\documentclass[12pt]{book}
\usepackage{psfrag}
\usepackage{epsfig}
\usepackage{bm}
\usepackage{amsmath}
\usepackage{mathrsfs}
\usepackage{amssymb}
\usepackage{axodraw}
\input psfig.sty
\usepackage{epsf}
\usepackage{graphicx}
\usepackage{indentfirst}

\newcommand{\gb}{{\cal O}}

\newcommand{\BR}{{\cal B}}

\newcommand{\dedx}{dE/dx}
\newcommand{\eff}{\varepsilon}

\newcommand{\gev}{\,\mbox{GeV}}
\newcommand{\mev}{\,\mbox{MeV}}
\newcommand{\pip}{\pi^+}
\newcommand{\pim}{\pi^-}
\newcommand{\piz}{\pi^0}

\newcommand{\kap}{K^+}
\newcommand{\kam}{K^-}

\newcommand{\etap}{\eta^{\prime}}
\newcommand{\omegap}{\omega^{\prime}}
\newcommand{\etac}{\eta_c}
\newcommand{\etacp}{\eta_c^{\prime}}
\newcommand{\hc}{h_c}
\newcommand{\chicz}{\chi_{c0}}
\newcommand{\chico}{\chi_{c1}}
\newcommand{\chict}{\chi_{c2}}
\newcommand{\chicj}{\chi_{cJ}}
\newcommand{\psp}{\psi^{\prime}}
\newcommand{\psip}{\psi(2S)}
\newcommand{\pspp}{\psi^{\prime \prime}}
\newcommand{\psipp}{\psi(3770)}
\newcommand{\jpsi}{J/\psi}
\newcommand{\DD}{D^+ D^-}
\newcommand{\EE}{e^+e^-}
\newcommand{\MM}{\mu^+\mu^-}
\newcommand{\TT}{\tau^+\tau^-}
\newcommand{\GG}{\gamma\gamma}
\newcommand{\pp}{\pi^+\pi^-}
\newcommand{\kk}{K^+K^-}
\newcommand{\ppkk}{\pi^+\pi^-K^+K^-}
\newcommand{\etappp}{\eta\pi^+\pi^-}
\newcommand{\kskl}{K^0_SK^0_L}
\newcommand{\kstk}{K^* \overline{K}}
\newcommand{\KKSC}{K^{*+}K^-}
\newcommand{\KKSN}{K^{*0}\overline{K^0}}
\newcommand{\zz}{\bar{p} p}
\newcommand{\LL}{\ell^+\ell^-}
\newcommand{\ogpi}{\omega\pi^0}
\newcommand{\omegapi}{\omega\pi^0}
\newcommand{\rhopi}{\rho\pi}
\newcommand{\rhopin}{\rho^0 \pi^0}
\newcommand{\rpi}{\rho\pi}
\newcommand{\OP}{\omega\pi^0}
\newcommand{\RP}{\rho\pi}
\newcommand{\RET}{\rho\eta}
\newcommand{\RETp}{\rho\eta^{\prime}}
\newcommand{\OET}{\omega\eta}
\newcommand{\OETp}{\omega\eta^{\prime}}
\newcommand{\FET}{\phi\eta} 
\newcommand{\FETp}{\phi\eta^{\prime}} 
\newcommand{\ccbar}{c\bar{c}}
\newcommand{\ccb}{c\overline{c}}
\newcommand{\QQb}{Q\overline{Q}}
\newcommand{\qqb}{q\overline{q}}
\newcommand{\nnb}{n\overline{n}}
\newcommand{\ppbar}{p\bar{p}}
\newcommand{\BBb}{B\overline{B}}
\newcommand{\KKb}{K\overline{K}}

\newcommand{\ppb}{p\overline{p}}

\newcommand{\LLb}{\Lambda \overline{\Lambda}}
\newcommand{\ddbar}{D\bar{D}}
\newcommand{\ddb}{D\overline{D}}
\newcommand{\DDb}{D\overline{D}}

\newcommand{\kskp}{K^0_S K^+ \pi^- + c.c.}
\newcommand{\kskn}{K^{*0}(892)\overline{K^0}+c.c.}
\newcommand{\VP}{1^-0^-}
\newcommand{\PP}{0^-0^-}
\newcommand{\fddd}{{\cal F} (\ddb \ra f)}
\newcommand{\fndd}{\overline{\cal F} (\ddb \ra f)}
\newcommand{\jpsipp}{J/\psi \pi^+\pi^-}
\newcommand{\ppjpsi}{\pi^+\pi^- J/\psi}

\newcommand{\PPJP}{\pi^+\pi^- J/\psi}
\newcommand{\ra}{\rightarrow}

\newcommand{\jpsito}{J/\psi \rightarrow }

\newcommand{\pspto}{\psi^{\prime} \rightarrow }
\newcommand{\psppto}{\psi^{\prime \prime} \rightarrow }
\newcommand{\EETO}{e^+e^-\to}
\newcommand{\M}{\frac{\sqrt{3}}{2}M}
\newcommand{\beq}{\begin{equation}}
\newcommand{\eeq}{\end{equation}}
\newcommand{\beqn}{\begin{eqnarray}}
\newcommand{\eeqn}{\end{eqnarray}}
\newcommand{\beqns}{\begin{eqnarray*}}
\newcommand{\eeqns}{\end{eqnarray*}}
\newcommand{\bfg}{\begin{figure}}
\newcommand{\efg}{\end{figure}}
\newcommand{\bitm}{\begin{itemize}}
\newcommand{\eitm}{\end{itemize}}
\newcommand{\bnum}{\begin{enumerate}}
\newcommand{\enum}{\end{enumerate}}
\newcommand{\btbl}{\begin{table}}
\newcommand{\etbl}{\end{table}}
\newcommand{\btbu}{\begin{tabular}}
\newcommand{\etbu}{\end{tabular}}
\newcommand{\psia} {\psi(4040)}
\newcommand{\psib} {\psi(4160)}
\newcommand{\psic} {\psi(4415)}

\newcommand{\bfk} {{\bf k}}
\newcommand {\rb}[1]{\raisebox{1.5ex}{#1}}

\newcommand {\tmea} {t_{\textrm{mea}}}
\newcommand {\texp} {t_{\textrm{exp}}}

\newcommand {\eseed} {E_{\textrm{seed}}}
\newcommand {\ethree} {E_{3\times3}}
\newcommand {\efive} {E_{5\times5}}

\newcommand {\odedx} {O_{\textrm{dE/dx}}}
\newcommand {\otof}  {O_{\textrm{TOF}}}
\newcommand {\oemc}  {O_{\textrm{EMC}}}
\newcommand {\omuc}  {O_{\textrm{MUC}}}
\newcommand {\oseq}  {O_{\textrm{seq}}}

\def\amplitd#1{\left|\mathscr{A}_{#1}\right|}
\newcommand{\amaz}{\amplitd{0}}
\newcommand{\amatwo}{\amplitd{2}}

\def\eref#1{(\ref{#1})}
\def\Journal#1#2#3#4{{#1} {\bf #2}, #3 (#4)}

\def\CTP{Commun. Theor. Phys.}
\def\IJMPA{Int. J. Mod. Phys. A}
\def\NCA{Nuovo Cimento}
\def\NIM{Nucl. Instrum. Methods}
\def\NIMA{Nucl. Instrum. Methods A}
\def\NPB{Nucl. Phys. B}
\def\NPA{Nucl. Phys. A}
\def\PL{Phys. Lett.}
\def\PLA{Phys. Lett. A}
\def\PLB{Phys. Lett. B}
\def\plb{Phys. Lett. B}
\def\PRL{Phys. Rev. Lett.}
\def\PRD{Phys. Rev. D}
\def\PRC{Phys. Rev. C}
\def\PRP{Phys. Rep.}
\def\ZPC{Z. Phys. C}
\def\EPJC{Eur. Phys. J. C}

\def\HEPNP{HEP \& NP}

\def\HEPNP{HEP \& NP}

\def\prpr{Phys. Rev.}

\def\nimnim{Nucl. Instrum. Methods}
\def\cpccpc{Comput. Phys. Commun.}
\def\hepnp{HEP \& NP}

\def\jpjp{Journal of Physics}

\def\Journal#1#2#3#4{{#1}, {#4}, {\bf #2}: {#3}}

\def\PRDKL#1#2#3{\Journal{\prpr}{D#1}{#2}{#3}}

\def\NIMKL#1#2#3{\Journal{\nimnim}{#1}{#2}{#3}}
\def\NIMAKL#1#2#3{\Journal{\nimnim}{A#1}{#2}{#3}}
\def\CPCKL#1#2#3{\Journal{\cpccpc}{#1}{#2}{#3}}
\def\HEPNPKL#1#2#3{\Journal{\hepnp}{#1}{#2}{#3}}

\def\JPGKL#1#2#3{\Journal{\jpjp}{G#1}{#2}{#3}}
\def\hepref#1{arXiv:{#1}}
\def\Figs {Figures}

\newcommand {\BegFig} {\begin{figure}%
                        \begin{center}%
                        }%
\newcommand {\EndFig} {\end{center}%
                       \end{figure}%
                       }%
\newcommand {\BegTab} {\begin{table}%
                        \begin{center}%
                        }
\newcommand {\EndTab} {\end{center}%
                       \end{table}%
                       }%

\usepackage{fancyheadings}
\renewcommand{\chaptermark}[1]{\markboth%
{\thechapter.\ #1}{}}

\long\def\singleMnfA#1#2#3#4{
\begin{figure}[p]
   \epsfxsize10cm
   \centerline{\epsffile{figures/#1}}
\vspace{-0.5cm}
   \caption[.]{\label{fig:#2}#3}
\vspace{-0.1cm}
}
\long\def\singleMnfB#1#2#3#4{
   \epsfxsize10cm
   \centerline{\epsffile{figures/#1}}
\vspace{-0.5cm}
   \caption[.]{\label{fig:#2}#3}
\end{figure}
}
\long\def\inst#1{\par\nobreak\kern 4pt\nobreak
    {\itshape #1}\par\vskip 10pt plus 3pt minus 3pt}
\RequirePackage{xspace}
\def\bes3{\mbox{\slshape B\kern-0.1em{E}\kern-0.1em S-III}}
\def\qqbar {\ensuremath{q\overline q}\xspace}
\def\ccbar {\ensuremath{c\overline c}\xspace}
\def\ssbar {\ensuremath{s\overline s}\xspace}
\def\Kbar    {\kern 0.18em\overline{\kern -0.18em K}{}\xspace}
\def\Kb      {\ensuremath{\Kbar}\xspace}
\def\KK      {\ensuremath{K\Kbar}\xspace}
\def\Kz      {\ensuremath{K^0}\xspace}
\def\Kzb     {\ensuremath{\Kbar^0}\xspace}
\def\KzKzb   {\ensuremath{\Kz {\kern -0.16em \Kzb}}\xspace}
\def\Ku      {\ensuremath{K^+}\xspace}
\def\Kub     {\ensuremath{K^-}\xspace}
\def\Kp      {\ensuremath{\Ku}\xspace}
\def\Km      {\ensuremath{\Kub}\xspace}
\def\Kpm     {\ensuremath{K^\pm}\xspace}
\def\Kmp     {\ensuremath{K^\mp}\xspace}
\def\Ks     {\ensuremath{K_S}\xspace}
\def\Kl     {\ensuremath{K_L}\xspace}
\def\KsKs   {\ensuremath{\Ks {\kern -0.16em \Ks}}\xspace}
\def\KlKl   {\ensuremath{\Kl {\kern -0.16em \Kl}}\xspace}
\def\KsKl   {\ensuremath{\Ks {\kern -0.16em \Kl}}\xspace}
\def\KlKs   {\ensuremath{\Kl {\kern -0.16em \Ks}}\xspace}
\def\Dbar    {\kern 0.18em\overline{\kern -0.18em D}{}\xspace}
\def\DD    {\ensuremath{D\Dbar}\xspace}
\def\Dz      {\ensuremath{D^0}\xspace}
\def\Dzb     {\ensuremath{\Dbar^0}\xspace}
\def\DzDzb   {\ensuremath{\Dz {\kern -0.16em \Dzb}}\xspace}
\def\DsP      {\ensuremath{D_S^+}\xspace}
\def\DsM     {\ensuremath{D_S^-}\xspace}
\def\DspDsm    {\ensuremath{\DsP {\kern -0.16em \DsM}}\xspace}
\def\Dp      {\ensuremath{D^+}\xspace}
\def\Dm     {\ensuremath{D^-}\xspace}
\def\DpBm    {\ensuremath{\Dp {\kern -0.16em \Dm}}\xspace}
\def\Bbar    {\kern 0.18em\overline{\kern -0.18em B}{}\xspace}
\def\Bb      {\ensuremath{\Bbar}\xspace}
\def\BB      {\ensuremath{B\Bbar}\xspace}
\def\Bz      {\ensuremath{B^0}\xspace}
\def\Bzb     {\ensuremath{\Bbar^0}\xspace}
\def\BzBzb   {\ensuremath{\Bz {\kern -0.16em \Bzb}}\xspace}
\def\Bu      {\ensuremath{B^+}\xspace}
\def\Bub     {\ensuremath{B^-}\xspace}
\def\Bp      {\ensuremath{\Bu}\xspace}
\def\Bm      {\ensuremath{\Bub}\xspace}
\def\Bpm     {\ensuremath{B^\pm}\xspace}
\def\Bmp     {\ensuremath{B^\mp}\xspace}
\def\BpBm    {\ensuremath{\Bu {\kern -0.16em \Bub}}\xspace}
\def\Bs      {\ensuremath{B_s}\xspace}
\def\Bsb     {\ensuremath{\Bbar_s}\xspace}
\newcommand{\optbar}[1]{\shortstack{{\tiny (\rule[.4ex]{1em}{.1mm})}
  \\ [-.7ex] $#1$}}
\def\BorBbar    {\kern 0.18em\optbar{\kern -0.18em B}{}\xspace}
\def\DorDbar    {\kern 0.18em\optbar{\kern -0.18em D}{}\xspace}
\def\KorKbar    {\kern 0.18em\optbar{\kern -0.18em K}{}\xspace}
\def\cp                {\ensuremath{C\!P}\xspace}

\def\bepc2{BEPC-II}
\mathchardef\Upsilon="7107
\def\Y#1S{\ensuremath{\Upsilon{(#1S)}}\xspace}
\newcommand{\scs}{\scriptscriptstyle}
\def\st{\scriptstyle}

\def\ra{\rightarrow}

\def\ko{K^0}

\def\al{\alpha}

\def\be{\begin{equation}}
\def\ee{\end{equation}}
\def\bea{\begin{eqnarray}}
\def\eea{\end{eqnarray}}

\begin{document}
\title{\bf \Huge Physics at \bes3 }
\vspace{3cm}
\author{
\hspace*{0cm} {\bf Editors} \\  
\vspace{2.0cm}
         \hspace*{0cm}   {\bf Kuang-Ta Chao and Yifang Wang}\\
\vspace{10.0cm}
{\bf IHEP-Physics-Report-\bes3-2008-001-v1}}

\vspace{1cm}

\date{}

\maketitle

\cleardoublepage

\vfill
\newpage
\large
\centerline{\bf \Huge Physics at \bes3 }
\vspace{3cm}
\centerline{\Large \bf Editors: } 
\centerline{Kuang-Ta Chao$^1$ and Yifang Wang$^2$ } 
\vspace{2cm}
\centerline{\small \it $^{1}$ Peking University, Beijing 100871}
\centerline{\small \it $^{2}$ Institute of High Energy Physics, P.O.Box 918, Beijing 100049}

\large
\vspace{2.0cm}
\centerline{\Large \bf Working Group and Conveners}
\vspace{1.0cm}
\noindent {\large \bf Part One: The \bes3 experiment and its physics}\\
 \hspace*{3cm} Conveners: Jian-Ping Ma, Ya-Jun Mao \\
\noindent {\large \bf Part Two: $e^+e^-$ Collision at $\sqrt{s} = 2-5$ GeV }\\
 \hspace*{3cm} Conveners: Yuan-Ning Gao, Xiao-Yuan Li \\
\noindent {\large \bf Part Three:  Light hadron physics}\\
 \hspace*{3cm} Conveners: Xiao-Yan Shen, Bing-Song Zou\\
\noindent {\large \bf Part Four: Charmonium Physics} \\
 \hspace*{3cm} Conveners: Cong-Feng Qiao, Changzheng Yuan\\
\noindent {\large \bf Part Five: Charm Physics}  \\
 \hspace*{3cm} Conveners: Hai-Bo Li, Zhi-Zhong Xing\\
\noindent {\large \bf Part Six: Tau Physics} \\
 \hspace*{3cm} Conveners: Antonio Pich, Changzheng Yuan\\

\newpage
\large
\centerline{\Large \bf Authors }
\vspace{.5cm}
\noindent\author{D.~M.~Asner$^{7}$, T.~Barnes$^{3,24}$, J.~M.~Bian$^1$, I.~I.~Bigi$^{17}$,
N.~Brambilla$^{11}$, I.~R.~Boyko$^{12}$, V. Bytev$^{12}$, K.~T.~Chao$^{21}$,
J.~Charles$^9$, H.~X.~Chen$^{1}$, J.~C.~Chen$^{1}$, Y.~Chen$^{1}$, Y.~Q.~Chen$^{15}$,
H.~Y.~Cheng$^{14}$, D.~Dedovich$^{12}$, S.~Descotes-Genon$^{20}$,
C.~D.~Fu$^{26}$, X.~Garcia~i~Tormo$^{3}$, Y.-N. Gao$^{26}$,  K.~L.~He$^{1}$,
Z.~G.~He$^{21}$, J.~F.~Hu$^{25}$, H.~M.~Hu$^{1}$, B.~Huang$^{1}$, Y.~Jia$^{1}$, H.-Y. Jin$^{31}$,
S.~Jin$^{1}$, Y.~P.~Kuang$^{26}$, H.~Lacker$^{13}$, H.~B.~Li$^{1}$,
J.~L.~Li$^{10}$, W.~D.~Li$^{1}$, X.~Y.~Li$^{15}$, B.~J.~Liu$^{1}$,
H.~H.~Liu$^{1}$, J.~Liu$^{1}$, H.~L.~Ma$^{1}$,
J.~P.~Ma$^{15}$, Y.~J.~Mao$^{21}$, X.~H.~Mo$^{1}$,  
S.~L.~Olsen$^{1,27}$,
A.~Pich$^{29}$,
A.~Pineda$^{4}$, R.~G.~Ping$^{1}$,  C.~F.~Qiao$^{10}$,
G.~Qin$^{1}$, H~.Qin$^1$, J.~M.~Roney$^{30}$, G.~Rong$^{1}$, L.~Roos$^{18}$, 
X.~Y.~Shen$^{1}$,  J.~Soto$^{6}$, A.~Stahl$^{2}$, S.~S.~Sun$^{1}$, 
S.~T$^\prime$Jampens$^{22}$, A.~Vairo$^{11}$ ,P.~Wang$^{1}$,
Y.~F.~Wang$^{1}$,
Y.~K.~Wang$^{28}$, N.~Wu$^{1}$, Y.~L.~Wu$^{15}$,
Z.~Z.~Xing$^{1}$, G.~F.~Xu$^1$, M.~Xu$^{25}$, M.~Yang$^1$, M.~Z.~Yang$^{16}$, Y.~D.~Yang$^{8}$,
C.~Z.~Yuan$^{1}$,
D.~H.~Zhang$^{1}$, D.~Y.~Zhang$^1$, J.~Y.~Zhang$^{1, 5}$, Z.~X.~Zhang$^1$, X.~M.~Zhang$^{1}$, X.~Y.~Zhang$^{23}$,
Y.~J.~Zhang$^{21}$, Q.~
Zhao$^{1}$,  A.~Zhemchugov$^{12}$,
H.~Q.~Zheng$^{21}$, Y.~H.~Zheng$^{10}$, M.~Zhong$^{26, 32}$, S.-L. Zhu$^{21}$, Y.~S.~Zhu$^{1}$,
V.~Zhuravlov$^{12}$, 
B.~S.~Zou$^{1}$, J.~H.~Zou~$^{23}$ 
\vspace{0.2cm}\\
\noindent {\small \it $^{1}$ Institute of High Energy Physics, Beijing 100049.\\
$^2$ III. Physikalisches Institut, RWTH Aachen University 52056 Aachen, Germany.\\
$^3$ High Energy Physics Division, Argonne National Laboratory 9700 South Cass Avenue, Ar-\\
\hspace*{0.25cm} gonne, IL 60439, USA.\\
$^4$ Dept. of Physics, Univiversitat Autonoma de Barcelona, E-08193 Bellaterra, Barcelona, Spain.\\
$^{5}$ China Center for Advanced Science and Technology (CCAST), Beijing 100080.\\
$^6$ Dept. d'Estructura i Constituents de la Materia, Universitat de Barcelona Diagonal 647, E-\\ 
\hspace*{0.25cm} 08028 Barcelona, Catalonia, Spain.\\
$^7$ Carleton University, 1125 Colonel By Drive, Ottawa, Ontario, K1S 5B6, Canada.\\
$^8$ Institute of Particle Physics, Central China Normal University, Wuhan 430079.\\
$^9$ Centre de Physique Th\'eorique, CNRS Luminy, Case 907, F-13288 Marseille  Cedex 9, France.\\
$^{10}$ Graduate University of the Chinese Academy of Sciences, Beijing 100039. \\
$^{11}$ Dipartimento di Fisica dell'Universita di Milano and INFN, Milano, Italy.\\
$^{12}$ JINR, 141980 Dubna, Moscow region, Russian Federation. \\
$^{13}$ Institute f\"ur Physik, Humboldt Universit\"at zu Berlin, 
Newtonstr. 15, 12489 Berlin, Germany 
$^{14}$ Institute of Physics, Academia Sinica, Taipei 115.\\
$^{15}$ Institute of Theoretical Physics of the Chinese Academy of Sciences, Beijing 100080. \\
$^{16}$ Department of Physics, Nankai University, Tianjin 300071. \\
$^{17}$ Department of Physics, University of Notre Dame du Lac Notre Dame, IN 46556, USA. \\
$^{18}$ Laboratoire de Physique Nucl\'eaire et Hautes Energies, Universit\'e Pierre  et Marie Curie Paris\\ 
\hspace*{0.25cm} 6, Universit\'e Denis Diderot Paris 7, CNRS/IN2P3,  F-75252 Paris, France. \\
$^{19}$ Physics Division, Oak Ridge National Laboratory, Oak Ridge, TN 37831, USA.\\
$^{20}$ Laboratoire de Physique Th\'eorique, Universit\'e Paris-Sud 11, CNRS, F-91405 Orsay Cedex,\\
\hspace*{0.25cm} France.\\
$^{21}$ Peking University, Beijing 100871.\\
$^{22}$ Laboratoire d'Annecy-le-vieux de Physique des Particules, Universit\'e de Savoie, CNRS/IN2P3, \\
\hspace*{0.25cm} F-74941 Annecy-le-Vieux, France.\\ 
$^{23}$ Department of Physics, Shandong University, Jinan, Shandong 250100. \\
$^{24}$ Department of Physics and Astronomy, University of Tennessee, Knoxville, TN 37996, USA.\\ 
$^{25}$ University of Science and Technology of China, Hefei 230026.\\
$^{26}$ Center for High Energy Physics, Tsinghua University, Beijing 100084.\\
$^{27}$ University of Hawaii, Honolulu, HI 96822, USA. \\
$^{28}$ Department of Physics,  Wuhan University, Wuhan 430072.\\
$^{29}$ Departament de Fisica Teorica, IFIC, Universitat de Valencia --CSIC, Apt. de Correus 22085, \\
\hspace*{0.25cm} E-46071 Valencia, Spain.\\
$^{30}$ Dept. of Physics, University of Victoria, P.O. Box 3055, Victoria, British Columbia Canada,\\ 
\hspace*{0.25cm} V8W 3P6. \\
$^{31}$ Institute of Modern Physics, Zhejiang University, Hangzhou 310027.
$^{32}$ Department of Physics, National University of Defense Technology,
Hunan, Changsha 410073.} }
\maketitle

\vfill

\newpage
\cleardoublepage
\newpage
\begin{center} \Large{\bf Abstract } \end{center}
\addcontentsline{toc}{chapter}{Abstract}

There has recently been a dramatic renewal of interest in the
subjects of hadron spectroscopy and charm physics.
This renaissance has been driven in part by experimental reports
of  \DzDzb mixing and the discovery of narrow $D_{sJ}$ states and
a plethora of
charmonium-like $XYZ$ states at the $B$ factories,
and the observation of an intriguing proton-antiproton threshold
enhancement and the possibly related $X(1835)$ meson state at
BESII.    At the same time, lattice QCD is now coming of age, 
and we are entering a new era when precise, quantitative 
predictions from lattice QCD
can be tested against experimental measurements.
For example,
the High Precision QCD (HPQCD) and United Kingdom QCD
(UKQCD)  collaboration's recent
high-precision, unquenched calculation of
$f_{D^+}= 208\pm 4$ MeV has been found to agree
with the CLEO-c collaboration measurement of
$f_{D^+}= 223\pm 17\pm8 $~MeV -- a precision level of $\sim 8$\%.
Intriguingly, this agreement does not extend to $f_{D_s}$, where
the HPQCD~+~UKQCD result
$f_{D_s} = 241\pm 3$~MeV is more than three standard deviations
below the current world average experimental value
$f_{D_s} = 276\pm 9$~MeV.   Precision improvements, especially
on the experimental measurements, are called for and will
be of extreme interest.

The \bes3 experiment at BEPCII in
Beijing, which will start operation in summer 2008,
will accumulate huge data samples of $10\times10^9$ $J/\psi$,
$3\times 10^9$  $\psi(2S)$ , 30 million
\DD or 2 million \DspDsm-pairs per running year, respectively,
running in the $\tau$-charm theshold region.  Coupled with
currently available results from CLEO-c, \bes3
will make it possible to study in detail, and with unprecedentedly
high precision, light
hadron spectroscopy in the decays of charmonium states and  charmed
mesons. In addition, about 90 million \DD pairs will be collected at \bes3
in a three-year run at the $\psi(3770)$ peak. Many high precision
measurements, including
CKM matrix elements related to charm weak decays, 
decay constants $f_{D^+}$ and $f_{D_S}$, Dalitz decays of
three-body $D$ meson decays, searches for $CP$ violation
in the charmed-quark sector, and absolute
decay branching fractions, will be accomplished.
\bes3 analyses are likely to be essential in deciding if recently
observed signs of mixing in the \DzDzb  meson
system are actually due to new physics or not.  \bes3 measurements
of $f_{D^+}$ and $f_{D_s}$ at the $\sim 1\%$ precision level
will match the precision of lattice QCD calculations and
provide the opportunity to probe the charged Higgs sector 
in some mass ranges that will be inaccessible to the LHC.
With modern techniques and
huge data samples, searches for rare,
lepton-number violating, flavor violating and/or invisible decays of
$D$-mesons,
charmonium resonances, and tau-leptons will be possible.
Studies of $\tau$-charm physics could
reveal or indicate the possible presence of new physics in the low
energy region.

This physics book provides detailed discussions on
important topics in $\tau$-charm physics that will be
explored during the next few years at \bes3 . Both theoretical and
experimental issues are covered, including
extensive reviews of recent theoretical
developments and experimental techniques.  Among the subjects covered
are:  innovations in  Partial Wave Analysis (PWA), theoretical
and experimental techniques
for Dalitz-plot analyses,
analysis tools to extract absolute branching fractions
and measurements of decay
constants, form factors, and $CP$-violation and \DzDzb-oscillation
parameters.
Programs of QCD studies and near-threshold tau-lepton physics
measurements are also discussed.

\normalsize
\tableofcontents
\clearpage{\pagestyle{empty}\cleardoublepage}


\clearpage{\pagestyle{empty}\cleardoublepage}

\clearpage{\pagestyle{empty}\cleardoublepage}

\setcounter{page}{0}
\pagenumbering{arabic}
\setcounter{chapter}{0}
\setcounter{part}{0}

%
\vfill
\newpage

\newpage
\part[The \bes3 experiment and its physics]{The \bes3 Experiment and Its
Physics\\
\vspace*{2cm}
 {\centering  \Large Conveners \\  
  Jian-Ping Ma, Ya-Jun Mao}\\
 \vspace*{1cm}
\Large Contributors \\
D.~M.~Asner, C.~D.~Fu, K.~L.~He, J.~F.~Hu, B.~Huang, W.~D.~Li, B.~J.~Liu,
J.~P.~Ma, Y.~J.~Mao, X.~H.~Mo, R.~G.~Ping,
G.~Qin,  S.~S.~Sun, Y.~F.~Wang, N.~Wu, M.~Xu, X.~Y.~Zhang, H.~Q.~Zheng, Y.~H.~Zheng}
\label{part:one}
\chapter[Physics goal of \bes3]{Physics Goal of \bes3 \footnote{ By Jian-Ping Ma}}
\label{chapter:goal}

The Standard Model (SM) has been successful at describing all
relevant experimental phenomena and, thus, has been generally accepted
as the fundamental theory of elementary particle physics.
Despite its success, the SM leaves many unanswered questions.
These can be classified into two main categories:
one for subjects related to possible new physics at unexplored energy 
scales and the other for nonperturbertive physics, mostly related
to Quantum Chromodynamics.
\par
The SM describes particle physics up to energies of around
$100$~GeV. It is expected that new particles and new interactions
will appear at some higher energy scale, say $1$~TeV. Those new particles
and new interactions are presumably  needed to solve some inconsistencies
within the SM and for the ultimate unification of all interactions. Such 
physics issues all belong to the first category, and will be addressed by 
experiments at the LHC, which will start
operation in 2008, and at the ILC, which is currently  being planned. 
\par
The second category of unanswered questions includes those
about nonperturbative effects.
QCD, the fundamental theory of the strong interactions, is well tested
at short distances, but at long distances nonperturbative effects 
become important and these are not well understood.  These
effects are very basic to the field of particle physics and
include {\it  e.g.}, the structure of hadrons and the
spectrum of hadronic states.  
Lower energy facilities with high luminosity can address these questions.
Among these, the Beijing Electron Positron Collider II (BEPCII),
which will operate in the $2$ GeV to $4.6$ GeV energy range,
will be an important contributor.  This is because it spans 
the energy range where {\it both} short-distance and 
long-distance effects can be probed.
\par
The BEPCII energy range includes the threshold of charmonia. The
discoveries of the low-lying charmonium states and of 
open-charmed hadrons were instrumental for 
the acceptance of quarks as truly dynamical
entities in general, and of the SM in particular. The surprising
discoveries of the narrow $D_{sJ}$ mesons, several hidden charm 
resonances
around 4 GeV region, and  the $X(1835)$ at BESII during the past few
years have considerably enhanced the interest in the study of the
spectroscopy of hadrons with and without {\em open} charm. The high
energy physics community has realized that comprehensive studies of
$e^+e^-$ annihilation in the charm-tau threshold
region can teach us novel and
unique lessons on hadronization and the interplay of perturbative
and nonperturbative dynamics. This has great value both in
its own right and for its contributions to improve the discovery 
potential for new
physics in the decays of $B$ mesons studied at 
LHCb and the $B$-factories.
The significance of physics around the threshold of charmonia
is also illustrated by the fact that the
Super-$B$ factories being designed at Frascati and KEK both include
the capability of running in the 4 GeV region.
\par
Theoretical studies of physics at the energy scale 
accessible to BEPCII continue to be actively pursued.
To provide a good understanding of physics at this scale,
theoretical tools derived from QCD have been invented.
For charmonia,
one can use the nonrelativistic QCD (NRQCR) and potential 
nonrelativistic
QCD (pNRQCD) models to make
theoretical predictions for physics involving both 
short- and long-distance effects,
where a factorization of the two different kinds of effects can  be 
accomplished  and predictions  that do not depend 
on the assumptions of any  particular model can be made. 
For charmed hadrons, one can 
at least partly rely on heavy quark effective theory (HQET)
for their study.
For physics involving long-distance effects only,
one can employ QCD sum-rule methods, or
lattice QCD and make predictions from first principles.
It is a fortunate coincidence that
the most powerful tool for the quantitative treatment of nonperturbative
dynamics, namely lattice QCD, is reaching a new level of sophistication with
uncertainties in calculations  of charmed quark dynamics
that are approaching the 1~percent level.
In addition to the theoretical tools derived from QCD,
many phenomenological models have been invented to deal with 
nonperturbative effects, especially
those at the 1~GeV scale or lower, such as  light hadron 
spectroscopy, decays
of charmonia and $D$-mesons into light hadrons, etc.
Many theoretical predictions obtained with the above-mentioned 
methods exist and call for tests from 
experiment.  The BEPCII/\bes3  facility will be ideal for
carrying out the task of confirming and validating
these approaches. 
\par
Fruitful physics results  have been produced with the earlier
Beijing Spectrometers (BESI and BESII) at BEPC. The precise
measurement of the $\tau$-lepton mass, performed by
BESI almost twenty years ago, remains the world's best 
measurement of this fundamentally important quantity.
The $R$-value measurements from BESII
have made an important improvement to the 
prediction  of the mass of the still undiscovered Higgs boson.  BESII
also observed an anomalous $p\bar p$ threshold mass enhancement  
in the radiative decay $J/\psi \to \gamma p\bar p$, an
observation that has stimulated many theoretical speculations.
The observation of non $D$-$\bar D$ decays of 
the $\psi(3770)$ by BESII also confounds theoretical
expectations.  Violation of the notorious $12\%$ rule 
has been observed in different $J/\psi$ and $\psi'$
decay channels. There are many other results that
could be mentioned.  At the same time, 
important results in this energy range
have also been obtained by the CLEO-c collaboration in 
the U.S., the most important of which include: the discovery of 
the $^1P_1$ state of charmonia (the
$h_c$), and a measurement of the  $D$-meson decay constant 
with an $8\%$ precision. 
With their large data sample
of $e^+e^- \to D\bar D$ events at the $\psi(3770)$, 
they are able to
measure absolute hadronic branching ratios 
with improved precision, {\it e.g.}, $B(D^0 \to K^-
\pi^+)$ and $B(D^+ \to K^- \pi^+ \pi^+)$
have been measured with errors below the $5\%$ level.
It is expected that these
phenomena will continue be be studied in 
\bes3 with even higher precision which
will provide a better understanding of non perturbative physics.
\vskip20pt
\par
The upgraded  BEPCII/\bes3 
is a unique and powerful facility for studying physics 
in the energy range up to 4~GeV, with a research program
that covers charmonium physics, $D$-physics, spectroscopy of light
hadrons
and $\tau$-physics.  It will  also enable searches for 
possible new interactions.   The upgraded collider will 
reach a luminosity of $L=10^{33}$cm$^{-2}$s$^{-1}$. 
At the peak of the $J/\psi$, BEPCII will produce $10^{10}$ $J/\psi$
events per year. These will provide \bes3 with the world largest data 
sample for  studying $J/\psi$ mesons and their decays.  With 
one-year-long runs at the design luminosity  we can expect
the following data samples:
\par
\begin{center}
\begin{tabular}{c||c|c|c|c}
\hline
     &  CMS Mass  & Peak Lum. & $ \sigma$  &
  No. of Events  \\ \hline
   $J/\psi$  & 3.097 & 0.6 & 3400  & $ 10\times 10^9$  \\ \hline
   $\tau^+\tau^-$ & 3.670  & 1.0 & 2.4 & $12\times 10^6$ \\ \hline
   $\psi(2S) $ & 3.686  & 1.0 & 640 & $3.2\times 10^9$ \\ \hline
   $D^0 \bar D^0 $ & 3.770  & 1.0 & 3.6 & $18\times 10^6$ \\ \hline
   $D^+ D^- $ & 3.770  & 1.0 & 2.8 & $14\times 10^6$ \\ \hline
   $D_s D_s $ & 4.030  & 0.6 & 0.32 & $1.0\times 10^6$ \\ \hline
   $D_s D_s $ & 4.170  & 0.6 & 1.0 & $2.0 \times 10^6$ \\ \hline
\end{tabular}
\end{center}
\par
It is evident from the table that there will be huge data samples of 
$J/\psi$ and $\psi(2S)$ events,
and large numbers of $D$ and $D_s$ meson decays. 
These will not only enable high
precision measurements, they will also provide the potential
for discovering phenomena that have been overlooked at previous
facilities because of statistical limitations. With these 
data samples, \bes3 will have opportunities to
search for new physics in rare decays of charmonia, charmed mesons, $\tau$
leptons and to probe $D^0 - \bar D^0$ oscillations and $CP$ asymmetry.
This physics yellow book gives detailed and comprehensive reviews of the
relevant experimental and theoretical issues and the tools 
that are available or needed to address them.
A brief summary of physics goals is given here.
\par\vskip20pt
\par
$\bullet$ {\bf Charmonium Physics }
\par
The total decay widths of the $J/\psi$ and $\psi'$ will be measured 
at a precision level that is better than $1\%$.
The $J/\psi$ has many different decay modes.  In two-body decays,
either of the final-state particles can be a pseudoscalar, a scalar,
a vector, an axial vector 
or a tensor meson. With a  $10^{10}$ $J/\psi$ event sample, 
these decay modes can be measured much more precisely than before.
Historically, there are some notorious problems related to decays of
charmonia.
Among them the most well known problems are the $\rho\pi$ puzzle, 
{\it i.e.} violations of the $12\%$-rule,  and  non-$D-\bar D$ decays
of the $\psi(3770)$. With \bes3's huge data samples, more 
detailed experimental information  will be forthcoming that will 
hopefully provide guidance leading to solutions of these  problems.
Transitions between various charmonium states will be 
measured with unprecedented precision.
With the possibility of running at higher energies, the recently 
discovered 
$Y(4260)$ could be accessed at BEPCII, and
this would offer \bes3 opportunities to study this unconventional
charmonium state.
\par
With such  huge data samples, it will be possible to detect some 
Cabbibo-suppressed $J/\psi$ decay channels. In these channels, 
the charmed  quark decays via
the weak interaction, while the anticharm quark 
combines with another quark to  form a $D$-meson. 
This process will provide the possibility for detecting 
effects of new physics 
at BEPCII, if, for example,  
branching ratios of those decays are found to be
larger than SM predictions.
Also, one can search for evidence of flavor-changing neutral currents.
This an area where \bes3 can make unique explorations for
physics beyond the SM.

\par\vskip20pt

$\bullet$ {\bf Light Hadron Spectroscopy and Search for New Hadronic States}
\par
Using $J/\psi$-decays, one can study light hadron spectroscopy
and search for new hadronic states.   The large $J/\psi$ sample
makes BEPC a ``glue'' factory, since the charmed- and 
anticharmed quark constituents of the $J/\psi$ almost
always annihilate into gluons.
This is very useful for glueball searches and for
probing the gluon contents of light hadrons, including
the low-lying scalar mesons.
\par
QCD predicts the existence of glueballs and lattice
QCD predicts their masses.  For example,  the $0^{++}$ glueball 
is predicted to have a mass that is between
$1.5$ and $1.7$~GeV. But to date the existence of these various 
glueballs  has still not been experimentally confirmed. 
Also, since QCD is a relativistic quantum field theory,
any hadron should have some gluon content if symmetries allow. These
gluon contents, especially those in scalar mesons, are crucial inputs
to the understanding of the properties of the light hadrons, 
such as the $f_0(1500,1700)$ scalar mesons.
The rich gluon environment in $J/\psi$ decays is an ideal 
place to study these issues.
\par
Recently, evidence  for exotic hadrons, {\it i.e.}
mesons that cannot be classified as a $q\bar{q}$ state
of the traditional quark model, have been 
seen experimentally.   In  principle,
QCD allows for the existence of  exotic hadrons. 
With high-statistics data samples, comprehensive  searches for
exotic states can be performed and the quantum numbers of any
candidates that are found can be determined.
\par
In BESII, an anomalous near-threshold mass enhancement 
is seen in the $p\bar{p}$ system produced in the radiative
decay process $J/\psi \rightarrow \gamma p \bar p$; similar
enhancements are seen in other baryonic systems. 
Various explanations for these enhancements have been proposed, 
{\it e.g.}, there may be resonances just below the 
mass thresholds.  However, a satisfactory and conclusive explanation 
has still not emerged. With  \bes3 data these enhancements
can be studied more in detail and, hopefully, a satisfactory
explanation can be established.

\par\vskip20pt
$\bullet$ {\bf ${\boldmath D}$-Physics}
\par
At BEPC, $D^+$ and $D^0$ mesons will be produced through the decays of
the $\psi(3770)$, and $D_s$ mesons
can be produced through $e^+e^-$ annihilation at $s$ around $(4.03{\rm
GeV})^2$.
The decay constants
$f_D$ and $f_{D_s}$ can be measured from purely leptonic decays with
expected systematic errors
of $1.2\%$ and $2.1\%$, respectively. Inclusive and
exclusive semileptonic
decays of $D$-mesons will also be studied to test various theoretical
predictions.
Moreover, through the study of the decays $D^0\to K^- e^+ \nu_e$ and 
$D^0\to
\pi^- e^+ \nu_e$
one can extract the CKM matrix elements $V_{cs}$ and $V_{cd}$ with an
expected systematic error of around $1.6\%$.
\par
With \bes3 it will be possible to measure
$D$-$\bar D$ mixing and search for
$CP$-violation. Theoretical predictions 
for mixing and $CP$-violation are unreliable; \bes3 can provide new 
experimental information about them.
\par
Rare- or forbidden decays can provide strict tests of the SM and
have the potential of uncovering
the effects of new physics beyond the SM.  With \bes3, they can be 
studied systematically.
Significant improvements of their branching ratio
measurements are expected.
The upper limits on branching ratios 
for unseen modes can be improved by factors 
of about $10^{-2}$.

\par\vskip20pt
$\bullet$ {\bf $\tau$-Physics}
\par
$\tau$-physics will also be studied at \bes3, where several important 
measurements can be made. 
Experimental studies of inclusive hadronic  decays
can provide precise determinations of the strange quark mass and the CKM 
matrix element $V_{us}$, while the study of leptonic decays can test 
the universality of the electroweak
interaction and give a possible hint of new physics. The measurement
precision of the
Michel parameter will be improved by a factor of between 2 and 4; 
the  $\tau$-mass will be measured with a precision
of $\delta m_\tau \sim 0.09$MeV, a factor of~3 
improvement on the BESI  result.

\par\vskip20pt
Beside presenting these physical goals and opportunities at the
BEPCII collider  with the  \bes3 detector,
this yellow book also presents some useful tools that
are relevant to \bes3 analyses.

\chapter[The \bes3 detector and offline software]{The \bes3 detector and offline
software \footnote{By Wei-Dong Li, Ya-Jun Mao and Yi-Fang Wang}}
\label{sec:detecors}

\section{Overview of the \bes3 Detector}

\begin{figure}[htbp]
\centerline{\includegraphics[angle=90,width=14cm,height=10cm]{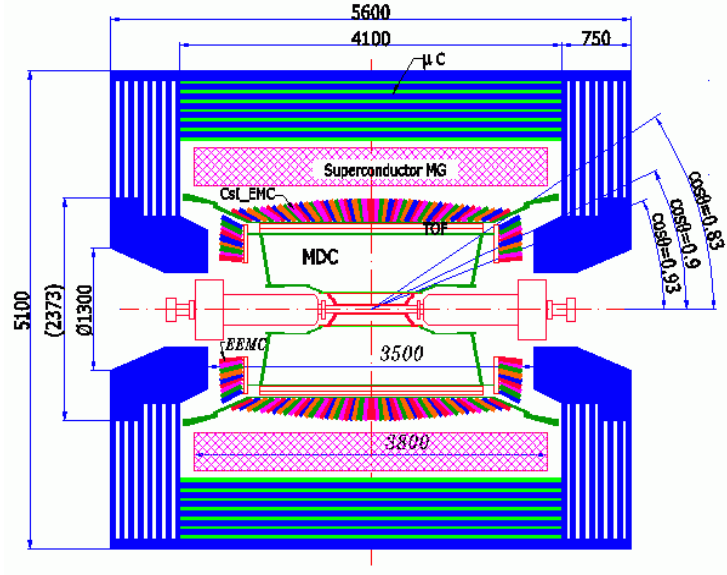}}
\caption{An Overview of the \bes3 Detector.}
\label{fig:r99bes}
\end{figure}

The \bes3 detector is designed to fulfill the physics requirements 
discussed in this report, and the technical requirements for a high 
luminosity, multi-bunch collider. Detailed descriptions of the \bes3 
detector can be found in Ref.~\cite{ref:bes3_det_tdr}. 
Figure~\ref{fig:r99bes} shows a schematic view of the
\bes3 detector, which consists of the following components:
\begin{itemize}
\item A Helium-gas based drift chamber with a single wire resolution 
that is better than 120 $\mu$m and 
a $dE/dX$ resolution that is better than 6\%. The 
momentum resolution in the  1T magnetic field is better than 0.5\% 
for charged tracks with a momentum of 1~GeV/c. 
\item A CsI(Tl) crystal calorimeter with an energy resolution 
that is better than 2.5\% and position resolution better 
than 6~mm for 1~GeV electrons and gammas.
\item A Time-of-Flight system with an intrinsic timing resolution 
better than 90 ps.
\item A super-conducting solenoid magnet with a central field of 1.0 
Tesla.
\item A 9-layer RPC-based muon chamber system with a spatial resolution 
that is better than 2~cm.
\end{itemize}
Details of each sub-detector and their performance, together with the 
trigger system, are discussed in subsequent sections. 
\section{\bes3 Offline Software}
\label{offline Software and Framework}

The \bes3 Offline Software System (BOSS) uses the C++ language and 
object-oriented techniques and runs primarily on the 
Scientific Linux CERN (SLC) operating system
The entire data processing and physics analysis software system
consists of  five functional parts: framework, simulation, 
reconstruction, calibration, and analysis tools.

The BOSS framework is based on the Gaudi package~\cite{part1:gaudi}, which provides 
standard interfaces and utilities for event simulation, data processing and 
physics analysis. The framework employs Gaudi's event data service as the data manager
and the event data conversion service for conversions between persistent data and 
transient objects. Three types of persistent event data have been defined in the 
BOSS system: raw data, reconstructed data
and Data-Summary-Tape (DST) data. 
Both reconstructed data and DST data are in ROOT format for easy 
management and usage.
Different types of algorithms can access data from Transient Event Data 
Store (TEDS) via the event data service. The detector's material and 
geometrical information 
are stored in GDML files, which can be retrieved by algorithms through 
corresponding services. 

The BOSS framework also provides abundant services and utilities for various needs. 
For instance, the magnetic field service provides the value of the field at 
any space point 
within the detector. The navigation service helps users to trace reconstructed tracks 
back to their Monte Carlo origins. Using 
the particle property service, the particles' 
properties can be accessed by various software components. A performance 
analysis tool 
is instrumented to profile the execution of the code and a time 
measurement tool has been
developed to facilitate code benchmarking. A pileup algorithm at 
the digital level can 
be used to mix a random trigger event with a simulated signal event 
so that the background 
simulation can be properly implemented. 

The software is managed by CMT~\cite{part1:cmt}, which 
can define a package, maintain the dependence between different packages and produce 
executables and libraries. 

\subsection{Simulation}

The \bes3 detector simulation, based on the GEANT4 
package~\cite{part1:geant4}, consists 
of four parts: event generators, detector description, particle tracking, 
and detector response. Event generators are discussed in
Sect.~\ref{part1:sec:generator}. 
A unique description of the detector geometry and materials, used by both the 
simulation and reconstruction package, has been developed based on
XML~\cite{part1:xml}. Particle 
tracking and their interactions with detector materials are handled by 
the GEANT4 package, 
while detector responses are modeled by the so-called digitization code, 
which takes 
into account detector components, including readout electronics, as well 
as the realistic situations such as noise, dead channels, etc. 
A simulation of the trigger system is also implemented. 

\subsection{Reconstruction}

The \bes3 reconstruction package consists mainly of the following 
four parts:
a) a track-finding algorithm and a Kalman-Filter-based track-fitting 
algorithm 
to determine the momentum of charged particles;
b) a particle identification algorithm based on $dE/dx$ and Time-Of-Flight
(TOF) measurements;
c) a shower- and cluster-finding algorithm for electromagnetic calorimeter
energy and position measurements; 
d) a muon track finder. 
In addtion, an event timing algorithm that determines the corresponding 
beam bunch crossing has been developed, and a secondary vertex and 
track refitting algorithm has been implemented.

\subsection{Calibration and database}

The calibration software consists of a
calibration framework and calibration algorithms. The framework
provides a standard way to obtain the calibration data objects 
for reconstruction and other algorithms. The calibration constants for
each sub-detector are produced by the associated calibration
algorithm, and are then stored in a ROOT file and a database along with other
information such as the run information, trigger condition, software 
and hardware version number, etc. 
The central database, which contains calibration data 
as well as some information 
from the online  and slow-control databases, will be distributed to
\bes3 collaborating institutions, and will also be available by remote 
access.  All of the databases at different sites will be synchronized via 
the nextwork, and updated periodically.

\section{Main Drift Chamber}

The main drift chamber (MDC), one of
the most important sub-detectors of \bes3,  can determine precisely the
momentum of a charged particle by measuring points along its trajectory in 
a well known magnetic field.  It can also determine the particle type 
by measuring the specific energy deposits ($dE/dx$) in the chamber.

\begin{figure}[htbp] 
\begin{center}
\includegraphics[width=8cm,height=6cm]{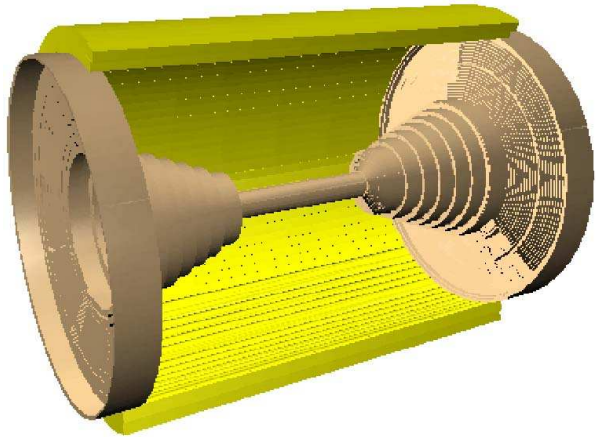}
\end{center}
\caption{An overview of the BESIII Main Drift Chamber}
\label{fig:mdc_view}
\end{figure}

The MDC is the innermost component of the \bes3 detector, and
 consists of inner and outer chambers without any intervening wall. 
The inner chamber can be replaced at some future data in the event
that it suffers severe damage due to high backgrounds. The absence 
of a chamber wall between the inner and outer chambers
eliminates a potential major source of  multiple scattering.

The MDC covers the polar angle $|\cos\theta| < 0.93$, with an inner 
radius of 60~mm, outer radius of 800~mm, and a maximum length 
of 2400~mm, as shown in Fig.~\ref{fig:mdc_view}. 
There are 43 cylindrical layers of drift cells that are
coaxial with the beam pipe, 
8 in the inner chamber and 35 in the outer chamber. 
All 8 layers in the inner chamber are stereo; 16 stereo 
layers and 19 axial layers  are interleaved in the 
outer chamber. 
In total, there are 6796 signal wires.  
The average half-width of a drift
cell is about 6~mm in the inner chamber and 8.1~mm in the outer chamber.

Aluminum wires ($\phi~110~\mu m$) are used for field shaping
 and gold-plated tungsten wires ($\phi~25~\mu m$) for
signals. A helium-based gas mixture (He/C$_3$H$_8$=60/40) is used as the working 
gas to reduce the effect of the multiple scattering, while keeping 
reasonable $dE/dx$ resolution. 

\begin{figure}[htbp]
\centerline{\includegraphics[width=10cm,height=5cm]{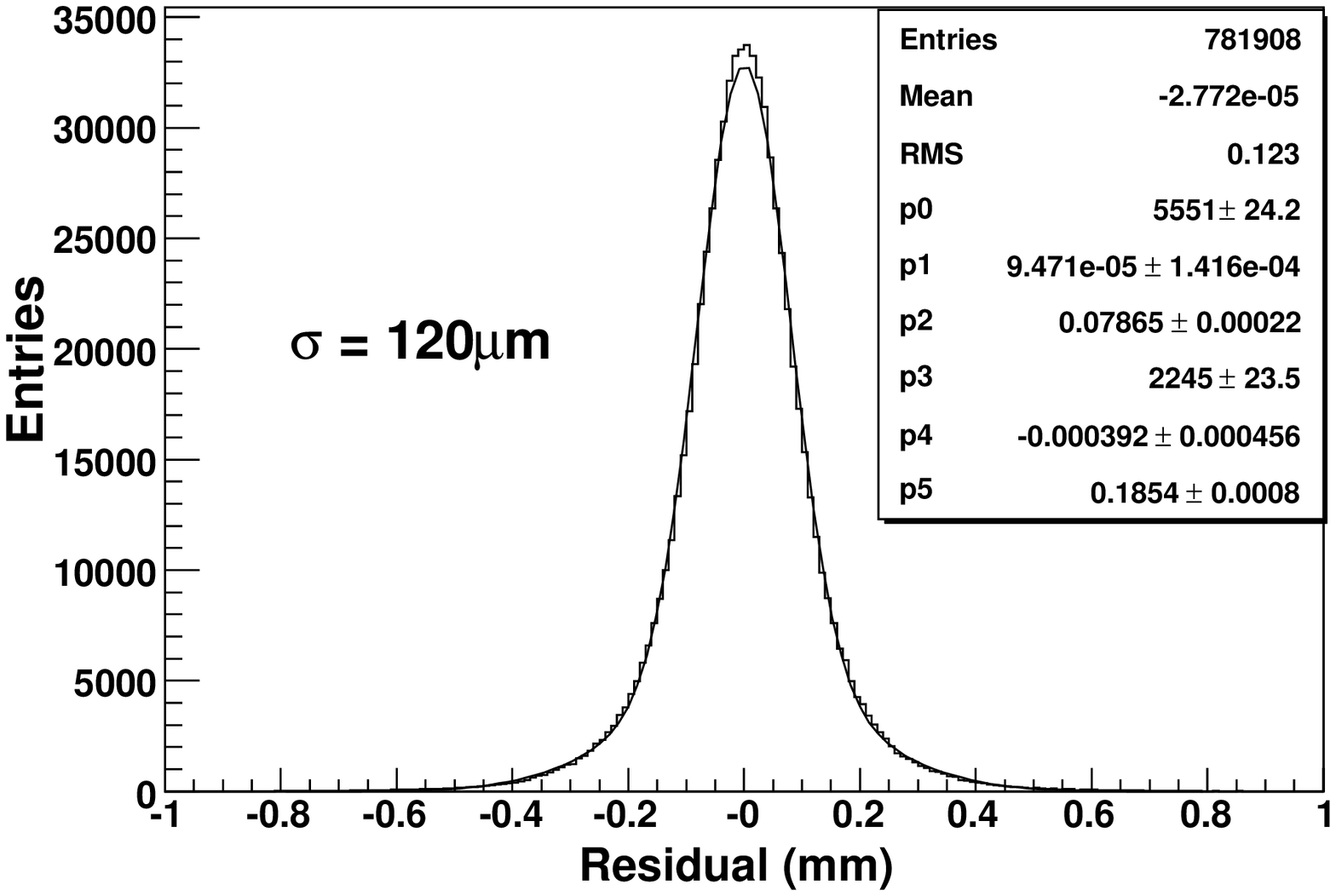}}
\centerline{\includegraphics[width=5cm,height=10cm,angle=-90]{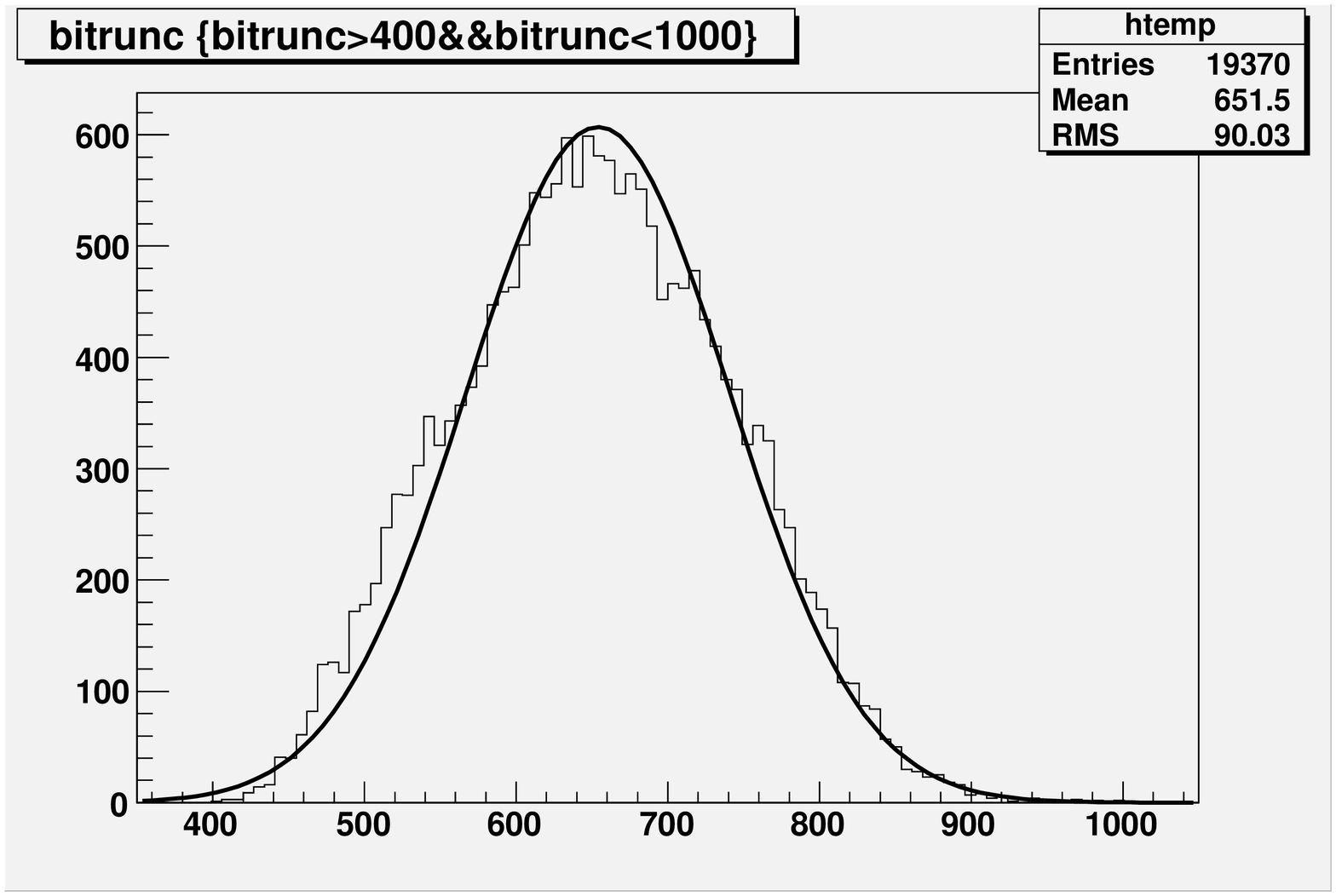}}
\caption{Upper plot: Single wire resolution of MDC from a cosmic-ray test; 
lower plot:  truncated mean of $dE/dx$ measurements from the same 
test.}
\label{fig:mdc_reso}
\end{figure}

A superconducting solenoid magnet provides an axial
1.0~Tesla magnetic field throughout the tracking volume. 
The single-wire resolution in 
the $R-\phi$ plane is designed to be better than 120~$\mu m$, 
the resolution  in $z$-direction at the vertex, measured 
with the  stereo wires, 
is 2~mm, the $dE/dX$ resolution from a truncated mean 
Landau distribution is better than
6\%, and the corresponding momentum resolution is better than 
$\sigma_{p_t}/p_t = 0.32\% p_t \bigoplus 0.37\% /\beta$,
where the first term comes from the trajectory measurement and 
the second term from  multiple scattering. 
Figure~\ref{fig:mdc_reso} shows
the single wire resolution of the MDC and 
truncated mean $dE/dX$ measurements from a cosmic ray test.

\subsection*{MDC Simulation}

The XML descriptions of the geometry and materials of 
the MDC are based on the
GEANT4 package. In particular, the tube class is used to describe and
build endplates and axial layers, while the hype class is used for stereo
layers and the twisttube class for stereo cells.  
During the track simulation, steps in the same cell are treated as 
one hit, and the digitization relies heavily on
calibration parameters via the calibration service function. 
Effects such 
as wire efficiency and resolution as a function of drift distance 
for each wire, noise in each layer, misalignment etc., 
have been modeled with parameters that can be tuned via a calibration 
service function.

\subsection*{MDC Reconstruction and calibration}

The MDC tracking algorithm starts with the
formation of track segments from hits using pre-calculated
patterns. It then links the found axial segments to circular
tracks and applies a circular fit using the least-square method. 
Stereo segments are subsequently added to track candidates
followed by an iterative helix fit. Finally, after collecting 
additional hits that might possibly belong to the track,
a track refitting procedure
based on the  Kalman-filter technique is performed.
From a Monte Carlo simulation, we determine that
this algorithm can maintain a tracking efficiency of more than 
98\% for  $p_T>$150 MeV/c tracks,
even in the presence of severe backgrounds. 
The $dE/dx$ reconstruction algorithm calculates
the energy loss of each charged particle through the chamber after
applying various corrections to the measured charge amplitudes, 
and then gives the probablity of each particle identification
hypothesis.
A GEANT4-based algorithm is developed to extrapolate a MDC track
into outer sub-detectors, taking into account the magnetic field and
the ionization loss of charged particle in the detector. 
The associated error matrix at a given space point is calculated 
taking into account multiple scattering effects. 

\subsection*{MDC calibration}

The MDC will be calibrated using $J/\psi \rightarrow \mu^+\mu^-$ events 
for both position and $dE/dX$ measurements.  Since
the production cross section at $J/\psi$ peak is huge, sufficient
statistics can be obtained in a short run period. The resulting 
calibration 
constants for the $x-t$ relations, timing, alignment, absolute efficiency 
of  wires, etc. for each run period are stored 
in the database for use by reconstruction algorithms. 
There is also a proposal to calibrate the MDC by turning off the magnetic 
field, so that straight tracks can be used for calibration in order to 
determine precisely the position of each 
wire~\cite{part1:calib_wulh}. 

\section{Time-Of-Flight System}

The Time-of-Flight (TOF) sub-detector, made of plastic scintillator 
bars and read out by fine-mesh phototubes,
is placed between the drift chamber and the electromagnetic calorimeter
and measures the flight time of charged 
particles in order to identify the particle-type. It also provides 
a fast trigger and helps reject cosmic-ray backgrounds. 

\begin{figure}[htbp]
\centerline{\includegraphics[width=8cm,height=6cm]{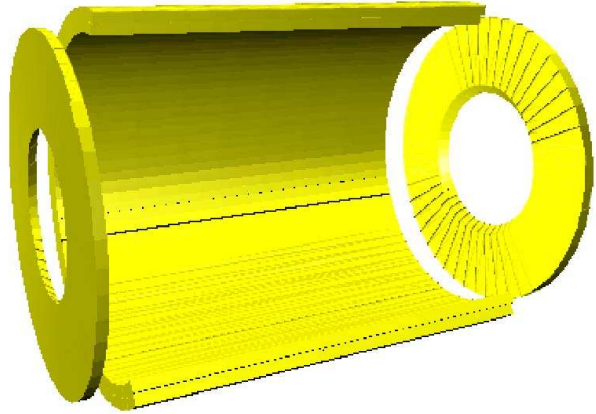}}
\caption{The \bes3 Time Of Flight System}
\label{fig:tof_bes3}
\end{figure}

The \bes3 TOF consists of two parts: the barrel and endcap as shown in 
Fig.~\ref{fig:tof_bes3}.
The solid angle coverage of the barrel TOF is $|\cos\theta| <0.83$, 
while that of the endcap is  $0.85< |\cos\theta| < 0.95$. The Barrel 
TOF consists of two layers of 88 plastic scintillator elements
arranged in a cylinder of mean radius  $\sim$870~mm.
Each scintillator bar 
has a length of 2380~mm, a thickness of 50~mm and a width of 50~mm;
it is read out at each end by a fine-mesh PMT.
Each endcap TOF array 
consists of 48 fan-shaped elements with an inner radius of
410~mm and an outer radius of 890~mm; these are read out from one end of 
the scintillator by a single fine-mesh PMT.

\begin{figure}[htbp]
\centerline{\includegraphics[width=5cm,height=11cm]{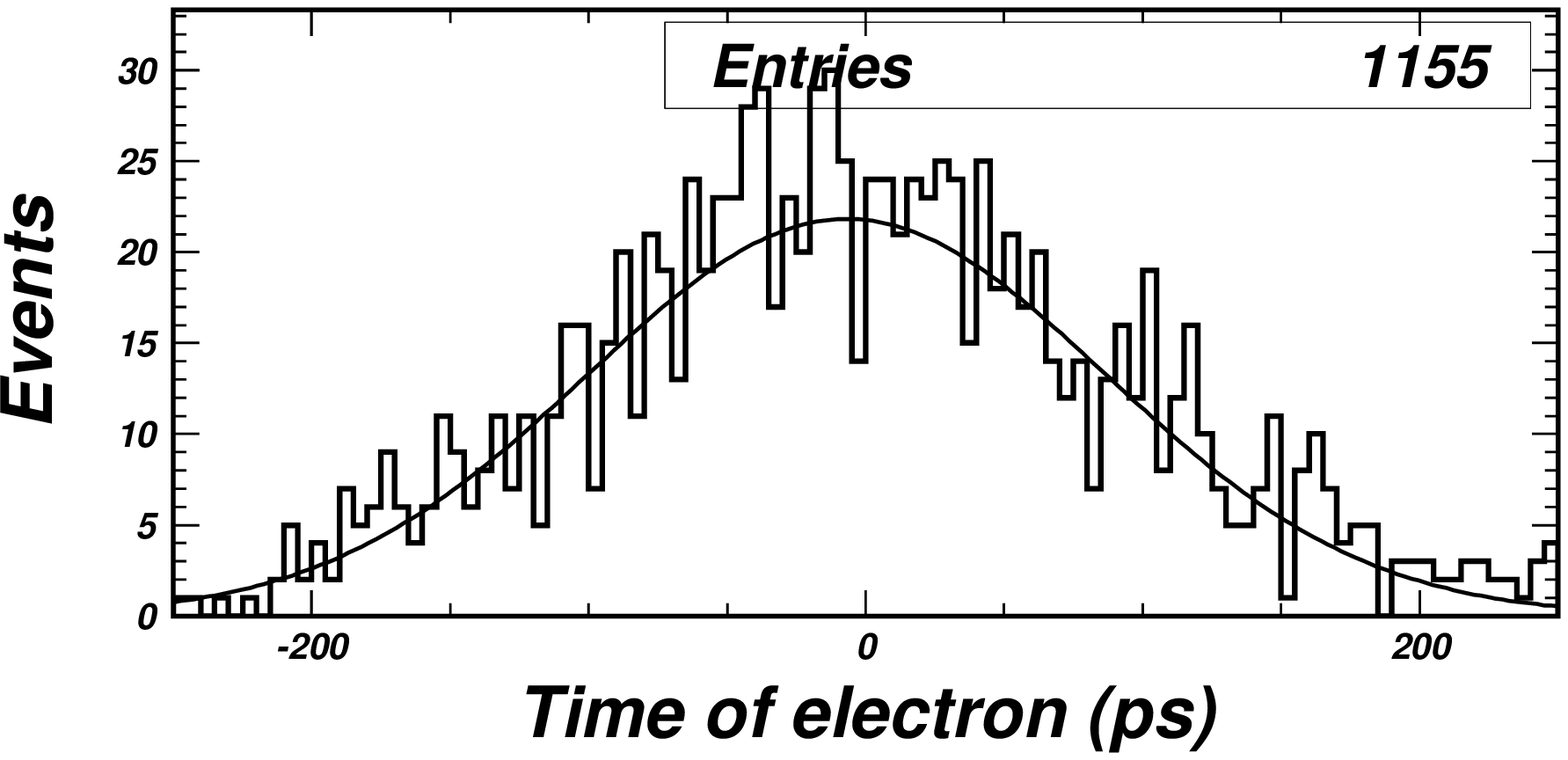}
\includegraphics[width=5cm,height=11cm]{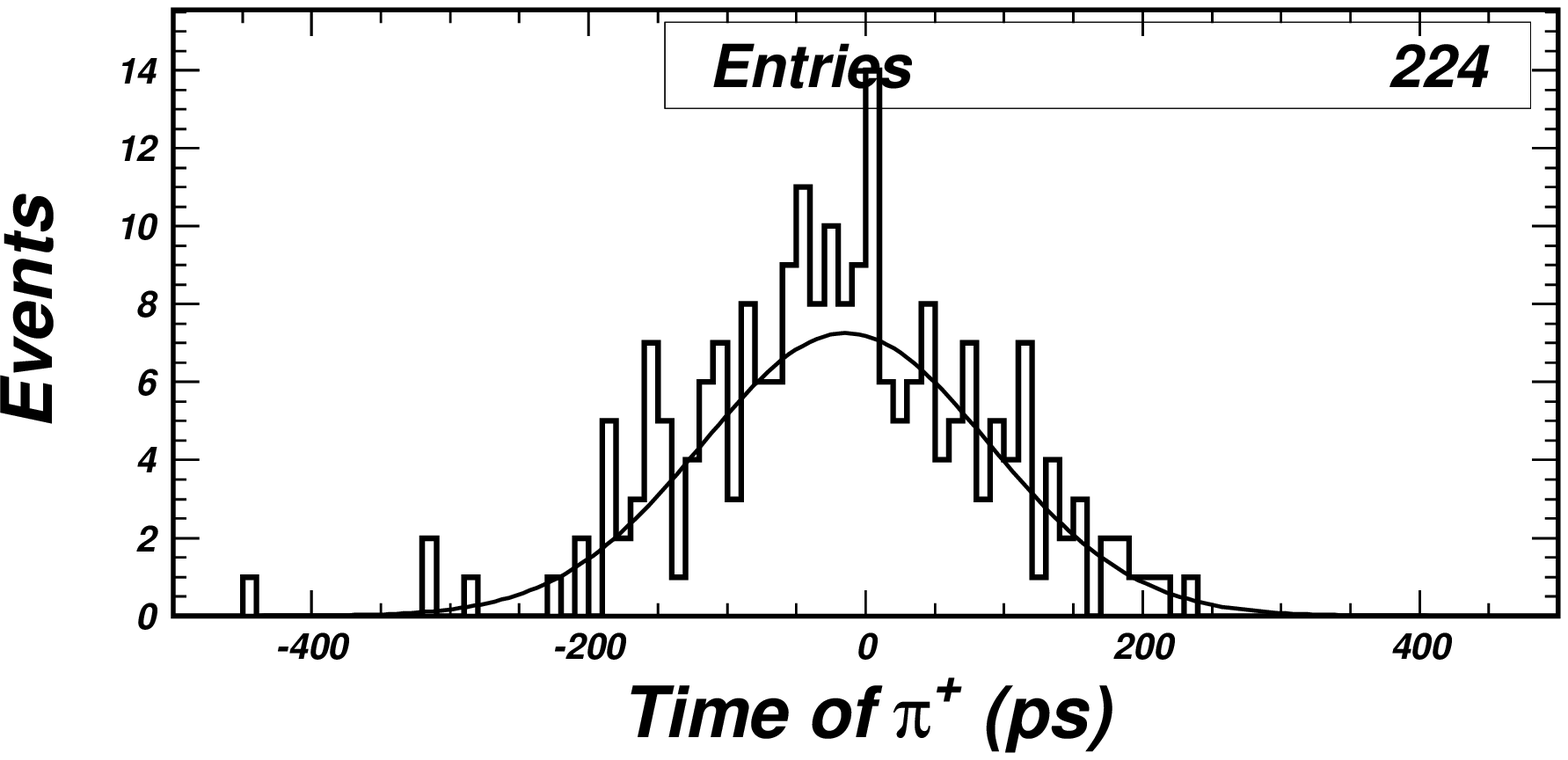}
\includegraphics[width=5cm,height=11cm]{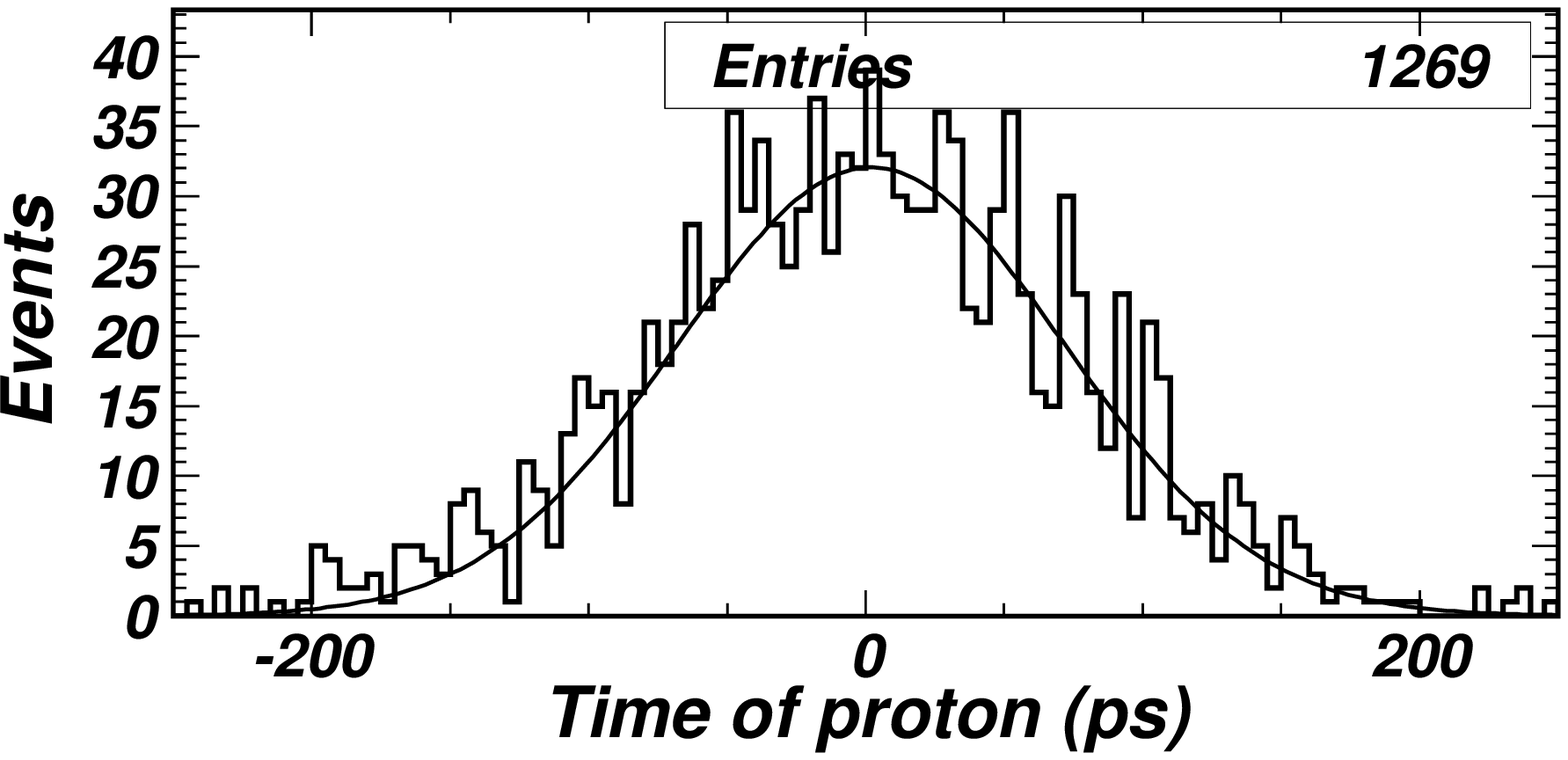}}\vspace{-4.0cm }
\caption{Time resolution of TOF counters from a beam test}
\label{fig:tof_reso}
\end{figure}

For this system, among all parameters, the time resolution is key.
This mainly depends on the following
contributions: the intrinsic TOF time resolution caused by the
characteristics of the scintillator and the PMT, time resolution and 
jitter in the readout electronics, the beam-bunch length and 
the bunch timing  uncertainty, 
the vertex and pathlength resolution of the track, 
time-walk effects, etc. The design intrinsic time 
resolution for a barrel counter is 90~ps and for an endcap counter
is 70~ps, which have been demonstrated in beam tests as shown in 
Fig.~\ref{fig:tof_reso}. The total time resolution 
for the double-layer barrel TOF and the end-cap is expected to be
about 100~ps. 

\subsection*{TOF Simulation}

The TOF simulation takes into account the scintillator, 
wrapping materials and photomultiplier tubes based on the GEANT4 package. 
A fast simulation model has been built that converts the
energy loss of a particle in the scintillator into photons,
propagates them (not individual photons but the pulse and its shape) 
to the PMTs and generates an electronic signal.  
A discriminator is then applied to the pulse to yield ADC and TDC
output. This algorithm has been tested with results from beam tests
and further tuning will be needed when data are available.
A full simulation that traces each optical photon can be activated to 
understand the details of the timing measurement.

\subsection*{TOF Reconstruction}

The TOF reconstruction starts from tracks
extrapolated to the TOF and matched with a particular TOF module.
The travel time of a charged particle from the interaction
point to the TOF is then calculated, after a weighted average
of results from PMTs at both ends of the scintilator bar and applying various 
corrections such as the effective light velocity in the scintillator, 
the light attenuation length, etc. The $dE/dx$ measurement is also 
obtained for both charged and neutral particles. Energy loss in 
the TOF can  then be added to that in the EMC in order 
to improve the shower energy resolution.

\subsection*{TOF Calibration}

The TOF calibration will be performed using $J/\psi$ decays to leptons,
both for timing and energy. The resulting calibration 
constants, such as effective velocity, attenuation length, muon energy
loss, etc., are stored in the database for use by the 
reconstruction algorithms. The TOF performance and status are monitored
regularly by a laser-fiberoptics  pulsing system.

\section{Electromagnetic Calorimeter}

\begin{figure}[htbp] 
\centerline{\includegraphics[width=8cm,height=6cm]{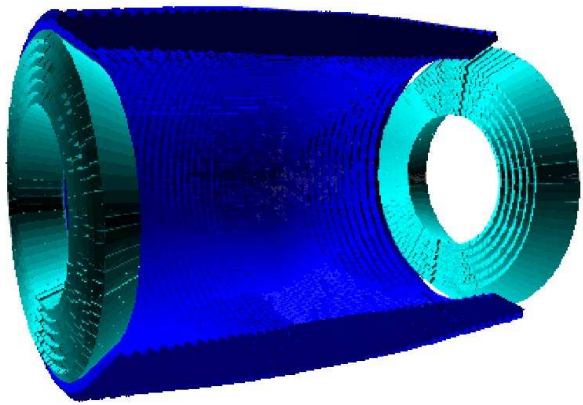}}
\caption{The BESIII Electromagnetic Calorimeter}
\label{fig:emc_bes3}
\end{figure}

The Electro-Magnetic Calorimeter (EMC) 
measures the energies and positions of electrons and photons
precisely, and plays an
important role in the \bes3 detector.
The calorimeter is comprised of one barrel and two endcap sections
as shown in Fig.~\ref{fig:emc_bes3}. The barrel has an inner radius 
of 940~mm and a length of 2750~mm, and covers the polar angle of 
$|\cos\theta|<0.83$. 
The endcaps have an inner radius
of 500~mm are placed at $z=\pm 1380$~mm from the collision point,
and  cover the polar angle range $0.85<|\cos\theta|<0.93$. 
The total acceptance is 93\% of 4$\pi$. A small gap of
about 50~mm between the endcaps and the barrel is required for
mechanical support structures, cables and cooling pipes. 
In the barrel, there
are a total of 44 rings of crystals along the $z$ direction,
each with 120 crystals. All crystals except for those in two rings at the
center of the detector point to $z = \pm$ 50~mm with a
slight tilt angle of 1.5$^o$ in the $\phi$ direction. Each endcap
consists of 6 rings that are  split into two tapered half-cylinders. 
All crystals point to $z = \pm 100$~mm with a tilt of 1.5$^o$ in the 
$\phi$
direction. The entire calorimeter has 6272 CsI(Tl) crystals and
a total weight of about 24 tons. 

The calorimeter is designed to have an energy measurement range for
electrons or photons from 20~MeV to 2~GeV, with an energy resolution
of about 2.3\%/$\sqrt{{E({\rm GeV})}}\bigoplus1$\%. The
design position resolution 
for an electromagnetic shower is
$\sigma_{xy}\leq 6~{\rm mm}/\sqrt{E({\rm GeV})}$ and the
electronics noise for each crystal 
is required to be less than 220~keV.

\subsection*{EMC Simulation}

In the simulation,
the EMC detector description is based on XML including crystals, 
casing, silicon photodiodes, preamplifier boxes, cables and 
the support system. The G4Trap class is used
for the barrel crystals and G4IrregBox, a class implemented by \bes3,
for the endcap crystals. The sensitive detector flag is set for crystals
and photodiodes. Hit information is recorded in the sensitive
detectors, and then the energy deposits summed,
 photon statistics computed, and the resulting photodiode
response is converted into electronics signals. 
The waveform for the electronics signals in the
time domain is obtained via an inverse Laplace transform;
the sampling and peak searching
process by the flash ADC is simulated to yield energy and time
information. Gaussian-type electronic noise is added to each bin
and the background is produced by summing the
waveforms. 

\subsection*{EMC Reconstruction}

\begin{figure}[htbp]
\centerline{\includegraphics[width=10cm,height=9cm]{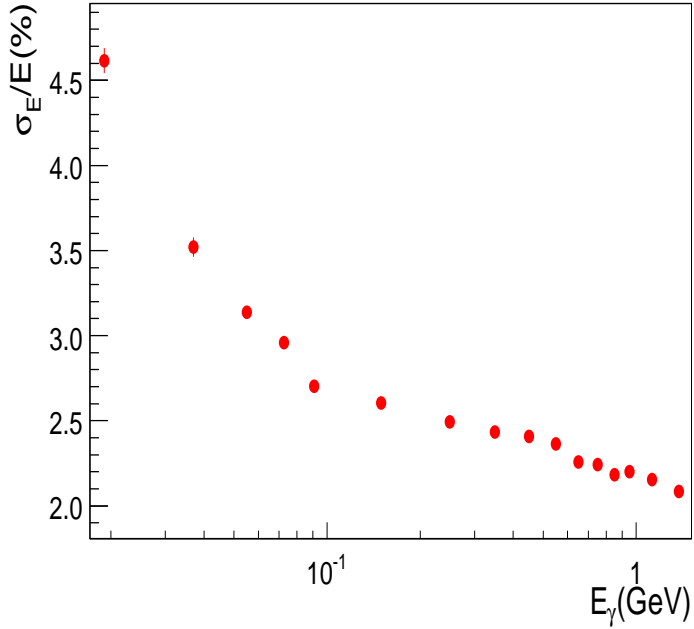}}
\caption{Energy resolution of EMC showers.}
\label{fig:emc_reso}
\end{figure}

Shower reconstruction in the EMC consists of three concatenated steps: 
The ADC value of each crystal is converted into energy using the
corresponding calibration constants. Clusters in both the barrel and 
end-caps are formed around the crystals with
local maximum energy deposits, called seeds. 
The  shower energies are computed from the energy sums
and the positions from energy-weighted first moments.
A splitting function is invoked to split the cluster
into several showers if multiple seeds are found in one cluster.
Matched energy deposits found in the TOF are added 
to the associated showers to improve 
energy resolution, particularly, for low energy photons. 
Figure~\ref{fig:emc_reso} shows the expected energy resolution for 
electromagnetic showers. 

\subsubsection*{EMC Calibration}

The EMC high energy response 
will be calibrated with Bhabha electrons at energies
of 1.55~GeV or more and the low energy response with 
$\pi^0\to\gamma\gamma$ decays. Each crystal has to be 
recalibrated periodically, and 
monitored frequently by a LED 
light pulser. Corrections due to temperature variations can be applied.
Calibration constants are stored in the database for use by the 
reconstruction algorithms. 

\section{Muon Identifier}

The \bes3 muon system is designed to distinguish
muons from hadrons by the characteristic hit
patterns they produce when penetrating the
return yoke of the \bes3 magnet.  The muon counter
is made of resistive plate chambers~(RPC) sandwiched 
by iron absorbers.
 
\begin{figure}[htbp]
\centerline{\includegraphics[width=8cm,height=6cm]{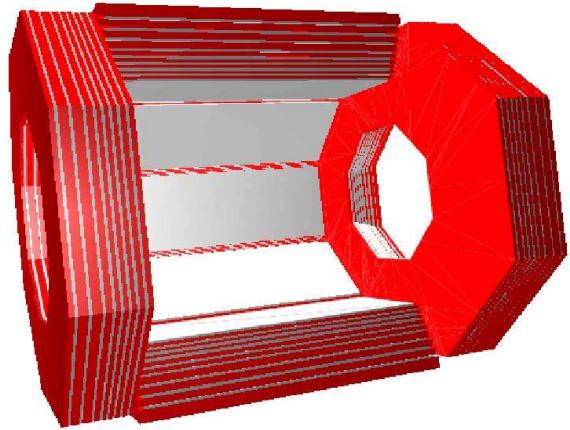}}
\caption{The BESIII Muon Identifier}
\label{fig:muon_bes3}
\end{figure}

The barrel part of the muon identifier is organized into
octants, each of which consists of 9 layers of muon counters
inserted into gaps in the iron, as shown in 
Fig.~\ref{fig:muon_bes3}. Each endcap is divided into quadrants,
each consisting of 8 RPC layers.  Proceding radially outward, the 
thicknesses of the iron absorbers are 3~cm, 3~cm, 3~cm, 4~cm,
4~cm, 8~cm, 8~cm, 8~cm and 15~cm. The muon counter
insertion gap between neighboring slabs is 4~cm.
The width of the RPC pickup strip is optimized at 4~cm, and 
only the $z$ orientation of odd gaps and azimuthal
orientation of even gaps are read out in the barrel, while in endcap, 
the $x$ orientation for odd gaps and $y$ orientation for even gaps are 
read out.

\subsection*{MUC Simulation}

The GDML technology is applied in the MUC simulation and
software objects are created for each component of the detection
system, including their materials, shapes, positions, sizes, etc.
At the lowest level, bakelite and gas objects are used to form
an RPC. A set of strip objects form a read-out plane. Objects of
RPCs, read-out planes and aluminium boxes form a muon counter module.
Iron slabs and modules are finally installed in the proper position to
build the muon identifier. The MUC digitization is
rather simple, an algorithm is developed to
select fired strips based on the distances from the tracks to strips.
The detection efficiency can be set for each pad 
based on test results; actual values will be assigned 
when data are available. 
Noise is simulated by a Poisson distribution with a noise level
determined from measurements made during the chamber construction. 
Again, actual values
will be assigned when detector is in operation. 

\subsection*{MUC Reconstruction}

The MUC reconstruction algorithm
starts with two searches that collect hits (fired strips)
in the two orientations. The two
collections are then combined to form track candidates 
and matched with tracks reconstructed in the MDC. 
For low momentum muons, it
may happen that too few layers in muon counters are fired. 
A subsequent search for tracks that looks through the remaining hits,
using  MDC tracks as seeds, is performed. 

\begin{figure}[htbp]
\centerline{\includegraphics[width=10cm,height=9cm]{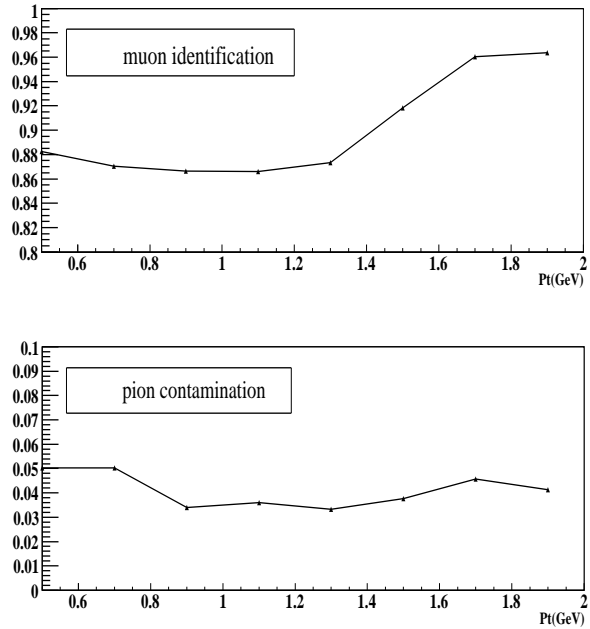}}
\caption{Muon/pion identification as a function of the transverse momentum}
\label{fig:muon_reso}
 \end{figure}

For each muon track candidate, a number of parameters are
calculated for muon/hadron identification, such as the depth of 
the track in the muon identifier, the maximum number of hits among 
layers the track penetrates, the match between the MUC stand-alone 
track and the MDC track, the $\chi^2$ of the MUC stand-alone track, etc. 
These parameters, together with the track momentum and exit 
angle from the MDC, are used as input to an Artificial Neural
Network for muon/hadron analysis. Figure~\ref{fig:muon_reso} shows the
performance of current muon identifier from simulated single muons
and pions in the momentum range between 0.5 and 1.9~GeV/c. 
In general, we are able to
reject pions to a level of $\sim4\%$ while keep 90\% of
real muons.

\subsection*{MUC Calibration}

The MUC Calibration is developed within the BOSS
framework. The main task is to study RPC detection efficiencies as
a function of area. In addition, the cluster size and noise level
are also  studied. Results are stored in a database for use
by reconstruction algorithms. 

\section{Trigger}

    The trigger system is required to select interesting physics 
events with a high efficiency and suppress backgrounds to a 
level that the data acquisition (DAQ) system can sustain. 
The maximum throughput of the \bes3 DAQ system is designed to be
4000Hz, hence the trigger system should be able to  
reduce various backgrounds and Bhabha events to a level less than 
2000~Hz while maintaining a high efficiency for signal events, 
which have a rate as high as
2000~Hz at the $J/\psi$ peak.

    A two-level scheme has been adopted for the \bes3 trigger system:
a level-1 hardware trigger with a level-2 software event filter.  
Signals from different sub-detectors are split into 
two, one for digitization and storage in the pipeline of the front-end 
electronics (FEE), and the other for the level-1 hardware 
trigger. Signals from sub-detectors are received and 
processed by the appropriate electronics modules in VME crates 
to yield basic trigger primitives such as the number of short and long 
tracks in the drift chamber, the number of fired scintillator bars
in TOF, the number of clusters and their topology in the electromagnetic 
calorimeter, etc. Information from these primitives are assembled
by the global trigger logic (GTL), which  generates an L1 strobe when
the trigger condition is satisfied. 

    Trigger conditions are pre-determined based on Monte Carlo 
simulations and will be adjusted based on the actual beam background
conditions.
Table~\ref{tab:trigger_eff} shows the trigger efficiencies for various
signal and backgrounds based on Monte Carlo simulations using the current
trigger table.

\begin{table}[t]
\begin{center}
\begin{tabular}{|c|c|c|}
\hline\hline
Events  &  Number of events simulated & Efficiency(\%) \\
\hline\hline
J/$\psi \rightarrow e^+e^-$   &    50,000  &  100.0 \\
\hline
J/$\psi \rightarrow \mu^+mu^-$   &  50,000  &  99.9 \\
\hline
J/$\psi \rightarrow$ Anything   &    10,000  &  97.7 \\
\hline
$\psi' \rightarrow$ Anything    &    10,000  & 99.5 \\
\hline
$\psi" \rightarrow  D\bar{D} $ Anything & 10,000   & 99.9 \\
\hline
J/$\psi \rightarrow \omega\eta \rightarrow 5\gamma$ &  10,000 & 97.9 \\
\hline
J/$\psi \rightarrow \gamma\eta \rightarrow 3\gamma$ &   10,000 & 92.8 \\
\hline
J/$\psi \rightarrow K^+K^-\pi^0 $  & 10,000  & 97.4 \\
\hline
J/$\psi \rightarrow \pi^0 p\bar{p} $ & 10,000  & 97.9 \\
\hline
J/$\psi \rightarrow p\bar{p} $ & 10,000  & 95.8 \\
\hline
Beam backgrounds    & 1,000,000 & 0.005 \\
\hline
cosmic-ray backgrounds &  100,000  & 9.4 \\
\hline\hline
\end{tabular}
\end{center}
\caption{Estimated trigger efficiencies for different types of events.}
\label{tab:trigger_eff}
\end{table}

Clearly, the efficiencies for most signals with topologies
containing  multiple charged tracks and photons
are satisfactory, even at very low momentum. The rejection power 
for beam backgrounds, which is estimated to have a maximum level of 
40~MHz, is about $5\times 10^{-5}$, resulting in a
background trigger rate of 
below 2000~Hz, even for extreme conditions.   The trigger rate for
cosmic-ray backgrounds is about 90 Hz.

\chapter{Analysis Tools}
\label{sec:tools}
\section[Monte Carlo Generators]{Monte Carlo Generators\footnote{By
Rong-Gang Ping}}
\label{part1:sec:generator}

\subsection{Introduction}
Precision measurements will be a central theme and
challenge for the \bes3 physics research program, and these will 
require high precision Monte Carlo (MC) generators.

High quality and precise MC simulations will be
essential for minimizing experimental systematic
uncertainties.  They are
used to determine detection efficiencies 
and to model backgrounds.  Thus,
the MC generators must simulate the 
underlying processes being studied as precisely as possible. 
Recently, high-precision generators 
({\it e.g.} KKMC, Bhlumi etc.) based on 
Yennie-Frautchi-Suura exponentiation 
have been developed for the QED processes 
$e^+e^-\to ff$ ($f:$ fermion). 
The official ``precision tags'' of these
generators are at the order of 1\% or less.
Generators that incorporate dynamical information into 
hadron decays have also
been developed, most notably EvtGen, which was produced by the 
BaBar and CLEO collaborations to model $B$ meson decays. 
These developments provide us with the luxury of
being able to choose which among the existing
generators is most suitable for simulating physics processes
in the tau-charm threshold  region.

However, most of these generators were originally intended for 
energies above the tau-charm threshold region. In general,
MC generators are  process- and model-dependent.  Only
a few MC generators cover the full energy scale of high energy
physics experiments.   So, at tau-charm threshold energies, the 
migration of the MC generators that were originally developed 
for higher energy scales requires careful tuning of
parameters and further comparisons with data.  
In addition, the comprehensive
generation of exclusive charmonium decays
requires that more models are included
in the EvtGen framework.

In this section, we present a general description of the \bes3
generator framework, and give brief introductions to the \bes3 event
generators, such as KKMC, BesEvtGen, various QED generators, and 
some inclusive generators. For details, the reader is directed to
to the generator guides and/or related publications.

\subsection{Generator framework}

\begin{figure}[htbp]

\begin{center}
\hspace*{-.cm} \epsfysize=6cm \epsffile{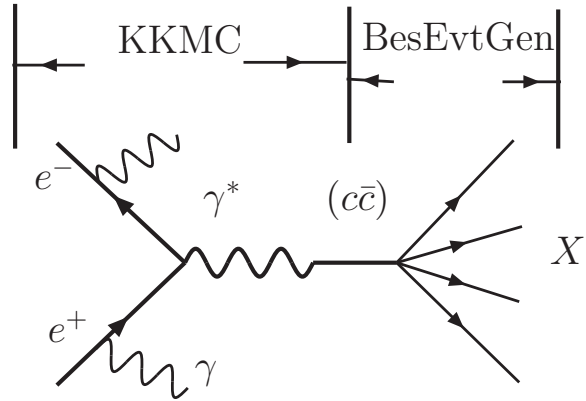}
\parbox{1\textwidth}{\caption{Illustration of \bes3 generator framework.
\label{part1:fig:fw}}}
\end{center}
\end{figure}

The default generator framework for \bes3 uses {\bf KKMC +
BesEvtGen} to generate charmonium decays. Charmonium production
via  $e^+e^-$ annihilation is illustrated 
in Fig.~\ref{part1:fig:fw}.
The incoming positrons and electrons can radiate real photons
via initial state radiation (ISR) before they annihilate into a 
virtual photon.  Corrections for these radiative processes 
are crucial in $e^+e^-$ annihilation experiments, especially
for measurements performed  near a  
resonance or a production threshold (see chapter 5). In order
to achieve precise results, generators for
$e^+e^-$ collision must carefully take ISR into account. 
The KKMC generator is used to simulate  $c \bar{c}$ production 
via $e^+e^-$ annihilation with the inclusion of ISR effects 
with high  precision; it also includes the effects of the beam
energy spread. The subsequent charmonium 
meson decays are generated with BesEvtGen.

It should be noted that the events are generated in the
centre-mass-system (cms) of the $e^+e^-$ beam. 
However,  the $e^+e^-$ beams at BEPCII
are not aligned exactly back to back; there is a crossing angle
between the two beam of about 22~mrad. Thus, the produced charmonium
state is not at rest and instead moves along the 
$x-$direction with a small
momentum. As a result, the generated events have to be boosted to the 
laboratory system before proceding through the detector simulation. 
This boost is implemented outside of the generator framework.

\subsection{\bes3 Generators}
Early generators used at \bes3 were those used for BESII, which
includes about 30 generators.  These are now obsolete and we do
not recommend their use.\footnote{Currently, the truth tables of these
generators are not available in the simulation} In what follows, we
focus on the generators currently used in the \bes3 generator
framework.
\subsubsection{KKMC}
KKMC~\cite{part1:kkmc} is an event generator for the precise
implementation of the Electroweak Standard Model 
formulae for the processes $e^+e^-\to f\bar f+n\gamma$ 
($f=\mu,~\tau,~d,~u,~s,~c,~b$) at 
centre-of-mass energies from the $\tau^+\tau^-$ 
threshold up to 1~TeV. KKMC was originally designed
for LEP, SLC, and is also suitable for future Linear Colliders,
$b,~c,~\tau-$ factories etc.

The most important features of KKMC are: the
implementation of ISR-FSR\footnote{FSR stands for final state radiation.}
interference;
second-order subleading corrections; and the exact matrix
element for two hard photons. Effects due to photon emission
from initial beams and outgoing fermions are calculated in QED up to
second order, including all interference effects, within 
the Coherent Exclusive Exponentiation (CEEX) scheme that is based on
Yennie-Frautschi-Suura exponentiation. Electroweak corrections are
included at first order with higher-order extensions using the
DIZET 6.21 library. Final-state quarks hadronize according to the
parton shower model using PYTHIA. Decays of the $\tau$ lepton are
simulated using the TAUOLA library, which takes spin
polarization effects into account.  The code and information on 
the program are available at
the KKMC web page~\cite{part1:webkkmc}.

In the \bes3 generator framework, KKMC is used to generate
charmonium states with the inclusion of ISR effects and the beam
energy spread. The resonances supported by KKMC include the
$J/\psi,~\psi(2S),~\psi(3770),~\psi(4030),~\psi(4160),~\psi(4415)$ and
secondary resonances, such as the 
$\rho,~\rho',~\rho'',~\omega,~\omega',~\phi$
and $\phi'$. Although KKMC supports the simulation of the decay of
these particles, the models in BesEvtGen are more powerful;
FSR effects can be included at the BesEvtGen level by
means of the PHOTOS package.

\subsubsection{BesEvtGen}
BesEvtGen~\cite{part1:besevtgen} is a generator for tau-charm physics
that has been developed from EvtGen,\footnote{The version is
V00-11-07} which was originally designed to study $B$ physics. 
EvtGen has a powerful interface that allows for the generation of
events for a given decay using a model that is easily created by
the user; it also provides access to other generators, such as PYTHIA
and PHOTOS.

The EvtGen interface uses  dynamical information to generate a
sequential decay chain through an accept-reject algorithm,
which is based on the amplitude probability combined with
forward and/or backward spin-density matrix information. The
EvtGen interface is designed to have the functionality to automatically
calculate these spin-density matrices. As an illustration of how the event
selection algorithm works, we consider the sequential decay
$J/\psi\to\rho^0\pi^0,~\rho^0\to \pi^+\pi^-$ and $\pi^0\to
\gamma\gamma$.

The first chain of the decay is selected based on the probability
\begin{equation}
P_\psi=\sum_{\lambda_\psi,\lambda_\rho}|M^{J/\psi\to\rho^0\pi^0}_{\lambda_\psi,\lambda_\rho}|^2,
\end{equation}
where $M$ stands for the amplitude for $J/\psi\to\rho^0\pi^0$ with the
helicity indexes $\lambda_\psi$ and $\lambda_\rho$. After decaying
the $J/\psi$, one has the forward spin-density matrix
\begin{equation}
{\mathcal
D}^{\rho^0}_{\lambda_\rho,\lambda_\rho'}=\sum_{\lambda_\psi}
M^{J/\psi\to\rho^0\pi^0}_{\lambda_\psi,\lambda_\rho}
[M^{J/\psi\to\rho^0\pi^0}_{\lambda_\psi,\lambda_\rho'}]^*.
\end{equation}
To generate the $\rho^0\to\pi^+\pi^-$ decay, proceed as with the
$J/\psi$, including also ${\mathcal
D}^{\rho^0}_{\lambda_\rho,\lambda_\rho'}$
\begin{equation}
P_\rho={1\over \textrm{Tr}{\mathcal
D}^{\rho^0}}\sum_{\lambda_\rho,\lambda_\rho'}{\mathcal
D}^{\rho^0}_{\lambda_\rho,\lambda_\rho'}A^{\rho^0\to\pi^+\pi^-}_{\lambda_\rho}[A^{\rho^0\to\pi^+\pi^-}_{\lambda_\rho'}]^*.
\end{equation}
To decay the $\pi^0$ with the full correlations between all
kinematic variables in the decay, the EvtGen interface automatically
calculates the backward spin-density matrix by
\begin{equation}
\tilde{\mathcal
D}^{\rho^0}_{\lambda_\rho,\lambda_\rho'}=A^{\rho^0\to\pi^+\pi^-}_{\lambda_\rho}[A^{\rho^0\to\pi^+\pi^-}_{\lambda_\rho'}]^*,
\end{equation}
then the spin-density matrix for the $\pi^0$ is
\begin{equation}
{\mathcal
D}^{\pi^0}=\sum_{\lambda_\psi,\lambda_\rho,\lambda_\rho'}\tilde{\mathcal
D}^{\rho^0}_{\lambda_\rho,\lambda_\rho'}M^{J/\psi\to\rho^0\pi^0}_{\lambda_\psi,\lambda_\rho}
[M^{J/\psi\to\rho^0\pi^0}_{\lambda_\psi,\lambda_\rho'}]^*,
\end{equation}
which is a constant for a spin-0 particle. Thus the $\pi^0$ decay is
selected by the probability
\begin{equation}
P_{\pi^0}={1\over \textrm{Tr}{\mathcal
D}^{\pi^0}}\sum_{\lambda_1,\lambda_2}{\mathcal
D}^{\pi^0}A^{\pi^0\to\gamma\gamma}_{{\lambda_1},\lambda_2}[A^{\pi^0\to\gamma\gamma}_{{\lambda_1},\lambda_2}]^*=
\sum_{\lambda_1,\lambda_2}A^{\pi^0\to\gamma\gamma}_{{\lambda_1},\lambda_2}[A^{\pi^0\to\gamma\gamma}_{{\lambda_1},\lambda_2}]^*.
\end{equation}

BesEvtGen incorporates baryons up to spin=3/2, and has about 30
models for simulating exclusive decays. The
amplitudes for these models are constructed using the helicity
amplitude method, and constrained by imposing parity
conservation. One of the most powerful models is DIY, which can
generate any decays using user-provided amplitudes.
Other useful models are those that generate decays using the
histogram distributions, such as MassH1, MassH2 and 
Body3.\footnote{These correspond to generating events 
according to a 1-D diagram, a Dalitz plot 
or the generation of 3-body decays according to a Dalitz plot plus two
angular distribution plots.} 
BesEvtGen provides access to the
PYTHIA and Lundcharm inclusive generators that can be used,
for example, to generate unknown decays of a given resonance.

\subsubsection{QED generators}

\begin{itemize}
\item Bhlumi and Bhwide

The generators Bhlumi~\cite{part1:bhlumi} and Bhwide ~\cite{part1:bhwide} 
are used to generate the Bhabha scattering process 
$e^+e^-\to e^+e^- + n\gamma$.
These are full energy scale generators even though
they were  originally designed for the high energy
LEP1/SLC and LEP2  colliders.  The Bhlumi generator is
suitable for generating  low angle Bhabha events
($\theta<6^\circ$), while the Bhwide generator is
intended  for wide angle
Bhabha events ($\theta>6^\circ$). Here ``suitable" means that
when these generators work within their suitable region,
they will achieve the tagged precision level; outside of that 
region their precision  will be poorer.
The precision of  Bhlumi is quoted as $0.11\%$ at the LEP1 energy
scale and 0.25\% for LEP2 experiments.  These estimates are based on
comparison with other MC calculations~\cite{part1:bhlumi}. 
The precision of the Bhwide is quoted as 0.3\% at the $Z$ boson peak and 
1.5\% at LEP2 energies.

 Bhlumi
is a multiphoton Monte Carlo event generator for low angle
($\theta<6^\circ$) Bhabha events that
provides four-momentum vectors  of the outgoing
electron, positron and photons. The first
${\mathcal{O}}(\alpha^1)_{YFS}$ version is described in
Ref.~\cite{part1:jadach}. The matrix elements are based on
Yennie-Frautschi-Suura (YFS) exponentiation. They include
exact first-order photonic corrections and leading-log
corrections at second order.  The other higher-order and 
subleading contributions are included in an approximate form.

Bhwide is a wide angle ($\theta>6^\circ$) generator for Bhabha
scattering. The theoretical formulation is based on
${\mathcal{O}}(\alpha)$ YFS exponentiation, with
${\mathcal{O}}(\alpha)$ virtual corrections from 
Ref.~\cite{part1:corr1}.
The YFS exponentiation is realized via Monte Carlo methods based on
a Bhlumi-type Monte-Carlo algorithm, but with some important
extensions: (1) QED interference between the electron and positron
lines are reintroduced as they are important for large
angle Bhabha scattering; (2) the full YFS form factor for the
$2\to2$ process, including all $s-,~t-$ and $u-$channels is
implemented; (3) the exact ${\mathcal{O}}(\alpha)$ matrix element
for the full Bhabha process is included.

Users of these generators are required to specify the
centre-mass-system energy, as well as other cuts on the electrons,
hard photons and soft photons in their job option file. 
Precisions are not given by authors for tau-charm 
energies,  but they are
the most precise generators we have for Bhabha processes.
\item Babayaga~\cite{part1:babayaga} is a 
Monte Carlo event generator for
$e^+e^-\to e^+e^-,~\mu^+\mu^-,~\gamma\gamma$ and $\pi^+\pi^-$ processes 
for energies below 12 GeV. It has a high-precision 
QED calculation of the Bahbha process and is intended 
for precise luminosity determinations for $e^+e^-$ 
$R$ measurements in the the hadronic resonance region. The
calculation is based on the matching of exact next-to-leading order
corrections with a parton shower algorithm. The accuracy of the
approach is demonstrated by a comparison with existing independent
calculations and through a detailed analysis of the main components
of theoretical uncertainty, including two-loop corrections, hadronic
vacuum polarization and light pair contributions. The theoretical
accuracy of Babayaga is quoted as 0.1\%~\cite{part1:bbyaversion}. The
current version of BABAYAGA used at \bes3 is 
V3.5~\cite{part1:bbyaversion}.

To use the Babayaga generator, the user is required to specify the
cms energy of the $e^+e^-$ system, together with the cuts on the
electron, positron and photons.
\end{itemize}

\subsubsection{Inclusive generators}
The PYTHIA programs are commonly used inclusive event generators
for high-energy  $e^+e^-,~pp$ and $ep$ collisions.
Historically, the family of event generators from the Lund group
started with JETSET in 1978; the PYTHIA program followed a few years
later. The version currently available is PYTHIA 6.4. The code and
further information can be found on the Pythia web 
page~\cite{part1:pythia}.

The Lundcharm model was especially adjusted by BESII
for simulating  $J/\psi$ and $\psi(2S)$ inclusive 
decays. $C-$ and $G-$parity constraints were imposed and comparisons
with experimental data were performed~\cite{part1:cjc}.  As
a result, \bes3  officially decided to
use this modified Lundcharm model to generate $J/\psi$ and
$\psi(2S)$ inclusive decays in the BesEvtGen framework.

An advantage of generating inclusive MC events
with Lundcharm running in the EvtGen 
framework is that the decay widths in the Lundcharm model 
can be controlled by the user. Thus, branching fractions and models for
known decays can be specified in the EvtGen decay dictionary, 
while unknown decays are generated with the Lundcharm model. 
When the Lundcharm model is called, a complete decay chain is 
generated, but only the first step of the $J/\psi$ or $\psi(2S)$ 
decay is read out and returned to the EvtGen interface. This 
interface has the 
functionality to check whether that particular decay is included 
among the EvtGen exclusive decay models. Only decays not included in 
the exclusive decay models are allowed to proceed; the decays of 
subsequent daughter
particles proceed via EvtGen Models.

EvtGen also allows access to the PYTHIA model to generate the
QED inclusive decays with the model "PYCONT".  At tau-charm energies,
the area law of the Lundcharm model should be implemented to
constrain the decays, however this has not yet been done in 
BesEvtGen.

\subsection{Summary and outlook}
We present a general description on the generator framework and
the event generators currently used at \bes3, which include KKMC,
BesEvtGen, Bhlumi, Bhwide, Babayaga and inclusive generators.
A cosmic ray generator, CORSIKA~\cite{part1:corsika}, is being
migrated.  Though some event generators for QED processes, such as
$ e^+e^- \to \mu^+\mu^-$ and $\tau^+\tau^-$ in KKMC, and
$e^+e^- \to\gamma\gamma,~\mu^+\mu^-$ and $\pi^+\pi^-$ in Babayaga, are
available, they still don't satisfy the \bes3 requirements for 
hadronic cross-section measurements. The migration of other 
generators is necessary, for example, MCGPJ ($\pi^+\pi^-,~K^+K^-$ 
and  $K^0_SK^0_L$ are available) and PHOKHARA
($\pi^+\pi^-,~\pi^+\pi^-\pi^0,~\pi^+\pi^-\pi^+\pi^-,~K^+K^-,~K^0_SK^0_L,~p\bar
p,~n\bar n$ and $\Lambda\bar \Lambda$ are available) with precision
levels of $(0.1\sim 0.2)\%$.


\section[Luminosity Measurements at \bes3]{Luminosity Measurements at
\bes3\footnote{By X.~H.~Mo, C.~D.~Fu, K.~ L.~He}}
\label{part1:sec:lumi}

\subsection{Introduction}\label{xct_intd}

With the large data samples that will be collected at \bes3
--- typically a few fb$^{-1}$--- unprecedented statistical precision 
will be achieved in the analyses of many channels.  Thus, 
non-statistical factors and effects will be the limiting factors 
on the precision of many experimental results.
Foremost among these limiting factors
will be the luminosity, which will be an 
input to many precision measurements, including the $\tau$ mass
measurement, $R$-values, $J/\psi$, $\psi^{\prime}$ and $\psi^{\prime 
\prime}$ total widths etc.  For these quantities, the luminosity error 
will directly translate into errors on their measured values. 
Thus, precision  luminosity measurements are a very important
aspect of the \bes3 physics program.

In $e^+e^-$ colliding beam experiments, generic physics analyses
commonly require the precise knowledge of the relative luminosity 
accumulated  on and off a resonance peak  so backgrounds from 
continuum production can be reliably subtracted. 
Some  analyses  make  internal consistency checks 
by dividing the full dataset into independent subsets
of comparable size. Here again there will be strong reliance
on the accuracy and stability of the
relative luminosity measurements. In addition, 
events used to calculate  the
luminosity, such as $e^+e^-$, $\mu^+\mu^-$ and $\gamma\gamma$, 
are interesting in their own right because of their
salient topologies that make them useful for online 
performance monitoring.

\begin{figure}[hbt]
\centerline{\psfig{file=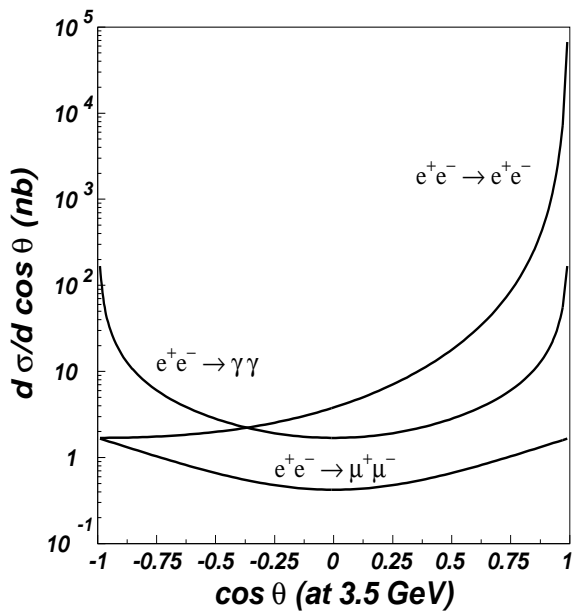,height=8.0cm ,width=7.5cm}}
\caption{\label{qedxct} Differential cross sections for the QED
processes: $e^+e^- \to e^+e^-$, $e^+e^- \to \mu^+\mu^-$, and $e^+e^- \to
\gamma\gamma$. The
center-of-mass energy  is 3.5 GeV.}
\end{figure}

In principle, any known process can be used to
measure luminosity.  However,
in order to obtain precise results, one usually selects a
process that has a characteristic experimental topology 
and a cross section that can be precisely calculated.
The QED processes
$e^+e^- \to e^+e^-$, $e^+e^- \to \mu^+\mu^-$, and
$e^+e^- \to \gamma\gamma$ satisfy these criteria 
and are commonly used for luminosity measurements. Their
differential cross sections, shown in Fig.~\ref{qedxct}, 
have the forms:
$$ \left. \frac{d\sigma}{d\Omega} \right|_{e^+e^- \to e^+e^-}
= \frac{\alpha^2}{4s} \left( \frac{ 3+\cos^2 \theta}{ 1-\cos
\theta}\right)^2~~, $$
$$ \left. \frac{d\sigma}{d\Omega} \right|_{e^+e^- \to \mu^+\mu^- }
= \frac{\alpha^2}{4s} \left( 1+\cos^2 \theta \right)~~, $$
$$ \left. \frac{d\sigma}{d\Omega} \right|_{e^+e^- \to \gamma\gamma }
= \frac{\alpha^2}{4s} \left( \frac{ 1+\cos^2 \theta}{
(E_b/p)^2-\cos^2 \theta} \right)^2~~. $$

The experimental response of the detector to each of these
reactions is quite distinct: their detection
efficiencies rely on  charged particle
tracking ($e^+e^-$ and $\mu^+\mu^-$), calorimetry ($e^+e^-$ and 
$\gamma\gamma$), muon identification ($\mu^+\mu^-$), and trigger 
algorithms. The expected theoretical cross
sections are calculable in QED;  at BEPCII energies weak interaction
effects are negligible. The primary theoretical obstacle in all
cases is the computation of electromagnetic radiative 
corrections in
a way that accommodates the experimental event selection criteria 
with adequate precision. This is usually accomplished by
Monte Carlo event generators that include diagrams with a
varying number of virtual and real radiative photons. In general,
the more accurate the theoretical calculation  the 
smaller the uncertainty on the luminosity measurement.

\begin{figure}[hbt]
\centerline{\psfig{file=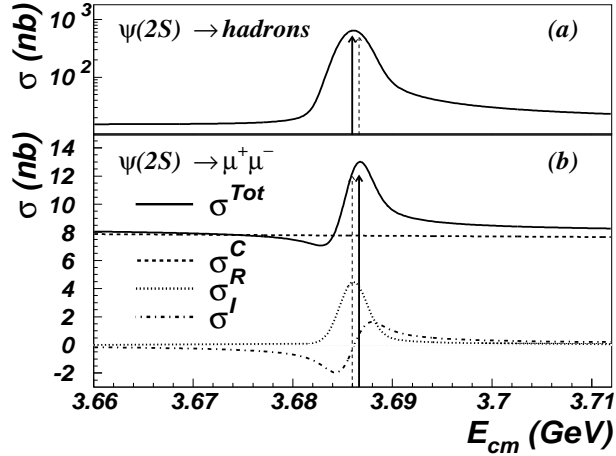,height=6.0cm,width=8.cm}}
\caption{\label{cmphup} Cross sections in the vicinity of $\psi(2S)$
for (a)~inclusive hadrons  and (b)~$\mu^+\mu^-$ (b) final states.
The solid line and arrow indicates the observed peak position
and the dashed line and arrow the actual position of the resonance
peak. In (b), the dashed line indicates the QED continuum~($\sigma^C$),
the dotted line he resonance~($\sigma^R$), the
dash dotted line the interference~($\sigma^I$), and
solid line the total cross section~($\sigma^{Tot}$).}
\end{figure}

Other factors that cannot be neglected are  interference effects
in the vicinity of resonance peaks. Such effects not only distort
the cross sections in the peak regions but 
can also shift the resonance peak position,
as the example shown in Fig.~\ref{cmphup} 
demonstrates. Resonance-continuum cross section ratios 
for the $J/\psi$, $\psi^{\prime}$, and $\psi^{\prime \prime}$ regions are 
listed in Table~\ref{rescon}. From the table we can see
that in the continuum region all three of the QED processes listed above
can be used to determine the luminosity, while only 
$\gamma\gamma$ and $e^+e^-$ can be used for the $\psi^{\prime}$, 
and only $\gamma\gamma$ is suitable for the $J/\psi$, if
high accuracy  is not to be compromised.

\begin{table}[htb]
\caption{\label{rescon}Resonance-continuum cross section ratios
in the $J/\psi$, $\psi^{\prime}$, and $\psi^{\prime \prime}$ 
peak regions.} \center
\begin{tabular}{c|ccc} \hline \hline
Res./Con. &  $J/\psi$   & $\psi^{\prime}$     & $\psi^{\prime \prime}$               \\  \hline
  $\mu^+\mu^-$   &  15.3      &  0.625     & $<1.28\times 10^{-5}$ \\
  $e^+e^-$   &  0.700     &  0.027     & $6.0\times 10^{-5}$   \\
  $\gamma\gamma$   & $<6\times 10^{-3}$
                       & $<5.8\times 10^{-3}$
                            & $<5.8\times 10^{-3}$  \\
\hline \hline
\end{tabular}
\end{table}

\subsection{Event selection and Algorithm}\label{xct_evt}

A detailed Monte Carlo simulation has been performed to develop
the event selection criteria listed in Table~\ref{lumevetsel}.
This is preliminary and is used as an example to study systematic 
errors  and identify what  further studies are needed. 

\begin{table}[htb]
\caption{\label{lumevetsel}Selection criteria for $e^+e^-$, $\gamma\gamma$, and
$\mu^+\mu^-$ final states.}
\center
\begin{tabular}{l|ccc} \hline \hline
      Description         & $e^+e^-$          & $\gamma\gamma$        & $\mu^+\mu^-$        \\ \hline
   \# neutral tracks      &        &  $\ge 2$     &              \\
   \# charged tracks      & $\ge 2$ ($<n$,
                            $n$ decided    &  $\le 1$     &  =2          \\
                          & by detector state) &          &              \\
   $ |\cos\theta|$        & $<0.8$         & $<0.8$       & $<0.8$       \\
Track momentum            & $>0.5E_b$      &              & (0.5-1.15)$E_b$ \\
Track acollinearity   & $<10^{\circ}$  &              & $<10^{\circ}$ \\
$ \cos\theta_1 \times \cos\theta_2\dagger$
                          & $<0.02$        & $<0.001$     & $<0$          \\
Shower Energy             & (0.5-1.1)$E_b$ & $>0.5E_b$    & (0.1-0.35) GeV\\
Shower acoplanarity   & $>8^{\circ}$   & $<2^{\circ}$ & $>5^{\circ}$  \\
Vertex \& TOF             &                &              & $|t1-t2|<3$ ns \\
\hline \hline
\end{tabular}\\
$\dagger$ : $\cos\theta_1$ and $\cos\theta_2$ are the $\cos \theta$
values of the two tracks with the largest momentum; the selection on their 
product  is equivalent to a back-to-back requirement.
\end{table}

\subsection{Systematic uncertainty}\label{part1:xct_sun}

Typical luminosity measurements at BESI and BESII are summarized in
Table~\ref{lumbes}, where the uncertainties are around 2-3\%.
Comparative results from other experiments are collected in
Table~\ref{lumother}. To achieve the goal of the \bes3 experiment
to have better than $1\%$ precision, effects such as backgrounds, 
trigger efficiency, errors in the MC simulation, 
data-taking stability and radiative corrections must
 be considered. 
Since \bes3  has not yet started to take data,  our systematic
study is based on BESII and CLEOc experience
as well as \bes3 simulations.

\begin{table}[htb]
\caption{\label{lumbes}Luminosity measurement uncertainties
using  $e^+e^-$ final states at BESI and BESII. The online 
luminosity is  measured at small angles
while offline luminosity is measured with large angle Bhabha events in the
Barrel Shower Counter.}
\center
\begin{tabular}{c|cccc} \hline \hline
Energy region &  $J/\psi $    & $\psi^{\prime} $      & $\psi^{\prime
\prime} $      &  $R$-value   \\ \hline
Method        &  online      & offline      & offline       & offline      \\
Uncertainty   &  $ 6 \% $    & $3.2 \% $    & $1.83 \% $    &  $< 3 \% $   \\
Reference     &\cite{part1:jbberr} &\cite{part1:jbberr} &\cite{part1:ppbberr} &\cite{part1:rbberr} \\
\hline \hline
\end{tabular}
\end{table}

\begin{table}[htb]
\caption{\label{lumother}Luminosity errors from other experiments}
\center
\begin{tabular}{c|cccc} \hline \hline
Exp. Group &  $E_{c.m.} $ & Mode                & Error    & Ref.  \\ \hline
  CLEO     &  10 GeV      & $e^+e^-$, $\mu^+\mu^-$, $\gamma\gamma$ & $1.0 \%$ & \cite{part1:lumcleo} \\
DA$\Phi$NE &  1-3 GeV     & $e^+e^-$               & $0.6 \%$ & \cite{part1:lumdafne} \\
\hline \hline
\end{tabular}
\end{table}

\subsubsection{Background analysis}


Cosmic-rays usually dominate the background in the $\mu$-pair
sample. Tight track quality requirements minimize this contamination
with almost no loss in efficiency. The remaining cosmic ray background
can be estimated with two independent variables, impact parameter,
{\it i.e.}, the distance of closest approach to the beam-axis in
the plane transverse to the beam, and time-of-fight. At CLEOc the
background estimates determined by these techniques are 
$(0.5\pm0.1)\%$  and $(0.6\pm0.1)\%$,
respectively~\cite{part1:lumcleo}. 
According to a KORALB~\cite{part1:koralb} Monte Carlo study, 
background from $\tau$-pair decays is $0.07\%$, and
from $e^+e^- \mu^+\mu^-$ events~\cite{part1:grteemm} is $<0.002\%$.

The background in the photon-pair event sample
 from Bhabha events where both charged tracks are missed
is estimated to be $(0.1\pm0.1)\%$~\cite{part1:grteemm}.
For Bhabha events, 
$\tau$-pairs contribute 0.03\%, $e^+e^- e^+e^-$ final states
yield $(0.05\pm0.05)\%$ \cite{part1:grteeee}, and cosmic-rays are
estimated to be $(0.1\pm0.1)\%$.
The backgrounds estimated by CLEOc are at the 0.1\% 
level for all three background sources;
this is also what is expected at \bes3.

\subsubsection{Trigger efficiency}\label{sxct_trig}

A two-level filtering technique is applied in the \bes3 data
acquisition system. The first level is the hardware trigger. The
second level  is a software trigger with special  algorithms
developed to select physics events
of different types~\cite{part1:fucdfilter}.

\begin{table}[htb]
\doublerulesep  2 pt \center \caption{\label{trigeff}Trigger
Efficiencies of \bes3 at $\sqrt{s}/2=1.89$ GeV.} \vskip 0.2cm
\begin{tabular}{ccccc} \hline \hline

 Event type &  $e^+e^-$            & $\mu^+\mu^-$              &
$\gamma\gamma$      \\ \hline
 Condition  & charge3           & charge1,2\&3       & common,neutralA\&B \\
 Level-1   &($99.95\pm 0.02)\%$  & $(92.2\pm 0.3)\%$   & $(99.78\pm 0.06)\%$ \\
 Level-2   &($99.999\pm 0.001)\%$& $(98.79\pm 0.06)\%$ & $(99.5\pm 0.3)\%$ \\
 Combined  &($99.95\pm 0.03)\%$  & $(91.1\pm 0.4)\%$   & $(98.9\pm 0.4)\%$ \\  \hline \hline
\end{tabular}
\end{table}
\noindent
The hardware trigger efficiencies are obtained using a method
similar to that of BESII~\cite{part1:trig,part1:trigref} and the online
software trigger efficiencies are given in Ref.~\cite{part1:fucdfilter}. 
The combined
results are summarized in Table~\ref{trigeff}.  A
conservative estimate of the uncertainty due to 
the trigger efficiency
is taken as 0.5\% for all three processes.

\subsubsection{Monte Carlo simulation}

The uncertainty due to Monte Carlo simulation can be analyzed by
comparing the data distributions with those from MC simulation.
Many different methods have been proposed to qualify the
difference~\cite{part1:lumcleo,part1:lumdafne}. We consider the method 
used in Ref.~\cite{part1:lumdafne} to be reasonable and 
will adopt it for use in \bes3.

\subsubsection{ Stability }
Since the detector environment will change during the running 
period, the measured luminosity may not be stable, even after
calibration.  This was checked by CLEO, BES and DA$\Phi$NE and no
obvious influence on the luminosity measurement was observed. Hence, no 
error is assigned for 
this for now.

\subsubsection{Theoretical accuracy}
The theoretical accuracy is actually constrained by the accuracy of the 
ISR calculation.
In the CLEOc analysis, radiative photons are generated
above some photon energy cutoff $k_0=E_{\gamma}/E_b$, and all
diagrams with softer photons are subsumed into the two body final
state. Two generators were used for $e^+e^-$ and $\mu^+\mu^-$ scattering, and
one for $\gamma\gamma$ events. The BK$ee$ program~\cite{part1:berendk} generates
$e^+e^-$ final states with zero or one radiative photon, yielding a
cross section that is accurate to the order of $\alpha^3$. Higher-order
corrections are available in the Bhlumi program~\cite{part1:sjadach},
which uses Yennie-Frautschi-Suura exponentiation to generate
multiple photons per event and yields a cross section with an
accuracy to the order of $\alpha^4ln^2(|t|/m_e^2)$, where $t$ is the typical
momentum-transfer. Similar to BK$ee$, the BK$\gamma\gamma$~\cite{part1:berendka}
and BKJ~\cite{part1:berendkj} Monte Carlo programs generate events with up
to one radiative photon and yields an order-$\alpha^3$ cross section
for $\gamma\gamma$ and $\mu$-pair respectively. Up to three radiative photons
in $\mu^+\mu^-$ events are possible (two from initial state radiation and
one from final state radiation) with FPAIR~\cite{part1:berendv}, which has
an order-$\alpha^4$ cross section accuracy. A photon cutoff of
$k_0=0.01$ is used for BK$ee$, BK$\gamma\gamma$, and BKJ, and $k_0=0.001$ for
Bhlumi and FPAIR. The generators used in the CLEOc analysis and the
predicted cross sections at $E_b=5.29$ GeV are listed in
Table~\ref{grtlum}.

\begin{table}[htb]
\caption{\label{grtlum}Generators for $e^+e^-$, $\gamma\gamma$, and $\mu^+\mu^-$ final
states.} \center
\begin{tabular}{l|ccc} \hline \hline
    Item             & $e^+e^-$          & $\gamma\gamma$          & $\mu^+\mu^-$        \\ \hline
\multicolumn{4}{c}{ CLEOc: cross section at $E_b$=5.29 GeV }
\\ \hline
$\alpha^3$ generator & BK$ee$         & BK$\gamma\gamma$        & BKJ             \\
cross section (nb)   & $8.45\pm 0.02$ &$1.124\pm 0.002$&$0.429\pm 0.001$ \\ \hline
$\alpha^4$ generator & BHLUMI         &                & FPAIR           \\
cross section (nb)   & $8.34\pm 0.02$ & $-$            &$0.427\pm 0.001$ \\ \hline
\multicolumn{4}{c}{ BESII: suitable for any energy region }
\\ \hline
$\alpha^3$ generator &  Radee         & Radgg          & Radmu \\
\hline
\multicolumn{4}{c}{ \bes3: suitable for any energy region }
\\ \hline

$\alpha^4$ generator & \multicolumn{3}{c}{ KKMC, Babayaga } \\
\hline \hline

\multicolumn{4}{c}{ DA$\Phi$NE: test at 1 GeV $<\sqrt{s}<$ 3 GeV }
\\ \hline
$\alpha^3$ generator &  Bhagenf       &                &     \\
                     &$\sigma_{\rm eff}(1.0195\mbox{GeV})=(430.7 \pm 0.3)$ nb &  &   \\ 
                     &  Mcgpj         &                &     \\
                     &  Bhwide        &                &     \\
$\alpha^4$ generator &  Babayaga      &                &     \\
                     &$\sigma_{\rm eff}(1.0195\mbox{GeV})=(431.0 \pm 0.3)$ nb &  &   \\
\hline\hline
\end{tabular}
\end{table}

Another check to ensure the correctness of theoretical calculations
is to compare the results from different generators. At DA$\Phi$NE,
the event generators \texttt{Babayaga}~\cite{part1:Calame2000,part1:Calame2001}
and \texttt{Bhagenf}~\cite{part1:Drago1997}, both developed for large 
angle Bhabha scatterings and based on the cross section calculation
in Ref.~\cite{part1:berendk}, have been 
evaluated, as well as their systematic uncertainties. The cross
sections calculated with the two generators, including the event
reconstruction efficiency, are listed in Table~\ref{grtlum}.

The error given in the cross section is mainly due to the
Monte Carlo statistics. The theoretical uncertainty claimed by the
authors is 0.5\% in both cases. The radiative corrections due to the
treatment of initial and final state radiation in \texttt{Bhagenf}
and \texttt{Babayaga} have been compared with two other event
generators: \texttt{Bhwide}~\cite{part1:bhwide} developed for LEP
and \texttt{Mcgpj}~\cite{part1:Fedotovich:2003ww} developed for VEPP-2M,
which are all based on the cross section calculated in
Ref.~\cite{part1:Arbuzov:1997pj}. In Ref.~\cite{part1:KLOEnote} a detailed
comparison has been performed and the agreement is within 0.5\%,
supporting the claims from the authors of the \texttt{Bhagenf}
and \texttt{Babayaga} generators.

At BESII, generators with cross section accuracy
up to order-$\alpha^3$ are used. At \bes3, a generic Monte Carlo generator
KKMC~\cite{part1:kkmcpaper} has been adopted; this can provide cross
sections with an accuracy up to order-$\alpha^4$. Further checks,
such as the shape of the photon energy spectrum, can be made when 
data are available.

\subsection{Summary}\label{part1:xct_sum}

Table~\ref{errlum} summarizes the errors in the luminosity
measurement at CLEO~\cite{part1:lumcleo} and estimated errors for \bes3.
Improvements at \bes3 includes:
\begin{itemize}
\item Statistics : the 0.2\% statistical uncertainty corresponds to 
250,000 Monte Carlo events:  1,000,000 events will accommodate a
statistical uncertainty at the level of 0.1\%, which can be easily 
realized at \bes3;
\item Backgrounds : The same level of background as seen at CLEO, namely 
0.1\%, is expected at \bes3;
\item Trigger efficiency : as discussed in Sect.~\ref{sxct_trig}, this
uncertainty can be conservatively taken to be 0.5\%; 
\item consistency between data and Monte Carlo: 1.0 \% ;
\item radiative corrections : 0.5\%.
\end{itemize}

\begin{table}[htb]
\caption{\label{errlum}Relative Error (\%) in luminosity for
CLEO and \bes3.}
\center
\begin{tabular}{c|ccc|ccc} \hline \hline
Exp.Group             &\multicolumn{3}{c|}{CLEO}& \multicolumn{3}{c}{BES} \\ \hline
Source                &$e^+e^-$  &$\gamma\gamma$   &$\mu^+\mu^-$  &$e^+e^-$
&$\gamma\gamma$
&$\mu^+\mu^-$  \\ \hline
Monte Carlo Statistic & 0. 2  &  0. 2  & 0. 2  & 0. 1  &  0. 1  & 0. 1  \\
Backgrounds           & 0. 1  &  0. 1  & 0. 1  & 0. 1  &  0. 1  & 0. 1  \\
Trigger Efficiency    & 0. 5  &  0. 1  & 1. 3  & 0. 5  &  0. 5  & 0. 5  \\
Detector Modeling     & 1. 1  &  0. 9  & 1. 4  & 1. 0  &  1. 0  & 1. 0  \\
Radiative Corrections & 1. 3  &  1. 3  & 1. 0  & 0. 5  &  0. 5  & 0. 5  \\
Summed in Quadrature  & 1. 8  &  1. 6  & 2. 2  & 1. 3  &  1. 3  & 1.
3
\\ \hline
combine               &\multicolumn{3}{c|}{1.1\%}& \multicolumn{3}{c}{0.8\%} \\
\hline \hline
\end{tabular}
\end{table}

\section[Particle Identification]{Particle Identification\footnote{K.~L.~He,
J.~F.~Hu, B.~Huang, G.~Qin, S.~S.~Sun, Y.~H.~Zheng}}
\label{sec:pid}

\subsection{Introduction}
\label{sec:pid:introduction}

Particle identification (PID) will play an essential role in most
\bes3 physics analyses~\cite{ref:bes3_det_tdr}. Good $\mu/\pi$
separation is required for precision $f_{D}/f_{D_{s}}$ measurements.
Excellent electron identification will help to improve the 
accracy of the CKM elements $|V_{\textrm{cs}}|$ and $|V_{\textrm{cd}}|$. 
The identification of hadronic ($\pi/K/p$) particles 
will be a commonly used tool in \bes3 physics analysis, and often 
the most crucial item to be considered.

Each part of the \bes3 detector executes its 
specific functions and
provides information that can be used to determine the
particle identity.  The PID capabilities 
are quite different for each sub-detector 
and for each different momentum range. To
improve the PID performance, a powerful technique is required to
combine the available information in the most optimal way, 
especially when  some of the inputs are highly correlated. 
In recent years, various PID algorithms, such as the likelihood
method~\cite{ref:pid_likelihood}, the Fisher
discriminator~\cite{ref:pid_kfd}, the H-Matrix
estimator~\cite{ref:pid_hmatrix}, artificial neural
networks~\cite{ref:pid_ann}, and the boosted decision
tree~\cite{ref:pid_bdt}, have been developed. 
Most of the PID requirements in the \bes3
physics program involve high quality
 $e/\pi$, $\mu/\pi$ and $\pi/K$ separation.

\subsection{The PID system of \bes3}
\label{sec:pid_sys}

The \bes3 detector~\cite{ref:bes3_det_tdr} consists of
a main drift chamber (MDC), Time-Of-Flight (TOF) counters, a CsI(Tl)
crystal calorimeter, and a muon identifier. All of them contribute 
to particle identification. 

\subsubsection{The $dE/dx$ measurements}
\label{sec:pid_dedx}

The main drift chamber (MDC) consists of 43 layers of sensitive
wires and operates with a $60\%/40\%$
$\textrm{He}/\textrm{C}_{3}\textrm{H}_{8}$ gas mixture. The expected
momentum resolution $\sigma_{p}/p$ is about 0.5\% at 1 GeV/c. The
energy loss in the drift chamber provides valuable information
for particle identification. The normalized pulse height, which is
proportional to the energy loss of incident particles in the drift
chamber, is a function of $\beta\gamma=p/m$, where $p$ and $m$ are
the momentum and mass of a charged particle.
\figurename~\ref{fig:tof_dedx}(a) shows the normalized pulse height
variation with momentum for different particle species. From the figure,
it is evident that electrons,
muons and pions with momenta around 0.2~GeV/c cannot be well 
separated  by $dE/dx$ pulse height measurements. 
Similarly, the $dE/dx$ pulse heights will not
discriminate electrons from kaons in the 0.5--0.6 GeV/c
momentum range.

There are a number of factors that can effect the $dE/dx$
performance~\cite{ref:bes3_dedx}: the number of hits, the average
path lengths in a cell, space charge and saturation effects,
electric field non-uniformities etc.  After
calibration, the $dE/dx$ measurement resolution is expected to be
between 6\% and 7\%. Using $dE/dx$ information, 3$\sigma$ $K/\pi$ 
separation can be achieved for momentum below 0.6 GeV/c; 
good $e/\pi$ separation can be obtained for all momenta
above 0.4 GeV/c.

\subsubsection{The TOF counter}
\label{pid_tof}

Radially outside of the MDC is the TOF system, which is 
specialized  for particle
identification. It consists of a two-layer barrel array and one
layer endcap array. There are two readout PMTs on each barrel
scintillator and one on each endcap scintillator. The expected time
resolution for the two layers combined 
is between 100 and 110~ps for $K$'s and $\pi$'s,
providing 2$\sigma$ K/$\pi$ separation up to 0.9 GeV/c.

The TOF system measures the flight time of
charged particle. The velocity ($\beta c$) and mass ($m$) of the
charged particle can be calculated from
\begin{equation}
\displaystyle\beta = \frac{L}{c\times\tmea},\quad
m^{2}=p^{2}\times\frac{1-\beta^{2}}{\beta^{2}}, \label{eq:tof_m2}
\end{equation}
where $\tmea$ is the measured time-of-flight, $L$ and $p$ are the
corresponding flight path and momentum of the charged particle given
by MDC measurements, and $c$ is the velocity of light in vacuum.
The typical mass square distributions for electrons, pions, kaons
and protons in different momentum ranges are shown in
\figurename~\ref{fig:tof_dedx}(b).

The PID capability relies on good time resolution ($\sigma_{t}$) of the
TOF system.  $\sigma_{t}$ depends on the pulse height, hit
position, and the beam status.  Usually the value of $\sigma_{t}$
varies for different TOF counters due to different performance 
of the scintillator, PMT, and electronics. Since the TOF
measurements are correlated due to the common event start time, the
weighted time-of-flight for two layers is obtained by a correlation
analysis discussed below and in in Ref.~\cite{ref:bes3_tof}.

\BegFig
\includegraphics[width=14cm]{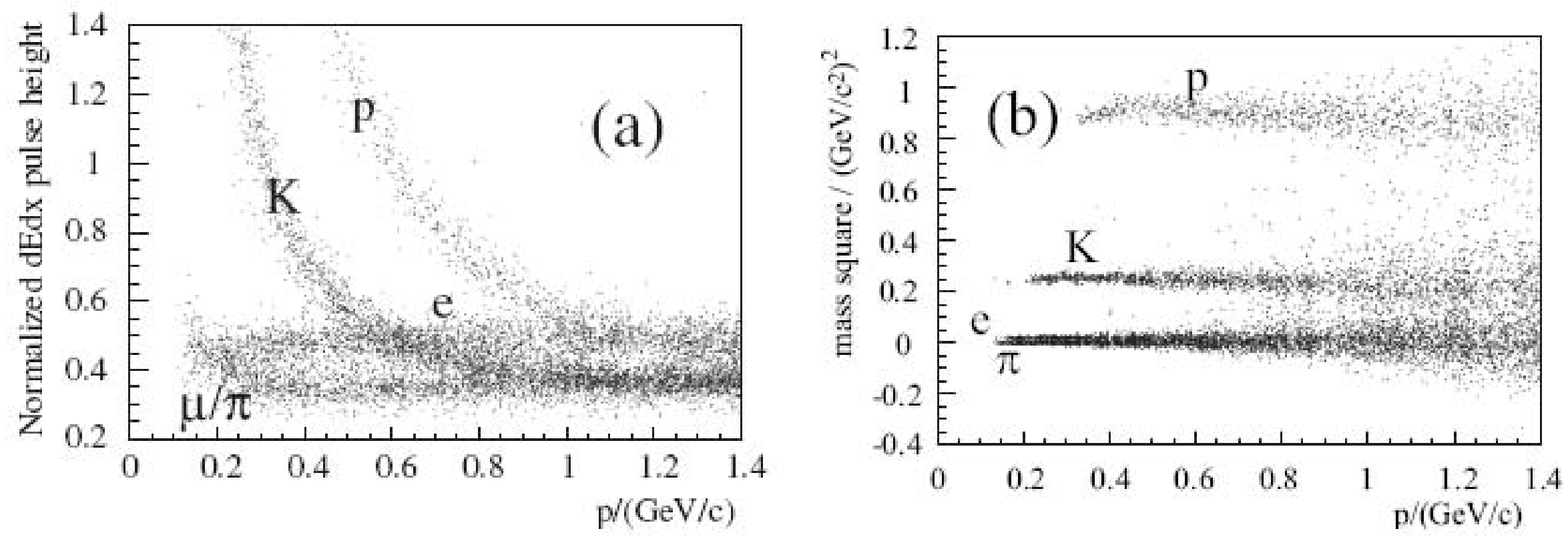}
\caption{(a) The normalized pulse heights ($dE/dx$) vs. momentum of
charged particles; (b) The mass square distribution from TOF
measurements.} \label{fig:tof_dedx} \EndFig

\subsubsection{The CsI(Tl) Calorimeter}
\label{sec:pid_emc}

The CsI(Tl) crystal electromagnetic calorimeter (EMC) contains 6240
crystals, and is used to measure the energy of photons precisely.
The energy and spatial resolutions are 2.5\% and 0.6~cm at
1~GeV, respectively. The characteristics of an electromagnetic shower is
distinctive for the electron, muon and hadron, thus the energy deposited 
and the shape of the shower in the calorimeter can be used as 
discrimination variables for PID.

The energy deposited 
by minimum ionizing
charged particles passing at normal incidence
through the EMC crystals without interacting is about 0.165~GeV.
Electrons and positrons lose all of their 
energies in
the calorimeter by producing electromagnetic showers, the ratio
of deposited energy to the track momentum ($E/p$) will be
approximately unity. Sometimes the energy deposited by hadrons will
have an $E/p$ ratio higher than that of the expected by ionization
due to the nuclear interactions in the CsI material.
\figurename~\ref{fig:energy_deposit}(a) shows the energy deposited 
{\it versus}
momentum for $e$, $\mu$ and $\pi$ in the EMC.

\BegFig
\includegraphics[width=14cm]{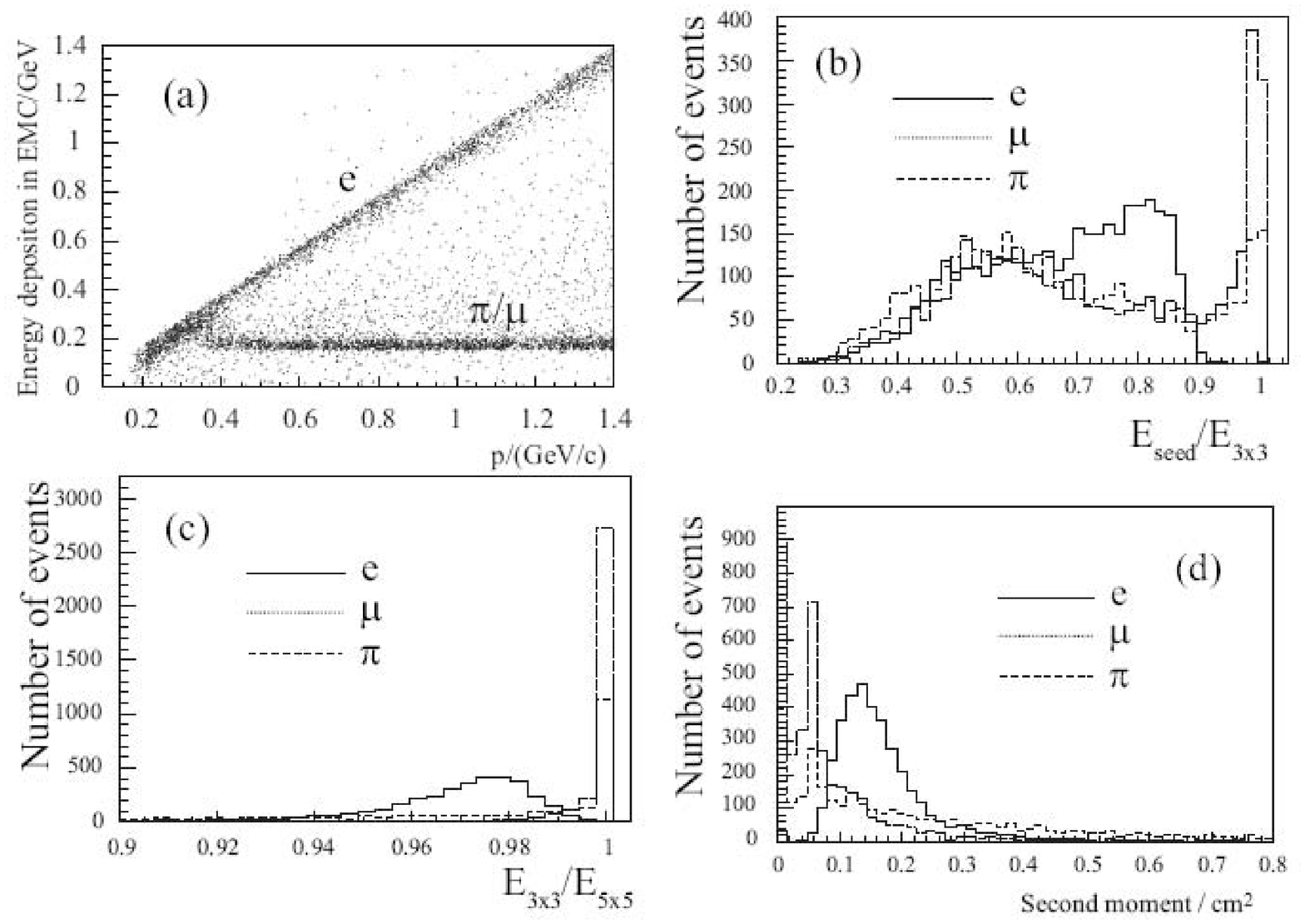}
\caption{(a) Energy deposit in the EMC vs. the momentum for $e$, $\mu$
and $\pi$; (b) ratio of $\eseed/\ethree$ for $e$, $\mu$ and $\pi$;
(c) ratio of $\ethree/\efive$ for $e$, $\mu$ and $\pi$; (d)
second-moment  distributions for $e$, $\mu$ and $\pi$.}
\label{fig:energy_deposit} \EndFig

The ``shape'' of the shower can be characterized by the three energies:
$\eseed$, the energy deposited in the central crystal; $\ethree$,
the energy deposited in the central $3\times3$ crystal array; and
$\efive$, the energy deposited in the central $5\times5$ crystal
array.  The ratios of $\eseed/\ethree$ and $\ethree/\efive$ for $e$,
$\mu$ and $\pi$ at 1 GeV/c are plotted in
\Figs.~\ref{fig:energy_deposit}(b) and~\ref{fig:energy_deposit}(c),
respectively.

The second-moment $S$ is defined as
\begin{equation}
\displaystyle S=\frac{\sum_{i}{E_{i}\cdot
d_{i}^{2}}}{\sum_{i}{E_{i}}},
\end{equation}
where $E_{i}$ is the energy deposited in the  $i^{\rm th}$ crystal, and
$d_{i}$ is the distance between the  $i^{\rm th}$ crystal and the center
position of reconstructed shower. The original idea of $S$ was
developed by the Crystal Ball experiment~\cite{ref:cb_secondmoment}
to distinguish clusters generated by $\pi^{0}$'s and $\gamma$'s. The
$S$ distributions for $e$, $\mu$ and $\pi$ at 1 GeV/c are shown in
\figurename~\ref{fig:energy_deposit}(d).

\subsubsection{The muon system}
\label{sec:pid_muc}

The magnet return yoke has nine layers of Resistive Plate Chambers
(RPC) in the barrel and eight layers in the endcap to form a muon
identifier. The spatial resolution is about 16.6~mm.

\BegFig
\includegraphics[width=6cm]{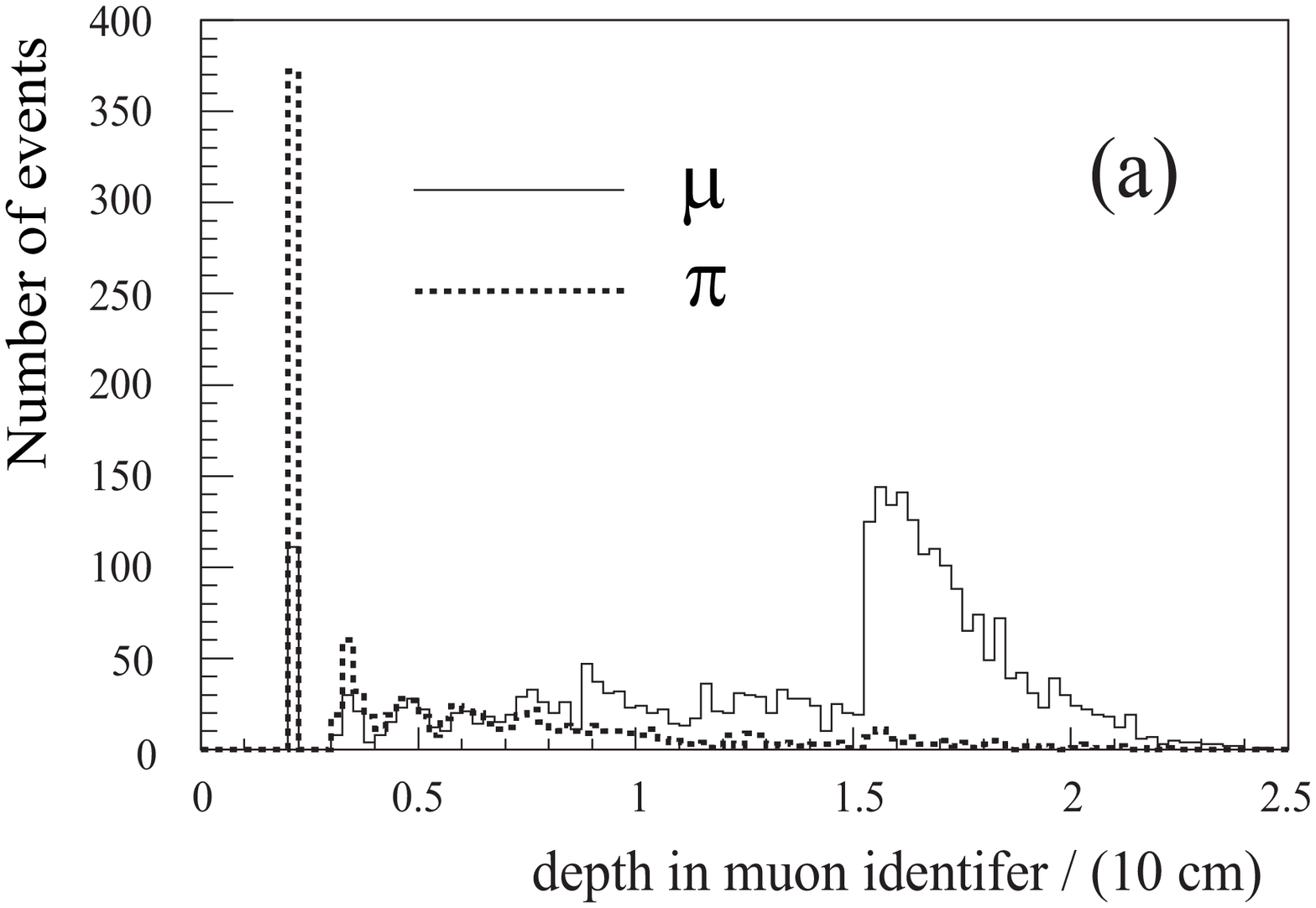}\quad
\includegraphics[width=6cm]{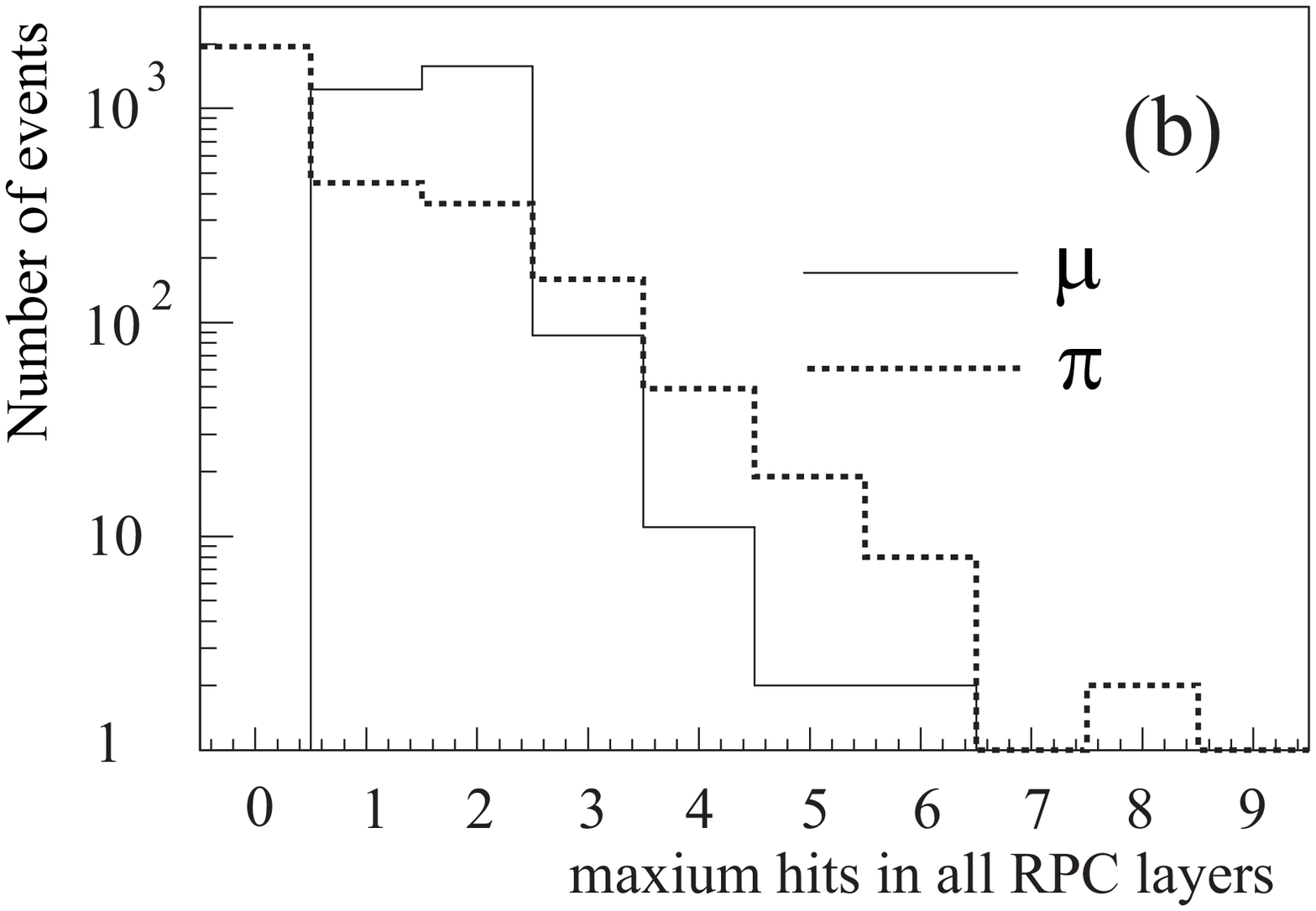}
\caption{(a)The travel depth of $\mu$ and $\pi$ in muon counter; (b)
The maximum number of hits for $\mu$ and $\pi$ in all RPC layers.}
\label{fig:muon vars} \EndFig

An electron's energy is exhausted in the calorimeter and  cannot
reach the muon counter. On the other hand
most of the hadrons pass through the material of
calorimeter and magnet coil, and are absorbed somewhere in the iron yoke.
Muons have a  strong penetrating probability and usually 
produce one hit in each layer. Hadrons can produce many hits in 
the layer near to where an interaction occurs. The distances 
between muon hits and
the extrapolated positions of an MDC track are used to reduce
hadron contamination to a low level, since the hits generated
by secondary muons from  $\pi/K$ decay will not match the
inner track very well. \figurename~\ref{fig:muon vars} shows the
distributions of penetration depth and the maximum number of hits in all
RPC layers for $\mu$'s and charged $\pi$'s with momentum  
in the range 0.8--1.5~GeV/c.

\subsection{The Likelihood Method}

Relative likelihoods (likelihood ratios) provide the most
powerful discrimination between particle identification 
hypotheses, and  the statistical significance
gives a measure of the consistency between data and the selected
hypotheses.

\subsubsection{Probability Density Functions}

The response of a detector to each particle species is given by a
probability density function (PDF), which, written as
${\cal{P}}(x;p, H)$, describes the probability that a particle of
species $H=e^{\pm},\mu^{\pm},\pi^{\pm}, K^{\pm},p,\bar{p}$ leaves a
signature $x$ described by a vector of measurements($dE/dx$, TOF,
$e/p$, ...). ${\cal{P}}(x;p,H)dx$ is the probability for the
detector to respond to a track of momentum $p$ and type $H$ with a
measurement in the range $(x, x+dx)$. As with any PDF, the integral
over all possible values is unity, $\int{\cal{P}}(x;p,H)dx=1$. Note
that the momentum is treated as part of the hypothesis for the PDF
and therefore is placed to the right of semicolon. Drift chamber
momentum measurements are usually of sufficient precision that they
can be treated as a given quantity. In borderline cases when the
precision is marginally sufficient, the PDF is 
sometimes smeared by the assumption
that the momentum is perfectly measured. 

The vector $x$ may describe a single measurement in one detector,
several measurements in one detector, or several measurements in
several detectors. The measurements may be correlated for a single
hypothesis. An example of correlated measurements within a single
device is $E/p$ and the shower shape of electrons in the EMC. An example
of correlated measurements in different detectors is the energy
deposited in the EMC and the muon chambers by charged
pions. In many cases the correlations are reasonably
small and the overall PDF can be determined as a product of the PDFs
for indivdual detectors. For example, the specific ionization
deposited by a charged track as it traverses the drift chamber has
almost no influence on time-of-flight measurements in the TOF.

The challenge of PID analysis is to determine the PDFs and their
correlations (if any) as well as to understand the uncertainties of these
distributions.

\subsubsection{Likelihood}

Given the relevant PDFs, the likelihood that a track with
measurement vector $x$ is a particle of species $H$ is denoted by
${\cal{L}}(H;p,x)$. Although the functional forms of
the PDFs and the
corresponding likelihood function are identical,
the difference between ${\cal{L}}(H;p,x)$ and ${\cal{P}}(x;p, H)$ is
subtle: probability is a function of the measurable quantities $(x)$
for a fixed hypothesis $(p, H)$; likelihood is a function of
particle type $(H)$ for a fixed momentum $p$ and the measured value
$(x)$. Therefore, an observed track for which $x$ has been measured
has a likelihood for each particle type. Competing particle type
hypotheses should be compared using the ratio of their likelihoods.
Other variables having a one-to-one mapping onto the likelihood
ratio are equivalent. Two commonly used mappings of the likelihood
ratios are the difference of log-likelihoods and a normalized likelihood
ratio, sometimes called the likelihood fraction. For example, to
distinguish between the $K^{+}$ and $\pi^{+}$ hypotheses for a track
with measurements $x_{\textrm{obs}}$, these three quantities would
be written as:
\begin{equation}
{\cal{L}}(K^{+};p_{\textrm{obs}},
x_{\textrm{obs}})/{\cal{L}}(\pi^{+};p_{\textrm{obs}},
x_{\textrm{obs}}) \label{eq_likeli_ratio}
\end{equation}
\begin{equation}
\log\left({\cal{L}}(K^{+};p_{\textrm{obs}},
x_{\textrm{obs}})\right)-\log\left({\cal{L}}(\pi^{+};p_{\textrm{obs}},
x_{\textrm{obs}})\right)\label{eq_likeli_log}
\end{equation}
\begin{equation}
\frac{{\cal{L}}(K^{+};p_{\textrm{obs}},
x_{\textrm{obs}})}{{\cal{L}}(K^{+};p_{\textrm{obs}},
x_{\textrm{obs}})+{\cal{L}}(\pi^{+};p_{\textrm{obs}},
x_{\textrm{obs}})}\label{eq_likeli_frac}
\end{equation}
It can be shown rigorously that the likelihood ratio
(Eq.~\eqref{eq_likeli_ratio} and its equivalents
Eq.~\eqref{eq_likeli_log} and Eq.~\eqref{eq_likeli_frac})
discriminate between hypotheses most powerfully. For any particular
cut on the likelihood ratio, there exists no other set of cuts or
selection procedure that gives a higher signal efficiency for the
same background rejection.

There has been an implicit assumption made so far that there is
perfect knowledge of the PDF describing the detector. In the real
world, there are often tails on distributions due to track
confusion, nonlinearities in detector response, and many other
experimental sources that are imperfectly described by the PDFs. 
While deviations from the expected distribution can be determined 
from control samples of real data, the tails of these distributions 
are often associated with fake or badly 
reconstructed tracks. That is why additional consistency
tests should be made. 

\subsubsection{Weighted Likelihood}

In the case (such as particle identification) that the {\it a priori}
probabilities of competing hypotheses are known numbers,
${\cal{P_{A}}}(H)$, likelihood can be used to calculate the expected
purities of given selection criteria. Consider the case of $K/\pi$
separation, the fraction of kaons in a sample with measurement
vector $x$ is given by
\begin{equation}
{\cal{F}}(K;x)=\frac{{\cal{L}}(K;x)\cdot{\cal{P_{A}}}(K)}
{{\cal{L}}(\pi;x)\cdot{\cal{P_{A}(\pi)}}+{\cal{L}}(K;x)\cdot{\cal{P_{A}}}(K)}
\label{eq_wt_likeli}.
\end{equation}
This can be considered as a weighted likelihood ratio where the
weighting factors are {\it a priori} probabilities. The ${\cal{F}}(K;x)$
are also called {\it posteriori} probabilities, 
relative probabilities, or
conditional probabilities, and their calculation according to
Eq.~\eqref{eq_wt_likeli} is an application of Bayes' theorem. The
purity, {\it i.e.}, the fraction of kaons in a sample selected with,
say, ${\cal{F}}(K;x)>0.9$, is determined by calculating the number
of kaons observed in the relevant range of values of ${\cal{F}}$ and
normalizing to the total number of tracks observed there, {\it
e.g.},
\begin{equation}
\displaystyle \textrm{fraction}({\cal{F}}_H>0.9)=\frac
{\int_{0.9}^{1}\frac{dN}{d{\cal{F}}(H;x)}{\cal{F}}(H;x)d{\cal{F}}(H;x)}
{\int_{0.9}^{1}\frac{dN}{d{\cal{F}}(H;x)}d{\cal{F}}(H;x)}  ,
\end{equation}
where the integration variable is the value of ${\cal{F}}(H;x)$.

\subsubsection{An example of TOF and $dE/dx$ PID}

At \bes3, TOF and $dE/dx$ are most essential for hadron separation.
In the TOF detector, the time-of-flight $t$ is measured with
a Gaussian resolution $\sigma_{t}$ that is assumed to be a
constant($\sim 100$~ps). Similarly, the energy loss in the drift chamber
($dE/dx$) also has a Gaussian distribution with a resolution of
$\sigma_{E}\sim6.5\%$. If all incident particles are known to be
either pions, kaons and protons at some fixed momentum, then the
distributions of $t$ and $dE/dx$ will consist of the superposition of
three Gaussian distributions, centered at the central values of 
($t_{\pi}$, $t_{K}$,$t_{p}$ ) and ($(dE/dx)_{\pi}$, $(dE/dx)_{K}$,
$(dE/dx)_{p}$) for pions, kaons and protons. The PDF for the pion
hypothesis is the normalized probability function
\begin{equation}
\begin{matrix}
\displaystyle{\cal{P}}(t;\pi)=\frac{1}{\sqrt{2\pi}\sigma_{t}}
\exp\left[-\frac{1}{2}\left(\frac{t-t_{\pi}}{\sigma_{t}}\right)^{2}\right]\\
\displaystyle{\cal{P}}(\textrm{dE/dx};\pi)=\frac{1}{\sqrt{2\pi}\sigma_{E}}
\exp\left[-\frac{1}{2}\left(\frac{\textrm{dE/dx}-\textrm{(dE/dx)}_{\pi}}{\sigma_{E}}\right)^{2}\right] 
.
\end{matrix}
\end{equation}
The PDFs for the kaon and proton hypotheses have similar forms. Using 
the
observed time of flight $t$ and $dE/dx$ information, the likelihoods
for pion, kaon and proton can be constructed from
\begin{equation}
\begin{matrix}
{\cal{L}}(\pi)={\cal{L}}(\pi;t,\textrm{dE/dx})={\cal{P}}(t;\pi)\cdot{\cal{P}}(\textrm{dE/dx};\pi) 
 ,\\
{\cal{L}}(K)={\cal{L}}(K;t,\textrm{dE/dx})={\cal{P}}(t;K)\cdot{\cal{P}}(\textrm{dE/dx};K) 
 ,\\
{\cal{L}}(p){\cal{L}}(p;t,\textrm{dE/dx})={\cal{P}}(t;p)\cdot{\cal{P}}(\textrm{dE/dx};p)  
.
\end{matrix}
\end{equation}

We consider $K/\pi$ separation in a sample that consists of 80\%
pions and 20\% kaons. Using the observed time of flight $t$ and
energy loss in the drift chamber, it is possible to calculate the
relative probabilities of pions and kaons for these measured $t$ and 
$dE/dx$ values:
\begin{equation}
\begin{matrix}
 \displaystyle {\cal{F}}(\pi)=
\frac{{\cal{P_A}}(\pi){\cal{L}}(\pi)}{{\cal{P_A}}(\pi){\cal{L}}(\pi)+{\cal{P_A}}(K){\cal{L}}(K)} 
,\\
 \displaystyle {\cal{F}}(K)=
\frac{{\cal{P_A}}(K){\cal{L}}(K)}{{\cal{P_A}}(\pi){\cal{L}}(\pi)+{\cal{P_A}}(K){\cal{L}}(K)} 
.
\end{matrix}
\end{equation}
By  construction, ${\cal{F}}(\pi)+{\cal{F}}(K)=1$.  The
calculation of relative probabilities are illustrated in
\figurename~\ref{fig_lh_tofpid}. As shown in
\figurename~\ref{fig_lh_tofpid}, the $K/\pi$ separation at 0.6 GeV
is better than it is at 1 GeV. \BegFig
\includegraphics[width=14cm]{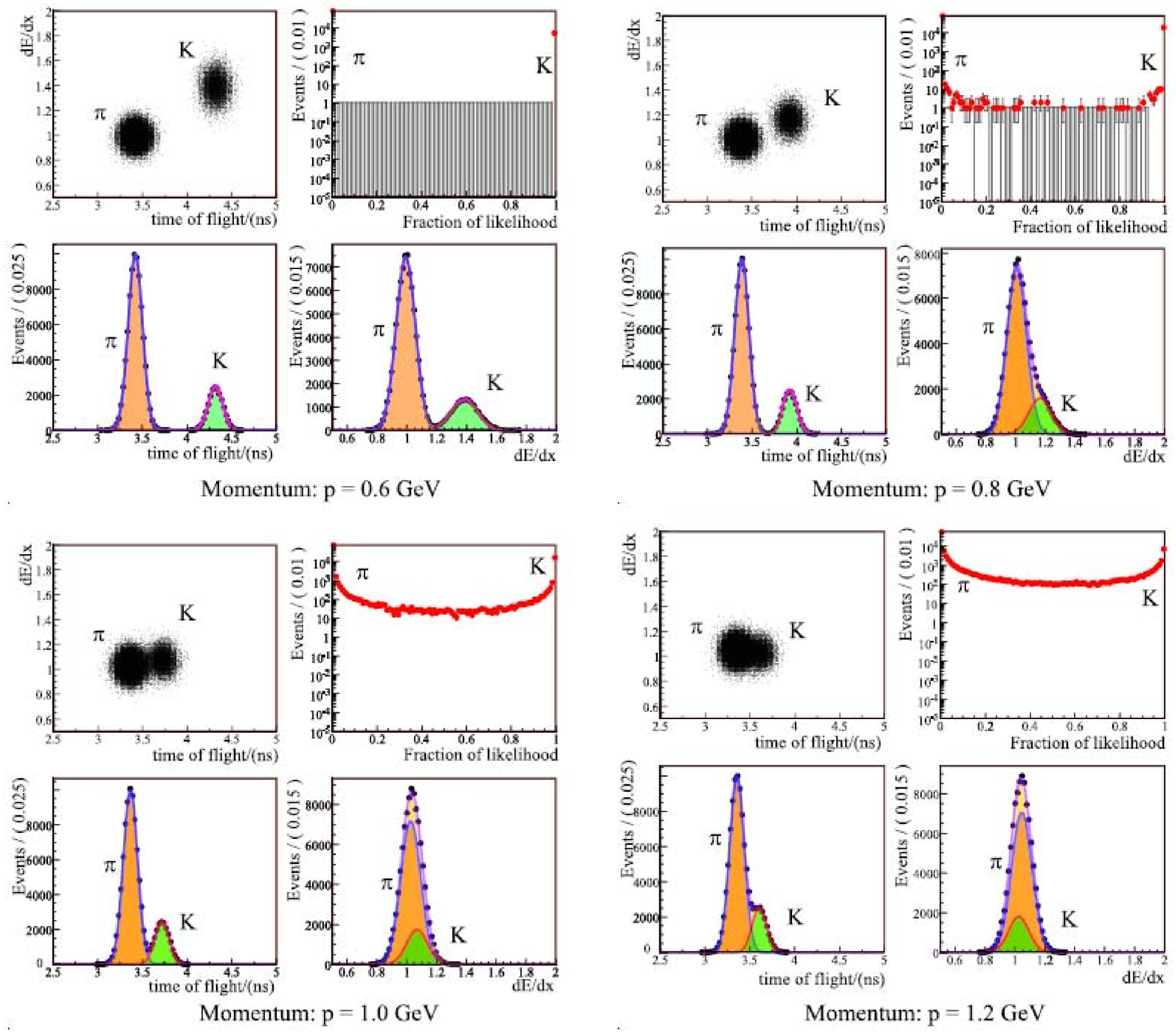}
\caption{The relative likelihood constructed by combining the TOF
and $dE/dx$ information for track momenta of 0.6, 0.8 , 1.0 and
1.2 GeV/c. The time of flight distribution is calculated for a 1.0 m
flight distance.} \label{fig_lh_tofpid} \EndFig

\subsubsection{Cell analysis}

In the example presented above no correlations are assumed
between the particle identification provided by TOF and that
provided by $dE/dx$. This is an acceptable approach if 
the TOF is a purely
passive detector and there are no other sources of correlation. An
approach that takes into account all correlations explicitly is cell
analysis. Basically, you make a multi-dimensional histogram of all
relevant variables and compute the fraction of tracks that land in
each cell for each hypothesis. You can then use these fractions as
the likelihood. The result is optimal with all correlations
completely accounted for, if the cells are small enough.

The trouble with this approach is that as the number of variables
becomes larger, the number of cells quickly gets out of control. It
becomes impossible to find enough ``training events" to map out the cell
distributions with adequate statistics. Still, it is a viable
approach for a small number of variables and is well suited to a
problem such as combining $E/p$ and event shape in calorimeter. This
would, in principle, involve three variables: $E/p$, shape, and 
the dip angle,
and one might get by with relatively large cells. A judicious choice
of cells that uses our knowledge of the underlying physics can
greatly reduce the number of cells needed. {\it e.g.}, the dip angle
might be eliminated as a variable if a dip-corrected shape variable
could be
devised. If groups of highly correlated variables can be treated
together, we might be able to construct a set of relatively
uncorrelated likelihoods. It may be necessary to combine information
from several detectors to construct some of these variables.

\subsection{A correlation analysis of TOF PID}
\label{sec:corr_tofpid}

A charged particle passing through the barrel array of
scintillators  will produce signals in one or two layers of the TOF
counter, corresponding to two or four 
time-of-flight measurements. 
However, at \bes3 the problem of averaging 
more
than one TOF measurement is complicated because the different
measurements are correlated due to the common event start time. A better
choice would be a weighted average of the different measurements.

\subsubsection{General algorithm}

For the covariance matrix elements given by
$(V_{t})_{ij}=\langle\delta t_{i}\delta t_{j}\rangle$, where $\delta
t_{i} = t_{i}-\overline{t}$, $\overline{t}$ is the average of
$t_{i}$,  the definition of the standard deviation is 
\begin{equation}
\sigma^{2}_{\overline{t}}=\sum_{ij}{w_{i}w_{j}(V_{t})_{ij}},
\label{eq:sigma_weight}
\end{equation}
\noindent
where $\overline{t}=\sum_{i}w_{i}t_{i}$ and $\sum_{i}{w_{i}}=1$.
Using standard Lagrange multiplier techniques, we obtain: 
\begin{equation}
\displaystyle
w_{i}=\frac{\sum_{k}(V_{t}^{-1})_{ik}}{\sum_{jk}(V_{t}^{-1})_{jk}}.
\label{eq:weight_solution}
\end{equation}

\subsubsection{Errors and correlations of TOF measurements}

The TOF time resolution ($\sigma_t$) can be factorized into
the product $\sigma_{t}(Q)\cdot\sigma_{t}(z)$, where
$\sigma_{t}(Q)$ and $\sigma_{t}(z)$ are functions of the pulse
height $Q$ and the hit position $z$~\cite{ref:tof_correction}. The
function $\sigma_{t}(Q)$ is complicated, and 
needs further study based on 
on real data. On the other hand, the $z$-dependent time
resolution $\sigma_{t}(z)$ is known and 
similar for electrons, muons and
hadrons~\cite{ref:tof_correction}.
\figurename~\ref{fig:z_error_onelayer}(a) shows a typical variation
of $\sigma_{t}(z)$ from one readout unit as a function of $z$ for
Bhabha events. The time resolution becomes poorer when the hit
position is far from the readout end.

For a given barrel TOF counter,
the $\tmea$ in the left-end and the right-end readout units can be
decomposed as
\begin{equation}
  t_{1} = t_{c} + (t_{D})_{1},\quad
  t_{2} = t_{c} + (t_{D})_{2},
\label{eq:t12}
\end{equation}
where $t_{1}$ and $t_{2}$ represent the $\tmea$'s in two readout
PMTs, $t_{c}$ represents the common part of $t_{1}$ and
$t_{2}$ including the common start time, and 
$(t_{D})_{1}$ and $(t_{D})_{2}$ represent the uncorrelated
parts of $t_{1}$ and $t_{2}$.  The covariance matrix for $t_{1}$ and
$t_{2}$ can be expressed as
\begin{equation}
V_{t}=\begin{pmatrix}
\sigma^{2}_{1} &\sigma^{2}_{c} \\
\sigma^{2}_{c} &\sigma^{2}_{2}
\end{pmatrix},
\label{eq:two_end_cov}
\end{equation}
where $\sigma_{1}$ and $\sigma_{2}$ are the time resolution in the
left-end and the right-end readout units, and $\sigma_{c}$ is the
uncertainty in $t_{c}$. 
According to the definition of the covariance
matrix, we have the following expressions
\begin{equation}
\begin{matrix}
\displaystyle \sigma_{1}^{2}=\langle\delta t_{1}\delta
t_{1}\rangle=\langle\delta t_{c}\delta
t_{c}\rangle+\langle\delta(t_{D})_{1}\delta(t_{D})_{1}\rangle,\\
\displaystyle \sigma_{2}^{2}=\langle\delta t_{2}\delta
t_{2}\rangle=\langle\delta t_{c}\delta
t_{c}\rangle+\langle\delta(t_{D})_{2}\delta(t_{D})_{2}\rangle, \\
\displaystyle \sigma_{c}^{2}=\langle\delta t_{1}\delta
t_{2}\rangle=\langle\delta t_{c}\delta t_{c}\rangle,
\end{matrix}
\label{eq:cov_matrix_elem}
\end{equation}
where we have used the fact that the correlations $\langle\delta
t_{c}\delta(t_{D})_{1}\rangle=0$, $\langle\delta
t_{c}\delta(t_{D})_{2}\rangle=0$ and
$\langle\delta(t_{D})_{1}\delta(t_{D})_{2}\rangle\approx 0$.

To get $\sigma_{c}$ conveniently, we define two new time
variables
\begin{equation}
\displaystyle t_{+}=\frac{t_{1}+t_{2}}{2},\quad \displaystyle
t_{-}=\frac{t_{1}-t_{2}}{2}. \label{eq:tpm}
\end{equation}
The fluctuations of $t_{+}$ and $t_{-}$ can be expressed as
\begin{equation}
\begin{matrix}
\displaystyle \sigma_{+}^{2} = \langle\delta t_{+}\delta
t_{+}\rangle =
\frac{\sigma_{1}^{2}+\sigma_{2}^{2}}{4}+\frac{\sigma_{c}^{2}}{2},\\
\displaystyle \sigma_{-}^{2} = \langle\delta t_{-}\delta
t_{-}\rangle=\frac{\sigma_{1}^{2}+\sigma_{2}^{2}}{4}-\frac{\sigma_{c}^{2}}{2},
\end{matrix}
\label{eq:res_tpm}
\end{equation}
where $\sigma_{+}$ and $\sigma_{-}$ are the time resolution of
$t_{+}$ and $t_{-}$. The value of $\sigma_{c}$ can be directly 
extracted as
$\sigma_{c}=\sqrt{\sigma^{2}_{+}-\sigma^{2}_{-}}$.
\figurename~\ref{fig:z_error_onelayer}(b) shows the distributions of
$\sigma_{+}(z)$, $\sigma_{-}(z)$ and $\sigma_{c}(z)$, where
$\sigma_{c}(z)$ is approximately a constant.
Substituting the expression of Eq.~\eqref{eq:two_end_cov}
into Eqs.~\eqref{eq:sigma_weight}$-$\eqref{eq:weight_solution}, we
get
\begin{equation}
\displaystyle
w_{1}=\frac{\sigma^{2}_{2}-\sigma^{2}_{c}}{\sigma^{2}_{1}+\sigma^{2}_{2}-2\sigma^{2}_{c}},\quad
w_{2}=\frac{\sigma^{2}_{1}-\sigma^{2}_{c}}{\sigma^{2}_{1}+\sigma^{2}_{2}-2\sigma^{2}_{c}},
\label{eq:sol_w12}
\end{equation}
and
\begin{equation}
 \displaystyle \sigma^{2}_{\overline{t}} =
\frac{\sigma^{2}_{1}\cdot\sigma^{2}_{2}-\sigma^{4}_{c}}{\sigma^{2}_{1}+\sigma^{2}_{2}-2\sigma^{2}_{c}}.
\label{eq:sol_sigma}
\end{equation}
The resulting $\sigma_{ \overline{t}}$ as a function of $z$ is shown in
\figurename~\ref{fig:z_error_onelayer}.  

\BegFig
\includegraphics[width=9cm]{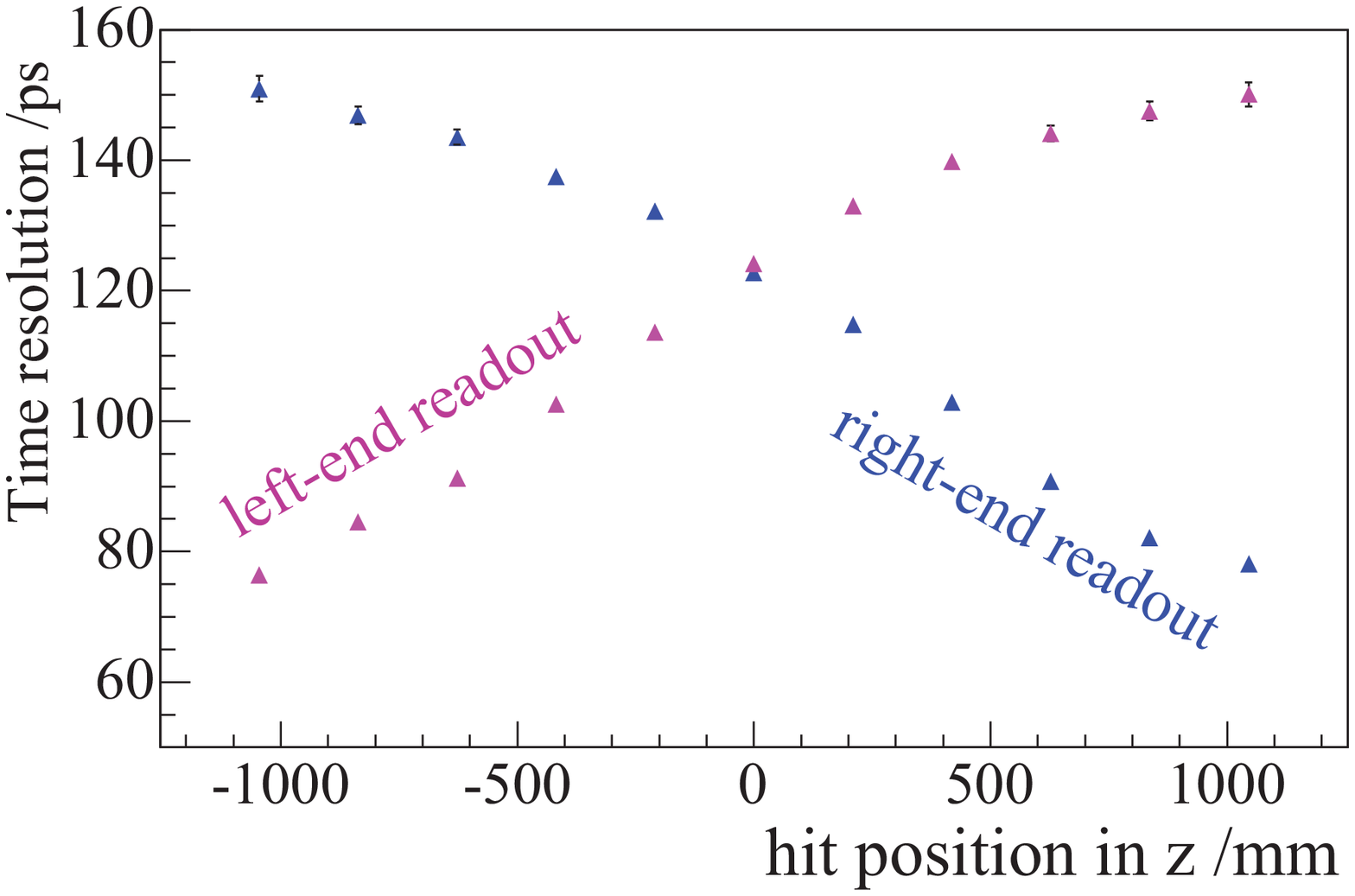}\\
\includegraphics[width=9cm]{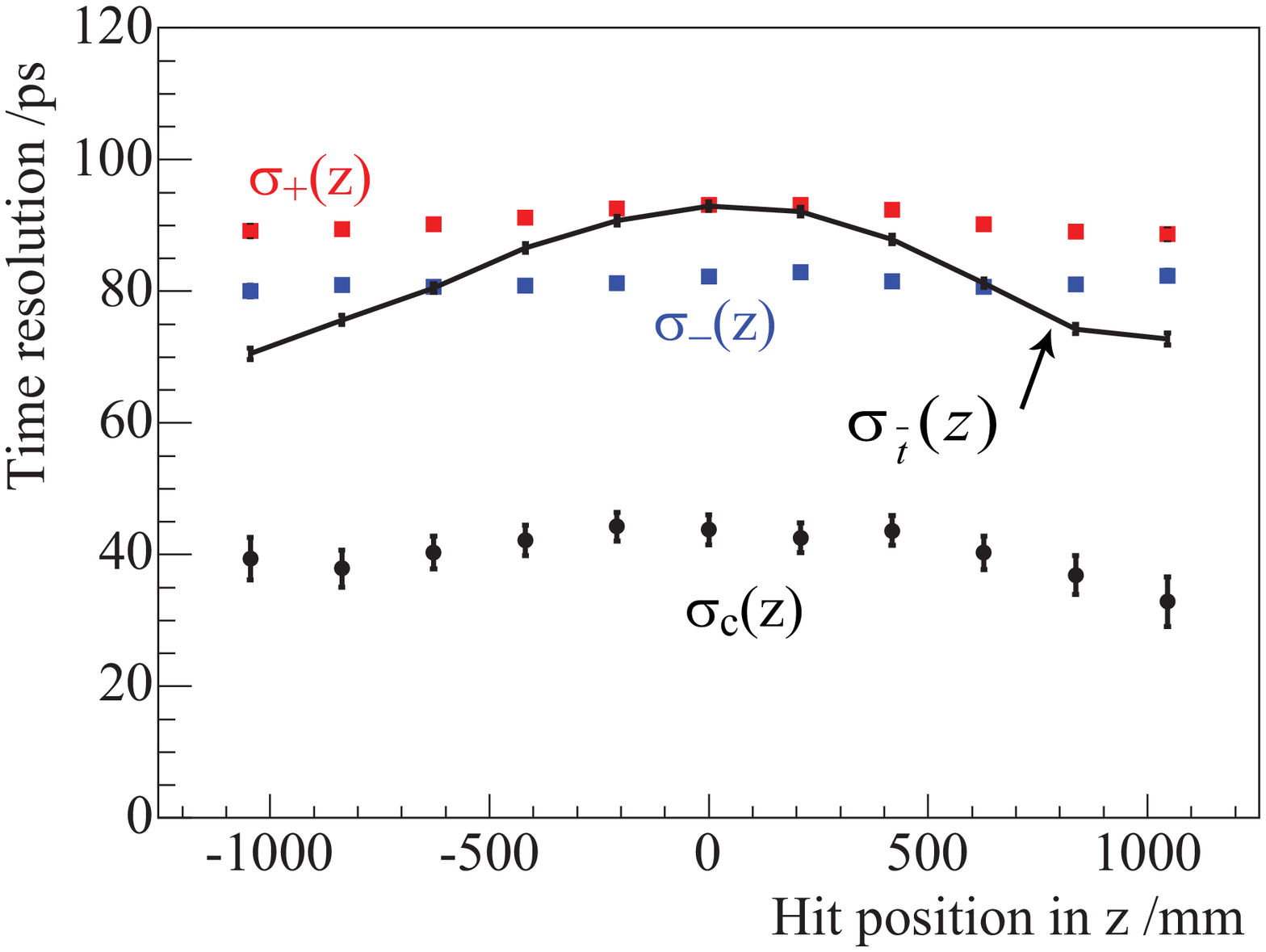}\quad
\includegraphics[width=9cm]{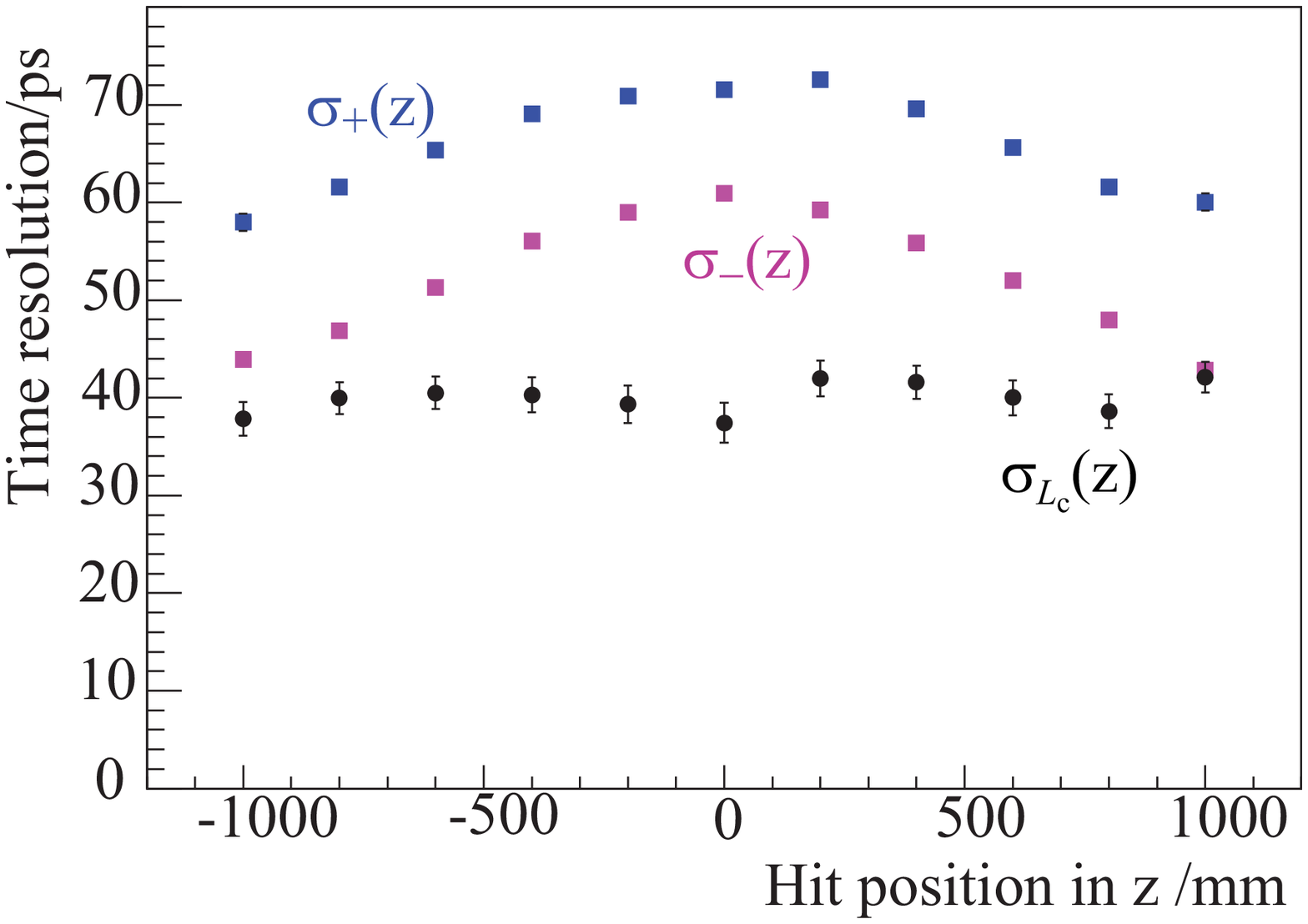}

\caption{(a)~The variation of $\sigma_{t}(z)$ for the left-end and
the right-end readout unit in a barrel TOF counter; (b)~Time resolution
of $t_{+}$, $t_{-}$, $t_{c}$ and the weighted time $\bar{t}$ for a
one-layer TOF measurement; (c)~The correlations between the two 
TOF layer measurements, where
$\sigma_{L_{c}}(z)=\sqrt{\sigma^{2}_{+}(z)-\sigma^{2}_{-}(z)}$.}
\label{fig:z_error_onelayer} \EndFig

\subsubsection{Combining the time-of-flight from two-layer
measurements}

Similar to the method adopted for the one-layer
measurement, we can construct the covariance matrix for
the two-layer
measurement case as follows
\begin{equation}
V_{t}=\begin{pmatrix}
      \sigma^{2}_{L_{1}} &\sigma^{2}_{L_{c}} \\
      \sigma^{2}_{L_{c}} &\sigma^{2}_{L_{2}}
      \end{pmatrix} ,
\label{eq:two_layer_cov}
\end{equation}
where $\sigma_{L_{c}}$ is the correlation between two-layer
measurements.
Substituting $t_{1}$, $t_{2}$ with $t_{L1}$, $t_{L2}$ in
Eqs.~\eqref{eq:t12} and \eqref{eq:tpm}, we get the corresponding
errors and correlations. 
The weighted time-of-flight of two-layer measurements is easily
obtained by applying the covariance matrix of
Eq.~\eqref{eq:two_layer_cov} in Eq.~\eqref{eq:sol_w12}. The
resulting $\tmea-\texp$ are shown in
\figurename~\ref{fig:delt_tof}(a). 

\BegFig
\includegraphics[width=10cm]{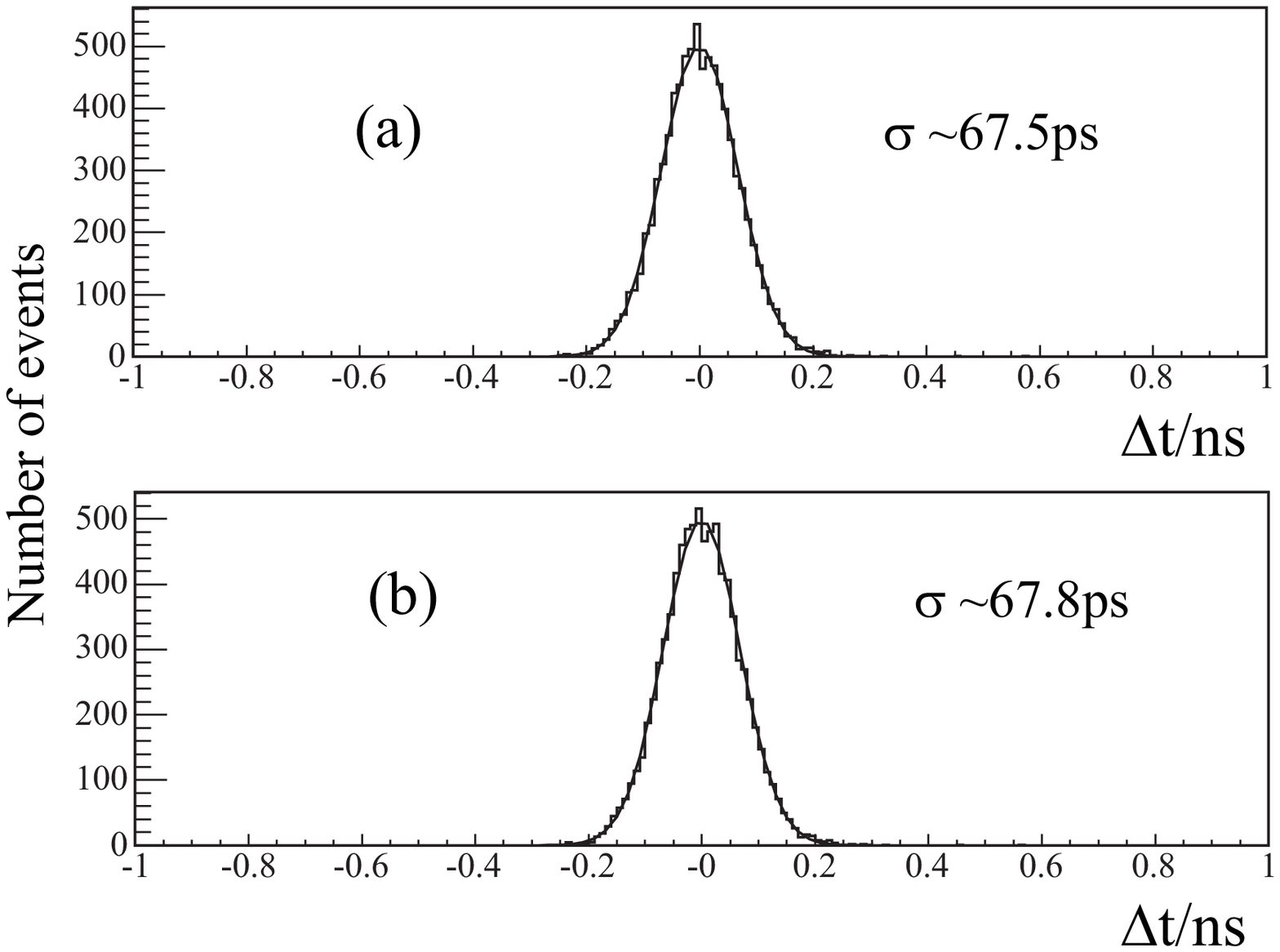}
\caption{$\Delta t=\tmea-\texp$ distributions: (a)~$\bar{t}$ is
weighted by $\bar{t}_{L_{i}}(i=1,2)$, $\bar{t}_{L_{i}}$'s are the
average time in each layer and are weighted by $t_{E_{i}}(i=1,2)$,
$t_{E_{i}}$'s are the TOF measurements in each end of readout units;
(b)~$\bar{t}$ is directly weighted by $t_{E_{i}}(i=1,2,\ldots,4)$,
$t_{E_{i}}$'s are the TOF measurements in the four-end of readout
units.} \label{fig:delt_tof} \EndFig

The apparatus of barrel TOF array can be considered as providing
four independent measurements of the time-of-flight for
a charged particle. The covariance matrix of TOF measurements can be
constructed as
\begin{equation}
V_{t}=\begin{pmatrix}
       \sigma^{2}_{1} &\sigma^{2}_{c} &\sigma^{2}_{c}
       &\sigma^{2}_{c} \\
       \sigma^{2}_{c} &\sigma^{2}_{2} &\sigma^{2}_{c}
       &\sigma^{2}_{c} \\
       \sigma^{2}_{c} &\sigma^{2}_{c} &\sigma^{2}_{3}
       &\sigma^{2}_{c} \\
       \sigma^{2}_{c} &\sigma^{2}_{c} &\sigma^{2}_{c}
       &\sigma^{2}_{4}
       \end{pmatrix}.
\label{eq:four_end_cov}
\end{equation}
In Eq.~\eqref{eq:four_end_cov}, $\sigma_{i}(i=1,2,\ldots,4)$ are the
resolutions of all readout units, the correlations ($\sigma_{c}$)
between the two-end of readout units in each layer,
and the correlation between two-layer measurements are in fact the same.
Employing the covariance matrix Eq.~\eqref{eq:four_end_cov} in
Eqs.~\eqref{eq:sigma_weight}$-$\eqref{eq:weight_solution}, the weight
factors $w_{i}(i=1,2,\ldots,4)$ can be easily calculated. The
resulting $\tmea-\texp$ distribution is shown in
\figurename~\ref{fig:delt_tof}(b).

As shown in \figurename s~\ref{fig:delt_tof}(a) and
\ref{fig:delt_tof}(b), the resulting time resolutions from two
weighting methods are consistent. The standard weighted method
adopted in the TOF calibration/reconstruction software
will be in two steps: combining the two-end TOF measurements in each
layer; calculating the weighted time from the two-layer
measurements.

\subsection{Applying the ANN technique in PID algorithm at \bes3}
\label{bes3_pid}

If the variables are not highly correlated, multiplying  the
likelihood associated with each variable should suffice. If
correlations are simple enough, a change of variables or a cell
analysis may suffice. If the variables are highly correlated, neural
nets and other opaque boxes might construct near-optimal
discrimination variables. The PDFs for the resulting variables can 
be used as the basis for a likelihood analysis. Using the same
formalism for neural network outputs as for conventional likelihood
analyses allows modular design of analysis software with no loss of
information and optimal discrimination between hypotheses.

At present, a class of Multilayer Perceptrons
(MLP)\cite{ref:root_nn} neural network has been applied to 
the \bes3 PID
algorithm and is implemented in ROOT~\cite{ref:root}. The PID
variables described in Section~{\ref{sec:pid_emc}} and
Section~\ref{sec:pid_muc} are correlated each other. With no loss of
information, a cell analysis may not be sufficient for the
likelihood method to get an optimal result. Since the
correlations of PID variables among sub-detectors are reasonably
small and can be ignored, the neural networks can be configured 
sequentially. 

\subsubsection{ Brief description of the artificial neural network  }
\label{sec:ann}

An artificial neural network~\cite{ref:pid_ann} is a computational
structure inspired by the study of biological neural processing.
Feed-forward neural networks, also known as multilayered
perceptrons, are most popular and widely used. The output of a
feed-forward neural network trained by minimizing, for example, the
mean square error function, directly approximates the Bayesian
posterior probability without the need to estimate the
class-conditional probabilities separately. A feed-forward neural
network (NN) is shown schematically in
\figurename~\ref{fig:mlp_neural}. Such networks provide a general
framework for estimating non-linear functional mapping between a set
of input variables {\bf x}($x_{1}$, $x_{2}$, $\ldots$, $x_{N}$) and
an output variable $O({\bf x})$ (or a set of output variables)
without requiring a prior mathematical description of how the output
formally depends on the inputs.

\BegFig
\includegraphics[width=12cm]{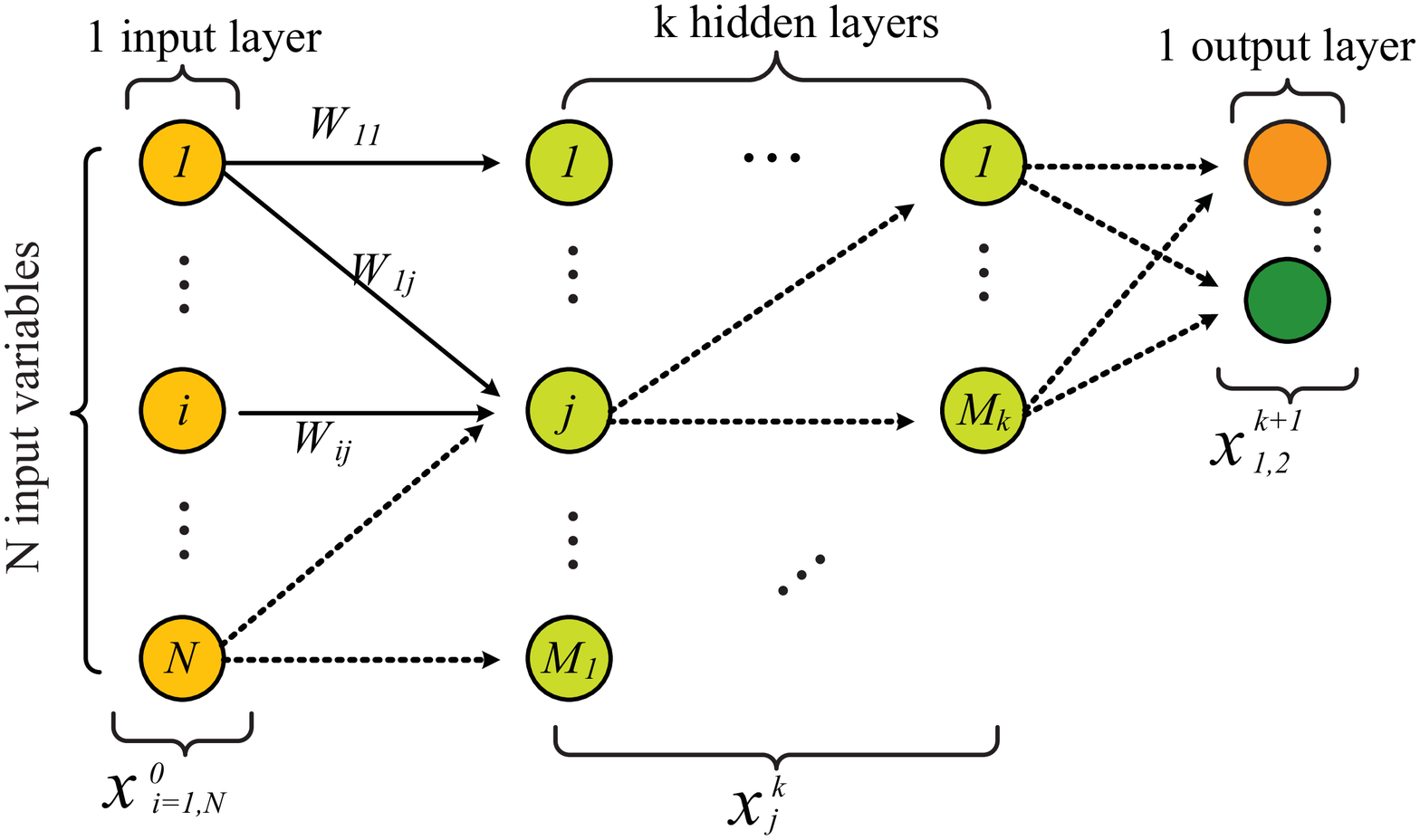}
\caption{The schematic structure of a multilayered perceptrons'
neural network: the input layer contains $N$  neurons as input
variables ($x^{0}_{i=1, 2, \ldots, N}$); the output layer contains
(two) neurons for signal and background event classes; in between
the input and output layers are a number of $k$ hidden layers with
arbitrary number of neurons ($x^{k}_{j=1, 2, \ldots, M_{k}}$). }
\label{fig:mlp_neural} \EndFig

The network is made of neurons characterized by a bias and weighted
links in between, the links are called synapses. A layer of
neurons makes independent computations on the data, and so it
receives and passes the results to another layer. The next layer may
in turn make its independent computations and pass on the results to
yet another layer. Finally, the processed results of the network can
be determined from the output neurons. As indicated in the sketch,
all neuron inputs to a layer are linear combinations of the neuron
output of the previous layer. For a given neuron $j$ in layer $k$,
we have the following equation
\begin{equation}
\displaystyle x^{k}_{j}= A\left(w^{k}_{0j}+
\displaystyle \sum_{i=1}^{M_{k-1}}{w^{k}_{ij}\cdot x^{k-1}_{i}}\right),
\label{eq:nn_transfer}
\end{equation}
where $x^{k-1}_{i}(i=1, 2, \ldots, M_{k-1})$ represents the input
signal from the previous layer $k-1$, $M_{k-1}$ is the total number
of neurons in layer $k-1$, $w^{k}_{ij}$'s represent the synaptic
weights of neuron $j$, the bias term $w^{k}_{0j}$ (not shown in
\figurename~\ref{fig:mlp_neural}) is acquired by adding a new
synapse to neuron $j$ whose input is $x^{k-1}_{0j}=1$. The transfer
from input to output within a neuron is performed by means of an
``activation function'' $A(x)$. In general, the activation function
of a neuron can be zero (deactivated), one (linear), or non-linear.
For a hidden layer, a typical activation function used in
Eq.~\eqref{eq:nn_transfer} is a sigmoid
\begin{equation}
\displaystyle A(x)=\frac{1}{1+e^{-x}}. \label{eq:nn_sigmoid}
\end{equation}
The transfer function of the output layer is usually linear. As a
consequence: a neural network without a hidden layer should give
identical discrimination power as a linear discriminant analysis
like the Fisher discriminator. In case of one hidden layer, the
neural network computes a linear combination of sigmoids.

The number of parameters (the synaptic weights $w^{k}_{ij}$'s in
Eq.~\eqref{eq:nn_transfer}) need to grow only as the complexity of
the problem grows. The parameters are determined by minimizing an
error function, usually the mean square error between the actual
output $O^{p}$ and the desired (target) output $t^{p}$,
\begin{equation}
\displaystyle E=\frac{1}{2N_{p}}\sum_{p=1}^{N}(O^{p}-t^{p})^{2},
\label{rq:nn_error}
\end{equation}
with respect to the parameters. Here $p$ denotes a feature vector or
pattern. The stochastic optimization algorithm used in learning
enables the model to be improved a little bit for each data point in
the training sample. Neural networks provide a very practical tool
because of the relatively small computational times required in
their training. Their fast convergence as well as their robustness in
supervised learning of multilayered perceptrons are due to the efficient
and powerful algorithms that have been developed in recent years.

\subsubsection{The configuration of PID networks}

The PID variables selected from each sub-detector together with the
incident momentum and the transverse momentum have been grouped and
trained separately, each sub-detector (the barrel part and the
endcap part) has one output. In this step, the neural network for each
sub-detector is quite simple. Almost all sub-networks are configured
with one hidden layer containing 2$N$ hidden neurons, where $N$ is
the number of the input neurons. A total of 50,000 single track events for 
each particle species with momentum ranging from 0.1 GeV/c to 1.6 GeV/c
and $-0.83<\cos\theta<0.83$ are trained for this neural network, where
$\theta$ is the incident polar angle. The output values are
constrained to be 1, 2, 3, 4 and 5 for electron, muon, pion, kaon
and proton, respectively. The training results for each sub-detector
are shown in
\Figs~\ref{fig:detector_performance}(a)--\ref{fig:detector_performance}(d).

The muon and pion bands are merged into one single peak (around
2.5) in the $dE/dx$, TOF and EMC outputs. The  EMC
and MUC information is not very applicable
for kaon and proton identification;
the EMC output
does provide some muon- pion discrimination. The MUC output
can separate muons from hadrons quite clearly.

\BegFig
\includegraphics[width=14cm]{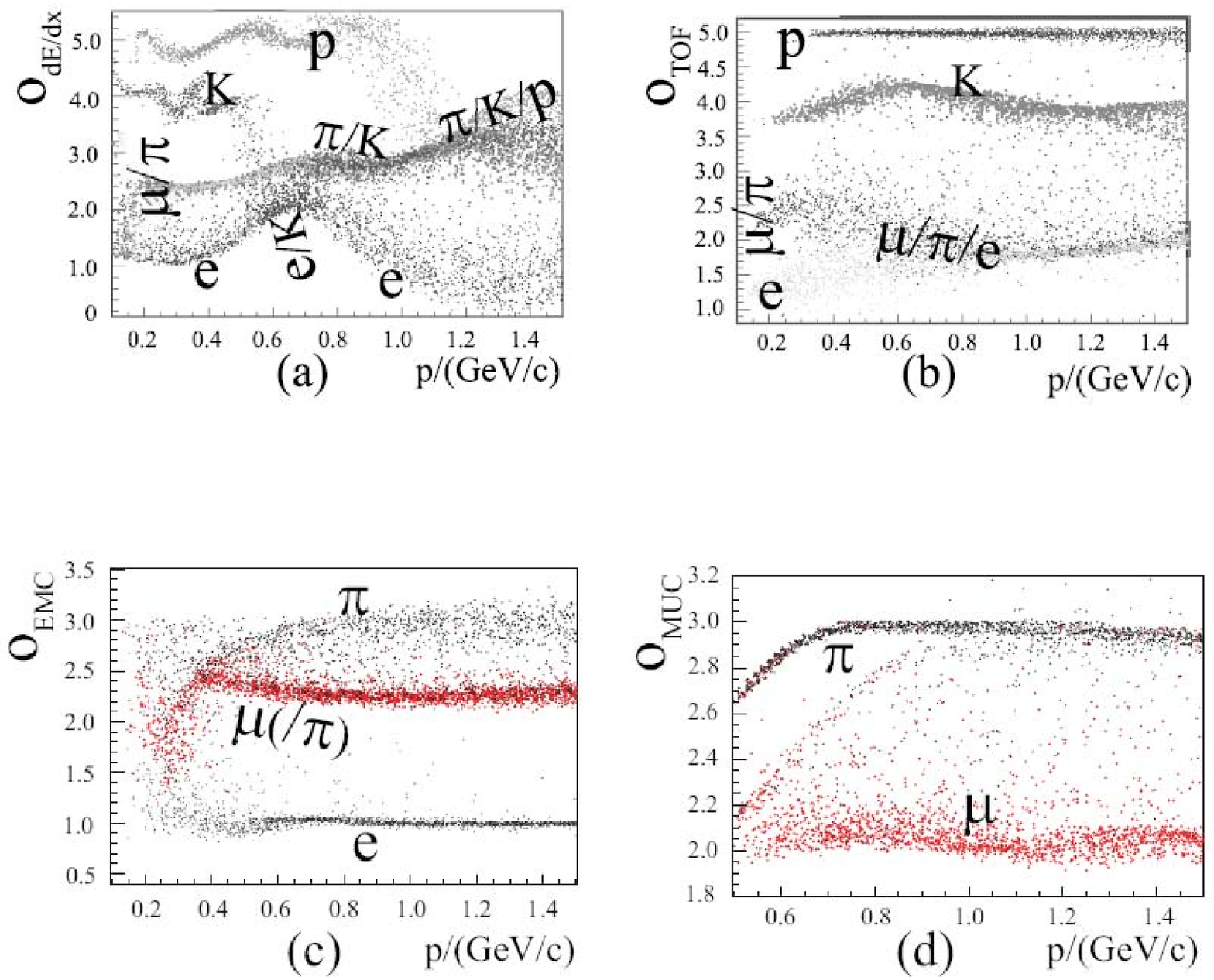}
\caption{NN outputs of sub-detectors. (a)~$dE/dx$ output; (b) TOF
output; (c) EMC output; (d) MUC output.}
\label{fig:detector_performance} \EndFig

The neural network outputs from the sub-detectors can be combined in
several ways to get near-optimal discrimination variables. For
example, the probability density functions (PDF) for the resulting
variables can be used as the basis for a likelihood analysis, or can
be used as the input variables for a sequential network. At present,
a conventional likelihood analysis based on the neural network output
variables and a sequential neural network analysis are applied in parallel
to the \bes3 PID algorithm. The sequential neural networks consists of
two input momentum variables and four input PID variables. The
momentum variables are the incident momentum and the transverse
momentum. The PID variables include the neural outputs from $dE/dx$
($\odedx$), TOF ($\otof$), EMC ($\oemc$) and MUC ($\omuc$) system
(the barrel part and the endcap part separately). The neural network is
configured with one hidden layer of ten hidden neurons.
Electron, muon, and hadron separations are studied with several
simulated Monte Carlo samples through different configurations of
networks. Cuts are put on the output of final discrimination
variables (the output of sequential network $\oseq$).

\subsubsection{Muon identification}

Muon candidates are required to have some response in the 
$\mu-$identifier.
The sequential neural network is trained with two PID variables: the
$\omuc$ and the $\oemc$. The $\mu-$ID abilities are studied in
different momentum partitions by comparing the discrimination
results from $\omuc$ and $\oseq$. \Figs~\ref{fig:pid_eff}(a) and
\ref{fig:pid_eff}(b) show the variations of the muon identification
efficiency and pion contamination rate as functions of incident
track momentum, where the track momentum is required to be greater
than a cut-off threshold ($\sim$500 MeV/c). Above 0.8 GeV/c, the
muon identification efficiency is around 90\%, and the pion
contamination rate is about 5\%. Additional information from the EMC
may help improve the $\mu-$ID ability.

As experienced in the BaBar experiment~\cite{ref:babar_muid}, the
additional variables, {\it e.g.}, the goodness of muon track fit and the
goodness of the muon track matching to the extrapolation position
from inner track system, may help reduce the background
contamination rates. These inputs will be studied at \bes3 in
the future.

\subsubsection{Electron identification}

\figurename~\ref{fig:pid_eff}(c) shows the variations of electron
identification efficiency and pion misidentification rate in
different momentum intervals as a function of cuts on $\oemc$. Above 0.6
GeV/c, one can see that the electron-ID efficiency is greater than
95\% while the pion contamination rate can be as low as $\sim 10^{-3}$.
On the other hand, with  $\oemc$ alone,
the $e/\pi$ separation is quite poor for low momentum 
tracks ({\it i.e.} less than 0.4 GeV/c).

\BegFig
\includegraphics[width=5cm]{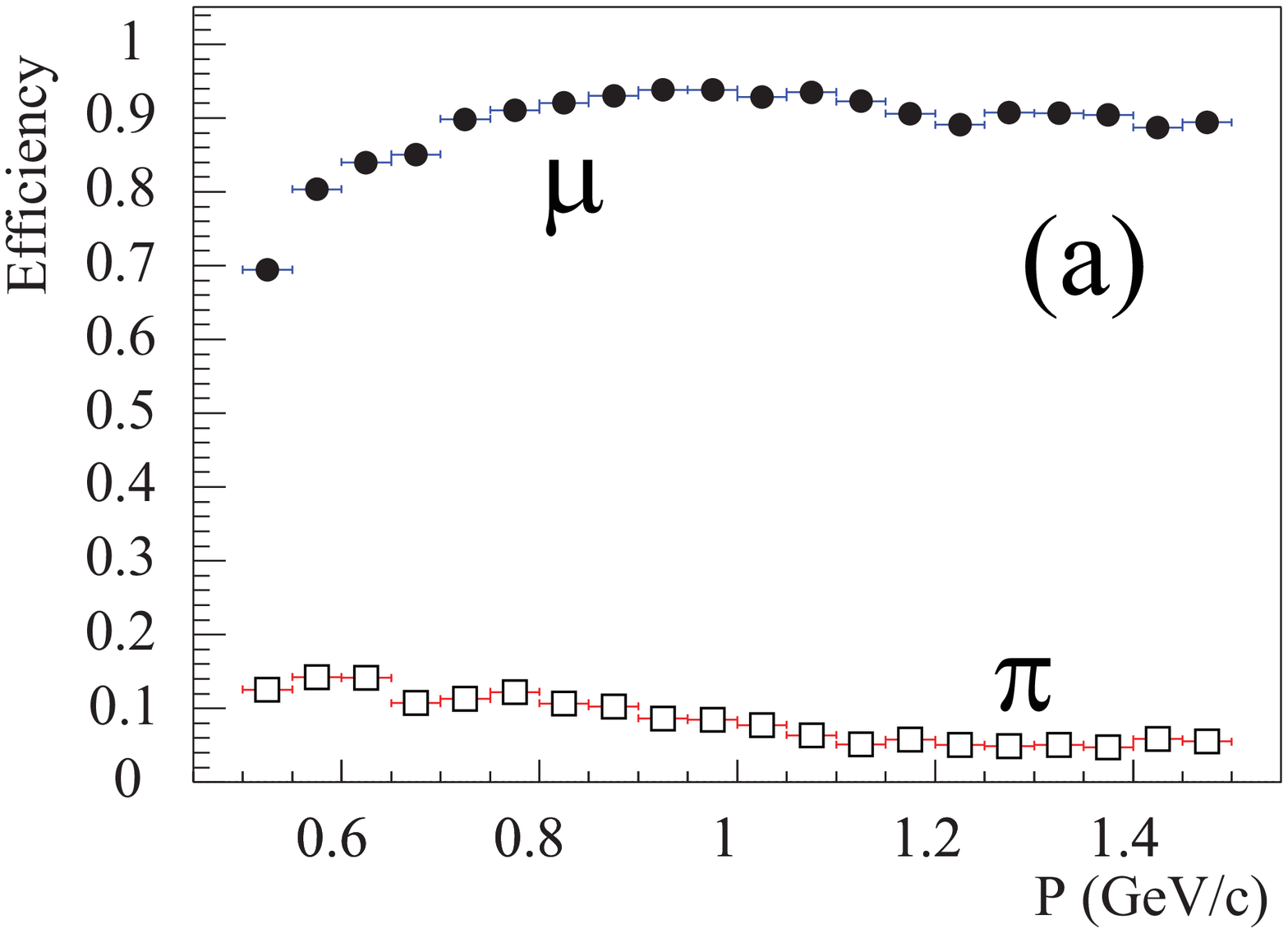}\quad
\includegraphics[width=5cm]{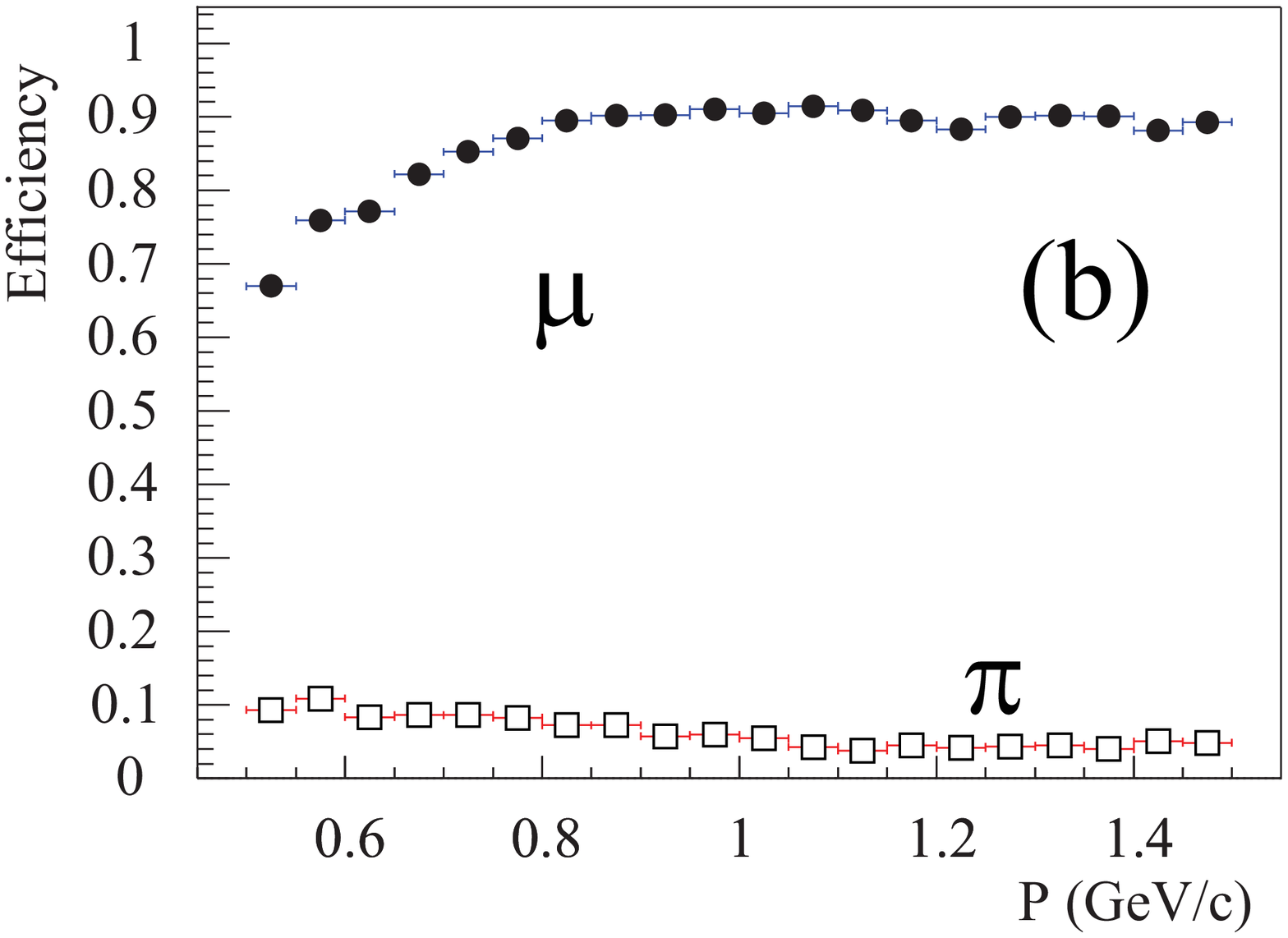} \quad
\includegraphics[width=5cm]{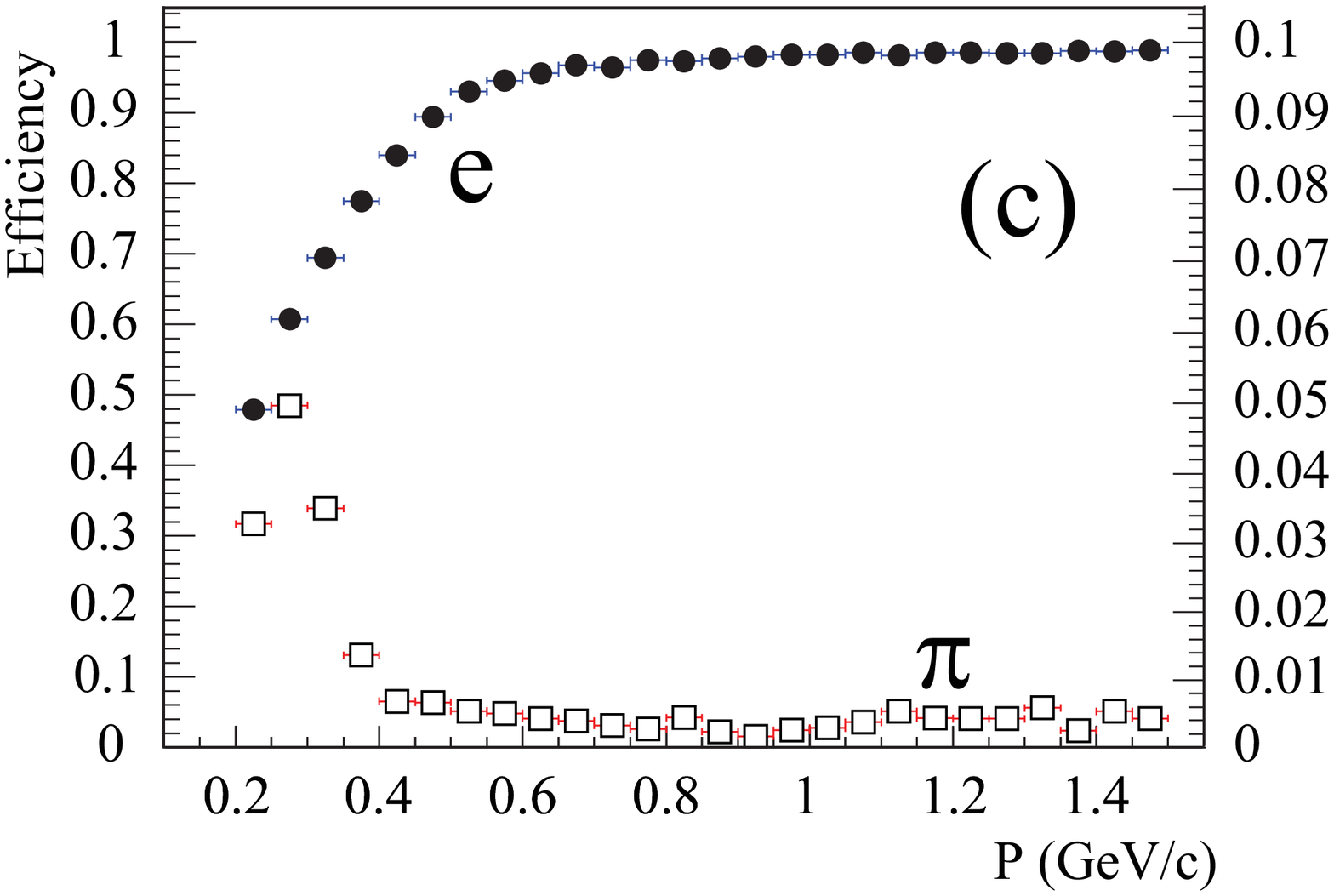}\\
\includegraphics[width=5cm]{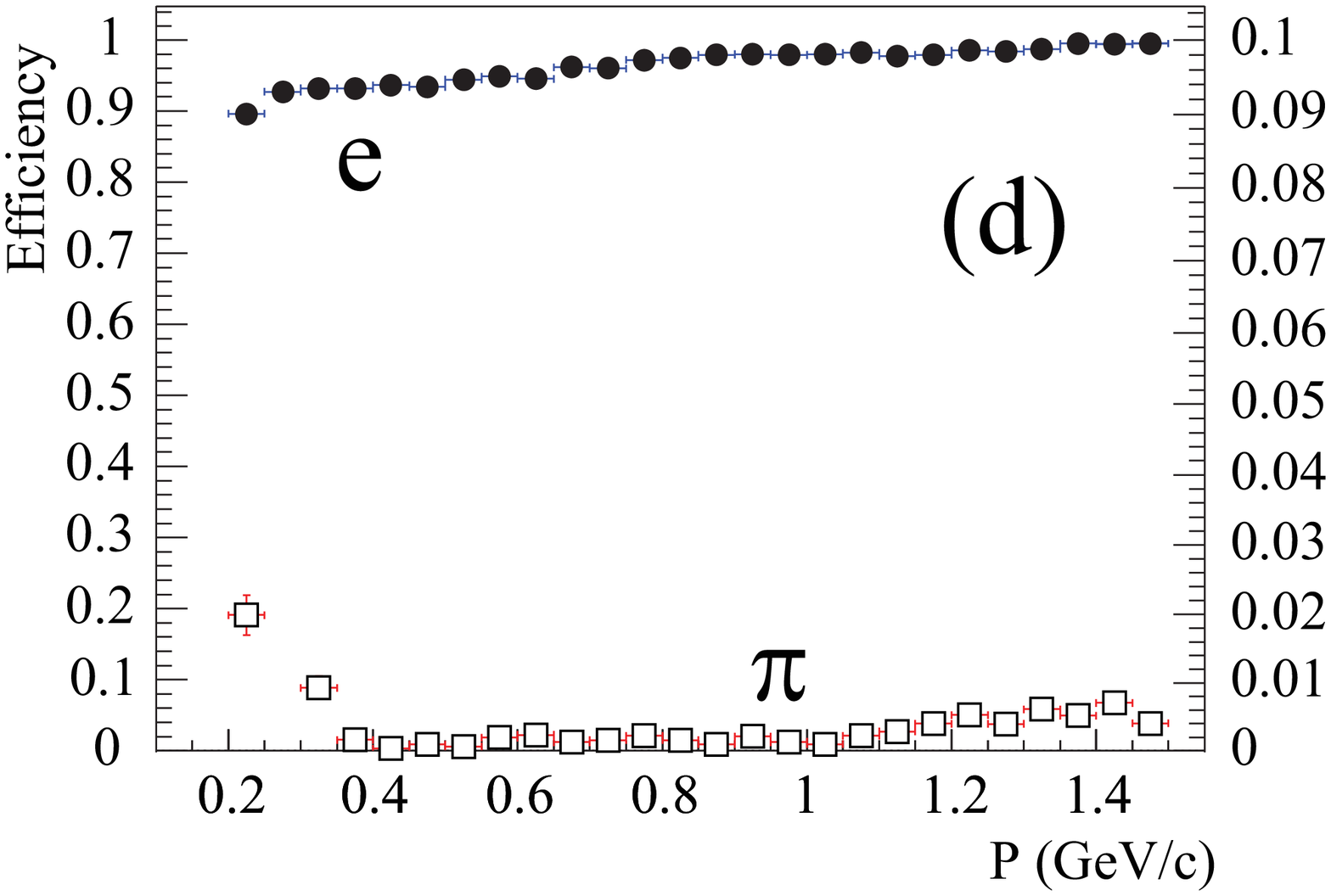} \quad
\includegraphics[width=5cm]{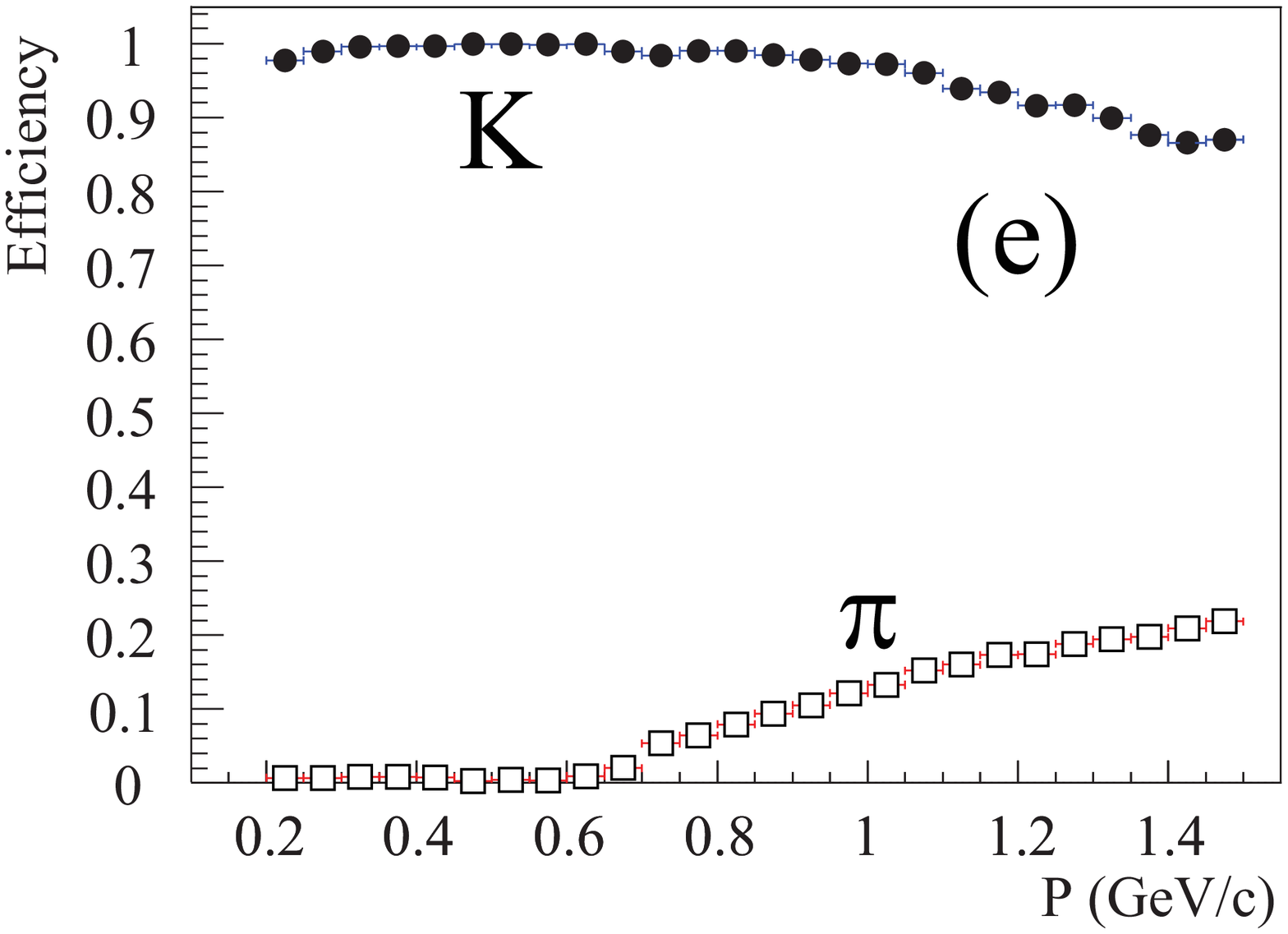}\quad
\includegraphics[width=5cm]{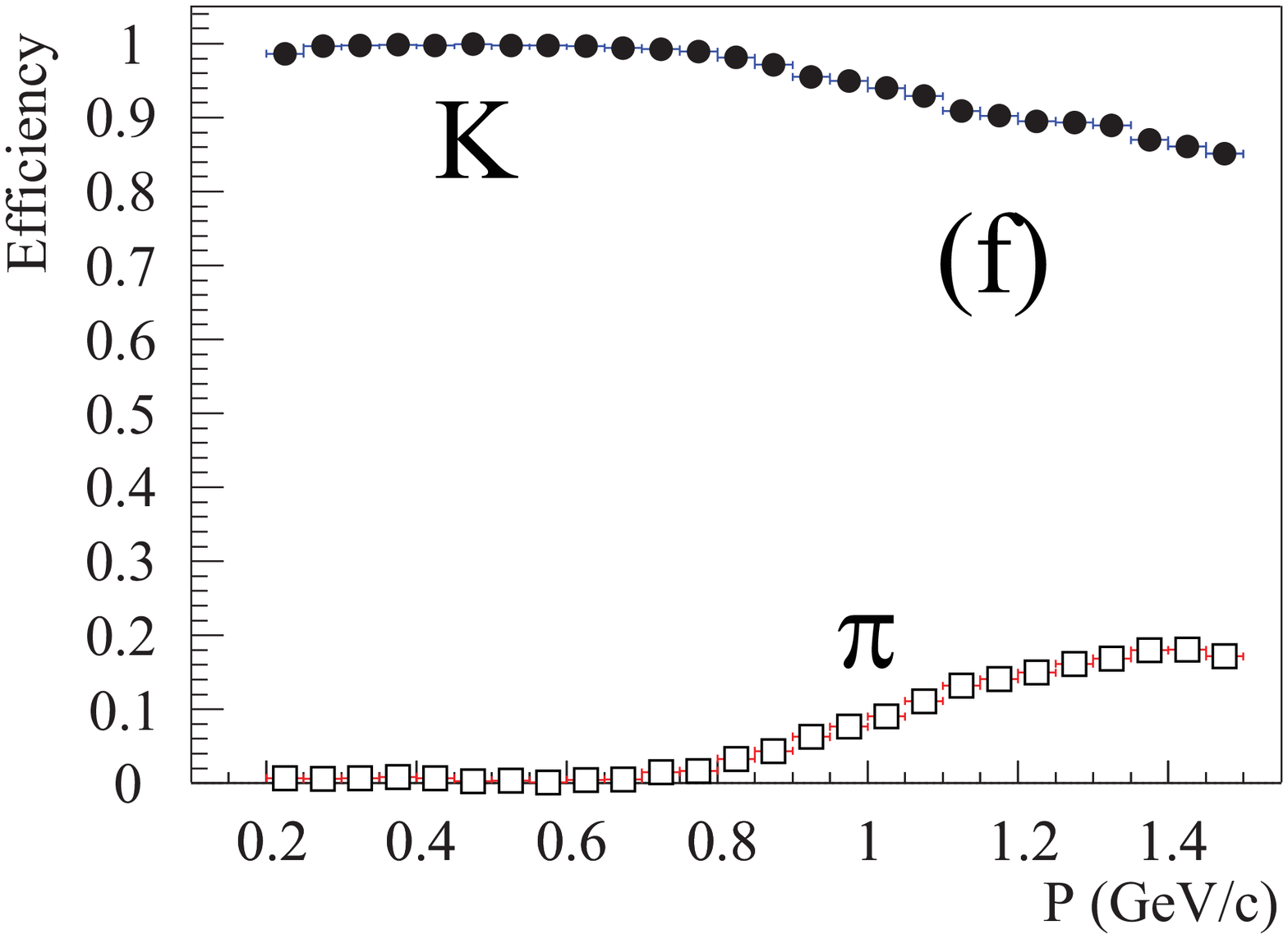}
\caption{PID efficiencies and contamination rates in different momentum
partitions. (a)~$\mu/\pi$ separation with $\omuc$; (b)~$\mu/\pi$
separation with $\omuc$ and $\oemc$; (c)~$e/\pi$ separation with
$\oemc$; (d)~$e/\pi$ separation with $\oemc$, $\otof$ and $\odedx$;
(e)~$K/\pi$ separation with the likelihood method; 
(f)~$K/\pi$ separation
with neural networks. In (c) and (d), the pion contamination rates are
enlarged by a factor of ten.} \label{fig:pid_eff} \EndFig

Both $\odedx$ and $\oemc$ are good discrimination variables to
separate electrons from pions above 0.4 GeV/c. $\otof$ can
separate $e/\pi$ quite effectively below 0.3 GeV/c. The neural network
trained with $\odedx$, $\otof$ and $\oemc$, shown in
\figurename~\ref{fig:pid_eff}(d), offers a nearly
uniform acceptance and background contamination between 0.25 GeV/c
and 1.6 GeV/c. It is interesting to note that the acceptance hole
between 0.2~GeV/c and 0.4 ~GeV/c almost vanishes after 
the application of an appropriate cut, 
even though no detector has clear discrimination power
for electrons in this region. 
The system is obviously making the inference that the
particle has to be an electron if it is not one of the others. This
is the combined contribution from the sub-detectors.

\subsubsection{ $\pi/K$ separation}

It is  generally believed that proton identification 
will be extremely good using the 
TOF and $dE/dx$ information at \bes3. Hence, only the
$K/\pi$ separation is focused on here. As discussed in Section
{\ref{sec:pid_sys}}, the $dE/dx$ can identify $K$'s and $\pi$'s very 
effectively below 0.6~GeV/c; the two-layer TOF can 
separate $K/\pi$ up to momenta of 0.9~GeV/c.

Traditionally, a likelihood method that combines the TOF and $dE/dx$
information is applied to hadron identification. To
construct the PDFs, data are divided into several bins in momentum
and $\cos\theta$ partitions to obtain the corresponding resolutions
and offsets. \figurename~\ref{fig:pid_eff}(e) shows the variation
of kaon identification efficiency and pion contamination rate as a
functions of momentum. In the real world, there are often tails on
distributions due to track confusion, nonlinearities in detector
response, and other experimental sources that are imperfectly
described in the PDFs. In light of this, it is helpful to apply 
the NN technique to hadron separation.

The network is
trained with two PID variables: $\otof$ and $\odedx$. The
PID ability is studied as a function of cuts on the output of 
the sequential
neural network $\oseq$.   The results are shown in
\figurename~\ref{fig:pid_eff}(f). Both the likelihood method and the
neural network method give similar results. Below 1~GeV/c, one
can see that the kaon-ID efficiency is greater than 95\% while the
pion contamination rate is less than 10\%. The $K/\pi$ separation is
extremely good for low momentum tracks ({\it i.e.}, less than 0.8 GeV/c).

\subsection{Future PID algorithms for \bes3}

The shapes of the PID variables from the EMC and MUC systems are 
complicated and there may exist non-linear correlations between 
between.
It is difficult for the likelihood method to construct the PDFs
analytically and handle these correlations properly. From our studies,
good electron-ID and muon-ID can be easily achieved from the neural
network at \bes3 with full detector information. In a simple
application, {\it e.g.}, for hadron separation, we get similar results
from the the neural network and the likelihood methods, where the PID
variables in TOF and $dE/dx$ systems are quasi-Gaussian, and the
correlation between two-layer TOF measurements is approximately
linear. Hence a flexible configuration of PID neural 
networks is employed
for both simple and complicated applications.

There are still a lot of factors that have to be taken into account
while applying the artificial neural network techniques to
particle identification at \bes3. For example, one or several input
variables may have to be removed due to the imperfect consistency
between data and Monte Carlo simulation. Impurities in the training
samples may introduce additional systematical uncertainties. More
detailed studies are needed in the future. Now the likelihood and
the artificial neural network PID algorithms are 
being studied in parallel.
The final \bes3 PID algorithm will definitely be a combination
of all the methods such as, for example, using the likelihood
method to combine the neural network outputs from sub-detectors. 

\setcounter{MaxMatrixCols}{32}

\section[Kinematic Fitting]{Kinematic Fitting \footnote{ By Kanglin He, Beijiang Liu, Min Xu, Xueyao Zhang}}
\label{sec:kfit}

\subsection{Introduction}
Kinematic fitting is a mathematical procedure in which one
uses the physical law governing a particle interaction or decay to
improve the measurements describing the process. For example,
consider the decay chains, $\psi(3770)\rightarrow D^{0}\overline{D^{0}}$, 
where $\overline{D^{0}}$ decays to the $CP$ eigenstate $K_{S}^{0}\pi^{0}$ 
and $D^{0}$ decays to the hadronic mode $K^{-}\pi^{+}$. 
There are several constraints that can be applied: 
(1) the $\pi^{+}\pi^{-}$ pair from $K^{0}_{S}$ decay must 
come from a common space point($2\times2-3$ = 1 constraint); 
(2) the momentum vector of $\pi^{+}\pi^{-}$ 
pair must be aligned with the position vector of the decay 
vertex relative to the interaction point($3-1$ = 2 constraints); 
(3) the mass of the $\gamma\gamma$ pair has to be equal to the 
$\pi^{0}$ mass(1 constraint); 
(4) energy and momentum are conserved in the  $D\overline{D}$ 
production  (4 constraints); 
and (5) the mass of $K^{0}_{S}\pi^{0}$ has to be 
equal to the mass of $K^{-}\pi^{+}$ (1 constraint). 
When the tracks are refit with these 9 constraints using 
the general algorithm discussed in next section, 
their parameters are forced to satisfy the constraints, thereby 
improving the mass and momentum resolution of the $D^{0}$ and 
the $\overline{D^{0}}$. 
These resolution improvements will translate to a larger signal 
to background ratio and frequently elevate marginal signals to 
statistical significant results.
The importance of kinematic fitting to data analysis
is demonstrated by its use in virtually all
modern high energy  physics experiments.

\subsection{General algorithm}

 The fitting technique is straightforward and is based on the well known
Lagrange multiplier method~\cite{part1:ref:fit1}. It is assumed that the
constraint equations can be linearized and summarized in two
matrices, {\bf{D}} and {\bf{d}}. Let $\alpha$ represent the
parameters for a set of $n$ tracks. It has the form of a column vector
\begin{equation}\label{part1:eq:alpha}
\alpha=\begin{pmatrix}
          \alpha_{1}\\
          \alpha_{2}\\
          \vdots \\
          \alpha_{n}
        \end{pmatrix}.%
\end{equation}
\noindent
Initially the track parameters have the unconstrained values $\alpha_{0}$,
obtained from the reconstruction. The $r$ constraint functions can be written generally as
\begin{equation}\label{eq:constraints}
\begin{matrix}
{\bf{H}}(\alpha)\equiv 0,    &\textrm{where }{\bf{H}}= \begin{pmatrix}
                                                   H_{1} &H_{2}&\cdots&H_{r}
                                                \end{pmatrix}.%
\end{matrix}%
\end{equation}
Expanding \eqref{eq:constraints} around a convenient point
$\alpha_{A}$ yields the linearized equations
\begin{equation}
0=\displaystyle\frac{\partial H(\alpha_{A})}{\partial\alpha}(\alpha-\alpha_{A})+H(\alpha_{A})={\bf{D}}\delta\alpha+{\bf{d}},
\end{equation}
where $\delta\alpha=\alpha-\alpha_{A}$. Thus we have
\begin{equation}
\begin{matrix}
   {\bf{D}}= \begin{pmatrix}
               \displaystyle\frac{\partial H_{1}}{\partial\alpha_{1}}
              &\displaystyle\frac{\partial H_{1}}{\partial\alpha_{2}}
              &\cdots
              &\displaystyle\frac{\partial H_{1}}{\partial\alpha_{n}} \\
              &  &  &  \\
               \displaystyle\frac{\partial H_{2}}{\partial\alpha_{1}}
              &\displaystyle\frac{\partial H_{2}}{\partial\alpha_{2}}
              &\cdots
              &\displaystyle\frac{\partial H_{2}}{\partial\alpha_{n}} \\
               \vdots &\vdots &\ddots &\vdots \\
               \displaystyle\frac{\partial H_{r}}{\partial\alpha_{1}}
              &\displaystyle\frac{\partial H_{r}}{\partial\alpha_{2}}
              &\cdots
              &\displaystyle\frac{\partial H_{r}}{\partial\alpha_{n}}
             \end{pmatrix},\quad
    &{\bf{d}}=\begin{pmatrix}
                H_{1}(\alpha_{A})\\
                H_{2}(\alpha_{A})\\
                \vdots \\
                H_{r}(\alpha_{A})
             \end{pmatrix},%
\end{matrix}%
\end{equation}
or $\displaystyle D_{ij}=\frac{\partial H_{i}}{\partial \alpha_{j}}$ and
$d_{i}=H_{i}(\alpha_{A})$. The constraints are incorporated using the method
of Lagrange multipliers in which the $\chi^{2}$ is written as a
sum of two term
\begin{equation}
\displaystyle \chi^{2}=(\alpha-\alpha_{0})^{T}V_{\alpha_{0}}^{-1}(\alpha-\alpha_{0})+2\lambda^{T}({\bf{D}}\delta\alpha+{\bf{d}}),
\end{equation}
where $\lambda$ is a vector of $r$ unknown Lagrange multipliers.
Minimizing the $\chi^{2}$ with respect to $\alpha$ and $\lambda$ yields two
vector equations that can be solved for parameters $\alpha$ and their 
covariance matrix:
\begin{equation}
     \begin{matrix}
        \displaystyle V^{-1}_{\alpha_{0}}(\alpha-\alpha_{0})+{\bf{D}}^{T}\lambda = 0, \\
        \displaystyle {\bf{D}}\delta\alpha+{\bf{d}}=0.
     \end{matrix}%
\end{equation}
The solution can be written as:
\begin{equation}
   \begin{matrix}
      \displaystyle \alpha = \alpha_{0}-V_{\alpha_{0}}{\bf{D}}^{T}\lambda, \\
      \displaystyle \lambda = V_{D}({\bf{D}}\delta\alpha_{0}+{\bf{d}}), \\
      \displaystyle V_{\alpha}=V_{\alpha_{0}}-V_{\alpha_{0}}{\bf{D}}^{T}V_{D}{\bf{D}}V_{\alpha_{0}},
   \end{matrix}%
   \label{eq:kf:sol}
\end{equation}
where $\displaystyle V_{D}=\left({\bf{D}}V_{\alpha_{0}}{\bf{D}}^{T}\right)^{-1}$ is the $r\times r$ constraint covariance matrix and
\begin{equation}
      \displaystyle \chi^{2}=\lambda^{T}V^{-1}_{D}\lambda=\lambda^{T}({\bf{D}}\delta\alpha_{0}+{\bf{d}}).
    \label{eq:kf:sol_chi2}
\end{equation}
Note that the $\chi^{2}$  can be written as a sum of $r$ distinct terms, one for each constraint. 
It can be shown that the new covariance matrix $V_{\alpha}$ has diagonal elements
that are smaller than the initial 
covariance matrix $V_{\alpha_0}$. In general, the nonlinearities of 
the constraint equations requires that the 
kinematic fitting procedure be applied iteratively until satisfactory convergence is achieved. 
Track parameters and their errors, covariance matrices, fit information and other quantities can be obtained after fitting.

The constraints ``pull'' the tracks away from  
their unconstrained values. 
The ``pull'' of the $i^{\rm th}-$track parameter is defined as:
\begin{equation}
\displaystyle \textrm{(pull)}_{i}=\frac{\alpha_{i}-\alpha_{0i}}{\sqrt{(V_{\alpha_{0}})_{ii}-(V_{\alpha})_{ii}}}.
\end{equation}
This is an important variable to test the track parameter and its error.
The resulting $\chi^{2}$ that is
obtained with $r$ constraints is distributed like a standard $\chi^{2}$ 
with $r$ degrees of freedom, 
if Gaussian errors apply. Of course, since track errors are only 
approximately Gaussian, the actual distribution will 
have more events in the tail than predicted by theory.  Still, knowledge 
of the distribution allows one to define reasonable $\chi^{2}$ cuts.

It is useful to compute how far the parameters have to move to
satisfy a particular constraint $j$. The initial ``distance from
satisfaction'' can be characterized by the quantity
$({\bf{D}}\delta\alpha_{0}+{\bf{d}})_{j}$ and the number of standard
deviations away from the satisfying the constraint is easily
calculated to be
\begin{equation}
  \displaystyle \sigma_{j}=\frac{D_{ji}\delta\alpha_{0i}+d_{j}}{\sqrt{(V^{-1}_{D})_{jj}}}.
\end{equation}
This information can be used to provide criteria for rejecting
background in addition to the overall $\chi^{2}$.

For kinematic fitting, it is important to choose a track representation 
that uses physically meaningful 
quantities and is complete. We adopt the 7-parameter $W$ format, 
defined as $\alpha_{W}=(p_{x},p_{y},p_{z},E,x,y,z)$,
a 4-momentum and a point where the 4-momentum is evaluated,
in the \bes3 kinematic fitting software package. It is 
straight-forward  to transfer the parameters of neutral tracks
and their covariance to the $W$ representation.
The $W$ format is conveient for transporting
particles in a magnetic field, and well suited for vertex
fitting. It has been noted that the $W$ format also have enough
information to represent the general decays of particles.

\subsection{Common Kinematic Constraints}
In this section, the constraints that are commonly encountered in high
energy physics are described. If multiple constraints are desired,  
one just extends the matrices by adding additional rows, one row per 
constraint.
This allows many constraints to be used simultaneously in the fit.

\subsubsection{Invariant mass and energy-momentum constraints}
The invariant mass constraint equation that forces a track to have an 
invariant mass $m_{c}$ is
\begin{equation}
E^{2}-p^{2}_{x}-p^{2}_{y}-p^{2}_{z}-m^{2}_{c}=0.
\end{equation}
Processes where invariant mass constraints 
are frequently applied in high energy physics analyses are 
$\pi^{0}/\eta\rightarrow \gamma\gamma$, $\eta\rightarrow 
\pi^{+}\pi^{-}\pi^{0}$, $K_{S}^{0}\rightarrow \pi^{+}\pi^{-}$, and
$\Lambda\rightarrow p \pi^{-}$, etc. All of these involve decays into 
several specific daughter particles. 
Since the detector resolutions for neutral particles are poorer
than those for charged particles, the 
invariant mass constraints 
for $\pi^{0}/\eta\rightarrow\gamma\gamma$ are 
almost always applied in data analyses.

In most $J/\psi$ and $\psi(2S)$ analyses, 
the final state daughter particles are 
fully reconstructed and are required 
to  satisfy energy-momentum conservation:
\begin{equation}
\begin{matrix}
             p_{x}-p_{cx}=0,\\
             p_{y}-p_{cy}=0 ,\\
             p_{z}-p_{cz}=0, \\
             E-E_{c}=0.
\end{matrix}
\end{equation}
Energy-momentum constraints are the most commonly
used  analysis tool. It is 
helpful for improving the momentum, 
energy and mass resolution, and for suppressing combinatorial 
backgrounds.

We consider the analysis of  
$J/\psi\rightarrow\rho\pi\rightarrow\pi^{+}\pi^{-}\pi^{0}\rightarrow\pi^{+}\pi^{-}\gamma\gamma$
as an example to illustrate the kinematic fitting procedure for 
energy-momentum constraints:
\begin{equation*}
\begin{matrix}
p_{x1}+p_{x2}+p_{x3}+p_{x4}=0,\\
p_{y1}+p_{y2}+p_{y3}+p_{y4}=0,\\
p_{z1}+p_{z2}+p_{z3}+p_{z4}=0,\\
E_{1}+E_{2}+E_{3}+E_{4}=M_{J/\psi},
\end{matrix}
\end{equation*}
where the $J/\psi$ is assumed to be at rest in the laboratory frame. 
Initially, the track parameters have the 
values $\displaystyle \alpha_{0}=\begin{pmatrix} \alpha^{0}_{1}\\ \alpha^{0}_{2}\\\alpha^{0}_{3}\\\alpha^{0}_{4}\end{pmatrix}$, where
$\alpha^{0}_{i}=\begin{pmatrix}
                 p^{0}_{x_{i}}\\
                 p^{0}_{y_{i}}\\
                 p^{0}_{z_{i}}\\
                 E^{0}_{i}
                 \end{pmatrix}$,
$i$=1, 2, 3, 4,                 
represent the four momentum vectors of $\pi^{+}$, $\pi^{-}$ and two photons. 
 The initial track covariance matrices are denoted as $V_{i0}$, $i$=1,2,3,4. 
The track parameters $\alpha_{i}$ can be expanded about these values (i.e., $\alpha_{iA}=\alpha_{i}^{0}$) giving for {\bf D} and {\bf d}:
\begin{equation*}
\begin{matrix}
{\bf D}=\begin{pmatrix}
        1 &0 &0 &0 &1 &0 &0 &0 &1 &0 &0 &0 &1 &0 &0 &0 \\
        0 &1 &0 &0 &0 &1 &0 &0 &0 &1 &0 &0 &0 &1 &0 &0 \\
        0 &0 &1 &0 &0 &0 &1 &0 &0 &0 &1 &0 &0 &0 &1 &0 \\
        0 &0 &0 &1 &0 &0 &0 &1 &0 &0 &0 &1 &0 &0 &0 &1
        \end{pmatrix},\quad
&{\bf d} = \begin{pmatrix}
    p^{0}_{x1}+p^{0}_{x2}+p^{0}_{x3}+p^{0}_{x4}\\
    p^{0}_{y1}+p^{0}_{y2}+p^{0}_{y3}+p^{0}_{y4}\\
    p^{0}_{z1}+p^{0}_{z2}+p^{0}_{z3}+p^{0}_{z4}\\
    E^{0}_{1}+E^{0}_{2}+E^{0}_{3}+E^{0}_{4}-M_{J/\psi}
    \end{pmatrix}.
\end{matrix}
\end{equation*}
The updated track parameters, covariance matrices and $\chi^{2}$ can be 
obtained by applying Eqs.~\eqref{eq:kf:sol} and~\eqref{eq:kf:sol_chi2}.
If one wants to apply the additional 
$\pi^{0}\rightarrow\gamma\gamma$ mass constraint:
\begin{equation*}
(E_{3}+E_{4})^{2}-(p_{x3}+p_{x4})^{2}-(p_{y3}+p_{y4})^{2}-(p_{z3}+p_{z4})^{2}=M^{2}_{\pi^{0}} ,
\end{equation*}
the derivative matrices {\bf D} and {\bf d} will take the forms:
\begin{equation*}
{\bf D}=\begin{pmatrix}
        1 &0 &0 &0 &1 &0 &0 &0 &1 &0 &0 &0 &1 &0 &0 &0 \\
        0 &1 &0 &0 &0 &1 &0 &0 &0 &1 &0 &0 &0 &1 &0 &0 \\
        0 &0 &1 &0 &0 &0 &1 &0 &0 &0 &1 &0 &0 &0 &1 &0 \\
        0 &0 &0 &1 &0 &0 &0 &1 &0 &0 &0 &1 &0 &0 &0 &1 \\
        0 &0 &0 &0 &0 &0 &0 &0
        &-2p^{0}_{x}
        &-2p^{0}_{y}
        &-2p^{0}_{z}
        &2E^{0}
        &-2p^{0}_{x}
        &-2p^{0}_{y}
        &-2p^{0}_{z}
        &2E^{0}
        \end{pmatrix},
\end{equation*}
and
\begin{equation*}
{\bf d} = \begin{pmatrix}
    p^{0}_{x1}+p^{0}_{x2}+p^{0}_{x3}+p^{0}_{x4}\\
    p^{0}_{y1}+p^{0}_{y2}+p^{0}_{y3}+p^{0}_{y4}\\
    p^{0}_{z1}+p^{0}_{z2}+p^{0}_{z3}+p^{0}_{z4}\\
    E^{0}_{1}+E^{0}_{2}+E^{0}_{3}+E^{0}_{4}-M_{J/\psi} \\
    (E^{0})^{2}-(p^{0}_{x})^{2}-(p^{0}_{y})^{2}-(p^{0}_{z})^{2}-M^{2}_{\pi^{0}}
    \end{pmatrix},
\end{equation*}
where $p^{0}_{x}=p^{0}_{x3}+p^{0}_{x4}$, $p^{0}_{y}=p^{0}_{y3}+p^{0}_{y4}$, $p^{0}_{z}=p^{0}_{z3}+p^{0}_{z4}$ and $E^{0}=E^{0}_{3}+E^{0}_{4}$.

\subsubsection{Reconstruction of $K^{0}_{S}$ and $\Lambda$ decay vertex}

\begin{figure}\centering
\includegraphics[width=8.0cm]{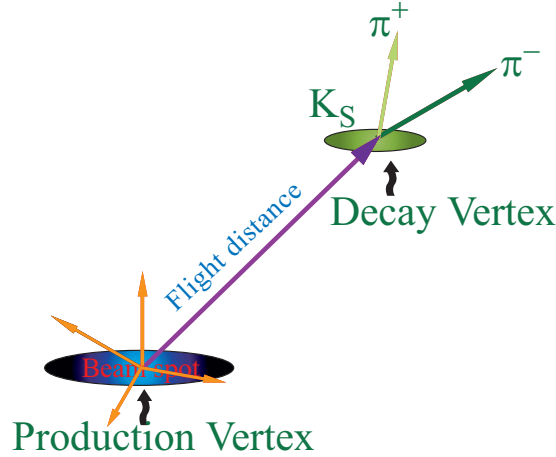}
\caption{ \label{fig::ksdecay}A $K^{0}_{S}$ travels a certain
distance (``the flight distance'') before decaying into its
daughters. These daughters are subsequently measured by the tracking
system.}
\end{figure}

To introduce the subject of decay vertex reconstruction, consider 
\figurename~\ref{fig::ksdecay},  which shows
a $K^{0}_{S}$ that decays to $\pi^{+}\pi^{-}$ at a secondary vertex 
after being produced in the beam interaction
region. An accurate determination of the lifetime requires that both
the beginning and endpoint of the $K^{0}_{S}$ flight vector be
determined accurately. The endpoint is determined by vertex fitting,
and its measurement accuracy is controlled purely by the tracking
error of the daughter particles. The beginning point is determined
by the beam spot size augmented, perhaps, 
as shown in \figurename~\ref{fig::ksdecay}, 
by other tracks produced in interaction point (IP), or by the
average of preliminary vertex for lot of events.

The motion of a neutral track before its decay is a simple linear
equation. We convert the  flight distance ($s$) measured from 
the production point $(x_{p}, y_{p}, z_{p})$ to 
the decay point $(x_{d}, y_{d}, z_{d})$, to the
proper decay time $c\tau$ using 
$s=\beta ct=\gamma\beta c\tau =(p/m)c\tau$, yielding the new
equations
\begin{equation}\label{eq:ctau}
      \begin{pmatrix}
           x_{p} \\
           y_{p} \\
           z_{p}
        \end{pmatrix} =
        \begin{pmatrix}
           \displaystyle x_{d}-\frac{p_{x}}{m}c\tau \\
            ~\\
           \displaystyle y_{d}-\frac{p_{y}}{m}c\tau \\
            ~\\
           \displaystyle z_{d}-\frac{p_{z}}{m}c\tau
        \end{pmatrix}.
\end{equation}
\noindent
The lifetime $c\tau$ is determined with Eq.~\eqref{eq:ctau} representing  
constraint conditions. We can apply
\begin{equation}
   \chi^{2}=(\alpha-\alpha_{0})^{T}V^{-1}_{\alpha_{0}}(\alpha-\alpha_{0})+2\lambda^{T}({\bf{D}}\delta\alpha+{\bf{E}}\delta c\tau +{\bf{d}})
\end{equation}
to solve for $c\tau$ and its error, while at the same time improving the track parameters and the start point~\cite{part1:ref:ctau}. 
Cascade decay vertices, such as, $\Xi^{-}\rightarrow\Lambda\pi^{-}$, then 
$\Lambda\rightarrow p\pi^{-}$ can also 
be reconstructed by applying a similar technique~\cite{part1:ref:xi_rec}.

\begin{figure}\centering
\includegraphics[width=7.5cm]{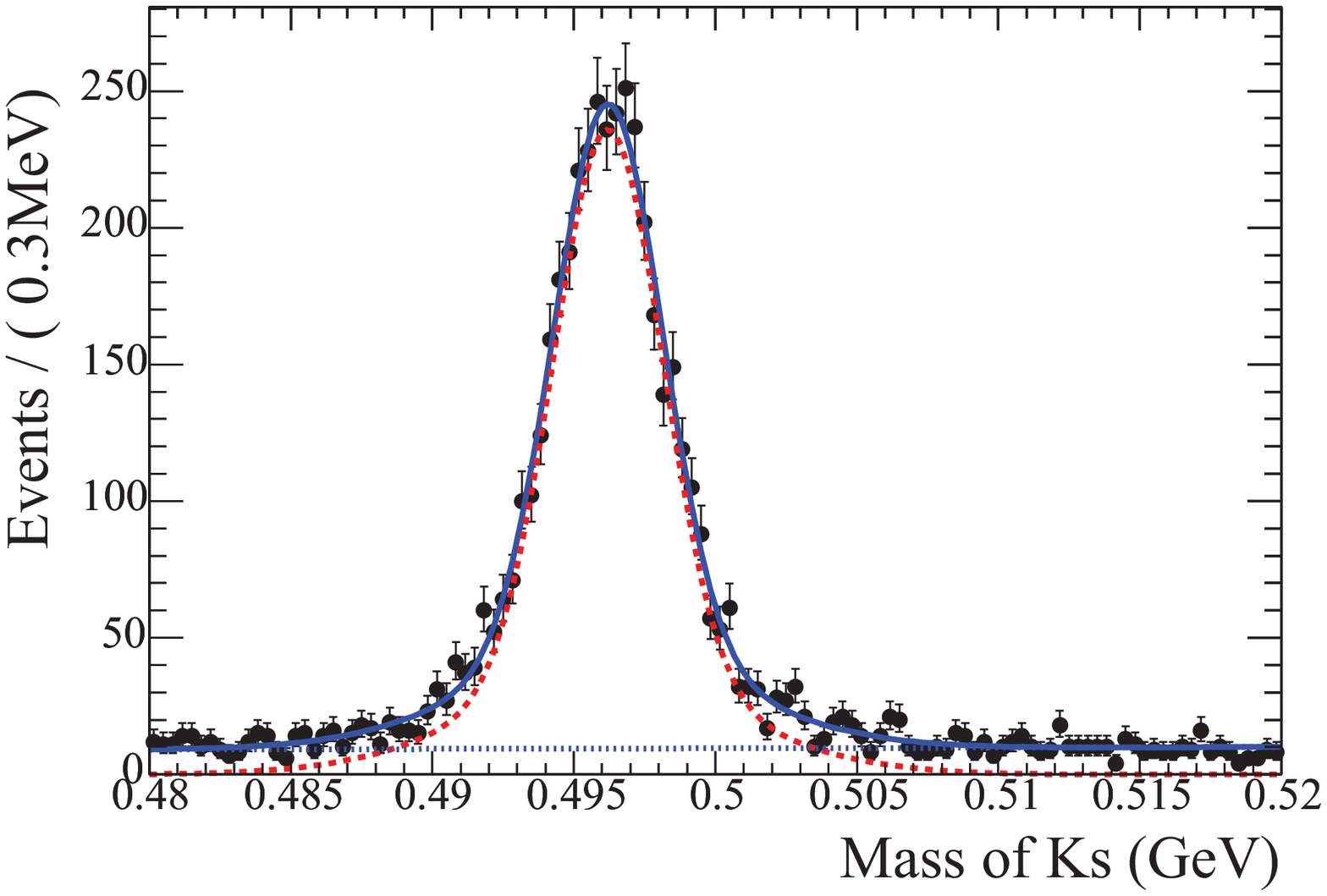}\quad
\includegraphics[width=7.5cm]{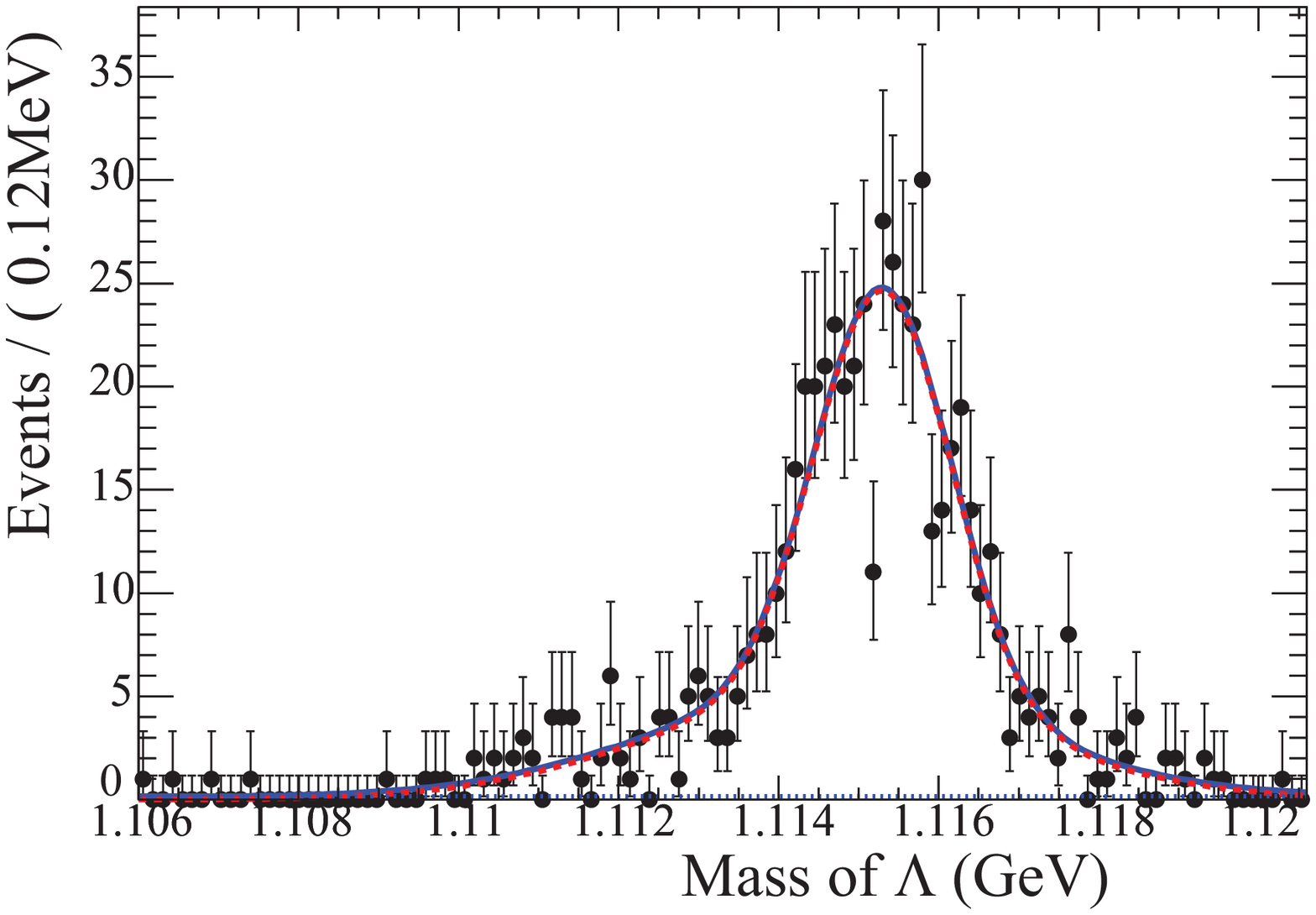}
\caption{The mass distribution of reconstructed $K^{0}_{S}$'s and
$\Lambda$'s from a MC simulation. The $K^{0}_{S}$ and 
$\Lambda$ mass resolutions are obtained from the weighted 
widths from  double Gaussian fits to the histograms.} 
\label{fig::ks_lam}
\end{figure}

A cut on the ratio of decay length to its error, $s/\sigma_{s}$, 
is useful to suppress  combinatorial backgrounds. 
\figurename~\ref{fig::ks_lam} 
shows the mass distribution of $K^{0}_{S}$ ($s/\sigma_{s}>2$) 
and $\Lambda$ ($s/\sigma_{s}>1$) after the secondary vertex 
reconstruction. 
The mass resolutions are $\sim3$~MeV for $K_{S}$, and 
$\sim1.2$~MeV for $\Lambda$.


\subsubsection{Kinematic constraints for charm tag reconstructions}

The general technique used in charm physics studies is referred to as tagging. 
At the peak of the $\psi(3770)$, 
charmed $D$ mesons are pair produced with no accompanying particles. 
Fully reconstructing one $D$ from a 
subset of tracks in an event
guarantees the remaining tracks must 
originate from the recoiling $\overline{D}$.

The total energy constraints $E=E_{c}(E_{\textrm{beam}})$ can be applied to improve the mass resolution 
of charm tags by the fact that each $D$ carries 
one-half the total cms energy, 
if the cms is at rest in the laboratory frame. In this case,
the reconstructed $D$ mass  is the famous,  so-called 
``beam constrained mass.''  At \bes3, The produced $\psi(3770)$ 
is  not at rest since 
the electron and positron beam collides with a finite 
crossing angle (22 mrad). 
The beam energy constraint can still be applied by simply 
boosting the tracks to the rest frame of the $\psi(3770)$.

The charmed partcle tags can be reconstructed in an alternative way that 
requires the mass of the reconstructed 
$D$ tag to be equal to the mass of recoiling $\overline{D}$. 
In this case we have
\begin{equation}
E^{2}_{D}-p^{2}_{Dx}-p^{2}_{Dy}-p^{2}_{Dz}=(E_{\psi}-E_{D})^{2}-(p_{\psi x}-p_{Dx})^{2}-(p_{\psi y}-p_{Dy})^{2}-(p_{\psi z}-p_{Dz})^{2},
\end{equation}
where $(p_{Dx}, p_{Dy}, p_{Dz}, E_{D})$ and $(p_{\psi x}, p_{\psi y}, p_{\psi z}, E_{\psi})$ denote the energy-momentum 
vectors of reconstructed $D$ tag and produced $\psi(3770)$. In the
$\psi(3770)$ rest frame, the equal mass 
constraint is totally equivalent to the beam energy constraint.

In Dalitz-plot or partial-wave analyses, an effective recoil mass can be 
calculated 
using $E_{\textrm{recoil}}=E_{\psi(3770)}-E_{\textrm{tag}}$  and 
$\vec{p}_{\textrm{recoil}}=\vec{p}_{\psi(3770)}-\vec{p}_{\textrm{tag}}$. 
One can perform a kinematic fit in which the mass of the charm  tag is 
constrained to the $D$ mass and the recoil mass is allowed to vary. 
The signal can then be seen in the recoil mass plot as a peak  near the 
$D$ mass. With this type of constraint, each event has the
same amount of phase space for its decay throughout the recoil  mass plot. 
This has the advantage that the kinematic boundaries 
of phase space are the same for 
the signal and sideband regions of the recoil mass plot.

\subsection{Applying Kalman filter techniques to kinematic fitting}

In 1960, R.~E.~Kalman published his famous paper describing a recursive 
solution to the discrete-data linear filtering problem~\cite{part1:ref:kalman}. 
It was introduced to the high energy physics world  in a paper by 
Fr\"{u}hwirth~\cite{part1:ref:fruhwirth_kalman}. The Kalman filter is a
set of mathematical equations that provides an efficient
computational (recursive) means to estimate the state of a process,
in a way that minimizes the mean of the square error. 
It is interesting to derive the kinematic fit formalism in the 
Kalman filter framework~\cite{part1:ref:kalman_kine}. 
This derivation is also relevant to the addition of exact constraints to 
a fit. At \bes3, there are two typical kinematic fitting problems:
\begin{itemize}
\item[1)] constraints with a covariance matrix;
\item[2)] constraints with virtual particles.
\end{itemize}
that have to be processed by applying the Kalman filter technique.

\subsubsection{Constraint with a covariance matrix}
The exact constraints can be regarded as a measurement equation with
infinite precision. Most kinematic constraints are of the exact type.
But at \bes3, the contribution from beam energy spread should be 
considered if
the data are taken off the peak of narrow resonance like the $J/\psi$
and $\psi(2S)$. In general, we call the ``measurement
equation'' the constraint with a covariance matrix.

Synchrotron radiation and the replacement of the radiated energy by
the accelerating cavities generate an energy spread for each beam
that results in an essentially Gaussian distribution in  beam
energies $E_{b}^{\prime}$ centered on the
nominal value ($E_{b}$)
\begin{equation}
\displaystyle G(E_{b},E_{b}^{\prime})=\frac{1}{\sqrt{2\pi}\Delta}
\exp\left[-\frac{(E_{b}-E_{b}^{\prime})^{2}}{2\Delta^{2}}\right],
\end{equation}
where $\Delta$ represents the beam energy spread.

Suppose that the electron and positron beams collide with a crossing
angle $2\theta$ in the $x$-$z$ plane (at \bes3, 
$\theta=11~\textrm{mrad}$).  In tha case
we have the measurement equation:
\begin{equation}
\begin{pmatrix} p_{x}&p_{y}&p_{z}&E\end{pmatrix} =
\begin{pmatrix}\displaystyle \sqrt{s}\tan\theta &0 &0
&\displaystyle \frac{\sqrt{s}}{\cos\theta}\end{pmatrix}
\end{equation}
where $E=2E_{b}$, $\delta E = \sqrt{2}\Delta$ and
$\delta\sqrt{s}=\displaystyle \sqrt{2}\Delta\cos\theta$. The corresponding
covariance matrix is given by
$2\cdot\Delta^{2}\cdot
\begin{pmatrix}\sin^{2}\theta &0 &0 &\sin\theta \\
0 &0 &0 &0 \\ 0 &0 &0 &0\\ \sin\theta &0 &0 &1
\end{pmatrix}$.
Clearly, the correlation between $p_{x}$ and $E$ are included.

\subsubsection{Constraint with virtual particles}

At \bes3, three kinds of constraints with virtual particles will commonly 
be encountered in data analysis. 
(In the following discussion, we suppose that a virtual particle is 
represented by a 4-momentum vector 
$q=(q_{x}, q_{y}, q_{z}, W)$ and all  detected particles are represented 
by a 4-momentum vector $p=(p_{x}, p_{y}, p_{z}, E)$.)

\begin{itemize}
\item No detection information available; the virtual particle is 
connected to kinematic fitting with some mass constraints. 
Energy-momentum conservation gives:
    \begin{equation*}
    \begin{matrix}
    p_{x}+q_{x}=p_{cx},\\
    p_{y}+q_{y}=p_{cy},\\
    p_{z}+q_{z}=p_{cz},\\
    E+W=E_{c}.
    \end{matrix}
    \end{equation*}
    In the analysis of $J/\psi\rightarrow \bar{p}n\pi^{+}$, where the $n$ 
is onot detected, 
the mass of the virtual particle $n$ provides an additional constraint. 
In  the analysis of $\psi(3770)\rightarrow D\bar{D}$, 
where one $D$ is reconstructed in one of its decay modes $X$ and 
$\bar{D}$ is not required to be reconstructed,
an equal mass constraint $m_{X}=m_{\textrm{missing}}$ can be employed to 
improve the $D$ tag mass resolution.
\item Only position information is available; the virtual particle is 
connected to the kinematic fitting with a set of 
measurement equations. Some particles, e.g., $K^{0}_{L}$, will register 
reliable position information when it interacts in the detector. 
In this situation, we may construct ``measurement equations'' as 
additional constraints to the energy-momentum conservation:
    \begin{equation*}
    \begin{matrix}
    x_{v}+\lambda q_{x}=x_{\textrm{clus}},\\
    y_{v}+\lambda q_{y}=y_{\textrm{clus}},\\
    z_{v}+\lambda q_{z}=z_{\textrm{clus}},
    \end{matrix}
    \end{equation*}
    where $(x_{v},y_{v},z_{v})$ is the position of interaction or decay 
vertex, $(x_{\textrm{clus}},y_{\textrm{clus}},z_{\textrm{clus}})$ is 
the position vector of the neutral cluster with the covariance matrix 
$V_{\textrm{clus}}$, and $\lambda$ is the flight path. 
The value of the
parameter $\lambda$ is not very interesting, it can be easily substituted 
with 
    \begin{equation*}
    \displaystyle \lambda=\frac{q^{T}V^{-1}_{\textrm{clus}}\Delta x}{q^{T}V^{-1}_{\textrm{clus}}q},
    \end{equation*}
    where $\Delta x=(x_{v}-x_{\textrm{clus}},
y_{v}-y_{\textrm{clus}},z_{v}-z_{\textrm{clus}})$.
\item The virtual particle is built by merging a set of tracks. 
Frequently, one wants to build virtual particles 
with a vertex constraint(e.g. $K^{0}_{S}\rightarrow \pi^{+}\pi^{-}$), or 
with a mass constraint (e.g., $\pi^{0}\rightarrow\gamma\gamma$). 
The pre-fitted $K^{0}_{S}$'s and $\pi^{0}$'s can be directly applied in 
physics analyses.
\end{itemize}

\subsection{Limitations of Kinematic Fitting}


The precision of kinematic fitting is governed by the model of 
constraint/measurement equations and the model of track parameters and 
their covariance matrices. 
To understand the power and limitations of the kinematic fitting, one has 
to understand these models well.

Since the natural widths of the narrow resonance  such as the $J/\psi$ and 
$\psi(2S)$ are much smaller than the detector resolution, 
the imperfection of energy-momentum constraints could be ignored. For 
broad resonances above charm threshold or for data-taking off of 
resonance peaks,  effects due to beam energy spread and initial state 
radiation must be taken into account.

Unlike the real world, the errors of track parameters are assumed to be 
Gaussian in the kinematic fitting procedure. In some experiments,
the track errors are poorly understood, making kinematic fittng
be of little use.   On the other hand, 
Kalman fitted tracks at \bes3 have better-understood tracking errors
and are well suited for kinematic fitting. 
As is known, it is quite difficult to obtain ``ideal'' track error in an 
experiment since there are many approximations in the 
Kalman track fitting error models, 
such as the wire resolution, fluctuations of energy losses and 
multiple scattering.  Tight $\chi^{2}$ cuts can cause large systematic 
uncertainties into an  analysis because of a long tail in the $\chi^{2}$ 
distribution.

For neutral tracks, the energy deposited in the crystal calorimeter 
is distributed asymmetrically.  If one is 
to avoid large inefficiencies in kinematic fits, the Gaussian error model 
for neutrals has to be modified. 
A solution may be carried out in the near future by application of the 
method of dynamic noise adjustment~\cite{part1:ref:dna}.

\section[Partial Wave Analysis]{Partial Wave Analysis\footnote{By Ning Wu, 
Hanqing Zheng}}
\label{sec:pwa}

\newcommand{\svec}[1]{ \stackrel{\rightarrow}{#1} }
\newcommand{\dvec}[1]{ \stackrel{\leftrightarrow}{#1} }
\newcommand{\lvec}[1]{ \stackrel{\leftarrow}{#1} }
\newcommand{\swav}[1]{ \stackrel{\sim}{#1} }
\newcommand{\tdot}[1]{ \stackrel{\cdot}{#1} }
\newcommand{\uhat}{ \hat U }
\newcommand{\ehat}{ \hat U_{\epsilon} }
\newcommand{\mhat}[1]{ \hat U_{\epsilon_{#1}} }
\newcommand{\define}{ \stackrel{\triangle}{=} }
\renewcommand{\theequation}{\thesection.\arabic{equation}}

\subsection{Introduction}

Partial Wave Analysis (PWA) is widely used in  high
energy experimental physics. It is a useful method for analyzing 
the correlation between the momenta of final state particles 
in order to determine the masses, widths and
spin-parities of intermediate resonances. The basis of PWA is relativistic
kinematics. 

It is known that all quantum states form a
Hilbert space that is a representation of 
inhomogeneous Lorentz space. Physicists 
are used to studying 
quantum states from the point of view of group theory. 
In this context, the quantum state
of a fundamental particle corresponds to an irreducible
representation of the Poincare group. The quantum state of a composite 
particle consisting of more fundamental constituents is represented by a 
state vector of the irreducible representation of the Poincare group 
reduced is reduced from
the direct product of the states of its component particles. Hence, the
Poincare group and its representation theory is the basis of the 
kinematic theory of relativistic particles. \\

By using the method of group representation and applying 
analysis techniques that exploit
the symmetries of the system, the form of the decay
matrix element can be changed to a new form where
the angular-dependent part
of the matrix element is expressed by a $D$-function, and the 
energy-dependent part is kept in a reduced matrix 
element~\cite{part1_pwa_1,part1_pwa_2,part1_pwa_3}.
In this new form, the angular information of the decay matrix
element is separated from the energy information. 
This property is quite useful in partial wave
analysis,  since the angular-dependence of the
decay matrix contains the information of the spin-parity of 
the decaying particles, and the energy dependence of
the decay matrix contains information about the interactions of its 
constituents,
or pole positions of intermediate states.
In the PWA technique, both the angular and energy information
of the decay matrix are utilized, and the
spin-parity and pole position of the resonance 
can be determined simultaneously.

In this report, we will briefly discuss various kinematic theories
of decay processes, such as the helicity formalism, the tensor formalism
etc. We then discuss how to apply the PWA technique
to sequential decay processes, and how to use PWA  to determine
the mass, width, spin-parity and branching ratios of a resonance.

\subsection{Decay Amplitude}

The helicity formalism is widely used in  PWA.
Since the  helicity formalism is based on  first
principles of quantum theory, it is considered to be
the standard method for determining the spin-parity of a resonance.

The concept of helicity was first proposed by
Chou and Shirokov in 1959~\cite{part1_pwa_4}.  Soon afterwards
Jacob, Wick and others systematically proposed the
helicity formalism~\cite{part1_pwa_1,part1_pwa_5,part1_pwa_6}. 
Subsequently, Chung found a way
to express the helicity coupling amplitudes
$F^J_{\lambda\nu}$ that is useful for
PWA~\cite{part1_pwa_7,part1_pwa_8,part1_pwa_9,part1_pwa_10,part1_pwa_11,part1_pwa_12,part1_pwa_13}. 
\\

The helicity operator is defined as the projection
of the spin operator along the direction of motion
\begin{equation}  \label{2.1}
h = \frac{\svec{J} \cdot \svec{P}}{|\svec{P}| }
= \svec{J} \cdot \hat{P},
\end{equation}
where $\hat{P}$ is the momentum unit vector.
The above definition only holds for a moving particle.
For particles at rest, the unit vector $\hat{P}$ generally has
no definition, so in this case the helicity
operator is defined as the projection of the spin operator
along the $z$-axis
\begin{equation}  \label{2.2}
h = J_3.
\end{equation}
If a moving particle is obtained by a Lorentz boost,
the unit vector  $\hat{P}$ in Eq.~(\ref{2.1})
is taken to be along the direction of the motion of
the particle before the boost. \\

One-particle
helicity states are taken to be the common eigenstate
of the operators $P_{\mu}$, $\svec{J}^2$ and $h$. These
are denoted by
\begin{equation}  \label{2.3}
| p \lambda \rangle,
\end{equation}
and satisfy
\begin{equation}  \label{2.4}
P_{\mu} | p \lambda \rangle
= p_{\mu} | p \lambda \rangle,
\end{equation}
\begin{equation}  \label{2.5}
\svec{J}^2 | p \lambda \rangle
= s(s+1) | p \lambda \rangle,
\end{equation}
\begin{equation}  \label{2.6}
h | p \lambda \rangle
= \lambda | p \lambda \rangle.
\end{equation}
\\

A two-particle helicity state is defined as
\begin{equation}  \label{2.7}
| p; JM \lambda_1 \lambda_2 \rangle
= \frac{N}{2 \pi}
\int {\rm d}U D^J_{M \lambda}(\alpha\beta\gamma)^*
R(\alpha\beta\gamma) \psi_{p \lambda_1 \lambda_2},
\end{equation}
where $R(\alpha\beta\gamma)$ is the rotation operator,
$D^J_{M \lambda}(\alpha\beta\gamma)$ is
the traditional
$D$-function~\cite{part1_pwa_14,part1_pwa_15,part1_pwa_16,part1_pwa_17},
$\psi_{p \lambda_1 \lambda_2}$ is the two-particle
direct product state in the canonical rest frame,
$N=\sqrt{\frac{2J+1}{4 \pi}}$ is the normalization
factor, and
\begin{equation}  \label{2.8}
{\rm d}U = \sin \beta
{\rm d}\alpha {\rm d}\beta {\rm d}\gamma ;
\end{equation}
the integration region is
\begin{equation}  \label{2.9}
-\pi < \alpha <  \pi, ~~
0 < \beta < \pi, ~~
-\pi < \gamma <  \pi.
\end{equation}
\\

We consider a two-body decay process.
Suppose that the spin-parity of the parent
particle is $J^{\eta_J}$, and spin-parities
of the two daughter particles are
$s^{\eta_s}$ and $\sigma^{\eta_{\sigma}}$, 
then, in the center of mass system of the parent
particle, the decay amplitude is
\begin{equation}  \label{2.10}
M^J_{\lambda\nu}(\theta, \varphi, m)
\propto D^{J~*}_{m,(\lambda-\nu)}
(\varphi, \theta, 0)
F^J_{\lambda\nu},
\end{equation}
where $\theta$ and $\varphi$ are polar and azimuthal
angle of one of the daughter particles in the center of
mass frame, $m$ is the magnetic quantum number of
the parent particle, $\lambda$ and $\nu$ are helicities
of the two daughter particles, and
$F^J_{\lambda\nu}$ is called the helicity coupling
amplitude. In $J/\psi$ hadronic or radiative
decay processes, parity conservation holds. In this
case, $F^J_{\lambda\nu}$ has the following symmetry
property
\begin{equation}  \label{2.11}
F^J_{\lambda \nu} =
\eta_J \eta_s \eta_{\sigma} (-)^{J - s - \sigma}
F^J_{-\lambda -\nu}.
\end{equation}
In the helicity formalism, all of the angular dependence
of the decay amplitude is contained in the $D$-function
as shown in Eq.~(\ref{2.10}); the helicity
coupling amplitudes $F^J_{\lambda\nu}$ are independent
of all angular information and only dependent on the energy of
the system. Details on how to calculate $F^J_{\lambda\nu}$
can be found in the
literature~\cite{part1_pwa_7,part1_pwa_8,part1_pwa_9,part1_pwa_10,part1_pwa_11,part1_pwa_12,part1_pwa_13}.
\\

In experimental physics analyses, most decays that are encountered
are sequential decays that include some intermediate resonant states. 
As an example, 
consider the following sequential decays
\begin{equation}  \label{2.12}
a \to b + c, ~~~ b \to d + e,
\end{equation}
where $b$ is an intermediate resonant state.
The decay amplitude for this sequential decay is
\begin{equation}  \label{2.13}
M^{s_a}_{\lambda\mu}(\theta_1, \varphi_1, m)
\cdot BW(s, M_b, \Gamma_b)
\cdot
M^{s_b}_{\nu\sigma}(\theta_2, \varphi_2, \mu),
\end{equation}
where $s_a$ and $s_b$ are the spin quantum numbers of
particles $a$ and $b$, respectively, $\lambda$, $\mu$,
$\nu$ and $\sigma$ are the  helicities of particles
$b$, $c$, $d$ and $e$, respectively, $m$ is the magnetic
quantum number of particle $a$ in its rest frame,
$M_b$ and $\Gamma_b$ are the mass and width of particle
$b$, $\theta_1$ and $\phi_1$ are the polar and azimuthal
angles of particle $b$ in the rest frame of particle $a$,
and $\theta_2$ and $\phi_2$ are the polar and azimuthal
angles of particle $d$ in the rest frame of particle $b$.
In this decay process, the final stable particles
are $c$, $d$ and $e$.  Particle $b$ is a resonant state
described by the Breit-Wigner amplitude
$BW(s, M_b, \Gamma_b)$. \\

In addition to the helicity formalism, one can sometimes
express the decay amplitude in the LS coupling formalism.
We again consider the two-body decay process
$a \to b + c$, and suppose that two final state
particles $b$ and $c$ are massive. Then, in the rest frame
of parent particle $a$, the decay amplitude in the LS coupling
formalism is
\begin{equation}  \label{2.14}
M^{s_a}_{ls}(\theta, \varphi; m_1, m_2, m)
\propto \sum_{m m_s}
\langle \theta \varphi m_1 m_2 |
l m s m_s \rangle
\langle l m s m_s | M | s_a m \rangle,
\end{equation}
where $s_a$ and $m$ are spin quantum number and magnetic
quantum number of particle $a$, $l$ is the relative orbital
angular momentum quantum number between particles $b$ and $c$,
$s$ is the total spin quantum number of particles $b$ and $c$,
$m_1$ and $m_2$ are magnetic quantum numbers of particles
$b$ and $c$, respectively,
and  $m$ is the magnetic quantum number
of particle $a$. Applying the Wigner-Eckart theorm, we have
\begin{equation}  \label{2.15}
\langle l m s m_s | M | s_a m \rangle
= \langle l m s m_s | s_a m \rangle
G^J_{ls}
\end{equation}
where $G^J_{ls}$ is the reduced matrix element. Using the
relation
\begin{equation}  \label{2.16}
\langle \theta \varphi m_1 m_2 |
l m s m_s \rangle =
\langle s_b m_1 s_c m_2 |
s m_s \rangle \cdot
Y^l_m(\theta, \varphi),
\end{equation}
where $Y^l_m(\theta, \varphi)$ is a spherical harmonic function,
we can convert Eq.~(\ref{2.14}) into
\begin{equation}  \label{2.17}
M^{s_a}_{ls}(\theta, \varphi; m_1, m_2, m)
\propto G^J_{ls} \langle s_b m_1 s_c m_2 |
s m_s \rangle \sum_m
\langle l m s m_s | s_a m \rangle
Y^l_m(\theta, \varphi).
\end{equation}
\\

It can be shown that the LS coupling amplitude $G^J_{ls}$
and helicity coupling amplitude $F^J_{\lambda\nu}$ are
related as
\begin{equation}  \label{2.18}
F^J_{\lambda\nu} = \sum_{ls}
\sqrt{\frac{2l+1}{2J+1}}
\langle l 0 s \delta | J \delta \rangle
\langle s_b \lambda s_c -\nu | s \delta \rangle
G^J_{ls}.
\end{equation}
Thus, we have
\begin{equation}  \label{2.19}
\sum_{\lambda \nu} |F^J_{\lambda\nu}|^2
= \sum_{ls} |G^J_{ls}|^2.
\end{equation}
\\

Another formalism used for PWA analysis is the so-called tensor
formalism, which was first proposed by Zemach 
in 1965~\cite{part1_pwa_18,part1_pwa_19}.
The original method proposed by Zemach was non-relativistic,
and all tensors were evaluated in their respective
rest frames. In this formalism, the decay amplitude 
and all the angular dependence is
expressed directly in terms of the 4-momentum vectors
of the initial and  final
state particles~\cite{part1_pwa_18,part1_pwa_19,part1_pwa_8,part1_pwa_9,part1_pwa_20,part1_pwa_21,part1_pwa_22}. 
As an example, 
consider $J/\psi$ decay.
The general form for the decay amplitude for $J/\psi$ hadronic
decay is
\begin{equation}  \label{2.20}
A = \psi_{\mu}(m) A^{\mu}
 = \psi_{\mu}(m) \sum_i \Lambda_i U_i^{\mu},
\end{equation}
where $\psi_{\mu}(m)$ is the polarization vector of the $J/\psi$,
$m$ the magnetic quantum number of the $J/\psi$ in its rest frame,
and $U_i^{\mu}$ the $i^{\rm th}$ partial wave amplitude with coupling
strength determined by a complex parameter $\Lambda_i$.
For $J/\psi$ radiative decays, the general form of the decay
amplitude is
\begin{equation}  \label{2.21}
A = \psi_{\mu}(m) e_{\nu}^*(m') A^{\mu\nu}
 = \psi_{\mu}(m) e_{\nu}^*(m') \sum_i \Lambda_i U_i^{\mu\nu},
\end{equation}
where $e_{\nu}(m')$ is the photon polarization four-vector, and
$m'$ the photon's helicity. In the tensor formalism, the main
task is to calculate the partial wave amplitude $U_i^{\mu}$
or $U_i^{\mu\nu}$. Details on how to calculate them can be
found in the
literature~\cite{part1_pwa_18,part1_pwa_19,part1_pwa_21,part1_pwa_22}. \\

\subsection{Partial Wave Analysis}

Once the decay amplitude is known, the next task is
the calculation of the differential cross-section of the decay.
Suppose all decay processes
and all respective decay amplitudes are known for 
a given channel with $N$ different decay modes,
then the total differential cross section is
\begin{equation}  \label{3.1}
\frac{{\rm d}\sigma}{{\rm d} \Phi}
= \sum_{m,\lambda}
|\sum_{i=1}^N \sum_{\mu} A_i(m,\lambda, \mu)|^2
+BG,
\end{equation}
where$A_i(m,\lambda, \mu)$ denotes the decay amplitude for the 
$i^{\rm th}$ decay
mode, $m$ is the helicity of the parent particle,
$\lambda=(\lambda_1, \lambda_2, \cdots)$ denotes
the set of helicities of final state particles,
$\mu = (\mu_1, \mu_2, \cdots)$ denotes the set
of helicities of intermediate resonances,
${\rm d} \Phi$ is the element of phase space,
and $BG$ represents the non-interfering background.
\\

In the tensor formalism, the  total differential
decay rate is expressed in a different way.
For $J/\psi$  non-radiative decay, it is expressed
as
\begin{equation}  \label{3.2}
\frac{{\rm d}\sigma}{{\rm d} \Phi}
= \sum_{i,j} P_{ij} \cdot F_{ij},
\end{equation}
where
\begin{equation}  \label{3.3}
 P_{ij} =  P_{ji}^*
= \Lambda_i \Lambda_j^*,
\end{equation}
and
\begin{equation}  \label{3.4}
F_{ij} =  F_{ji}^*
= \frac{1}{2} \sum_{\mu=1}^2 U_i^{\mu} U_j^{\mu *}.
\end{equation}
For $J/\psi$ radiative decay, Eqs.~(\ref{3.2}) and
(\ref{3.3}) still can be used, but Eq.~(\ref{3.4}) changes to
\begin{equation}  \label{3.5}
F_{ij} =  F_{ji}^*
= - \frac{1}{2} \sum_{\mu=1}^2
U_i^{\mu \nu} g^{\bot \bot }_{\nu \nu'} U_j^{\mu \nu' *}.
\end{equation}
\\

The normalized probability density function (PDF)
that is used to describe the decay process is
\begin{equation}  \label{3.6}
f(x,\alpha)
= \frac{{\rm d}\Gamma / {\rm d} \Phi}{\Gamma}
W(\Phi),
\end{equation}
where $x$ represents a set of quantities that are measured
experimentally, $\alpha$ represents some unknown parameters
that have to be determined by the fit, and $W(\Phi)$ represents
effects of detection efficiency. The total decay width,
$\Gamma$, is given by
\begin{equation}  \label{3.7}
\Gamma = \int W(\Phi)
\frac{{\rm d}\sigma}{{\rm d}\Phi}
{\rm d}\Phi.
\end{equation}
In PWA, the decay width is determined
by Monte Carlo integration,
\begin{equation}  \label{3.8}
\Gamma =  \frac{1}{N_{mc}} \sum_{j=1}^{N_{mc}}
\left (\sum_{m,\lambda}
|\sum_{i=1}^N \sum_{\mu} A_i(m,\lambda, \mu)|^2
+BG \right )_j ,
\end{equation}
where $N_{mc}$ is the total number of Monte Carlo
events, and the subscript $j$ indicates that the quantity is
evaluated for the $j^{\rm}$-th Monte Carlo event. It is necessary
that these Monte Carlo events are obtained from a complete
detector simulation and pass all of the selection 
conditions applied to the actual data sample.\\

The maximum likelihood method is
utilized in the fit. The likelihood function is
given by the adjoint probability density for all the
data, 
\begin{equation}  \label{3.9}
{\cal L} = \prod_{i=1}^{N_{events}}
f(x,\alpha),
\end{equation}
where  $N_{events}$ is the total number of events in the
channel. In the data analysis, the goal is to find the set of values,
$\alpha$, that minimize $S$, which is defined as 
\begin{equation}  \label{3.10}
S = - {\rm ln} {\cal L}.
\end{equation}
Different spin-parity intermediate resonances have different
angular distributions, and different PDFs are used to fit
to invariant mass spectra the angular distributions.
Because the better fits will have smaller values of $S$, we use the
information provided by the value of $S$ to determine 
the solution that gives the best fit and in this way 
discrimate between different hypotheses for the
spin-parity of a given resonance. \\

\subsection{Mass, Width and Spin-parity}

From the decay amplitude [Eq.~(\ref{2.10})],
we know that different spin-parity hypotheses
for an intermediate resonance
give different angular distributions.
Because the helicity coupling amplitude $F^J_{\lambda\nu}$ is
a slowly varing function of energy [see Eq.~(\ref{2.13})], we know
that the invariant mass spectrum is mainly dominated by
the mass and width of a resonance. 
Since the PWA fits both the
angular distributions and the invariant mass spectrum
simultaneously,  we can determine the mass, width and
spin-parity.\\

The spin parity of a resonance is determined mainly from the
fit to the angular distributions. As an example, we discuss
how to determine the spin-parity of the $\sigma$ particle in
$J/\psi \to \omega \pi \pi$ decays~\cite{part1_pwa_23,part1_pwa_24}. 
We know that the possible spin-parity 
values for a resonance in the $\pi\pi$
system produced in $J/\psi \to \omega \pi \pi$ decay
can only be $0^{++}$, $2^{++}$, $4^{++}$ etc. 
Each of these hypothesis is fit to the data and the 
one with the smallest $S$ value is selected; 
the spin-parity of the selected hypothesis
will be that of the $\pi\pi$  resnoance. \\

The mass and width of a resonance is determined 
in a diffferent
way, still using the $\sigma$ particle as an example.
Assuming the spin-parity of the $\sigma$ particle
is known, we first keep the masses and widths of all resonances
in this channel fixed except the mass of the $\sigma$ particle and
perform the likelihood fit.  The value of the mass of the $\sigma$
particle is changed step by step and the value of the likelihood
function of the corresponding fit is minimized.  
The mass value corresponding
to the minimum value of $S$ is the
measured mass of the $\sigma$ particle. 
A similar way is used to determine the width of the $\sigma$ particle.
This technique is called mass and width scanning. 

\subsection{Applications}

PWA is a powerful tool for the study of  hadron spectroscopy, in
particular, for the study of the structure in a spectrum of 
sequential decays. It has been widely
in BES physics analyses of $J/\psi$ and $\psi'$ decays,
and many meaningful results have been obtained. 
Here, we give a few examples to show how it works.

\begin{figure}[htbp]
\centerline{\epsfig{file=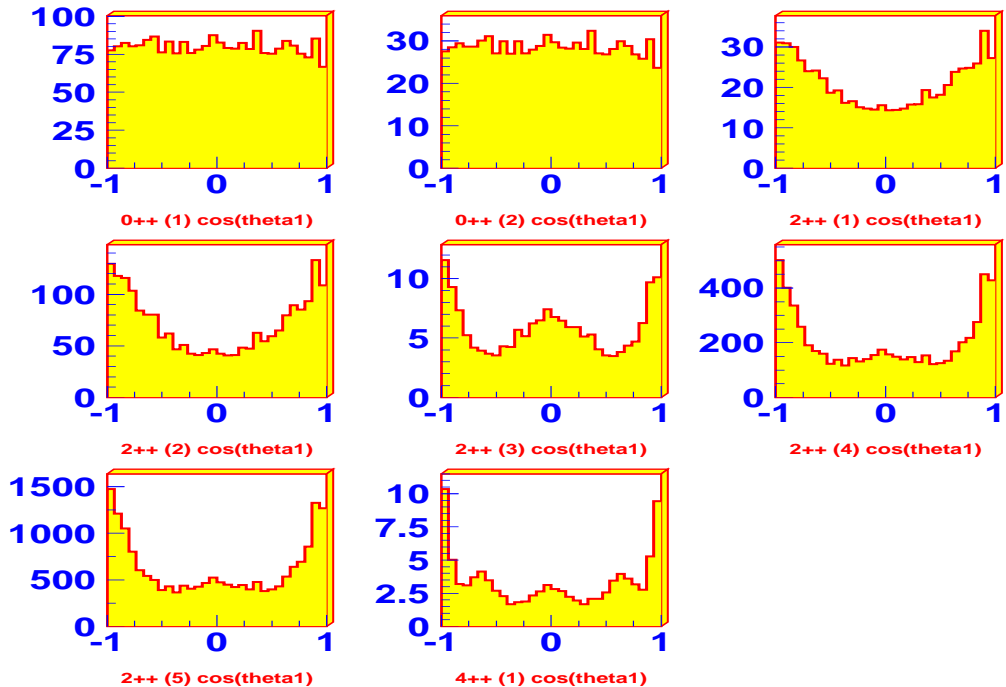,height=4.0in,width=6.0in}}
\caption[]{ Angular distributions for different spin-parity states. }
\label{w01}
\end{figure}

\begin{figure}[htbp]
\centerline{\epsfig{file=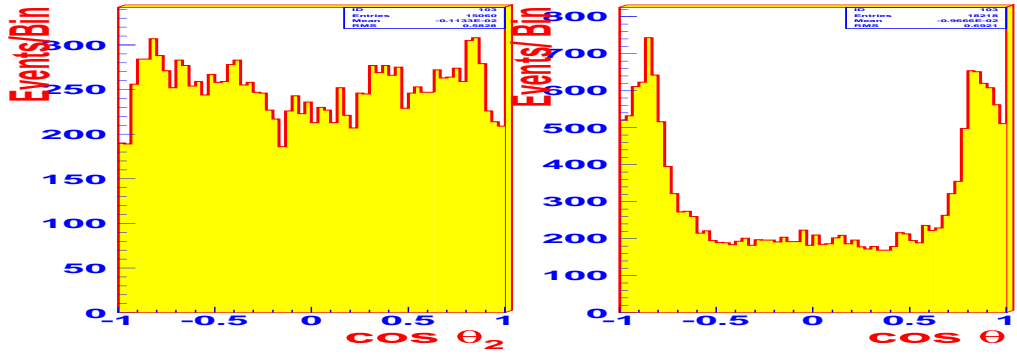,height=2.0in,width=6.0in}}
\caption[]{ LEFT: angular distribution for the $\sigma$ particle,
    RIGHT: angular distribution for the $f_2(1270)$ }
\label{w02}
\end{figure}

First, we discuss the $J/\psi \to \omega \pi \pi$
channel~\cite{part1_pwa_23,part1_pwa_24}. As stated above,
the possible spin-parities for a resonance in the $\pi\pi$
spectrum are $0^{++}$, $2^{++}$, $4^{++}$ etc. For a
$0^{++}$ resonance, there are two independent helicity
paramenters. For  $2^{++}$ and $4^{++}$ resonance, there
are five independent helicity paramenters. Different parameters
correspond to different angular distributions, especially for
the pion polar angle distribution in the $\pi\pi$
center of mass frame. Figure~\ref{w01} shows typical
pion polar angle distributions  in
the $\pi\pi$ center of mass frame. From the figure we can see
that the behavior of the angular distribution for different
spin-parities are quite different. For a $0^{++}$ resonance,
it is relatively flat; in contrast, for a $2^{++}$ resonance, 
it is concave, while for a $4^{++}$ resonance, it is 
severely concave with a complex superstructure.  
Figure~\ref{w02} compares the angular
distributions of the $\sigma$ particle 
(left panel) with the $2^{++}$ $f_2(1270)$ meson
(right panel) as a comparison.  
It is
clear that the angular distribution
of the $\sigma$ particle is quite similar to 
$0^{++}$ expectations, and the angular distribution 
of $f_2(1270)$ is that expected for a standard
$2^{++}$ state.

\begin{figure}[htbp]
\centerline{\epsfig{file=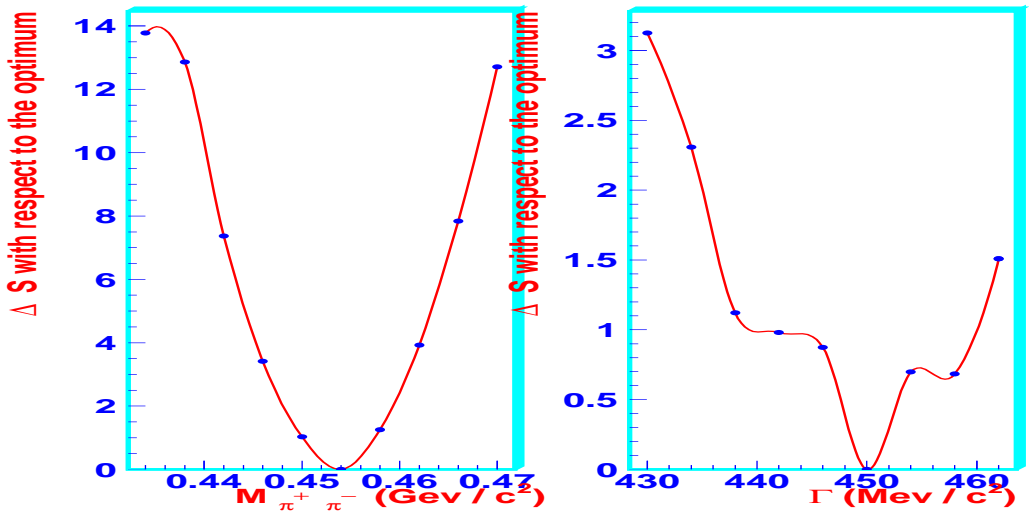,height=3.0in,width=6.0in}}
\caption[]{ Mass and width scans for the $\sigma$ particle. }
\label{w03}
\end{figure}

The mass and width of the $\sigma$ particle 
is determined by  mass and width scans; the corresponding change in $S$
is shown in Fig. \ref{w03}.  In both scan curves, minima are
clearly seen, and these correspond to the measured mass and width
of the $\sigma$ particle. \\

Using a similar method, we can determine masses, widths and spin-parities
of other resonances.
\\

An interesting example of a PWA is the $J/\psi \to \gamma VV$ 
class of decays,
such as $J/\psi \to \gamma \rho \rho$, $J/\psi \to \gamma \omega \omega$,
$J/\psi \to \gamma K^*(892) \bar K^*(892)$ etc., where 
important information is contained
in the distribution of the $\chi$ angle, which is the angle between
the decay plans of the two vector mesons. If the resonance is a
$0^{-+}$ meson, the distribution should follow
\begin{equation}  \label{5.1}
\frac{{\rm d} N}{{\rm d} \chi}
\sim \sin^2 \chi,
\end{equation}
while the expectation for a $0^{++}$ meson is
\begin{equation}  \label{5.2}
\frac{{\rm d} N}{{\rm d} \chi}
\sim 1 + \alpha \cos^2 \chi.
\end{equation}
These two distributions are quite different: 
for a $0^{-+}$ meson the number of events increases 
with increasing  $\chi$,
while for a $0^{++}$ meson  the number of events decreases
with increasing  $\chi$. 

Sometimes, angular distributions are completely determined by
the spin-parity of the resonance. One simple example is
$J/\psi \to \rho \pi$ with $\rho \to \pi \pi$, where the
differential angular distribution is
\begin{equation} \label{5.3}
\frac{{\rm d}\sigma}{{\rm d} \Omega} \sim
\sin^2 \theta_2 [ \cos^2 \varphi_2
+ \cos^2 \theta_1 \sin^2 \varphi_2 ],
\end{equation}
where $\theta_2$ and $\varphi_2$ are the pion polar angle and azimuthal 
angle distributions in the $\rho$ center of mass system, and $\theta_1$ is
the polar angle of the $\rho$ in the $J/\psi$ center of mass system.
In the $\rho$ center of mass system, the angular distribution
for the  polar angle $\theta_2$ is
\begin{equation} \label{5.4}
\frac{{\rm d} N}{{\rm d} \cos \theta_2} \sim
\sin^2 \theta_2,
\end{equation}
while that for the azimuthal angle $\phi_2$ is
\begin{equation} \label{5.5}
\frac{{\rm d} N}{{\rm d} \varphi_2} \sim
1 + 2 \cos^2 \varphi_2.
\end{equation}
On the right hand side of Eq.~(\ref{5.3}), there are no free 
parameters and, 
so, the angular distribution is completely specified by
the spin-parity of the $\rho$.
This is a special case;  in most cases,
the angular distributions are dependent of some unknown parameters,
and the magnitude of these parameters  effects the behavior of
angular distribution. \\

Relativistic effects also influence the angular
distributions of the decay particles.  As an example,
we discuss the case of $J/\psi$
decaying into two spin half particles:  $J/\psi \to \Sigma \bar\Sigma$,
$J/\psi \to \Lambda \bar \Lambda$, $J/\psi \to e^+ e^-$,
and $J/\psi \to \mu^+ \mu^-$.
The differential angular distribution for this class of decay processes is
\begin{equation} \label{5.6}
\frac{{\rm d} \sigma}{{\rm d} \Omega} \sim
1 + \alpha \cos^2 \theta,
\end{equation}
where $\theta$ is the pole angle of a daughter particle in the
$J/\psi$ center of mass system, and
\begin{equation} \label{5.7}
\alpha = \frac{|F^{1}_{\frac{1}{2} -\frac{1}{2}}|^2
    - 2 |F^{1}_{\frac{1}{2} \frac{1}{2}}|^2}
    {|F^{1}_{\frac{1}{2} -\frac{1}{2}}|^2
    + 2 |F^{1}_{\frac{1}{2} \frac{1}{2}}|^2}.
\end{equation}
For the case that the $J/\psi$ decays into a baryon and anti-baryon pair,
the velocity of the baryon is non-relativistic, so the parameter $\alpha$
can take any value between -1 and 1. But for $J/\psi$ decaying into
$e^+ e^-$ or $\mu^+ \mu^-$, the final state particle is relativistic.
In this case, the helicity coupling amplitude
$F^{1}_{\frac{1}{2} \frac{1}{2}}$ vanishes and $\alpha=1$.
\\

\subsection{Discussions}

Partial wave analysis is a powerful tool for the study of hadron
spectroscopy: it can simultaneously determine
the mass, width, branching ratio and
spin-parity of a resonance.  However, 
the theoretical calculation of the formulae
used in the PWA analysis can be quite complicated.
 Moreover, there are many practical difficulties in the application
of PWA.  For example, a typical PWA requires a 
enormous amounts of CPU time and lots of memory;  at 
some point, it becomes impractical. 
New computing methods are needed to
facilitate PWA at \bes3, where the statistics 
for many interesting channels will be huge. 

For known background sources, Monte Carlo techiniques can be 
used to simulate their behavior and fit the data.  Alternatively,
one can write out the theoretical
formula for the relative decay amplitude and directly fit it 
as part of the PWA.  In other cases, however,
the origins of backgrounds 
are not completely known, which introduces uncertainties
into the PDFs  used to model them.   Often,
these are quite similar to phase space and a non-interfering
constant amplitude can be used in the fit.  However, this is
not always the case, and additional free parameters may have
to be added to the likelihood, thereby consuming additional
memory and CPU time.

A commonly used PWA technique is a bin-by-bin fit.
In this case,  the parameter space is divided into many small bins.
In each bin, if the its size is small enough, one can
approximately assume that the amplitude and phase are constant. 
By analysing the data bin-by-bin, one can obtain the magnitudes of
the amplitude and phase of a resonance in each bin, which
gives direct measurements of the variation of the amplitude 
and phase with mass.
Thus, a bin-by-bin fit can enable one to determine the phase motion of
a resonance. Precise measurements of the phase motion is important for
theoretical analyses. \\

In physics analyses, most intermediate resonances 
that are encountered are relatively narrow and their mass 
positions are far from thresholds.  Sometimes, however,
we have to deal with wide resonances that are close to 
threshold. In these cases, the traditional 
Breit-Wigner function, which is an approximate description
that is only valid for narrow resonances far away from the
threshold, is not applicable.  To date, there does not exist 
a mature description for the shape of a wide near-threshold 
resonance that is widely
accepted.  For these cases, when different
Breit-Wigner forms are used  in the fit, the 
 masses and widths that are determined can be quite different. 
In fact, in these cases,  the 
masses and widths  derived directly from the Breit-Wigner function are 
not  the physical masses and 
widths of the resonances. The quantity with the most  physical 
significance is  the pole position, and the
physical mass and width of the resonance should be calculated
from it.  In   analyses of broad,
near-threshold resonances, we find
that while the masses and widths derived from different Breit-Wigner
forms are completely different, the pole positions
are approximately the same. Therefore, in these
cases, it is best to use pole positions to describe these resonances. \\

\section[Dalitz-plot Analysis Formalism]{Dalitz-plot Analysis Formalism\footnote{By David Asner}}


%
Originally the primary application of Dalitz-plot analyses was to determine
the spin and parity of light mesons. Recently Dalitz-plot analyses have 
emerged as a powerful tool in the study of $D$ and $B$ mesons.

Charmed meson decay dynamics have been studied extensively over the last
decade. 
Recent studies of multi-body decays of charmed mesons probe a variety
of physics including doubly Cabibbo suppressed 
decays~\cite{cleoa,cleob,Link:2004mx}, 
searches for $CP$ 
violation~\cite{cleob,Kopp:2000gv,cleoacp,cleopipipi0,babarD2KKpi}, 
$T$ violation~\cite{focust}, 
$D{^0}\hbox{--}\overline D{^0}$ mixing~\cite{cleoc,mixing}, 
the properties
of established light 
mesons~\cite{e687pipipi,e791dspipipi,focuspipipi,cleoksetapi0}, 
the properties of 
$\pi\pi$~\cite{cleoa,focuspipipi,e791pipipi},
$K\pi$~\cite{e791kpipi,e791kpipia}, and $KK$~\cite{babardtokkk}
$S$-wave states, and 
the dynamics of four-body final 
states~\cite{Link:2003pt,focuskkpipi}.

Recently $B$ meson decay dynamics have been studied.
Multi-body decays of $B$ mesons also probe a variety
of physics including, charmless 
$B$-decays~\cite{Garmash:2004wa,Wang:2005fc,Aubert:2004bt,Aubert:2005ce,Aubert:2005kd}, 
measurements of the Cabibbo-Kobayashi-Maskawa (CKM) angle $\gamma/\phi_3$
\cite{Poluektov:2004mf,Abe:2004gu,Abe:2005ct,BabarKspipiBW,Aubert:2005yj},
searches for direct $CP$ 
violation~\cite{Aubert:2004bt,Aubert:2005ce,Garmash:2005ji}, charm 
spectroscopy~\cite{Abe:2003zm,Abe:2004cw}, the properties
of established light mesons
\cite{Garmash:2004wa,Aubert:2005ce,Aubert:2005kd}, 
the properties of $KK$~\cite{Garmash:2004wa,Aubert:2005kd} 
and $K\pi$~\cite{Garmash:2004wa,Aubert:2004bt,Aubert:2005ce} 
$S$-waves,
and the three-body production of 
baryons~\cite{Wang:2005fc,Abe:2004sr}.  Time-dependent Dalitz-plot (TD) 
analyses have been used to determine the CKM angle
$\alpha/\phi_2$ with $B\to \pi^+\pi^-\pi^0$~\cite{Aubert:2004iu}
and to resolve the two-fold ambiguity in the CKM angle $\beta/\phi_1$ with
$B\to D\pi^0, D\to K^0_S\pi^+\pi^-$~\cite{Bondar:2005gk,Abe:2005zz}. 
A TD analysis of 
$B^0 \to D^{*\pm}K^0\pi^\mp$~\cite{Aubert:2004at} is sensitive 
$\gamma/\phi_3$.
Future studies could improve sensitivity to new physics in TD
analyses of $b \to s$ penguin decays~\cite{Aubert:2005kd}.

Additionally, partial wave analyses have been used to study the
dynamics of charmonium decays to hadrons, following the formalism
presented in Refs.~\cite{Zou:2002ar,Dulat:2005in}, in radiative 
decays~\cite{Bai:2003ww,Bai:2000ss,Bai:1999mm,Bai:1999mk}
and in decays to all hadronic final 
states~\cite{Ablikim:2005kp,Ablikim:2005jy,Ablikim:2004qn,Ablikim:2004st}. 
Multi-body decays of charmonium to all hadronic final states
can be analyzed with the Dalitz-plot analysis technique.
Studies of the $\pi\pi$, $K\pi$ and $KK$ $S$-wave in charmonium
decays probe most of the phase space accessible in $B$ decays.
Thus, Dalitz-plot analyses of charmonia could lead to reduced 
systematic errors in many $B$ analyses.


Weak nonleptonic decays of $B$ and charmed mesons are expected to 
proceed dominantly through resonant two-body decays in several 
theoretical models~\cite{theory};
see Ref.~\cite{JDJackson} for a review of resonance phenomenology.
These amplitudes are typically calculated with the Dalitz plot analysis 
technique~\cite{dalitz}, which uses
the mininum number of independent observable quantities. 
For the three-body decay of a spin-0 particle to all pseudo-scalar final 
states, $D, B \to abc$, the decay rate is 
\begin{equation}
\Gamma = \frac{1}{(2\pi)^3 32 \sqrt{s^3}}\left|\cal M\right|^2dm^2_{ab}dm^2_{bc},
\label{eqn:gamma}
\end{equation}
where $m_{ij}$ is the invariant mass of $i-j$ and the coefficient of the 
amplitude includes all kinematic factors.
The scatter plot in $m^2_{ab}$ versus $m^2_{bc}$ is called a Dalitz plot. 
If $\left|\cal M \right|^2$ is constant the allowed region of the plot 
will be populated uniformly with events.
Any variation in $\left|\cal M \right|^2$ over the Dalitz plot
is due to dynamical rather than kinematical effects.
It is straightforward to extend the formalism beyond three-body final states.
For $N$-body final states, phase space has dimension $3N-7$. Other cases
of interest include one vector particle or a fermion/anti-femion pair 
(e.g. $B \to D^*\pi\pi$, $B \to \Lambda_c p \pi$, $B \to K \ell \ell$)
in the final state. For the former case phase space has dimension $3N-5$ and 
for the latter two $3N-4$.


\medskip

The amplitude of the process,  $R \rightarrow rc, r \rightarrow ab$
where $R$ is a  $D$, $B$, or $q \bar q$ meson and $a$,$b$,$c$ are 
pseudo-scalars, is given by 
\begin{eqnarray}
& {\cal M}_r(J,L,l,m_{ab},m_{bc}) = \sum_\lambda \left<ab|r_\lambda\right>T_r(m_{ab}) \left<cr_\lambda|R_J\right> 
\label{eqn:a} \\ \nonumber
 & =  Z(J,L,l,\vec{p},\vec{q}) B^{R}_L(|\vec{p}|) B^{r}_L(|\vec{q}|) T_r(m_{ab}),
\end{eqnarray}
where the sum is over the helicity states $\lambda$ of the intermediate resonance particle $r$, 
$a$ and $b$ are the daughter
particles of the
resonance $r$, $c$ is the spectator particle, 
$J$ is the total angular momentum of $R$, 
$L$ is the orbital angular momentum between $r$ and $c$,
$l$ is the orbital angular momentum between $a$ and $b$ equivalent to the
spin of $r$, 
$\vec{p}$ and $\vec{q}$ are the three-momenta of $c$ and $a$, 
respectively, in the $r$ rest frame, $Z$ 
describes the angular distribution of final state particles, $B^{R}_L$ and $B^{r}_L$
are the barrier factors for the production of $rc$ and $ab$, respectively,
with angular momentum $L$, and
$T_r$ is the dynamical function describing the resonance 
$r$. 
The amplitude for modeling the Dalitz plot is a phenomenological object.
Differences in the
parameterizations of $Z$, $B_L$ and $T_r$, as well as the set of 
resonances $r$, complicate the comparison of results from 
different experiments. 

Usually the resonances are modeled with a
Breit-Wigner form although
some more recent analyses have used the $K$-matrix 
formalism~\cite{Kmatrix,Au:1986vs,Anisovich:2002ij}
with the $P$-vector approximation~\cite{aitchison} to describe the 
$\pi\pi$ $S$-wave. 

The nonresonant (NR) contribution to $D\to abc$ is parameterized as 
constant  ($S$-wave) with no variation in magnitude or phase across the 
Dalitz plot.
The available phase space is much greater for $B$ decay and the nonresonant 
contribution to $B\to abc$ requires a more sophisticated parameterization.
Theoretical models of the NR 
amplitude~\cite{Fajfer:1998yc,Cheng:2002qu,Fajfer:2004cx,Cheng:2005ug} 
do not reproduce the
distributions observed in the data. Experimentally, several 
parameterizations have been used~\cite{Garmash:2004wa,Aubert:2005kd}.

\subsection{Barrier Factor $B_L$}
The maximum angular momentum $L$ in a strong decay is limited by the linear
momentum $\vec{q}$. Decay particles moving slowly with an impact parameter 
(meson radius) $d$ of 
order 1~fm have difficulty generating sufficient angular momentum to 
conserve the spin of the resonance.  
The Blatt-Weisskopf~\cite{Blatt-Weisskopf,Hippel-Quigg} 
functions $B_L$, given in Table~\ref{tab:barrier},
weight the reaction amplitudes to account for this spin-dependent effect.
These functions are normalized to give $B_L=1$ for 
$z=(\left|{\vec q}\right|d)^2=1$.
Another common formulation $B^\prime_L$, also in Table~\ref{tab:barrier}, 
is 
normalized to give $B^\prime_L=1$ for $z=z_0=(\left|{\vec q_0}\right|d)^2$
where $q_0$ is the value of $q$ when $m_{ab}=m_r$. 

\begin{table}[htbp]
\centering
\begin{tabular}{ccc} \hline
$L$  & $B_L(q)$  & $B^\prime_L(q,q_0)$  \\ 
\hline
0  & 1 & 1  \\ \\
1 & $\sqrt{\frac{2z}{1+z}}$  & $\sqrt{\frac{1+z_0}{1+z}}$ \\  \\
2 & $\sqrt{\frac{13z^2}{(z\!-\!3)^2\!+\!9z}}$ &  $\sqrt{\frac{(z_0\!-\!3)^2\!+\!9z_0}{(z\!-\!3)^2\!+\!9z}}$ \\  \\
&  where $z = (\left|{\vec q}\right|d)^2$ & and $z_0= (\left|{\vec q_0}\right|d)^2$\\  \hline
\end{tabular}
\caption
{Blatt-Weisskopf barrier factors.}
\label{tab:barrier}
\end{table}

\subsection{Angular Distributions}
The tensor, or Zemach formalism~\cite{Zemach,Filippini:yc} and
the helicity formalism~\cite{helicity,Filippini:yc} yield identical
descriptions of the angular distributions for the decay process 
$R\to rc, r\to ab$
for reactions where $a$, $b$ and $c$ are spin-0 and the initial state is
unpolarized.
In this scenario, the angular distributions for $J=0,1,2$ are 
given in Table~\ref{tab:angular}.
For polarized initial states, the helicity formalism~\cite{helicity} is 
used to 
determine the distinct angular distribution for each helicity state 
$|\lambda|$.
The angular distributions for $J=1,2$ for a polarized initial 
are given in Table~\ref{tab:angularpolarized}.
The sign of the helicity cannot be determined from the Dalitz plot alone 
when $a$, $b$ and $c$ are spin-0.
For final-state particles with non-zero spin 
(e.g. radiative charmonium decays), the helicity formalism
is required.

For the decays of pseudoscalars to three pseudoscalars the formalism
simplifies considerably as the angular distribution $Z$ depends only 
on the spin $l$ of resonance $r$. 
Since $J=0$ and $L=l$, only the first three
rows of Table~\ref{tab:angular} are required.

\begin{table}
\centering
\begin{tabular}{lc} \hline
$J\to l+L$  & Angular Distribution\\ \hline
$0 \!\to\! 0\!+\!0$  & uniform  \\ 
$0 \!\to\! 1\!+\!1$  & $(1\!+\!\zeta^2)\cos^2\theta$  \\ 
$0 \!\to\! 2\!+\!2$  & $(\zeta^2\!+\!\frac{3}{2})^2(\cos^2\theta \!-\! 1/3)^2$  \\ 
$1 \!\to\! 0\!+\!1$  & uniform \\ 
$1 \!\to\! 1\!+\!0$  & $1\!+\!\zeta^2\cos^2\theta$ \\ 
$1 \!\to\! 1\!+\!1$  & $\sin^2\theta$  \\ 
$1 \!\to\! 1\!+\!2$  & $1\!+\!(3\!+\!4\zeta^2)\cos^2\theta$ \\ 
$1 \!\to\! 2\!+\!1$  & $(1\!+\!\zeta^2)[1\!+\!3\cos^2\theta\!+\!9\zeta^2(\cos^2\theta \!-\! 1/3)^2]$ \\ 
$1 \!\to\! 2\!+\!2$  &  $(1\!+\!\zeta^2)\cos^2\theta\sin^2\theta $ \\ 
$2 \!\to\! 0\!+\!2$  & uniform \\ 
$2 \!\to\! 1\!+\!1$  & $3\!+\!(1\!+\!4\zeta^2)\cos^2\theta$ \\ 
$2 \!\to\! 1\!+\!2$  & $\sin^2\theta$ \\ 
$2 \!\to\! 2\!+\!0$  & $1\!+\!\frac{\zeta^2}{3}\!+\!\zeta^2\cos^2\theta \!+\! \zeta^4(\cos^2\theta \!-\! 1/3)^2$ \\ 
$2 \!\to\! 2\!+\!1$  & $1\!+\!\frac{\zeta^2}{9}\!+\!(\frac{\zeta^2}{3}\!-\!1)\cos^2\theta \!-\!\zeta^2(\cos^2\theta\!-\!1/3)^2$ \\ 
$2 \!\to\! 2\!+\!2$  & $1\!+\!\frac{\zeta^2}{9}\!+\!(\frac{\zeta^2}{3}\!-\!1)\cos^2\theta\!+\!\frac{(16\zeta^4 \!+\! 21\zeta^2 \!+\!9)(\cos^2\theta \!-\! 1/3)^2}{3}$ \\ \hline
\end{tabular}
\caption
{Angular distributions for each $J,L,l$ for {\it unpolarized} initial states
where $\theta$ is the  angle between particles $a$ and $c$ in the rest 
frame of resonance $r$,  $\sqrt{1+\zeta^2}$ is a relativistic correction 
with $\zeta^2 = E^2_r/m^2_{ab} - 1$,  and $E_r = 
(m^2_{R}+m^2_{ab}-m^2_c)/2m_{R}$.}
\label{tab:angular}
\end{table}

\begin{table}
\centering
\begin{tabular}{lc} \hline
$J\to l+L$  & Angular Distribution\\ \hline
$1 \!\to\! 1\!+\!0$  & $F_0 \gamma^2 \cos^2\theta + F_1 \sin^2\theta$ \\ 
$1 \!\to\! 1\!+\!1$  & $F_1\sin^2\theta$  \\ 
$1 \!\to\! 1\!+\!2$  & $F_0(2\gamma/3)^2\cos^2\theta+F_1(1/9)\sin^2\theta$\\ 
$1 \!\to\! 2\!+\!1$  & $2F_0 \gamma^4(\cos^2\theta-1/3)^2$ \\ 
& $ + F_1 \gamma^2[2/9\!+\!2/3\cos^2\theta\!-\!2(\cos^2\theta-1/3)^2]$ \\
$1 \!\to\! 2\!+\!2$  &  $F_1 \gamma^2\cos^2\theta\sin^2\theta $ \\ 
$2 \!\to\! 1\!+\!1$  & $F_0(2\gamma^2/3)\cos^2\theta^2 + F_1(1/2)\sin^2\theta$ \\ 
$2 \!\to\! 1\!+\!2$  & $F_1\sin^2\theta$ \\ 
$2 \!\to\! 2\!+\!0$  & $F_0(4\gamma^4/3+4\gamma^2/3+1/3)(\cos^2\theta \!-\! 1/3)^2 $\\ & $+ F_1 \gamma^2[4/9+4/3\cos^2\theta - 4(\cos^2\theta\!-\!1/3)^2]$ \\ 
& $+ F_2[8/9 - 4/3 \cos^2\theta + (\cos^2\theta\!-\!1/3)^2]$ \\ 
$2 \!\to\! 2\!+\!1$  & $F_1 \gamma^2[1/9\!+\!1/3\cos^2\theta -(\cos^2\theta\!-\!1/3)^2]$\\ & $ + F_2[8/9 - 4/3\cos^2\theta + (\cos^2\theta\!-\!1/3)^2]$ \\ 
$2 \!\to\! 2\!+\!2$  & $3F_0 (4\gamma^2/9\!-\!1/9)^2 (\cos^2\theta\!-\!1/3)^2$\\ & $  + F_1 \gamma^2[1/9\!+\!4/3\!-\!(\cos^2\theta\!-\!1/3)^2]/9$ \\ & $+ F_2 [8/9\!-\!4/3\cos^2\theta + (\cos^2\theta\!-\!1/3)^2]/9$ \\ \hline
\end{tabular}
\caption
{Angular distributions for $J\neq 0,L\neq 0,l$ for {\it polarized} 
initial states
where $\cos\theta$ is the angle between particles $a$ and $c$ in the rest frame of resonance $r$,
$\gamma = E_r/m_{ab}$, and $E_r = (m^2_{R}+m^2_{ab}-m^2_c)/2m_{R}$.
$F_\lambda$ denotes the fraction of the initial state in helicity
state $\lambda$. For unpolarized initial states setting $F_\lambda$=1 recovers
the angular distributions obtained from the Zemach formalism shown 
in Table~\ref{tab:angular}.}
\label{tab:angularpolarized}
\end{table}


\medskip
\subsection{The Dynamical Function $T_R$}
The dynamical function $T_r$ is derived from the $S$-matrix formalism. 
In general, the amplitude for a final state $f$ to couple to an initial 
state $i$ is
$S_{fi} = \left<f|S|i\right>$,
where the scattering operator $S$ is unitary
and satisfies $SS^\dagger=S^\dagger S=I$. 
The transition operator 
${\hat T}$ is defined by separating the probability that $f=i$ yielding,
%
\begin{equation}
S = I + 2iT = I + 2i\left\{\rho\right\}^{1/2}{\hat T}\left\{\rho\right\}^{1/2},
\label{eqn:s}
\end{equation} 
where $I$ is the identity operator,  ${\hat T}$ is Lorentz invariant 
transition operator,
$\rho$ is the diagonal phase space matrix  where $\rho_{ii}=2q_i/m$ and 
$q_i$ is the momentum of $a$ in the $r$ rest frame
for decay channel $i$.
In the single channel $S$-wave scenario,
$S=e^{2i\delta}$ 
satisfies unitarity and implies 
\begin{equation}
{\hat T}=\frac{1}{\rho}e^{i\delta}\sin\delta.
\label{eqn:t}
\end{equation} 
transition operator.

There are three common formulations of the dynamical function. The
Breit-Wigner formalism
is the simplest formulation
 - the first term in a Taylor expansion about a $T$ matrix pole.
The $K$-matrix formalism~\cite{Kmatrix} is more general 
(allowing more than one $T$ matrix pole
and coupled channels while preserving unitarity). The Flatt\'e distribution\cite{flatte} is used to parameterize resonances near threshold and is equivalent to a one-pole,
two-channel $K$-matrix.

\subsection{Breit-Wigner Formulation}
The common formulation of a Breit-Wigner resonance decaying to spin-0 
particles $a$ and $b$ is
\begin{equation}
T_r(m_{ab}) = \frac{1}{m^2_r - m^2_{ab} - im_r \Gamma_{ab}(q)}
\label{eqn:bw}
\end{equation}
where the ``mass dependent'' width $\Gamma$ is

\begin{equation}
	\Gamma = \Gamma_r \left(\frac{q}{q_r}\right)^{2L+1}
\left(\frac{m_r}{m_{ab}} \right) B^\prime_L(q,q_0)^2
\label{eqn:gammaab}
\end{equation}
where $B^\prime_L(q,q_0)$ is  the Blatt-Weisskopf barrier factor from 
Table~\ref{tab:barrier}.
A Breit-Wigner parameterization best describes isolated, non-overlapping 
resonances far from the threshold of additional decay channels. 
For the $\rho$ and $\rho(1450)$ a more complex parameterization 
as suggested by Gounaris-Sakarai~\cite{gs} is often 
used~\cite{Aubert:2004bt,Abe:2005ct,Aubert:2005yj,Aubert:2004iu}.

Unitarity can be violated when the dynamical function is parameterized as the sum
of two or more overlapping Breit-Wigners.
The proximity of a threshold to the resonance shape distorts the line
shape from a simple Breit-Wigner. 
This scenario is described by the Flatt\'e formula and is discussed below.


\subsection{$K$-matrix Formalism}
The $T$ matrix can be described as
\begin{equation}
{\hat T} = (I - i{\hat K}\rho)^{-1} {\hat K},
\label{eqn:t3matrix}
\end{equation}
where 
${\hat K}$ is the Lorentz invariant
$K$-matrix describing the 
scattering process and 
$\rho$ is the phase space factor.

Resonances appear as a sum of poles in the $K$-matrix
\begin{equation}
{\hat K}_{ij} = \sum_\alpha \frac{\sqrt{m_\alpha \Gamma_{\alpha i}(m) m_\alpha \Gamma_{\alpha j}(m) }}
{(m^2_\alpha - m^2)\sqrt{\rho_i\rho_j}}.
\end{equation}
The $K$-matrix is real by construction, thus the associated $T$-matrix 
respects unitarity.

For the special case of a single channel, single pole we obtain
\begin{equation}
K = \frac{m_0\Gamma(m)}{m^2_0 - m^2},
\end{equation}
and
\begin{equation}
T = K(1-iK)^{-1} =  \frac{m_0\Gamma(m)}{m^2_0 - m^2 -im_0\Gamma(m)},
\end{equation}
which is the relativistic Breit-Wigner formula.
For the special case of a single channel, two poles we have
\begin{equation}
K = \frac{m_\alpha\Gamma_\alpha(m)}{m^2_\alpha - m^2} 
+ \frac{m_\beta\Gamma_\beta(m)}{m^2_\beta - m^2} ,
\end{equation}
and in the limit that $m_\alpha$ and $m_\beta$ are far apart relative to the widths we
can approximate the $T$ matrix as the sum of two Breit-Wigners,
$T(K_\alpha+K_\beta) \approx T(K_\alpha) + T(K_\beta)$,
\begin{equation}
T \approx  \frac{m_\alpha\Gamma_\alpha(m)}{m^2_\alpha - m^2 -im_\alpha\Gamma_\alpha(m)} 
+  \frac{m_\beta\Gamma_\beta(m)}{m^2_\beta - m^2 -im_\beta\Gamma_\beta(m)}.
\label{eqn:t2matrix}
\end{equation}
In the case of two nearby resonances 
Eq.~\ref{eqn:t2matrix} is not valid and exceeds
unity (and hence $T$ violates unitarity).

This formulation, 
which applies to $S$-channel production in two-body scattering
$ab \to cd$, 
can be generalized to describe the production of resonances
in other processes, such as the decay of charmed mesons. 
The key assumption here
is that the two-body system described by the  $K$-matrix does {\it not} 
interact with the rest of the final state~\cite{aitchison}. The quality of 
this assumption varies with 
the production process and is appropriate for scattering experiments like $\pi^- p \to \pi^0\pi^0 n$, radiative decays such as $\phi, J/\psi \to \gamma \pi\pi$ and semileptonic decays such as $D \to K\pi\ell\nu$. This assumption may be of limited validity
for production processes such as $p{\overline p} \to \pi\pi\pi$ or $D \to \pi\pi\pi$.
In these scenarios the two-body Lorentz invariant amplitude, ${\hat F}$, is given as
\begin{equation}
{\hat F}_i = (I - i {\hat K}\rho)_{ij}^{-1} {\hat P}_j = 
({\hat T}{\hat K}^{-1})_{ij}{\hat P}_j ,
\label{eqn:tmatrix}
\end{equation}
where 
$P$ is the production vector that parameterizes the resonance
production in the open channels. 

For the $\pi\pi$ $S$-wave,
a common formulation of the 
$K$-matrix~\cite{Anisovich:2002ij,focuspipipi,Aubert:2005yj} is
\begin{equation}
K_{ij}\!(s)\!=\!\left\{ \sum_\alpha \frac{g_i^{(\alpha)}g_j^{(\alpha)}}{m_\alpha^2\!-\!s}\!+\!f_{ij}^{sc}\frac{1 \!-\! s_0^{sc}}{s\!-\!s_0^{sc}}\right\}
\times\frac{s\!-\!s_A/2m^2_{\pi}}{(s\!-\!s_{A0})(1\!-\!s_{A0})}.
\label{eqn:kmatrix}
\end{equation}
The factor $g_i^{(\alpha)}$ is the real
coupling constant of the $K$-matrix pole
$m_\alpha$ to meson channel $i$; the parameters $f_{ij}^{sc}$ and $s_0^{sc}$
describe a smooth part of the $K$-matrix elements; the multiplicative factor 
$\frac{s-s_A/2m_\pi^2}{(s\!-\!s_{A0})(1\!-\!s_{A0})}$ 
suppresses a false
kinematical singularity near the $\pi\pi$ threshold - the Adler zero; and 
the number 1 has units ${\rm GeV}^2$.

The production vector, with $i=1$ denoting $\pi\pi$, is 
\begin{equation}
P_{j}\!(s) = \left\{ \sum_\alpha \frac{\beta_\alpha g_j^{(\alpha)}}{m_\alpha^2\!-\!s}+ f_{1j}^{pr}\frac{1 \!-\! s_0^{pr}}{s\!-\!s_0^{pr}}\right\}
 \times  \frac{s\!-\!s_A/2m^2_{\pi}}{(s\!-\!s_{A0})(1\!-\!s_{A0})}\,,
\label{eqn:pvector}
\end{equation}
where the free parameters of the Dalitz plot fit are the complex production 
couplings $\beta_\alpha$,
and the production vector background parameters $f_{1j}^{pr}$ and $s_0^{pr}$.
All other parameters are fixed by scattering experiments. 
Reference~\cite{Au:1986vs} 
describes the $\pi\pi$ scattering data with a 4 pole, 2 channel 
($\pi\pi$, $KK$) model while Ref.~\cite{Anisovich:2002ij} describes the scattering data with
5 pole, 5 channel ($\pi\pi$, $KK$, $\eta\eta$, $\eta^\prime\eta^\prime$ and 
$4\pi$) model. The former has been implemented by CLEO~\cite{cleopipipi0}
and the latter by FOCUS~\cite{focuspipipi} and BaBar~\cite{Aubert:2005yj}.
In both cases only the $\pi\pi$ channel was analyzed. A more complete
coupled channel
analysis would simultaneously fit all final states accessible by rescattering.

\subsection{Flatt\'e Formalism}
The scenario where another channel opens close to the resonance position can be
described by the Flatt\'e formulation
\begin{equation}
{\hat T(m_{ab})}\!=\!\frac{1}{m^2_r-m_{ab}^2-i(\rho_1g_1^2+\rho_2g_2^2)},\,\,\hfill\,\,g_1^2+g_2^2 = m_r\Gamma_r.
\label{eqn:flatte}
\end{equation}
This situation occurs in the $\pi\pi$ S-wave where the $f_0(980)$ is near the $K{\overline K}$ threshold and in the $\pi\eta$ channel where the $a_0(980)$ also lies near $K{\overline K}$ threshold. For the $a_0(980) $ resonance 
the relevant coupling constants are $g_1=g_{\pi\eta}$ and $g_2 = g_{KK}$
and the phase space terms are $\rho_1=\rho_{\pi\eta}$ and 
$\rho_2=\rho_{KK}$, where
\begin{equation}
\rho_{ab} = 
\sqrt{\left(1-(\frac{m_a-m_b}{m_{ab}})^2\right)\left(1+(\frac{m_a-m_b}{m_{ab}})^2\right)} 
. \label{eqn:a980phase}
\end{equation}
For the $f_0(980)$ the relevant coupling constants are 
$g_1 = g_{\pi\pi}$ and $g_2 = g_{KK}$ 
and the phase space terms are $\rho_1=\rho_{\pi\pi}$ and 
$\rho_2=\rho_{KK}$. The charged
and neutral $K$ channels are usually assumed to have the same coupling 
constant but separate phase space factors due to $m_{K^+} \neq m_{K^0}$
resulting in
\begin{equation}
\rho_{KK} \!=\!\frac{1}{2}\left(\sqrt{1 \!-\! \left(\frac{2m_{K^\pm}}{m_{KK}}\right)^2} \!+\! \sqrt{1 \!-\! \left(\frac{2m_{K^0}}{m_{KK}}\right)^2}\;\right)\,.
\label{eqn:f980phase}
\end{equation}


\subsection{Branching Ratios from Dalitz Fits}
The fit to the Dalitz plot distribution using either the Breit-Wigner or the
$K$-matrix formalism factorizes into a resonant contribution to the amplitude ${\cal M}_j$
and a complex coefficient, $a_j e^{i\delta_j}$, 
where $a_j$ and $\delta_j$ are real. 
The definition of a rate of a single process, given a set of
amplitudes $a_j$ and phases $\delta_j$ is the square of 
the relevant matrix
element (see Eq.~\ref{eqn:gamma}). 
In this spirit, the fit fraction is usually defined as
the integral over the Dalitz plot ($m_{ab}$ vs $m_{bc}$) 
of a single amplitude
squared divided by the integral over the Dalitz plot
of the square of the coherent sum of all amplitudes,
\begin{equation}
{\rm Fit\, Fraction}_j =  \frac{\int \left| 
a_j e^{i\delta_j}{\cal M}_j
\right|^2 dm^2_{ab}dm^2_{bc}}
{ \int \left| \sum_k  
a_k e^{i\delta_k}
{\cal M}_k \right|^2
dm^2_{ab}dm^2_{bc}} ,
\label{eqn:FF}
\end{equation}
where ${\cal M}_j$ is defined by Eq.~\ref{eqn:a} and 
described in Ref.~\cite{Kopp:2000gv}.
The sum of the fit fractions for all components
will in general not be unity due to interference.

It should be noted that when the $K$-matrix description
  in Eq.~\ref{eqn:tmatrix} is used to describe a wave (e.g. $\pi\pi$ 
$S$-wave) then
  ${\cal M}_j$ refers to the entire wave.  In these circumstances, it
  may not be straightforward to separate it into a sum of individual
  resonances unless these are narrow and well separated, in which
  case Eq.~\ref{eqn:t2matrix} can be used.


\subsubsection{Reconstruction Efficiency}
The efficiency for reconstructing an event as a function of position
on the Dalitz plot is in general non-uniform. 
Typically, a signal Monte Carlo sample generated with a uniform 
distribution in phase
space is used to determine the efficiency. 
The variation in efficiency across the Dalitz plot varies with experiment and decay mode.
Most recent analyses
utilize a full
GEANT~\cite{GEANT} detector simulation.


Finite detector resolution can usually be safely neglected as most resonances are
comparatively broad. Notable exceptions where detector resolution effects must be
modeled are $\phi \to K^+K^-$, $\omega \to \pi^+\pi^-$, 
and $a_0 \to \eta \pi^0$. One approach is to convolve the resolution function
in the Dalitz-plot variables $m^2_{ab}$, $m^2_{bc}$ with the function that
parameterizes the resonant amplitudes. 
In high statistics data samples resolution effects near the phase space boundary
typically contribute to a poor goodness of fit.
The momenta
of $a$,$b$ and $c$ $can$ be 
recalculated with a $R$ mass constraint. This forces the 
kinematical boundaries of the Dalitz plot to be strictly respected. 
If the three-body mass is not constrained, then the
efficiency (and the parameterization of background) may also depend on the
reconstructed mass. In fits to multi-body decays of charmonia and bottomonia
it is not appropriate to constrain the mass due to the finite natural width of 
the parent.

\subsubsection{Background Parameterization}
The contribution of background to the charm and $B$ samples varies by
experiment and final state. 
The background naturally falls into five categories:
(i) purely combinatoric background containing no resonances,
(ii) combinatoric background containing intermediate resonances, such as a real $K^{*-}$
or $\rho$, plus additional random particles,
(iii) final states containing identical particles as in $D^0 \to K^0_S\pi^0$
background to $D^0 \to \pi^+\pi^-\pi^0$ and $B \to D\pi$ background to $B\to K\pi\pi$,
(iv) mistagged decays such as a real $\overline{D}^0$ or $\overline{B}^0$ incorrectly identified as $D^0$ or $B^0$ and
(v) particle misidentification of the decay products such as
 $D^+\to \pi^-\pi^+\pi^+$ or $D_s^+\to K^-K^+\pi^+$ reconstructed as
 $D^+\to K^-\pi^+\pi^+$.


The contribution from combinatoric background with intermediate resonances 
is distinct from the 
resonances in the signal because the former do $not$ interfere with the
latter 
since they are not from true resonances. Additionally, processes
such as $\psi' \to \gamma \chi_{c2} \to \gamma(\gamma J/\psi) \to \gamma\gamma(\pi\pi)$ and $\psi' \to \pi^0 J/\psi, J/\psi \to \pi\pi$, do $not$ interfere since electromagnetic and hadronic
transitions proceed on different time scales.
Similarly, $D^0 \to \rho \pi$ and $D^0 \to K^0_S \pi^0$ do not interfere
since strong and weak transitions proceed on different time scales.
The usual identification tag of the initial particle as a
$D{^0}$ or a $\overline D{^0}$
is the charge of the distinctive slow
pion in the decay sequence $D^{*+}\!\to\!D{^0}\pi_s^+$ or
$D^{*-} \to \overline D^0 \pi_s^-$.  
Another possibility
is the identification or ``tagging'' of one of the $D$ mesons from
$\psi(3770)\!\to\!D^0\overline D{^0}$ as is done for $B$ mesons 
from $\Upsilon(4S)$.
The mistagged background is subtle and may be mistakenly enumerated in the $signal$ fraction
determined by a $D^0$ mass fit. Mistagged decays
contain true $\overline D^{0}$'s or $\overline B^{0}$'s and so 
the resonances in the mistagged sample exhibit interference 
on the Dalitz plot. 

%


\part[$e^+e^-$ Collision at $\sqrt{s} = 2-5$ GeV]{$e^+e^-$ Collision at $\sqrt{s} = 2-5$ GeV\\
\vspace*{2cm}
\Large Conveners: Yuan-Ning Gao, Xiao-Yuan Li\\
 \vspace*{1cm}
Contributors \\
H.~M.~Hu, X.~Y.~Li, J.~P.~Ma, G.~Rong, P.~Wang, D.~H.~Zhang}
\label{part:two}
\chapter[Physics Processes and Radiative Corrections]{Physics Processes and
Radiative Corrections}
\label{sec:processes}

\section[Physics Processes at BEPC-II]{Physics Processes  at BEPC-II
\footnote{by J.~P.~Ma}}

To help make this document self-contained, we list here basic 
$e^+e^-$ continuum processes that can be explored
at BEPC-II.  In addition to their own  interest, 
continuum amplitudes have an 
important influence in the resonance region, {\it e.g.}
in  the $J/\psi$, $\psi'$ and $\psi''$ peak regions. A careful 
understanding of
continuum processes is necessary for detailed studies 
at these resonances.
Since the cms energy of BEPC-II is far below the mass of 
the $Z$-boson,
weak interactions will not play an important role. 
Therefore, the basic processes are only 
QED- and QCD-related.

There are many QED-processes in $e^+e^-$ collisions.
We limit ourself to those that contain only 
two particles in the  final state.
These processes are $e^+ e^- \to \mu^+ \mu^-$
or $\tau^+\tau^-$,
$e^+e^- \to e^+ e^-$ and $e^+e^- \to \gamma\gamma$.
We first consider the process
\begin{equation}
   e^- (p_1) + e^+ (p_2) \to  \mu^- (p_3) + \mu^+ (p_4),
\end{equation}
where the four-momenta are indicated in the brackets. We define
the quantities
\begin{equation}
  s= (p_1 +p_2)^2,\ \ \ \
  \cos \theta = \frac {\vec p_1 \cdot \vec p_3}{ \vert \vec p_1 \vert \vert
\vec p_3 \vert}.
\end{equation}
Since the mass of the electron is tiny, we neglect it, 
but we keep terms involving the mass of the $\mu$. The differential 
cross-section for $e^+e^-\to \mu^+\mu^-$ is given by:
\begin{eqnarray}
\frac {d\sigma}{d\Omega} &=& \frac{\alpha^2}{4s} \sqrt{ 1 -\frac{4
m_\mu^2}{s} }
\left ( 1 + 4 \frac{m^2_\mu}{s} + \left (1 -\frac{4 m_\mu^2}{s} \right )
\cos^2 \theta \right ) ,
\nonumber\\
\sigma &=& \frac{4\pi \alpha^2}{3 s}  \sqrt{ 1 -\frac{4 m_\mu^2}{s} } \left
( 1+ 2\frac{m^2_\mu}{s} \right ).
\end{eqnarray}
One gets the result for $e^+e^- \to \tau^+ \tau^-$
by replacing $m_\mu$ with $m_\tau$.

For the process
\begin{equation}
   e^- (p_1) + e^+ (p_2) \to  \gamma (p_3) + \gamma (p_4),
\end{equation}
the differential cross section reads:
\begin{equation}
\frac {d\sigma}{d\Omega} = \frac{\alpha^2}{s} \frac{ 1+\cos^2 \theta}{\sin^2
\theta}.
\end{equation}
The differential cross-section for Bhabha scattering
\begin{equation}
  e^- (p_1) + e^+ (p_2) \to  e^- (p_3) + e ^+ (p_4)
\end{equation}
reads:
\begin{equation}
\frac{ d\sigma }{d\cos\theta} = \frac{\pi\alpha^2}{2 s}
\frac {(3+\cos^2\theta )^2}{(1-\cos\theta )^2 }.
\end{equation}
These formulae summarize all two-body QED processes at 
BEPC-II.  They all can be found in standard text books.

In addition to QED processes, 
there are also hadronic processes. Because the energy
of BEPC-II
is rather low, the most interesting continuum process is
\begin{equation}
 e^- + e^+ \to hadrons.
\end{equation}
If $s$ is sufficiently large, the $e^-e^+$ pair 
initially  annihilates into quarks and gluons.
If we assume that the transmission probability of quarks 
and gluons into hadrons is unity,  
the total cross-section can be calculated  
using perturbative QCD, in the sense that there are
no infrared- or  collinear divergences.
The total cross-section is usually expressed 
in terms of the  famous $R$-value, which is defined as:
\begin{equation}
R (s)= \frac{ \sigma(e^+ e^- \to hadrons)} { \sigma (e^+ e^- \to \mu^+
\mu^-)}.
\end{equation}
To the one-loop correction level, $R$ is given by:
\begin{equation}
  R(s) = N_c \sum_{q} Q_q^2 \left [ 1 + \frac{\alpha_s (\mu) }{\pi}
+{\mathcal O }(\alpha_s^2) \right ]
    \theta (s-4m_q^2).
\end{equation}
Here $\mu$ is the renormalization scale, 
which is taken to be $\mu^2=s$ in order to
minimize higher-order corrections. Two-loop level 
calculations of $R$ can be found in
Ref.~\cite{part2_ma_r-two-loop}.
\par
It should be noted that the theoretical basis for the $R$ 
calculation is the operator product expansion (OPE). 
Therefore, there are addition power-corrections to
the perturbative prediction given in the above. 
These power-corrections can be important for \bes3. 
Part of these corrections can be obtained from QCD sum 
rules, with a result that
reads~\cite{part2_ma_itep}:
\begin{eqnarray}
\int_0^\infty ds \exp(-s/M^2) R^{I=1} (s) &=& \frac{3}{2} \left [ 1
+\frac{\alpha_s (M)}{\pi}
         +{\mathcal O }(\alpha_s^2)
\right.
\nonumber\\
 &&   \left.  + \frac{\pi^2}{3 M^4} \langle 0 \vert \frac{\alpha_s(M)}{\pi}
G_{\mu\nu}^a G^{a,\mu\nu} \vert 0 \rangle
      -\frac{448\pi^3}{81 M^6} \vert \langle 0 \vert \bar q q\vert 0\rangle
\vert^2 +\cdots \right ],
\nonumber
\end{eqnarray}
where the scale parameter $M$ should be taken as a typical hard scale. 
In general, these power corrections to $R$
are expected to be small.
\par
Other QCD processes, such as jet production, inclusive-, and exclusive 
hadron production have been studied
at $e^+ e^-$ colliders with cms energies that are much higher than 
those of BEPC-II.  However,  since \bes3 will primarily concentrate 
on  physics in the resonance regions, these topics may not be
heavily pursued.


\section[Radiative corrections]{Radiative corrections\footnote{By Ping Wang}}
\label{sec:radiativeinee}

 All measurements made 
by $e^{+}e^{-}$ colliding experiments must apply 
radiative correction in order to get
results that can be compared with results from
other types of experiments.

\subsection{First order perturbation and exponentiation}
The lowest-order perturbation term takes account of
initial state radiation of a single photon. The 
cross section is expressed by the Bonneau and 
Martin formula~\cite{part2_rad_BM}
\begin{eqnarray}
\sigma(W)=\sigma_0(W) \{1+\frac{2\alpha}{\pi}(\frac{\pi^2}{6}
-\frac{17}{36})+\frac{13}{12} \beta   \nonumber \\
\nonumber \\
+  \beta \int^E_0 \frac{dk}{k} [(1-\frac{k}{E}
+ \frac{k^2}{2E^2}) \frac{\sigma_0(W-k)}{\sigma_0(W)}-1]\},
\label{MB}
\end{eqnarray}
where 
\begin{equation}
\beta=\frac{\alpha}{\pi}\log(\frac{s}{m_e^2}-1).
\label{beta1}
\end{equation}
In Eq.~(\ref{MB}), the upper limit of 
integration, $k_{max}$, has been set to $E$, corresponding to 
the fact that an electron can lose all its energy to 
radiation.   Soft photon emission is contained in the $dk/k$
factor, which is just the classical result corrected for energy 
conservation by the cross section $\sigma_0(W-k)$. It is
convenient to rewrite Eq.~(\ref{MB}) with the soft photon part
displayed separately from the ``hard'' terms 

\begin{eqnarray}
\sigma(W)=\sigma_0(W)[1+\delta] + \beta \int_0^E \frac{dk}{k}
[\sigma_0(W-k)-\sigma_0(W)] \nonumber \\
\nonumber \\
-\frac{\beta}{E} \int^E_0 dk (1-\frac{k}{2E}) \sigma_0(W-k);
\label{1photon}
\end{eqnarray}
where
$$
\delta = \frac{2\alpha}{\pi} (\frac{\pi^2}{6}-\frac{17}{36})
+\frac{13}{12} \beta
$$
is a small number that changes slowly with energy. (At  
the $J/\psi$, $\beta=0.076$, $\delta=0.085$.) The last term 
in Eq.~(\ref{1photon}) is small compared with the first two.

If instead of integrating over the entire
photon energy range, we only include photons up to a maximum
energy of $k_{max}$, and we assume that the variation of
$\sigma_0(s)$ with  $s$ is very slow, we have
\begin{equation}
\sigma(s) \approx \sigma_0(s) (1+ \delta_1 + 
\beta \ln (\frac{k_{max}}{E})),
\label{1storder}
\end{equation} 
where 
$$
\delta_1 = \frac{3}{4} \beta + 
\frac{2\alpha}{\pi}(\frac{\pi^2}{6}-\frac{1}{4}).
$$
The variable $k_{max}$ is usually determined by 
the experimental event selection criteria. However
in some experiments there is a natural 
cut off. For example, in the case of
a resonance, the width
of the resonance $\Gamma$ provides 
such a natural cut off.
Another example is the $\tau$-pair cross section near
threshold. Here $k_{max} = E-2m_{\tau}$. 

\begin{figure}[hbt]
\includegraphics[width=12cm,height=6cm]{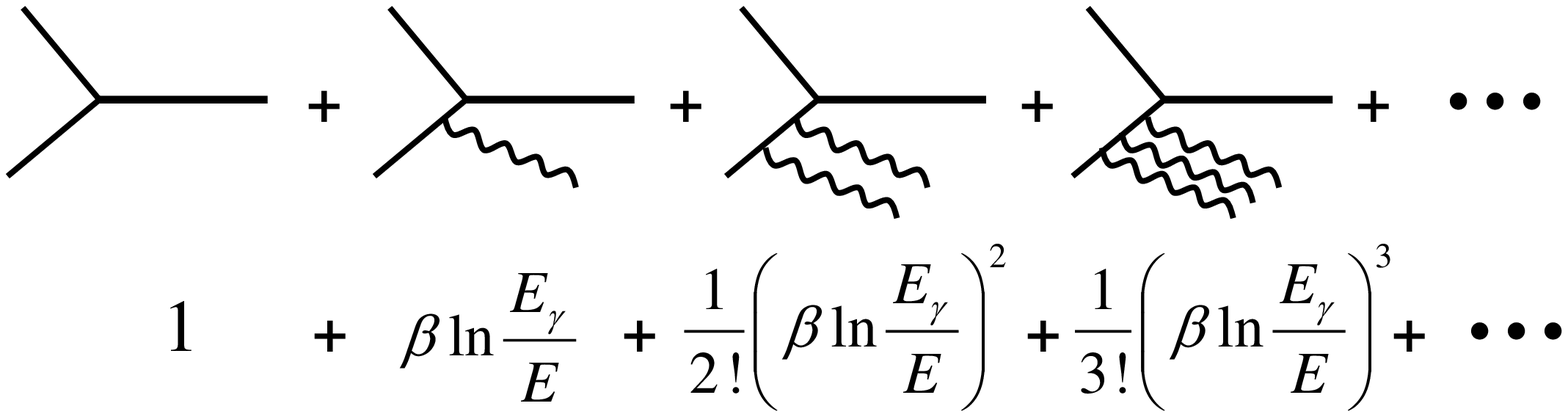}
\caption{\label{rad3} The summation of multiphoton emission.}
\end{figure}

For a narrow resonance like the $J/\psi$ 
$$
\beta \ln (\frac{\Gamma}{M}) \approx -81\%,
$$ 
which is very large for a perturbative expansion.
Similarly large values obtain for the $\psi^\prime$.
For energies near the $\tau$-pair threshold, 
$\beta\ln (\frac{E-2m_{\tau}}{E})$ can also be large. 
In these circumstances, the first-order perturbation 
expansion of Eq.~(\ref{1storder}) is not enough, and
higher-order terms, including 
multi-photon emission,  must be included.
This is done by 
\begin{eqnarray}
\sigma & = & \sigma_0 (1+\delta_1+\delta_2)
(1 + \beta \ln (\frac{k_{max}}{E}) \nonumber \\
\nonumber \\ 
&& + \frac{1}{2!}\beta^2 \ln^2 (\frac{k_{max}}{E})
+ \frac{1}{3!}\beta^3 \ln^3 (\frac{k_{max}}{E}) 
+ \cdots ) \nonumber \\
  \nonumber \\
& = & exp (\beta \ln \frac{k_{max}}{E} )
= (\frac{k_{max}}{E})^\beta.
\label{expon} 
\end{eqnarray}
That is, the bremsstrahlung spectrum is modified to become
\begin{equation}
\frac{d \sigma}{dk} = \sigma_0(1 + \delta + \cdots )\beta 
k^{(\beta-1)},
\end{equation}
where the $1/k$ factor
is replaced by $k^\beta (1+\delta_1+\cdots)/k$.
This is illustrated in Fig.~(\ref{rad3}). The summation
can be justified if the emitted photon is collinear with 
the incoming electron,\footnote{In this section we use
electron to designate either he incoming electron or positron.} 
in which case  the electron 
is almost on the mass shell after the emission. 
This means that the electron,
after emitting the photon, is undisturbed
and, so, each successive photon is independent. 

To accommodate the $s$ dependence of $\sigma_0(s)$  
in  multi-photon emission, we introduce
a factor
$$
\exp [-\beta \ln(E/k)] = (k/E)^\beta
$$
into the integrand in Eq.~(\ref{1photon}). The 
radiatively corrected cross section then becomes
\begin{equation}
\sigma(W) = \beta \int^E_0\frac{dk}{k}(\frac{k}{E})^\beta 
\sigma_0(W-k) + \delta \sigma_0(W).
\label{expo1}
\end{equation}
This is called exponentiation. It was originally 
conjectured by Schwinger in 1949~\cite{part2_rad_Schwinger}, 
and later developed by 
Yennie, Frautschi and Suura in 1961~\cite{part2_rad_YFS}.

The exponentiation in Eq.~(\ref{expo1}) 
is done after phase space integration over photon
momenta, which is appropriate for inclusive distributions. This
method is called ``inclusive'' exponentiation,
in contrast to ``exclusive exponentiation''
in Monte Carlo simulations as discussed below.
\subsection{Structure function approach}
A long-standing question in radiative correction methodology
was what part of the lowest-order radiative corrections
should one exponentiate?
This was rigorously solved based on quantum field theory
by Kuraev and Fadin~\cite{part2_rad_rad.1,part2_rad_rad.2,part2_rad_rad.3}. 
In their approach,  single-photon annihilation
is regarded as a Drell-Yan process. In the leading
logarithmic approximation ({\it i.e.} when only the terms 
containing $\log(\frac{s}{m_e^2})$ 
are retained), the cross section can be expressed
in the form 

\begin{equation}
\sigma(s) = \int \int dx_1 dx_2 D(x_1,s) D(x_2,s), 
\tilde{\sigma}(x_1 x_2 s),
\label{sig}
\end{equation}
where 
$$
\tilde{\sigma}(s) = \frac{\sigma_{B}(s)}{|1-\Pi(s)|^2}.
$$
In the above expression, $\sigma_{B}(s)$ is the Born-order
cross section, and $\Pi(s)$ is the vacuum polarization
factor.  The function $D(x,s)$ in Eq.~(\ref{sig})
is called the structure function. It
satisfies the Lipatov-Altarelli-Parisi
equation~\cite{part2_rad_LAP}:
\begin{equation}
D(x,Q^2) = \delta(x-1) + \int_{m_e^2}^{Q^2} \frac{\alpha(Q^{\prime 2})}
{2\pi} \frac{dQ^{\prime 2}}{Q^{\prime 2}} \int_x^1 \frac{dz}{z}
P(z)D(\frac{x}{z},Q^{\prime 2}),
\label{Lipatov}
\end{equation}
where $\alpha(Q^2)$ is the running coupling constant given by
$$
\alpha(Q^2)=\alpha/[1-\frac{\alpha}{3\pi} \ln\frac{Q^2}{m^2}],
$$
and $P(z)$ is the regularized $e \rightarrow e + \gamma$
splitting function.
\begin{equation}
P(z) = \frac{1+z^2}{1-z} - \delta(1-z)\int_0^1 \frac{dx (1+x^2)}{1-x}.
\label{splitting}
\end{equation}
\noindent
The structure function has a clear and intuitive meaning:
it represents the probability density for finding ``inside''
the parent electron a virtual electron with momentum fraction $x$
and virtuality $Q^2$. ($Q$ is the four-momentum of the virtual 
electron.)

By solving Eq.~(\ref{Lipatov}) for $D(x,s)$
and substituting it into Eq.~(\ref{sig}),
the radiatively corrected cross section can be expressed as
\begin{eqnarray}
\sigma(s)\;=\int^{1-s_m/s}_0 dx \; \tilde{\sigma}(s(1-x))F(x,s) ,
\label{rad}
\end{eqnarray}
where $\sqrt{s}$ is the cms energy of the colliding beam,
$\sqrt{s_m}$ is
the cut-off of the invariant mass in the event selection,
\begin{equation}
F(x,s)\;=\;\beta x^{\beta-1}\delta^{V+S}+\delta^{H} ,
\label{Fexp}
\end{equation}
with $\beta$ expressed in Eq.(\ref{beta1}), and 
\begin{equation}
\delta^{V+S}\;=\;1+\frac{3}{4}\beta+\frac{\alpha}{\pi}
\left(\frac{\pi^{2}}{3}-\frac{1}{2}\right)+\beta^{2}
\left(\frac{9}{32}-\frac{\pi^{2}}{12}\right) ,
\label{V+S}
\end{equation}
\begin{eqnarray}
\delta^{H}\; & = & \;-\beta\left(1-\frac{x}{2}\right) \nonumber \\
              &   & \nonumber \\
& & +\frac{1}{8}\beta^{2}\left[4(2-x)\ln\frac{1}{x}-
\frac{1+3(1-x)^{2}}{x}\ln(1-x)-6+x\right] .
\end{eqnarray}
Here the conversion of  bremsstrahlung photons to
real $e^{+}e^{-}$ pairs is
included in the cross section, which is the usual
experimental situation.  Thus there is cancellation between the
contributions of virtual and real $e^{+}e^{-}$ pairs in the leading
term~\cite{part2_rad_rad.3}. 

In quantum field theory,
it can be rigorously proved that Eq.~(\ref{rad}) summarizes all 
of the leading-log (LL) terms such as
$$
(\frac{\alpha}{\pi} \log(\frac{s}{m^2_e}))^N
$$
for $N$ from 1 to $\infty$.
This can be improved to include the next-to-leading-log (NLL)
terms~\cite{part2_rad_Zline}, such as
$$
(\frac{\alpha}{\pi})^{(N+1)} (\log(\frac{s}{m^2_e}))^N.
$$
Since for the BEPC-II energy scale, $\frac{s}{m^2_e}$ is large,
these leading terms give the main contribution
to the initial-state radiative correction.
The expression for $F(x,s)$ in Eq.~(\ref{Fexp}) also incorperates the
non-leading  first-order terms from explicit calculation.
This method has an accuracy of $0.1\%$.

The master formula of Eq.~(\ref{rad}) is universally applicable.
For example, it can be used for resonances, $R$-value 
measurements, and threshold cross section determinations 
such as those used for the $\tau$
mass measurement. One only needs to substitute the Born order
cross section for each case.

For resonances,
\begin{equation}
\sigma_{B}(s)\;=\frac{12\pi \Gamma^0_{ee} \Gamma_{f}}{(s-M^{2})^{2}
+\Gamma^{2} M^{2}} ,
\end{equation}
where $M$ and $\Gamma$  are the mass and total width of the resonance;
$\Gamma^0_{ee}$ and $\Gamma _{f}$   are the partial  widths
for the $e^{+}e^{-}$  mode  and the detected final
state,  respectively.
Since the decay of a
quarkonium $1^{--}$ state to an $e^{+}e^{-}$ pair is via a
virtual photon, there is always vacuum polarization associated with
the process. So, the experimentally measured $e^+e^-$ partial width,
denoted explicitly as $\Gamma_{ee}^{exp}$, is related to
$\Gamma^0_{ee}$ by the expression
\begin{equation}
\Gamma_{ee}^{exp}=\frac{\Gamma^0_{ee}}{|1-\Pi(M^2)|^2}.
\label{gee}
\end{equation}
The vacuum polarization gets contributions
from leptons and hadrons. The latter is treated in Ref.~\cite{part2_rad_vac}.
In order to fit for resonance parameters, the contribution from
hadronic resonances to the vacuum polarization should not
include the resonance being measured.  The
Particle Data Group adopts the convention of
Refs.~\cite{part2_rad_Tsai,part2_rad_Luth}
that $\Gamma_{ee}$ means $\Gamma_{ee}^{exp}$.
So, for the cross section of inclusive hadrons, the vacuum
polarization correction should not be applied. 

For the near-threshold $\tau$ mass measurement, the 
Born-order cross section is
$$
\sigma_B(s)=\frac{4\pi\alpha^2}{3s} \frac{v(3-v^2)}{2}
F_c(v)F_r(v),
$$
with
$$
F_c=\frac{\pi\alpha/v}{1-\exp(-\pi\alpha/v)},
$$
and
\begin{eqnarray}
F_{r}= & 1+(\alpha/\pi v) \{(1+v^2)[\ln \frac{1+v}{2} \ln \frac{1+v}{1-v}
+2\ell(\frac{1-v}{1+v}) - \frac{\pi^2}{3} +
2\ell (\frac{1+v}{2}) \nonumber \\
 \nonumber \\
& -2\ell(\frac{1-v}{2}) - 4\ell (v) + \ell(v^2) ] +
[\frac{11}{8}(1+v^2) -3v + \nonumber \\
 \nonumber \\
& \frac{1}{2}\frac{v^4}{3-v^2} ]
\ln(\frac{1+v}{1-v}) +6v\ln(\frac{1+v^2}{2})
- 4v\ln v + \frac{3}{4}v\frac{(5-3v^2)}{(3-v^2)} \},
\end{eqnarray}
where
$$
\ell (x) = - \int^x_0\ln(1-t) dt/t,
$$
and $s_m=4m_{\tau}^2$ in Eq.(\ref{rad}). In these equations,
$v=\sqrt{1-4m_{\tau}^2/s}$ is the velocity of $\tau$.

Equation~(\ref{rad}) was first derived by Kuraev and 
Fadin~\cite{part2_rad_rad.1}, 
and it has been reproduced and improved by by Altarelli and
Martinelli, as well as by Nicrosini and Trentadue~\cite{part2_rad_rad.2}.
Berends and his cowokers
have done an explicit second-order calculation to check Kuraev and
Fadin's results~\cite{part2_rad_rad.3}. As a result of these efforts, 
Eq.~(\ref{rad}) has been fully checked and is now established as
the correct and accurate formula for radiative corrections.

After the publication of Kuraev and Fadin's work, 
a group of SLAC researchers
(G.~Alexander, P.~Drell, V.~Luth {\em et al.})
used their formula and reanalyzed
all $\psi$ and $\Upsilon$ family states.
They corrected the values of the total widths and
leptonic partial widths of these 
states~\cite{part2_rad_Luth}. The results and the
method of this paper have been adopted by the Particle Data Group.
Since the mid-1980s, it has been the accepted way
to treat radiative corrections by all experiments,
including BESI and BESII ({\it e.g.} the $J/\psi$ and $\psi^{\prime}$ 
scan, as well as $\tau$ mass measurement 
papers)~\cite{part2_rad_Zhu,part2_rad_Mo,part2_rad_tau}. 
It has also been used for $Z$ line-shape fitting in order to
extract the parameters of the $Z$ by the LEP and SLC 
groups~\cite{part2_rad_Zline}.

The correctness and precision of this approach
has been fully checked.

\subsection{Initial and final state radiation}
Direct calculations show that the vertex corection plus 
soft bremsstrahlung with photon energy up to 
$\Delta E \ll E$ is 

\begin{equation}
\frac{3}{4} \beta - \frac{1}{4} +\frac{\pi^2}{6} 
-\beta \ln(\frac{E}{\Delta E}),
\label{vertex}
\end{equation}
where $\beta$ is given by Eq.~(\ref{beta1}).

\begin{figure}[hbt]
\begin{minipage}{7cm}
\includegraphics[width=6cm,height=6cm]{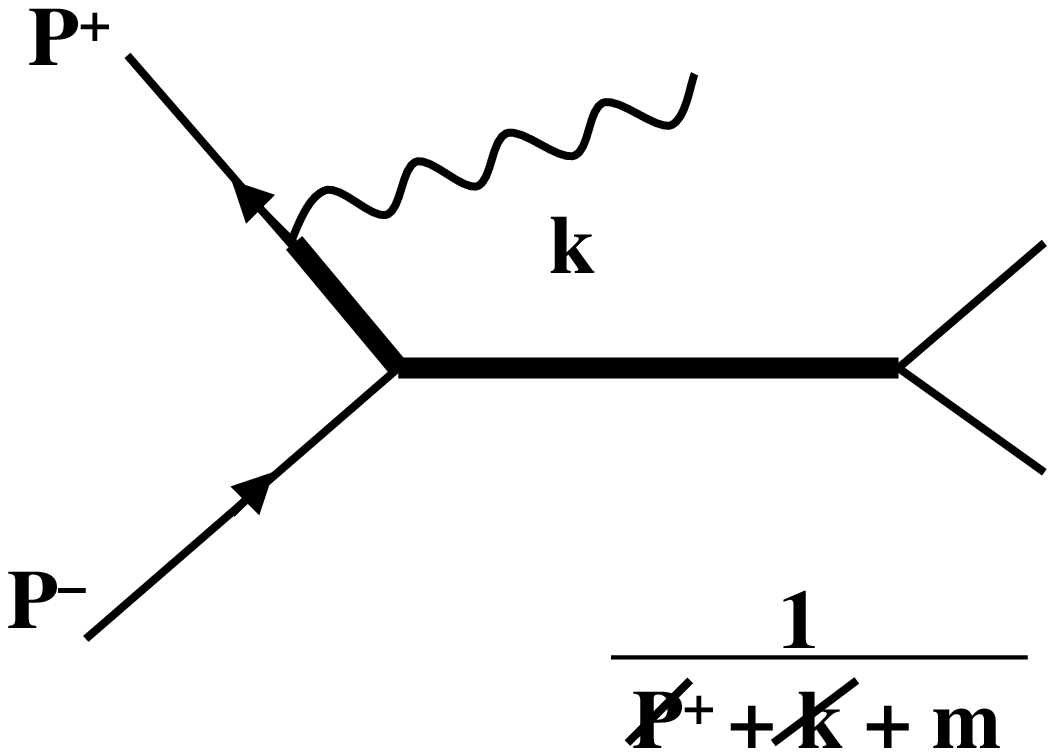}
\end{minipage}
\begin{minipage}{7cm}
\includegraphics[width=6cm,height=6cm]{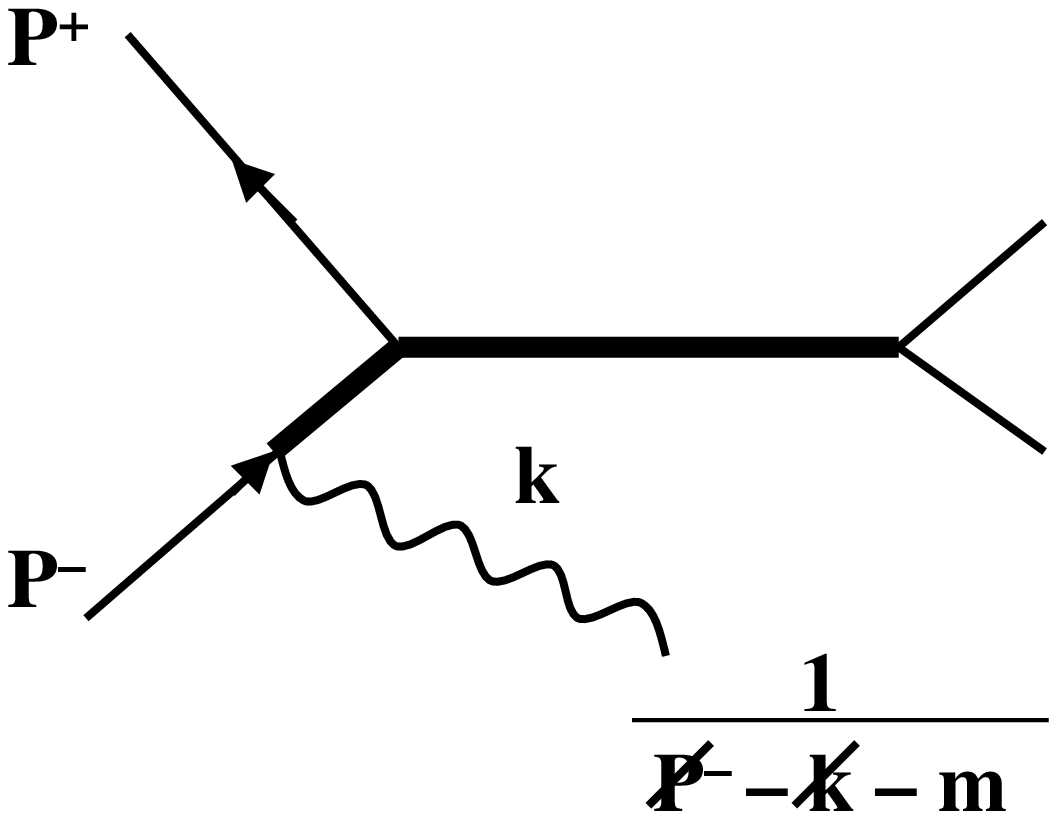}
\end{minipage}
\caption{\label{figrad} The fermion propagator 
with the emission of a photon by a fermion.}
\end{figure}

The bremsstrahlung spectrum is 
\begin{equation}
\beta \frac{1}{2} [1+(1-x)^2] x^{-1} dx,
\label{brem}
\end{equation}
where $x=k/E$ with $k$ the energy of the radiated photon.
Here the factor $\beta$
is due to the radiation of photons along the direction
of the incoming electron.
It comes from (the square of) the fermion propagator 
with the emission of a photon by the fermion
as shown in Fig.~\ref{figrad}.

The fermon propagator has the form 
\begin{eqnarray}
\frac{1}{\rlap /p -\rlap /k -m_l} 
= \frac{\rlap /p -\rlap /k + m_l}{(p-k)^2 - m^2} 
= \frac{\rlap /p -\rlap /k + m_l}{p \cdot k} \nonumber \\ 
\nonumber \\
= \frac{\rlap /p -\rlap /k + m_l}
{E_{l}E_{\gamma} - |p_l| E_{\gamma} \cos \theta}.
\label{div}
\end{eqnarray}
For $E \gg m_e$, $E_{l} \approx |p_l|$. So, at $\theta =0$,
{\it i.e.} when the photon is emitted along the 
direction of the fermion, 
Eq.~(\ref{div}) becomes nearly divergent.
The term with the factor $\beta$ comes from 
the integration of
the photon angle relative to the fermion around 
$\theta = 0$.  
At the GeV mass scale, $\beta$ is at the order of 
$0.1$, and far greater than the QED fine structure 
constant $\alpha$. The appearance of terms with the 
factor $\beta$ is called the colliner divergence  
(it diverges as $m_e/E \rightarrow 0$), 
or the mass singularity.

Kinoshita, Lee and Nauenberg observed that 
if one sums up all the states with the same energy
(degenerate states) the mass singularity cancels out 
to all orders in perturbation theory. In $e^+e^-$
experiments, the initial state is preselected by 
the machine so that both the $e^+$ and $e^-$ have a definite
energy, $E_{beam}$. Thus, the mass singularity due to 
initial state radiation remains. The final state 
can be summed and, after doing so, we expect their mass
singularity will cancel to all orders in perturbation
theory. This can be seen mathematically by
integrating the bremsstrahlung spectrum 
in Eq.~(\ref{brem}) from 
$\Delta E \ll E$ to $E$, we obtain
\begin{eqnarray}
\int^1_{\Delta E/E} \beta (x^{-1} -1 + \frac{1}{2} )dx \nonumber \\ 
\nonumber \\
\approx \beta (\ln \frac{E}{\Delta E} - \frac{3}{4}).
\label{intbre}
\end{eqnarray}
Adding Eq.~(\ref{intbre}) to Eq.~(\ref{vertex}), the radiative
correction becomes 
$$
\frac{2\alpha}{\pi} (-\frac{1}{4} + \frac{\pi^2}{6} ),
$$
which no longer has a mass singularity. 

The mass singularity from initial state bremsstrahlung 
$\beta$ and the vertex correction do not cancel each 
other when the photon spectrum is integrated because 
as the bremsstrahlung is emitted from the initial
electron or positron, the virtual photon four-momentum 
changes from $s$ to $s(1-k/E)$. Usually the cross section
depends on $s$. For example, in the point-like coupling
$e^+e^- \rightarrow \mu^+\mu^-$, the cross section is
inversely proportional to $s$. In such a case, we
have to multiply the integrand of Eq.~(\ref{intbre}). 
by a factor 
$$
(1-x)^{-1}.
$$ 
In this case, the mass singularity of 
the resulting expression does not cancel that from
vertex correction. In the language of Lee and 
Nauenberg we do not expect any cancellation for
the initial state because we are not summing over all
degenerate initial states. The machine preselects
the electron and positron to each have energy $E_{beam}$, and 
no other degenerate state participates in the interaction. 

This is very important when dealing with hadronic
final states that are usually too complicated
to be radiatively corrected. The above indicates
that we can ignore the radiative corrections to the 
final states to an accuracy of $\frac{\alpha}{\pi}$.

Nevertheless, in some MC generators with $0.1\%$
precision, final state radiation is included. 
This is discussed below.

\subsection{MC generators: inclusive process}

The KKMC Montr Carlo program~\cite{part2_rad_Jadach2}
generates radiatively corrected $e^+e^-$ annihilations 
to $\mu^+\mu^-$, $\tau^+ \tau^-$ final states
as well as inclusive hadronic final states formed from
$u$, $d$, $s$, $c$, $b$
quark pairs. 

To realize the Monte Carlo simulation, the authors of 
KKMC introduce ``coherent exclusive exponentiation.'' 
The exclusivity means that the exponentiation procedure
is done at the level of the fully differential (multi-photon) 
cross section, before any phase-space integration 
over the photon momenta is done. 
This is in contrast to the
``inclusive exponentiation'' of Eq.~(\ref{expo1}) where
the exponentiation is done after the integration of
the photons'  phase space.

In the Monte Carlo event, multiple photons are emitted
from the initial state electron;
each photon may have a momentum with a finite 
angle from the incoming electron.  The algorithm of coherent
exclusive exponentiation treats radiative 
corrections to infinite order not only including
initial-state-radiation, but also 
interference between initial-final state radiation 
and narrow resonances. Full scale next-to-leading-log
corrections are included.
For quarks and $\tau$ leptons, the simulation of 
hadronization is done by PYTHIA (in an earlier version
JETSET was used).  Beam polarization and spin effects
in $\tau$ decays, both longitudinal and transverse, 
are taken into account.  $\tau$ decays are done by TAUOLA.
A precision level of $0.2\%$ is achieved.

The program language is FORTRAN 77, with 
popular extensions such as long 
variable names, long source lines, etc.
There is a make file that controls the compiling. 
The program is written with  the possible
future translation into an object-oriented
language such as C++ in mind. The program source code 
is organized into modules, also called pseudo-classes,
which have the structure of the C++ class, as far 
as this is possible to do in FORTRAN 77.  

For most users of KKMC in $e^+e^-$ experiments, 
a set of input parameters must be specified. 
This is done in the file user.input.
For most of the parameters, the users are advised to use
the default values, or if changed, with special caution,
{\it e.g.} in consultation with the program authors. There are
also some parameters which are irrelevant to the
$\tau$-charm threshold physics, {\it e.g.} those for
beamstrahlung and weak interactions.
Only a small subset of the parameters stored in xpar(10000)
need to be set by the users.

\subsection{MC generators: exclusive process}
Currently there are two MC generators for exclusive processes:
BABAYAGA and MCGPJ.

\subsubsection{BABAYAGA}

BABAYAGA~\cite{part2_rad_bby} uses the following 
differential cross section master formula
\begin{eqnarray}
\sigma(s)= & \int dx_1 dx_2 dy_1 dy_2 \int d\Omega \nonumber \\
 \nonumber \\ 
& D(x_1,Q^2) D(x_2,Q^2) D(y_1,Q^2) D(y_2,Q^2)
\frac{d\sigma_0(x_1 x_2 s)}{d \Omega}.
\label{PS}
\end{eqnarray}
Here $D(x,Q^2)$ is the strucure function which 
satisfies Eq.~(\ref{Lipatov}).
Equation~(\ref{PS}) accounts for 
photon radiation emitted
by both initial state and final state fermions. 
Although this formula allows for the inclusion 
of the universal virtual photon and real photonic 
corrections up to all orders of perturbation theory, 
strictly speaking, the radiation included
in Eq.~(\ref{PS})  
is collinear with the emitting fermions.

The algorithm employed by BABAYAGA goes beyond
the collinear approximation by solving 
Eq.~(\ref{Lipatov}) by a Monte Carlo algorthm, the 
so-called Parton-Shower algorithm. This algorithm 
exactly solves Eq.~(\ref{Lipatov}) with the iterative 
solution
\begin{eqnarray}  
D(x,Q^2) & = & \Pi(Q^2,m^2) \delta(1-x) \nonumber
 \\ \nonumber \\
& + & \frac{\alpha}{2\pi} \int^s_{m^2} \Pi(Q^2,s^\prime)
\frac{d s^\prime}{s^\prime} \Pi(s^\prime, m^2)
\int^{x_+}_{0} dy P(y) \delta(x-y)  \nonumber \\ 
\nonumber \\
& + & (\frac{\alpha}{2\pi})^2 
\int^{Q^2}_{m^2} \Pi(Q^2,s^\prime) \frac{d s^\prime}{s^\prime}
\Pi(s^{\prime\prime},m^2) \nonumber \\
\nonumber \\
& \times & \int_0^{x_+} dx_1 \int_0^{x_+} dx_2 P(x_1) P(x_2)
\delta(x- x_1 x_2) \nonumber
 \\ \nonumber \\
 & + & \hspace{0.2cm}3\hspace{0.2cm}{\rm photons}.
\label{PSD}
\end{eqnarray}
In Eq.~(\ref{PSD}),
$$
\Pi(s_1,s_2) = \exp[-\frac{\alpha}{2} \pi \int^{x_1}_{x_2}
\frac{d s^\prime}{s^\prime} \int^{x_+}_{0} dz P(z)]
$$
is the Sudakov form factor~\cite{part2_rad_Sudakov}, which represents the 
probability that the electron evolves from virtuality
(virtuality is the squared four-momentum of the 
off-mass-shell particle)
$s_2$ to virtuality $s_1$ with no emission of
photons of energy fraction greater than (an infrared
regulator) $\epsilon = 1 - x_+$. The solution in 
Eq.~(\ref{PSD}) accounts for ``soft + virtual'' and real
photons radiation up to all orders of perturbation
theory in the leading-logarithmic approximation.
The Sudakov form factor exponentiates the leading 
logarithmic contribution of the $\cal{O}(\alpha)$
``soft + virtual'' cross section, as well as the dominant
contribution coming from the infrared cancellation 
between the virtual box and the initial-final state 
interference of the bremsstrahlung diagrams. 

In the implementation of the parton shower algorithm,
BABAYAGA simulates a shower of photons emitted by 
the electron according to Eq.~(\ref{PSD}). 
When the algorithm stops, we are left with the
energy fraction $z_1$ of each photon, distributed 
according to the splitting function of
Eq.~(\ref{splitting}), as well as the virtualities
of the electron at each branching and the 
remaining energy fraction $x$ of the electron after the
showering. The $x$ variable is distributed according 
to the structure function $D(x,Q^2)$. By means of 
these quantities, an approximate branching kinematics 
is obtained. 

The main advantage of the parton shower algorithm 
with respect to the collinear treatment of the 
electron evolution is the possiblity of going
beyond the strictly collinear approximation and 
generating transverse momentum $p_\perp$ of the 
electrons and photons at each branching.
In fact, in BABAYAGA the kinematics
of the branching process 
$$
e(p) \rightarrow e^\prime(p^\prime) + \gamma(q)
$$
is as 
\begin{eqnarray}
p = (E,\vec{0},p_z), \nonumber
\\ \nonumber \\
p^\prime = (zE,\vec{p}_\perp,P^\prime_z), \nonumber
\\ \nonumber \\
q = ((1-z)E, -\vec{p}_\perp,q_z).
\end{eqnarray}
Once the variables $p^2$, $p^{\prime 2}$ and $z$ are
generated by the parton shower algorithm,
the on-shell condition $q^2=0$, together
with  longitudinal momentum conservation, 
allows one to obtain an expression for the $p_\perp$
value:
$$
p_\perp^2 = (1-z)(z p^2 - p^{\prime 2}).
$$
To first order in $p^2/E^2 \ll 1$, $p_\perp^2 \ll 1$.
All the photons generated by BABAYAGA can have transverse
momentum, which is in contrast with the other Monte Carlo
generator for exclusive processes, MCGPJ. In the latter only
one photon can be emitted at a large angle from one of 
the fermions, while other photons are emitted
within a very small angle from the fermions.   

In BABAYAGA, up to 10 photons are emitted from each
incoming and outgoing particles in the branching process. 
Altogether up to 40 photons can be produced in each event. 
But only the 2  most energetic ones are output. 

BABAYAGA is written only for two-body final states.
Besides $e^+e^-$, $\mu^+\mu^-$, $\gamma\gamma$,
hadronic final states include only $\pi^+\pi^-$. 
For this
mode,  final state radiation is not considered.
In the output, there are two final state particles
and two photons. 
The accuracy level is $0.5\%$. The programming language 
is FORTRAN.

Unlike KKMC, which stores input parameters in an array 
xpar with dimension of 10000, the input file of BABAYAGA
has only 27 parameters. Their meaning is clearly explained 
in the file input.txt.

\subsubsection{High precision MC generator for the Bhabha process}
BABAYAGA@NLO~\cite{part2_rad_bbyn} is a Monte Carlo program that
simulates the Bhabha process at flavor factories
from the $\phi$  to the $\Upsilon$'s.
The calculation is based on the matching of exact
next-to-leading-order corrections with a parton shower
algorithm.  

A corrected differential cross section in the
parton shower approach can be written as
\begin{eqnarray}
d\sigma^\infty = \Pi(Q^2,\epsilon) F_{SV} \{d\sigma_0 
+ \sum_{n=1}^\infty \frac{d\sigma_0}{n!} \nonumber
\\ \nonumber \\
\times \prod_{i=1}^n [\frac{\alpha}{2\pi} P(x_i) I(x_i) dx_i
\theta(x_i-\epsilon) F_{i,H}] \},
\end{eqnarray}
where $F_{SV}$ and $F_{i,H}$ are identical to 1 in the pure
parton shower case. To go beyond the leading 
logarithm approximation and preserve the summation
of the higher-order corrections, this program
sets
\begin{eqnarray}
F_{SV} = 1 + \frac{d\sigma^{\alpha, ex}_{SV}
-d\sigma^{\alpha, PS}_{SV}}{d\sigma_0}, \nonumber
\\ \nonumber \\
F_{i,H} = 1 + \frac{d\sigma^{\alpha, ex}_{i,H}
-d\sigma^{\alpha, PS}_{i,H}}{d\sigma^{\alpha, PS}_{i,H}}, 
\end{eqnarray}
where $d\sigma^{\alpha, ex}_{SV}$ are the complete
expressions for the soft and virtual $\cal{O}(\alpha)$
cross sections and the real one-photon emission 
cross section; $d\sigma^{\alpha, PS}_{SV}$
$d\sigma^{\alpha, PS}_{i,H}$ 
are the parton-shower approximations
for the soft and virtual 
cross sections and the real one photon emission 
cross section. 

Its accuracy level is 0.1\%. 
Currently, this is the most accurate
Monte Carlo program for the Bhabha process. This
program is ideally suited for luminosity
measurements at \bes3.

\subsubsection{MCGPJ}
MCGPJ is another Monte Carlo generator for
exclusive processes~\cite{part2_rad_MCGPJ}.
The current version simulates $e^+e^-$, $\mu^+\mu^-$, 
$\pi^+\pi^-$, $K^+K^-$  and $K_SK_L$ final states. 

This program takes higher-order
radiative corrections into account
by means of the structure function
formalism. It involves a convolution of the Born-order
cross section with the electron  structure 
function, which describes the leading effects due to
emission of photons in the collinear region as well 
as the radiation of $e^+e^-$ pairs. The collinear region is
defined as a part of the angular phase space inside four narrow
cones surrounding the directions of the motion of the
initial and final state particles. The emitted photons
are  inside these cones, which have an opening angle $2\theta_0$.
The angle $\theta_0$ has to  obey the restriction
$$
\frac{m_e}{E_{cm}} \ll \theta_0 \ll 1.
$$   
This serves as an auxiliary parameter and usually its value
is taken to be
$$
\theta \sim \sqrt{\frac{2m_e}{E_{cm}}}.
$$
Although this program introduces an additional parameter $\theta_0$, 
the physical results are independent of its value. (As is
the case for the other auxiliary parameter $\epsilon$, 
---the infrared cutoff (or so-called infrared regulator)---
that is used in every Monte Carlo program invloving radiative 
corrections.)  
The calculation takes into account
hard photons emitted in the four narrow cones,
which give the main contribution the the radiative correction.
In addition, one hard photon is allowed to be emitted 
at a large angle that is outside the narrow cones. 
This is included
by incorporating an explicit $\cal{O}(\alpha)$ calculation. 
The accuracy level of the program is $0.2\%$.
It is written in C++.

For the modes other than the Bhabha process, MCGPJ has higher 
accuracy than BABAYAGA.  Unlike BABAYAGA, MCGPJ
can be extended to three-body and multi-body 
final states.


\subsection{MC generators: the radiative return process}

The usual way to measure $R$ values is by energy-scan 
experiments conducted on $e^+e^-$ colliders. But an 
alternative option has been suggested which 
uses radiative-return at flavor factories~\cite{part2_rad_rr}. 
These colliders operate at fixed energies and
with enormous luminosities. This peculiar feature 
of a factory  allows the use of the radiative return, {\it i.e.}
the reaction
$$
e^+(p_1) e^-(p_2) \rightarrow \gamma(k_1) + \gamma^{*}(Q)
(\rightarrow hadrons) 
$$
to explore the hadronic cross section
in a wide range of $Q^2$ in a single experiment.

Nominally, invariant masses of hadronic systems everywhere
between the production threshold for the respective mode
and the operating cms energy of the flavor factory are accessible.
In practice, in order to isolate the reaction, 
it is useful to consider only events with a hard photon,
which significantly lowers the accessible mass range. 

A dedicated Monte Carlo program PHOKHARA~\cite{part2_rad_Henryk}
has been developed for such processes. It includes 
next-to-leading-order corrections from virtual and real 
photon emission. 
In the generated events, one or two
photons are emitted from the incident $e^+$ or $e^-$,
or the final state particles. 
Two-body and multi-body hadrnic final states are included.
The present version has nine modes, including
$\mu^+\mu^-$, $\pi^+\pi^-$, $K^+K^-$, nucleon-anti-nucleon,
$\pi^+\pi^-\pi^0$, $\pi^+\pi^-\pi^+\pi^-$,
$\pi^+\pi^-\pi^0\pi^0$ and $\Lambda \bar{\Lambda}$.
Final-state radiation and interference between 
initial- and final-state radiation
are included for two-body final states.

The programing language is FORTRAN. It is based on the 
calculation of Feynman diagrams with one or two emitted 
photons and one-loop corrections.

\subsection{Summary}
In $e^+e^-$  annihilation experiments, 
radiative corrections are crucial for achieving
precision results in virtually all types of
measurements. Sometimes they
can change the observed numerical results in a profound way, 
particularly measurements of resonance properties 
and cross section determinations  near the production
threshold. Thanks to steady progress of theoretical techniques,
we can now calculate the integrated cross section using
 formulae derived with the structure function approach.
This approach sums up the contributions of leading 
logarithmic terms to infinite order of $\alpha$ in 
the QED perturbative expansion.
For differential cross sections, there are
different methods to include higher order terms
in $\alpha$. These schemes are implemented by 
Monte Carlo programs for both inclusive and exclusive processes. 
The latest Monte Carlo programs achieve 0.1\% precision
levels for the Bhabha process (BABAYAGA@NLO), and 0.2\% 
precision levels for other 
inclusive and exclusive processes (KKMC, MCGPJ and PHOKHARA).
They are for general use at
favor factory experiments such as \bes3.  

\newpage


\chapter[Hadronic fragmentation]{Hadronic fragmentation\footnote{By Hai-Ming Hu}}
\label{chapter:fragmentation}

Quantum Chromdynamics (QCD) is the unique candidate theory
for the strong interactions. However, hadronization processes occur
in the nonperturbative regime where there are no reliable first
principle calculations. Therefore, phenomenological models 
have been built and are widely used in experiments.
\begin{figure}[htbp]
\begin{center}
\includegraphics[width=9cm,height=8cm]{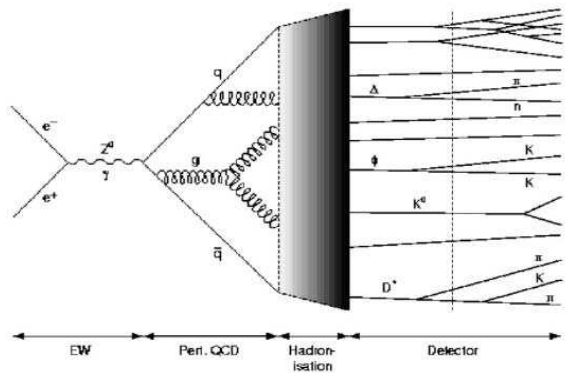}
\caption{Schematic illustration of hadron
production in $e^+e^-$ annihilation.}\label{hadpic}
\end{center}
\end{figure}

Hadron production in inclusive 
$e^+e^-$ annihilation processes is illustrated in
Fig.~\ref{hadpic}.
There are two successful hadronization models at high energies: 
the cluster model 
(HERWIG)~\cite{part2_r_e_herwig1,part2_r_e_herwig2} and
the Lund string fragmentation model
(JETSET)~\cite{part2_r_e_jetset1,part2_r_e_jetset2}. They
agree well with data at the $Z^0$ energy scale,  but were not
intended for simulating processes at intermediate and low energy
scales.  The approximate conditions that are incorporated
into these models are not appropriate in the BEPC-II energy region.

\section{String fragmentation in the Lund model}
~~~~The transverse and longitudinal momentum of hadrons
relative to directions of the primary partons 
produced in $e^+e^-$ annihilations are governed by different 
mechanisms.  In the Lund model~\cite{part2_r_e_bobook}, 
the transverse  momentum is described as a quantum tunneling 
effect, and the longitudinal
momentum is described as fragmenting string in $(1+1)$
dimensional ({\it i.e.} time $t$ and longitudinal distance $x$) 
phase  space.

The Lund model uses a semi-classical massless
relativistic string to model the QCD color field. The
foundations of the Lund model are universal, including 
relativity, causality and quantum mechanics. The picture of the
string may be summarized as: the color force field between the
$q\bar{q}$ pair is confined to a narrow tube with potential
density $\kappa$; quarks $q$ and $\bar{q}$ located at the ends of
string; a gluon $g$ is treated as a transverse excitation of
string. The string fragmentation is pictured as: the $q$ and
$\bar{q}$ stretch field force as they move along
in opposite directions in the center-of-mass system; new
$q\bar{q}$ (or $q\bar{q}q'\bar{q}')$ pairs tunnel out from the
quantum vacuum, and the string breaks at the production 
vertix, forming two srings that terminate on the 
produced quarks.   
The fragmentation of the string is causally 
disconnected, and follows a
scaling behavior.  The produced $q\bar{q}$ 
(or $q\bar{q}q'\bar{q}'$) may form mesons and baryons if they carry
with the correct flavor quantum numbers, otherwise they just
behave as vacuum fluctuations and do not lead to any observable
effects. 


\begin{figure}[htbp]
\begin{center}
\includegraphics[width=7cm,height=5cm]{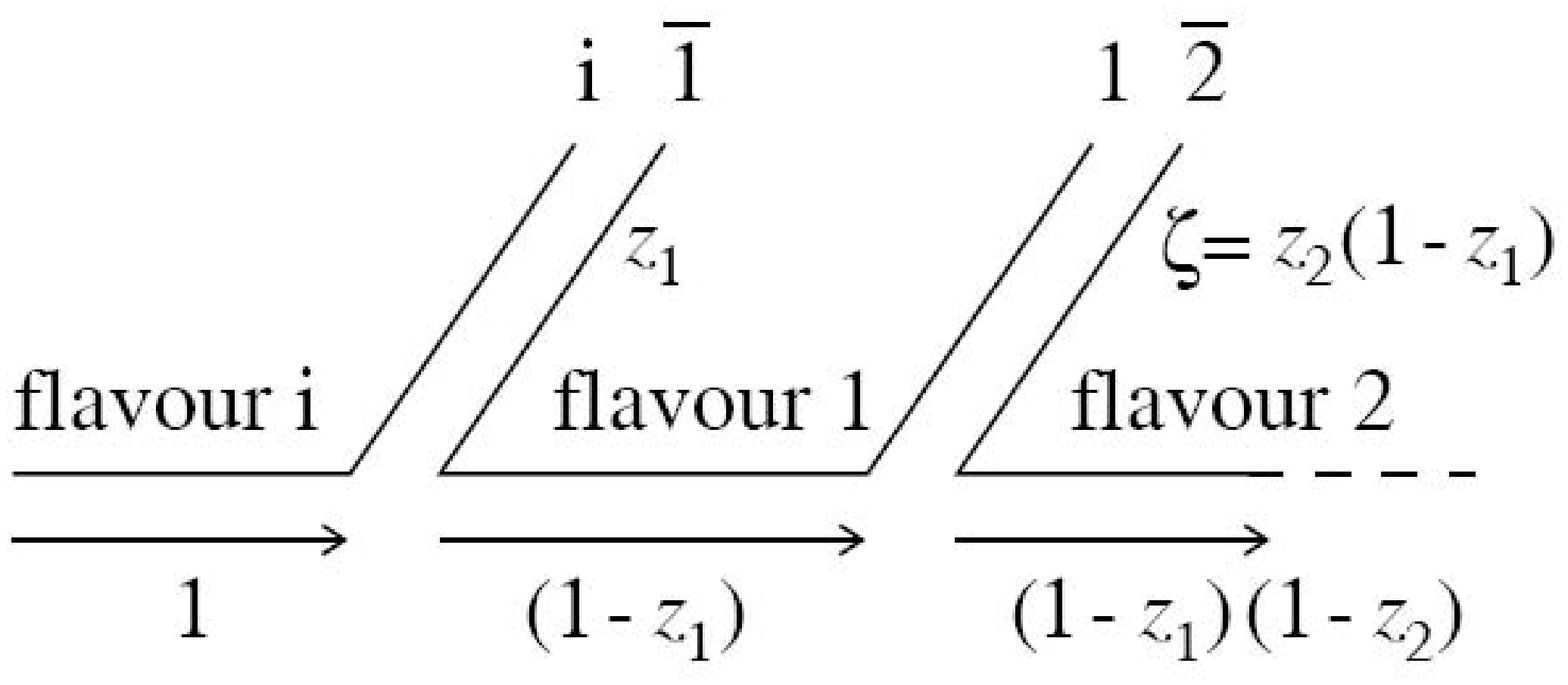}
\includegraphics[width=7cm,height=5cm]{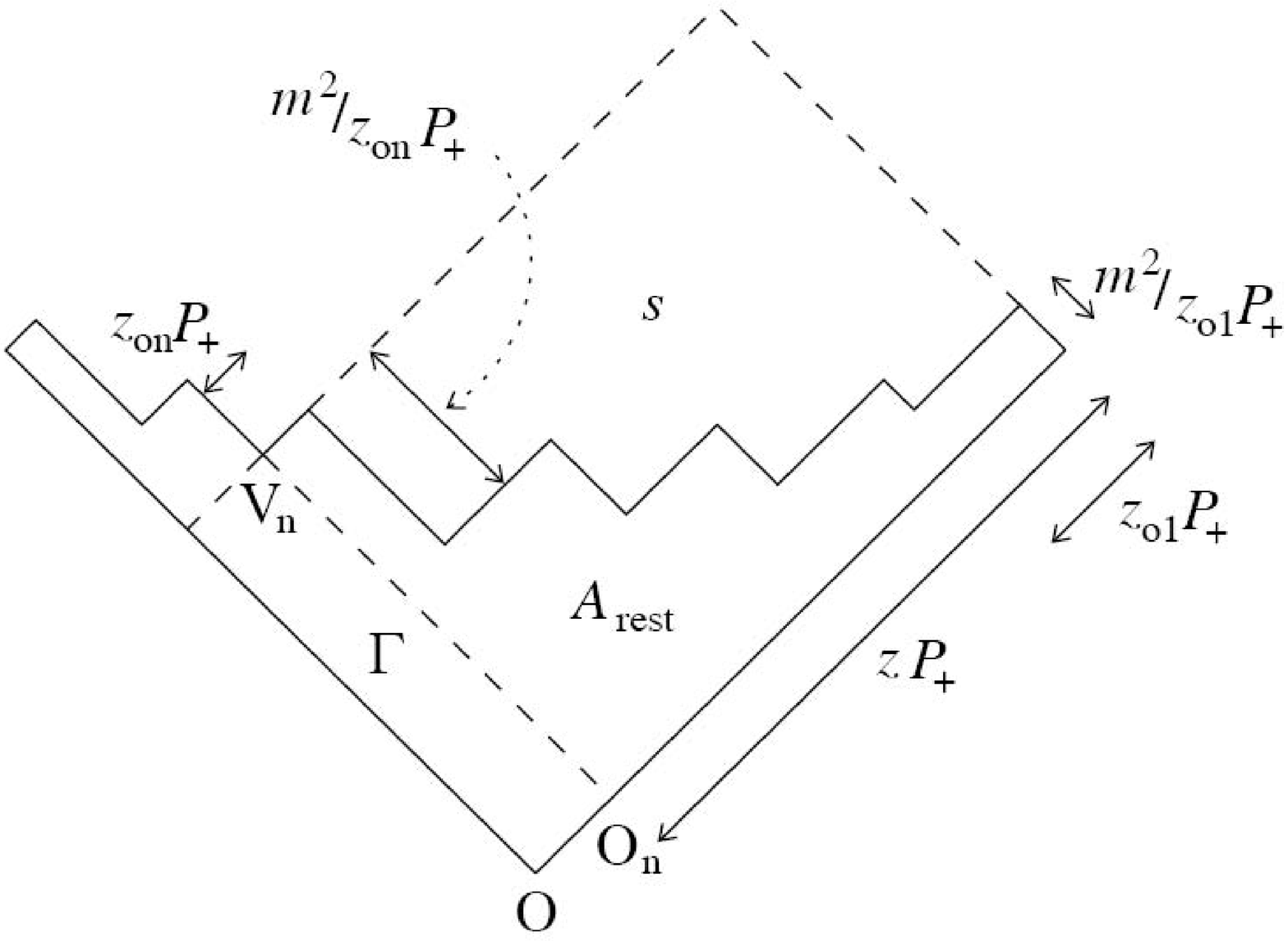}
\put(-345,100){\bf (I)} \put(-160,100){\bf (II)}
\label{hadronization}\caption{(I): an iterative cascade chain, where
$i$ is the flavor of the initial quark, and $1\bar{1}$, $2\bar{2}$
are the flavors of step-produced quark pairs, $z_j$ is the
light-cone momentum of the $j-$th particle; (II): an $n$-particle
cluster in the infinite iterative processes, where the fragmentation
starts from the right end.}\label{part2:iterative}
\end{center}
\end{figure}
In the Lund model, $n$ hadrons are produced via string fragmentation in
an iterative way~\cite{part2_r_e_bobook} [see Figure 
\ref{part2:iterative}(I)]. The Lund
model based JETSET program is a successful generator 
for high energies, such as at $Z^0$ scale, 
in which the hadron production is simulated according
to the fragmentation function in iterative and inclusive ways:
\begin{equation}\label{fz}
f(z)=\frac{N}{z}(1-z)^a\exp(-bm^2/z),
\end{equation}
where $m$ is the mass of hadron,
$z$ is the fraction of light-cone momentum, $a$ and $b$ are two
dynamical parameters and $N$ is the normalization constant. In the
deduction of $f(z)$, right-left symmetry for  string
fragmentation is assumed
(fragmentation starting from the right end and
left end of the string are equal) and the cluster with energy
$\sqrt{s}$ is produced from a total system for which an infinitely large
energy is assumed. Thus, the remaining string with area $\Gamma$
continues to have infinitely large energy after each step of the
fragmentation [see Fig. \ref{part2:iterative}(II)], which allows the
effects of hadronic masses to be ignored. At the $Z^0$ scale, 
where the total cms energy is much larger than the sum of
the masses of the produced hadrons, the above
assumptions are a good approximation. In the BEPC-II energy region,
hadronization finishes within a few fragmentation steps; Table
\ref{part2:pn} shows values for $P(n)$, the probability
for producing $n$ hadrions. The momentum of the first $n-1$
hadrons are determined by $f(z)$, while the last one is set
by the conservation of the energy-momentum. Thus, at low energies,
the momentum of a large fraction of the produced hadrons violate the Lund 
model assumptions.  This problem is avoided by having the string fragment 
in an exclusive way using the Lund area law.
\begin{table}[htbp]
\begin{center}
\caption{The probability $P(n)$ of the string fragmenting into $n$
preliminary hadrons.}
\begin{tabular}{|c|c|c|c|c|c|}\hline
$E_{ECM}$(GeV)& $P(2)$  &  $P(3)$ &$P(4)$&$P(5)$&$P(6)$\\\hline
2.200 &$0.1745$&$0.6507$&$0.1593$&0.0148&0.0005\\\hline 2.600
&$0.1416$&$0.6435$&$0.1922$&$0.0217$&$0.0010$\\\hline 3.070
&$0.1125$&$0.6256$&$0.2286$&$0.0316$&$0.0017$\\\hline
3.650 &$0.0930$&$0.6446$&$0.2925$&$0.0502$&$0.0034$\\
\hline
\end{tabular}
\label{part2:pn}
\end{center}
\end{table}

\section{The Lund area law}


In Lund model, the production of the $n$-hadron state takes the
following steps
\begin{displaymath}
e^+e^-\Rightarrow q\bar{q}\Rightarrow {\rm string} \Rightarrow
m_1+m_2+\cdots +m_n.
\end{displaymath}
Similar to the expression in quantum field theory, the total matrix
element of the hadronization is written
as~\cite{part2_r_e_luarlw2001,part2_r_e_hadron03}
\begin{equation}
{\cal M}\equiv {\cal M_{\rm QED}}(e^+e^-\rightarrow q\bar{q}){\cal
M}_{\rm LUND}(q\bar{q}\rightarrow m_1,m_2,\cdots m_n).
\end{equation}
The first step is described by the usual QED matrix element ${\cal
M}_{QED}$. The dynamics of the second step, which  governs a string
fragmenting into a particular $n-$particle state with masses
$m_1,m_2,\cdots ,m_n$ is factorized as
\begin{equation}
{\cal M}_{\rm LUND}(q\bar{q}\rightarrow m_1,m_2,\cdots m_n) = C_n
{\cal M}_{\bot} {\cal M}_{//},
\end{equation}
where ${\cal M}_{\bot}$ and ${\cal M}_{//}$ describe the transverse
and longitudinal momentum distributions (relative to
the direction of the initial $q\bar{q}$ momentum), respectively,
and $C_n$ is a dimensionless normalization constant.

\begin{figure}
\begin{center}
\includegraphics[width=7cm,height=5cm]{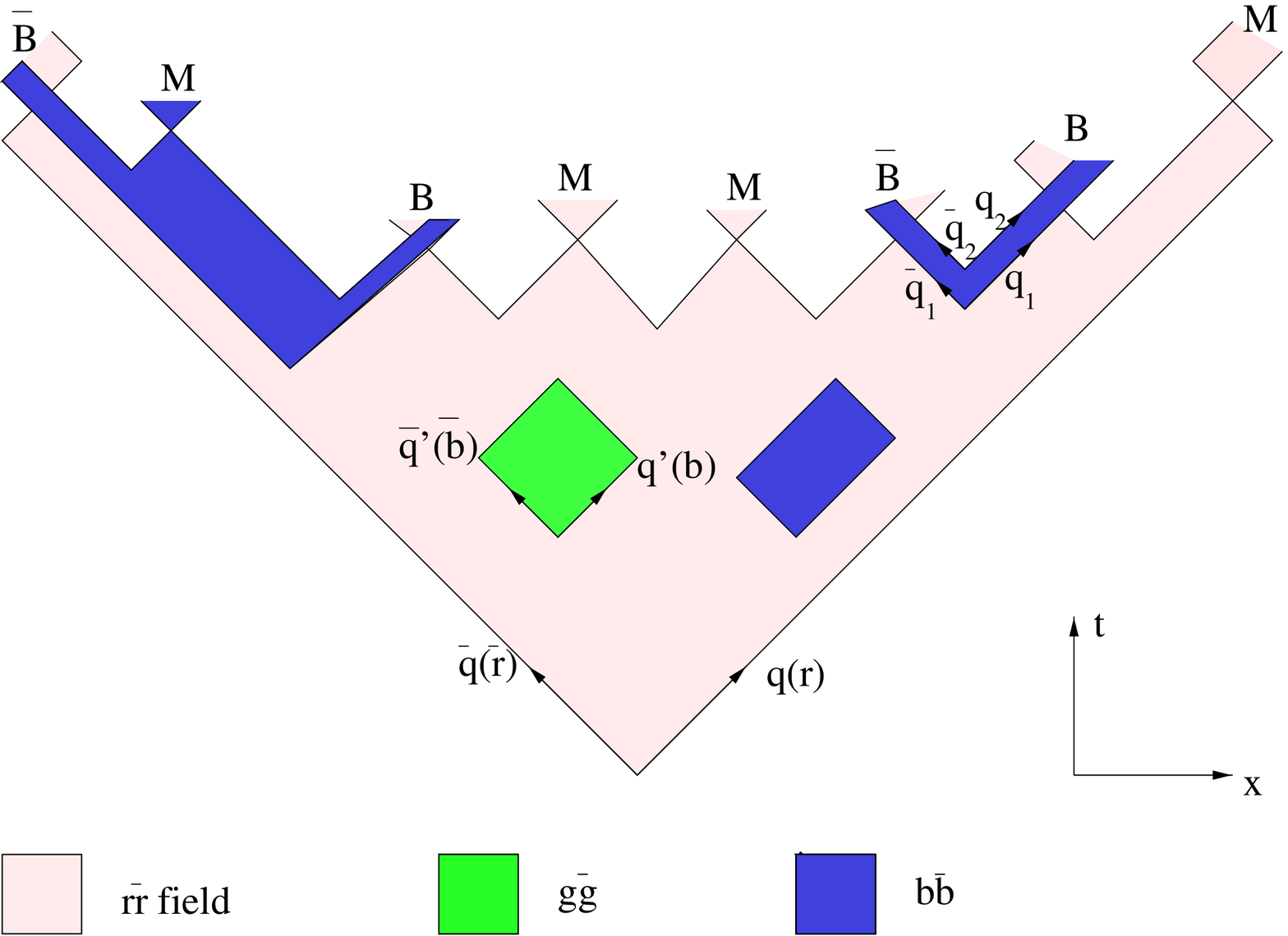}
\includegraphics[width=8cm,height=6cm]{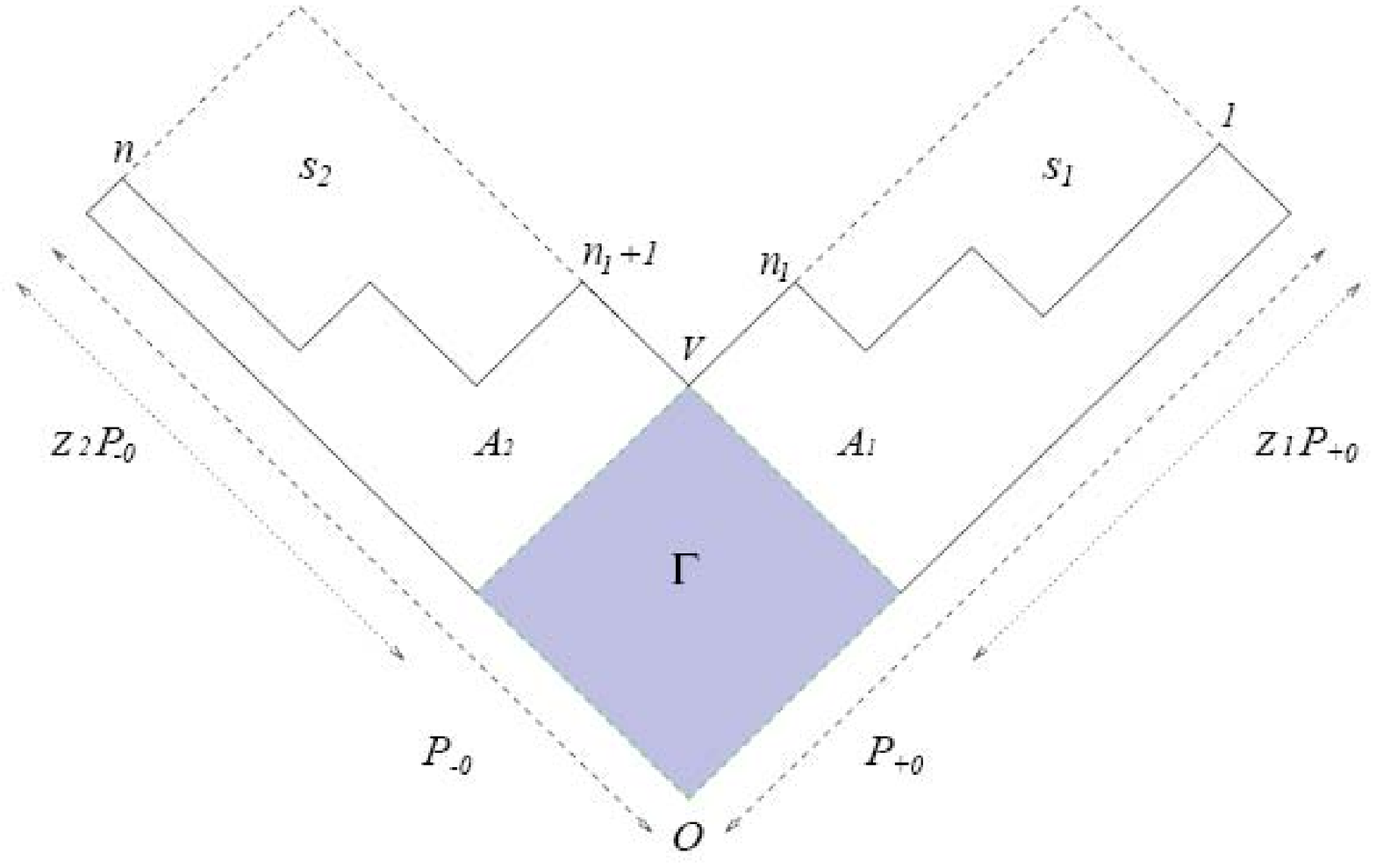}
\caption{(a) String fragmentation in time-longitudinal phase-space
by a set of new pairs ($q\bar{q}$ or $q\bar{q}q\bar{q}$)
production, hadrons (mesons $M$ and baryons $B$) form at 
the vertices; (b)
the vertex $V$ divides the $n$-body string fragmentation into two
clusters that contain $n_1$ and $n_2$ hadrons with squared
invariant masses $s_1$ and $s_2$.}\label{lundarea}
\end{center}
\end{figure}

The origin of transverse momentum is from quantum fluctuations in 
the ground state of the string, which implies a Gaussian distribution,
\begin{equation}
{\cal M}_{\bot}=\exp(-\sum_{j=1}^n\vec{k}_{j }^{~2}),
\end{equation}
where the relation between transverse momentum $\vec{p}_{\bot}$ and
the dimensionless vector $\vec{k}$ is
\begin{equation}
\vec{k}_{j}\equiv \frac{\vec{p}_{\bot j}}{2\sigma},
\end{equation}
and the variance is $\sigma^2\equiv <\vec{p}_{\bot}^{~2}>$. The
longitudinal matrix is described by the Lund area law
\begin{equation}
{\cal M}_{//}=\exp(i\xi {\cal A}_n), \label{matrix}
\end{equation}
where ${\cal A}_n$ is the area enclosed by the quark-antiquark
light-cone lines of $n$-particles (see Fig.~\ref{lundarea}). The
parameter $\xi$ is given by
\begin{equation}
\xi=\frac{1}{2\kappa} +i\frac{b}{2},
\end{equation}
where $\kappa$ is the string tension constant and $b$ is
a flavor-independent dynamical parameter.
The phase of the hadronic state is given by the real part of $\xi$,
and the imaginary part corresponds to the production rate of the
hadronic state. 

\section{Solutions of the Lund area law}
~~~~In the BEPC-II energy region, the emitted gluons are usually soft,
which leads to a small broadening of the two-jet system. The quark
and the accompanying gluons behave as a single quark-jet. 
Gluon emission does not change the topological shapes of the final
states and may be neglected.

The probability for the string to fragment into $n-$particles state
can be written as
\begin{equation}
d\wp_n (m_1,m_2,\cdots m_n)=\delta (1-\sum_{j=1}^n\frac{m_{\bot
j}^2}{sz_j})\delta (1-\sum_{j=1}^n z_j)\delta
(\sum_{j=1}^n\vec{k}_j) \overline{\sum}|{\cal M}_{LUND}|^2d\Phi_n,
\label{lundml}
\end{equation}
where $z_j$ is the light-cone momentum $z_j\equiv(E_j\pm
p_{zj})/(E_0\pm P_{z0})$, $m_{\bot j}$ is the transverse mass
\begin{equation}
m_{\bot j}=\sqrt{m_j^2+p_{\bot j}^2}=\sqrt{m_j^2+4\sigma^2
\vec{k}_{j}^2} ,~~~~~~(j=1,2,\cdots n),
\end{equation}
and $d\Phi_n$ is the $n-$particle dimensionless phase-space element
\begin{equation}
d\Phi_n=\prod_{j=1}^n d^2\vec{k}_{j} \frac{dz_j}{z_j}.
\end{equation}
The summation over the fragmentation hadrons gives the following
factor
\begin{equation}
\overline{\sum}\Longrightarrow B_n\prod\limits_j^n(VPS)_j(SUD)_j,
\end{equation}
where $B_n$ is the combinatorial number of string fragmentations,
$(VSP)$ is the ratio of vector meson to pseudoscalar meson 
production, which is suppressed from
its spin-state counting value $3:1$, and $(SUD)$ reflects 
the suppression of strange meson production relative to 
that for ordinary mesons.

When integrated over the kinematic variables of the $n$ hadrons,
the Lund area law [Eq.~(\ref{lundml})] has the following
solutions for~\cite{part2_r_e_luarlwbohu,part2_r_e_luarlwhu}:\\
\noindent  $\bullet$ String $\Rightarrow$ 2 hadrons
\begin{equation}
\wp_2=\frac{C_2}{\sqrt{\lambda (s,m_{\bot 1}^2,m_{\bot
2}^2)}}[\exp(-b{\cal A}_2^{(1)}+\exp(-b{\cal A}_2^{(2)})];
\end{equation}

\noindent $\bullet$ String $\Rightarrow$ 3 hadrons
\begin{equation}
d\wp_3=\frac{C_3}{\sqrt{\Lambda}}\exp(-b{\cal A}_3)d{\cal A}_3;
\end{equation}
The area law is  Lorentz invariant and factorizes
so that the total system may be divided into two
subsystems, each of which contains $n_1$ and $n_2$ hadrons $(n_1, n_2
= 2~{\rm or}~3)$ [see Fig.~\ref{lundarea}(b)]. Applying the analytical
results for each subsystem, one gets
\begin{equation}
\label{e3e4} d\wp_n(s;s_1,s_2)=\frac{ds_1ds_2}{\sqrt{\lambda
(s,s_1,s_2)}} [\exp(-b\Gamma^{(1)})+exp(-b\Gamma^{(2)} )]
 \wp_{n_1}(s_1)\wp_{n_2}(s_2).
\end{equation}
Multi-body fragmentation can be treated in a similar way.


\section{Preliminary multiplicity}
~~~~The Lund area law can give the multiplicity distribution of the
fragmentation hadrons. In fact, the dimensionless $n$-particles
(neutral and charged) partition function can be defined
as~\cite{part2_r_e_luarlw2001}
\begin{equation}
Z_n=\int d\Phi_n exp(-b{\cal A}_n).
\end{equation}
The relation between $Z_n$ and the multiplicity distribution is
\begin{equation}
P_n=\frac{Z_n}{\sum Z_n},
\end{equation}
which has the approximate form
\begin{equation}
P_n=\frac{\mu^n}{n!}\exp[c_0+c_1(n-\mu)+c_2(n-\mu)^2].
\end{equation}
The quantity $\mu$ can be written in the energy-dependent form
\begin{equation}
\mu =\alpha +\beta \exp(\gamma \sqrt{s}),
\end{equation}
where the parameters $c_0$, $c_1$, $c_2$, $\alpha$, $\beta$ and $\gamma$
have to be tuned with experimental data.



\section{Parameter tuning}

The Lund area law is incorporated into the JETSET program, mainly
as the subroutine LUARLW, which contains many phenomenological
parameters. Their values are unknown and must be tuned by
comparing with data to ensure good agreement for important
distributions, such as energy, momentum, multiplicities, polar
angle etc. When comparing data distributions with Monte
Carlo events, the statistics of all kinds of events (hadronic
events, QED backgrounds and beam associated backgrounds) in Monte
Carlo sample and data have to be the same, and this requirement 
can be met by
\[
{\rm data \Leftrightarrow Monte Carlo
~~sample}\left\{\begin{array}{cc}
{\rm beam-gas} & BG^* \\
\mu^+\mu^- & L\cdot\sigma_{\mu\mu}\\
e^+e^- & L\cdot\sigma_{ee} \\
\tau^+\tau^- & L\cdot\sigma_{\tau\tau} \\
\gamma\gamma & L\cdot\sigma_{\gamma\gamma} \\
{\rm hadrons} & L\cdot\sigma_{had} .
\end{array}
\right.
\]
The beam associated sample, $BG^*$, can be obtained from the raw data
according to the methods described in Ref.~\cite{part2_r_e_bgstar}.

\begin{figure}
\begin{center}
\includegraphics[width=3.5cm,height=3.5cm]{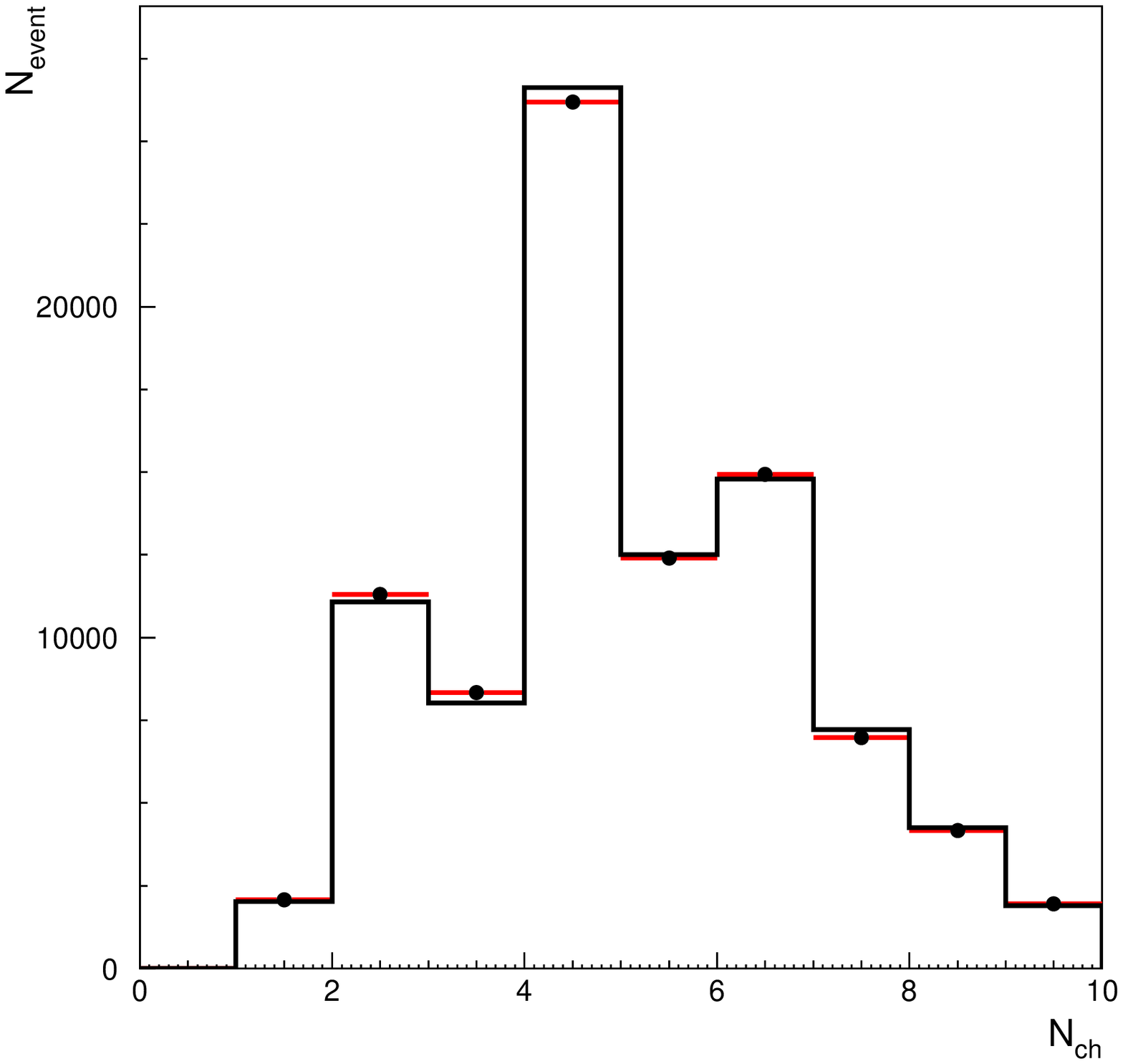}
\includegraphics[width=3.5cm,height=3.5cm]{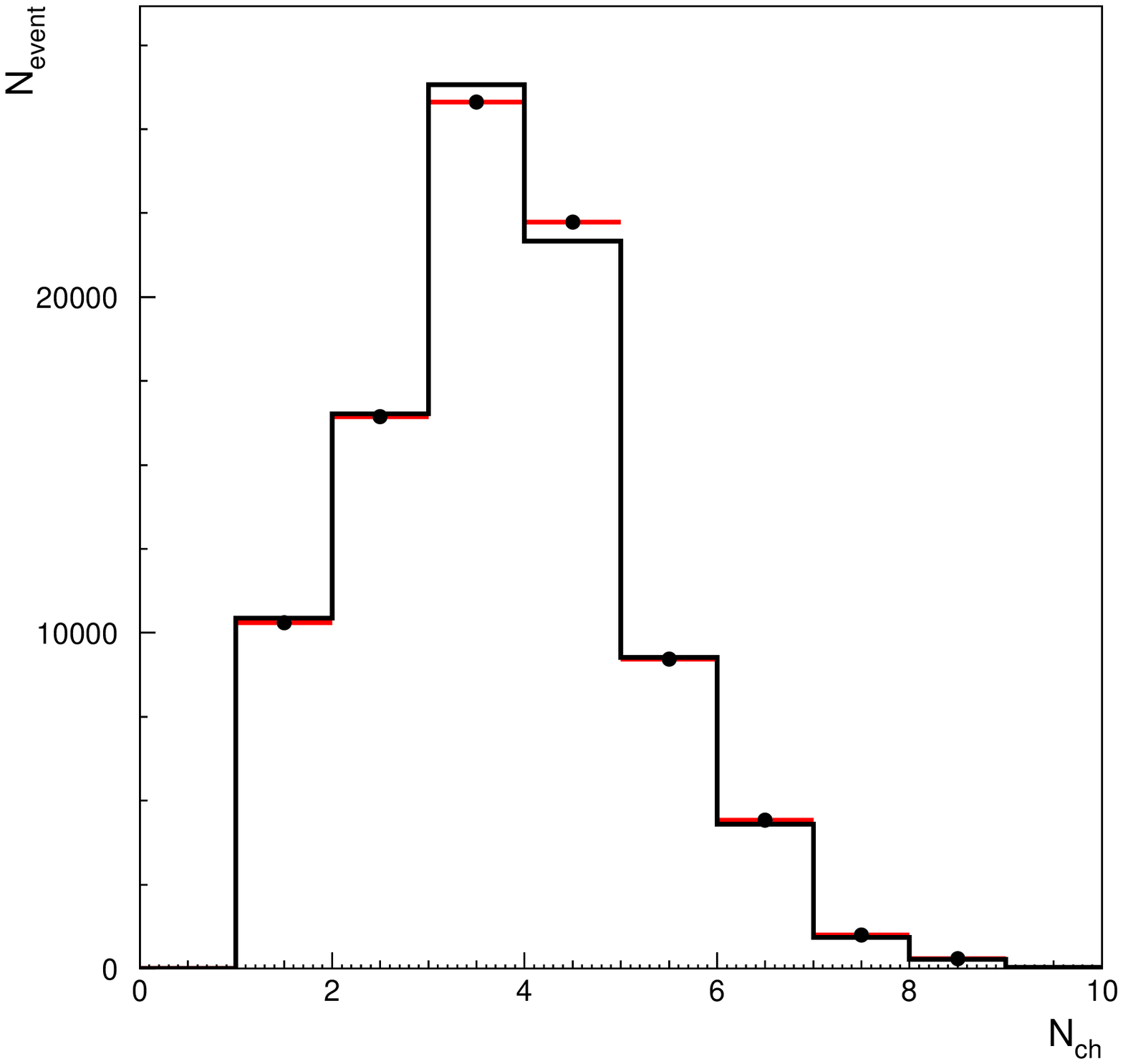}
\includegraphics[width=3.5cm,height=3.5cm]{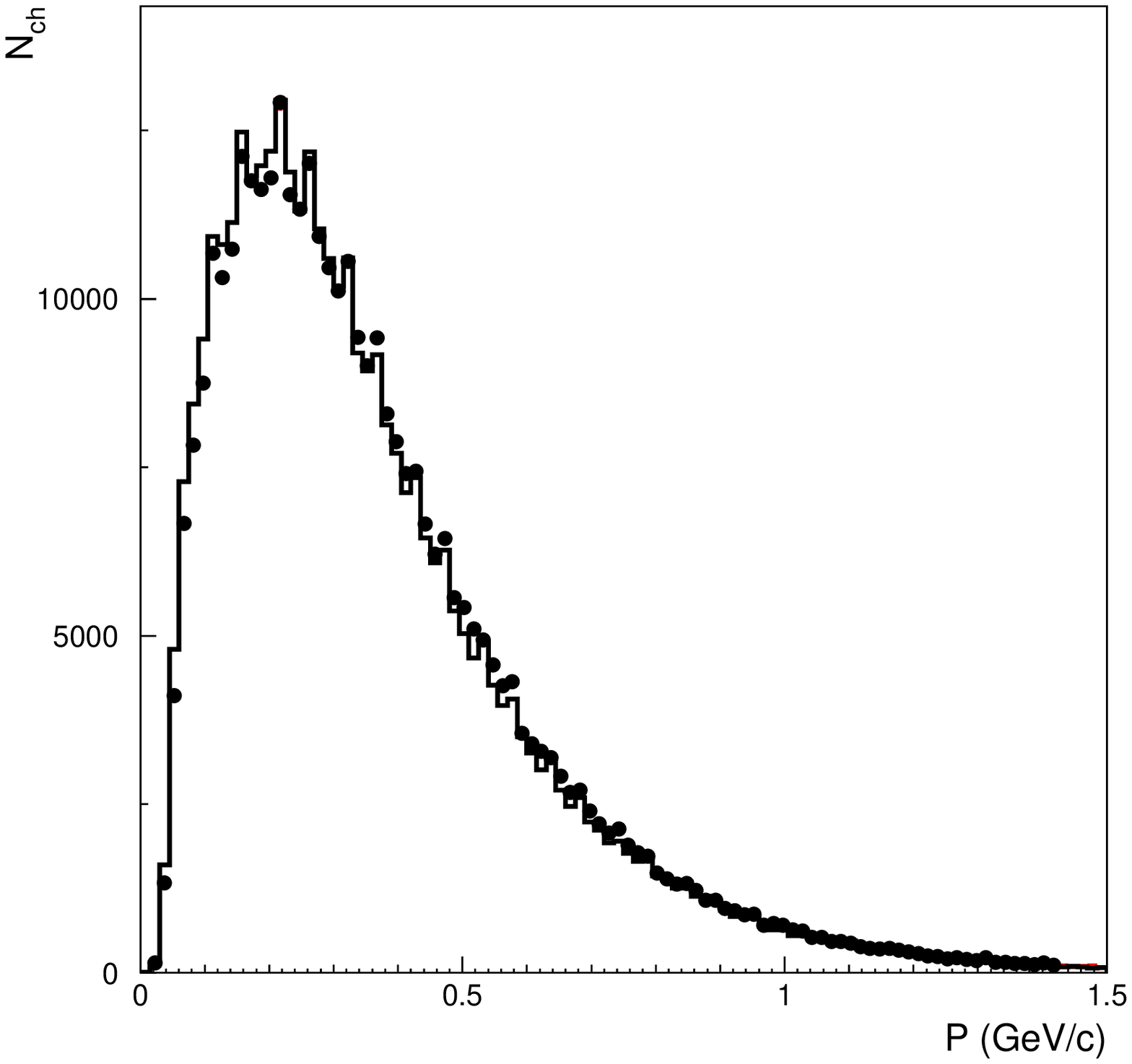}
\includegraphics[width=3.5cm,height=3.5cm]{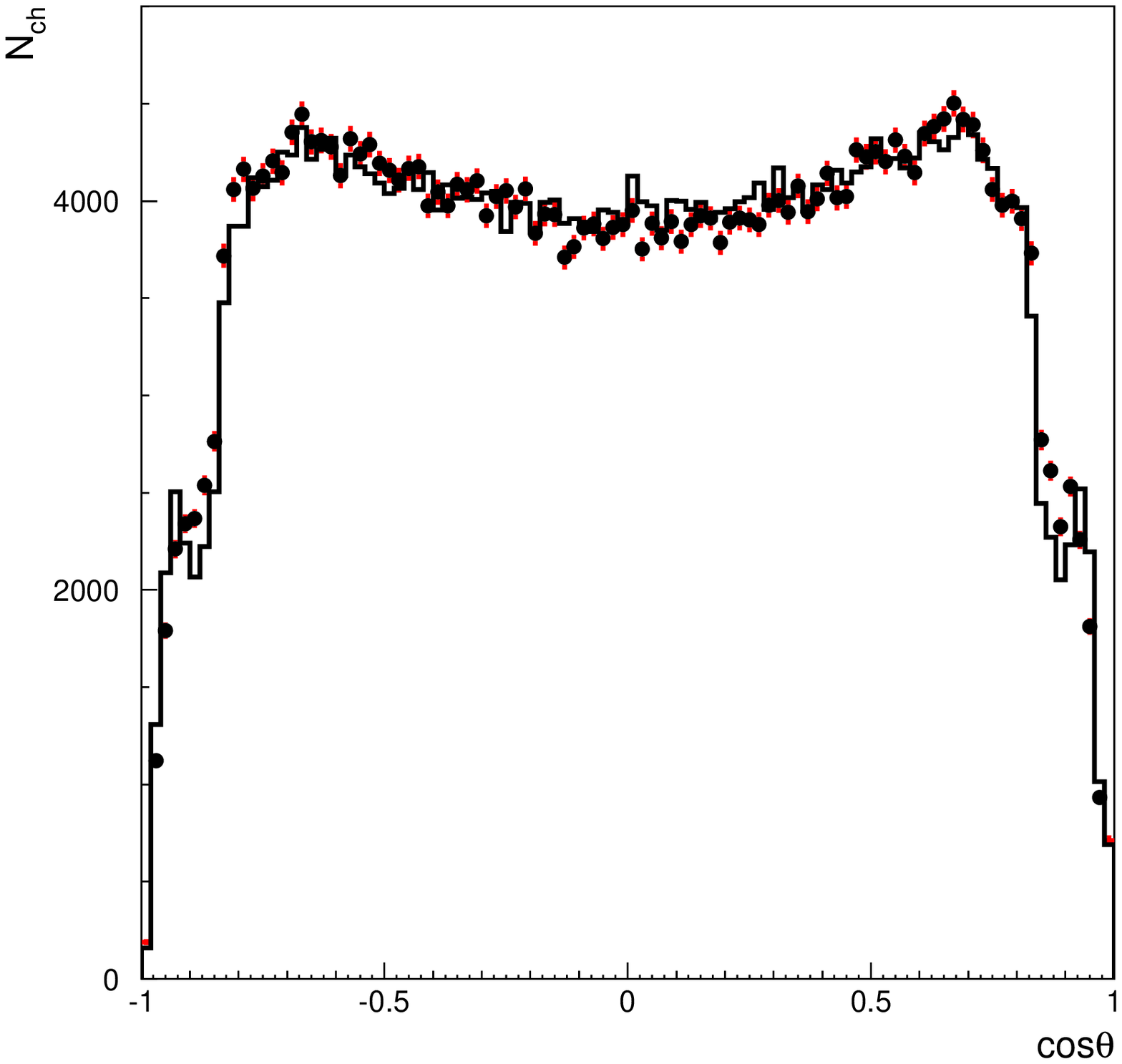}
\includegraphics[width=3.5cm,height=3.5cm]{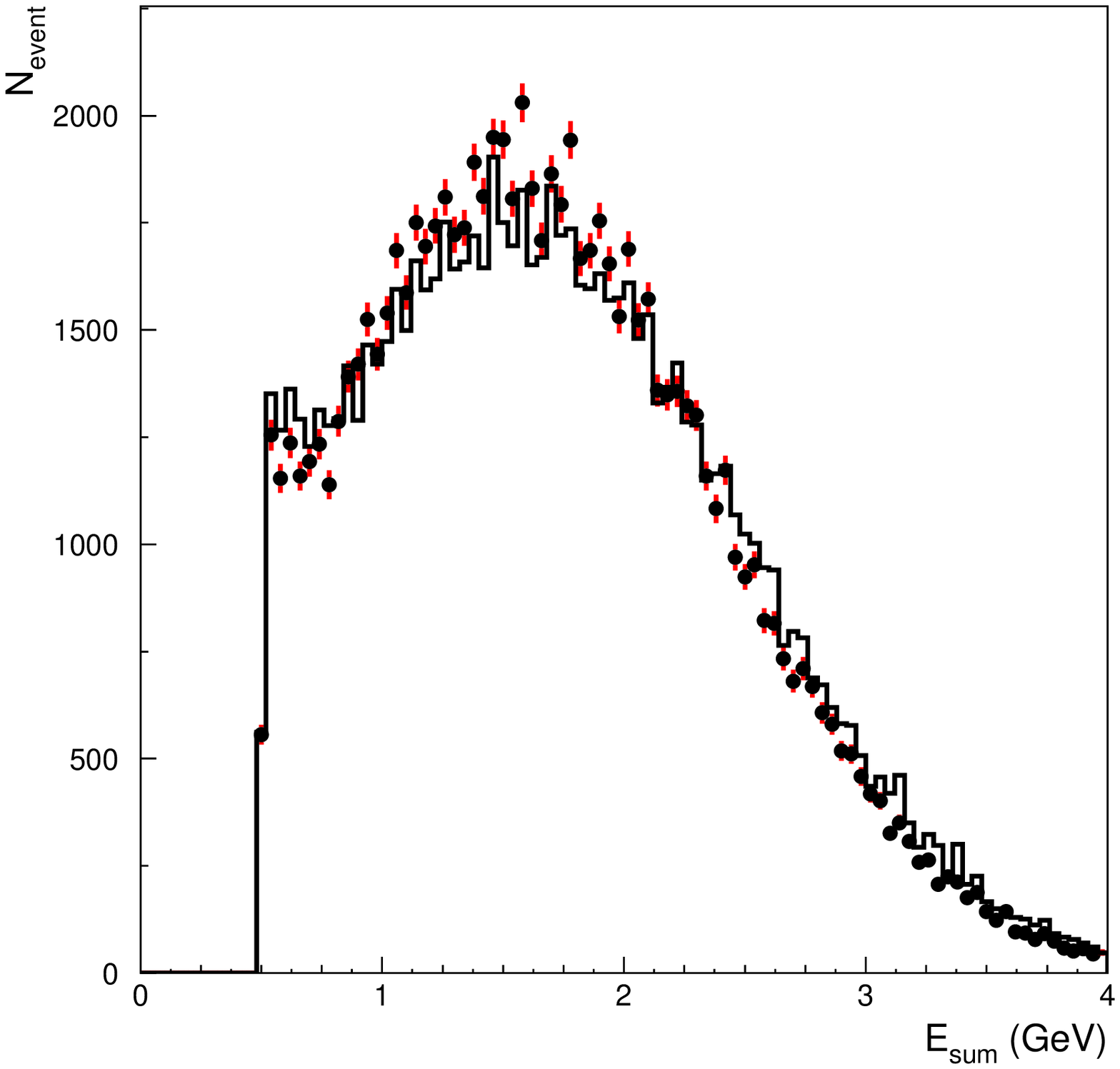}
\includegraphics[width=3.5cm,height=3.5cm]{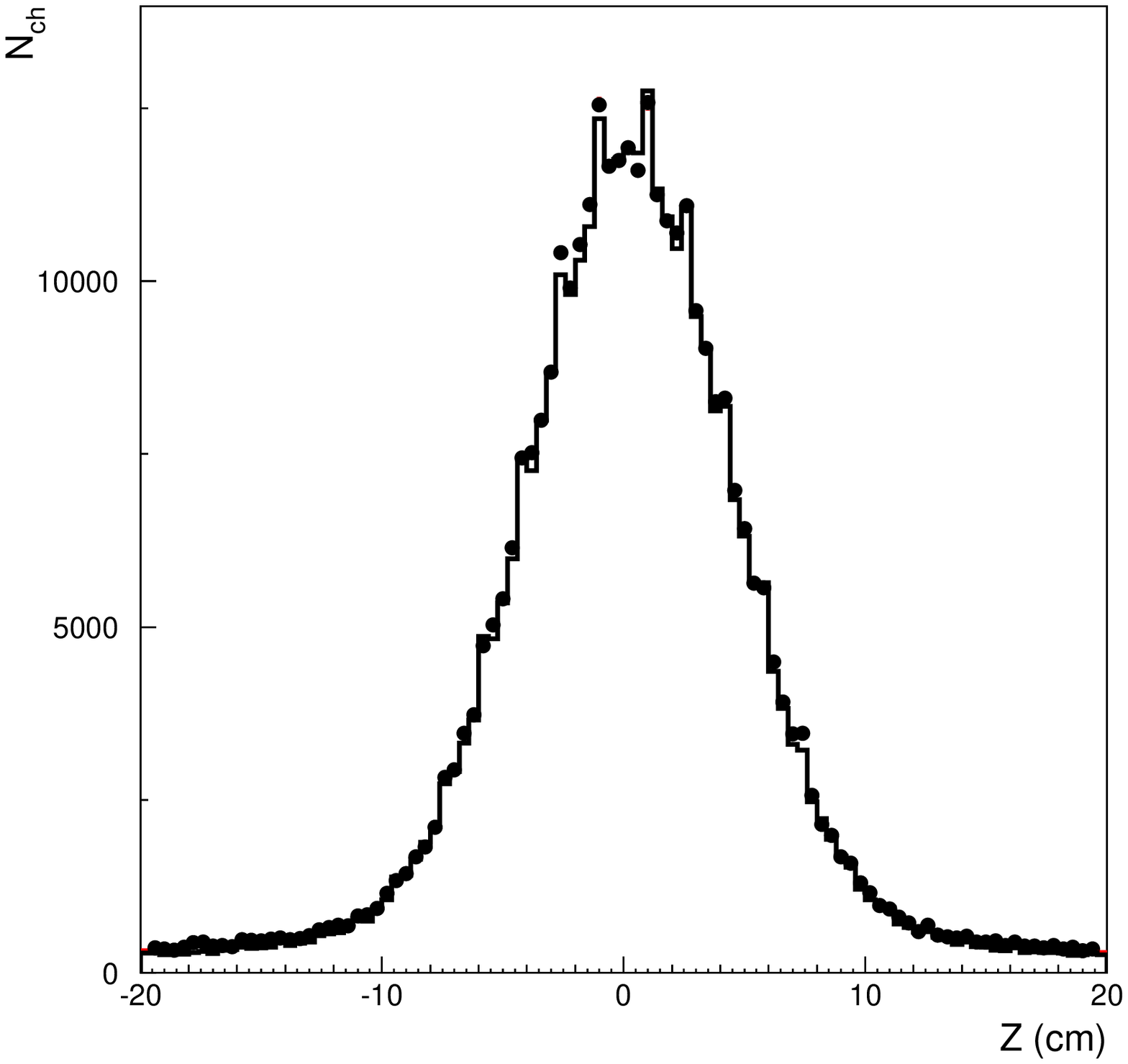}
\includegraphics[width=3.5cm,height=3.5cm]{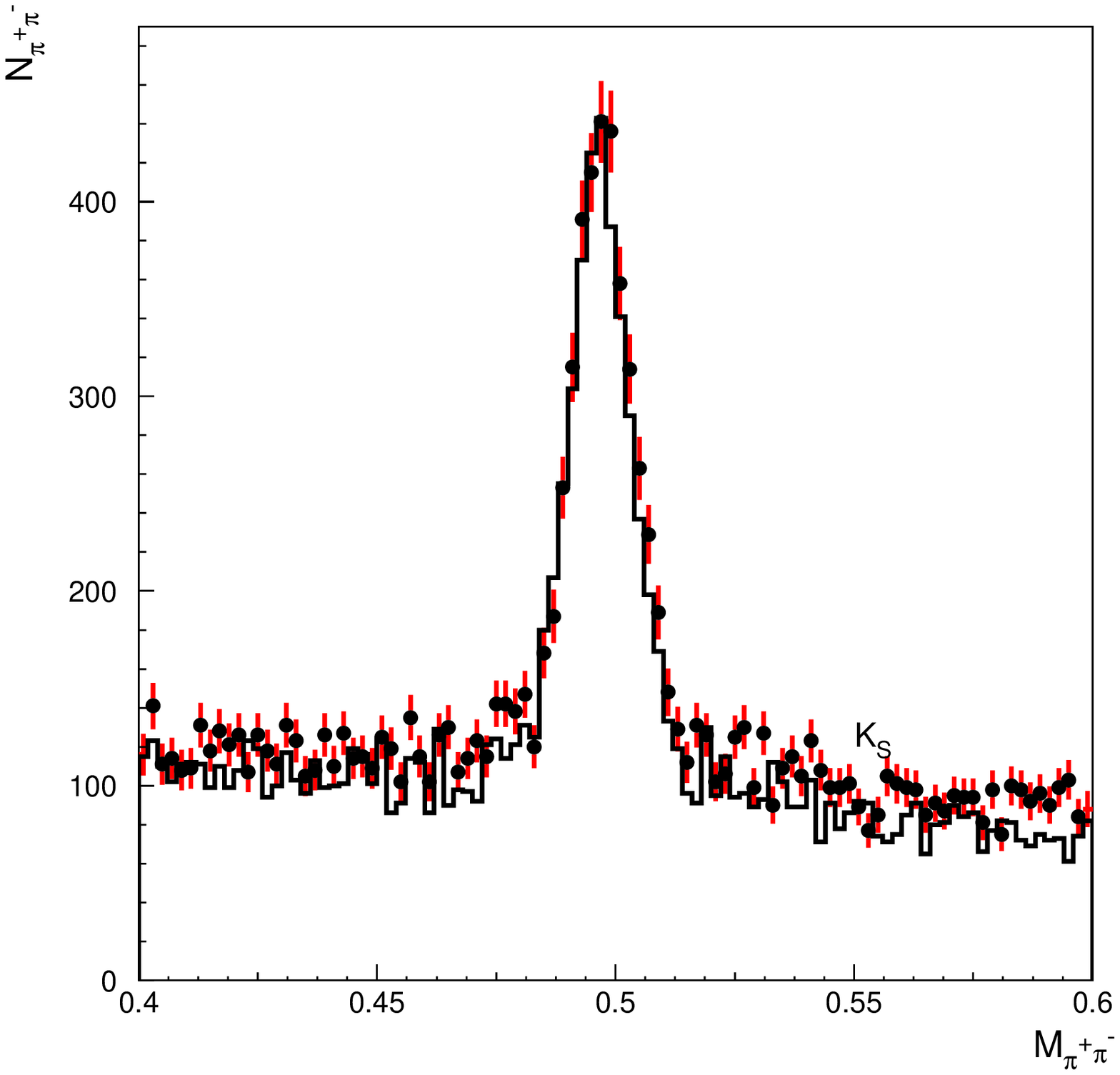}
\includegraphics[width=3.5cm,height=3.5cm]{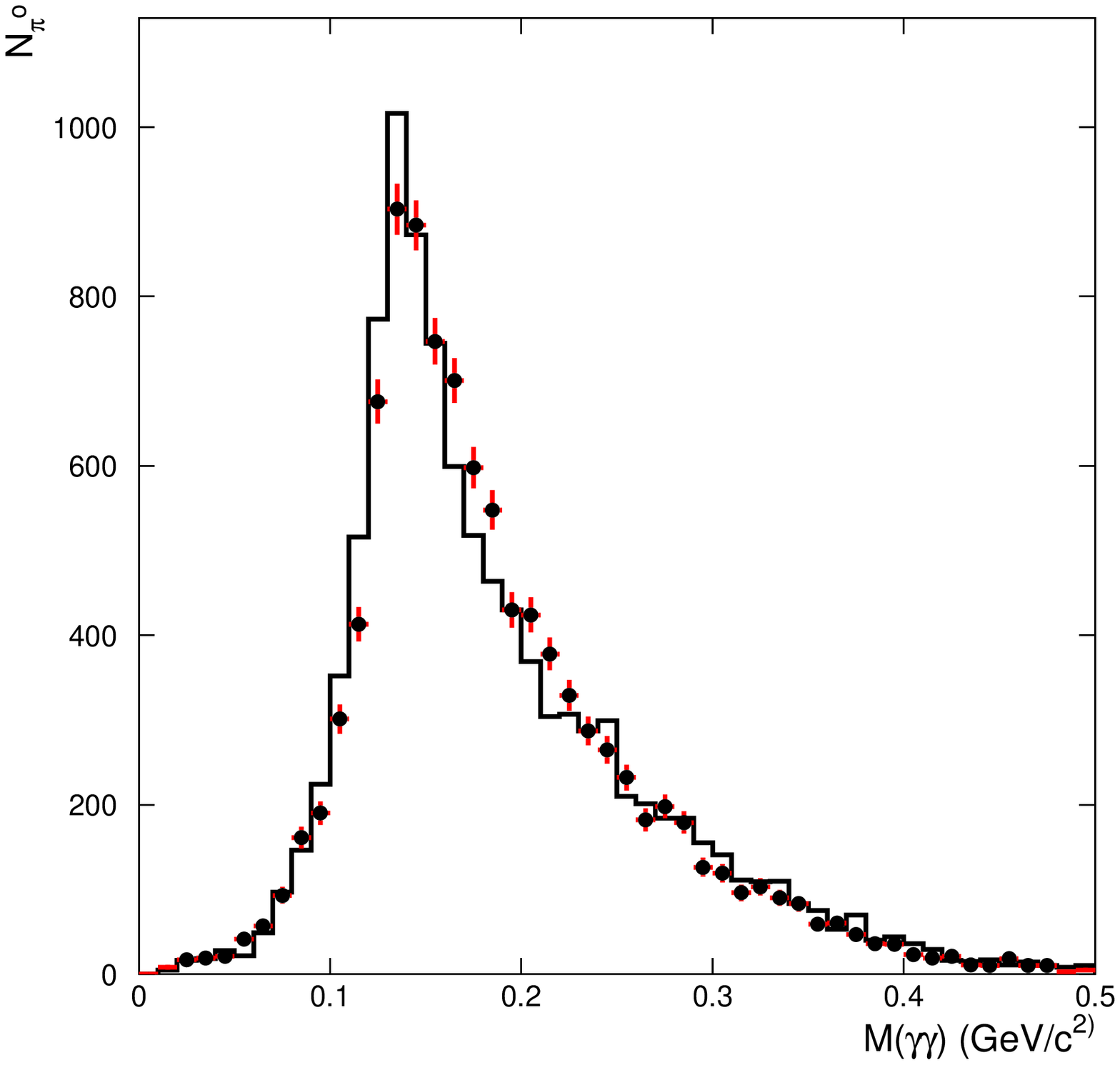}
\caption{Comparisons of data with simulated LUARLW events for
distributions of:  multiplicity of
charged tracks, multiplicity of  good tracks, momentum $p$,
polar-angle $\cos\theta$, energy deposited in BSC, vertex position of
charged tracks, invariant mass of $K_S\to\pi^+\pi^-$ and
$\pi^0\to\gamma\gamma$, at $\sqrt{s}=3.65$~GeV.}\label{datalundarea}
\end{center}
\end{figure}

The distributions to be compared are those that are directly
related to the hadronic criteria: multiplicity, space position,
momentum, polar-angle, energy deposit, the ratio of $\pi/K$, the
fractions of the short life-time particle $K_S$ and $\Lambda$
(which will influence their secondary vertexes), the time of
flight etc. The main parameters to be tuned are:
$PARJ(1-3)$, $PARJ(11-17)$ in JETSET~\cite{part2_r_e_jetset} and
the parameters in
LUARLW~\cite{part2_r_e_luarlwbohu,part2_r_e_luarlwhu}. Some
comparisons of the sensitive distributions between \bes3 data and
LUARLW at detector level are illustrated in
Fig.~\ref{datalundarea}.


\chapter[R values and precision test of the Standard Model]{R values  and
precise test of the Standard Model \footnote{By Hai-Ming Hu and Xiao-Yuan 
Li}}
\label{sec:r_value}

\def\amu{a_\mu}
\def\damu{\delta \amu}
\def\amuh{a_\mu^{{\mathrm had}}}
\def\MZ{M_Z}
\def\GZ{\Gamma_Z}
\def\MW{M_W}
\def\Mt{m_t}
\def\swsqeffl{\sin^2\theta^{\rm lept}_{\rm eff}}
\def\a{\alpha}
\def\ai{\alpha^{-1}}
\def\aiz{\alpha^{-1}(\MZ)}
\def\dalf{\Delta\alpha}
\def\dalp{\delta \dalf}
\def\das{\Delta\alpha(s)}
\def\daz{\Delta\alpha(\MZ)}
\def\dah{\Delta\alpha^{(5)}_{\rm had}}
\def\dahs{\Delta\alpha^{(5)}_{\rm had}(s)}
\def\dahz{\Delta\alpha^{(5)}_{\rm had}(\MZ^2)}
\def\dahzE{\Delta\alpha^{(5)}_{\rm had}(-\MZ^2)}
\def\dah0{\Delta\alpha^{(5)}_{\rm had}(-s_0)}

\newcommand{\gamu}{\gamma_{\mu}}
\newcommand{\gv}{\mbox{GeV}}
\newcommand{\mv}{\mbox{MeV}}
\newcommand{\tv}{\mbox{TeV}}
\newcommand{\adp}{\frac{\alpha_s}{4 \pi} }
\newcommand{\Gmu}{G_{\mu}}
\newcommand{\mz}{M^2_Z}
\newcommand{\mw}{M^2_W}

\newcommand{\dal}{\Delta \alpha}
\newcommand{\dro}{\Delta \rho}
\newcommand{\sinth}{\sin^2 \Theta\: .}
\newcommand{\costh}{\cos^2 \Theta\: .}
\newcommand{\sinf}{\sin^2 \Theta_f}
\newcommand{\cosf}{\cos^2 \Theta_f}
\newcommand{\sini}{\sin^2 \Theta_i}
\newcommand{\cosi}{\cos^2 \Theta_i}
\newcommand{\sinW}{\sin^2 \Theta_W}
\newcommand{\cosW}{\cos^2 \Theta_W}
\newcommand{\sing}{\sin^2 \Theta_g}
\newcommand{\cosg}{\cos^2 \Theta_g}

\newcommand{\wz}{\sqrt{2}}
\newcommand{\ha}{\frac{1}{2}}
\newcommand{\daa}{\frac{\delta \alpha}{\alpha}}
\newcommand{\dee}{\frac{\delta e}{e}}
\newcommand{\epm}{e^+e^-}
\newcommand{\pipi}{\pi^+\pi^-}
\newcommand{\epmb}{e^+e^- \rightarrow b\bar{b}}
\newcommand{\epmf}{e^+e^- \rightarrow f\bar{f}}
\newcommand{\epmm}{e^+e^- \rightarrow \mu^+\mu^-}
\newcommand{\sha}{\sigma(e^+e^- \rightarrow {\rm hadrons})}
\newcommand{\noi}{\noindent}
\newcommand{\bit}{\begin{itemize}}
\newcommand{\eit}{\end{itemize}}
\newcommand{\bary}{\begin{array}}
\newcommand{\eary}{\end{array}}
\newcommand{\crn}{\nonumber \\}
\newcommand{\bet}{\begin{center} \begin{tabular}}
\newcommand{\ent}{\end{tabular} \end{center}}
\newcommand{\bb}{\begin{thebibliography}}
\newcommand{\eb}{\end{thebibliography}}
\newcommand{\ci}[1]{\cite{#1}}
\newcommand{\bi}[1]{\bibitem{#1}}  
\newcommand{\lab}[1]{\label{#1}}
\newcommand{\re}[1]{(\ref{#1})}
\newcommand{\SM}{Standard Model }
\newcommand{\STi}{Slavnov-Taylor identity }
\newcommand{\WTi}{Ward-Takahashi identity }
\newcommand{\STis}{Slavnov-Taylor identities }
\newcommand{\WTis}{Ward-Takahashi identities }

\newcommand{\eps}{\epsilon}
\newcommand{\veps}{\varepsilon}
\newcommand{\gapprox}{\raisebox{-.2ex}{$\stackrel{\textstyle>}
{\raisebox{-.6ex}[0ex][0ex]{$\sim$}}$}}
\newcommand{\lapprox}{\raisebox{-.2ex}{$\stackrel{\textstyle<}
{\raisebox{-.6ex}[0ex][0ex]{$\sim$}}$}}
\newcommand{\MSS}{${\mathrm{MS}}$ }
\newcommand{\MOM}{${\mathrm{MOM}}$ }
\newcommand{\OS}{${\mathrm{OS}}$ }
\newcommand{\MSb}{$\overline{\mathrm{MS}}$ }
\newcommand{\MSs}{$\overline{\mathrm{MS}}$ scheme}
\newcommand{\MSbm}{\overline{\mathrm{MS}} }
\newcommand{\alms}{\alpha_{\bar{\rm MS}}}
\newcommand{\dams}{\delta \alpha_{\bar{\rm MS}}}   
\newcommand{\alos}{\alpha_{\rm OS}}
\newcommand{\daos}{\delta \alpha_{\rm OS}} 
\newcommand{\daaos}{\left.\frac{\delta \alpha}{\alpha}\right|_{\rm OS}}
\newcommand{\daams}{\left.\frac{\delta \alpha}{\alpha}\right|_{\bar{\rm 
MS}}}
\newcommand{\stt}{s_\theta^2}
\newcommand{\ctt}{c_\theta^2}
\newcommand{\ctf}{c_\theta^4}
\newcommand{\cts}{c_\theta^6}
\newcommand{\ctit}{c_\theta^{-2}}
\newcommand{\Lnqmone}{L_{qM_1}}
\newcommand{\Lnqmtwo}{L_{qM_2}}


Measurements of the total cross section for $e^+e^-$--annihilation
into hadrons are indispensable input for the determination of the
non--perturbative hadronic contribution to the running of the  QED
fine  structure constant, an essential input parameter in precision
electroweak measurements. A number of excellent reviews on this subject
are available~\cite{part2:r:recent}. Some of the material from these
review papers is summarized here. In Sect.~\ref{part2:sec:r:1} the
definition of $R_{\rm had}$, its experimental determination and the
present status of $R_{\rm had}$ measurements at low energy are presented.
Evaluations of $\dahz$, the non-perturbative hadronic
contribution to the running of the fine structure constant, are
given in Sect.~\ref{part2:sec:r:2}. Section~\ref{part2:sec:r:3} contains
a discussion of the choice of input parameters for Standard Model tests,
in particular, the effective fine structure constant at
the scale $\sqrt{s}=M_{\rm Z}$.

\section{$e^+e^- \rightarrow {\rm
hadrons}$ cross sections and the $R_{had}$ value}
\label{part2:sec:r:1}
$R_{\rm had}(s)$, by the proper definition, is the {\em ratio of
the total cross sections} according to following equation,

\be R_{\rm had}(s)=\frac{\sigma_{tot} (\epm \ra \gamma^* \ra
hadrons)} {\sigma (\epm \ra \gamma^* \ra \mu^+ \mu^-)}\;.
\label{Rdef}
\ee
\noindent
Usually, however, experiments do not determine $R_{\rm had}$ as the
ratio of the total cross sections given in  Eq.~\ref{Rdef}.
Rather, the hadronic experimental cross section is first corrected
for QED 
effects~\cite{part2:r:BonneauMartin,part2:r:Tsai,part2:r:BerendsKleiss,part2:r:Eidelman78}, 
which include bremsstrahlung as well as vacuum
polarization corrections. The latter account for the running of
the fine structure constant $\alpha(s)$.  After these corrections
are applied, the resulting $\sigma_{tot}$ is divided by the Born cross
section $\sigma_0 (\epm \ra \gamma^* \ra
\mu^+\mu^-)=\frac{4\pi\alpha ^2}{3s}$.  As a result, the ``working''
definition of $R_{\rm had}(s)$ is:

\bea
R_{\rm had}(s)=\frac{\sigma_{tot} (\epm \ra \gamma^* \ra 
hadrons)_{exp}^{corr}} {\sigma_0 (\epm \ra \gamma^* \ra \mu^+
\mu^-)}\;.
\eea
\noindent
Note that, the experimental cross section $\sigma (\epm \ra
\gamma^* \ra \mu^+ \mu^-)$ never appears here.  They are,
however, used by some authors to check how good the normalizations are
(see {\it e.g.}, Ref.~\cite{part2:r:BerBoe}).

Some general comments concerning the $R_{\rm had}$ determination
are in order.
\begin{itemize}
\item {\bf Exclusive vs Inclusive}

Usually, for energies below $ \sim 2~\gv$ exclusive cross sections
are measured for individual channels, while at higher energies the
hadronic final states are treated inclusively.  In exclusive
measurements, one directly 
measures the total and differential cross
sections for all exclusive reactions that are
kinematically allowed in that energy region. 
Having measured the exclusive cross sections, one can determine
the total cross sections and the value of $R$ by simply summing
them. This is, of course, not at all trivial since one has to be
sure that there is neither double counting nor missing final
states and that correlations between different
channels are properly taken into account~\cite{part2:r:Eidel}. There is,
in fact, still a systematic difference between the sum of
exclusive channels and the inclusive $R_{\rm had}$ measurements in
the energy range [1.4--2.1] $~\gv$ ~\cite{part2:r:Hagiwara}. In view of
the many channels in this energy region,
inclusive measurement should be pursued as   
much as possible~\cite{part2:r:Jeger06}.

\item{\bf  Energy scan vs Radiative return }

Measurements of hadronic cross sections have usually been
performed via energy scans, {\it i.e.}, by systematically varying
the $e^+e^-$ beam energies. This traditional way of measuring 
the hadronic cross section has one disadvantage --- it needs
dedicated running periods. On the other hand, modern flavor
factories, such as the Frascati $\phi$-factory DA$\Phi$NE or the
PEP-II and KEKB $B$-factories, are designed for a fixed
cms energy $\sqrt{s}$.  An energy scan for the
measurement of hadronic cross sections would, therefore,
not be very efficient
and an alternative way, the radiative-return method, has been
proposed.  The radiative-return method (for a
brief theoretical review see Ref.~\cite{part2:r:Czyz06} and references
cited therein)  relies on the
observation that the cross section of the reaction $e^+e^- \to
\mathrm{hadrons} +\mathrm{photons}$, with photons emitted from the
initial leptons, factorizes into a function $H$, which is fully calculable
within QED, and the cross section of the reaction $e^+e^-\to
\mathrm{hadrons}$ \bea &&\kern-30ptd\sigma(e^+e^- \to
\mathrm{hadrons} + \gamma\mathrm{'s})(s,Q^2) =
 \nonumber \\
 \kern+30pt &&H \cdot
 d\sigma(e^+e^-\to \mathrm{hadrons})(Q^2) \ ,
\label{master}
\eea
where $Q^2$ is the invariant mass of the hadronic system. Thus
from the measured $Q^2$ dependence of the
reaction $e^+e^- \to \mathrm{hadrons} + \mathrm{photons}$
at fixed $\sqrt{s}$, one can
evaluate $\sigma(e^+e^-\to \mathrm{hadrons})$, if the function
$H$ is known. As is evident from Eq.~(\ref{master}), the radiative
return method allows for the extraction of the hadronic cross   
section from the production energy threshold of a given hadronic
channel almost to the operating cms energy of a given experiment
($\sqrt{s}$). The smaller cross section for the radiative process
in comparison to the corresponding
process without photon emission has to be
compensated by higher luminosities.  This requirement is met by
the meson factories (DAPHNE, KEKB and PEPII).  These
 were all built for purposes other than hadronic cross section
measurements, but
their huge luminosities provide data samples that are large enough
for precise measurements of interesting hadronic channels and
give  information on rare channels that are not
easily accessible in scan experiments.

The radiative-return method has been successfully applied by
{\small KLOE} to measure the pion form factor below
$1~\gv$~\cite{part2:r:Aloisio04} and by {\small BaBar} for
the timelike proton-antiproton form factor and for several exclusive
higher multiplicity final states in the mass range from threshold
to $4.5~\gv$~\cite{part2:r:Aubert04}. The combination of KLOE and
BaBar data  covers the entire mass range below $\sim 4.5~\gv$.
For an extensive review
of the recent radiative-return results from both collaborations
see Ref.~\cite{part2:r:Denig06}.  This method has the advantage
of having the same normalization for each energy point.  It
requires a good theoretical understanding of the radiative      
corrections, a precise determination of the angle and energy of 
the emitted photon, and the full control of backgrounds,
especially from events where a photon is emitted in the final state
{\small (FSR)}. The Karlsruhe-Katowice group computed the
radiative corrections up to {\small NLO} for different exclusive
channels, implementing them in the event generator {\small
PHOKHARA}~\cite{part2:r:Rodrigo:2001jr,part2:r:Kuhn:2002xg,part2:r:Rodrigo:2001kf,part2:r:Czyz:2002np,part2:r:Czyz:PH03}.
The current precision level for the
$\pi^+\pi^-\gamma$ final state is $0.5\%$.

\item{\bf Status of $R_{\rm had}$ at low energy}

During the last thirty years the $R_{\rm had}$ ratio has been
measured by several experiments. Figure~\ref{plot:RHAD} gives an
updated summary of the $R_{\rm had}$ measurements by different
experiments and the current precision in different e$^+$e$^-$
cms energy regions by Burkhardt and
Pietrzyk~\cite{part2:r:Burk-Piet05}.

\begin{figure}[htbp]
\begin{center}
\includegraphics[width=9cm]{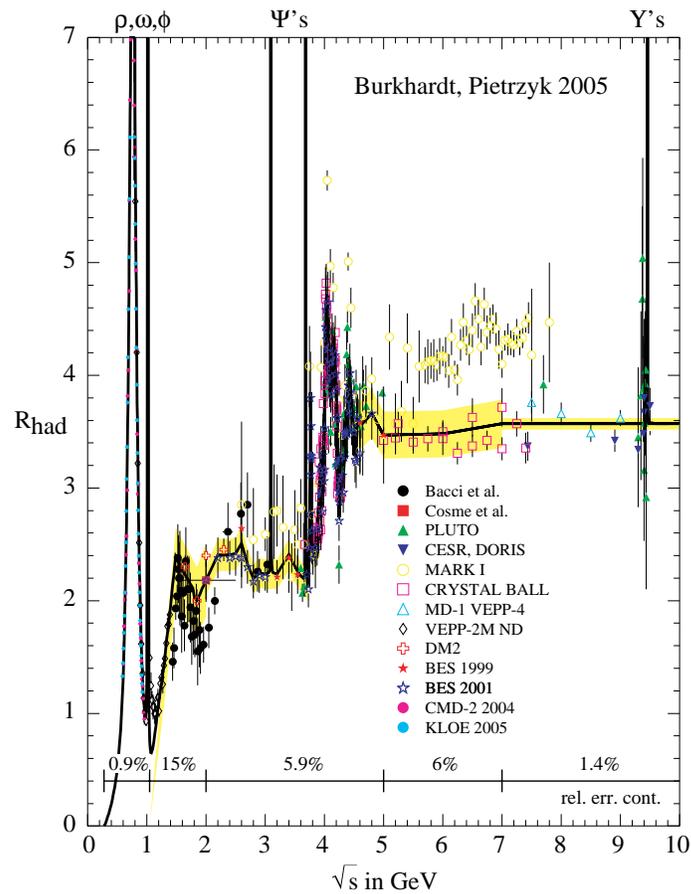}
\vspace{-0.5cm} \caption{$R_{\rm had}$ $versus$ cms energy.
Measurements are shown with statistical errors. The relative   
uncertainty assigned to the parameterization is shown as a band and
given with numbers at the bottom (from Ref.~\cite{part2:r:Burk-Piet05}).}
\label{plot:RHAD}
\end{center}
\end{figure}

\begin{itemize}

\item {\small\bf The $\pi^+ \pi^-$ threshold region.}

The experimental data are poor below about $400~ \mv$ because the 
cross section is suppressed near the threshold. The most effective
way to measure the threshold in the time-like region is provided
by radiative-radiation events, where the emission of an
energetic photon allows one to compensate in part for
the experimental difficulties associated with the detection
of two  pions that are nearly at rest.

\item {\small\bf The $\rho$ peak region.}

The $\pi^+\pi^-$ region between 0.5 and $1 ~\gv$ has been studied
by different experiments. {\small CMD-2}~\cite{part2:r:Akhmetshin03} and
{\small SND}~\cite{part2:r:Achasov05} performed an energy scan at the 
$e^+e^-$ collider {\small VEPP--2M} ($\sqrt{s}\in $ [0.4--1.4]
$\gv$) with $\sim 10^6$ and $\sim 4.5\times 10^6$ events,
respectively, and with systematic errors 
that range from 0.6\% to 4\% in
the relative cross-sections, depending on the dipion mass value.
The pion form factor has also been measured by {\small KLOE} using
ISR, and results from {\small BaBar} are expected soon.  {\small KLOE}'s
published results~\cite{part2:r:Aloisio04} are based on an integrated
luminosity of 140~pb$^{-1}$ and has  relative errors of 1.3\%
in the [0.6--0.97]~$\gv$ energy region (which are dominated by 
systematics).
{\small KLOE} subsequently collected more than 2~fb$^{-1}$ at
the $\phi$ meson peak, which corresponds to
 $\sim 2 \times 10^{7}$  $\pi^+\pi^-\gamma$ events
in the $\rho$ peak region.
{\small BaBar}~\cite{part2:r:Aubert04} has already collected over 
300~fb$^{-1}$ at the $\Upsilon(4S)$ peak, and plans to collect about  
1~ab$^{-1}$ by the end of data taking.

The results of these four experiments ({\small CMD-2, SND, KLOE,
BaBar}) in the next few years will probably allow us to know the
$\pi^+\pi^-$ cross-section for most of the $\rho$ region with a
relative accuracy that is better than 1\% (considering both
statistical and systematic errors). In summary~\cite{part2:r:Burk-Piet05}:

\begin{itemize}
\item very minor change have been introduced by
the recent {\small CMD2}, {\small KLOE} and {\small SND} measurements;
\item previous measurements in the $\rho$ region are already
quite precise.
\end{itemize}

\item {\small\bf The 1.05--2.0 GeV energy region.}

The [1.05--2.0 GeV] region is the most poorly known. As can be
seen in Fig.~\ref{plot:Fraction} in  Sect.~\ref{part2:sec:r:2},
the $R_{\rm had}$ measurements in this energy region
contributes about  40\% of the
uncertainty of the total dispersion integral for  
$\Delta^{(5)}_{had}(m_Z^2)$~\cite{part2:r:Burk-Piet05}.
It also provides most of the contribution to the uncertainty of
$a_{\mu}^{\mbox{$\scriptscriptstyle{\rm HLO}$}}$ above 1 $\gv$.
New $R_{\rm had}$ measurements in this energy region
will be very valuable.

\item {\small\bf  The high energy region}

In the high energy region, we distinguish the $J/\psi$ and the
$\Upsilon$ resonances and the background inclusive measurements of
the total hadronic cross section that are usually presented in
terms of $R_{\rm had}$ values.

For the narrow resonances $\omega$, $\phi$, the $J/\psi$ family (6
states) and the $\Upsilon$ family (6 states) we can safely use the
Breit-Wigner parameterization:
\be
\sigma_{BW}(s)=\frac{12 \pi}{M_R^2}\frac{\Gamma_{ee}}{\Gamma_R}
\frac{M_R^2 \Gamma_R \Gamma(s) }{(s-M_R^2)^2+M_R^2 \Gamma^2(s)},
\label{BWR}
\ee
which for zero-width resonances has the form
\be
\sigma_{NW}(s)=\frac{12 \pi^2}{M_R} \Gamma_{ee}\: \delta(s-M_R^2) .
\label{NWA}
\ee  
   
The masses, widths and the electronic branching fractions
for these resonances are listed in
the Review of Particle Properties~\cite{part2:r:PDG}.

In the region from the $J/\psi$ to the $\Upsilon$ the earlier
$R_{\rm had} $-measurements are from
Mark~I~\cite{part2:r:Bacino78,part2:r:Siegrist82},
$\gamma \gamma 2$~\cite{part2:r:Bacci79}, DASP~\cite{part2:r:DASP},
PLUTO~\cite{part2:r:Criegee82}, LENA~\cite{part2:r:Niczyporuk82}, Crystal 
Ball
(CB)~\cite{part2:r:Edwards90} and MD-1~\cite{part2:r:Blinov91}.

\begin{itemize}
\item {\small\bf The $2.0--5.0~\gv$ energy region}

In this energy region the earlier results have a precision of
15\%$\sim$20\%~\cite{part2:r:Siegrist82,part2:r:Bacci79,part2:r:Criegee82}.
In 2001, {\small BES-II} published the results of $R_{\rm had}$ 
measurements at 85 different cms. energies between 2 and
$4.8~\gv$ with an average precision of 6.6\%~\cite{part2:r:BES01},
an additional 6~points were published in
1999~\cite{part2:r:BES99}. These represent a substantial improvement.
The {\small BES-II} results were used in the 2001 evaluation of
$\dahz$~\cite{part2:r:Burk-Piet01,part2:r:Jeger01a} and its
uncertainty was reduced to 5.9\%.

\item {\small\bf The $5.0--7.0~\gv$ energy region}

There is a longstanding annoyance: the values of $R_{\rm had}$
for the energy region $5.0 -7.8~\gv$
from the Mark-I experiment~\cite{part2:r:Siegrist82}
are  higher than theoretical expectations, {\it i.e.}
$ 4.4 \pm 0.4 $ as opposed to $\sim 3.4$ (see
Fig.~\ref{plot:RHAD}). In 1990, the average value of $R_{\rm had}$  
in the $e^+e^-$ cms energy region between 5 and 7.4~$\gv$ was
reported by the Crystal Ball
Collaboration~\cite{part2:r:Edwards90} to be $3.44 \pm 0.03 \pm
0.018$ which is much more in line with expectation of
perturbation QCD assuming 5 quarks, and in agreement with other
experiments PLUTO, LENA and MD-1. These results were used in 
a 1995 evaluation of
$\dahz$~\cite{part2:r:Eidel-Jeger,part2:r:Burk-Piet95}. They,
together with the $R_{\rm had}$ measurements in the energy region
between 2 and 5~$\gv$ from {\small
BESII}~\cite{part2:r:BES01,part2:r:BES99}, were the two major
changes in the history of the determination of $\dahz$. Although the
results of the Crystal Ball are preferred by theorists, these data  
were never published, leaving some room for doubt. {\small CLEO}
may have ability to settle this problem~\cite{part2:r:Dytman04}.

\item {\small\bf The $7.0--12.0~\gv$ energy region}

In this energy region, pQCD calculations are not
expected to be very reliable, although the
calculations are in good agreement with the existing data.
Improved measurements of $R_{\rm had}$ are necessary to avoid
dependence on pQCD.  A
{\small CLEO} measurement in the $\Upsilon(4S)$ continuum
region at $\sqrt{s} = 10.52~\gv$,
$ R_{\rm had} = 3.56 \pm 0.01 \pm
0.07$~\cite{part2:r:Ammar98},  is one of the most accurate
$ R_{\rm had}$ measurements ever made with both 
statistical and systematic errors that are quite small.
Recently the {\small CLEO} collaboration has taken data
for $R$ measurements at energies of  6.96, 7.38, 8.38, 9.4,
10.0, 10.33~{\rm and}~11.2~$\gv$.
These data are under analysis and results can be
expected soon~\cite{part2:r:Dytman04}.

\item {\small\bf The energy region above $12.0~\gv$}
The values of $R_{\rm had}$ in this region are described by a
parameterization based on third-order QCD~\cite{part2:r:Burk-Piet01}

\end{itemize}
\end{itemize}
\end{itemize}
As we discuss below,  hadronic cross section measurements are
crucial for the precise evaluation of the hadronic contributions
to running of the electromagnetic coupling $\alpha_{QED}$. This
requires a better knowledge of the hadronic cross section
over the entire energy range, starting from the $ 2m_{\pi}$ threshold.
The optimal exploitation of a high energy linear collider,
such as the ILC, for
precision physics requires a reduction
of the errors on the  low energy $R_{had}$ measurements by
a factor of something like ten.

\section{Hadronic Vacuum Polarization}
\label{part2:sec:r:2}
The running of the electromagnetic coupling with momentum
transfer, $\alpha(0)\rightarrow\alpha(s)$, caused by fermion-pair
loop insertions in the photon propagator, is customarily written
as
\begin{eqnarray}
\alpha(s) & = & \frac{\alpha(0)}{1-\Delta\alpha(s)}
          ~ = ~
         \frac{\alpha(0)}{1
                            -\Delta\alpha_{\mathrm{e\mu\tau}}(s)
                            -\Delta\alpha_{\mathrm{top     }}(s)
                            -\Delta\alpha_{\mathrm{had     }}^{(5)}(s)
                          }\,,
\label{eq:sm:alpha}
\end{eqnarray}
where $\alpha(0)=1/137.036$~\cite{part2:r:Mohr-Taylor}.  The contribution
from leptons, $\alpha_{\mathrm{e\mu\tau}}(s)$, is calculated 
up to third order:
$\Delta\alpha_{\mathrm{e\mu\tau}}(M_{\rm Z}^2)=3149.7686 \times
10^{-5}$ with negligible uncertainty~\cite{part2:r:MS98}.  Since heavy
particles decouple in QED, the top-quark contribution is small:
$\Delta\alpha_{\mathrm{top}}(M_{\rm Z}^2)=-0.00007(1)$ as
calculated by the TOPAZ0 and ZFITTER programs
as a function of the pole  mass of
the top quark, $m_t$.  The running electromagnetic coupling is
insensitive to new particles with high masses.  For light-quark
loops, diagrammatic calculations are not viable since at these low
energy scales perturbative QCD is not applicable.  Therefore, the
total contribution from the five light quark flavors to the hadronic
vacuum polarization, $\dahz$, is more accurately obtained from  
a dispersion integral over the measured hadronic cross-section in
electron-positron annihilations at low center-of-mass energies.
The relevant vacuum polarization amplitude satisfies the
convergent dispersion relation~\cite{part2:r:Jeger01a}
         
\bea Re \Pi'_{\gamma}(s)-\Pi'_{\gamma}(0)=\frac{s}{\pi} Re
\int_{s_0}^{\infty} ds' \frac{Im \Pi'_{\gamma}(s')}{s'(s'-s -i  
\veps)}, \nonumber \eea and, using the optical theorem (unitarity), one
has
\bea Im \Pi'_{\gamma}(s)=\frac{s}{e^2}
                    \sigma_{tot} (\epm \ra \gamma^* \ra {\rm hadrons})(s)
\;. \nonumber \eea
\noindent
In terms of the cross section ratio

\bea R_{\rm had}(s)=\frac{\sigma_{tot} (\epm \ra \gamma^* \ra {\rm
hadrons})}
     {\sigma (\epm \ra \gamma^* \ra \mu^+ \mu^-)}\;, \nonumber
\eea
where $\sigma (\epm \ra \gamma^* \ra \mu^+
\mu^-)=\frac{\displaystyle 4\pi \al ^2}{\displaystyle 3s}$ is the tree
level value, we obtain

\bea \Delta \al _{\rm hadrons}^{(5)}(M_Z^2) &=& -\frac{\al
M_Z^2}{3\pi}Re\int_{4m_{\pi}^2}^{\infty}
ds\frac{R(s)}{s(s-M_Z^2-i\veps)}\;. \eea

The dispersion integral can be evaluated either by direct
integration between measured data points or by using a  
parameterization of the measured cross section of $\epm \ra {\rm
hadrons}$. In the first approach one tries to rely on the
experimental data as much as possible and directly integrates
over the experimental cross section measurements,
joining adjacent data point with straight lines (the trapezoidal rule).
In this approach one can take into account uncertainties of
separate measurements in a straightforward
manner~\cite{part2:r:Eidel-Jeger}.
In the second approach one fits
the experimental points to some model and integrates the
resulting parameterization of the data. This procedure inevitably
leads to some model-dependence and it is not clear how experimental
errors, especially systematic uncertainties, are taken into
account~\cite{part2:r:Eidel-Jeger}.
     
Detailed evaluations of this dispersive integral directly from the
experimental data have been carried out by many
authors~\cite{part2:r:Hagiwara,part2:r:Jeger06,part2:r:Burk-Piet05,
part2:r:Burk-Piet01,part2:r:Jeger01a,part2:r:Eidel-Jeger,
part2:r:Burk-Piet95,part2:r:BJVP89,part2:r:Swartz,part2:r:Jeger01b}.
There are also several evaluations of $\dahz$ that are
more theory driven~\cite{part2:r:deltaalpha}.

An important conclusion from studies before 1989,  described in
Ref.~\cite{part2:r:BJVP89}, is that the independent programs and
parameterization methods gave nearly identical results. Differences
in central values obtained from the use of the trapezoidal rule
between many data points, partially smoothed functions or broad
averages were negligible compared to the experimental uncertainties.
The uncertainty in the result obtained from the
integration is, therefore, almost entirely due to the experimental
errors in the determination of $R_{\rm had}(s)$. 


The result of Refs.~\cite{part2:r:Eidel-Jeger} and
\cite{part2:r:Burk-Piet95},
\begin{eqnarray}
\Delta\alpha_{\rm had}^{(5)} = 2804 \, (65) \times 10^{-5},
\label{eq:dalhad:exp:old}
\end{eqnarray}
was used by the {\small LEP} Collaborations and the {\small LEP}
Electroweak Working Group as the input parameter to constrain the
Standard Model until summer of 2000~\cite{part2:r:LEPEWWG01}.

After {\small BES-II} published its sequence of substantially
improved total cross section measurements between $2 ~{\rm and}~ 5
~\gv$~\cite{part2:r:BES99}~\cite{part2:r:BES01}, some of these analyses 
were
updated to include the the
{\small BES-II} as well as measurements from {\small
CMD-2}~\cite{part2:r:Akhmetshin03}, obtaining~\cite{part2:r:Burk-Piet01}:

\begin{eqnarray}
\Delta\alpha_{\rm had}^{(5)} = 2761 \, (36) \times 10^{-5} ,
\label{eq:dalhad:exp:new}
\end{eqnarray}
and~\cite{part2:r:Jeger01a}:

$$ \Delta\alpha_{\rm had}^{(5)} = 2757 \, (36) \times 10^{-5} . $$
The factor-of-two reduction in the quoted uncertainty from
$70 \times10^{-5}$ in 
Refs.~\cite{part2:r:Eidel-Jeger,part2:r:Burk-Piet95} 
to $36 \times 10^{-5}$
in Refs.~\cite{part2:r:Burk-Piet01,part2:r:Jeger01a}
is mainly due to the {\small BES-II} data.

\begin{figure}[htbp]
\begin{center}
\includegraphics[width=10cm]{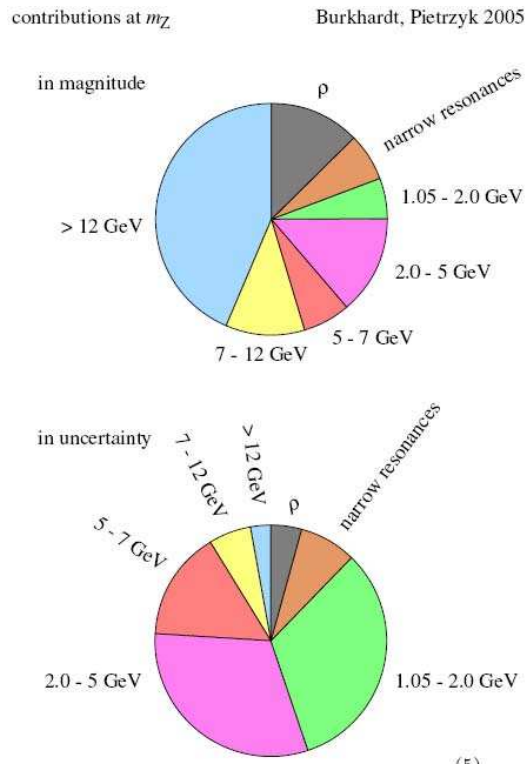}
\caption{Relative contributions to $\dahz$ in magnitude (top) and
uncertainty (bottom) (from Ref.~\cite{part2:r:Piet06}).} 
\label{plot:Fraction}
\end{center}
\end{figure}

\begin{figure}[htbp]
\begin{center}
\includegraphics[width=9cm]{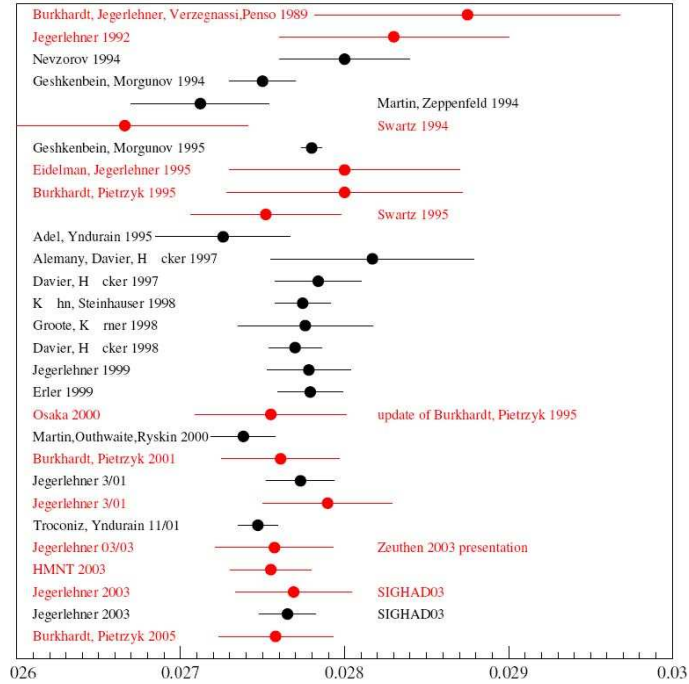}
\caption{Comparison of different estimates of $\dahz$. Estimates based
on dispersion integration of the experimental data are shown with
red solid dots and estimates relying on additional theoretical
assumptions are shown as black solid dots (from 
Ref.~\cite{part2:r:Piet06}). } \label{plot:comparison}
\end{center}
\end{figure}

The  estimates of $\dahz$ from 
Refs.~\cite{part2:r:Burk-Piet01,part2:r:Jeger01a}, listed in
Eq.~\ref{eq:dalhad:exp:new}, were  used as input 
by the {\small LEP} Collaborations and the {\small LEP} Electroweak 
Working Group until the summer of 2004.
A more recent update,
$ \Delta\alpha_{\rm had}^{(5)} = 2758 \, (35) \times 10^{-5} $
\cite{part2:r:Burk-Piet05}, includes the {\small
KLOE}~\cite{part2:r:Aloisio04} measurements.
Figure~\ref{plot:Fraction} (from Ref.~\cite{part2:r:Piet06}) 
illustrates the
relative contributions from different $e^+e^-$ cms energy
regions to $\dahz$, both in magnitude (top) and 
uncertainty (bottom). The region
between $1.05-2~\gv$ gives a significant contribution to the
uncertainty despite its small contribution to the magnitude.
A comparison of estimates of $\dahz$ performed during 90's is shown
in Fig.~\ref{plot:comparison}(from Ref.~\cite{part2:r:Piet06}).

\begin{figure}[htbp]
\vspace*{-4mm} \centering
\includegraphics[width=9cm]{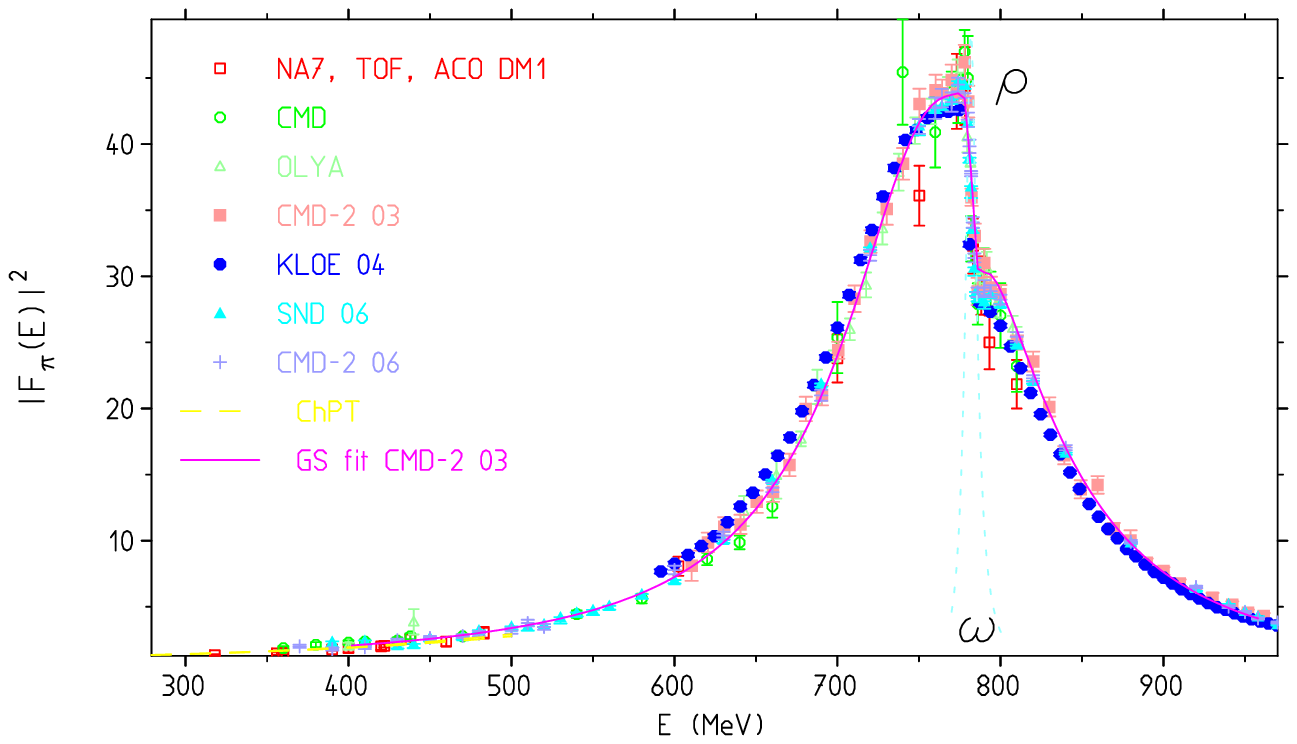}
\includegraphics[width=9cm]{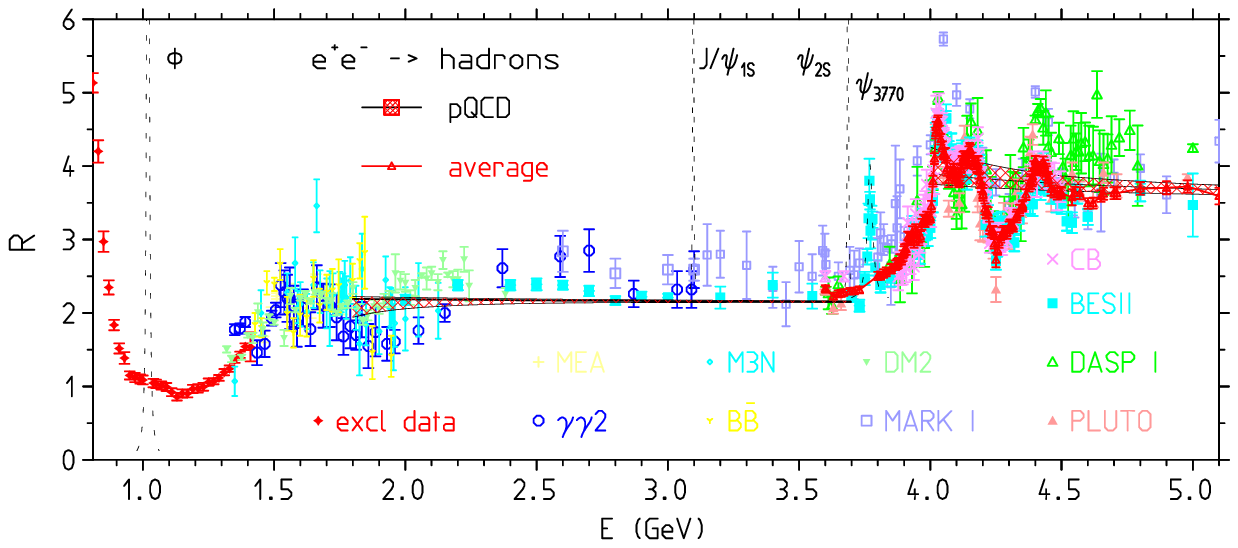}
\includegraphics[width=9cm]{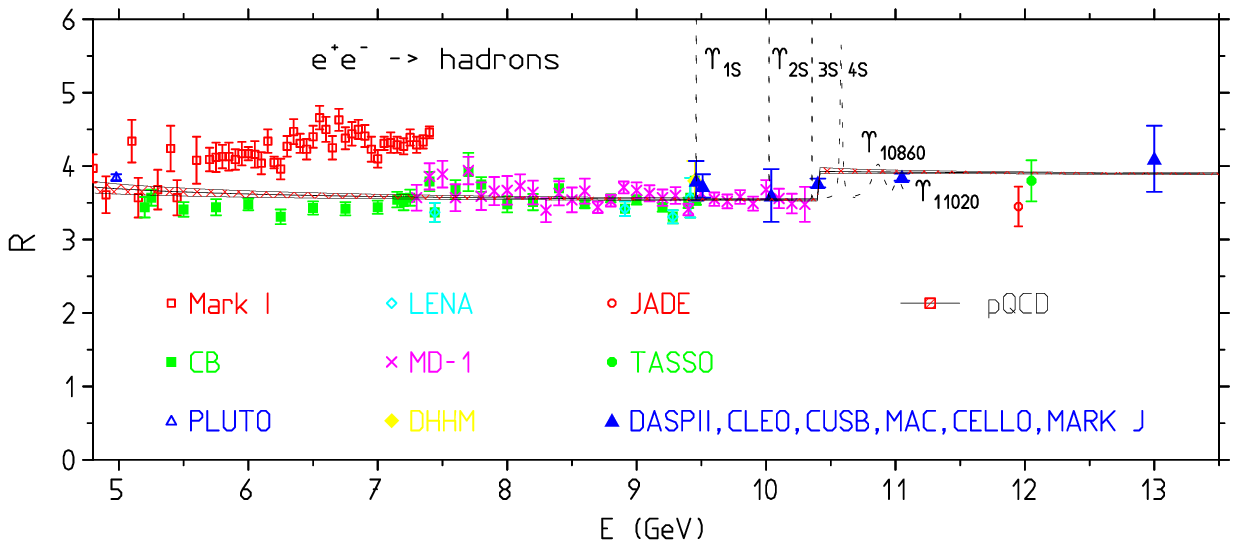}   
\caption{A compilation of the presently available experimental
hadronic $\epm$--annihilation data (from Ref.~\cite{part2:r:Jeger06}).}
\label{fig:Rdata}
\end{figure}

\begin{figure}[htbp]
\vspace*{-00mm} \centering
\includegraphics[width=10cm]{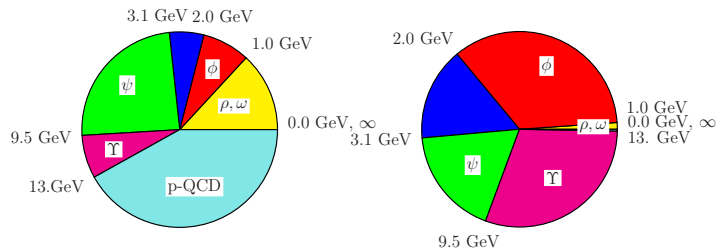}
\caption{$\dahz$: contributions (left) and 
squared~errors (right) from
different $\sqrt{s}$ regions (from Ref.~\cite{part2:r:Jeger06}).} 
\label{fig:distR}
\vspace*{-4mm}
\end{figure}

Another recent update~\cite{part2:r:Jeger06} based on a compilation of
the data shown in Fig.~\ref{fig:Rdata} (from Ref.~\cite{part2:r:Jeger06}),
yields $ \Delta\alpha_{\rm had}^{(5)} = 0. 027607 \pm  0.000225 $
or $ 1/\alpha^{(5)}(m^2_{\rm Z})=128.947 \pm 0.035$.
Contributions from various energy regions and the origin of the
errors in this estimate are shown in Fig.~\ref{fig:distR} (from
Ref.~\cite{part2:r:Jeger06}).
More details are given in Table~\ref{tab:alphacont} (from
Ref.~\cite{part2:r:Jeger06}).

\begin{table*}[t]
\vspace*{-4mm}
\centering \caption{Contributions for $\dahz \times 10^4$ (direct
integration method) with relative (rel) and absolute (abs) error
in percent (from Ref.~\cite{part2:r:Jeger06}).} \label{tab:alphacont}
\begin{tabular}{crrr}
\hline\noalign{\smallskip}
   Energy range      & $\dahz \times 10^4$ &
  {rel~[\%]}  & {abs~[\%]}\\
\noalign{\smallskip}\hline\noalign{\smallskip}
   $\rho,\omega$ ($E<2M_K$)        &   36.23 [$\:    $ 13.1](0.24) &   0.7  
&   1.1 \\
        $2M_K<E<2~\gv$             &   21.80 [$\;\;\,$  7.9](1.33) &   6.1  
&  34.9  \\
     $2~\gv<E<M_{J/\psi}$          &   15.73 [$\;\;\,$  5.7](0.88) &   5.6  
&  15.4  \\
   $M_{J/\psi}<E<M_{\Upsilon}$     &   66.95 [$\:    $ 24.3](0.95) &   1.4  
&  18.0  \\
   $M_{\Upsilon}<E<E_{\rm cut}$    &   19.69 [$\;\;\,$  7.1](1.24) &   6.3  
&  30.4  \\
   $E_{\rm cut}<E$ pQCD            &  115.66 [$\:    $ 41.9](0.11) &   0.1  
&   0.3  \\
   $E < E_{\rm cut}$ data          &  160.41 [$\:    $ 58.1](2.24) &   1.4  
&  99.7 \\
           total                   &  276.07         [100.0](2.25) &   0.8  
& 100.0  \\
\noalign{\smallskip} \hline
\end{tabular}
\end{table*}  

In summary, to reach higher precision, more
experimental effort is required to measure 
$\sha$ precisely both at low and 
intermediate energies~\cite{part2:r:Jeger06}.

\section{Input parameters and their uncertainties }
\label{part2:sec:r:3}
In 2005 and 2006,  results from 
``final'' fits of electroweak measurements 
taken at the $Z$ resonance by the {\small SLC} and 
{\small LEP} experiments were reported~\cite{part2:r:LEPEWWG05a}.
The mass and width of the $Z$ boson, $M_Z$
and $\Gamma_Z$, and its couplings to fermions, for example the
$\rho$ parameter and the effective electroweak mixing angle for
leptons, were precisely determined:
\begin{center}
\begin{tabular}{lll}
$M_{\rm Z}$       & = &           $91.1875  \pm 0.0021 ~{\rm GeV}$\\
$\Gamma_{\rm Z}$       & = &            $2.4952  \pm 0.0023 ~{\rm GeV}$\\
$\rho_l$ & = &            $1.0050  \pm 0.0010$      \\
$\sin^2\theta^{\rm lept}_{\rm eff}$ & =&            $ 0.23153 \pm
0.00016$ \,.
\end{tabular}
\end{center}

Using SM radiative corrections, 
the large and diverse set of {\small LEP} and {\small SLC}
measurements provide many stringent tests of 
the Standard Model and tight constraints on its 
free parameters.
The masses of the $W$ boson and the top quark are
predicted to be: $M_{\rm W}=80.363 \pm 0.032~{\rm GeV}$ and
$m_t=173^{+13}_{-10}~{\rm GeV}$, in good
agreement with the subsequent direct measurements of these 
quantities at {\small LEPII} and the {\small
TEVATRON}~\cite{part2:r:LEPEWWG05b}, 
thereby successfully testing the Standard
Model at the level of its radiative corrections. 

Note that the SM perturbation calculations are 
{\sl approximations} that are obtained by truncation of
the perturbation series. The accuracy of the finite-order
approximation for each $Z$-pole observable depends on the input
parameter set that is used in the calculations.  A 
natural
choice is the QED-like parameterization in terms of
the parameters
\begin{center}
\begin{equation}
\alpha,~~~~ \alpha_s,~~~~ M_{\rm W },~~~~M_{\rm Z}, ~~~~m_{\rm
H},~~{\rm and}~~m_f,
\end{equation}
\end{center}
where $M_{\rm  W} $ and $M_{\rm Z} $ are the masses of the $W$
and $Z$ bosons, $m_{\rm H}$ is the mass of the Higgs boson,  
$m_f$ are the masses of all known fundamental fermions $f$, in
particular $m_t$ is the top quark mass, and $\alpha$ and  
$\alpha_s$ just are the coupling constants of the electromagnetic
and strong interactions.
Loop corrections induce a running of the
coupling constants with increasing momentum transfer 
(or $s$),  The running of the strong coupling, $\alpha_s(s)$, is
even larger than that for the QED coupling $\alpha.$  
Since the $Z^0$ resonance is  sufficiently dominant for  
$Z$-pole observables, the relevant coupling constants become simply
$\alpha(m^2_Z)$ and $\alpha_s(m^2_Z)$.

Within the SM, the mass of the $W$ 
that is directly measured  at
the {\small TEVATRON} and {\small LEPII}, 
is related to $M_{\rm Z}$ and the Fermi constant 
$\Gmu$ through radiative  corrections.
A very precise value for the latter,
$\Gmu=1.16637(1)\cdot10^{-5}~\gv^{-2}$~, is derived from
measurements of the muon lifetime plus two-loop corrections.
The 9~ppm precision on $\Gmu$ greatly exceeds the relative
precision with which $M_{\rm W}$ will be measured at any time in the
foreseeable future. This motivates our substitution of
$\Gmu$ for $M_{\rm W}$ as an input parameter for SM
calculations.

Therefore,  the fine-structure constant $\alpha$, the
Fermi coupling constant $G_{\mu}$, and the mass of Z boson $M_{\rm
Z}$ are the preferred choice for input parameters for precise 
calculations of SM radiative
corrections, since these are the most precisely measured:
\begin{itemize}
\item the fine-structure constant in the Thomson limit determined
from the $e^{\pm}$ anomalous magnetic moment, the quantum Hall 
effect, and other measurements is~\cite{part2:r:Mohr-Taylor}
\begin{equation}
\begin{array}{lcl}
\alpha^{-1}(0)&=&137.03599911(46),\\
\frac{\displaystyle \delta\alpha}{\displaystyle \alpha}&=&
3.6\times
10^{-9};\\
\end{array}
\end{equation}
\item the Fermi constant determined from the muon
lifetime 
formula is~\cite{part2:r:Marciano-Sirlin,part2:r:Ritbergen-Stuart}
\begin{equation}
\begin{array}{lcl}
G_{\mu}&=&1.16637(1),\\
\frac{\displaystyle \delta G_{\mu}}{\displaystyle G_{\mu}}&=&
8.6\times
10^{-6};\\
\end{array}
\end{equation}
\item the $Z$ boson mass determined from the $Z$-lineshape scan
at {\small LEP I} is~\cite{part2:r:LEPEWWG05a}
\begin{equation}
\begin{array}{lcl}
M_{\rm Z}&=&91.1875(21),\\
\frac{\displaystyle \delta M_{\rm Z}}{\displaystyle M_{\rm Z}}&=&
2.4\times
10^{-5};\\
\end{array}
\end{equation}
\item the effective fine-structure constant at the 
$M_Z$ scale is~\cite{part2:r:Burk-Piet05}
\begin{equation}
\begin{array}{lcl}
\alpha^{-1}(M_{\rm Z})&=&128.940(48),\\
\frac{\displaystyle \delta \alpha(M_{\rm Z})}{\displaystyle
\alpha(M_{\rm Z})}&=& (1.6 \sim 6.8) \times
10^{-4}.\\
\end{array}
\end{equation}
\end{itemize}
The relative uncertainty of $\alpha(M_{\rm Z})$ is roughly one
order of magnitude worse than that of $M_{\rm Z}$, making it one
of the limiting factors in the calculation of precise SM
predictions.

Note that $\dal$ enters in electroweak precision physics typically
when calculating the weak mixing parameter $\sini$
from $\al$, $\Gmu$ and $M_Z$ via~\cite{part2:r:Jeger91}
\be
\sini\:\cosi\:
=\frac{\pi \al}{\sqrt{2}\:G_\mu\:M_Z^2} \frac{1}{1-\Delta r_i}, \ee
where \begin{eqnarray} \Delta r_i &=&\Delta r_i({\al ,\: \Gmu ,\: M_Z 
,}\:m_H,
\:{m_{f\neq t},\:m_t}) \end{eqnarray} includes higher-order 
corrections
that can be calculated in the SM or in alternative models.

The value of $\Delta r_i$ depends on the definition of $\sini$. 
The various
definitions coincide at tree level and only differ by 
quantum-loop induced effects. From the weak gauge boson masses, 
the electroweak gauge couplings and 
the neutral current couplings of the charged
fermions we obtain
\begin{eqnarray}
\sinW &=& 1-\frac{M_W^2}{M_Z^2},\\
\sing &=& e^2/g^2=\frac{\pi \al}{\sqrt{2}\:G_\mu\:M_W^2},\\
\sinf &=& \frac{1}{4|Q_f|}\;\left(1-\frac{v_f}{a_f}
\right)\;,\;\;f\neq \nu\;,
\end{eqnarray}
respectively. For the most
important cases the general form of $\Delta r_i$ reads
\begin{eqnarray}
\Delta r_i &=& \dal - f_i(\sini)\:\Delta\rho + \Delta
r_{i\:\mathrm{remainder}},
\label{der}
\end{eqnarray}
where:
\begin{itemize}
\item The large term $\Delta \alpha$ is due to the photon vacuum
polarization
\begin{center}
\begin{equation}
\Delta\alpha = \Pi^{\gamma \gamma}_1(0) - \Pi^{\gamma
\gamma}_1(M^2_Z).
\end{equation}
\end{center}
This is a universal term that influences  predictions for $M_W$,
$A_{LR}$, $A^f_{FB}$, $\Gamma_f$, etc. These terms can be   
calculated safely in perturbation theory.

\item $\Delta \rho$ is the famous correction to the
$\rho$--parameter that is defined as the neutral to charged
current ratio
\begin{center}
\begin{equation}
\Delta \rho = \rho-1 = \frac{\displaystyle G_{NC}}{\displaystyle
G_{\mu}} = \frac{\displaystyle \Pi^{ZZ}_1(0)}{\displaystyle
M^2_{\rm Z }}-\frac{\displaystyle \Pi^{WW}_1(0)}{\displaystyle
M^2_{\rm W}}.
\end{equation}
\end{center}

$\dro$  exhibits the leading top mass correction

\begin{eqnarray} \dro \simeq \frac{\sqrt2 \Gmu}{16 
\pi^2}\:3m_t^2\;\;;\;\;\;m_t
\gg m_b , \end{eqnarray} 
that allowed {\small LEP} experiments to obtain 
a rather good indirect estimate of the top quark mass prior to 
its discovery at the {\small TEVATRON}.

Note that in Eq.~(\ref{der}) $f_W=c_W^2/s_W^2 \simeq 3.35$ is
substantially enhanced relative to $f_f=1$.

\item The ``remainder'' term, although sub-leading, is very
important for the interpretation of the precision experiments at
{\small LEP} and includes part of the leading Higgs mass
dependence. For a heavy Higgs particle we obtain the simple
expression

\begin{eqnarray}
\Delta r^{\rm Higgs}_i \simeq \frac{\sqrt2 \Gmu \mw}{16
\pi^2}\: \left\{c^H_i\: (\ln \frac{m_H^2}{\mw}- \frac56) \right\}
\;\;\;\;\;m_M \gg M_W ,
\end{eqnarray}
where, for example, $c^H_f=(1+9 \sinf)/(3\sinf)$ and
$c^H_W=11/3$.
\end{itemize} 
The uncertainty $\delta \Delta \alpha$ implies uncertainties
$\delta M_W$, $\delta \sinf$ that are given by

\begin{eqnarray}
\frac{\delta M_W}{M_W} &\sim& \ha \frac{\sinW}{\cosW-\sinW}
\;\delta \dal \sim 0.23 \;\delta \dal, \\
\frac{\delta \sinf}{\sinf} &\sim& ~~\frac{\cosf}{\cosf-\sinf}   
\;\delta \dal \sim 1.54 \;\delta \dal\; .
\end{eqnarray}

\begin{figure}[htbp]
\begin{center}
\includegraphics[width=8cm]{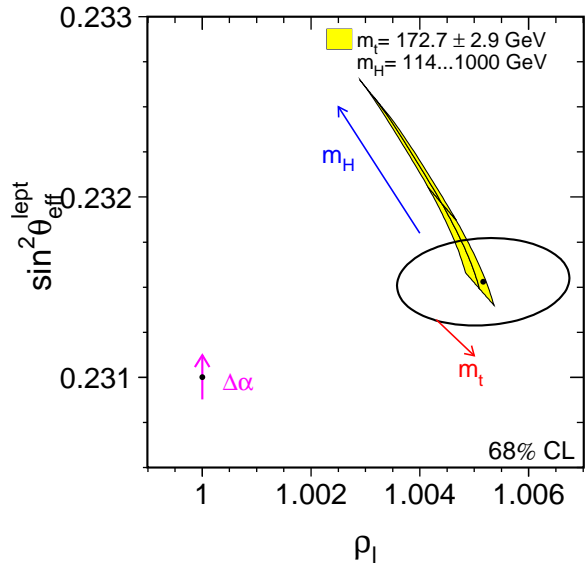}
\vspace{-0.5cm} \caption{Contour curve of 68\% probability in the
($\rho_l$, $\sin^2\theta^{\rm lept}_{\rm eff}$) plane.  The
prediction of a theory based on electroweak Born-level formulae
and QED with running $\alpha$ is shown as the dot, with the arrow
representing the uncertainty due to the hadronic vacuum   
polarization $\Delta \alpha^{(5)}_{\rm had}(M^2_{\rm Z})$. The
same uncertainty also affects the SM prediction, shown as the
shaded region drawn for fixed $\Delta \alpha^{(5)}_{\rm
had} (M^2_{\rm Z})$ while $m_t$ and $m_H$ are varied in the ranges
indicated (from Ref.~\cite{part2:r:LEPEWWG05a}). }
\label{fig:coup:qed-ew-cor}
\end{center}
\end{figure}

The effects of the uncertainty due to $\dahz$ on the SM
prediction for the $\rho$ parameter and $\sin^2\theta^{\rm
lept}_{\rm eff}$ can be seen in Fig~\ref{fig:coup:qed-ew-cor}.
While the SM prediction for the $\rho$ parameter is not 
effected by the uncertainty in $\dahz$, the uncertainty on the
prediction of $\swsqeffl$ within the SM due to the uncertainty
on $\dahz$ is nearly as large as the accuracy of the experimental
measurement of $\swsqeffl$. The present error in the effective
electromagnetic coupling constant, $\delta \Delta\alpha(M^2_{\rm 
Z}) = 35 \times 10^{-5}$~\cite{part2:r:Burk-Piet05}, dominates the
uncertainty of the theoretical prediction of $\sin^2 \!\theta_{\rm
eff}^{\rm lept}$, inducing an error $\delta(\sin^2 \!\theta_{\rm
eff}^{\rm lept}) \sim 12 \times 10^{-5}$, which is not much smaller
than the experimental value $\delta(\sin^2 \!\theta_{\rm eff}^{\rm
lept})^{\rm EXP} = 16 \times 10^{-5}$ determined by {\small LEP-I}
and {\small SLD}~\cite{part2:r:LEPEWWG05a}. This observation underlines
the importance of a precise cross-section measurement of
electron-positron annihilation into hadrons at low cms
energies.

Moreover, since measurements of the effective {\small EW} mixing   
angle at a future linear collider will improve the precision by about
an order of magnitude~\cite{part2:r:Weiglein:2004hn}, a much smaller 
value of $\delta\Delta\alpha(M^2_{\rm Z})$ will be required. It is
therefore crucial to assess all viable options to further reduce
this uncertainty.

Table~\ref{tab:future} (from 
Ref.~\cite{part2:r:Jeger01a,part2:r:DAFNE})
shows that an uncertainty $\delta \Delta\alpha_{\rm had}^{(5)}
\sim 5 \times 10^{-5}$, needed for precision physics at a future
linear collider, requires the measurement of the hadronic cross
section with a precision of $O(1\%)$ from threshold up to the
$\Upsilon$ region.
\begin{table}[h]
\begin{center}
 \renewcommand{\arraystretch}{1.4}
 \setlength{\tabcolsep}{1.6mm}
\begin{tabular}{|c|c|c|c|}
\hline $\delta \Delta\alpha_{\rm had}^{(5)} \times 10^{5} $ &
$\delta(\sin^2 \!\theta_{\rm eff}^{\rm lept}) \times 10^{5}$
& Request on $R$\\
\hline \hline
35   &  12.5 & Present \\
\hline
7   &   2.5 & $\delta R/R \sim 1\%$ for $\sqrt{s} \leq M_{J/\psi}$\\
\hline
5   &   1.8 & $\delta R/R \sim 1\%$ for $\sqrt{s} \leq M_{\Upsilon}$ \\
\hline
\end{tabular}
\caption{\label{tab:future} Values of the uncertainties $\delta 
\Delta\alpha_{\rm had}^{(5)}$ (first column) and the errors
induced by these uncertainties on the theoretical {\small SM}
prediction for $\sin^2 \!\theta_{\rm eff}^{\rm lept}$ (second
column). The third column indicates the corresponding requirements
on the $R$ measurement.}
\end{center}
\end{table}

In the SM, the Higgs mass $m_H$ is the only relevant unknown
parameter and by confronting the calculated with the
experimentally determined value of $\sini$ one obtains 
important indirect constraint on the Higgs mass. The uncertainty
$\delta \Delta \alpha$ thus obscures the indirect
bounds on the Higgs mass obtained from electroweak precision
measurements. As is mentioned in Sect.~\ref{part2:sec:r:2}, the current 
uncertainty in the 
$1.05 \sim 2.0~\gv$ energy region is 15\%. Improving the precision
from 15\% (see Fig.~\ref{plot:RHAD}) to 5\% would
change the total uncertainty on $\dahz$ from 0.00035 to 0.00027.
The change in the fitted value of the Higgs mass would be small;
however changing $R_{\rm had}$ by $\pm 1 \sigma$ in this cms
energy region would shift the central value of the fitted Higgs
mass by $ ^{+16}_{-9}~\gv$. Therefore more precise measurements in
this cms energy region are important.

\begin{figure}[htbp]
\begin{center}
\includegraphics[width=7cm]{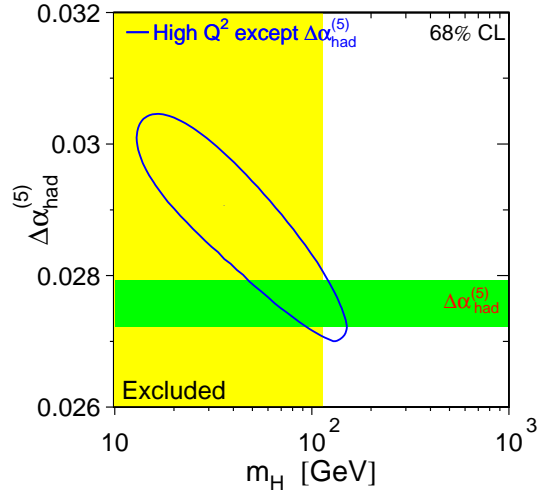}
\vspace{-0.5cm} \caption{ Contour curve of 68\% probability in the
$(\dahz,m_H)$ plane, based on all measurements other than that
of $\dahz$.  The direct measurements of the excluded   
observable is shown as the horizontal bands of width $\pm1$   
standard deviation.  The vertical band shows the 95\% confidence
level exclusion limit on $m_H$ of $114.4~\gv$ derived from the
direct search at {\small LEPII}~\cite{part2:r:LEPEWWG06}.}
\label{fig:msm:mhah}
\end{center}
\end{figure}

The importance of the external $\dahz=0.02758\pm
0.00035$\cite{part2:r:Burk-Piet05} determination for the constraint on
$m_H$ is shown in Fig.~\ref{fig:msm:mhah}.  Without the external
$\dahz$ constraint, the fit results are
$\dahz=0.0298^{+0.0010}_{-0.0017}$ and $m_H=29^{+77}_{-15}~\gv$,
with a correlation of $-0.88$ between these two fit results.

\begin{figure}[htbp]
\begin{center}
\includegraphics[width=8cm]{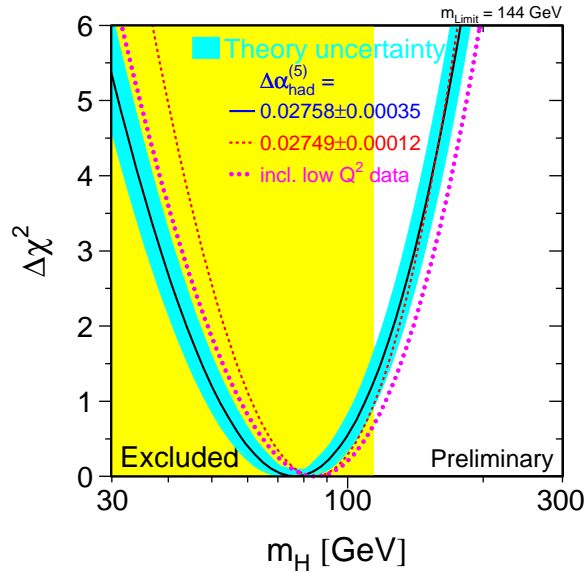}
\vspace{-0.5cm} \caption{The black line is the result of the Electroweak
Working Group fit using all data~\cite{part2:r:LEPEWWG06}; the band 
represents an estimate of the
  theoretical error due to missing higher-order corrections.  The vertical
  band shows the 95\% CL exclusion limit on $m_H$ from the direct 
  search.}
\label{fig:blueband}
\end{center}
\end{figure}

The latest global fit of the {\small LEP} Electroweak Working  
Group, which employs the complete set of {\small EW} observables, 
leads to the predicted value $m_H = 91^{+45}_{-32}$ GeV, with a 95\%
confidence level upper limit of 186~GeV~\cite{part2:r:LEPEWWG06}
(see Fig.~\ref{fig:blueband}). This upper limit increases to
219~GeV when the {\small LEP-II} direct search lower
limit of 114~GeV is included.
%

%
%

\section{Measurement of $R$ value at \bes3}
 
\begin{center}
\begin{figure}[htbp]
\begin{center}
\includegraphics[width=11cm,height=8cm]{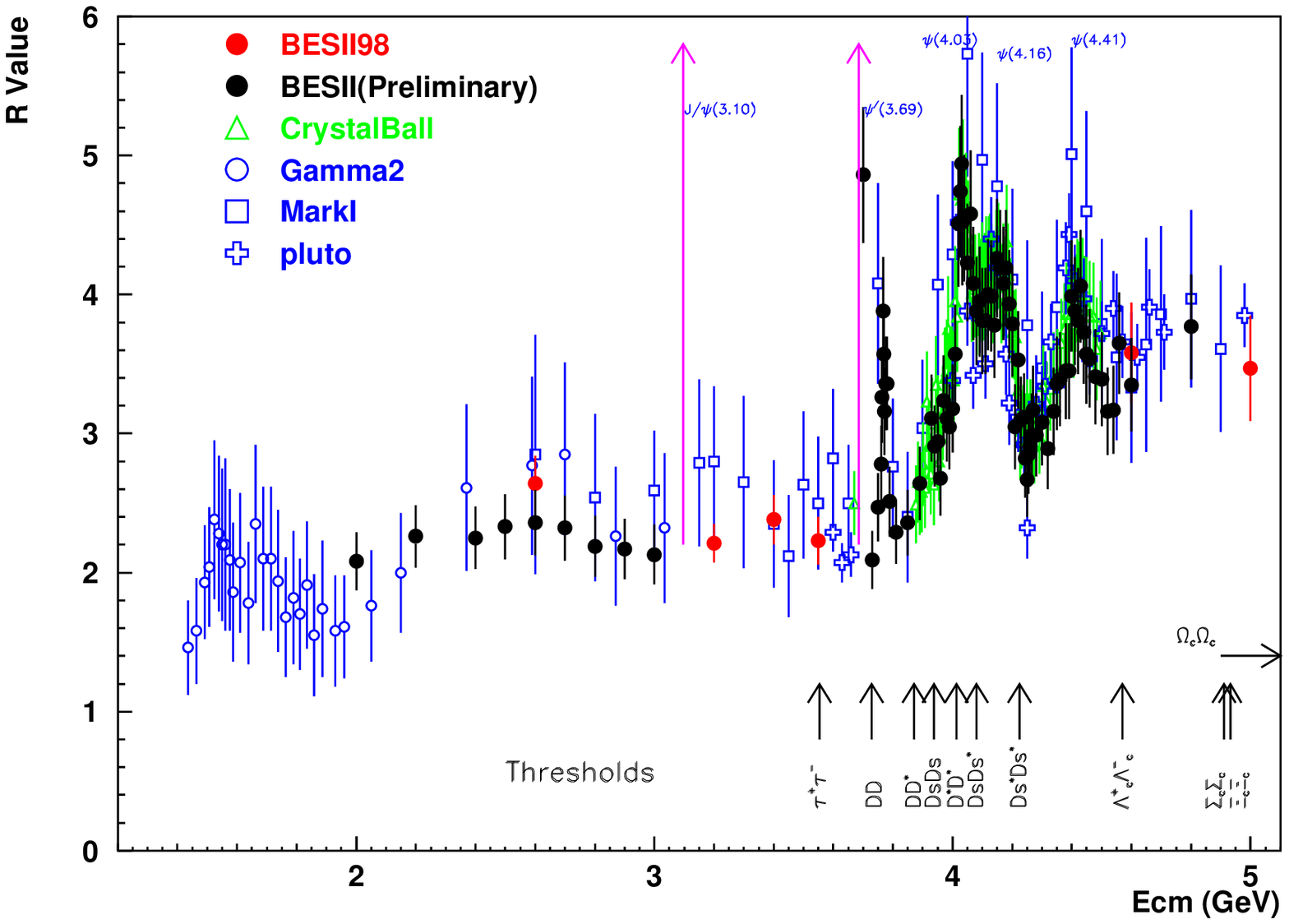}
\includegraphics[width=11cm,height=8cm]{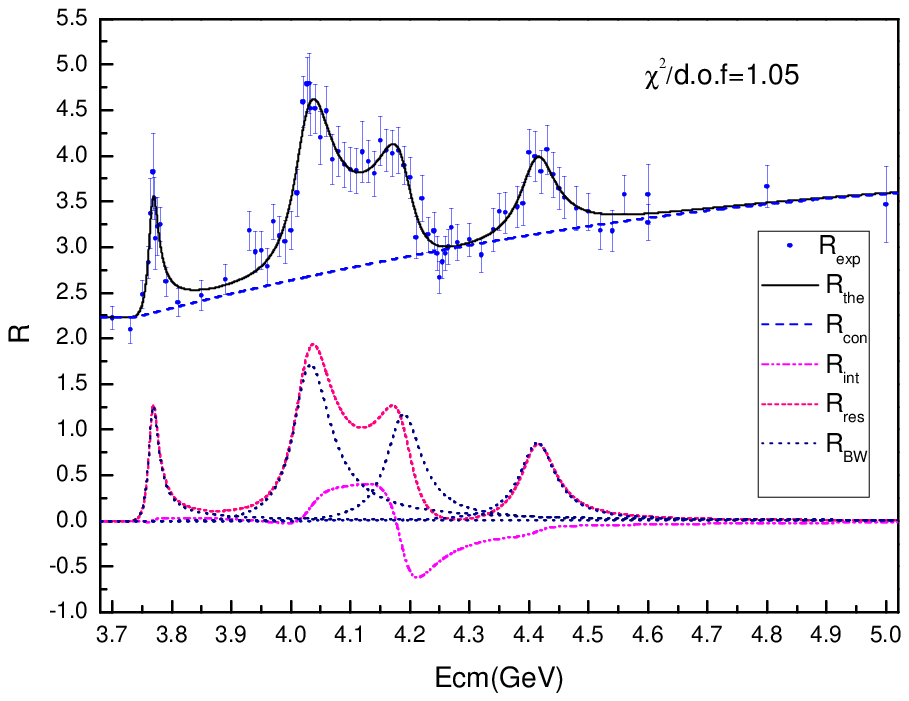}
\end{center}
\vspace{-5mm}\caption{Upper: $R$ values measured between 2-5 GeV.
Lower: the charmoniuum resonance structures measured with BESII.}
\label{r99bes}
\end{figure}  
\end{center}

~~~~The $R$ value is defined in Eq.~\ref{Rdef}.  Experimentally, it
is determined 
from ~\cite{part2_r_e_cb86,part2_r_e_besr98,part2_r_e_besr99}
\begin{equation}
R_{exp}=\frac{ N^{obs}_{had} - N_{bg}} { \sigma^0_{\mu\mu}  L
\epsilon_{trg} \bar\epsilon_{had}(1+\delta)}, \label{rexp}
\end{equation}
where $N^{obs}_{had}$ is the number of observed hadronic events,
$N_{bg}$ is the number of the residual backgrounds, $L$ is the    
integrated luminosity, $\bar\epsilon_{had}$ is
the average detection efficiency for hadronic
events~\cite{part2_r_e_luarlw2001}, $\epsilon_{trg}$ is the trigger
efficiency~\cite{part2_r_e_trigger}, $(1+\delta)$ is the correction
factor of the initial state radiation (ISR),
and $\sigma^0_{\mu\mu}$ is the theoretical Born cross section for
$e^+e^-\to\mu^+\mu^-$. Figure~\ref{r99bes} shows the measured $R$
values below 5 GeV and the higher charmoniuum resonance structures
measured by BES-II.

\subsection{Hadronic event selection and background subtraction}
~~~~In the BEPC energy region, the event types in the raw data are the
QED processes ($e^+e^-\to e^+e^-$, $\mu^+\mu^-$, $\tau^+\tau^-$,
$\gamma\gamma$, etc.), hadronic processes (including continuum  
and resonance states) and beam-associated backgrounds. The     
observed final-state particles are $e$, $\mu$, $\pi$, $K$, $p$ and
$\gamma$.

Different types of final states can be identified with specialized
selection criteria. At high energies ($E_{cm}>2$ GeV), there are many
unknown and unobservable hadronic production channels, one cannot
measure $R$  by summing up all of the exclusive cross   
sections. The strategy for selecting an inclusive sample of
hadronic events is to first subtract the backgrounds, and then 
judge the remaining events by specific
criteria~\cite{part2_r_e_hadsel04}:
(1) if an event is classified as $e^+e^-$ or $\gamma\gamma$ or
$\mu^+\mu^-$, it is rejected; (2) the residual backgrounds are
further removed by the hadronic criteria; (3) an energy deposit cut
is used to remove most of the beam-associated backgrounds.

In the data analysis, the hadronic events are classified by the 
number of their charged tracks and the hadronic criteria may be  
divided into track level and event level. Hadronic events with
two or more good charged tracks ($n_{gd}\geq 2$) are selected
by the criteria described in Refs.~\cite{part2_r_e_besr98}
and \cite{part2_r_e_besr99}.  For events with a single good
charged track ($n_{gd}= 1$),  one or more $\pi^0$'s are   
required~\cite{part2_r_e_hadsel04}.

The number of observed hadronic events, $N^{obs}_{had}$, is
obtained by fitting the $z$-vertex distribution 
 of the selected events with a Gaussian signal function
and a polynomial background~\cite{part2_r_e_besr99}.  The number of 
the residual QED backgrounds $N_{bg}$ is estimated statistically from
\begin{equation}
N_{bg}=L\sum\limits_{\tiny QED}\epsilon_{\tiny
QED}\cdot\sigma_{\tiny QED},
\end{equation}
where $\sigma_{\tiny QED}$ is the theoretical cross section ofor the
QED process and $\epsilon_{\tiny QED}$ is the corresponding
residual efficiency~\cite{part2_r_e_cb86}.

The fraction of the hadronic events lost in the reconstruction and
data analysis is compensated by the Monte Carlo-determined
hadronic event efficiency
\begin{equation}
\bar\epsilon_{had}=\frac{N_{obs}^{MC}}{N_{gen}^{MC}},
\end{equation}
where $N^{MC}_{gen}$ is the number of the inclusive hadronic events
produced by the generator, and $N^{MC}_{obs}$ is the observed number of
events passing the detector simulations and the event selection
requirements.

\subsection{Scan of the continuum and resonances}
~~~~The measured $R$ values for the 2-5~GeV energy region 
are shown in Fig.~\ref{r99bes}. The region between
2.0~and~3.73~GeV is called  the
continuum region (except for the two narrow resonances $J/\psi$ and
$\psi'$). The $R$ values in this region are approximately smooth    
and change very slowly with energy, as predicted by QCD. The
energy range above 
the open-charmed meson pair-production threshold (3.73~GeV), is called 
the resonance region. 
There,  previously observed resonances are 
are the $\psi(3773)$, $\psi(4040)$, $\psi(4190)$ and
$\psi(4415)$~\cite{part2_r_e_seth}\cite{part2_r_e_besresfit}; 
the $Y(4260)$ resonance was only recently 
observed~\cite{part2_r_e_y4200} and only in its $\pi^+\pi^- J/\psi$
decay mode;  $Y(4260)$ decays to open-charmed mesons have yet 
to be seen.  They might be observed at \bes3 
in a scan with small energy steps and   
large statistics at each point. 

The measured $R$ value 
have to be  corrected by the
ISR factor $(1+\delta)$, which depends on the $R$ value and   
the resonance parameters. In general, the measurement of 
$R$ value in the resonance region has to be an iterative
procedure~\cite{part2_r_e_besresfit} and the following issues  
should be taken into account:

\noindent $\bullet$~~\underline{Breit-Weigner amplitude:}
The relativistic Breit-Wigner amplitude for $e^+e^-$ $\to$ resonance
$\to$ final state $f$ is
\begin{equation}
{\cal T}_r^{f}(W)
=\frac{M_r\sqrt{\Gamma_r^{ee}\Gamma_r^f}}{W^2-M_r^2+ iM_r\Gamma_r}
e^{i\delta_r}, \label{rlaf}
\end{equation}
where $W\equiv E_{cm}\equiv\sqrt{s}$ is the cms energy,
the index $r$ represents the resonance being considered, $M_r$ is 
the nominal resonance mass, $\Gamma_r$ is the full width, 
$\Gamma_r^{ee}$ is  
the electron width, $\Gamma_r^{f}$ is the hadronic width for the 
decay channel $f$, and $\delta_r$ is the phase.

High mass charmoniuum states can decay into several two-body
final states $f$. According to the Eichten model~\cite{part2_r_e_eichten} and
existing experimental data~\cite{part2_r_e_barnes}, 
the allowed decay channels
(including their conjugate states) are:
\begin{eqnarray}
\psi (3770)&\Rightarrow& D\bar{D};\nonumber\\
\psi (4040)&\Rightarrow& D\bar{D},D^{\ast}\bar{D}^{\ast},
                         D\bar{D}^{\ast},
                         D_s\bar{D}_s;\nonumber\\
\psi (4140)&\Rightarrow& D\bar{D},D^{\ast}\bar{D}^{\ast},
                         D\bar{D}^{\ast},
                         D_s\bar{D}_s,D_s\bar{D}_s^{\ast};\nonumber\\
\psi (4415)&\Rightarrow& D\bar{D},D^{\ast}\bar{D}^{\ast},
                         D\bar{D}^{\ast},
                         D_s\bar{D}_s,D_s\bar{D}_s^{\ast},
D_s^{\ast}\bar{D}_s^{\ast}, D\bar{D_{1}}, D\bar{D^{*}_{2}}.\nonumber
\end{eqnarray}
\noindent
The total squared inclusive amplitude of the resonances is the
incoherent sum over all different decay channels $f$,
\begin{equation}
|{\cal T}_{res}|^2=\sum_{f} |\sum_{r} {\cal T}_{r}^{f}(W)|^2.
\end{equation}
The resonance cross section can be expressed in terms of $R$ as
\begin{equation}
R_{res} = \frac{12\pi}{s} \{|{\cal T}_{\psi(2S)}|^2 + |{\cal 
T}_{res}|^2\}, \label{eqres}
\end{equation}
where the contribution from the high-mass
tail of $\psi(2S)$ is included.

\noindent $\bullet$~~\underline{Hadronic width:}
The hadronic width of a broad resonance depends on the energy. The
calculation of $\Gamma_r^{f}(W)$ involves strong interactions,
and phenomenological models have to be used. 
The decay of a resonance can
be viewed as quantum mechanical  barrier 
penetration~\cite{part2_r_e_gaocsl}, and this
predicts the energy-dependence of the
hadronic width to be \cite{part2_r_e_blatt,part2_r_e_galtieri}
\begin{equation}
\Gamma_{r}^{f}(W)=\hat{\Gamma}_r\sum\limits_L
\frac{Z_{f}^{2L+1}}{B_L},\label{widtqm}
\end{equation}
where $L$ is the orbital angular momentum of the decaying final
state, $\hat{\Gamma}_r$ is a parameter to be determined by fitting
experimental data, $Z_{f}\equiv \rho P_{f}$, $\rho$ is the radius of
the interaction (of the order of a few fermis) and $P_{f}$ is the decay
momentum. The energy-dependent functions $B_L$ are 
given in Ref.~\cite{part2_r_e_blatt}.

\noindent $\bullet$~~\underline{Continuum backgrounds:}
Contributions from continuum backgrounds that originate from
initial light-quark pairs ($u\bar{u}$, $d\bar{d}$ and $s\bar{s}$)
are well described by pQCD for cms energies above 2~GeV. 
Because of the proximity of the
production threshold, open-$c\bar{c}$ continuum backgrounds can
only be described by phenomenological models or experimental
expressions. The DASP group assumes that the continuum charm backgrounds
are the two-body states $D\bar{D}$, $D\bar{D}^{\ast}$,
$D^{\ast}\bar{D}^{\ast}$, $D_{s}\bar{D_{s}}$,
$D_{s}\bar{D_{s}}^{\ast}$ and $D_{s}^{\ast}\bar{D_{s}}^{\ast}$. The
cross sections for these continuum backgrounds are given in 
Ref.~\cite{part2_r_e_dasp78}. One can assume that there are many possible 
continuum states above the open-charm threshold, and the inclusive cross
section is expected to vary smoothly with energy.  For simplicity, it
is parameterized as a second-order polynomial~\cite{part2_r_e_besresfit}.
The shape of the continuum backgrounds can be calculated using the 
LUND area law [Eq.~(\ref{lundml})],  but the normalization constants 
have to be determined from data. Note that both the
DASP-form and the polynomial provide smooth backgrounds shapes, while
the LUND area law can be used to describe more complex threshold 
behaviors.

\noindent $\bullet$~~\underline{Fitting function:}
The iterative fits can be done with a least squares method
using MINUIT~\cite{part2_r_e_minuit}
with a $\chi^2$ is defined as
\begin{equation}
\chi^2 = \sum\limits_i \frac{[f_c\widetilde{R}_{exp}(W_i) -
\widetilde{R}_{the}(W_i)]^2} {[f_c\Delta
\widetilde{R}^{(i)}_{exp}]^2}+\frac{(f_c-1)^2}{\sigma^{2}_{c}},  
\label{fitf2}
\end{equation}
where the experimental and the corresponding theoretical quantities
are
\begin{equation}
\widetilde{R}_{exp}=\frac{ N^{obs}_{had} - N_{bg} } {
\sigma^0_{\mu\mu}  L  \epsilon_{trg} \bar\epsilon_{had}} ~~~~~~{\rm
and}~~~~~~~ \widetilde{R}_{the}=(1+\delta) R_{the} . \label{rlike}
\end{equation}
Here $\Delta \widetilde{R}^{(i)}_{exp}$ is the statistical and
non-common systematic errors of $\widetilde{R}_{exp}(W_i)$,
$\sigma_{c}$ is the common error, and the
scale factor $f_c$ reflects the
influence of the common error. If interference between 
the continuum
and resonance states is ignored, $R_{the}$ is given by
\begin{equation}
R_{the}=R_{con}+R_{res}.\label{rthe}
\end{equation}

The measurement of the resonance parameters and $R$ is performed
iteratively: the initial values of the parameters are used as the
inputs to calculate the radiative correction factors $(1+\delta)$ at
all energy-points; then these values 
are updated in the next iteration; the fitting
continues until $\chi^2$ satisfies the convergence conditions and
the correct error matrix is given, and the measured $R$ values 
and resonance parameters are obtained.

\subsection{$R$ values from radiative return}


~~~~Energy scans were used to collect data at different cms
energies to measure  $R$ values at  BEPC/BESII, and the effects 
of ISR are corrected by
the factor $(1+\delta)$. The luminosity of BEPC-II will be 100   
times higher than that at BEPC, and the photon resolution at 
\bes3 is significantly improved.  This will make it possible 
to measure $R$ values with the ISR (radiative return) events:
\begin{displaymath}
e^+e^-\to\gamma^{hrd}+\gamma^{sft}+{\rm hadrons},
\end{displaymath}
where $\gamma^{hrd}$ indicates an observed hard radiated photon with
large momentum from the initial $e^+$ or $e^-$, and $\gamma^{sft}$
indicates possible observed or unobserved soft photons. 
In terms of $x$, the fraction of the beam particle's
energy that is transferred to the ISR photon, the
effective energy for producing the final-state hadrons 
is $s'=s(1-x)$. The cross section for the ISR process 
$e^+e^-\to\gamma^{hrd}+f$ with a
particular final state $f$ is related to the cross section
$\sigma_f(s)$ for the direct annihilation as~\cite{part2_r_davier}
\begin{equation}
\frac{d\sigma_f(s,x)}{dx}=W(s,x)\sigma_f[s(1-x)],
\end{equation}
where $x=2E_{\gamma}^*/\sqrt{s}$, $E_{\gamma}^*$ is the radiated
photon 
fractional energy in the $e^+e^-$ frame (laboratory system) and
$\sqrt{s}$ the nominal center-of-mass energy of the collider. The
radiator function $W(s,x)$ has been computed including radiative
correction to an accuracy that is better than
$1\%$~\cite{part2_r_hepex0502025,part2_r_hepex0512023}
\begin{equation}
W(s,x)=\beta [(1+\delta)x^{\beta-1}-1+\frac{x}{2}],
\end{equation}
where $\beta=2\alpha/\pi(\ln(s/m_e^2)-1)$ and $\delta$ takes into
account vertex and self-energy corrections.

The production of ISR photons is strongly concentrated along the
incident beam direction, 
which means that most of the ISR events can not be
identified and lost in data analysis. In order to accumulate a data
sample within a small window of lower effective energy $\Delta
s'$ and with large enough statistics for a particular channel $f$, it
is necessary to run BEPCII at a fixed nominal energy $\sqrt{s}$
(such as a resonance peak) for a long period, as was done with PEP-II at
SLAC.

Typically, many photons are observed in an event, 
most of which are produced either
via hadronic processes ({\it e.g.}
$\pi^0\to\gamma\gamma$ etc.), by
interactions between produced hadrons
and the material of the detector or by soft ISR. 
In order to be able to distinguish the hard ISR photon
from soft ISR photons and
hadronically produced photons, 
the measured energy region
$s'$ must be substantially lower than $s$. 
For example, PEP-II runs at
nominal energy $\sqrt{s}=10.5$ GeV, 
while the the energy region for
the {\small BaBar} $R$ measurements
is below 5 GeV.

The main focus of radiative-return
 $R$ measurements has been the study of exclusive
hadronic channels. The interesting channels at low energies are:
$\pi^+\pi^-$, $K^+K^-$, $p\bar{p}$, $\pi^+\pi^-\pi^0$, $4\pi$,
$5\pi$, $6\pi$, $\pi\pi\eta$, $K\bar{K}\pi$, $K\bar{K}\pi\pi$,
$2K2\bar{K}$, $K\bar{K}\eta$.  Measurements of the leptonic
process $e^+e^-\to\gamma\mu^+\mu^-$ provides the ISR luminosity.
Thus the Born cross section $\sigma_s(s')$ is obtained
from~\cite{part2_r_davier}
\begin{equation}
\sigma_f(s')=\frac{\Delta N_{\gamma
f}\epsilon_{\mu\mu}(1+\delta_{FSR}^{\mu\mu})}{\Delta N_{\gamma
\mu\mu}\epsilon_{f}(1+\delta_{FSR}^{f})}\sigma_{\mu\mu}(s'),\label{sigmaf}
\end{equation}
where $\Delta N_{\gamma f}$ is the number of detected $\gamma f$
events in the bin of width $\Delta s'$ centered at $s'$,
$\epsilon_f$ is the detected efficiency for the final state $f$, and
$\delta_{FSR}^{f}$ is the fraction of times
the photon is emitted by one of the final state 
particles.  The quantities for $\gamma\mu^+\mu^-$ have  similar
meanings. It is important to correctly compute the $s'$ value for
each event as it has to be obtained from the momenta of
the produced particles, 
since the photon energy is rather insensitive to $s'$ in the
low-energy regime of interest.

The $R$ measurement for the inclusive hadronic cross section with ISR
data needs more detailed study. One of the main difficulties,
based on BaBar's experience,  is that the resolution deteriorates
rapidly for low recoil masses~\cite{part2_r_davier}.

Assuming that the luminosity of BEPC-II is 
$10^{33}{\rm cm}^2s^{-1}$ at 4 GeV, one can 
collect about $10^8$ hadronic events in a one-year run
($\sim 10^7$s). Table~\ref{dsdn} shows the estimated number of the
produced hadronic events in a series of effective energy intervals
$\Delta\sqrt{s'}$. The hadronic efficiency depends on $\sqrt{s'}$,
it was typically about 0.50--0.68 for $N_{gd}\geq1$-prong events
in the solid angle $|\cos\theta|<0.86$ for BESII,  but the
efficiency will fall rapidly with decreasing $\sqrt{s'}$.
\begin{table}[ht]
\begin{center}
\caption{Estimated numbers of the produced hadronic events in a
series of effective energy intervals in a one year ($10^7$
seconds) BEPC-II run at $\sqrt{s}=4$GeV.}
\begin{tabular}{|c|c|c|c|c|c|c|c|}\hline
$\sqrt{s'}$& $\Delta N_{had}^{gen}$  && $\sqrt{s'}$ &$\Delta
N_{had}^{gen}$&&$\sqrt{s'}$&$\Delta N_{had}^{gen}$\\\hline
0.4--0.5&$4.8\times10^5$&&1.3--1.4&$3.0\times10^5$&&2.2--2.3&$3.6\times10^5$\\\hline
0.5--0.6&$5.3\times10^5$&&1.4--1.5&$3.0\times10^5$&&2.3--2.4&$3.6\times10^5$\\\hline
0.6--0.7&$12.1\times10^5$&&1.5--1.6&$2.7\times10^5$&&2.4--2.5&$3.7\times10^5$\\\hline
0.7--0.8&$31.8\times10^5$&&1.6--1.7&$3.3\times10^5$&&2.5--2.6&$3.7\times10^5$\\\hline
0.8--0.9&$10.7\times10^5$&&1.7--1.8&$3.2\times10^5$&&2.6--2.7&$3.8\times10^5$\\\hline
0.9--1.0&$3.2\times10^5$&&1.8--1.9&$2.8\times10^5$&&2.7--2.8&$3.8\times10^5$\\\hline
1.0--1.1&$7.5\times10^5$&&1.9--2.0&$2.5\times10^5$&&2.8--2.9&$3.9\times10^5$\\\hline
1.1--1.2&$2.4\times10^5$&&2.0--2.1&$3.3\times10^5$&&2.9--3.0&$4.2\times10^5$\\\hline
1.2--1.3&$2.2\times10^5$&&2.1--2.2&$3.4\times10^5$&&&\\\hline
\end{tabular}
\label{dsdn}
\end{center}
\end{table}

\subsection{Systematic errors}
~~~~The two largest error sources for the $R$ measurement are the
error on the number of 
observed hadronic events $N_{had}^{obs}$ and the error
associated with hadronic event detection 
efficiency $\bar\epsilon_{had}$. Since the
systematic errors of these two terms are estimated by comparing the
difference between data and MC, the equivalent number of hadronic
events can be defined as
\begin{equation}
\tilde{N}_{had}^{obs}=\frac{N_{had}}{\bar\epsilon_{had}}
=N_{gen}^{MC}\frac{N_{had}}{N_{obs}^{MC}}.
\end{equation}
In Eq.~(\ref{rexp}), the main systematic error for the $R$ value
measurement is estimated as
\begin{equation}\label{r_error}
\frac{\Delta R}{R}\cong\sqrt{(\frac{\Delta
\tilde{N}_{had}}{\tilde{N}_{had}})^2+(\frac{\Delta L}{L})^2 +
(\frac{\Delta \epsilon_{trg}}{\epsilon_{trg}})^2 + (\frac{\Delta
(1+\delta)}{(1+\delta )})^2}.
\end{equation}

In addition, the error of the tracking efficiency $\sigma_{trk}$,
which reflects the difference of the track reconstruction between
data and MC, will cause an additional error $\Delta\epsilon_{trk}$. The
probability that $n_{er}$ charged tracks are incorrectly reconstructed in
an $n_{gd}$-prong event ($n_{er}\leq n_{ch}$) can be considered to
roughly obey a binomial distribution
$B(n_{er};n_{gd},\sigma_{trk})$. With a multiplicity distribution 
of the form $P(n_{gd})$, $\Delta\epsilon_{trk}$ is estimated to  
be~\cite{part2_r_e_hadsel04}
\begin{equation}
\Delta\epsilon_{trk}
=\sum_{n_{gd}}P(n_{gd})B(n_{er};n_{gd},\sigma_{trk}).
\end{equation}
For measurements of the inclusive cross section, only those cases
where {\it all} $n_{gd}$ tracks in an event are incorrectly reconstructed 
($n_{er}=n_{gd}$) will contribute to the error on $R$.

\chapter[Experimental tests of QCD]{Experimental
tests of QCD\footnote{By Hai-Ming Hu}}
\label{chapter:qcd_test}

~~~~Quantum Chromodynamics (QCD) provides a means to apply
perturbative techniques to quark-gluon evolution processes with
large momentum
transfer~\cite{part2_r_e_yndurain,part2_r_e_yuldokshitzer}. However,
the treatment of hadronization (the transition from quarks and
gluons to hadrons at nonperturbative scales) has not been solved
mathematically. At present, some assumptions about 
hadronization are commonly made. 
Local parton-hadron duality
(LPHD)~\cite{part2_r_e_azimov} predicts that
the parton distributions
are simply renormalized in the hadronization process without 
any shape
changes. The modified leading-logarithmic approximation (MLLA)
takes into account the soft partons and strict transverse momentum
ordering in subsequent perturbative series. All distributions
derived from LPHD/MLLA contain a few free parameters that have
to be determined from experiment, and LPHD/MLLA itself still has
to be experimentally tested. Some distributions that have not yet 
been predicted by LPHD/MLLA can be measured in experiments.  
Experimental 
tests of OCD-motivated models are very helpful for providing  
understanding of the strong interaction and 
for giving guidance to the
development of nonperturbative QCD techniques.

%

\section{Inclusive distributions}

The different types of hadronic  production measurements can be 
classified as total cross sections, and
exclusive, inclusive, semi-inclusive and single-particle
distributions according to how many and what kinds of the final
particles are measured~\cite{part2_r_e_humble}. Note that
the meanings of these terms may not be exactly the same in different
references. In the following, some possible measurements are
listed.

In general,  inclusive distributions are measured as a function of
the kinematic variables $(s,p_{//},p_\bot)$. 
Two questions are addressed: (1)
how do the distributions change with the cms energy
$\sqrt{s}$? and (2) how do these distributions change with momentum
$(p_{//},p_\bot)$ when $\sqrt{s}$ is fixed?  The answer to the second
question depends on the type of the initial state and the
properties of measured particles in the final state.

Feynman hypothesized that at sufficiently high energies, 
where quark and hadron masses can be neglected, 
quark fragmentation depends only on the quark 
flavor and a dimensionless scaling variable
$x=2p/E_{cm}$. This scaling property means that the distributions
are  functions only of the scaling variable $x$ and transverse
momentum $p_\bot$ at large energies. Thus, the cross section
$sd\sigma /dx$ averaged over $p_\bot$ should be
$s$-independent~\cite{part2_r_e_patrick}. The scaling assumption is
found to be a good approximation in high energy reactions. 
It has also been
tested at BES-II, but further tests are needed to understand where
and how scaling fails.  This should provide 
some insight into pQCD and non-pQCD.

Commonly, inclusive distributions are measured as a functon
of Feynman $x$ or $\xi = -\ln x$, rapidity $y$ or $\eta$, and $p_\bot$.

\noindent$\bullet$~\underline{$x$ distribution}\\
Inclusive differential cross sections can be expressed in terms
of the structure functions $F_1$ and $F_2$~\cite{part2_r_e_drelletal},
\begin{equation}
\frac{d^2\sigma}{dxd\cos\theta}=\frac{3}{4}\sigma_0x\beta[2F_1(x,s)+
\frac{x\beta}{2}F_2(x,s)\sin^2\theta],
\end{equation}
where $\sigma_0$ is the zeroth-order QED cross section. The
relationships between $F_1$ \& $F_2$ and the longitudinal and transverse
structure functions $F_T$ \& $F_L$ are
\begin{eqnarray}
F_T(x,s)&=&2F_1(x,s),\\
F_L(x,s)&=&2F_1(x,s)+xF_2(x,s).
\end{eqnarray}

\noindent$\bullet$~\underline{$\xi$ distribution}\\
The inclusive $\xi$ distribution can be 
derived from 
MLLA/LPHD~\cite{part2_r_e_schmelling,part2_r_e_fongwebber1,
part2_r_e_fongwebber2}
\begin{equation}\label{xidis}
\frac{1}{\sigma}\frac{d\sigma^h}{d\xi}=2K_{LPHD}\times f_{MLLA}(\xi
,Q_0,\Lambda_{eff}),
\end{equation}
where $f_{MLLA}(\xi ,Q_0,\Lambda_{eff})$ is the $\xi$ distribution 
at the parton level
and $K_{LPHD}$ is a factor 
that connectsthe parton level and hadron level in 
the context of LPHD.
The $\xi$ range for Eq.~(\ref{xidis}) is $0\leq\xi\leq Y\equiv
\ln (\sqrt{s}/2\Lambda_{eff})$. In the neighborhood of the peak
value $\xi^*$, the $\xi$ distribution can be replaced with a
distorted Gaussian form~\cite{part2_r_e_dokshitzer,part2_r_e_fong}:
\begin{equation}
DG(\xi;N,\bar{\xi},\sigma,s_k,k)=\frac{N}{\sigma\sqrt{2\pi}}
\exp[\frac{1}{8}k-\frac{1}{2}s_kk\delta-\frac{1}{4}(2+k)\delta^2
+\frac{1}{6}s_kk\delta^3+\frac{1}{24}k\delta^4],
\end{equation}
where $\delta=(\xi-\bar{\xi})/\sigma$, $\bar{\xi}$ is the average
value of $\xi$, $\sigma$ is the width of the $\xi$ distribution, and
$s_k$ and $k$ are the skewness and kurtosis, respectively. 
For a normal Gaussian
function, both $s_k$ and $k$ are zero. MLLA/LPHD predicts that the
peak position of $\xi$ depends on the energy as
\begin{equation}
\xi^*=0.5Y+\sqrt{cY}+c+\mathcal{O}(1/\sqrt{Y}),
\end{equation}
where $c$ is a function of
the number of colors $N_c$ and the number of the active quarks $n_f$.
The energy dependence of $\xi^*$ may be described 
as~\cite{part2_r_e_akwrawy}
\begin{equation}
\xi^*=A\ln s+B,
\end{equation}
where $A$ and $B$ are free parameters that are determined from
experiment.
\begin{figure}[htbh]
\centering{
\includegraphics[width=8.0cm,height=7.0cm]{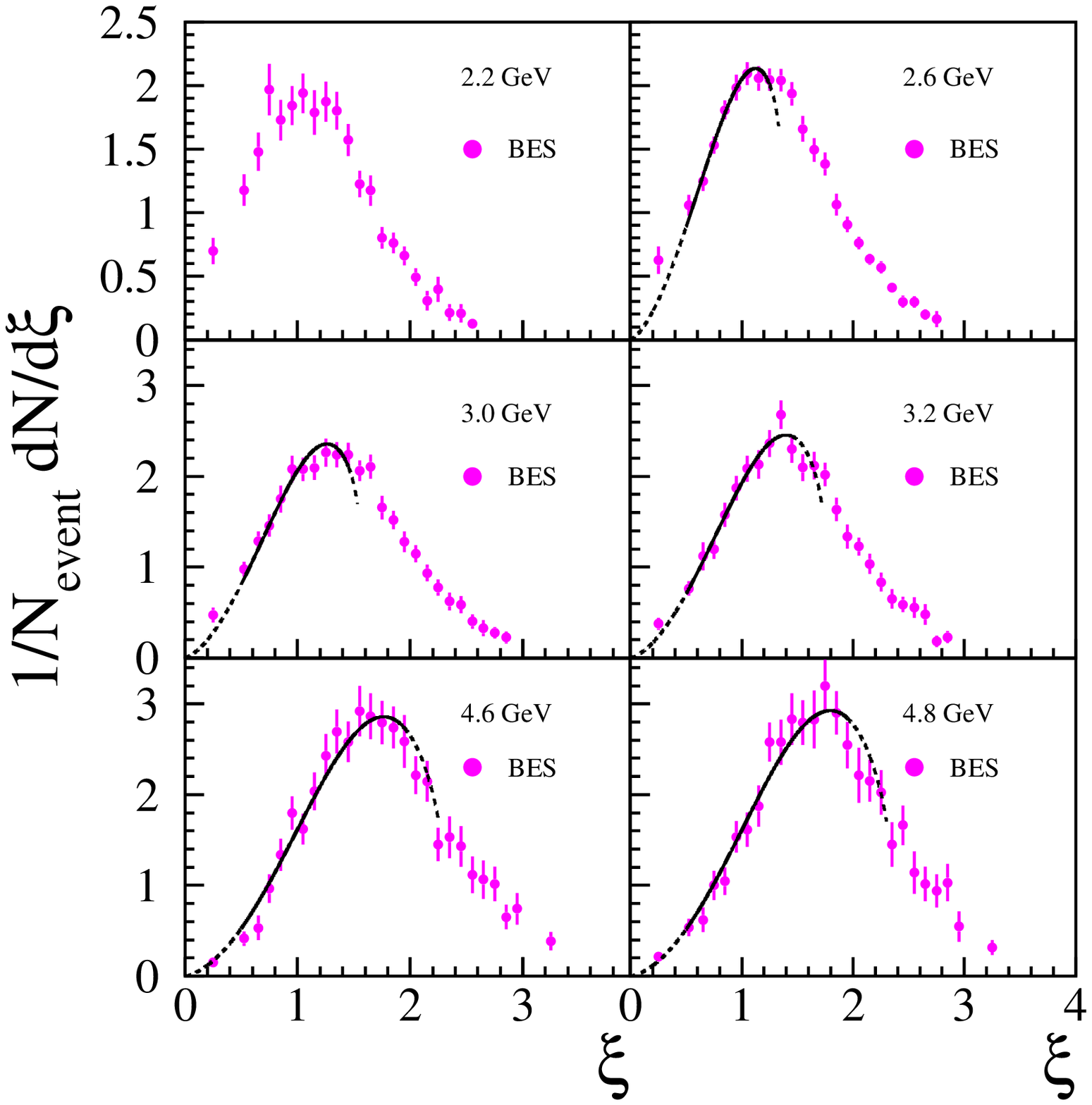}
\put(-185,180){(a)}
\includegraphics[width=8.0cm,height=7.0cm]{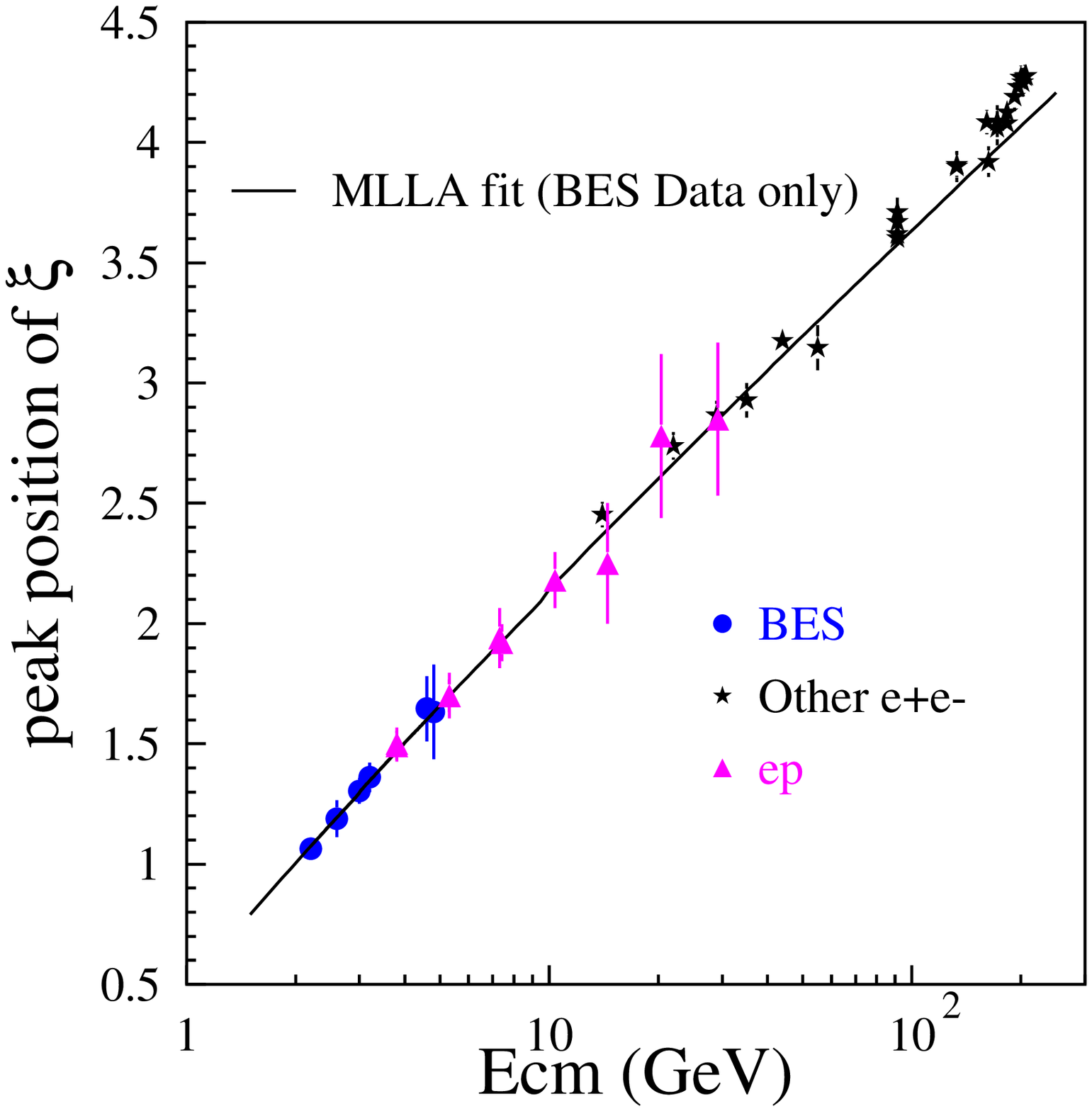}
\put(-175,180){(b)}\\
\includegraphics[width=7.5cm,height=6.5cm]{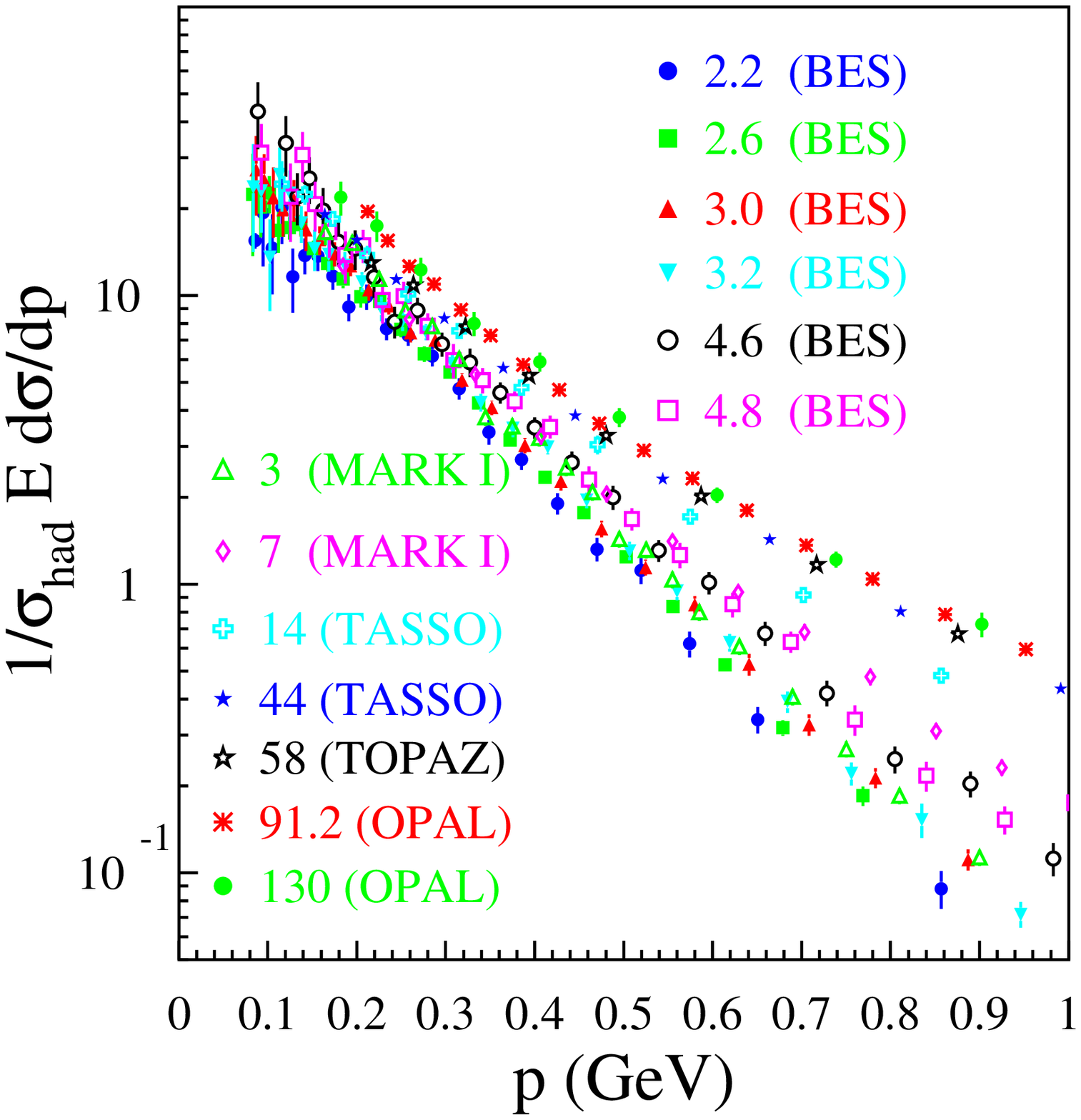}
\put(-175,170){(c)} }
\includegraphics[width=7.5cm,height=6.5cm]{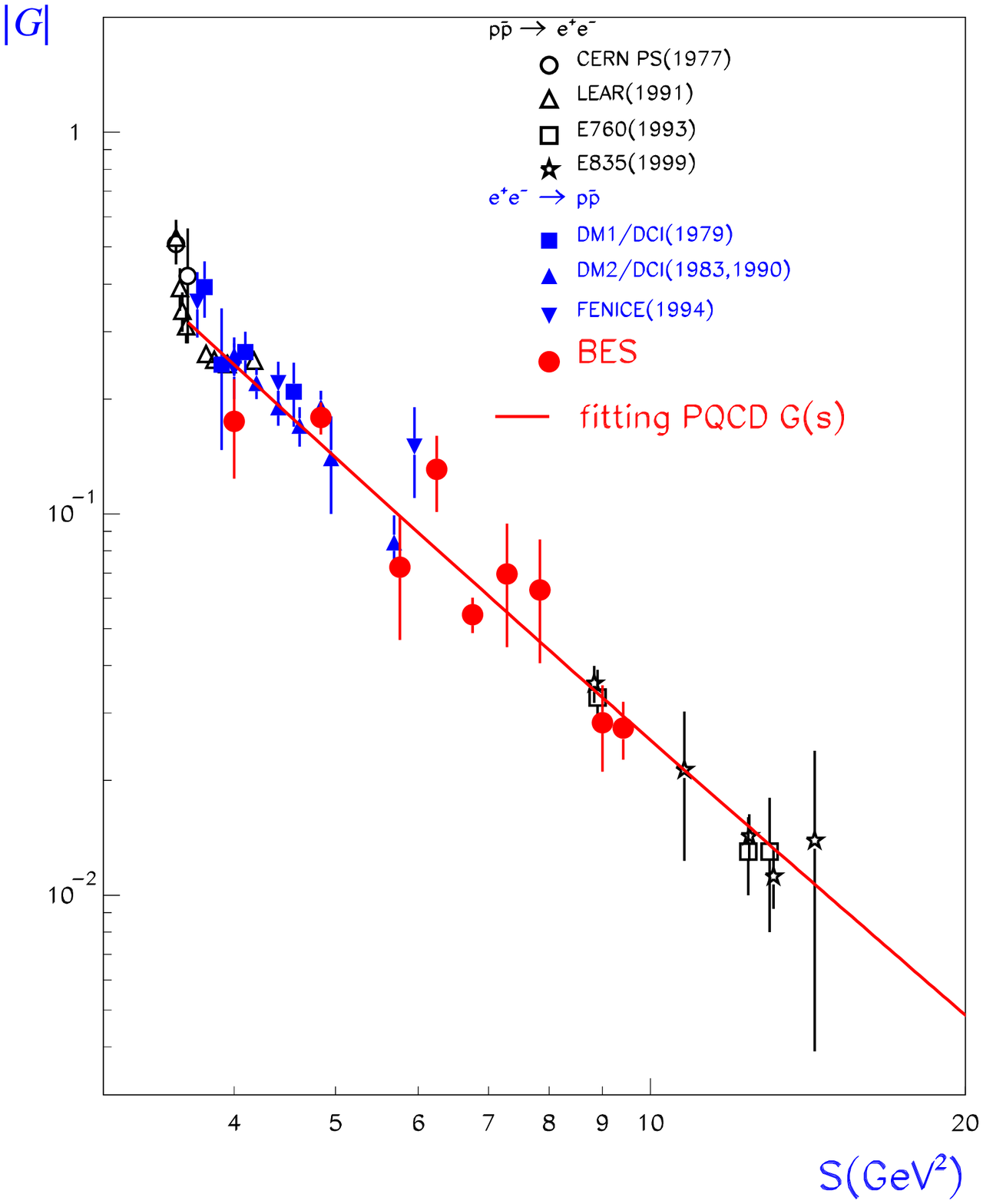}
\put(-175,170){(d)} \caption{(a) Distributions of $\xi$; (b)
energy dependence of the peak position of $\xi$; (c) distribution of
momentum $p$; (d) $p\bar{p}$ form factors.} \label{inclus}
\end{figure}

\section{Exclusive cross sections and form factors}
Exclusive cross sections can be written as functions of form
factors that embody the influence of the strong interaction
on the properties of electro-magnetic
interaction vertices.  Precise
measurements of hadronic form factors helps promote the
understanding to the strong interaction. In particular, in the BEPC
energy region, $e^+e^-\to\pi^+\pi^-$, $\pi^+\pi^-\pi^0$,
$\pi^+\pi^-\pi^+\pi^-$, $\pi^+\pi^-\pi^0\pi^0$,
$\pi^+\pi^-K^+K^-$, $p\bar{p}$, $\Lambda\bar{\Lambda}$ etc. have
large production cross sections and it is  important 
that they are properly modeled by
the Monte Carlo generators.

\noindent $\bullet$~~\underline{$e^+e^-\to N\bar{N}$ }\\
Nucleons, the proton $p$ and neutron $n$, are the two most common
components of matter. Their electromagnetic form
factors describe their internal structure and dynamics.

In the process $e^+e^-\to p\bar{p}$, a pair of
spin-1/2 baryons with internal structure are produced. The current
contains two independent form factors, the Dirac $F_1(q^2)$ and
Pauli $F_2(q^2)$ form factors~\cite{part2_r_e_drell}, defined as
\begin{equation}
\langle N(p')|J^\mu(0)|N(p)\rangle=e\bar{u}(p')[F_1(q^2)\gamma^\mu
+\frac{i}{2M}F_2(q^2)\sigma_{\mu\nu}q_\nu]u(p),
\end{equation}
where $M$ is the mass of nucleon and $q^2=(p-p')^2=s$ is the squared
momentum transfer at the photon-nucleon vertex. $F_1(q^2)$ and
$F_2(q^2)$ are normalized so that for the proton $F_1^p(0)=1$ and
$F_2^p(0)=\kappa_p=\mu_p-1$, and for neutron $F_1^n(0)=0$ and
$F_2^n(0)=\kappa_n=\mu_n$, where $\kappa_p$ and $\kappa_n$ are the
anomalous part of proton and neutron magnetic moments. The electric
$G_E(s)$ and magnetic $G_M(s)$ form factors are defined by the
combinations
\begin{equation}
G_E=F_1+\tau F_2, \,\,\,\, G_M=F_1+F_2.
\end{equation}

The differential and total cross sections are given
by~\cite{part2_r_e_botov,part2_r_e_dafnenf}
\begin{eqnarray}
\frac{d\sigma}{d\Omega}&=&\frac{\alpha^2\beta}{4s}C[|G_M(s)|^2(1+\cos^2\theta
) +\frac{1}{\tau}|G_E(s)|^2\sin^2\theta ] \label{diffcroxt}\\
\sigma&=&\frac{4\alpha^2\pi\beta}{3s}[|G_M(s)|^2+\frac{1}{2\tau}|G_E(s)|^2],\label{totcroxt}
\end{eqnarray}
where $\tau =s/4M^2$, $\theta$ is the scattering angle,
$\beta=\sqrt{1-1/\tau}$ is the velocity and $C$ is the Coulomb
correction factor. For the neutron $C=1$, and for the proton
\begin{equation}
C=\frac{y}{1-exp(-y)}, \,\,\, y=\pi\alpha M/\beta\sqrt{s}.
\end{equation}
The electronic $G_E(s)$ and magnetic $C_M(s)$ are two independent
form factors with  normalizations $G_E(0)=1$ and $G_M(0)=\mu_p$.
Figure~\ref{inclus}(d) shows the measured form factor of the proton
under the assumption that $G_E(s)=C_M(s)$. The measurement of the 
angular distribution of 
Eq.~(\ref{diffcroxt}) determines $G_E(s)$ and
$C_M(s)$ simultaneously, provided  the data sample is sufficiently large.


\noindent $\bullet$~~\underline{$e^+e^-\to\pi$ mesons}\\
In 1988, the $\rho(1600)$ entry in the PDG tables was 
replaced by two new resonances: the
$\rho(1450)$ and $\rho(1700)$.  It has been suggested that this
assignment can be validated by a theoretical analysis on the
consistency of  the $2\pi$ and $4\pi$ electromagnetic form factors.
In any case, detailed experimental data on the cross section for
$e^{+}e^{-}\to \pi^{+}\pi^{-}\pi^{+}\pi^{-}$ will make possible the
accurate determination of the parameters of 
the $\rho$-meson and its radial recurrences.

The venerable vector-meson-dominance model (VMD) has 
been modified to expand the
contributions of the lowest-lying vector-mesons 
to include their high-mass
recurrencies~\cite{part2_r_e_php1,part2_r_e_php2}. Previous 
analyses~\cite{part2_r_e_php1}
used different decay channels to obtain the relevant parameters.
Reference~\cite{part2_r_e_wenghu4piff} uses the cross section for
$e^{+}e^{-}\to \pi^{+}\pi^{-}\pi^{+}\pi^{-}$ derived from the
extended VMD model to fit {\small BaBar} experimental data.
Cross sections for the processes $e^+e^-\to
\pi^+\pi^-,~\omega\pi^0,~\eta\pi^+\pi^-$, $\pi^+\pi^-\pi^+\pi^-$ and
$\pi^+\pi^-\pi^0\pi^0$ can be found 
in Refs.~\cite{part2_r_e_php1,part2_r_e_php2}.

\noindent $\bullet$~~\underline{Higher order corrections for 
$e^+e^-\to\pi^+\pi^-(\gamma)$}\\
More precision measurements of the pion form factor will require
a careful consideration of the radiative corrections, including the
initial state (IS) and the final state (FS) contributions. 
Reference~\cite{part2_r_e_hoefer} presents field-theory-based 
calculations for $e^+e^- \to \pi^ + \pi^- (\gamma)$ production 
with higher order radiative corrections. The Born cross section 
for $e^+e^- \to \pi^ + \pi^-$ can be written as
\begin{equation}
\frac{d\sigma_0}{d\Omega}=\frac{\alpha^2\beta_\pi^3}{8s}\sin^2\theta
|F_\pi(s)|^2,~~~{\rm
or}~~~\sigma_0(s)=\frac{3\pi\alpha^2\beta_\pi^3}{3s}|F_\pi(s)|^2,
\end{equation}
where $\theta$ is the angle between $\vec{p}_{\pi^-}$
and $\vec{p}_{e^-}$ and $\beta_\pi=\sqrt{1-4m_\pi^2/s}$. The form
factor $F_\pi(s)$ encodes the substructure of the pion with the
charge normalization constraint $F_\pi(0)=1$ (classical limit). For
real radiation $e^+e^-\to\pi^+\pi^-\gamma$,  the invariant mass
square of the pion pair is $s'=(p_{\pi^+}+p_{\pi^-})^2$. 
Reference~\cite{part2_r_e_hoefer} presents the differential cross section, 
the total cross section and the pion form factor with higher-order 
corrections. 
Discussions for $e^+e^-\to K_SK_L(\gamma)$ and $K^+K^-(\gamma)$
can be found in Ref.~\cite{part2_r_e_arbuzov}. The Monte Carlo generator
MCGPJ~\cite{part2_r_e_fedotovich}  simulates events with
two-body final states, including $e^+e^-\to\pi^+\pi^-(\gamma)$,
$K^+K^-(\gamma)$, $\mu^+\mu^-(\gamma)$ and $e^+e^-(\gamma)$.

\section{Multiplicity distribution}
~~~~The charged particle multiplicity distribution is a basic quantity for
any reaction.
Precise
measurements need reliable Monte Carlo and efficiency
matrices to translate from measured  to physical
(theoretical) quantities.  Experiments often measure 
the multiplicity
distribution of the charged particles with the definition:
\begin{equation}
P(n_c)=\frac{\sigma_{n_c}}{\sum\sigma_{n_c}},
\end{equation}
where $\sigma_n$ is the topological cross section 
for $n$-particle
production; the average multiplicity is given by
\begin{equation}
\langle n_c \rangle =\sum  n_c P(n_{c}).
\end{equation}

The observed multiplicity distribution $P_{obsv}(n_c)$ is not the
physical distribution $P_{true}(n_c)$. They are related by the
migration matrix 
$M(n_c',n_c)$~\cite{part2_r_e_jean,part2_r_e_siegrist},
\begin{equation}
P_{obsv}(n_c')=\sum_{n_c}M(n_c',n_c)P_{true}(n_c).\label{pn1}
\end{equation}
The matrix element $M(n_c',n_c)$ corresponds to the probability
for events
with $n_c$ produced charged particles to end up with $n_c'$ 
detected charged tracks.
This can be determined from MC as
\begin{equation}
M(n_c',n_c)=\frac{N_{MC}(n_c',n_c)}{\sum_{n_c}N_{MC}(n_c',n_c)},
\end{equation}
where $N_{MC}(n_c',n_c)$ is the number of events with $n_c$
generated and $n_c'$ observed charged particles.
Equation~(\ref{pn1}) can be rewritten in matrix notation as
\begin{equation}
P_{obsv}=MP_{true},~~~~{\rm or}~~~P_{true}=M^{-1}P_{obsv},
\end{equation}
where $P_{true}$ and $P_{obsv}$ are vectors whose elements are
$P_{true}(n_c)$ and $P_{obsv}(n_c')$.

QCD-motivated models interpret the jet evolution as 
a branching process,
and predict mean multiplicities to increase with energy as
\begin{equation}
\langle n_c \rangle =a+b\cdot \exp\{c[\ln(s/Q^2_0)]^{1/2}\}.
\end{equation}
The NLO-based QCD evolution relation for  fragmentation
predicts~\cite{part2_r_e_malaza}
\begin{equation}
\langle
n_c\rangle=a[\alpha_s(s)]^b\exp[c/\sqrt{\alpha_s(s)}][1+d\sqrt{\alpha_s(s)}],
\end{equation}
where
\begin{equation}
b=\frac{1}{4}+\frac{10n_f}{27\beta_0},~~~~c=\frac{\sqrt{96\pi}}{\beta_0},
\end{equation}
with $n_f$ the number of the active quarks, $\beta_0=11-(2/3)n_f$;
$a$ and $d$ free parameters. NLO predicts the second-order 
moment to be
\begin{equation}
R_2\equiv\frac{\langle n_c(n_c-1)\rangle}{\langle
n_c\rangle^2}=\frac{11}{8}[1-c\sqrt{\alpha_s(s)}],
\end{equation}
where
\begin{equation}
c=\frac{1}{\sqrt{6\pi}}\frac{4455-40n_f}{1782}.
\end{equation}

The average multiplicity derived from MLLA/LPHD can be written
as~\cite{part2_r_e_lupia}
\begin{equation}
\langle n_c\rangle =c_1\frac{4}{9}N_{LA}+c_2,
\end{equation}
where
\begin{equation}
N_{LA}=\Gamma (B)(\frac{z}{2})^{(1-B)}I_{1+B}(z),
\end{equation}
and $z=\sqrt{48Y/\beta_0}$, $Y=\ln (\sqrt{s}/2Q_0)$, $B=a/\beta_0$,
$a=11+2n_f/27$. Here $\Gamma$ is the Gamma function and $I_x$ is the
modified Bessel function of order $x$. 
The factor $Q_0$ is the energy scale
parameter, its value is taken to be $0.27$ GeV in
Ref.~\cite{part2_r_e_lupia}.

A commonly used expression for the average multiplicity 
is~\cite{part2_r_e_mattic}
\begin{equation}
\langle n_c \rangle = a +b\ln s +c \ln^2s,
\end{equation}
where $a$, $b$ and $c$ are free parameters. The energy variation
of the multiplicity distribution can be studied by measuring its
variance
\begin{equation}
D_c=\sqrt{\langle n_c^2\rangle -\langle n_c\rangle}.
\end{equation}
In hadron production experiments, the 
forward and backward correlations
\begin{equation}
\langle n_B\rangle = a+b n_F~~~~{\rm or}~~~~\langle n_F\rangle =c+d
n_B
\end{equation}
can give some useful information.

Feynman scaling predicts that the hadronic cross section
$\sigma_{n_c}(s)$ satisfy a scaling law for large $\langle n_c
\rangle$~\cite{part2_r_e_kno}:
\begin{equation}
\langle n_c\rangle\frac{\sigma_{n_c}(s)}{\sum_{n_c}\sigma_{n_c}(s)}
=\langle n_c\rangle P(n_c)\to\Psi (z),
\end{equation}
which means  $\langle n_c \rangle P(n_c)$ depends on $n_c$ through
$z\equiv n_c/\langle n_c\rangle$.  Here, $\Psi (z)$ is a energy
independent function. The scaling law has been tested at high
energies~\cite{part2_r_e_mattic}, but the approximation 
of the scaling
assumption has not been tested in BEPC energies with high precision.
\begin{figure}
\centering{
\includegraphics[width=5cm,height=4.5cm]{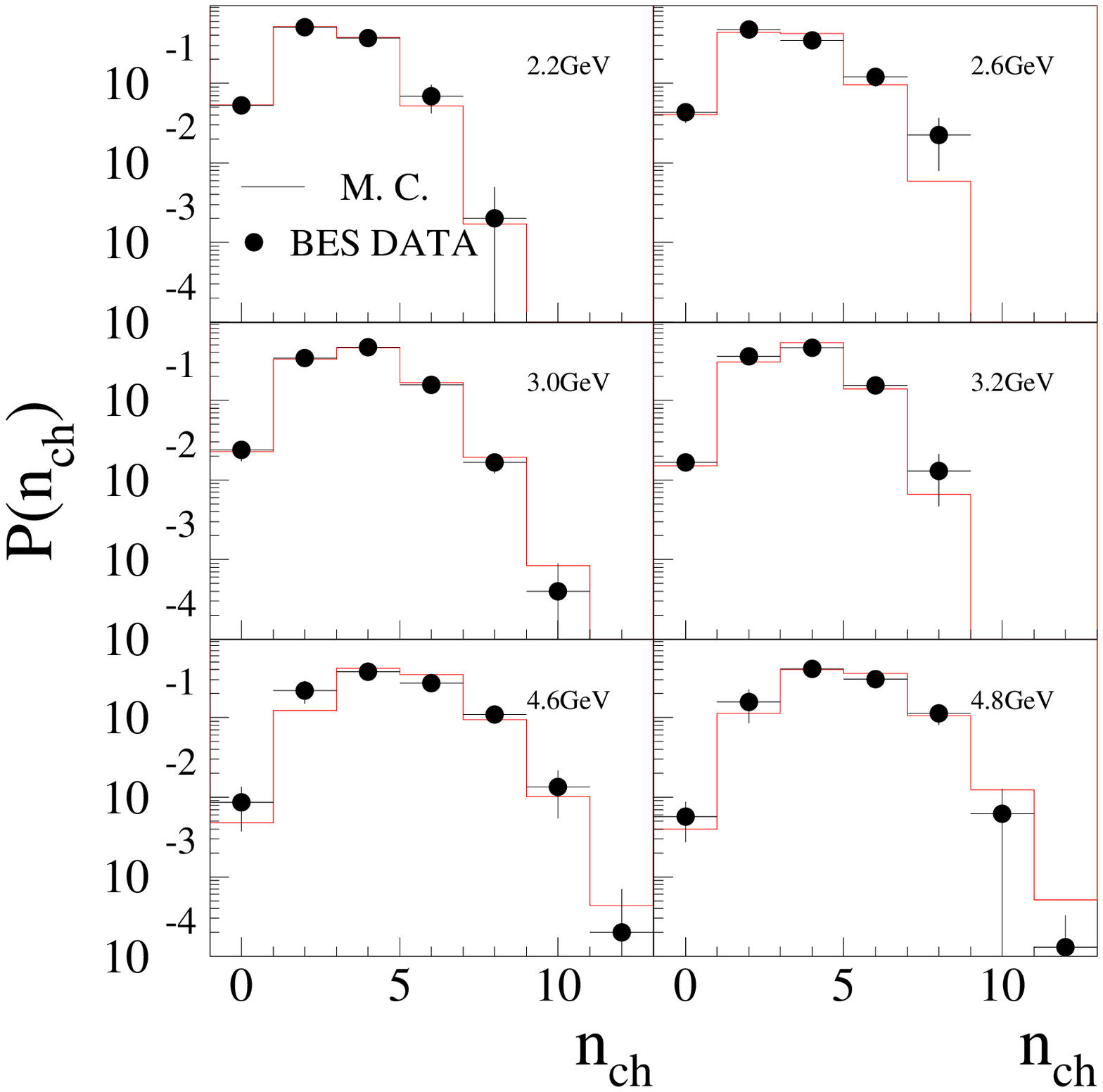}
\includegraphics[width=5cm,height=4.8cm]{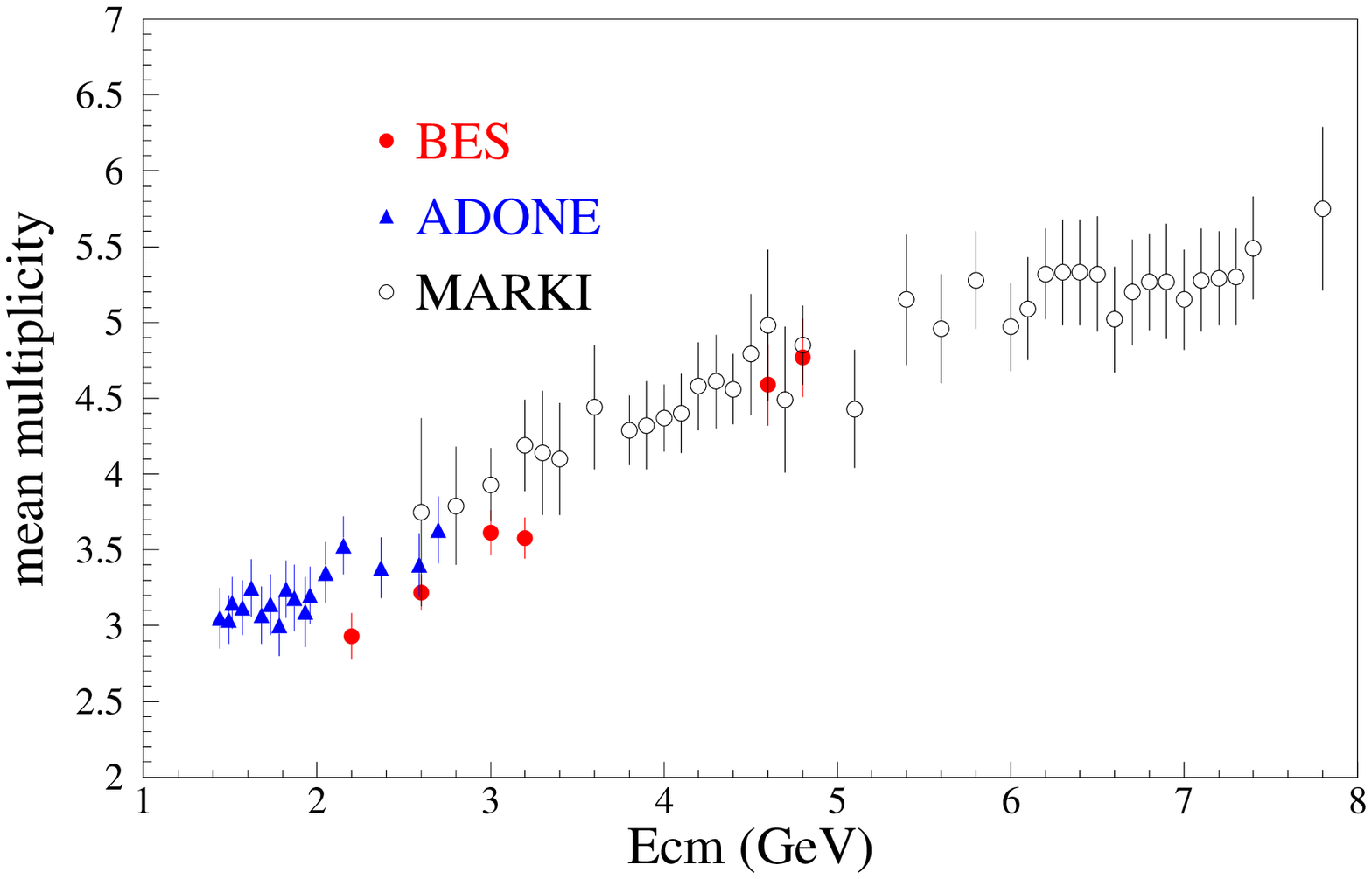}
\includegraphics[width=5cm,height=4.4cm]{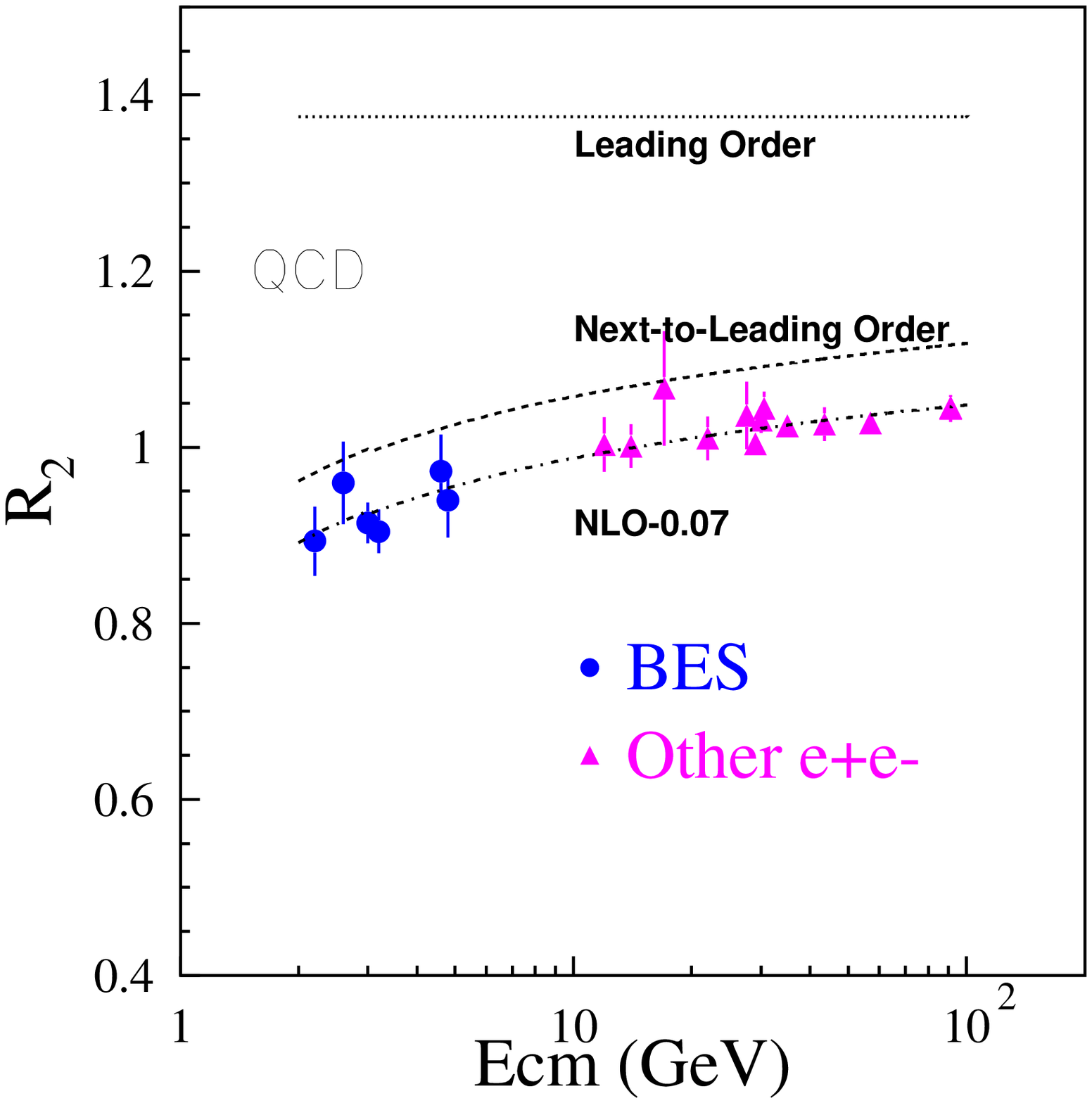}
}\caption{Left: multipicity distributions; middle: the
energy dependence of average multiplicity; right: a comparison of
data and QCD predictions for $R_2$.} \label{multy}
\end{figure}


\section{Kinematical and dynamical correlations}
In inclusive measurements, only the kinematics of one particle are
measured; the kinematics of all the others are averaged. Thus, a great
amount of information is lost in the summation. Measurements of
correlation effects is a more valid way to extract dynamical
information from experimental data and test hadronization models.

The simplest correlation function is the two-particle correlation.
In order to separate pseudo-correlations from the true one,
experiments measure the inclusive correlation
function~\cite{part2_r_e_carruther}
\begin{equation}
C_2(x_1,x_2)=C_S(x_1,x_2)+C_L(x_1,x_2),
\end{equation}
where
\begin{eqnarray}
C_S(x_1,x_2)&=&\sum_{n_c}P(n_c)C_2^{(2)}(x_1,x_2),\nonumber\\
C_L(x_1,x_2)&=&\sum_{n_c}P(n_c)\Delta
\rho^{(n_c)}(x_1)\Delta\rho^{(n_c)}(x_2) .
\end{eqnarray}
Here $x$ can be any kinematical observable. $C_S$ is the
average of the semi-inclusive correlation functions and is more
sensitive to the dynamical correlations. $C_L$ arises from mixing
different topological particle densities.  Related distributions
are defined as
\begin{equation}
\rho_1^{(n_c)}(x)=\frac{1}{\sigma_{n_c}}\frac{d\sigma}{dx},
\end{equation}
\begin{equation}
\rho_2^{(n_c)}(x_1,x_2)=\frac{1}{\sigma_{n_c}}\frac{d^2\sigma}{dx_1dx_2},
\end{equation}
\begin{equation}
C_2^{(n_c)}(x_1,x_2)=\rho_2^{(n_c)}(x_1,x_2)-\rho_1^{(n_c)}(x_1)\rho_1^{(n_c)}(x_2),
\end{equation}
\begin{equation}
\Delta\rho^{(n_c)}(x)=\rho_1^{(n_c)}(x)-\rho_1(x).
\end{equation}
Discussions about distribution
and correlation functions can  be found in
Refs.~\cite{part2_r_e_mueller,part2_r_e_kendall,part2_r_e_carruthers}.

\section{Topological event shapes}
A typical event produced in a collision process has several 
final-state particles  and
kinematical parameters are needed to summarize global
correlations among these particles. 
To describe the complicated topologies encountered in
multihadronic events, a number of event measures 
have been introduced, including the 
event-shape parameters sphericity and thrust.
These quantities are intended to provide a global view of the
properties of a given process, wherein the full information content
of the event is condensed into one or two measured
numbers.   QCD-motivated models
give quantitative predictions for sphericity and thrust
that agree with high energy
measurements; these
have not yet been studied at intermediate energies.

The SPEAR Group introduced a method based on the inertia tensor in
classical dynamics to make the first studies of jet structure in $e^+e^-$
annihilation events~\cite{part2_r_e_spear,part2_r_e_jetset2}.\\
$\bullet$~\underline{Sphericity $S$}\\
Sphericity is a geometrical parameter reflecting the degree
of isotropy in configurations of final-state hadrons. The sphericity tensor 
is defined as~\cite{part2_r_e_bjorken}
\begin{equation}
S^{\alpha\beta}\equiv\frac{\sum_ip_i^{~\alpha}p_i^{~\beta}}{\sum_i
|\vec{p}_i|^{~2}},~~~~\alpha , \beta =1,2,3,
\end{equation}
where $\vec{p}_i$ is the momentum of final state hadron $i$
and the summation runs over all observed hadrons in an event.
The three eigenvalues of $S^{\alpha\beta}$ are ordered as
$\lambda_1\geq\lambda_2\geq\lambda_3$ with $\lambda_1 +\lambda_2
+\lambda_3=1$. ``Sphericity'' is defined as
\begin{equation}
S=\frac{3}{2}(\lambda_1+\lambda_2),
\end{equation}
so that $0 < S < 1$. Sphericity is essentially a measure of the
summed $p_{\bot}$ with respect to the event axis. A severe two-jet event
corresponds to $S\approx 0$ and an isotropic event to $S\approx 1$.
In practice, the special direction $\vec{n}_S$, which minimizes
the value of $S$, is found:
\begin{equation} \label{sph}
S={\rm min}
\frac{3}{2}\frac{\sum_i|\vec{n}_S\cdot\vec{p}_i|^2}{\sum_i\vec{p}_i^2}.
\end{equation}


\noindent$\bullet$~\underline{Trust $T$}\\
Trust is a parameter that reflects the degree of anisotropy of the
three-momenta of final state particles and is defined as
\begin{equation}\label{tru}
T={\rm
max}\frac{\sum_i|\vec{n}_T\cdot\vec{p}_i|}{\sum_i|\vec{p}_i|}.
\end{equation}
where the vector $\vec{n}_T$ defines the ``thrust axis'', 
{\it i.e.} the direction that maximizes the value of $T$.
It is often useful to divide an event into two hemispheres by a
plane through the origin and perpendicular to the trust axis. The
allowed range of $T$ is $0.5 < T < 1$;  a severe two-jet event
corresponds to $T\approx 1$ and an isotropic event to $T\approx
0.5$.  The quantity $\tau = 1-T$ is often used to
replace $T$.

pQCD predicts that average trust values can be expressed
as~\cite{part2_r_e_Ellis}
\begin{equation}
<\tau >=<\tau^{pert}>+<\tau^{pow}>,
\end{equation}
where  $<\tau^{pert}>$ is the value calculated by pQCD, and
$<\tau^{pow}>$ is the contribution from  power corrections. 
$<\tau^{pert}>$ has been calculated based on NLO to be
\begin{equation}
<\tau^{pert}>=A\tilde{\alpha}_s+(B-2A)\tilde{\alpha}_s^2,
\end{equation}
where $\tilde{\alpha}_s=\alpha_s/2\pi$, and the coefficients $A$
and $B$ are obtained from the integral of the NLO matrix 
element to be $A=2.103$ and $B=40.99$~\cite{part2_r_e_Seymour}.

\section{Bose-Einstein correlations (BEC)}
~~~~In quantum mechanics, the wave function of identical bosons
is symmetric under the exchange of any two 
identical bosons. This property  leads
to a special statistical correlation, {\it i.e.} the so-called
Bose-Einstein correlation, which exists in boson system even in
the absence of any interactions. This symmetry leads an 
interference term that
contains information on the space-time extent of 
hadronic sources.

\subsubsection{Correlation function}
~~~~The manifestation of BEC is that the possibility of finding two
identical bosons in a small phase-space volume element is larger than 
that  for two different particles. The space-time properties of hadronic 
source can be inferred from measurements of Bose-Einstein correlation
functions~\cite{part2_r_e_boal}. Suppose that a particle can be emitted 
at the space-time point $x=(\vec{r},t)$ from an extended hadronic 
source with the probability amplitude $f_c(x)$, and the wave function of
the particle is a plane wave $\psi(x)\sim\exp(ipx)$. If the hadronic
source is coherent, the probability for observing a stable and free
particle with 4-momentum $p=(\vec{p},E)$ reads
\begin{equation}
P_c(p)=|\int\psi(x)f_c(x)d^4x|^2.
\end{equation}
The joint probability for observing two identical bosons with
momenta $p_1$ and $p_2$ can be written as
\begin{equation}
P_c(p_1,p_1)=\frac{1}{2}|\int d^4x_1d^4x_2
\psi(x_1,x_2)f_c(x_1)f_c(x_2)|^2.
\end{equation}
The spatial distribution of the hadronic source and the momentum
distribution are related by a Fourier transformation and, so, the spatial
distribution can be derived from the measured momentum distribution.

In experiments, correlation functions with slightly different
definitions are used, such as~\cite{part2_r_e_boal}
\begin{equation}
R_2(p_1,p_2)=\frac{P_c(p_1,p_1)}{P_c(p_1)P_c(p_2)}-1,
\end{equation}
where one can see that $R_2=0$ if there is no 
correlation between the two
particles. Another form that is commonly used is~\cite{part2_r_e_Utyuzh}
\begin{equation}
C_2(Q^2)=\frac{N_2(p_i,p_j)}{N_1(p_i)N(p_j)},
\end{equation}
which is the ratio of the two-identical-boson distribution to the
product of two single boson distributions. Following a suggestion in
Ref.~\cite{part2_r_e_biyajima}, 
Bose-Einstein correlation up to $5^{th}$ order can
be used in experimental studies. The correlation function of 
order $q$:
\begin{equation}
R_q(Q_{q\pi}^2)=\frac{N_q(Q_{q\pi}^2)}{N_q^{BG}(Q_{q\pi}^2)},
\end{equation}
is defined as the ratio of the distribution of like-charged
$q$-tuplets ($q=2,3,4,5$) $N_q(Q_{q\pi}^2)$ and a reference
distributin (background) $q$-tuplets $N_q^{BG}(Q_{q\pi}^2)$ obtained
from  random event-mixing. The variable $Q_{q\pi}^2$ is defined
as the sum over all permutations:
\begin{equation}
Q_{q\pi}^2=Q_{12}^2+Q_{13}^2+Q_{23}^2+\cdots +Q_{(q-1)q}^2,
\end{equation}
of the squared 4-momentum difference $Q_{ij}^2=(p_i-p_j)^2$ of
particle $i$ and $j$.

Gyulassy suggests a simple procedure for incorporating the Coulomb
final-state interaction between two
charged pions~\cite{part2_r_e_gyulassy}:
\begin{equation}
[R(\vec{p}_1,\vec{p}_2)+1]_{theory+Coulomb}= W(\vec{p}_1,\vec{p}_2)
[R(\vec{p}_1,\vec{p}_2)+1]_{theory},
\end{equation}
where the Gamov factor $W(\vec{p}_1,\vec{p}_2)$ is the square modulus of
the relativistic Coulomb wave function at the 
origin~\cite{part2_r_e_gamov}
\begin{equation}
W(\vec{p}_1,\vec{p}_2)=\frac{2\pi\eta}{\exp(2\pi\eta)-1} ,
\end{equation}
with $\eta=\alpha m_{\pi}/|\vec{p}_1-\vec{p}_2|$. The Gamov factor
suppresses the correlation function at small relative
momentum.

\subsubsection{Parametrization of the source distribution}
~~~~Measurements of the BEC have two aims: the study the quantum
effects of the particles and the space-time distribution of 
hadronic sources. For the latter, some models for the 
sources have been suggested~\cite{part2_r_e_boal}.

\noindent$\bullet$~\underline{Gaussian source}:\\
The most extensively used parametrization is a static hadronic
source with a Gaussian form
\begin{equation}
\rho (\vec{r})=\frac{1}{\sqrt{a_x^2a_y^2a_z^2\pi^3R^6}}
\exp\{-[(x/a_x)^2+(y/a_y)^2+(z/a_z)^2]/R^2\},
\end{equation}
where $(a_x,a_y,a_z)$ are dimensionless constants that allow
for a nonspherical source, and $R$ represents the scale of the source.

\noindent$\bullet$~\underline{Source with finite lifetime}:\\
If one considers a source with a finite lifetime $\tau$ and space
dimension $r_0$, its distribution can be written
as~\cite{part2_r_e_koonin,part2_r_e_yano}
\begin{equation}
\rho(\vec{r},t)=\frac{1}{\pi^2r_0^3\tau}\exp(-r^2/r_0^2-t^2/\tau^2).
\end{equation}

The free parameters in the models can be determined by fitting the
data, and the possible space-time distributions of the hadronic
sources can be deduced. It is expected that the following subjects
may be done for charged bosons: (1) two-body correlations: the
inflections of multi-body correlation and the final-state
electromagnetic and strong interactions;  (2) the multiplicity
dependence of BE correlations; (3) the space-time form of hadronic
sources; and (4) BE correlations in resonance decays. This 
latter measurement requires
large hadronic samples and excellent particle
identification.

\section{Possible fractal structure of final state
phase-space}
~~~~There has been a flaw in the study of the spectrum of
the final state particles, {\it i.e.} one usually pays attention to
averaged distributions only and attributes fluctuations
to the statistical effects associated with
having a finite number of particles. 
Events with abnormally high particle density condensed
in small phase-space volumes
have been observed in several types of high
energy 
reactions~\cite{part2_r_e_burnett,part2_r_e_adamus,part2_r_e_alner,part2_r_e_adamovich,
part2_r_e_braunschweig,part2_r_e_behrend,part2_r_e_abreu,part2_r_e_akrawy,part2_r_e_decamp}.

Bialas and Peschanski~\cite{part2_r_e_bialas,part2_r_e_peschanski} 
suggested that spikes observed in experimental distributions 
could be the manifestation of ``intermittency'' in hadron physics.
They argue that if intermittency occurs in
particle production, large fluctuations are not only expected, they
should also exhibit self-similarity with respect to the size of the
phase-space volume. Various
efforts~\cite{part2_r_e_carruthers89,part2_r_e_wuliu}
have been undertaken to understand the physics behind these experimental
findings.

The important questions to address are: do the anomalous
fluctuations have intrinsic dynamical origins? is the
phase-space of the final state isotropic or not? is the
phase-space continuous or fractal? does the approximate
intermittency observed at very high energies also exist at
intermediate energies? can intermittency be explained 
theoretically (such as by cascading , BEC, etc.)?

The study of these phenomena
has two aspects: (1) Experimentally, one
measures fractal moments   and the Hurst index. 
In a one-dimension case, phase-space variables can be chosen 
as the rapidity, or the transverse momentum, or azimuthal angle. 
Combinations of any two of these can be used for two-dimensional
analyses. (2) A theoretical qestion is whether or not the 
asymptotic fractal behavior in the perturbative evolution 
of partons persists after hadronization processes?

\chapter[Lineshapes of charmonium resonances]{Lineshapes of charmonium
resonances\footnote{By Gang Rong and Da-Hua Zhang}}
\label{sec:resonance_parameters}

\section{Resonance production and $1^{--}$ charmonium states}
\label{sec:gang_para_1}

In $e^+e^-$ annihilation, 
$1^{--}$ resonances of the quarkonium family 
can be formed directly in the $s$-channel. Each of
these can have a number of different decay modes. 
The total width of a resonance $\Gamma_{\rm tot}$ is 
the sum of its partial widths for each decay mode, {\it i.e.} 
$\Gamma_{\rm tot}=\sum_i \Gamma_i$. 
For a resonance $R$  decaying to a final state 
$F$, the cross section for the $R\to F$ production is given by
the relativistic Breit-Wigner formula~\cite{part2:gang:bwformula}
$$\sigma_{(e^+e^- \rightarrow R \to F)}(E) = 
  \frac {12\pi \Gamma_{ee}\Gamma_{F} } 
        {(E^2-M_R^2)^2+\Gamma_{\rm tot}^2 M_{R}^2 }, $$
\noindent
where $\Gamma_{ee}$ is the leptonic width, $\Gamma_{F}$ is the partial
width for the decay $R \rightarrow F$, $M_R$ is the mass of the
resonance $R$, and $E$ is the center-of-mass energy. 
The corresponding non-relativistic Breit-Wigner formula is given by
$$\sigma_{(e^+e^- \rightarrow R \to F)}(E) = 
\frac{3\pi}{M_R^2}
               \frac
{\Gamma_{ee}\Gamma_{F}}{(E-M_R)^2+\Gamma_{\rm tot}^2/4}. $$
\noindent
The $1^{--}$ quarkonium resonances 
are generally described by the
parameters $M_R$, $\Gamma_{\rm tot}$, 
the hadronic width $\Gamma_{\rm h}$ 
for the decay $R\rightarrow {\rm hadrons}$, and 
their leptonic width $\Gamma_{ee}$. 

There are six established $1^{--}$ charmonium resonances,
the $J/\psi$, $\psi(2S)$, $\psi(3770)$, $\psi(4040)$,
$\psi(4160)$ and $\psi(4415)$. These are generally 
classified as the $1^3S_1$,  $2^3S_1$,
$1^3D_1$-dominant , $3^3S_1$, 
$4^3S_1$-dominant, and $5^3S_1$ 
$c\bar c$ charmonium states~\cite{part2:gang:y_b_ding_prd51_5064}, 
respectively. They are directly produced in  
$e^+e^-$ annihilation in the energy region between 3.0 and 4.5~GeV.
Using the well measured resonance parameters of 
the $J/\psi$ and $\psi(2S)$ as input,
charmonium models predict the masses, total widths 
and leptonic widths of these other $1^{--}$ charmonium states.

In addition to the above-listed charmonium resonances, 
a $1^{--}$ resonance called the $Y(4260)$ was
discovered by the BaBar experiment in the 
$e^+e^-$ radiative-return production of
$\pi^+\pi^- J/\psi$ final states.
In addition to uncertainties in the interpretation of the
$Y(4260)$,  there are on-going debates about the 
classifications of some of the of the other $1^{--}$ states.
For example, 
Close and Page~\cite{part2:gang:F_Close} suggest that the
$\psi(4040)$ and $\psi(4160)$ are strong mixtures of 
a ground-state hybrid-charmonium state 
with mass $M\sim4.1$ GeV 
and the conventional $\psi(3S)$ charmonium
meson. 

Theoretical classifications of the charmonium states 
are based on 
the measured resonance parameters and their decay products.
To help clarify the situation, more  precise 
measurements of these
resonance parameters, and studies their decay final 
states are needed.

More precisely measured resonance parameters 
of the $\psi(3770)$ are needed  for
a better understanding of the  nature of  this state too.
At present, the $\psi(3770)$ is considered to be a mixture of 
the $1^3D_1$ and $2^3S_1$ charmonium states, an interpretation 
that is based on the measured leptonic width. 
However, there are still open
questions about $\psi(3770)$ production and decay. 
According to QCD-based models,
more than $97\%$ of the $\psi(3770)$ decays should proceed into 
$D\bar D~(D^0\bar D^0~{\rm and}~D^+D^-) $ final states.
However, the BES Collaboration recently measured the 
branching fraction for 
$\psi(3770) \rightarrow {\rm non-}D\bar D$ to be
$B[\psi(3770) \rightarrow {\rm non-}D\bar D]
=(16.4\pm 7.3 \pm 4.2)\%$~\cite{part2:gang:prl97_121801_y2006} and
$B[\psi(3770) \rightarrow {\rm non-}D\bar D]
=(14.5\pm 1.7 \pm 5.8)\%$~\cite{part2:gang:plb641_y2006_145}.
These two independent results
were obtained from two different data samples 
with different analysis methods.
Both of the measurements indicate a branching fraction that 
is larger than  the $3\%$ QCD  expectation.
If the branching fraction is really 
significantly larger than $3\%$,
there may be some new process involved.
To resolve this puzzle, more accurate 
measurements of the $\psi(3770)$ resonance parameters
are needed.  This will require
more careful studies of the line-shapes for the
formation and for subsequent inclusive hadonic decays,
inclusive $J/\psi$ decays, and other specific exclusive 
decays.

In addition  to the conventional charmonium states,
there may be other $1^{--}$ states
in the energy region between 3.0 and 4.5 GeV,
such as hybrid charmonia, glueballs, and/or 
4-quark states.
To understand better the nature of the 
charmonium states and to
search for other types of $1^{--}$ 
mesons, more precise measurements 
of the resonance parameters and careful analysis 
of the line-shapes of the
resonances for different decay 
modes are essential. 

\section{Key points for precision measurements}
\label{sec:gang_para_2}

Parameters of $1^{--}$ resonances can  be directly determined  
from the energy dependence of the cross sections for 
production via $e^+e^-$ annihilation. These parameters can also be 
measured in othe production experiments such as $p \bar p$ 
annihilation~\cite{part2:gang:e760}.

The MARK-I~\cite{part2:gang:mark1_jpsi, part2:gang:mark1_psipp}, 
FRAG~\cite{part2:gang:frag_jpsi}, FRAM~\cite{part2:gang:fram_jpsi},
DASP~\cite{part2:gang:dasp_jpsi}, MARK-II~\cite{part2:gang:mark2_psipp}, 
DELCO~\cite{part2:gang:delco_psipp}, BES~\cite{part2:gang:bes_jpsi} 
and BES-II~\cite{part2:gang:bes2_psip,part2:gang:prl97_121801_y2006}
experiments previously
measured these parameters using $e^+e^-$ annihilation;
 E760~\cite{part2:gang:e760} measured the masses and 
the total widths of the $J/\psi$ and $\psi(2S)$ 
in $p \bar p$ annihilation by analyzing cross section scan
data.

An examination of the analyses of the cross section scan data from 
previous experiments shows some potential problems with the way
the resonance parameters have been extracted.
First, other than BES-II measurements of the resonance
parameters of the $\psi(2S)$ and 
$\psi(3770)$~\cite{part2:gang:prl97_121801_y2006},
none of the previous  $e^+e^-$ experiments considered
the effects of photon vacuum polarization corrections to continuum
hadron production on the total and leptonic  widths.  
If BES-II had not considered these effects,
their measured
total width of the $\psi(2S)$ resonance would be reduced 
by about 40~keV~\cite{part2:gang:prl97_121801_y2006}, 
while the leptonic width would be larger
by about $4\%$. 
Moreover, most previous experiments assumed
that the detection efficency for the observation of inclusive
hadronic events is an almost linear
function of  the $e^+e^-$ cms energy.
This assumption systematically 
increases the inferred leptonic width of a resonance.
In addition, none of the previous experiments considered the effect 
of possible interference 
between the amplitude for continuum hadron production 
and the resonance amplitude.
This effect can also systematically shift the 
measured resonance parameters by a
significant amount.

To measure the resonance parameters more precisely, one has
to determine accurately the detection efficencies for 
inclusive hadronic events at different cms energies.
In addition, photon vacuum polarization corrections 
for continuum hadron production in the resonance region
have to be applied and the
possible interference between the amplitude 
for continuum hadron production and the amplitude for
the electromagnetic decays of the resonance
has to be considered. Moreover, 
better methods for dealing with inital state radiative (ISR) and 
photon vacuum polarization corrections should be employed. 

To determine detection efficencies 
precisely at different cms energies, 
Monte Carlo generators that include both ISR and photon vacuum 
polarization corrections have to be used.  The generator 
should not only include 
charmonium production and decays, but also
all relevant sub-processes. 
It should also allow for the possibility of interference 
between  the  continuum and resonance amplitudes and
simulate the line-shapes of the 
narrow resonances exactly.
 Recently, Zhang~{\it et al.}~\cite{part2:gang:zhang_rong_chen} developed 
a generator with these properties for use in the measurement of the 
$\psi(3770)$ and $\psi(2S)$ resonance parameters and branching 
fractions for $\psi(3770)\rightarrow D^0\bar D^0, D^+D^-$ and 
$\psi(3770)\rightarrow 
{\rm non-}D\bar 
D$~\cite{part2:gang:prl97_121801_y2006,part2:gang:plb641_y2006_145}. 
This generator was developed with the aim of
making precise measurements
of the resonance parameters and search for new $1^{--}$ states with
the \bes3 detector at the BEPC-II collider.

\section{An example for measuring $\psi(2S)$ and $\psi(3770)$ parameters}
\label{sec:gang_para_3}

Among the six resonances 
that are seen via direct production in $e^+e^-$
annihilation between 3.0 and 4.5~GeV, the
$\psi(2S)$ and $\psi(3770)$ are closest 
in mass to each other. To get reliably
measured parameters for these two resonances,  simultaneous
measurements with a common data set are necessary. 
In this section, 
we take the $\psi(2S)$ and $\psi(3770)$ measurements
performed with BES-II as an example to 
illustrate how  the simultaneous determination of the
two resonances' parameter values can be done.
The method for measuring the parameters of the $1^{--}$ resonances
in  more complicated cases is similar.
The data sets used for these measurements were taken 
in March,~2003. The total integrated luminosity of
the data sets is $\sim 5$ pb$^{-1}$ collected over the 
 3.66 to 3.88~GeV energy range. 

\begin{figure}
{\centering
\includegraphics[width=14.0cm,height=7.0cm]
{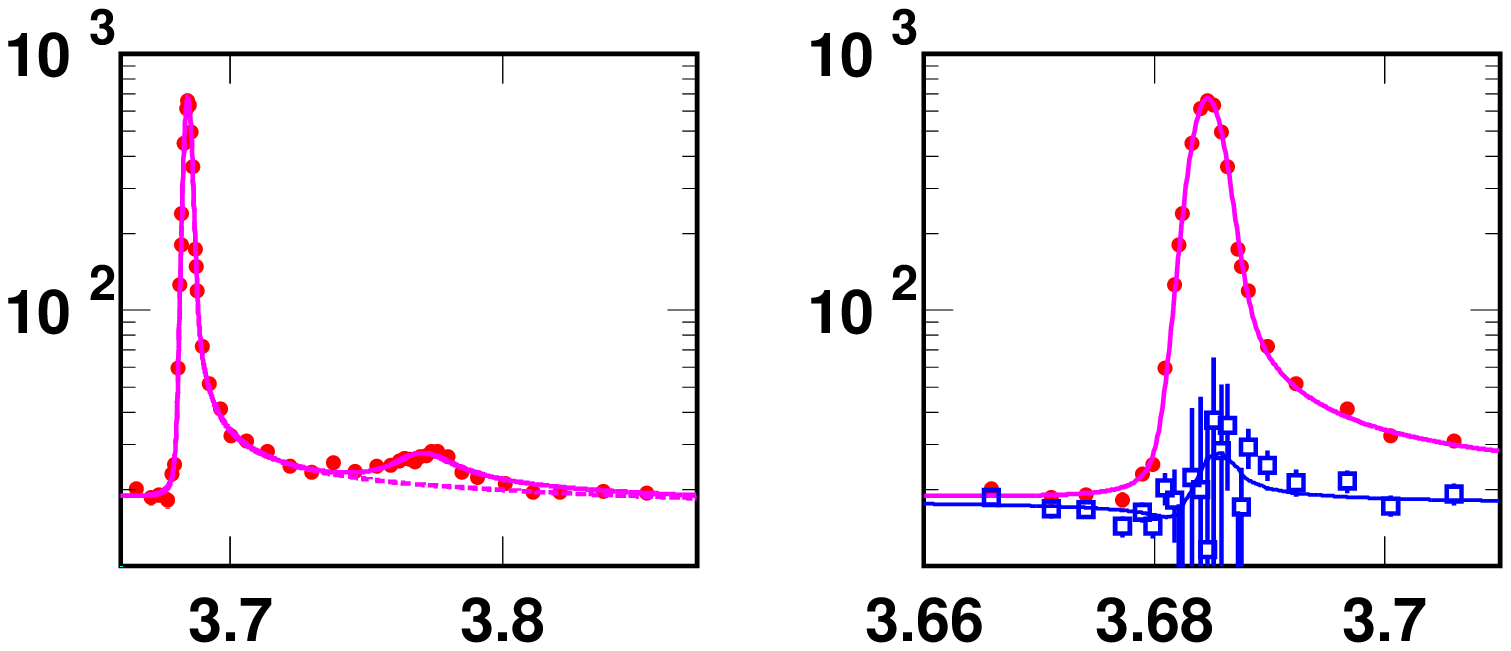}
\put(-300,-10.0){\bf\large $E_{\rm cm}$ [GeV]}
\put(-110,-10.0){\bf\large $E_{\rm cm}$ [GeV]} 
\put(-410,80){\rotatebox{90}{\bf\large $\sigma^{\rm obs}_{\rm had}$~~~[nb]}}
\put(-200,80){\rotatebox{90}{\bf\large $\sigma^{\rm obs}_{\rm had}$~~~[nb]}}
\put(-240,82.0){\large (a)}
\put(-37,82.0){\large (b)} 
\put(-19.0,36.0){1}
\put(-19.0,20.0){2}
\caption{The hadronic cross sections versus
the CMS energy
(see text).}
\label{xsct_had_ddbar1}
}
\end{figure}

\begin{figure}
{\centering
\includegraphics[width=14.0cm,height=8.0cm]
{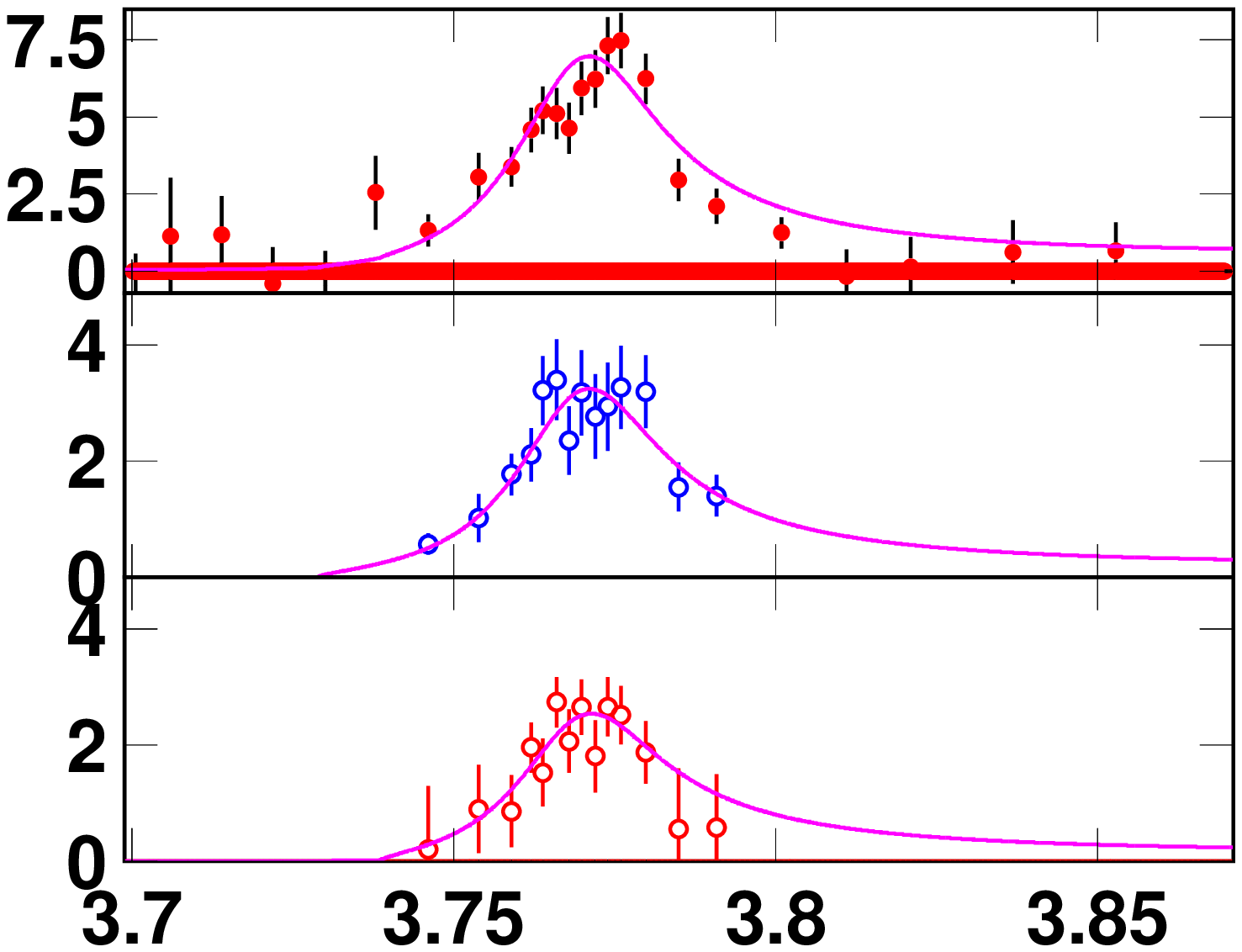}
\put(-200,-12){\bf\large $E_{\rm cm}$~~ [GeV]}
\put(-365,85){\rotatebox{90}{\large $\sigma^{\rm obs}$~~ [nb]} }
\put(-95,190.0){(a)}
\put(-95,120.0){(b)} 
\put(-95,55.0){(c)} 
\put(-160,190.0){hadrons}
\put(-160,120.0){$D^0\bar D^0$}
\put(-160,55.0){$D^+D^-$}
\caption{The observed cross sections versus the cms
energy.
}
\label{xsct_had_ddbar2}
}
\end{figure}

The $\psi(3770)$ and $\psi(3686)$ resonance parameters
can be extracted from a fit to the observed hadronic cross sections or
from a fit to both the observed hadronic cross sections 
and the observed $D \bar D$ ($D^0\bar D^0$ and $D^+D^-$) cross sections.
The observed hadronic cross sections are determined from
\begin{equation}
\sigma^{\rm obs}_{\rm had} =\frac{N^{\rm obs}_{\rm had}}
                {L~\epsilon_{\rm had}~\epsilon_{\rm had}^{\rm trig}
                },
\end{equation}
where $N^{\rm obs}_{\rm had}$ is the number of
the observed hadronic events,
$L$ is the integrated luminosity,
$\epsilon_{\rm had}$ is the efficiency for the
detection of inclusive hadronic events 
and $\epsilon_{\rm had}^{\rm trig}$ is the trigger efficiency
for recording hadronic events in the online data acquisition system.
The observed cross sections for $D^0 {\bar D}^0$ (or $D^+D^-$)
production are determined from
\begin{equation}
 \sigma_{D^0{\bar D}^0({\rm or~}D^+D^-)}^{\rm obs} =
                       \frac {N_{D^0_{\rm tag}} ({\rm or~}N_{D^+_{\rm tag}})
}
                       {2 \times L \times B \times \epsilon },
\end{equation}
where $N_{D^0_{\rm tag}}$ ($N_{D^+_{\rm tag}}$) is
the number of reconstructed $D^0$ ($D^+$) events 
obtained from an analysis of the
$K^{\mp}\pi^{\pm}$ and $K^{\mp}\pi^{\pm}\pi^{\pm}\pi^{\mp}$
(or $K^{\mp}\pi^{\pm}\pi^{\pm}$) invariant mass spectra as
discussed in detail in Ref.~\cite{part2:gang:xsct_ddbar_bes};
$B$ is the branching fraction
for the relevant decay mode,
and $\epsilon$ is its detection and trigger efficiency.
Figure~\ref{xsct_had_ddbar1} shows the observed cross sections
(points with errors) 
for inclusive hadronic event production, while
Figs.~\ref{xsct_had_ddbar2}(b) and~(c), respectively,
display the observed cross sections (circles with errors) 
for $D^0\bar D^0$ and $D^+D^-$ production.

The determination of the resonance parameters
is accomplished by simultaneously fitting the observed cross sections
for $\psi(2S)$, $\psi(3770)$, $D^0\bar D^0$ and $D^+D^-$
to functions 
that describe the combined $\psi(2S)$, $\psi(3770)$ resonance shapes,
the tail of the $J/\psi$ resonance and the 
non-resonant hadronic background,
as well as the partial $\psi(3770)$ resonance shapes
for $\psi(3770) \rightarrow {D^0 \bar D^0}$ and
$\psi(3770) \rightarrow {D^+D^-}$.
Assuming that there are no additional structures and effects,
we use a pure $P$-wave Born-order Breit-Wigner function
with energy-dependent total widths
to describe $\psi(3770)$ 
and $D\bar D$ ($D^0\bar D^0$ and $D^+ D^-$ ) production
via $\psi(3770)$ decay.
The $\psi(3770)$ resonance shape is taken to be
{\begin{small}
\begin{equation}
\sigma^{\rm B}_{\psi(3770)}(s^{'}) =
    \frac{12 \pi \Gamma_{ee}^{\psi(3770)}
            \Gamma_{\rm tot}^{\psi(3770)}(s^{'})}
{{(s^{'}-{M^{\psi(3770)}}^2)^2 + 
[M^{\psi(3770)}\Gamma_{\rm tot}^{\psi(3770)}(s^{'})]^2}},
\end{equation}
\end{small} } 
\noindent
\hspace{-0.4cm}
while the $D\bar D$ resonance shapes are taken as
{\begin{small}
\begin{equation}
\sigma^{\rm B}_{D\bar D}(s^{'}) =
    \frac{12 \pi \Gamma_{ee}^{\psi(3770)}\Gamma_{D\bar D}(s^{'})}
{{(s^{'}-{M^{\psi(3770)}}^2)^2 + [M^{\psi(3770)}
   \Gamma_{\rm tot}^{\psi(3770)}(s^{'})]^2}}, 
\end{equation}
\end{small} } 
\noindent
\hspace{-0.5cm}
where 
$M^{\psi(3770)}$ and $\Gamma_{ee}^{\psi(3770)}$ are
the mass and leptonic width of the $\psi(3770)$ resonance, respectively;
$\Gamma_{D\bar D}$ is the partial width 
of $\psi(3770)$ decay into $D \bar D$;  
$\Gamma_{\rm tot}^{\psi(3770)}(s^{'})$ and 
$\Gamma_{D\bar D}(s^{'})$ have
energy dependence defined as
{\begin{small}
\begin{equation}
 \Gamma_{\rm tot}^{\psi(3770)}(s^{'})=\Gamma_{D^0\bar D^0}(s^{'})+
          \Gamma_{D^+D^-}(s^{'})+\Gamma_{{\rm non}-D\bar D}(s^{'}),
\end{equation}
\end{small} } 
where
\begin{eqnarray}
\Gamma_{D^0\bar D^0}(s^{'}) & = &
\Gamma_0~ \theta_{00}
           \frac{(p^{}_{D^0})^3} {(p^0_{D^0})^3}
            \frac{1+(rp_{D^0}^0)^2} {1+(rp_{D^0})^2}B_{00},
\end{eqnarray}
\begin{eqnarray}
\Gamma_{D^+D^-}(s^{'}) & = &
\Gamma_0~ \theta_{+-}
           \frac{(p^{}_{D^+})^3} {(p^0_{D^+})^3}
            \frac{1+(rp_{D^+}^0)^2} {1+(rp_{D^+})^2}B_{+-},
\end{eqnarray}
and
\begin{equation}
\Gamma_{{\rm non-}D\bar D}(s^{'}) = \Gamma_0~\left[ 1
   - B_{00}
   - B_{+-}\right] .
\end{equation}
Here $p^0_D$ and $p_D$ are
the momenta of the $D$ mesons
produced at the peak of the $\psi(3770)$ and at
the actual cms energy $\sqrt{s^{'}}$, respectively;
$\Gamma_0$ is the total width of
the $\psi(3770)$ at its peak,   
$B_{00}=B(\psi(3770)\rightarrow D^0\bar D^0)$ and
$B_{+-}=B(\psi(3770)\rightarrow D^+D^-)$
are the branching fractions for
$\psi(3770)\rightarrow D^0\bar D^0$ and
$\psi(3770)\rightarrow D^+D^-$, respectively,
$r$ is the interaction radius of the $c\bar c$,
and $\theta_{00}$ and $\theta_{+-}$ are step   
functions that account for the thresholds of $D^0\bar D^0$ and
$D^+D^-$ production, respectively. 
In the fit we take $\Gamma_0$, $B_{00}$, $B_{+-}$
and $r$ as free parameters.

The non-resonant background shape is taken as
{\small
\begin{eqnarray}
\nonumber  \sigma^{nrsnt}_h(s) & =\int_{0}^{\infty}ds'' G(s,s'')
  \int^1_0 dx  \frac{ R_{uds}(s') {\sigma^B_{\mu^+\mu^-} }(s')} 
  {|1-\Pi(s')|^2} F(x,s)  & \\
   & + f_{D\bar D}\left[ (\frac{p_{D^0}}{E_{D^0}})^3\theta_{00}
                  + (\frac{p_{D^+}}{E_{D^+}})^3\theta_{+-} \right]
 \sigma^B_{\mu^+\mu^-}(s), 
\label{non-reson}
\end{eqnarray}
}
\noindent
\hspace{-0.3cm} 
where 
$G(s,s'')$ is a Gaussian function
that describes the beam-energy spread,
$F(x,s)$ is the sampling function~\cite{part2:gang:kuraev},
$1/{|1-\Pi(s(1-x))|}^2$ is the vacuum polarization correction
function~\cite{part2:gang:berends} including the contributions from all 
$1^{--}$ resonances, 
the QED continuum hadron spectrum
as well as the contributions from the lepton pairs ($e^+e^-$,
$\mu^+\mu^-$ and $\tau^+\tau^-$)~\cite{part2:gang:zhang_rong_chen};
$\sigma^B_{\mu^+\mu^-}(s)$ is the Born cross section
for $e^+e^-\rightarrow \mu^+\mu^-$, 
$E_{D^0}$ and $E_{D^+}$ are the energies of $D^0$ and $D^+$ mesons
produced at the nominal energy $\sqrt{s}$, 
$f_{D\bar D}$ is a parameter to be fitted, 
and $R_{uds}(s')$ is the $R$ value for light hadron production via
direct one-photon $e^+e^-$ annihilation. 

In the fit we leave $R_{uds}(s')$ as a   
free parameter, assuming that its value is independent of the energy, 
and fix the $J/\psi$ resonance parameters
at their PDG values~\cite{part2:gang:pdg04}.
Figure~\ref{xsct_had_ddbar1}(a)
shows the observed cross sections and the results of the fit,
where the solid line shows the fit to the data and
the dashed line represents the contributions 
from the $J/\psi$, $\psi(2S)$ and continuum hadron production.
To examine directly the contribution from the vacuum
polarization corrections to the Born hadronic cross section due to one
photon annihilation, we subtract the contributions of
the  $\psi(2S)$
and $\psi(3770)$ as well as the $J/\psi$ from
the observed cross sections to get the expected cross sections
for continuum hadron production 
corrected for radiative effects
given by Eq.\eref{non-reson}.
The squares with errors in Fig.~\ref{xsct_had_ddbar1}(b) show the
extracted cross sections, 
where the errors are the original absolute errors
of the total observed cross sections 
shown in Fig.~\ref{xsct_had_ddbar1}(a).
The blue curve (lower curve)
in Fig.~\ref{xsct_had_ddbar1}(b)
shows the fit to the expected cross sections for continuum hadron
production corrected for the radiative effects 
as given in Eq.~\eref{non-reson}.
Figure~\ref{xsct_had_ddbar2}(a) shows the
observed cross sections for 
inclusive hadronic event production, where the contributions from the 
$J/\psi$ and $\psi(2S)$ radiative tails as well as  continuum hadron 
production  are removed.
Figures~\ref{xsct_had_ddbar2}(b) and~(c)
display the observed cross sections for $D^0\bar D^0$ and $D^+D^-$
production together with the fits to the data, respectively.

The results from the fit are summarized in Table~\ref{tbl_psipp_prmt},
where the first error is statistical and second systematic.
From the fit we obtain  $R_{uds}$  
in the region between 3.660 and 3.872 GeV to be
          $$R_{uds} = 2.262 \pm 0.054 \pm 0.109,$$
where the errors are, respectively, statistical and the systematic.
The fit yields a non-$D\bar D$ branching fraction for 
the $\psi(3770)$ of
$$B(\psi(3770) \rightarrow {\rm non}-D\bar D)= (16.4 \pm 7.3 \pm 4.2)\%.$$ 
The fit has $\chi^2/{\rm d.o.f} = 65.4/64 = 1.02$.
\begin{table}
\centering   
\caption{    
The measured $\psi(3770)$ and $\psi(2S)$ parameters,
where M is the mass, 
$\Gamma_{\rm tot}$ the total width [$\Gamma_{\rm tot}=\Gamma_0$ for
$\psi(3770)$], $\Gamma_{ee}$ the partial leptonic width 
and $\Delta {\rm M}$ the measured mass
difference of the $\psi(3770)$ and the $\psi(2S)$.
} \label{tbl_psipp_prmt}
\vspace{0.2cm}
\begin{tabular}{ccccc} \hline \hline
Res. & M~(MeV)  &  $\Gamma_{\rm tot}$~(MeV)  & $\Gamma_{ee}$~(eV) &
$\Delta {\rm M}$ (MeV) \\ \hline
$\psi(3770)$  & $3772.2 \pm 0.7 \pm 0.3$  & $26.9\pm 2.4 \pm 0.3$ &
$251 \pm 26 \pm 11$ &  
          \\
$\psi(2S)$  & $3685.5 \pm 0.0 \pm 0.3$  & $0.331\pm 0.058 \pm 0.002$ &
$2330 \pm 36 \pm 110$ &
$86.7 \pm 0.7$  \\
\hline \hline
\end{tabular}
\end{table}  

The continuum background shape effects the measured
total and leptonic widths of the resonances
from the line-shape analysis.
If we use 
{\small    
\begin{eqnarray}
\sigma^{nrsnt}_h(s) & =
\nonumber h~\sigma^B_{\mu^+\mu^-}(s) + f_{D\bar D}\times \hspace{30mm} & \\
\hspace{0mm}
 & \left[(\frac{p_{D^0}}{E_{D^0}})^3\theta_{00}
                 +(\frac{p_{D^+}}{E_{D^+}})^3\theta_{+-} \right]
 \sigma^B_{\mu^+\mu^-}(s),
\label{non-reson-1}
\end{eqnarray}
}
\noindent
\hspace{-1.0mm}in the fit to the data (where $h$ is
a free parameter), we obtain   
$\Gamma^{\rm tot}_{\psi(2S)}=290 \pm 59 \pm 5$ keV,
$\Gamma^{ee}_{\psi(2S)}=2.378 \pm 0.036 \pm 0.103$ keV,
$\Gamma^{\rm tot}_{\psi(3770)}=27.3 \pm 2.5 \pm 1.1$ MeV and
$\Gamma^{ee}_{\psi(3770)}=256 \pm 27 \pm 13$ eV,
with almost unchanged measurements of the resonance masses.
This fit has $\chi^2/{\rm d.o.f} = 75.3/64 = 1.18$.
This indicates that the vacuum polarization corrections
to the Born order cross sections for the continuum hadron production
cannot be ignored in
precision measurements of the resonance parameters of the narrow
resonances like $J/\psi$ and $\psi(2S)$ (and also the $\Upsilon(1S)$
etc.) in $e^+e^-$ cross section scan experiments. 
Ignoring the effects of the vacuum polarization corrections
on continuum hadron production 
in the analysis of the cross section scan data    
taken in the $\psi(2S)$ resonance region would decrease the $\psi(2S)$ total
width by about 40 keV.

\section{Resonance measurements at \bes3}
\label{sec:gang_para_4}

As mentioned above, precise measurements of the $1^{--}$
charmonium resonance parameters are important 
for the understanding of the
dynamics of  charmonium resonance production in $e^+e^-$ annihilation. 
Before clarifying the situation of the understanding the exact natures of
$\psi(4040)$ and $\psi(4160)$, one has to measure precisely and accurately
their resonance parameters. 
At present, one still does not understand why the $\psi(3770)$ decays to
non-$D\bar D$ with a such large branching fraction. To 
resolve this ``puzzle,"
one has to measure the parameters of the 
$\psi(3770)$  and $\psi(2S)$ more precisely. 
In this section, we discuss how well
the branching fractions and parameters of the 
$\psi(3770)$ can be measured at \bes3.

According to our experience with the
measurement the $\psi(3770)$ and $\psi(2S)$ 
parameters,  the branching fraction for
$\psi(3770)\rightarrow {\rm non}-D\bar
D$, and the $R$ value for the range from 3.65 to 
3.88~GeV~\cite{part2:gang:prl97_121801_y2006,part2:gang:hep_ex_0612056,
part2:gang:plb641_y2006_145,part2:gang:prl97_262001_y2006}, 
we expect that the total systematic 
uncertainty on the observed cross section for inclusive
hadronic event production can be reduced to the level of $\sim 2.5\%$ 
at \bes3.  In light of this, one should consider 
collecting  cross section scan
data with a statistical precision of $\sim 1.5\%$.

\begin{figure}
{\centering   
\includegraphics[width=14.5cm,height=11.5cm]
{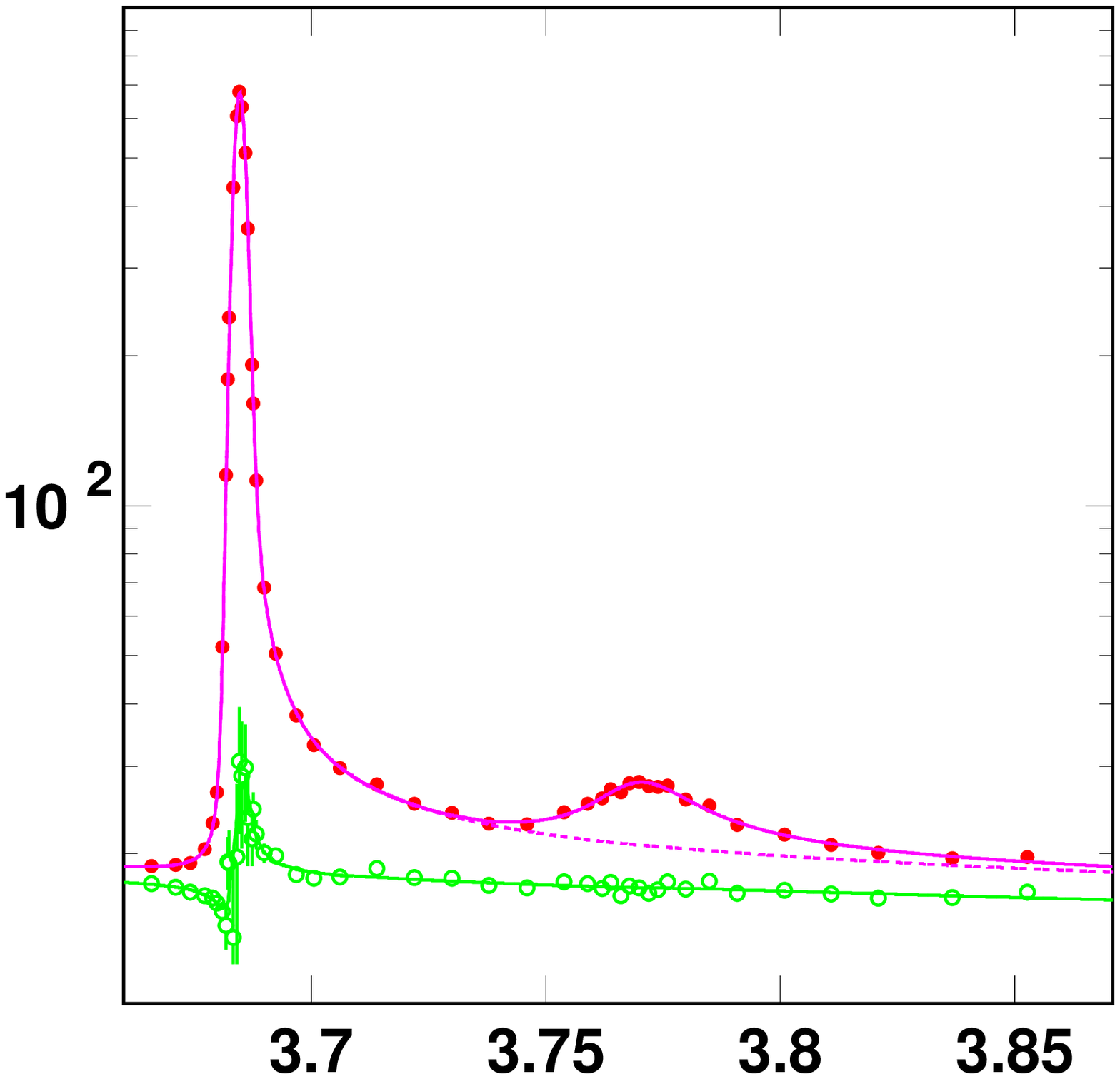}
\put(-220,1){\bf\large $E_{\rm cm}$~~ [GeV]}
\put(-399,120){\rotatebox{90}{\large $\sigma^{\rm obs}$~~ [nb]} }
\caption{The observed cross sections versus the cms
energy.
}
\label{xsct_had_psip_psipp_mc}
}
\end{figure}

Assuming that 
the $\psi(2S)$ and $\psi(3770)$ are the only two
resonances in the 3.65 to 3.88~GeV energy range, 
we did a Monte Carlo study to determine how well we can measure their
resonance parameters and the branching
fraction for $\psi(3770) \rightarrow {\rm non}-D\bar D$. 
The events were generated at the same 49 energy points where we
collected cross section scan data with 
the BES-II detector in March, 2003.
The total integrated luminosity is about 60 pb$^{-1}$.
Following the same procedure that was used to deal with 
the cross section scan data discussed above, we obtain  
$\psi(2S)$ and $\psi(3770)$ resonance
oparameters and the branching fraction for
$\psi(3770) \rightarrow {\rm non}-D\bar D$.
Figure~\ref{xsct_had_psip_psipp_mc} shows the observed cross section
as a function of cms energy, where the dots with errors
show the cross section for inclusive hadronic event production; the 
circles with errors represent the cross section for continuum hadron 
production with ISR and photon vacuum polarization corrections applied; 
the red (green) solid line gives the best fit to the observed 
cross sections for
inclusive hadronic event production (continuum hadronic event
production); while the red dashed line shows the total contributions from
continuum hadronic event production, 
$\psi(2S)$ production and $J/\psi$ production. 
Tables~\ref{tbl_psip_prmt_mc} and~\ref{tbl_psipp_prmt_mc}
summarize, respectively, the results of the
simulated measurements of 
the $\psi(2S)$ and the $\psi(3770)$ resonance parameters, 
where the errors are statistical
and systematic, respectively. As a comparison, we also list the values of
the parameters input to the Monte Carlo simulation. 
The measured $R_{\rm uds}$ value (``measured") along with the input 
$R_{\rm uds}$ value (``input") are listed in Table~\ref{tbl_r_mc}, where 
the
errors are statistical and systematic errors, respectively. 

\begin{figure}
{\centering   
\includegraphics[width=14.5cm,height=8.5cm]
{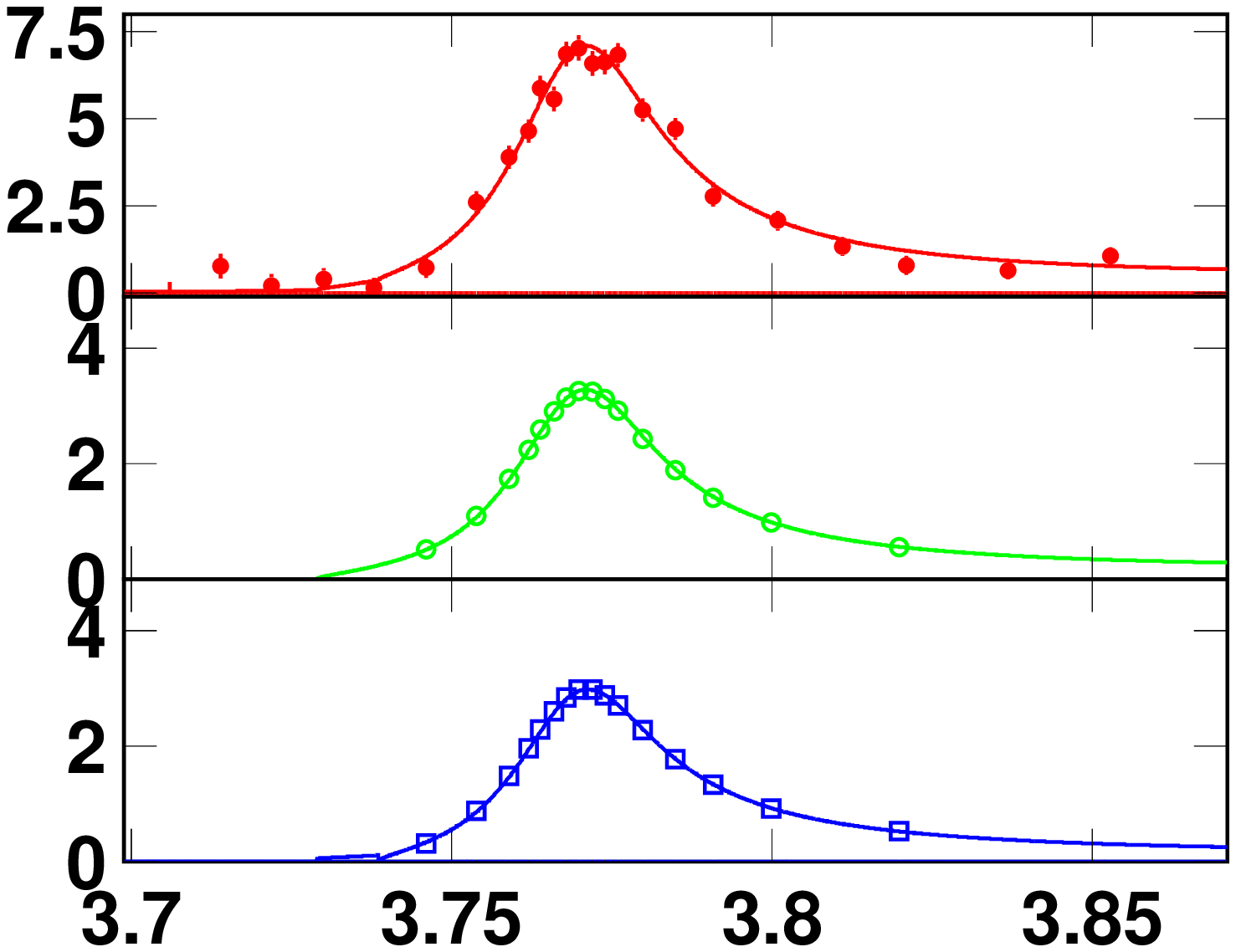}
\put(-220,-12){\bf\large $E_{\rm cm}$~~ [GeV]}
\put(-381,80){\rotatebox{90}{\large $\sigma^{\rm obs}$~~ [nb]} }
\put(-145,200.0){(a)}
\put(-145,130.0){(b)} 
\put(-145,65.0){(c)} 
\put(-120,200.0){hadrons}
\put(-120,130.0){$D^0\bar D^0$}
\put(-120,65.0){$D^+D^-$}
\caption{The observed cross sections versus the cms
energy.
}
\label{xsct_had_ddbar2_mc}
}
\end{figure}
\begin{table}
\centering
\caption{
The input and measured $\psi(2S)$ parameters,
where M is the mass,
$\Gamma^{\rm tot}$ the total width
and $\Gamma^{ee}$ the partial leptonic width, where ``input" means the
value of the parameter used in the Monte Carlo simulation and
``measured" means its measured values.
}
\label{tbl_psip_prmt_mc}
\vspace{0.2cm}
\begin{tabular}{ccccc} \hline \hline
input or & M~(MeV)  &  $\Gamma_{\rm tot}$~(keV)  & $\Gamma_{ee}$~(keV)
 & non-rsnc \\
measured            &                      &       &       & BCK shape \\ \hline
input  & $3686.09$  & $337$ & $2.33$ & --  \\
measured  & $3686.08 \pm 0.01 \pm 0.1$  &
$349.6\pm 13.0 \pm 2$  & $2.33 \pm 0.01 \pm 0.05$
 & Eq.~\eref{non-reson} \\
measured  & $3686.06 \pm 0.01 \pm 0.1$  &
$289.1\pm 11.0 \pm 2$  & $2.42 \pm 0.01 \pm 0.05$
 & Eq.~\eref{non-reson-1} \\
 \hline \hline

\end{tabular}
\end{table}
\begin{table}
\centering   
\caption{    
The input and measured $\psi(3770)$ parameters,
where M is the mass, 
$\Gamma^{\rm tot}$ the total width [$\Gamma^{\rm tot}=\Gamma_0$ for
$\psi(3770)$] and $\Gamma^{ee}$ the partial leptonic width, where
``input" means the 
value of the parameter used in the Monte Carlo simulation and
``measured" means its measured value.
} 
\label{tbl_psipp_prmt_mc}
\vspace{0.2cm}
\begin{tabular}{cccc} \hline \hline
input/measured & M~(MeV)  &  $\Gamma^{\rm tot}$~(MeV)  & $\Gamma^{ee}$~(eV) \\ \hline
input  & $3772.3$  & $26.9$ & $251$  \\
measured  & $3772.1 \pm 0.2 \pm 0.1$  & 
$26.8\pm 0.5 \pm 0.1$ &       
$255 \pm 7 \pm 6$ \\
 \hline \hline             
\end{tabular}
\end{table}  
\begin{table}
\centering
\caption{
The input and measured $R_{\rm uds}$ values in the resonance region from
3.65 to 3.88~GeV.
}
\label{tbl_r_mc}
\vspace{0.2cm}
\begin{tabular}{cc} \hline \hline
$R_{\rm uds}$  [measured] & $R_{\rm uds}$ [input] \\ \hline
$2.254\pm 0.012 \pm 0.056$ & $2.260$ \\
\hline \hline
\end{tabular}
\end{table}
\begin{table}
\centering
\caption{
The input and measured branching fraction for 
$\psi(3770)\rightarrow {\rm non}-D\bar D$,
where "input" means the
value of the parameter used in the Monte Carlo simulation and
"measured" means its measured value.
}
\label{tbl_bf_psipp_mc}
\vspace{0.2cm}
\begin{tabular}{ccc} \hline \hline
input/measured & $B(\psi(3770)\rightarrow D\bar D)$ [$\%$] &
$B(\psi(3770)\rightarrow {\rm non}-D\bar D)$ [$\%$] \\ \hline
input  & $90\%$ & $10\%$ \\
measured  & $88.8 \pm 2.4 \pm 2.0$ & $11.2\pm 2.4 \pm 2.0$ \\
 \hline \hline
\end{tabular}
\end{table}
%


In Table~\ref{tbl_psip_prmt_mc}, the ``non-rsnc BCK shape" 
entry indicates which
non-resonant background shape, as given in 
Eqs.~\eref{non-reson} and~\eref{non-reson-1},
is used in the data reduction.
From the Monte Carlo simulation, we find that
the total width of the $\psi(2S)$ resonance
is reduced by $\sim 20\%$ and 
the partial leptonic width is increased by $\sim 4\%$
if we do not consider the effect of the photon vacuum polarization
corrections on continuum hadron production in the data reduction.

Figures~\ref{xsct_had_ddbar2_mc}(b) and~(c)
show the observed cross sections for $D^0\bar D^0$ and $D^+D^-$
production with the best fits to the data, respectively. The observed cross
sections for the decays $\psi(3770)\rightarrow {\rm hadrons}$ is displayed
in Fig.~\ref{xsct_had_ddbar2_mc}(a). The fit yields the branching fractions
for $\psi(3770)\rightarrow D \bar D$ and 
$\psi(3770)\rightarrow {\rm non}-D \bar D$  summarized in
Table~\ref{tbl_bf_psipp_mc} along with the branching fractions used in the
Monte Carlo simulation. The errors listed in the table are the statistical
and systematic, respectively.
The fit also gives the measured branching fractions for 
$\psi(3770) \rightarrow D^0 \bar D^0$ and 
$\psi(3770) \rightarrow D^+ D^-$ to be
$$B[\psi(3770) \rightarrow D^0 \bar D^0]= (46.3\pm 1.3 \pm 1.0)\%, $$
and
$$B[\psi(3770) \rightarrow D^+ D^-]= (42.5\pm 1.2 \pm 0.9)\%; $$
the corresponding  MC input values are
$B[\psi(3770) \rightarrow D^0 \bar D^0]= 46.8\%$ and
$B[\psi(3770) \rightarrow D^+ D^-]= 43.2\%$,  respectively.

From analyzing the Monte Carlo sample generated at the 49 energy points, we
notice that the line-shape of the observed cross sections are smooth.
However, if there are states other than the $\psi(2S)$ and
$\psi(3770)$ in this energy region, the observed cross sections would
deviate from these expectations (as shown by the lines in
Fig.~\ref{xsct_had_psip_psipp_mc} and
Fig.~\ref{xsct_had_ddbar2_mc}). This provides a method to search for 
new $1^{--}$ states in the energy region from 3.0 to 4.5 GeV.

For measurements of the parameters of the
other resonances in the energy region
from 3.87 to 4.50~GeV, the method used for
the measurements of the $\psi(2S)$ and $\psi(3770)$ parameters
can be used.  However, since the $\psi(4040)$,
$\psi(4160)$ and $Y(4260)$ resonances overlap each other,
the possibility of interference between
them has to be included.\footnote{At present, the 
leptonic width of
the $Y(4260)$ resonance is not measured.}
To get reliable  resonance parameters, 
point-to-point detection efficiencies should be used
to extract the observed cross sections, not
the almost straight line detection efficiency curves
used in most of the previous $e^+e^-$ experiments. 
An improved formalism for dealing with the ISR correction 
should be used.  In this way 
precise and accurate measurements of
the resonance parameters can be made.  These considerations
are also important for $J/\psi$ resonance parameter determinations.
 
\section{Summary}  
\label{sec:gang_para_5}

For a better understanding of
the dynamics of charmonium resonance production in
$e^+e^-$ annihilation, precise measurements of the parameters of the
charmonium resonances in the energy region from 3.0 to 4.5~GeV
are needed. 
At present,  most available measurements of the parameters 
of the $1^{--}$ charmonium resonances may be biased 
since most previous $e^+e^-$ experiments
did not consider the effects of the photon vacuum polarization corrections
on continuum hadron production, did not consider the
possible interference between the continuum hadron amplitude and the
resonance amplitude, and used almost linear
detection efficiency curves to extract the observed cross sections.
For these reasons, the parameters of the resonances should be
remeasured with the \bes3 detector.
Such measurements could provide useful input into the debate 
about the interpretation
of the $\psi(4040)$ and $\psi(4160)$ states. 

Precision measurements of the
parameters of the $\psi(2S)$ and $\psi(3770)$ would address the 
question of non-$D\bar D$ decays of $\psi(3770)$. A careful analysis of 
the line-shape of the observed cross sections for inclusive hadronic 
event production, $D\bar D$ (including $D^*$ and $D_S^+$) production,
other exclusive final states (such as $J/\psi \pi\pi$, ~$J/\psi 
\eta$, ~$J/\psi \eta'$, 
~$J/\psi \omega$, ~$\phi K^+K^-$, ~$\phi \pi^+\pi^-$,
~$\chi_{cJ}\rho$, ~$\chi_{cJ}\omega$~...) and inclusive final states 
(~$J/\psi X$, ~$\eta X$, ~${\eta'} X$ ...) 
at different energies in the range from 3.7 to 4.5~GeV will provide the
opportunity to search for new, non-conventional
resonances. These can be done well with the \bes3
detector.

\part[Light hadron physics]{Light Hadron Physics\\
\vspace{2cm}
{\Large Conveners\\
  Xiao-Yan Shen, Bing-Song Zou\\}
\vspace{1cm} 
{\Large Contributors\\
 Jian-Ming Bian, Vladimir Bytev, Ying Chen, Hong-Ying Jin, Shan Jin, Hu Qin, Xiao-Yan Shen, 
 Ning Wu, Guo-Fa Xu, Zhen-Xia Zhang, Qiang Zhao, Alexey Zhemchugov, Han-Qing Zheng, Shi-Lin Zhu, Bing-Song Zou\\}}

\label{part:three}
\def\NCA{\em Nuovo Cimento}
\def\NIM{\em Nucl. Instrum. Methods}
\def\NIMA{{\em Nucl. Instrum. Methods} A}
\def\NPB{{\em Nucl. Phys.} B}
\def\PLB{{\em Phys. Lett.}  B}
\def\PRL{\em Phys. Rev. Lett.}
\def\PRD{{\em Phys. Rev.} D}
\def\ZPC{{\em Z. Phys.} C}
\def\sl{\llap{$/$}}
\newcommand{\Dpp}{D^+}
\newcommand{\MeVcc}{\ensuremath{\hbox{MeV}/c^2}}
\newcommand{\half}{\ensuremath{{1\over 2}}}

\newcommand{\script}{\scriptscriptstyle}
\newcommand{\begineqn}{\begin{equation}}
\newcommand{\ba}{\begin{array}{c}}
\newcommand{\ea}{\end{array}}
\newcommand{\bqa}{\begin{eqnarray}}
\newcommand{\eqa}{\end{eqnarray}}
\newcommand{\p}{\partial}
\newcommand{\e}{\,\,\mbox{##}}
\newcommand{\no}{\nonumber\\}

\newcommand{\ds}{D^{\ast}}
\newcommand{\dss}{D^{\ast \ast}}
\newcommand{\D}{D_{s}}
\newcommand{\Ds}{D^{\ast}_{s}}
\newcommand{\at}{a_{2}(1320)}
\newcommand{\aoo}{a_{1}(1260)}
\newcommand{\az}{a_{0}(1450)}
\newcommand{\bo}{b_{1}(1235)}
\newcommand{\ft}{f_{2}(1270)}
\newcommand{\fo}{f_{1}(1285)}
\newcommand{\fz}{f_{0}(1370)}
\newcommand{\ho}{h_{1}(1170)}
\newcommand{\kt}{K_{2}^{\ast}(1430)}
\newcommand{\kll}{K_{1}(1270)}
\newcommand{\kz}{K_{0}^{\ast}(1430)}
\newcommand{\kh}{K_{1}(1400)}
\newcommand{\fts}{f_{2}^{'}(1525)}
\newcommand{\fos}{f_{1}(1510)}
\newcommand{\fzs}{f_{0}(1370)}
\newcommand{\hos}{h_{1}(1380)}
\newcommand{\kq}{K (1460)}
\newcommand{\ksq}{K^{\ast} (1410)}
\newcommand{\pq}{\pi (1300)}
\newcommand{\euq}{\eta_{u} (1295)}
\newcommand{\esq}{\eta_{s} (1490)}
\newcommand{\rqb}{\rho (1450)}
\newcommand{\oq}{\omega (1420)}

 


Our knowledge about the structure of matter and the 
nature of the interactions
between its constituent components follows a hierarchy that 
closely tracks the evolution 
of experimental measurements from low to higher and higher energies. 
For example, with an electron beam of a given energy, one can
access the microscopic structure of matter at length scales
corresponding to the de-Brogie wave length of the electrons
that are being used:  $\lambda=2\pi\times
197.3$ MeV$\cdot$fm/$E_\gamma$, were $E_\gamma$ is the
energy of the photon that is exchanged in the experimentally
observed process.  As the energy of the beams that are available get 
higher and higher, the sizes of objects that can be probed get smaller
and smaller.

Using this relation, theoretical prescriptions for phenomena 
observed at different microscopic scales can be classified in 
terms of the corresponding
energy scales.  At  momentum transfers of several MeV, which
is much below $\Lambda_{QCD}\sim 200$~MeV 
the fundamental scale of Quantum Chromodynamics (QCD),   
explicit chiral-symmetry
breaking is revealed by the pattern of pseudoscalar meson 
masses and the $\pi N$ $\sigma$-terms.  Chiral perturbation 
theory can, in principle, provide a rigorous solution for 
physics that involve soft ({\it i.e.} low-energy) pions. 
At the very high energy (short distance) extreme 
({\it i.e.} energies much greater than $\Lambda_{QCD}$), partonic 
and gluonic degrees of freedom are
revealed in deep inelastic scattering measurements. 
In this energy region perturbative expansions of QCD are valid and
rigorous solutions can be pursued.  However, in the intermediate
region between these two energy extremes, excitations of meson 
and baryon resonances reveal effective constituent degrees 
of freedom. These, together with 
their mutual strong interactions, give rise to the
complex spectroscopy of hadrons. Since neither 
perturbative expansions of QCD nor chiral perturbation theory 
is valid at these length scales, it remains a great challenge to
physicists to describe the underlying physics on the basis of
first principles.   This challenge is
the focus of a large amount of recent experimental and 
theoretical effort. 
 
Hadron spectroscopy has been the major platform for probing many 
of the dynamical aspects of strong interactions in the non-perturbative 
regime. It also connects fundamental approaches such as lattice QCD 
calculations to phenomenological tools such as the quark model, 
QCD sum rules, etc., and provide insights into 
non-perturbative properties of QCD. 
 
Quarks, the basic building blocks of hadrons, are bound together 
by the color force mediated by the exchanges of gluons to form 
color-singlet hadrons.  The underlying dynamics are described 
by the widely tested QCD Lagrangian~\cite{de-r-g-g}: 
\bea 
{\cal L}&=& i\sum_i \bar{q}_i(x)[\partial_\mu-ig_s\sum_a\frac 12 
\lambda^a A_\mu^a(x)]\gamma^\mu q_i(x) \nonumber\\ 
&&-\frac 14\sum_a F^a_{\mu\nu}(x) F^{\mu\nu a}(x) -\sum_i 
\bar{q}_i(x) m_i q_i(x) \ , 
\eea 
where $g_s$ is the strong coupling constant, $i$ is the 
index for quark-flavor of the  $q_i(x)$  Dirac spinor in
three-dimensional color space and $\lambda^a$ denote the 
$3\times 3$ Gell-Mann matrices with $a=1,\dots, 8$. Here, a 
$q\bar{q}$ pair is the minimal configuration of quarks and 
antiquarks that can form a color-singlet meson, while 
$qqq$ is the minimal configuration that can form a 
color-singlet baryon. 
 
Note that the QCD quark-gluon interaction conserves flavor, and the 
interaction strength is flavor-independent. The only dependence on 
flavor in the QCD Lagrangian is through the quark-mass terms.

In the $u$, $d$ \& $s$ light quark sector, the mass 
differences are relatively small: $m_d-m_u\simeq 3$ MeV and 
$m_s-m_d=150$ MeV. Therefore, the strong interactions have an 
approximate global $SU(3)$ flavor symmetry, where quarks (antiquarks) 
are assigned to a triplet
representation {\bf 3} ($\bar{\bf 3}$). Mesons 
made of $q\bar{q}$ are then irreduciable representations given by 
the following product decomposition: 
\be 
{\bf 3}\otimes\bar{\bf 3}={\bf 1}+{\bf 8} \ , 
\ee 
and baryons are: 
\be 
{\bf 3}\otimes{\bf 3}\otimes{\bf 3}={\bf 1}_a+{\bf 8}_\lambda 
+{\bf 8}_\rho +{\bf 10}_s, 
\ee 
where the subscripts $a$, $s$, $\lambda$ and $\rho$ denote 
antisymmetric, symmetric, and the two mixed symmetries for the 
two-body substates within the three-quark system. 
 
$SU(3)$-flavor symmetry implies the existence of flavor nonets 
with the same $J^P$ but different  charges in the meson spectrum. 
For example, there are eight  pseudoscalars with masses below 
$\sim$500~MeV: $\pi^0$, $\pi^\pm$, $K^0$, 
$\bar{K^0}$, $K^\pm$ and $\eta$, 
and one at about 1~GeV: $\eta^\prime$. These states 
are collectively identified as the $0^-$ flavor nonet of 
the meson spectrum. 
 
Similarly, for the lowest-mass baryons, one expects the 
existence of flavor singlets, octets and decuplets. In
the spectrum of the lowest-lying baryon states, one identifies eight 
baryons that correspond to
an octet with $J^P=1/2^+$: $p(uud)$, $n(udd)$, 
$\Sigma^+(uus)$, $\Sigma^0(uds)$, $\Sigma^-(dds)$, $\Lambda(uds)$, 
$\Xi^0(uss)$ and $\Xi^-(dss)$, with masses  in the
$0.9\sim 1.3$~GeV range, and ten states of a decuplet with 
$J^P=3/2^+$: $\Delta^{++}(uuu)$, $\Delta^+(uud)$, $\Delta^0(udd)$, 
$\Delta^-(ddd)$, $\Sigma^+(uus)$, $\Sigma^0(uds)$, 
$\Sigma^-(dds)$, $\Xi^0(uss)$, $\Xi^-(dss)$, and $\Omega^-(sss)$, 
with masses in the $1.2\sim 1.7$~GeV range.  The flavor-singlet 
baryon with $J^P = 1/2^+$ has a higher mass that is
produced by the spatial and spin 
degrees of freedom.

By relating the breaking of $SU(3)$-flavor symmetry to the mass 
term in the QCD Lagrangian, an early successful prediction of the 
quark model was that the hadrons within an $SU(3)$-flavor 
multiplet will differ in mass linearly with the 
number of $s$-quarks that they contained.  Namely, one 
has 
\bea 
\Omega^-(1672)-\Xi(1530)&\simeq & 
\Xi(1530)-\Sigma(1385)\nonumber\\ 
&\simeq & 
\Sigma(1385)-\Delta(1232) \nonumber\\ 
\phi(1020)-K^*(890) &\simeq & K^*(890)-\rho(770)  \ . 
\eea 
A compact expression is given by the Gell-Mann-Okubo mass 
relation: 
\be 
\Sigma+3\Lambda = 2(N+\Xi) \ . 
\ee 
 
The main hypotheses of the 
QCD-motivated naive quark 
model~\cite{isgur-karl-model,part3:chao} are:
\begin{description}
\item[i)]  Spontaneous breaking of chiral 
symmetry leads to the presence of massive 
constituent quarks within a hadron as effective degrees of 
freedom. 
\item[ii)] Hadrons can be viewed as a quark system in which the 
gluon fields generate effective potentials that depend on the 
relative positions and spins of the massive quarks. 
\end{description} 
These two hypotheses inspire a non-relativistic treatment as
a leading-order approach.  Meanwhile, for quarks that have
constituent masses that are comparable to the QCD scale, the creation 
of constituent quark pairs will be suppressed.  Thus, the 
lowest-lying meson states are $q\bar{q}$ and 
lowest-lying baryon states
are $qqq$.  By treating the gluon fields as effective 
potentials, the hadron wavefunctions only depend on 
quark variables. 
 
By incorporating these simple hypotheses 
in a framework that accommodates the color, 
flavor, spin, and spatial degrees of freedom in the wavefunction 
solutions for bound quark systems, the quark model provides an 
efficient and evidently successful classification for a large 
number of hadrons that are observed experimentally~\cite{close-book}. 
 
 
In the light hadron sector, the nonrelativistic approach of the 
naive quark model unavoidably suffers the problem that the 
excitation energies of the excited hadron states ({\it i.e.}
resonances) are comparable to the 
constituent masses, thereby making the nonrelativistic approach 
questionable. Although it is still unclear why such a 
nonrelativistic model works so well for the light flavor 
 $u$-, $d$- and $s$-quark systems, there is no doubt about its 
extensive range of empirical successes. Various investigations 
are making 
progress on providing a firmer basis for the quark model 
phenomenology. For instance, recent progress in lattice QCD is 
providing evidence for effective degrees of freedom that 
can be identified as constituent quarks inside the nucleon. 
 
Another tantalizing aspect of the low-energy properties of QCD that needs 
understanding is the possible 
existence of hadrons with structures that
are more complex than the traditional $q\bar{q}$ or $qqq$ 
configurations of the quark model.  QCD is a non-Abelian theory 
that does not appear to forbid formations of so-called `exotics`, 
{\it i.e.} color-singlet constitutent arraangements that are
more complex than the conventional $q\bar{q}$ or $qqq$ hadrons. 
These include, for example: glueballs, made only of gluons; hybrids, made 
of both quarks and gluons as effective constituents; multiquark states;
{\it etc.}. 
These forms of hadrons have still evaded any explicit confirmations from 
experiment or theory. This 
reflects poorly on what we know about the fundamental strong force 
in the nonperturbative regime. Such states, if they exist, will enrich the 
population of states in the spectroscopy of hadrons, and deepen our 
understanding of the properties of QCD. As a result, they have 
received a tremendous amount af
experimental and theoretical effort.

Applications of various types of QCD-motivated
phenomenological models to the study of hadron
spectroscopy can be found in the literature.
These include: the
semirelativistic flux-tube model~\cite{ckp,stancu};
the instanton model~\cite{instanton-model}; 
the Goldstone boson exchange model~\cite{goldstone-boson-exchange};
the diquark model~\cite{diquark}; etc.
Masses of the non-strange $P$-wave baryons have also 
been examined by means of a mass operator analysis in large
$N_c$ QCD~\cite{large-nc}. Moreover, 
there continue to be attempts at formulating 
relativistic versions of the quark model~\cite{re-qm}. 
QCD-inspired approaches for understanding
the dynamics of hadron EM and strong decays and hadron-hadron
interactions have also been extensively described in the literature.
Generally speaking, those approaches take into
account different specific ingredients of the quark interactions 
and shed new light on corresponding aspects of non-perturbative QCD. 
(See Ref.~\cite{capstick-roberts} for a recent review.) The ultimate
prescription for understanding non-perturbative QCD dynamics in the light
quark sector may come from  future lattice QCD studies.
 
The study of light hadron spectroscopy is is a major activity at many 
hadron facilities. The primary focus of the
CLAS collaboration at Hall B at Jefferson  
Laboratory are the properties of baryons, including those of the
ground states and a nearly complete set of excited baryon   
resonances. It has the capability to do
polarized-beam asymmetry measurements and, eventually,
experiments with polarized targets will also be performed.  With
its primary electron energy upgraded to 12 GeV, measurements of the
baryon resonance form-factors from low to high momentum-transfers
will be carried out. One of the most important physics goals of
CLAS is to look for the ``missing baryon resonances", {\it i.e.}
states that are predicted by the quark model, but which have
so far not been seen in the $\pi$-$N$
scattering data.  Similar projects are also underway at ELSA, ESRF,
MAMI and SPring-8, via nucleon photo- and/or electro-excitations.
In contrast to these, the study of baryon resonances in charmonium
decays at \bes3 will have the advantages of reduced non-resonant 
background
and the ability to disentangle individual resonances 
thanks to the natural isospin
filter provided by the initial charmonia. We discuss
experiments on light meson spectroscopy at other experimental
facilities such as BaBar, Belle,  BNL, Fermilab, CLEO-c/Cornell, FAIR, 
{\it etc.}, below.

\chapter{Meson spectroscopy}
\label{sec:meson}

\section{Introduction} 
 
The ultimate goal of the study of hadron spectroscopy is to understand 
the dynamics of the constituent interactions.  For light
hadron systems, perturbative QCD in not applicable,  and
the path towards this ultimate goal has to be via
phenomenological methods or lattice QCD (LQCD) 
calculations. In the past decade, LQCD has experienced drastic 
improvements aided in part by the continued rapid development of computing 
resources. But a lot of technical difficulties persist
in the simulation of fully non-perturbative QCD processes; 
{\it e.g.}, unquenched calculations are still unavailable. In contrast, 
the development of phenomenological methods has  made
progress over the whole range of 
modern sciences, and especially in the study of hadron spectroscopy 
(see Ref.~\cite{godfrey-napo} and references therein).  For these 
approaches,
experimental data  is required to provide necessary constraints on the 
various parameters introduced by the theory. 
 
In this Chapter, we will 
discuss how a $q\bar{q}$ meson can be constructed in the quark 
model~\cite{close-book,godfrey-napo}. In addition to ``conventional"
$q\bar{q}$ mesons, it is known that the non-Abelian property 
of QCD also suggests the possible existence of bound states that
are made completely of gluons, {\it i.e.} so-called ``glueballs", 
and/or a gluon continuum. Furthermore, it may also be possible to form 
multiquark mesons, such as $qq\bar{q}\bar{q}$, and so-called ``hybrid" 
mesons, which contain both $q\bar{q}$ and a gluon ($g$) as 
constituents, $q\bar{q}g$.  Such unconventional states, if they 
exist, will enrich greatly the meson spectrum and shed light on 
the dynamics of strong QCD. Thus, searching for those 
unconventional mesons in experiment is a topical subject 
for modern intermediate- and  high-energy 
physics~\cite{Scalarnonet,amsler-tornqvist}. 
 
\section{Conventional meson spectrum}


For conventional $q\bar{q}$ mesons, the goal of
 phenomenological studies is, on 
the one hand, to find an empirically efficient way to describe the 
meson spectrum, and on the other hand maintain the general properties 
of QCD. In the 
quark model framework, the starting point for the study
of the meson spectrum is the construction of 
the Hamiltonian for color-singlet $q\bar{q}$ systems: 
\be\label{hamiltonian-qcd} 
H=\int d{\bf x}\sum_i q^\dag_i({\bf x})\beta 
\left(m_i-\frac{\Delta}{2m_i}\right)q_i({\bf x}) +\frac 12 \int 
d{\bf x} d{\bf y} V_0({\bf x}-{\bf y})\sum_{ija} q^\dag_i({\bf 
x})\frac{\lambda^a}{2} q_i({\bf x})q^\dag_j({\bf 
y})\frac{\lambda^a}{2} q_j({\bf y}) \ , 
\ee 
where $i$ and $j$ are flavor indices, $\lambda^a$ are the 
Gell-Mann Matrices for the SU(3)-color interactions, and $V_0({\bf 
x}-{\bf y})$ is a central potential, {\it i.e.} it only depends on $|({\bf 
x}-{\bf y})|$. Note that this Hamiltonian is independent of flavor, 
as required by QCD. Also, one can see that a particular assumption 
for the form of the potential $V_0$ can reflect properties of QCD,
such as in one-gluon-exchange Coulomb approximation $V_0({\bf 
x})=\alpha_s/|{\bf x}|$ at short separation distances, with
an increasing confinement behavior at larger distances.

The potential part of the Hamiltonian operating on a color-singlet 
$q\bar{q}$, $|(q\bar{q})_{\bf 1}\rangle$, gives 
\be 
\langle (q\bar{q})_{\bf 1}| \hat{V}|(q\bar{q})_{\bf 1}\rangle = 
-\frac 43 V_0 \ , 
\ee 
where $\hat{V}$ denotes the second term of 
Eq.~(\ref{hamiltonian-qcd}). Therefore, one can simplify the 
Hamiltonian for a color-singlet $q_i\bar{q}_j$ system to be: 
\be\label{hamiltonian-2} 
H=\frac{{\bf p}_x^2}{2m_i} +\frac{{\bf p}_y^2}{2m_j} -\frac 43 
V_0({\bf x}-{\bf y}) \ , 
\ee 
where ${\bf p}_x$ (${\bf p}_y$) and ${\bf x}$ (${\bf y}$) are the 
momentum and position of quark $i$ ($j$), respectively. 
 
In the center-of-mass frame, the Hamiltonian can be written as 
\be 
H=\frac{{\bf P}^2}{2M} +\frac{{\bf p}^2}{2m} + V({\bf r}) \ , 
\ee 
where ${\bf P}={\bf p}_x+{\bf p}_y$ is the c.m. momentum of the 
$q\bar{q}$ system, ${\bf r}={\bf x}-{\bf y}$ and ${\bf 
p}=({\bf p}_x-{\bf p}_y)/2$ are the relative distance and momentum 
between these two quarks, and $M\equiv m_i+m_j$ \& $m\equiv m_i 
m_j/M$ are the total and reduced masses. Since $V({\bf r})$ is 
assumed to be spin-independent, Eq.~(\ref{hamiltonian-2}) is 
invariant under separate orbital and spin rotations. By defining 
the radial quantum number $N$, the orbital angular momentum $L$, and 
the total spin $S=0,1$, one can classify the eigenstates of 
Eq.~(\ref{hamiltonian-2}) (including the spin wavefunctions) as $N 
^{2S+1}L_J$ with the total angular momentum $J=L$ for $S=0$ or 
$J=|L-1|, \ L, \ L+1$ for $S=1$. Given an explicit form for $V({\bf 
r})$, one in principle is able to produce a $q\bar{q}$ 
spectrum to compare with experimental data. For instance, the 
ground state will correspond to $N=1, L=0$ with $S=0$ or 1, {\it i.e.} a 
spin singlet $1 ^1S_0$ and a spin triplet $1 ^3S_0$. In the 
charmonium spectrum -- a suitable example for spin-independent 
potential quark model -- one can identify these as the $\eta_c(2980)$ and 
$J/\psi(3097)$. Table~\ref{nonet} lists a classification
based on this scheme for some of 
the low-lying mesons in the light quark sector~\cite{pdg2006}. 
\begin{table} 
\begin{center} 
\caption{ Quark model classification for some of those low-lying 
$q\bar{q}$ states. $f^\prime$ and $f$ denote the octet-singlet 
mixings. It should be noted that the nature of 
the $f_0(1710)$ and 
$f_0(1370)$ is still subject to debate.  } 
\begin{tabular}{cc|c|c|c|c}\hline 
&& $I=1$ &   $I=1/2$ & $I=0$ & $I=0$ \\ 
$N^{2S+1}L_J$ & $J^{PC}$ & $u\bar{d}, d\bar{u}, 
(d\bar{d}-u\bar{u})/\sqrt{2}$ & 
 $u\bar{s}, s\bar{u}, d\bar{s},s\bar{d}$ & $f^\prime$ & $f$ \\ 
\hline $1 ^1S_0$ & $0^{-+}$ & $\pi$ & $K$ & $\eta$ & $\eta^\prime$ 
\\ 
$1 ^3 S_1$ & $1^{--}$ & $\rho(770)$ & $K^*(892)$ & $\phi(1020)$ & 
$\omega(782)$ \\ 
$1 ^1 P_1$ & $1^{+-}$ & $b_1(1235)$ & $K_{1B}$ & $h_1(1380)$ & 
$h_1(1170)$ \\ 
$1 ^3P_0$ & $0^{++}$ & $a_0(1450)$ & $K^*_0(1430)$ & $f_0(1710)$ & 
$f_0(1370)$ \\ 
$\dots$ & &&&&\\ 
 \hline 
\end{tabular} 
\end{center} 
\label{nonet} 
\end{table}

There are also spin-dependent forces between the quarks that 
result in fine and hyperfine structures in the hadron 
spectrum. In fact, without spin-dependent forces, the 
spectrum obtained from the Hamiltonian of 
Eq.~(\ref{hamiltonian-2}) cannot match the pattern observed in 
experiment. On the other hand, relativistic effects will break 
the invariance of Hamiltonian under separate orbital and spin 
rotations. Therefore, it is natural to introduce spin-dependent 
forces that are phenomenologically analogous to those in the hydrogen 
atom. 
 
De Rujula, Georgi and Glashow first illustrated 
that the spin-dependent forces  in the quark potential 
originate from the short-range QCD one-gluon-exchange 
(OGE)~\cite{de-r-g-g}. In a nonrelativistic expansion, in addition 
to a colored Coulomb-type potential, the OGE generates a
Breit-Fermi interaction: 
\be 
H_{BF}=k\alpha_s\sum_{i<j} S_{ij} \ , 
\ee 
where $k=-4/3$ and $-2/3$ for a $q\bar{q}$ singlet and 
$q q$ in a $\bar{\bf 3}$, respectively, and 
\bea 
S_{ij}& =& \frac{1}{|{\bf r}|}-\frac{1}{2m_i m_j}\left(\frac{{\bf 
p}_i\cdot {\bf p}_j}{|{\bf r}|} +\frac{{\bf r}\cdot({\bf 
r}\cdot{\bf p}_i){\bf p}_j}{|{\bf r}|^3}\right) 
-\frac{\pi}{2}\delta({\bf r})\left(\frac{1}{m_i^2}+\frac{1}{m_j^2} 
+\frac{16{\bf S}_i\cdot{\bf S}_j}{3m_i m_j}\right)\nonumber\\ 
&& -\frac{1}{2|{\bf r}|^3}\left\{\frac{1}{m_i^2}({\bf r}\times{\bf 
p}_i)\cdot{\bf S}_i -\frac{1}{m_j^2}({\bf r}\times{\bf 
p}_j)\cdot{\bf S}_j \right.\nonumber\\ 
&&\left. + \frac{1}{m_i m_j}\left[2({\bf r}\times{\bf 
p}_i)\cdot{\bf S}_j -2({\bf r}\times{\bf p}_j)\cdot{\bf S}_i 
-2{\bf S}_i\cdot{\bf S}_j + 6\frac{({\bf S}_i\cdot{\bf r})({\bf 
S}_j\cdot{\bf r})}{|{\bf r}|^2}\right]\right\}+ \cdots \nonumber\\ 
\eea 
 
\noindent 
Although it is still questionable as to whether or not
one can apply the OGE 
picture to the hadron spectrum, these results turn out 
to be in good agreement with 
experimental observations.

We do not review the Goldstone boson exchange model developed in 
the past decade, but refer readers to recent 
reviews~\cite{goldstone-boson-exchange} for details. A review of 
constituent quark model approaches for $q\bar{q}$ system can be 
found in Ref.~\cite{godfrey-napo}.

According to the constituent quark model (CQM), mesons and baryons
are composed of $q\bar q$ and $qqq$ respectively. The CQM provides a
convenient framework for the classification of hadrons and most of the
experimentally observed hadron states fit into this scheme quite
neatly. Any states beyond the CQM are labelled as ``non-conventional"
hadrons.

However, the CQM is only a phenomenological model. It is not derived
from the underlying theory of the strong interaction---{\it i.e.} Quantum
Chromodynamics (QCD). Hence the CQM spectrum is not necessarily
the same as the physical spectrum of QCD, which remains ambiguous
and elusive after decades of intensive theoretical and
experimental exploration. No one has been able to either prove or
exclude the existence of non-conventional states rigorously,
since no one has yet been able to solve the confinement issue in QCD.

Hadron physicists generally take a modest and practical attitude.
If one supposes that these non-conventional states exist, then the
important issues are: (1) How to determine their characteristic
quantum numbers and estimate their masses, production
cross-sections and decay widths reliably? (2) How and in which
channels can we distinguish their signals from background and 
identify them unambiguously?

There are three classes of non-conventional hadrons. The first
class are mesons with "exotic" $J^{PC}$ quantum numbers. The
possible angular momentum, parity and charge conjugation parity of
a neutral $q\bar q $ meson can only be $J^{PC}=0^{++}, 0^{-+}, 1^{++},
1^{--}, 1^{+-}, \cdots$;  $q\bar q $ mesons cannot
have $J^{PC}=0^{--}, 0^{+-}, 1^{-+}, 2^{+-}, 3^{-+}, \cdots
$. Any state with these ``exotic" quantum numbers will clearly be beyond
the CQM. We want to emphasize that these would be ``exotic" only in the 
context 
of the CQM. One can construct color-singlet local operators to verify
that these quantum numbers are allowed in QCD. ``Exotic" quantum
numbers provide a powerful signatures for experimental searches 
for non-conventional meson states. In contrast, the $qqq$ baryons of
the CQM exhaust all possible $J^P$, i.e., $J^P={1\over 2}^{\pm},
{3\over 2}^{\pm}, {5\over 2}^{\pm}, \cdots$.

The second class of unconventional hadrons 
are those with exotic flavor content.  One example 
of this would be the purported  $\Theta^+$ pentaquark. This was 
studied in
the $K^+ n$ channel with the quark content $uudd\bar s$. Such a state,
should it exist, would clearly be beyond the CQM. 
Exotic flavor content is thus an important
aspect of experimental searches for non-QPM states.

The third class are hadrons that have non-exotic quantum numbers but
do not fit  into the CQM spectrum.  Take, for example, the 
$J^{PC}=0^{++}$ scalar mesons. Below 2~GeV there exists the $\sigma,
f_0(980), f_0(1370), f_0(1500), f_0(1710), f_0(1790)$ and the $f_0(1810)$
candidate scalar meson states. Within the CQM, only two scalars 
are expected in this mass range, at least 
if we ignore radial excitations. Including radial excitations, the CQM 
could accommodate no more than four scalars. Clearly there is a 
serious overpopulation of the scalar meson spectrum. If all of the states 
listed above are genuine and distinct, 
the content of some of them cannot be simply $q\bar q$.  The
overpopulation of the spectrum provides another useful tool for
the experimental search for non-conventional states.

Glueballs are hadrons comprised of gluons. Quenched lattice QCD
simulations suggest that the scalar glueball is the lightest
of these states, with a mass in the $1.5\sim 1.7$~GeV
range. Glueballs with  other quantum numbers are expected to 
have higher masses.  In the large $N_c$ limit, glueballs
decouple from the conventional $q\bar q$ mesons \cite{witten}.
In the real world with $N_c=3$, glueballs can
mix with any nearby $q\bar q$ mesons that have the same quantum numbers,
which renders the experimental identification of scalar glueballs
very difficult. In the following we discuss all of the non-conventional
hadrons according to their underlying quark gluon structure.

\section{Glueball spectrum} 
 
Glueballs are bound states of at least 2 or 3 gluons in a color 
singlet due to the non-Abelian property of QCD: 
\bea 
gg: & {\bf 8}\otimes {\bf 8} &={\bf 1} + {\bf 8} + \dots 
\nonumber\\ 
ggg: & {\bf 8}\otimes{\bf 8}\otimes{\bf 8} &=({\bf 8}\otimes{\bf 
8})_{\bf 8}\otimes{\bf 8}={\bf 1} + \dots , 
\eea 
where the charge conjugation is $C=+$ for $gg$ states and $C=-$ 
for $ggg$ states. Assuming that gluons inside glueballs are 
massive, for $gg$ with orbital angular momentum $L=0$, 
states of $0^{++}$ and $2^{++}$ can be formed, with $0^{++}$ being the 
ground state. For $ggg$, the lowest states are $0^{-+}$, 
$1^{--}$ and $3^{--}$. 
 
Both $gg$ and $ggg$ can form states with quantum numbers that 
cannot be produced by $q\bar{q}$ quark model states. Such 
states, called ``oddballs," would be ``smoking guns" in searching 
for glueball candidates. (Experimental evidence for such a state
is still unavailable.) 
Possible quantum numbers for oddballs are: 
$0^{--}$, $0^{+-}$, $1^{-+}$, $2^{+-}$, $3^{-+}\dots$. However, 
for $gg$ states, if the gluons inside are massless, $J=\mbox{odd}$ 
states would be forbidden by Yang's theorem~\cite{part3:ref:yang} though 
they  could exist in $ggg$ sector.

There is no explicit correlation between the $gg$ and $ggg$ 
ground state masses,  although the $0^{++}$ is expected to be lighter 
than the $0^{-+}$. Theoretical predictions for the glueball masses 
vary significantly among different approaches. 
Early phenomenological models find rather light masses for the scalar 
glueball, {\it e.g.}, $M(0^{++})=0.65\sim 1.21$ GeV in the bag 
model~\cite{carlson,donoghue,chanowitz-sharpe}, and 
$M(0^{++})=1.15$ GeV in a potential model~\cite{cornwall-soni}. 
Other QCD-motivated approaches produce larger masses for the scalar 
such as $M(0^{++})=1.52$ GeV in a flux-tube 
model~\cite{isgur-paton}, and $M(0^{++})\simeq 1.5$ GeV in QCD sum rule 
calculations~\cite{novikov,shifman,latorre,narison,kisslinger,faessler,huang-jin-zhang}. 
 
Over the past twenty years, extensive numerical studies have been 
carried out aimed at calculating the glueball spectrum in LQCD. Although 
the earliest LQCD predictions~\cite{lattice-v1,lattice-v2} for the 
glueball masses varied significantly,  the more recent predictions for 
several lightest glueballs have converged to a similar mass region 
despite 
the use of different approaches~\cite{mt,mp,ukqcd,part3:chen}. The 
lowest glueball state in these calculations
is the $J^{PC}=0^{++}$ state (scalar) with 
a mass of about $1.5\sim 1.7$ GeV, and the mass ratios of the 
tensor and pseudoscalar to the scalar are about 1.4 and 1.5, 
respectively. The latest results~\cite{part3:chen} on the glueball 
spectrum from a larger and finer lattice are listed in 
Table~\ref{spectrum1} and shown in Fig.~\ref{spectrum2}. 
\begin{table} 
\begin{center} 
\caption{  The predicted glueball spectrum in physical units. In 
column 2, the first error is the statistical uncertainty coming 
from the continuum extrapolation, the second one is the 1\% 
uncertainty resulting from the approximate anisotropy. In column 
3, the first error comes from the combined uncertainty of $r_0 
M_G$, the second from the uncertainty of $ r_0^{-1}=410(20)\,{\rm 
MeV}$ } 
\begin{tabular}{ccc}\hline\hline 
$J^{PC}$& $r_0 M_G$ &   $M_G\,({\rm MeV}$) \\ 
\hline 
$0^{++}$ & 4.16(11)(4)   &    1710(50)(80)     \\ 
$2^{++}$ & 5.83(5)(6)    &    2390(30)(120)    \\ 
$0^{-+}$ & 6.25(6)(6)    &    2560(35)(120)    \\ 
$1^{+-}$ & 7.27(4)(7)    &    2980(30)(140)     \\ 
$2^{-+}$ & 7.42(7)(7)    &    3040(40)(150)     \\ 
$3^{+-}$ & 8.79(3)(9)    &    3600(40)(170)     \\ 
$3^{++}$ & 8.94(6)(9)    &    3670(50)(180)     \\ 
$1^{--}$ & 9.34(4)(9)    &    3830(40)(190)     \\ 
$2^{--}$ & 9.77(4)(10)   &    4010(45)(200)     \\ 
$3^{--}$ & 10.25(4))(10) &    4200(45)(200)     \\ 
$2^{+-}$ & 10.32(7)(10)  &    4230(50)(200)     \\ 
$0^{+-}$ & 11.66(7)(12)  &    4780(60)(230)     \\ 
\hline\hline 
\end{tabular} 
\end{center} 
\label{spectrum1} 
\end{table} 
 
\begin{figure}[htbp]
\centerline{\epsfig{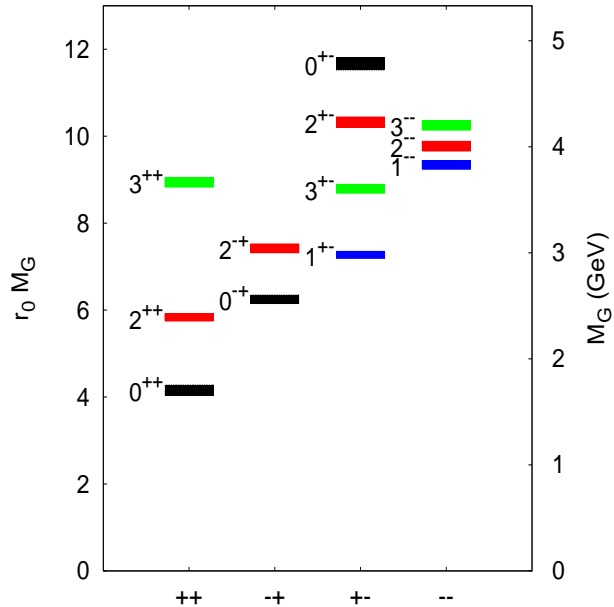}}
\caption{The mass spectrum of 
glueballs in pure $SU(3)$ gauge theory. The masses are given 
both in terms of $r_0$ ($r_0^{-1}=410\,{\rm MeV}$) and in ${\rm 
GeV}$. The thickness of each colored box indicates the statistical 
uncertainty of the mass.}  
\label{spectrum2} 
\end{figure} 
 
The calculations of the glueball spectrum are mostly from quenched 
lattice QCD. Therefore, an open question remains: {\it How 
large is the systematic uncertainty associated with the 
use of the quenched approximation?} A recent preliminary 
analysis of the scalar glueball mass based on the MILC 
dynamical gauge configuration indicates that the scalar 
glueball mass of the dynamic lattice simulation 
is not much effected by this approximation~\cite{Irving}. \\


\subsection{Glueball signatures} 
 
Using the spectrum of the $q\bar{q}$ nonet of pseudoscalar 
($0^{-}$) and vector ($1^{-}$) mesons as a reference, 
one expects the scalar nonet to lie in the 1$\sim$2~GeV
mass window~\cite{godfrey-napo,Scalarnonet,amsler-tornqvist}. 
The established $a_0(1450)$ and $K_0^*(1430)$ scalar states in this mass 
region are rather naturally assigned as the $I=1$ and $I=1/2$ 
members, respectively, leaving at issue the
identification of the $I=0$ 
members of the nonet. For $I=0$, there are 
a number of candidate states reported in the 
literature~\cite{cb-96,wa102,bes-phipipi,bes-wkk,bes-wphi}, namely, 
the $f_0(1370)$, $f_0(1500)$, $f_0(1710)$, and recent observations of $f_0(1790)$ and
$f_0(1810)$ at BES-II.  As mentioned above, this over-population may
already be a signal for the existence of 
scalars beyond the conventional quark model classification. 
 
There also exists a scalar nonet with masses below 1~GeV, {\it i.e.} 
the $f_0(980)$, $a_0(980)$, $\sigma(600)$, and $\kappa(800)$, which 
are candidates for multiquark or molecule 
states~\cite{jaffe,jaffe-low}. This nonet is discussed in detail 
in Chapter~\ref{sec:softpion-physics}.

Recent LQCD results~\cite{mp,ukqcd,part3:chen} indicate that the mass 
region between 1~and~2~GeV is extremely interesting for the ground 
state scalar glueball ($J^{PC}=0^{++}$) search,  especially in light
of the observation of a number of $f_0$ states with similar 
masses ($f_0(1370)$, $f_0(1500)$ and 
$f_0(1710)$)~\cite{bugg-physrept}. However, since glueball 
properties are not expected to be drastically different from those
of conventional $q\bar{q}$ mesons, one is faced with the 
difficulty of distinguishing the scalar glueball from conventional 
$q\bar{q}$ states. Nonetheless, more than two $f_0$ states with 
similar masses implies that mixing of a pure gluonic scalar 
(glueball) with $I=0$ members of the nearby $q\bar{q}$ nonet can be 
occurring~\cite{close-amsler,close-kirk,close-zhao,weingarten,
cheng-chua-liu,li-shen-yu,he-li,faessler}. 
This greatly complicates the task of identifying the scalar 
glueball both experimentally and theoretically. In contrast, signals for 
oddballs would be decisive evidence for the existence of glueballs. 
Unfortunately, to date concrete experimental identification of an
oddball is still unavailable. 
 
Theoretical expectations for a glueball with conventional quantum 
numbers have been widely discussed in the 
literature~\cite{Scalarnonet}. Although  these are
not sufficient in most cases for distinguishing a glueball candidate from 
a conventional $q\bar{q}$ state, they are still useful for 
providing guidance for further efforts. In the following
we briefly review some of these expectations. \\
 
{\bf 1). Flavor-blindness of glueball decays \\} 
 
The  predicted
ratios for flavor-singlet glueball decay branching fractions are: 
\be 
\frac{1}{P.S.}\Gamma(G\to \pi\pi : 
K\bar{K}:\eta_8\eta_8:\eta_1\eta_8:\eta_1\eta_1)=3:4:1:0:1 \ , 
\ee 
where P.S. denotes the phase space factor, and $\eta_1$ and 
$\eta_8$ are the $I=0$ flavor-singlet and 
flavor-octet members of the SU(3) 
nonet. It can be shown that this relation holds for $\pi\pi : 
K\bar{K} : \eta\eta : \eta\eta^\prime : \eta^\prime\eta^\prime$ 
after taking into account the singlet-octet mixing: 
\bea 
\eta &=& \eta_8\cos\theta 
-\eta_1\sin\theta\nonumber\\ 
\eta^\prime &=&\eta_8\sin\theta +\eta_1\cos\theta . 
\eea 
 
The most significant feature for a pure glueball decays to $PP$ is 
the vanishing branching ratio for $G\to \eta\eta^\prime$. However, 
the observation of a
vanishing rate for $X \to\eta\eta^\prime$ does not 
necessarily allow one to conclude 
unambiguously that $X=G$, since interference 
between different components in a $q\bar{q}$ state could also lead 
to a vanishing $\eta\eta^\prime$ branching 
ratio~\cite{close-amsler,close-kirk,close-zhao}.\\ 
 
{\bf 2). Glueball couplings to $\gamma\gamma$ \\}
 
Since gluons are electrically neutral, glueball production in 
$\gamma\gamma$ collisions and glueball decays into $\gamma \gamma$
final states are suppressed.   However, a small branching fraction  
for $X\to \gamma\gamma$ does not necessarily prove 
that $X$ is a glueball because of effects of possible interference 
with mixed components. For instance, in Ref.~\cite{cfl},
the branching fractions for $f_0 \to 
\gamma\gamma$ are found to be 
$f_0(1710):f_0(1500):f_0(1370) \simeq 3:1:12$, which indicates 
smaller $\gamma\gamma$ couplings for 
$f_0(1500)$ and $f_0(1710)$ than that for $f_0(1370)$.
 
Alternatively, the scalar's $q\bar{q}$ component can be probed in 
$e^+ e^-$ annihilation via two virtual photon intermediate states. 
VEPP and DA$\Phi$NE have access to the production of scalars in the
1$\sim 2$~GeV mass range, while BEPCII will be able to access 
heavier scalars, up to about 3 GeV~\cite{brodsky-goldhaber-lee}.\\ 
  
{\bf 3). Glueball production in heavy quarkonium radiative decays\\}
 
The $J/\psi$ radiative decay process is gluon rich and ideal for 
searches of glueballs as intermediate hadronic resonances in 
the $J/\psi\to \gamma G\to \gamma + \mbox{hadrons}$ process. This
will provide access to 
all isoscalar states with charge conjugation $C=+$ and forbids all 
$C=-$ states. These allowed quantum numbers include the low-lying 
glueballs and hybrids for which the production phase space is 
generally large.  Thus, a systematic study of $J/\psi$ 
radiative decays with high statistics at \bes3 will be extremely 
important for clarifying some long-standing puzzles. 
 
Information about intermediate resonances can be obtained by 
measuring their hadronic and/or radiative decays. Two measures are 
proposed in the literature to quantify the gluonic contents of the 
resonances. One is called ``Stickiness", and defined as
\be 
S= \frac{\Gamma(J/\psi\to \gamma X)\times P.S.(X\to 
\gamma\gamma)}{\Gamma(X\to \gamma\gamma)\times 
P.S.(J/\psi\to\gamma X)}, 
\ee 
by Chanowitz~\cite{chanowitz-84}. This is intended to measure 
the color-to-electric-charge ratio with phase space factored 
out, and maximize the effects from the glue dominance inside a 
glueball. If $X$ is a glueball, one would expect that its 
production is favored in $J/\psi$ radiative decays, while its 
decays into $\gamma\gamma$ are strongly suppressed. Therefore, a 
glueball should have large stickiness in comparison to a 
conventional $q\bar{q}$ state. 
 
Cakir and Farrar~\cite{cakir-farrar} propose another quantitative 
measure of the gluonic content of a resonance by calculating its 
branching ratio to gluons, i.e. $b_{R\to gg}$. This quantity can 
be related to the production branching ratio of a resonance $R$ in heavy 
quarkonium radiative decays. Its value is expected to be
in the range of $b_{R\to 
gg}= O(\alpha_s^2)\simeq 0.1-0.2$ for a $q\bar{q}$ state, and 
$\sim 1$ for a glueball. Interestingly, for most of the 
well established $q\bar{q}$ states, $b_{R\to gg}$ is found to be rather 
small, while for the scalar glueball candidates $f_0(1500)$ and 
$f_0(1710)$ the value is rather large, in particular for the 
$f_0(1710)$~\cite{cfl}. 
 
These two measures, however, still cannot provide unambiguous 
evidence for glueballs.  ``Stickiness" works in a world where 
there is no glueball and $q\bar{q}$ mixing (quenched 
approximation in LQCD), {\it i.e.} for pure glueballs. In the
event that  glueball-$q\bar{q}$ mixing occurs (unquenched), 
interference between glueball and $q\bar{q}$ components can violate 
the expectation of large stickiness for glueball 
states. In this sense, the scheme of Ref.~\cite{cakir-farrar,cfl} 
seems to be more applicable because it directly measures the 
coupling of a resonance to gluons. \\
  
{\bf 4). Chiral suppression \\} 
 
A criterion first pointed out by Carlson~\cite{carlson-81} and 
recently developed by Chanowitz~\cite{chanowitz-2005} is the 
chiral suppression mechanism for $J=0$ glueballs. Due to the fact 
that in pQCD the amplitude is proportional to the current quark 
mass in the final state,   $J=0$ glueballs should have larger 
couplings to {\it e.g.} $K\bar{K}$ than to $\pi\pi$. For $J\neq 
0$, the decay amplitude is flavor symmetric. Combined with the 
quenched LQCD prediction, this suggests that the $f_0(1710)$ is a 
gluon-dominant state while the $f_0(1500)$ is dominated by
$s\bar{s}$~\cite{weingarten}. However, because of the complexity 
of the gluon hadronization and the unclear 
extent of $G-q\bar{q}$ mixing, the 
observation of a relatively large branching ratio to $K\bar{K}$ for a 
candidate does not conclusively establish it as being a 
glueball~\cite{chao-he-ma,zhang-jin}. \\

{\bf 5). Charmonium hadronic decays \\}
 
Decays of charmonium states to light hadrons have great advantages for
systematic analyses for both light mesons and 
baryons. For instance,  $SU(3)$-flavor symmetry breaking can be 
investigated in the decays $J/\psi\to VP, VS, VT$; $\chi_{cJ}\to 
VV$, $PP$, $SS$; $\eta_c\to VV$, $PP$, etc. The 
decays $J/\psi\to V f_1$ also provide access to the axial vectors 
$f_1(1285)$ and $f_1(1420)$. An important issue here is the study 
of the properties of the final state mesons and the search for evidence
for nonconventional states, 
such as: the $f_0(980)$ and $a_0(980)$ as either four-quark 
states~\cite{jaffe} or $K\bar{K}$ molecules; a scalar glueball 
component of the
$f_0(1370)$, $f_0(1500)$ \& $f_0(1710)$; and hybrids. 
Specifically, for states $X$ recoiling from an $\omega$ or $\phi$ in 
$J/\psi\to \omega X$ and $\phi X$, one can gain information about 
the flavor components of the resonance $X$. Since the $\omega$ and $\phi$ 
are ideally  mixed, i.e. $|\omega\rangle = 
|u\bar{u}+d\bar{d}\rangle/\sqrt{2}$ 
and $|\phi\rangle = |s\bar{s}\rangle$, the flavor content of a 
resonance $X$ can be probed based on the OZI rule~\cite{part3:ozi}. 
However, it should cautioned that so far at least the role played by the 
empirical OZI rule has not been fully understood in the charmonium 
energy region. For systems that are recoiling from light 
mesons, large OZI-violations have been found in some QCD-inspired 
calculations~\cite{isgur-geiger,lipkin-zou,li-bugg-zou}. Dynamical 
studies of possible OZI rule violations should be carried out; these
may be of essential importance for understanding the production of 
various final state mesons~\cite{close-zhao,zzm,zhao-chic,zhao-etac}.\\ 
 
{\bf 6). Other glueball-favored processes \\}
 
Glueball signals have also been searched for in other reactions, such as 
$p\bar{p}$ annihilation, and central collisions of the type $pp\to pp G$. 
In $p\bar{p}$ reactions, because of the competition of 
$q\bar{q}$ meson formation
via quark and antiquark rearrangement, 
identification of glueball signals will be contaminated. 
Similarly, competition from $q\bar{q}$ production in $pp$ 
central collisions will mix with possible glueball signals. 
 
It is proposed by Close and Kirk~\cite{close-kirk-filter} that 
in $pp$ central collisions, the production of $S$-wave 
resonances (e.g. $0^{++}$ and $2^{++}$ glueballs, 
or $S$-wave tetraquark states, or $K\bar{K}$ molecules) 
will be favored due to the small recoil transverse momentum difference 
(d$P_T$) of the final state protons, while $q\bar{q}$ 
production will be favored in the larger d$P_T$ region (d$P_T\ge 
O(\Lambda_{QCD})$). Therefore, different kinematic regions serve 
as a production filter for $S$-wave resonances. Following 
this, an experimental analysis by WA102~\cite{wa102-filter} 
revealed a clear azimuthal dependence as a function of $J^{PC}$ and 
$P_T$ at the proton vertices, and the scalars appeared to divide into 
two classes: $f_0(980)$, $f_0(1500)$ and $f_0(1710)$ strongly peak 
at small $\phi$ angle (corresponding to small d$P_T$) and 
$f_0(1370)$ at large $\phi$.

The above criteria cannot individually provide indisputable 
evidence for glueball candidates with conventional quantum 
numbers, especially in the case of a scalar glueball. However, the 
combined effect of all of the above expectations and criteria 
might be useful for placing  bounds on the glueball 
and $q\bar{q}$ contents of a state and, thereby, provide some insight 
into  the complex issue of strong QCD. In fact, because of the
limitations imposed by the use of the 
quenched approximation in present LQCD calculations, 
the effects arising from an unquenched approach need to be 
investigated in detail. Phenomenological progress along this 
direction was proposed by Tornqvist~\cite{tornqvist-95} who 
emphasized the difference between bare states (quark model bound 
states that do not decay) and dressed ones (physical states that 
decay to hadrons). For the case of scalars, Boglione and 
Pennington~\cite{boglione-pennington} showed that a pure glue or 
$q\bar{q}$ state (e.g. unquenched bare state studied by LQCD) is 
so different from the dressed hadrons that reliable calculations 
of the hadron properties are crucial for extracting the scalars 
and for making sense of the experimental measurements.

\subsection{Glueball candidates}
\label{sec:glueball-cands}  

\subsubsection{Scalar glueball candidates} 

The abundance of isoscalar scalars in the $1\sim 2$~GeV, 
mass region, {\it i.e.} the $f_0(1370)$, $f_0(1500)$, and 
$f_0(1710)$ (the $f_0(1790)$ and $f_0(1810)$ 
should be confirmed in further experiments), 
makes them natural scalar glueball candidates. In the following, 
we briefly review the available experimental information about these 
states and examine the theoretical expectations of their nature. 
Controversies will be identified. \\
 
{\bf 1). $f_0(1370)$\\} 
 
The $f_0(1370)$ is broader than the $f_0(1500)$ and 
$f_0(1710)$, and is strongly coupled to $\pi\pi$. Its decays into 
$K\bar{K}$ were also observed by the Crystal Barrel $p\bar{p}$ 
annihilation experiment~\cite{cb-96}, and confirmed by the WA102 $pp$ 
scattering experiment~\cite{wa102} in the $\pi\pi$, 
$K\bar{K}$ and $\eta\eta$ decay channels. 
Its absence in $\eta\eta^\prime$ can be mainly due to the limited phase
space and does not necessarily mean that it has a large glueball component. 
In light of the large value of the ratio 
$BR_{f_0(1370)\to \pi\pi}/BR_{f_0(1370)\to K\bar{K}}= 2.17\pm 
0.9$~\cite{wa102}, the $f_0(1370)$ 
seems to be a likely candidate
for a $n\bar{n}$ scalar meson ($n=u$ or $d$). 
 
Recent data from BESII 
also show a strong $f_0(1370)$ signal in 
$J/\psi \to \phi \pi^+ \pi^-$ \cite{bes-phipipi}. 
Figure \ref{phipp}(a) shows the $K^+ K^-$  
invariant mass distribution for $J/\psi \to K^+K^-\pi^+\pi ^-$,
where a strong $\phi$ signal is evident. The spectrum of
$\pi^+ \pi^-$ invariant masses recoiling against the $\phi$ is 
shown in Fig.~\ref{phipp}(b), where, in addition to a 
strong $f_0(980)$ peak, 
there is a broad enhancement around 1370~MeV. 

\begin{figure}[htbp]
\centerline{\epsfig{file=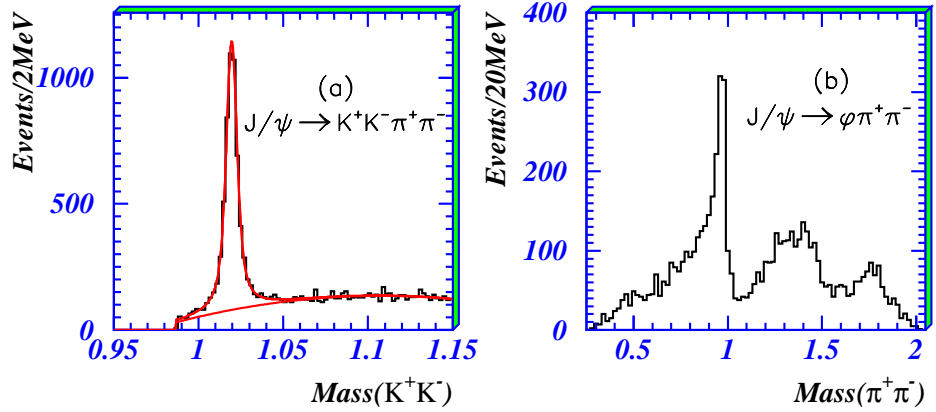,width=5.0in}}
\caption{ (a) The $K^+K^-$ invariant mass distribution for
$J/\psi \to K^+K^-\pi ^+\pi ^-$.
(b) The $\pi^+\pi^-$ invariant mass for events selected within $\pm 15$
MeV of the $\phi$.}
\label{phipp}
\end{figure} 
 
No significant $f_0(1370)$ signal
is observed in $J/\psi \to \phi K^+ K^-$
\cite{bes-phipipi}, $\omega K^+ K^-$ \cite{bes-wkk} or
$\omega \pi^+ \pi^-$ \cite{bes-wpipi}.
Figures~\ref{phikk}(a) and (b) show the $\phi$ signal and $K^+ K^-$ 
invariant mass recoiling against a $\phi$ for 
$J/\psi \to K^+K^-K^+K^-$ decays.

\begin{figure}[htbp]
\centerline{\epsfig{file=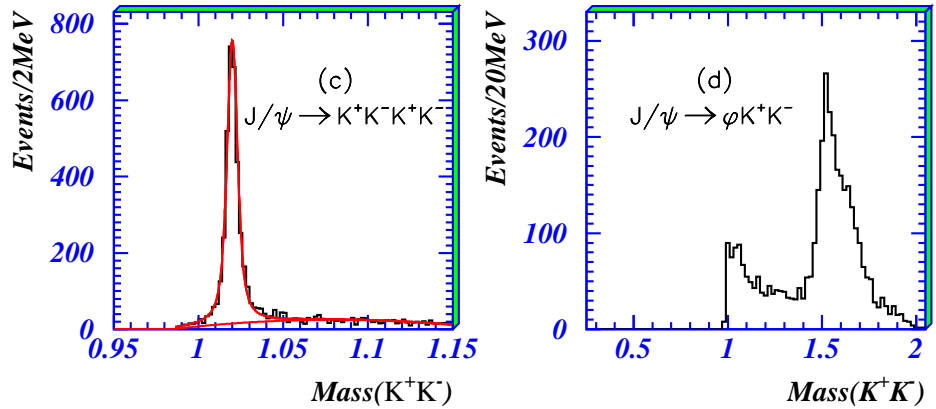,width=5.0in}}
\caption{ (a) The $K^+K^-$ invariant mass distribution for
$J/\psi \to K^+K^-K^+K^-$.
(b) The $K^+K^-$ invariant mass for events selected within $\pm 15$
MeV of the $\phi$.}
\label{phikk}
\end{figure}

A partial wave analysis (PWA) was performed to $J/\psi \to \phi \pi ^+\pi 
^-$ 
and $\phi K^+K^-$ using
relativistic tensor expressions for the amplitudes as
documented in Ref. \cite{pwa-zou}.
The full angular dependence of the decays of the $\phi$ and 
$\pi ^+\pi ^-$ or $K^+K^-$ resonances is fitted, including 
correlations between them.
The line-shape of the $\phi$ is not fitted, because the
$\phi$ is much narrower than the experimental resolution.
The $\phi \pi ^+\pi ^-$ and $\phi K^+K^-$ data are fitted
simultaneously, and resonance masses and widths are constrained to be
the same for both sets of data. The PWA results show that the peak 
around 1370~MeV in the $\pi^+ \pi^-$ invariant mass spectrum 
in Fig. \ref{phipp}(b) comes from a dominant
$f_0(1370)$ term that
interferes with an $f_2(1270)$ and a smaller $f_0(1500)$. 
The $\phi K\bar K$ data contains a strong
peak due to $f_2'(1525)$, with a shoulder on its upper side 
that can be fitted by interference between $f_0(1500)$ and $f_0(1710)$.
The mass and width of the $f_0(1370)$ are determined to be: 
$M=1350 \pm 50$~MeV and $\Gamma = 265 \pm 40$~MeV.

The absence of an $f_0(1370)$ signal in $J/\psi\to \omega \pi\pi$ 
indicates
that the branching ratio for $J/\psi\to 
\phi f_0(1370)$ is larger than that for $J/\psi\to \omega 
f_0(1370)$. This certainly raises concerns about the $f_0(1370)$ 
interpretation as an $n\bar{n}$ ( $n$ = $u$ or $d$) meson since, in the context
of the OZI rule, one would 
expect that an $n\bar{n}$ meson would more likely recoil against 
an $\omega$ than against a $\phi$.\\ 
  
{\bf 2). $f_0(1500)$ \\}
 
The $f_0(1500)$ was observed in many experiments, such as pion 
induced interactions $\pi^- p$, $p \bar{p}$ annihilation \cite{ppb1,ppb2}, 
$pp$ central collisions \cite{ppcentral1,ppcentral2} and 
$J/\psi$ radiative decays~\cite{bugg1,bes-4pi}. Most of the data on 
the $f_0(1500)$ comes from the Crystal 
Barrel collaboration, which resolved two scalar states in this mass 
region, and determined their decay branching ratios to a number of 
final states, including $\pi^0 \pi^0$, $\eta \eta$, $\eta 
\eta^\prime$, $K_L K_L$ and $4\pi^0$. It has also been noted that
it is absent  in the glueball-suppressed processes 
$\gamma \gamma \to K_s K_s$ and $\pi^+ \pi^-$.   All of these 
features favor the interpretation of the 
$f_0(1500)$ as a non-$q \bar{q}$ state.

If the $f_0(1500)$ is a scalar glueball, it 
should be copiously produced in $J/\psi$ radiative decays. The 
 $J/\psi\to\gamma\pi^+\pi^-$ process was analyzed previously by the
Mark\,III~\cite{mark3-gpipi}, DM2~\cite{dm2-gpipi} and 
BES\,I~\cite{bes1charge} experiments, where there was evidence 
for the $f_2(1270)$ and an additional $f_2(1720)$. In addition, a high 
mass shoulder to the $f_2(1270)$, at about 1.45~GeV, was 
seen. A revised amplitude analysis of the Mark\,III data found
this shoulder to be a scalar at $\sim$ 1.43~GeV, and, in 
addition, found that the peak at $\sim$ 1.7~GeV to be scalar 
rather than tensor~\cite{dunwoodie}. The 
$J/\psi\to\gamma\pi^0\pi^0$ process was also studied by the 
Crystal Ball~\cite{cball} and BES\,I experiments~\cite{bes1}, but 
no partial wave analyses were performed on this channel. 
 
Recently, BES reported the results on $J/\psi $ radiative decays 
to $\pi^+\pi^-$ and $\pi^0\pi^0$ based on a sample of 58M $J/\psi$ 
events taken with the BES\,II detector \cite{bes2-gpipi}.
Figure~\ref{gpipimass} shows the $\pi^+\pi^-$ mass spectrum for the
selected events, together with the corresponding background
distributions and Dalitz plot.
There is a strong $\rho^0(770)$ peak mainly due to background from
$J/\psi\to\rho^0\pi^0$. A strong $f_2(1270)$ signal, a shoulder on the
high mass side of $f_2(1270)$, an enhancement at $\sim$ 1.7~GeV,
and a peak at $\sim$ 2.1~GeV are clearly visible.
The lightly shaded histogram in Fig. \ref{gpipimass} corresponds to the
dominant background $J/\psi\to\pi^+\pi^-\pi^0$.
Other backgrounds are shown as the dark shaded
histogram in Fig.~\ref{gpipimass}. Figure~\ref{gpi0pi0mass} shows the 
$\pi^0\pi^0$ mass spectrum and  Dalitz plot.
The shaded histogram corresponds to the sum of estimated backgrounds
determined using  branching ratios from Ref.~\cite{pdg2006}. 
In general, the $\pi^+\pi^-$ and $\pi^0\pi^0$
mass spectra exhibit similar structures above 1.0~GeV.

\begin{figure}[htbp]
    \centerline{
    \psfig{file=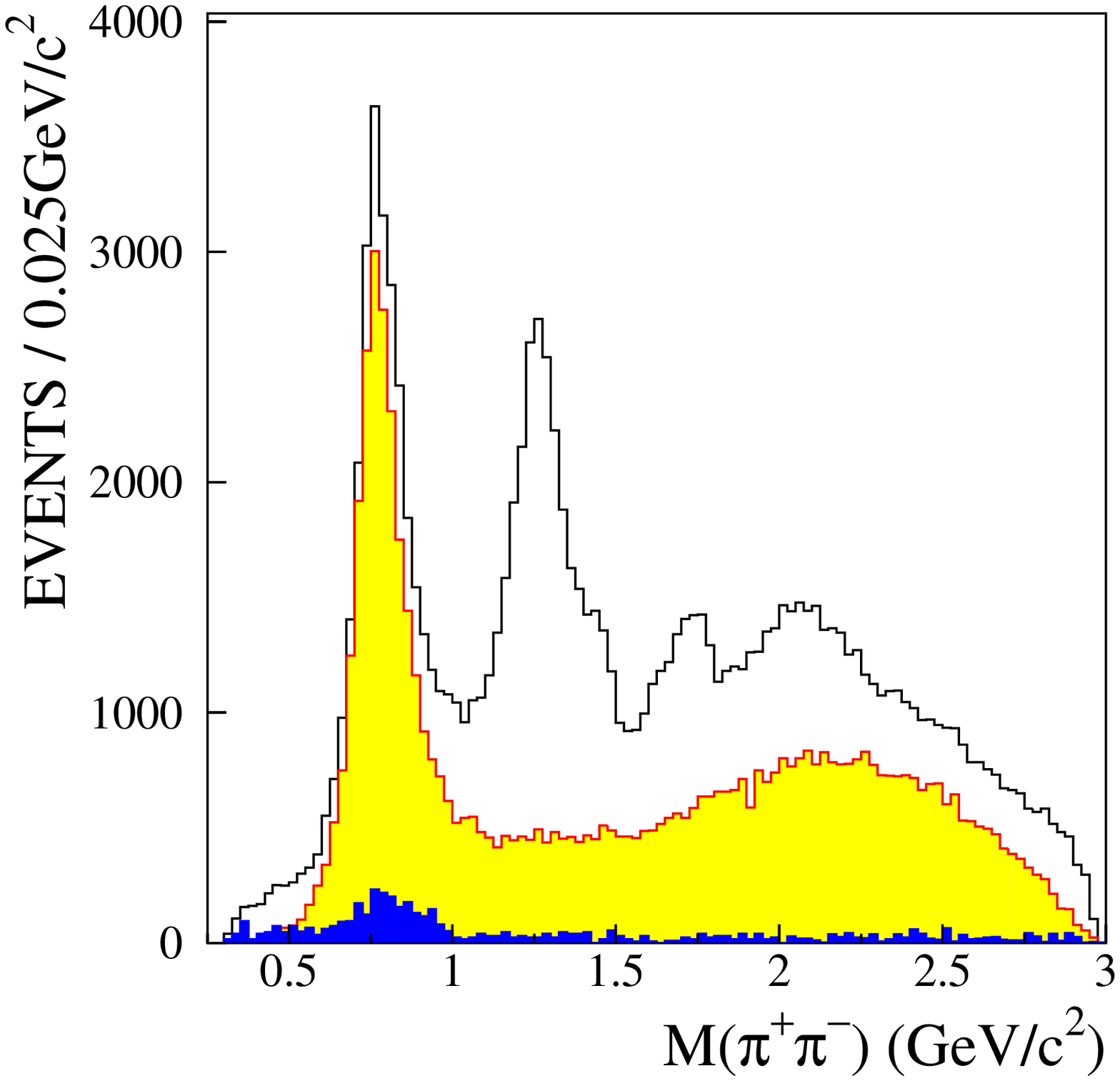,width=6.5cm,height=5.5cm}
    \psfig{file=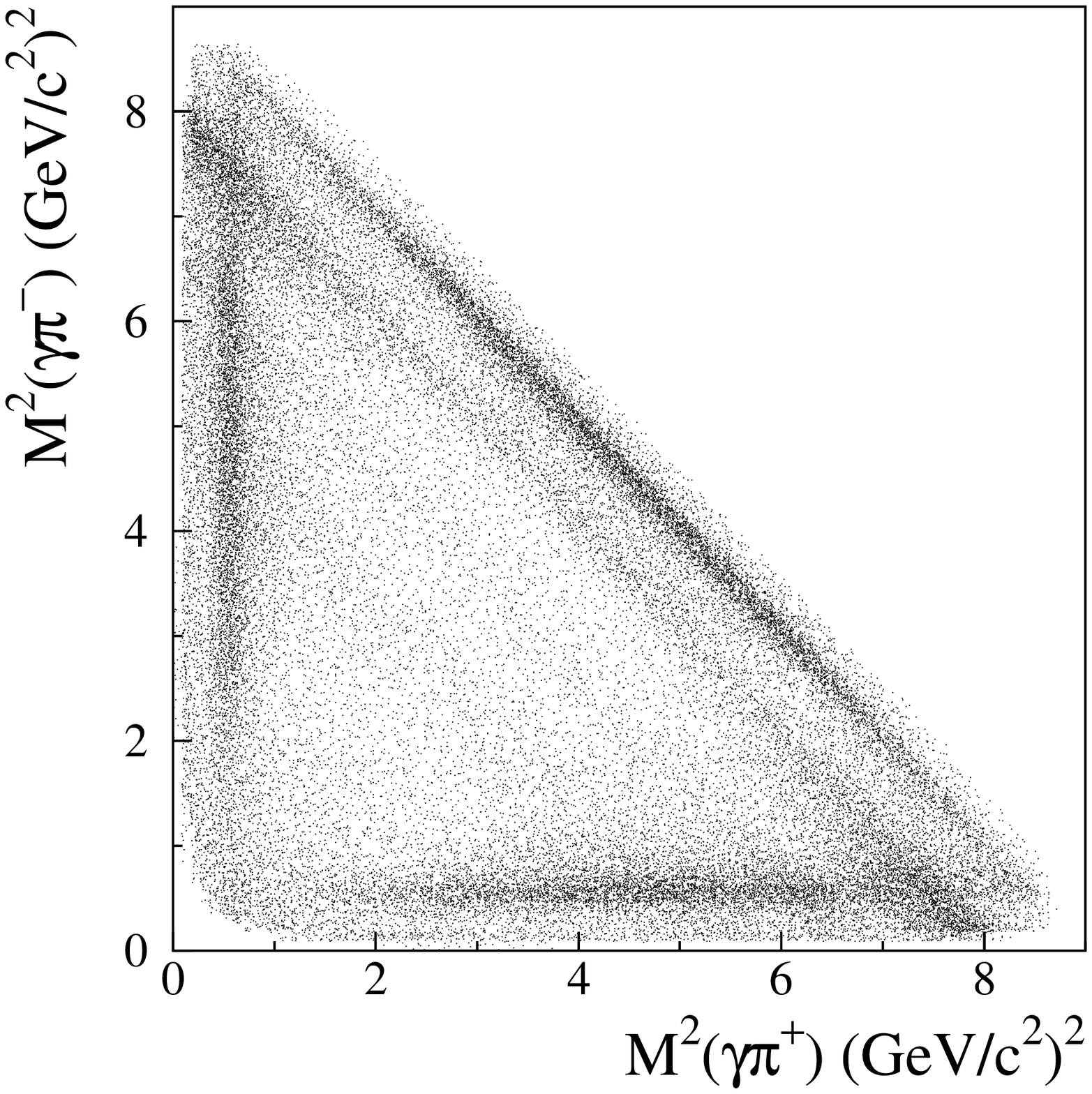,width=6.5cm,height=5.5cm}}
   \caption{Invariant mass spectrum of $\pi^+\pi^-$ and the Dalitz
plot for $J/\psi\to\gamma\pi^+\pi^-$, where the lightly and dark shaded
histograms in the upper panel correspond to $J/\psi\to\pi^+\pi^-\pi^0$
and other estimated backgrounds, respectively.}
   \label{gpipimass}
   \end{figure}

\begin{figure}[htbp]
    \centerline{
    \psfig{file=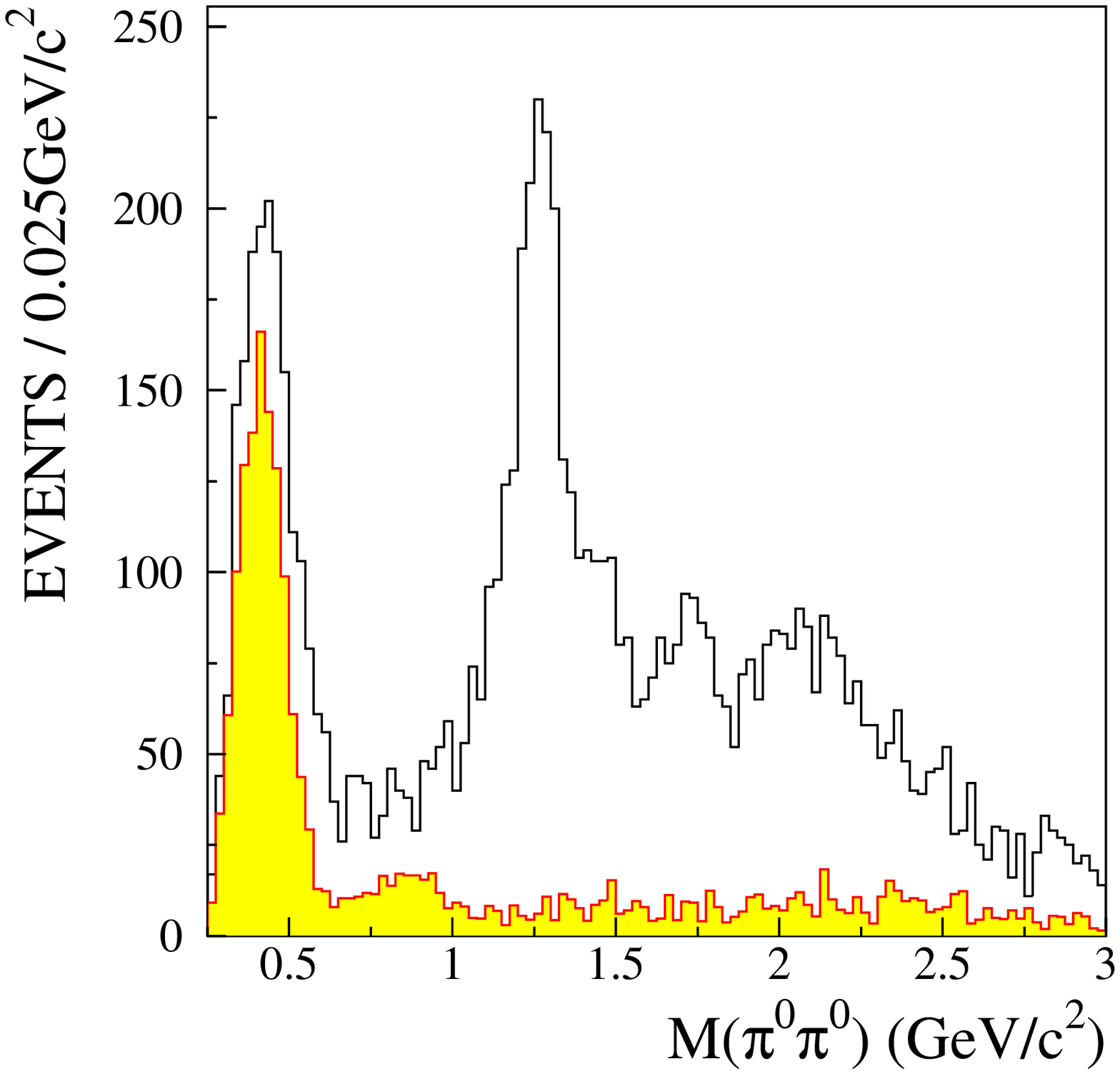,width=6.5cm,height=5.5cm}
    \psfig{file=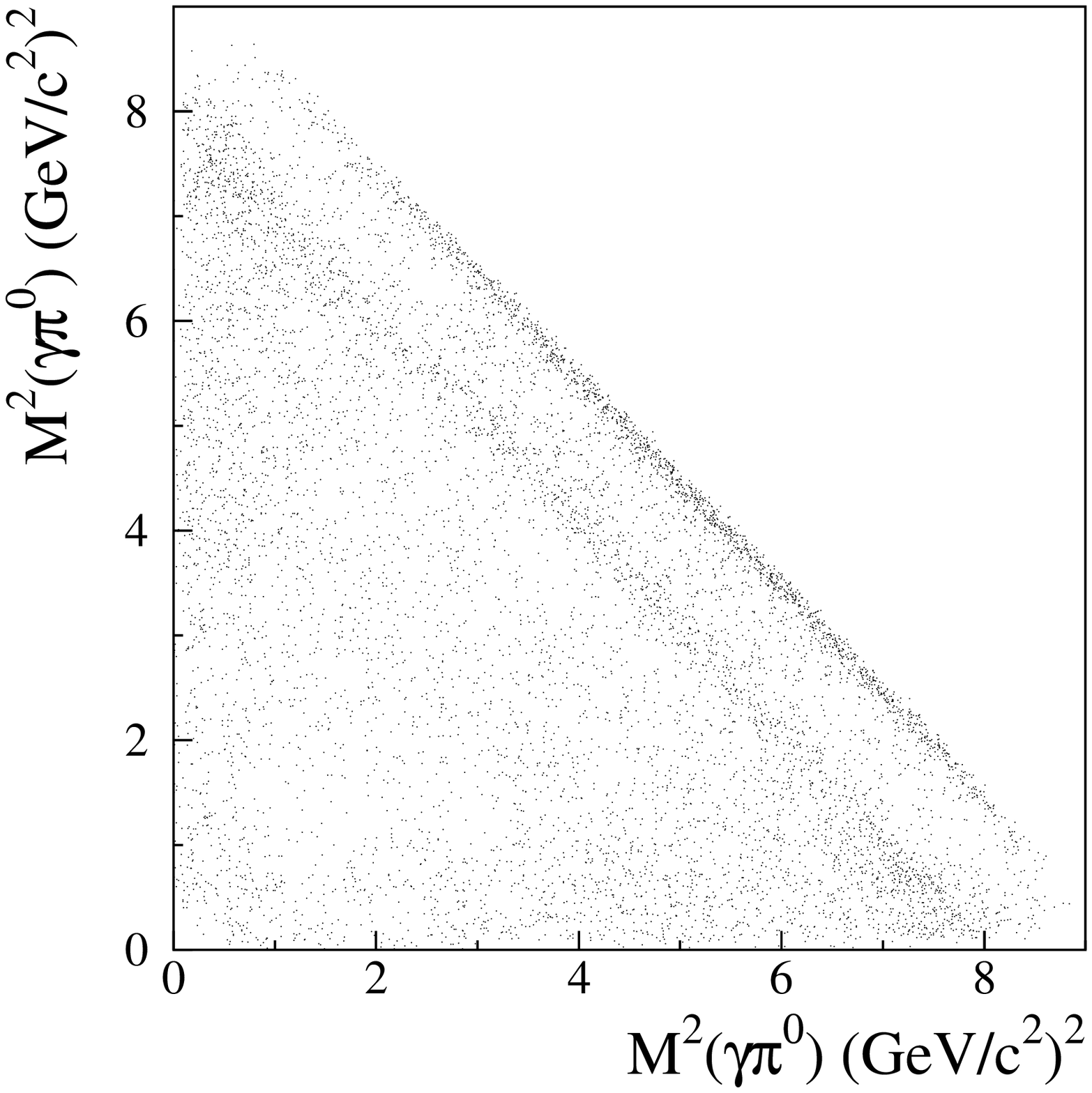,width=6.5cm,height=5.5cm}}
   \caption{Invariant mass spectrum of $\pi^0\pi^0$ and the Dalitz
plot for $J/\psi\to\gamma\pi^0\pi^0$, where the shaded histogram
in the upper panel corresponds to the estimated backgrounds.}
   \label{gpi0pi0mass}
   \end{figure}

Partial wave analyses 
(PWA) of the 1.0~to~2.3~GeV $\pi \pi$ mass range
were carried out using the relativistic covariant tensor 
amplitude method. 
There are conspicuous peaks due to $f_2(1270)$ and two 
$0^{++}$ states in the 1.45 and 1.75~GeV mass regions. The 
first $0^{++}$ state,  which is
considered to be the $f_0(1500)$,
has a mass of $1466\pm 6\pm 20$~MeV, a 
width of $108{^{+14}_{-11}}\pm 25$~MeV, and a branching 
fraction ${\cal B}(J/\psi\to \gamma f_0(1500) \to\gamma 
\pi^+\pi^-) = (0.67\pm0.02\pm0.30) \times 10^{-4}$.
Spin 0 is strongly preferred over spin 
2. Figure~\ref{fm1cr} shows the $\pi^+ \pi^-$ and  $\pi^0 \pi^0$ invariant 
mass distributions for $J/\psi \to \gamma \pi^+ \pi^-$ and 
$\gamma \pi^0 \pi^0$, where the PWA fit projection is shown as a 
histogram. 
 
\begin{figure}[htbp] 
    \centerline{ 
    \psfig{file=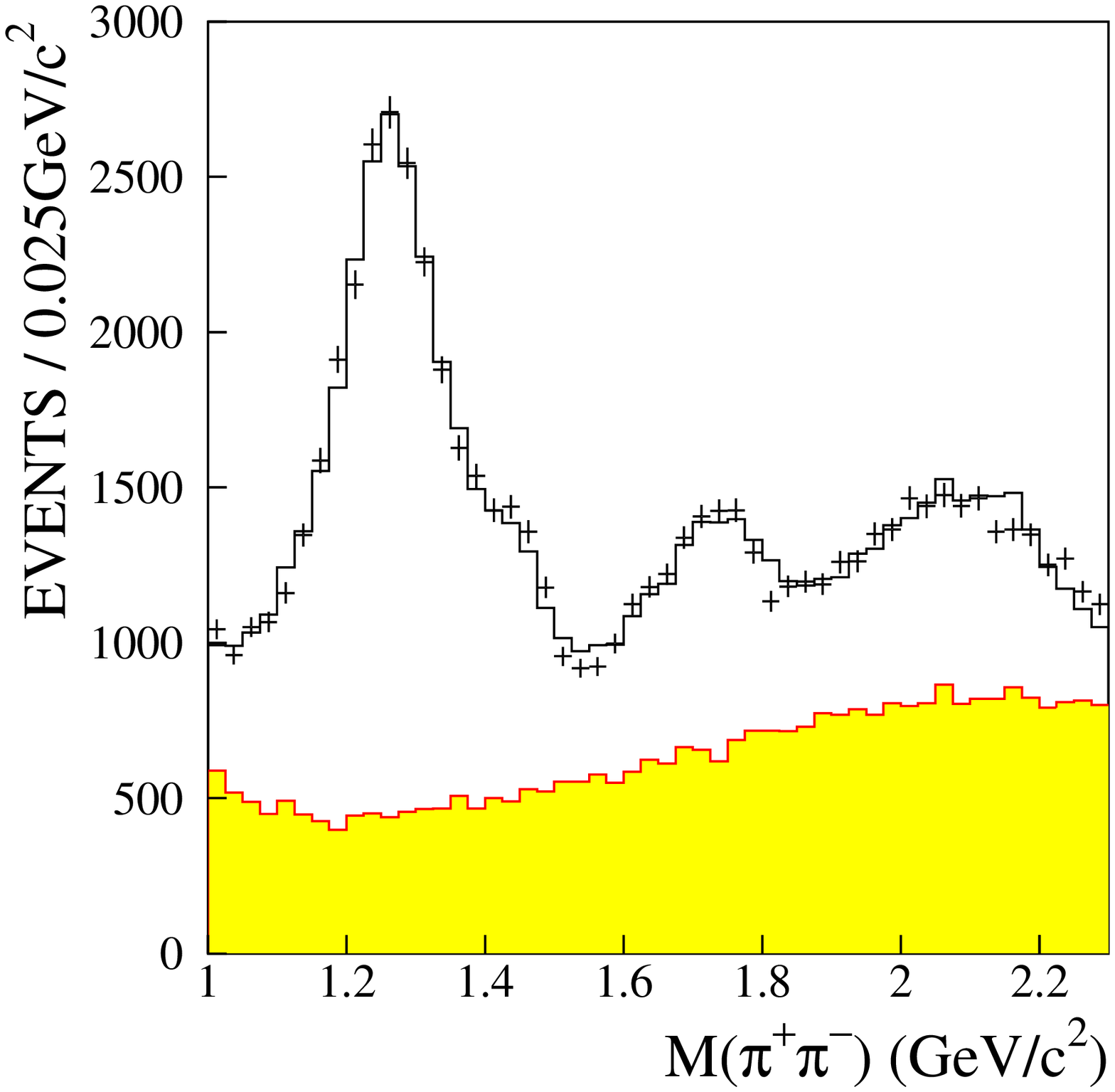,width=6.5cm,height=5.5cm} 
    \psfig{file=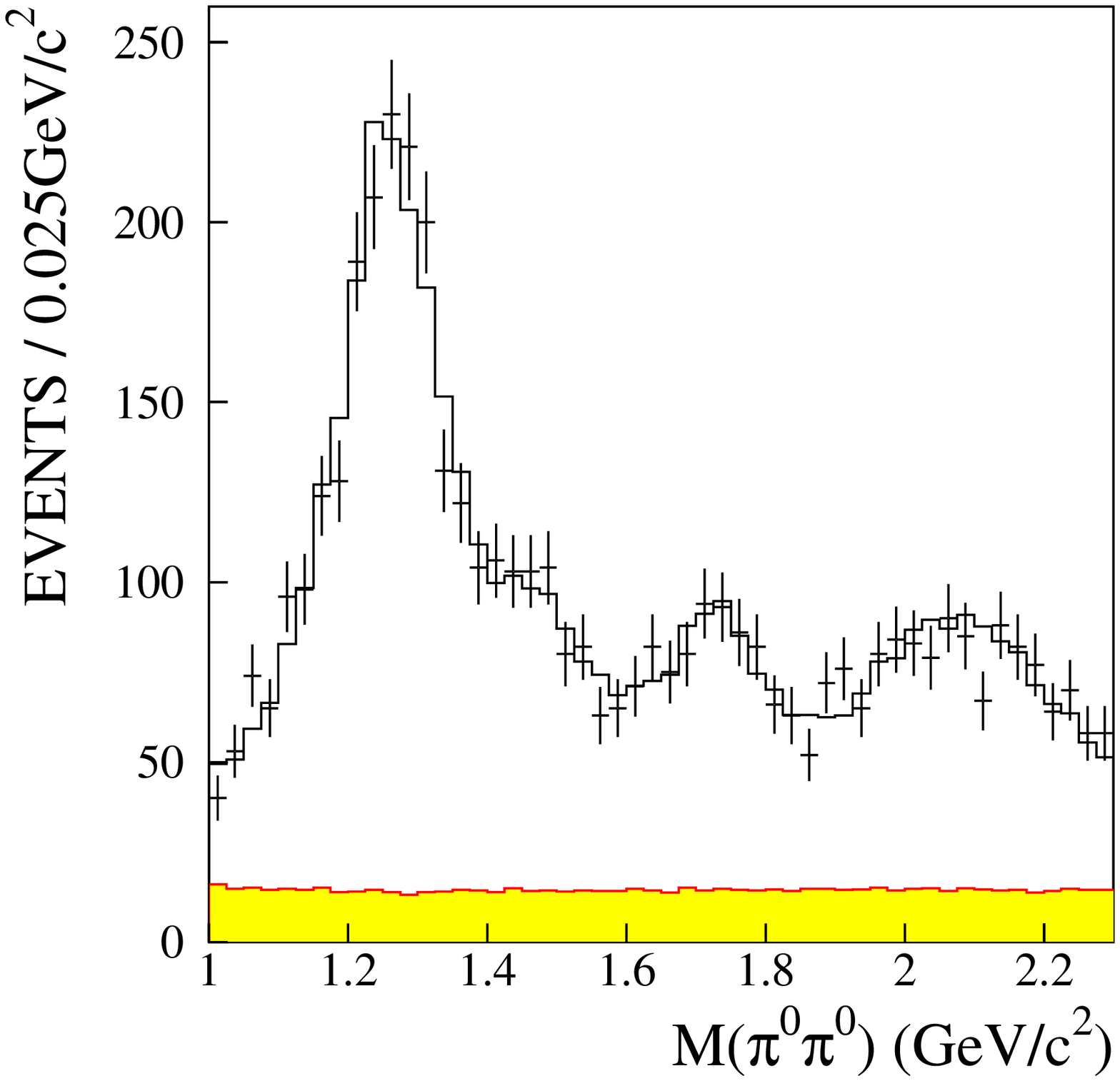,width=6.5cm,height=5.5cm}} 
    \caption{The $\pi^+\pi^-$ invariant mass distribution from 
$J/\psi \to \gamma \pi^+\pi^-$. The crosses are data, the full 
histogram shows the PWA fit projection, and the shaded histogram 
corresponds to the background.} 
    \label{fm1cr} 
    \end{figure} 
 
In contrast to the $f_0(1370)$, the $f_0(1500)$ is not directly
observed in $J/\psi$ hadronic decays, such as 
$J/\psi\to \phi K\bar{K}, \ \omega K\bar{K}, \
\phi\pi\pi, \ \omega\pi\pi$ at BES~\cite{bes-phipipi,bes-wkk}. In most
of the above production channels, the $f_0(1500)$ turns to have a
larger branching fraction to $\pi\pi$ than $K\bar{K}$. However, recent 
Belle data on $B^+\to K^+ K\bar{K}$ and 
$K^+\pi\pi$~\cite{f01500-belle} suggest 
the existence of a
$0^{++}$ scalar at 1.5 GeV with a larger branching fraction to $K\bar{K}$ 
than to $\pi\pi$.

Searching for the $f_0(1500)$ in more decay modes and 
studying its spin-parity will be crucial for clarifying its nature. \\

{\bf 3). $f_0(1710)$ \\} 
 
The $f_0(1710)$ is a main competitor of the $f_0(1500)$ for the 
lightest $0^{++}$ glueball assignment, primarily due to its large 
production 
rate in gluon rich processes such as $J/\psi$ radiative decays, 
$pp$ central production etc., as well as the predictions of lattice QCD.
However, there has been a long history of uncertainty about the 
properties of the $f_0(1710)$. Table~\ref{f01710}~\cite{pdg2006}
lists results on the $f_0(1710)$ from different 
experiments before the BESII era. Apparently, different experiments 
give different masses, widths and spin-parities. The latest analysis
of the MarkIII $J/\psi \to \gamma K \bar K$ data by 
Dunwoodie~\cite{dunwoodie} 
favors $J^P = 0^+$ over an earlier assignment of $2^+$.  The 
latest central production data of WA76 and WA102 
($pp \to p(K\bar{K})p$) also favor $0^+$ \cite{wa76,wa102-2} as does 
the BESI $J/\psi \to \gamma 4\pi$ data~\cite{bes-4pi}.
The spin-parity of $f_0(1710)$ in these observed processes is crucial for 
clarifying whether the $f_0(1710)$ is a $q \bar q$ or a non-$q \bar q$ 
state.  If $J=0$, the $f_0(1710)$ and $f_0(1500)$ might represent 
the glueball and the $q \bar q$ state, or more likely each is a 
mixture of both. However, if $J=2$, it will be difficult to assign 
a glueball status to an $f_2(1710)$, since this mass would be at odds 
with  all current lattice gauge calculations. 
 
To study the structure around 1.7~GeV,
partial wave analyses (PWA) of $J/\psi \to \gamma K \bar K$, 
$\gamma \pi \pi$, $\omega \pi^+ \pi^-$, $\omega K^+ K^-$, $\phi 
\pi^+ \pi^-$ and $\phi K^+ K^-$ were carried out using
the $5.8 \times 10^7 J/\psi$ events collected at BESII. 
 
Figure~\ref{gkkb} shows the $K^+K^-$ and $K^0_SK^0_S$ mass spectra for
selected $J/\psi \to \gamma K \bar K$ events, together with the 
corresponding background distributions. These two mass
spectra agree closely below 2.0 GeV. The resonant structure in the
1.7~GeV mass region  is clearly visible in both decay
modes, in addition there is a strong signal for the $f'_2(1525)$.

A partial wave analyses, using relativistic covariant
tensor amplitudes constructed from Lorentz-invariant combinations of
the 4-vectors and the photon polarization for $J/\psi$ initial states
with helicity $\pm 1$, was carried out for $K \bar K$ masses below 
2.0 GeV. Both bin-by-bin and global fits were performed and consistent
results were obtained. The analyses show that spin 0 is strongly preferred 
over spin 2 for the resonance around 1.7~GeV. 
The $f_0(1710)$ peak mass is $1740\pm 
4^{+10}_{-25}$~MeV  and the width is
$166{^{+5}_{-8}}{^{+15}_{-10}}$ MeV. 
 
\begin{figure} 
 \centerline{\epsfig{file=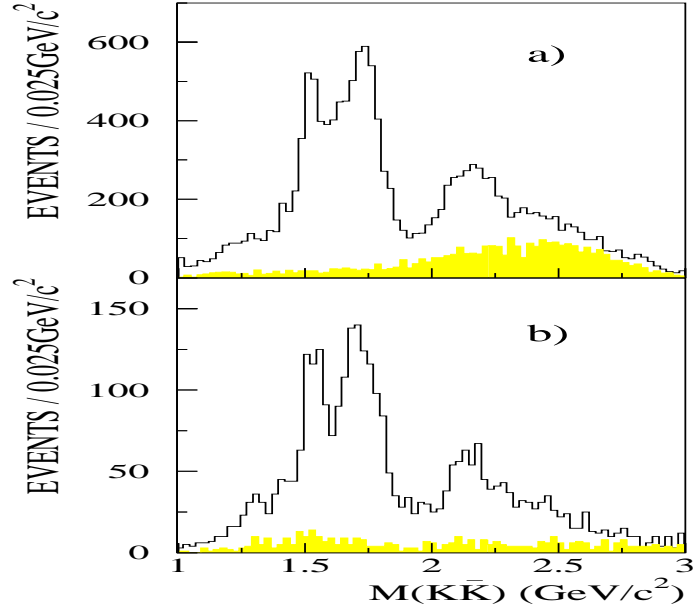,width=90mm,height=8.0cm}} 
\caption{Invariant mass spectra of a) $K^+K^-$, 
b) $K^0_SK^0_S$ for $J/\psi \to \gamma K \bar K$ events, where the 
shaded histogra correspond to the estimated background 
contributions.} 
  \label{gkkb} 
\end{figure} 
 
In $J/\psi \to \omega K^+ K^-$ decays, the $K^+ K^-$ invariant mass 
shows a conspicuous signal for the $f_0(1710)$, as is shown in Fig. 
\ref{wkk}. 
A partial wave analysis determines its mass and width to be: 
$M = 1738 \pm 30$~MeV and $\Gamma = 125 \pm 20$~MeV. 

\begin{figure}[htbp]
\begin{center}
\epsfig{file=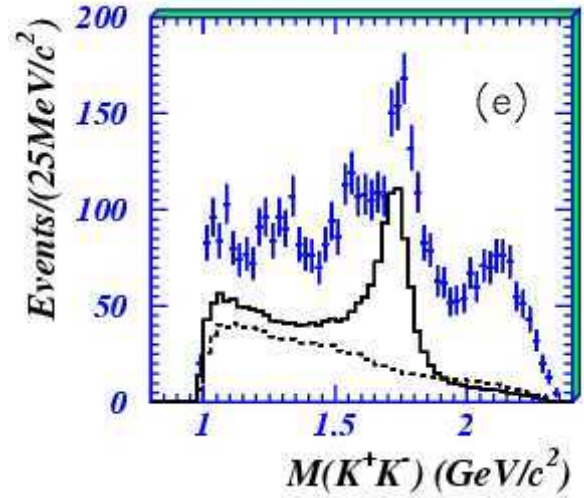,width=8.0cm}
\caption{$K^+K^-$ invariant 
mass spectrum in $J/\psi \to \omega K^+ K^-$. The crosses are data 
and the histogram is PWA fit projection for the $0^{++}$ amplitude.} 
\label{wkk} 
\end{center}
\end{figure} 

In the spectrum of $\pi^+ \pi^-$ invariant masses
recoiling against the $\phi$
in $J/\psi\to \phi \pi^+\pi^-$ decays, shown in Fig. \ref{phipp}, a 
peak at around 1.79~GeV is evident. No peak in this region is seen
in the $K^+K^-$ spectrum for the $J/\psi \to \phi K^+ K^-$ channel. 
A simultaneous PWA fit to $J/\psi \to \phi K^+ K^-$ and 
$\phi \pi^+ \pi^-$ shows that this peak has 
a mass and width of $M = 1790 ^{+40}_{-30}$~MeV and 
$\Gamma= 270^{+60}_{-30}$~MeV, and spin 0 is preferred over spin 
2. This state, the $f_0(1790)$, is distinct from the $f_0(1710)$, seen in 
the $J/\psi \to \gamma K^+ K^-$ and $\omega K^+ K^-$ channels, because
of its different mass, width and decay or production branching
fractions to $\pi\pi$ and $KK$.
 
For $J/\psi \to \gamma \pi \pi$, the production of a $0^{++}$ state is 
also observed in the 1.7~GeV mass region. Its mass and width are 
$1765^{+4}_{-3}\pm13 $~MeV and $145\pm8\pm69$~MeV. 
This $0^{++}$ state may be the $f_0(1710)$ that is observed in 
$J/\psi \to \gamma K \bar K$ and $\omega K^+ K^-$, it may be 
the $f_0(1790)$ which is seen 
in $J/\psi \to \phi \pi^+ \pi^-$, or it may be a superposition 
of these two states. 
 
\begin{table}
\begin{center} 
\caption{ Results of $f_0(1710)$ from different experiments before the BESII era.} 
\begin{tabular}{c|c|c|c|c} 
\hline 
&&&&\\ 
{ Process }&{ Collaboration}&{$M($MeV$)$}&{$\Gamma($MeV$)$}&{$J^{PC}$}\\ 
&&&&\\ 
\hline 
&&&&\\ 
{    $J/{\psi}~{\rightarrow}~\gamma \eta \eta$}&C.~B.(82)&{$1640 
\pm 
50$}&$200^{+100}_{-70}$&$2^{++}$\\ 
&&&&\\ 
{ 
$\pi^{-}p~{\rightarrow}~K^{0}_{S}K^{0}_{S}n$}&BNL(82)&$1771^{+77}_{-53}$&$200^{+156}_{-9}$&$0^{++}$\\ 
&&&&\\ 
{   $\pi^{-}N~{\rightarrow}~K^{0}_{S}K^{0}_{S}n$}&FNAL(84)&$1742 
\pm 
15$&$57\pm38$&-----\\ 
&&&&\\ 
{   $\pi^{-}p~{\rightarrow}~\eta\eta N$}&GAMS(86)&$1755 \pm 
8$&$<50$&$0^{++}$\\ 
&&&&\\ 
{   $J/{\psi}~{\rightarrow}~\gamma K^{+} K^{-}$}&MARK3(87)&$1720 
\pm 
14$&$130\pm20$&$2^{++}$\\ 
&&&&\\ 
{   $J/{\psi}~{\rightarrow}~\gamma K^{+} K^{-}$}&DM2(88)&$1707 \pm 
10$&$166\pm33$&-----\\ 
&&&&\\ 
{   $pp~{\rightarrow}~p(K^+K^-)p$}&WA76(89)&$1713 \pm 
10$&$181\pm30$&$2^{++}$\\ 
~~~~{${\rightarrow}~p(K^0_S K^0_S)p$}&&$1706\pm 10$&$104\pm 30$&\\ 
&&&&\\ 
{   $J/{\psi}~{\rightarrow}~\gamma K\bar{K}$}&MARK3(91)&$1710 \pm 
20$&$186\pm30$&$0^{++}$\\ 
&&&&\\ 
{   $p\bar{p}~{\rightarrow}~\pi^0\eta\eta$}&E760(93)&$1748 \pm 
10$&$264\pm25$&$even^{++}$\\ 
&&&&\\ 
{   $J/{\psi}~{\rightarrow}~\gamma 4\pi$}&MARK3 data&$1750 \pm 
15$&$160\pm40$&$0^{++}$\\ 
&D. Bugg(95)&&&\\ 
&&&&\\ 
{   $J/{\psi}~{\rightarrow}~\gamma K\bar{K}$}&MARK3 
data&$1704^{+16}_{-23}$&$124^{+52}_{-44}$&$0^{++}$\\ 
&Dunwoodie&&&\\ 
&&&&\\ 
{   $pp~{\rightarrow}~p(K\bar{K})p$}&WA102(99)&$1730 \pm 
15$&$100\pm25$&$0^{++}$\\ 
&&&&\\ 
{   $J/{\psi}~{\rightarrow}~\gamma 
4\pi$}&BES(2000)&$1740^{+20}_{-25}$&$135^{+40}_{-25}$&$0^{++}$\\ 
 
&&&&\\ 
\hline 
\end{tabular} 
\end{center}
\label{f01710} 
\end{table} 

More recently, BESII reported the observation of 
a near-threshold enhancement in the $\omega \phi$
invariant mass spectrum from the OZI suppressed decays of $J/\psi
\to \gamma \omega \phi$ \cite{bes-gwf}.
Figure~\ref{gwf}(a) shows the $K^+K^-\pi^+\pi^-\pi^0$ invariant
mass distribution for events with $K^+K^-$ invariant mass within the
$\phi$ mass range ($|m_{K^+K^-}-m_{\phi}|<15$~MeV) and
the $\pi^+\pi^-\pi^0$ mass within the $\omega$ mass range
($|m_{\pi^+\pi^-\pi^0}-m_{\omega}|<30$~MeV);  a structure
peaked near the $\omega\phi$ mass threshold is observed. 
The peak is also evident as a band along the upper
right-hand edge of the Dalitz plot in Fig.~\ref{gwf}(b).
No evidence of an enhancement near the
$\omega\phi$ threshold is observed for events from either the
$\omega$ or $\phi$ sidebands. Moreover,
studies using inclusive and exclusive Monte-Carlo samples
show that the $\omega\phi$ threshold enhancement is not 
due to background.

The significance of the $\omega\phi$ threshold enhancement is more
than $10\sigma$. From a partial wave analysis with covariant helicity
coupling amplitudes,  a spin-parity of $0^{++}$ with the $\omega\phi$
system in an $\mathcal{S}$-wave, is favored. The mass 
and width
of the enhancement are determined to be $M = 1812^{+19}_{-26}$ (stat)
$\pm$ 18 (syst) MeV and $\Gamma = 105 \pm 20$ (stat) $\pm$ 28
(syst) MeV, and the product branching fraction is
$\mathcal{B}(J/\psi\to\gamma X)\cdot \mathcal{B}(X\to\omega\phi)$ =
(2.61 $\pm$ 0.27 (stat) $\pm$ 0.65 (syst)) $\times$ $10^{-4}$.  The
mass and width of this state are not compatible with any known scalars
listed in the Particle Data Book~\cite{pdg2006}.

\begin{figure}[htbp]
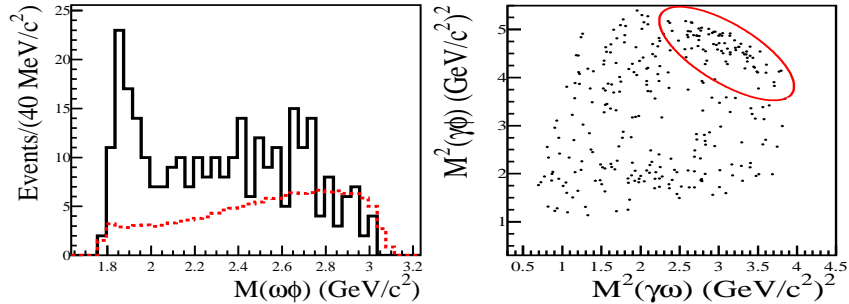

    \centerline{
    \psfig{file=Part3/figures/mwf.epsi,width=5.5cm,height=4.0cm}
    \psfig{file=Part3/figures/daliz.epsi,width=5.5cm,height=4.0cm}}
  \caption{(a) The $K^+K^-\pi^+\pi^-\pi^0$ invariant mass distribution
 for the $J/\psi\to \gamma \omega\phi$ candidate events. The dashed curve
indicates the acceptance varying with the $\omega\phi$
invariant mass. (b) Dalitz plot.}
    \label{gwf}
    \end{figure}

These states have attracted considerable interest and
stimulated much speculation about their underlying 
nature~\cite{bicudo,bugg-06,chao-x1810,hllz,liba,zhao-zou-x1810}. 
They raise essential questions about their production 
mechanisms in $J/\psi$ hadronic and radiative 
decays~\cite{zhao-zou-x1810}. 
 
In additional to the above, recent analyses suggest the existence 
of a broad scalar $f_0(1200-1600)$ with a half width 500-900 
GeV~\cite{anisovich-05,mo}. The existence of such a broad 
state would lead to a rearrangement of the 
assigments of particles to the scalar nonets, where 
$\sigma(600)$ and $\kappa(800)$ are no longer physical states. In 
contrast to two nonets, one above and one below 1~GeV, the new 
arrangement might be: i) $f_0(980), \ f_0(1300), \ a_0(980)$, 
$K_0(1415)$ for the $q\bar{q}$ with radial quantum number $n=1$; 
and ii) $f_0(1500), \ f_0(1750), \ a_0(1520)$,  $K_0(1820)$ for 
$n=2$. The broad $f_0(1200-1600)$ is then regarded as a 
descendant of the scalar glueball. 
 
The light-meson spectroscopy of scalar states in the mass range of
1$\sim$2~GeV/, which has long been a source of controversy, remains
very complicated.  Overlapping states interfere with each
other differently in different production and decay channels.
Therefore, high statistics and high precision experimental data 
that includes many production and decay channels are needed to 
sort out the properties of these scalar states.

In brief, in the 1$\sim$2 GeV mass region, there are at least three 
isoscalar scalars, the $f_0(1370)$, $f_0(1500)$ and $f_0(1710)$,
that are 
well established experimentally.  However, the behaviour of these 
states in different processes has a number of unexpected features, 
which raise questions about their nature. These include: 
What  are the constituent structures of these scalars? 
Are any one of these scalars a glueball state? 
Is the glueball a pure and/or mixed state? 
What and how does the present experimental information 
tell us about scalar production and decay mechanisms? $\dots$ \\
 

\subsubsection{Pseudoscalar glueball candidates} 

The Particle Data Book~\cite{pdg2006}
lists five $0^{-+}$ states above 1~GeV: 
the $\eta(1295)$, $\eta(1405)$, 
$\eta(1475)$, $\eta(1760)$, and $\eta(2225)$. The former three are 
well established in a variety of experimental observations (see review 
by Amsler and Masoni~\cite{pdg2006}), while the latter two states 
need further confirmation. 

Signals for the $\eta(1295)$ were observed in $\pi^- 
p$~\cite{adams01,fukui91c,manak00a}, $\bar{p} p$ 
annihilation~\cite{abele98,anisovich01,amsler04b}, and $J/\psi$ 
radiative decays~\cite{augustin92} in the $\eta\pi\pi$ spectrum 
either via $a_0(980)\pi$ or directly coupled to $\eta\pi\pi$. 
There is no clear signal for the $\eta(1295)$ in $K\bar{K}\pi$, which 
could be a hint for its non-strange $q\bar{q}$ nature. 
 
Historically, there was only one pseudoscalar, the so-called 
$E/\iota(1440)$, observed in $p\bar{p}$ 
annihilation~\cite{baillon67} and $J/\psi$ radiative 
decays~\cite{scharre80,edwards82e,augustin90}. After about 1990, more 
and more observations revealed the existence of two resonance 
structures around 1.45 GeV in 
the $a_0(980)\pi$, $K\bar{K}\pi$ and 
$K^*\bar{K}$ 
spectra~\cite{rath89,adams01,bai90c,augustin92,bertin95-97,cicalo99,nichitiu02}. 
The lower-mass state, the $\eta(1405)$, has large couplings to 
$a_0(980)\pi$ and $K\bar{K}\pi$, while the higher-mass state, the 
$\eta(1475)$, favors $K^*\bar{K}$. 

The $\eta(1405)$ was confirmed 
by MARKIII~\cite{bolton92b}, Crystal Barrel~\cite{amsler95f} and 
DM2~\cite{augustin90} in its decays into $\eta\pi\pi$; its
production has been seen
in both $J/\psi$ radiative decays and $\bar{p}p$ annihilations. 
 In contrast, although the $\eta(1475)$ has been observed in $K\bar{K}\pi$ 
($K^*\bar{K}$)~\cite{rath89,adams01,bai90c,augustin92,bertin95-97,cicalo99,nichitiu02}, 
signals for it are not seen in $\eta\pi\pi$. A study 
of $K\bar{K}\pi$ and $\eta\pi\pi$ production in $\gamma\gamma$ 
collisions~\cite{acciarri01g} showed that the $\eta(1475)$ appeared in 
$K\bar{K}\pi$, but not in $\eta\pi\pi$, while the $\eta(1405)$ was
not seen in either channel. 
 
A reasonable assignment for these three states, the $\eta(1295)$, 
$\eta(1405)$ and $\eta(1475)$, was proposed by Close and 
Kirk~\cite{close-kirk-97}, and Barnes {\it et 
al.}~\cite{barnes97}, who suggested that the $\eta(1295)$ is the 
radial excitation of the $\eta^\prime$. Because of its degeneracy with the
$\pi(1300)$, the $\eta(1295)$ should be dominantly $n\bar{n}$ and hence 
strongly coupled to $\eta\pi\pi$ (it has not been seen in 
$K\bar{K}\pi$). As a result, its $I=0$ partner should be mainly 
$s\bar{s}$ due to ideal mixing. Notice that the $\eta(1405)$ is 
not seen in $\gamma\gamma$ while the $\eta(1475)$ appears in 
$K\bar{K}\pi$ but not $\eta\pi\pi$. This observation leads to the 
identification of the $\eta(1475)$ as a $s\bar{s}$ state and the 
$\eta(1405)$ as a $0^{-+}$ glueball 
candidate~\cite{close-kirk-97,barnes97,cfl}. 
 
The above classification was questioned by Klempt who pointed out
that the absence of  $\eta(1295)$ production in $\gamma\gamma$ collisions 
made it hard to assign it to a $q\bar{q}$ state~\cite{klempt-ichep-04}.
Applying a quark model approach by Barnes {\it et
al.}~\cite{barnes97}, Klempt showed that the three $\eta$
resonances could be due to a single state. The wavefunction
overlap between the initial and final states can give rise to different
invariant mass distributions for $\eta^*\to a_0(980)\pi$,
$\sigma\eta$ and $K^*\bar{K}$ and, thus, result in the different peak
positions that have been interpreted as signals for different
states. In this scheme, the $\eta(1440)$ exists as the radial
excitation of the $\eta$ in the range from 1.3 to 1.5 GeV, while that
for the $\eta^\prime$ is identified as the $\eta(1760)$.

To clarify the nature of these $\eta$ resonances,
higher-statistics searches in $\gamma\gamma$ collisions and $J/\psi$
radiative decays have been strongly urged. In a recent analysis by the
BESII experiment~\cite{bes-04-ggv}, the $\eta(1295)$ is observed in
$J/\psi\to \gamma \eta(1295)\to \gamma (\gamma \rho)$, but absent in
$J/\psi\to \gamma(\gamma\phi)$. Meanwhile, another state, the
$\eta(1424)$, is seen in $J/\psi\to \gamma X$ with $X\to
\gamma\rho$, but is absent in its decays into $\gamma\phi$.
This seems to favor the interpretation by Klempt, but is still not
conclusive.

The $\eta(1760)$ was reported by the MARK III collaboration in
$J/\psi$ radiative decays and was found to decay to
$\omega\omega$~\cite{ww-mark} and $\rho\rho$~\cite{rr-mark}.  It was
also observed by the DM2 collaboration in $J/\psi$ radiative decays in
the $\rho\rho$ decay mode with a mass of $M$ = 1760 $\pm$ 11~MeV
and a width of $\Gamma$ = 60 $\pm$ 16 MeV~\cite{rr-dm} and in the
$\omega\omega$ decay mode~\cite{ww-dm}. The BESI experiment reported
its $\eta\pi^+\pi^-$ decay with a mass of $M$ = 1760 $\pm$ 35
MeV, but without a determination of its width~\cite{bes-etapipi}.  Also,
the possible production of a pseudoscalar $\phi\phi$ threshold enhancement
in $\pi^-p$ scattering has been reported~\cite{etkin}. The $\eta(1760)$
was suggested to be a $3^1S_0$ pseudoscalar $q\bar{q}$ meson, but some
authors suggest a mixture of glueball and $q\bar{q}$ or a
hybrid~\cite{page-li,wu}. Recently, in Ref.~\cite{liba-2}, it was argued that
the pseudoscalar glueball may be in the 1.5 to 1.9~GeV mass
region, and that it would also have Vector Vector decay modes. 

The decay channel $J/\psi\to\gamma\omega\omega$, $\omega\to
\pi^+\pi^-\pi^0$ was analyzed by BESII, using a sample of 
5.8 $\times$ 10 $^7$ $J/\psi$ events \cite{bes2-gww}. 
The histogram in Fig.~\ref{gww}(a) shows the  $2(\pi^+ \pi^- \pi^0)$
invariant mass distribution of events when both $
\pi^+ \pi^- \pi^0$ masses are within the $\omega$ range ($|m_{
\pi^+\pi^-\pi^0 }-m_{\omega}| <$40 MeV). 
The $\omega\omega$
invariant mass distribution peaks at 1.76 GeV, just above the
$\omega\omega$ mass threshold. The phase space 
distribution and the acceptance versus $\omega \omega$ invariant mass
are also shown in the figure. The corresponding Dalitz plot is
shown in Fig. \ref{gww}(b). The shaded histogram in Fig.~\ref{gww}(c)
indicates the background.

\begin{figure}[htbp]
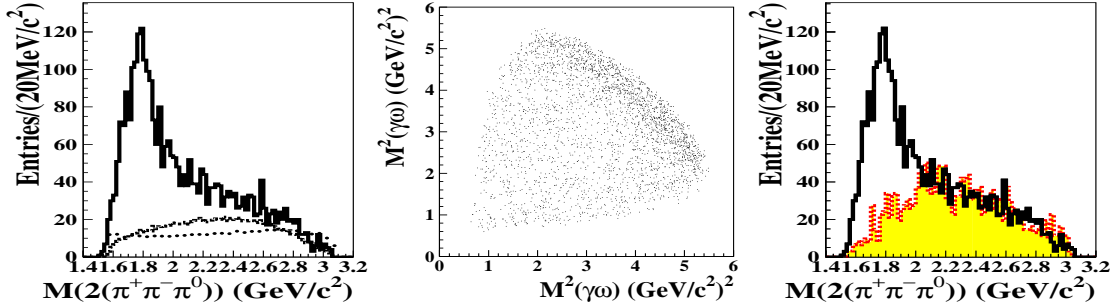

    \centerline{
    \psfig{file=Part3/figures/fig2a.epsi,width=4.8cm,height=4.0cm}
    \psfig{file=Part3/figures/fig2b.epsi,width=4.8cm,height=4.0cm}
    \psfig{file=Part3/figures/fig2c.epsi,width=4.8cm,height=4.0cm}}
    \caption{ (a) The  $2(\pi^+ \pi^- \pi^0)$
invariant mass distribution for candidate events. The dashed
curve is the phase space invariant mass distribution, and the dotted curve
shows the acceptance versus the $\omega\omega$
invariant mass. (b) The Dalitz plot. (c) The shaded histogram is the
background.} 
\label{gww}
\end{figure}

Analysis of angular correlations indicates
that the $\omega\omega$ system below 2~GeV is predominantly
pseudoscalar. A partial wave analysis confirms the predominant
pseudoscalar nature, together with small $0^{++}$ and $2^{++}$
contributions, and yields a pseudoscalar mass $M$ = 1744 $\pm$ 10 (stat)
$\pm$ 15 (syst) MeV, a width $\Gamma$ = $244^{+24}_{-21}$
(stat) $\pm$ 25 (syst) MeV, and a product branching fraction
Br($J/\psi\to\gamma\eta(1760)$) $\cdot$ Br($\eta(1760)\to
\omega\omega$) = (1.98 $\pm$ 0.08 (stat) $\pm$ 0.32 (syst)) $\times$
$10^{-3}$. The $\eta(1760)$ observed here is broader than the one 
listed in the PDG~\cite{pdg2006}. 

To identify the pseudoscalar glueball and clarify the pseudoscalar 
spectrum, further studies of the $\eta(1295)$, $\eta(1405)$,
$\eta(1475)$, $\eta(1760)$ and other high mass $0^{-+}$ states are needed
with a high statistics data sample.\\

\subsubsection{Tensor glueball candidates}
 
Lattice QCD predicted the $2^{++}$ tensor glueball to be the second 
lowest-mass glueball state with a mass around 2.3 GeV, which makes it 
interesting to search for it in experiments. 
 
Mark III first presented signals for a narrow state ($\Gamma \sim 
20$ MeV), the so-called $\xi(2230)$ or $f_J(2220)$, at 2.2 GeV in 
$J/\psi\to \gamma 
K^+K^-$~\cite{mark-III}, and later in $\gamma K_s^0 
K_s^0$~\cite{mark-III-2}. However, there was no clear signal seen 
at DM2 in the same decay channels~\cite{dm2-xi}. In hadron scattering 
experiments, the GAMS Collaboration found a structure 
in the $\eta\eta^\prime$ invariant mass spectrum
at $m=2220\pm 10 $~MeV with a width of $\Gamma\sim 80$ MeV in $\pi^- p\to 
\eta\eta^\prime n$ interactions at 38~GeV and 100~GeV~\cite{gams}, 
while  the LASS group reported a structure at 
2.2 GeV in the $K_s^0 K_s^0$ invariant mass spectrum 
for $K^- p\to K_s^0 K_s^0\Lambda$~\cite{lass}. 
  
Further evidence for this state was 
reported by BESI based on a $\sim 8 \times 
10^{6} \ J/\psi$ event sample. 
Structures near 2220~MeV were reported in $\pi^+\pi^-$, 
$K^+ K^-$, $K_s^0 K_s^0$, $p\bar{p}$, and 
$\pi^0\pi^0$~\cite{bes-96a,bes-98a}. In addition, stringent limits
were placed 
on the two-photon coupling of the $\xi(2230)$~($f_J(2220)$)
by CLEO collaboration studies of the reactions 
$\gamma \gamma \to K_s K_s$~\cite{cleo97} and 
$\gamma \gamma \to \pi^+ \pi^-$~\cite{cleo98}.
The copious production of $\xi(2230)$ in $J/\psi$
radiative decays, its narrow width and small two-photon
coupling suggested it be the lightest tensor glueball candidate.
However, the $\xi(2230)$ was neither seen in the inclusive $\gamma$ 
spectrum by the Crystal Ball collaboration \cite{cbl} nor in $p \bar p$ 
annihilations
to $K_s^0K_s^0$~\cite{evangelista-97}, $\eta\eta$ and
$\pi^0\pi^0$~\cite{cb-97}.

The $\xi(2230)$~($f_J(2220)$) was not observed in the mass spectrum of
$K^+K^-$, $\pi^+\pi^-$ or $p \bar p$ in the higher-statistics
$5.8\times 10^{7}$ $J/\psi$ event sample at BESII.
However, preliminary results from
a partial wave analysis (PWA) of $J/\psi \to \gamma K^+K^-$
at BESII showed that it is difficult to exclude
the existence of the $\xi(2230)$; a $4.5 \sigma$
significance signal with mass, width and 
product branching fraction consistent with the
BESI results was found~\cite{liaohb}.
More careful studies are needed to draw firm conclusions on
the $\xi(2230$ at \bes3.

No other tensor glueball candidates have been observed in
radiative $J/\psi$ decays in the mass range favored by LQCD. 
If the $\xi(2230)$~($f_J(2220)$) does indeed not exist, one of the 
following must be true: 1) The LQCD
prediction on the tensor glueball mass is unreliable; 2) The tensor
glueball production rate in an particular
exclusive mode is very low;
3) The glueball width is wide.  Thus, in order
to confirm whether or not a tensor glueball exists,
more experimental and theoretical efforts are needed.
Reliable theoretical predictions on the production rates of glueballs in
$J/\psi$ radiative decays and  their decay properties 
would be useful.

\subsection{Hunting for glueballs at \bes3}

The peak luminosity of BEPCII is designed to be
$10^{33}$ $cm^{-2}s^{-1}$ at the center-of-mass energy around the
$\psi(3770)$ peak; scaling from this we determine an expectation
for the luminosity at the $J/\psi$ peak that is 
about 60\% of the peak design value. If the average luminosity is 
assumed to be half of the peak
luminosity and the effective running time for data accumulation is
$10^7$~s/year, the  3400 nb peak cross section for $J/\psi$ production
translates into about 10~billion $J/\psi$ events accumulated
in a one year data run.   Compared to previous
exposures, this data sample is huge.  (This is nearly 200 times
as large as the BESII 58 million $J/\psi$ event sample.) 
Moreover, the new \bes3 detector has  much better
performance characteristics than those of
previous $e^+e^-$ detectors that operated at the $J/\psi$ peak.
The large data sample and excellent detector peformance 
will make possible studies of light hadron spectroscopy
and searches for new hadron states with sensitivities
that far exceed those of previous measurements.

Glueballs are expected to be copiously produced in radiative $J/\psi$
decays. The inclusive photon spectrum from radiative $J/\psi$ decays can be
used to search for new states,  {\it e.g.,} glueballs,
especially when these states have a
relatively large production rate in radiative $J/\psi$ decays.

The study of exclusive $J/\psi$ radiative and hadronic decays,
as well as two-photon processes will also provide important information
for the identification of the glueballs. As suggested by Close,
the decays of $J/\psi \to \gamma \gamma \rho$ and $\gamma \gamma \phi$
can act as flavor filters that can be used to tag the flavor content
of mesons that decay to $\gamma\gamma$.

As briefly summarized in Section~\ref{sec:glueball-cands}, 
the existence of at least three $I=0$ scalar mesons
in the $1\sim 2$~GeV mass range --- the $f_0(1370)$, $f_0(1500)$
and $f_0(1710)$ --- is well established. However the nature 
of these states remains a mystery. Recently, two 
additional  scalar meson candidates, the
$f_0(1790)$ and $f_0(1810)$, were reported by BESII and
require comfirmation.

With a high statistics $J/\psi$ data sample taken with a high performance
detector, the confirmation of these new states 
and the determination of their
quantum numbers via PWA, their masses, widths and
decay branching ratios, as well as systematic studies
of the $f_0(1370)$, $f_0(1500)$ and $f_0(1710)$ produced
in both  radiative and hadronic decays to
$PP$ ($P$ denotes a pseudoscalar meson), $VV$ ($V$ denotes a
vector meson), {\it etc} is necessary and possible.
This will help clarify the scalar  situation. In addition,
studies
of these states in high statistics two-photon data samples will be crucial
for determining their gluonic content through the determination of their
two-photon couplings.

Five pseudoscalar states above 1 GeV: the $\eta(1295)$, $\eta(1405)$,
$\eta(1475)$, $\eta(1760)$, and $\eta(2225)$, are listed in the
PDG \cite{pdg2006}. Of these, some are well established, while
others need further confirmation. The nature of these states is still
controversial, as is discussed above in Sect.~\ref{sec:glueball-cands}.
With \bes3 data,
the decays $J/\psi \to \gamma \eta \pi \pi$, $\gamma \eta' \pi \pi$,
$\gamma K \bar K \pi$, $\gamma V V$ {\it etc.,} and their corresponding
hadronic decays can be investigated, which will help identify the
pseudoscalar glueballs and 
eventually understand the pseudoscalar spectrum.\\

\subsubsection{The tensor glueball candidate in  
$J/\psi \to \gamma \eta \eta$ and $\gamma \eta \eta'$}

In order to investigate the \bes3 sensitivity for searching for the
$2^{++}$ glueball candidate $f_J(2220)$, we did a simulation
study of $J/\psi \to \gamma \eta \eta$ and $\gamma \eta \eta'$ decays
where the $\eta$ decays
to $\gamma \gamma$ or $\pi^+ \pi^- \pi^0$ and 
the $\eta'$ to $\eta \pi^+ \pi^-$.
A full Monte-Carlo simulation, based on a \bes3 detector
model in the GEANT4 MC framework, is used.

The final states of the examined channels include five photons or
five photons plus multi-prong charged pions.
In the event selection, all of the candidate events are required to 
satisfy the following common criteria for charged tracks and
photons:  1) all charged tracks are required to be
within the polar angle region of $|\cos\theta|<0.93$, have points of 
closest approach that are within 1 cm of the beam axis
and 5~cm of the center of the interaction point; 
2) a total net charge is zero;
3) each candidate photon is required to have an energy deposit in the
electromagnetic calorimeter that is greater than 40~MeV, to be isolated 
from charged tracks by more
than $20^\circ$ in both the $x-y$ and $r-z$ planes and an angle between
any  other photon in the event that is greater than $7^\circ$. 
Candidate $\eta$ mesons are reconstructed
via their decays to $\gamma \gamma$ or $\pi^+\pi^-\pi^0$,
and $\eta'$ mesons are reconstructed from the decay chain
$\eta' \to \eta \pi^+\pi^-$,
$\eta \to \gamma \gamma$.
A four-constraint (4C) energy-momentum conservation kinematic fit
is applied using the hypothesis of each decay mode. When the number of
photons in an event exceeds the minimum, all combinations are tried, and the
combination with the smallest $\chi^2$ is retained. \\

a). $J/\psi \to \gamma \eta \eta$ \\

In the simulation of $J/\psi \to \gamma \eta \eta$,
we include the processes:
$J/\psi \to \gamma f_0(1500)$, $\gamma f_0(1710)$,
$\gamma f_2(1910)$,
$\gamma f_0(2100)$, $\gamma f_2(2100)$ and $\gamma f_J(2220)$. The
input masses and widths of the resonances are 
taken from ref.~\cite{pdg2006} and listed in
Table~\ref{resonances}.

\begin{table}[htbp]
\begin{center}
\caption{\label{resonances} The masses and widths of the input resonances}
\begin{tabular}{|c|c|c|c|c|c|c|}
\hline

&$f_0(1500)$&$f_0(1710)$&$f_2(1910)$&$f_0(2100)$&$f_2(2150)$&$f_J(2220)$\\
\hline
Mass(MeV)&1507.0&1714.0&1915.0&2103.0&2156.0&2231.0 \\
\hline
Width((MeV)&109.0&140.0&163.0&206.0&167.0&23.0\\
\hline
\end{tabular}
\end{center}
\end{table}
\noindent
In ref.~\cite{pdg2006}, the following branching fractions are
reported:
$$Br(J/\psi \to \gamma f_0(1500))Br(f_0(1500) \to \eta \eta)=
1.84 \times 10^{-5}$$
$$Br(J/\psi \to \gamma f_0(1710))Br(f_0(1710) \to \eta \eta)=
2.88 \times 10^{-5}.$$
Branching ratios not listed in ref.~\cite{pdg2006} are taken to be:
$$Br(J/\psi \to \gamma f_2(1910))Br(f_2(1910) \to \eta \eta) \sim
1 \times 10^{-5}$$
$$Br(J/\psi \to \gamma f_0(2100))Br(f_0(2100) \to \eta \eta) \sim
1 \times 10^{-5}$$
$$Br(J/\psi \to \gamma f_2(2150))Br(f_2(2150) \to \eta \eta) \sim
1 \times 10^{-5}$$
$$Br(J/\psi \to \gamma f_J(2220))Br(f_J(2220) \to \eta \eta) \sim
1 \times 10^{-5}.$$

The  further requirements $|M_{\gamma \gamma}-M_{\eta}|<0.035$ GeV
or $|M_{\pi^+\pi^-\pi^0}-M_{\eta}|<0.030$~GeV are applied
for $J/\psi \to \gamma \eta \eta$, $\eta \to \gamma \gamma$ or
$\eta \to \pi^+\pi^-\pi^0$, respectively.
Table~\ref{reso-eff} lists the mass resolutions and efficiencies for each
resonance region for both of the
$\eta$ decay modes after the application of the
above-listed selection criteria. For the case of
$\eta \to \pi^+\pi^-\pi^0$, there are charged pions in the final states
and, therefore, the mass resolution is better than that for the
$\eta \to \gamma \gamma$ case where there are only neutral tracks.
However, the selection efficiency for this mode is very low.

\begin{table}[htbp]
\begin{center}
\caption{\label{reso-eff}
Mass resolutions and efficiencies for $J/\psi \to \gamma \eta
\eta$}
\begin{tabular}{|c|c|c|c|}
\hline
process  & $\eta$ decay mode & mass resolution (MeV)& efficiency(\%)
\\
\hline
$J/\psi \to \gamma f_0(1500) \to \gamma \eta \eta$
& $\eta \to \gamma \gamma$ & $18.4 \pm 0.2$ & 25.7 \\ \cline{2-4}
& $\eta \to \pi^+\pi^-\pi^0$ & $10.9 \pm 0.4$ & 1.56 \\ \hline
{$J/\psi \to \gamma f_0(1710) \to \gamma \eta \eta$}
& $\eta \to \gamma \gamma$ & $20.1 \pm 0.2$ & 25.9 \\ \cline{2-4}
& $\eta \to \pi^+\pi^-\pi^0$ & $9.7 \pm 0.6$ & 1.71 \\ \hline
{$J/\psi \to \gamma f_2(1910) \to \gamma \eta \eta$}
& $\eta \to \gamma \gamma$ & $21.3 \pm 0.3$ & 25.9 \\ \cline{2-4}
& $\eta \to \pi^+\pi^-\pi^0$ & $10.7 \pm 0.5$ & 1.73 \\ \hline
{$J/\psi \to \gamma f_0(2100) \to \gamma \eta \eta$}
& $\eta \to \gamma \gamma$ & $22.0 \pm 0.2$ & 26.5 \\ \cline{2-4}
& $\eta \to \pi^+\pi^-\pi^0$ & $10.5 \pm 0.4$ & 1.37 \\ \hline
{$J/\psi \to \gamma f_2(2150) \to \gamma \eta \eta$}
& $\eta \to \gamma \gamma$ & $22.2 \pm 0.2$ & 26.5 \\ \cline{2-4}
& $\eta \to \pi^+\pi^-\pi^0$ & $9.8 \pm 0.4$ & 1.44 \\ \hline
{$J/\psi \to \gamma f_J(2230) \to \gamma \eta \eta$}
& $\eta \to \gamma \gamma$ & $22.8 \pm 0.3$ & 26.7 \\ \cline{2-4}
& $\eta \to \pi^+\pi^-\pi^0$ & $11.0 \pm 0.3$ & 2.09 \\ \hline

\end{tabular}
\end{center}
\end{table}

The main backgrounds for $J/\psi \to \gamma \eta \eta \to 5 \gamma$
come from $J/\psi \to \omega \eta, \omega \to \gamma \pi^0$ and
$J/\psi\to\gamma\eta,\eta\to\pi^0\pi^0\pi^0$. However, these tend to
accumulate in the high mass region of the $\eta \eta$ invariant mass
spectrum.

Figure~\ref{g2eta-5g-1} shows the $\eta \eta$ invariant mass spectrum
for the incoherent sum of the generated signals and backgrounds
when both $\eta$ candidates decay into $\gamma\gamma$. 
The generated signal and background
events are normalized to a $10^{10}$ $J/\psi$ event sample.

\begin{figure}[htbp]
\begin{center}
\epsfig{file=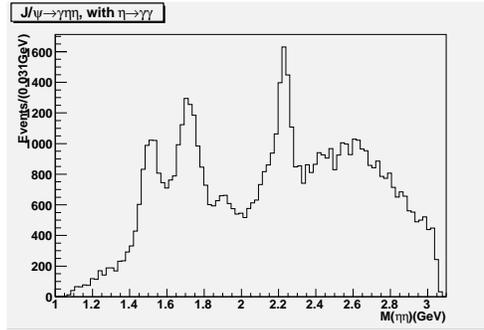,width=6.5cm}
\caption{The $\eta \eta$ invariant mass spectrum in $J/\psi \to
\gamma \eta \eta \to 5 \gamma$. The generated signals and backgrounds
are normalized to $10^{10} J/\psi$ decays and are added incoherently.
\label{g2eta-5g-1}}
\end{center}
\end{figure}

For $J/\psi \to \gamma \eta \eta \to 5 \gamma 2(\pi^+\pi^-)$, the main
contamination to the $\eta \eta$ spectrum is from $J/\psi \to
2(\pi^+\pi^-)3 \pi^0$. The $\eta \eta$ invariant mass spectrum
for the incoherent sum of the generated signals and backgrounds
with $\eta$ decaying into $\pi^+\pi^-\pi^0$ is shown in
Fig.~\ref{g2eta-3pi-1}. The generated signal and background events
are normalized to $10^{10}$ total $J/\psi$ events.

\begin{figure}[htbp]
\begin{center}
\epsfig{file=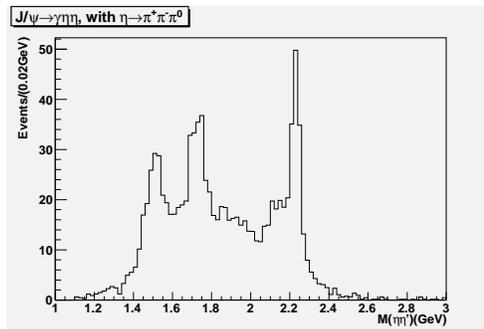,width=6.5cm}
\caption{The $\eta \eta$ invariant mass spectrum in $J/\psi \to
\gamma \eta \eta \to 5 \gamma 2(\pi^+\pi^-)$. The generated signals
and backgrounds
are normalized to $10^{10} J/\psi$ events and are added incoherently.
\label{g2eta-3pi-1}}
\end{center}
\end{figure}


A significant $f_J(2220)$ signal is seen in both the $\eta \to
\gamma\gamma$ and  $\pi^+\pi^-\pi^0$ modes.
The Breit-Wigner fit to the $\eta \eta$ invariant mass spectra 
yields a statistical significance for the $f_J(2220)$ 
signal that is larger than $7\sigma$ in both cases.

If we take the product branching fraction for 
$J/\psi \to \gamma f_J(2220),
f_J(2220) \to \eta \eta$ to be $0.5 \times 10^{-5}$ and that for
$J/\psi \to \gamma f_J(2150), f_J(2150) \to \eta \eta$ to be
$1 \times 10^{-5}$, the statistical significance of 
the $f_J(2220)$ is found to be larger than $7\sigma$ for the
both the $\eta \to 2 \gamma$ and 
$\eta \to \pi^+\pi^-\pi^0$ modes.
The $\eta \eta$ invariant mass spectra are shown in Fig. \ref{g2eta-5g-2}
for $\eta \to 2\gamma$ and Fig. \ref{g2eta-3pi-2} for
$\eta \to \pi^+ \pi^- \pi^0$, respectively.
If we assume the product branching fraction for
$J/\psi \to \gamma f_J(2220),
f_J(2220) \to \eta \eta$ is $ 0.5 \times 10^{-6}$ and that for
$J/\psi \to \gamma f_2(2150), f_2(2150) \to \eta \eta$ is
$3 \times 10^{-5}$, the statistical significance of the
$f_J(2220)$ signal
is larger than $7~\sigma$ for the $\eta \to 2 \gamma$ and $4.3~\sigma$ for
$\eta \to \pi^+\pi^-\pi^0$ modes.

\begin{figure}[htbp]
\centering
\includegraphics[width=0.5\textwidth]
{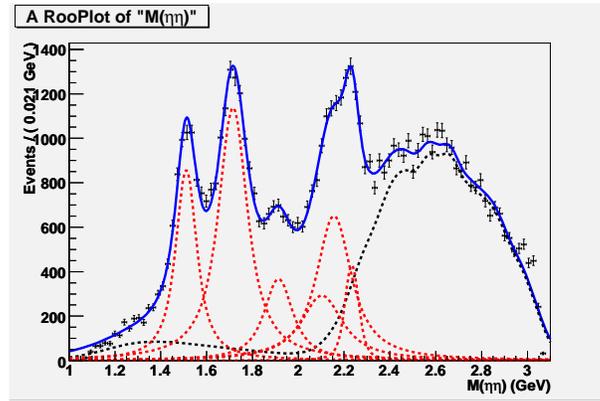}
\caption{Fit result after adding all resonances together.
($J/\psi\to\gamma\eta\eta,
    \eta\to\gamma\gamma,\eta\to\gamma\gamma$ channel).
\label{g2eta-5g-2}}
\end{figure}

\begin{figure}[htbp]
\centering
\includegraphics[width=0.5\textwidth]
{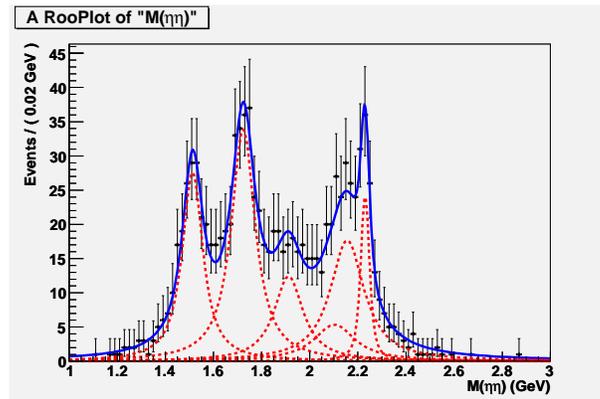}
\caption{Fit result after adding all resonances
together ($J/\psi\to\gamma\eta\eta,
    \eta\to\pi^+\pi^-\pi^0,\eta\to\pi^+\pi^-\pi^0$ channel).
\label{g2eta-3pi-2}}
\end{figure}

Table~\ref{input-output} shows a comparison of the input
resonances' masses, widths and branching ratios with those obtained
from the fit, when the product branching fraction for 
$J/\psi \to \gamma
f_J(2220)$, $f_J(2220) \to \eta \eta$ is taken to be $0.5 \times 10^{-5}$
and $J/\psi \to \gamma f_2(2150)$, $f_2(2150) \to \eta \eta$ to be
$1 \times 10^{-5}$. The fitted results are consistent with the input
values.

From the simulation, one can see that if the branching ratio of
$f_J(2220)$ is large enough compared with that for the
nearby resonance, {\it e.g.} the $f_2(2150)$, the $f_J(2220)$
can be clearly seen in the mass spectrum.
If $f_J(2220) \to \eta \eta$ is $ 0.5 \times 10^{-6}$ and
$J/\psi \to \gamma f_2(2150), f_2(2150) \to \eta \eta$ is
$3 \times 10^{-5}$, then the broad $f_2(2150)$ may 
interfere with the observation
of the $f_J(2220)$. In this case, a partial wave
analysis (PWA) that determines the contribution of each
spin-parity component will be needed to distinguish the
$f_J(2220)$ and $f_2(2150)$. \\

\begin{table}[htbp]
  \centering
  \begin{tabular}{|c|c|c|c|c|c|c|}
\hline
          &       \multicolumn{2}{c|}{Mass(MeV)}     &
\multicolumn{2}{|c|}{Width(MeV)}     &
\multicolumn{2}{|c|}{Branching ratio($\times 10^{-7}$)}  \\
\hline
          &     input  & fit   &    input  & fit   &
input  & fit                     \\
\hline
   $f_0(1500)$   &   $1507.0$    &   $1512.8\pm 3.8$   &    $109.0$    &
$97.4\pm 8.2$       &  $14.43$   &  $14.67\pm 0.94$
\\
\hline
   $f_0(1710)$   &   $1714.0$    &   $1723.8\pm 3.5$   &    $140.0$    &
$117.2\pm 10.4$       &  $22.58$   &  $22.75\pm 1.47$
\\
\hline
   $f_2(1910)$   &   $1915.0$    &  $1912.2\pm 1.1$    &    $163.0$    &
$143.8\pm 8.2$       &  $7.8$   &   $7.98\pm 0.93$
\\
\hline
   $f_0(2100)$   &   $2103.0$    &   fixed   &    $206.0$    &    fixed
&  $7.8$   &  $6.59\pm 2.62$            \\
\hline
   $f_2(2150)$   &   $2156.0$    &   fixed   &    $167.0$    &    fixed
&  $7.8$   &  $7.98\pm 2.72$            \\
\hline
   $f_J(2220)$   &    $2231.1$   &  $2230.8\pm 2.7$    &    $23.0$     &
$27.6\pm 7.2$       &   $3.9$  &  $4.57\pm 0.71$
\\
\hline
  \end{tabular}
\caption{Input and output comparison of mass, width and branching ratio in
the simulation. Here, the masses and widths of $f_0(2100)$ and $f_2(2150)$
are
fixed. ($J/\psi\to\gamma\eta\eta,
    \eta\to\pi^+\pi^-\pi^0,\eta\to\pi^+\pi^-\pi^0$ channel).
\label{input-output}}
\end{table}

b). $J/\psi \to \gamma \eta \eta'$ \\

According to some theoretical calculations, the $f_J(2220)$ may have a
large decay branching fraction to $\eta \eta'$. Therefore we performed
a full Monte-Carlo simulation to investigate the
sensitivity of the observation of $f_J(2220)$ in this decay channel.
Here, only the case where the $\eta'$ is reconstructed through
$\eta' \to \eta \pi^+ \pi^-$ and the $\eta$ through $\eta \to \gamma \gamma$
is considered, corresponding to a final state with
five photons and one $\pi^+ \pi^-$ pair.
To select the $\eta$ and $\eta'$, $|M_{\gamma
\gamma}-M_{\eta}|<0.035$~GeV
and $|M_{\pi^+\pi^-\eta}-M_{\eta'}|<0.040$~GeV are required.
The resonances included in the simulation are the same as
those used  the in $J/\psi \to \gamma \eta \eta$ simulation.
Table~\ref{g2etap-reso-eff} lists the mass resolutions and efficiencies
in each resonance region for $J/\psi \to \gamma \eta \eta'$.

\begin{table}[htbp]
\begin{center}
\caption{Mass resolutions and efficiencies for $J/\psi \to \gamma \eta
\eta'$.\label{g2etap-reso-eff}}
\begin{tabular}{|c|c|c|}
\hline
process  & mass resolution (MeV)& efficiency(\%)
\\
\hline
$J/\psi \to \gamma f_0(1500) \to \gamma \eta \eta'$
& $14.2 \pm 0.4$ & 4.5 \\ \hline
$J/\psi \to \gamma f_0(1710) \to \gamma \eta \eta'$
& $16.7 \pm 0.5$ & 4.8 \\ \hline
$J/\psi \to \gamma f_2(1910) \to \gamma \eta \eta'$
& $17.2 \pm 0.4$ & 4.9 \\ \hline
$J/\psi \to \gamma f_0(2100) \to \gamma \eta \eta'$
& $18.6 \pm 0.4$ & 5.6 \\ \hline
$J/\psi \to \gamma f_2(2150) \to \gamma \eta \eta'$
& $18.1 \pm 0.4$ & 5.7 \\ \hline
$J/\psi \to \gamma f_J(2220) \to \gamma \eta \eta'$
& $19.0 \pm 0.5$ & 6.1 \\ \hline

\end{tabular}
\end{center}
\end{table}

The main backgrounds for $J/\psi \to \gamma \eta \eta' \to 5 \gamma
\pi^+ \pi^-$ 
come from $J/\psi \to \omega \eta, \omega \to \pi^+\pi^- \pi^0$,
$\eta \to \gamma \gamma$, $J/\psi \to \omega \pi^0$, $\omega \to
\pi^+ \pi^- \pi^0$ and $J/\psi \to \omega \eta', \omega \to \pi^+\pi^-
\pi^0, \eta \to \gamma \gamma$.

According to the Particle Data Book~\cite{pdg2006},
$$Br(J/\psi \to \gamma f_0(1500))Br(f_0(1500) \to \eta \eta') \sim
1.8 \times 10^{-5}$$
$$Br(J/\psi \to \gamma f_0(1710))Br(f_0(1710) \to \eta \eta') \sim
1.8 \times 10^{-5}.$$
\noindent
Other branching fractions not listed in ~\cite{pdg2006} 
are taken to be:

$$Br(J/\psi \to \gamma f_2(1910))Br(f_2(1910) \to \eta \eta') \sim
1 \times 10^{-5}$$
$$Br(J/\psi \to \gamma f_0(2100))Br(f_0(2100) \to \eta \eta') \sim
1 \times 10^{-5}$$
$$Br(J/\psi \to \gamma f_2(2150))Br(f_2(2150) \to \eta \eta') \sim
1 \times 10^{-5}$$
$$Br(J/\psi \to \gamma f_J(2220))Br(f_J(2220) \to \eta \eta') \sim
1 \times 10^{-5}.$$

Figure~\ref{g2etap} shows the $\eta \eta'$ invariant mass spectrum
for the incoherent sum of the generated signals and backgrounds, with the
branching fraction assumptions listed above.
The generated signal and background
events are normalized to $10^{10}$  $J/\psi$ events.
The background level in the 2220~MeV region is very low and the
$f_J(2220)$ signal can be clearly seen.

\begin{figure}[htbp]
\begin{center}
\epsfig{file=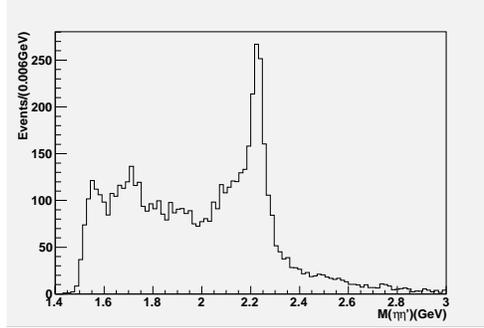,width=6.5cm}
\caption{The $\eta \eta' $ invariant mass spectrum in $J/\psi \to
\gamma \eta \eta' \to 5 \gamma \pi^+\pi^-$. The generated signals
and backgrounds
are normalized to $10^{10} J/\psi$ events and are added incoherently.
\label{g2etap}}
\end{center}
\end{figure}

If a different branching fraction is assumed for the  $f_2(2150)$, say
$$Br(J/\psi \to \gamma f_2(2150))Br(f_2(2150) \to \eta \eta') \sim
3 \times 10^{-5},$$ the resulting $\eta \eta'$ invariant mass spectrum
for the incoherent sum of the generated signals and backgrounds
looks like that shown in Fig.~\ref{g2etap-2}.

\begin{figure}[htbp]
\begin{center}
\epsfig{file=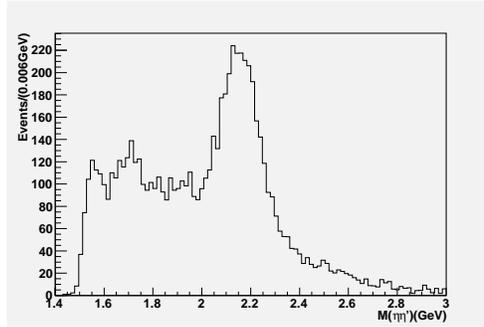,width=6.5cm}
\caption{The $\eta \eta$ invariant mass spectrum in $J/\psi \to
\gamma \eta \eta' \to 5 \gamma \pi^+\pi^-$. The generated signals
and backgrounds
are normalized to $10^{10} J/\psi$ events and are added incoherently.
\label{g2etap-2}}
\end{center}
\end{figure}

In summary, the decays of $J/\psi \to
\gamma \eta \eta$ and $\gamma \eta \eta'$ are studied based on
a full Monte-Carlo simulation of \bes3. 
From the simulation, we determine that the
sensitivity of searches for the tensor
glueball $f_J(2220)$ at\bes3 not only depends on the production rate
of $f_J(2220)$ in $J/\psi$ radiative decays and its decay branching ratios
to the examined final states, but also on the production
and decays of nearby resonances.  In some cases, a partial wave
analysis will be needed to resolve ambiguities. \\

\subsubsection{Inclusive photon spectrum}

The inclusive photon spectrum from radiative $J/\psi$ decays can be used
to search for new states, especially when these states have a
relatively large production rate in radiative $J/\psi$ decays, as
is expected for glueballs.

The Crystal Ball experiment presented an inclusive photon
spectrum for radiative $J/\psi$ decays, where the $\eta$, $\eta'$
and a peak corresponding to a recoil mass of around 1440 MeV are clearly
seen \cite{cbl}. However no clear signal for
$J/\psi \to \gamma f_J(2220)$ is observed and no numerical upper
limit on inclusive $f_J(2220)$ production in radiative $J/\psi$
decays is set because of the uncertainties in the photon efficiency
as a function of energy.

A Monte-Carlo study of inclusive radiative $J/\psi$ decays at \bes3
has been
performed. For the selected events, the charged tracks are required to be
within the polar angle region of $|\cos\theta|<0.93$ and to come from
the interaction point. The total net charge of the charged tracks
is required to be zero.
The candidate photon is required to have an energy deposit in the
electromagnetic calorimeter that is greater than 40 MeV and have polar 
and azimuthal opening angles between it
and any charged track that are greater than 
$20^\circ$.  In order to reject photons from $\pi^0$ decays,
it is required that the invariant mass of it and any other photon 
in the event should be greater than 0.2~GeV.
A pairing with an invariant mass between 0.5 and 0.7~GeV
is considered to be an $\eta$ candidate and is also removed.

Figure~\ref{inclusive} shows the results of
a Monte-Carlo study of inclusive radiative
$J/\psi$ decays at \bes3.
Here, the branching ratio of $J/\psi \to \gamma f_J(2220)$ is assumed
to be $Br(J/\psi \to \gamma f_J(2220))=2.5 \times 10^{-3}$.

\begin{figure}[htbp]
\begin{center}
\epsfig{file=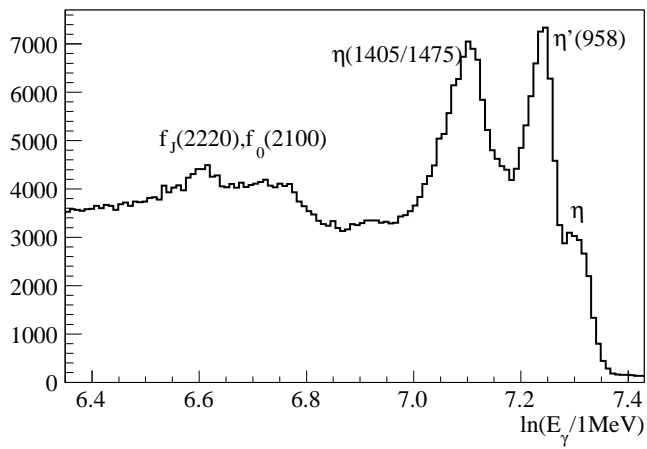,width=10.0cm}
\caption{The inclusive photon spectrum for $J/\psi$ radiative
         decays.\label{inclusive}}
\end{center}
\end{figure}

\section{Hybrid Mesons}

\subsection{Theoretical models for hybrid mesons}

\subsubsection{Large $N_C$ expansion}

Hybrid mesons are hypothesized to be formed from a $q\bar q$ pair plus 
one explicit
gluon field $G$. In the large $N_c$ limit, the amplitude for
creating a hybrid meson from the vacuum has the same $N_c$-order
as that for creating a $q\bar q$ meson \cite{cohen}. If kinematics
and other conservation laws allow, the production cross section for
hybrid mesons is expected to be roughly the same as that of
ordinary mesons. At least it is not suppressed in the large $N_c$
limit. In the same limit, hybrid mesons and ordinary mesons 
with the same quantum numbers can mix freely.  Thus, the
identification of hybrid mesons can be very difficult unless they have
exotic quantum numbers. That is why considerable efforts have been
devoted to the search for $1^{-+}$ hybrid mesons.\\

\subsubsection{Flux tube model}

The flux tube model is based on intuition gained from the strong-coupling
limit of lattice QCD~\cite{IP1,IP2}. In this picture, a meson is
described as a quark-antiquark pair linked by a color flux
tube. The quarks move adiabatically in an effective potential
generated by the dynamics of the flux tube. 
The flux tube can rotate along its axis, but the orbital angular momentum
along the flux tube is zero. When the flux tube is in its ground state, the
excitations of the quark-antiquark pair yields the conventional meson 
spectrum.

Hybrid mesons are defined as excitations of the color flux
tube. The lowest-lying exotic hybrid meson is predicted to have
quantum numbers $J^{PC}=1^{-+}$ and a mass around 1.9 GeV~\cite{IP2}, 
consistent with  predictions from
lattice QCD.  In the flux tube model, the
decay of a hybrid is triggered by the breaking the
flux tube~\cite{IP1}. In this breaking process, a quark-antiquark
pair is  created with spin $S_{q\bar q}=1$,
orbital angular momentum $L_{q\bar q}=1$ and total angular
momentum $J_{q\bar q}=0$, a process called `$^3P_0$ pair
creation'. In this picture, a state with spin $S=0$
can not decay into two $S=0$ states \cite{page}. 
In the flux tube model, the quark-antiquark pair of the lowest-lying 
$1^{-+}$ hybrid is in a spin singlet and, thus, cannot decay
into a pair of spin zero mesons, such as $\pi\pi$ and $\pi\eta$.
In addition, when a single
flux tube  breaks into two flux tubes (two
mesons), the relative coordinate of the two final flux tubes
(the line that connects the centers of the two flux tubes) is parallel
to the original one (denoted as ${\bf r}$). As a result, the two 
final-state mesons that materialize from the two tubes
cannot absorb the unit of string angular momentum about the ${\bf
r}$ axis \cite{IP1,page} as relative orbital angular
momentum. Therefore a $1^{-+}$ hybrid (with one excited phonon
polarized along the flux tube) cannot decay into two ground
states, such as $\pi\pi$, $\pi\eta$ and $\pi\rho\cdots$. The
preferential decay modes are those with one excited meson, such as
$b_1\pi$, $f_1\pi\cdots$. This selection rule can be violated when
the two final ground states have different spatial wave functions
({\it i.e.,} different spatial size). One calculation shows
that the partial width for $1^{-+}\rightarrow\rho\pi$ can be large
and compatible with $\pi_1(1600)$ being a hybrid if the $\pi$
shrinks to a point~\cite{close}.

The original flux tube model (the IKP model) was modified by the
introduction of a new decay vertex that is constructed using
the heavy quark expansion of the Coulomb-gauge QCD Hamiltonian to
identify relevant operators~\cite{swanson1,swanson2}. This new
model (the PSS model), 
which is an extension of IKP, states
that the decay amplitude for a hybrid meson vanishes when the
daughter mesons are identical. This means that not only
$S$- plus $S$-wave final states are forbidden but also
$P$- plus $P$-wave final states, and the preferred decay channels 
are $S$- plus $P$-wave  pairs. The predicted
partial widths for $1^{-+}$ hybrid meson to each channel
differ as shown in Table \ref{fluxtube} \cite{swanson2}, 
where the mass of the $1^{-+}$ hybrid is taken to be 1.6~GeV. It
should be noted that in PSS the width to $\rho\pi$ is larger than that
for $f_1\pi$. An extensive study of the decay patterns of  
hybrid mesons with other quantum numbers in the flux tube  
model have been performed by Page, Swanson and 
Szczepaniak~\cite{swanson2}.

In Tables~\ref{part3:tab:t1} through ~\ref{part3:tab:t16}, 
collected from ref.~\cite{swanson2}, the 
dominant widths for hybrid decays $H \rightarrow A B$ for various
$J^{PC}$ hybrids in a partial wave $L$ are presented.
In each table, column 1 indicates the $J^{PC}$ of the
hybrid, column 2 the decay mode and column 3, $L$. 
In columns 4, 5, 6 and 8, the predictions of this model are indicated. Column
6 uses the `standard parameters' used throughout ref.~\cite{swanson2} and 
are
defined in the Appendix of ref.~\cite{swanson2}. Column 5 uses the same parameters, 
except that all
hybrids are assumed to be $0.2$ GeV heavier (and the $c\bar{c}$ hybrids
$0.3$ GeV heavier to put them above the $D^{**}D$ thresholds
at approximately $4.3$ GeV). Column 4 uses
so-called `alternative parameters.' Comparisons of columns
4 and 6 can, therefore,
be used to estimate the parameter sensitivity of
the predictions. For hybrid decays to two ground state $S$-wave mesons,
the `reduced width' is indicated in column 8. This is the width divided by
the dimensionless ratio
$(\beta_A^2-\beta_B^2)^2/(\beta_A^2+\beta_B^2)^2$,
where $\beta$ is the inverse radius of the SHO wave function
\protect\cite{page95hybrid}.
This gives a measure of how strong the decay is with the 
wave function dependence explicitly removed. In column 7, the IKP
model predictions for the `standard parameters' are given. Thus,
columns 6
and 7 should be compared to see differences between the PSS  
and IKP model predictions.

\begin{table}
\begin{center}
\caption{Decay widths of the $1^{-+}$ hybrid meson from the two
flux tube model.\label{fluxtube}}
\begin{tabular}{c|c|c|c|c|c}
  \hline
&$b_1\pi$&$\rho\pi$&$f_1\pi$&$\eta(1295)\pi$&$K^*K$\\
PSS(MeV)&24 &9&5&2&0.8\\
IKP(MeV)&59 &8&14&1&0.4\\
  \hline
\end{tabular}
\end{center}
\end{table}

\begin{table}
\caption{$2^{-+}$ Isovector Hybrid Decay Modes from Ref.
\cite{swanson2}.}
\begin{tabular}{lccccccc}
\hline
\hline
         &                &   &  alt & 2.0 GeV hybrid & standard & IKP & reduced \\
\hline
$2^{-+}$ & $\rho\pi  $     & P &    9 &  16  &  13  &  12  &  57  \\
         & $K^{\ast} K$    & P &    1 &   5  &   2  &   1  &  17  \\
         & $\rho\omega$    & P &    0 &   0  &   0  &   0  &  20  \\
         & $ f_2(1270)\pi  $     & S &   19 &  10  &  9   &  14  &      \\
         &                 & D &   .1 &  .2  & .05  &  11  &      \\
         & $ f_1(1285)\pi  $     & D &   .1 &  .3  & .06  &  \O  &      \\
         & $ f_0(1370)\pi  $     & D &   .02&  .08 & .01  &  .6  &      \\
         & $ b_1(1235)\pi  $     & D & \O   & \O   & \O   &  20  &      \\
         & $ a_2(1320)\eta  $    & S &  --  &  7   &  --  &  --  &      \\
         &                 & D &  --  & .01  &  --  &  --  &      \\
         & $ a_1(1260)\eta  $   & D &   0  & .05  &   0  &   0  &      \\
         & $ a_0(1450)\eta  $    & D & --   &  0   &  --  &  --  &      \\
         & $ K_2^{\ast}(1430) K $       & S & --   & 11   & --   &  --  &      \\
         &                 & D & --   &  0   & --   &  --  &      \\
         & $ K_1(1270) K $       & D & 0    & .01  &  0   & .02  &      \\
         & $ K_0^{\ast}(1430) K $       & D & --   & 0    & --   &  --  &      \\
         & $ K_1(1410) K $       & D & --   & 0    & --   &  --  &      \\
         & $ \rho_1450) \pi $    & P &  .8  &  12  &   3  &   2  &      \\
         & $ K^{\ast}(1410) K   $    & P &   -- &   1  &  --  &  --  &      \\
         & $ \Gamma$       &  &  30   &  63  &  27  & 59   &      \\
\hline

\end{tabular}
\label{part3:tab:t1}
\end{table}

\begin{table}
\caption{$1^{-+}$ Isovector Hybrid Decay Modes from Ref.
\cite{swanson2}.\label{part3:tab:t2}}
\begin{tabular}{lccccccc}
\hline
\hline
         &                &   &  alt & high mass & standard & IKP & reduced \\
\hline

$1^{-+}$ & $ \eta \pi    $ & P &    0 &   .02&   .02& .02  &  99  \\
         & $ \eta^{'} \pi$ & P &    0 &   .01&   .01& 0    &  30  \\
         & $ \rho \pi    $ & P &    9 &  16  &  13  &  12  &  57  \\
         & $ K^{\ast} K  $ & P &    1 &   5  &   2  &   1  &  17  \\
         & $ \rho \omega $ & P &    0 &   0  &   0  &   0  &  13  \\
         & $ f_2(1270)\pi  $     & D & .2   &  .5  &  .1  & \O   &      \\
         & $ f_1(1285)\pi  $     & S &  18  &  10  &  9   &  14  &      \\
         &                 & D &   .06&  .2  & .04  &   7  &      \\
         & $ b_1(1235)\pi      $ & S &  78  &  40  &  37  &  51  &      \\
         &                 & D &   2  &   3  &   1  &  11  &      \\
         & $ a_2(1320)\eta  $    & D & --   & .02  &  --  &  --  &      \\
         & $ a_1(1260)\eta  $    & S & 5    &   7  &   3  &   8  &      \\
         &                 & D & 0    & .01  &  0   &  .01 &      \\
         & $ K_2^{\ast}(1430) K $       & D & --   & 0    & --   &  --  &      \\
         & $  K_1(1270) K $       & S &  4   & 7    &  2   &   6  &      \\
         &                 & D & 0    & .2   & 0    & .04  &      \\
         & $  K_1(1410) K $       & S & --   & 33   & --   &  --  &      \\
         &                 & D & --   &  0   & --   &  --  &      \\
         & $ \pi(1300) \eta $    & P & --   &   5  &  --  &  --  &      \\
         & $ \eta_u(1295) \pi $    & P & 3    &  27  &  11  &   8  &      \\
         & $ K(1460) K    $    & P & --   &  .8  &  --  &  --  &      \\
         & $ \rho(1450) \pi $    & P &  .8  &   12 &    3 &   2  &      \\
         & $ K^{\ast}(1410) K   $    & P &  --  &   1  &  --  &  --  &      \\
         & $\Gamma $       &   & 121  & 168 & 81    &  117 &      \\
\hline

\end{tabular}
\end{table}

\begin{table}
\caption{$1^{--}$ Isovector Hybrid Decay Modes from Ref.
\cite{swanson2}.\label{part3:tab:t3}}
\begin{tabular}{lccccccc}
\hline
\hline
         &                &   &  alt & 2.0 GeV hybrid & standard & IKP &
reduced \\
\hline
$1^{--}$ & $ \omega \pi  $ & P &   9  &  16  &  13  &  12  &  57  \\
         & $\rho\eta     $ & P &   4  &   9  &   6  &   4  &  30  \\
         & $\rho\eta^{'} $ & P &  .1  &   1  &   .2 &  .1  &   1  \\
         & $ K^{\ast} K  $ & P &   3  &   9  &   5  &  3   &  34  \\
         & $ \at\pi  $     & D &  .5  &   2  &   .3 &  16  &      \\
         & $ \aoo\pi  $     & S &   78 &  41  &  37  &  51  &      \\
         &                 & D &   .4 &   .8 &   .2 &  11  &      \\
         & $ \ho\pi  $     & S &      &      &   \O &      &      \\
         &                 & D &      &      &  \O  &      &      \\
         & $ \bo\eta     $ & S &      &      &  \O  &      &      \\
         &                 & D &      &      &  \O  &      &      \\
         & $ \kt K $       & D & --   & 0    & --   &  --  &      \\
         & $ \kll K $       & S &  6   &  12  &  4   &  11  &      \\
         &                 & D &  0   & .01  & 0    &  0   &      \\
         & $ \kh K $       & S & --   &  17  & --   &  --  &      \\
         &                 & D & --   & 0    & --   &  --  &      \\
         & $ \oq \pi     $ & P &  1   &  14  &  4   &  4   &      \\
         & $ \ksq K      $ & P &  --  &   3  &  --  &  --  &      \\
         & $\Gamma $       &  &  103  &  121 &  70  &  112 &      \\
\hline

\end{tabular}
\end{table}

\begin{table}
\caption{$2^{+-}$ Isovector Hybrid Decay Modes from Ref.
\cite{swanson2}.\label{part3:tab:t4}}
\begin{tabular}{lccccccc}
\hline
\hline
         &                &   &  alt & 2.0 GeV hybrid & standard & IKP &
reduced \\
\hline

$2^{+-}$ & $ \omega \pi  $ & D &  .5  &   1  &   1  &   1  &   4  \\
         & $\rho\eta     $ & D &   .1 &  .6  &  .2  &   .1 &  1   \\
         & $\rho\eta^{'} $ & D &   0  &  .02  &   0  &   0  &  0  \\
         & $ K^{\ast} K  $ & D &  .04 &  .2  & .08  & .04  & .6   \\
         & $ \at\pi  $     & P &  .7  &  .9  &  .4  & 130  &      \\
         &                 & F &   0  &  .02 &   0  &  .2  &      \\
         & $ \aoo\pi  $     & P &  3   &  4   &  2   & 45   &      \\
         &                 & F & .01  &  .02 &  0   & .3   &      \\
         & $ \ho\pi  $     & P &  2   &   2  &  1   & 69   &      \\
         &                 & F & .01  & .03  & .01  & .5   &      \\
         & $ \bo\eta  $    & P & .02  &  .5  & .01  & .8   &      \\
         &                 & F &  0   &   0  &   0  &  0   &      \\
         & $ \kt K $       & P & --   & .04  & --   &  --  &      \\
         &                 & F & --   &   0  & --   &  --  &      \\
         & $ \kll K $       & P &  0   &  .03 & 0    & .6   &      \\
         &                 & F &  0   & 0    &  0   &  0   &      \\
         & $ \kh K $       & P & --   & .3   & --   &  --  &      \\
         &                 & F & --   &  0   & --   &  --  &      \\
         & $ \pq \pi  $    & D & .08  &  1   &  .2  & .2   &      \\
         & $ \oq \pi  $    & D & .02  &  .4  &  .04 & .04  &      \\
         & $ \ksq K      $ & D &  --  &  .01 &  --  &  --  &      \\
         & $\Gamma$        &   & 7    &  11 &   5   &  248 &      \\
\hline

\end{tabular}
\end{table}

\begin{table}
\caption{$0^{-+}$ Isovector Hybrid Decay Modes from Ref.
\cite{swanson2}.\label{part3:tab:t5}}
\begin{tabular}{lccccccc}
\hline
\hline
         &                &   &  alt & 2.0 GeV hybrid & standard & IKP &
reduced \\
\hline

$0^{-+}$ & $ \rho \pi    $ & P &  37  &  63  &  51  &  47  & 230  \\
         & $ K^{\ast} K  $ & P &   5  &  18  &  10  &   5  &  69  \\
         & $ \rho \omega $ & P &      &      &  \O  &      &      \\
         & $ f_2(1270)\pi  $     & D &   1  &   3  & .6   & 8    &      \\
         & $ \fz\pi  $     & S &  62  &  40  & 30   & 62   &      \\
         & $ \at\eta  $    & D &  --  &  .1  & --   & --   &      \\
         & $ \az\eta  $    & S &  --  &  4   & --   & --   &      \\
         & $ \kt K $       & D & --   &   .02& --   &  --  &      \\
         & $ \kz K $       & S & --   & 44   & --   &  --  &      \\
         & $ \rqb \pi $    & P &  3   &  47  & 10   & 10   &      \\
         & $ \ksq K   $    & P &  --  &   5  & --   & --   &      \\
         & $\Gamma $       &  &  108  & 224  & 102  & 132  &      \\
\hline

\end{tabular}
\end{table}

\begin{table}
\caption{$1^{+-}$ Isovector Hybrid Decay Modes from Ref.
\cite{swanson2}.\label{part3:tab:t6}}
\begin{tabular}{lccccccc}
\hline
\hline
         &                &   &  alt & 2.0 GeV hybrid & standard & IKP &
reduced \\
\hline

$1^{+-}$ & $\omega\pi  $   & S &   23 &  19  &  26  &  38  & 118  \\
         &                 & D &   .3 &  .8  &  .4  &  .3  &  2   \\
         & $\rho\eta     $ & S &   15 &  21  &  25  &   22 & 118  \\
         &                 & D &  .07 &  .3  &  .1  &  .06 &  .6  \\
         & $\rho\eta^{'} $ & S &   3  &  8   &  5   &   4  & 25  \\
         &                 & D &  0   &  .01 &  0   &  0   &   0  \\
         & $ K^{\ast} K  $ & S &  27  &  52  & 47   & 36   & 339  \\
         &                 & D &  .02 &  .1  & .04  & .02  & .3   \\
         & $ \at\pi  $     & P & 19   &  26  & 10   & 49   &      \\
         &                 & F &  0   &   .02&  0   &  .1  &      \\
         & $ \aoo\pi  $     & P &  9   &   10 &  5   &   29 &      \\
         & $ \az\pi  $     & P &  3   &   6  &  1   &  26  &      \\
         & $ \ho\pi  $     & P &  \O  &  \O  &  \O  & 95   &      \\
         & $ \bo\eta  $    & P &  \O  &  \O  &  \O  &  1   &      \\
         & $ \kt K $       & P & --   &  1   & --   &  --  &      \\
         &                 & F & --   &  0   & --   &  --  &      \\
         & $ \kll K $       & P &  .04 &  .6  & .02  &  5   &      \\
         & $ \kz K $       & P & --   &  .4  & --   &  --  &      \\
         & $ \kh K $       & P & --   &  .4  & --   &  --  &      \\
         & $ \oq \pi  $    & S &  16  &  82  &  58  & 79   &      \\
         &                 & D & .01  &  .2  &  .02 & .02  &      \\
         & $ \ksq K   $    & S &  --  & 110  &  --  &  --  &      \\
         &                 & D &  --  &  .01 &  --  &  --  &      \\
         & $\Gamma$        &   &  115 & 338  & 177  & 384  &      \\
\hline

\end{tabular}
\end{table}

\begin{table}
\caption{$0^{+-}$ Isovector Hybrid Decay Modes from Ref.
\cite{swanson2}.\label{part3:tab:t7}}
\begin{tabular}{lccccccc}
\hline
\hline
         &                &   &  alt & 2.0 GeV hybrid & standard & IKP &
reduced \\
\hline

$0^{+-}$ & $ \aoo\pi  $     & P & \O   &  \O  & \O   & 309  &      \\
         & $ \ho\pi  $     & P & 47   &  45  & 24   &  37  &      \\
         & $ \bo\eta  $    & P & .6   &  12  &  .4  &  .3  &      \\
         & $ \kll K $       & P &  .7  &  10  & .4   &  7   &      \\
         & $ \kh K $       & P & --   &   1  & --   &  --  &      \\
         & $ \pq \pi  $    & S & 60   & 246  &  222 &  312 &      \\
         & $ \kq K    $    & S & --   & 115  & --   &  --  &      \\
         & $\Gamma $       &   & 108  & 429  & 247  &  665 &      \\
\hline

\end{tabular}
\end{table}

\begin{table}
\caption{$1^{++}$ Isovector Hybrid Decay Modes from Ref.
\cite{swanson2}.\label{part3:tab:t8}}
\begin{tabular}{lccccccc}
\hline
\hline
         &                &   &  alt & 2.0 GeV hybrid & standard & IKP &
reduced \\

\hline

$1^{++}$ & $ \rho \pi    $ & S &  23  &  19  &  26  &  38  &  116 \\
         & $             $ & D &   1  &   3  &   2  &   1  &   8  \\
         & $ K^{\ast} K  $ & S &  14  &  26  &  24  &  18  &  170 \\
         & $             $ & D &  .04 &  .3  &   .09& .04  &   .6 \\
         & $ \rho \omega $ & S &   0  &   0  &   0  &   0  &  47  \\
         &                 & D &   0  &   0  &   0  &   0  &  .03  \\
         & $ f_2(1270)\pi  $     & P & 4    &  5   & 2    & 75   &      \\
         &                 & F & .01  & .03  & 0    &  .3  &      \\
         & $ f_1(1285)  \pi  $     & P &  7   &  9   & 4    &   62 &      \\
         & $ \fz \pi  $      & P &  \O  & \O   & \O   &   4  &      \\
         & $ \bo\pi $      & P &  \O  &  \O  & \O   &      &      \\
         & $ \at\eta  $    & P &  --  &  .9  &  --  &  --  &      \\
         &                 & F &  --  &  0   &  --  &  --  &      \\
         & $ \aoo\eta  $    & P &  .2  &  3   & .09  &   1  &      \\
         & $ \az\eta  $    & P &  --  &  \O  &  --  &  --  &      \\
         & $ \kt K $       & P & --   &  .4  & --   &  --  &      \\
         &                 & F & --   &   0  & --   &  --  &      \\
         & $ \kll K $       & P &  .07 &   1  & .05  &  1   &      \\
         & $ \kz K $       & P & --   &   0  & --   &  --  &      \\
         & $ \kh K $       & P & --   &   .7 & --   &  --  &      \\
         & $ \rqb \pi $    & S & 14   & 80   & 50   & 66   &      \\
         &                 & D & .02  & .6   & .05  & .04  &      \\
         & $ \ksq K   $    & S &  --  & 55   & --   & --   &      \\
         &                 & D &  --  & .01  & --   & --   &      \\
         & $\Gamma$        &  &  63  & 204  & 108  & 269   &      \\
\hline
\end{tabular}
\end{table}

\begin{table}
\caption{Isoscalar Hybrid Decay Modes from Ref. \cite{swanson2}.
\label{part3:tab:t9} }
\begin{tabular}{lccccccc}
\hline
\hline
         &                &   &  alt & 2.0 GeV hybrid & standard & IKP &
reduced \\

\hline
$2^{-+}$ & $ K^{\ast} K  $ & P & 1    &   5  &   2  & 1    &  17  \\
         & $ \at \pi $     & S &  52  & 31   &  25  & 45   &      \\
         &                 & D &  .2  & .6   &  .1  & 22   &      \\
         & $ \aoo\pi  $     & D &  .5  &  1   &  .3  & \O   &      \\
         & $ \az\pi  $     & D &  .02 &  .1  & .01  &  .6  &      \\
         & $ f_2(1270)\eta  $    & S &  --  &  8   & --   &  --  &      \\
         &                 & D &  --  & .02  & --   &  --  &      \\
         & $ f_1(1285)  \eta  $    & D &  --  & .02  & --   &  --  &      \\
         & $ \fz\eta  $    & D &  --  &   0  & --   &  --  &      \\
         & $ \kt K $       & S & --   & 11   & --   &  --  &      \\
         &                 & D & --   &  0   & --   &  --  &      \\
         &                 & G & --   &  0   & --   &  --  &      \\
         & $ \kll K $       & D & 0    & .01  &  0   & 0    &      \\
         & $ \kz K $       & D & --   & 0    & --   &  --  &      \\
         & $ \kh K $       & D & --   & 0    & --   &  --  &      \\
         & $ \ksq K   $    & P & --   &  1   &  --  &  --  &      \\
         & $\Gamma$        &   & 54   &  58  &  27  &  69  &      \\
\hline
$1^{-+}$ & $\eta^{'} \eta$ & P &  0   &  0   &  0   &   0  & 10   \\
         & $ K^{\ast} K  $ & P & 1    &  5   &  2   &   1  & 17   \\
         & $ \at\pi  $     & D &  .4  &   1  &  .2  &  \O  &      \\
         & $ \aoo\pi  $     & S &  59  &  30  &  28  &  38  &      \\
         &                 & D &  .3  &  .6  &  .2  &  34  &      \\
         & $ f_2(1270)\eta  $    & D &  --  &  .05 & --   &  --  &      \\
         & $ f_1(1285)  \eta  $    & S &  --  &    8 & --   &  --  &      \\
         &                 & D &  --  &   .01& --   &  --  &      \\
         & $ \kt K $       & D & --   & 0    & --   &  --  &      \\
         & $ \kll K $       & S &  4   & 7    &  2   &   7  &      \\
         &                 & D & 0    & .2   & 0    & 0    &      \\
         & $ \kh K $       & S & --   & 33   & --   &  --  &      \\
         &                 & D & --   &  0   & --   &  --  &      \\
         & $ \pq \pi  $    & P & 8    &  65  &  27  &  27  &      \\
         & $ \euq \eta$    & P & --   &  6   &  --  &  --  &      \\
         & $ \kq K    $    & P & --   &  .8  &  --  &  --  &      \\
         & $ \ksq K   $    & P & --   &   1  &  --  &  --  &      \\
         & $\Gamma $       &   &  73  & 158  &  59  & 107  &      \\
\hline
\end{tabular}
\end{table}

\begin{table}
\caption{Isoscalar Hybrid Decay Modes from Ref. \cite{swanson2}.
\label{part3:tab:t10} }
\begin{tabular}{lccccccc}
\hline
\hline
         &                &   &  alt & 2.0 GeV hybrid & standard & IKP &
reduced \\

\hline
$0^{-+}$ & $ K^{\ast} K  $ & P & 5    & 18   & 10   &   5  & 69   \\
         & $ \at\pi  $     & D &  2   &  6   &  1   & 16   &      \\
         & $ \az\pi  $     & S & 145  & 114  &  70  & 175  &      \\
         & $ f_2(1270)\eta  $    & D &  --  &.2    & --   &  --  &      \\
         & $ \fz\eta  $    & S &  --  & 23   & --   &  --  &      \\
         & $ \kt K $       & D & --   &   .02& --   &  --  &      \\
         & $ \kz K $       & S & --   & 44   & --   &  --  &      \\
         & $ \ksq K   $    & P &      &   5  &      &      &      \\
         & $\Gamma $       &   & 152  & 210  &  81  & 196  &      \\
\hline
$1^{--}$ & $ \rho \pi    $ & P & 28   &  47  & 38   & 35   & 172  \\
         & $ \omega \eta $ & P & 3    &   9  &  6   &  4   &  29  \\
         & $\omega\eta^{'}$& P & .1   &   1  &  .2  & .3   &   .8  \\
         & $ K^{\ast} K  $ & P & 3    &  9   &   5  &  3   &  35  \\
         & $ \bo\pi  $     & S & \O   &  \O  &  \O  &      &      \\
         &                 & D &      &      &  \O  &      &      \\
         & $ \ho\eta  $    & S &      &      &      &  \O  &      \\
         & $ \kt K $       & D & --   & 0    & --   &  --  &      \\
         & $ \kll K $       & S &  6   &  12  &  4   &  11  &      \\
         &                 & D &  0   & .01  & 0    &  0   &      \\
         & $ \kh K $       & S & --   &  17  & --   &  --  &      \\
         &                 & D & --   & 0    & --   &  --  &      \\
         & $ \rqb  \pi  $  & P &  2   &  35  &  8   &  7   &      \\
         & $ \oq \eta $    & P & --   &  .6  &  --  &  --  &      \\
         & $ \ksq K   $    & P & --   &   3  &  --  &  --  &      \\
         & $\Gamma $       &   &  42  & 134  &  61  &  60  &      \\
\hline
\end{tabular}
\end{table}

\begin{table}
\caption{Isoscalar Hybrid Decay Modes from Ref. \cite{swanson2}.
\label{part3:tab:t11} }
\begin{tabular}{lccccccc}
\hline
\hline
         &                &   &  alt & 2.0 GeV hybrid & standard & IKP &
reduced \\
\hline
$2^{+-}$ & $ \rho \pi    $ & D & 1    &   4  &  2   &  2   &  11  \\
         & $ \omega \eta $ & D & .1   &  .5  &  .2  &   .1 &   1  \\
         & $\omega\eta^{'}$& D &  0   &  .03  &   0  &   0  & 0  \\
         & $ K^{\ast} K  $ & D & .04  &  .2  &  .08 &  .04 & .6   \\
         & $ \bo\pi  $     & P &   4  &  5   & 2    & 164  &      \\
         &                 & F &  .02 &  .07 & .01  &   .8 &      \\
         & $ \ho\eta $     & P & .2   &  .7  & .1   &  6   &      \\
         & $ \kt K $       & P & --   & .04  & --   &  --  &      \\
         &                 & F & --   &   0  & --   &  --  &      \\
         & $ \kll K $       & P &  0   &  .03 & 0    & .6   &      \\
         &                 & F &  0   & 0    &  0   &  0   &      \\
         & $ \kh K $       & P & --   & .3   & --   &  --  &      \\
         &                 & F & --   &  0   & --   &  --  &      \\
         & $ \rqb \pi $    & D &  .02 & .8   &  .06 &  .05 &      \\
         & $ \oq \eta $    & D &  --  &  0   &  --  &  --  &      \\
         & $ \ksq K   $    & D & --   &  .01 &  --  &  --  &      \\
         & $\Gamma$        &   & 5    &  12  &   4  & 166  &      \\
\hline
$1^{+-}$ & $ \rho \pi    $ & S & 70   &  57  &  77  & 114  & 350  \\
         &                 & D & .8   &   2  &   1  &   1  &   6  \\
         & $ \omega \eta $ & S & 15   &  22  &  25  & 22   & 119  \\
         & $             $ & D & .07  &  .3  &  .1  & .06  &  .6  \\
         & $\omega\eta^{'}$& S & 4    &  8   &  5   & 15   & 24   \\
         & $             $ & D & 0    &  .02 &   0  &  0   & 0   \\
         & $ K^{\ast} K  $ & S & 27   &  52  &  47  & 36   & 339  \\
         & $             $ & D &  .02 &   .1 &  .04 & .02  &  .3  \\
         & $ \bo\pi  $     & P &  \O  &  \O  &  \O  & 231  &      \\
         & $ \ho\eta  $    & P &  \O  &  \O  & \O   &  9   &      \\
         & $ \kt K $       & P & --   &  1   & --   &  --  &      \\
         &                 & F & --   &  0   & --   &  --  &      \\
         & $ \kll K $       & P &  .04 &  .6  & .02  &  5   &      \\
         & $ \kz K $       & P & --   &  .4  & --   &  --  &      \\
         & $ \kh K $       & P & --   &  .4  & --   &  --  &      \\
         & $ \rqb \pi  $   & S &  42  & 240  &  150 & 199  &      \\
         &                 & D & .01  &  .4  &  .04 & .03  &      \\
         & $ \oq \eta $    & S &  --  & 38   &  --  &  --  &      \\
         &                 & D &  --  &  0   &  --  &  --  &      \\
         & $ \ksq K   $    & S &  --  & 110  &  --  &  --  &      \\
         &                 & D &  --  &  .01 &  --  &  --  &      \\
         & $\Gamma$        &   &  158 & 529  & 305  & 632  &      \\
\hline
\end{tabular}
\end{table}

\begin{table}
\caption{Isoscalar Hybrid Decay Modes from Ref. \cite{swanson2}.
\label{part3:tab:t12}}
\begin{tabular}{lccccccc}
\hline
\hline
         &                &   &  alt & 2.0 GeV hybrid & standard & IKP &
reduced \\

\hline
$0^{+-}$ & $ \bo\pi  $     & P &  110 &  119 &  56  &  85  &      \\
         & $ \ho \eta $    & P &  4   &  17  &  3   &  2   &      \\
         & $ \kll K $       & P &  .7  &  10  & .4   &  7   &      \\
         & $ \kh K $       & P & --   &   1  & --   &  --  &      \\
         & $ \kq K    $    & S &  --  & 115  &  --  &  --  &      \\
         & $\Gamma$        &   & 115  & 262  &  59  &  94  &      \\
\hline
$1^{++}$ & $ K^{\ast} K  $ & S & 17   &  26  &  24  &  18  & 170  \\
         &                 & D &  .04 &  .3  &  .09 &  .04 &  .6  \\
         & $ \at\pi  $     & P &   10 &  14  &   5  & 179  &      \\
         &                 & F &  .01 & .06  &  .01 &  .4  &      \\
         & $ \aoo\pi  $     & P &   28 & 30   &   14 & 232  &      \\
         & $ \az\pi  $     & P &  \O  &  \O  &   \O &   6  &      \\
         & $ f_2(1270)\eta  $    & P &  --  &   1  & --   &  --  &      \\
         &                 & F &  --  &   0  & --   &  --  &      \\
         & $ f_1(1285)  \eta  $    & P &  --  &   2  & --   &  --  &      \\
         & $ \fz\eta  $    & P & \O   &  \O  & \O   &  --  &      \\
         & $ \kt K $       & P & --   &  .4  & --   &  --  &      \\
         &                 & F & --   &   0  & --   &  --  &      \\
         & $ \kll K $       & P &  .07 &   1  & .05  &  1   &      \\
         & $ \kz K $       & P & --   &   0  & --   &  --  &      \\
         & $ \kh K $       & P & --   &   .7 & --   &  --  &      \\
         & $ \ksq K   $    & S & --   & 55   &  --  &  --  &      \\
         &                 & D &  --  & .01  &  --  &  --  &      \\
         & $\Gamma$        &   &  55  & 130  &  43  & 436  &      \\
\hline
\end{tabular}
\end{table}

\begin{table}
\caption{$s \bar s$ Hybrid Decay Modes from Ref. \cite{swanson2}.
\label{part3:tab:t13}}
\begin{tabular}{lccccccc}
\hline
\hline
         &                &   &  alt & high mass & standard & IKP & reduced \\
\hline
$2^{-+}$  & $K^{\ast} K  $ & P &  6   & 13   &  11  &   8  &  82  \\
         & $ \kt K $       & S &  28  &  29  & 21   & 44   &      \\
         &                 & D &  .03 &  .5  &  .02 &  1   &      \\
         & $ \kll K $       & D &  .2  &  .5  &  .1  &   10 &      \\
         & $ \kz K $       & D & .02  &   .3 &   .01&   .2 &      \\
         & $ \kh K $       & D &  .06 &    .5&  .03 &  .6  &      \\
         & $ \fts\eta $    & S &  --  &  20  &  --  &  --  &      \\
         &                 & D &  --  &  .2  &  --  &  --  &      \\
         & $ \fos\eta $    & D &  --  &  .03 &  --  &  --  &      \\
         & $ \fzs\eta $    & D &  .01 &   .08&  0   &  .1  &      \\
         & $ \ksq K   $    & P &  2   &  27  &  6   &  5   &      \\
         & $\Gamma$        &   &  36  &  91  &  38  &  69  &      \\
\hline
$1^{-+}$  & $ \eta^{'}\eta$& P &  0   &  0   &   0  &   0  &  44  \\
          & $ K^{\ast} K $ & P &  6   & 13   &  11  &   8  &  82  \\
         & $ \kt K $       & D &  .07 & 1    &  .04 &  \O  &      \\
         & $ \kll K $       & S &  14  & 10   &  11  &  14  &      \\
         &                 & D &   3  & 8    &   2  &  21  &      \\
         & $ \kh K $       & D &  83  & 76   & 61   & 121  &      \\
         &                 & D &  .03 &  .2  & .02  &  .4  &      \\
         & $ \fts\eta $    & D &  --  & .04  &  --  &  --  &      \\
         & $ \fos\eta $    & S &  --  & 21   &  --  &  --  &      \\
         &                 & D &  --  & .02  &  --  &  --  &      \\
         & $ \kq K    $    & P &  1   &  45  & 4    &  3   &      \\
         & $ \esq \eta$    & P &  --  &  15  & --   &  --  &      \\
         & $ \ksq K   $    & P &  2   &  27  &  6   &  5   &      \\
         & $\Gamma$        &   & 109  & 216  &  95  & 172  &      \\
\hline
$0^{-+}$ & $ K^{\ast} K  $ & P & 26   & 52   &  46  &  33  & 330  \\
         & $ \kt K $       & D &   .4 &   6  &  .2  &  1   &      \\
         & $ \kz K $       & S &  113 &   117&  83  &  174 &      \\
         & $ \fts\eta $    & D &  --  &  .2  &  --  &  --  &      \\
         & $ \fzs\eta $    & S &  72  &  105 &  64  & 109  &      \\
         & $ \ksq K   $    & P &   7  &  110 &  22  &  18  &      \\
         & $\Gamma$        &   & 218  & 390  & 215  & 335  &      \\
\hline
$1^{--}$ & $ K^{\ast} K  $ & P & 13   & 26   &  23  &  16  & 165  \\
         & $ \phi   \eta $ & P & 2    &  19  &  11  &   3  &  89  \\
         & $ \phi\eta^{'}$ & P & .01  &  2   &  .1  &  .02 &  .5  \\
         & $ \kt K $       & D &  .1  & 2    & .07  &   2  &      \\
         & $ \kll K $       & S &  23  & 16   &  18  &  24  &      \\
         &                 & D &   .2 &  .6  &  .1  &  2   &      \\ 
         & $ \kh K $       & S &  43  & 40   &  32  &  63  &      \\ 
         &                 & D &  .1  & .6   &  .04 &  .7  &      \\ 
         & $ \hos\eta $    & S &      &      & \O   &      &      \\ 
         &                 & D &      &      & \O   &      &      \\ 
         &                 & D &  .07 &  .6  & .04  &  .3  &      \\ 
         & $ \ksq K   $    & P &  3   &   55 &  11  &   9  &      \\ 
         & $\Gamma$        &   &  84  & 155  &  95  & 120  &      \\ 
\hline
\end{tabular}
\end{table}

\begin{table}
\caption{$s \bar s$ Hybrid Decay Modes from Ref. \cite{swanson2}.
\label{part3:tab:t14}}
\begin{tabular}{lccccccc}
\hline
\hline
         &                &   &  alt & high mass & standard & IKP & reduced
\\
\hline
$2^{+-}$ & $ K^{\ast} K  $ & D &  1   &  3   &   2  &   1  &  13  \\
         & $ \phi \eta   $ & D & .06  &  .8  &  .3  &  .08 &   2  \\
         & $ \phi\eta^{'}$ & D & 0    &  0  &  0   &  0   &   0  \\
         & $ \kt K $       & P &  .3  & 1    &  .2  &  32  &      \\
         &                 & F &   0  & .03  &  0   & .01  &      \\
         & $ \kll K $       & P &  .2  &  .3  &  .1  &  17  &      \\
         &                 & F &  .04 &  .2  &  .02 &  .6  &      \\
         & $ \kh K $       & P &  3   &   8  &   2  &   28 &      \\
         &                 & F &  0   &   0  &  0   &  0   &      \\
         & $ \hos\eta $    & P &  .3  &  2   & .2   &  9   &      \\
         &                 & F &   0  &  0   &  0   &  0   &      \\
         & $ \ksq K   $    & D &  .04 &   2  &  .1  & .08  &      \\
         & $\Gamma$        &   &  5   &  18  &   5  &  79  &      \\
\hline
$1^{+-}$ & $ K^{\ast} K  $ & S & 20   & 19   &  34  &  42  & 247  \\
         &                 & D &  .6  &  2   &   1  &   .6 &   7  \\
         & $ \phi   \eta $ & S &  11  & 63   &  66  &  28  & 523  \\
         &                 & D &  .03 &  .5  &  .2  &  .04 &   1  \\
         & $ \phi \eta^{'}$& S &  2   &  19  &  8  &  3  & 61  \\
         & $             $ & D &  0   &  .02 &   0  &  0   &   0  \\
         & $ \kt K $       & P &  8   & 35   &  5   &  10  &      \\
         &                 & F &  0   & .02  &  0   &  .01 &      \\
         & $ \kll K $       & P &  4   &  5   &  2   &  122 &      \\
         & $ \kz K $       & P &  3   & 14   &  2   &   18 &      \\
         & $ \kh K $       & P &  3   &  8   &  2   &   4  &      \\
         & $ \hos\eta $    & P &  \O  &  \O  &  \O  & 14   &      \\
         & $ \ksq K   $    & S &   39 &  206 &  181 & 201  &      \\
         &                 & D &   .02&    1 &  .06 & .04  &      \\
         & $\Gamma$        &   &  91  & 373  & 301  & 443  &     \\
\hline
$0^{+-}$ & $ \kll K $       & P &  66  & 95   &  43  &  165 &      \\
         & $ \kh K $       & P &  10  &  30  &   6  &  36  &      \\
         & $ \hos \eta$    & P &  8   &  42  &  5   &  4   &      \\
         & $ \kq K    $    & S & 46   & 323  & 205  & 221  &      \\
         & $\Gamma$        &   & 130  & 490  &  259 & 426  &     \\
\hline
$1^{++}$ & $ K^{\ast} K  $ & S & 10   &  9   &  17  &  21  & 123  \\
         &                 & D &  1   &  4   &   2  &   1  &  15  \\
         & $ \kt K $       & P &   3  &  13  &   2  & 27   &      \\
         &                 & F &   0  &  .05 &   0  & .01  &      \\
         & $ \kll K $       & P &   7  &  11  &   5  &  37  &      \\
         & $ \kz K $       & P &   \O &  \O  &   \O &  2   &      \\
         & $ \kh K $       & P &   6  &  16  &   3  &  29  &      \\
         & $ \fts\eta $    & P &  --  &  2   &  --  &  --  &      \\
           &               & F &  --  &  0   &  --  &  --  &      \\
         & $ \fos\eta $    & P &  --  &  4   &  --  &  --  &      \\
         & $ \fzs\eta $    & P &  \O  &  \O  &  \O  &  2   &      \\
         & $ \ksq K   $    & S & 19   & 103  & 90   & 100  &      \\
         &                 & D & .05  & 2    &  .1  & .08  &      \\
         & $\Gamma$        &   &  46  & 164  & 119  & 219  &      \\
\hline
\end{tabular}
\end{table}

\begin{table}
\caption{$c \bar c$ Hybrid Decay Modes from Ref. \cite{swanson2}.
\label{part3:tab:t15}  }
\begin{tabular}{lccccccc}
\hline
\hline
         &                &   &  alt & high mass & standard & IKP & reduced \\
\hline
$2^{-+}$ & $ D^{\ast}D $   & P &  .5   &  .1  & .8    &  4   &  19  \\
         & $\dss(2^{+})D$  & S & --   &   9   &  --  &  --  &      \\
         &                 & D & --   &  .2  &  --  &  --  &      \\
         & $\dss(1^{+}_L)D$& D & --   &   .2  &  --  &  --  &      \\
         & $\dss(0^{+})D$  & D & --   &   .2  &  --  &  --  &      \\
         & $\dss(1^{+}_H)D$& D & --   &   .2 &  --  &  --  &      \\
         & $\Gamma$        &   & .5   &  10    & .8   &  4   &       \\
\hline
$1^{-+}$ & $ D^{\ast}D $   & P &  .5  &  .1  & .8    &  4   &  19  \\
         & $\dss(2^{+})D$  & D & --   &   .5  &  --  &  --  &      \\
         & $\dss(1^{+}_L)D$& S & --   &   1.2  &  --  &  --  &      \\
         &                 & D & --   &   2.5 &  --  &  --  &      \\
         & $\dss(1^{+}_H)D$& S & --   &   25 &  --  &  --  &      \\
         &                 & D & --   &   0  &  --  &  --  &      \\
         & $\Gamma $       &   & .5   &  29  & .8   &  4   &      \\
\hline
$0^{-+}$ & $ D^{\ast}D $   & P &  2   &   .3  & 3   & 16   & 76  \\
         & $\dss(2^{+})D$  & D & --   &   2.5  &  --  &  --  &      \\
         & $\dss(0^{+})D$  & S & --   &   25 &  --  &  --  &      \\
         & $\Gamma $       &   & 2    &  28  & 3    &  16  &      \\
\hline
$1^{--}$ & $ D^{\ast}D $   & P &  1   &  .2  &  1.5    &  8   &  38  \\
         & $\dss(2^{+})D$  & D & --   & 1     &  --  &  --  &      \\
         & $\dss(1^{+}_L)D$& S & --   & 7   &  --  &  --  &      \\
         &                 & D & --   &  .3   &  --  &  --  &      \\
         & $\dss(1^{+}_H)D$& S & --   & 10   &  --  &  --  &      \\
         &                 & D & --   & .2   &  --  &  --  &      \\
         & $\Gamma$        &   & 1    & 19   & 1.5   & 8   &      \\
\hline
$2^{+-}$ & $ D^{\ast}D $   & D & .2   &  .2  &  .3   &  1   &   7 \\
         & $\dss(2^{+})D$  & P & --   &  .5   &  --  &  --  &      \\
         &                 & F & --   & .02  &  --  &  --  &      \\
         & $\dss(1^{+}_L)D$& P & --   &  0 &  --  &  --  &      \\
         &                 & F & --   &  0   &  --  &  --  &      \\
         & $\dss(1^{+}_H$)D& P & --   &  3  &  --  &  --  &      \\
         &                 & F & --   &  0   &  --  &  --  &      \\
         & $\Gamma $        &   & .2   & 4   & .3   &  1    &      \\
\hline
$1^{+-}$ & $ D^{\ast}D $   & S &  .3   &   .1 &  .5   &  8   &   12 \\
         &                 & D & .1   &    .1&  .1  &  .5  &  4  \\
         & $\dss(2^{+})D$  & P & --   & 13   &  --  &  --  &      \\
         &                 & F & --   & .01  &  --  &  --  &      \\
         & $\dss(1^{+}_L)D$& P & --   & 2    &  --  &  --  &      \\
         & $\dss(0^{+})D$  & P & --   & 8    &  --  &  --  &      \\
         & $\dss(1^{+}_H)D$& P & --   & 2.5   &  --  &  --  &      \\
        & $\Gamma $        &   & .4   &  26  &  .6   & 8.5  &      \\
\hline
\end{tabular}
\end{table}

\begin{table}
\caption{$c \bar c$ Hybrid Decay Modes from Ref. \cite{swanson2}.
\label{part3:tab:t16}}
\begin{tabular}{lccccccc}
\hline
\hline
         &                &   &  alt & high mass & standard & IKP & reduced
\\
\hline

$0^{+-}$ & $\dss(1^{+}_L)D$& P & --   & 25  &  --  &  --  &      \\
         & $\dss(1^{+}_H)D$& P & --   & 15  &  --  &  --  &      \\
         & $\Gamma $       &   & --   & 40  &  --  &  -- &       \\
\hline
$1^{++}$ & $ D^{\ast}D $   & S & .2   &  .1  &  .3   &   1  &  6  \\
         &                 & D & .2   &  .2  &  .3   &   .3  &  8  \\
         & $\dss(2^{+})D$  & P & --   &  5  &  --  &  --  &      \\
         &                 & F & --   & .03   &  --  &  --  &      \\
         & $\dss(1^{+}_L)D$& P & --   & 5   &  --  &  --  &      \\
         & $\dss(0^{+})D$  & P & --   & \O   &  --  &  --  &      \\
         & $\dss(1^{+}_H)D$& P & --   & 5   &  --  &  --  &      \\
         & $\Gamma$        &   &  .4  & 15  & .6   & 1.3  &      \\
\end{tabular}
\end{table}

\subsubsection{QCD sum rules}

The mass and decay width of the $1^{-+}$ hybrid meson has been
studied using QCD sum rules. Within this framework, one considers
a two-point correlator
\begin{equation}
\Pi_{\mu\nu}(q^2)=i\int \mathrm{d}^4x\, {\mathrm{e}}^{iqx} \langle
0|T\{j_{\mu}(x),j^+_{\nu}(0)\}|0 \rangle ,
\end{equation}
where $j_{\mu}(x)=\bar q(x)T^a\gamma_{\nu}igG^a_{\mu\nu}q(x)$ is
the interpolating current for the $1^{-+}$ isospin vector hybrid
meson.

The spectral density
$\rho_{v}(s)=\frac{1}{\pi}\textbf{Im}\Pi_{v}(s)$ can be expressed
in terms of the hybrid meson observables such as its mass etc:
\begin{equation}
\frac{1}{\pi}\textbf{Im}\Pi_{v}(s)=\sum_R
M_R^6f_R^2\delta\left(s-M_R^2\right)+{\rm QCD~continuum}.
\end{equation}
It can also be related to the correlator $\Pi_{v}(q^2)$ at the
scale $-q^2$ via the dispersion relation
\begin{equation}
 \Pi_{v,s}(q^2)={(q^2)^n
\int^\infty_0ds\frac{\rho_{v}(s)}{s^n(s-q^2)}+\sum^{n-1}_{k=0}a_k(q^2)^k},
\end{equation}
where the $a_k$ are appropriate subtraction constants.

After invoking a Borel transformation to enhance the
lowest-lying resonance in the spectral density, we have the QCD
sum rules
\begin{equation}\label{bor}
R_k\left(\tau,s_0\right)=\int^{s_0} s^k{\mathrm{e}}^{-s\tau}
\rho_{v}(s)\mathrm{d}s~;~k=0,1,2,\ldots ,
\end{equation}
where the quantity $R_k$ represents the QCD prediction, and $s_0$
is the threshold parameter.

The sum rules for a $1^{-+}$ hybrid meson have been obtained by various
authors. The prediction for the hybrid mass is sensitive to the
threshold $s_0$, and the sum rule in the leading order of $\alpha_s$
expansion is unstable. When the next-to-leading-order correction
is included, the sum rule becomes  more stable.  An
upper bound on the $1^{-+}$ hybrid mass is predicted to be 2.0 GeV 
\cite{jin}.

The decay widths of the $1^{-+}$ hybrid can be obtained by
a three-point correlator
\begin{equation}
\Pi(p,q)=i\int \mathrm{d}^4x\mathrm{d}^4y\, {\mathrm{e}}^{ipx+iqy}
\langle 0|T\{j_A(x)j_B(y)j_\mu(0)\}|0 \rangle,
\end{equation}
where $j_A(x)$ and $j_B(y)$ are operators that annihilate the
final states $A$ and $B$ respectively.

When $A$ and $B$ are two pseudoscalars, $\Pi(p,q)=F_1
(p+q)_\mu+F_2(p-q)_\mu$. Only $F_2$ is relevant to the process
$1^{-+}\rightarrow AB$ and vanishes at the leading order
\cite{Govaerts}. At the next leading order, it was estimated
\cite{narison1} to be: $\Gamma(1^{-+}\rightarrow \pi\eta^\prime)\sim
3$MeV, $\Gamma(1^{-+}\rightarrow \pi\eta)\sim
0.3$MeV, which is quite consistent with 
flux tube model predictions.

However, the channel $\pi\rho$ is not narrow in the QCD sum rules
approach. By using the three-point function at the symmetric
point, the width of $1^{-+}\rightarrow \pi\rho$ was predicted 
to be in
the 250-600 MeV range~\cite{narison1,Narison2}. Later, it
was pointed out that the calculation at the symmetric point
receives large contamination from  higher resonances and the
continuum~\cite{zhusl}. By using the light-cone QCD
sum rules and a double Borel transformation, the width is
reduced to be $40\pm 20$~MeV~\cite{zhusl}. The similar
channel, $K^*K$, is suppressed by the kinematic phase space. The
$f_1\pi$ channel is very broad ($\sim 100$ MeV) in the QCD sum 
rule approach.

It is interesting to compare QCD sum rule predictions
with those of the flux tube model, since the bases of these two
approaches are contradictory. The former is based on an
$\alpha_s$ expansion while the latter is based on the strong
coupling expansion. From Table~\ref{flux-sm}, the decay patterns from 
these two approaches 
are seen to be very similar although the partial decay widths for
each channel are quite different:
\begin{equation}
\Gamma(f_1\pi)>\Gamma(\rho\pi)>\Gamma(K^*K)>\Gamma(\eta^\prime\pi)
>\Gamma(\eta\pi).
\end{equation}   

\begin{table}\label{tab2}
\begin{center}
\caption{Comparison of the decay widths of the $1^{-+}$ hybrid
meson for the flux tube model and QCD sum rule approaches.
\label{flux-sm}}
\begin{tabular}{c|c|c|c}
  \hline
& QCDSR & Flux Tube Model& PDG ($\pi_1(1600)$)\\
$b_1\pi$(MeV)&unstable&$40\pm 20$ &seen\\
$f_1\pi$ (MeV)&100 &$10\pm5$ &seen\\
$\rho\pi$(MeV)& $40\pm 20$&9&seen\\
$K^*K$ (MeV)& 8&0.6&no\\
$\eta^\prime\pi$ (MeV)&3&small&seen\\
$\eta\pi$ (MeV)&0.3&small&no\\
  \hline
\end{tabular}
\end{center}
\end{table}


\subsubsection{Comments}

It is important to note that the gluon inside the hybrid meson can
easily split into a  $q\bar q$ pair. Therefore, tetraquarks can
always have the same quantum numbers as hybrid mesons,
including exotic ones.  The discovery of a candidate hadron with
$J^{PC}=1^{-+}$ does not ensure that it is an exotic hybrid meson. One
has to exclude the tetraquark possibility based on its mass, decay
width, decay patterns {\it etc}. This argument holds for 
the $\pi_1(1400)$ and $\pi_1(1600)$ hybrid candidates.

The flux tube model predicts that hybrid mesons prefer to decay into a
pair of mesons with $L=1$ and $L=0$. Heavy hybrid mesons tend to decay
into one $P$-wave heavy meson and one pseudoscalar meson according
to a light-cone QCD sum rule calculation~\cite{zhu-qcd}. A lattice
QCD simulation suggests the string breaking mechanism may play an
important role for the decays of 
heavy quarkonium hybrids~\cite{bali}. 
When the string between the heavy quark and antiquark breaks, 
light mesons are created. As a result,
the preferred final states are a heavy quarkonium plus light
mesons. However, readers should exercise caution  with these
so-called ``selection rules." None of them have been tested
experimentally because no  $1^{-+}$ hybrid candidates have
yet been established unambiguously. 

\subsection{Signatures of the hybrid mesons}

Hybrid mesons are color-singlet composites of constituent quarks and
gluons, such as $q \bar q g$  bound states.  Evidence for the
existence of hybrid mesons would be direct proof of the
existence of the gluonic degree of freedom and the validity of
QCD. The conventional wisdom is that it would be more fruitful to
search for low-lying hybrid mesons with exotic quantum numbers
than to search for glueballs. Hybrids have the additional
attraction that, unlike glueballs, they span complete flavor
nonets and, thus, provide many possibilities for experimental
detection. In addition, the lightest hybrid multiplet includes at
least one $J^{PC}$ exotic state.

In searches for hybrids, there are two primary methods to distinguish them
from conventional states. One  is to look for an excess of
observed states over the number predicted by the quark model. The
drawback to this method is that it depends on a good understanding
of the hadron spectrum in a mass region where it is still rather
murky. At present, phenomenological models have not been
tested experimentally to the extent to where a given state can be
reliably ruled out as a conventional meson. The situation is
further muddled by the expectation of
mixing between conventional $q \bar q$
states and hybrids with the same $J^{PC}$ quantum numbers. The
other approach is to search for the states with quantum numbers
that cannot be accommodated in the quark model. The discovery of
exotic quantum numbers would be definite evidence of something
new.


Some experimental  searches for an isovector $1^{-+}$ hybrid
have claimed positive evidence for the existence of such a state.
Evidence for an exotic $\pi \eta$ resonance 
in the charge exchange reaction $\pi^- p \to \eta \pi^0 n$ 
was claimed by the GAMS collaboration~\cite{gams-exotic}.  These 
findings were, however, found to be
ambiguous in later analyses~\cite{gams-exotic-2}. Evidence for an 
exotic $P$-wave $\pi \eta$ state was also reported 
by the VES experiment~\cite{ves-exotic}.
The observation of a $\pi \eta$ resonance, 
with a mass and width that coincide with those of the $a_2(1320)$,
was claimed by a KEK group~\cite{kek-exotic}.
However, here feedthrough from the dominant $D$-wave into
the $P$-wave  cannot been excluded.  E852 at BNL reported 
the observation of an isovector
$1^{-+}$ state with a mass and width of 
$(1370 \pm 16 ^{+50}_{-30})$~MeV
and $(385 \pm 40^{+65}_{-105})$~MeV, respectively, produced
in $\pi^- p \to \eta\pi^-p$ at 18 GeV/c~\cite{e852-1}.
In these studies, the $\pi \eta$ $P$-wave is seen to have a
forward-backward asymmetry,
which is evidence for interference 
between even and odd $\pi \eta$ partial waves.  Subsequently,
the Crystal Barrel Collaboration found
evidence for an $I^G (J^{PC})= 1^- (1^{-+})$ exotic state~\cite{cbl-exotic1}
with mass and width of $(1400 \pm 20 \pm 20)$~MeV and
$(310 \pm 50^{+50}_{-30})$~MeV, respectively, in the reaction 
$\bar p n \to \pi^-\pi^0 \eta$ produced by stopping antiprotons in
liquid deuterium. The partial wave analysis of data on $p \bar p$
annihilation at rest in liquid hydrogen (LH$_2$) into $\pi^0\pi^0 \eta$ by
the Crystal Barrel shows that the inclusion of a $\pi \eta$ $P$-wave in 
the fit gives supporting evidence for an $1^{-+}$ exotic state with
parameters compatible with the previous findings~\cite{cbl-exotic2}.
Another isovector $1^{-+}$ meson, the $\pi_{1}(1600)$, was
observed in $\rho\pi$~\cite{e852-rho}, $\eta^{\prime}\pi$~\cite{e852-etap}, and
$f_{1}\pi$~\cite{e852-Kuhn} final states.
The latter experiment also revealed a higher state,
$\pi_{1}(2000)$~\cite{e852-Kuhn}, and $f_{1}\pi$ decays of
the $\pi_{1}(1600)$ and the $\pi_{1}(2000)$ are measured.
Such a rich spectrum of exotic mesons is somewhat puzzling, 
since lattice QCD~\cite{lattice-exotic}
and flux-tube model~\cite{Isgur-exotic,Barnes95-exotic} calculations 
predict only one low-mass $\pi_{1}$ meson. 

In the flux-tube model, the lightest $1^{-+}$ isovector hybrid is 
predicted
to decay primarily into $b_{1}\pi$ \cite{Isgur-exotic}. The $f_{1}\pi$ 
branch is also
expected to be large and many other decay modes are suppressed. However,
few experiments have addressed the $b_{1}\pi$ and $f_{1}\pi$ decay channels.
The VES collaboration reported a broad $1^{-+}$\ peak in $b_{1}\pi$
decay~\cite{VES-exotic-2}, and Lee, \textit{et al.}~\cite{Lee-2} 
observed significant $1^{-+}$ strength in $f_{1}\pi$\ decay. In neither 
case, however, was a definitive resonance interpretation of the $1^{-+}$ 
wave possible.   Preliminary results from
a later VES analysis show the excitation of 
the $\pi_{1}(1600)$~\cite{Dorofeev}.  The
E852 experiment at BNL reported the observation of a strong excitation of 
the exotic $\pi_{1}(1600)$ in the $(b_{1}\pi)^{-}$ decay channel, and
confirmed the exotic $\pi_{1}(2000)$ in the reaction $\pi^{-}
p\rightarrow\pi^{+}\pi^{-}\pi^{-}\pi^{0}\pi^{0}p$ ~\cite{e852-lu}.

\subsection{Monte-Carlo simulation of $1^{-+}$ exotic state 
$\pi_1(1400)$}

From simple counting of the powers of the electromagnetic and strong 
coupling constants, one obtains:
\begin{equation}                                                    
 \Gamma(J/\psi\to MH)>\Gamma(J/\psi\to MM')\approx\Gamma(J/\psi\to MG),
\end{equation}
where $M$ stands for an ordinary $q \bar q$ meson, $G$ for a glueball and 
$H$ for a hybrid state.
Therefore, hadronic $J/\psi$ decays  provide a good place to
search for hybrid states.
A full Monte-Carlo simulation of the decays $J/\psi \to \rho \eta \pi^0$,
with $\rho \to \pi^+\pi^-$ and $\eta \to \gamma \gamma$, was done, 
and a partial wave
analysis was carried out to study the \bes3
sensitivity to an $1^{-+}$ exotic state 
in the $\eta \pi^0$ final state.  The simulation is based on
GEANT4 and the \bes3 detector design.

For $J/\psi \to \rho \eta \pi^0$, we generated 
$\rho \pi_1(1400)$ as well as the expected
non-exotic processes
$J/\psi \to \rho a_0(980)$, $\rho a_2(1320)$ and
$\rho a_2(1700)$, with the decay braching
fractions:
\begin{eqnarray*}
Br(J/\psi\to\rho a_0(980))\sim 4.38\%\\
Br(J/\psi\to\rho\pi_1(1400)) \sim14.57\%\\
Br(J/\psi\to\rho a_2(1320))\sim21.39\%\\
Br(J/\psi\to\rho a_2(1700))\sim41.64\% ,\\
\end{eqnarray*}
and considered
the angular distributions of different spin-parities and the interference
between them. The main background to $J/\psi\to\rho\eta\pi^0$ comes from
$J/\psi\to\gamma\rho^+\rho^-$, which were also
generated.

To select candidate events, we require two good charged tracks
with zero net charge and at least four good photons. A good charged track 
is one that is within the polar angle region $|cos\theta|<0.93$ and has
points of closest approach that are within 1~cm of the beam axis
and within 5~cm of the center of the  interaction region.
The two charged tracks are required to
consist of an unambiguously identified $\pi^+\pi^-$ pair.
Candidate photons are required to have an energy
deposit in the electromagnetic calorimeter that is
greater than 50 MeV and to be isolated
from charged tracks by more than $20^{\circ}$ in both the x-y and
r-z planes; at least four photons
are required. A four-constraint (4C) energy-momentum conservation
kinematic fit is performed to the $\pi^+\pi^-\gamma\gamma\gamma\gamma$
hypothesis and the $\chi^{2}_{4C}$ is required to be less than 15.
For events with more than four selected photons, the combination with
the smallest $\chi^{2}$ is choosen.
The photons from the decays of $\pi^0$ and $\eta$ are selected 
based on the
combination with the smallest $\delta$, where

\begin{equation}
\delta=\sqrt{{(M_{\eta}-M{\gamma_i\gamma_j})^2}+
{(M_{\pi^0}-M{\gamma_k\gamma_l})^2}}.
\end{equation}

All of the generated $J/\psi \to \rho \eta \pi^0$ events are subjected to 
the selection criteria described above. The points with error bars in 
Fig.~\ref{exotic1-zzx}(a) show the $\eta \pi^0$ invariant mass spectrum
for the surviving events and the shaded area shows the background.
The signal and background events are normalized to $1.5 \times 10^8$
$J/\psi$ events.  
The mass resolutions and efficiencies in the 1.4~GeV mass region
are about 10 MeV and 27.2\% respectively.

 
All of the signal events are added by taking into account 
possible interference
between them, and the backgrounds are added incoherently. 
A partial wave analyses is applied to these events using
the covariant helicity coupling amplitude method
to construct the amplitudes. 
The relative magnitudes and phases of the
amplitudes are determined by a maximum likelihood fit.
Each resonance is represented by a
constant-width Breit-Wigner functions of the form 
$$
BW_{X}=\frac{m\Gamma}{s-m^2+im\Gamma},
$$
where $s$ is the square of the two-particle invariant mass, $m$ and 
$\Gamma$ are
the mass and width of intermediate resonance $X$, respectively. 
The background events from $J/\psi\to\gamma\rho^+\rho^-$ are
given the opposite log likelihood in the fit to cancel the background
events in the data. In this analysis, the masses and widths of $a_0(980)$
and $a_2(1700)$ are fixed to PDG values and those for $a_2(1320)$
and $\pi_1(1400)$ are allowed to float.

Figure~\ref{exotic1-zzx} (a) shows a comparison of the $\eta \pi^0$ 
invariant 
mass spectrum for the generated events and that from PWA projections. The
consistency is reasonable.  Figure~\ref{exotic1-zzx} (b) indicates the 
contributions from each component.
The angular distributions for the generated events and that from 
the PWA are shown in Figs.~\ref{exotic2-zzx}.

\begin{figure}[htbp]
\begin{center}
\epsfig{file=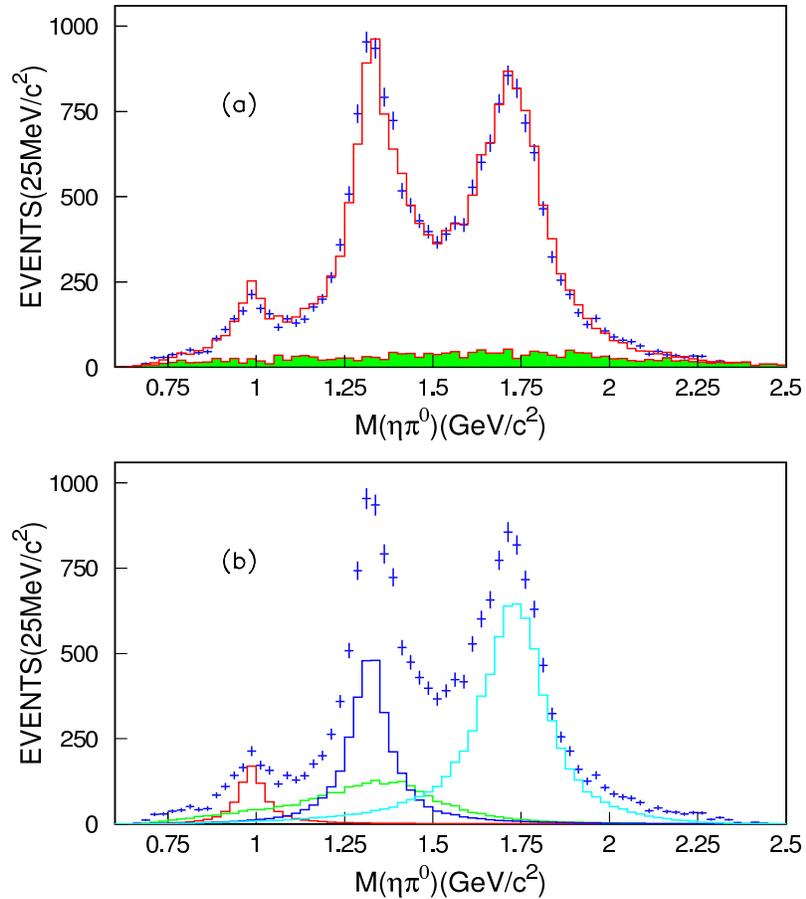,width=10.5cm,height=12.0cm}
\caption{The $\eta \pi^0$ invariant mass spectrum for
$J/\psi\to \rho \eta \pi^0$. The signals are Monte-Carlo events 
generated by taking into
account the angular distributions of each resonance and the interference
between them, as described in the text; the backgrounds are added to the
signals incoherently. The sum of the signals and backgrounds are shown
as the point with error bars. 
(a). The comparison of the generated mass spectrum and PWA projection from
all contributions. The background is shown as the shaded region. 
(b). The contribution to the PWA projections of the different 
resonant components.
\label{exotic1-zzx}}
\end{center}
\end{figure}

\begin{figure}[htbp]
\begin{center}
\epsfig{file=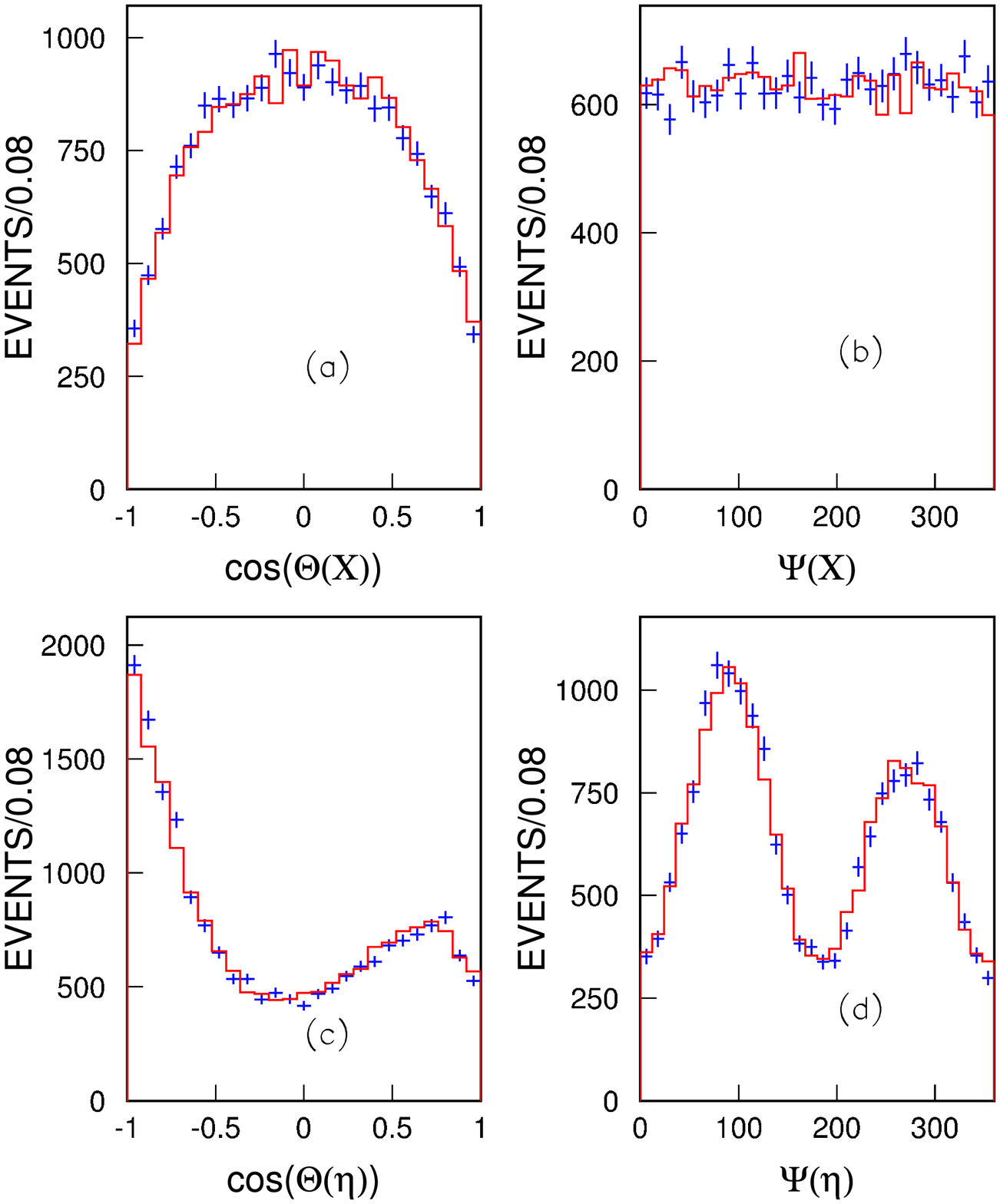,width=12.5cm,height=10.0cm}
\caption{Angular distributions for $J/\psi\to \rho \eta \pi^0$. The 
points with error
bars indicate the distributions for signal and background events and the
histograms show the PWA results.\label{exotic2-zzx}}
\end{center}
\end{figure}

A comparison of the input and output 
values for the masses, widths and branching fractions is shown in 
Table~\ref{exotic3-zzx}, where the output masses, widths and branching frations
that are obtained from the PWA analysis agree with the input values 
resonably well.

\begin{table}
\caption{Input and output comparison of masses, widths and branching
 fractions. Here the masses and widths of $a_0(980)$ and $a_2(1700)$ 
 are fixed to PDG values.  \label{exotic3-zzx}}
 \begin{center}
 \begin{tabular}{c|c|c|c|c|c}
 \hline
 \hline
& & $a_0(980)$ & $a_2(1320)$ & $\pi_1(1400)$ &  $a_2(1700)$ \\
\hline
Mass (MeV) & input & 0.985 & 1.318 &1.376 &1.732   \\
                 & output& fixed & $1.320\pm0.002$&$1.380\pm0.008$ & fixed\\
\hline
Width (MeV) & input &0.08&107 &360 &0.194 \\
                 & output&fixed &$112\pm 4$& $376\pm16$& fixed\\
\hline
Fraction  & input &4.38 &21.39 &14.57&43.36 \\
                 & output&$4.55\pm0.30$&$19.50 \pm $& $14.53\pm$& $41.64\pm$\\
\hline
  \end{tabular}
  \end{center}
 \end{table}

The Monte-Carlo simulation indicates that the partial wave
analysis is able to separate components with similar masses
but with different spin-parities
({\it i.e.}, $a_2(1320)$ and $\pi_1(1400)$),
when the statistics are large and the 
detector performance
is good. With a large statistics sample, it will also be possible to
measure the phase motion, which would give additional convincing
evidence for the existence of a resonance.

\section{Multiquarks}

\subsection{Multiquark candidates}

When N $(N\ge 4)$ quarks-antiquarks are confined within a single
MIT bag, a {\it multiquark} state is formed. The color structure within
a multiquark state is complicated and not unique. It always has a
component that is the product of two (or more) color-singlet
hadrons. A special mechanism is needed to prevent the multiquark from
falling apart easily and thereby becoming extremely broad when it lies 
above threshold fr decay into conventional hadrons.

There are several well known multiquark candidates. The first one
is the $H$ dibaryon suggested by Jaffe decades ago~\cite{dibaryon}.
On-going doubly strange hypernuclei search experiments have pushed
its binding energy to be less than several MeV, about to the point where
the existence of the $H$ existence is now rather dubious. Jaffe also 
suggested that the
low-lying scalar nonet are tetraquarks because of their low
and inverted mass pattern~\cite{sigma}. The third one is the
rapidly
fading $\Theta^+$ pentaquark. Interested readers may consult 
a recent review~\cite{IJMPA}. 

In general, a multiquark state is expected to have a broad width
since it can easily fall apart into mesons and/or baryons
when  its mass is above
the mass threshold for producing these hadrons. Multiquark
states may only be experimentally observable 
when their masses are near these mass thresholds
--- either below or just above them --- otherwise the widths
of the multiquark states might be too wide to be 
experimentally distinguishable from non-resonant background.
The $J/\psi$ meson serves a unique role for searches of 
new hadrons and studies of  light hadron spectroscopy. 
Recently, a number of new structures have been 
observed in $J/\psi$ decays, including a strong near-threshold mass 
enhancements in the $p\bar p$ invariant mass spectrum from 
$J/\psi\rightarrow\gamma p\bar p$ decays~\cite{gpp}, the $p \bar \Lambda $ and the 
$K^- \bar \Lambda$ mass spectra in  $J/\psi \rightarrow p K^- \bar \Lambda$ 
decays~\cite{pkl}, the $\omega\phi$ mass spectrum in the double-OZI suppressed 
decay $J/\psi\to\gamma\omega\phi$~\cite{bes-gwf}, and two new resonances, 
the $X(1835)$, in $J/\psi\to\gamma\pi^+\pi^-\eta'$ decays~\cite{x1835} as well as
a very broad $1^{--}$ resonant structure in the $K^+K^-$
invariant mass spectrum from
$J/\psi \rightarrow K^+K^-\pi^0$ decays.  All of these new structures are
possibly multiquark states. \\

\subsubsection{Observation of a strong near-threshold mass enhancement in
the $p\bar p$ invariant mass spectrum}

The BES Collaboration observed an anomalous strong $p\bar{p}$
mass enhancement near the
$m_p + m_{\bar p}$ mass threshold in $J/\psi\rightarrow\gamma p\bar p$ 
decays~\cite{gpp}.  It can be fit with an $S$- or $P$- wave Breit-Winger
resonance function; in the case of the $S$-wave fit, the mass is
below $m_p + m_{\bar p}$ at 
$1859^{+3}_{-10}$$^{+5}_{-0.5}$~MeV and the width is smaller
than 30~MeV at the 90$\%$ C.L. (see Fig. ~\ref{gpp_mfit}).

\begin{figure}[htpb]
\begin{center}
 \vspace{9.0cm}
\includegraphics{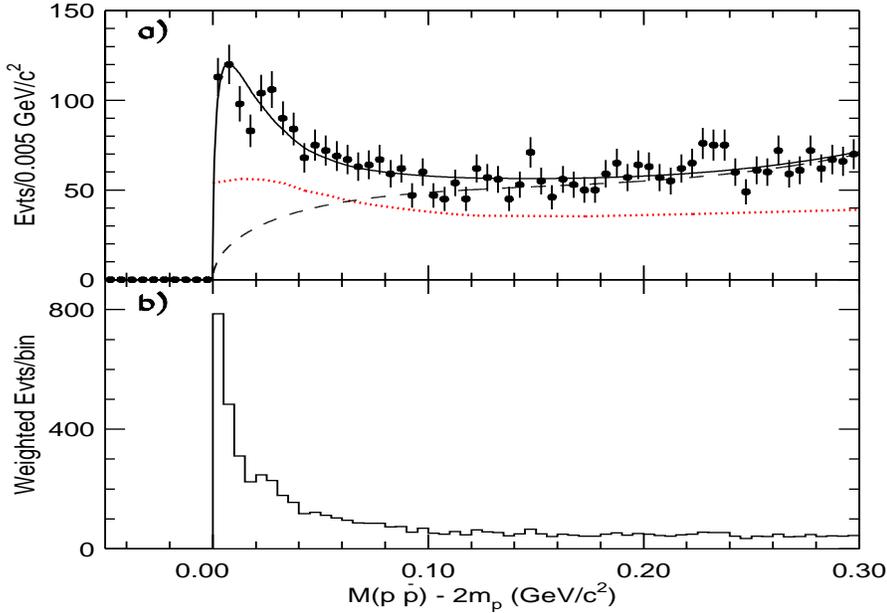}

\caption{\label{gpp_mfit}
(a) The near-threshold $M_{p\bar p}-2m_p$ distribution for
the $J/\psi\to \gamma p \bar p$ events.  The dashed
curve is the background from $\pi^0 p\bar{p}$
final states.  The dotted curve
indicates how the acceptance varies with $p \bar p$ invariant
mass.
(b) The  $M_{p\bar p}-2m_p$ distribution with events weighted
by $q_0/q$.}
\end{center}
\end{figure}

It is interesting to note that such a strong mass threshold
enhancement is not observed in $p \bar p$ cross section 
measurements, also not in B decays~\cite{B-ppbar}, 
nor in radiative $\psi(2S) or \Upsilon(1S) \rightarrow 
 \gamma p \bar{p}$ decays \cite{part3_bes_radpsi2s,part3_cleo-gpp},  and not in 
$J/\psi \rightarrow \omega p \bar{p}$ \cite{part3_bes_wppbar}. 
These non-observations disfavor a purely $p \bar{p}$ FSI interpretation 
of the strong mass threshold enhancement which is, as of now, 
uniquely observed in the $J/\psi \rightarrow\gamma p \bar{p}$ 
decay process.

This surprising experimental observation has stimulated a number of 
theoretical speculations~\cite{ppbar,part3:ref:theory,gao,yan,fsi1,fsi2}.  
Among these, the most intriguing is that it is an example
of a $p\bar p$ bound state, 
sometimes called {\it baryonium}~\cite{ppbar,baryonium,yan}, which has
been the subject of many experimental searches~\cite{richard}.

It is worth noting that if the observed $p \bar p$ mass enhancement is due to a 
resonance, the decay branching fraction (BF) to $p\bar p$ is larger than 
expected  for a conventional $q\bar{q}$ meson.  From Ref.~\cite{gpp}, 
$B(J/\psi\to\gamma X)\cdot B(X\to p \bar p)$ = $(7.0\pm 0.4(stat)^{+1.9}_{-0.8}(syst))
\times 10^{-5}$.   Measurements of the inclusive photon spectrum in $J/\psi$ 
decays by the Crystal Ball group limit the branching fraction for
radiative production of a 
narrow resonance around 1.85 GeV to
be below $2\times 10^{-3}$ (an estimate that is based on
data reported in Ref.~\cite{cbl}). 
Thus, if a resonance is responsible for producing the observed 
enhancement, 
it would have a branching fraction to $p\bar p$ that is larger 
than 4$\%$, which would be the $largest$ $p\bar p$ branching fraction 
of all known mesons~\cite{pdg2006}. Since decays to $p \bar p$ are 
kinematically possible only for a small portion of the high-mass tail 
of the resonance and have very limited phase space, a large $p \bar p$ 
branching fraction implies an  unusually strong coupling to $p \bar p$, as 
expected for a $p \bar p$ bound state. \\


\subsubsection{Observation of X(1835) in $J/\psi\to \gamma \pi^+\pi^-\eta'$}

The baryonium interpretation of the $p\bar p$ mass enhancement requires
a new resonance with a mass around 1.85~GeV, which would be 
supported by the observation of a resonance in other decay channels.
Possible strong decay channels for a $p\bar p$ bound state, suggested in 
Ref.~\cite{gao,yan}, include $\pi^+\pi^-\eta'$. 

An analysis of 
$J/\psi\to\gamma\pi^+\pi^-\eta'$ with $\eta' \to \eta \pi^+\pi^-$
and $\eta' \to \gamma \rho$ is reported
in Ref.~\cite{x1835}. 
For both the $\eta' \to \eta \pi^+\pi^-$
and $\eta' \to \gamma \rho$ data samples,
the  $\pi^+\pi^-\eta'$ invariant
mass spectra for the selected events exhibit
peaks at a mass around 1835~MeV.
Figure~\ref{sum} is the $\eta' \pi^+\pi^-$ mass spectrum
for both $\eta'$ decay modes combined, where
a distinct peak near 1835~MeV is observed.
The combined spectrum is
fitted with a Breit-Wigner (BW) function convolved with a Gaussian mass
resolution function (with $\sigma = 13$~MeV) to represent the
$X(1835)$
signal plus a smooth polynomial background. The mass and width
obtained from the fit are
$M=1833.7\pm 6.1$~MeV and $\Gamma=67.7\pm 20.3$~MeV,
respectively.
The signal
yield from the fit is $264\pm 54$ events with a confidence level of
45.5$\%$ ( $\chi^2/d.o.f.$ = 57.6/57) and $-2\ln L=58.4$.  A fit to the mass
spectrum without a BW signal function returns $-2\ln L = 126.5$. The change
in
$-2\ln L$ with $\Delta(d.o.f.) = 3$  corresponds to a statistical
significance
of 7.7~$\sigma$ for the signal.

Using MC-determined selection efficiencies of $3.72\%$ and $4.85\%$ for the
$\eta'\to\pi^+\pi^-\eta$ and $\eta'\to\gamma\rho$ modes, respectively,
we determine a product branching fraction of
\begin{center}
$B(J/\psi\to\gamma X(1835))\cdot B(X(1835)\to\pi^+\pi^-\eta') =
(2.2\pm 0.4)\times 10^{-4}.$
\end{center}

\begin{figure}[htbp]
\centerline{\epsfig{file=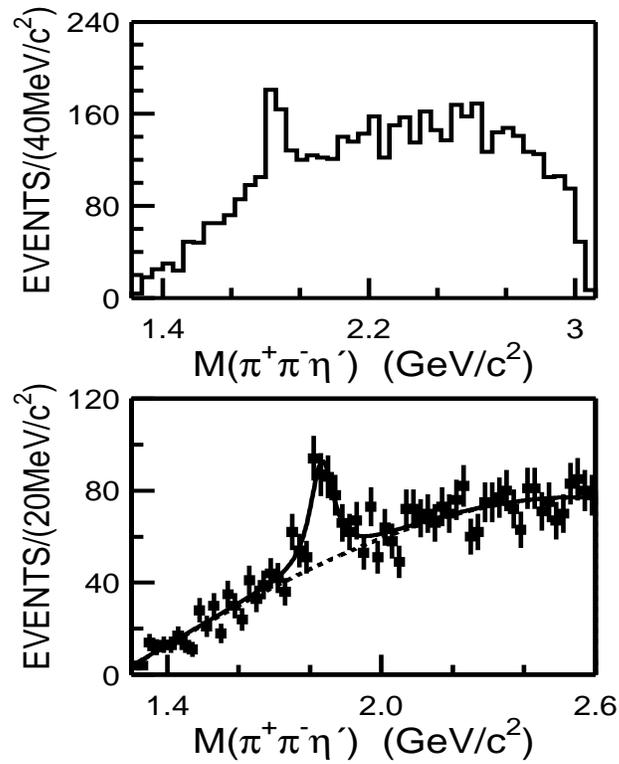,width=8cm,height=10cm}}

\caption{\label{sum}   The $\pi^+\pi^-\eta'$ invariant mass distribution for
           selected events from both the
$J/\psi\to\gamma\pi^+\pi^-\eta'(\eta'\to\pi^+\pi^-\eta,
\eta\to\gamma\gamma)$
           and $J/\psi\to\gamma\pi^+\pi^-\eta' (\eta'\to\gamma\rho)$ analyses. 
The bottom
           panel shows the fit (solid curve) to the data (points with error 
           bars); the dashed curve indicates the background function. }
\end{figure}

The mass and width of the 
$X(1835)$ are not compatible with any known meson resonance~\cite{pdg2006}.
We examined the possibility that the $X(1835)$ is responsible for the 
$p\bar{p}$ mass threshold enhancement observed in radiative
$J/\psi\to\gamma p \bar{p}$ decays \cite{gpp}.  
Subsequent to the publication of Ref.~\cite{gpp}, it was pointed out 
that the $S$-wave BW function used for the fit should 
be modified to include the effect of final-state-interactions (FSI) 
on the shape of the $p\bar{p}$ mass spectrum~\cite{fsi1,fsi2}. 
Redoing the $S$-wave BW fit to 
the $p\bar{p}$ invariant mass spectrum of Ref.~\cite{gpp} with
the zero Isospin and $S$-wave FSI factor of Ref.~\cite{fsi2}
included,  yields a mass  
$M = 1831 \pm 7$~MeV and a width $\Gamma < 153$~MeV (at the 90$\%$ 
C.L.);  values that are in good agreement with the 
mass and width of $X(1835)$ seen in $\pi\pi\eta '$.  
Moreover, according to 
Ref.~\cite{yan}, the $\pi\pi\eta'$ decay mode is expected to be a
strong sub-threshold decay channel for 
a $p \bar p$ bound state. Thus, the $X(1835)$ resonance is a prime 
candidate for
the source of the $p\bar{p}$ mass threshold enhancement in 
$J/\psi\to \gamma p\bar{p}$ process. 
In this case, the $J^{PC}$ and $I^G$ of the $X(1835)$ could only be $0^{-+}$ 
and $0^+$, which can be tested in future experiments.  Also in this context, 
the relative $p \bar p$ decay strength is quite  strong:
$B(X\to p\bar p)/B(X\to \pi^+\pi^-\eta')\sim 1/3$ (The product BF 
determined from
the fit that includes FSI effects on the $p\bar{p}$ mass spectrum is within
the systematic errors of the result reported in Ref.~\cite{gpp}.)
Since decays to $p \bar p$ are kinematically allowed only for a small portion 
of the high-mass tail of the resonance and have very limited phase space, 
the large $p \bar p$ branching fraction implies an unusually strong coupling 
to $p \bar p$, as expected for a $p \bar p$ bound state~\cite{baryonium,zhu}.  
However, other possible interpretations of the $X(1835)$ that have no relation
to the $p \bar p$ mass threshold enhancement are not excluded. \\


\subsubsection{Observation of $p\bar \Lambda$ mass threshold enhancement in
$J/\psi\to pK^-\bar\Lambda$}

An analysis of  $J/\psi\to pK^-\bar\Lambda$ 
is described in detail in Ref.~\cite{pkl}.
The $p \bar\Lambda$ invariant mass spectrum for selected events is shown 
in Fig.~\ref{x208}(a), where an enhancement is evident near the 
$m_{\Lambda} + m_p$ mass 
threshold.  No corresponding structure is seen in a sample of
$J/\psi \rightarrow p K^- \bar\Lambda$ Monte-Carlo events generated with a uniform 
phase space distribution.  The $pK^- \bar\Lambda$ Dalitz plot is shown in 
Fig.~\ref{x208}(b), where, in addition to bands for the well established 
$\Lambda^*(1520)$ and $\Lambda^*(1690)$,  a significant $N^*$ band 
near the $K^-\bar\Lambda$ mass threshold, and a $p \bar\Lambda$ mass 
enhancement in the right-upper part of the Dalitz plot,
isolated from both the $\Lambda^*$ and $N^*$ bands, are evident.


\begin{figure}[htbp]
\centerline{\epsfig{file=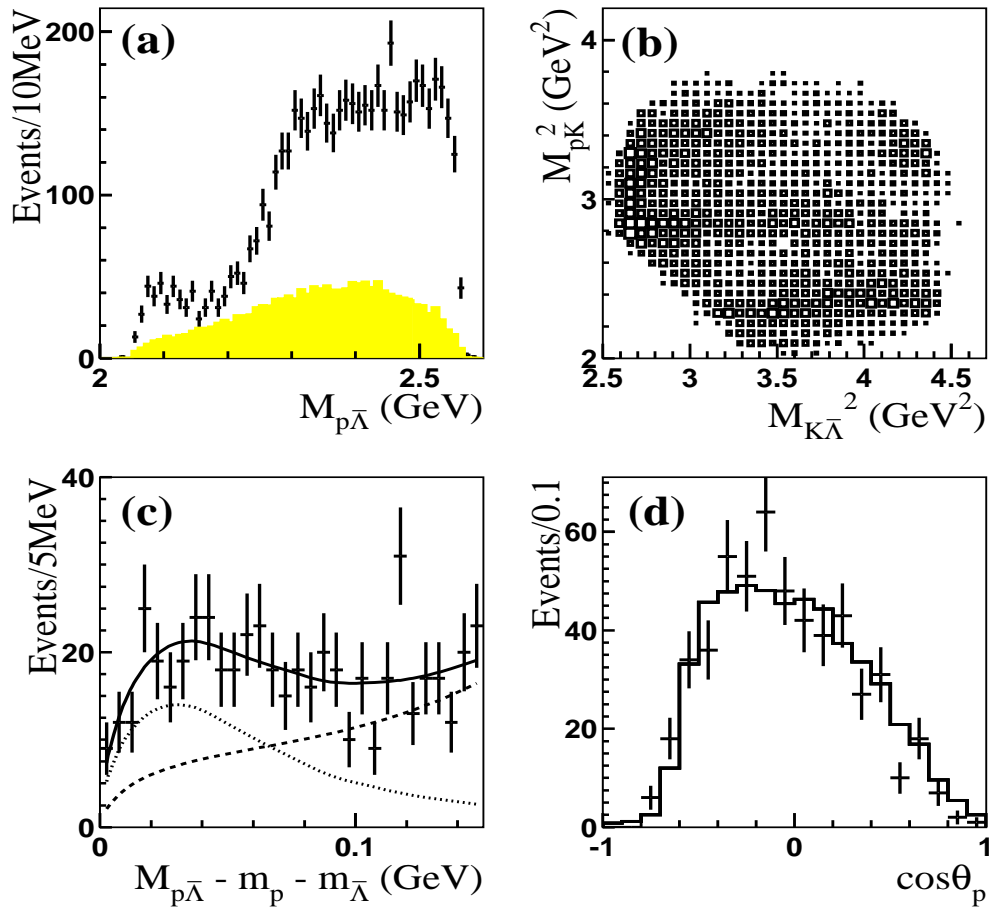,width=13cm,height=12cm}}

\caption{\label{x208}  (a) The points with error bars indicate the measured 
          $p\bar \Lambda$ mass spectrum from $/\psi\to K^+\bar{p}\Lambda$ 
          decays; the shaded histogram indicates  
          phase space MC events (arbitrary normalization).
	  (b) The Dalitz plot for the selected event sample.
	  (c) A fit (solid line) to the data. The 
	  dotted curve indicates the Breit-Wigner signal and the 
          dashed curve the phase space `background'.
	  (d) The $\cos \theta_p$ distribution under the enhancement,
           the points are data and the histogram is the MC
           (normalized to the data).}
\end{figure}

The $p\bar{\Lambda}$ invariant mass enhancement is fitted 
with an acceptance-weighted $S$-wave Breit-Wigner 
function, together with a function describing 
the phase space contribution, as shown in Fig.~\ref{x208}(c).
The fit
gives a peak mass of $m=2075\pm 12$~MeV and a width $\Gamma=90 \pm 35$~MeV.  
The enhancement deviates from the shape of the phase space contribution with a 
statistical significance of about $ 7\sigma$. 

The fit yields $N_{res}=238 \pm 57$ signal events, corresponding to
a product branching fraction

\begin{center}

$ BR(J/\psi \rightarrow K^-X) BR (X\rightarrow p \bar\Lambda) 
=(5.9\pm 1.4)\times 10^{-5}$.

\end{center}

Searches for the same enhancement in $K\pi$ and $K\pi\pi$ modes in the 
$J/\psi, \psi' \rightarrow KK\pi, KK\pi\pi$ decays would help to confirm 
the
presence of this anomalous peak and understand its nature, and in particular
distinguish whether or not
it is due to a conventional $K^*$ meson, a 
multiquark state, or a  baryon-antibaryon resonance.
If its decay widths to $K\pi$ and $K\pi\pi$ modes are much smaller than
to $p \bar\Lambda$, its interpretation as a conventional $K^*$ meson
would be disfavored.\\


\subsubsection{Observation of $K^-\bar\Lambda$ mass threshold 
enhancement in $J/\psi\to pK^- \bar\Lambda$}

In the Dalitz plot of Fig.~\ref{x208}(b), a clear band
is also observed near the $K^- \bar\Lambda$ mass threshold.
We have performed a preliminary partial wave analysis (PWA)
of these data and determined that 
the mass of this threshold structure (dubbed $N_X^*$) is in the range 
between 1500 to 1650~MeV, with a width between
70 and 110~MeV and a favored spin-parity of $1/2^-$. It
has a product branching fraction $B(J/\psi \rightarrow \overline{p} N_X^*)
B(N_X^* \rightarrow K^- \bar \Lambda)$ 
that is larger than $2\times 10^{-4}$.


Considering the fact that the $N_X^*$ mass is below or very close
to the $K^- \bar\Lambda$ mass threshold, the phase space available to the
$K^-\bar\Lambda$ final state is very small. Thus, the large branching 
fraction to  $K^-\bar\Lambda$ 
indicates that he $N_X^*$ has very strong coupling to $K^-\bar\Lambda$,
suggesting that it could be the $K^-\bar\Lambda$ resonant state
predicted by the chiral-$SU(3)$ quark model~\cite{zhangzy-hf}.\\


\subsubsection{Observation of a broad $1^{--}$ resonant structure
 in the $K^+K^-$ mass spectrum in $J/\psi \to K^+K^-\pi^0$}

A broad peak is observed at low $K^+K^-$ invariant masses in 
$J/\psi \to K^+K^-\pi^0$ decays; the 
analysis is described in detail in Ref.~\cite{kkpi0}. 
The Dalitz plot for the
selected $K^+K^-\pi^0$ events is shown in
Fig.~\ref{fig:kkpi}(b), where there is a broad $K^+K^-$ 
(diagonal) band in 
addition to prominent $K^*(892)$ and $K^*(1410)$ signals.  This band
corresponds to the broad low-mass
peak observed around 1.5~GeV in the
$K^+K^-$ invariant mass projection shown in Fig.~\ref{fig:kkpi}(c).


\begin{figure}[htbp]
\centerline{\epsfig{file=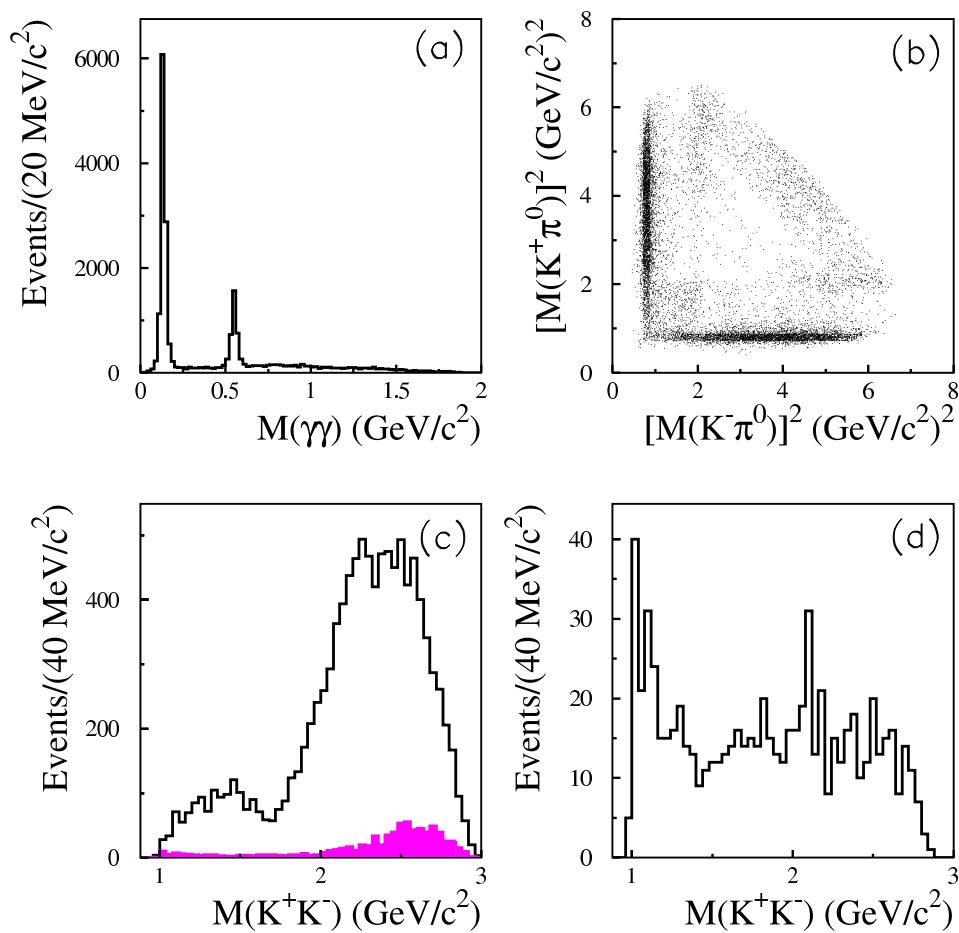,width=5.0in}}

\caption{  (a) The $\gamma\gamma$ invariant mass distribution
    for $J/\psi\to \gamma\gamma K^+K^-$ events.  (b) The
    Dalitz plot for $K^+K^-\pi^0$ candidate events.  (c) The $K^+K^-$
    invariant mass distribution for the $K^+K^-\pi^0$ candidate events; 
    the solid histogram is data and the shaded histogram is the background
    (normalized to the data).  (d) The $K^+K^-$ invariant mass 
    distribution
    for the $\pi^0$ mass sideband events (not normalized).}
 \label{fig:kkpi}
\end{figure}

A partial wave analysis shows that the $J^{PC}$ of this structure is
$1^{--}$. Its pole position is determined to be
$(1576^{+49}_{-55}$$^{+98}_{-91})$~MeV - $i
(409^{+11}_{-12}$$^{+32}_{-67})$~MeV, and the 
product branching fraction is
$B(J/\psi\to X \pi^0)\cdot B(X\to K^+K^-)$=
$(8.5\pm0.6^{+2.7}_{-3.6})\times 10^{-4} $, where the first errors are
statistical and the second are systematic.  These parameters are not
compatible with any known meson resonances~\cite{pdg2006}.

To understand the nature of this broad $1^{--}$ peak, it is important
to search for a similar structure in $J/\psi\to K_S K^\pm \pi^\mp$
decays to determine its isospin. It would also be intriguing to search for
it in $K^*K, KK\pi$ decay modes.  In the mass region covered by the $X$, 
there are
several other $1^{--}$ states, such as the $\rho(1450)$ and
$\rho(1700)$, but the width of the $X$ is much broader than the widths
of any of these other meson states. This may be an indication that the $X$ 
has a different nature than these other mesons.  For example, the very 
broad width is suggestive of a multiquark state.

\subsection{Simulation of the $X(1835)$ in $J/\psi$ decays
at \bes3/BEPCII}

The numerous near-threshold enhancements observed
at BESII motivated us to simulate the production of $X(1835)$ in 
$J/\psi\to \gamma X(1835)$,
$J/\psi\to \omega X(1835)$ and $J/\psi\to \phi X(1835)$, with
$X(1835) \to \eta^\prime \pi^+\pi^-$, at \bes3. \\


\subsubsection{$J/\psi \to \gamma X(1835)$}

In the simulation, we only focused on the decay mode of $X(1835)\to
\eta^\prime \pi^+\pi^-$ with $\eta^\prime \to \eta\pi^+\pi^-$ and $ \eta\to
\gamma\gamma$. Therefore, the final-state composition is $3\gamma
2(\pi^+\pi^-)$. The main backgrounds come from $J/\psi\to
\eta^\prime \pi^+\pi^-\pi^0$, $\omega \eta^\prime$,
$b_1\pi$ and $\phi f_2^\prime$, of which $J/\psi\to \eta^\prime
\pi^+\pi^-\pi^0$ is dominant.

Charged tracks are selected by requiring that they are within
the polar angle region $|\cos\theta|<0.93$ and within the vertex region of
$r_{xy}<1$~cm and $|z|<$5~cm.  Four good charged tracks with
zero net charge are required. Candidate photons 
are
required to have an energy deposit in the
EMC that is greater than 40 MeV, to be isolated from 
charged tracks by more
than $20^\circ$ in both the $x-y$ and $r-z$ planes and 
the opening angle with any other photon in the event
that is greater than $7^\circ$. 
Finally, the four charged tracks and photons in the event are
kinematically fitted using four energy and momentum conservation
constraints (4C) to the $J/\psi \to 3 \gamma 2(\pi^+ \pi^-)$ hypothesis.
The fit is repeated using all permutations and the
combination with the best fit to $3\gamma 2(\pi^+\pi^-)$ is retained.
In order to further suppress the backgrounds and improve the mass
resolution, a 5C kinematic fit is imposed to constrain the invariant mass
of two of the photons to the $\eta$ mass. 
The $\eta'$ is selected by requiring $|m_{\eta \pi^+\pi^-}-m_\eta'| 
< 0.04$ GeV.

About 11,000 $J/\psi \to \gamma X(1835)$, $X(1835)\to
\eta^\prime \pi^+\pi^-$, $\eta^\prime \to \eta\pi^+\pi^-$ and $ \eta\to
\gamma\gamma$ simulated events
pass the above-listed  selection criteria.  According to the 
product branching fraction reported in 
Ref.~\cite{x1835},  this number of events is equivalent to 
a $3 \times 10^9$  $J/\psi$ event sample.
We also generated and selected events from the
background channels mentioned above and normalized these
to the same number of $J/\psi$ events.

The \bes3  $\eta' \pi^+\pi^-$ mass 
resolution near the resonance mass of 1835~MeV
is about 3~MeV and the selection effficiency 
for the above-listed selection criteria is about 10\%.

Figure~\ref{x-radiative} shows the 
$\eta' \pi^+ \pi^-$ invariant mass spectrum 
for  the selected  $J/\psi\to \gamma X(1835)$, 
$X(1835) \to \eta' \pi^+ \pi^-$ signal and background events.
Compared with the BESII results ((Fig.~\ref{sum}), the 
background level at \bes3 is much lower 
and the $X(1835)$ signal more distinct due to much better 
energy and momentum resolutions.

\begin{figure}[htbp]
\begin{center}
\epsfig{file=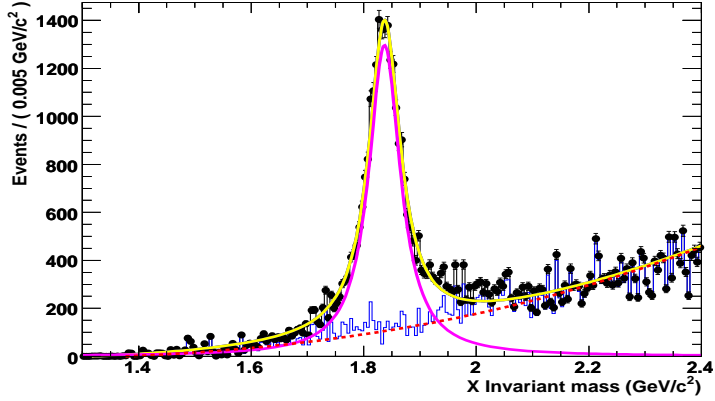,width=10.5cm,height=6.0cm}
\caption{The $\eta^\prime \pi^+\pi^-$ invariant mass for
$J/\psi\to \gamma \eta^\prime \pi^+\pi^-$. The generated signals
and backgrounds are normalized to $3 \times 10^{9} J/\psi$ events 
and are added incoherently.}
\label{x-radiative}
\end{center}
\end{figure}

A fit with a Breit-Wigner signal and a second order polynomial 
background function yields
mass, width and branching fractions for the $X(1835)$ that
are consistent with the MC-input parameters.\\

\subsubsection{$J/\psi \to \omega X(1835)$}

A sample of $J/\psi \to \omega X(1835)$ events was generated with $\omega$
decaying into $\pi^+\pi^-\pi^0$ and $X(1835)$ to $\eta'\pi^+\pi^-$.
The final states contain four $\gamma$'s and three $(\pi^+\pi^-)$ pairs. 
Potential
background channels are $J/\psi\to\omega K_s K \pi$, $KK 4\pi$, 
$\phi 4\pi$, $\phi K^\star K$, $\omega K^\star K$, and $\pi^0 6\pi$.
 
The selection of the good charged tracks and photons are similar to
those for the $J/\psi \to \gamma X(1835)$ analysis described above. 
Events with six 
good charged tracks and at least four good photons are retained.
The 4C kinematic fit is applied using the
 $J/\psi \to 4\gamma 3(\pi^+\pi^-)$ hypothesis.

After the final selection, the efficiency for 
$J/\psi \to \omega X(1835)$, $\omega \to \pi^+\pi^-\pi^0$, 
$X(1835) \to \eta'\pi^+\pi^-$, and $\eta' \to \eta \pi^+\pi^-$ is about
2.4\% and the mass resolution at 1835~MeV is 3~MeV.

Figure \ref{x-omega} shows the invariant mass spectrum of
$\eta' \pi^+ \pi^-$ for the $J/\psi\to \omega X(1835)$, $\omega \to
\pi^+\pi^-\pi^0$, $X(1835) \to \eta' \pi^+ \pi^-$ signal 
and background events. 

\begin{figure}[htbp]
\begin{center}
\epsfig{file=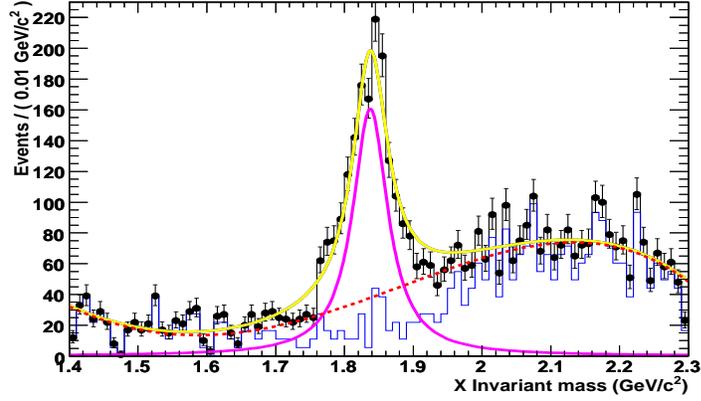,width=10.5cm,height=6.0cm}
\caption{The $\eta^\prime \pi^+\pi^-$ invariant mass for
$J/\psi\to \omega \eta^\prime \pi^+\pi^-$. The generated signals
and backgrounds are normalized to $3 \times 10^{9} J/\psi$ events
and are added incoherently.}
\label{x-omega}
\end{center}
\end{figure}

\subsubsection{$J/\psi \to \phi X(1835)$}

For $J/\psi \to \phi X(1835)$, $\phi \to K^+ K^-$ and $\eta' \to \eta
\pi^+\pi^-$, the final states include two photons, two charged kaons
and four charged pions. The numbers of generated signal events
and background events
that pass the event selection criteria, are listed in 
Table~\ref{phi-x},
where one can see that the background level is very low. 
The invariant mass spectrum of
$\eta' \pi^+ \pi^-$ for the $J/\psi\to \phi X(1835)$, $\phi \to
K^+K^-$, $X(1835) \to \eta' \pi^+ \pi^-$ for signal and background
events are shown in Fig.~\ref{x-phi}, where the $X(1835)$ is almost free 
of 
the backgrounds.
The selection effciency for this channel is about 2.6\% and the 
$\pi^+\pi^-\eta'$ invariant mass 
resolution near 1835~MeV is about 3 MeV.

\begin{table}[htbp]
\begin{center}
\caption{$J/\psi\rightarrow\phi X$ and its background channels} 
\label{phi-x}
\begin{tabular}{|c|r|r|r|r|r|r|r|} \hline

                    & $\phi X(1835)$ & $KK 4\pi$ & $\omega K_s K\pi$
&$\phi f_2^\prime$ & $\phi 4\pi$ &$\omega K^\star K$ & $\pi^0
6\pi$\\ \hline
 Entries & 169318 &54798 &89891 &79874 &45959 &90866 &127540 \\
 \hline
 $N_{ch}=6$ &21272&15611&15025&1254&11405&26310&48359\\\hline
$N_{\pi^+}=2$ &&&&&&&\\
$N_{\pi^-}=2$ &18024&13021&1036&1063&9266&2224&516\\\hline
$N_{K^+}=1$ &&&&&&&\\
$N_{K^-}=1$ &17492&12409&184&1031&8972&175&9\\ \hline
Vertex&5434&3816&18&464&2671&91&8\\\hline
4C-fit&2369&206&2&42&156&0&0\\\hline
5C-fit&1914&0&0&3&1&0&0\\\hline\hline
Scale&1914&0&0&8&86&0&0\\\hline
\end{tabular}
\end{center}
\end{table}

\begin{figure}[htbp]
\begin{center}
\epsfig{file=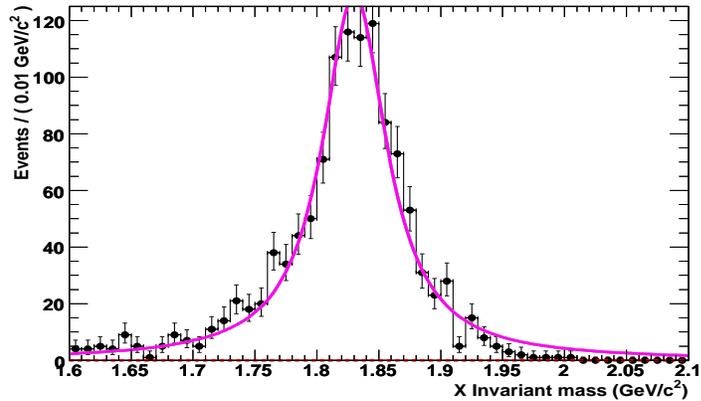,width=10.5cm,height=6.0cm}
\caption{The $\eta^\prime \pi^+\pi^-$ invariant mass for
$J/\psi\to \phi \eta^\prime \pi^+\pi^-$. The generated signals
and backgrounds are normalized to $3 \times 10^{9} J/\psi$ events
and are added incoherently.}
\label{x-phi}
\end{center}
\end{figure}

\section{Molecular states}

Molecular states are bound states of two (or more) color-singlet
hadrons. In other words, there are two (or more) MIT bags.
Color-singlet hadrons --- usually $\pi$-mesons ---
 are exchanged between these bags to produce the attractive force.  There
have been speculations that the $f_0(980)$ and $a_0(980)$ are $K \bar K$
molecules since they are only 10~MeV below $K\bar K$ threshold
\cite{isgur}. The $\Lambda (1405)$ is sometimes postulated as a $K N$
molecule.  More recently, the $X(3872)$ meson, which has a mass 
that is very nearly
equal to $m_{D^0} + m_{D^{*0}}$, has been hypothesized to be a
$D^0\bar{D^{*0}}$ molecular state~\cite{tornqvist_deuson}.

\chapter{Baryon spectrum}
\label{sec:baryon}

\section{Baryons as $qqq$ states} 
 
The classical picture for baryons is that each baryon is comprised 
of three quarks. Based on this picture, the non-relativistic 
constituent quark model (NRCQM) provides an explicit classification for 
light baryons in terms of group symmetry.  Until now, all 
established baryons listed in the PDG tables~\cite{pdg2006}. 
can be ascribed to three-quark ($qqq$) 
configurations.
 
Here we briefly review the main ingredients of the model
and indicate how the many deviations that have been 
observed in experiment might be exposing complicated aspects of
strong QCD dynamics. The understanding of these deviations and 
searches  for new non-NRCQM
states and phenomena is the main motivation for 
further studies of baryon spectroscopy. 
 
In the light quark ({\it i.e.} $u$, $d$, and $s$) sector, the total 
wavefunction of a baryon consists of four parts: i) the spatial 
wavefunction $\psi$; ii) the flavor wavefunction $\phi$; iii) the spin 
wavefunction $\chi$; and iv) the color wavefunction $\phi_c$. For such 
systems of fermions, the Pauli principle requires that the total 
wavefunction is antisymmetric under exchange of any two quarks. 
Therefore, the total wavefunction must be anti-symmetrized. Note 
that the color wavefunction of
a normal baryon state is a color singlet state and, thus, 
the color wavefunction is always 
anti-symmetric under exchange of any two 
quarks. As a consequence, one needs to {\it symmetrize} the rest 
part of the total wavefunction, i.e. spatial, spin and isospin, to 
obtain an overall antisymmetric baryon wavefunction. Since the 
representations of the permutation group $S_3$ are also 
representations of the $SU(2)$, $SU(3)$ and $SU(6)$ groups, it is 
convenient 
to construct the baryon wavefunctions on the basis of $S_3$ group 
symmetry.

In general, there are four representations of $S_3$: the totally 
symmetric basis $e^s$, the totally antisymmetric basis $e^a$, and 
the two 
mixed-symmetry bases $e^\lambda$ and $e^\rho$, which are defined 
under the permutation transformations: 
\begin{equation} 
P_{12}\left( \begin{array}{c} 
e^\lambda\\ 
e^\rho 
\end{array} \right) = 
\left(\begin{array}{cc} 
1 & 0\\ 
0 & -1 
\end{array}\right) 
\left(\begin{array}{c} 
e^\lambda\\ 
e^\rho 
\end{array}\right) \ , 
\end{equation} 
and 
\begin{equation} 
P_{13}\left( \begin{array}{c} 
e^\lambda\\ 
e^\rho 
\end{array} \right) = 
\left(\begin{array}{cc} 
-\frac 12 & -\frac{\sqrt{3}}{2}\\ 
-\frac{\sqrt{3}}{2} & \frac 12 
\end{array}\right) 
\left(\begin{array}{c} 
e^\lambda\\ 
e^\rho 
\end{array}\right) \ , 
\end{equation} 
where $P_{12}$ and $P_{13}$ are permutation operators for exchange 
of $1 \leftrightarrow 2$ and $1\leftrightarrow 3$, respectively.

The wavefunctions of spin ($SU(2)$), flavor ($SU(3)$) and combined
spin-flavor ($SU(6)$) are also representations of the $S_3$ group, 
and can be expressed as: 
\begin{eqnarray} 
SU(2) \ & {\bf 2}\otimes{\bf 2}\otimes{\bf 2} &= {\bf 4}_s +{\bf 
2}_\rho + 
{\bf 2}_\lambda , \\ 
SU(3) \ & {\bf 3}\otimes{\bf 3}\otimes{\bf 3} &= {\bf 10}_s +{\bf 
8}_\rho + 
{\bf 8}_\lambda + {\bf 1}_a ,\\ 
SU(6) \ & {\bf 6}\otimes{\bf 6}\otimes{\bf 6} &= {\bf 56}_s +{\bf 
70}_\rho + {\bf 70}_\lambda + {\bf 20}_a , 
\end{eqnarray} 
where the subscripts denote the corresponding $S_3$ basis for each 
representation, and the bold numbers denote the dimension of the 
corresponding representation. Note that we cannot construct an 
antisymmetric spin state with identical spin 1/2 fermions. 
 
Thus, the spin-flavor wavefunctions can be expressed as $|{\bf 
N}_6, ^{2S+1} {\bf N}_3\rangle $, where ${\bf N}_6$ and ${\bf 
N}_3$ denote the $SU(6)$ and $SU(3)$ representation and $S$ stands for 
the total spin of the wavefunction. For example, the spatial 
ground states with a spatial part that is trivially symmetric 
necessarily have a  spin-flavor ($\chi$-$\phi$) part 
that is also symmetric, which can 
be made up of either $\phi^s\chi^s$ or $(\phi^\rho\chi^\rho 
+\rho^\lambda\chi^\lambda)/\sqrt{2}$; {\it i.e.},  $|{\bf 56}, 
^{\bf 4} {\bf 10}\rangle^s$ or $|{\bf 56}, ^{\bf 2} {\bf 
8}\rangle ^s$. These are, respectively,  
the $J^P=3/2^+$ decuplet that contains the 
$\Delta(1232)$ and the $J^P=1/2^+$ octet that contains the nucleon. 
 
In the NRCQM, it is assumed that the total orbital angular 
momentum of the three-quark system is conserved, which means that 
the spatial wavefunction is also a representation of the $S_3$ 
group. Combining the spatial wavefunctions, baryons can be
constructed in the $SU(6)\otimes O(3)$ symmetry limit. A useful 
classification, one used by the PDG~\cite{pdg2006}, is to assign the 
baryons 
into harmonic oscillator bands that have the same number of 
quanta of excitation $N$. Each band consists of a number of 
supermultiplets, specified by (${\bf N}_6$, $L^P_N$) with $L$ the 
total quark orbital angular momentum and $P$ the total parity. The 
$N=0$ bands consists of the spatial ground state 56-plet 
(56,$0^+_0$). The $N=1$ band contains the $L=1$ negative-parity 
baryons with masses below 1.9~GeV and consists of the (70,$1^-_1$) 
multiplet, which, with a total of $70\times 3$ spin-flavor states, is 
composed of two decuplets with $J^P=1/2^-$ and $3/2^-$, two nonets 
with $J^P=1/2^-$ and $3/2^-$, and three octets with $J^P=1/2^-$, 
$3/2^-$ and $5/2^-$. The $N=2$ band contains five 
supermultiplets: (56,$0^+_2$), (70,$0^+_2$), (56,$2^+_2$), 
(70,$2^+_2$), and (20,$1^+_2$). 
The PDG~\cite{pdg2006} gives quark-model 
assignments for some of the known baryons in terms of the 
flavor-spin $SU(6)$ basis listed in Table~\ref{baryon1}. Here 
for simplicity the $SU(6) \otimes O(3)$ configurations are assumed 
unmixed. In reality, not only do the $\Lambda$ singlet and octet 
states mix,  states with same $J^P$ but different $L$-$S$ 
combinations can mix as well~\cite{Isgur1,Capstick1}. 
 
\begin{table}
\begin{center}
\caption{Quark-model assignments for some of the known baryons 
in terms of the flavor-spin SU(6) basis. Only the dominant
representation is listed. (Copied from the PDG,
(Ref~\cite{pdg2006})}
\begin{tabular}{cccccccc}
\hline
\hline
$J^P$  & $(D, L^P_N)$ & $S$  & & & Octet members & & Singlets \\
\hline
$1/2^+$ & $(56,0^+_0)$ & 1/2 & $N(938)$ & $\Lambda(1116)$ 
&$\Sigma(1193)$ & $\Xi(1318)$ & \\
$1/2^+$ & $(56,0^+_2)$ & 1/2 & $N(1440)$ & $\Lambda(1600)$ 
&$\Sigma(1660)$ & $\Xi(?)$ & \\
$1/2^-$ & $(70,1^-_1)$ & 1/2 & $N(1535)$ & $\Lambda(1670)$ 
&$\Sigma(1620)$ & $\Xi(?)$ & $\Lambda(1405)$ \\
$3/2^-$ & $(70,1^-_1)$ & 1/2 & $N(1520)$ & $\Lambda(1690)$ 
&$\Sigma(1670)$ & $\Xi(1820)$ &$\Lambda(1520)$ \\
$1/2^-$ & $(70,1^-_1)$ & 3/2 & $N(1650)$ & $\Lambda(1800)$ 
&$\Sigma(1750)$ & $\Xi(?)$ & \\
$3/2^-$ & $(70,1^-_1)$ & 3/2 & $N(1700)$ & $\Lambda(?)$ 
&$\Sigma(?)$ & $\Xi(?)$ & \\
$5/2^-$ & $(70,1^-_1)$ & 3/2 & $N(1675)$ & $\Lambda(1830)$ 
&$\Sigma(1775)$ & $\Xi(?)$ & \\
$1/2^+$ & $(70,0^+_2)$ & 1/2 & $N(1710)$ & $\Lambda(1810)$ 
&$\Sigma(1880)$ & $\Xi(?)$ & $\Lambda(?)$\\
$3/2^+$ & $(56,2^+_2)$ & 1/2 & $N(1720)$ & $\Lambda(1890)$ 
&$\Sigma(?)$ & $\Xi(?)$ & \\
$5/2^+$ & $(56,2^+_2)$ & 1/2 & $N(1680)$ & $\Lambda(1820)$ 
&$\Sigma(1915)$ & $\Xi(2030)$ & \\
$7/2^-$ & $(70,3^-_3)$ & 1/2 & $N(2190)$ & $\Lambda(?)$ 
&$\Sigma(?)$ & $\Xi(?)$ &$\Lambda(2100)$ \\
$9/2^-$ & $(70,3^-_3)$ & 3/2 & $N(2250)$ & $\Lambda(?)$ 
&$\Sigma(?)$ & $\Xi(?)$ & \\
$9/2^+$ & $(56,4^+_4)$ & 1/2 & $N(2220)$ & $\Lambda(2350)$ 
&$\Sigma(?)$ & $\Xi(?)$ & \\
\hline 
  & &   & & & Decuplet members & &  \\
\hline
$3/2^+$ & $(56,0^+_0)$ & 3/2 & $\Delta(1232)$ &$\Sigma(1385)$ 
& $\Xi(1530)$ & $\Omega(1672)$ & \\
$3/2^+$ & $(56,0^+_2)$ & 3/2 & $\Delta(1600)$ &$\Sigma(?)$
& $\Xi(?)$ & $\Omega(?)$ & \\
$1/2^-$ & $(70,1^-_1)$ & 1/2 & $\Delta(1620)$ &$\Sigma(?)$
& $\Xi(?)$ & $\Omega(?)$ & \\
$3/2^-$ & $(70,1^-_1)$ & 1/2 & $\Delta(1700)$ &$\Sigma(?)$
& $\Xi(?)$ & $\Omega(?)$ & \\
$5/2^+$ & $(56,2^+_2)$ & 3/2 & $\Delta(1905)$ &$\Sigma(?)$
& $\Xi(?)$ & $\Omega(?)$ & \\
$7/2^+$ & $(56,2^+_2)$ & 3/2 & $\Delta(1950)$ &$\Sigma(2030)$
& $\Xi(?)$ & $\Omega(?)$ & \\
$11/2^+$ & $(56,4^+_4)$ & 3/2 & $\Delta(2420)$ &$\Sigma(?)$
& $\Xi(?)$ & $\Omega(?)$ & \\
\hline
\hline
\end{tabular}
\end{center}
\label{baryon1}
\end{table} 
 
All members of the spatial ground state 56-plet (56,$0^+_0$) are 
experimentally well-established. Their static properties, such 
as masses, magnetic moments etc., are well reproduced by the most 
basic versions of the quark model with either harmonic-oscillator or 
linear confinement terms, which are generally spin-independent, 
plus a  spin-spin 
interaction that produces the octet-decuplet mass splitting. 
 
The situation for excited states is much more complicated. In 
order to reproduce the properties of the excited baryon states, 
various quark models with different dynamical ingredients
have been proposed and developed.
See Ref.~\cite{Capstick1} for a recent review. Among these models, two 
of the most detailed and most quoted ones are the one gluon exchange 
model (OGE) and the Goldstone boson exchange model (GBE). 
 
The OGE models originated from the renowned 1975 paper of De 
Rujula, Georgi and Glashow (DGG)~\cite{DGG}. In addition to
the confinement 
potential, the basic ingredient of the DGG model is a 
flavor-independent but spin-dependent interaction 
that is modeled on the
short-range, one-gluon-exchange (OGE) of QCD.  In a non-relativistic 
approximation, the OGE leads to a potential : 
\begin{equation} 
V^{OGE}({\bf r}_{ij}) = V^{OGE}_{cen}({\bf r}_{ij}) + 
V^{OGE}_{ten}({\bf r}_{ij})+V^{OGE}_{LS}({\bf r}_{ij}), 
\end{equation} 
where $V^{OGE}_{cen}({\bf r}_{ij})$,  $V^{OGE}_{ten}({\bf 
r}_{ij})$ and $V^{OGE}_{LS}({\bf r}_{ij})$ are central, tensor and 
spin-orbit forces, respectively.  These have the forms 
\begin{eqnarray} 
V^{OGE}_{cen}({\bf r}_{ij}) &=& \frac{\alpha_S}{4} 
(\lambda^c_i\cdot\lambda^c_j)\left\{ \frac{1}{r_{ij}} - 
\frac{\pi}{2}\delta({\bf r}_{ij}) \left[\frac{1}{m_i^2} + 
\frac{1}{m_j^2} + \frac{4}{3 m_i m_j}({\bf \sigma}_i\cdot {\bf 
\sigma}_j)\right]\right\}, \\ 
V^{OGE}_{ten}({\bf r}_{ij}) &=& -\frac{\alpha_S}{4} 
(\lambda^c_i\cdot\lambda^c_j)\frac{1}{4 m_i m_j} \frac{1}{3r^3_{i 
j}}(3{\bf \sigma}_i\cdot {\bf\hat r}_{ij}{\bf \sigma}_j\cdot 
{\bf\hat r}_{ij} - {\bf \sigma}_i\cdot {\bf \sigma}_j), \\ 
V^{OGE}_{LS}({\bf r}_{ij}) &=& -\frac{\alpha_S}{4} 
(\lambda^c_i\cdot\lambda^c_j)\frac{m_i^2+m_j^2+4m_i m_j}{8m_i^2 
m_j^2}\frac{1}{3r^3_{i j}} [{\bf L}\cdot ({\bf \sigma}_i + {\bf 
\sigma}_j)], 
\end{eqnarray} 
where $\lambda^c_i$ ($i=1,\cdots,8$) are the color-$SU(3)$ 
unitary spin matrices. The most detailed and phenomenologically 
successful extension of the DGG model was made by Isgur and his 
collaborators, first non-relativistically~\cite{Isgur1,Capstick1}, 
and subsequently in a
relativized formulation~\cite{Capstick1,Capstick2}. In their model 
calculations, the $V^{OGE}_{LS}({\bf r}_{ij})$ term is neglected 
based on the argument that it is cancelled by an inevitable Thomas 
precession term that is generated by confinement \cite{Isgur2}. 
Otherwise, the inclusion of this term would
spoil the agreement with the observed spectrum. 
Results from a recent OGE quark 
model calculation~\cite{Capstick2} for the excitation spectrum of 
the nucleon is shown in Fig.~\ref{fig-qm1}, 
where it is compared with experimental 
observations.  Many baryon electromagnetic couplings and strong 
decay couplings have also been calculated within the context of 
this model~\cite{Capstick1}. 
 
\begin{figure}[htpb] 
\vspace{0cm} 
\hspace{0cm}\includegraphics[scale=0.8]{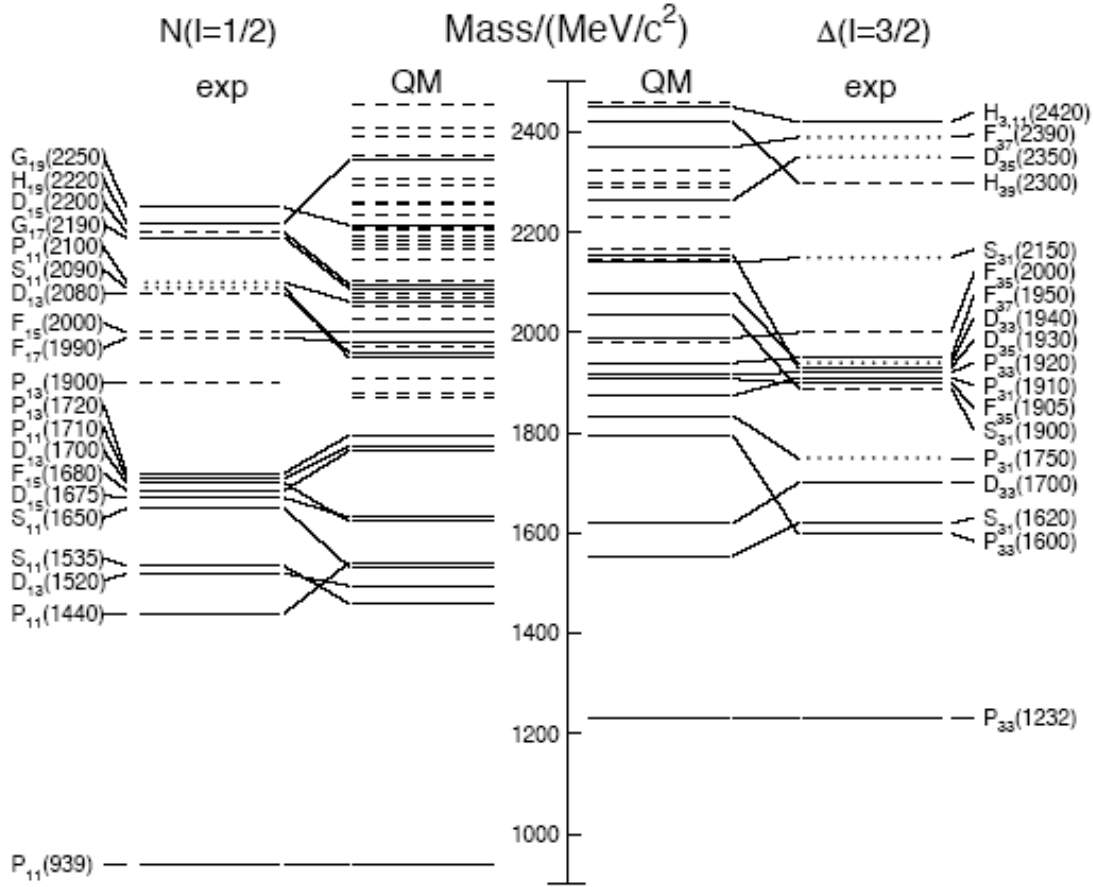} 
\caption{The nucleon excitation spectrum. The 
positions of the excited state identified in experiment are compared to those 
predicted by a modern quark model calculation. Left hand side: 
isospin $I=1/2$ N-states, right hand side: isospin $I=3/2$ 
$\Delta$-states. Experimental: (columns labelled `exp'), three and 
four star states are indicated by full lines (two-star dashed 
lines, one-star dotted lines). At the very left and right of the 
figure the spectroscopic notation of these states is given. Quark 
model \cite{Capstick2}: (columns labelled `QM'), all states for 
the N=1,2 bands, low lying states for N=3,4,5 bands. Full lines: 
at least tentative assignment to observed states, dashed lines: so 
far no observed counterparts. (Copied from PDG \cite{pdg2006}) } 
\label{fig-qm1} 
\end{figure} 
 
The GBE models originated from the 1984 work 
of Manohar and Georgi~\cite{Manohar}.
The basic idea here is that constituent 
quarks with internal structure are a consequence of the the 
spontaneous breaking of the approximate chiral symmetry of QCD and, 
thus, couple to the chiral meson fields. Glozman and 
Riska~\cite{CCQM} have popularized this idea by making an extensive 
analysis of the baryon spectrum using a model based on a hyperfine 
interaction arising solely from the exchange of a pseudoscalar 
octet instead of the gluons of the OGE model. 
In their model, the spin-dependent 
hyperfine interaction is flavor-dependent: 
\begin{equation} 
H_{HF}\propto \sum_{i<j}V(\vec r_{ij})\lambda^F_i\cdot\lambda^F_j 
\vec\sigma_i\cdot\vec\sigma_j \, ,
\end{equation} 
where the $\lambda^F_i$ are flavor $SU(3)$ Gell-Mann matrices. It is 
the flavor-dependent factor $\lambda^F_i\cdot\lambda^F_j$ of the 
GBE interaction that produces parity orderings for the $N^*$, 
$\Delta^*$ and $\Lambda^*$ spectra that are different than
those of the OGE models, and in 
agreement with observation. After further extending this 
model to include the exchange of a nonet of vector mesons and a 
scalar meson~\cite{glozman}, the low-lying $N^*$, $\Delta^*$ and 
$\Lambda^*$ spectra can be well reproduced as shown in 
Fig.~\ref{fig-glozman}. A problem for the model is whether or not 
the large number of parameters introduced to fit the spectra can 
also reproduce the relevant strong decays of the excited baryons 
and baryon-baryon interactions. 
 
\begin{figure}[htpb] 
\vspace{0cm} \hspace{0cm}\includegraphics[scale=0.9]{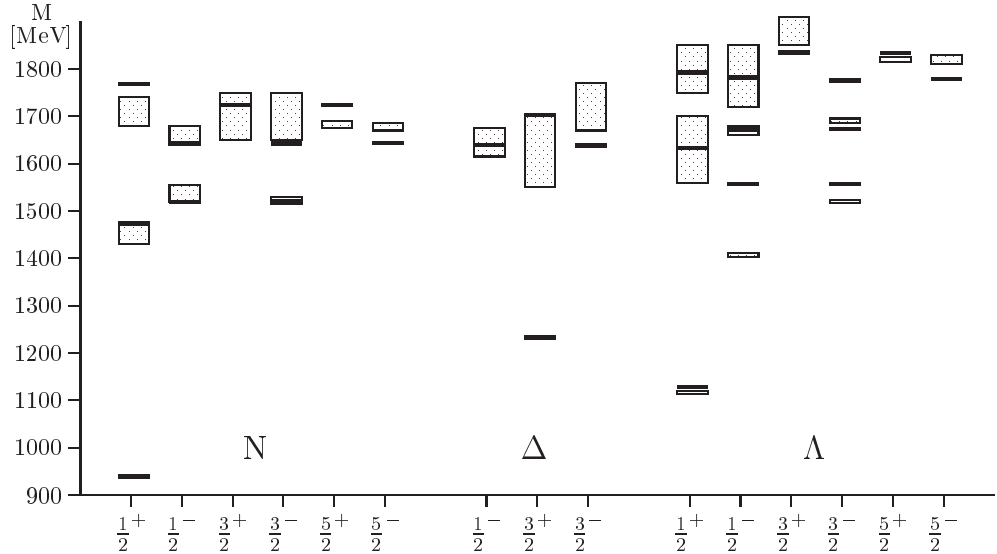} 
\caption{Energy levels of low-lying $N^*$, $\Delta^*$ and 
$\Lambda^*$ baryons from Ref.\cite{glozman}, compared to the range 
of central values for resonances masses from the PDG \cite{pdg2006}. } 
\label{fig-glozman} 
\end{figure} 
 
In order to describe simultaneously the baryon spectrum, baryon 
decays and baryon-baryon interactions, some hybrid models 
that include both OGE and GBE interactions have been 
developed~\cite{Faessler}. Until now, these hybrid models 
have been mainly  applied  to studies of dibaryon systems 
using parameters determined from fits to 
ground state baryon masses, the deuteron binding energy, etc. 
 
Besides the OGE and GBE interactions, an alternative 
flavor-spin-dependent hyperfine interaction that derives from
instanton effects was proposed~\cite{Metsch}. This interaction 
acts only on scalar, isoscalar pairs of quarks in relative 
$S$-wave states: 
\begin{equation} 
<q^2;S,L,I|H_{HF}|q^2;S,L,I>~\propto~{\cal W}~ 
\delta_{S,0}\delta_{L,0}\delta_{I,0}, 
\end{equation} 
where $\cal W$ is the radial matrix element of the contact 
interaction. In this approach, baryons are described by the 
homogeneous Bethe-Salpeter equation with three- and 
two-particle instantaneous interaction kernels, which implement 
confinement and the flavor-spin-dependent interaction from 
instanton effects to account for the major mass splittings. Its 
predictions for the baryon spectrum are of similar quality as OGE 
models. But it can give a description of the mass spectrum while 
implementing relativistic covariance both in the quark dynamics 
and in the calculation of currents needed for decay observables. 
 
Other quite well known models based on the three-quark picture for baryons 
include the MIT bag model~\cite{MIT-bag}, the cloudy bag 
model~\cite{cloudy-bag} and
the algebraic model~\cite{Bijker}. The MIT bag 
model is relativistic and confines three valence quarks to the 
interior of baryons by a bag pressure term with a parameter 
$B$ that is used 
to set the scale of the baryon masses. The cloudy bag model 
incorporates chiral invariance by allowing a cloud of pion fields 
to couple to the confined quarks only at the surface of the MIT 
bag. The algebraic model is also called the collective model. Its 
approach to the dynamics is not the usual solution of some 
Schr\"odinger-like equation, but rather bosonic quantization of 
the spatial degree of freedom that has a $Y$-shaped string-like 
configuration with possible vibrations and rotations. For more 
details of these models, we refer readers to the corresponding 
original papers~\cite{MIT-bag,cloudy-bag,Bijker} or 
Ref.~\cite{Capstick1} for a recent review. 
 
In the following sections, we  review some of the outstanding 
problems in baryon spectroscopy in the context of the
$qqq$ picture for baryons and 
introduce new pictures of baryons that go
beyond this simplest three-quark configuration.
We then summarize  the accomplishments of BESI and BESII in the 
area of baryon spectroscopy and discuss the prospects 
for \bes3.

\section{Outstanding problems in baryon spectroscopy} 
 
Although the quark model has achieved a number of significant 
successes in the 
interpretation of many of the static properties of nucleons and  
excited resonances, our present knowledge on baryon spectroscopy 
is still in its infancy~\cite{pdg2006}. Many very fundamental 
issues in baryon spectroscopy are still not well 
understood~\cite{Capstick1}. 
 
On the theoretical side, an unsolved fundamental problem 
that still persists is: 
{\it What are the
proper effective degrees of freedom for describing the internal 
structure of baryons?} Several pictures based on various effective 
degrees of freedom are shown in Fig.~\ref{fig1}. 
 
\begin{figure}[htpb] 
\vspace{0cm} 
\hspace{0cm}\includegraphics[scale=0.5]{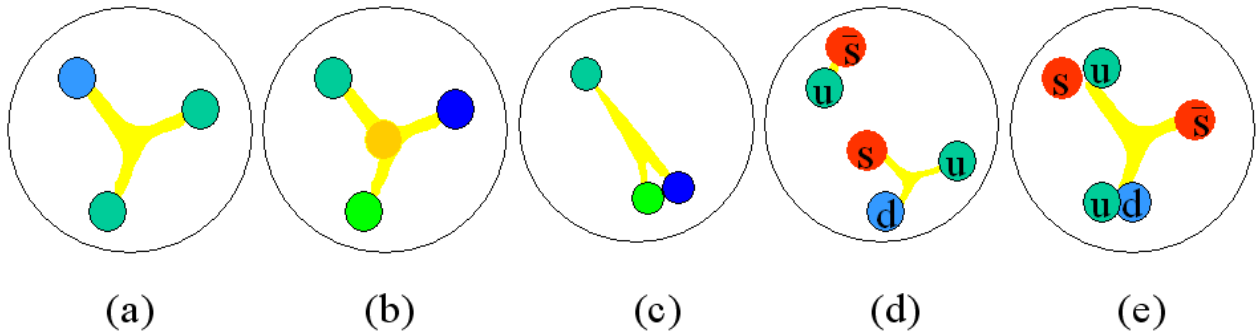} 
\caption{Various pictures for internal quark-gluon structure of 
baryons: (a) $qqq$, (b) $qqqg$ hybrid, (c) diquark, d) meson-baryon 
state, (e) pentaquark with diquark clusters. } \label{fig1} 
\end{figure}

The classical and simple $qqq$ configuration of the
constituent quark model, shown in 
Fig.~\ref{fig1}(a), has been very successful at explaining the 
static properties such as masses and magnetic moments of the 
spatial ground states of the flavor $SU(3)$ octet and decuplet 
baryons. It predicted the $\Omega^-$ baryon to have a mass around 
1670~MeV, as was subsequently discovered experimentally. 
However, its predictions for 
the spatially excited baryons have not been as successful, as 
illustrated in Fig.~\ref{fig-qm1}. In the simple 
$qqq$ constituent quark model, 
the lowest spatially excited baryon is expected to be a ($uud$) 
$N^*$ state with one quark in orbital angular momentum $L=1$ 
state, and, hence, with negative parity. 
Experimentally~\cite{pdg2006}, the lowest negative parity 
$N^*$ resonance is found to be the $N^*(1535)$, 
which is heavier than two other spatially excited 
baryons: the $\Lambda^*(1405)$ and $N^*(1440)$. In the classical 
$qqq$ constituent quark model, the $\Lambda^*(1405)$ with spin-parity 
$1/2^-$ is supposed to be a ($uds$) baryon with one quark in 
orbital angular momentum $L=1$ state and about 130 MeV heavier 
than its $N^*$ partner, the $N^*(1535)$; the $N^*(1440)$ with 
spin-parity $1/2^+$ is supposed to be a ($uud$) state with one 
quark in radial $n=1$ excited state and should be heavier than the
$L=1$ excited ($uud$) state $N^*(1535)$ (based on the result that for 
a simple harmonic oscillator potential the state energy is 
$(2n+L+3/2)\hbar\omega$).  Thus,
for the three lowest spatially excited 
baryons, the classical quark model picture has already failed. 
 
The second outstanding problem for the classical $qqq$ model is 
that in many of its forms it predicts a substantial number of 
`missing $N^*$ states' around 2~GeV, which have so far not
been observed~\cite{Capstick1}. Since the predicted 
number of excited states decreases with the number of the
effective degrees of freedom, it is argued that
the missing $N^*$ states problem favors the 
diquark picture as shown in Fig.~\ref{fig1}(c), which has fewer
degree of freedom and predicts fewer $N^*$ states~\cite{diquark-zbs}. 
For example, in diquark models, the two quarks forming the diquark 
are constrained to be in the relative $S$-wave, and, thus, cannot 
combine with the third quark to form (20,$1^+_2$)-multiplet baryons. 
Experimentally, not a single (20,$1^+_2$)-multiplet baryon has yet to
be identified~\cite{pdg2006}.  However, the non-observation of these 
missing $N^*$ states does not necessarily mean that they do not 
exist. In the limit that the $\gamma$ or $\pi$ couples to only one 
quark in the nucleon in  $\gamma N$ or $\pi N$ reactions, the 
(20,$1^+_2$)-multiplet baryon cannot be produced~\cite{Zhaoq}. 
As for higher-order effects, they can have couplings to 
$\pi N$ and $\gamma N$, but maybe these are too weak to be seen
in presently available $\pi N$ and $\gamma N$ 
experiments~\cite{Capstick1,Zhaoq}. Other production processes should be 
explored.  Moreover, the diquark models have been successful in only 
a few, very limited areas. 
 
The third outstanding problem for the classical $qqq$ quark model is 
that in deep inelastic scattering and Drell-Yan experiments the 
number of $\bar d$ quarks in the proton is found to be more 
than the number of $\bar u$ quarks by about 12\%~\cite{Garvey}. 
It  is argued that this  favors a 
mixture of meson-baryon states as shown in Fig.~\ref{fig1}(d). 
In this picture, the $\bar d$ over $\bar u$ excess in the 
proton is explained by a mixture of $n\pi^+$ with the $\pi^+$ 
composed of $u\bar d$~\cite{Thomas}; the $N^*(1535)$ and 
$\Lambda^*(1405)$ are identified as quasi-bound $K\Sigma$ 
and $\bar KN$ states, respectively~\cite{Weise}. The extreme of this 
picture is that only the ground state baryon-octet $1/2^+$ and 
baryon-decuplet $3/2^+$ are dominated by $qqq$, while all excited 
baryons are generated by meson-baryon coupled channel 
dynamics~\cite{Lutz,Oset1}. However, a mixture of pentaquark 
components with diquark clusters, as shown in Fig.~\ref{fig1}(e),
could also explain these properties~\cite{zou1,zr,liubc,zhusl-zbs}. 
 
Another unsolved fundamental question is: {\it Even if we know the 
effective degrees of freedom in the baryon, how do we deal with the 
interactions between them?} In general, in most fields of physics, 
two-body forces dominate and three-body forces are treated as a 
residual interaction. In QCD, however, the three-body force between 
three quarks is expected to be the ``primary" force that reflects the 
$SU(3)_c$ gauge symmetry; a point of view  
that is strongly supported by a recent lattice 
calculation~\cite{Takahashi1}. An earlier 
constituent quark model calculation~\cite{Capstick3} also suggested 
that the three-quark potential is directly responsible for the 
structure and properties of baryons. It is much more complicated to 
deal with a three-body force than the usual two-body force. 
Furthermore, the center of the $Y$-shaped gluon field could act as 
an additional, vibrational degree of freedom that makes the baryon 
behave as a $qqq$-gluon hybrid as shown by Fig.~\ref{fig1}(b). 
Fortunately, the gluonic excitation energy is found to be about 
1~GeV for a typical hadronic scale, which is substanially larger 
than typical quark excitation energies~\cite{Takahashi2}. This 
large gluonic excitation energy explains 
the great success of the simple $qqq$ quark model for the 
spatial ground state baryons. On the other hand, the 
three-body $Y$-shaped gluon field interaction is obtained in the quenched 
approximation~\cite{Takahashi1}; some people believe that the effective 
interaction field between the constituent quarks should be a meson 
field rather than a gluon field~\cite{Manohar,CCQM}. Thus, the 
three-body force may also be generated by various couplings of 
meson fields. 
 
There are also various phenomenological models for hybrid 
baryons~\cite{Barnes1}. The earliest one is the bag model which places 
relativisitic quarks and gluons in a spherical cavity and allows 
them to interact via QCD forces such as one-gluon exchange, 
the color Compton effect, etc.~\cite{Barnes2}. The lightest 
hybrid baryon was predicted to be an ``extra" $1/2^+$ $N^*$ $P11$ 
state with a mass of about $1.5\sim 1.6$ GeV, which is suggestive
of the $N^*(1440)$ Roper resonance. 
This is supported by QCD sum rule 
calculations~\cite{Kisslinger} that also predict the lightest 
$1/2^+$ hybrid to be around 1.5~GeV and conclude that the Roper 
is largely a hybrid. However, more recent flux tube model 
calculations~\cite{Page} predict a larger mass for the lightest 
hybrid baryons, about 1.9 GeV,  with a twofold degenerate 
pair of $1/2^+$ and $3/2^+$ $N^*$ hybrids. This 
is closer to lattice QCD 
predictions~\cite{Takahashi2}. 
 
In reality, a baryon state around 2~GeV could be a mixture 
of all five of the configurations shown in Fig.~\ref{fig1}. 
As for the  existence or non-existence of genuine pentaquark states, 
we refer the reader to recent reviews \cite{klempt}.

On the experimental side, our present knowledge of baryon spectroscopy 
has come almost entirely from partial-wave analyses of $\pi N$ total, 
elastic, and charge-exchange scattering data from more than twenty 
years ago~\cite{pdg2006}. However, recently a new generation of 
experiments on $N^*$ physics with electromagnetic probes has been 
started at new facilities such as CEBAF at JLAB, ELSA at Bonn, 
GRAAL at Grenoble and SPRING8 at JASRI. Some nice results have 
already been produced~\cite{CLAS,ELSA,GRAAL,Bonn}.  However, a 
problem for 
these experiments is that above 1.8 GeV there are many broad 
resonances with various possible quantum numbers that overlap 
each other and are very difficult to disentangle. Moreover 
resonances with weak couplings to $\pi N$ and $\gamma N$ will not 
show up in these experiments.

\section{Baryon Spectroscopy at BESI and BESII} 
 
In 2000, BESII started a baryon resonance research program
that has been focused on the study of excited $N^*$ 
baryons~\cite{Zou1}.  This takes advantage of the fact that 
$J/\psi$ and $\psi'$ decays provide excellent opportunities for 
studying  excited nucleon ($N^*$)
and hyperon ($\Lambda^*$, $\Sigma^*$ \&  $\Xi^*$)
resonances~\cite{Zou2}. 
The  Feynman graph for the production of excited nucleons and 
hyperons in $\psi$ decays is 
shown in Fig.~\ref{fig:1}, ($\psi = J/\psi$ or $\psi'$). 
 
\begin{figure}[htbp] 
\vspace{-1.2cm} 
\includegraphics[scale=0.6]{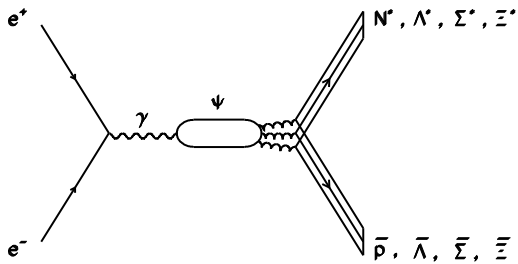} \vspace{-1.5cm} 
\caption{$\bar pN^*$, $\bar\Lambda\Lambda^*$, $\bar\Sigma\Sigma^*$ 
and $\bar\Xi\Xi^*$ production from $e^+e^-$ collision through 
$\psi$ meson.} \label{fig:1} 
\end{figure} 
 
In comparison to other facilities, the BES baryon program has advantages
in at least three important aspects: 

\noindent 
(1) isospin conservation ensures that the 
$J/\psi\to\bar NN\pi$ ($\bar NN\pi\pi$) decay processes produce
pure isospin=1/2 $\pi N$ ($\pi\pi N$) systems.  In contrast,
such systems produced in $\pi N$ 
and $\gamma N$ experiments are mixture of isospin=1/2 and 3/2, and 
analyses of these final states
suffer from the complications of the isospin decomposition; 

\noindent
(2) decays of $\psi$ mesons to final states containing baryons 
proceed via three  or more virtual gluons, which is a 
favorable environment for producing hybrid 
($qqq$-gluon) baryons, and for looking for ``missing" $N^*$ 
resonances, such as members of a possible (20,$1^+_2$)-multiplet 
baryons, that have weak couplings to both $\pi N$ and $\gamma N$, 
but a strong coupling to $g^3N$; 

\noindent
(3) In addition to $N^*$  $\Lambda^*$ \& $\Sigma^*$ baryons,  $\psi$
decays can access doubly strange
$\Xi^*$ baryons. Many 
QCD-inspired models~\cite{Capstick2,CCQM} are expected to be more 
reliable for baryons containing two strange quarks because of the heavier 
quark mass. More than thirty $\Xi^*$ resonances are predicted 
to exist, while currently only two such states are 
experimentally well established. 
The theory for these states is left 
essentially unchallenged because of this paucity of data. 
 
BESI started data-taking in 1989 and was upgraded in 1998 to BESII. 
BESI collected 7.8 million $J/\psi$ events and 3.7 million $\psi'$ 
events. BESII collected 58 million $J/\psi$ events and 14 
million $\psi'$ events. 
 
\begin{figure}[htbp] 
\includegraphics[scale=0.25]{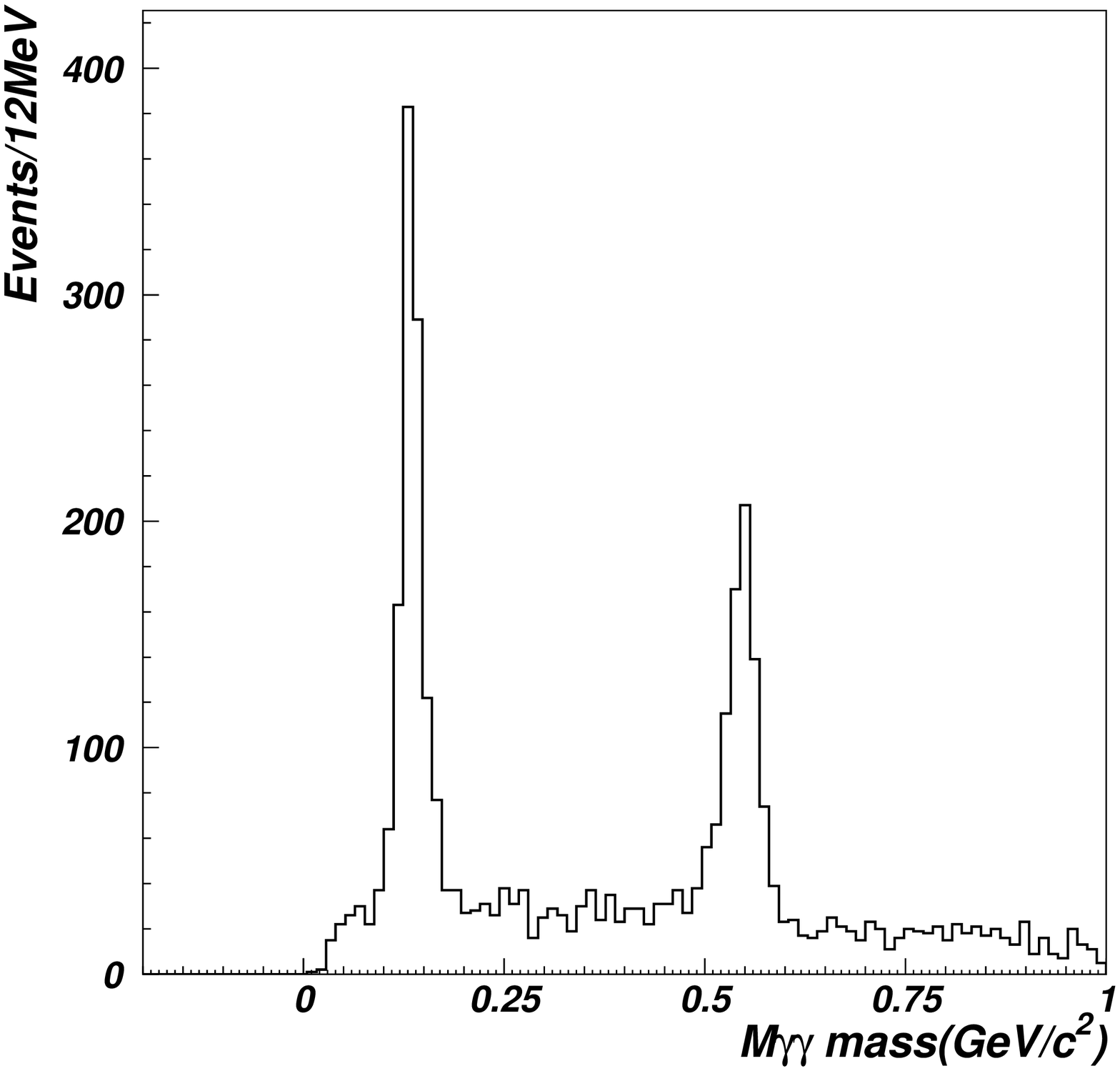} 
\includegraphics[scale=0.25]{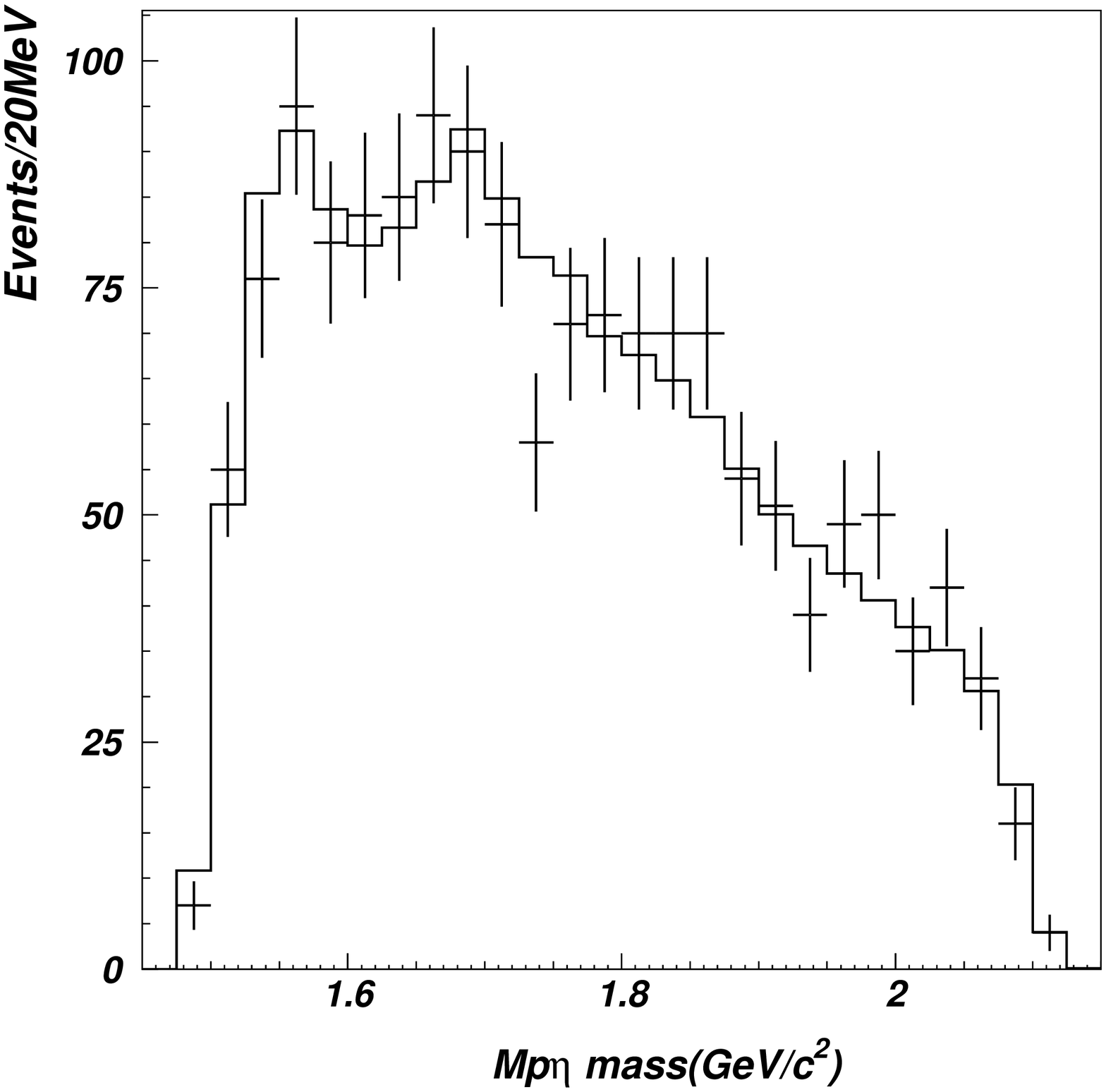}\vspace{-0.5cm} 
\caption{\label{fig2} left: $\gamma\gamma$ invariant mass for 
$J/\psi\to\bar pp\gamma\gamma$; right: $p\eta$ invariant mass 
spectrum for $J/\psi\to\bar pp\eta$. BESI data} 
\end{figure}

From the 7.8 million $J/\psi$ event sample collected at BESI 
between 1990 and  1991, events of the type 
$J/\psi\to\bar pp\pi^0$ and $\bar pp\eta$ were 
selected and reconstructed with the $\pi^0$ and $\eta$ 
detected via their $\gamma\gamma$ decay mode~\cite{Zou1}. The 
$\gamma\gamma$ invariant mass, shown in 
Fig.~\ref{fig2}~(left), exhibits 
two distinct peaks corresponding to the $\pi^0$ and 
$\eta$. 
The $p\eta$ invariant mass spectrum, shown in 
Fig.~\ref{fig2}~(right),  has two peaks:
one at 1540~MeV and another at 1650~MeV. A partial 
wave analysis was applied to the $J/\psi\to\bar pp\eta$ 
channel~\cite{Zou1}, using the effective Lagrangian 
approach~\cite{Nimai,Olsson} with the Rarita-Schwinger 
formalism~\cite{Rarita,Fronsdal,Chung,Liang} and the 
extended automatic Feynman Diagram Calculation (FDC) 
package~\cite{Wang}.  The results indicate a
definite signal for a $J^{P}=\frac{1}{2}^-$ component at $M = 
1530\pm 10$~MeV with $\Gamma =95\pm 25$~MeV, 
which is very close to the $\eta N$ threshold. 
In addition, there is a distinct resonance at
$M = 1647\pm 20$~MeV with $\Gamma = 145^{+80}_{-45}$~MeV
and a preferred $J^P$ value of $\frac{1}{2}^-$. 
These two $N^*$ resonances are probably
the well established $S_{11}(1535)$ and 
$S_{11}(1650)$ states.  In the higher $p\eta$($\bar{p}\eta$) 
mass region, there is evidence for a structure around 1800~MeV; 
however, with the limited statistical precision of BESI, a determination
of its quantum numbers was not possible.  The 
$p\pi^0$ invariant mass spectrum from $J/\psi\to p\bar p\pi^0$, 
shown in Fig.~\ref{fig3}~(left), has clear peaks around 1500~MeV 
and 1670~MeV, and some weak structure around 2~GeV. 
 
\begin{figure}[htbp] 
\hspace{-1cm}\includegraphics[scale=0.3]{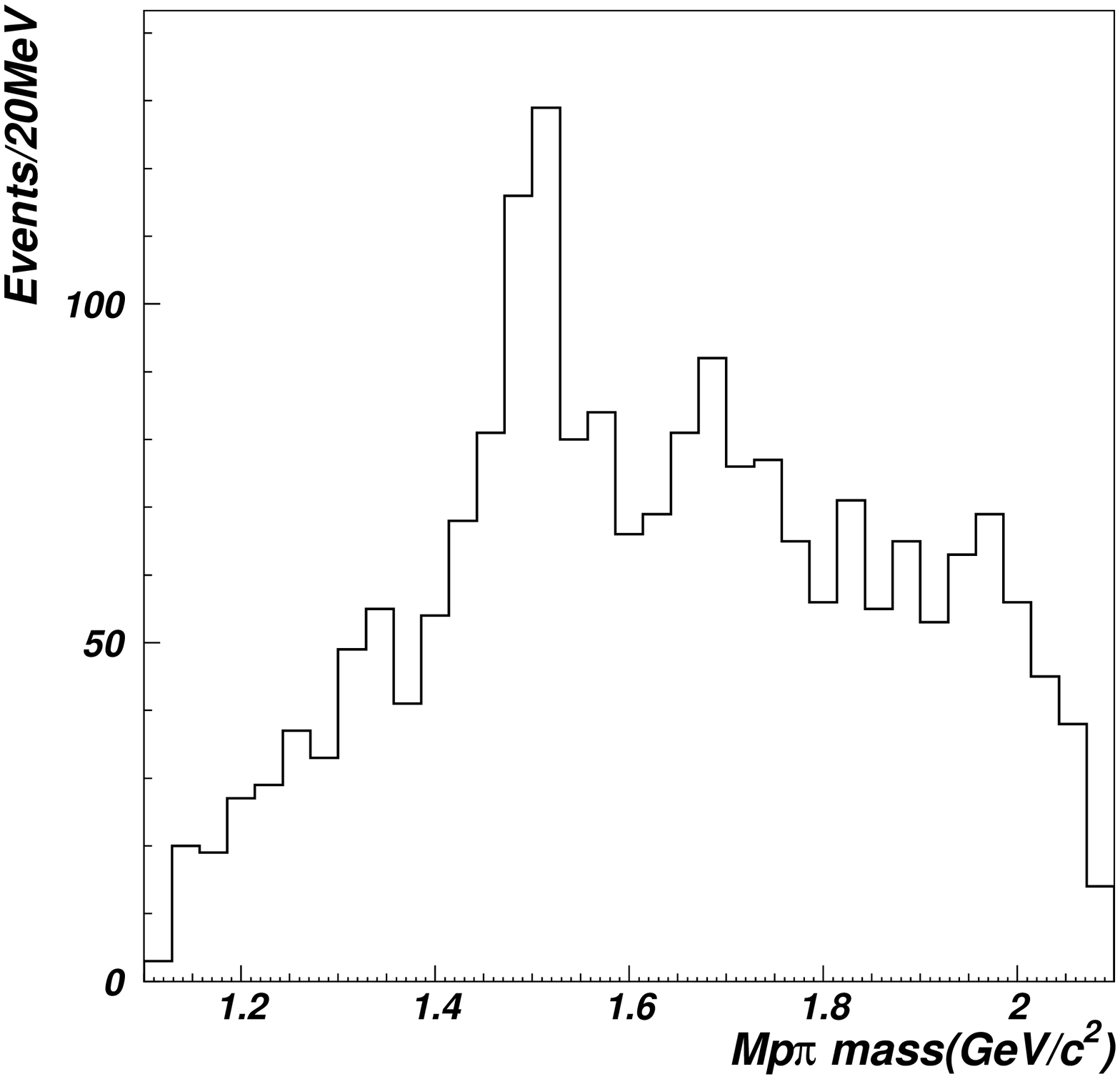} 
\vspace{-0.5cm}\includegraphics[scale=0.35]{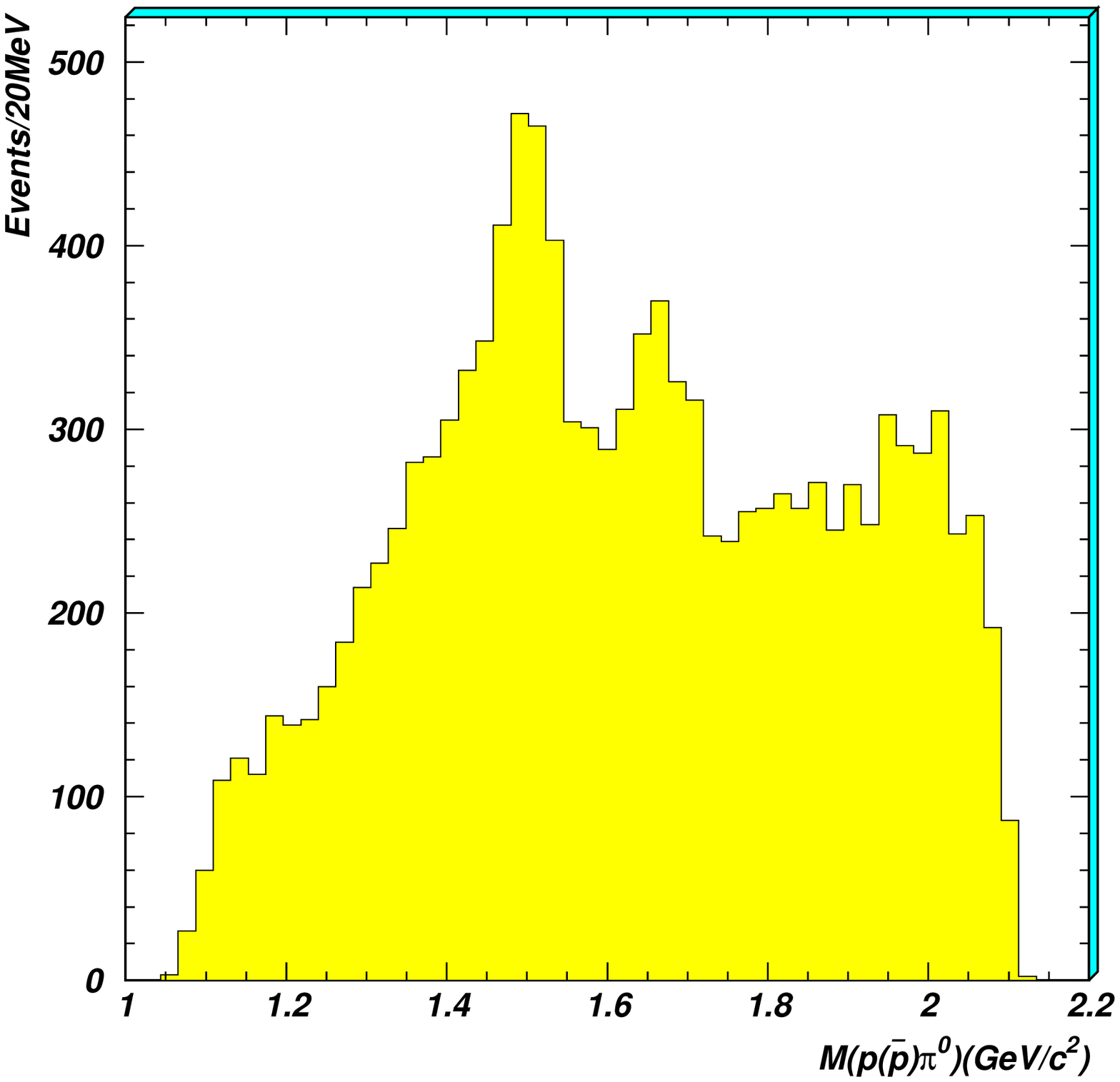} 
\caption{\label{fig3} $p\pi^0$ invariant mass spectrum for 
$J/\psi\to\bar pp\pi^0$ from BESI (left) and preliminary BESII 
data (right)} 
\end{figure} 
 
With the 58 million $J/\psi$ event sample collected in 
the more efficient BESII detector, an order-of-magnitudee 
increase in the number of reconstructed 
events for each channel is obtained. 
Results for $J/\psi$ decays to 
$p\bar p\pi^0$, $p\bar n\pi^-+c.c.$, $pK^-\bar\Lambda + c.c.$ and 
$\Lambda\bar\Sigma\pi$+c.c.  are shown in 
Figs.~\ref{fig3},~\ref{fig3.2},~\ref{fig4}~\&~\ref{fig5}, 
respectively. 
These are useful channels for studies of $N^*$, $\Lambda^*$ and 
$\Sigma^*$ resonances. 
 
For the $J/\psi\to p\bar p\pi^0$ channel, the $N\pi$ 
invariant mass spectrum from the BESII sample, shown in
Fig.~\ref{fig3}~(right), 
looks similar to that of the BESI data shown in 
Fig.~\ref{fig3}~(left), but with much higher statistics. 
 
\begin{figure}[htbp] 
\includegraphics[scale=0.35]{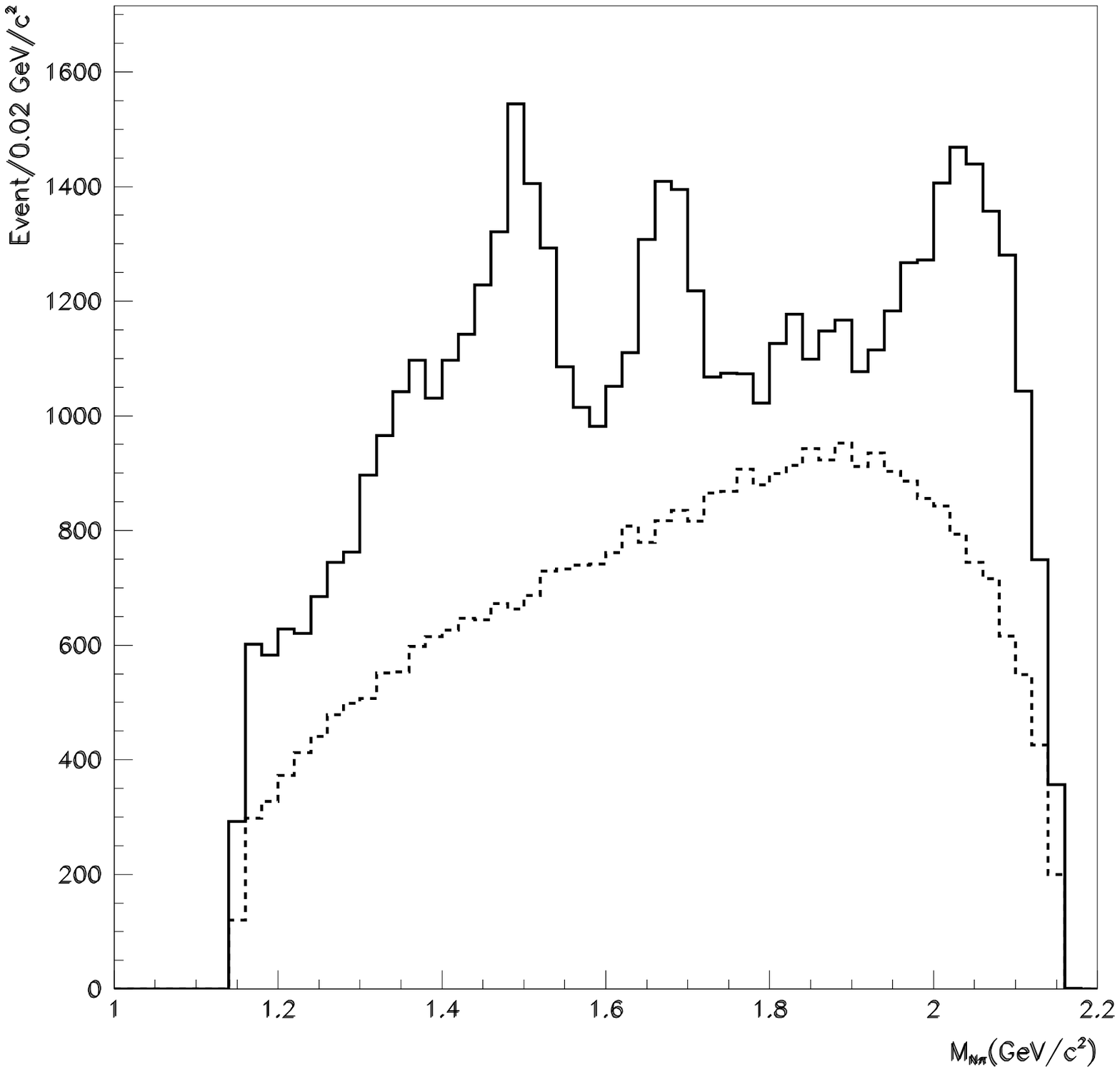} 
\includegraphics[scale=0.6]{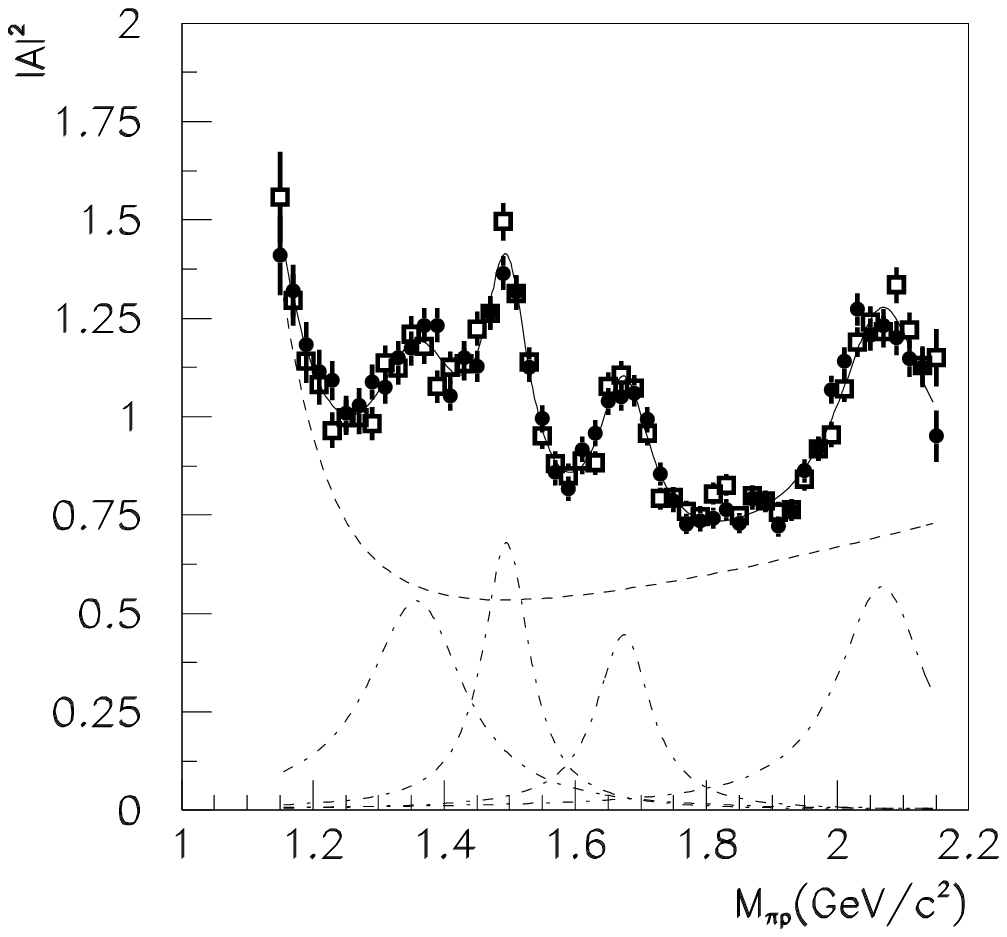}\vspace{-0.8cm} 
\caption{\label{fig3.2} The $p\pi^-$ and $\bar p\pi^+$ invariant 
mass spectra for $J/\psi\to p\pi^-\bar n$ (left) and $\bar p\pi^+ 
n$ (right), compared with phase space distribution and data 
divided by Monte Carlo phase space vs $p\pi$ invariant mass for 
$J/\psi\to \bar p\pi^-\bar n$ (solid circle) and  $J/\psi\to\bar 
p\pi^+n$ (open square).} 
\end{figure}

In the analysis of the $J/\psi\to p\bar n\pi^-$ channel, 
only the proton and $\pi^-$ are  detected.  After the application
of some selection requirements to reduce backgrounds, the 
missing mass spectrum shows a very clean peak
corresponding to the missing antineutron. In the 
$p\pi^-$ invariant mass spectrum, shown in Fig.~\ref{fig3.2} 
(left), in addition to the two well known $N^*$ peaks at 
1500~MeV and 1670~MeV,   $N^*$ peaks 
around 1360~MeV and 2030~MeV are evident. The 
charge-conjugate channel $\bar p\pi^+n$ gives results 
that are very similar, as shown in Fig.~\ref{fig3.2}~(right). 
 
To investigate the behavior of the amplitude squared, 
the invariant mass distribution has to be 
corrected for phase-space and  efficiency effects. The 
corrected results are shown in Fig.~\ref{fig3.2}~(right). At 
low $p\pi$ invariant mass, the tail from the nucleon pole term, 
predicted from theoretical considerations~\cite{Okubo,Liang2}, is 
clearly seen. In addition, four peaks, around 1360~MeV, 
1500~MeV, 1670~MeV and 2065~MeV, are evident. Note that 
the well known lowest-mass nucleon 
resonance, the $\Delta(1232)$, which dominates 
the low mass $\pi N$  and $\gamma N$ scattering data, 
does not show up here because of the isospin filtering 
effects of $J/\psi$ decay. While the peaks around 1500~MeV 
and 1670~MeV correspond to the well known second and third 
resonance peaks observed in $\pi N$ and $\gamma N$ scattering 
data, the two peaks around 1360~MeV and 2065~MeV have not been 
previously observed in $\pi N$ invariant mass spectra. The 1360~MeV 
peak could be from the $N^*(1440)$,  which has a pole
position near 1360~MeV~\cite{pdg2006,Manley,Dytman} and 
is usually buried under a huge 
$\Delta(1232)$ peak in $\pi N$ and $\gamma N$ experiments; 
The peak near 2065~MeV may be one
(or more) of the long-sought-for 
``missing" $N^*$ resonance(s). For the decay $J/\psi\to\bar 
NN^*(2065)$, the orbital angular momentum of $L=0$ is much 
preferred due to the centrifugal barrier suppression factor 
for $L\geq 1$. For $L=0$, the spin-parity of the $N^*(2065)$ would 
be limited to be $1/2^+$ and $3/2^+$. This could be the reason that the 
$N^*(2065)$ shows up as a peak in $J/\psi$ decays and 
not in $\pi N$ and $\gamma N$ production experiments, 
where all $1/2^\pm$,  $3/2^\pm$, 
$5/2^\pm$ and $7/2^\pm$ $N^*$ resonances around 2.05~GeV 
are allowed and can overlap 
and interfere with each other. A simple Breit-Wigner 
fit~\cite{bes2} gives a mass and width for the $N^*(1440)$ peak of 
$1358\pm 6 \pm 16$~MeV and $179\pm 26\pm 50$~MeV; for the new 
$N^*(2065)$ the fitted mass and width are $2068\pm 3^{+15}_{-40}$~MeV and 
$165\pm 14\pm 40$~MeV. A partial wave analysis indicates 
that the $N^*(2065)$ peak contains both spin-parity $1/2^+$ and 
$3/2^+$ components~\cite{bes2}. 
 
\begin{figure}[htbp] 
\includegraphics[scale=0.25]{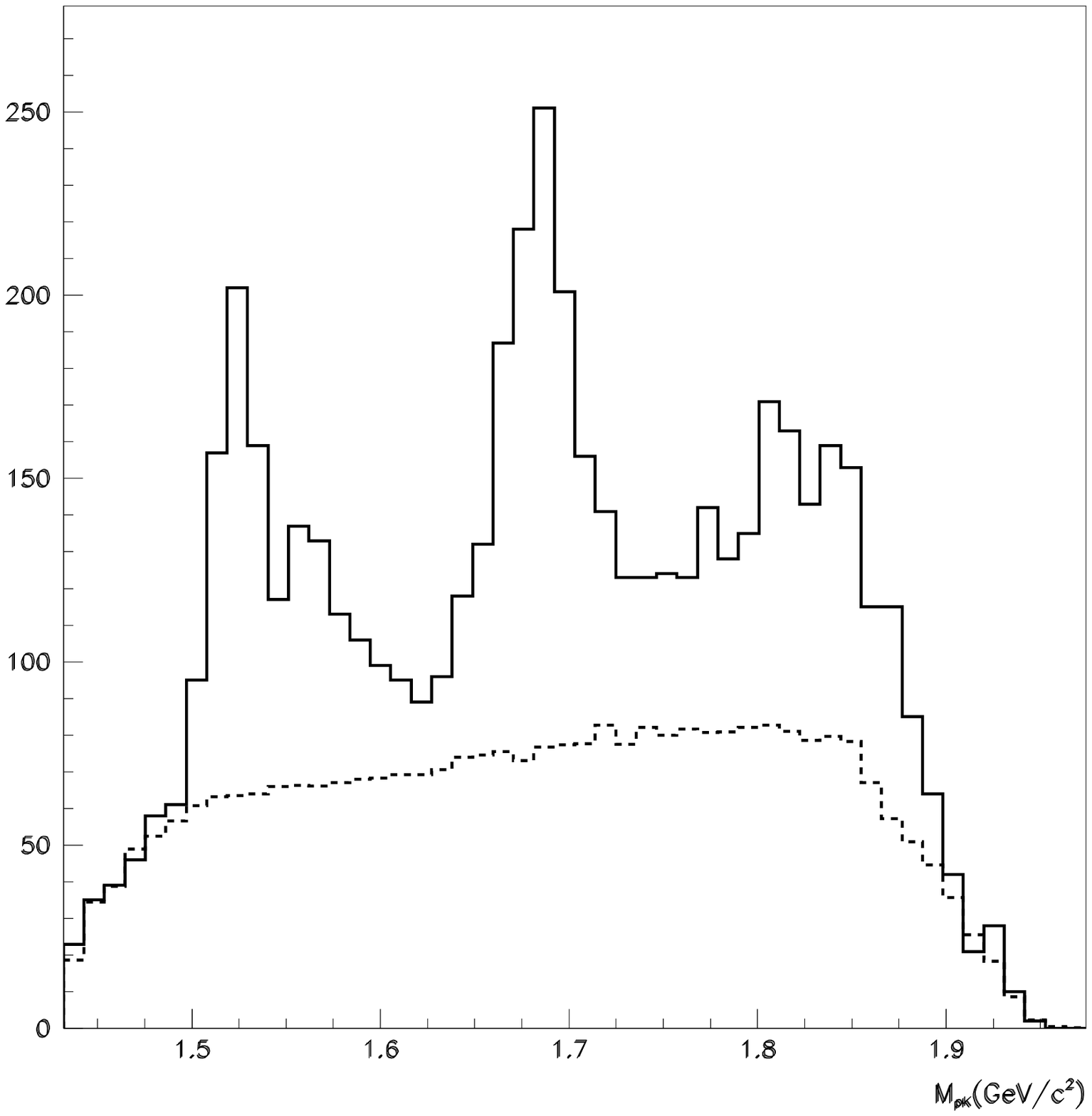} 
\includegraphics[scale=0.25]{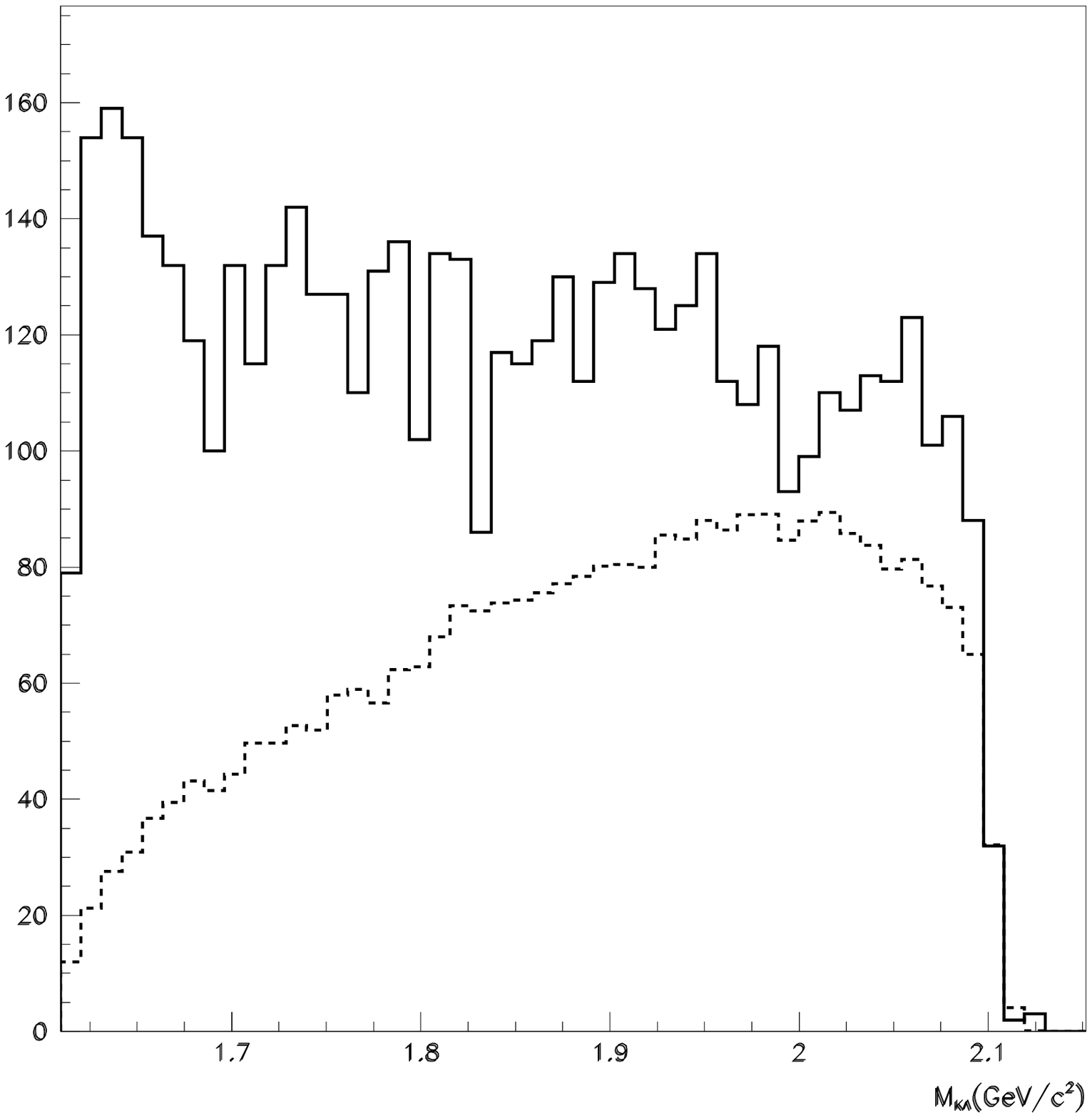} 
\includegraphics[scale=0.25]{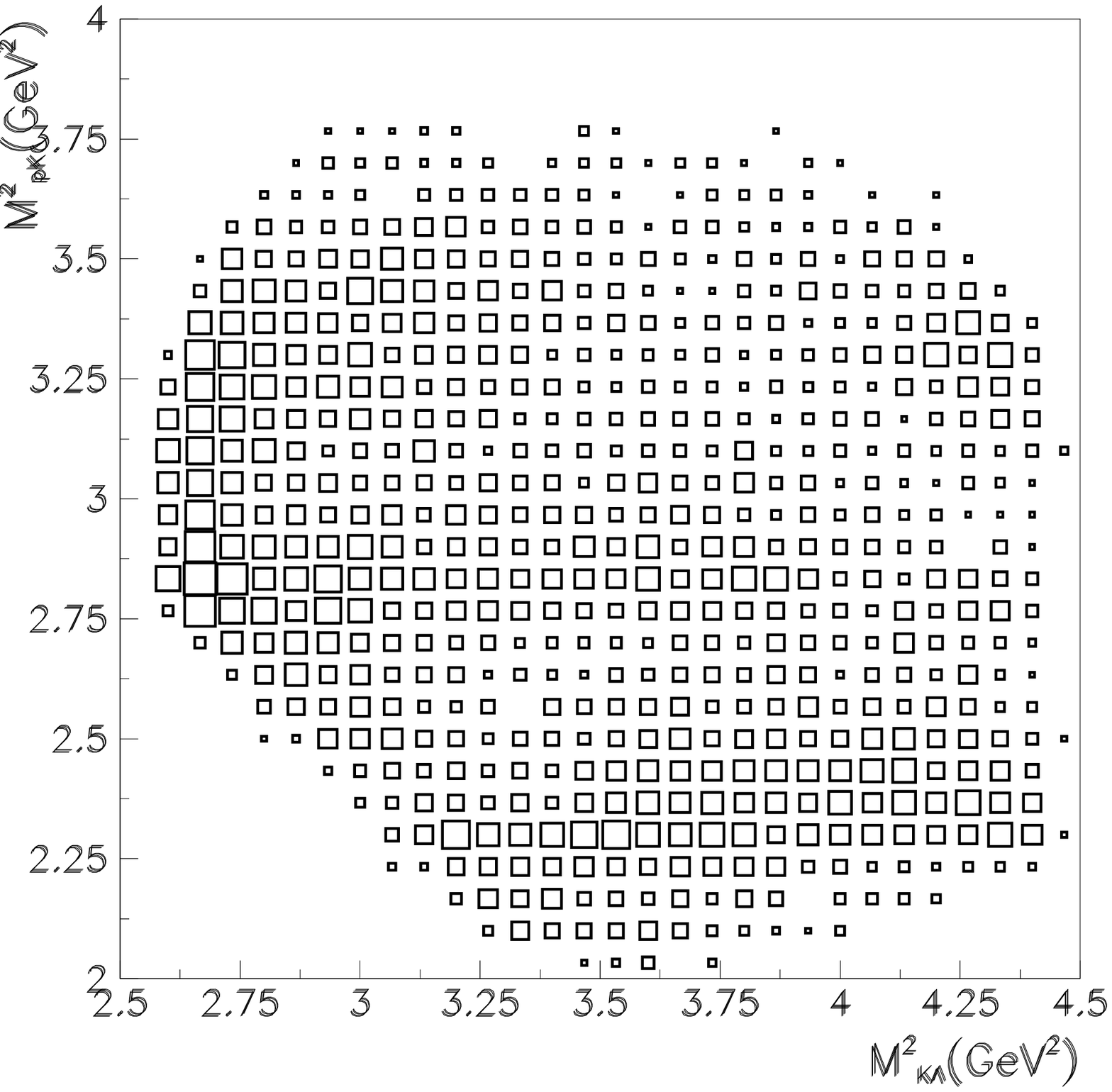} 
\caption{\label{fig4} $pK$ (left) and $K\Lambda$ 
(middle) invariant mass spectra for $J/\psi\to 
pK^-\bar\Lambda$+c.c., compared with phase space distribution; 
right: the Dalitz plot for $J/\psi\to pK^-\bar\Lambda$+c.c.} 
\end{figure}

In the BESII $J/\psi\to pK^-\bar\Lambda$ and $\bar pK^+\Lambda$ 
data~\cite{yang-zbs}, there are clear $\Lambda^*$ peaks at 1.52~GeV, 
1.69~GeV and 1.8 GeV in the $pK$ invariant mass spectrum
shown in Fig.~\ref{fig4}~(left), and $N^*\to K\Lambda$ peaks 
near the $K\Lambda$ threshold at 1.9~GeV  and another near 
2.05~GeV (Fig.~\ref{fig4}~(center)). 
The $N^*$ peak near the $K\Lambda$ threshold is most
probably due to the $N^*(1535)$, which is known to have large 
coupling to $K\Lambda$~\cite{Oset1,liubc}. The SAPHIR experiment 
at ELSA~\cite{ELSA} also observed an $N^*$ peak around 1.9~GeV in 
the $K\Lambda$ invariant mass spectrum from photo-production, and 
a fit~\cite{Mosel} to the data reveals a large $1/2^-$ near-threshold 
enhancement that is mainly due to the $N^*(1535)$. The $N^*$ peak 
at 2.05~GeV is compatible with the $\pi N$ peak observed in
$J/\psi\to N\bar N\pi$ decays. 
Fits to the Dalitz plot distribution (Fig.~\ref{fig4}~(right)) 
prefer a spin-parity for the $N^*(2050)$ of $3/2^+$.
 
\begin{figure}[htbp] 
\begin{center} 
\includegraphics[scale=0.35]{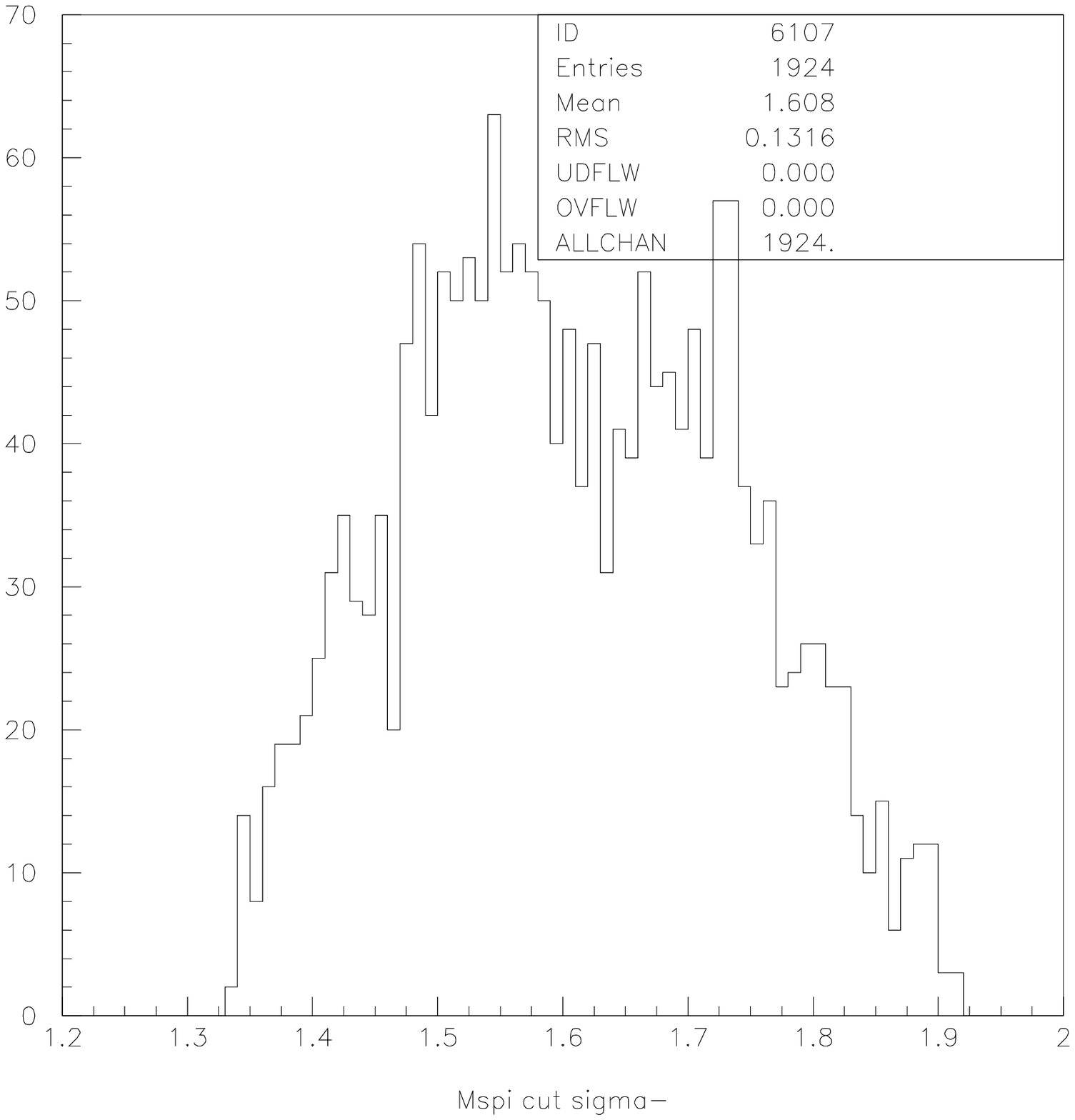} 
\includegraphics[scale=0.35]{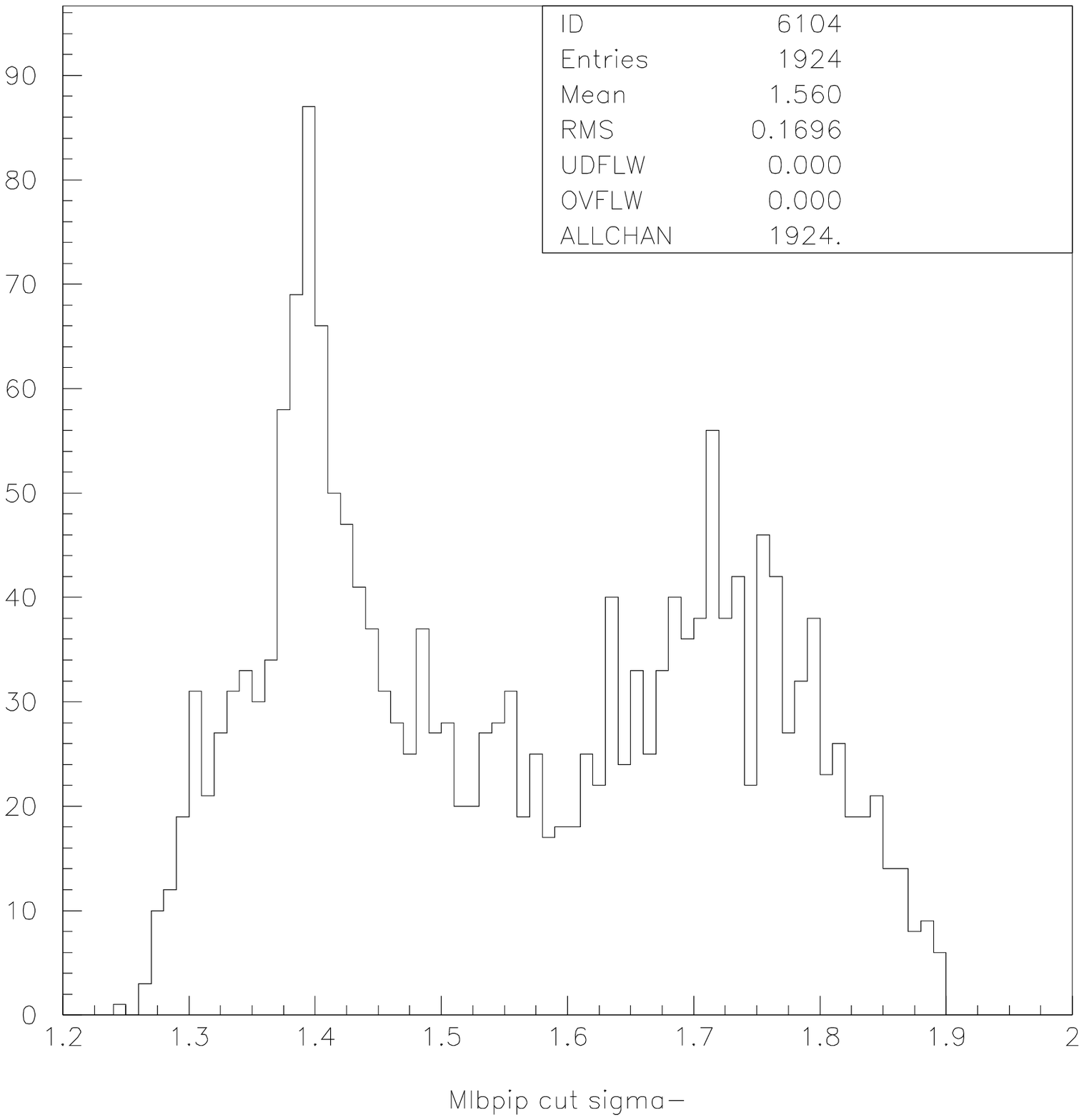} 
\includegraphics[scale=0.35]{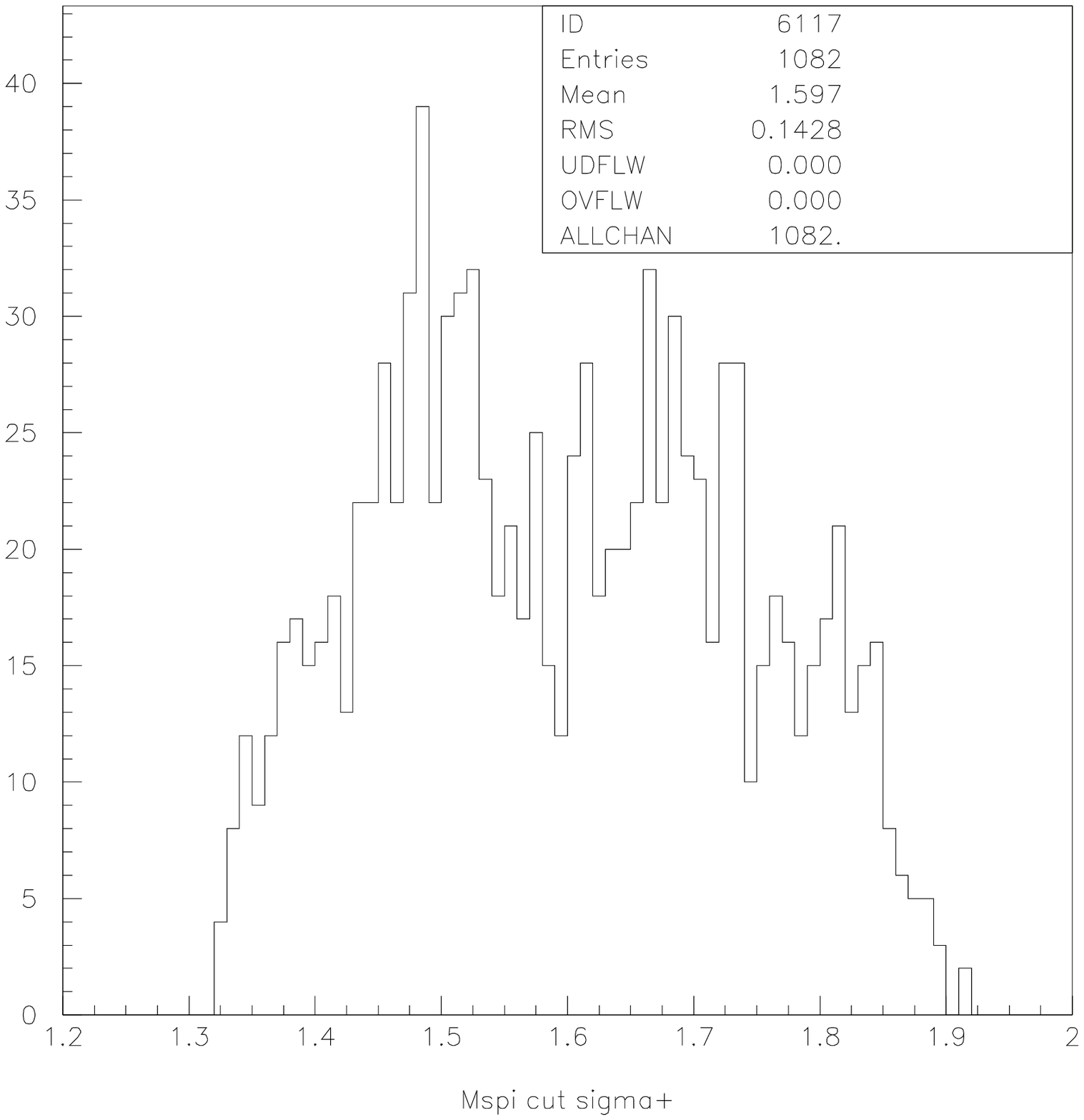} 
\includegraphics[scale=0.35]{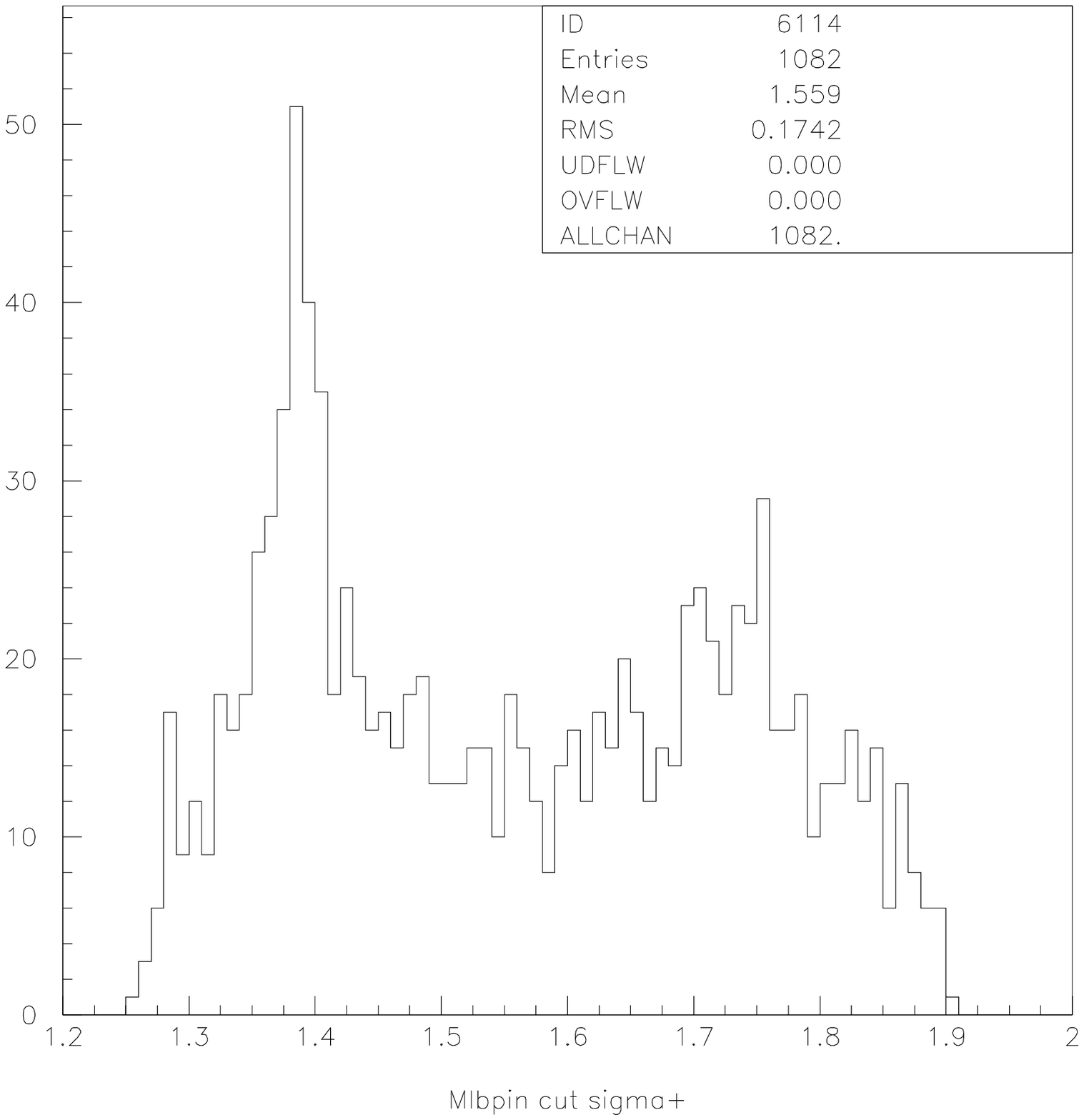} 
\end{center} 
\caption{\label{fig5} $\bar\Sigma\pi$ (left) and $\Lambda\pi$ 
(right) invariant mass spectrum for $J/\psi\to 
\Lambda\bar\Sigma^+\pi^-$ (up) and $J/\psi\to 
\Lambda\bar\Sigma^-\pi^+$ (down), respectively. 
These are preliminary data from BESII~\cite{zoubs4}.} 
\end{figure} 
 
In $J/\psi\to\Lambda\Sigma\pi$ decays~\cite{zoubs4}, 
 $\Lambda^*$ peaks in the $\Sigma\pi$ invariant mass spectra
(Fig.~\ref{fig5}~(left)) are seen at 1.52~GeV, 1.69~GeV and 1.8~GeV.
These are 
similar to the $\Lambda^*$ peaks seen in the 
$pK\Lambda$ channel, although less distinct.  In
the $\Lambda\pi$ invariant mass spectra from 
$J/\psi\to\Lambda\Sigma\pi$, shown in Fig.~\ref{fig5}~(right), 
there is a very clear peak around 1.385~GeV corresponding to the 
well established $\Sigma(1385)$ resonance 
and there is an additional $\Sigma^*$ peak around 1.72~GeV. 
 
\subsection{Partial Wave Analysis for Baryon Resonances}

In order to get more useful information about the properties of the 
baryon resonances that are being produced, such as their 
$J^{P}$ quantum numbers, masses, widths, production and decay rates, 
etc.,  partial wave analyses (PWA) are necessary. A brief introduction to
PWA is given below with specific emphasis on baryonic final states; 
a more generalized description on PWA 
can be found in section \ref{sec:pwa}. 
 
The basic procedure for  partial wave analysis uses the standard 
maximum likelihood method and consists of three main steps:

\noindent
 (1) construct amplitudes $A_i$ for each allowed partial 
wave $i$; 
 
\noindent
(2) form the total transition probability for each event 
from linear combinations of the individual
partial wave amplitudes:
$ w= |\sum_i c_iA_i|^2 $, where $c_i$ are free parameters 
to be determined from the fit; 
 
\noindent
(3) determine the $c_i$ parameters as well as resonance
mass and width parameters by  maximizing the likelihood 
function ${\it L}$:
\begin{equation} 
 {\it L}=\prod_{n=1}^N\frac{w_{data}}{\int w_{MC}}  , 
\end{equation} 
\noindent
where $N$ is the number of reconstructed data events and 
$w_{data}$, $w_{MC}$ are evaluated for data and Monte Carlo 
events, respectively. 
 
For the form of the partial wave amplitudes, we use the 
effective Lagrangian approach~\cite{Nimai,Olsson} with the 
Rarita-Schwinger formalism~\cite{Rarita,Fronsdal,Chung}. In this 
approach, there are three basic elements for constructing 
amplitudes: particle spin wave functions, propagators and the
effective vertex couplings; the amplitude can be written out 
using Feynman rules for tree diagrams. 
 
For example, for $J/\psi\to\bar NN^*(3/2+)\to\bar 
N(k_1,s_1)N(k_2,s_2)\pi(k_3)$, the amplitude can be expressed as 
\begin{equation} 
A_{3/2+}=\bar u(k_2,s_2)k_{2\mu}P^{\mu\nu}_{3/2}(c_1g_{\nu\lambda} 
+c_2k_{1\nu}\gamma_\lambda 
+c_3k_{1\nu}k_{1\lambda})\gamma_5v(k_1,s_1)\psi^\lambda ,
\end{equation} 
where $u(k_2,s_2)$ and $v(k_1,s_1)$ are 1/2-spinor wave functions 
for the $N$ and $\bar N$, respectively, and $\psi^\lambda$ 
is the spin-1 wave 
function ({\it i.e.} the polarization vector) for the $J/\psi$. The 
$c_1$, $c_2$ and $c_3$ terms correspond to three possible 
couplings of the $J/\psi\to\bar NN^*(3/2^+)$ vertex. The $c_1$, 
$c_2$ and $c_3$ can be taken as constant parameters or some 
smooth vertex form-factor modulation 
can be included if necessary. The spin $3/2$ 
propagator $P^{\mu\nu}_{3/2}$ for $N^*(3/2^+)$ resonances is 
\begin{equation} 
P^{\mu\nu}_{3/2}= \frac{\gamma\cdot p+ M_{N*}}{M^2_{N*}-p^2 
-iM_{N*}\Gamma_{N*}}\left[g^{\mu\nu}-\frac{1}{3}\gamma^\mu\gamma^\nu 
-\frac{2p^\mu p^\nu}{3M^2_{N*}} 
+\frac{p^\mu\gamma^\nu-p^\nu\gamma^\mu}{3M_{N*}}\right],
\end{equation} 
with $p=k_2+k_3$. Other partial wave amplitudes can be constructed 
similarly~\cite{Rarita,Liang}.

\section{Prospects for baryon spectroscopy at BESIII} 
 
Recent empirical indications of a positive strangeness 
magnetic moment and positive strangeness radius of the proton 
suggest that the five-quark components in baryons may be largely in 
colored diquark cluster configurations rather than in 
``meson cloud'' configurations or in the form of a sea of 
quark-antiquark pairs~\cite{zou1,zr}. The diquark cluster picture 
also gives a natural explanation that the excess of $\bar d$ over 
$\bar u$ in the proton is due to the presence of a $[ud][ud]\bar d$ 
component. More precise measurements and analyses of 
the strange form factors are needed to examine the relative 
importance of the meson-cloud components and $q^2q^2\bar q$ 
components in the proton. 
 
For baryons, the spatial excitation energy  
could be larger than that needed to pull a $q\bar q$ pair from 
the gluon field around a quark to form diquark clusters 
that contain  a valence 
quark. As a result, five-quark components could be dominant for some 
excited baryon states. 
 
The diquark cluster picture for the five-quark components in baryons 
also gives a natural explanation for the long-standing 
mass-reversal problem of the $N^*(1535)$, $N^*(1440)$ and 
$\Lambda^*(1405)$ 
resonances, as well as the unusual decay pattern of the $N^*(1535)$ 
resonance, {\it i.e.} the large coupling to $K\Lambda$. 
These could  be the consequence of a large 
$|[ud][us]\bar s>$ component~\cite{zou1,liubc}. 
 
The diquark cluster picture predicts the existence of $SU(3)$ nonet
partners of the $N^*(1535)$ and $\Lambda^*(1405)$, {\it i.e.}, an 
additional $\Lambda^*~1/2^-$ around 1570~MeV, a triplet 
$\Sigma^*~1/2^-$ around 1360~MeV and a doublet $\Xi^*~1/2^-$ 
around 1520~MeV~\cite{zhusl-zbs}. There is, in fact, 
some evidence for all of these in BES $J/\psi$ data. 
Figure~\ref{fig4}~(left) 
shows the $pK$ invariant mass spectrum for $J/\psi\to 
pK^-\bar\Lambda$+c.c. and Fig.~\ref{fig5}~(right)
shows the $\Lambda\pi$ 
invariant mass spectrum for 
$J/\psi\to\Lambda\bar\Sigma^+\pi^-$~\cite{zoubs4}. 
In the $pK$ invariant mass spectrum, beneath
the narrow $\Lambda^*(1520)~3/2^-$ peak is a quite obvious 
broader peak around 1570~MeV. A preliminary partial wave 
analysis~\cite{yanghx2001} indicates its spin-parity 
to be $1/2^-$.  Such a $\Lambda^*(1570)~1/2^-$ resonance 
would fit neatly into the new 
scheme for a $1/2^-$ $SU(3)$ baryon nonet. In the $\Lambda\pi$ 
invariant mass spectrum of Fig.~\ref{fig5}~(right), 
there are signs of a broad structure
under the $\Sigma^*(1385)~3/2^+$ peak.  No partial wave 
analysis has yet been performed for this channel, but there is 
good reason to expect that there may be $1/2^-$ component 
underneath the $\Sigma^*(1385)~3/2^+$ peak. 
 
According to the PDG~\cite{pdg2006}, the branching fractions for 
$J/\psi\to\bar\Sigma^-\Sigma^*(1385)^+$ and 
$J/\psi\to\bar\Xi^+\Xi^*(1530)^-$ are $(3.1\pm 0.5)\times 10^{-4}$ 
and $(5.9\pm 1.5)\times 10^{-4}$, respectively. These two 
processes are $SU(3)$-violating decays since the $\Sigma$ and $\Xi$ 
belong to an $SU(3)$ $1/2^+$ octet while $\Sigma^*(1385)$ and 
$\Xi^*(1530)$ belong to an $SU(3)$ $3/2^+$ decuplet. 
For comparison, the $SU(3)$-violating decay 
$J/\psi\to\bar p\Delta^+$ has a
branching fraction that is less than $1\times 10^{-4}$,
while the $SU(3)$-allowed 
decay $J/\psi\to\bar pN^*(1535)^+$  has a branching fraction 
of $(10\pm 3)\times 10^{-4}$.  Thus the branching fractions for 
$J/\psi\to\bar\Sigma^-\Sigma^*(1385)^+$ and 
$J/\psi\to\bar\Xi^+\Xi^*(1530)^-$ are anomalously large. A 
possible explanation for this anomaly is that there 
are substantial $1/2^-$ components under the $3/2^+$ peaks and 
the two branching ratios were obtained assuming a pure $3/2^+$ 
contribution. This possibility could be easily checked with the 
high statistics \bes3 data in near future. 
 
\begin{figure}[htbp] 
\centerline{\psfig{file=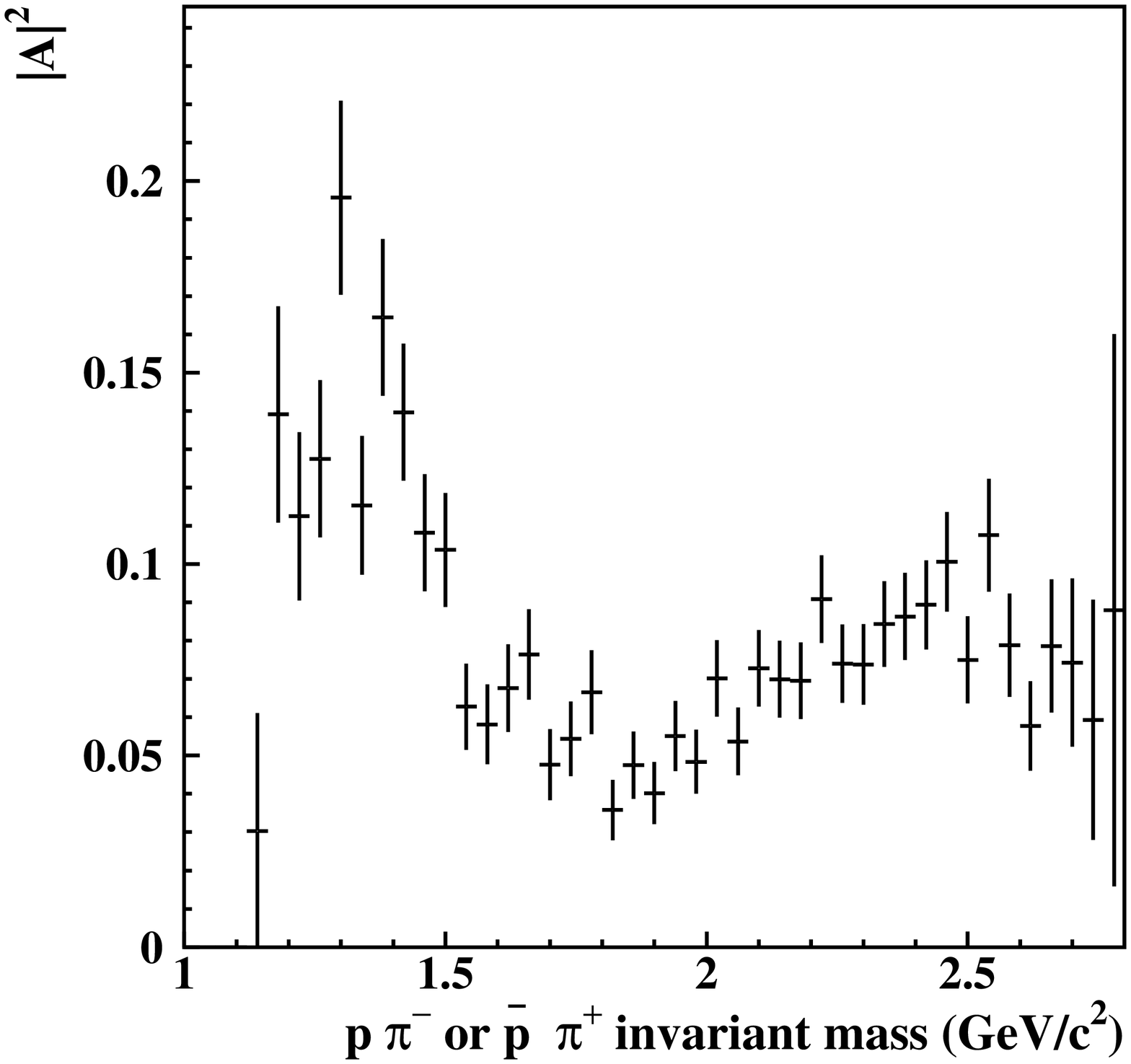,width=6cm} 
            \psfig{file=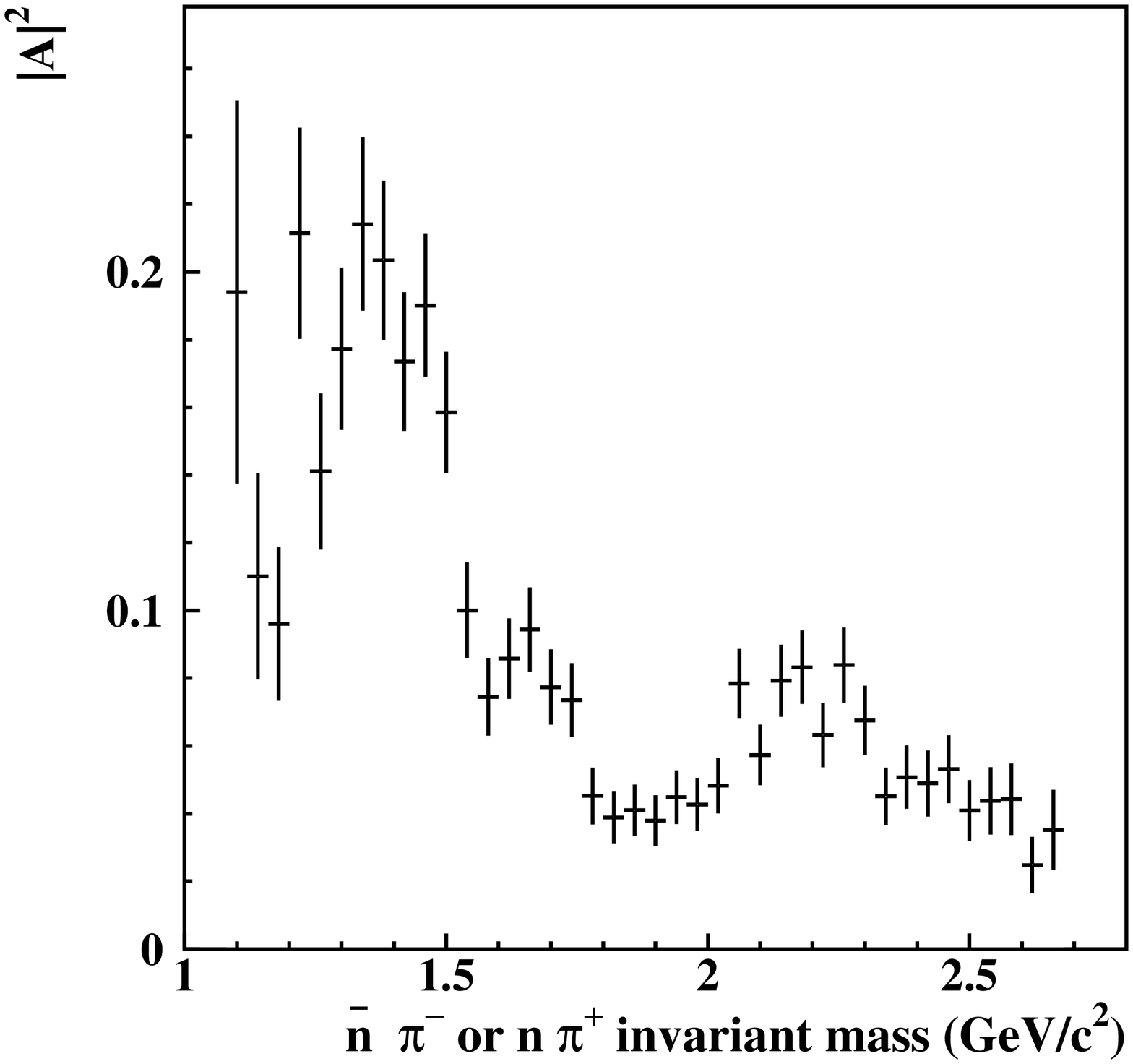,width=6cm}} 
\caption{\label{am} Data corrected by MC simulated efficiency and 
phase space versus $p \pi^-$ (or $\bar{p} \pi^+$) and $\bar{n} 
\pi^-$ (or $n \pi^+$) invariant mass for $\psi' \to p \bar{n} 
\pi^-+c.c.$ candidate events~\cite{psip2}.} 
\end{figure} 
 
\begin{figure}[htbp] 
\includegraphics[scale=0.295]{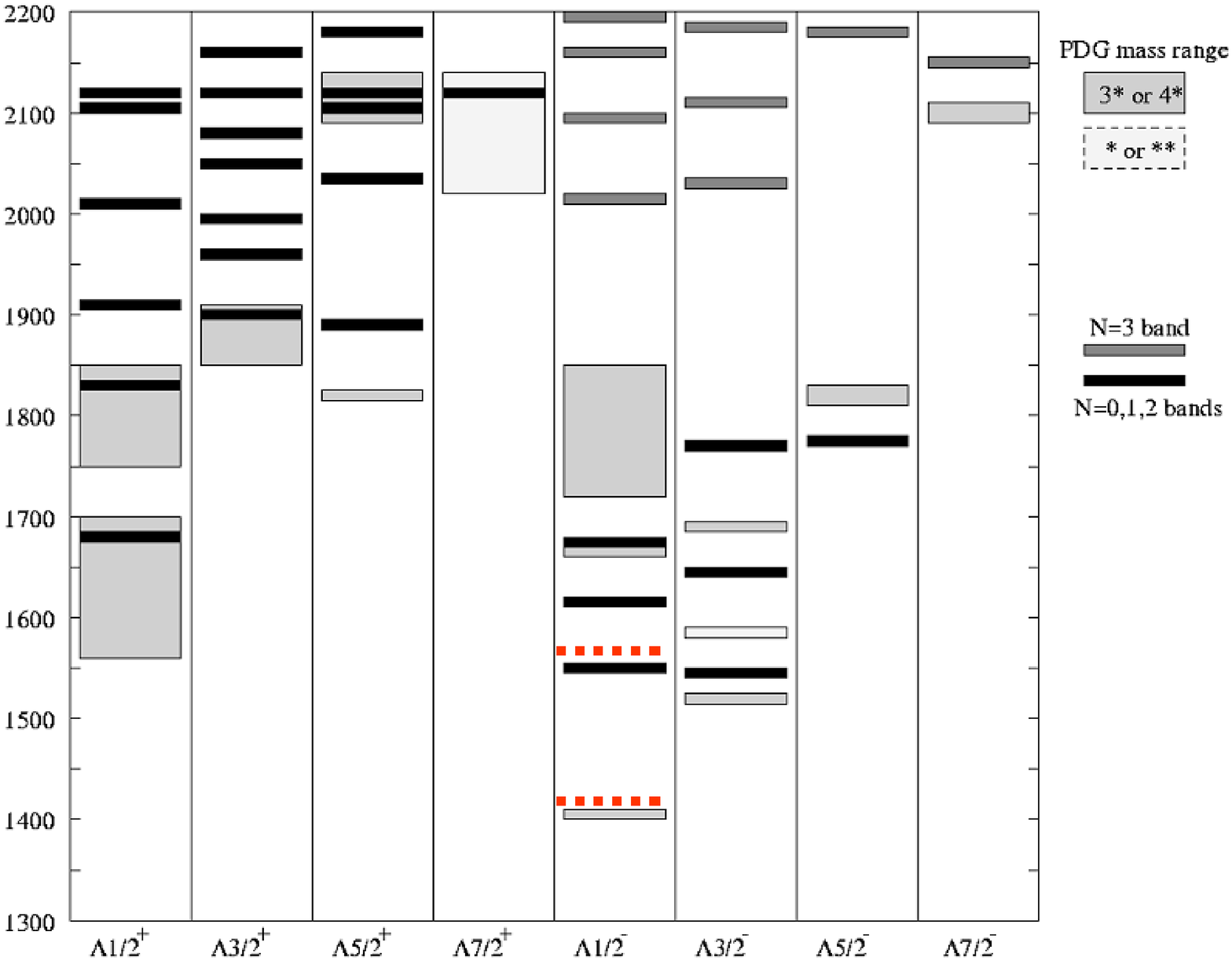} 
\includegraphics[scale=0.32]{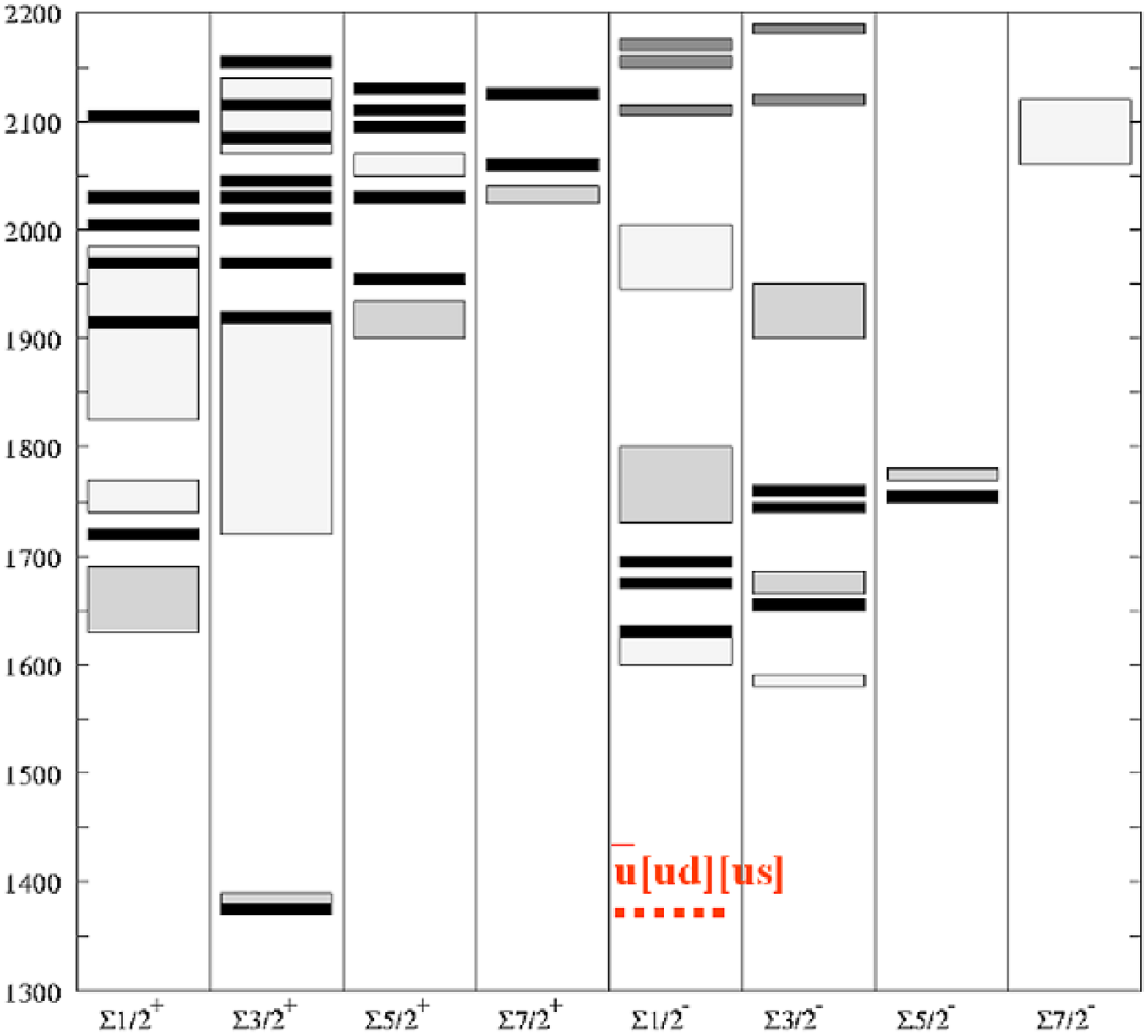} 
\includegraphics[scale=0.5]{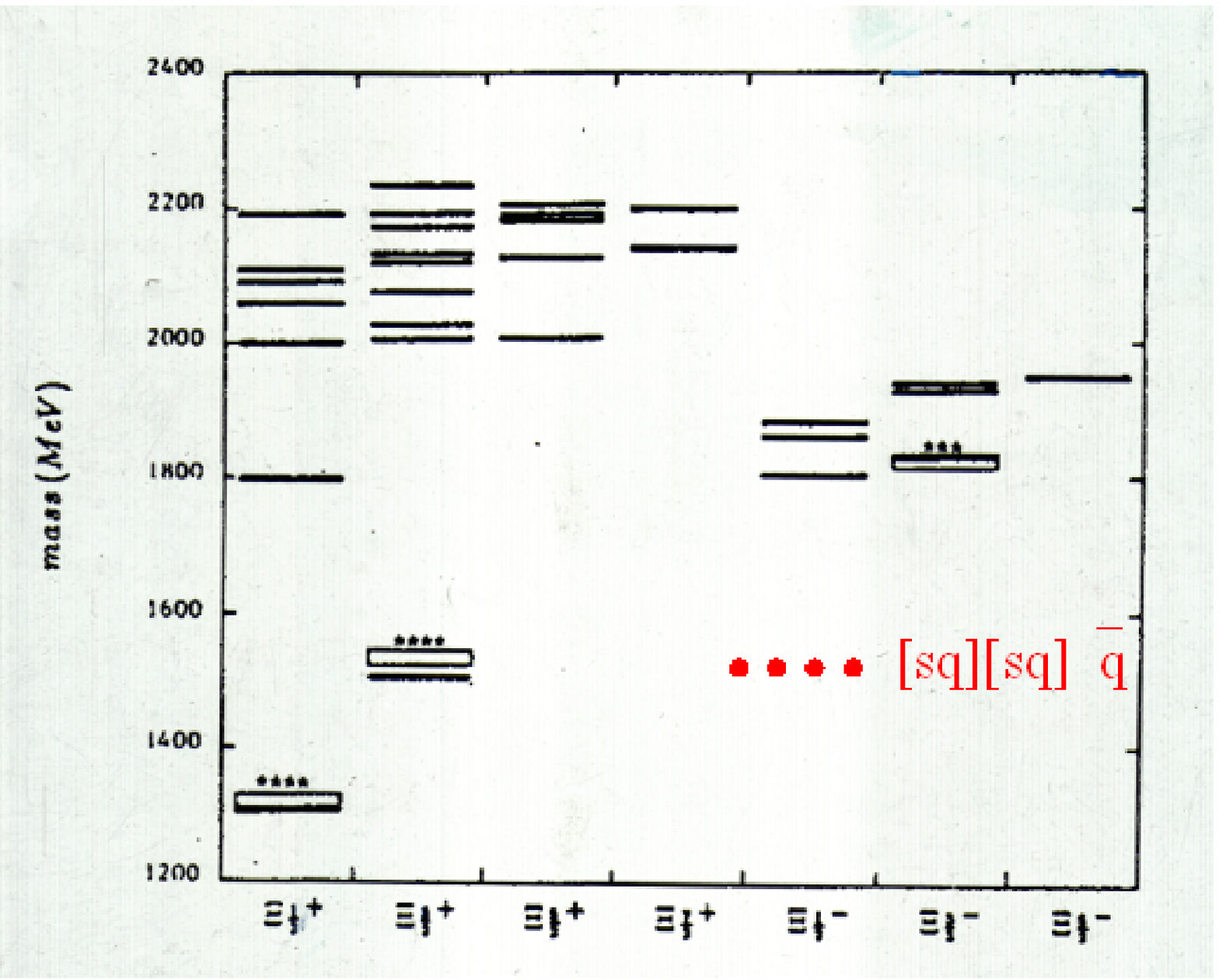} 
\caption{\label{fig6} Predicted $\Lambda^*$,$\Sigma^*$,$\Xi^*$ 
mass spectra (solid line) by the classical 3q quark model 
\cite{isgur-zbs,Capstick1} and the predicted lowest $1/2^-$ 
$\Lambda^*$,$\Sigma^*$,$\Xi^*$ states (dashed line) with 
pentaquark configuration~\cite{zhusl-zbs}, compared with observed 
states (indicated by boxes). } 
\end{figure} 
 
With two orders-of-magnitude higher statistics
expected at \bes3, numerous
important  baryon spectroscopy issues can be studied with 
both $J/\psi$ and $\psi'$ decays. Data from the $\psi'$ will 
significantly extend the mass range for the study of baryon 
spectroscopy. For example, in 
a sample of $\psi'\to p \bar{n} \pi^-+c.c.$ 
events collected at BESII~\cite{psip2}, obvious 
structures at $M_{N\pi}> 2$~GeV in the $N\pi$ invariant mass 
spectra are evident (Fig.~\ref{am}). However,
the statistics are insufficient for drawing
any conclusions about high mass $N^*$ 
resonances~\cite{psip2,psip1}. 
Determinations of the properties of these high mass 
$N^*$ resonances can be done with the
huge \bes3 data samples. 
More importantly, many of
the ``missing" $\Lambda^*$,$\Sigma^*$ and $\Xi^*$ 
hyperon resonances are expected to be produced and observable in 
the high statistics \bes3 $J/\psi$ and $\psi'$ data samples. 
Figure~\ref{fig6} 
shows the $\Lambda^*$,$\Sigma^*$,$\Xi^*$ mass spectra 
(solid line) predicted by the classical $qqq$ quark 
model`\cite{isgur-zbs,Capstick1} and the predicted lowest $1/2^-$ 
$\Lambda^*$,$\Sigma^*$,$\Xi^*$ states (dashed line) with 
pentaquark configurations~\cite{zhusl-zbs}, compared with observed 
states (indicated by boxes). 
The classical $qqq$ quark model~\cite{isgur-zbs} predicts 
more than 30 $\Xi^*$ resonant states between 1.78~and 2.35~GeV as 
shown in Fig.~\ref{fig6}. None of these have yet been established. 
These $\Xi^*$ states cannot be produced in $J/\psi$ decays  
because of the limited phase space, but all of them can be 
produced in $\psi'$ decays. 
\bes3  $\psi'$ data will enable us to complete the 
$\Lambda^*$, $\Sigma^*$ and $\Xi^*$ spectrum and 
distinguish between various 
models for their internal structure, such as the simple $qqq$ 
quark structure and more complicated structures
in which pentaquark components dominate. 
 
\begin{table}[htb] 
\caption{ Measured $J/\psi$ decay branching ratios (BR$\times 
10^3$) for channels involving baryon anti-baryon and meson(s) 
\cite{pdg2006,bes2}} \label{part3:table:1} 
\renewcommand{\arraystretch}{1.2} 
\begin{tabular}{cccccc} 
\hline $p\bar n\pi^-$  & $p\bar p\pi^0$  &  $p\bar p\pi^+\pi^-$ & 
$p\bar p\eta$ & $p\bar p\eta'$ & $p\bar p\omega$ \\ 
\hline $2.4\pm 0.2$ & $1.1\pm 0.1$ & $6.0\pm 0.5$ & $2.1\pm 0.2$ & 
$0.9\pm 0.4$ & $1.3\pm 0.3$ \\ 
\hline $\Lambda\bar\Sigma^-\pi^+$ & $pK^-\bar\Lambda$ & $pK^-\bar\Sigma^0$ 
& $\bar pp\phi$ & $\Delta(1232)^{++}\bar p\pi^-$ & $pK^-\bar\Sigma(1385)^0$ \\ 
\hline $1.1\pm 0.1$ & $0.9\pm 0.2$ & $0.3\pm 0.1$ 
& $0.045\pm 0.015$ & $1.6\pm 0.5$ & $0.51\pm 0.32$ \\ 
\hline 
\end{tabular}\\ 
\end{table}

The measured $J/\psi$ decay branching ratios for channels 
involving baryon anti-baryon plus meson(s) are listed in Table 
\ref{part3:table:1}. With $10^{10}$ $J/\psi$ events, all
of these channels will have large enough event samples 
to support partial wave analyses. 
Among these 
channels, the $\Sigma\bar\Lambda\pi +c.c.$ channels will be given 
high priority in order to
pin down the lowest $1/2^-$ $\Sigma^*$ and 
$\Lambda^*$ resonances as well as other, 
higher excited $\Sigma^*$ and 
$\Lambda^*$ states. Another very important channel is 
$K^-\Lambda\bar\Xi^+ + c.c.$ which will be the best channel for finding 
the lowest $1/2^-$ $\Xi^*$ resonance and other ``missing" 
$\Xi^*$ states that decay via  $\Xi^*\to K\Lambda$. 
This channel should be 
rather easily reconstructed in \bes3. One can select events 
containing a $K^-$ and a $\Lambda$ with $\Lambda\to p\pi^-$
and  then use the narrow peak in the $K^-\Lambda$ 
recoil mass spectrum to 
identify the undectected $\bar\Xi^+$. 
 
For $10^9$ $\psi^\prime$ events, the $K^-\Lambda\bar\Xi^+ + c.c.$ 
and $p\bar p\phi$ final states, which 
have very limited phase space in $J/\psi$ decays, 
will be given high priority. The 
$K^-\Lambda\bar\Xi^+ + c.c.$ channel will provide
opportunities to discover many of the
``missing" $\Xi^*$ resonances, while the $p\bar p\phi$ 
channel should allow us to find $N^*$ resonances 
that have a large 
coupling to $N\phi$ \cite{huangf} and, thus, large five-quark 
components. 
 
After analyzing the easier three-body final states, 
four-body and even five-body 
channels could also be investigated. Among these, 
$\Delta(1232)^{++}\bar p\pi^-$ in $p\bar p\pi^+\pi^-$ and 
$\Delta(1232)^{++}\bar\Sigma^-K^-$ in $p\bar\Sigma^-\pi^+K^-$ are 
very good channels for finding ``missing" $\bar\Delta^{*--}$ states
decaying to $\bar p\pi^-$ and $\bar\Sigma^-K^-$. The 
spectrum of isospin 3/2 $\Delta^{++*}$ resonances is of special 
interest since it is the most experimentally accessible system 
that is comprised of three identical valence quarks.  It has recently
been proposed that the lowest 
$1/2^-$ baryon decuplet contains large 
vector-meson-baryon molecular components~\cite{xiejj}. In this new 
scheme, the $\Xi^*(1950)$ is predicted to be
a $1/2^-$ resonance with a
large coupling to $\Lambda K^*$. The 
$\psi^\prime\to\bar\Xi\Lambda K^*$ 
process will provide a very good place to look for 
a ``missing" $\Xi^*$ 
with a large $\Lambda K^*$ coupling. 
 
In summary, \bes3 data can play a very important role in 
the study of 
excited nucleons and hyperons, {\it i.e.}, 
the very poorly understood $N^*$, $\Lambda^*$, 
$\Sigma^*$, $\Xi^*$ and $\Delta^{*++}$ resonances.

\chapter{Physics of soft pions and the lightest scalars at \bes3}
\label{sec:softpion-physics}

A brief introduction to the physics related to soft pions, light
scalar mesons and the final state intereaction theorem for \bes3 
experiments is given.

\section{Partially conserved vector current (PCAC) and soft pions}
\subsection{PCAC}
 Strong interaction physics is described by Quantum Chromodynamics
(QCD), where $\bar\psi\psi$ has vacuum quantum numbers
 and a non-vanishing vacuum expectation value: $<\bar\psi\psi>\neq
 0$. As a consequence, the  pseudo-Goldstone bosons, $\pi^a$,
 couple to the axial-vector current, $<0|A_\mu^a|\pi^a>\neq
0$. If isospin is a good quantum number then,
 \begineqn \label{pcac1}
<0|A_\mu^a |\pi^b>=f_\pi\delta^{ab}p_\mu\ ,
 \end{equation}
where $f_\pi=93$MeV is the $\pi$ decay constant as determined from
$\pi\rightarrow l\nu$ decay. For an on-shell pion Eq.~(\ref{pcac1})
becomes
 \begineqn <0|\partial^\mu
A_\mu^a |\pi^b>=f_\pi\delta^{ab}m_\pi^2\ ,\,\,\,\mbox{or} \,\,\,
<0|\partial^\mu A_\mu^a |\pi^b>=f_\pi m_\pi^2<0|\phi^a|\pi^b>\ .\end{equation}
Generalizing to the operator form, we have \begineqn
\partial^\mu A_\mu^a =f_\pi m_\pi^2\phi^a\ .
\end{equation} This  is the celebrated Partial  Conservation of  Axial
Current (PCAC) relation~\cite{PCAC}.
The most important applications of PCAC are the derivations of
various soft pion theorems. In the following we briefly discuss a
few of them.

\subsection{Adler's theorem with one soft pion}\label{Adler}

 Adler's theorem states that:
\textit{~~~In order to calculate the matrix element for
a strong interaction process involving
one soft pion: $i\rightarrow f+\pi$, one only needs to consider
the process without the pion, $i\rightarrow f$, and then insert the pion
into any of the external lines, using derivative coupling
theory. }
 A diagramatic explanation to the Adler's theorem is given in 
Fig.~\ref{zheng-fig1}.

\begin{figure}[htbp]
\centerline{\epsfig{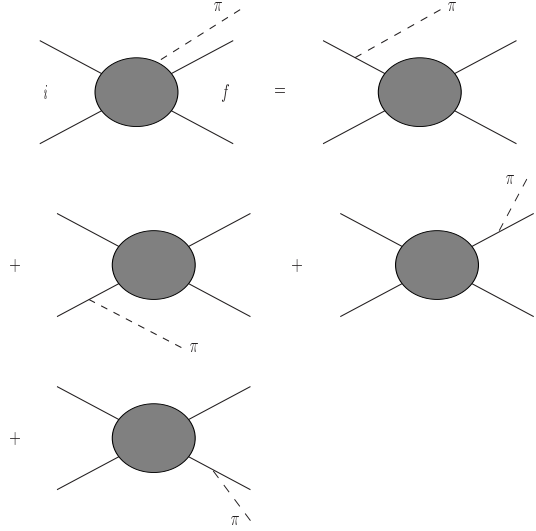}}
\caption{Diagramatic explanation of the Adler's theorem.}  
\label{zheng-fig1} 
\end{figure} 

\noindent Notice that the theorem as stated above is only exact when
all four components of the pion momentum $k^\mu_\pi\to 0$
simultaneously ($\pi$ off-shell). In practice it means that any quantity
$k\cdot p$, where $p$ is any momentum involved in the process $i\to f
+\pi$, can be considered to be small. A simple proof of the
theorem can be found in Ref.~\cite{coleman}.

An important application of the soft pion theorem is to $\pi\pi\to
\pi\pi$ scatterings. Where, since there is no 3$\pi$ coupling,
\begineqn
T(p_1,p_2,p_3\to 0,p_4)=0\ .
\end{equation} 
In terms of the Mandelstam variables, when
$p_3\to 0$, one has $s=(p_1+p_2)^2 =m_\pi^2$,
$t=(p_1-p_3)^2=m_\pi^2$, $u=(p_1-p_4)^2 =m_\pi^2$, and we find that the 
Adler zero for $\pi\pi$ scattering corresponds to $s=t=u=m_\pi^2$. This
zero is not far from $\pi\pi$ scattering threshold. Hence the
threshold parameters for $\pi\pi$ scatterings are suppressed due to
the existence of this $T$-matrix zero.

An important process in $J/\psi$ physics where the soft pion theorem
cn be applied is $\psi'\to J/\psi\pi\pi$ decay. 
The decay branching
ratio of $\psi'(3685)$ into $J/\psi$ and light hadrons is $(55.7\pm
2.6)\%$~\cite{pdg2006}, including ${\rm Br}(\psi'\rightarrow
J/\psi\,\,\, \pi^+\pi^-)=(30.5\pm 1.6)\%$ and ${\rm
Br}(\psi'\rightarrow J/\Psi\,\,\, \pi^0\pi^0)=(18.2\pm 1.2)\%$.
Therefore the $\psi'\rightarrow J/\psi  \pi\pi$ process is a
dominant $\psi'$ decay channel.  The transition amplitude for
$\psi'\rightarrow J/\psi \pi\pi$ is
 \begineqn
M=<\psi\pi\pi|\psi'>=<P,\epsilon ;p_{1a},p_{2b}|P',\epsilon'>\ .
 \end{equation}
Using PCAC one can obtain, after some algebraic
manipulation~\cite{cahn},
 \bqa
<\psi 
\pi\pi|\psi'>&=&-(p_1^2-m_\pi^2)(p_2^2-m_\pi^2)(\phi^a\phi^b)\nonumber\\
&=&f_\pi^{-2}p_1^\mu p_2^\nu (\overline {A}_\mu^a\overline
{A}_\nu^b)-if_\pi^{-1}(p_1^2+p_2^2-m_\pi^2)<\psi|\tilde\Sigma^{ab}|\psi'>\
,
 \eqa
where $\overline A$ indicates that the pion pole part of the axial
vector current, $A$, has been subtracted. Both the 
term in the parentheses, and the inner product term
({\it i.e.} the Sigma term) are regular functions when $p_\mu\to
0$~\cite{cahn}. It is evident from the formula that when
$p_1^\mu\to 0$ and $p_2$ is on-shell or vice versa, $M\to 0$. This is
the Adler zero condition.

\subsection{The linear $\sigma$ model and chiral 
shielding}\label{SLsigmaM}
Historically,  the $\sigma$ meson was first introduced in association
with the $SU(2)\times SU(2)$ linear $\sigma$ model in an attempt to
describe the spontaneous breaking of chiral symmetry. The model has
the advantage of a natural realization of PCAC and current algebra.
In the linear sigma model lagrangian there are  iso-triplet
$(\pi_1,\pi_2,\pi_3)$ fields and a scalar iso-singlet $\sigma$
field~\cite{Lee}:
 \bqa\label{LsigmaM}
\mathcal L &=& \mathcal L_s+c\sigma \ ,\nonumber \\%
\mathcal L_s &=& {1\over 2}[(\partial_\mu \sigma)^2+(\partial_\mu \pi)^2]-{m^2\over 2}[\sigma^2+\pi^2]%
-{\lambda\over 4}[\sigma^2+\pi^2]^2\ .
\eqa%
The last term in the first equation indicates that chiral symmetry
is not exact, $c=f_\pi m_\pi^2$ is a small quantity. If $c=0$,  the
lagrangian is invariant under the $SU_L(2)\times SU_R(2)$
chiral rotations: \bqa\label{chiralrotation}
&&\vec{\pi} \rightarrow \vec{\pi}+\vec{\alpha} \times\vec{\pi}-\vec{\beta} \sigma \ ,\nonumber \\%
&&\sigma \rightarrow \sigma+\vec{\beta} \cdot \vec{\pi}\ .
\eqa%
When $m^2<0$, chiral symmetry is spontaneously broken; the minimum
of the effective potential is taken at $<\sigma>=f_\pi$ and we can
redefine another scalar field with vanishing vacuum expectation
value $s$ as,
\be%
\sigma=s+f_\pi\ ,\ \ <s>=0\ .
\end{equation}%
The new lagrangian expressed in terms of the shifted field is
\be\label{Lsigmash}
\mathcal L = {1\over 2}[(\partial_\mu \pi)^2-{m_\pi}^2\pi^2]+{1\over 2}[(\partial_\mu s)^2-{m_\sigma}^2s^2]%
-\lambda f_\pi s(s^2+\pi^2)-{\lambda\over 4}[s^2+\pi^2]^2 \ .
\end{equation}%
From the above we find
 \begineqn {m_\sigma}^2=m^2_\pi+2\lambda f_\pi^2\ .
\end{equation} In addition we determine the $\sigma\pi\pi$ coupling 
constant,
\begineqn%
g_{\sigma\pi\pi}={{m_\sigma^2-m_\pi^2}\over f_\pi}=2\lambda f_\pi\ .
\end{equation}%
Here the coupling constant $g_{\sigma\pi\pi}$ is proportional to the mass
square of the $\sigma$ meson and, thus, the interaction becomes very
strong in the large $m_\sigma$ limit.
As a result, there does not exist a trivial
decoupling limit when $m_\sigma\to\infty$. 
As a consequence,
the decay width of $\sigma\to \pi\pi$ (in the chiral limit) at tree
level is proportional to the $m_\sigma^3$:
 \bqa
\Gamma(\sigma\rightarrow \pi  \pi )&=&  {{3m^3_{\sigma}} \over 32\pi
f^2_\pi } \sqrt{1-{4m_\pi^2 \over m_{\sigma}^2}} \ .\eqa
 From
this formula one sees that when $m_\sigma =600MeV$,
$\Gamma_\sigma\simeq660$MeV, {\it i.e.,} its large width already exceeds
its mass. The appearance of a large width  indicates the failure of
perturbation calculations and also the Breit--Wigner description
of the resonance,  since the latter is only a narrow width
approximation~\cite{ZengGao}.
 The
$\pi\pi$ scattering amplitudes can be calculated at tree level using
the linear sigma model, as shown in Fig.~\ref{zheng-fig2}.

\begin{figure}[htbp]
\centerline{\epsfig{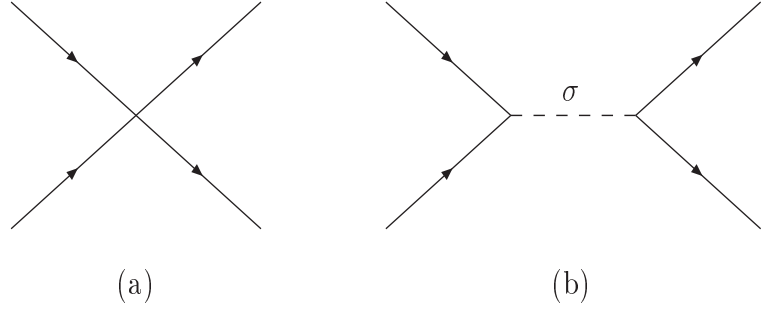}}
\caption{First-order Feynman graphs for $\pi\pi$ scatterings in the linear
$\sigma$ model. The crossed
channel diagrams are not shown.}  
\label{zheng-fig2} 
\end{figure} 

There are only two types of diagrams:
the 4$\pi$ contact interaction (Fig.~\ref{zheng-fig2}(a)), and 
the $\sigma $ exchange diagram  (Fig.~\ref{zheng-fig2}(b)).
One has,
\bqa\label{Iamp}
&&T^{I=0}(s,t,u) = 3A(s,t,u)+A(t,u,s)+A(u,s,t)\ ,\nonumber \\%
&&T^{I=1}(s,t,u) = A(t,u,s)-A(u,s,t)\ ,\nonumber \\%
&&T^{I=2}(s,t,u) = A(t,u,s)+A(u,s,t)\ ,%
\eqa%
where
 \begineqn\label{Astu}
A(s,t,u)=-2\lambda(1+{{2\lambda v^2}\over{{s-m_\sigma}^2}})%
=-{{{m_\sigma}^2-{m_\pi}^2}\over{{s-m_\sigma}^2}}{{s-{m_\pi}^2}\over
{f_\pi}^2}\ .
\end{equation}%
In the above, the first term on the $r.h.s.$ of the first equality
corresponds to the background contact $\lambda \phi^4$ interaction.
To preserve the boundedness of the vacuum energy from below, $\lambda$
must be positive and, thus, the  $\lambda\phi^4$ contact interaction is
repulsive. The second term is from the $\sigma$ pole and provides an
attractive force at low energies.  The explicit calculation described at
the conclusion of Sec.~\ref{Adler} confirms this; the two
terms in Eq.~(\ref{Astu}) cancel almost exactly at threshold, which
is as expected from the low energy soft pion theorem.

The partial wave projection of the full amplitude is,
 \begineqn
T_J^I(s)={1\over 32\pi(s-4m_\pi^2)} \int^{0}_{4m_\pi^2-s} dt\,
P_J(1+{2t\over s-4m_\pi^2}) T^{I}(s,t,u)\ .
 \end{equation}
  From
Eq.~(\ref{Astu}) one gets partial wave amplitudes $T^I_J$  as
follows:
 \bqa\label{LsMR} { T}_0^0&=&\frac{\lambda}{16\pi
}\left(\frac{3(m_\sigma^2-m_\pi^2)}{m_\sigma^2-s}
+\frac{2(m_\sigma^2-m_\pi^2)\log[\frac{m_\sigma^2+s-4m_\pi^2}{m_\sigma^2}]}{s-4m_\pi^2}-5\right)\ ,\nonumber\\
{ T}_0^2&=&-{\lambda\over 8\pi}\left(1-\frac{2\lambda
v^2}{s-4m_\pi^2}
\log[\frac{m_\sigma^2+s-4m_\pi^2}{m_\sigma^2}]\right)\ ,\nonumber\\
{ T}_1^1&=&\frac{\lambda^2v^2(16m_\pi^2-4s-2(4m_\pi^2-2m_\sigma^2-s)
\log[\frac{m_\sigma^2+s-4m_\pi^2}{m_\sigma^2}])}{8\pi(s-4m_\pi^2)^2}\
.
 \eqa
It is not very easy to recognize the $T$-matrix zero from the
above partial wave amplitudes. Taking the limit of $m_\sigma\to\infty$ 
while keeping $\lambda$ fixed, we find that the zeros are located at
$s=m_\pi^2/2$, $4m_\pi^2$  and $2m_\pi^2$ for $IJ=00$, $11$ and $20$,
respectively. The positions of these zeros also
receive small corrections from higher orders. In the limit of
$m_\sigma\to\infty$ and $f_\pi$ fixed, the partial wave amplitudes from
Eq.~(\ref{LsMR}) become
 \bqa
{T}_0^0=\frac{2s-m_\pi^2}{32\pi f_\pi^2}\ ,\,\,
{T}_0^2=\frac{2m_\pi^2-s}{32\pi f_\pi^2}\ ,\,\,
{T}_1^1=\frac{s-4m_\pi^2}{96\pi f_\pi^2}\ ,
 \eqa
which agree with the lowest order results from chiral perturbation
theory, and are, in fact, predictions from current algebra.

The lesson  we learn from the above is that it is very difficult to
identify the $\sigma$ pole (if there is one) in
experiments, since the background contribution cancels that 
from the $\sigma$
near threshold, as dictated by the soft pion theorem.
In the language of Ref.~\cite{achasov06}, this is called `chiral
shielding'. Furthermore, the $\sigma$ meson has a large width which
makes it difficult to distinguish from backgrounds.

\subsection{Why should there be a $\sigma$ resonance?}

The chiral shielding of the  $\sigma$  has probably been known  since the
invention of the linear $\sigma$ model. It implies, on one hand,
the cancelation between a positive $\sigma$ contribution and a
negative background contribution to the $\pi\pi$ scattering phase
shift, a fact that has led many physicists to argue for the possible 
existence
of the $\sigma$ resonance~\cite{tornqvist,negative phase}. On the
other hand, since no firm evidences for the $\sigma$ meson had been
found experimentally, it was thought that the $\sigma$ meson is not
necessary for describing chiral symmetry breaking and this, 
in turn, led to
the invention of the non-linear
realization of chiral symmetry, from which chiral
perturbation theory was constructed~\cite{GL}. The latter has
been very successful at describing low energy strong interaction
physics without referring to a $\sigma$ meson. Indeed the chiral
cancelation mechanism in the linear realization only provides at
most a self-consistent description to the data and is insufficient
to provide solid proof for the existence of the $\sigma$ pole.
However, the background contribution from the nearby left-hand cut can
be estimated using chiral perturbation theory, and 
the background contribution to the $\pi\pi$ scattering phase
shift is found to be negative. This proves that the $\sigma$ pole is
essential for the application of chiral perturbation theory to
experimental data~\cite{XZ00}.

The existence of the $\sigma$ pole has been reported in production
experiments, such as in $D$-meson decays in E791~\cite{Aitala}
and $J/\psi$ and $\psi(2S)$ decays
at BES~\cite{bessigma1,bessigma2} .  In Ref.~\cite{Aitala}, a
sample of $1172\pm 61$ $D^-\to \pi^-\pi^+\pi^+$ decays (the ratio
$\Gamma(D^+\to\pi^-\pi^+\pi^+)/\Gamma(D^+\to K^-\pi^+\pi^+)$ is
about 3\%), is used to provide evidence for the existence of a scalar 
resonance with mass and width  $M=478^{+24}_{-23}\pm 17$MeV,
$\Gamma=324^{+42}_{-40}\pm 21$MeV, with a corresponding pole position
of $M=489$MeV, $\Gamma=346$MeV. As is mentioned in the following section,
BES data also show a strong broad scalar
$\pi\pi$ resonance in $J/\psi$ and $\psi(2S)$ decays.  

Furthermore, it is found that the broad
scalar resonance accounts for approximately half of all decays. 
However, the 
results from the BES collaboration~\cite{bessigma1} and the E791
collaboration~\cite{Aitala} look somewhat different, reflecting the
uncertainties in parameterization of a light and broad resonance.

\section{Beyond the Breit--Wigner description of a broad resonance}

In production experiments, different processes have
shown evidence for the $\sigma$ meson. However, since the $\sigma$
pole has a large width, it is difficult to separate it unambiguously
from background contributions. It has been ponted out
that in phase shift analyses, crossing symmetry plays a very important 
role and, thus, is useful for determining the pole
location~\cite{ZhouJHEP}. The mass and width are found to be
$m_\sigma=470\pm 50$MeV and $\Gamma_\sigma=570\pm 50$MeV,
respectively, which is in good agreement with the Roy equation
determination of the $\sigma$ pole: $m_\sigma=441^{+16}_{-8}$MeV and
$\Gamma_\sigma=544^{+18}_{-25}$MeV~\cite{CCL}.
The evidence for such a light and broad resonance requires careful
attention to the issue on how to parameterize it.

\subsection{How to describe a light and broad resonance}\label{param}

For a resonance as broad as the $\sigma$ meson, its lineshape's
peak mass value can be drastically different from its pole mass
location~\cite{xiaoly06}. The simple form for a
resonance propagator is
 \begineqn\label{sigmapropagator}
  \triangle(s)=\frac{1}{s-M^2+iM\Gamma(s)}\ ,
   \end{equation}
where $\Gamma(s)$ is the momentum dependent width, which is proportional
to the square of the coupling constant, $g^2$ and also proportional
to the phase space factor, $\rho$. In Eq.~(\ref{sigmapropagator}),
$M$ is a mass parameter (the lineshape mass), when $\sqrt{s}=M$ the
real part of the propagator vanishes. The pole mass of the
resonance, denoted as $m_{pole}$, however, is defined as the zero of
the denominator of $\triangle(s)$:
 \begineqn
m^2_{pole}-M^2+iM\Gamma(m_{pole}^2)=0\ .
 \end{equation}
 The complex root of this equation is located on the un-physical 
sheet\footnote{The kinematic factor, $\rho$ is
of square root cut structure where the resonance couples to two
massive particles. The un-physical sheet is reached by
 changing the sign of $\rho$.} as
 dictated by the requirement of micro-causality.
 Apparently, $m_{pole}$ is usually complex. It may be further written as
 $m_{pole}\equiv m-i\Gamma/2$. When the coupling constant $g$ 
is large, $m$ and $\Gamma$ can be totally different from $M$ and 
$\Gamma(M^2)$.   Taking the
$\sigma$ resonance as an example, one estimate gives $m_{pole}\simeq
450$MeV, but with a lineshape mass as large as  1~GeV. The latter in
general is not a good definition of a broad resonance, because when
the coupling strength becomes strong, the real part of the
self-energy may contain a non-negligible $s$
dependence~\cite{tornqvist}. One of the frequently used simple
parametrization forms for the $\sigma$ meson is
 \begineqn\label{pro2}
 \triangle(s)=\frac{1}{s-M^2+i\rho(s)G}\ ,
  \end{equation}
  where $G$ is a constant to be fixed by the fit.  However, as
  pointed out in Ref.~\cite{Zhengkappa},  Eq.~(\ref{pro2}) does not
  describe a single pole.  Instead, it contains in addition a virtual
  pole for equal-mass scattering, or an additional resonance pole for
  unequal-mass scattering. These additional poles are located 
  in the small $|s|$ region and should not, in fact, exist if we 
  trust chiral perturbation theory predictions for small $|s|$.
  The spurious pole(s) hidden in Eq.~(\ref{pro2}) can be
  simply subtracted out in which case Eq.~(\ref{pro2}) can be
  rewritten as
  \begineqn\label{pro2'}
\triangle(s)=\frac{1}{s-M^2+i\rho(s)s G}\ .
  \end{equation}
  This equation, called the "Red Dragon" by Minkowski and 
  Ochs~\cite{MO99}, 
  depicts very well the smooth but
  steady rise of the low energy $\pi\pi$ phase shift in the $IJ=00$ 
  channel.

  A parametrization frequently found in the literature introduces an 
  exponential form-factor
  aimed at suppressing the resonance contribution at large momentum 
  separations
  (see, for example, Ref.~\cite{zoubugg,tornqvist}). This has the form,
 \begineqn\label{tornq}
 \Gamma(s)=\alpha(2s-m_\pi^2)\frac{\sqrt{s-4m_\pi^2}}{\sqrt{s}}e^{-(s-M_r^2)/4\beta^2}\
 ,
 \end{equation}
 where $\alpha$ and $\beta$ are free parameters and $M_r$ is the resonance
 mass. Other parametrization forms can also be found
  in the literature~\cite{arantes}. The
BES experiment~\cite{bessigma1} tested both Eq.~(\ref{pro2'}) and
Eq.~(\ref{tornq}). Despite the very different parameterizations,
they were not able, at their level of experimental precision, 
to distinguish between them based on the quality of fit. On
the other hand, there exist ambiguities associated with the
description of a broad resonance. There are only a few 
general rules that govern the form of the propagator: first,
chiral suppression of the $\sigma$ coupling to pion fields at low
energies has to be taken into account; second, spurious
singularities hidden in the propagator, if any, have to
lie very far from the physical region where they can be
considered as purely background contributions; third, the
propagator must obey  real analyticity. The latter means
$\Delta(s-i\epsilon)=\Delta(s+i\epsilon)^*$. Aside from these 
basic rules, it is not
known how to separate clearly background contributions from a
broad pole contribution. There  exist some freedom to absorb part
of the background contributions into the $\sigma$ propagator and
vice versa. Hence the differences between two parameterizations only
reflect the different definition of background contributions.
Nevertheless, the $\sigma$ pole location itself is a physical
quantity and should not, in principle, depend on the 
parameterization form that is used.
\subsection{The $\sigma$ pole in the BESII experiment}
In an analysis of the $J/\psi\to\omega\pi^+\pi^-$~\cite{bessigma1} 
process, 
the BESII collaboration found strong evidence for the existence of 
the $\sigma$, as shown in Figs.~\ref{bessigma11} and ~\ref{bessigma12}. 
The pole mass and width are found to be
$m_\sigma=541\pm 39$MeV and $\Gamma_\sigma=504\pm 84$MeV, respectively. 
Different
parameterizations were tested in the analysis. Nevertheless, within
the current experimental precision and the lack of the knowledge of
the production vertex, it is difficult to distinguish between
different parametrization forms.

\begin{figure}[htpb]
\centerline{\psfig{file=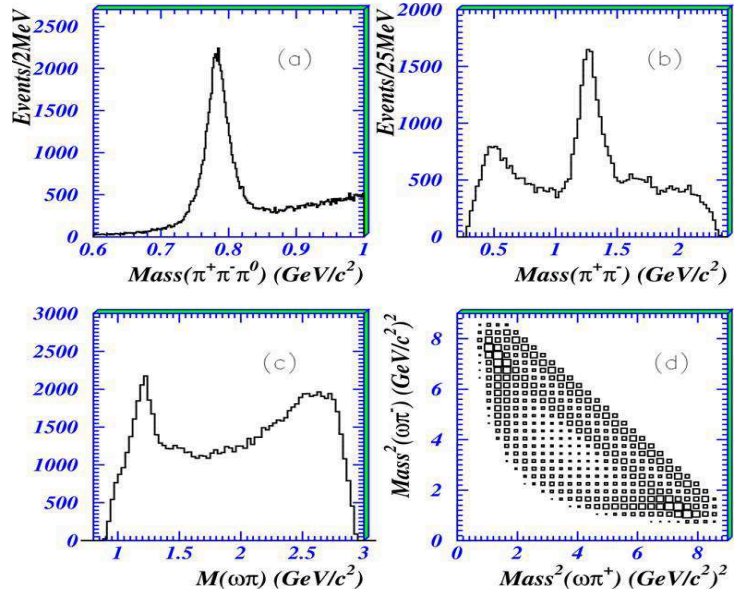,width=10.cm,height=8.cm}}
\caption{(a.) Distribution of $\pi ^+\pi ^- \pi ^0$ mass.
  (b.) Distribution of the $\pi^+ \pi^-$ invariant mass recoiling
  against the $\omega$. (c.) Distribution of $\omega \pi$ invariant mass.
  (d.) Dalitz plot.}
\label{bessigma11}
\end{figure}%

\begin{figure}[htpb]
\centerline{\psfig{file=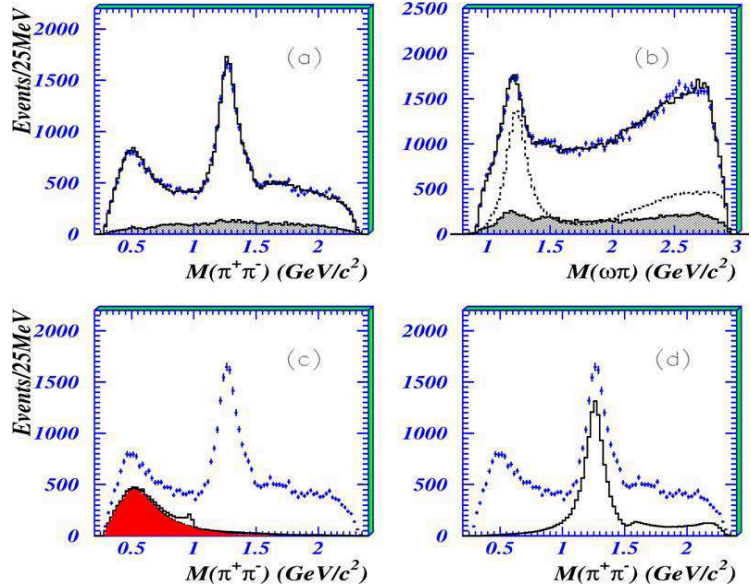,width=10.cm,height=8.cm}}
\caption{ Mass projections of data compared with the fit (histograms); 
          the shaded region
            shows background estimated from sidebands.
            (a) and (b): $\pi \pi$ and $\omega \pi$ mass; the
            dashed curve in (b) shows the fitted $b_1(1235)$ signal
            (two charge combinations).
            (c) and (d): mass projections of $0^{++}$ and $2^{++}$
            contributions to $\pi^+ \pi^-$ from the fit;
            in (c), the shaded area shows the $\sigma$ contribution
            alone, and the full histogram shows the coherent sum
            of $\sigma$ and $f_0(980)$.}
\label{bessigma12}
\end{figure}%

In an early study of the $\psi'\to J/\psi\,\pi^+\pi^-$ process by the BES
collaboration, it is found that the data can
be well described without assuming the existence 
of the $\sigma$~\cite{besppjpsi}.
However, the $\sigma$ pole, if it exists in $J/\psi\to \omega\pi\pi$
process, has to be present in $\psi'\to J/\psi\,\pi^+\pi^-$ as well.
In Ref.~\cite{bessigma2}, this process is re-investigated and
the data are described by the cancelation
between the  sigma pole and a negative background, in
agreement with the picture given in Section~\ref{SLsigmaM}. More
importantly, the sigma pole location, although with large
uncertainties, is found to be stable and consistent with the result
given in Ref.~\cite{bessigma1}, in agreement with the concept of pole
universality discussed below. 
Figure~\ref{figLiGang} shows the experimental results.
\begin{figure*}
\centerline{\psfig{file=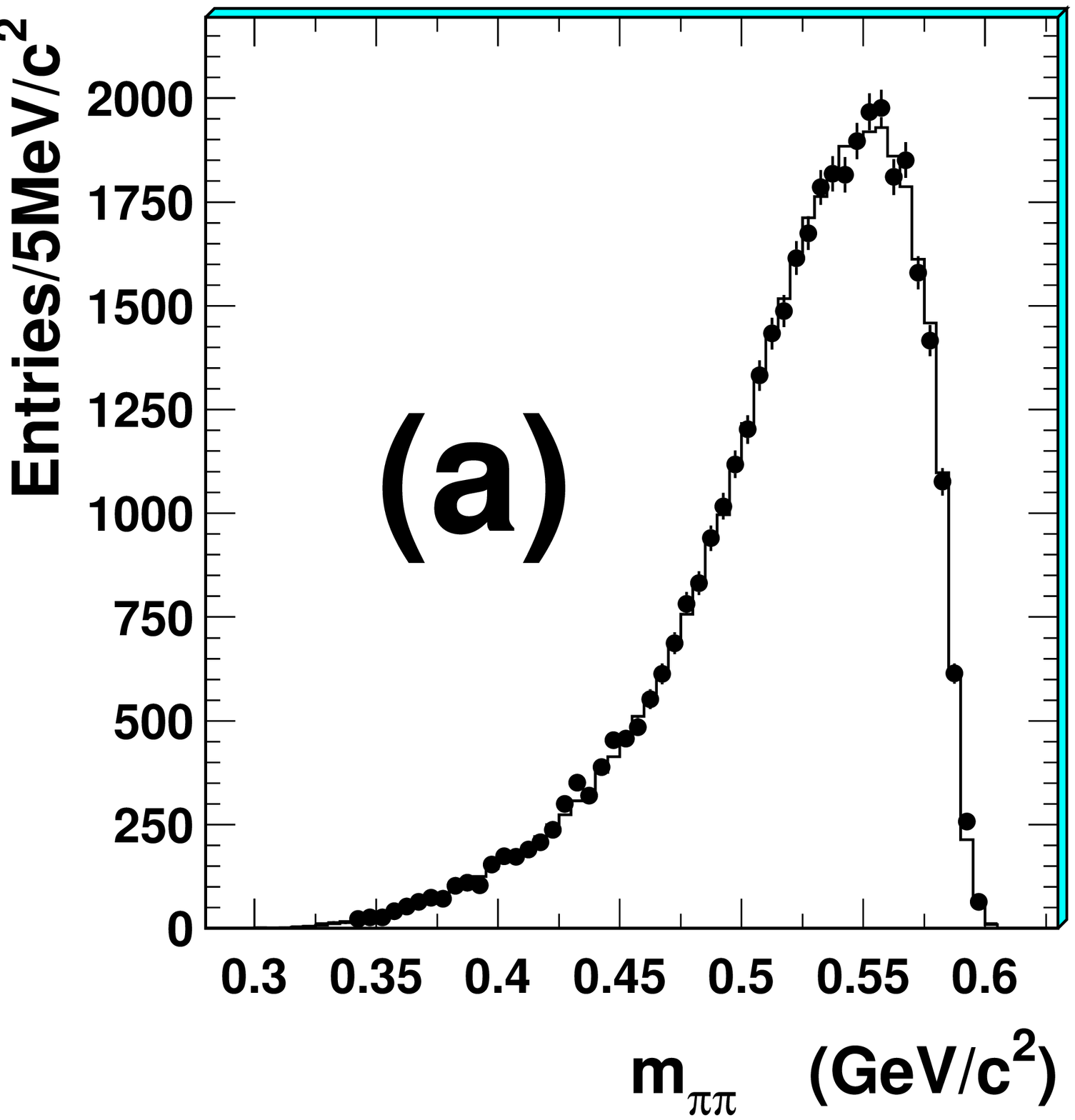, width=5.cm,height=4.5cm}
\psfig{file=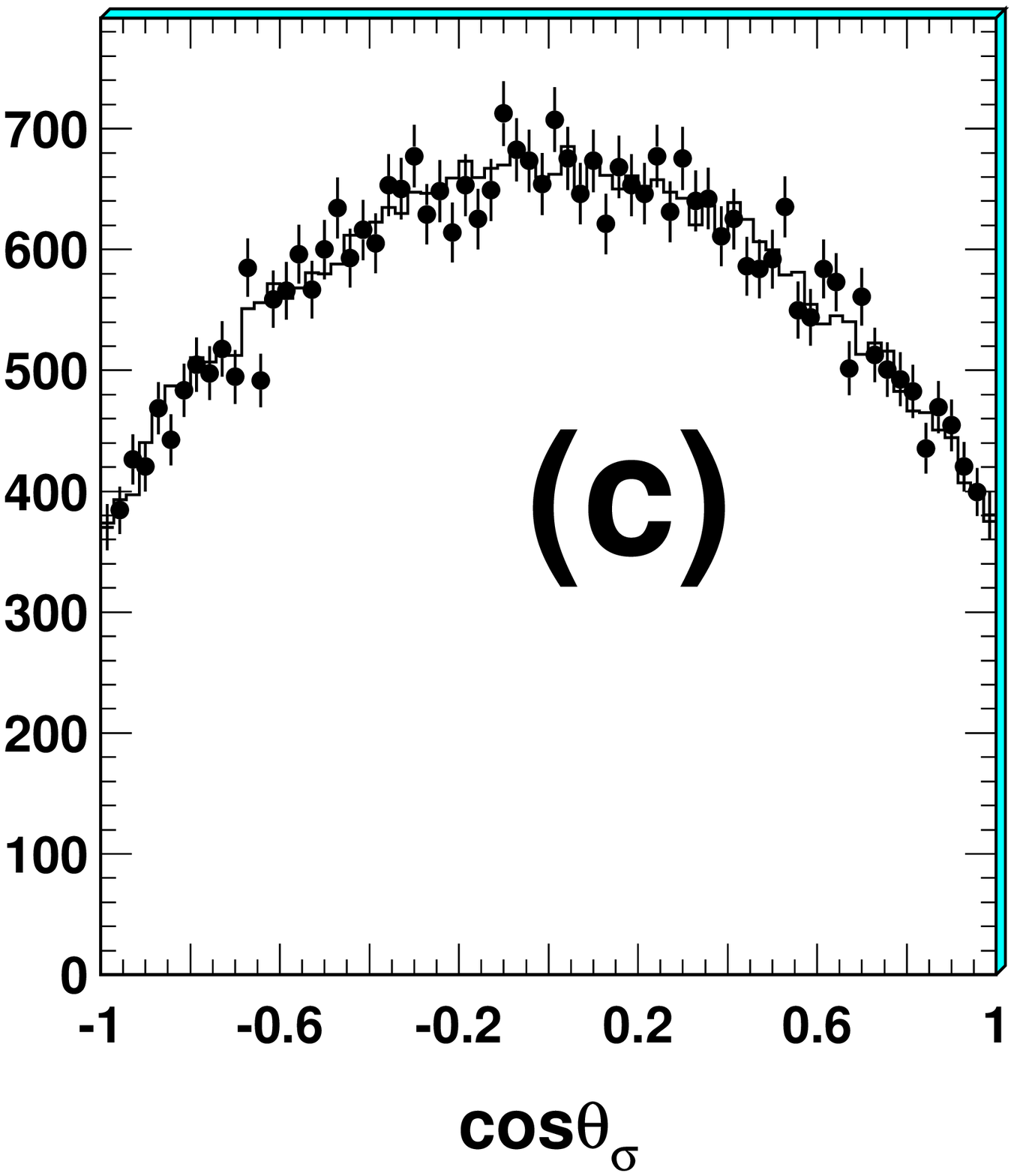, width=5cm,height=4.5cm}
\psfig{file=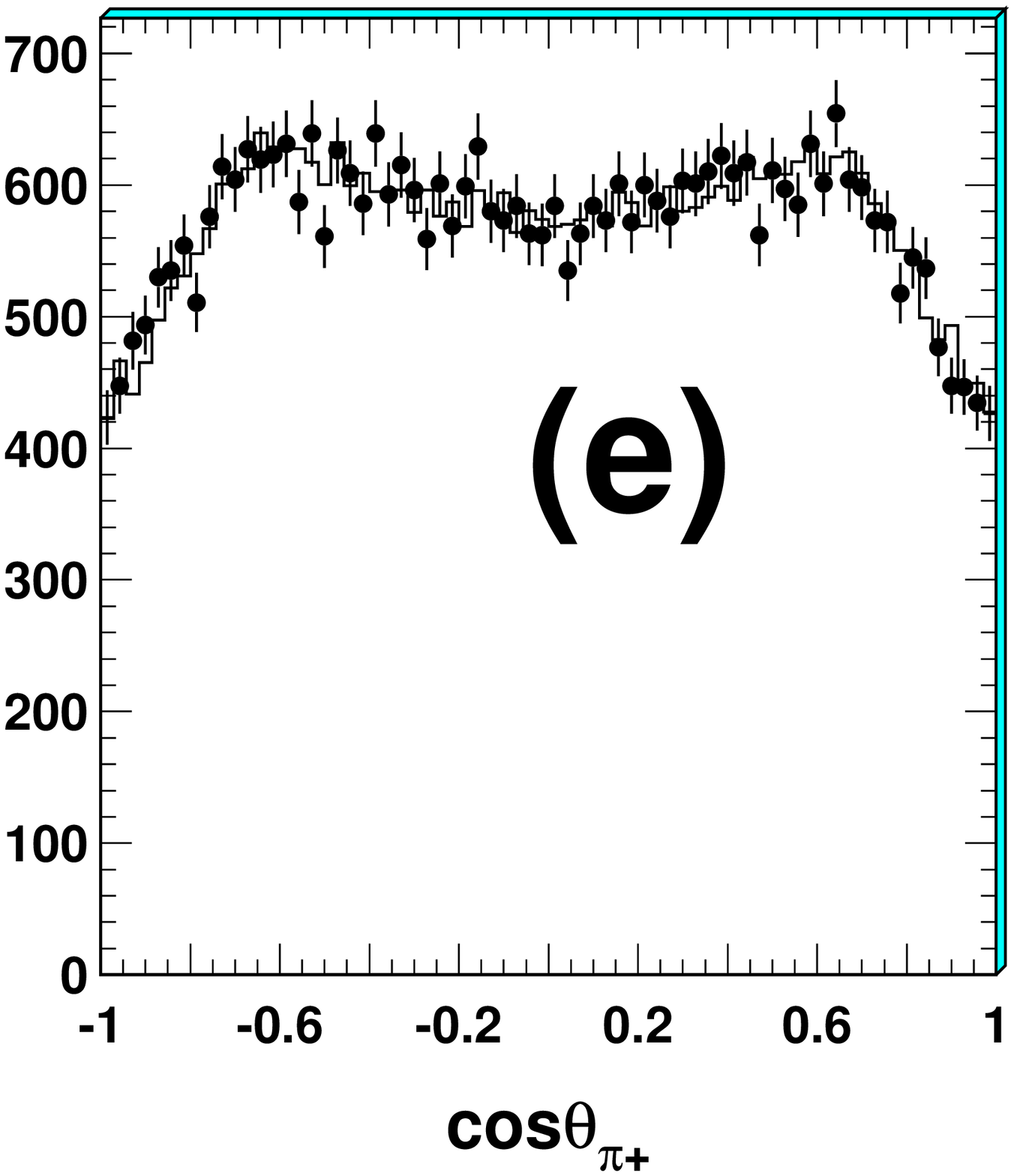, width=5cm,height=4.5cm}} \centerline{
\psfig{file=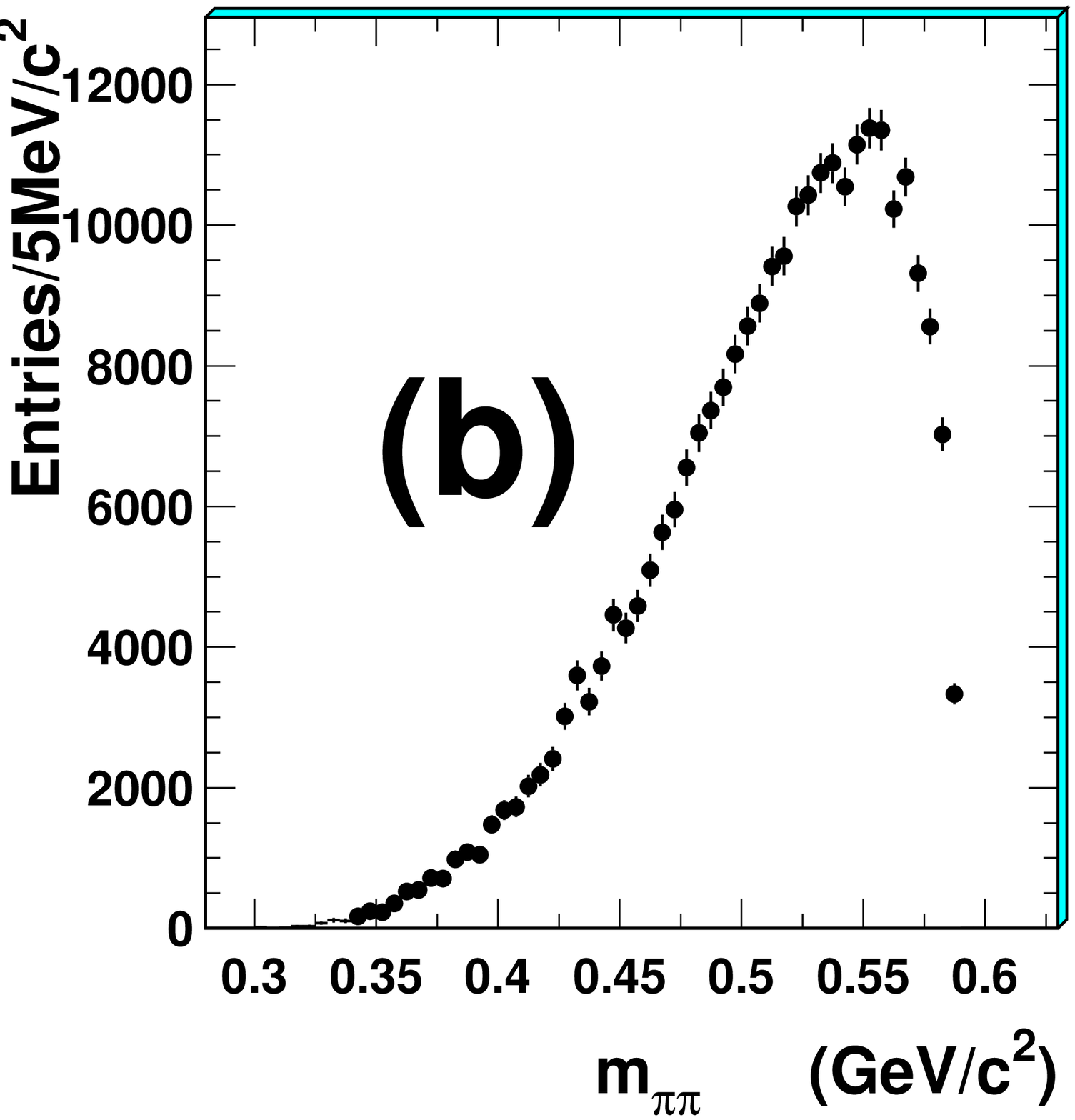, width=5cm,height=4.5cm}
\psfig{file=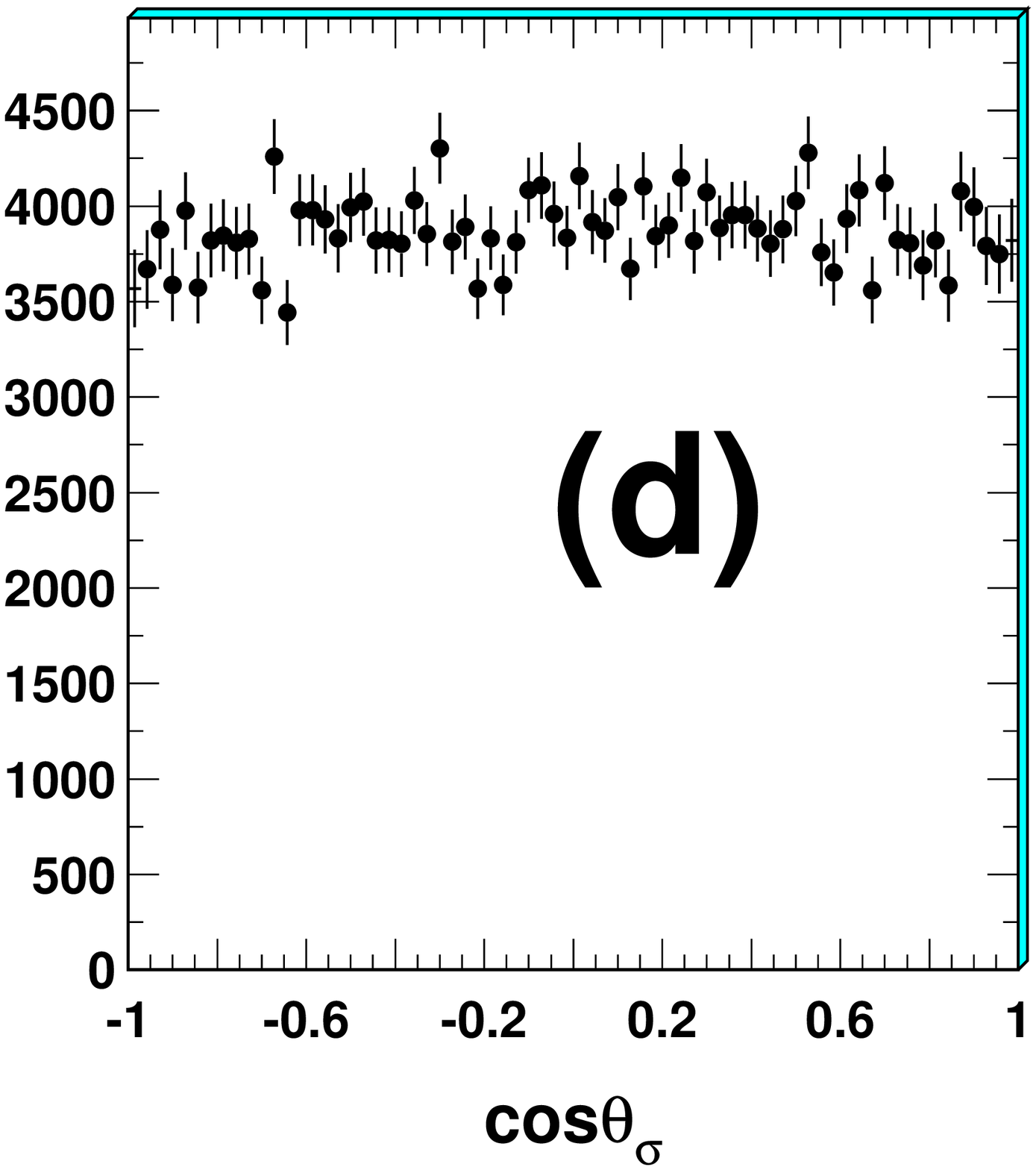, width=5cm,height=4.5cm}
\psfig{file=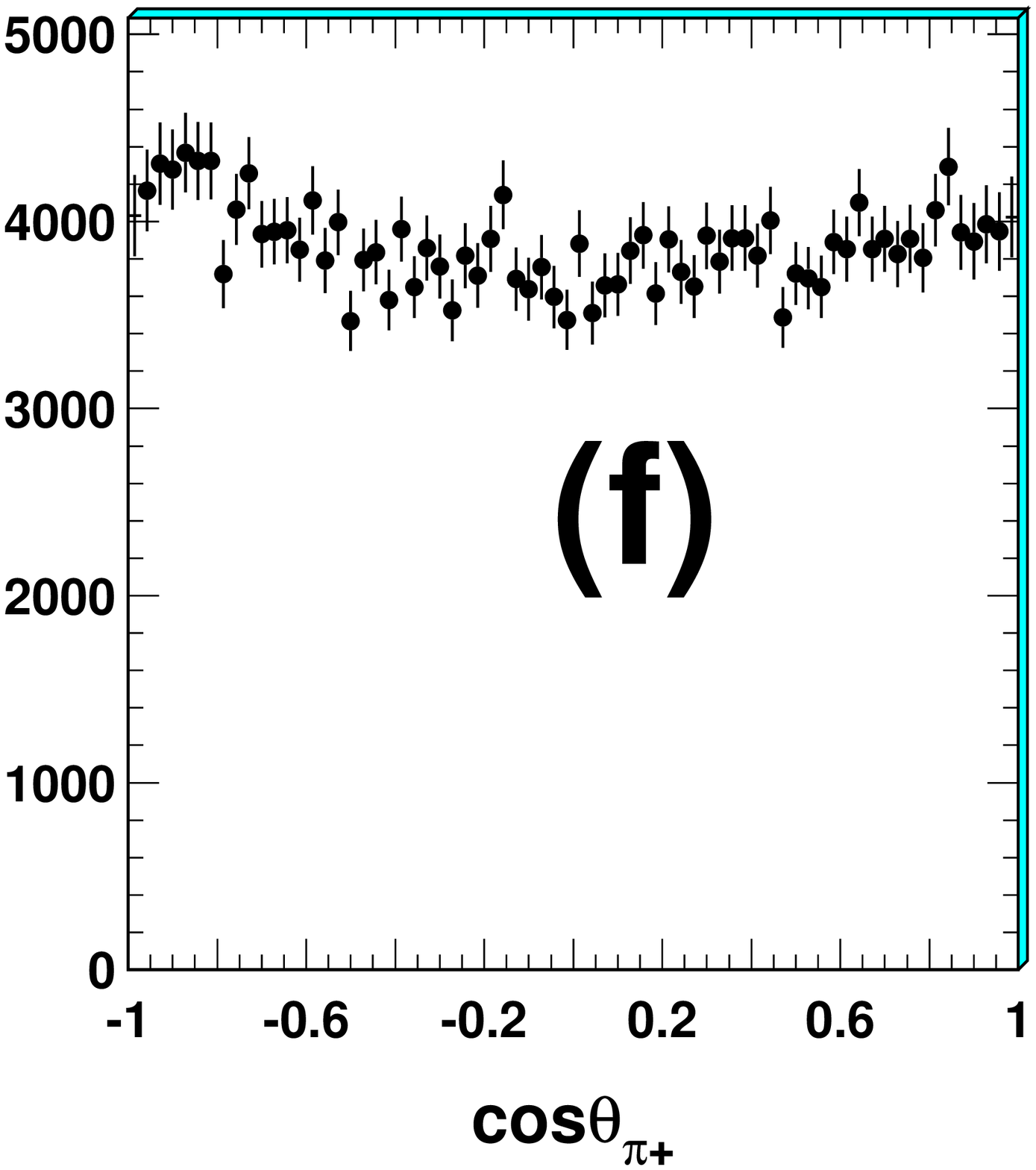, width=5cm,height=4.5cm}}
\caption{Fit results of $\psi(2S)\rightarrow\pi^+\pi^-J/\psi$~(P.K.U.
ansatz). Dots with error bars are data and the histograms are the
fit results.
 (a) and (b) are the  $\pi^+\pi^-$ invariant mass,
 (c) and (d) the cosine of the $\sigma$ polar angle in the lab frame,
 and (e) and (f) the cosine of the $\pi^+$ polar angle in the $\sigma$
 rest frame. (From Ref.~\cite{bessigma2})
} \label{figLiGang}
\end{figure*}


It is generally expected that for a
certain resonance, its pole location should be unique, no matter what
process it appears in. 
Although resonance parameters, such as mass and width, obtained using 
different parametrizations as described in Sec. \ref{param} can be
very different, their pole positions are the same. This is demonstrated 
by experimental results obtained by the 
BES experiment~\cite{bessigma1,bessigma2,beskappa}.

\section{Final state interaction theorem and Omn\'es solution}

\subsection{  The final state interaction  theorem}

Assuming there exist a group of eigenstates of the strong interaction
and total angular momentum $J$ ($J$ is a good quantum number), ¡¡$|1>,
|2>, ..., |n>$, the strong interaction $S$ matrix has the diagonal
form $S_{0}=diag( e^{2i\delta_1},\cdots,e^{2i\delta_n} )$. After
introducing the ``weak'' interaction, the $S$ matrix elements are
no longer diagonal. The non-diagonal elements are assumed weak,
$i\Sigma\propto O(\epsilon)$. In the zero-th order approximation,
$S=S_0$. In the first-order approximation
 \begineqn
  S=S_0+i\Sigma \ .
 \end{equation}
Unitarity requires $SS^+=S^+S=1$, or
 \begineqn 1\simeq S_0^+S_0+i(\Sigma
S_0^+-S_0\Sigma^+)+O(\epsilon^2)\ .
 \end{equation}
This leads to $\Sigma S_0^+=S_0\Sigma^+$.
 From time reversal invariance
 $\Sigma_{mn}=\Sigma_{nm}$ we get
 \begineqn
\Sigma_{mn}=|\Sigma_{mn}|e^{i(\delta_n+\delta_m)}\ .
 \end{equation}
For a ``weak" two body final state amplitude,¡¡we have $A_{i\to
f}=|A_{i\to f}|e^{i\delta_f}$ where $\delta_f$ is the phase shift of
the two body final-state interaction. This is called Watson's
theorem for final state interactions~\cite{watson}. Examples are
$\gamma\gamma\to 2\pi$, $K_s\to 2\pi$, $\Sigma^-\to n\pi^-$ and
$\Lambda^0\to \pi N$, etc..

\subsection{$D^+\to K^-\pi^+l\nu_l$ decays}
 An important process for future \bes3 experimentation, for which the 
final state interaction
theorem applies, is $D^+\to K^-\pi^+l\nu_l$.  Since the 
leptons are spectators to the strong interactions, the strong phase 
is generated from $\pi K$ rescattering, when the invariant mass of the 
$\pi K$  system is less than the $\pi\eta'$ threshold, are exactly the 
same as the one that appears in $\pi K$ elastic scattering, according to 
the final state interation theorem.  The $D^+\to K^-\pi^+l \nu_l$ process 
should be $P$-wave
dominant (via $D^+\to \bar K^{*0} l \nu_l$), but the FOCUS
Collaboration found evidence for a small, even-spin $K\pi$ amplitude
that interferes with the dominant $\bar K^{*0}$ component
~\cite{Focus1}.  The data can be described by $\bar K^{*0}$
interference with either a constant amplitude or a broad spin 0
resonance. Thus, a careful reanalysis at \bes3 that extracts the $S$-wave
component and the phase is  very important for further investigation
of the $\kappa$ pole problem. The final state interaction
theorem is also confirmed
in Ref.~\cite{Focus2}. Edera and Pennington give a review on the
related experimental analyses in Ref.~\cite{pennington2}.

\subsection{$D^+\to K^-\pi^+\pi^+$ decays}
In Ref.~\cite{Aitala:2002kr},  an isobar model is used to
parameterize the partial wave amplitude.   In this model, the decay
amplitude ${\cal A}$ is described by a sum of quasi two-body terms
$D\to R+k,~R\to i+j$, in each of the three channels $k=1,2,3$:
 \begineqn\label{eq:isobarmodel}
   {\cal A} =d_0 e^{i\delta_0} +
        \sum_{n=1}^N d_{n} e^{i\delta_{n}}
        ~{F_{\script R}(p,r_{\script R},J)\over
        m_{\script R_n}^2-s_{ij}-im_{\script R_n}\Gamma_{\script R_n}(s_{ij})}
        \times
        F_{\script D}(q,r_{\script D},J)
        ~M_J(p,q)\ ,\nonumber\\
 \end{equation}
where $s_{ij}$ is the squared invariant mass of the $ij$ system, $J$
is the spin, $m_{\script R_n}$ the mass and $\Gamma_{\script R_n}(s_{ij})$
the width of the $n$th resonance,
$F_{\script R}$ and $F_{\script D}$ are form factors, with effective
radius parameters $r_{\script R}$ and $r_{\script D}$, for all $R_n$ and
for the parent $D$ meson, respectively,  $p$ and $q$ are momenta of
$i$ and $k$, respectively, in the $ij$ rest frame, and $M_J(p,q)$ is a
factor introduced to describe spin conservation in the decay. For
more detailed discussions on this expression see
Ref.~\cite{Aitala}. 

The complex coefficients $d_n
e^{i\delta_n}~(n=0,N)$ are determined by the $D$ decay dynamics and
are parameters estimated by a fit to the data. The first, the
non-resonant ($NR$) term describes direct decay to ${i+j+k}$ with no
intermediate resonance, and $d_0$ and $\delta_0$ are assumed to be
independent of $s_{ij}$. For $D^+ \to K^- \pi^{+}_{a}\pi^{+}_{b}$ decays,
${\cal A}$ is Bose-symmetrized with respect to interchange of $\pi^+_a$ and
$\pi^+_b$. In Ref.~\cite{Aitala:2002kr} it is noticed  that the $NR$
term is small, and that a further term, parameterized as a new
$J=0$ resonance $\kappa(800)$, with $m_{\script R}=(797\pm 19\pm 43)$MeV
and $\Gamma_{\script R}=(410\pm 43\pm 87)$MeV, gives a much better
description of the data.

Apparently the above parametrization forms are not satisfactory for
the purpose of searching for broad resonances and for testing
final state phases. In Ref.~\cite{meadows}, an improved approach to
the above analysis is presented that uses a generalized isobar picture
of two body interactions. While higher $K\pi$ waves are described by
sums of known resonances, the $s$-wave amplitude and phase are
determined bin-by-bin in $K\pi$ mass. The phase variation is found
to be not that of the $K^-\pi^+$ elastic scattering obtained by the LASS
Collaboration~\cite{LASS}. The applicability of the Watson theorem in
three-body hadronic decays is examined in Ref.~\cite{meadows}. In 
three-body $D^+\to K^-\pi^+\pi^+$ decays, the 
$K^-\pi^+$ pairs form both isospin 1/2 and 3/2
components. It is not clear, however, how to estimate the $I={3\over
2}$ $S$-wave component. In the literature it is often assumed, based
on a simple spectator-quark model for $D$ decay to $K\pi\pi$,
that the $K\pi$ system has only $I=1/2$. However, it is found that
if $I=\half$ dominates, the Watson theorem does not describe
these data very well.  This question is re-examined in
Ref.~\cite{pennington2}. It is suggested that in $D\to K\pi\pi$
decays there exists a different mixture of $I=1/2$ and $I=3/2$
$S$-wave interactions than in elastic scattering. Applying Watson's
theorem to this generalized isobar model allows one to estimate the
$I=3/2$ $K\pi$ $S$-wave component, and it is found to be larger than
in either hadronic scattering or semi-leptonic processes.

Another very interesting process for future \bes3 experimentation is
$\psi'\to J/\psi \pi\pi$. This is a three body final state. But due 
to color transparency arguments, one can, to a very good approximation,
neglect the final state interactions between the $J/\psi$ and the pions.
In this case, the problem is reduced to a two-body final state. The
kinematics constrains the $\pi\pi$ system to be solely elastic and,
thus, the $\psi'\to J/\psi\, \pi\pi$ channel provides another good
opportunity to extract the final state phase of $S$-wave $\pi\pi$
interactions. However, in this channel the phase is obtained by
interference between $S$- and $D$-waves and the latter is very small.
So, high statistics are needed, which will be available
at \bes3. The measured phase shift difference
$\delta_0^0-\delta_0^2$ will be important complementary information
to other determinations, since it covers the energy region around 
500~MeV, which is not covered by previous experiments.

Precision experiments also require more  theoretical investigations 
into three-body final state interactions, such as, for example, 
in the $D\to K\pi\pi$ process. Corrections to the Watson theorem 
for three-body decays are
discussed theoretically in Ref.~\cite{caprini}.

\subsection{More on final state interactions in the $J/\psi\to \omega\pi\pi$ process}
The final state phases in $J/\psi\to \omega\pi\pi$ process is also
discussed in Ref.~\cite{buggA0}. The final state is three body and
a complicated pattern of final state interactions is involved.
Nevertheless the $\omega\pi$ two-body interaction may be subtracted
using a suitable parametrization 
({\it i.e.}, the generalized isobar model). In
the low $\pi\pi$ invariant mass region, the $\pi\pi\to 4\pi$ final
state interaction is negligible. Hence the remaining part of the
decay amplitude characterizing the $\pi\pi$ final state
interactions, denoted as $A_0$, may be related to the Watson theorem
by fixing the final state phase: \begineqn
A_0(s)=|A_0(s)|e^{i\delta_\pi}\ ,\end{equation}
 which is similar to the scalar
form factor \begineqn F_0(s)=|F_0(s)|e^{i\delta_\pi} \ .\end{equation} That means the
ratio $R(s)=A_0(s)/F_0(s)$ is real when $s<4M_K^2$. Furthermore,
since the $\sigma$ pole both in $A_0$ and $F_0$ cancel each other
and the cut in $R$ is distant (starting from $4M_K^2$), $R$ has to
be a slowly varying function, at least at low energies ({\it i.e.}, when
$s<<4M_K^2$). This can be seen immediately if one writes down a
dispersion relation of $R$ with one subtraction,
 \begineqn\label{Rintegral}
  R(s)=R_0+R_1s+\frac{(s-4M_K^2)^2}{\pi}\int_{4M_K^2}\frac{\mathrm{Im}R(t)}{(t-4M_K^2)^2(t-s)}\
  .
 \end{equation}
The scalar form factor can be determined from theory~\cite{anan}.  The
 ratio $R$ is plotted in Fig.~\ref{figbugg} where one
sees that $R$ does not show any curvature-like structure, hence the
integral in Eq.~(\ref{Rintegral}) is small and $R\simeq R_0+R_1s$.
Furthermore, the scalar amplitude may be parameterized as
$A_0(s)=R_0(1-s/s_0)F_0(s)$ and $s_0\simeq
1.65$GeV$^2$~\cite{leutwylerAcores}. On the other hand, $J/\psi$
decays to $\omega\pi\pi$, $\omega K\bar K$, $\phi\pi\pi$ and $\phi
K\bar K$ are studied in Ref.~\cite{meissner06}, using the assumption
that $A_0/F_0\simeq const$. The observed energy dependence, shown
in Fig.~\ref{figbugg}, seems to suggest that the final-state
interaction between the $\omega$ and $\pi$ and also the three-body final 
state interaction can be safely removed. This conclusion seems to disagree
with the conclusion reported by E791 based on
their analysis of $D\to K\pi\pi$ decays (bear
in mind that the processes are different). Hence it would be
interesting for future \bes3 experiments and related
theoretical studies to improve the understanding of 
three body final state interactions.
\begin{figure}[hbt]
 \centerline{%
 \epsfig{file=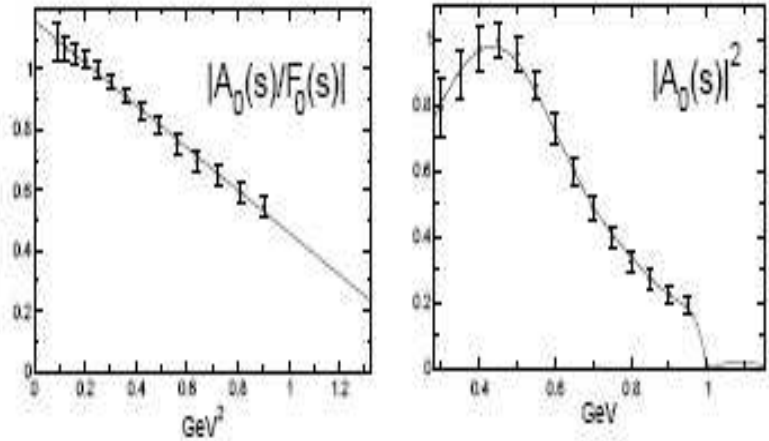,
         width=0.45\textheight,
         height=0.3\textheight,angle=0}}
 \caption{%
 \label{figbugg}The comparison between the scalar form factor and
 the scalar amplitude extracted from BES $J/\Psi\to \omega\pi\pi$
 process. Data are from Ref.~\cite{buggA0}.}
\end{figure}


\subsection{The Omn\'es solution}

Consider a pion form factor, or a form-factor-like quantity, $A(s)$,
which is an analytic function on the entire cut $s$ plane. Neglecting
inelasticity effects and assuming that the single channel unitarity
relation holds over the entire physical region: $4m_\pi^2<s<\infty$,
the spectral function of the form-factor satisfies
 \begineqn\label{ImA}
  {\rm Im}A=\rho AT^*\ ,
 \end{equation}
where $T$ is the $\pi\pi$ (partial wave) scattering amplitude. 
Equation~(\ref{ImA}) has a simple solution, called the Omn\'es solution:

  \begineqn\label{omnes}
A(s+i\epsilon)=P_n(s)\exp\{{s\over\pi}\int^\infty_{4m_\pi^2}\frac{\delta_\pi(s')ds'}{s'(s'-s-i\epsilon)}\}\
,
 \end{equation}
where $P_n(s)$ is a (real analytic) polynomial, representing the
possible zeros of $A(s)$ on the complex $s$ plane, and $\delta_\pi$
is the $\pi\pi$ scattering phase shift.

\section{On the nature of the lightest scalar resonances}
Issues related to the
lightest scalar resonances, $\sigma$, $\kappa$, $f_0(980)$ and
$a_0(980)$ have attracted much interest.  Important among these
is the $\kappa$, the other ({\it i.e.} in addition to the $\sigma$) 
broad resonance among these states.

The E791 experiment
at Fermilab reported the evidence for the $\kappa$ in the $D^+ \to K^+ 
\pi^+ \pi^-$ \cite{E791kappa}, with a mass and width of 
$797 \pm 19 \pm 43$ MeV/c$^2$ and $410 \pm 43 \pm 87$ MeV/c$^2$.
The $\bar{K}^*(892)^0K^+\pi^-$ channel in
$J/\psi\to K^+K^-\pi^+\pi^-$ was studied at BES~\cite{beskappa}
and a clear low mass enhancement in the
invariant  mass spectrum of $K^+\pi^-$ is observed.
Two independent partial
wave analyses were performed and different parametrizations of
$\kappa$ tried.  Both analyses favor the interpretation
of the low mass enhancement as the $\kappa$.
The average mass and width of the $\kappa$ in the two analyses are
878 $\pm$  23$^{+64}_{-55}$  MeV/$c^2$ and
499 $\pm$ 52$^{+55}_{-87}$  MeV/$c^2$, respectively,
corresponding to a pole at
$ ( 841 \pm 30^{+81}_{-73} ) - i( 309 \pm 45^{+48}_{-72} ) $ MeV/$c^2$.
Figure \ref{beskappa} shows the $K \pi$ invariant mass
spectrum that recoils against a $K^*(892)$. The crosses are data and
histograms represent the PWA fit projection. The shaded area shows 
the $\kappa$ contribution.

\begin{figure}[htbp]
\centerline{\psfig{file=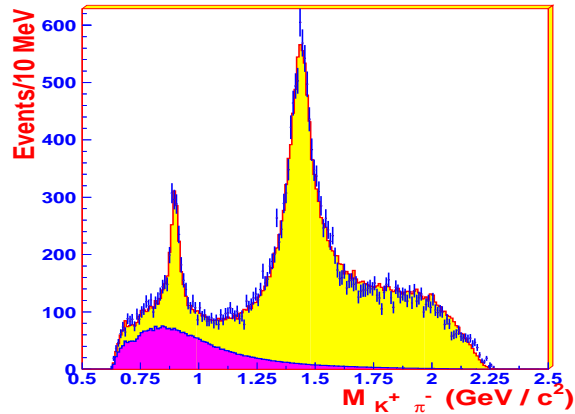,width=8.cm,height=6.cm}}
  \caption{\label{beskappa}
    The $K\pi$ invariant mass recoiling against a $K^*$.
    The crosses are data and histograms represent the PWA fit projection.
    The shaded area shows the $\kappa$ contribution.}
\end{figure}

These results should be compared with a recent
phase shift analysis based on LASS~\cite{LASS} data:
$M_\kappa=694\pm 53$MeV, $\Gamma_\kappa=606\pm
59$MeV~\cite{Zhoukappa}, and the Roy--Steiner equation
analysis~\cite{Moussallam}: $M_\kappa=658\pm 13$MeV,
$\Gamma_\kappa=557\pm 24$MeV.

 An immediate question of interest
concerning these scalars is whether or not they comprise the lowest
scalar octet, and, if the answer is yes, how to explain the
large difference between their masses and widths. A widely accepted
understanding to such question has not yet emerged, but the key
must be related to a proper treatment to the strong interaction
dynamics especially with respect to unitarity. Large $SU_f(3)$-breaking
effects must also be taken into account. First of all,
 it should be stressed that a naive quark model description of the
 lightest scalar meson cannot be successful. As emphasized by
 T\"ornqvist~\cite{Tornqvist99}, this is simply because in naive 
quark models, chiral symmetry, which is crucial for scalars, 
is absent. Indeed,
 if the `$\sigma$' resonance really has something to do with the
$\sigma$ particle in the linear $\sigma$ model
Eq.~(\ref{Lsigmash}), then from (Eq.~\ref{chiralrotation}) it is
realized that the $\sigma$ is the chiral partner of the pion field.
The latter is a collective excitation -- that would be a Goldstone
boson. Hence the $\sigma$ must share many properties of the pion
field, and it is well known that it is very difficult to understand the 
pion field in terms of simple quark models.

The $\kappa$ meson contains an $s$-quark component,
 while the $a_0$ does not. In
the naive $q\bar{q}$ picture this would indicate a heavier $\kappa$
than $a_0$. In a phenomenological approach, Jaffe relates the
unusual properties of the light scalar mesons to the assumption that
the lightest scalar mesons are mainly composed of $(qq)(\bar q\bar
q)$~\cite{Jaffescalar}. With this, he observes that the spectrum of
the flavor nonet is inverted as compared to a standard $q\bar q$
nonet, and contains a light isospin singlet, a strange doublet, and
a heavy triplet plus singlet with hidden strangeness. This picture
agrees well with the observed spectrum of a light $\sigma$, a
strange $\kappa$, and the heavier $a_0(980)$ and $f_0(980)$. It also
explains why the $a_0(980)$ and $f_0(980)$ strongly  couple to
$K\bar K$ and $\pi\eta$. However, the large width of the $\sigma$
and $\kappa$ are still waiting for an explanation in this approach.
On the other hand, the mass splitting between the $\sigma$ and
$a_0(980)$ may also be explained by instanton effects~\cite{'t
Hooft} and are  investigated in the literature~\cite{instanton}.

The $\gamma\gamma\to \pi^+\pi^-,\,\,\pi^0\pi^0$ process has been
studied in Ref.~\cite{pennington3} using a dispersive calculation
and the decay width $\Gamma(\sigma\to 2\gamma)$ is extracted. The
result is compared to quark model calculations with different
assumptions about the content of the sigma meson and it indicates that the
extracted $\sigma$ coupling to two photons is very different from
that for gluonium or even tetraquark descriptions to the $\sigma$
resonance.

In Ref.~\cite{narison06},  QCD spectral sum rules (QSSR)
calculations based on different proposals ($q \bar q $, $q q
\bar q\bar q$ and gluonium) are performed and discussed. It is found 
that, in
the I=1 and I=1/2 channel the unusual wrong splitting between the
mass  of $a_0(980)$ and $\kappa$ and also the width of $a_0(980)$
can be understood from QSSR within a $q \bar q$ assignment.
However the large width of $\kappa$ can not be explained either
in $q \bar q$ or $q q \bar q\bar q$ pictures. In the $I=0$ channel, the
important role of QCD trace anomaly was emphasized, and based on this
the observed masses of the $\sigma$ and $f_0(980)$ can  be explained
as a maximal mixing between a low lying gluonium and a conventional
$q \bar q$ state, based on analysis of BES results on 
$J/\psi\to\phi\pi\pi$, $\omega\pi\pi $,   $f_0K^+K^-$ and other
experimental information. Furthermore, with the aid of
experimental information on OZI-violating $J/\psi\to\phi\pi\pi$,
$D_s\to 3\pi$ decays and especially $J/\psi\to \gamma S$, it is found
that most of the vacuum scalars below 2~GeV, other than the $f_0(1370)$, 
contain a large gluonium component.

A lattice study of the vacuum scalars has also been performed, and
it suggests the existence of a low-lying sigma~\cite{kunihiro06}. It
is pointed out that the physics content of the $\sigma$, i.e., a
tetraquark, a hybrid with a glue ball or a $q\bar q$ collective
state, is obscure: the disconnected diagram gives the dominant
contribution~\cite{kunihiro06}. Investigations have also been made
based on Lagrangian models at the meson or quark levels. In
Ref.~\cite{Scalarnonet}, the lightest scalars are depicted as a nonet
with a complicated structure. Near the center they are $(qq)_{\bar
3}(\bar q\bar q)_3$ in an $S$-wave, with some $q \bar q $ in a $P$-wave, 
but
further out they rearrange as $(q\bar q)_{1}(q\bar q)_1$ and finally
as meson-meson states. In Ref.~\cite{beveren06} a review is given
of studies on lightest scalars based on a coupled-channel model
of potential scattering. The potential contains a confining part of
harmonic oscillator type as well as a $^3 P_0$ transition potential
characterizing the phase transition from a $q \bar q $ pair to a
meson pair. In such a model calculation, $T$ matrix zeros are found
to be close to those predicted by current algebra. Reasonable fit
results are found for phase shifts, using only a small set of
parameters and the $\sigma$ and $\kappa$ resonances are predicted
dynamically. Variations of Pad\'e approximations to $\chi$PT
amplitudes are also studied in the literature to explore the  nature
of the lightest scalars that are generated dynamically~\cite{OOP}.

The substructure  of the $f_0(980)$ resonance is also a long
standing issue. Even its small peak width may be explained as the
Flatt\'e effect, which means that although it manifests itself as 
a narrow peak, the
$f_0(980)$ may have a large decay width~\cite{Zou}. One
interpretation of the quark content of the $f_0(980)$ is that it
mainly has an $s \bar s$ component, while another explanation is 
that it is  a $K \bar{K}$ molecule. The latter would require
 \begineqn {\Gamma(\phi\rightarrow f_0
\gamma) \over \Gamma(\phi\rightarrow a_0 \gamma)}\sim 1 \ .
 \end{equation}
However, recent KLOE
results indicate~\cite{Kloe}
 \begineqn {\Gamma(\phi\rightarrow f_0
\gamma) \over \Gamma(\phi\rightarrow a_0 \gamma)}=6.1\pm 0.6\ .
 \end{equation}
  If, on the other hand, it is explained as a $q\bar{q}$ state, it
cannot explain the large branching ratio ($\geq 10^{-4}$). Another
interpretation of the lightest scalar resonances is that they are
$qq\bar{q}\bar{q}$ states. In such a scheme the mass relations are
explained~\cite{maiani} without considering the large widths and
strong interaction corrections. In Ref.~\cite{bugg06} some decay
modes of these resonances are estimated using both $q\bar{q}$ and
$qq\bar{q}\bar{q}$ picture and it was found that  neither picture
can give the correct branching ratio.

To conclude, the nature of the lightest scalars remains
mysterious. Future \bes3 experimentation in this area will be
highly prized since it could provide information helpful
for  understanding strong interaction physics at low energies.

\chapter{Two-photon physics}
\label{sec:two_photon}

\newcommand{\Li}{\mbox{Li}_2}
\newcommand{\vecc}[1]{\mbox{\boldmath $#1$}}
\newcommand{\Ree}{\mbox{Re}}
\newcommand{\dd}{\mbox{d}}

\newcommand\nn{\nonumber}

\def\Li#1#2{{\mathrm{Li}}_{#1}\left(#2\right)}
\def\order#1{{\mathcal O}\left(#1\right)}

\newcommand{\br} [1]{ \left( #1 \right) }
\newcommand{\brs}[1]{ \left[ #1 \right] }
\newcommand{\brf}[1]{ \left\{ #1 \right\} }

\section{Introduction}

Two-photon processes have always held a special fascination for physicists
because of the opportunity they provide to study the conversion of pure 
photons into matter (leptons and hadrons). The
original calculation for these processes 
is reported in Ref.~\cite{part3:ref:Zemach}, and the first detailed 
consideration of possible applications of these process was 
first presented independently by two 
groups~\cite{Balakin:1970jv,Brodsky:1970vk}.

Two-photon collisions can be accessed in electron-positron
or electron-electron colliders via the process
$$e^\pm+e^-\to e^\pm+e^- +\gamma+\gamma\to e^\pm+e^- +X,$$
where $X$ describes any final state. Note that intermediate photons can
be virtual or almost real. In the case of almost real photons,
one has the possibility of studying
processes of type $\gamma\gamma\to X$.
Not all final states can be accessed in two-photon processes;
since each photon has odd $C$ parity, only 
final states $X$ with even $C$ parity are produced.

An interesting feature of this process is that
its cross-section rises slowly with the energy and becomes
comparable with the $e^+e^-$
annihilation channel at $\sqrt{s}/2 \approx 1$ 
GeV,
(see Fig.~\ref{twophoton_fig1})~\cite{BGMS1975}.
\begin{figure}[htbp]
\centerline{\epsfig{file=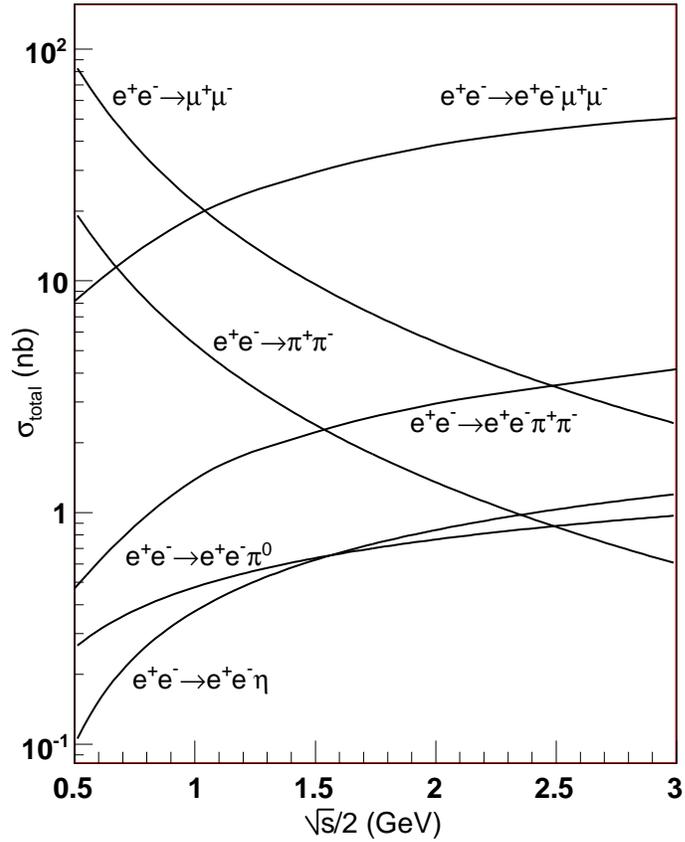,width=10.0cm}}
\caption {Cross sections for two-photon production of lepton and hadron 
pairs.}
\label{twophoton_fig1}
\end{figure}
For example, the cross section for $e^+e^-\to\pi^+ \pi^-$ pair 
in the single-photon annihilation
channel is proportional to $\alpha^2/q^2$, where $q$ is the virtuality
of the intermediate photon.  Athough the two-photon process has an
additional factor of $\alpha^2$ it has a logarithmic reinforcement 
that increases with the energies of initial-state particles~\cite{BG:37}:
$$\sigma_{e^\pm e^-\to e^\pm e^- h}=\frac{\alpha^4}{18\pi^2m_\pi^2}
\ln\frac{sm_\rho^2}{m_e^2m_\pi^2}\ln\frac{sm_\rho^6}{m_e^6m_\pi^2}
\biggr(\ln\frac{s}{m_\pi^2}\biggl)^2 .$$
This is because the two-photon  cross-section is
enhanced when the virtualities of interacting photons
goes to zero (see Fig.~\ref{twophoton_fig2}).
\begin{figure}[htbp]
\centerline{\epsfig{file=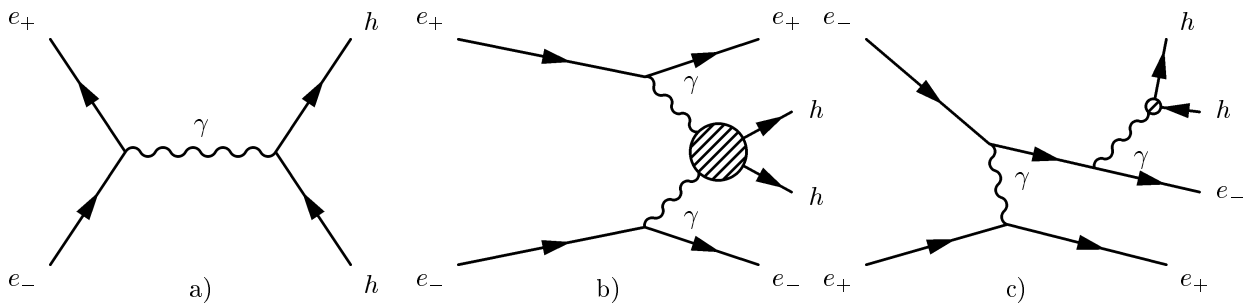}}
\caption {Feynman diagrams for hadron pair production
in $e^+e^-$ collisions. The contribution from 
diagram b) is enhanced by the intermediate photon poles.}
\label{twophoton_fig2}
\end{figure}
In this case, they are almost real and, from kinematic restrictions, are 
emitted
along initial particle directions as illustrated in 
Fig.~\ref{twophoton_fig3}.
\begin{figure}[htbp]
\centerline{\epsfig{file=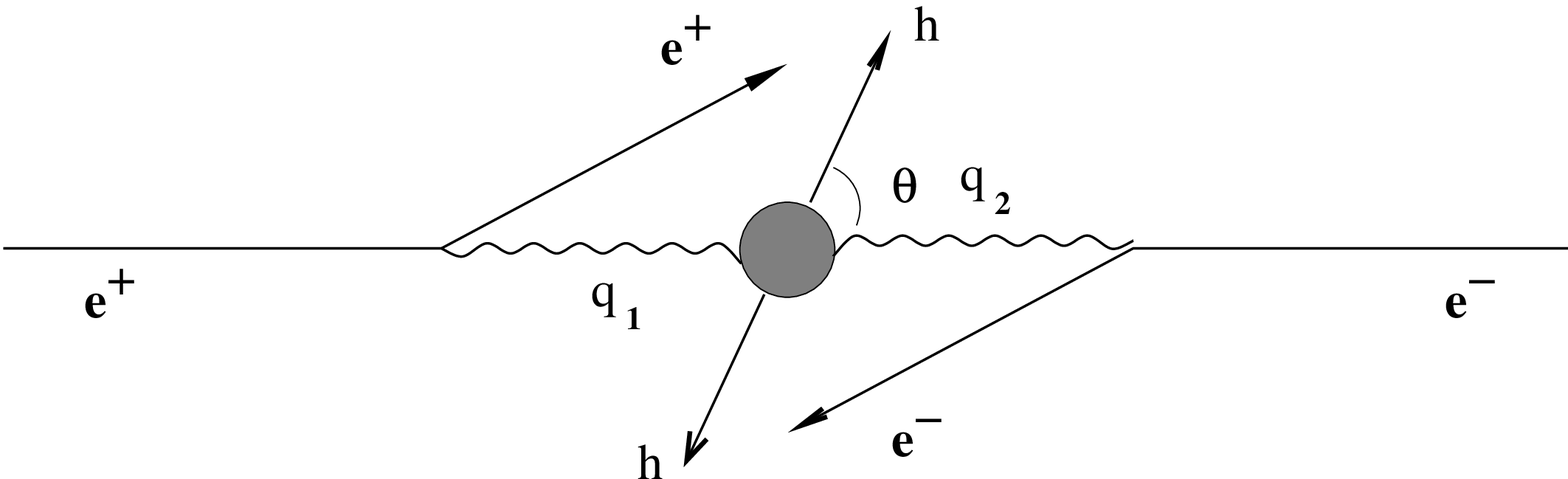,width=12.5cm}}
\caption{Kinematics of the two-photon process.}
\label{twophoton_fig3}
\end{figure}

Keeping this enhancement in mind and taking into account only leading 
terms (the so-called Equivalent 
Photon Approximation (EPA)) the cross section has the form 
\begin{eqnarray}
\frac{\dd\sigma^{(0)}}{\dd s\dd\Gamma}=2\biggl(\frac{\alpha}{\pi}\biggr)^2
\frac{1}{s}\ln^2\frac{E}{m_e}
f(\frac{\sqrt{s}}{2E})\frac{\dd\sigma_{\gamma\gamma\to X(s)}}{\dd\Gamma},
\nonumber \\
f(\gamma)=-(2+\gamma^2)^2\ln\gamma-(1-\gamma^2)(3+\gamma^2),
\nonumber
\end{eqnarray} 
where $E$ is the energy of initial electron.
We can see that  the cross-section for the two-photon process splits
into two different parts: one part is connected with emission of the
two real photons by the incoming beam particles and the other part is
final state production by the two photons. This formula contains the
essence of two-photon physics.

The cost of this
simplicity is the underestimation of the cross-section in some
specific kinematic regions, and loss of  information 
about the scattered particle's
angular dependence,  and deep-virtual scattering, when one of the photons 
has a large $|q^2|$.
On the other hand, events with highly virtual  intermediate photons
are quite rare, since for these there is no $q^2$ pole enhancement.
These are called "single-tag" events
({\it i.e.}, events where one photon has a nonzero virtuality $q^2$)
and are useful for studying resonance production by off-shell photons.

The main research object in two-photon physics is the 
determination of the amplitudes
for $\gamma\gamma\to X$ processes, both for off-mass-shell and for
almost-real photons.  Another way to describe
two-photon cross sections is in terms of five structure functions.
Three of them can represent the cross section
$\sigma_{a,b}$ for  scalar ($a,b=S$) and transverse photons
($a,b=T$). The other structure functions $\tau_{TT}$
and $\tau_{TS}$ correspond to transitions where
each of the photons flip their spin while conserving
the total helicity~\cite{BGMS1975}:
\begin{eqnarray}
\dd\sigma=
\frac{\alpha^2}{16\pi^4q_1^2q_2^2}\sqrt{\frac{(q_1q_2)^2-q_1^2q_2^2}{(p_1p_2)^2-m_1^2m_2^2}}
\biggl(
4\rho^{++}_1\rho^{++}_2\sigma_{TT}
+2|\rho_1^{+-}\rho_2^{+-}|\tau_{TT}\cos(2\tilde{\phi})
\nonumber \\
+2\rho^{++}_1\rho^{00}_2\sigma_{TS}
+2\rho^{00}_1\rho^{++}_2\sigma_{ST}
+\rho^{00}_1\rho^{00}_2\sigma_{SS}
-8|\rho^{+0}_1\rho^{+0}_2|\tau_{TS}\cos(\tilde{\phi})
\biggr)
\frac{\dd^3 p_1'\dd^3p_2'}{E_1 E_2}.
\nonumber
\end{eqnarray}
Here $\rho_{i}^{ab}$ are the density matrices of the
virtual photon in the $\gamma\gamma$-helicity basis.

At large $q_1$ or $q_2$ the two-gamma cross-section $\sigma_{ab}$
can be calculated in the context of perturbative QCD.
For small virtualities  of the intermediate photons
({\it i.e.}, $q_i^2\ll W^2$, where $W$ is the invariant mass
of the state $X$),  the situation is not so simple. 
However, it is
possible to use different low-energy models.
For example, when it is assumed that the  $W$ and
$q_i$ dependence factorize~\cite{Schuler:1996qr}, one has the form: 
\begin{equation}
\sigma_{ab}(W,q_i^2)=h_a(q_1^2)h_b(q_2^2)\sigma_{\gamma\gamma}(W^2);
\nonumber
\end{equation}
this approximation is called the "improved EPA".
For more details, see the original papers 
of Refs.~\cite{Brodsky:1971ud,BGMS1975};  a recent review can be found in 
Ref.~\cite{Penn1}.
A summary of recent experimental work can be found in Ref.~\cite{Whaley}.
A very interesting review of two-photon event
generators (for LEP1 energies)
can be found in Ref.~\cite{Lonnblad:1995vh}.

\section{Experimental status of two-photon processes}

Lepton pair production in two-photon processes can be calculated
in the framework of QED.  While di-lepton production does not have much
bearing on new physics searches, measurements at large cm energies,
where radiative corrections are essential,
can provide precise tests of QED.
At  \bes3 energies, di-lepton
production by two photons is not 
particularly interesting, although it has to be taken into account as a
possible background source for some channels.

Of more interest is two-photon production of hadronic final states.
During the past few years a number of experiments 
using single-tag photons were carried out, primarily
to measure the photon structure function
$F_2^\gamma$.  These were done at cm energies well above the
 \bes3 range.  Also, high energy 
$\gamma\gamma\to jet$ or $\gamma\gamma\to 2jet$ processes have been studied,
and measurements of the $p_T$ distributions for inclusive $\pi$ 
production and cross-sections for charm and beauty production reported.

In the energy range accessible at \bes3, hadron production via two 
photon collisions has been 
studied by several experiments. 
At the MARK-II experiment~\cite{MARK2_1} 
the $\gamma\gamma\to \pi^+\pi^-$ cross section was measured
for invariant masss from $0.35$ to $1.6$~GeV, 
and $f_2(1270)$ and $f_0(1010)$ production was observed.
In addition, the
decay width of the $\eta'(958)$ to two photons
was measured~\cite{MARK2_2}.


The CLEO-II collaboration detected
two-photon production of pairs of charged pions and kaons
with invariant masses in the range between $1.5$ and $5.0$ GeV.
The data show a $\sim 40\%$ discrepancy from leading-order 
(LO) QCD predictions~\cite{part3:ref:CLEO}.

The CELLO Collaboration measured 
$\gamma\gamma\to \pi^+\pi^-$ production over
the energy range $0.75-1.9$ GeV \cite{CELLO}.

The Belle collaboration reported the observation of $f_0(980)$ 
production in the $\gamma\gamma\to \pi^+\pi^-$ reaction~\cite{BELLE}.
They find a line-shape for  the resonance 
that differs  from the standard Breit-Wigner form
and a total $f_0(980)$  width that  is much narrower 
than the PDG value. The authors of
Ref.~\cite{Achasf0} argue that this effect can be explained by 
destructive interference between the $f_0(980)$
and coherent non-resonant $\pi^+\pi^-$ background.

Neutral pion pair production by two photons has been measured 
by the Crystal Ball collaboration~\cite{CrB}. 
The $\gamma\gamma\to \pi^0\pi^0$ 
process provides a very crucial test of 
Chiral Perturbation Theory (ChPT)~\cite{chpt_test}. 
A key feature is that this process appears only in 
one-loop calculations in 
ChPT and some theoretical predictions conflict 
with the Crystal Ball data. This can be used to estimate
the two-loop contribution and determine constraints of
higher-order renormalization constants. Because of the very 
high luminosity of BEPCII it should be possible to obtain 
new and more precise data of neutral pion pair production with
\bes3.

In  the photon-photon center of mass energy
$0.5<W<2$ GeV  one can measure resonant structures in the hadron spectrum.
In the $\pi\pi$ final state there are many, perhaps too many, 
scalar resonances: $f_0(400-1200)$, $f_0(980)$,
$f_0(1370)$, $f_0(1500)$, $f_0(1710)$, etc.  Some are not well
established and some authors doubt their existence~\cite{doubtf0}.
In the \bes3 experiment, higher statistics can be obtained over the 
same angular acceptance than at (for example) the Belle experiment.
By combining data from different channels, such as $\pi^0\pi^0$ and
$\pi^+\pi^-$, it will be possible to separate the contributions 
to the cross sections of states with definite spin. 
Accurate measurements of this channel can provide
the possibility of determining whether or not any of the above-noted
candidate scalar states  are glueballs, based on
non-observation at a high level of sensitivity.

Searches for events with higher hadron multiplicity in the final state are 
also of interest. 
As an example, one can search for three pion final states
with invariant mass below $2$~GeV.
The ARGUS collaboration presented a clear $a_{2}(1370)$ signal 
resonance but did not observe the pseudoscalar resonance 
$\pi_2(1300)$ or any significant enhancements near the
$\pi_2(1670)$~\cite{ARGUS_3pi}. However, this measurement 
contradicts results from the Crystal Ball~\cite{CRB_3_pi} and  
CELLO~\cite{CELLO_3_pi} experiments. \bes3 can shed light on 
this puzzle.

Single-tag data (when one of the photons is considerably off-shell) migh
possibly be measured with \bes3, although here the cross section is much 
lower than for two-photon processes with real photons. Such measurements 
can be used to determine the photon structure function at small $q^2$.

The total hadronic cross-section can be measured by \bes3 as well.
In this case corrections for the angular acceptance of the detector
will be necessary. 

\section{Measurement of two-photon processes at \bes3}

The combination of the very high luminosity of the BEPCII accelerator 
together 
with the excellent \bes3 detector performance offers opportunities 
for  precision measurements
of hadron production by two photons.

To study the \bes3 potential for $\gamma\gamma$ physics, the Monte-Carlo
program GALUGA 2.0~\cite{Schuler_2} was used with small modification.
Since the production of charged and neutral pion pairs in two-photon 
reactions is an important topic for \bes3, relevant models based on 
the point-like pion approximation and on ChPT~\cite{BGMS1975,Bijnens} 
were implemented in the program.

The luminosity function has been calculated for several working
energies of the BEPCII machine (see Fig. \ref{twophoton_lumi}). 
Given that the integrated luminosity of  BEPCII will be of order 
\mbox{5 $fb^{-1}$}/year, one can see that relatively high statistics  
two-photon event samples are
expected at \bes3, especially at low energies ($m_{\pi} < W < 
m_{f_{0}}$).

\begin{figure}[htbp]
\centerline{\epsfig{file=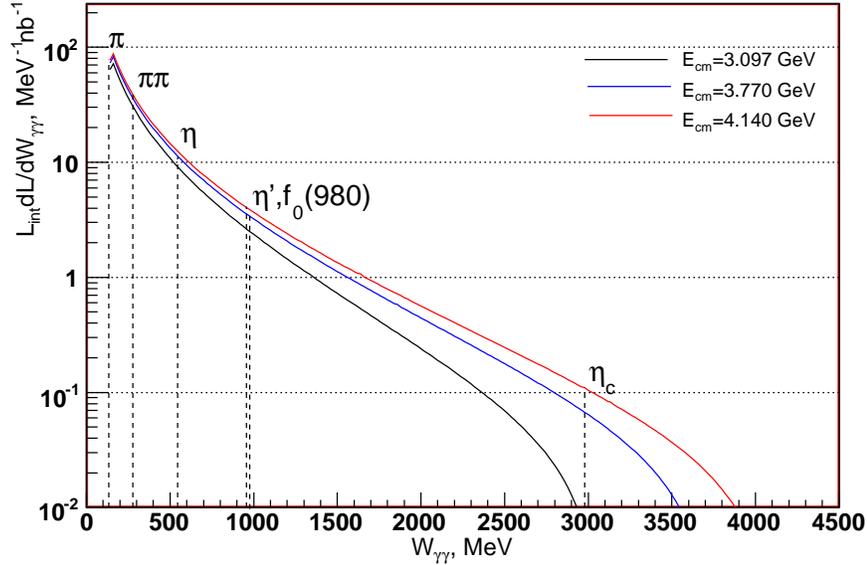, width=12.5cm}}
\caption { The photon-photon flux at \bes3 for several beam energies, 
for an integrated luminosity of 1 $fb^{-1}$.} 
\label{twophoton_lumi} 
\end{figure}


One can, in principle at least, tag two-photon events by 
the detection of one or both of
scattered electrons (single tag or double tag mode). Tagging allows one 
to suppress backgrounds significantly, but at the cost of
a dramatic reduction in experimental
statistics. In addition, the requirement of 
the detection of a scattered electron 
into a specific energy and angular range can distort the measurement 
results. 

The main source of background in the \bes3 experiment will be 
hadronic decays of charmonium with one or more particles undetected. To
suppress this background, at least single tagging is needed.  Studies 
using the no-tag mode are probably only possible
at $E_{cm}$ = 3.77 GeV, or at other non-narrow resonances.   
Background contributions from $D$ and $\Psi(3770)$ decays 
are small and, probably, easily 
suppressed.  However this has yet to be confirmed by 
a detailed background study. 
Since no special tagging system is available in the \bes3 detector,
only scattered electrons at the angular range $|cos(\Theta)|<0.93$
can be used.

The number of two-photon-produced  hadronic final states
has been estimated for $E_{cm}$=3.77 GeV and an 
integrated luminosity $L_{int}=5 fb^{-1}$ ($\approx$ 1 year of data
taking). Results of the calculation are provided in 
\tablename~\ref{twophoton_stat}.

\begin{table}
\centering
\begin{tabular}{|c|c|c|c|c|} \hline
Mode  & Total cross-section [nb] & No tag $\times 10^{6}$& Single tag $\times 10^{3}$ & Double tag \\ \hline %
$\pi^{+}\pi^{-}$ & 2.38   &  11.9  &  236.5  & 5860 \\
$\pi^{0}\pi^{0}$ & 0.062  &  0.31  &  25.5 & 885    \\
$\pi^{0}$        & 0.67   &  3.35  &  7.85   & 90   \\
$\eta$           & 0.24   &  1.20  &  32.8   & 490  \\
$\eta'$          & 0.37   &  1.85  &  113.0  & 2255 \\
$a_{0}(980)$     & 0.33   &  1.65  &  7.87   & 990   \\
$f_{0}(980)$     & 0.046  &  0.23  &  9.5    & 140  \\
$\eta_{c}$       & 0.0016 &  0.008 &  1.9    & 225  \\
\hline %
\end{tabular}
\caption{ The total number of two-photon-produced events
for  an integrated luminosity of 5 $fb^{-1}$ at
$E_{cm}$=3.77~GeV.}
 \label{twophoton_stat}
\end{table}

The relative increase in the number of heavier mesons, like
$\eta, \eta'$, in the tagged modes with respect to 
the decreasing total cross 
sections is due to the increase of the production cross
section with increasing $W$ (see Fig.~\ref{twophoton_res}). 
As a result,
the scattering angle of electrons and, consequently, the number of
double-tag events for these mesons will likely be larger 
than those for $\pi^{0}$ production.

\begin{figure}[htbp]
\centerline{\epsfig{file=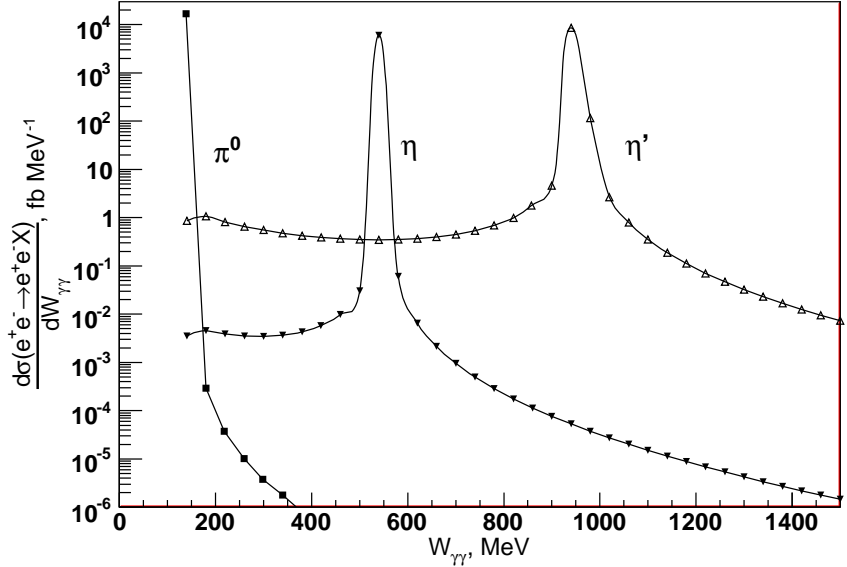, width=12.5cm}}
\caption { Differential cross section $\frac{d\sigma}{dW}$ of meson
 formation in two photon reactions $e^{+}e^{-}\rightarrow e^{+}e^{-}X$ ($E_{cm}=3.77 GeV$).} 
\label{twophoton_res} 
\end{figure}

The production of $\pi^{+}\pi^{-}$ and $\pi^{0}\pi^{0}$ pairs in two
photon processes at \bes3 has been studied in detail for
$E_{cm}$=3.77 GeV. In Figs.~\ref{twophoton_pipi1},~\ref{twophoton_pipi2},
 \& ~\ref{twophoton_pi0pi0} 
one can see the momentum distribution of
the produced pions. Only events where both pions 
are within the \bes3  angular 
acceptance ($|cos(\Theta)|<0.93$) are selected. The plots are normalized
by the total number of events. The no-tag-mode events 
are indicated by solid line, the
dotted line corresponds to double-tag-mode events.

\begin{figure}[htbp]
\centerline{\epsfig{file=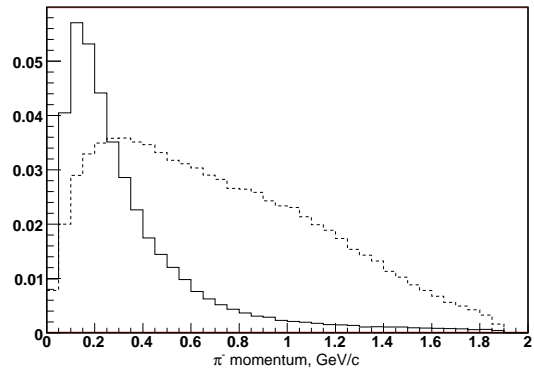, width=8.0cm}}
\caption { Momentum distribution of $\pi^{+}$ mesons from
  $e^{+}e^{-}\rightarrow e^{+}e^{-}\pi^{+}\pi^{-}$.  } 
\label{twophoton_pipi1} 
\end{figure}
\begin{figure}[htbp]
\centerline{\epsfig{file=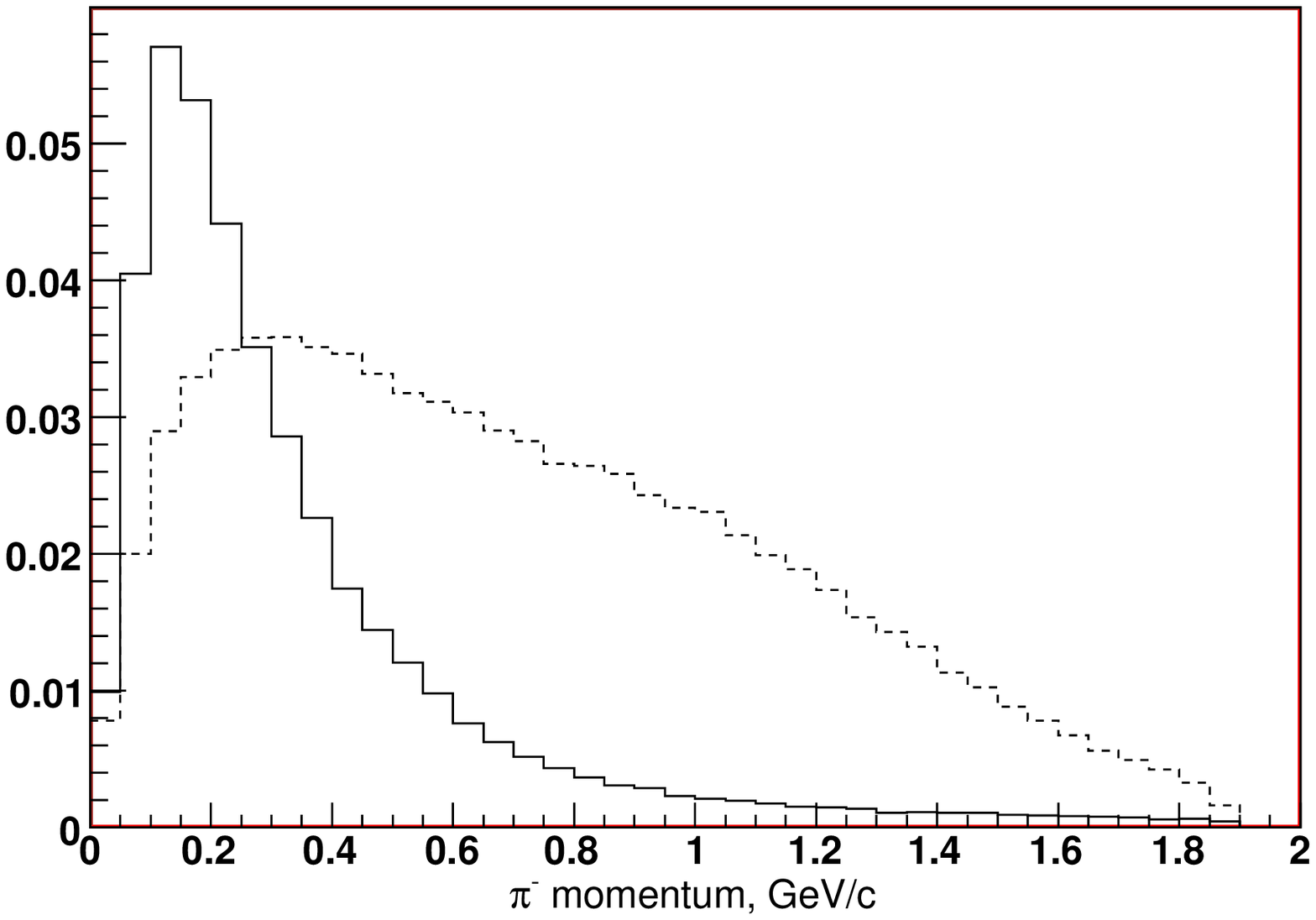, width=8.0cm}}
\caption { Momentum distribution of $\pi^{-}$ mesons from
  $e^{+}e^{-}\rightarrow e^{+}e^{-}\pi^{+}\pi^{-}$.} 
\label{twophoton_pipi2} 
\end{figure}
\begin{figure}[htbp]
\centerline{\epsfig{file=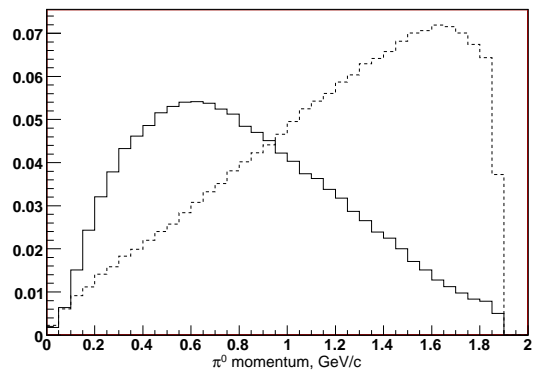, width=8.0cm}}
\caption { Momentum distribution of $\pi^{0}$ mesons from
  $e^{+}e^{-}\rightarrow e^{+}e^{-}\pi^{0}\pi^{0}$. } 
\label{twophoton_pi0pi0}
\end{figure}

\section{Summary}
A feasibility study for measurements of
hadron production in two-photon reactions at
\bes3 has been carried out. The two-photon invariant mass region
accessible at the \bes3 energies extends up to 3~GeV. 
For data-taking at the $J/\psi$ and $\psi^{\prime}$ resonances,
electron tagging of two-photon events will be necessary for reducing
backgrounds from hadronic charmonium decays. At $E_{cm}$ = 3.77 GeV, 
however,
where backgrounds from $\Psi(3770)$ and $D$ decays is expected to be 
small, no-tag data will be useful. The
double-tag requirement will provide the 
most effective
background suppression, however the 
data yield would be too small except for a few
channels (production of $\pi^{+}\pi^{-}$, $\eta'$). Single-tag
measurements are more promising: for example, the estimated 
numbers of events should be sufficient for precision measurements
of neutral-pion pair production. In general, \bes3 offers some good
opportunities for precision measurements of 
the production of low-mass hadronic
systems in two-photon collisions.

\part[Charmonium Physics]{Charmonium Physics\\
\vspace*{2cm}
\Large Conveners\\
 Cong-Feng Qiao, Changzheng Yuan.\\
 \vspace*{1cm}
\Large Contributors\\ 
T.~Barnes, N.~Brambilla, H.~X.~Chen, Y.~Q.~Chen,
X.~Garcia~i~Tormo, Z.~G.~He, Y.~Jia, Y.~P.~Kuang, H.~B.~Li,
J.~L.~Li, H.~L.~Ma, J.~P.~Ma, X.~H.~Mo, S.~L.~Olsen,
A.~Pineda, R.~G.~Ping, C.~F.~Qiao, G.~Rong, J.~Soto,
A.~Vairo, P.~Wang, C.~Z.~Yuan, D.~H.~Zhang,
~X.~M.~Zhang and Y.~J.~Zhang}
\label{part:four}

\chapter{Theoretical Frameworks of Charmonium Physics}
\label{sec:cc_theory}


The discovery of the $J/\psi $ in 1974 had a huge impact on the
development of the theory of strong interactions and the Standard
Model. Still today, mesons made from two heavy quarks, {\it i.e.}
heavy quarkonia, play a 
prominent role in investigations of QCD dynamics both within and 
beyond the Standard Model~\cite{nora_Brambilla:2004wf}.
These are multi-scale systems that probe all of the energy regimes 
of QCD: from the hard region, where expansions in the coupling constant 
are legitimate, to the low-energy region, where nonperturbative
effects dominate. Heavy quark-antiquark states  are thus an ideal,
and to some extent unique, laboratory where our understanding of
nonperturbative QCD and its interplay with perturbative QCD can be
tested in a controlled framework. In correspondence with the
hierarchy of energy scales in quarkonia,  a hierarchy of
nonrelativistic effective field theories (NR EFT) may be
constructed, each one with fewer degrees of freedom that are left 
dynamical and thus simpler. Some of these physical scales are large and  
may be treated with perturbation theory. These features make 
mesons that are made from two heavy quarks accessible inside QCD.

{\if Here we will focus on the lowest energy EFT, called 
pNRQCD~\cite{cyq_Pineda:1997bj,cyq_Brambilla:1999xf}, that can be
constructed for heavy quarkonium mesons. We will give a brief
introduction to the main concepts underlying this new EFT and
discuss a wide range of its applications. The focus and the main
interest is clearly on charmonium physics, but  results obtained
for  bottomonium physics are also discussed both for their own
interest and their impact on charmonium physics. Due to the simplicity
of pNRQCD, several model independent predictions are possible.
Therefore, on one hand the progress in our understanding of 
NREFTs~\cite{cyq_Brambilla:2004jw}  makes it possible to move beyond
phenomenological models and to  provide a systematic description
from QCD of all aspects of heavy-quarkonium physics. On the other
hand, the recent progress (coming from $B$-factories, CLEO, Fermilab
and BES) in the measurement of several heavy-quarkonium
observables makes it meaningful to address problems associated
with their precise theoretical determination.
\par
In this situation  \bes3 offers a unique opportunity to advance
our understanding of strong interactions and our control of some
parameters of the Standard Model. \fi }

\section[Non-Relativistic QCD effective theory]
{Non-Relativistic QCD effective theory\footnote{By Yu-Qi Chen}}
\label{sec:cc_nrqcd}








Charmonia are bound states of the $c\bar c$ pair by the strong
interaction.
The charm quark mass $m_c$ is  so  large 
 that the  motion of the charm quark inside the bound state is
slow and charmonium can be approximately regarded as a
non-relativistic bound state. $v^2 \sim 0.3 $ for charmonia and
0.1 for bottomnia  by potential model calculations or by Lattice
simulation,  where $v$ is the relative velocity between the $c$
and $\bar{c}$. 
For a small value of $v$, there are several distinct energy scales
in charmonium. The 3-momentum is order $m_cv$. The binding energy
is order $m_cv^2$. Numerically, they are around 800 MeV and 500
MeV, respectively.
 In the limit of $v^2
\ll 1$, those energy scales satisfy a hierarchy relation $m_c \gg
m_cv \gg m_cv^2$, with which the effects at energy scale $m_c$ can
be integrated out explicitly.  The resultant theory, being
expressed as a Pauli two-component field theory, is just the {\it
nonrelativistic QCD} (NRQCD) effective
theory~\cite{cyq_Caswell:1985ui,cyq_Lepage:1992tx,cyq_Bodwin:1994jh}. NRQCD is
equivalent to the full QCD but it simplifies QCD by taking a
nonrelativistic expansion and by reducing the number of the energy
scales. Essentially, NRQCD can be regarded as an effective theory
which expands the full QCD in powers of $v$. It turns out to be
very useful in dealing with charmonium relevant processes such as
spectroscopies, annihilation decays, and inclusive productions.
Starting from NRQCD, one may integrate out further the effects at
energy scale $m_cv$ and obtain an effective theory contains only
energy scale $m_cv^2$, called as {\it potential NRQCD}
(pNRQCD)~\cite{cyq_Pineda:1997bj,cyq_Brambilla:1999xf,cyq_Brambilla:2004jw},
which will be reviewed in the next subsection.

\subsection{NRQCD effective theory}

QCD fully describes the strong interactions of the quarks and
gluons. QCD lagrangian for heavy quark is given by:
\begin{equation}
{\cal L} \;=\; \bar{\Psi}(x) \, (\,i \not\!D -m_c\,) \Psi (x) \;,
\end{equation}
where $ iD^\mu = i\partial^\mu +g A^{a\mu}\,T^a  $ is the
covariant derivative. It looks simple but it turns out to be
complicated. As a relativistic quantum field theory, it describes
heavy quark and antiquark together and quantum effects from all
energy scales.
It can be simplified in the nonrelativistic limit $v \ll 1 $. First,
in this limit, the heavy quark pair creation and annihilation
effects is suppressed.  The quark and the antiquark are decouple.
All relativistic effects can be expanded as power series of $v$.
Second, those quantum effects arising from high energy scales $m$ or
above are perturbatively calculable since QCD is a theory with
asymptotical freedom. This can be done by taking the hard cut-off
energy scale $\Lambda$ less than $m_c$. Quantum effects above
$\Lambda$ can be expressed as contributions from  a sum of certain
local operators in powers of $v$ since in this region the internal
particles are far off-shell and can only propagate in a short
distance which is less than $1/m_c$. Consequently, taking  $\Lambda
< m_c$, adding certain local operators in the lagrangian, and making
expansion in terms of $v$, one obtains an effective theory, which
reproduces the results of the full QCD at low energy scale. The
coefficients of those local operators are called as the Wilson short
distance coefficients which can be calculated by matching the QCD
full theory and that of the effective theory.

In NRQCD effective theory, the quark and the antiquark fields are
described by the two component Pauli field instead of the Dirac
four component field in QCD. Up to order $v^4$, the NRQCD
effective lagrangian
reads~\cite{cyq_Caswell:1985ui,cyq_Lepage:1992tx,cyq_Bodwin:1994jh}
\allowdisplaybreaks
\begin{eqnarray}
{\cal L}_{\rm NRQCD}&=& {\cal L}_l\,+\,{\cal L}_0 \,+\,{\delta
\cal L} \;, \label{NRQCD}
\end{eqnarray}
where ${\cal L}_l$ is the usual lagrangian that describes gluons
and light quarks;
\begin{eqnarray}
 {\cal L}_{0}&=&\psi^{\dagger} \left(\,i D_0 + \, \frac{\mathbf{D}^2}{2 m_c}\,
 \right)\psi \,+\,\chi^{\dagger} \left(\,i D_0 + \, \frac{\mathbf{D}^2}{2 m_c}\,
 \right)\chi \;\,
\label{L0} \end{eqnarray}
is the leading order  NRQCD effective lagrangian; and
\begin{eqnarray}
{\delta \cal L} &=&
 {c_1 \over 8 m_c^3}\,
 \psi^{\dagger}  \, (\mathbf{D}^2)^2 \psi
 \,+\,
  \frac{c_2}{8m_c^2} \,
 \psi^{\dagger} \,g(\, \mathbf{D \cdot E} - \mathbf{E \cdot  D} \,)\psi \,
 + {c_3\over 2m_c}\, \psi^{\dagger} \,g \mathbf{\mbox{\boldmath $\sigma$} \cdot B}\,\psi
 \nonumber
\\
&&  + \,i {c_4 \over 8 m_c^2}\, \psi^{\dagger}\, g\,
\mathbf{\mbox{\boldmath $\sigma$} \cdot \left(D \times E -E \times
D\right) } \, \psi
\,+\,
{\rm charge\;\, conjugate\;\, terms} \label{delL}
\end{eqnarray}
%
are $v^2$ correction terms  to ${\cal L}_0$. Here $\psi$ and $\chi$
are the Pauli spinor fields of quark and antiquark, respectively.
 Gauge invariance implies
that gluon field appears in the lagrangian always only through the
gauge-covariant derivatives $i D_0$, $i{\bf D}$ and the QCD field
strength ${\bf E} $, ${\bf B}$.

In $ {\cal L}_0 $, both the $i D_0$ term and the ${\bf D}^2/2m_c$
term contribute to the same order of the quarkonium energy in
Coulomb gauge. In QED,  neglecting ${\bf A}$ field, variation to the
$ {\cal L}_0 $ gives to the Sch{\o}dinger equation of the hydrogen
atom. No Pauli matrix $\sigma^i$ in $ {\cal L}_0 $ implies  a
symmetry of  the heavy quark spin. It means  that spin-flip of the
heavy quark does not change the energy of the system at leading
order and the spin-dependent effects are suppressed. Consequently,
at leading order the states $J/\psi$ and $\eta_c$ are degenerate and
the states $\chi_{c0},\chi_{c1},\chi_{c2} $, and $h_c$  are also
degenerate. This symmetry is violated after including the next
leading order corrections since the spin dependent term with
$\sigma^i$ appearing in $ {\delta \cal L} $. The velocity scaling
rule~\cite{cyq_Caswell:1985ui,cyq_Lepage:1992tx,cyq_Bodwin:1994jh} can be used to
count the relative importance of each term in $ {\delta \cal L} $.
Those Wilson short-distance coefficients $c_i$'s has been calculated
in pQCD. They can be expanded as power series of $\alpha_s$ at
energy scale $m_c$. Other coefficients can be determined by matching
the full QCD and the NRQCD effective theory. Some relationships
between these coefficients follows from the reparameterization
invariance~\cite{cyq_Luke:1992cs,cyq_Chen:1993sx,cyq_Sundrum:1997ut,cyq_Finkemeier:1997re,cyq_Chen:2002yr,cyq_Brambilla:2003nt}.
Especially, the coefficient of the kinetic energy term in ${\cal
L}_0$ is not renormalization.

\subsection{Inclusive Charmonium Annihilation Decays }

An important decay mode of charmonium is the annihilation decay,
in which the $c\bar{c} $ pair annihilates into a light quark
antiquark pairs or gluons, or photon(s). Those light partons
eventually form light hadrons by hadronization. The inclusive
process that $c\bar{c} $ annihilates into all light hadrons is
infrared safety and hence perturbatively calculable although the
hadronization process is infrared sensitive. When a $c\bar{c}$
pair annihilates, the total energy $2m_c$ is released. Thus it
occurs at distance of $1/(2 m_c)$. The typical size of a
charmonium state is order of $1/(m_c v)$. Thus in the limit of
$v\ll 1$,  the effects happened at these two distinct distance
scales are separated well. NRQCD factorization
formula~\cite{cyq_Bodwin:1994jh} provide a systematic framework to
analyze the annihilation decay rates. In the formula, the
annihilation decay rate is factorized into a sum of the products
of the short-distance coefficients and the long-distance matrix
elements~\cite{cyq_Bodwin:1994jh}:
\begin{equation}
\Gamma ( H ) \;=\; {1 \over 2 M_{H} } \; \sum_{mn} {C_{mn}(\mu)
\over m_c^{d_{mn}-N-1}}
        \langle H | {\cal O}_{mn} | H \rangle^{(\mu)} \;,
\label{fact-Gam}
\end{equation}
where $M_H$ is the  mass of the charmonium state $H$ and $d_{mn}$
is the mass dimension of the operator ${\cal O}_{mn}$. The matrix
elements $ \langle H | {\cal O}_{mn} | H \rangle $ are expectation
values in the quarkonium state $H$ of local 4-fermion operators
that have the structure
\begin{equation}
{\cal O}_{mn} \;=\; \psi^\dagger {{\cal K}'}_m^\dagger \chi \;
    \chi^\dagger {\cal K}_n \psi \;,
\label{O_mn}
\end{equation}
where  ${\cal K}_n$ and ${{\cal K}'}_m^\dagger$ are products of a
color matrix ($1$ or $T^a$), a spin matrix, and a polynomial in the
gauge covariant derivative ${\bf D}$. The color singlet dimension 6
and 8 operators are:
\begin{eqnarray}
 &&   {\cal O}_1(^3S_1) = \psi^{\dag} {\mbox{\boldmath $\sigma$}} \chi \, \chi^{\dag} {\mbox{\boldmath $\sigma$}} \psi,
\nonumber
\\
&&
 {\cal O}_1(^1S_0) = \psi^{\dag} \chi \, \chi^{\dag}  \psi, \nonumber
 \label{o-j-psi}
\end{eqnarray}
\begin{eqnarray}
&&
 {\cal O}_1(^3P_0) = \psi^{\dag} \mathbf{D}\cdot
 {\mbox{\boldmath
 $\sigma$}} \chi \, \chi^{\dag}
    \mathbf{D} \cdot {\mbox{\boldmath
 $\sigma$}} \psi, \nonumber
\\
&& {\cal O}_1(^3P_1) = \psi^{\dag} \mathbf{D}\times
 {\mbox{\boldmath
 $\sigma$}} \chi \, \chi^{\dag}
    \mathbf{D} \times {\mbox{\boldmath
 $\sigma$}} \psi, \nonumber
\\
&& {\cal O}_1(^3P_2) = \psi^{\dag}  \sigma^{(i} D^{j)} \chi \,
\chi^{\dag}  \sigma^{(i} D^{j)} \psi,\nonumber
\\
&&
 {\cal O}_1(^1P_1) = \psi^{\dag} \mathbf{D} \chi \, \chi^{\dag}
    \mathbf{D} \psi,
 \label{o-chi-cj}
\end{eqnarray}
The corresponding color octet operators could be obtained by
inserting $T^{a}$ between $\psi^{\dag}$ and $\chi$ and between
$\chi^{\dag}$ and $\psi$. In NRQCD factorization formula, the
quantum number of the state $H$ is not necessarily the same with
that of the operator $O$. When $H$ and $Q$ hold the same quantum
number, $H$ decays via the leading Fock space. Otherwise, $H$ decays
via the higher Fock space with electric or magnetic transition. It
differs from the conventional color-singlet model in which the $H$
decays only via the leading Fock space. The relative importance of
each matrix element can be estimated by the velocity scaling
rule~\cite{cyq_Bodwin:1994jh}. Suppose $E$ and $M$ be the total number of
the electric and magnetic transitions; $D$ the derivative number
contain in the operator $O$, then the matrix element $\langle H |
{\cal O}_{mn} | H \rangle$ scales like $v^{3+D+E+3M/2}$. This rule
can be used to estimate the order of magnitude of the NRQCD matrix
elements.

 To reduce the number of the NRQCD matrix elements, some comments can be
given below:

 (1) In the
electromagnetic annihilation decay mode, the $c\bar{c}$ pair
annihilates to the vacuum. Thus one needs to insert a vacuum state
$|0\rangle \langle 0 |$ between those two two-fermion operators.
It corresponds to take the vacuum saturation approximation, in
which the NRQCD matrix elements are proportional to the square of
the wavefunctions at the origin for $S-$wave and its derivative
for $P-$wave and so on. The approximation validates up to order
$v^4$~\cite{cyq_Bodwin:1994jh} for a  $S-$wave NRQCD matrix elements.

(2) In annihilation processes of charmonium decay and production,
the relativistic correction is typically large. The Gramm-Kapustin
relation~\cite{cyq_Gremm:1997dq} relates the leading order matrix
elements to the relativistic correction ones. However, the pole
mass enter into the relation so that some ambiguities are
involved.

(3) With pNRQCD, the color-octet matrix elements are related to
the corresponding color-singlet one by some universal factor which
are state and flavor
independent~\cite{cyq_Pineda:1997bj,cyq_Brambilla:1999xf,cyq_Brambilla:2004jw}.
This will be reviewed in detail in the next subsection.

Thus up to certain orders of $v$, there are only a few independent
long-distance matrix elements, which are nonperturbative in nature.
They may be determined by fitting experimental data, by Lattice
simulation, or by pNRQCD. This feature makes NRQCD factorization
formula be useful in analyzing the inclusive or electromagnetic
charmonium annihilation decays.

 The short-distance coefficients  for annihilation decay rates
$C_{mn}(\mu) $ can be expanded as a power series of
$\alpha_s(\mu)$. They can be determined by matching perturbative
calculations of $c \bar{c}$ scattering amplitudes.  For
annihilation decays, $\mu\sim m$. The $\mu$ dependence in the
short-distance coefficients cancels that in the matrix elements. A
general matching prescription is the threshold expansion
method~\cite{cyq_Braaten:1996jt}.

In sections \ref{sec:LightHadronDecay} and \ref{sec:LightHadronhc}, we will discuss the light hadron decays of the
$J/\psi$, $\psi'$, $\eta_c$, $\chi_{cJ}$ and $h_{c}$ in framework of
NRQCD factorization method.

\section[pNRQCD and its applications]
{pNRQCD and its applications\footnote{By Nora Brambilla, Yu Jia,
and Antonio Vairo}}
\label{sec:cc_pnrqcd}

\def\bfnabla{\mbox{\boldmath $\nabla$}}
\def\bfsigma{\mbox{\boldmath $\sigma$}}
\def\lQ{\Lambda_{\rm QCD}}
\def\siml{{\ \lower-1.2pt\vbox{\hbox{\rlap{$<$}\lower6pt\vbox{\hbox{$\sim$}}}}\ }}
\def\simg{{\ \lower-1.2pt\vbox{\hbox{\rlap{$>$}\lower6pt\vbox{\hbox{$\sim$}}}}\ }}
\newcommand{\MS}{\overline{\rm MS}}
\def\lla{\langle\!\langle}
\def\rra{\rangle\!\rangle}

\def\als{\alpha_{\rm s}}
\def\siml{{\ \lower-1.2pt\vbox{\hbox{\rlap{$<$}\lower6pt\vbox{\hbox{$\sim$}}}}\ }}



\if

The $J/\psi $ discovery in 1974 had  great  impact on the
development of the theory of strong interactions and the Standard
Model. Systems made by two heavy quarks occupy  still today a
prominent place \cite{nora_Brambilla:2004wf} in our investigation
of the QCD dynamics within and beyond the Standard Model. They are
multi-scale systems probing all the energy regimes of QCD, from
the hard region, where an expansion in the coupling constant is
legitimate, to the low energy region, where nonperturbative
effects dominate. Heavy quark-antiquark states  are  thus an ideal
and to some extent unique laboratory where our understanding of
nonperturbative QCD and its interplay with perturbative QCD may be
tested in a controlled framework. In correspondence with  the
hierarchy of energy scales in quarkonia, a hierarchy of
nonrelativistic effective field theries (NR EFT) may be
constructed, each one with  less degrees of freedom left dynamical
and thus simpler. Some of these physical scales are large and  may
be treated in perturbation theory. These features make  two heavy
quark systems accessible inside QCD.

Here we will focus on the lowest energy EFT, called pNRQCD
\cite{cyq_Pineda:1997bj,cyq_Brambilla:1999xf},  that can be
constructed for heavy quarkonium. We will give a brief
introduction to the main concepts underlying this new EFT and
discuss a wide range of applications. The focus and the main
interest is clearly on charmonium physics, but  results obtained
in  bottomonium physics are also discussed for their potential
interest and impact on charmonium physics. Due to the simplicity
of pNRQCD  several model independent predictions become possible.
Therefore, on one hand the progress in our understanding of NREFTs
\cite{cyq_Brambilla:2004jw}  makes it possible to move beyond
phenomenological models and to  provide a systematic description
from QCD of all aspects of heavy-quarkonium physics. On the other
hand, the recent progress (coming from B-factories, CLEO, Fermilab
and BES) in the measurement of several heavy-quarkonium
observables makes it meaningful to address the problem of their
precise theoretical determination.
\par
In this situation  BESIII offers a unique opportunity to advance
our understanding of strong interactions and our control of some
parameters of the Standard Model.

\fi


\if

The description of hadrons containing
two heavy quarks is a rather challenging problem,
which adds to the complications of the bound state in field theory
those coming from a nonperturbative QCD low-energy dynamics.
A  simplification is provided  by  the nonrelativistic nature
suggested by the large mass of the heavy quarks
and  manifest in the spectrum pattern.
Systems made by two heavy quarks are  characterized
 by three energy scales, hierarchically ordered by the heavy quark velocity
in the center of mass frame
$v \ll 1$: the mass $m$ (hard scale), the momentum transfer $mv$ (soft scale),
which is proportional to the inverse of the typical size of the system $r$,
 and  the binding energy $mv^2$ (ultrasoft scale), which is proportional to the
inverse of the typical time of the system. In bottomonium $v^2
\sim 0.1$, in charmonium $v^2 \sim 0.3$. In perturbation theory
$v\sim \als$. Feynman diagrams will get contributions from all
momentum regions associated with these scales. Since these
momentum regions depend on $\als$, each Feynman diagram
contributes to a given observable with a series in $\als$ and a
non trivial counting. For energy scales close to $\lQ$, the scale
at which nonperturbative effects become dominant, perturbation
theory breaks down and one has to rely on nonperturbative methods.
 Regardless of this, the non-relativistic hierarchy
$ m \gg mv \gg mv^2 $ will persist  also below  the  $\lQ$  threshold.

The wide span of energy scales involved makes also a lattice
calculation in full QCD extremely challenging. We may, however,
take advantage of the existence of a hierarchy of scales by
substituting QCD with simpler but equivalent EFTs. Lower energy
EFTs may be constructed by systematically integrating out modes
associated to  energy scales not relevant for the two quark
system. Such integration  is made  in a matching procedure that
enforces the equivalence between QCD and the EFT at any  given
order of the expansion in $v$. Any prediction of the EFT is
therefore a prediction of QCD with an error of the size of the
neglected order in $v$.
By integrating out the hard modes one  obtains Nonrelativistic QCD
 \cite{nora_Caswell:1985ui} (cf. the contribution by Yu-Qi Chen in this Book).
 In such EFT soft
and ultrasoft scales are left dynamical and still their mixing  complicates
 calculations and power counting.
In the last few years the problem of systematically treating the
remaining dynamical scales in an EFT framework has been addressed
by several groups \cite{nora_group} and has now reached a good
level of understanding. We give here an introduction to  the
simplest EFT you can write down for two heavy quark systems, where
only ultrasoft degrees of freedom remain dynamical. This is
potential NRQCD\cite{cyq_Pineda:1997bj,cyq_Brambilla:1999xf}
 \footnote{for an alternative and equivalent EFT (in the case in which $\lQ$ is
the smallest scale) see \cite{nora_vnrqcd}.}.

\fi

\subsection{pNRQCD}

pNRQCD
\cite{cyq_Pineda:1997bj,cyq_Brambilla:1999xf,cyq_Brambilla:2004jw}
is the EFT for two heavy quark systems that follows from NRQCD by
integrating out the soft scale $mv$. Here the role of the
potentials and the quantum mechanical nature of the problem are
realized in  the fact that the Schr\"odinger equation appears as
zero order problem for the two quark states.
We may distinguish two situations:
1) weakly coupled pNRQCD when $mv  \gg \lQ$,
where the matching from NRQCD to pNRQCD may be performed in
 perturbation theory;
2) strongly coupled pNRQCD when $ mv \sim \lQ$,
where the matching has to be nonperturbative.
Recalling that $r^{-1}
  \sim mv$, these two situations correspond  to systems with inverse
typical radius smaller than or  of the same order as  $\Lambda_{\rm QCD}$.

\subsubsection{Weakly coupled pNRQCD}
The effective degrees of freedom that remain dynamical  are: low
energy $Q\bar{Q}$ states (that can be decomposed into  a singlet
and an octet field under colour transformations) with energy of
order  $ \Lambda_{\rm QCD},mv^2$ and
    momentum   ${\bf p}$ of order $mv$,
plus   low energy (ultrasoft)  gluons
with energy  and momentum of order
$\lQ,mv^2$.
All the  gluon fields are multipole  expanded (i.e.  expanded  in
the quark-antiquark distance $r$).
The Lagrangian is then given by terms of the type
\begin{equation}
{c_k(m, \mu) \over m^k}  \times  V_n(r
\mu^\prime, r\mu)   \times  O_n(\mu^\prime, mv^2, \lQ)\; r^n  .
\end{equation}
where the matching coefficients $c_k$ are inherited from NRQCD and
contain the logs in the quark masses, while  the pNRQCD potential
matching  coefficients $V_n$  encode the non-analytic behaviour in
$r$. At  leading order in the multipole expansion, the singlet
sector of the Lagrangian gives rise to equations of motion of the
Schr\"odinger type.
Each term in  the pNRQCD Lagrangian has a definite power counting.
The bulk of the interaction is carried by potential-like terms,
nut non-potential interactions, associated with the propagation of
low energy degrees of freedom are present as well. These
retardation (or non-potential) effects start at the
next-to-leading order  (NLO) in the multipole expansion and are
systematically encoded in the theory and  typically related to
nonperturbative effects \cite{cyq_Brambilla:1999xf}. There is a
systematic procedure to calculate corrections in $v$ to physical
observables: higher order perturbative (bound state) calculations
in this framework become viable. In particular the EFT can be used
for a very efficient resummation of
 large logs (typically logs of the ratio of energy and momentum scales)
 using the renormalization group  (RG) adapted to the case of correlated
scales \cite{nora_rg,nora_vnrqcd};
 Poincar\'e invariance is not lost, but shows up
in some exact relations among the matching coefficients
\cite{cyq_Brambilla:2003nt}.

\subsubsection{Strongly coupled pNRQCD}
In this case the
matching to pNRQCD is nonperturbative.
Away from threshold
(precisely when heavy-light meson pair and heavy hybrids
develop a mass gap of order $\lQ$ with respect to the energy of the $Q\bar{Q}$ pair),
the quarkonium singlet field $\rm S$ remains as the only low energy dynamical
degree of freedom in the pNRQCD Lagrangian (if no ultrasoft pions are considered),
which reads
\cite{nora_Brambilla:2000gk,nora_Pineda:2000sz,cyq_Brambilla:2004jw}:
\begin{equation}
\quad  {L}_{\rm pNRQCD}= {\rm Tr} \,\Big\{ {\rm S}^\dagger
   \left(i\partial_0-{{\bf p}^2 \over
 2m}-V_S(r)\right){\rm S}  \Big \} .
\end{equation}
The matching potential $V_S(r)$ is a series in the expansion in the inverse of the quark masses:
static, $1/m$ and $1/m^2$ terms have been calculated, see
\cite{nora_Brambilla:2000gk,nora_Pineda:2000sz}.
They involve NRQCD matching coefficients and low energy nonperturbative parts given in terms
of Wilson loops and field strengths insertions in the Wilson loop.
In this regime   we recover the quark potential singlet model from
 pNRQCD.  However the potentials are  calculated from QCD in the formal nonperturbative
matching procedure. An actual evaluation of the low energy part
requires lattice evaluation \cite{nora_Bali:2000gf} or QCD vacuum
models calculations \cite{nora_vac}.

\subsection{Applications}

It is important to establish when $\lQ$ sets in, i.e. when we have
to resort to non-perturbative methods. For low-lying resonances,
it is reasonable, although not proved, to assume $mv^2 \simg \lQ$.
The system is weakly coupled and we may rely on perturbation
theory, for instance, to calculate the potential. The theoretical
challenge here is performing higher-order calculations and the
goal is precision physics. For high-lying resonances, we assume
$mv \sim \lQ$.  The system is strongly coupled and the potential
must be determined non-perturbatively, for instance, on the
lattice. The theoretical challenge here is providing a consistent
framework where to perform lattice calculations and the progress
is measured by the advance in
 lattice computations.

For what concerns systems close or above the open flavor
threshold, a complete and satisfactory understanding of the
dynamics has not been achieved so far. Hence, the study of these
systems is on a less secure ground than the study of states below
threshold. Although in some cases one may develop an EFT owing to
special dynamical conditions (as for the $X(3872)$ interpreted as
a loosely bound $D^0 \, \bar{D}^{*\,0}$ $+$ ${\bar D}^0 \,
D^{*\,0}$ molecule), the study of these systems largely relies on
phenomenological models. The major theoretical challenge here is
to interpret the new states in the charmonium region discovered at
the B-factories in the last years.

\subsection{QCD potentials}

pNRQCD allows us to properly define the QCD potentials
 and give a well defined procedure to properly calculate them.
In this modern  description  the potentials are  matching
coefficients of the EFT and as such depend on the scale of the
matching. In weakly coupled pNRQCD the soft scale is bigger than
$\lQ$ and so the singlet and octet potentials  have  to be
calculated in the perturbative matching. In
\cite{nora_Brambilla:1999qa} a determination of the singlet
potential at three loops leading log has been obtained inside the
EFT which gives the way to deal with the well known infrared
singularity arising in the potential at this order. From this,
$\als$ in the $V$ regularization can be obtained, showing at this
order and  for this regularization a dependence on the infrared
behaviour of the theory. The finite terms in the singlet static
potential at three loops are not yet known but has been estimated
\cite{nora_Chishtie:2001mf}. Recently also the logarithmic
contribution at four loops has been calculated
\cite{nora_Brambilla:2006wp}.
 The three loop renormalization group improved
calculation of the static singlet potential has been compared to
the lattice calculation and found in good agreement up to about
0.25 fm \cite{nora_Pineda:2002se}. The static octet potential is
known at two loops  \cite{nora_Kniehl:2004rk} and again agrees
well with the lattice data \cite{nora_Bali:2003jq}.

At a scale $\mu$ such that $mv \sim \lQ \gg \mu \gg mv^2$,
confinement sets in and the potentials become admixture of
perturbative terms, inherited from NRQCD, which encode high-energy
contributions, and non-perturbative objects. Strongly coupled
pNRQCD gives us the general form of the  potentials obtained in
the nonperturbative matching to QCD in the form of Wilson loops
and Wilson loop chromoelectric and chromomagnetic field strengths
insertions
 \cite{nora_Brambilla:2000gk,nora_Pineda:2000sz}, very well suited for lattice calculations.
These will be in general acomplex valued functions.  The real part
controls the spectrum and the imaginary part controls the decays.

The real part of the potential has been one of the first
quantities to be calculated on the lattice (for a review see
\cite{nora_Bali:2000gf}). In the last year, there has been some
remarkable progress. In \cite{nora_Koma:2006si}, the $1/m$
potential has been calculated for the first time. The existence of
this potential was first pointed out in the pNRQCD framework
\cite{nora_Brambilla:2000gk}. A $1/m$  potential is typically
missing in potential model calculations. The lattice result shows
that the potential has a $1/r$ behaviour, which, in the charmonium
case, is of the same size as the $1/r$ Coulomb tail of the static
potential and, in the bottomonium one, is about 25\%. Therefore,
if the $1/m$ potential has to be considered part of the
leading-order quarkonium potential together with the static one,
as the pNRQCD power counting suggests and the lattice seems to
show, then  the leading-order quarkonium potential would be,
somewhat surprisingly, a flavor-dependent function. In
\cite{nora_Koma:2006fw}, spin-dependent potentials have been
calculated with unprecedented precision. In the long range, they
show, for the first time, deviations from the flux-tube picture of
chromoelectric confinement \cite{nora_Brambilla:1998bt}. The
knowledge of the potentials in pNRQCD could provide an alternative
to the direct determination of the spectrum in NRQCD lattice
simulations: the quarkonium masses would be determined by solving
the Schr\"odinger equation with the lattice potentials. The
approach may present some advantages: the leading-order pNRQCD
Lagrangian, differently from the NRQCD one, is renormalizable, the
potentials are determined once for ever for all quarkonia, and the
solution of the Schr\"odinger equation provides also the
quarkonium wave functions, which enter in many quarkonium
observables: ~ decay widths,~ transitions,~ production
cross-sections. The  existence of a   power counting inside the
EFT  selects the leading and the subleading terms in
quantum-mechanical perturbation theory. Moreover,  the quantum
mechanical  divergences (typically encountered in perturbative
calculations involving iterations of the potentials, as in the
case of the  iterations of spin delta potentials)  are absorbed by
NRQCD matching coefficients. Since a  factorization between the
hard (in the NRQCD matching coefficients) and  soft  scales (in
the Wilson loops or nonlocal gluon correlators) is realized and
since the  low energy objects are only  glue dependent,
confinement investigations, on the lattice and in QCD vacuum
models become feasible \cite{nora_vac,nora_Brambilla:1998bt}.

The potentials evaluated on the lattice once used in the
Schr\"odinger equation produce the spectrum.
The calculations involve only   QCD parameters  (at some scale and in some
scheme).

\subsection{Precision determination of Standard Model parameters}

\subsubsection{$c$ and $b$ mass extraction}
The lowest  heavy quarkonium states are   suitable systems to
extract a precise determination of the mass of the heavy quarks
$b$ and $c$. Perturbative determinations of the $\Upsilon(1S)$ and
$J/\psi$ masses have been used to extract the $b$ and $c$ masses.
The main uncertainty in these determinations comes from
nonperturbative nonpotential contributions (local and nonlocal
condensates) together with possible effects due to subleading
renormalons. These determinations  are competitive with those
coming from different systems and different approaches (for the
$b$ mass see e.g. \cite{nora_El-Khadra:2002wp}). We report some
recent determinations in Table~\ref{Tabmasses}.

\begin{table}[h]
\addtolength{\arraycolsep}{0.2cm} \caption{Different recent
determinations of ${\overline m}_b({\overline m}_b)$ and
${\overline m}_c({\overline m}_c)$ in the $\MS$ scheme from the
bottomonium and the charmonium systems. The displayed results
either use direct determinations or non-relativistic sum rules.
Here and in the text, the $^*$ indicates that the theoretical
input is only partially complete at that order.}
\begin{center}
\begin{tabular}{|c|c|c|}
\hline
reference & order &  ${\overline m}_b({\overline m}_b)$ (GeV)
\\
\hline
\cite{nora_Pineda:2001zq} & NNNLO$^*$ & $4.210 \pm 0.090 \pm 0.025$
\\
\cite{nora_Brambilla:2001qk} & NNLO +charm & $4.190 \pm 0.020 \pm 0.025$
\\
\cite{nora_Eidemuller:2002wk} & NNLO& $4.24 \pm 0.10$
\\
\cite{nora_Penin:2002zv} & NNNLO$^*$& $4.346 \pm 0.070$
\\
\cite{nora_Lee:2003hh} & NNNLO$^*$ & $4.20 \pm 0.04$
\\
\cite{nora_Contreras:2003zb} & NNNLO$^*$ & $4.241 \pm 0.070$
\\
\cite{nora_Pineda:2006gx} & NNLL$^*$ & $4.19 \pm 0.06$
\\
\hline
\hline
reference  & order &  ${\overline m}_c({\overline m}_c)$ (GeV)
\\
\hline
\cite{nora_Brambilla:2001fw} & NNLO & $1.24 \pm 0.020$
\\
\cite{nora_Eidemuller:2002wk}  & NNLO & $1.19 \pm 0.11$
\\
\hline
\end{tabular}
\end{center}
\label{Tabmasses}
\end{table}

A recent analysis  performed by the QWG \cite{nora_Brambilla:2004wf}
and based on all the previous determinations indicates
 that at the moment the mass extraction from  heavy quarkonium
involves  an error of about 50 MeV both for the  bottom
($1\% $ error) and in the charm ($4 \% $ error) mass. It would be
very important to be able to further reduce the error on the heavy quark masses.

\subsubsection{Determinations of  $\als$.}
Heavy quarkonia leptonic and non-leptonic inclusive and radiative decays
 may provide means to extract $\alpha_s$. The present
  PDG determination of $\alpha_s$ from bottomonium pulls down
  the global $\alpha_s$ average  noticeably \cite{nora_Brambilla:2004wf}.
  Recently, using the most recent CLEO data on radiative $\Upsilon(1S)$ decays
   and dealing with the octet contributions within weakly coupled pNRQCD,
a new determination of $\als(M_{\Upsilon(1S)})= 0.184^{+0.014}_{-0.013}$
has been obtained \cite{nora_al}, which corresponds to
$\als(M_Z)= 0.119^{+0.006}_{-0.005}$ in agreement
  with the central value of the PDG and with competitive errors.
A similar extraction of $\als$ from the inclusive photon spectrum
for radiative $J/\psi$ decays may be possible (cf. the
contribution by X.~Garcia~i~Tormo and J.~Soto in this Book in
section~\ref{sec:cc_inclusive_rad}).

\subsubsection{Top-antitop production  near threshold at ILC.}
In \cite{nora_Pineda:2006gx,nora_Hoang:2001mm} the total cross
section for top quark pair production close to threshold in e+e-
annihilation is investigated at NNLL in the weakly coupled EFT.
Here we see how the summation of the large logarithms in the ratio
of the energy scales significantly reduces the scale dependence.
Studies  like these will make  feasible a precise extractions of
the strong coupling, the top mass and the top width at a future
ILC.

\if

We will consider separately systems with a small interquark radius
(low-lying states) and systems with a radius comparable or bigger
than the confinement scale $\lQ^{-1}$ (high-lying states). It is
difficult to say to which group a specific resonance may belong,
since there are no direct measurements of the interquark radius.
Electric dipole transitions or quarkonium dissociation in a
medium, once a well founded theory treatment of such processes
will be given, may give a clear cut procedure. At the moment one
uses the typical EFT approaches assuming  that a particular scales
hierarchy holds  and checking then a posteriori that the
prediction and the error estimated inside such framework are
consistent with the data.

Low-lying $Q\bar{Q}$ states are assumed to realize the hierarchy:
$ m \gg  mv \gg mv^2 \simg \lQ$, where $mv$ is the typical scale
of the inverse distance between the heavy quark and antiquark and
$mv^2$ the typical scale of the binding energy. At a scale $\mu$
such that $mv \gg \mu \gg mv^2$ the effective degrees of freedom
are $Q\bar Q$ states (in color singlet and octet configurations),
low-energy gluons and light quarks.

\begin{table}[ht]
\caption{Different perturbative determinations of the $B_c$ mass compared with the experimental
value and a recent lattice determination.}
\label{TabBc}
\begin{center}
\begin{tabular}{|c|cccc|}
\hline
\multicolumn{5}{|c|}{$B_c$ mass ~(MeV)}\\
\hline
\cite{nora_Acosta:2005us} (expt) &\cite{nora_Allison:2004be} (lattice) & \cite{nora_Brambilla:2000db} (NNLO)
& \cite{nora_Brambilla:2001fw} (NNLO)& \cite{nora_Brambilla:2001qk} (NNLO)\\
\hline
$6287\pm 4.8 \pm 1.1 $ &$6304\pm12^{+12}_{-0}$ & 6326(29) & 6324(22) & 6307(17)
\\
\hline
\end{tabular}
\end{center}
\end{table}

Once the heavy quark masses are known, one may use them to extract
other quarkonium ground-state observables. The $B_c$ mass has been
calculated at NNLO  in
\cite{nora_Brambilla:2000db,nora_Brambilla:2001fw,nora_Brambilla:2001qk}.
These values agree well with the unquenched lattice determination
of \cite{nora_Allison:2004be}, which shows that the $B_c$ mass is
not very sensitive to non-perturbative effects. This is confirmed
by a recent measurement of the $B_c$ in the channel  $B_c \to
J/\psi \, \pi$ by the CDF collaboration at the Tevatron; they
obtain with 360 pb$^{-1}$ of data $M_{B_c} = 6285.7 \pm 5.3 \pm
1.2$ MeV \cite{nora_Acosta:2005us}, while the latest available
figure based on 1.1 fb$^{-1}$ of data is $M_{B_c} = 6276.5 \pm 4.0
\pm 2.7$ MeV (see
http://www-cdf.fnal.gov/physics/new/bottom/060525.blessed-bc-mass/).
See Table \ref{TabBc}.

The bottomonium and charmonium ground-state hyperfine splitting
has been calculated at NLL in \cite{nora_Kniehl:2003ap}. Combining
it with the measured $\Upsilon(1S)$ mass, this determination
provides a quite precise prediction for the $\eta_b$ mass:
$M_{\eta_b} = 9421 \pm 10^{+9}_{-8} ~{\rm MeV}$, where the first
error is an estimate of the theoretical uncertainty and the second
one reflects the uncertainty in $\als$. Note that the discovery of
the $\eta_b$ may provide a very competitive source of $\als$ at
the bottom mass scale with a projected error at the $M_Z$ scale of
about $0.003$. Similarly, in \cite{nora_Penin:2004xi}, the
hyperfine splitting of the $B_c$ was calculated at NLL accuracy:
$M_{B_c^*}  - M_{B_c} = 65 \pm 24^{+19}_{-16}~{\rm MeV}$.

High-lying $Q\bar{Q}$ states are assumed to realize the hierarchy:
$ m \gg  mv \sim \lQ \gg mv^2$. A first question is where the
transition from low-lying to high-lying takes place. This is not
obvious, because we cannot measure directly $mv$. Therefore, the
answer can only be indirect and, so far, there is no clear
agreement in the literature. A weak-coupling treatment for the
lowest-lying bottomonium states ($n=1$, $n=2$ and also for the
$\Upsilon(3S)$) appears to give positive results for the masses
at NNLO in \cite{nora_Brambilla:2001fw} and at N$^3$LO$^*$ in
\cite{nora_Penin:2005eu}. The result is more ambiguous for the
fine splittings of the bottomonium $1P$ levels in the NLO analysis
of \cite{nora_Brambilla:2004wu} and positive only for the
$\Upsilon(1S)$ state in the  N$^3$LO$^*$ analysis of
\cite{nora_Beneke:2005hg}.

Masses of high-lying quarkonia may be accessed by lattice calculations.

Allowed magnetic dipole transitions between charmonium and
bottomonium ground states have been considered at NNLO  in
\cite{nora_Brambilla:2005zw,nora_Vairo:2006js}. The results are:
$\Gamma(J/\psi \to \gamma \, \eta_c) \! = (1.5 \pm
1.0)~\hbox{keV}$ and $\Gamma(\Upsilon(1S) \to \gamma\,\eta_b)$ $=$
$(k_\gamma/39$ $\hbox{MeV})^3$ $\,(2.50 \pm 0.25)$ $\hbox{eV}$,
where  the errors account for uncertainties (which are large in
the charmonium case) coming from higher-order corrections. The
width $\Gamma(J/\psi \to \gamma\,\eta_c)$ is consistent with
\cite{nora_Yao:2006px}. Concerning $\Gamma(\Upsilon(1S) \to
\gamma\,\eta_b)$, a photon energy $k_\gamma = 39$ MeV corresponds
to a $\eta_b$ mass of 9421 MeV.

In the weak-coupling regime, the magnetic-dipole hindered transition
$\Upsilon(2S) \to \gamma\,\eta_b$
at leading order \cite{nora_Brambilla:2005zw} does not agree with the experimental upper bound
\cite{nora_Artuso:2004fp}, while the ratios for different $n$ of the radiative decay widths
$\Gamma(\Upsilon(nS) \to \gamma\,X)$ are better consistent with
the data if $\Upsilon(1S)$ is assumed to be a weakly-coupled bound state
and $\Upsilon(2S)$ and $\Upsilon(3S)$ strongly coupled ones \cite{nora_GarciaiTormo:2005bs}.

The radiative transition $\Upsilon(1S)\to\gamma\,X$ has been
considered in \cite{nora_Fleming:2002sr,nora_GarciaiTormo:2005ch}.
The agreement with the CLEO data of \cite{nora_Nemati:1996xy} is
very satisfactory (cf. the contribution by N.~Brambilla, Y.~Jia
and A.~Vairo in this Book).


The imaginary part of the potential provides the NR\-QCD decay matrix elements
in pNRQCD.

For excited states, they typically factorize in a part, which is the wave function in
the origin squared (or its derivatives), and in a part which contains
gluon tensor-field correlators
\cite{nora_Brambilla:2001xy,nora_Brambilla:2002nu,nora_Brambilla:2003mu,nora_Vairo:2003gh}.
This drastically reduces the number of non-perturbative parameters needed; in pNRQCD,
these are wave functions at the origin and universal gluon tensor-field correlators,
which can be calculated on the lattice.
Another approach may consist in determining the correlators on one set
of data (e.g. in the charmonium sector) and use them to make predictions
for another (e.g. in the bottomonium sector). Following this line in
\cite{nora_Brambilla:2001xy,nora_Vairo:2002nh}, at NLO in $\als$, but at leading
order in the velocity expansion, it was predicted
${\Gamma_{\rm had}(\chi_{b0}(2P))}/{\Gamma_{\rm had}(\chi_{b2}(2P))} \approx$
 $4.0$ and ${\Gamma_{\rm had}(\chi_{b1}(2P))}/$ ${\Gamma_{\rm
had}(\chi_{b2}(2P))} \approx$ $0.50$. Both determinations turned
out to be consistent, within large errors, with the CLEO III data
\cite{nora_Brambilla:2004wf}.

For the lowest resonances, inclusive decay widths are given in
weakly coupled pNRQCD by a convolution of perturbative corrections
and nonlocal nonperturbative correlators. The perturbative
calculation embodies large contributions and requires large logs
resummation. The ratio of electromagnetic decay widths was
calculated for the ground state of charmonium and bottomonium at
NNLL order in \cite{nora_Penin:2004ay}. In particular, they
report: $\Gamma(\eta_b\to\gamma\gamma) /\Gamma(\Upsilon(1S)\to
e^+e^-) = 0.502 \pm 0.068 \pm 0.014$, which is a very stable
result with respect to scale variation. A partial  NNLL$^*$ order
analysis of the absolute width of $\Upsilon(1S) \to e^+e^-$ can be
found in \cite{nora_Pineda:2006ri}.

The SELEX collaboration at Fermilab reported evidence of five resonances that
may possibly be identified with doubly charmed baryons  \cite{nora_Ocherashvili:2004hi}.
Although these findings have not been confirmed by other experiments (notably
by FOCUS, BELLE and BABAR) they have triggered a renewed
theoretical interest in doubly heavy baryon systems.

Low-lying $QQq$ states are assumed to realize the hierarchy: $ m \gg mv \gg
\lQ$, where $mv$ is the typical inverse distance between the two heavy quarks
and $\lQ$ is the typical inverse distance between the centre-of-mass of the two heavy quarks
and the light quark.
At a scale $\mu$ such that $mv \gg \mu \gg \lQ$ the effective
degrees of freedom are $QQ$ states (in color antitriplet and sextet
configurations), low-energy gluons and light quarks. The most suitable EFT at
that scale is a combination of pNRQCD and HQET
\cite{nora_Brambilla:2005yk,nora_Fleming:2005pd}. The hyperfine splittings of the doubly heavy
baryon lowest states have been calculated at NLO in $\als$ and at LO in
$\lQ/m$ by relating them to the hyperfine splittings of the $D$ and $B$ mesons (this
method was first proposed in \cite{nora_Savage:di}). In \cite{nora_Brambilla:2005yk}, the
obtained values are: $M_{\Xi^*_{cc}}-M_{\Xi_{cc}} = 120 \pm 40$ MeV
and $M_{\Xi^*_{bb}}-M_{\Xi_{bb}} = 34 \pm 4$ MeV, which are
consistent with the quenched lattice determinations of
\cite{nora_Flynn:2003vz,nora_Lewis:2001iz,nora_AliKhan:1999yb,nora_Mathur:2002ce}.
Chiral corrections to the doubly  heavy baryon masses, strong decay widths and
electromagnetic decay widths have been considered in \cite{nora_Hu:2005gf}.

Also low-lying $QQQ$ baryons can be studied in a weak coupling framework.
Three quark states can combine in four color configurations: a singlet,
two octets and a decuplet, which lead to a rather rich dynamics
\cite{nora_Brambilla:2005yk}. Masses of various $QQQ$ ground states have been
calculated with a variational method in \cite{nora_Jia:2006gw}: since baryons made of three
heavy quarks have not been discovered so far, it may be important for future searches
to remark that the baryon masses turn our to be lower
than those generally obtained in strong coupling analyses.
For $QQQ$ baryons with a typical distance of the order $\lQ$ inverse, the form of the
static, $1/m$ and spin dependent nonperturbative potentials have been obtained in pNRQCD
\cite{nora_Brambilla:2005yk}. Up to now only the static potential has been evaluated on the lattice
\cite{nora_Bali:2000gf,nora_Suganuma:2004zx}.

\fi

\subsection{Gluelump Spectrum and exotic states}

The gluelumps are states formed by a gluon and two hravy quarks in
a octet configuration at small interquark distance
\cite{nora_Foster:1998wu}. The mass of such nonperturbative
objects are typically measured on the lattice. The tower of
hybrids static energies \cite{nora_Juge:2002br}
 reduces to the gluelump masses for small interquark distances.
In pNRQCD \cite{cyq_Brambilla:1999xf,nora_Bali:2003jq}
 the full structure of the gluelump spectrum has been studied,
obtaining model independent predictions on the shape, the pattern,
the degeneracy and the multiplet structure of the hybrid static
energies for small $Q\bar{Q}$ distances that well match and
interpret the existing lattice data. These studies may be
important both to  elucidate the confinement mechanism (the
gluelump masses control the behaviour of the nonperturbative glue
correlators appearing in the spectrum and in the decays) and in
relation to the exotic states recently observed at the
B-factories. The $Y(4260)$  in the charmonium sector may be
identified with an hybrid state inside such picture
\cite{nora_Vairo:2006pc}.

\subsection{Outlook}
pNRQCD makes possible to investigate a wide range of heavy
quarkonium observables in a controlled and systematic fashion and
therefore it makes possible  to learn about one of the most
elusive sectors of the Standard Model: low-energy QCD. Among the
most interesting and open theory challenges there are:  the
construction of   an EFT close to threshold to understand the new
exotic states, the taming of  quarkonium  production and the
development of  an EFT approach   to quarkonium suppression in
media and quarkonium-nuclei interaction.

The many new data coming in the last few years from B-Factories,
CLEO, BES, HERA and the Tevatron experiment have given us glimpses
of interesting physics to be explored. With the new theory tools
discussed here and with the impressive number of  produced and
detected charmonium states, BES-III will make the difference in
this field allowing  to carry on important investigation inside
the Standard Model and beyond the Standard Model (see
\cite{nora_Brambilla:2004wf} Chapter Beyond the Standard Model and
\cite{nora_Sanchis-Lozano:2005fj}).

\chapter[Charmonium Spectroscopy]
{Charmonium Spectroscopy}
\label{sec:cc_spectrum}
\def\be{\begin{equation}}
\def\ee{\end{equation}}
\def\bd{\begin{displaymath}}
\def\ed{\end{displaymath}}
\def\ba{\begin{eqnarray}}
\def\ea{\end{eqnarray}}
\def\ccbar{$c\bar c$ }
\def\ra{\rightarrow }
\def\rt{\rightarrow }
\def\etac{\eta_c }
\def\etacp{\eta^{\prime}_c }
\def\jpsi{J/\psi }
\def\psp{\psi^{\prime} }
\def\psip{\psi^{\prime} }
\def\pipi{\pi^+\pi^-}
\def\mumu{\mu^+\mu^-}
\def\hc{h_c }

\def\C{\rm C}
\def\D{\rm D}
\def\F{\rm F}
\def\I{\rm I}
\def\J{\rm J}
\def\L{\rm L}
\def\M{\rm M}
\def\P{\rm P}
\def\S{\rm S}
\def\T{\rm T}
\def\X{\rm X}


\section[Charmonium spectroscopy]
{Introduction\footnote{By T. Barnes}}

Historically~\cite{ted_Aubert:1974js,ted_Augustin:1974xw,ted_Appelquist:1974zd,
ted_DeRujula:1974nx,ted_Appelquist:1974yr,ted_Eichten:1974af}
charmonium has played an important r\^ole in the study of the
strong interaction, including the search for exotica, and in the
development of our understanding of the forces between quarks. The
spectrum of relatively narrow charmonium states that cannot decay
into open-charm modes is experimentally very clear; this has made
it possible to precisely measure the masses of states that in the
quark model are identified with N,L multiplets, and in addition
the effects of the weaker spin-dependent forces (spin-spin,
spin-orbit and tensor) in splitting these multiplets into states
of definite S and J can also be quantified. Not only are the
effects of the Breit-Fermi Hamiltonian (which is due to one gluon
exchange, ``OGE") evident in the spectrum, but the Lorentz nature
of the confining interaction itself has also been established (as
scalar rather than timelike vector).

In future studies of charmonium spectroscopy at \bes3S it should be
possible to discover many charmonium states that are expected
theoretically but have not yet been observed. Some of the as yet
unknown states are expected to be very narrow resonances. In
addition one can presumably identify novel exotica that are
currently unknown or poorly understood, such as charm molecules
and charmonium hybrids. Our understanding of the known charmonium
states can be greatly improved through more precise measurements
of transitions involving these states, including both radiative
and strong couplings. These studies have the potential to greatly
improve our understanding of strong decays in particular. In this
section we will review the current status of our understanding of
charmonium and closely related systems, and will discuss some ways
in which \bes3 can contribute to our understanding of QCD through
future experimental studies of charmonium.

\section{Conventional charmonium states}


The spectrum of the known ``conventional" charmonium states, by
which we mean states that appear to be in relatively good
agreement with the predictions of simple $c\bar c$ potential
models, is summarized in Table~\ref{Table_expt_spectrum}. The
table, which is abstracted from the 2006 PDG \cite{ted_PDG2006},
gives the usual quark model N,L multiplet assignment, the quark
model spectroscopic assignment, the PDG name, and the mass and
width of each state.

\begin{table}[htbp]
\caption{Experimental spectrum of reasonably well established
``conventional" \ccbar states. This table gives the usual quark
model multiplet and spectroscopic assignment for each state and
the PDG label for the state, followed by their masses and widths.
Partner states that are expected theoretically but not yet
identified experimentally are indicated by a dash.}
\label{Table_expt_spectrum}
\begin{center}
\begin{tabular}{ccc|lc}
\hline
\hline
& & & &
\\
Mult.
& Spec.
& \quad Expt. state \quad
& \qquad Mass [MeV] \quad
& Width [MeV]
\\
& & & &
\\
\hline
1S
&  $1^3{\rm S}_1 $
&  $J/\psi(1S) $
& \quad $ 3096.916 \pm 0.011 $  \qquad
& $ 0.0934 \pm 0.0021$
\\

&  $1^1{\rm S}_0 $
&  $\eta_c(1S) $
& \quad $ 2980.4 \pm 1.2   $
& $ 25.5 \pm 3.4   $
\\
\hline
1P
&  $1^3{\rm P}_2 $
&  $\chi_{c2}(1P) $
& \quad $ 3556.20 \pm 0.09 $
& $2.06\pm 0.12$
\\

&  $1^3{\rm P}_1 $
&  $\chi_{c1}(1P) $
& \quad $ 3510.66 \pm 0.07 $
& $ 0.89 \pm 0.05$
\\

&  $1^3{\rm P}_0 $
&  $\chi_{c0}(1P) $
& \quad $ 3414.76 \pm 0.35   $
&  $ 10.4 \pm 0.7$
\\

&  $1^1{\rm P}_1 $
&  $h_c(1P) $
& \quad $ 3525.93 \pm 0.27   $
& $ < 1$
\\
\hline
2S
&  $2^3{\rm S}_1 $
&  $\psi(2S)$
& \quad $ 3686.093 \pm 0.034 $
& $ 0.337 \pm 0.013 $
\\

&  $2^1{\rm S}_0 $
&  $\eta_c(2S) $
& \quad $ 3638 \pm 4   $
& $14\pm 7$
\\
\hline
1D
&  $1^3{\rm D}_3 $
&  -
&
&
\\

&  $1^3{\rm D}_2 $
&  -
&
&
\\

&  $1^3{\rm D}_1 $
&  $\psi(3770) $
& \quad  $ 3771.1 \pm 2.4  $
& $23.0\pm 2.7$
\\

&  $1^1{\rm D}_2 $
&  -
&
&
\\
\hline
2P
&  $2^3{\rm P}_2 $
&  $\chi_{c2}(2P) $
& \quad $ 3929 \pm 5 \pm 2 $
& $29\pm 10\pm 2$
\\

&  $2^3{\rm P}_1 $
&  -
&
&
\\

&  $2^3{\rm P}_0 $
&  -
&
&
\\

&  $2^1{\rm P}_1 $
&  -
&
&
\\

\hline
3S
&  $3^3{\rm S}_1 $
&  $\psi(4040) $
& \quad $ 4039 \pm 1      $
& $80\pm 10$
\\

&  $3^1{\rm S}_0 $
&  -
&
&
\\
\hline
2D
&  $2^3{\rm D}_3 $
&  -
&
&
\\

&  $2^3{\rm D}_2 $
&  -
&
&
\\

&  $2^3{\rm D}_1 $
&  $\psi(4160) $
& \quad  $ 4153 \pm 3 $
& $103 \pm 8 $
\\

&  $2^1{\rm D}_2 $
&  -
&
&
\\
\hline
4S
&  $4^3{\rm S}_1 $
&  $\psi(4415) $
& \quad $ 4421 \pm 4   $
& $62\pm 20$
\\

&  $4^1{\rm S}_0 $
&  -
&
&
\\
\hline
\hline
\end{tabular}
\end{center}
\end{table}

\begin{table}[htbp]
\caption{Theoretical predictions for the spectrum of \ccbar states
in a nonrelativistic potential model (NR) and the Godfrey-Isgur
relativized potential model (GI). (This table is abstracted from
Ref.\cite{ted_Barnes:2005pb}.)} \label{Table_spectrum}
\begin{center}
\begin{tabular}{cr|c|cc}
\hline
\quad Mult. \quad
& State \phantom{x}
& Input mass
& \multicolumn{2}{c}{Theor. mass}
\\
&
& (NR model)
& \phantom{xx}NR\phantom{xx}
& \phantom{xx}GI \phantom{xx}
\\
\hline
1S &  $J/\psi(1^3{\rm S}_1) $ & 3097 & 3090 & 3098\\
   &  $\eta_c(1^1{\rm S}_0) $ & 2979 & 2982 & 2975\\
\hline
2S &  $\psi'(2^3{\rm S}_1) $  & 3686 & 3672 & 3676\\
   &  $\eta_c'(2^1{\rm S}_0)$ & 3638 & 3630 & 3623\\
\hline
3S &  $\psi(3^3{\rm S}_1) $   & 4040 & 4072 & 4100\\
   &  $\eta_c(3^1{\rm S}_0) $ &      & 4043 & 4064\\
\hline
4S &  $\psi(4^3{\rm S}_1) $   & 4415 & 4406 & 4450\\
   &  $\eta_c(4^1{\rm S}_0) $ &      & 4384 & 4425\\
\hline
1P &  $\chi_2(1^3{\rm P}_2) $ & 3556 & 3556 & 3550\\
   &  $\chi_1(1^3{\rm P}_1) $ & 3511 & 3505 & 3510\\
   &  $\chi_0(1^3{\rm P}_0) $ & 3415 & 3424 & 3445\\
   &  $h_c(1^1{\rm P}_1) $    &      & 3516 & 3517\\
\hline
2P &  $\chi_2(2^3{\rm P}_2) $ &      & 3972 & 3979\\
   &  $\chi_1(2^3{\rm P}_1) $ &      & 3925 & 3953\\
   &  $\chi_0(2^3{\rm P}_0) $ &      & 3852 & 3916\\
   &  $h_c(2^1{\rm P}_1) $    &      & 3934 & 3956\\
\hline
3P &  $\chi_2(3^3{\rm P}_2) $ &      & 4317 & 4337\\
   &  $\chi_1(3^3{\rm P}_1) $ &      & 4271 & 4317\\
   &  $\chi_0(3^3{\rm P}_0) $ &      & 4202 & 4292\\
   &  $h_c(3^1{\rm P}_1) $    &      & 4279 & 4318\\
\hline
1D &  $\psi_3(1^3{\rm D}_3) $ &      & 3806 & 3849\\
   &  $\psi_2(1^3{\rm D}_2) $ &      & 3800 & 3838\\
   &  $\psi(1^3{\rm D}_1) $   & 3770 & 3785 & 3819\\
   & $\eta_{c2}(1^1{\rm D}_2)$&      & 3799 & 3837\\
\hline
2D &  $\psi_3(2^3{\rm D}_3) $ &      & 4167 & 4217\\
   &  $\psi_2(2^3{\rm D}_2) $ &      & 4158 & 4208\\
   &  $\psi(2^3{\rm D}_1) $   & 4159 & 4142 & 4194\\
   & $\eta_{c2}(2^1{\rm D}_2)$&      & 4158 & 4208\\
\hline
1F &  $\chi_4(1^3{\rm F}_4) $ &      & 4021 & 4095\\
   &  $\chi_3(1^3{\rm F}_3) $ &      & 4029 & 4097\\
   &  $\chi_2(1^3{\rm F}_2) $ &      & 4029 & 4092\\
   &  $h_{c3}(1^1{\rm F}_3) $ &      & 4026 & 4094\\
\hline
2F &  $\chi_4(2^3{\rm F}_4) $ &      & 4348 & 4425\\
   &  $\chi_3(2^3{\rm F}_3) $ &      & 4352 & 4426\\
   &  $\chi_2(2^3{\rm F}_2) $ &      & 4351 & 4422\\
   &  $h_{c3}(2^1{\rm F}_3) $ &      & 4350 & 4424\\
\hline
1G &  $\psi_5(1^3{\rm G}_5) $ &      & 4214 & 4312\\
   &  $\psi_4(1^3{\rm G}_4) $ &      & 4228 & 4320\\
   &  $\psi_3(1^3{\rm G}_3) $ &      & 4237 & 4323\\
   & $\eta_{c4}(1^1{\rm G}_4)$&      & 4225 & 4317\\
\hline
\hline
\end{tabular}
\end{center}
\end{table}

This spectrum of states can be described surprisingly well using a
simple \ccbar potential model. In these models one typically
assumes a zeroth-order (spin-independent) potential that combines
a OGE color Coulomb term and a linear confining interaction, \be
V_0^{(c\bar c)}(r) = -\frac{4}{3}\frac{\alpha_s}{r} + br. \ee This
is augmented by the spin-dependent Breit-Fermi Hamiltonian due to
OGE, and an inverted spin-orbit term that arises from the
(assumed) scalar nature of the confining interaction, \be
V_{spin-dep} = \frac{32\pi\alpha_s}{9 m_c^2}\, \vec {\S}_c \cdot
\vec {\S}_{\bar c}\, \delta(\vec x\,)\, + \frac{1}{m_c^2}\, \bigg[
\Big( \frac{2\alpha_s}{r^3} - \frac{b}{2 r} \Big) \, \vec {\L}
\cdot \vec {\S} + \frac{4\alpha_s}{r^3} \, \T \ \bigg] . 
\label{eqn:spin-dependent} \ee Two
examples of the spectra predicted by this type of model, a fully
nonrelativistic potential model ``NR" and the relativized
Godfrey-Isgur model \cite{ted_Godfrey:1985xj} ``GI"  are shown in
Fig.\ref{ccbar_spectrum}. (Here and in much of our discussion of
charmonia we will refer to the \ccbar potential model of
Ref.\cite{ted_Barnes:2005pb}, since potential models give
predictions for many of the observed properties of charmonium
resonances. Much of the original literature is also cited in this
reference, and should be reviewed for a more complete discussion.
We also note that several excellent and very extensive reviews of
charmonia have appeared
\cite{ted_Quigg:2004nv,ted_Galik:2004mi,ted_Brambilla:2004wf,ted_Seth:2005ak,ted_Swanson:2006st,ted_Rosner:2006sv,ted_Seth:2006si}
which discuss recent developments in the field in detail.)

\begin{figure}[htbp]
\centerline{
\includegraphics[width=14cm,angle=0,clip=]{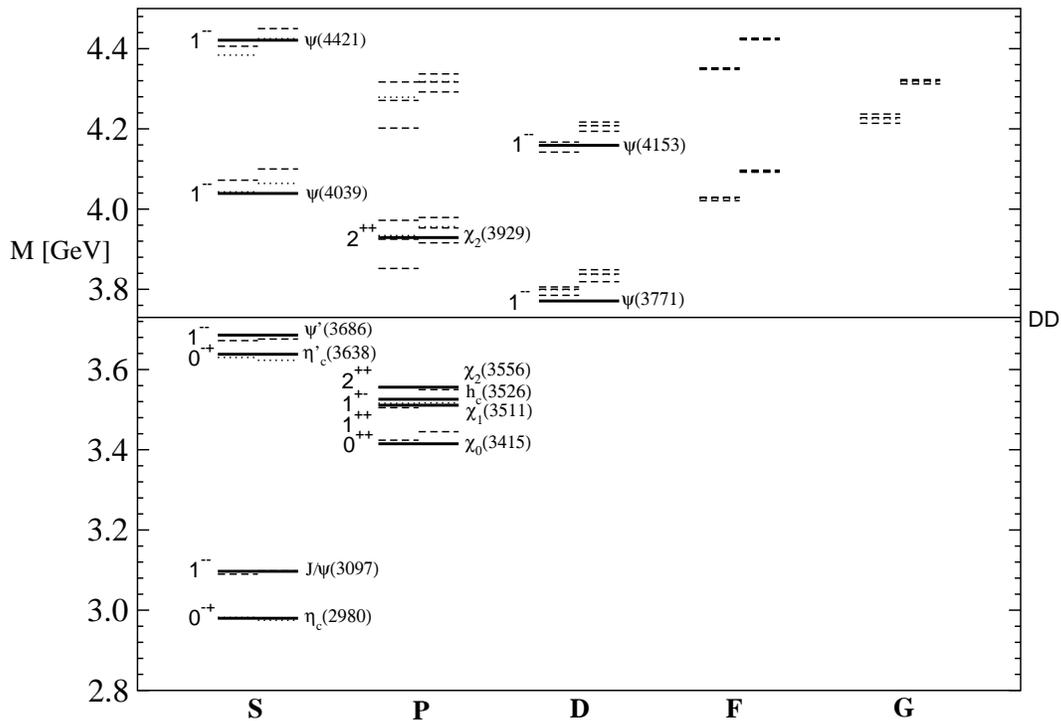}
} \caption{Predicted and observed spectrum of charmonium states
(Table~\ref{Table_spectrum}). The solid lines are experiment for
reasonably well-established charmonium states, and the particle
labels give the actual 2006 PDG masses in MeV rather than the
usual rounded PDG mass labels. The broken lines are theory (NR
model left, GI right). In the theoretical spectra, spin triplet
levels are dashed and spin singlets are dotted. The DD open-charm
threshold at 3.73 GeV is also shown.} \label{ccbar_spectrum}
\end{figure}

The experimental and theoretical masses (in the two models of
Table~\ref{Table_spectrum}) are shown in
Fig.~\ref{ccbar_spectrum}. Evidently there is reasonably good
agreement between the current experimental spectrum and the
potential model predictions. Comparison between the
nonrelativistic and relativized models shows some discrepancies,
notably in the scales of 2P and 3P multiplet splittings and in the
positions of the higher-L multiplets. The immediate experimental
tasks suggested by this spectrum are the identification of the
remaining 2P states, the 3P states and the higher-L levels.
Identification of the remaining 2P states (those with C=(+)) can
be carried out at \bes3 using E1 radiative transitions from the
higher $1^{--}$ levels $\psi(4040)$ and $\psi(4160)$).
Identification of the 3P levels will be more complicated because
hybrid charmonia are expected at similar masses; we shall see that
this may be possible using $\psi(4415)$ radiative transitions.
Finally, the important problem of the identification of the
higher-L states is a difficult problem. It may be possible to
identify one such state, the ${}^3$F$_2$ $c\bar c$, in radiative
transitions from the $\psi(4160)$. Experimentally accessing the
other higher-L $c\bar c$ states is an unsolved problem.

It is also possible to predict the spectrum of charmonia,
especially the lowest-lying states in each sector of state space,
using lattice gauge theory (LGT); for some recent references see
Refs.{\cite{ted_Okamoto:2001jb,ted_diPierro:2003bu,ted_Umeda:2003ia,ted_Dudek:2006gb}}.
This approach is very attractive in that it does not require the
assumptions of potential models, and can in addition be applied to
the study of novel states such as the $\J^{\P\C}$-exotic
charmonium hybrids, where potential models are of unknown
accuracy. To date the results of LGT for the charmonium spectrum
are very similar to the predictions of potential models for the
lighter states, which are experimentally well established. Current
LGT predictions for the more controversial higher-mass states
unfortunately have rather large errors (100~MeV is typical at
present), so they cannot yet usefully be compared to experimental
candidates or potential model calculations. Future LGT
calculations of the charmonium spectrum should be very interesting
in this regard. One very interesting question that LGT can
hopefully address will be the importance of quenching (the effect
of closed $q\bar q$ loops on the spectrum and couplings of
charmonia). The closely related question of the importance of
closed virtual continuum channels on the properties of heavy
quarkonia and hadrons more generally is a very important and
controversial issue.


\subsection{Particle widths}

Charmonium states that are below their open-charm decay threshold
(all states below $2m_D = 3.73$~GeV, and in addition the lightest
$2^{-+}$ and $2^{--}$ 1D states) must decay through annihilation
of the $c\bar c$ pair. The total widths in these decays have
traditionally been described as annihilation into gluons, using
the corresponding formulas for positronium annihilation into
photons but with $\alpha_s$ vertices and combinatoric color
factors.

These calculations are dubious for several reasons. First, the
assumption of free gluons final states and pQCD dynamics is
presumably a bad approximation, since the actual glueball spectrum
has widely spaced discrete levels and a mass gap that is about
half as large as charmonium itself. Second, when higher-order pQCD
radiative corrections are incorporated, it is often found that
they are so large as to make the numerical predictions clearly
unreliable. Finally, the use of positronium formulas is a
nonrelativistic approximation, which is not well justified for the
charmonium system.

Despite these many caveats these pQCD predictions for charmonium
annihilation total widths remain interesting as rough guidelines,
and some of the qualitative predictions do seem to be satisfied.
For example, since $\alpha_s$ is not large at the $c\bar c$ mass
scale one expects that the annihilation decay widths of negative
C-parity charmonia (which must decay into at least three gluons)
should be rather smaller than the decay widths of similar positive
C-parity charmonia (which can decay into two gluons). Inspection
of the total widths of well established \ccbar states below
3.73~GeV in Table~\ref{Table_expt_spectrum} shows that this
guideline is indeed well satisfied.

Similar positronium annihilation rate formulas for other states
can be adapted to give strong (gluonic and $q\bar qg$) total
annihilation width estimates for other heavy quarkonia. These lead
to the expectation that the two as yet unknown D-wave \ccbar
states $h_{c2}$ ($2^{-+}$ ${}^1\D_2$) and $\psi_2$ ($2^{--}$
${}^3\D_2$) should be quite narrow, with strong annihilation
widths of perhaps $\sim 1$~MeV~\cite{ted_Barnes:2003vb}. These
predictions of very narrow widths make the discovery of the
$h_{c2}$ and $\psi_2$ two of the most important outstanding
experimental goals of charmonium spectroscopy. We note in passing
that their 1D partner state $\psi_3$ ($3^{--}$ ${}^3\D_3$) should
have a comparably small total width; it can decay to DD, but the
L=3 angular momentum barrier should make this a narrow state as
well.

Discovery of these narrow states may be feasible at \bes3.
Production of the three narrow 1D \ccbar states may be possible
using E1 radiative transitions from the appropriate 2P \ccbar
states; once formed, these narrow 1D states may then be detected
through their large E1 branching fractions to the 1P \ccbar
multiplet. (Ref.\cite{ted_Barnes:2005pb} gives E1 radiative
partial width estimates for all these transitions.)


\section[Issues for \bes3]
{Issues for \bes3\footnote{By S.L. Olsen}}

\bes3 is well suited to address the remaining experimental questions
that are related to the low-mass ({\it i.e.} below open-charm
threshold) charmonium spectrum.  These include
precise determinations of the masses and widths of the $\etac$
and $\etacp$, and the relation of
the $\hc$ mass relative to the center-of-gravity mass of the
$\chi_{cJ}$ states.  In addition, we advocate a precision 
measurement of the partial width $\Gamma(\jpsi\rt\gamma\etac)$. 

\subsection{Measurements related to the $\etac$}

Although the $\etac$ is the ground  state $c\bar{c}$ meson and, 
in some sense,  the most fundamental of the charmonium mesons, 
its properties  are very poorly understood,
both theoretically and experimentally.  

\subsubsection{Mass of the $\etac$}
In spite of the
fact that the PDG lists more than 20 measurements of the $\etac$
mass~\cite{ted_PDG2006}, the world average value, 
$2979.8\pm 1.2$~MeV has a precision
that is poorer, by an order-of-magnitude or more than the
listed mass values for the $\jpsi$, $\psp$, $h_c$, $\chi_{c1}$ or
$\chi_{c2}$.  The agreement between different measurements is poor;
the PDG group's fit to the measured values has a confidence level of
0.6\% (see Fig.~\ref{fig:etac_M_G_PDG}).  Moreover, the mass value
itself is somewhat mysterious.  From 
Eqn.~\ref{eqn:spin-dependent},
the potential model expectation for the $\jpsi$-$\etac$ mass splitting is
\be
M(\jpsi) - M(\etac) =
\frac{32\pi\alpha_s(m_c)}{9 m^2_c}|\Psi(0)|^2,
\ee
where $|\Psi(0)|$, the wave function at the origin,
can be determined from the $\jpsi\ra e^+e^-$ partial
width via
\be
\Gamma(\jpsi\ra e^+e^-) =
\frac{16\pi\alpha^2 q^2_c|\Psi(0)|^2}{M^2_{\jpsi}}
(1 - \frac{16\alpha_s}{3\pi} + ...).
\ee
Using the world average value
$\Gamma(\jpsi\ra e^+e^-) = 5.55\pm0.14$~keV,
$\alpha_s(M^2_{\jpsi})\simeq 0.33$,
and $m_c\simeq M_{\jpsi}/2$, we find a
prediction for the mass splitting of
$M(\jpsi)-M(\etac)\sim 60$~MeV, which is about
half the measured value of $116.5\pm 1.2$~MeV.
This discrepancy is also seen in lattice QCD 
calculations~\cite{choe:2003slo}.

\begin{figure}[htbp]
\centerline{
\includegraphics[width=14cm,angle=0,clip=]{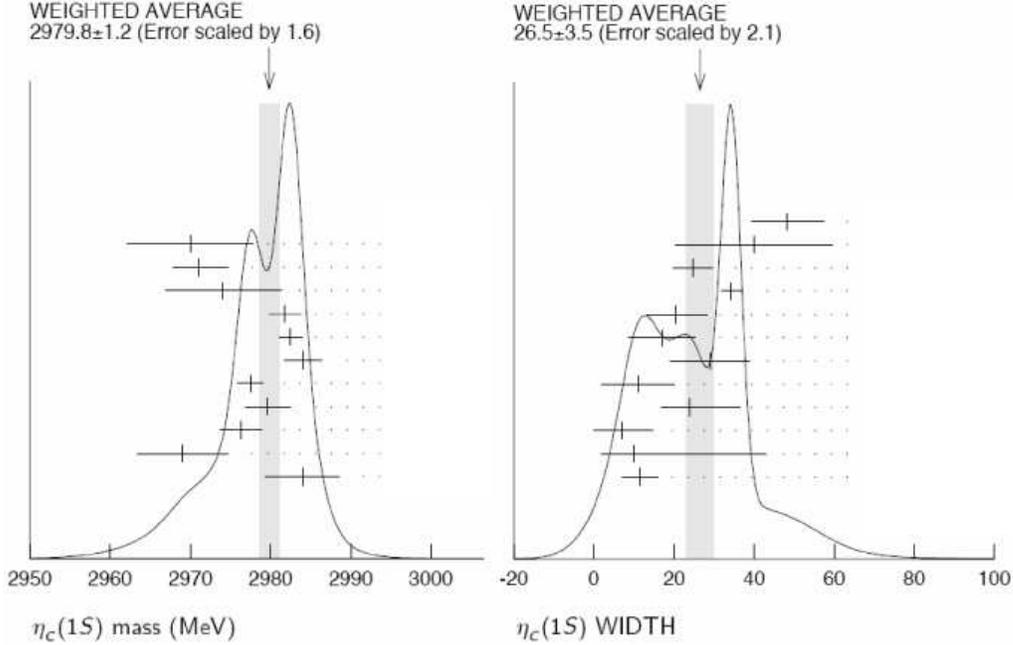}
}  
\caption{
The distribution of measurements of the $\eta_c$ mass (left)
and width (right) from PDG-2006.}  
\label{fig:etac_M_G_PDG}
\end{figure}

\subsubsection{Width of the $\etac$}
The experimental situation on the $\eta_c$ total width is even more
confused.  The right panel of Fig.~\ref{fig:etac_M_G_PDG} shows
the range of $\eta_c$ total width measurements listed in the PDG
ables.  The PDG fit for a world average value of
$\Gamma = 26.5\pm 3.5$~MeV has a confidence level of only 0.1\%,
and, so, the errors on the measurements going into
the averaging process are scaled by a factor of 2.1. Here
there is 
a notable discrepancy between measurements using $\etac$ mesons
produced from the radiative decay $\jpsi\ra\gamma\etac$, which
tend to give low values, and those where the $\etac$ is produced
directly by two-photons, {\it i.e.} $\gamma\gamma\ra\etac$, or
in $B$ mesons decays via $B\ra K\etac$, which give higher values.

\subsubsection{The $\Gamma(\psi\rt\gamma\etac)$ partial width} 

In the potential model, radiative decays of the $\jpsi$ 
and $\psip$ to the $\etac$ states, 
{\it i.e.,} $\psi(nS)\rt\gamma\eta_c(n'S)$, 
proceed primarily via magnetic dipole (M1) 
transitions, with an amplitude given by
\be
\Gamma(\psi(nS)\ra \gamma \eta_c(n'S)) =
\frac{4}{3}\alpha q^2_c\frac{k^3_{\gamma}}{m^2_c}\vert \int dr~r^2~
R_{n^{\prime}0}(r)R_{n0}(r)~j_0(\frac{k_{\gamma}r}{2})\vert^2 .
\label{fig:M1_width}
\ee
Here $q_c$ and $m_c$ are the charge and mass of the $c$-quark
and $k_{\gamma}$ is the transition photon energy.  For
$\jpsi\rt\gamma\etac$, $n=n'=1$ and, in the limit of
zero recoil momentum, {\it i.e.} $k_{\gamma}r\simeq 0$, the
overlap integral is unity and the $\etac$ line shape is
modulated by a factor of $k^3_{\gamma}$. 

Taking $m_c=M_{\jpsi}/2$ and using  the PDG world
average values for the $\etac$ mass and $\jpsi$ 
mass and total width,
we determine the predicted partial width value from 
Eqn.~\ref{fig:M1_width} to be
\be
\Gamma(\jpsi\rt\gamma\etac) = 2.90 ~ {\rm keV~~~~Prediction},
\ee
which would correspond to the branching fraction
${\mathcal B}(\Gamma(\jpsi\rt\gamma\etac) = 3.1\%$.
The PDG value for this branching fraction
is based solely on a single measurement from
the Crystal Ball group~\cite{xtal-ball_gaiser:1986slo} and is 
${\mathcal B}(\Gamma(\jpsi\rt\gamma\etac) = 1.3\pm 0.4\%$.
The corresponding partial width is
\be
\Gamma(\jpsi\rt\gamma\etac) = 1.2\pm 0.4~{\rm keV~~~~Crystal~Ball~ 
measurement},
\ee
which is far below the predicted value.

This very dramatic discrepancy between potential model predictions
and the measured value of the $\jpsi\rt\gamma\etac$ partial width
makes it urgent that \bes3 remeasure this quantity 
reliably and with high precision.   In addition to its obvious
theoretical significance, such a measurement has important engineering
consequences.  The extraction of all $\etac$ branching fraction 
determinations that use $\etac$ mesons produced by the 
$\jpsi\rt\gamma\etac$ process (as is the case for most
of the $\etac$ branching fraction measurements listed
in the PDG) rely on this measurement.

\subsubsection{Experimental considerations}

With a sample of  $\sim 10^{9}$ $\jpsi$ events accumulated
at the resonance peak, we can expect many thousands of detected
$\gamma\etac$ events for each of the dominant $\etac$
decay modes.  This will provide the opportunity to measure
the mass and width in each mode with a statistical precision that is
about and order-of-magnitude better than the current world averages. 
Since the effects of interference between the $\etac$ amplitude 
and that for possible coherent continuum production of
the same final state will be an
important source of systematic error,  the measurements should
be done using as many different $\etac$ decay modes as possible.  

Additional data accumulated at the peak of the $\psip$ would
be very useful.  The transition gamma-ray energies for the
$E1$ decays $\psip\rt\gamma\chi_{c2}$ and $\psip\rt\gamma\chi_{c1}$, 
namely $127.60\pm0.09$~MeV and $166.07 \pm 0.08$~MeV, respectively, 
are very similar to that of the gamma in $\jpsi\rt\gamma\etac$ 
and, thanks to precision measurements from Fermilab 
experiment E835~\cite{andreotti:2005slo}, these energy values
are very well known.  Moreover,  the $\chi_{c1,2}$ and the $\etac$ 
have many decay final states in common and with similar
branching fractions.   Therefore, since the $\chi_{c1,2}$ states 
are narrow and the branching fractions for $\psip\rt\gamma\chi_{c1,2}$
are larger than that for $\jpsi\rt\gamma\etac$, a sample of
$\sim 10^{8}$ events taken at the $\psip$ peak will allow for
measurements of these two calibration channels that would have 
precision levels comparable to those
of the $\etac$ measurements with the same final state, 
made with $\sim 10^{9}$ $\jpsi$ events. 
Such complementary measurements would be useful for
controlling the experimental systematic
effects to levels that are smaller than the statistical errors.

For the $\psip\rt\gamma\etac$ transition, $n=2$ and $n^{\prime}=1$,
the radial wave functions in Eqn.~\ref{fig:M1_width} are different
and the $\etac$ recoil momentum is not small.   In
this case, the the line shape is modulated by an {\it additional}, 
model-dependent factor that the CLEO group estimated to
be $\sim k^4_{\gamma}$~\cite{slo_cleoc_psip2gammaetac}.  Therefore,
although $\psip$ data will provide checks and calibrations
for measurements done with $\jpsi$ data,
this strong,  model-dependent modulation of the line shape 
probably precludes using  the $\psp\rt\gamma\etac$
data itself to make precise model-independent measurements of the
 mass and width of the $\etac$.

The ${\mathcal B}(\jpsi\rt\gamma\etac)$ measurement has to
be done inclusively, {\it i.e.}, the $\etac$ signal has to
be measured from the $E_{\gamma}\simeq 114$~MeV peak 
in the inclusive photon spectrum, independently (as much as possible) 
of any specific $\etac$ decay mode.  Since the Crystal Ball
group was able to identify and measure such a peak, this
should be doable in \bes3, where the electromagnetic
calorimeter has better granulatity and a factor-of-two better
gamma-ray energy resolution.  A difficulty here is 
that the low energy of the transition gamma makes it
not useful as an event trigger.  For $\jpsi$ running,
trigger conditions have to be established that insure all
significant $\etac$ decay modes satisfy them with high
efficiency.   

Trigger biases can be avoided by using tagged $\jpsi$
events from the $\pi^+\pi^-\jpsi$ decays of the $\psip$
that are triggered only by the tracks of the transition
$\pi^+\pi^-$ pair.  With a 
$\sim 10^{8}$ event sample taken at the peak of the $\psip$
resonance, we can expect about $\sim 10^{7}$ tagged $\jpsi$
decays, including nearly $10^{5}$ with monochromatic
gammas from the $\jpsi\rt\gamma\etac$ transition
detected in the electromagnetic calorimeter.

\subsection{Mass and width of the $\etacp$}

Among the below-threshold charmonium states,
the $\etacp$ meson has the most poorly measured mass and width.
The PDG average for the mass, $3637\pm 4$~MeV, has a fit confidence
level of 2.1\%.  The average width value, based on two measurements,
is $14\pm 7$~MeV.

At \bes3, $\etacp$ mesons will be produced during $\psip$ running 
via the radiative M1 decay $\psp\rt \gamma\etacp$. According to
the relation given in Eqn.~\ref{fig:M1_width}, the partial width   
$\Gamma(\psp \ra \gamma \etacp)$ can be estimated by   
scaling the measured the measured partial width for $\jpsi \ra\gamma\etac$
($1.19 \pm 0.34$~keV by $k^{\prime 3}_{\gamma}/k^3_{\gamma}$, where
$k^{\prime}_{\gamma} = 48 \pm 4$~MeV ($k_{\gamma}= 114.3 \pm 1.2$~MeV) is 
gamma energy
for the $\psp \gamma\etacp$ ($\jpsi \gamma\etac$) transition.  This
gives an expected partial width
$\Gamma(\psp\ra\gamma\etacp)=87\pm 25$~eV, and branching fraction
${\mathcal B}(\psp\ra \gamma\etacp) = (2.6 \pm 0.7)\times 10^{-4}$.
Thus, the expected $\etap$ mesons production rate in \bes3 
will be low and, because of the low energy of
the transition gamma ray,  detection
will be difficult.  Precise $\etacp$ mass and width measurements   
will be a challenge for \bes3.

\subsection{The $\hc$ mass}

The $\hc$ meson was observed
by CLEO in both inclusive and exclusive processes.  In the 
inclusive process $\psp\ra\pi^0\hc$ the $\hc$ decay products
are not detected and the signal is a
distinct peak in the $\pi^0$ recoil mass distribution.
Figure~\ref{fig:hc_CLEO} shows the very clean signal seen by CLEO 
for the exclusive process $\psp\ra\pi^0\hc$ with 
$\hc\rt\gamma\etac$~\cite{part4_CLEO-ch_c05}.
Evidence for the $\hc$ was also reported by Fermilab experiment E835 in the process
$\bar{p}p\ra\hc\ra\gamma\etac\ra\gamma\gamma\gamma$~\cite{part4_E835}.
CLEO measures $M(\hc)=3525.35\pm0.27\pm0.2$~MeV; E835
reports $3525.8\pm 0.2 \pm 0.2$~MeV.

\begin{figure}[htbp]
\centerline{
\includegraphics[width=9cm,angle=0,clip=]{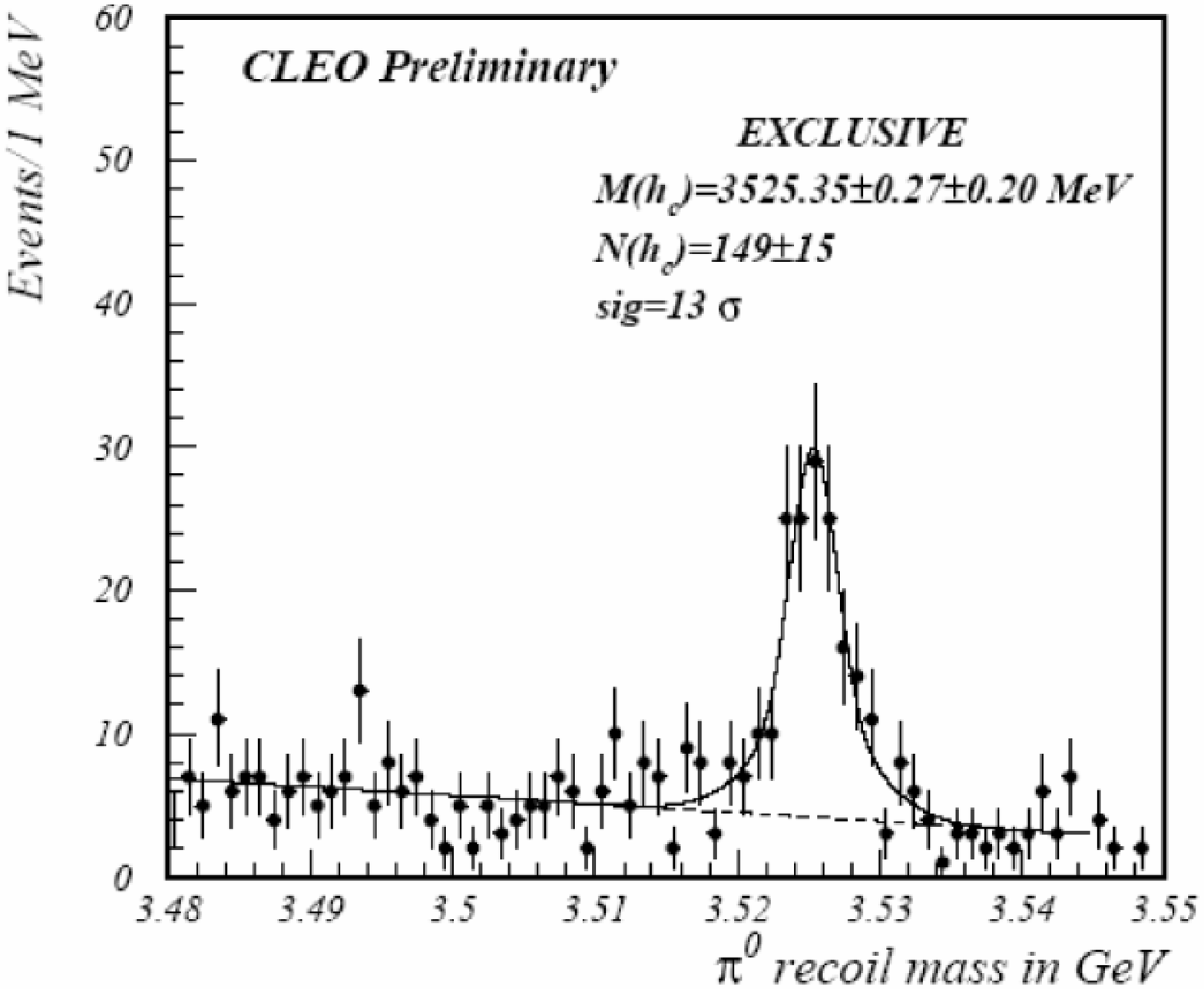}
}  
\caption{\label{fig:hc_CLEO}
The exclusive signal for $\psp\ra\pi^0 h_c$; $h_c\ra \gamma\etac$
from CLEO.
}
\end{figure}

Naively, if the radial dependence of the
$\vec {\S}_c \cdot \vec {\S}_{\bar c}$ term   
in Eq.~\ref{eqn:spin-dependent} is
truly a $\delta$-function, one would expect the mass
of the $\hc$ to be equal to the
center-of-gravity mass of the $\chi_{cJ}$ states:
\be
M_{c.o.g.}(\chi_{c}) = \frac{1}{9}(M(\chi_{c0})
+ 3M(\chi_{c1}) + 5M(\chi_{c2}))= 3525.30\pm 0.08~{\rm MeV},
\ee
where PDG masses for the $\chi_{cJ}$ states are used.
A more detailed calculation~\cite{part4_CO}, discussed
below in Sect.~\ref{sec:transition_hadron_hc}, predicts the mass
to be a few MeV below $M_{c.o.g.}(\chi_c)$.
The $M(\hc)$ measurements seem to prefer the naive
expectations.  However, the current precision 
leaves room for at least a factor of two
improvement from \bes3.  The cleanliness of the
CLEOc $h_c$ signal in the exclusive decay channel
(see Fig.~\ref{fig:hc_CLEO}) indicates that improved
statistics afforded by the large $\psip$ sample expected 
for \bes3 will result in a commensurate improvement
in the precision of the $h_c$  mass measurement.


\section[Novel states]
{Novel states incorporating a \ccbar subsystem\footnote{By T. Barnes}}

\subsection{Charm Molecules}
\label{sec:novel_barnes_x3872}

Molecules are weakly bound states of more than one hadron, of
which by far the best known examples are nuclei and hypernuclei.
There are also speculations that certain unusual hadrons may also
be similar weakly bound states of mesons, such as the $f_0(980)$
and $a_0(980)$ ``KKbar-molecule" candidates. One of the earliest
suggestions for a mesonic molecule in the charm sector pertains to
the $\psi(4040)$; it was suggested by Voloshin {\it et al.}
\cite{ted_Voloshin:1976ap} that this state might be a $\D^*\D^*$
molecule, because the mass is close to the mass of two $\D^*$
mesons, and this state couples very strongly to $\D^*\D^*$ final
states. (See also
Refs.\cite{ted_DeRujula:1976qd,ted_Iwao:1980iw}.) It has since
been realized that this $\D^*\D^*$ dominance is also expected for
a conventional $3{}^3$S$_1$ \ccbar charmonium state
\cite{ted_LeYaouanc:1977ux,ted_Barnes:2005pb}.

The obvious characteristics of molecules are that they should have
masses just below the sum of the masses of their constituent
hadrons (to within a nuclear physics scale binding energy of
perhaps {\it ca.} 10~MeV, and that they should decay strongly to
their constituents (if allowed by width effects) or to final
states that those constituents would naturally couple strongly to.
Their most plausible quantum numbers are those of an S-wave
constituent pair, since the residual binding forces between
color-singlet hadrons are relatively weak and short relative  to
forces between quarks and gluons. (This argues against the
$\psi(4040)$ being a $\D^*\D^*$ molecule, since the binding forces
would be competing against a P-wave angular momentum barrier.)

Rather remarkably, a very strong candidate for a charmed meson
molecule was identified in recent years in B-meson decays, the
X(3872). This state was originally identified by Belle
\cite{ted_Choi:2003ue} in the final state $J/\psi \pi^+\pi^-$,
where the mass distribution of the $\pi^+\pi^-$ system is
consistent with, and believed to be dominantly due to, a $\rho^0$.

The mass and width (still only known as an upper limit) for this
state are \be \M(\X(3872))  = \ 3871.2 \pm 0.5\ \ {\rm MeV}, \ee
\be \Gamma(\X(3872) ) \ < \ 2.3\  {\rm MeV}, \ 90\% \ {\it c.l.}
\ee There is now clear evidence from the decay angular
distribution that the $\J^{\P\C}$ quantum numbers are $1^{++}$
\cite{ted_Abe:2005iy}. Comparison with the spectrum of expected
\ccbar states in Fig.\ref{ccbar_spectrum} shows that no $1^{++}$
\ccbar state is expected near this mass, although one could
clearly consider a low-mass $2{}^3$P$_1$ $\chi_1'$ state as a
possibility. This and other \ccbar possibilities were initially
considered, and the combined mass and narrow width of the X(3872)
were found to be inconsistent with any \ccbar assignment
\cite{ted_Barnes:2003vb}.

The proximity of the X(3872) to the mass of a (neutral)
$\D^0\D^{*0}$ pair (at an almost identical $3871.2 \pm 0.6$~MeV)
immediately suggested that this state might be a weakly bound
$\D^0\D^{*0}$ molecule (see for example
Refs.\cite{ted_Tornqvist:2003na,ted_Close:2003sg,ted_Swanson:2003tb}).
(A binding energy of {\it ca.} 1~MeV is allowed by the current
experimental mass errors.) Note that this $\D\D^*$ molecule would
have a much smaller charged $\D^+\D^{*-}$, $\D^-\D^{*+}$
component, since this basis state has a mass of $3879.3 \pm
0.6$~MeV, and hence is at a much higher mass on the scale of a
bound state with only 1~MeV binding.

Although the residual ``nuclear" forces that can lead to the
formation of hadronic molecules are usually not well understood,
in this case there is less ambiguity; the longest-ranged force,
which should be the most important for a weakly bound system, is
one pion exchange. Since the strength of the required $\D^*\D\pi$
coupling can be inferred from the partial width for $\D^*\to
\D\pi$, it is straightforward to calculate the effective $\D\D^*$
interaction due to one pion exchange. (There is however some
ambiguity in the treatment of the short-range truncation of this
force.) Studies of one pion exchange forces in the $\D\D^*$ system
by Tornqvist \cite{ted_Tornqvist:2003na} and Swanson
\cite{ted_Swanson:2003tb} showed that the $1^{++}$ state
experienced the strongest binding forces from one pion exchange,
and that they were numerically just strong enough to (perhaps)
form a $\D\D^*$ bound state.

One especially striking prediction of the (neutral D)
D$^0$D$^{*0}$ molecule model is that one should observe comparable
strength $J/\psi \omega$ and $J/\psi \rho^0$ decay modes
\cite{ted_Swanson:2003tb} (see also
Ref.\cite{ted_Tornqvist:2004qy}), due to the maximal isospin
breaking present in the initial state. This prediction appears to
have been confirmed by the evidence from Belle for the $\omega$ in
the $3\pi$ mode X(3872)$\to J/\psi \pi^+\pi^-\pi^0$
\cite{ted_Abe:2005ix}. The $3\pi$ invariant mass peaks at the
highest mass, as expected for a virtual $\omega$, and the ratio of
$2\pi$ to $3\pi$ branching fractions is close to unity,
\begin{equation}
\frac{\Gamma({\rm X(3872)} \to J/\psi \pi^+ \pi^- \pi^0)}
{\Gamma({\rm X(3872)} \to J/\psi \pi^+ \pi^-)} = 1.0 \pm 0.4 \pm 0.3,
\end{equation}
as predicted by Swanson in the DD$^*$ model.

There is also evidence from Belle for the radiative transition
X(3872)$\to \gamma\; J/\psi$ \cite{ted_Abe:2005ix}, with the width
ratio
\begin{equation}
\frac{\Gamma({\rm X(3872)} \to  \gamma\; J/\psi)}
{\Gamma({\rm X(3872)} \to J/\psi \pi^+ \pi^-)} = 0.14 \pm 0.05,
\label{X3872_rad}
\end{equation}
which should be useful in testing the details of different models
of the X(3872).

There are several interesting studies of charm molecules that
might be possible at BES. Detection of this $1^{++}$ state at all
at an $e^+e^-$ facility is nontrivial, but may be feasible through
E1 decays of the $\psi(4040)$ and $\psi(4160)$. (Both will couple
to the X(3872) through it's \ccbar component.) If the X(3872) can
be studied, a measurement of the total width of the X(3872) would
be very interesting, since the current upper limit is not far
above the estimate Swanson gives in the $\D\D^*$ molecule model
\cite{ted_Swanson:2003tb,ted_Swanson:2006st}. Measurements of the
branching fractions of the X(3872) into the various final states
predicted by the molecule model would then be very interesting,
especially the ratios of different isospin modes such as J/$\psi
\omega$ versus J/$\psi \rho^0$.

Given that the X(3872) does appear likely as a $\D\D^*$ molecule,
it is of interest to search for other molecules that are predicted
assuming similar interactions between charmed mesons. Swanson
\cite{ted_Swanson:2006st} notes that one attractive possibility is
a $0^{++}$ $\D^*\D^*$ molecule, which could be produced in
$\psi(4160)$ E1 decays and observed in the J/$\psi \omega$ final
state.

\subsection{Charmonium Hybrids}

\label{sec:novel_barnes_y4260}
The discovery of charmonium hybrids may be the most exciting goal of the current
and near future studies of the charmonium system. These states should be accessible
at BES.

Hybrids are hadrons in which the gluonic degree of freedom has
been excited. The nature of this gluonic excitation is not well
understood at present, and has been described by various models,
including valence gluon models such as the bag model and
nonperturbative systems such as the flux tube model. To date most
theoretical studies have considered hybrid mesons ($q\bar q$ with
a gluonic excitation) and hybrid baryons ($qqq$ with a gluonic
excitation). Although the models differ in detail regarding their
predictions for the spectrum of hybrids, there is general
agreement that hybrid mesons have the very attractive feature of
including states with $\J^{\P\C}$-exotic quantum numbers. These
are the quantum numbers $\J^{\P\C} = 0^{--}$ and the series
$0^{+-}, 1^{-+}, 2^{+-}, 3^{-+}, \dots$, which are forbidden to
conventional $q\bar q$ systems, but can all be formed from hybrid
basis states. The search for resonances with exotic quantum
numbers is therefore a principal goal of searches for hybrid
mesons. Hybrids basis states also span all the conventional $q\bar
q$ $\J^{\P\C}$ quantum numbers, and should give rise to a rich
overpopulation of the experimental spectrum relative to the
expectations of the naive $q\bar q$ quark model.

In view of the current uncertainty regarding the nature of hybrids
in models, the predictions of LGT studies are correspondingly very
important. The most recent studies of charmonium hybrids suggest a
mass scale for the lightest exotic charmonium hybrid, which
probably has $\J^{\P\C} = 1^{-+}$, of approximately 4.4~GeV (see
for example
Refs.\cite{ted_Dudek:2006ej,ted_Bernard:1997ib,ted_Mei:2002ip,ted_Liao:2002rj,ted_Bali:2003tp,ted_Luo:2005zg}).
In the flux tube model, which in comparison anticipates the
lightest charmonium hybrid multiplet near 4.2~GeV
\cite{ted_Barnes:1995hc}, this lightest hybrid multiplet is
predicted to contain three exotic and five nonexotic hybrid
states, including both a $1^{-+}$ exotic and a $1^{--}$ nonexotic.

The prediction of hybrids in this approximate mass region,
including a nonexotic $1^{--}$ state, is especially interesting in
view of the recent discovery of the  Y(4260). This remarkable new
state was reported by BaBar in initial state radiation (ISR) in
the reaction $e^+e^- \to \gamma_{ISR} J/\psi \pi^+\pi^-$
\cite{ted_Aubert:2005rm}, and has been confirmed by CLEO-c
\cite{ted_He:2006kg} and Belle \cite{ted_Abe:2006hf} in the same
process. There may also be evidence for an enhancement in $J/\psi
\pi^+\pi^-$ near 4.26~GeV in the decay B$\to$K$J/\psi \pi^+\pi^-$
(in both neutral and negative B/K charge states)
\cite{ted_Aubert:2005zh}. The mass and width reported for the
Y(4260) by BaBar \cite{ted_Aubert:2005rm} are
\begin{equation}
{\rm M} = 4259 \pm 8 {+2 \atop -6} \ {\rm MeV},
\label{Y4260_mass}
\end{equation}
\begin{equation}
\Gamma = 88 \pm 23 {+6 \atop -4} \ {\rm MeV},
\label{Y4260_width}
\end{equation}
with consistent results from CLEO-c, but a somewhat higher mass
and larger width from Belle.

The ISR production mechanism tells us that this state must be
$1^{--}$, but it cannot be a conventional $c\bar c$ state because
the $1^{--}$ states in this mass region are well established from
earlier $e^+e^-$ annihilation experiments. (The Y(4260) is
bracketed by the 2$^3$D$_1$ $\psi(4160)$ and the 4$^3$S$_1$
$\psi(4415)$, which have masses that are in excellent agreement
with the expectations of $c\bar c$ potential models.)

The Y(4260) (if a real resonance) evidently represents
``overpopulation'' of the expected quark potential model spectrum
of $1^{--}$ $c\bar c$ states. In addition, as most models of
hybrids have a vanishing $c\bar c$ wavefunction at contact, it has
long been speculated that they would have small $e^+e^-$ widths,
and thus make rather weak contributions to R. This overpopulation
of the spectrum and the fact that there is no enhancement visible
in R near this mass has led to suggestions that this state may be
a charmonium hybrid \cite{ted_Close:2005iz}.

Another indication that the Y(4260) may be a charmonium hybrid
follows from a LGT study by the UKQCD group
\cite{ted_McNeile:2002az} of the strong decay couplings of exotic
$b\bar b$ hybrids. This LGT study found strikingly large couplings
of $b\bar b$ hybrids to closed flavor modes (specifically to
$\chi_b S$, where $S$ is a light scalar isoscalar meson that would
decay to $\pi\pi$). This is sufficiently similar to the BaBar
observation of the Y(4260) in the single closed-charm mode $J/\psi
\pi^+\pi^-$ to be cited as additional possible evidence for a
hybrid interpretation.

The unusual $J/\psi \pi^+\pi^-$ mode and the UKQCD study suggest
searches of any other accessible closed-charm modes with $1^{--}$
quantum numbers, such as $J/\psi \eta$, $J/\psi \eta'$, $\chi_J
\omega$ and so forth. Ideally the light system should have quantum
numbers thought to couple strongly to pure glue, such as $0^{++}$
and $0^{-+}$.

In specific decay models, notably the flux-tube decay model
\cite{ted_Isgur:1985vy}, theorists have long anticipated that the
dominant open-flavor decay modes of hybrids would be a meson pair
with one internal S-wave (for charmonium hybrids, D, D$^*$, D$_s$,
D$_s^*$) and one internal P-wave (such as D$_J$ and D$_{sJ}$). In
the case of the Y(4260) this suggests dominance of the decay mode
DD$_1$(2430). This broad D$_1$ has a width of {\it ca.}~400~MeV,
and decays to D$^*\pi$, so this suggests a search for evidence of
the Y(4260) in DD$^*\pi$. Since this is a prediction of a decay
model in an untested regime (hybrids), one should be cautious and
search the more familiar two-body modes DD, DD$^*$, D$^*$D$^*$,
D$_s$D$_s$, D$_s$D$_s^*$ and D$_s^*$D$_s^*$ for evidence of the
Y(4260) as well. If there is evidence of a large DD$_1$(2430)
signal, the Y(4260) would then be quite convincing as a hybrid
having properties predicted by the flux tube model. If it appears
in some of these open charm modes such as DD$^*$ and D$_s$D$_s^*$
at rates comparable to or larger than $J/\psi\pi^+\pi^-$, one
might claim a hybrid but speculate that the flux tube decay model
was inaccurate in predicting hybrid decay modes. Finally, if the
Y(4260) does not appear in any other mode, one might be skeptical
about whether the $J/\psi\pi^+\pi^-$ signal is due to a resonance
at all; there are nonresonant possibilities, such as production of
DD$_1$ in $e^+e^- \to $ DD$_1$ followed by an inelastic FSI that
produces a broad $J/\psi\pi^+\pi^-$ enhancement due to the (very
broadened) onset of DD$_1$(2430) threshold events (which would
appear near 4.3~GeV).

In any case since it is clear from the reported mass that the
Y(4260) is not a conventional $c\bar c$ state, it will be very
important to establish the properties of this signal through the
accumulation of better statistics. This applies even more strongly
to the more recently reported states discussed in the following
section.



\section[The $XYZ$ mesons, recent experimenatl developments]
{The $XYZ$ mesons, recent experimental 
developments\footnote{By S.L. Olsen}}

In this section we briefly disuss some of the 
other so-called $XYZ$ mesons, concentrating  
mainly of recent recent experimental developments.

\subsection{The $X(3940)$ (and $X(4160)$)}

Belle observed the $X(3940)$ 
recoiling from the $\jpsi$ in double-charmonium production in the 
reaction $e^+e^-\to J/\psi +X$ 
at $E_{cm}\simeq 10.58$~GeV~\cite{slo_Belle_x3940}.
In addition to the $X(3940)$, Belle observed the well known $J=0$
charmonium states $\eta_c$, $\chi_{c0}$, and $\eta_c(2S)$ with properties 
consistent with PDG values.   

While a distinct signal for $X(3940)\rt D\bar{D^*}$ is also seen,
there is no evidence for the $X(3940)$ in either the $D\bar{D}$ 
or $\omega \jpsi$
decay channels.  If the $X(3940)$ has $J=0$, as seems to be the case
for mesons produced via this production mechanism,
the absence of a substantial
$D\bar{D}$ decay mode strongly favors $J^{P}=0^{-+}$,
for which the most likely charmonium assignment
is the $\eta^{\prime\prime}_c$, the $3^1S_1$ charmonium state.
The fact that the lower mass $\eta_c(1S)$ and $\eta_c(2S)$ are also 
produced in double charm production seems to support this assignment.  
The predicted width for a $3^1S_0$ state with a mass of 3943~MeV is
$\sim 50$~MeV~\cite{Eichten:2005ga}, which is in acceptable
agreement with the measured $X(3940)$ width.

However, there are problems with this assignment, the first being
that the measured mass
of the $X(3940)$, recently updated by Belle to be
$3942\pm 8$~MeV/$c^2$~\cite{slo_Belle_x4160}, is  below potential model
estimates of $\sim$4050~MeV/$c^2$ or higher~\cite{ted_Barnes:2005pb}.
A further complication is the recent observation by
Belle of a mass peak in the $D^*\bar{D^*}$ system
$m = 4156\pm 29$~MeV/$c^2$ and
total width of $\Gamma = 139^{+113}_{-65}$~MeV/$c^2$,
recoiling from a $J/\psi$ in the process $e^+e^-\rt J/\psi
D^*\bar{D^*}$~\cite{slo_Belle_x4160}.
Using similar arguments, this latter state, called the $X(4160)$ 
could also be attributed to  the $3^1S_0$ state.
But the $X(4160)$ mass is well above expectations for the $3^1S_0$
and well below those for the $4^1S_0$, which is predicted to be
near 4400~MeV/$c^2$~\cite{ted_Barnes:2005pb}.
Although the $X(3940)$ {\it or} the $X(4160)$ might conceivably fit a
charmonium assignment, it seems very unlikely that both of them could be
accommodated as $c\bar{c}$ states.

\subsection{The $Y(3940)$}

Belle observed the $Y(3940)$ via its $Y(3940)\rt \omega J/\psi$
decay in $B \rt K \omega J/\psi$ decays~\cite{slo_Belle_y3940}.  
This observation has  recently
been confirmed by BaBar~\cite{slo_BaBar_y3940}.
Belle reports a mass and width of $M=3943\pm 17$~MeV/c$^2$ and
$\Gamma=87\pm 34$~MeV/$c^2$ while BaBar reports the preliminary values of
$M=3914.3^{+3.8}_{-3.4} \pm 1.6$~MeV and
$\Gamma=33^{+12}_{-8} \pm 0.6$~MeV which are somewhat different from 
Belle's.

The measured product branching fractions agree:
${\cal B}(B\rt KY(3940)){\cal B}(X(3940)\rt\omega J/\psi)
= (7.1 \pm 3.4)\times 10^{-5}$ (Belle), and
${\cal B}(B\rt KY(3940)){\cal B}(X(3940)\rt\omega J/\psi)
= (4.9 \pm 1.1)\times 10^{-5}$ (BaBar).
These values together with an assumption that the
branching fraction ${\cal B}(B\rt K Y(3940))$ is less than
or equal to $1\times 10^{-3}$, the value that is typical
for allowed $B\rt K$+charmonium decays,  implies
a partial width $\Gamma(Y(3940)\rt \omega J/\psi) > 1$~MeV/$c^2$,
which is at least an order-of-magnitude higher
than those for hadronic transitions between any of the
established charmonium states.  The Belle group's 90\%
confidence level limit on
${\cal B}(X(3940)\rt\omega J/\psi) < 26\%$~\cite{slo_Belle_x3940}
is not stringent enough to rule out the possibility that 
the $X(3940)$ and the $Y(3940)$ are the same state.

\subsection{The $Z(3930)$}

The $Z(3930)$ is a peak reported by Belle in the spectrum of $D\bar{D}$ 
mesons produced in $\gamma\gamma$ collisions, with  mass and width
$M=3929\pm 6$~MeV/$c^2$ and
$\Gamma=29\pm 10 $ MeV/$c^2$~\cite{slo_Belle_z3930}. 
The $D \bar D$ decay mode makes it impossible
for the $Z(3930)$ to be the $\eta_c(3S)$ state.
The two-photon production process can only produce $D\bar{D}$
in a $0^{++}$ or $2^{++}$ state and for these, the $dN/d\cos\theta^*$
distribution, where $\theta^*$ is the angle between the $D$ meson 
and the incoming photon in the $\gamma\gamma$ cm, are quite distinct:
flat for $0^{++}$ and $\propto \sin^4\theta^*$ for $2^{++}$.  The
Belle measurement strongly favors the $2^{++}$ hypothesis
(see Fig.~\ref{fig:belle_z3930}), making
the $Z(3930)$
a prime candidate for the $\chi^{\prime}_{c2}$, the $2^3P_2$ charmonium 
state.
The predicted mass of the $\chi_{c2}(2P)$ is 3972~MeV/$c^2$ and the
predicted total width assuming the observed mass value is
$\Gamma_{\rm 
total}(\chi_{c2}(2P))=28.6$~MeV/$c^2$~\cite{slo_swanson:2006,ted_Barnes:2005pb,Eichten:2005ga},
in good agreement with the experimental measurement.  Furthermore,
the two-photon production rate for
the $Z(3930)$ is also consistent with expectations for
the $\chi^{\prime}_{c2}$~\cite{Barnes:1992sg}.

\begin{figure}[htbp]
\centerline{
\includegraphics[width=7cm,angle=0,clip=]{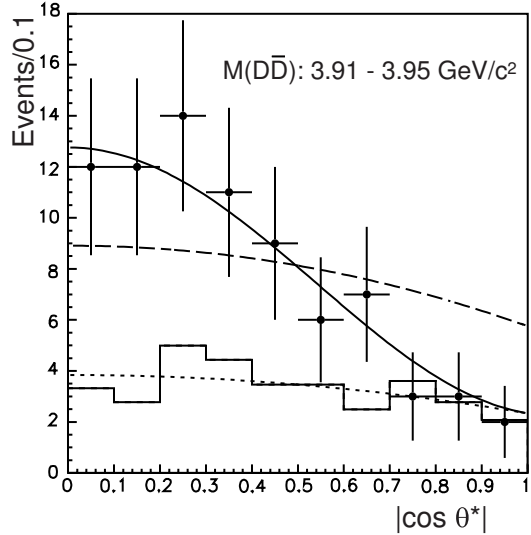}
}  
\caption{
Belle's $\chi_{c2}(2P)$ candidate~\cite{slo_Belle_z3930}:
$\cos \theta^*$, the angle of the $D$~meson relative to the
beam axis in the $\gamma\gamma$ center-of-mass frame for
events with $3.91 < m(D \bar D) < 3.95\,\mbox{GeV}$; the data (circles) 
are
compared with predictions for $J=2$ (solid) and $J=0$ (dashed).
The background level can be judged from the solid histogram
or the interpolated smooth dotted curve.
\label{fig:belle_z3930}}
\end{figure}

\subsection{$\pi^+\pi^- \psi'$ resonances at 4360~MeV/$c^2$ and
4660~MeV/$c^2$}

In addition to the $Y(4260)$, which is discussed at length
in Sect.~\ref{sec:novel_barnes_y4260} above,
BaBar also found a broad peak in the cross section for $e^+e^-\rt
\pi^+\pi^- \psi'$ that is distinct from the $Y(4260)$; its peak
position and width are not consistent with those of the
$Y(4260)$~\cite{slo_BaBar_y4325}.  The BaBar observation was
subsequently confirmed by a Belle group study that
used a larger data sample~\cite{slo_Belle_y4660}.
The Belle group was able to determine
that the $\pi^+\pi^-\psi'$ mass enhancement is produced by
two distinct peaks, one, the $Y(4360)$ with $M= 4361 \pm 13$~MeV/$c^2$
and $\Gamma = 74 \pm 18$~MeV/$c^2$ and a second, the $Y(4660)$ with
$M=4664 \pm 12$~MeV/$c^2$
and $\Gamma = 48 \pm 15$~MeV/$c^2$~\cite{slo_Belle_y4660}.  These
masses and widths are not consistent with any of the
established $1^{--}$ charmonium states, and no sign
of a peak at either of these masses is evident in the
$e^+e^-$ total annihilation cross section~\cite{slo_BES_R}
or in the exclusive cross sections  $e^+e^- \rt 
D\bar{D}$~\cite{belle_dd:2008},
$D\bar{D^*}$ or $D^*\bar{D^*}$~\cite{belle_dstrdstr:2006},
or $D\bar{D}\pi$ (non-$D^*$)~\cite{belle_ddpi:2007}, which
indicates that the $\pi^+\pi^- \psi'$ partial width for these
states is unusually large (at least by charmonium standards).
Moreover, as is evident in Fig.~\ref{fig:y4260_etc}, which
shows the recent Belle results~\cite{slo_Belle_y4260}
for $\pi^+\pi^- J/\psi$ (top)
and $\pi^+\pi^- \psi^{\prime}$ (bottom) with the same
horizontal mass scales,
there is no sign of either the $Y(4360)$ or $Y(4660)$ in the
$\pi^+\pi^- J/\psi$ channel; nor is there any sign of
the $Y(4260)$ peak in the $\pi^+\pi^- \psi^{\prime}$ mass spectrum.

\begin{figure}[htbp]
\centerline{
\includegraphics[width=8cm,angle=0,clip=]{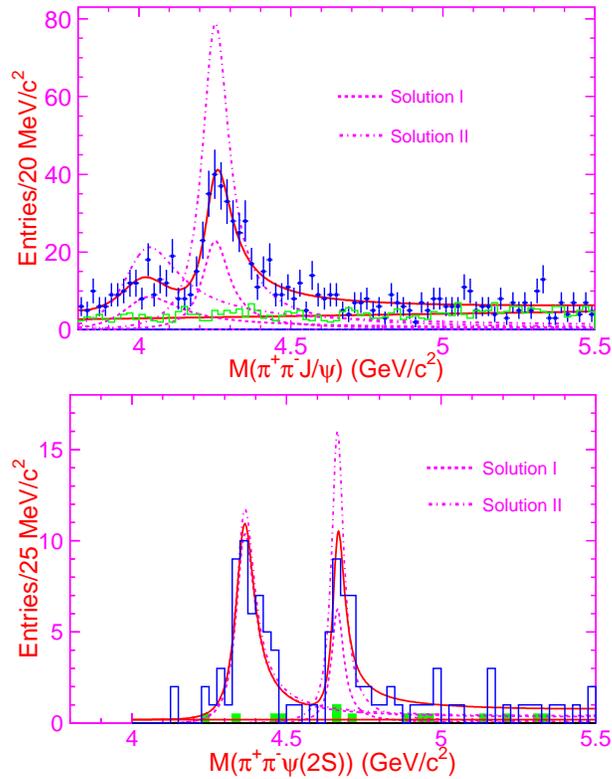}
}  
\caption{
The $\pi^+ \pi^- J/\psi$ (Top) and
$\pi^+\pi^- \psi^{\prime}$ (Bottom) invariant mass
distributions for the ISR processes
$e^+ e^- \rt  \gamma \pi^+ \pi^- J/\psi (\psi^{\prime})$,
from Refs.~\cite{slo_Belle_y4260,slo_Belle_y4660}.
The curves indicate the results of fits of interfering
Breit Wigner resonances to the data.
}  
\label{fig:y4260_etc}
\end{figure}

\subsection{The $Z^{\pm}(4430)\rt\pi^{\pm} \psi^{\prime}$}

In Summer 2007, the Belle group reported 
observed the relatively narrow enhancement in the $\pi^+ \psi^{\prime}$
invariant mass distribution in the $B\rt K\pi^{\pm}\psi'$ 
decay process shown in Fig.~\ref{fig:z4430}~\cite{slo_Belle_z4430}.
The fitted peak mass and width values are $M=(4433\pm 
5)$~MeV/$c^2$  and $\Gamma=(45^{+35}_{-18})$~MeV/$c^2$,
which is too narrow to
be caused by interference effects in the $K\pi$ channel.  The
$B$ meson decay rate to this state, which is called $Z^{\pm}(4430)$,
is similar to that for decays to the $X(3872)$ and $Y(3940)$, which
implies that the $Z^{\pm}(4430)$ has a substantial branching fraction
({\it i.e.} greater than a few percent) to $\pi^{\pm}\psi^{\prime}$ and,
thus, a partial decay width for this mode that is on the MeV scale.
There are no reports of a $Z^{\pm}(4430)$ signal in the $\pi^+J/\psi$
decay channel.

\begin{figure}[htbp]
\centerline{
\includegraphics[width=7cm,angle=0,clip=]{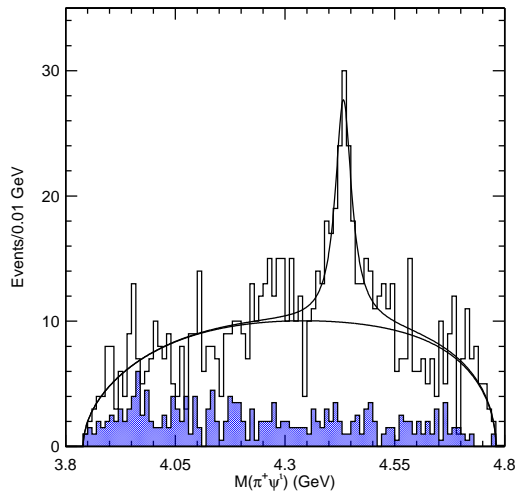}
}  
\caption{
The $\pi^{\pm} \psi^{\prime}$ invariant mass
distribution for $B\rt  K\pi^{\pm} \psi^{\prime}$ decays~\cite{slo_Belle_z4430}.
The shaded histogram is the estimated background. The
curve is the result of a fit to a relativistic $S$-wave
Breit Wigner signal function plus a phase-space-like
continuum term.}
\label{fig:z4430}
\end{figure}

Among the $XYZ$ exotic meson candidates, the $Z^{\pm}(4430)$ is unique
in that it has a non-zero electric charge, a feature that cannot
occur for $c\bar{c}$ charmonium states or $c\bar{c}$-gluon hybrid
mesons.  It is, therefore, a prime candidate for a multiquark meson.

\subsection{Evidence for corresponding states in the $s$- and $b$-quark sectors?}

The proliferation of meson candidates that are strongly coupled
to $c\bar{c}$ quark pairs but not compatible with a conventional
charmonium assignment leads one naturally to question whether or
not similar states exist that are strongly coupled to $s\bar{s}$
or $b\bar{b}$ quark pairs. There is some evidence that this, in fact,
may be the case.

\subsubsection{The $Y(2175)$}
In 2006, the BaBar group reported a resonance-like structure in the
$f_0(980)\phi$ invariant mass distribution produced in
$e^+e^-\rt \gamma_{ISR} f_0(980)\phi$ radiative-return
events~\cite{slo_BaBar_y2175}.  They report resonance parameters of
$M= (2170 \pm 10 \pm 15)$~MeV \&
$\Gamma = (58 \pm 16  \pm 20 )$~MeV. They see no signal for
this peak in a sample of $K^*(892)K\pi$
events that has little kinematic overlap with $f_0(980)\phi$, and
conclude that this structure, which they call the $Y(2175)$,   
has a relatively large branching fraction for $f_0(980)\phi$.

The similarities with the $Y(4260)$, both in production and    
decay properties,
leads naturally to the speculation that the $Y(2175)$ might be an
$s\bar{s}$ analogue of the $Y(4260)$,
{\it i.e.} it is the ``$Y_s(2175)$''.
On the other hand,
there is no compelling evidence against it being a
conventional $3^3S_1$ or  $2^3D_1$ $s\bar{s}$
``strangeonium''  state.  The study of the $Y(2175)$ in
other production
and decay modes would be useful for distinguishing
between different possibilities~\cite{slo_ding_y2175}.

\begin{figure}[htbp]
\centerline{
\includegraphics[width=6cm,angle=0,clip=]{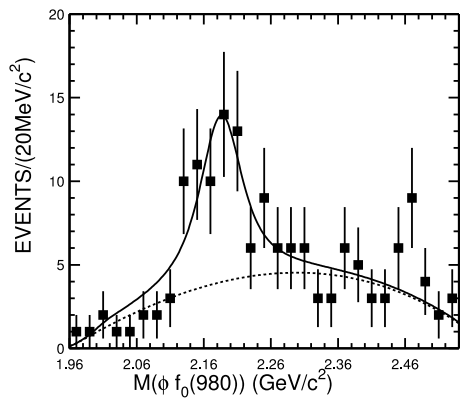}
}  
\caption{The
$M(f_0(980)\phi)$ distribution for $\jpsi\rt \eta f_0(980)\phi$ decays
in BESII.}
\label{fig:bes_y2175} 
\end{figure}

The BESII group made a first step in this program
by finding an $f_0(980)\phi$  mass peak with
similar parameters produced in
$\jpsi\rt\eta f_0(980)\phi$ decays
(see Fig.~\ref{fig:bes_y2175})~\cite{slo_Bes_y2175}.
The BESII fit yields a mass and width of
$M= (2186 \pm 10 \pm 6)$~MeV \&
$\Gamma = (65 \pm 23  \pm 17 )$~MeV, which are in
good agreement with BaBar's measurements.  

The next
steps will be finding it in other decay modes and
searching for counterpart states  with quantum numbers
other than $1^{--}$  that, perhaps, decay into final
states containing an $\eta^{\prime}$.  This will be
an important task for \bes3, and is
discussed in some detail elsewhere in
this report.

\subsubsection{Anomalous $\pipi\Upsilon(nS)$ production at the
$\Upsilon(5S)$}

Using a sample of 236 million $\Upsilon(4S)$ mesons,
BaBar~\cite{slo_BaBar_4s2pipi1s} observed $167 \pm 19$
and $97 \pm 15$ event signals for
$\Upsilon(4S)\rt\pipi\Upsilon(1S)$ and $\pipi\Upsilon(2S)$,
respectively, from which they infer partial widths
$\Gamma(\Upsilon(4S)\rt\pipi\Upsilon(1S)) = (1.8 \pm 0.4)$~keV.
$\Gamma(\Upsilon(4S)\rt\pipi\Upsilon(2S)) = (2.7 \pm 0.8)$~keV.
Belle~\cite{slo_Belle_4s2pipi1s}, with a sample of 464 million $\Upsilon(4S)$
events  reported a $44 \pm 8$ event signal for the transition
$\Upsilon(4S)\rt\pipi\Upsilon(1S)$, from which they infer a partial
width
$\Gamma(\Upsilon(4S)\rt\pipi\Upsilon(1S)) = (3.65 \pm 0.67 \pm 0.65)$~keV.
These partial widths are comparable in magnitude to those measured
for $\pipi$ transitions between the $\Upsilon(3S)$, $\Upsilon(2S)$
and $\Upsilon(1S)$, as discussed below in Sect.~\ref{sec:transition_hadron}.

In 2006, Belle had a one-month-long run at $e^+e^-$ cm energy of
10.87~GeV,
which corresponds to the peak mass of the $\Upsilon(5S)$.  The total
data sample collected was 21.7~fb$^{-1}$ and the
number of $\Upsilon(5S)$ events collected was 6.3 million.
Scaling from the $\Upsilon(4S)$ observations, they did not
expect to see any significant evidence for 
$\Upsilon(5S)\rt\pi^+\pi^-\Upsilon(nS)$ transitions in this data set.
Contrary to expectations based on the $\Upsilon(4S)$ measurements,
the Belle group found large  numbers of $\pipi\Upsilon(nS)$
events in this data sample: $325 \pm 20$ $\pipi\Upsilon(1S)$ events and
 $186 \pm 15$ $\pipi\Upsilon(2S)$ events (see
Figs.~\ref{fig:belle_5s2pipi1s_fig2a}(a) and (b))~\cite{slo_Belle_5s2pipi1s}.
(The $\Upsilon(2,3S)\rt\pipi\Upsilon(1S)$ signals
in Fig.~\ref{fig:belle_5s2pipi1s_fig2a}(a) are
produced by radiative-return transitions
$e^+e^-\gamma_{ISR}\Upsilon(2,3S)$.)

\begin{figure}[htbp]
\centerline{
\includegraphics[width=10cm,angle=0,clip=]{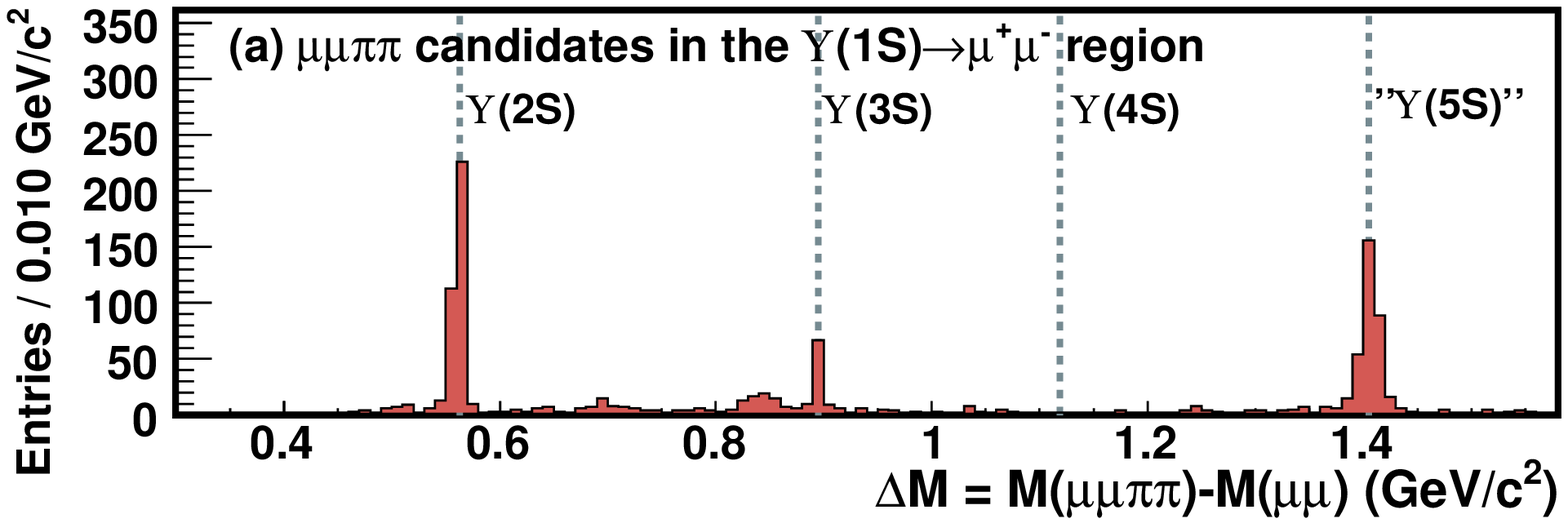}
}

\centerline{
\includegraphics[width=10cm,angle=0,clip=]{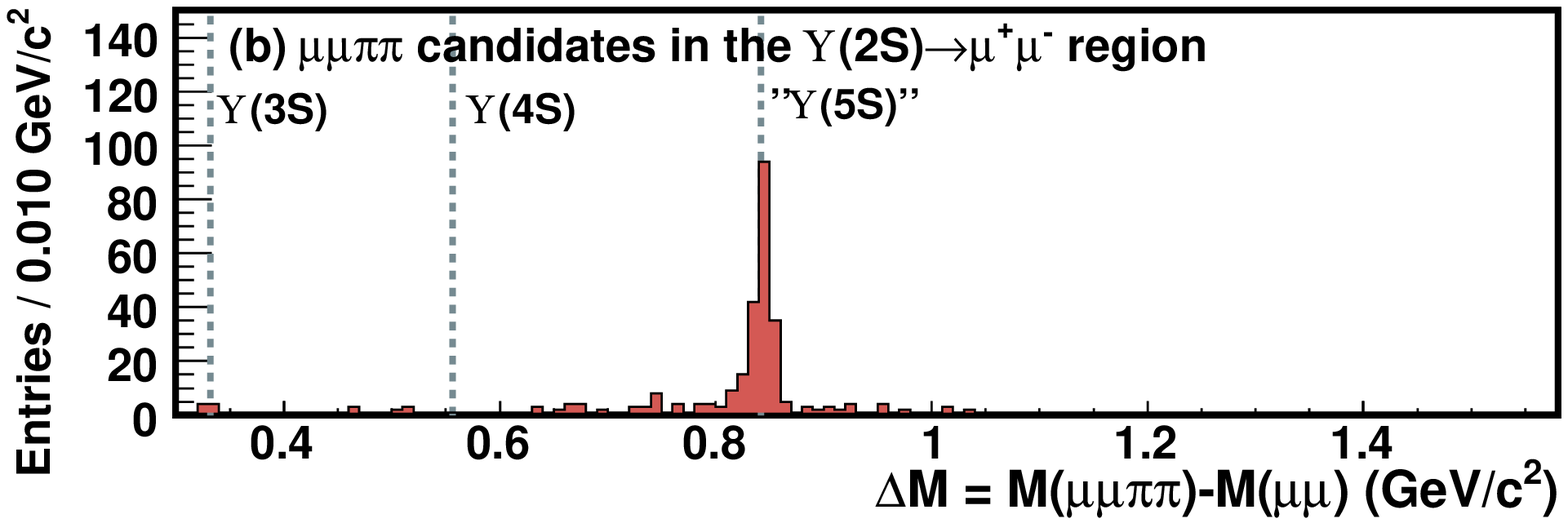}
}
  
\caption{ Belle's
$M(\mumu\pipi)-M(\mumu)$ mass difference distributions
for events with {\bf (a)} $M(\mumu)=\Upsilon(1S)$
and {\bf (b)} $M(\mumu)=\Upsilon(2S)$. Vertical dashed
lines show the expected locations for
$\Upsilon(nS)\rt \pipi\Upsilon(1,2S)$ transitions.}
\label{fig:belle_5s2pipi1s_fig2a}
\end{figure}

If one assumes that these events
are coming from $\Upsilon(5S)\rt\pipi\Upsilon(nS)$ transitions, the
inferred partial widths are huge:
$\Gamma(\Upsilon(5S)\rt\pipi\Upsilon(1S)) = (590 \pm 40 \pm 90)$~keV.
$\Gamma(\Upsilon(5S)\rt\pipi\Upsilon(2S)) = (850 \pm 70 \pm 160)$~keV,
more than two orders-of-magnitude higher than corresponding transitions
from the $\Upsilon(4S)$.

A likely explanation for these unexpectedly large partial
widths (and, in fact, the motivation for
Belle's pursuit of this subject) is that there
is a ``$Y_b$'', {\it i.e.} a $b\bar{b}$ counterpart of the
$Y(4260)$, that is overlapping the $\Upsilon(5S)$~\cite{slo_hou_yb}
and this
state is producing the $\pipi\Upsilon(1,2S)$ events that are seen.

\subsection{Summary}

There is a large (and growing) number of candidate charmonium-like
meson states that  have been observed
that do not seem to fit into the quark-antiquark classification scheme
of the constituent quark model.  

These states exhibit a  number of peculiar features:
\begin{itemize}
\item
Many of them have partial widths for decays to charmonium~+~light~hadrons
that are at the $\sim$MeV scale, which is much larger than is typical
for established $c\bar{c}$ meson states.
\item
They are relatively narrow although many of them are well above
relevant open-charm thresholds.
\item
There seems to be some selectivity: states seen to decay to final
states with a $\psip$ are not seen in the corresponding $\jpsi$
channel, and {\it vice versa}.
\item
The new $1^{--}$ charmonium states are not apparent in the 
$e^+e^-\rt$ charmed-meson-pair or the total hadronic cross sections.
\item
There are no evident changes in the properties of these states
at the $D^*D^{**}$ mass threshold.
\item
Although some states are near mass thresholds for pairs of open charmed
mesons, this is not a universal feature.
\item
There is some evidence that similar states exist in the $s$- and $b$-quark
sectors.

\end{itemize}

Attempts to explain these states theoretically have usually been
confined to subsets of the observed states.  For example,
the $X(3872)$ and $Z(4430)$ have been attributed to bound
molecular states of $D\bar{D^*}$ and $D^*\bar{D^{**}}$ mesons,
or as  diquark-antidiquark tetraquark states,
the $Y(4260)$ as a $c\bar{c}$-gluon hybrid, etc.
However, no single model seems able to deal with the whole system
and their properties in a compelling way.  In general, the predictions
of the various models have had limited success.

\subsection{Implications for \bes3}

This continues to be a data-driven field, with an increasingly
large number of new results continuing to come out from BaBar, Belle 
and BESII. Many of the observed states in the $c\bar{c}$ sector states can be
accessed at \bes3; examples of how this might be done 
are discussed above in Sects.~\ref{sec:novel_barnes_x3872} 
and~\ref{sec:novel_barnes_y4260}.  If  there is a 
corresponding spectroscopy for the $s\bar{s}$ sector, which seem likely to 
be the case, all of the low-lying versions of
those states should be accessible at \bes3 either in 
decays of $\psi$ resonances and/or by continuum $e^+e^-$ production. 

\begin{figure}[htbp]
\centerline{
\includegraphics[width=6cm,angle=0,clip=]{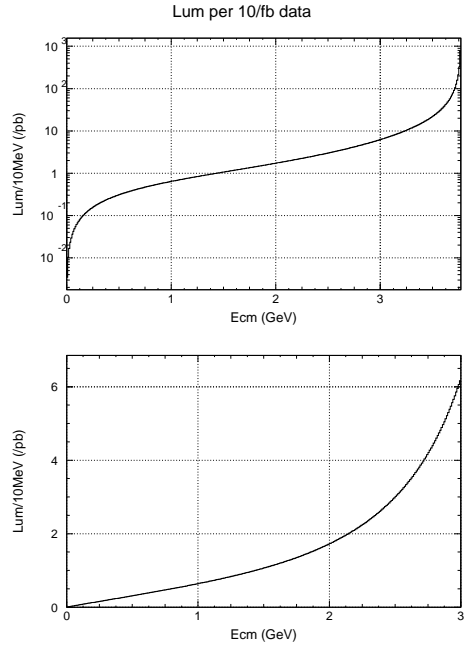}
}
\caption{The energy dependence of the luminosity associated with the initial state
radiation collected by \bes3 in a 10~fb$^{-1}$ data sample accumulated at 
the peak
of the $\psi(3770)$ resonance. The top  has a semi-log scale, the bottom 
one has a linear scale.}
\label{fig:slo_lum3770} 
\end{figure}

For the latter, dedicated energy scans might not be the best strategy.  Thanks
to BEPCII's  high luminosity, continuum $e^+e^-$ processes in the 
$2\sim 3$~GeV c.m. energy region can be accessed via initial-state-radiation
while running at the $\psi(3770)$ resonance, or higher.  It is
expected that \bes3 will collect about 5~fb$^{-1}$/yr at the
$\psi(3770)$, where many years will be invested. 
Figure~\ref{fig:slo_lum3770} shows the  energy dependence 
of the luminosity associated 
with the initial state
radiation collected by \bes3 in a 10~fb$^{-1}$ data sample accumulated 
at the peak of the $\psi(3770)$ resonance~\cite{slo_ycz_private}.  
Since the total cross section 
for $e^+e^-$ annihilation
into hadrons near $E_{cm}=2.5$~GeV is $\sim 30$~nb, thousands of hadronic 
events per
10~MeV c.m. energy bin will be collected in this energy region during 
a multiyear run at the $\psi(3770)$.

\chapter{Charmonium transitions}
\label{sec:transition}

\section[Hadronic transitions]
{Hadronic transitions\footnote{By Yu-Ping Kuang}}
\label{sec:transition_hadron}
\newcommand{\Tr}{{\mbox{\rm Tr}}}
\newcommand{\mn}{{\mu\nu}}
\renewcommand{\half}{{1\over 2}}

\subsection{QCD Multipole Expansion}

Hadronic transitions are important decay modes of heavy quarkonia
(bound states of heavy quarks $Q$ and $\overline{Q}$). For
instance, the branching ratio for $\psi^\prime\to J/\psi+\pi+\pi$
is approximately $50\%$. In general, let us consider the hadronic
transitions
\begin{eqnarray}                       
\Phi_I\to\Phi_F+h
\label{HT}
\end{eqnarray}
in which $\Phi_I$, $\Phi_F$ and $h$ stand for the initial state quarkonium,
the  final state quarkonium, and the emitted light hadron(s),
respectively.

In the $c\overline{c}$ and $b\overline{b}$ systems, the typical
mass difference $M_{\Phi_I}-M_{\Phi_F}$ is around a few hundred
MeV, so that the typical momentum of the light hadron(s) $h$ is
low. In this section, we consider only the single-channel approach
(For a coupled-channel approach, see
Refs.~\cite{part4_ZK91,part4_Kuang06} and references therein).
In this picture, the light hadron(s) $h$ are produced from gluons emitted
by the $Q$ and/or $\overline{Q}$ in the transition. The typical
momentum of the emitted gluons is also low, and, thus, perturbative
QCD does not work in these processes. Certain nonperturbative
approaches are needed for studying hadronic transitions. The {\it
QCD multipole expansion} (QCDME) is a feasible approach to
hadronic transitions.

Due to the nonrelativistic nature of the process, the heavy quarkonia
$n^{\sigma}L_J$ [labelled by the principal quantum number $n$, the
orbital angular momentum $L$, the total angular momentum $J$, and
the spin multiplicity $\sigma$ ($\sigma=1~{\rm or}~ 3$)], can be
considered as solutions of the Schr\"odinger equation within a given
potential model. The typical radius $a=\sqrt{\langle r^2\rangle}$
of the $c\overline{c}$ and $b\bar{b}$ quarkonia obtained in this
way is of the order of $~10^{-1}$ fm. For soft gluon emission
with gluon momenta $k$ satisfying $ak<1$, $ak$ can be a good
expansion parameter.
In classical electrodynamics, the coefficient of the $(ak)^l$ term
in the multipole expansion contains a factor
$\displaystyle\frac{1}{(2l+1)!}$. Hence such a multipole expansion
actually works better than expected simply from consideration of the
magnitude of $(ak)^l$. 
Note that the convergence of QCDME does not depend on the value of
the QCD coupling constant $g_s$. Therefore QCDME is a feasible
approach to the soft gluon emission in hadronic transitions
(\ref{HT}).

QCDME has been studied by many authors
\cite{part4_Gottfried,part4_BR,part4_Peskin,part4_VZNS,part4_Yan,part4_KYF}.
The gauge invariant formulation is given in Ref. \cite{part4_Yan}.
For a systematic review of this type of the theory and its
applications to hadronic transitions, see
Ref.~\cite{part4_Kuang06}. In this expansion, the general formula
for the $S$ matrix element between the initial state $|I\rangle$
and the final state $|F\rangle$ in the single-channel approach has
been given in Ref. \cite{part4_KYF}.
Explicit evaluation of the $S$ matrix
elements in various cases will be presented in the following
sections.

\subsection{Hadronic Transitions Between $\bm S$-Wave Quarkonia}

In the single-channel approach, the  amplitude for hadronic
transitions (\ref{HT}) is diagrammatically shown in
Fig.~\ref{HTfig}. In the figure, there are two complicated
vertices: the  vertex of {\it multipole gluon emissions} ({\bf
MGE}) from  the heavy quarks and the vertex of {\it hadronization}
({\bf H}) describing the conversion of the emitted gluons into
light hadron(s). The {\bf MGE} vertex is at the scale of the heavy
quarkonium, and depends on the properties of the heavy quarkonia.
The {\bf H} vertex is at the scale of the light hadron(s) and is
independent of the heavy quarkonia. In the following, we shall
treat them separately. To illustrate, we take the $\pi\pi$ and
$\eta$ transitions between $S$-wave quarkonia as examples.

\begin{figure}[h]
\centerline{
\includegraphics[width=6.5truecm,clip=true]{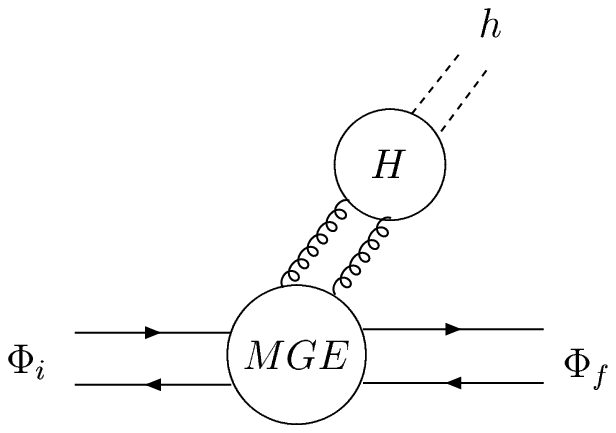}
}
\null\vspace{-0.4cm}
\caption{\footnotesize Diagram for a typical hadronic transition in the single-channel approach.}
\label{HTfig}
\end{figure}

\subsubsection{$\bm{\pi\pi}$ Transitions}

Consider the transition $n_I^3S_1\to n_F^3S_1+\pi+\pi$. These
processes are dominated by double electric-dipole transitions
(E1E1), whose transition amplitude can be obtained from
Refs.~\cite{part4_KYF,part4_Yan,part4_KY81}.
The $n_I^3S_1\to n_F^3S_1+\pi+\pi$ transition rate can be
expressed as \cite{part4_KY81}
\begin{eqnarray}                        
\Gamma(n_I^3S_1\to n_F^3S_1~\pi~\pi)=|C_1|^2G|f^{111}_{n_I0n_F0}|^2,
\label{GammaS-S}
\end{eqnarray}
where
\begin{eqnarray}                     
f^{LP_IP_F}_{n_Il_In_Fl_F}\equiv \sum_{K}\frac{\int R_F(r)r^{P_F}R^*_{KL}(r)r^2dr
\int R^*_{KL}(r^\prime)
r^{\prime P_I}R_I(r^\prime)r^{\prime 2}dr^\prime}{M_I-E_{KL}},
\label{f}
\end{eqnarray}
and the phase-space factor $G$ is given by \cite{part4_KY81}
\begin{eqnarray}                      
&&G\equiv\frac{3}{4}\frac{M_{\Phi_F}}{M_{\Phi_I}}\pi^3\int K\sqrt{1-
\frac{4 m_\pi^2}{M_{\pi\pi}^2}}
(M_{\pi\pi}^2-2m_\pi^2)^2~dM_{\pi\pi}^2,
\label{G}\\
&&K\equiv \frac{\sqrt{(M_{\Phi_I}
+M_{\Phi_F})^2-M_{\pi\pi}^2}\sqrt{(M_{\Phi_I}-M_{\Phi_F})^2-M_{\pi\pi}^2}}{2M_{\Phi_I}}.\nonumber
\label{K}
\end{eqnarray}
In (\ref{f}), $R_I$, $R_F$, and $R_{KL}$ are the radial wave functions
of the initial, final, and intermediate vibrational states, and
they can be calculated from the Schr\"odinger equation within a
given potential model.

There is only one overall unknown constant $C_1$ left in this
transition amplitude, and it can be determined by taking one
well measured hadronic transition rate as an input. So far, the
best-measured $S$-state to $S$-state $\pi\pi$ transition rate is
$\Gamma(\psi^\prime\to J/\psi~\pi\pi)$. The updated experimental
value is~\cite{ted_PDG2006}
\begin{eqnarray}                       
&&\Gamma_{\rm tot}(\psi^\prime)=277\pm 22~{\rm keV},\nonumber\\
&&B(\psi^\prime\to J/\psi~\pi^+\pi^-)=(31.8\pm 0.6)\%, \nonumber\\
&&B(\psi^\prime\to J/\psi~\pi^0\pi^0)=(16.46\pm 0.35)\%.
\label{input}
\end{eqnarray}

\begin{table}[h]

\caption{The value of $|C_1|^2$ and the predicted rates
$\Gamma(\Upsilon^\prime\to \Upsilon~\pi\pi)$,
$\Gamma(\Upsilon^{\prime\prime}\to \Upsilon~\pi\pi)$, and
$\Gamma(\Upsilon^{\prime\prime}\to\Upsilon^\prime~\pi\pi)$ ( in
keV) in the Cornell model and the BGT model. The corresponding
updated experimental values of the transition rates are taken from
Ref. \cite{ted_PDG2006} and listed for comparison.}
\label{table1}
\centering
\begin{tabular}{cccc}
\hline\hline
 &Cornell&BGT&Expt.\\
 \hline
$|C_1|^2\hspace{2.8cm}$&$83.4\times 10^{-6}$&$67.8\times 10^{-6}$&\\
$\Gamma(\Upsilon^\prime\to\Upsilon\pi\pi)$~(keV)~~&8.6~~~&7.8~~~&$~~~~~~8.89\pm1.17~~~~~~$\\
$\Gamma(\Upsilon^{\prime\prime}\to\Upsilon\pi\pi)$~(keV)&0.44&1.2~~~&$1.33\pm 0.22$\\
$\Gamma(\Upsilon^{\prime\prime}\to\Upsilon^\prime\pi\pi)$~(keV)&0.78&0.53&$0.98\pm 0.28$\\
\hline\hline
\end{tabular}
\end{table}

\null\vspace{-0.4cm} Using (\ref{input}) as an input to determine
$C_1$, we can predict all of the $S$-state to $S$-state $\pi\pi$
transition rates in the $\Upsilon$ system. Since the amplitude
(\ref{f}) depends on the potential model, the determined value of
$|C_1|$ is model dependent. Here we take the Cornell Coulomb plus
linear potential model \cite{part4_Cornell} and the
Buchm\"uller-Grunberg-Tye (BGT) potential model \cite{part4_BGT}
as examples to show the determined $|C_1|$ and the predicted rates
of $\Upsilon^\prime\to \Upsilon~\pi\pi$,
$\Upsilon^{\prime\prime}\to \Upsilon~\pi\pi$, and
$\Upsilon^{\prime\prime}\to\Upsilon^\prime~\pi\pi$. The results
are
listed in Table~\ref{table1}. 
We see that the predicted rates for the BGT model are close to the
experimental values.

Note that the phase space factor $G$ for
$\Upsilon^{\prime\prime}\to\Upsilon~\pi\pi$ is much larger than
that for $\Upsilon^\prime\to \Upsilon~\pi\pi$:
$G(\Upsilon^{\prime\prime}\to\Upsilon~\pi\pi)/
G(\Upsilon^\prime\to \Upsilon~\pi\pi)=33$. Thus, one might naively
expect that
$\Gamma(\Upsilon^{\prime\prime}\to\Upsilon~\pi\pi)>\Gamma(\Upsilon^\prime\to
\Upsilon~\pi\pi)$. However, as we see from the experimental values
in Table~\ref{table1},
$\Gamma(\Upsilon^{\prime\prime}\to\Upsilon~\pi\pi)
/\Gamma(\Upsilon^\prime\to \Upsilon~\pi\pi)\approx
1.33/8.89=0.15$. The reason why the predictions for this ratio are
close to the experimental value is that the contributions from
various intermediate states to the overlap integrals in the
summation in $f^{111}_{3010}$ [cf. Eq.~(\ref{f})] {\it drastically
cancel} each other due to the fact that the
$\Upsilon^{\prime\prime}$ wave function contains two nodes. This
is a {\it characteristic} of this type of intermediate state
model (QCS or bag model).

Improved theoretical studies of these processes in a systematic
relativistic coupled-channel theory are needed since the
$\psi^\prime$ and $\Upsilon^{\prime\prime}$ are close the the open
flavor thresholds.

Recently, the $\eta'_c$ state was found experimentally with a
mass $M_{\eta'_c}=3.638\pm0.004$ GeV \cite{ted_PDG2006}. It will be
interesting to measure the hadronic transition
$\eta'_c\to\eta_c\pi\pi$ at \bes3. A very crude estimate of the
transition rate was made in Ref.~\cite{part4_Voloshin} by only
estimating the phase space in the single-channel approach. Since
the $\eta'_c$ lies very close to the $c\bar c$ threshold, a more
sophisticated theoretical study of this transition that takes into
account the coupled-channel and relativistic corrections is
needed.

\subsubsection{$\bm\eta$ Transitions}

The transitions $~n_I^3S_1\to n_F^3S_1+\eta~$ have contributions
from E1M2 and M1M1 transitions, and is dominated by an E1M2
transition. Similar to $\pi\pi$ transitions, using the data
\cite{ted_PDG2006}
\begin{eqnarray}                   
\Gamma_{\rm tot}(\psi^\prime)=277\pm 22~{\rm keV},,
~~~~~~~~B(\psi^\prime\to J/\psi~\eta)=(3.09\pm 0.08)\%
\label{input'}
\end{eqnarray}
as inputs, we can predict the rates for
$\Upsilon^\prime\to\Upsilon~\eta$ and
$\Upsilon^{\prime\prime}\to\Upsilon~\eta$ to be
\begin{eqnarray}                       
\Gamma(\Upsilon(n_I^3S_1)\to\Upsilon~\eta)=\displaystyle\frac{\bigg|\displaystyle
\frac{f^{111}_{n_I010}(b\bar{b})}{m_b}\bigg|^2}
{\bigg|\displaystyle\frac{f^{111}_{2010}(c\bar{c})}{m_c}\bigg|^2}\frac{|{\bm q}(b\bar{b)}|^3}
{|{\bm q}(c\bar{c})|^3}
\Gamma(\psi^\prime\to J/\psi~\eta).
\label{Upsilon,eta}
\end{eqnarray}
where $\bm q(b\bar{b})$ and $\bm q(c\bar{c})$ are the momenta of
$\eta$ in $\Upsilon(n_I^3S_1)\to\Upsilon~\eta$ and $\psi^\prime\to
J/\psi~\eta$, respectively. Taking the BGT model as an example to
calculate the ratio of transition amplitudes in
(\ref{Upsilon,eta}), we obtain
\begin{eqnarray}                         
\Gamma(\Upsilon^\prime\to\Upsilon~\eta)=0.022~{\rm keV},~~~~~~
\Gamma(\Upsilon^{\prime\prime}\to\Upsilon~\eta)=0.011~{\rm keV},
\label{Upsilon'Upsilon",eta}
\end{eqnarray}
which are consistent with the present experimental bounds \cite{ted_PDG2006}
\begin{eqnarray}                  
\Gamma(\Upsilon^\prime\to\Upsilon~\eta)<0.064~{\rm keV},~~~~~~
\Gamma(\Upsilon^{\prime\prime}\to\Upsilon~\eta)<0.045~{\rm keV}.
\label{eta-expt}
\end{eqnarray}

We can also compare the ratios $R^\prime\equiv
\Gamma(\Upsilon^\prime\to\Upsilon~\eta)/\Gamma(\psi^\prime\to
J/\psi~\eta)$ and $R^{\prime\prime}\equiv
\Gamma(\Upsilon^{\prime\prime}
\to\Upsilon~\eta)/\Gamma(\psi^\prime\to J/\psi~\eta)$ with 
recent experimental measurements. Recently BESII has reported
accurate measurements of $\Gamma(\psi^\prime\to J/\psi~\eta)$ and
$\Gamma(\psi^\prime\to J/\psi~\pi^0)$ \cite{part4_besggJ}. With
the new BESII values and the bounds on
$\Gamma(\Upsilon^\prime\to\Upsilon~\eta)$ and
$\Gamma(\Upsilon^{\prime\prime}\to\Upsilon~\eta)$
\cite{ted_PDG2006}, the experimental bounds on $R^\prime$ and
$R^{\prime\prime}$ are \cite{part4_besggJ}
\begin{eqnarray}                       
R^\prime|_{expt}<0.0061,~~~~R^{\prime\prime}|_{expt}<0.0043.
\label{R'R"expt}
\end{eqnarray}
Using the BGT model to calculate the ratios $R^\prime$ and 
$R^{\prime\prime}$, we obtain
\begin{eqnarray}                    
R^\prime|_{BGT}=
0.0025,~~~~~~~~~~
R^{\prime\prime}|_{BGT}=
0.0013,
\label{R'R"}
\end{eqnarray}
which are consistent with the experimental bounds
(\ref{R'R"expt}). \bes3 can further improve the measurements of
$\Gamma(\psi^\prime\to J/\psi~\eta)$ and $\Gamma(\psi^\prime\to
J/\psi~\pi^0)$.

\subsection{$\bm{\pi\pi}$ Transitions of $\bm D$-Wave Charmonium}

The $\psi(3770)$ (or $\psi^{\prime\prime}$) is primarily the $1D$
state of the charmonium. The measured leptonic width of the
$\psi(3770)$ is $(0.24\pm 0.06)$ keV \cite{ted_PDG2006}. If we
simply regard the $\psi(3770)$ as a pure $1D$ state of charmonium, the
predicted leptonic width is smaller than the experimental value by
an order of magnitude. Therefore it is generally considered that the
$\psi(3770)$ is a mixture of charmonium states, i.e., the observed
$\psi'$ and $\psi(3770)$ states are
\cite{part4_KY90,part4_Kuang02,part4_Godfrey}.
\begin{eqnarray}                             
&&\psi^\prime=\psi(2S)\cos\theta+\psi(1D)\sin\theta,\nonumber\\
&&\psi(3770)=-\psi(2S)\sin\theta+\psi(1D)\cos\theta.
\label{mixing}
\end{eqnarray}
The mixing angle $\theta$ can be determined phenomenologically by
fitting the ratio of the leptonic widths of the $\psi^\prime$ and
$\psi(3770)$. The leptonic widths of the $\psi(2S)$ and $\psi(1D)$ are
proportional to the wave function at the origin $\psi_{2S}(0)$ and
the second derivative of the wave function at the origin
$\displaystyle
\frac{5}{\sqrt{2}}\frac{d^2\psi_{1D}(0)/dr^2}{2m_c^2}$,
respectively. Therefore the determination of $\theta$ depends on
the potential model. Here we take two potential models as
examples, namely the Cornell model \cite{part4_Cornell} and an
improved QCD-motivated potential model by Chen and Kuang (CK)
\cite{part4_CK92} that produces more successful phenomenological
results. The determined values of $\theta$ are
\begin{eqnarray}                    
&&{\rm Cornell}:~~\theta=-10^\circ,\nonumber\\
&&{\rm CK}:~~~~~~~\theta=-12^\circ.
\label{theta}
\end{eqnarray}

Since the $\psi(3770)$ lies above the $D\bar{D}$ threshold, it is
usually believed that the $\psi(3770)$ mainly decays into the open
channel $D\bar{D}$. However, the measured cross section
$\sigma(e^+e^-\to\psi(3770)\to D\bar{D})=5.0\pm 0.5~{\rm nb}$
\cite{part4_MARKIII} is smaller than the measured direct
production cross section
$\sigma(e^+e^-\to\psi(3770))=7.5\pm0.8~{\rm nb}$
\cite{part4_MARKII,part4_CrystalBall84}. Experiments have been
searching for non-$D\bar D$ decays of the $\psi(3770)$ 
that might account for this discrepancy for a long
time. In 2005, BESII detected the hadronic transition $\psi(3770)\to
J/\psi~\pi\pi$ \cite{part4_BES05}, which was later 
confirmed by CLEO-c~\cite{part4_CLEO-c06}. This was the first
experimentally observed non-$D\bar D$ decay mode of the
$\psi(3770)$. More non-$D\bar D$ decay channels have been searched
for.

Theoretical studies of the $\psi(3770)\to J/\psi~\pi\pi$ transition
were reported much earlier \cite{part4_KY90,part4_Kuang02}. This
transition is dominated by E1E1 gluon emission and its
rate is given by~\cite{part4_KY90}
\begin{eqnarray}                       
\Gamma\big(\psi(3770)\to J/\psi~\pi\pi\big)&=&|C_1|^2\bigg[\sin^2\theta ~G(\psi^\prime)
~|f^{111}_{2010}(\psi^\prime)|^2
+ \nonumber \\
&&\frac{4}{15}\bigg|\frac{C_2}{C_1}\bigg|^2\cos^2\theta ~H(\psi^{\prime\prime})
~|f^{111}_{1210}(\psi^{\prime\prime})|^2\bigg].
\label{3770rate}
\end{eqnarray}

This rate depends both on the value of $C_2/C_1$ and the potential
model (through the amplitudes $f^{111}_{2010}$, $f^{111}_{1210}$).
We use the Cornell model \cite{part4_Cornell} and the CK model
\cite{part4_CK92} as examples. Taking the possible parameter range
of~\cite{part4_KY90}
\begin{eqnarray}                       
1\le C_2/C_1\le 3,
\label{C_2/C_1}
\end{eqnarray}
we obtain the values of $\Gamma\big(\psi(3770)\to
J/\psi+\pi^++\pi^-\big)$ listed in Table~\ref{table2}. Note that
$S$-$D$ mixing only affects the rate at the few percent level
and, so, the rate is essentially 
that for $\Gamma(\psi(1D)\to J/\psi~\pi^+\pi^-)$.

\begin{table}[h]
\caption{The predicted transition rate $\Gamma(\psi(3770)\to
J/\psi+\pi^++\pi^-)$ (in keV) in the Cornell model and the CK
model with the updated input data (\ref{input}).} \label{table2}
\tabcolsep 0.5cm \centering
\begin{tabular}{cc}
\hline\hline
Model&$\Gamma(\psi(3770)\to J/\psi~\pi^+\pi^-)$~(keV)\\
\hline
Cornell& $26\--139$\\
CK& $32\-- 147$\\
\hline\hline
\end{tabular}
\end{table}

The BESII-measured branching ratio ${\mathcal B}(\psi(3770)\to
J/\psi+\pi^++\pi^-)$, based on 27.7 pb$^{-1}$ data 
at the $\psi(3770)$, is \cite{part4_BES05}
\begin{eqnarray}                       
B\big(\psi(3770)\to J/\psi+\pi^++\pi^-\big)=(0.34\pm 0.14\pm 0.09)\%,
\label{BES3770-BR}
\end{eqnarray}
which, using the $\psi(3770)$ total width of\cite{ted_PDG2006}
\begin{eqnarray}                     
\Gamma_{tot}\big(\psi(3770)\big)=23.0\pm 2.7~{\rm MeV},
\label{3770width}
\end{eqnarray}
gives the partial width value \cite{part4_BES05}
\begin{eqnarray}                      
\Gamma_{BES}\big(\psi(3770)\to J/\psi+\pi^++\pi^-\big)=80\pm32\pm21~ {\rm keV}.
\label{BES3770-Gamma}
\end{eqnarray}
This is in agreement with the theoretical predictions in
Table~\ref{table2}. Using the BESII data (\ref{BES3770-Gamma}) and
Eq.~(\ref{3770rate}) to determine $C_2/C_1$, we obtain
\begin{eqnarray}                      
C_2/C_1=2^{+0.7}_{-1.3}.
\label{BESC2/C1}
\end{eqnarray}
This shows that $C_2/C_1$ is, in fact, within the estimated range
(\ref{C_2/C_1}).

The CLEO-c  measurement of the branching ratio is \cite{part4_CLEO-c06}
\begin{eqnarray}                        
B\big(\psi(3770)\to J/\psi+\pi^++\pi^-\big)=\big(0.214\pm 0.025\pm 0.022\big)\%,
\label{CLEO3770-BR}
\end{eqnarray}
which corresponds to a partial width of
\begin{eqnarray}                       
\Gamma\big(\psi(3770)\to J/\psi+\pi^++\pi^-\big)=50.5\pm 16.9~{\rm keV}.
\label{CLEO3770-Gamma}
\end{eqnarray}
Considering the experimental errors, this is consistent with the
BESII result (\ref{BES3770-Gamma}). We can also determine $C_2/C_1$
from (\ref{CLEO3770-Gamma}) and (\ref{3770rate}), and the result
is
\begin{eqnarray}                        
C_2/C_1=1.52^{+0.35}_{-0.45}.
\label{CLEO-cC2/C1}
\end{eqnarray}
This is consistent with the value (\ref{BESC2/C1}) determined from
the BESII data, but with better precision.

Further improvement of the measurement of the transition rate of
$\psi(3770)\to J/\psi+\pi^++\pi^-$ at \bes3 
is needed to determine the fundamental parameter $C_2/C_1$
that  occurs in other hadronic transitions.

\subsection{Studying the $\bm{h_c}$ State}

\label{sec:transition_hadron_hc}

The spin-singlet $P$-wave state $h_c$ (or $\psi(1^1P_1)$) is of
special interest since the difference between its mass $M_{h_c}$
and the center-of-gravity of the $\chi_{cJ}$ states, $M_{\rm
c.o.g.}(\chi_{c})
=(5M_{\chi_{c2}}+3M_{\chi_{c1}}+M_{\chi_{c0}})/9=3525.30\pm 0.08$
MeV, gives useful information about the spin-dependent interactions
between the heavy quark and antiquark. If the spin-dependent
interaction is perturbative, it is shown in Ref.~\cite{part4_CO}
that $M_{h_c}$ will be a few MeV smaller than $M_{\rm
c.o.g.}(\chi_{c})$. There have been a number of experiments 
that searched for the $h_c$ state.

In $\bar{p}p$ collisions, $h_c$ can be directly produced. The E835
experiment recently found the $h_c$ state via the decay channel
$\bar{p}p\to h_c\to\eta_c\gamma$, and the measured resonance mass
is $M_{h_c}=3525.8\pm 0.2\pm 0.2$ MeV with a width
$\Gamma_{tot}(h_c)\lesssim 1$ MeV \cite{part4_E835}. The measured
production rate is consistent with the theoretical range given in
Ref.~\cite{part4_KTY88} (see, in addition, Ref.~\cite{part4_E835}).

At $e^+e^-$ colliders, the $h_c$ state cannot be produced
directly in the $s$-channel due to its $C$ and $P$ quantum numbers.
Because of the limited phase space, the best way to search for
the $h_c$ state at CLEO-c or \bes3 is through the isospin-violating
hadronic transition
\cite{part4_CrystalBall83,part4_KTY88,part4_Kuang02}
\begin{eqnarray}                      
\psi^\prime\to h_c+\pi^0.
\label{psi'-h_cpi^0}
\end{eqnarray}
Recently, CLEO-c has found the $h_c$ state via the process
$\psi'\to h_c\pi^0\to\eta_c\gamma\gamma\gamma$
\cite{part4_CLEO-ch_c05}. The measured resonance mass is
$M_{h_c}=3525.35\pm 0.27\pm 0.2$ MeV \cite{part4_CLEO-ch_c05}, which
is consistent with the E835 result at about the $1\sigma$ level.

Theoretical studies were given much earlier in
Refs.~\cite{part4_KTY88,part4_Kuang02,part4_GR02}. The process
$\psi^\prime\to h_c+\pi^0$ is dominated by E1M1 transitions. \if
Taking account
The transition amplitude is
\begin{eqnarray}               
{\cal M}_{E1M1}&=&i\frac{g_Eg_M}{6}\frac{1}{2m_c}\sum_{KL}\frac{\langle h_c|\bar x_k|KL\rangle
\langle KL|(s_c-s_{\bar c})_l|\psi^\prime\rangle+
\langle h_c|(s_c-s_{\bar c})_l|KL\rangle
\langle KL|\bar x_k|\psi^\prime\rangle}{M_{\psi^\prime}-E_{KL}}
\nonumber \\
&&\langle \pi^0|E_kB_l|0\rangle,
\label{E1M1amplitude}
\end{eqnarray}
where $\bm s_c$ and $\bm s_{\bar c}$ are spins of $c$ and
$\bar{c}$, respectively. Since $\pi^0$ is a pseudoscalar, the
hadronization factor $\langle\pi^0|E_kB_l|0\rangle$ is
nonvanishing only when $E_kB_l=\delta_{kl}\bm{E}\cdot\bm{B}/3
=\delta_{kl}F^a_{\mu\nu}\tilde F^{a\mu\nu}/12$ which is related to
the chiral anomaly \cite{part4_GTW}. Using the
Gross-Treiman-Wilczek formula \cite{part4_GTW} without making
further approximations, we obtain
\begin{eqnarray}                       
\langle\pi^0|\alpha_sF^a_{\mu\nu}\tilde{F}^{a\mu\nu}|0\rangle
=\frac{4\pi}{\sqrt{2}}\frac{m_d-m_u}{m_d+m_u}f_\pi m_\pi^2,
\label{GTWresult}
\end{eqnarray}
in which the factor $(m_d-m_u)/(m_d+m_u)$ reflects the violation
of isospin. To predict the transition rate with these expressions,
we should determine the relations between the effective coupling
constants $\alpha_E=\displaystyle\frac{g^2_E}{4\pi}$,
$\alpha_M=\displaystyle\frac{g^2_M}{4\pi}$ and the coupling
constant $\alpha_s$ appearing in Eq.~(\ref{GTWresult}). With
certain approximations, we can calculate the transition rates
$\Gamma(\psi^\prime\to J/\psi~\pi\pi)$ expressed in terms of
$\alpha_E$ \cite{part4_KY81,part4_KTY88}, so that $\alpha_E$ can
be determined by taking the input data (\ref{input}). The
determined $\alpha_E$ is approximately $\alpha_E\approx 0.6$ which
is just about the commonly estimated value of $\alpha_s$ at the
light hadron scale. So we approximately take
$\alpha_s\approx\alpha_E$. The determination of $\alpha_M$ is
quite uncertain. Here we take a possible range \cite{part4_KTY88}
\begin{eqnarray}                     
\alpha_E\le\alpha_M\le 3\alpha_E
\label{alpha_M}
\end{eqnarray}
to estimate the rate. \fi 
Taking account of the $S$-$D$ mixing
(\ref{mixing}) in the $\psi'$, the transition rate is
\cite{part4_Kuang02}
\begin{eqnarray}                    
\Gamma(\psi^\prime\to h_c\pi^0)&=&\frac{\pi^3}{143m_c^2}\bigg(\frac{\alpha_M}{\alpha_E}\bigg)
\bigg|\cos\theta\bigg(f^{110}_{2011}+f^{001}_{2011}\bigg)-\sqrt{2}\sin\theta
\bigg(f^{110}_{1211}+f^{201}_{1211}\bigg)\bigg|^2\nonumber\\
&&\times\frac{E_{h_c}}{M_{\psi^\prime}}
\bigg[\frac{m_d-m_u}{m_d+m_u}f_\pi m_\pi^2\bigg]^2|\bm q_\pi|.
\label{Gamma(psi'-h_cpi^0)}
\end{eqnarray}
Numerical results based on the CK potential model are \cite{part4_Kuang02}
\begin{eqnarray}                    
\Gamma(\psi^\prime\to h_c\pi^0)&=&0.06\bigg(\frac{\alpha_M}{\alpha_E}\bigg)~{\rm
keV},\nonumber \\
B(\psi^\prime\to h_c\pi^0)&=&(2.2\pm 0.2)\bigg(\frac{\alpha_M}{\alpha_E}\bigg)\times 10^{-4}.
\label{psi'-h_cpi^0result}
\end{eqnarray}
Calculations show that the dependence of the transition rate on
the potential model is mild \cite{part4_Kuang02,part4_KTY88}.

We know that the $\pi^0$ decays $99\%$ into two photons. Thus the
signal in (\ref{psi'-h_cpi^0}) is $\psi^\prime\to h_c\gamma\gamma$
with $M_{\gamma\gamma}=m_{\pi^0}$. If the momenta of the two
photons can be measured with sufficient accuracy, one can look for
monochromatic $\pi^0$s as the signal. From the branching
ratio in Eq. (\ref{psi'-h_cpi^0result}), we see that, taking account
of a $\sim 10\%$ detection efficiency, hundreds of signal events can be
observed in a data sample of 10 millions of $\psi^\prime$s. The
backgrounds are shown to be either small or can be clearly
excluded \cite{part4_Kuang02}. Once the two photon energies
$\omega_1$ and $\omega_2$ are measured, the $h_c$ mass can be
extracted from the relation
$M_{h_c}^2=M_{\psi^\prime}^2+m_{\pi^0}^2-2M_{\psi^\prime}(\omega_1+\omega_2)$.

To have a clearer signal, one can further look at the decay
products of the $h_c$. It has been shown that the main decay channel for
the $h_c$ is $h_c\to\eta_c\gamma$ \cite{part4_Kuang02}. So, the
cleanest signal would be $\psi'\to h_c\pi^0\to \eta_c\gamma\gamma\gamma$.
The branching ratio $B(h_c\to\eta_c\gamma)$ depends on the
hadronic width of the $h_c$. In Ref.~\cite{part4_Kuang02}, the
hadronic width of $h_c$ was studied both in the conventional
perturbative QCD (PQCD) and in the nonrelativistic QCD (NRQCD)
approaches. The predictions are \cite{part4_Kuang02}
\begin{eqnarray}                          
&&{\rm PQCD}:~~~~~~~~~~~~~~~~~~~~~~~~B(h_c\to\eta_c\gamma)=(88\pm 2)\%,
\label{PQCDB}\\
&&{\rm NRQCD}:~~~~~~~~~~~~~~~~~~~~~~B(h_c\to\eta_c\gamma)=(41\pm 3)\%.
\label{NRQCDB)}
\end{eqnarray}
These lead to the predictions
\begin{eqnarray}                          
{\rm PQCD}:~~~~~~B(\psi'\to h_c\pi^0)\times B(h_c\to\eta_c\gamma)&=&1.9\left(\frac{\alpha_M}{\alpha_E}\right)
\times 10^{-4} \nonumber \\
& = &(1.9\--5.8)\times 10^{-4},
\label{PQCDBB}
\end{eqnarray}
\begin{eqnarray}
{\rm NRQCD}:~~~~B(\psi'\to h_c\pi^0)\times B(h_c\to\eta_c\gamma)&=&0.9\left(\frac{\alpha_M}{\alpha_E}\right)\times 10^{-4}
\nonumber \\
&=&(0.9\--2.7)\times 10^{-4}.
\label{NRQCDBB}
\end{eqnarray}
CLEO-c measured
\cite{part4_CLEO-ch_c05}
\begin{eqnarray}                          
~~~~B(\psi'\to h_c\pi^0)\times B(h_c\to\eta_c\gamma)=(3.5\pm 1.0\pm 0.7)\times 10^{-4}
\label{CLEO-cBB}
\end{eqnarray}
which is within the  theoretically predicted
range (\ref{PQCDBB}) for the PQCD approach. However, considering the large
measurement errors in (\ref{CLEO-cBB}), the present CLEO-c value is also
consistent with the NRQCD prediction (\ref{NRQCDBB}). Future
improved measurements at \bes3 and CLEO-c 
can distinguish 
between these two different approaches to hadronic decays.

{\if

Comparing the PQCD prediction (\ref{PQCDBB}) with the CLEO-c
result (\ref{CLEO-cBB}), we can determine
\begin{eqnarray}                          
\frac{\alpha_M}{\alpha_E}=1.8\pm0.9.
\label{alpha_M/alpha_E}
\end{eqnarray}
This is really within the estimated range
(\ref{alpha_M}). Note that the $50\%$ uncertainty in
(\ref{alpha_M/alpha_E}) comes from the errors in (\ref{CLEO-cBB}).
Since $\alpha_M/\alpha_E$ is a fundamental parameter in the theory
of QCD multipole expansion, it is important to improve the
measurement of $B(\psi'\to h_c\pi^0)\times B(h_c\to\eta_c\gamma)$
at \bes3 and CLEO-c to determine this fundamental parameter to a
higher precision.

Since the hadronization factor (\ref{GTWresult}) in this $\psi'\to
h_c\pi^0$ process is obtained from the Gross-Treiman-Wilczek
relation without taking approximations, the agreement between
(\ref{PQCDBB}) and (\ref{CLEO-cBB}) implies that the above
theoretical approach to the heavy quark {\bf MGE} factor in
(\ref{E1M1amplitude}) is quite reasonable.

\fi}

CLEO-c has also studied the branching ratios for some exclusive
hadronic decay channels of $\eta_c$ \cite{part4_CLEO-ch_c05}. More
accurate measurement of the branching ratios of these exclusive
hadronic channels at \bes3 and CLEO-c can also be used to compare
the corresponding predictions in Ref.~\cite{part4_Kuang02} and test
the PQCD and NRQCD approaches.

In summary, there are many aspects that need improved
experimental studies of the $h_c$ state at \bes3 and
CLEO-c.
\begin{description}
\item{\bf i})~~~Because of the errors in the E835 and CLEO-c
experiments, we still cannot judge whether $M_{h_c}$ is larger or
smaller than $M_{c.o.g.}(\chi_{c})$. Improved
measurements of $M_{h_c}$, e.g. including exclusive channel
measurements, can clarify this issue. \item{\bf
ii})~~In order to determine the fundamental parameter
$\alpha_M/\alpha_E$ and to test the PQCD and NRQCD approaches
to the hadronic decays of $h_c$, improved measurements of
$B(\psi'\to h_c\pi^0)\times B(h_c\to\eta_c\gamma)$
are needed. \item{\bf iii})~Branching ratios for
various exclusive channels have been calculated in
Ref.~\cite{part4_Kuang02} and found to be different
in the PQCD 
and the NRQCD approachs. Improved measurements of the exclusive
channel branching ratios can distinguish between these two approaches.
\item{\bf iv})~~Some other decay modes of the $h_c$ state have
been discussed in Ref.~\cite{part4_KTY88}. For instance,
$\Gamma(h_c\to J/\psi~\pi\pi) = 4.12 \left(\displaystyle
\frac{\alpha_M} {\alpha_E}\right)$ keV, etc \cite{part4_KTY88}.
After the accumulation of a large enough $\psi^\prime$ sample, other
decay modes of $h_c$ may also be measured and the properties
of the $h_c$ state better understood.
\end{description}

\subsection{$\bm{\pi\pi}$ Transitions of $\bm P$-Wave Quarkonia}

Theoretical studies of the hadronic transitions
$\chi_b(2^3P_{J_I})\to\chi_b(1^3P_{J_F})\pi\pi$ have been reported in
Ref.~\cite{part4_KY81}. Recently, CLEO measured the transition
rate of $\Gamma(\chi_b(2^3P_{J_I})\to \chi_b(1^3P_{J_F})\pi\pi)$
\cite{part4_Cawlfield}, and the results are consistent with the
theoretical predictions \cite{part4_Kuang06}.

So far, no hadronic transitions of the $\chi_{cJ}$ states have
yet been observed. The $\chi_{cJ}$ decays that have been observed are 
mainly decays into light hadrons, and the hadronic
widths of the three $\chi_{cJ}$ states are rather different. The
$\chi_{c1}$ has the smallest hadronic decay rate \cite{ted_PDG2006},
and is, thus, the most promising of the three $\chi_{cJ}$
states for studying hadronic transitions. The main hadronic
transition process for $\chi_{c1}$ is expected
to be $\chi_{c1}\to\eta_c\pi\pi$
which is dominated by the E1-M1 transition. Its transition rate
has been computed in Ref.~\cite{part4_LK06}, and the obtained rate
in the two-gluon approximation is
\begin{eqnarray}                            
\Gamma(\chi_{c1}\to\eta_c\pi\pi)=\displaystyle\frac{4\alpha_E\alpha_M}{8505\pi
m_c^2} |f^{010}_{1110}+|f^{101}_{1110}|^2
\left(M_{\chi_{c1}}-M_{\eta_c}\right)^7, \label{ratechic1}
\end{eqnarray}
where
\begin{eqnarray}                                      
&&f^{LP_IP_F}_{n_Il_In_Fl_F}=\sum_K f^{LP_IP_F}_{n_Il_In_Fl_F}(K),\nonumber\\
&&f^{LP_IP_F}_{n_Il_In_Fl_F}(K)\equiv\displaystyle\frac{\int R^*_F(r'){r'}^{P_F}R_{KL}(r'){r'}^2dr'
\int R^*_{KL}(r){r}^{P_I}R_I(r){r}^2dr}{E_I-E_{KL}}.
\label{ff}
\end{eqnarray}
We use the CK potential model \cite{part4_CK92} as an example to
calculate the radial wave functions in (\ref{ff}).
To determine $\alpha_E$, we take the same approximation for
$\Gamma(\psi'\to J/\psi~\pi\pi)$ which is
\begin{eqnarray}                                
\Gamma(\psi'\to
J/\psi~\pi\pi)=\frac{8\alpha_E^2}{8505\pi}|f^{111}_{2010}|^2(M_{\psi'}-M_{J/\psi})^7,
\label{2S-1Spipi}
\end{eqnarray}
and determine $\alpha_E$ by taking Eqn.~(\ref{input}) as input. This
gives $\alpha_E=0.46$. In this case the predicted partial width is
\begin{eqnarray}                         
\null\hspace{-0.2cm}
\Gamma(\chi_{c1}\to\eta_c\pi\pi)=9.0\left(\frac{\alpha_M}{\alpha_E}\right){\rm
keV} =16.3\pm 8.1~{\rm keV}. \label{rate}
\end{eqnarray}
The result in the Cornell potential model is smaller than the
value in (\ref{rate}) by $14\%$ \cite{part4_LK06}, so that the
model dependence of the prediction is not significant. 
The total width of $\chi_{c1}$ is
$\Gamma(\chi_{c1})=0.89\pm 0.05$ MeV \cite{ted_PDG2006}. So that the
predicted branching fraction is
\begin{eqnarray}                             
B(\chi_{c1}\to\eta_c\pi\pi)=(1.82\pm1.02)\%.
\label{BR}
\end{eqnarray}

At $e^+e^-$ colliders, the $\chi_{c1}$ state can be produced 
at the $\psi'$ peak via $\psi'\to\gamma\chi_{c1}$ decay. 
In the rest frame of the $\psi'$, the momentum of 
the $\chi_{c1}$ is 171 MeV  which is only
$5\%$ of its mass $M_{\chi_{c1}}=3510.66$ MeV. Therefore we can
neglect the motion of $\chi_{c1}$, and simply take the branching
fraction (\ref{BR}) to estimate the event numbers in the experiments.

The detection of the process
\begin{eqnarray}                           
\psi'\to\gamma\chi_{c1}\to\gamma\eta_c\pi\pi
\label{chain}
\end{eqnarray}
can be performed in two ways, namely the {\it inclusive} and the
{\it exclusive} detections \cite{part4_CLEO}. In the inclusive
detection, only the photon and the two pions are detected, while
the $\eta_c$ is inferred from the missing energy and momentum. In the 
exclusive
detection, the photon, the two pions, and the decay products
of the $\eta_c$ are all detected. Reconstruction of the $\eta_c$ and
$\chi_{c1}$ from the measured final state tagging particles can
suppress backgrounds.

The inclusive detection requires measuring the momenta of the
photon and the pions to certain precision. It is difficult to do
this kind of analysis with the BESII data because the BESII
photon resolution is not good enough. At \bes3 and CLEO-c,
this kind of detection may be possible.

For \bes3, it is not expected to be difficult to accumulate 
a data sample of $\sim 10^8$ $\psi'$ events. 
CLEO-c is now running at the $\psi'$ peak again, and will
soon accumulate $\sim3\times 10^7$ $\psi'$ events. The branching ratio
of $\psi'\to\gamma\chi_{c1}$ is
$B(\psi'\to\gamma\chi_{c1})=(8.7\pm0.4)\%$ \cite{ted_PDG2006}.
Taking account of a $15\%$ detection efficiency, we obtain the
number of events of type~(\ref{chain}) at \bes3 and CLEO-c
\begin{eqnarray}                          
&&
{\it 
BES-III}:\hspace{1.2cm}N_{incl}(\psi'\to\gamma\chi_{c1}\to\gamma\eta_c\pi\pi)
=(2.38\pm1.43)\times 10^4;
\label{inclusiveBES III}\\
&&
{\rm CLEO-c}:\hspace{1.0cm}N_{incl}(\psi'\to\gamma\chi_{c1}\to\gamma\eta_c\pi\pi)
=(7.15\pm4.27)\times10^3.
\label{inclusiveCLEO-c}
\end{eqnarray}
These numbers are so large that it seems 
 the transition (\ref{chain}) should be 
clearly identified.

For the exclusive detection, suitable decay modes of the $\eta_c$
should be used for identifying the $\eta_c$ in the process~(\ref{chain}).
Some feasible decay modes that have reasonable branching fractions are
\cite{ted_PDG2006,part4_CLEO-ch_c05}
\begin{eqnarray}                    
&&\eta_c\to\rho\rho:~~~~~~~~~~~~~~~~~~~~B(\eta_c\to\rho\rho)=(2.0\pm0.7)\%,
\label{rhorho}\\
&&\eta_c\to K^*\bar{K}^*:~~~~~~~~~~~~~~~B(\eta_c\to K^*\bar{K}^*)=(1.03\pm0.26)\%,
\label{KK}\\
&&\eta_c\to\phi\phi:~~~~~~~~~~~~~~~~~~~~B(\eta_c\to\phi\phi)=(0.27\pm0.09)\%,
\label{phiphi}\\
&&\eta_c\to K^*(892)^0K^-\pi^+:~~~B(\eta_c\to K^*(892)^0K^-\pi^+)=(2.0\pm0.7)\%,
\label{KKpi}\\
&&\eta_c\to K^+K^-\pi^+\pi^-:~~~~~~~B(\eta_c\to K^+K^-\pi^+\pi^-)=(1.5\pm 0.6)\%,
\label{KKpipi}\\
&&\eta_c\to 2(\pi^+\pi^-):~~~~~~~~~~~~~B(\eta_c\to 2(\pi^+\pi^-)=(1.20\pm0.30)\%,
\label{2pipi}\\
&&\eta_c\to\eta\pi\pi\to\gamma\gamma\pi\pi:~~~~~~~
B(\eta_c\to\eta\pi\pi\to\gamma\gamma\pi\pi)=(1.9\pm0.7)\%,
\label{etapipi}\\
&&\eta_c\to K_SK^\pm\pi^\mp:~~~~~~~~~~~B(\eta_c\to K_SK^\pm\pi^\mp)=(1.9\pm0.5)\%,
\label{K_SKpi}\\
&&\eta_c\to K_LK^\pm\pi^\mp:~~~~~~~~~~~B(\eta_c\to K_LK^\pm\pi^\mp)=(1.9\pm0.5)\%.
\label{K_LKpi}
\label{modes}
\end{eqnarray}
The modes (\ref{K_SKpi}) and (\ref{K_LKpi}) have been used by
CLEO-c in the search for the $h_c$ state \cite{part4_CLEO-ch_c05}.

\begin{table}[h]
\caption{Predictions for the numbers of events in the exclusive
detection of the process (\ref{chain}) in the CK potential model
using the $\eta_c$ decay modes shown in
(\ref{rhorho})$\--$(\ref{K_LKpi}) at BESII, \bes3 , and CLEO-c.
The accumulated numbers of the $\psi'$ events are taken to be
$1.4\times 10^7$ for BESII, $10^8$ for \bes3, and $3\times
10^7$ for CLEO-c. The detection efficiency is taken to be $10\%$
for BES II, and $15\%$ for \bes3 and CLEO-c.}\label{table3}
\begin{center}
\begin{tabular}{cccc}
\hline\hline
modes&BESII&\bes3&CLEO-c\\
\hline
$\rho\rho$&$44\pm42$&$476\pm454$&$143\pm136$\\
$K^*\bar{K}^*$&$23\pm20$&$245\pm210$&$74\pm62$\\
$\phi\phi$&6$\pm$6&$64\pm60$&$19\pm18$\\
$K^*(892)^0K^-\pi^+$&$44\pm42$&$476\pm454$&$143\pm136$\\
$K^+K^-\pi^+\pi^-$&$34\pm34$&$358\pm358$&$107\pm107$\\
$2(\pi^+\pi^-)$&$26\pm23$&$286\pm243$&$85\pm73$\\
$\eta\pi\pi\to\gamma\gamma\pi\pi$&$42\pm41$&$452\pm439$&$136\pm132$\\
$K_SK^\pm\pi^\mp$&$42\pm36$&$452\pm391$&$136\pm118$\\
$K_LK^\pm\pi^\mp$&$42\pm36$&$452\pm391$&$136\pm118$\\
\hline\hline
\end{tabular}
\end{center}
\end{table}

For the exclusive detection, the requirement of the precision of
the photon momentum measurement in (\ref{chain}) is not as strict.
So it is possible to do this kind of analysis with the BESII data
except for those $\eta_c$ decay modes that have photons. Therefore
we also estimate the event numbers in BESII. The BESII data sample
contains $1.4\times 10^7$ $\psi'$ events. Considering the ability
of the BESII detector, we take a detection efficiency of $\sim 10\%$.
The predicted numbers of events for BESII, \bes3,
and CLEO-c are listed in Table~\ref{table3}. We see that the
exclusive detection of the process (\ref{chain}) can be well
studied at \bes3 and CLEO-c. Considering the theoretical
uncertainties, the number of events at BESII may be marginal.

As we have mentioned in the last subsection, inproved
measurements of $B(\psi'\to h_c\pi^0)\times B(h_c\to \eta_c\gamma)$
at \bes3 and CLEO-c will determine
$\alpha_M/\alpha_E$ more acurately.  In this case, the predictions 
for the numbers of events of process~(\ref{chain}) will be more
reliable.

\subsection{Summary}

We have seen that the theory of hadronic transitions based on
QCDME is quite successful (For details see
Ref.~\cite{part4_Kuang06}). Future studies will include 
hadronic transitions involving excited states that are close to 
or above
the open-flavor threshold. Therefore, the development of a systematic
theory for hadronic transitions that includes relativistic and
coupled-channel effects is needed.

Soon, the \bes3 group will have a high-quality detector plus the highest 
luminosity ever enjoyed by an $e^+e^-$ experiment in 
the charm threshold region.  Thus,
even transitions with small rates could  be detected and studied
by them.


\section[Radiative transition]
{Radiative transition\footnote{By Ted Barnes}}
\label{sec:transition_rad}

\subsection{E1 radiative transitions}

Radiative transitions are a very interesting feature of charmonium
physics. They are quite straightforward to evaluate in $c\bar c$
potential models, and (with sufficient statistics) provide a route
from the initial $1^{--}$ states produced in $e^+e^-$ annihilation
to C = $(+)$ charmonia.

The largest rates are for E1 (electric dipole) transitions, which
in the nonrelativistic quark model involve a simple matrix element
of $\vec x$. The results we will quote here use the expression

\be \Gamma_{\rm E1}( {\rm n}\, {}^{2{\S}+1}{\rm L}_{\J} \to {\rm
n}'\, {}^{2{\S}'+1}{\rm L}'_{{\J}'} + \gamma)
 =  \frac{4}{3}\,  e_c^2 \, \alpha \,
{\rm E}_{\gamma}^3 \, \frac{{\rm E}_f^{(c\bar c)}}{{\rm
M}_i^{(c\bar c)}}\, C_{fi}\, \delta_{{\S}{\S}'} \, |\,\langle {\rm
n}'\, {}^{2{\S}'+1}{\rm L}'_{{\J}'} | \; r \; |\, {\rm n}\,
{}^{2{\S}+1}{\rm L}_{\J} \rangle\, |^2 \ee where $e_c= 2/3$ is the
$c$-quark charge in units of $|e|$, $\alpha$ is the fine-structure
constant, E$_{\gamma}$ is the photon's energy, and the angular
matrix element $C_{fi}$ is \be C_{fi}=\hbox{max}({\L},\; {\L}')
(2{\J}' + 1) \left\{ { {{\L}' \atop {\J}} {{\J}' \atop {\L}} {{\S}
\atop 1}  } \right\}^2 . \ee This is the formula quoted by
Ref.\cite{ted_Kwong:1988ae}, except for our inclusion of a
relativistic phase space factor. (We note in passing that it is
also possible to evaluate these E1 transition rates using LGT;
preliminary results of this work have been presented by
Dudek~\cite{ted_Dudek:2006ej}.) We evaluate the matrix elements
$\langle {n'}^{2{\S}'+1}{\L}'_{{\J}'} |\; r \; |
n^{2{\S}+1}{\L}_{\J}  \rangle$ using the nonrelativistic
Schr\"odinger wavefunctions obtained in the model described in the
previous section.

\begin{table}[htbp]
\caption{ E1 radiative transitions of the low-lying narrow \ccbar
states in the NR and GI potential models, abstracted from
Ref.\cite{ted_Barnes:2005pb}. The masses are taken from
Table~\ref{Table_spectrum}; experimental masses were used (rounded
``input" column) if known, otherwise, theoretical values were
assumed.} \label{E1rad_a}
\begin{center}
\begin{tabular}{c | c c c c c c c c }
\hline Initial meson & Multiplets & Final meson &
\multicolumn{2}{c}{E$_{\gamma}$ (MeV)} &
\multicolumn{2}{c}{$\Gamma_{\rm thy}$~(keV)}
& $\Gamma_{\rm expt}$~(keV) & \\
&                &             & NR  & GI  & NR & GI &   \\
\hline \hline
\\
$\chi_2$ & 1P $\to$ 1S & $J/\psi$ & 429.  & 429.  & 424. & 313.  &
426. $\pm$ 51.
\\
\hline
\\
$\chi_1$ & 1P $\to$ 1S & $J/\psi$ & 390.  & 389.  & 314. & 239.  &
291. $\pm$ 48.
\\
\hline
\\
$\chi_0$ & 1P $\to$ 1S & $J/\psi$ & 303.  & 303.  & 152. & 114.  &
119. $\pm$ 19.
\\
\hline
\\
$h_c$ & 1P $\to$ 1S & $\eta_c$ & 504.  & 496.  & 498. & 352.  &
\\
\hline \hline
\\
$\psi' $ & 2S $\to$ 1P & $\chi_{2}$  & 128.  & 128.  &  38. & 24.
& 27. $\pm$ 4.
\\
& & $\chi_{1}$  & 171.  & 171.  &  54. & 29.  & 27. $\pm$ 3.
\\
& & $\chi_{0}$  & 261.  & 261.  &  63. & 26.  & 27. $\pm$ 3.
\\
\hline
\\
$\eta_c'$ & 2S $\to$ 1P & $h_c$ & 111.  & 119.  &  49. & 36.  &
\\
\hline \hline
\\
$\psi_3(1^3{\rm D}_3) $ & 1D $\to$ 1P & $\chi_2$ & 242.  & 282.  &
272. & 296.  &
\\
\hline
\\
$\psi_2(1^3{\rm D}_2) $ & & $\chi_2$ & 236.  & 272.  &  64. & 66.
&
\\
& & $\chi_1$ & 278.  & 314.  & 307. & 268.  &
\\
\hline
\\
$\psi(3770) $ & & $\chi_2$ & 208.  & 208.  &  4.9 & 3.3  & $<  21$
$(90\% \ c.l.)$ \cite{ted_Briere:2006ff}
\\
& & $\chi_1$ & 250.  & 251.  & 125. & 77.  &  $70 \pm 17 $
\cite{ted_Briere:2006ff}
\\
& & $\chi_0$ & 338.  & 338.  & 403. & 213.  & $172 \pm 30$
\cite{ted_Briere:2006ff}
\\
\hline
\\
$h_{c2}(1^1{\rm D}_2) $ & & $h_c$ & 264.  & 307.  & 339. & 344.  &
\\\hline
\end{tabular}
\end{center}
\end{table}

\begin{table}[htbp]
\caption{E1 radiative transitions of the broader $1^{--}$
charmonium states above 4 GeV in the NR and GI quark models
(evaluated as in Table~\ref{E1rad_a}).} \label{E1rad_b}
\begin{center}
\begin{tabular}{c | c c c c c c c c }
\hline Initial meson & Transition & Final meson &
\multicolumn{2}{c}{E$_{\gamma}$ (MeV)} &
\multicolumn{2}{c}{$\Gamma_{\rm thy}$~(keV)}
& $\Gamma_{\rm expt}$~(keV) & \\
&                &          &     NR    & GI   & NR & GI &   \\
\hline
\\
$\psi(4040) $ & 3S $\to$ 2P & $\chi_2'$ & 67.  &  119. &  14. &
48.  &
\\
& & $\chi_1'$ & 113.  & 145.  & 39. & 43.  &
\\
& & $\chi_0'$ & 184.  & 180.  & 54. &  22. &
\\
& \phantom{3S} $\to$ 1P & $\chi_2$ & 455.  & 508.  &  0.70 & 12.7
&
\\
& & $\chi_1$ & 494.  & 547.  &  0.53 & 0.85  &
\\
& & $\chi_0$  & 577.  & 628.  &  0.27 & 0.63  &
\\
\hline
\\
$\psi(4160) $ & 2D $\to$ 2P & $\chi_2'$ & 183.  & 210.   & 5.9  &
6.3 &
\\
& & $\chi_1'$ & 227.  & 234.   & 168. & 114. &
\\
& & $\chi_0'$ & 296.  & 269.   & 483. & 191.  &
\\
& \phantom{2D} $\to$ 1P & $\chi_2$ & 559.  & 590. & 0.79    &
0.027 &
\\
& & $\chi_1$ & 598.  &  628. & 14.    & 3.4  &
\\
& & $\chi_0$ & 677.  &  707. & 27.    & 35.  &
\\
& \phantom{2D} $\to$ 1F & $\chi_2(1^3{\rm F}_2)$ & 128.  &  101.
& 51.  & 17. &
\\
\hline
\\
$\psi(4415) $ & 4S $\to$ 3P & $\chi_2(3^3{\rm P}_2)$ & 97.  & 112.
&   68. &   66.  &
\\
& & $\chi_1(3^3{\rm P}_1)$ & 142.  & 131.  & 126. &   54.  &
\\
& & $\chi_0(3^3{\rm P}_0)$ & 208.  & 155.  & 0.003 &   25. &
\\
& \phantom{4S} $\to$ 2P & $\chi_2'$ & 421.  & 446. &   0.62 &
15.  &
\\
& & $\chi_1'$ & 423.  & 469.  &  0.49 &   0.92  &
\\
& & $\chi_0'$ & 527.  & 502.  &  0.24 &  0.39 &
\\
& \phantom{4S} $\to$ 1P & $\chi_2$ & 775.  & 804. &  0.61  &  5.2
&
\\
& & $\chi_1$ & 811.  & 841.  & 0.41  &  0.53  &
\\
& & $\chi_0$ & 887.  & 915.  & 0.18  &  0.13 &
\\\hline
\end{tabular}
\end{center}
\end{table}

Several very interesting features of E1 radiative transitions are
evident in Tables~\ref{E1rad_a} and \ref{E1rad_b}. First, the 1P
$\to$ 1S (Table~\ref{E1rad_a}) transitions are in very reasonable
agreement with experiment. It is notable that the predicted
radiative partial width for $h_c \to \gamma \eta_c$ is especially
large, which would be an interesting measurement provided that an
entry channel for the $h_c$ can be identified at BES.

The theoretical rates for the 2S $\to$ 1P transitions
(Table~\ref{E1rad_a}) appear too large by about a factor of two,
although the relativized model of Godfrey and Isgur
\cite{ted_Godfrey:1985xj} does not share this difficulty. We note
in passing that apparent good agreement between a pure-$c\bar c$
charmonium potential model and experiment may actually be
spurious; decay loop effects will contribute two-meson continuum
components to all these charmonium resonances, which may
significantly modify the predicted radiative transition rates.

E1 radiative transitions from the higher-mass charmonium states
are especially interesting. The 1${}^3$D$_1$ candidate
$\psi(3770)$ (Table~\ref{E1rad_a}) is predicted to have large
partial widths to $\gamma \chi_1$ and  $\gamma \chi_0$ (with
branching fractions of 0.5\% and 1.7\% respectively), but the
branching fraction to $\gamma \chi_2$ is predicted to be only
about $2 \cdot 10^{-4}$. This small number however follows from
the assumption that the $\psi(3770)$ is a pure ${}^3$D$_1$ state;
if there is a significant admixture of S-wave basis states in the
$\psi(3770)$, \be |\psi(3770)\rangle = \cos(\theta) \, |{}^3{\rm
D}_1 \rangle + \sin(\theta) \, | {2}^3{\rm S}_1 \rangle
\label{3770_mix} \ee one typically finds a much larger radiative
width to $\gamma \chi_2$
\cite{ted_Ding:1991vu,ycz_rosner,ted_Rosner:2004mi}. (See
Fig.~\ref{3770_rad}. The sign of the mixing angle $\theta$ depends
on the convention for the normalization of the ${}^3$D$_1$ and
${2}^3$S$_1$ basis states; note for example in Fig.1 of
Ref.\cite{ted_Rosner:2004mi} that a zero $\psi(3770) \to \gamma
\chi_2$ width requires a small negative mixing angle, whereas with
our conventions it would be positive.) Since the coupling of the
$\psi(3770)$ to $e^+e^-$ suggests a significant $2^3$S$_1$
component, a measurement of this radiative partial width at BES
will be especially useful as an independent test of the presence
of this amplitude in the $\psi(3770)$ wavefunction.

\begin{figure}
\centerline{
\includegraphics[width=10cm,angle=0,clip=]{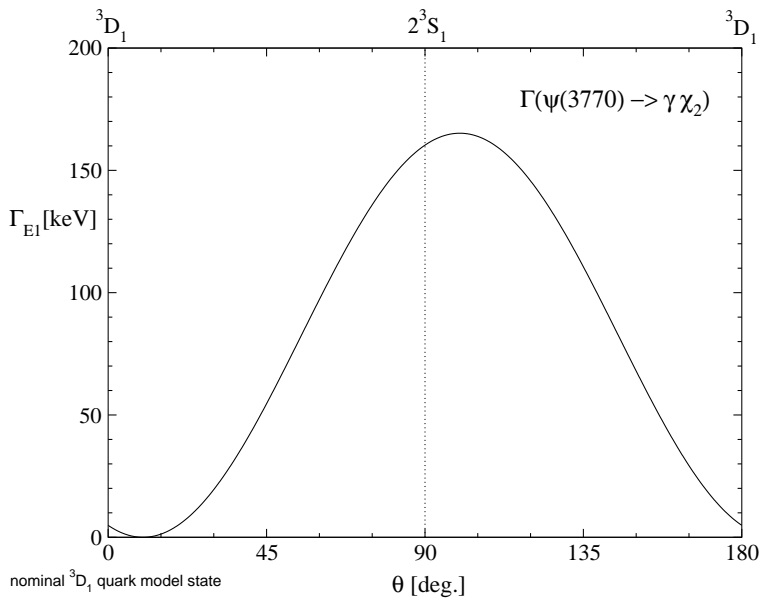}
} \caption{Predicted radiative partial width for the E1 transition
$\psi(3770)\to \gamma \chi_2$ as a function of the
${2}^3$S$_1$-${}^3$D$_1$ mixing angle $\theta$.} \label{3770_rad}
\end{figure}

We note in passing that if the dominant mechanism of ${}^3$D$_1$ -
${2}^3$S$_1$ basis state mixing in the $\psi(3686)$ and
$\psi(3770)$ is through virtual charm meson decay loops such as
DD, DD$^*$ and D$^*$D$^*$, the assumption of a $2\otimes 2$
orthogonal mixing matrix as in Eq.\ref{3770_mix} is incorrect. In
this case the $\langle {}^3$D$_1|\psi(3686)\rangle $ and $\langle
{2}^3$S$_1|\psi(3770)\rangle $ overlaps will no longer be simply
related, and radiative transition amplitudes will also receive
contributions due to photon emission from the two-meson continua.

Next we consider radiative decays of the higher-mass vectors
$\psi(4040)$ and $\psi(4160)$. As is evident in
Table~\ref{E1rad_b}, if the $\psi(4040)$ is dominantly a
${3}^3$S$_1$ state as we assume here, it should have very small E1
radiative widths to the triplet members of the 1P multiplet, with
branching fractions of at most about $10^{-6}$. Since other
components of the $\psi(4040)$ state vector (for example
${}^3$D$_1$ or charmed meson continua) may lead to significant
radiative couplings to 1P states, it will be very interesting for
BES to search for these radiative modes. The radiative widths of
the $\psi(4040)$ to the controversial 2P triplet states are
predicted to be much larger, with branching fractions of up to
$\sim 10^{-3}$. Since the 2P states should have large strong
branching fractions to DD ($\chi_0$ and $\chi_2$) and DD$^*$
($\chi_1$ and $\chi_2$) \cite{ted_Barnes:2005pb}, it may be
possible to identify these states in the DD and DD$^*$ invariant
mass distributions of the decays $\psi(4040) \to \gamma $DD and
$\gamma $DD$^*$.

Radiative decays of the $\psi(4160)$ share certain features with
the decays of both the $\psi(3770)$ and the $\psi(4040)$. First,
the coupling to the 2P multiplet is again predicted to be much
stronger than to the 1P multiplet, so radiative decays of the
$\psi(4160)$ can be used to search for 2P states. The branching
fraction to the ${2}^3$P$_0$ $\chi_0$ state in particular may be
as large as $\sim 0.5\%$ if it is at $\approx 3850$~MeV. Second,
just as for the $\psi(3770)$ a strong suppression of decays to all
${}^3$P$_2$ ``$\chi_2$" states is predicted, but this follows from
the assumption that the $\psi(4160)$ is a pure D-wave $c\bar c$
state; with an S-wave $\psi(4160)$ admixture, which is required to
explain the $\psi(4160)$ leptonic width, the coupling to
${}^3$P$_2$ states may be much larger; this should be searched for
at BES. Finally, an interesting feature of D-wave \ccbar E1 decays
is that one may reach the currently unknown 1F \ccbar multiplet
(specifically the state ${}^3$F$_2$); $\psi(4160)$ decays to
$\gamma DD$ should be appropriate for this. Unfortunately, the
${}^3$F$_2$ state is expected to be rather broad
\cite{ted_Barnes:2005pb}.

\subsection{Accessing the new states near 3.9 GeV through E1
transitions}

\begin{table}[htbp]
\caption{Theoretical E1 radiative partial widths of the
$\psi(4040)$ and $\psi(4160)$ into $C=(+)$ 2P $c\bar c$ states,
recalculated in the NR model with masses suggested by the new XYZ
states.} \label{ta2}
\begin{center}
{\begin{tabular}{@{}c|cccc@{}}
\hline
& & & & \\
Initial State & Final State & E1 Width  & E1 B.F.      &
\\
              &             & (keV)     &              &
\\\hline
& & & & \\
$\psi(4040)$ & $\chi_2'(3929)$  & $56.$ & $0.7 \cdot 10^{-3}$   &  \\
             & $\chi_1'(3940)$  & $25.$ & $0.3 \cdot 10^{-3}$   &  \\
             & $\chi_0'(3940)$  & $8.3$ & $0.1 \cdot 10^{-3}$   &  \\
&  &  &  & \\
&  &  &  & \\
$\psi(4160)$ & $\chi_2'(3929)$  & $9.9$  &  $0.1 \cdot 10^{-3}$  &  \\
             & $\chi_1'(3940)$  & $129.$ &  $1.3 \cdot 10^{-3}$  &  \\
             & $\chi_0'(3940)$  & $172.$ &  $1.7 \cdot 10^{-3}$  &  \\
& & & & \\
\hline
\end{tabular}}
\end{center}
\end{table}

The $\psi(4040)$ and $\psi(4160)$ can be used as $1^{--}$ entry
states for the study of the new XYZ states near 3.9~GeV. As shown
in Table~\ref{ta2}, both these states are expected to have
relatively large E1 branching fractions into the 2P $c\bar c$
multiplet, $\psi(4040,4160) \to \gamma \chi_J'$. (These E1 partial
widths were calculated as in Table~\ref{E1rad_b} and
Ref.\cite{ted_Barnes:2005pb}, but the final masses were lowered to
3929~MeV and 3940~MeV to accommodate the new XYZ resonances as 2P
candidates.) Note that these lower masses give significantly
larger E1 partial widths than in Table~\ref{E1rad_b}. Evidently,
studies of E1 transitions from the $\psi(4040)$ and $\psi(4160)$
at BES should allow the identification of the 2P resonances
through their hadronic decays. In this approach one would study
the invariant mass and angular distributions of the final charmed
mesons in the processes $e^+e^- \to \psi(4040,4160) \to \gamma
{\rm DD}$ and $\gamma {\rm DD}^*$.


\subsection{M1 radiative transitions}

M1 transitions between charmonium states in pure $c\bar c$ models
result from photon emission through the $H_I = - \vec \mu \cdot
\vec B$ magnetic moment interaction of the $c$ quark (and
antiquark), and as such are suppressed relative to E1 transitions
by the small factor of $1/m_c$ in the magnetic moment operator.
The M1 transition amplitude is proportional to the matrix element
of the spin operator, with a spatial factor that (without recoil
corrections) is simply the matrix element of unity. M1 transitions
are therefore nonzero only between states with the same L$_{c\bar
c}$ (and different S$_{c\bar c}$, since the C-parity must change).
If we assumed a spin-independent zeroth-order potential and
neglect recoil effects, M1 transitions between different radial
multiplets would vanish because the n$^3$S$_1$ and n$'^1$S$_0$
states have orthogonal spatial wavefunctions. One such transition
is actually observed in charmonium, $\psi{\, '} \to \gamma
\eta_c$, which must be due in part to the nonorthogonal $\psi{\,
'}$ and $\eta_c$ spatial wavefunctions and final meson recoil
effects.

The formula for M1 decay rates analogous to the E1 formula used in
the previous section is

\bd \Gamma_{\rm M1}( {\rm n}\, {}^{2{\S}+1}{\rm L}_{\J} \to {\rm
n}'\, {}^{2{\S}'+1}{\rm L}'_{{\J}'} + \gamma)
 =  \frac{4}{3}\,  e_c^2 \, \frac{\alpha}{m_c^2}\,
{\rm E}_{\gamma}^3 \, \frac{{\rm E}_f^{(c\bar c)}}{{\rm
M}_i^{(c\bar c)}}\, {2\J'+1\over 2\L +1}\, \delta_{{\L}{\L}'} \,
\delta_{{\S},{\S}'\pm 1} \, \ed \be \cdot |\,\langle {\rm n}'\,
{}^{2{\S}'+1}{\rm L}'_{{\J}'} |\,
{\rm n}\, {}^{2{\S}+1}{\rm L}_{\J} \rangle\, |^2 \ . \ee
Evaluating this formula for transitions from the $\psi$ and
$\psi{\, '}$ gives the results shown in
Table~\ref{Table_M1_rates}. A more detailed study of M1 radiative
decay rates, incorporating recoil corrections (which are
numerically important for transitions between multiplets such as
2S $\to$ 1S) appears in Ref.\cite{ted_Barnes:2005pb}.


An even larger discrepancy between experiment and theory is
evident in the ``hindered" M1 transition $\psi{\, '} \to \gamma
\eta_c$. Since this rate is only nonzero due to recoil effects
(not included here) and corrections to the naively orthogonal 1S
and 2S $c\bar c$ wavefunctions, the discrepancy is perhaps less
surprising than that found in the allowed 1S $\to$ 1S $\psi \to
\gamma \eta_c$ transition rate. In any case this is another
example of an M1 transition rate in charmonium in which experiment
and theory are clearly in disagreement. Since the experimental
rate is again only about $4\sigma $ from zero, it would be very
useful to improve the accuracy of this measurement at BES.

\begin{table}[h]
\caption{Theoretical and experimental M1 radiative partial widths
of the $\psi$ and $\psi{\, '}$, neglecting recoil effects.}
\label{Table_M1_rates}
\begin{center}
{\begin{tabular}{@{}c|cccc@{}}
\hline
 Initial meson & Final meson
& \phantom{}  &  $\Gamma_{thy.}$~(keV) & $\Gamma_{expt.}$~(keV) \\
\hline J$/\psi$   & $\eta_c $           && 2.9   &  $ 1.2 \pm 0.3$
\\
\hline $\psi{\, '}$   &  $\eta_c{\, '}$          && 0.21  &
\\
$\psi{\, '}$   &  $\eta_c$           && 4.6   &  $0.8 \pm 0.2 $
\\
$\eta_c{\, '}$   &  ${\rm J}/\psi $ && 7.9   &
\\
\hline
\end{tabular}}
\end{center}
\end{table}

A well-known problem is evident in the decay rate J$/\psi \to
\gamma \eta_c$, which is that the predicted rate in the
nonrelativistic potential model is about a factor of 2-3 larger
than experiment. Since this rate only involves the charm quark
magnetic moment, and hence only its mass, this discrepancy is a
surprise. The relativized Godfrey-Isgur model
\cite{ted_Godfrey:1985xj} predicts a somewhat smaller rate of
2.4~keV, which is still about a factor of two larger than
experiment. Since the errors are rather large, it would clearly be
very interesting to improve the experimental accuracy of this
surprising partial width. If this discrepancy is confirmed, it may
be an indication that pure-$c\bar c$ models are a rather
inaccurate description of charmonium, and that other components of
the state vector such as two-meson continua make comparable
important contributions to the M1 transition amplitudes. In view
of the inaccuracy of the theoretical 1S M1 transition rate, it
would also be interesting to test the 2S transition rate $\psi{\,
'} \to \gamma \eta_c{\, '}$ experimentally. Unfortunately, this
rate is predicted to be a rather small 0.21~keV in the
nonrelativistic $c\bar c$ model.

An even larger discrepancy between experiment and theory is
evident in the ``hindered" M1 transition $\psi{\, '} \to \gamma
\eta_c$. Since this rate is only nonzero due to recoil effects
(not included here) and corrections to the naively orthogonal 1S
and 2S $c\bar c$ wavefunctions, the discrepancy is perhaps less
surprising than that found in the allowed 1S $\to$ 1S $\psi \to
\gamma \eta_c$ transition rate. In any case this is another
example of an M1 transition rate in charmonium in which experiment
and theory are clearly in disagreement. Since the experimental
rate is again only about $4\sigma $ from zero, it would be very
useful to improve the accuracy of this measurement and to search
for the other M1 transitions at BES.

M1 decays between charmonium resonances have only been observed
between S-wave states. The rates between orbitally excited states
are typically predicted to be quite small, due to the small
splittings within excited-L multiplets. They are large enough
however to be observable given narrow initial states and large
event samples. For example, a hypothetical $^1$D$_2$ $c\bar c$
assignment for the X(3872) could be tested through a search for
its M1 decay to $\gamma \psi(3770)$, which would have a partial
width of about 1.2~keV in the nonrelativistic potential model. The
$h_c$ decay $h_c \to \gamma  \chi_0$ has similar phase space, and
is predicted to have a partial width of 0.8~keV. In contrast, the
smaller phase space of the M1 transition from the higher-mass
$\chi_2$ state leads to an expected partial width for $\chi_2 \to
\gamma h_c$ of only about 60~eV.

\section[Channels for measurement at \bes3]
{Channels for measurement at \bes3\footnote{By Changzheng Yuan}}
\label{sec:transition_channels}
\renewcommand{\dedx}{dE/dx}
\renewcommand{\BR}{\mathcal{B}}
\newcommand{\jpc}{J^{PC}}
\renewcommand{\eff}{\varepsilon}
\renewcommand{\gev}{\,\mbox{GeV}}
\renewcommand{\mev}{\,\mbox{MeV}}
\newcommand{\gevcc}{\,\mbox{GeV}/c^2}
\newcommand{\mevcc}{\,\mbox{MeV}/c^2}
\renewcommand{\ra}{\rightarrow}
\newcommand{\JPC}{J^{PC}}
\renewcommand{\psp}{\psi^{\prime}}
\renewcommand{\pspto}{\psi^{\prime}\to}
\renewcommand{\jpsi}{J/\psi}
\renewcommand{\jpsito}{J/\psi\to}
\renewcommand{\pspp}{\psi^{\prime\prime}}
\renewcommand{\psppto}{\psi^{\prime\prime}\to}
\renewcommand{\ppjpsi}{\pi^+\pi^-J/\psi}
\renewcommand{\hc}{h_c(^1P_1)}
\renewcommand{\etac}{\eta_c}
\renewcommand{\etacp}{\eta_c^{\prime}}
\renewcommand{\chicz}{\chi_{c0}}
\renewcommand{\chico}{\chi_{c1}}
\renewcommand{\chict}{\chi_{c2}}
\newcommand{\chicJ}{\chi_{cJ}}
\renewcommand{\ppkk}{\pi^+\pi^-K^+K^-}
\renewcommand{\EE}{e^+e^-}
\renewcommand{\EETO}{e^+e^-\to}
\renewcommand{\etappp}{\eta\pi^+\pi^-}
\newcommand{\ccg}{c\bar{c}g}
\renewcommand{\beqns}{\begin{eqnarray*}}
\renewcommand{\eeqns}{\end{eqnarray*}}
\renewcommand{\bfg}{\begin{figure}}
\renewcommand{\efg}{\end{figure}}
\renewcommand{\bitm}{\begin{itemize}}
\renewcommand{\eitm}{\end{itemize}}
\renewcommand{\bnum}{\begin{enumerate}}
\renewcommand{\enum}{\end{enumerate}}
\renewcommand{\btbl}{\begin{table}}
\renewcommand{\etbl}{\end{table}}
\renewcommand{\btbu}{\begin{tabular}}
\renewcommand{\etbu}{\end{tabular}}



\subsection{Introduction}

The BEPCII design peak luminosity is
$10^{33}$~cm$^{-2}$s$^{-1}$ for center-of-mass energies near the
$\pspp$ peak. Thus, the 
peak luminosity at the $\psp$, which is less than
100~MeV below the $\pspp$ peak,  should be about the same. Since the beam
energy spread of BEPCII will be around 1.4~MeV, the peak cross
section for $\psp$ production will be around 600~nb. Assuming that the
average luminosity is half of the peak luminosity and the
effective running time each year is around $10^7$~s, one can expect
as many as 3~billion $\psp$ events in a one year 
run~\cite{ycz_YZQ_hepnp}.  This data sample 
would be huge compared to those used by previous experiments, and the detector 
performance will also be much better.  The combined effect will be 
to make high precision measurements
of $\psp$ decays possible, and make searches for the modes
with very small branching fractions feasible.

The spectrum of charmonium states
below the open-charm threshold is shown in
Fig.~\ref{ccb_spec}.  Since the mass of $\psp$ is higher than those
of all of the other $n=1$ $S$- and $P$-wave
charmonium states, all these lower-mass charmonia
can, in principle, be accessed by  radiative and/or hadronic decays of
$\psp$. In the past, such processes have been fruitful for
both our theoretical and experimental understanding
of charmonium physics~\cite{ycz_QWG,ted_PDG2006}, 
However, because of the limited statistics of the old generation
experiments, and the poorer detector performance, not all of the
possible transition have been measured.  Some that are crucial 
for the further development of phenomenological models of charmonium 
physics have still not been observed.

\begin{figure*}[htbp]
\centerline{\psfig{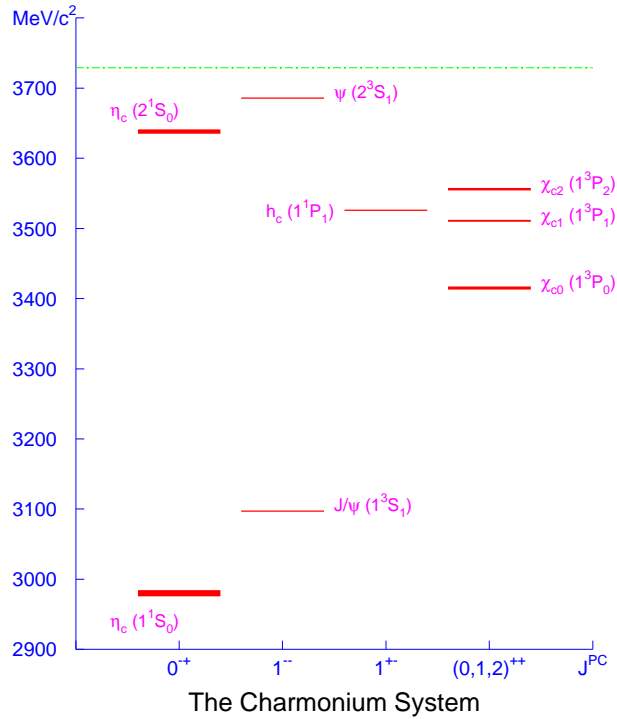}}
\caption{Charmonium spectroscopy below the open charm threshold.}
\label{ccb_spec}
\end{figure*}

In this section, we  list all the allowed radiative transitions
and hadronic transitions that can be studied with the large $\psp$
data sample that will be collected by \bes3. In addition,
we give an  overview of
the status of our studies and point out some topics where
additional theoretical effort is called for.

\subsection{Radiative transitions}

Since the $\JPC$ of the photon is $1^{--}$,
single photon  transitions can only
occur between two states of different $C$-parity. The 
transitions are
either electric- or magnetic-multipole
processes, depending on the spins and parities of the initial and
final states. In those cases where
the spins of the initial and final states
are $S_i$ and $S_f$, respectively, the total angular momentum
carried by the photon ($J_{\gamma}$) can be any integer between
$|S_i-S_f|$ and $S_i+S_f$. If the product of the parities of the
initial state ($\pi_i$) and final state ($\pi_f$) is equal to
$(-1)^{J_{\gamma}}$, the transition is an $E J_{\gamma}$ transition;
otherwise, if $\pi_i\cdot \pi_f=(-1)^{J_{\gamma}+1}$, it is an $M
J_{\gamma}$ transition. It is apparent that the electric multipole
transitions preserve the initial quark spin directions, while the 
magnetic multipole transitions are accompanied by spin flip of one of
the quarks.

In general, when more than one multipole transitions are allowed,
only the lowest one is important.   Nevertheless, for some charmonium
transitions, contributions of higher multipoles have been studied 
both theoretically and experimentally.

Radiative transitions between charmonium states have been
studied extensively by many authors both theoretically and
experimentally~\cite{ycz_PRD42p2293,ycz_PRD21p203,ycz_PRD26p2295,ycz_PRD25p2938,
ycz_PRD28p1692,ycz_PRD28p1132,ycz_besii_ggpsi,ycz_cleoc_tran}. The 
partial widths for some
of the transitions have also been calculated with lattice
QCD~\cite{ycz_dudek}.

\subsubsection{$\psp$ decays}

\bitm

 \item $\psp\to \gamma \chicJ$, $J=0,1,2$

These are the transitions between $S$-wave  and
$P$-wave spin triplets. For $\pspto \gamma \chicz$ there is only
an $E1$ transition, while for $\pspto \gamma \chico$ there can be $E1$
and $M2$ transitions, and in $\pspto \gamma \chict$ there can be
be $E1, M2$, as well as $E3$ transitions.

In general, it is believed that $\pspto \gamma \chicJ$ is
dominated by the $E1$ transition, but with some $M2$ (for $\chico$
and $\chict$) and $E3$ (for $\chict$) contributions due to
relativistic corrections. These contributions have been used to
explain the big differences between the calculated pure $E1$
transition rates and the experimental results~\cite{ycz_PRD21p203}.
They will also affect the angular distribution of the radiative
photon. Thus, measurements of the photon angular distribution can be
used to determine the contributions of the higher multipoles in
the transition.

Moreover, for $\pspto \gamma \chict$, the $E3$ amplitude is
directly connected with $D$-state mixing in the $\psp$, which has been
regarded as a possible explanation of the large leptonic
annihilation rate of $\pspp$~\cite{ycz_PRD28p1132}. Since recent
studies~\cite{ycz_rosner,ycz_wympspp,ycz_wymkskl} also suggest the $S$- and
$D$-wave mixing of $\psp$ and $\pspp$ may be the key to solve the
longstanding ``$\rho\pi$ puzzle'' and to explain $\pspp$
non-$D\bar{D}$ decays, more experimental information on multipole
amplitudes gains additional importance.

Decay angular distributions in $\pspto \gamma \chict$ were studied
by the Crystal Ball experiment using $\pspto\gamma\gamma
J/\psi$ events~\cite{ycz_Cball}; no significant contributions from higher multipoles
were found, but the errors were large due to
the limited statistics. In a recent analysis at
BESII~\cite{ycz_gpipi}, $\pspto \gamma \chict \to\gamma \pip \pim$ and
$\gamma K^+K^-$ decays were used for a similar study. The analysis
gave a magnetic quadrupole amplitude
$a_2'=-0.051^{+0.054}_{-0.036}$ and an electric octupole
amplitude $a_3'=-0.027^{+0.043}_{-0.029}$~\cite{ycz_amplitudes},
neither of which differs significantly from zero. The results are in
good agreement with what is expected for a pure $E1$ transition.
As for the $D$-state mixing of $\psp$, the results do not
contradict previous theoretical calculations by more than one
standard deviation~\cite{ycz_PRD30p1924}.

The contribution of these higher multipoles are of theoretical
interest, so further studies at \bes3, where a
much higher sensitivity for probing the higher multipoles
contribution would be possible, are anticipated.

 \item $\psp\to \gamma \etac$

This is a hindered $M1$ transition, as it occurs between $n=2$ and
$n=1$ states.

 \item $\psp\to \gamma \etacp$

This is an $M1$ transition, and analogous to the similar transition
between $\jpsi$ and $\etac$. However, the transition rate, which
is proportional to $E^3_{\gamma}$, is very
small since the mass difference between the $\psp$ and $\etacp$ is
not very large.

The study of the $\etacp$ in $\psp$ decays at \bes3 will challenge  the
ability of the experimentalists and the capabilities of the detector.

\eitm

\subsubsection{$\etacp$ decays}

The observation of these transitions will be very helpful for
understanding the $\etacp$ properties. From the experimental
point of view, these final states are observationally clean,
but the rates are small.

\bitm

 \item $\etacp\to \gamma \jpsi$

This is an $M1$ transition. It has been
calculated in Ref.~\cite{ycz_PRD30p1924}.

 \item $\etacp\to \gamma \hc$

This is an $E1$ transition. It was calculated some time ago
~\cite{ycz_PRD28p1692} to have a partial width of 16~keV.

 \eitm

\subsubsection{$\chicJ$ decays}

\bitm

 \item $\chicJ\to \gamma \jpsi$

These are the transitions between the $P$-wave and
$S$-wave spin triplets. For $\chicz\to \gamma \jpsi$, there is only
an $E1$ transition, while for $\chico\to \gamma \jpsi$, there are $E1$
and $M2$ transitions, and for $\chict\to \gamma \jpsi$, there could
be $E1, M2$, as well as $E3$ transitions.

Decay angular distributions for $\pspto \gamma \chict$
were studied by the Crystal Ball experiment using 
$\psp\to\gamma\gamma J/\psi$ events~\cite{ycz_Cball}; the 
no significant contribution 
from higher multipoles
was found, but the errors were large due to
the limited statistics. The $\chict\to\gamma\jpsi $ 
decay was also studied by E835 in $\ppb$ annihilation.

 \item $\chict\to \gamma \hc$

This can be an $M1$, $E2$ and/or $M3$ transition. There are no 
published calculations for this process.

\eitm

\subsubsection{$\hc$ decays}

\bitm

 \item $\hc\to \gamma \etac$

This $E1$ transition was the discovery mode for the $\hc$ state that
was used by CLEO~\cite{ycz_cleoc1p1}. The transition branching fraction is
expected to be large (more than 50\% of all $\hc$ decays),
which was confirmed by the CLEOc measurement. This should be
measured with higher precision at \bes3.

 \item $\hc\to \gamma \chicz$, $\gamma \chico$

$\hc\to \gamma \chicz$ is an $M1$ transition, and $\hc\to \gamma
\chico$ is an $M1$ and/or $E2$ transition. There are no calculations
of this transition in the literature.

Measurements of these transitions as well as $\chict\to \gamma
\hc$ will be difficult, since the rates may be small, and the photons
are very low energy.

\eitm

\subsubsection{$\jpsi$ decays}

\bitm

 \item $\jpsi\to \gamma \etac$

This is an $M1$ transition and a better measurement is needed 
to clarify the difference between the existing Crystal Ball measurement,
which is smaller than
theoretical predictions. Also, there are
some discrepancies between $\etac$
properties measured using $\jpsi$ decays 
and those derived  from experiments using other production modes, such
as $\gamma^*\gamma^*$ fusion, B decays etc., that have to be 
investigated and clarified.

This mode can be studied using either $\psp$ data sample, via
$\pspto \jpsi \pip\pim$, or using the $\jpsi$ data sample
collected at the $\jpsi$ resonance peak.

\eitm

All the radiative transitions between the charmonium states listed
above are indicated by arrows in Fig.~\ref{ccb_rad}.

\begin{figure*}[htbp]
\centerline{\psfig{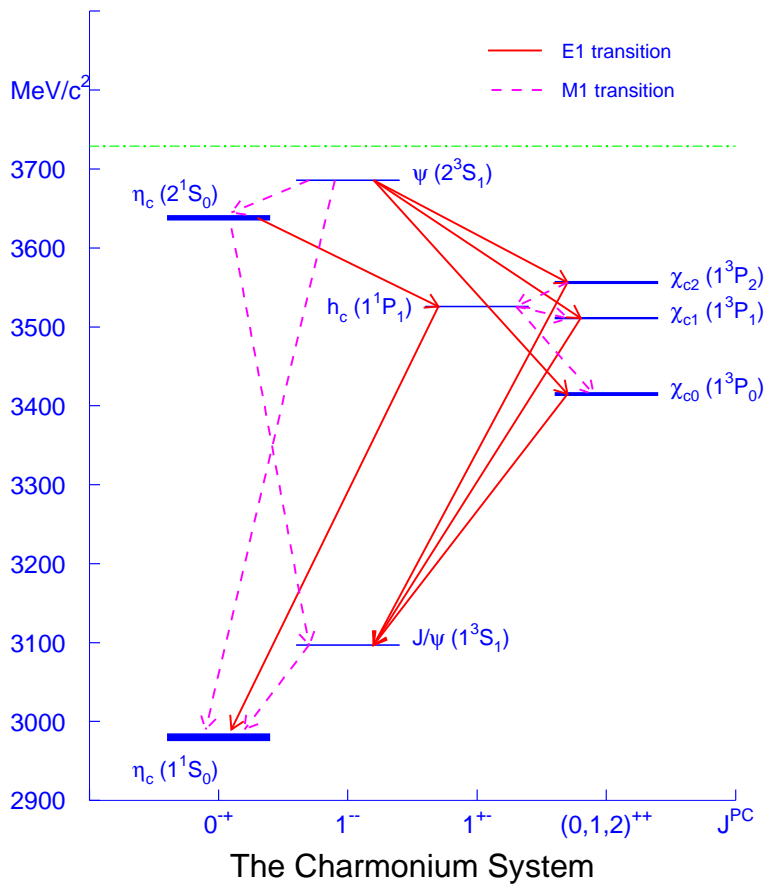}}
\caption{Radiative transitions between charmonium states below the
open charm threshold.} \label{ccb_rad}
\end{figure*}

\subsection{Hadronic transitions}

There are strong and electromagnetic transitions between two
charmonium states if the mass difference is large enough to
produce one or more $\pi$'s, 
and/or an $\eta$. $C$-parity conservation and
Parity conservation may forbid some of the transitions, and these
are pointed out below. The study of their usefulness for
searches for rare decays and potential signals for new 
physics is beyond the scope of this section.

Only the hadronic transitions of the $\psp$ have been well studied
experimentally. These  include the $\pi^+\pi^-\jpsi$, $\pi^0\pi^0\jpsi$,
$\eta\jpsi$ and $\pi^0\jpsi$ decay modes of the $\psp$. 
Extensive theoretical calculations have been
for these transitions as well. The other possible transitions
have not bee well studied.  Below we indicate  those where 
results are available.

It should be noted that, since the mass differences between the
charmonium states are not large, the light hadrons are generally
produced at very low momentum, this may provide some unique opportunities
to studying the physics of this energy domain.

\subsubsection{$\psp$ decays}

Since the mass difference between the $\psp$ and many of the
charmonium states are much larger than one $\pi$ mass, there
are a number of possible transitions.
All the kinematically allowed transitions are indicated in
Fig.~\ref{ccb_had_psp} and discussed below.

\begin{figure*}[htbp]
\centerline{\psfig{file=./Part4/transition/czy/cc_had_psp.epsi,width=10cm}}
\caption{Hadronic transitions of $\psp$ to other charmonium
states.} \label{ccb_had_psp}
\end{figure*}

\bnum

 \item {$\psp\to \etac +X$}

The mass difference between the $\psp$ and $\etac$ is 706~$\mevcc$, which is
greater than $5m_{\pi}$ and $m_{\eta}+m_{\pi}$; all of the possible
combinations are listed below. There are no reported
measurements for any of the
channels listed here, and only a very few theoretical considerations.

\bitm

  \item $\psp\to n\piz \etac$, $n=1,2,3,4,5$: $C$-violating, not allowed

  \item $\psp\to \pip\pim\etac$: $G$-parity violating, EM decays only, via
  $\rho^*$

  \item $\psp\to \pip\pim\piz\etac$: strong decays, via $\omega^*$

It was predicted that this mode would have a branching fraction at the 1\%
level~\cite{ycz_survival} in a model that was developed to explain the
``$\rhopi$ puzzle'' between $\psp$ and $\jpsi$ hadronic decays.
Using a 3~million $\psp$ data sample,  CLEOc~\cite{ycz_cleoc_etac3pi} 
established an upper
limit on the branching fraction of less than
$1.0\times 10^{-3}$ at 90\% C.L.

  \item $\psp\to 2(\pip\pim)\etac$: $G$-parity violating, EM decays only

  \item $\psp\to \pip\pim 2\piz \etac$: $G$-parity violating, EM decays only

  \item $\psp\to 2(\pip\pim)\piz\etac$: strong decays

  \item $\psp\to \pip\pim 3\piz \etac$: strong decays

  \item $\psp\to \eta\etac$: $C$-violating, not allowed

  \item $\psp\to \eta\piz\etac$: $C$-violating, not allowed

\eitm

\item {$\psp\to \jpsi +X$}

The mass difference between the $\psp$ and $\jpsi$ is 589~$\mevcc$,
which is greater than $4m_{\pi}$ and $m_{\eta}$, all the possible
combinations are listed below.
The channels in this category have been studied well both
experimentally and theoretically, thanks to the large decay branching
fractions and the distinct signature of the leptonic decays of the $\jpsi$.

\bitm

  \item $\psp\to \piz \jpsi$: $G$-parity violating, EM decays

This transition has been
observed (via $\piz\to \gamma\gamma$ and $\jpsi\to \ell^+\ell^-$) by
many experiments, most recently BESII and CLEO.

  \item $\psp\to \pip\pim\jpsi$: strong decays

This is one of the main transition modes of the $\psp$ with a
branching fraction that 
corresponds to  about one third of all decays. The $\pip\pim$ mass shows
the interesting feature of events 
clustering at high $\pip\pim$ masses, which
has been a hot topic of theoretical discussion 
that dates from the time of the discovery
of this decay mode and persisting until now.

The $\pi\pi$ are produced mainly in an $S$-wave, with the same 
$0^{++}$ quantum
number as the $\sigma$, the $D$-wave component was found to be small
by a BESI analysis based on a 4~million $\psp$ event sample.

The process has been analyzed in various models by many
authors~\cite{part4_Kuang06,ycz_shifman,ycz_yanml,ycz_sigma,ycz_guofk}; all the models fit
the data well.

  \item $\psp\to \piz\piz\jpsi$: strong decays

This is similar to $\pip\pim\jpsi$ mode. 
Isospin symmetry predicts that its
production rate should be
half of that of $\pip\pim\jpsi$. This was tested
with high precision by CLEOc  using a 3~million $\psp$
event sample.

An isospin violation  may exist, but it should be small, as indicated 
by the $\piz\jpsi$ and $\eta\jpsi$ rate difference.
This may be tested with more and higher precision data.

  \item $\psp\to \pip\pim\piz\jpsi$: $G$-parity violating, EM decays only

This rate can be roughly estimated from the $\piz\jpsi$ decay rate.

  \item $\psp\to 3\piz\jpsi$: $G$-parity violating, EM decays only

  \item $\psp\to 2(\pip\pim)\jpsi$: strong decays

Here the phase space is small, however
the rate may not be small since it is a
strong decay. The experimental detection is difficult, since the
$\pi^{\pm}$ momenta are low.

  \item $\psp\to \pip\pim 2\piz\jpsi$: strong decays

This is similar to $\psp\to 2(\pip\pim)\jpsi$, {\it i.e.} hard to detect.

  \item $\psp\to 4\piz \jpsi$: strong decays

This is similar to $\psp\to 2(\pip\pim)\jpsi$.  The detection of eight 
photons with energy near half of the $\piz$ mass will challenge the
Electromagnetic Calorimeter.

  \item $\psp\to \eta\jpsi$: strong decays

There are many measurements of this mode.
The ratio between the rate of this mode and the
isospin-violating $\piz\jpsi$ mode is used to measure the mass
difference of the $u$- and $d$-quarks, and the strength of 
electromagnetic contributions to $\psp$ hadronic transitions.

\eitm

\item {$\psp\to \chicJ +X$}

The mass difference between the $\psp$ and $\chicz$ is 271~$\mevcc$,
only slightly greater than $2m_{\piz}$ and lower than
$m_{\pip}+m_{\pim}$.  Since the width of the $\chicz$ is around
10~$\mevcc$, the decay $\psp\to \pip\pim\chicz$ could, in principle, 
be accessed via the low-mass tail of the $\chicz$. 

The mass difference between the $\psp$ and $\chico$
is 176~$\mevcc$, slightly greater than $m_{\piz}$;  the mass
difference between the $\psp$ and $\chict$ is 130~$\mevcc$, lower than
$m_{\piz}$. All the possible transitions are listed below.

\bitm

  \item $\psp\to n\piz \chicJ$, $n=1,2$: $C$-violating, not allowed

  \item $\psp\to \pip\pim\chicz$: $G$-parity violating, EM decays only, via
  $\rho^*$.  However the available phase space is very small and 
  only the low-mass tail of $\chicz$
  can be produced. There are no measurements and no theoretical
  calculations.

\eitm

\item {$\psp\to \hc +X$}

The mass difference between the $\psp$ and $\hc$ is 160~$\mevcc$,
slightly greater than $m_{\piz}$. The only kinematically allowed 
hadronic transition is $\psp\to \piz \hc$.

\bitm

  \item $\psp\to \piz\hc$: $G$-parity violating, EM decays only

This is the $\hc$ discovery mode that was exploited by
CLEO~\cite{ycz_cleoc1p1}. The product of the branching fraction and
that of $\hc\to \gamma \etac$ was determined in the same experiment.

More experimental effort is needed to understand this
transition, as well as to study better the properties of the $\hc$.

\eitm

\enum

\subsubsection{$\etacp$ decays}

Since the mass of $\etacp$ is only slightly below than that of
$\psp$, the mass difference between the $\etacp$ and many of the
charmonium states is also larger than one $\pi$ mass. As a result,
there are many possible transitions.
All of the kinematically allowed hadronic transitions are indicated in
Fig.~\ref{ccb_had_etacp} and discussed below.

\begin{figure*}[htbp]
\centerline{\psfig{file=./Part4/transition/czy/cc_had_ecp.epsi,width=10cm}}
\caption{Hadronic transitions of $\etacp$ to other charmonium
states.} \label{ccb_had_etacp}
\end{figure*}

\bnum

 \item {$\etacp\to \etac +X$}

The mass difference between $\etacp$ and $\etac$ is 658~$\mevcc$,
which is greater than $4m_{\pi}$ and $m_{\eta}$.  The possible
hadronic transitions are listed below.

\bitm

  \item $\etacp\to \piz \etac$: $P$-violating, not allowed

  \item $\etacp\to \pip\pim\etac$: strong decays, via
  $\sigma$

Voloshin~\cite{ycz_voloshin} pointed out that this decay is related to
the well studied $\psp\to \pip\pim \jpsi$ transition, and estimated
that the branching fraction  could be around 5-10\%, including
the neutral $\piz\piz$ mode. No experimental
information is currently available.

The study of this transition will very hard to do with
$\etacp$ mesons produced from radiative $\psp$ decays,
since in this process the $\etacp$ is produced with a very small 
branching fraction  and the photon energy is quite
low, which makes it hard to distinguish from background.

  \item $\etacp\to \piz\piz\etac$: strong decays, via
  $\sigma$

This is similar to, but harder than, the detection of $\etacp\to \pip\pim\etac$
decays.  Its observation will require the identification of
multi-photons in the event.

  \item $\etacp\to \pip\pim\piz\etac$: $G$-parity violating, EM decays only,
  also  high orbital angular momentum

This probably will not be detectable at \bes3.

  \item $\etacp\to 3\piz\etac$: $G$-parity violating, EM decays only,
  high orbital angular momentum

This will also be unlikely to be detected at \bes3.

  \item $\etacp\to 2(\pip\pim)\etac$: strong decays

Here the phase space small and this may not be detected at \bes3.

  \item $\etacp\to \pip\pim 2\piz \etac$: strong decays

This process hase very small phase space
and multiple low momentum charged and neutral pions.
It is unlikely that it will be detected at \bes3.

  \item $\etacp\to 4\piz \etac$: strong decays

Here the phase space is small and there are numerous
low-energy photons.   It will not be detectable at
\bes3.

  \item $\etacp\to \eta\etac$: $P$-violating, not allowed

\eitm

\item {$\etacp\to \jpsi +X$}

The mass difference between the $\etacp$ and $\jpsi$ is 541~$\mevcc$,
which is slightly greater than $4m_{\piz}$ and smaller than
$2(m_{\pip}+m_{\pim})$. Considering the uncertainty of the $\etacp$
mass is large and the width of $\etacp$ is probably at the $\sim 10$~$\mevcc$ level,
the high mass tail of the $\etacp$ could, in principle,
decay into $2(\pip\pim)\jpsi$. All
the possible combinations are listed below.

\bitm

  \item $\etacp\to n\piz \jpsi$, $n=1,2,3,4$: $C$-violating, not
  allowed

  \item $\etacp\to \pip\pim\jpsi$: $G$-parity~violating, EM decays only, 
via
  $\rho^*$

  \item $\etacp\to \pip\pim\piz \jpsi$: strong decays, via
  $\omega^*$

  \item $\etacp\to 2(\pip\pim)\jpsi$: $G$-parity violating, EM decays only,
  with very small phase space

  \item $\etacp\to \pip\pim2\piz \jpsi$: $G$-parity violating, EM decays only,
  with very small phase space

\eitm

The detection of the above modes maybe a bit easier than the
corresponding
$\etacp \to \etac$ transition modes since the $\jpsi$ tag is much
simpler, very narrow and quite distinct.

A naive estimate indicates that the rates for
the $\etacp\to \jpsi$ transition should
be smaller than the $\etacp\to \etac$ transitions, since the
former requires a quark spin flip. There has been no serious theoretical
effort expended on these estimations.

\item {$\etacp\to \chicJ +X$}

The mass difference between the $\etacp$ and $\chicz$ is 223~$\mevcc$,
slightly greater than $m_{\piz}$; the mass difference between
$\etacp$ and $\chico$ is 128~$\mevcc$, and that between $\etacp$
and $\chict$ is 82~$\mevcc$, smaller than $m_{\piz}$. The only
possible transition is $\etacp\to \piz \chicz$.

\bitm

  \item $\etacp\to \piz \chicz$: $G$-parity violating, EM decays only

\eitm

\item {$\etacp\to \hc +X$}

The mass difference between he $\etacp$ and $\hc$ is 112~$\mevcc$,
smaller than $m_{\piz}$. No hadronic transitions are allowed.

\enum

\subsubsection{$\hc$ decays}

The mass difference between the $\hc$ and many of the charmonium
states are also much larger than one $\pi$ mass, thus there are
many possible transitions.
All the allowed hadronic transitions are indicated in
Fig.~\ref{ccb_had_hc} and discussed  below.

\begin{figure*}[htbp]
\centerline{\psfig{file=./Part4/transition/czy/cc_had_hc.epsi,width=10cm}}
\caption{Hadronic transitions of $\hc$ to other charmonium
states.} \label{ccb_had_hc}
\end{figure*}

\bnum

 \item {$\hc\to \etac +X$}

The mass difference between the $\hc$ and $\etac$ is 546~$\mevcc$,
which is greater than $4m_{\pi}$ and about the same as $m_{\eta}$. All 
possible hadronic transitions are listed below.

\bitm

  \item $\hc\to n\piz \etac$, $n=1,2,3,4$: $C$-violating, not allowed

  \item $\hc\to \pip\pim\etac$: $G$-parity violating, EM decays only, via
  $\rho^*$

  \item $\hc\to \pip\pim\piz\etac$: strong decays, via $\omega^*$

  \item $\hc\to 2(\pip\pim)\etac$: $G$-parity violating, EM decays only

  \item $\hc\to \eta\etac$: $C$-violating, not allowed

\eitm

$\hc\to \pip\pim\etac$ and $\hc\to \pip\pim\piz\etac$ should be
looked for experimentally, $\hc\to 2(\pip\pim)\etac$ rate may be
too small to be detectable at \bes3.

\item {$\hc\to \jpsi +X$}

The mass difference between the $\hc$ and $\jpsi$ is 429~$\mevcc$,
greater than $3m_{\piz}$. All possible hadronic transitions are listed
below.

There is no experimental information currently available, neither
are there any theoretical calculations.

\bitm

  \item $\hc\to \piz \jpsi$: $G$-parity violating, EM decays only

  \item $\hc\to \pip\pim\jpsi$: strong decays, via $\sigma$

  \item $\hc\to \piz\piz\jpsi$: strong decays, via $\sigma$

  \item $\hc\to \pip\pim\piz\jpsi$: $G$-parity violating, EM decays only

  \item $\hc\to 3\piz \jpsi$: $G$-parity violating, EM decays only

\eitm

\item {$\hc\to \chicz +X$}

The mass difference between the $\hc$ and $\chicz$ is 111~$\mevcc$,
smaller than $m_{\piz}$. There are no kinematically
allowed hadronic transitions.

\enum

\subsubsection{$\chict$ decays}

The mass difference between the $\chict$ and many of the charmonium
states are also much larger than one $\pi$ mass.   Thus there are
many possible hadronic transitions.
All of the possibile transitions are indicated in
Fig.~\ref{ccb_had_chic2} and discussed below.

\begin{figure*}[htbp]
\centerline{\psfig{file=./Part4/transition/czy/cc_had_chic2.epsi,width=10cm}}
\caption{Hadronic transitions of $\chict$ to other charmonium
states.} \label{ccb_had_chic2}
\end{figure*}

\bnum

\item {$\chict\to \etac +X$}

The mass difference between the $\chict$ and $\etac$ is 576~$\mevcc$,
greater than $4m_{\pi}$ and $m_{\eta}$, all the possible
hadronic transitions are listed below.

\bitm

  \item $\chict\to \piz \etac$: $G$-parity violating, EM decays only

  \item $\chict\to \pip\pim\etac$: Strong decays, high orbital angular
  momentum

  \item $\chict\to \piz\piz\etac$: Strong decays, high orbital angular
  momentum

  \item $\chict\to \pip\pim\piz\etac$: $G$-parity violating, EM decays only

  \item $\chict\to 3\piz\etac$: $G$-parity violating, EM decays only

  \item $\chict\to 2(\pip\pim)\etac$: strong decays, but the phase space is
  very small and orbital angular momentum very high

  \item $\chict\to \pip\pim 2\piz \etac$: strong decays, phase space
  very small, orbital angular momentum very high

  \item $\chict\to 4\piz \etac$: strong decays, with small phase space
  and high orbital angular momentum

  \item $\chict\to \eta\etac$: strong decays, but with small phase space

\eitm

\item {$\chict\to \jpsi +X$}

The mass difference between the $\chict$ and $\jpsi$ is 459~$\mevcc$,
slightly greater than $3m_{\pi}$. All possible hadronic transitions
are listed below.

\bitm

  \item $\chict\to n\piz \jpsi$, $n=1,2,3$: $C$-violating, not
  allowed

  \item $\chict\to \pip\pim\jpsi$: $G$-parity violating, EM decays only, via
  $\rho^*$

  \item $\chict\to \pip\pim\piz\jpsi$: strong decays, via
  $\omega^*$

\eitm

\item {$\chict\to \chicz +X$, $\chico +X$}

The mass difference between the $\chict$ and $\chicz$ is 141~$\mevcc$,
slightly greater than $m_{\piz}$; the mass difference between
$\chict$ and $\chico$ is 46~$\mevcc$, smaller than $m_{\piz}$. The
only possible hadronic transition is $\chict\to \piz \chicz$.

\bitm

  \item $\chict\to \piz \chicz$: $P$-violating, not allowed

\eitm

\item {$\chict\to \hc +X$}

The mass difference between the $\chict$ and $\hc$ is 30~$\mevcc$,
smaller than $m_{\piz}$. There are no 
kinematically allowed hadronic transitions.

\enum

\subsubsection{$\chico$ decays}

The mass difference between the $\chico$ and many of the charmonium
states are also much larger than one $\pi$ mass.  Thus, there are
many possible transitions.
All the allowed transitions are indicated in
Fig.~\ref{ccb_had_chic1} and discussed below.

\begin{figure*}[htbp]
\centerline{\psfig{file=./Part4/transition/czy/cc_had_chic1.epsi,width=10cm}}
\caption{Hadronic transitions of $\chico$ to other charmonium
states.} \label{ccb_had_chic1}
\end{figure*}

\bnum

\item {$\chico\to \etac +X$}

The mass difference between the $\chico$ and $\etac$ is 531~$\mevcc$,
greater than $3m_{\pi}$, all the possible transitions are listed
below.

\bitm

  \item $\chico\to \piz \etac$: $P$-violating, not allowed

  \item $\chico\to \pip\pim\etac$: strong decays, via $\sigma$

A very rough measurement at BESII showed that it was
not observable using the BESII data sample.

  \item $\chico\to \piz\piz\etac$: strong decays, via $\sigma$

  \item $\chico\to \pip\pim\piz\etac$: $G$-parity violating, EM decays only

  \item $\chico\to 3\piz\etac$: $G$-parity violating, EM decays only

\eitm

\item {$\chico\to \jpsi +X$}

The mass difference between the $\chico$ and $\jpsi$ is 413~$\mevcc$,
slightly greater than $3m_{\piz}$ and smaller than
$m_{\pip}+m_{\pim}+m_{\piz}$ by 2~$\mevcc$. All the possible
transitions are listed below.

\bitm

  \item $\chico\to n\piz \jpsi$, $n=1,2,3$: $C$-violating, not
  allowed

  \item $\chico\to \pip\pim\jpsi$: $G$-parity violating, EM decays only, via
  $\rho^*$

  \item $\chico\to \pip\pim\piz\jpsi$: no phase space

\eitm

\item {$\chico\to \chicz +X$}

The mass difference between the $\chico$ and $\chicz$ is 95~$\mevcc$,
smaller than $m_{\piz}$. There are no 
kinematically allowed hadronic transitions.

\enum

\subsubsection{$\chicz$ decays}

The mass difference between the $\chicz$ and many of the charmonium
states are also much larger than one $\pi$ mass, thus there are
many possible transitions.
All the allowed transitions are indicated in
Fig.~\ref{ccb_had_chic0} and discussed below.

\begin{figure*}[htbp]
\centerline{\psfig{file=./Part4/transition/czy/cc_had_chic0.epsi,width=10cm}}
\caption{Hadronic transitions of $\chicz$ to other charmonium
states.} \label{ccb_had_chic0}
\end{figure*}

\bnum

\item {$\chicz\to \etac +X$}

The mass difference between the $\chicz$ and $\etac$ is 435~$\mevcc$,
greater than $3m_{\pi}$, all the possible hadronic transitions are listed
below.

\bitm

  \item $\chicz\to \piz \etac$: $G$-parity violating, EM decays only

  \item $\chicz\to \pip\pim\etac$: $P$-violating, not allowed

  \item $\chicz\to \piz\piz\etac$: $P$-violating, not allowed

  \item $\chicz\to \pip\pim\piz\etac$: $G$-parity violating, EM decays only

  \item $\chicz\to 3\piz\etac$: $G$-parity violating, EM decays only

\eitm

\item {$\chicz\to \jpsi +X$}

The mass difference between the $\chicz$ and $\jpsi$ is 318~$\mevcc$,
slightly greater than $2m_{\pi}$. All the possible hadronic transitions
are listed below.

\bitm

  \item $\chicz\to n\piz \jpsi$, $n=1,2$: $C$-violating, not
  allowed

  \item $\chicz\to \pip\pim\jpsi$: $G$-parity violating, EM decays only, via
  $\rho^*$

\eitm

\enum

\subsection{Summary}

In this section, we listed all the kinematically
allowed transitions between
the known charmonium states below the charm threshold, more
studies are needed for a better understanding of these transitions.

\section[Monte Carlo simulation of spin-singlet charmonium states]
{Monte Carlo simulation of spin-singlet charmonium
states\footnote{By Haixuan Chen}}
\label{sec:transition_simulation}


\subsection{$\psip\to\gamma\etac,\etac\to K^0_S K \pi$}

Using the BESIII Offline Software System (BOSS), a data sample of
about 150,000 Monte Carlo simulated 
$\psip\to\gamma\etac,\etac\to K^0_S K^{\pm} \pi^{\mp}$  
events has been analyzed. In order to extract
the $\etac$ signal, two charged tracks with net
charge zero and originating from the interaction region plus a 
$K_S \to \pi^+\pi^-$ decay with a
reconstructed secondary vertex are required. The photon
candidate with highest energy is regarded as the radiative photon
from the $\psip$. The $K_S$ selection includes the
requirements $m_{K_S} \in
[0.44,0.56]~\hbox{GeV}/c^2$ and $\cos \theta > 0.9$,
where $\theta$ is the angle between the radiative photon and the
PMISS direction of the $\etac\to K^0_S K \pi$ candidate. 
Figure~\ref{mkskpi_etac}
shows the $K_S K\pi$ invariant mass distribution, where the
$\etac$ signal is evident; a fit gives a mass resolution value of
$(6.78\pm0.05)~\hbox{MeV}/c^2$.  The output value of $m_{\etac}$
from the fit to the reconstructed $M_{K_S K \pi}$ peak is $2979.94\pm 0.06
~\hbox{MeV}/c^2$.   The difference from the input mass is $\triangle m_{\etac} =
(0.14\pm0.06)~\hbox{MeV}/c^2$, thereby
demonstrating that $m_{\etac}$ can be correctly reconstructed in
this channel.  The detection efficiency based on these event selection
requirements is $(14.60\pm0.07)\%$,  where uniform angular distributions
are assumed. The detection efficiency is presumed to be smaller if 
more realistic angular distributions are considered. 
Using the branching fractions
of $\psip\to\gamma\etac$ and $\etac\to K^0_S K^{\pm}
\pi^{\mp}$~\cite{ted_PDG2006}, and assuming a detection efficiency of
$10\%$,  $1800$  $\etac$ signal signal events are expected to be observed
in a $3\times 10^{8}$ $\psip$ event sample (which could be collected within
about 0.1 year of BESIII operation).

\begin{figure}[htbp]
\centerline{\psfig{file=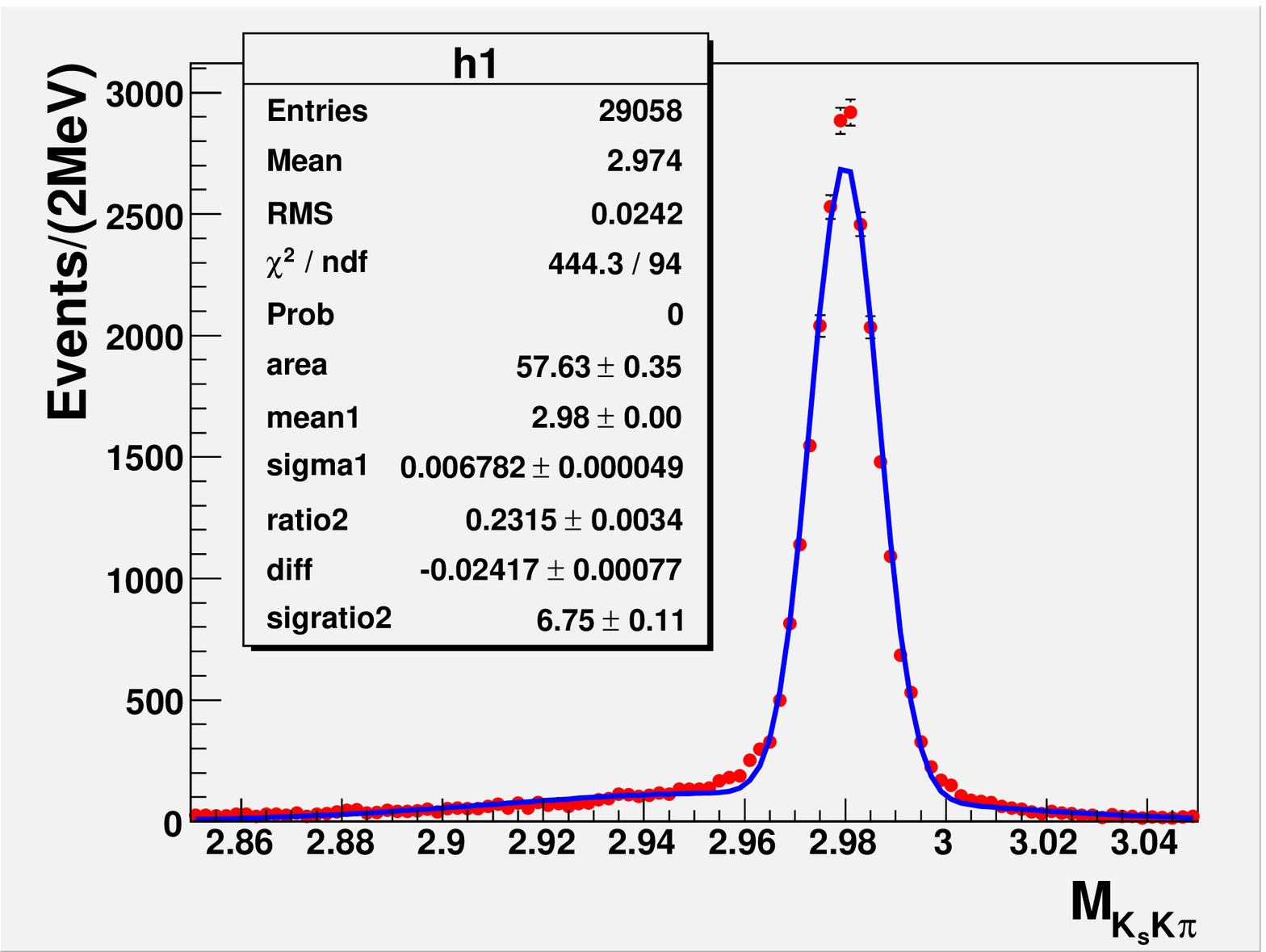,width=8.0cm}}
\caption{The $K_S K\pi$ nvariant mass distribution in the
$\etac$ mass region. The distribution is fitted using a double
Gaussian function.}
\end{figure}\label{mkskpi_etac}

\subsection{$\psip\to\gamma\etacp,\etacp\to K^0_S K \pi$}

About 150,000 Monte Carlo simulated 
$\psip\to\gamma\etacp,\etacp\to K^0_S K^{\pm} \pi^{\mp}$ events are
analyzed.  The same selection criteria as those used in analysis of
$\psip\to\gamma\etac$ are applied to the reconstructed data and 
the resulting $K_S K\pi$ invariant mass distribution is shown in
Fig.~\ref{mkskpi_etacp}.  An $\etacp$ signal with a
mass resolution of $(9.64\pm 0.11)~\hbox{MeV}/c^2$ is evident. 
The output value of $m_{\etacp}$ from a fit is $3628.41\pm 0.11
~\hbox{MeV}/c^2$, which corresponds to an input-output mass
difference of  $\triangle m_{\etacp} =(0.41\pm0.11)~\hbox{MeV}/c^2$.
This means that $m_{\etacp}$ can also be reconstructed correctly. The 
detection efficiency  based on these event selection criteria
is $(8.03\pm0.11)\%$ where uniform angular distributions are assumed. 
The detection efficiency is expected to be smaller for more realistic angular
distributions.  Assuming a detection efficiency of
$5\%$, about $900$ $\etacp\to K_S K\pi$ signal events are expected to be
observed in a $3\times 10^{9}$ $\psip$ event sample.

\begin{figure}[htbp]
\centerline{\psfig{file=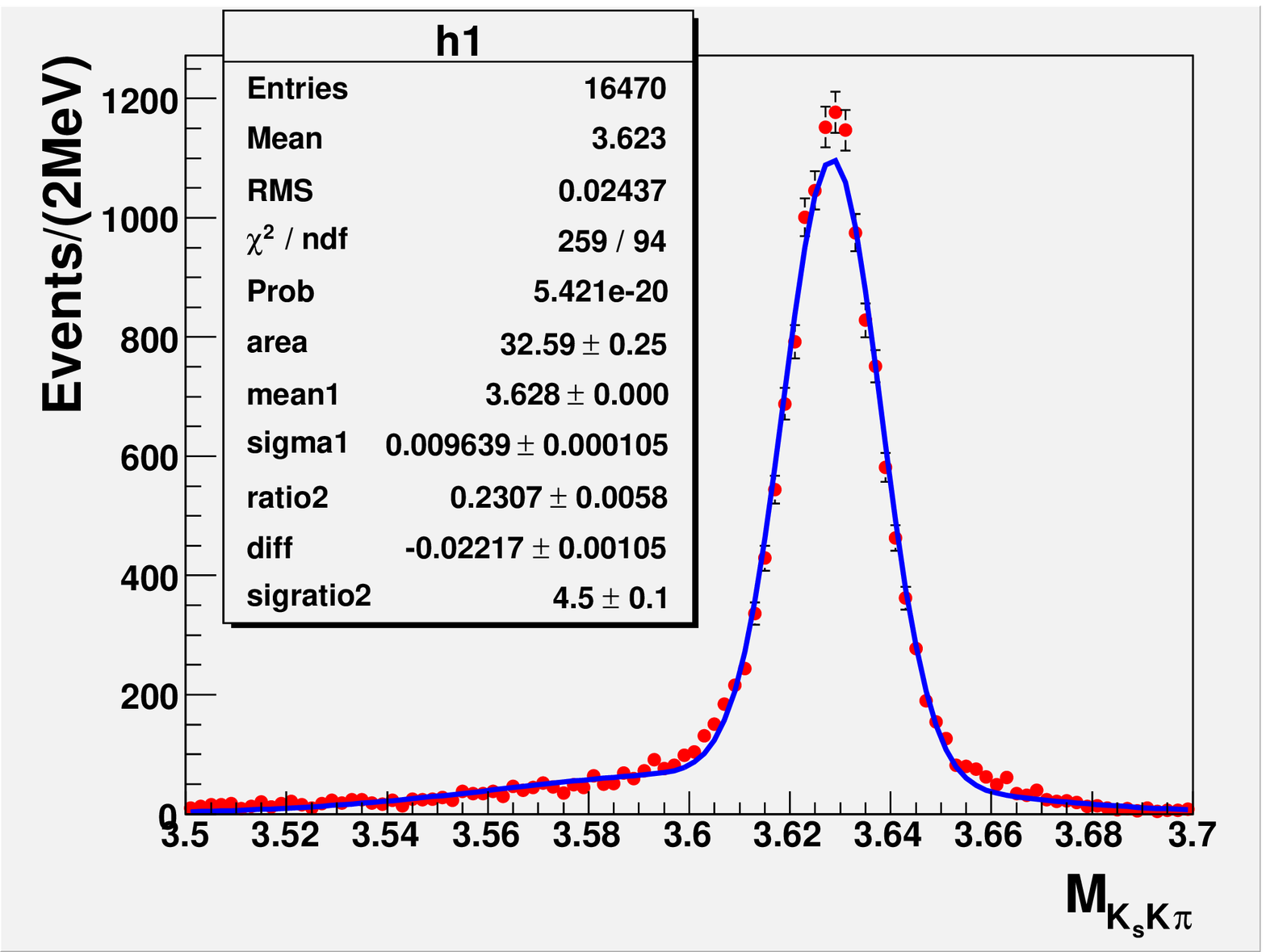,width=8.0cm}}
\caption{The $K_S K\pi$ invariant mass distribution in the
$\etacp$ mass region. The distribution is fitted with a double
Gaussian.}
\end{figure}\label{mkskpi_etacp}

\subsection{$\psip\to\pi^0 h_c,h_c\to \gamma\etac, \etac\to K^0_S K
\pi$}

About 50,000 Monte Carlo simulated events of the type 
$\psip\to\pi^0 h_c,h_c\to \gamma\etac, \etac\to K^0_S K^{\pm} \pi^{\mp}$ 
are analyzed. For the event selection, two charged tracks with net charge
zero and originating from the interaction region plus a $K_S\to\pi^+\pi^-$
candidate with a reconstructed secondary vertex are required. At least three
photon candidates are required in each event and the photon
candidate with highest energy is regarded as the radiative photon
from the $h_c\to\etac$ transition.  The two-photon combination with smallest
$|m_{\gamma\gamma}-m_{\pi^0}|$ value are regarded as the candidate
$\pi^0$. The $K_S$ selection includes the requirement $m_{K_S} \in
[0.44,0.56]~\hbox{GeV}/c^2$. The $\etac\to K_S K \pi$ signal requirements
are $m_{K_S K \pi} \in [2.6,3.4]~\hbox{GeV}/c^2$ and $\cos \theta >
0.95$, where $\theta$ is the angle between the
radiative photon from the $h_c\to\gamma\etac$ 
decay and the PMISS direction of the 
candidate $\etac\to K^0_S K \pi$ track combination. 
The resulting invariant mass distribution of  $\gamma K_S K\pi$
is shown in Fig.~\ref{mgkskpi_hc}. An $h_c$ signal is
observed with a resolution of $(11.3\pm
0.6)~\hbox{MeV}/c^2$. The output value of $m_{h_c}$ from the
fit to the peak is $3528.1\pm 0.7 ~\hbox{MeV}/c^2$, which
corresponds to an input-output difference of
$\triangle m_{h_c} = (2.0\pm0.7)~\hbox{MeV}/c^2$.
This indicates that $m_{h_c}$ is not
reconstructed perfectly in this mode, presumably
because of the presence of three 
relatively low energy photons. 
A possible reason is that the current version of BOSS's
modeling and reconstruction of the photon energy is 
not very accurate and this results
in a systematic error in the final result.
The detection efficiency for these event selection criteria is
$(8.71\pm0.33)\%$, where uniform angular distributions are assumed. 
The detection efficiency is expected to be smaller for more realistic angular
distributions.  Assuming a detection efficiency of
$5\%$, an $h_c \to\gamma\etac$, $\etac\to K_S K\pi$ signal of about 
$450$ detected events  is expected in a
$3\times 10^{9}$ $\psip$ event sample.

\begin{figure}[htbp]
\centerline{\psfig{file=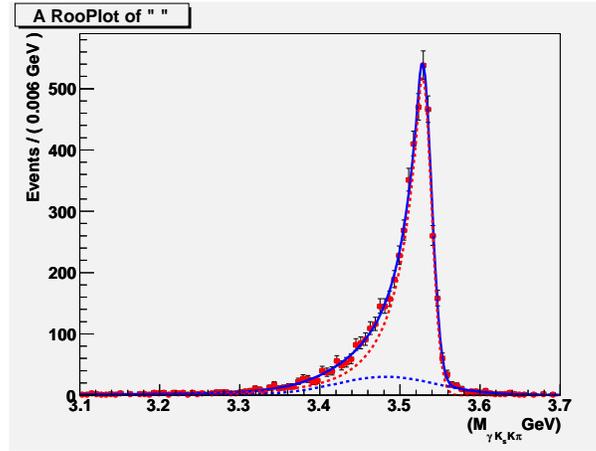,width=8.0cm}}
\caption{Invariant mass distribution of $M_{\gamma K_S K \pi}$ in
the $h_c$ mass region. The distribution is fitted using a CB
function plus double Gaussian.}
\end{figure}\label{mgkskpi_hc}

\chapter{Charmonium Leptonic and EM Decays}
\label{sec:cc_inclusive}


\section[Leptonic and EM Decays in Potential Models]
{Leptonic and EM Decays in Potential Models\footnote{By T.
Barnes}}
\label{sec:cc_lepton_barnes}
\def\be{\begin{equation}}
\def\ee{\end{equation}}
\def\bd{\begin{displaymath}}
\def\ed{\end{displaymath}}
\def\ba{\begin{eqnarray}}
\def\ea{\end{eqnarray}}
\def\ccbar{$c\bar c$ }
\def\C{\rm C}
\def\D{\rm D}
\def\F{\rm F}
\def\I{\rm I}
\def\J{\rm J}
\def\L{\rm L}
\def\M{\rm M}
\def\P{\rm P}
\def\S{\rm S}
\def\T{\rm T}
\def\X{\rm X}



\subsection{$e^+e^-$ widths of $1^{--}$ states}

Leptonic partial widths are immediately accessible at $e^+e^-$
machines, and they provide interesting (and currently rather
puzzling) information regarding the wavefunctions of $1^{--}$
charmonium states. In the nonrelativistic limit of an S-wave
quarkonium system the coupling to $e^+e^-$ through a virtual
photon involves the wavefunction at contact (the van Royen -
Weisskopf formula); for an S-wave $c\bar c$ system this partial
width is given by \cite{ted_VanRoyen:nq}, \be \Gamma_{c\bar
c}^{e^+e^-}({}^3{\rm S}_1) = \frac{16}{9}\, \alpha^2\;
\frac{|\psi(0)|^2} {{\rm M}_{c\bar c}^2} \ee where the radial
wavefunction is normalized to $\int_0^\infty r^2 dr\, |\psi(r)|^2
= 1$. for relativistic bound states the annihilation is nonlocal,
so a nonvanishing $e^+e^-$ coupling of D-wave (and higher-L
$1^{--}$ states) is also predicted. This width at leading
nonrelativistic order is proportional to the L$th$ derivative of
the $Q\bar Q$ wavefunction at contact, and for D-wave $c\bar c$
states is given by \be \Gamma_{c\bar c}^{e^+e^-}({}^3{\rm D}_1) =
\frac{50}{9}\, \alpha^2\; \frac{|\psi''(0)|^2} {{\rm M}_{c\bar
c}^2 m_c^4}. \ee This is typically much smaller than the leptonic
widths of $^3$S$_1$ states. Evaluating these widths using the
nonrelativistic quark model wavefunctions described here gives the
results quoted in Table~\ref{ted_Table_ee}.

\begin{table}[htbp]
\caption{Predictions of a nonrelativistic $c\bar c$ potential
model for $e^+e^-$ partial widths, together with current (2006)
PDG experimental values \cite{ted_PDG2006}.} \label{ted_Table_ee}
\begin{center}
{\begin{tabular}{@{}c|ccl@{}}
\hline
 \ State \ & \ Asst. \ &   \ $\Gamma^{e^+e^-}_{thy.}$(keV) \
& \ $\Gamma^{e^+e^-}_{expt.}$(keV)\
\\
\hline
& & & \\
J/$\psi$
& ${1}\,^3$S$_1$
&  12.13\phantom{1}
& $ 5.55 \pm 0.14\pm 0.02 $
\\
& & & \\
$\psi{\, '}$
& ${2}\,^3$S$_1$
&   \phantom{1}5.03\phantom{1}
& $ 2.48 \pm 0.06 $
\\
& & & \\
$\psi(3770)$
& ${1}^3$D$_1$
&   \phantom{1}0.056
& $ 0.242 {+ 0.027\atop -0.024} $
\\
& & & \\
$\psi(4040)$
& ${3}\,^3$S$_1$
&   \phantom{1}3.48\phantom{1}
& $ 0.86 \pm 0.07 $
\\
& & & \\
$\psi(4160)$
& ${2}^3$D$_1$
&   \phantom{1}0.096
& $ 0.83 \pm 0.07 $
\\
& & & \\
$\psi(4415)$
& ${4}\,^3$S$_1$
&   \phantom{1}2.63\phantom{1}
& $ 0.58 \pm 0.07 $
\\
& & & \\
\hline
\end{tabular}}
\end{center}
\end{table}

Note that the agreement between the model and experiment is not
especially good. This overestimate of the J/$\psi$ leptonic width
by roughly a factor of two seems to be a common difficulty in
naive potential models, and may in part be due to the use of the
nonrelativistic ``wavefunction at contact" approximation for this
decay rate. Non-valence components in the charmonium states, such
as $c\bar c g$ or D meson pairs from decay loops, may also
contribute to this inaccuracy. Concerns have also been expressed
that the leading-order QCD radiative corrections may be large
\cite{ted_Barbieri:1975ki}, which could significantly reduce the
overall scale of the leptonic widths. Of course the leading-order
pQCD corrections are prescription-dependent, so it is not clear
that this claim is reliable. In any case it is evident that a
simple change of scale alone will not resolve the discrepancies in
Table~\ref{ted_Table_ee}, since the nominally D-wave states
$\psi(3770)$ and $\psi(4160)$ both have much larger couplings that
would be expected for pure D-wave states.

The large experimental leptonic widths of the $\psi(3770)$ and
$\psi(4160)$ relative to predictions for pure D-wave $c\bar c$
states (see Table~\ref{ted_Table_ee}) may be due to admixtures of
S-wave $c\bar c$ components
\cite{ycz_rosner,ted_Rosner:2004mi}. As we shall discuss in
the next section, these mixing angles can be estimated through
measurements of the E1 radiative decay rates $\psi(3770) \to
\gamma \chi_2$ and (with more difficulty) $\psi(4160) \to \gamma
\chi_2$ and $\gamma \chi_2{\, '}$, since these transitions are
very sensitive to the presence of S-wave $c\bar c$ components.
Radiative width ratios such as $\Gamma(\psi(3770) \to \gamma
\chi_2)/\Gamma(\psi(3770) \to \gamma \chi_1)$ and the leptonic
width ratio $\Gamma(\psi(3770) \to e^+e^-)/\Gamma(\psi{\, '} \to
e^+e^-)$ provide two independent tests of S-D mixing.

At BES, in addition to testing the existing measurements of the
leptonic widths of the vector states cited in
Table~\ref{ted_Table_ee}, it will also be very important to
determine the leptonic widths of new $1^{--}$ hybrid candidates
such as the Y(4260) (to be discussed in the section on hybrid
charmonia). It has widely been anticipated that $1^{--}$ hybrids
should have suppressed leptonic widths relative to conventional
S-wave $1^{--}$ vector quarkonia; this may indeed be the case for
the Y(4260), since it is not apparent in the existing R
measurements.

\subsection{Two-photon couplings}

Although BES will not have adequate $\sqrt{s}$ to usefully exploit
two-photon production of charmonium resonances, this has been a
useful technique at higher-energy $e^+e^-$ facilities, which we
briefly mention here in the interest of completeness. This subject
has been reviewed elsewhere \cite{Barnes:1992sg}.

Two-photon resonance production involves the reaction $e^+e^- \to
e^+e^- R$, $R\to f$, where $R$ is a C = $(+)$ meson resonance and
$f$ is the exclusive final state observed. This process proceeds
dominantly through the two-photon coupling of the resonance,
$e^+e^- \to e^+e^- \gamma \gamma$, $\gamma \gamma \to R$, and
hence implicitly gives the two-photon partial width
$\Gamma_{\gamma\gamma}(R)$ times the branching fraction $B(R\to
f)$. One may also consider the case of one or both photons
significantly off mass shell $q_\gamma^2 \neq 0$, which gives the
generalized widths $\Gamma_{\gamma^*\gamma}(R)$ and
$\Gamma_{\gamma^*\gamma^*}(R)$. The interesting question for
off-shell widths is whether they are given by a vector dominance
formula with the mass of the relevant vector, in this case the
J$/\psi$.

In the limit of large quark mass (and hence a zero-range charm
quark propagator) the two-photon width of an S-wave charmonium
state is proportional to the wavefunction at contact squared. In
this approximation, higher-L states are produced with amplitudes
proportional to the L$^{th}$ derivative of the wavefunction at
contact \cite{ted_Ackleh:1991ws,ted_Ackleh:1991dy}. (In practice
the attractively simple contact approximation appears marginal at
best at the $c\bar c$ mass scale.) Both the $\eta_c$ and
$\eta_c{\, '}$ have been seen in two-photon collisions; the
$\gamma\gamma$ width of the $\eta_c$ is approximately 7~keV,
comparable to quark model expectations. Two-photon production of
the $\eta_c'$ has been observed by CLEO-c \cite{ted_Asner:2003wv},
but the $\eta_c{\, '}\to \gamma\gamma$ partial width is not known
because experimentally one measures the $\gamma\gamma$ width times
the branching fraction to an exclusive final state, and the
$\eta_c{\, '}$ absolute branching fractions are not known.

Higher-L $c\bar c$ two-photon widths are suppressed by the mass of
the charm quark, and therefore are not very well established; only
the P-wave $\chi_0$ and $\chi_2$ states and the $2{}^3$P$_2$
candidate $\chi_2(3929)$ have been observed. The theoretical ratio
of P-wave $\gamma\gamma$ widths within a multiplet (such as 1P or
2P) in the large quark mass limit is
$\Gamma_{\gamma\gamma}(^3$P$_0)/ \Gamma_{\gamma\gamma}(^3$P$_2) =
15/4$, however this may be modified significantly by QCD radiative
corrections and finite quark mass effects. Nonrelativistically the
$\gamma\gamma$ state produced by the $^3$P$_2$ should be pure
helicity two; relativisitic corrections are expected to give rise
to a small helicity-zero $\gamma\gamma$ component as well.
Although one can in principle produce higher-L $c\bar c$ states in
$\gamma\gamma$ collisions, in practice these rates fall rapidly
with increasing L for $c\bar c$ and heavier $Q\bar Q$ systems. As
an example, the two-photon width of a hypothetical $^1$D$_2$(3840)
$c\bar c$ state is predicted to be only 20~eV
\cite{Barnes:1992sg}.

%
%
%
%
%
%
%

\subsection[QCD radiative and relativistic corrections to EM decays]
{QCD radiative and relativistic corrections to EM decays\footnote{By
Y.J. Zhang}}

\begin{table}[htbp]
\caption[Present PDG values for the leptonic widths of the $J/\psi$,
         $\psi(2S)$, and $\psi(3770)$ states]
        {Present PDG \cite{zhang_Yao:2006px} values for the leptonic widths of the $J/\psi$,
         $\psi(2S)$, and $\psi(3770)$ states.}
\label{tab:charmED-1}
\renewcommand{\arraystretch}{1.2}
\begin{center}
\begin{tabular}{|c|c|c|c|c|}
\hline Resonance & $\Gamma_{\rm tot}$ ({\rm keV}) & $\Gamma_{ee}$
({\rm keV})  & ${\cal B}_{\mu\mu}$(\%) & ${\cal B}_{\tau\tau}$(\%)
\\ \hline%
$J/\psi$ & $93.4 \pm 2.1$& $5.55 \pm 0.14\pm 0.02$ & $5.93 \pm 0.06$ & --- \\
\hline
 $\psi$(2S) & $337 \pm 13$  & $2.48 \pm 0.06$ & 0.73 $\pm$
0.08  & 0.28$\pm$ 0.07   \\ \hline $\psi(3770)$ & $(23.0 \pm 2.7)
\times 10^3$ & $0.242 ^{+0.027}_{-0.024}$&---&---\\ \hline
\end{tabular}
\end{center}
\renewcommand{\arraystretch}{1.0}
\end{table}

Beyond the tree level result,  QCD radiative and relativistic
corrections in the electromagnetic decay of charmonium have been
considered. The one-loop QCD radiative correction to the leptonic
width of $^3S_1$ states has been calculated in
Refs.\cite{zhang_Barbieri:1975ki,zhang_Celmaster:1978yz}
\begin{eqnarray}
\Gamma_{Q\bar Q}^{e^+e^-}({}^3{ S}_1) = e_Q^2\, \alpha^2\;
\frac{|\psi(0)|^2} {m_Q^2}\left(1-\frac{16 \alpha_s}{3\pi}\right).
\end{eqnarray}
where $|\psi(0)|$ is the radial wavefunction at the origin, and it
is normalized by $\int r^2 |\psi(r)|^2=1$. For $m_e \ll m_c$ and
$m_\mu \ll m_c$, the masses of leptons are ignored here.

The two-loop QCD correction has been given in Ref.
\cite{zhang_Beneke:1997jm}, and these contributions lead to
\begin{eqnarray}
\Gamma_{Q\bar Q}^{e^+e^-}({}^3{ S}_1)&=&e_Q^2\, \alpha^2\;
\frac{|\psi(0)|^2} {m_Q^2}\left\{1-4C_F {\alpha_s(m)\over\pi}
+\right.\nonumber
\\&&\left.\left[-117.46+0.82n_f+{140\pi^2\over 27}
\ln\left({2m\over\mu_\Lambda}\right)\right]
\left({\alpha_s\over\pi}\right)^2\right\},
\end{eqnarray}
where $\alpha_s(m_c)\approx 0.35$ and $\alpha_s(m_b)\approx 0.21$.

Relativistic corrections to the leptonic width of $^3S_1$ states
have also been considered to ${\cal O}(v^4)$, and given in
Ref.\cite{zhang_Bodwin:2002hg}:
\begin{eqnarray}
\Gamma_{Q\bar Q}^{e^+e^-}({}^3{ S}_1) &=& {F_{ee}({}^3S_1) \over
m_Q^2}
        \Big| \langle 0| \chi^\dagger \bm{\sigma} \psi |{}^3S_1\rangle
\Big|^2 \nonumber\\
&&+{G_{ee}({}^3S_1) \over m_Q^4}
        {\rm \, Re \,}\left[ \langle{}^3S_1| \psi^\dagger \bm{\sigma} \chi
|0\rangle \cdot
        \langle 0| \chi^\dagger \bm{\sigma} (-\frac{i}{2}
\overleftrightarrow{\bf D})^2
\psi |{}^3S_1\rangle \right]\nonumber\\
&&+{H_{ee}^1({}^3S_1) \over m_Q^6} \langle{}^3S_1| \psi^\dagger
\bm{\sigma} (-\frac{i}{2}\overleftrightarrow{\bf D})^2 \chi
|0\rangle \cdot
        \langle 0| \chi^\dagger \bm{\sigma} (-\frac{i}{2}\overleftrightarrow{\bf D})^2
\psi |{}^3S_1\rangle\nonumber\\
&&+{H_{ee}^2({}^3S_1) \over m_Q^6}
        {\rm \, Re \,}\left[ \langle{}^3S_1| \psi^\dagger \bm{\sigma} \chi
|0\rangle \cdot
        \langle 0| \chi^\dagger \bm{\sigma} (-\frac{i}{2} \overleftrightarrow{\bf D})^4
\psi |{}^3S_1\rangle \right], \label{Gpsiee}
\end{eqnarray}
where the short-distance coefficients at leading order in $\alpha_s$
are
\begin{eqnarray}\label{3s1-ee-coeffs-1}
F_{ee}({}^3S_1)&=&{2\pi\over 3}e_Q^2\alpha^2,\\
G_{ee}({}^3S_1)&=&-{8\pi\over 9}e_Q^2\alpha^2,\\
H^1_{ee}({}^3S_1)+H^2_{ee}({}^3S_1)&=& {58\pi\over 54}e_Q^2\alpha^2.
\end{eqnarray}
The relativistic leading-order matrix element $ \Big| \langle 0|
\chi^\dagger \bm{\sigma} \psi |{}^3S_1\rangle \Big|^2 $ is related
to the radial wavefunction at the origin by
 \begin{eqnarray} \Big|
\langle 0| \chi^\dagger \bm{\sigma} \psi |{}^3S_1\rangle \Big|^2=
\frac{N_c}{2\pi}|\psi(0)|^2.
\end{eqnarray}
Relations of higher order matrix elements to leading-order matrix
element $\Big| \langle 0| \chi^\dagger \bm{\sigma} \psi
|{}^3S_1\rangle \Big|^2$ are
 \begin{eqnarray}{\cal
O}(v^2)\Big| \langle 0| \chi^\dagger \bm{\sigma} \psi
|{}^3S_1\rangle \Big|^2
 &\sim&
  \langle{}^3S_1| \psi^\dagger
\bm{\sigma} \chi |0\rangle \cdot
        \langle 0| \chi^\dagger \bm{\sigma} (-\frac{i}{2}
\overleftrightarrow{\bf D})^2 \psi |{}^3S_1\rangle
 \nonumber \\
 {\cal O}(v^4)\Big| \langle 0|
\chi^\dagger \bm{\sigma} \psi |{}^3S_1\rangle \Big|^2 &\sim&
\langle{}^3S_1| \psi^\dagger \bm{\sigma}
(-\frac{i}{2}\overleftrightarrow{\bf D})^2 \chi |0\rangle \cdot
        \langle 0| \chi^\dagger  \bm{\sigma} (-\frac{i}{2}\overleftrightarrow{\bf D})^2
\psi |{}^3S_1\rangle\nonumber \\ &\sim& \langle{}^3S_1| \psi^\dagger
\bm{\sigma} \chi |0\rangle \cdot
        \langle 0| \chi^\dagger \bm{\sigma} (-\frac{i}{2} \overleftrightarrow{\bf D})^4
\psi |{}^3S_1\rangle
\end{eqnarray}

\begin{table}[htbp]
\caption[Present PDG values for the two-photon widths of the
$\eta_c$,
         $\chi_{c0}$,  $\chi_{c2}$,and $\eta_c\prime$ states]
        {Present PDG \cite{zhang_Yao:2006px} values for the two-photon widths of the
$\eta_c$,
         $\chi_{c0}$,  $\chi_{c2}$,and $\eta'_c$ states.}
\label{tab:charmED}
\renewcommand{\arraystretch}{1.2}
\begin{center}
\begin{tabular}{|c|c|c|c|}
\hline Resonance & $\Gamma_{\rm tot}$ ({\rm
MeV})&$\Gamma_{\gamma\gamma}$ ({\rm keV})& ${\cal B}_{\gamma\gamma}
\times 10^4$
\\ \hline%
$\eta_c$ & $25.5 \pm 3.4$& $7.2 \pm 0.7 \pm 2.0$ & $2.8\pm 0.9$  \\
\hline
 $\chi_{c0}$ & $10.4 \pm 0.7$ & $2.87 \pm 0.39$ & $2.76 \pm 0.33$
   \\ \hline
   $\chi_{c2}$ & $2.06\pm0.12$ & $0.53 \pm 0.05$& $2.59 \pm 0.19$\\ \hline
$\eta_c'$ & $14 \pm 4$& seen& seen  \\
\hline
\end{tabular}
\end{center}
\renewcommand{\arraystretch}{1.0}
\end{table}

For the decay of  ${}^1S_0$ states into two photons, the one-loop
QCD radiative correction is given in
Refs.~\cite{zhang_Barbieri:1979be,zhang_Hagiwara:1980nv}:
\begin{equation}
\Gamma_{Q\bar Q}^{\gamma\gamma}({}^1{ S}_0) =2\pi
e_Q^4\alpha^2\left[1+\left({\pi^2\over
4}-5\right)C_F{\alpha_s\over\pi}\right]\frac{\Big| \langle
0|\chi^\dagger \psi |{}^1S_0\rangle \Big|^2 }{m_Q^2}.
\end{equation}
And the relativistic leading-order matrix element $\Big| \langle
0|\chi^\dagger \psi |{}^1S_0\rangle \Big|^2  $ is related to the
radial wavefunction at origin by
 \begin{eqnarray} \Big| \langle
0|\chi^\dagger \psi |{}^1S_0\rangle \Big|^2 =
\frac{N_c}{2\pi}|\psi(0)|^2.
\end{eqnarray}
The two-loop QCD radiative corrections are given in
Ref.~\cite{zhang_Czarnecki:2001zc}:
\begin{eqnarray} {{\Gamma_{Q \bar Q
}^{ e^+e^-}(^3S_1)} \over {\Gamma_{Q\bar Q}^{\gamma\gamma}({}^1{
S}_0)}} = {1\over 3e_Q^2} \left[ 1 - 0.62 \alpha_s(m_Q) +
\alpha_s(m_Q)^2 \left( 2.37 \ln \alpha_s(m_Q) - 1.8 \right) \right].
\label{resac}
\end{eqnarray}

Relativistic corrections to the two-photon width of $^1S_0$ states
have also been considered to ${\cal O}(v^4)$, and given in
Ref.\cite{zhang_Bodwin:2002hg}:
\begin{eqnarray}
\Gamma_{Q\bar Q}^{\gamma\gamma}({}^1{
S}_0)&=&{F_{\gamma\gamma}({}^1S_0) \over m_Q^2}
\Big| \langle 0|\chi^\dagger \psi |{}^1S_0\rangle \Big|^2 \nonumber\\
&&\hbox{} +{G_{\gamma\gamma}({}^1S_0) \over m_Q^4} {\rm \, Re
\,}\left[ \langle{}^1S_0| \psi^\dagger \chi |0\rangle \langle 0|
\chi^\dagger (-\frac{i}{2} \overleftrightarrow{\bf D})^2 \psi
|{}^1S_0\rangle \right]\nonumber\\
&&\hbox{} +{H_{\gamma\gamma}^1({}^1S_0) \over m_Q^6} \langle{}^1S_0|
\psi^\dagger(-\frac{i}{2}\overleftrightarrow{\bf D})^2 \chi
|0\rangle \langle 0| \chi^\dagger (-\frac{i}{2}
\overleftrightarrow{\bf D})^2 \psi
|{}^1S_0\rangle\nonumber\\
&&\hbox{} +{H_{\gamma\gamma}^2({}^1S_0) \over m_Q^6} {\rm \, Re
\,}\left[ \langle{}^1S_0| \psi^\dagger \chi |0\rangle \langle 0|
\chi^\dagger (-\frac{i}{2} \overleftrightarrow{\bf D})^4 \psi
|{}^1S_0\rangle \right], \label{Getagg}
\end{eqnarray}
where the short-distance coefficients at leading order in $\alpha_s$
are
\begin{eqnarray}\label{3s1-ee-coeffs}
F_{\gamma\gamma}({}^1S_0)&=&2\pi e_Q^4\alpha^2,\\
G_{\gamma\gamma}({}^1S_0)&=&-{8\pi e_Q^4\over 3}\alpha^2.,\\
H_{\gamma\gamma}^1({}^1S_0)+H_{\gamma\gamma}^2({}^1S_0)&=&
{136\pi\over 45}e_Q^4 \alpha^2.
\end{eqnarray}

The relative orders of the matrix elements are similar to that for
the $^3S_1$ states.

The two-photon decay of the spin one state $\chi_{c1}$ is forbidden
by the Landau-Yang theorem~\cite{zhang_landau,zhang_Yang:1950rg}.
The one-loop QCD radiative corrections to the decays of ${}^3P_0$
and ${}^3P_2$ states into two photons are given in Refs.~\cite{
zhang_Huang:1996bk,zhang_Huang:1996cs,zhang_Petrelli:1997ge,zhang_Barbieri:1981xz,zhang_Barbieri:1980yp}:
\begin{eqnarray}\label{eq:chic02TwoGamma}
\Gamma_{\chi_{c0}}^{\gamma\gamma}&=&\frac{27e^4_Q\alpha^2}{m^4}
|R^{\prime}_{\chi_{c0}}(0)|^2
\left[1+\frac{\alpha_s}{\pi}(\frac{\pi^2}{3}-\frac{28}{9})\right],\nonumber \\
\Gamma_{\chi_{c2}}^{\gamma\gamma}&=&\frac{36e^4_Q\alpha^2}{5m^4}
|R^{\prime}_{\chi_{c2}}(0)|^2\left(1-\frac{\alpha_s}{\pi}\frac{16}{3}\right).
\end{eqnarray}
The current PDG values are\cite{zhang_Yao:2006px}  $\Gamma^{\gamma
\gamma}_{\chi_{c0}}$=(2.87$\pm$0.39) keV, $\Gamma^{\gamma
\gamma}_{\chi_{c2}}$=(0.53$\pm$0.05) keV, then we can get
$\mathcal{R} \equiv\Gamma^{\gamma \gamma}_{\chi_{c2}}/\Gamma^{\gamma
\gamma}_{\chi_{c0}}=0.186\pm 0.02$. Recently it is given by the CLEO
Collaboration that $\Gamma^{\gamma
\gamma}_{\chi_{c0}}$=(2.53$\pm$0.37$\pm$0.26) keV, $\Gamma^{\gamma
\gamma}_{\chi_{c2}}$=(0.60$\pm$0.06$\pm$0.06) keV, and then
 $\mathcal{R} $=0.237$\pm$0.043$\pm$0.034.
\cite{zhang_Ecklund:2008hg}. From Eq.(\ref{eq:chic02TwoGamma}),
$\mathcal{R} $=$4/15\sim 0.267$ at leading-order in $\alpha_s$ and
$4/15(1-1.76 \alpha_s)\sim 0.149$ at next to leading-order in
$\alpha_s$ can be obtained.

The value of $|R^{\prime}_{\chi_{c2}}(0)|^2$ may be calculated using
potential models, and then the two-photon widths and $\mathcal{R}$
can be estimated with Eq.(\ref{eq:chic02TwoGamma}). The theoretical
predictions for two-photon widths of $\chi_{c2}$ and $\chi_{c0}$ and
the ratio $\mathcal{R}$ derived from them are listed in
Table.\ref{tab:zhang_theory}.
\begin{table}[h]
\renewcommand{\arraystretch}{1.2}
\begin{center}
\caption[ Potential model predictions for the two-photon widths of
$\chi_{c2}$ and $\chi_{c0}$ and the ratio $\mathcal{R}$ derived from
them]{ Potential model predictions for the two-photon widths of
$\chi_{c2}$ and $\chi_{c0}$ and the ratio $\mathcal{R}$ derived from
them. They are cited from Ref.~\cite{zhang_Ecklund:2008hg} }
\begin{tabular}{lccc}
\hline \hline Reference & $\Gamma_{\gamma\gamma}(\chi_{c2})$ (eV) &
$\Gamma_{\gamma\gamma}(\chi_{c0})$ (eV) & $\mathcal{R}$ \\ \hline
Barbieri~\cite{zhang_Barbieri:1975am} & 930 & 3500 & 0.27 \\
Godfrey~\cite{zhang_Godfrey:1985xj} & 459 & 1290 & 0.36 \\
Barnes~\cite{zhang_barnes1} & 560 & 1560 & 0.36 \\
Bodwin~\cite{zhang_Bodwin:1992ye} & 820$\pm$230 & 6700$\pm$2800 & 0.12$^{+0.15}_{-0.06}$ \\
Gupta~\cite{zhang_Gupta:1996ak} & 570 & 6380 & 0.09 \\
M\"{u}nz~\cite{zhang_Munz:1996hb} & 440$\pm$140 & 1390$\pm$160 & 0.32$^{+0.16}_{-0.12}$ \\
Huang~\cite{zhang_Huang:1996cs} & 490$\pm$150 & 3720$\pm$1100 & 0.13$^{+0.11}_{-0.06}$ \\
Ebert~\cite{zhang_Ebert:2003mu} & 500 & 2900 & 0.17 \\
Schuler~\cite{zhang_Schuler:1997yw} & 280 & 2500 & 0.11 \\ \hline
\hline
\end{tabular}
\label{tab:zhang_theory}
\end{center}
\renewcommand{\arraystretch}{1.0}
\end{table}

\section[Leptonic and EM Decays in Effective Field Theories]
{Leptonic and EM Decays in Effective Field Theories\footnote{By A.
Pineda}}
\label{sec:cc_inclusive_lepton}
\renewcommand{\be}{\begin{equation}}
\renewcommand{\ee}{\end{equation}}
\renewcommand{\bea}{\begin{eqnarray}}
\renewcommand{\eea}{\end{eqnarray}}
\def\lQ{\Lambda_{\rm QCD}}
\renewcommand{\als}{\alpha_s}
\renewcommand{\MS}{\overline{\rm MS}}

\def\lskip{\vglue\baselineskip\vglue-\parskip\noindent}


\subsection{Introduction}
The use of effective field theories allows one to tackle the
computation of heavy quarkonium observables directly from QCD.
The effective field theories available for the study of these systems are
NRQCD \cite{cyq_Caswell:1985ui,cyq_Bodwin:1994jh} and pNRQCD 
\cite{cyq_Pineda:1997bj,cyq_Brambilla:1999xf,nora_Brambilla:2000gk}.
They profit from the fact that a dimensionless variable
(the relative velocity between the quark and antiquark), $v \ll 1$, appears and
that the mass, $m_Q$, of the heavy quark is much larger than $\lQ$.
Therefore, a bunch of inequalities between the
different physical scales appearing in the system show up and effective field
theories are specially suitable to take profit of them.
The most obvious inequalities we have are $m_Q \gg m_Qv \gg m_Qv^2$ and
$m_Q \gg \lQ$, though some others may arise in some specific cases
as we will discuss next.

Trying to get a further understanding on the dynamics of the heavy
quarkonium one can distinguish between the weak and
strong coupling regime.
In short, we will denote by weak coupling regime the
situation where the soft scale, $m_Qv$, is much larger than $\lQ$
and the strong coupling regime the situation when $m_Qv \sim \lQ$.
To discern to which regime each bottomonium and charmonium state
belongs to is
one of main open questions in heavy quarkonium physics.
Whereas there is certain consensus that the bottomonium ground
state belongs to the weak coupling regime
\cite{nora_Brambilla:2001fw,nora_Pineda:2001zq,nora_Brambilla:2001qk,nora_Lee:2003hh,
lpt_Penin:2005eu,lpt_Beneke:2005hg}. The situation
is not that clear for higher excitations of bottomonium and
for charmonium. There are different claims by different groups.
To illustrate the point,
we can find the evaluations in Refs. \cite{nora_Brambilla:2001fw,
nora_Brambilla:2001qk}
where
they perform a weak coupling analysis for the first two
states of charmonium and for the first three states of bottomonium
claiming that a reasonable description can be obtained with
perturbation theory. On the other hand there is the
analysis of Ref. \cite{lpt_Beneke:2005hg}, where they claim that no convergence
is found in the perturbative series for these states. It should
be mentioned that different renormalon subtraction schemes has
been used in both approaches. It would be welcome to see
how things work out in other schemes. Another
point that should be stressed is that these analysis mainly focussed
on the determination of the heavy quarkonium spectrum. However,
if a given state belongs to the weak or strong coupling regime
should also be reflected in other observables of these states.
In particular this should happen for the inclusive electromagnetic
decays, being among the cleanest observables one can think of.
However, the situation for the inclusive electromagnetic
decays is quite less clear. The convergence of the
perturbative series is much worse than in the spectrum (actually
this has some side effects in finite order computation of
heavy quarkonium sum rules). A detailed discussion can be
found in Ref. \cite{lpt_Pineda:2006ri}.
Note that this bad convergence also affects to the bottomonium
ground state, not to mention higher excitations or charmonium.
Therefore, before making clear cut statements upon whether one
state belongs to the weak or strong coupling regime the
situation of the inclusive electromagnetic decay widths
should be clarified.

Another aspect that should be mentioned is that finite order
theoretical evaluations of the inclusive electromagnetic decay widths
appear to be very scale dependent. At this
respect the use of renormalization group techniques \cite{lpt_Pin2} have shown itself
to be a convenient tool to reduce the scale dependence of bottomonium
and charmonium observables \cite{lpt_Kniehl:2003ap,lpt_Penin:2004ay,
nora_Pineda:2006gx,lpt_Pineda:2006ri}.

In
the next section we elaborate in somewhat more detail upon the above
discussion.

\subsection{Weak coupling regime}

In the weak coupling regime the inclusive electromagnetic decay widths read
\begin{eqnarray}
\label{vector}
\Gamma(\Upsilon(nS) \rightarrow e^+e^-) =
16\pi \,
{C_A \over 3} \left[ { \alpha_{EM}\, e_Q \over M_{\Upsilon(nS)}} \right]^2
\left|\phi^{(s=1)}_n({\bf 0}) \right|^2
\left\{ c_1-d_1{M_{\Upsilon(nS)}-2m_Q \over 6m_Q} \right\}^2;
\end{eqnarray}
\begin{eqnarray}
\label{pseudo}
\Gamma(\eta_b(nS) \rightarrow \gamma\gamma) =
16 \pi\,  C_A \left[ { \alpha_{EM}\, e_Q^2 \over M_{\eta_b(nS)}} \right]^2
\left|\phi^{(s=0)}_n({\bf 0}) \right|^2
\left\{
c_0-d_0{M_{\eta_b(nS)}-2m_Q \over 6m_Q}
\right\}^2.
\end{eqnarray}

The coefficients $c_{0,1}$ are known with two loop accuracy
\cite{lpt_KalSar,lpt_HarBro,lpt_CzaMel1,lpt_BSS,lpt_CzaMel2}. $d_0=d_1=1$ at
lowest order.
The wave function corrections are also known
with two-loop accuracy. The corrections to the wave function at
the origin are obtained by taking the residue of the Green
function at the position of the poles
\begin{equation}
\left|\phi^{(s)}_n({\bf 0}) \right|^2
=
\left|\phi^{(0)}_n({\bf 0})\right|^2
\left(1+\delta \phi^{(s)}_n\right)
=
{\
\vbox{\hbox{\rlap{Res}\lower9pt\vbox
{\hbox{$\scriptscriptstyle{E=E_n}$}}}}\
} \!
 G_{s}({\bf 0},{\bf 0};E)
\,,
\end{equation}
where the LO wave function is given by
\begin{equation}
\left|\phi^{(0)}_n({\bf 0})\right|^2=
{1 \over \pi} \left({m_QC_F\als \over 2n }\right)^3.
\end{equation}
The corrections to $\delta \phi_n^{(s)}$
produced by $\delta V$ have been calculated with NNLO
accuracy \cite{lpt_Melnikov2,lpt_Penin} in the direct matching scheme. One can
also obtain them in the dimensional regularized $\MS$ scheme with
NNLL accuracy by incorporating the renormalization group improved
matching coefficients \cite{lpt_Pineda:2006ri}. Also a partial evaluation
of the matching coefficients at NNLL is available
\cite{nora_Pineda:2006gx,lpt_Pineda:2006ri}, see also
\cite{lpt_Pineda:2000gz,lpt_Pin2,lpt_Hoang:2002yy}. Therefore, the complete NNLO result
for the decay is at present available as well as a partial evaluation
with NNLL accuracy. One may try then to apply these results to
charmonium and bottomonium decay widths.

\begin{figure}[h!]
   \begin{center}
    \includegraphics[width=.75\textwidth]{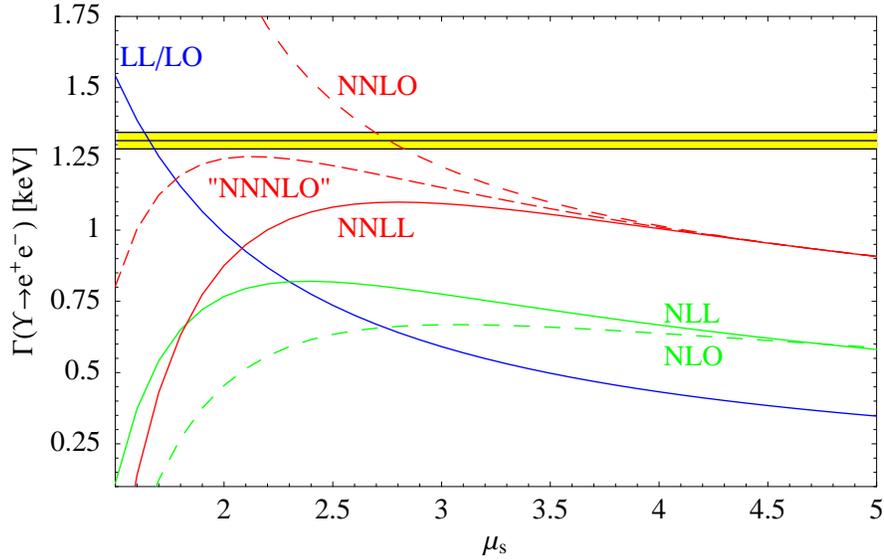}
   \caption{\small{Prediction for the $\Upsilon(1S)$ decay rate to
   $e^+e^-$.  We work in the RS' scheme. Figure from Ref. \cite{lpt_Pineda:2006ri}.}}
   \label{fig:decayee}
   \end{center}
\end{figure}

\begin{figure}[h!]
   \begin{center}
    \includegraphics[width=.75\textwidth]{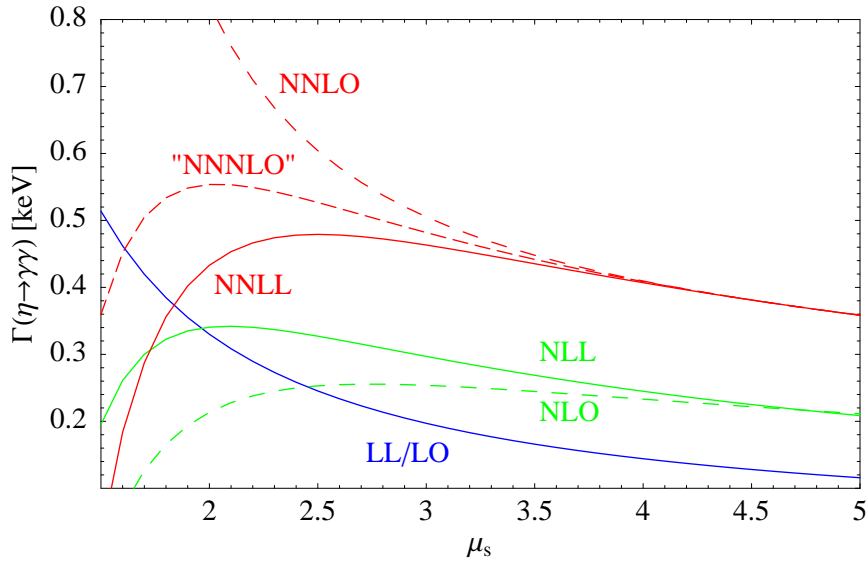}
   \caption{\small{Prediction for the $\eta_b(1S)$ decay rate to two
   photons.  We work in the RS' scheme. Figure from Ref. \cite{lpt_Pineda:2006ri}.}}
   \label{fig:decaygg}
   \end{center}
\end{figure}

The results for the vector and pseudoscalar bottomonium ground state
decay can be found in
Figs.~\ref{fig:decayee} and \ref{fig:decaygg} respectively.
One problem is the large magnitude of the two loop correction to the
matching coefficient of the electromagnetic current. This is puzzling
from the theoretical point of view, since the hard scale should be
perturbative in nature. On the
other hand one may argue that the result for the matching coefficient
is scheme dependent and that a better convergence is expected for the
full result where the scheme dependence disappears. This
is actually not so as we show in Figs. \ref{fig:decayee} and
\ref{fig:decaygg},
which we take from Ref. \cite{lpt_Pineda:2006ri}.
One can discuss whether the charm quark mass
is large enough but the convergence problem also appears in the case
of bottomonium. This may point out to the fact that this convergence problem
is fake.
One may try to introduce renormalization group techniques. This
significantly diminishes the
scale dependence of the theoretical prediction and improves the
convergence of the series, yet, the
absolute magnitude of the correction is large.
From the numerical analysis, we find that the NNLL corrections are
huge, especially for the $\eta_b(1S) \rightarrow \gamma\gamma$
decay. The result we obtain for this decay is compatible with the
number obtained in Ref. \cite{lpt_Penin:2004ay}. This is somewhat
reassuring, since in that reference the ratio of the spin-one
spin-zero decay was considered, which was much more scale independent,
as well as more convergent (yet still large) than for each of the
decays themselves.  This agreement can be traced back to the fact
that, for the spin-one decay, for which we can compare with
experiment, we find that the NNLL result improves the agreement with
the data.  Overall, the resummation of logarithms always significantly
improves over the NNLO result, the scale dependence greatly improves,
as well as the convergence of the series. On the other hand the
problem of lack of convergence of the perturbative series is not
really solved by the resummation of logarithms and it remains as an
open issue. Due to the lack of convergence no numbers or errors
are given. In this respect we
can not avoid to mention that, whereas the perturbative series in
non-relativistic sum rules is sign-alternating, is not
sign-alternating for the electromagnetic decays. Finally, we would
also like to remark the strong scale dependence that we observe at low
scales, which we believe to have the same origin than the one observed
in $t\bar{t}$ production near threshold in Ref. \cite{lpt_Pineda:2006ri}.

\begin{figure}[t]
\begin{center}
\epsfxsize=\textwidth
\epsffile{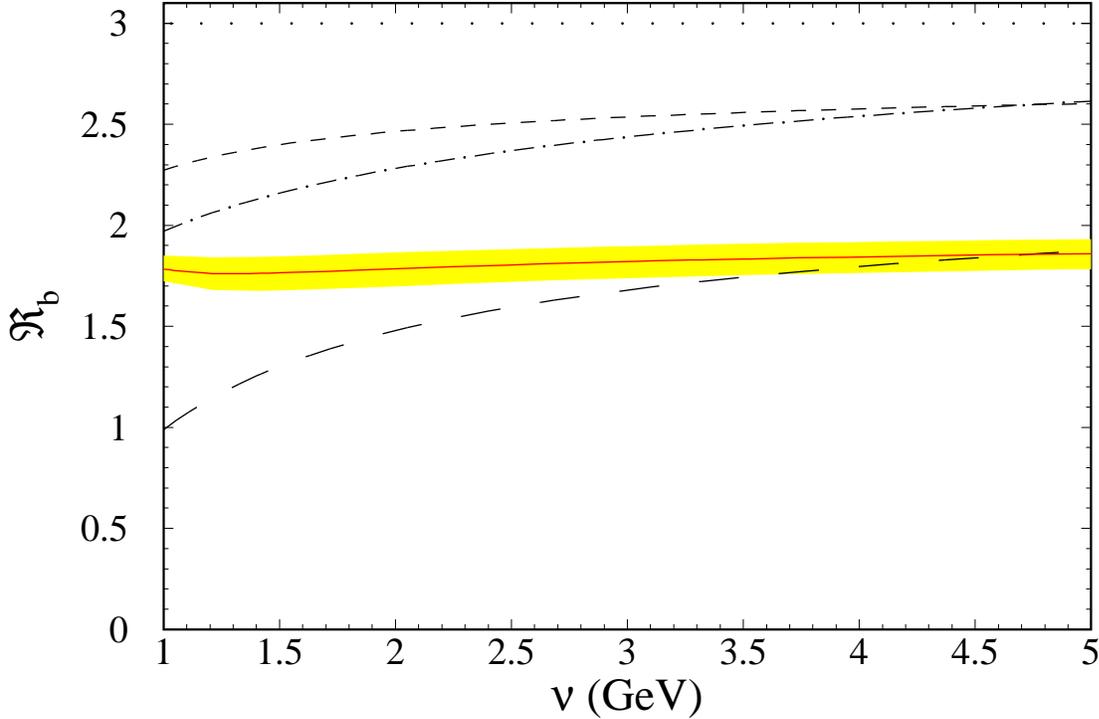}
\end{center}
\caption{\label{figb} The spin  ratio  as the function of the
renormalization scale $\nu$ in LO$\equiv$LL (dotted line), NLO (short-dashed
line), NNLO (long-dashed line), NLL (dot-dashed
line), and NNLL (solid line) approximation for the bottomonium
ground state with $\nu_h=m_b$. For the NNLL result the band reflects
the errors due to $\alpha_s(M_Z)=0.118\pm 0.003$.
Figure from Ref. \cite{lpt_Penin:2004ay}.}
\end{figure}

\begin{figure}
\begin{center}
\epsfxsize=\textwidth
\epsffile{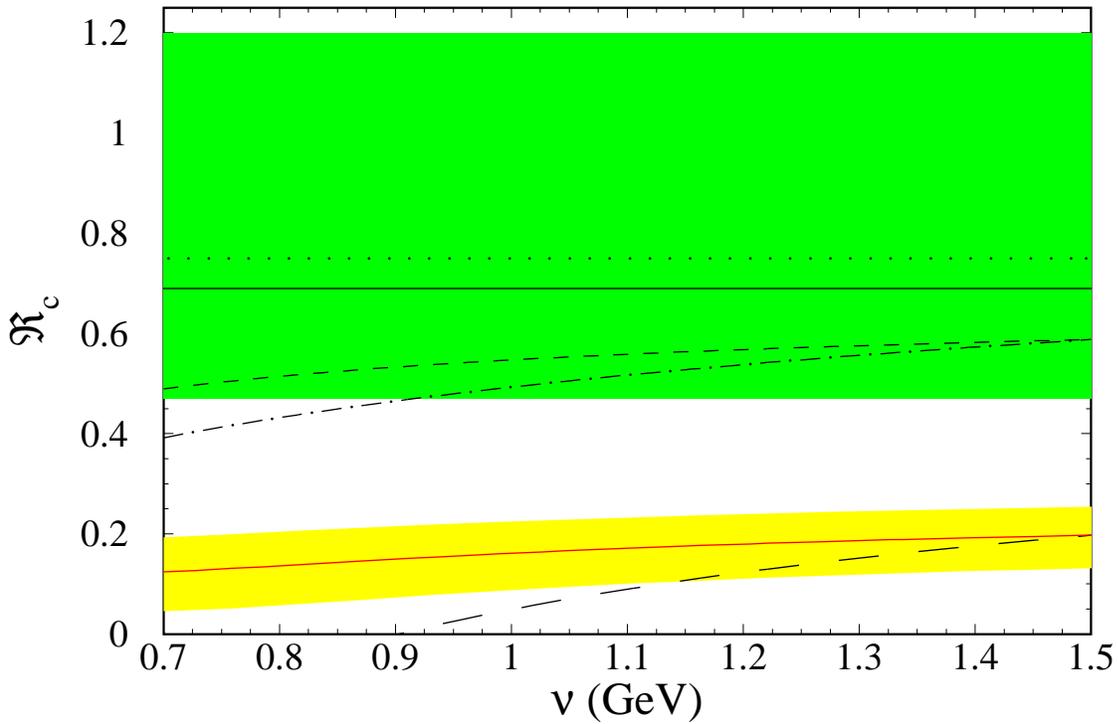}
\end{center}
\caption{\label{figc}  The spin  ratio  as the function of the
renormalization scale $\nu$ in LO$\equiv$LL (dotted line), NLO (short-dashed
line), NNLO (long-dashed line), NLL (dot-dashed
line), and NNLL (solid line) approximation for the charmonium
ground state with $\nu_h=m_c$.  For the NNLL result the lower (yellow)
band reflects
the errors due to $\alpha_s(M_Z)=0.118\pm 0.003$. The upper (green) band
represents the experimental error of the ratio where the central
value is given by the horizontal solid line.
Figure from Ref. \cite{lpt_Penin:2004ay}.}
\end{figure}

One may argue that maybe considering
the ratio between spin one and spin zero decays one may get a better
convergence (moreover for the renormalization group result we
do not have the complete expression at NNLL but we do for the ratio
\cite{lpt_Penin:2004ay}).
Actually this is so as we can see for the bottomonium
but for charmonium physics
the magnitude of the corrections is still very large. One may wonder whether
there are renormalon effects here \cite{lpt_BC}.

For bottomonium, the logarithmic expansion shows nice convergence and
stability (c.f. Fig.~\ref{figb})
despite the presence of ultrasoft contributions with
$\alpha_s$ normalized at a rather low scale $\nu^2/m_b$.
At the same time, the perturbative corrections are important
and reduce the leading order result by approximately $41\%$.
For illustration, at the
scale of minimal sensitivity, $\nu=1.295$~GeV, we have the following series:
\be
{\cal R}_b={\Gamma(\Upsilon(1S) \rightarrow
  e^+e^-)\over\Gamma(\eta_b(1S)\rightarrow\gamma\gamma)}
={1\over 3Q_b^2}\left(1-0.302-0.111\right)\,.
\ee
In contrast, the fixed-order expansion blows up at the scale
of the inverse Bohr radius.

For the charmonium, the NNLO approximation becomes negative at an
intermediate scale between $\alpha_sm_c$ and $m_c$ (c.f. Fig.~\ref{figc})
and the use of the
NRG is mandatory to get a sensible perturbative approximation. The NNLL
approximation has good stability against the scale variation but the
logarithmic expansion does not converge well.  This is the main factor
that limits the theoretical accuracy since the nonperturbative
contribution is expected to be under control. For illustration, at the
scale of minimal sensitivity, $\nu=0.645$ GeV, we obtain
\be
{\cal R}_c={\Gamma(J/\Psi(1S)\rightarrow
  e^+e^-)\over\Gamma(\eta_c(1S)\rightarrow\gamma\gamma)}
={1\over 3Q_c^2}\left(1-0.513-0.326\right)\,.
\ee
The central value of our NNLL result is $2\sigma$ below the experimental
value. The discrepancy may be explained by the large higher order
contributions. This should not be surprising because of the rather large value
of $\alpha_s$ at the inverse Bohr radius of charmonium.  For the charmonium
hyperfine splitting, however, the logarithmic expansion converges well and the
prediction of the renormalization group is in perfect agreement with the
experimental data \cite{lpt_Kniehl:2003ap}.
Thus one can try to improve the convergence of the series for the
production/annihilation rates by accurately taking into account the
renormalon-related contributions. One point to note is that with a potential
model evaluation of the wave function correction, the sign of the NNLO term is
reversed in the charmonium case \cite{lpt_CzaMel2}.
This is very interesting since,
actually, one can also see that the main effects of higher order
effects in the static potential is to make it to agree with the
lattice potential and therefore with the standard parameterizations
of the non-perturbative potential like Cornell, etc...
At the same time the
subtraction of the pole mass renormalon from the perturbative static potential
makes explicit that the potential is steeper and closer to lattice and
phenomenological potential models \cite{lpt_sum}. Therefore, the incorporation of
higher order effects from the static potential may improve the agreement with
experiment, and it may look foreseeable to obtain a description
of the ground state charmonium with mainly perturbative ingredients. However,
how to match the hard coefficient with the perturbative potential
is challenging since the logs of the potential
get entangled with those of the hard matching coefficient. In any case, if we estimate the theoretical uncertainty as the
difference of the NNLL and the NLL result at the soft scale $\alpha_sm_c$, the
theoretical and experimental values agree within the error bars.

\subsection{Final discussion}

In order to determine unambiguously  which bottomonium and
charmonium states belong to the weak or strong coupling regime
 it is necessary
to understand the lack of convergence of the perturbative series
for the inclusive electromagnetic decays. At this respect understanding
the following points would be of great help:

\begin{itemize}

\item A complete NNLL computation would be most welcome 
in order to have a better
estimate of the theoretical uncertanties.

\item A complete NNNLO computation would be most welcome in that it 
would provide a better
estimate of the theoretical uncertanties. Moreover it would help to asses
the importance of the resummation of logarithms.

\item Estimates of the non-perturbative correctionsare needed. 
They also depend on the
same chromoelectric gluonic correlator than the $\Upsilon(1S)$ mass.

\item The rearrangement (renormalon-based?) of the perturbative series
needs better a understanding.

\end{itemize}

Finally, we also note that a renormalization group analysis for 
non-relativistic
charmonium sum rules remains to be done. In this respect, 
there are new R measurements in the sub-charmonium region
with BES at BEPC and just above the charm threshold with
CLEO at CESR. These provide important input to the
determinations of the charm quark mass.

\chapter[Radiative decays]{Radiative decays}
\label{sec:radiative_chapter}

\section[Inclusive Radiative Decays]
{Inclusive Radiative Decays\footnote{By Xavier Garcia i Tormo and
Joan Soto}} \label{sec:cc_inclusive_rad}

\def\bfnabla{\mbox{\boldmath $\nabla$}}
\def\bfsigma{\mbox{\boldmath $\sigma$}}
\def\lQ{\Lambda_{\rm QCD}}
\def\als{\alpha_{\rm s}}
\def\siml{{\ \lower-1.2pt\vbox{\hbox{\rlap{$<$}\lower6pt\vbox{\hbox{$\sim$}}}}\ }}

\subsection{Introduction}

Inclusive radiative decays of heavy quarkonium systems to light
hadrons have been a subject of investigation since the early days
of QCD \cite{soto_Brodsky:1977du,soto_Koller:1978qg}. It was
thought for some time that a reliable extraction of $\als$ was
possible from the inclusive $\gamma gg$ decay normalized to the
inclusive $ggg$ decay. However, when the experimental data became
available for $J/\psi$ \cite{soto_Scharre:1980yn}, it turned out
that the photon spectrum, and in particular the upper end-point
region of it, appeared to be poorly described by the theory. The
situation was slightly better for the $\Upsilon (1S)$
\cite{soto_Nemati:1996xy}, where, at least, good agreement with
QCD was found in the central region \cite{soto_Wolf:2000pm}. In
fact, the whole photon spectrum for the $\Upsilon (1S)$ is now
well understood thanks to a number of theoretical advances which
have taken place in recent years (see
\cite{soto_GarciaiTormo:2005ch} and references therein). Here we
will mainly translate in a critical way these advances to the case
of the $J/\psi$.

We will stay in the effective theory framework of Non-relativistic
QCD (NRQCD) \cite{cyq_Caswell:1985ui,cyq_Bodwin:1994jh} and
Potential NRQCD (pNRQCD)
\cite{cyq_Pineda:1997bj,cyq_Brambilla:1999xf}, and our
terminology will follows that of \cite{cyq_Brambilla:2004jw}. Let
us remind the reader that heavy quarkonium systems enjoy the
hierarchies of scales $m\gg mv\gg mv^2$ and $m \gg \lQ$, where $m$
is the heavy quark mass, $v\ll 1$ the relative velocity of the
heavy quarks and $\lQ$ a typical hadronic scale. States fulfilling
$\lQ \lesssim mv^2$ are said to be in the weak coupling regime
(the binding is essentially due to a Coulomb-like potential)
whereas states fulfilling $\lQ \gg mv^2$ are said to be in the
strong Coupling regime (the binding is due to a confining
potential). States below the open flavor threshold and not too
deep are expected to be in the strong coupling regime whereas deep
states are expected to be in the weak coupling one. States above
(or very close to) the open flavor threshold are not expected to
be in either regime.

\subsection{The photon spectrum}

The contributions to the decay width can be split into direct
($^{dir}$) and fragmentation ($^{frag}$)

\be
\frac{d\Gamma}{dz}=\frac{d\Gamma^{dir}}{dz}+\frac{d\Gamma^{frag}}{dz}
\ee We will call direct contributions to those in which the
observed photon is emitted from the heavy quarks and fragmentation
contributions to those in which it is emitted from the decay
products (light quarks). This splitting is correct at the order we
are working but should be refined at higher orders. $z\in [0,1]$
is defined as $z=2E_\gamma /M$, $M$ being the heavy quarkonium
mass.

\subsubsection{Direct Contributions}\label{secdirect}

The starting point is the QCD formula \cite{soto_Rothstein:1997ac}
\begin{eqnarray}
{d \Gamma\over dz}&=&z{M\over 16\pi^2} {\rm Im} T(z) \nonumber \\
T(z)&=&-i\int d^4 x e^{-iq\cdot x}\left< V_Q (nS)
 \vert T\{ J_{\mu} (x) J_{\nu} (0)\} \vert
V_Q (nS)
 \right> \eta^{\mu\nu}_{\perp}
\label{gdz}
\end{eqnarray}
where $J_{\mu} (x)$ is the electromagnetic current for heavy
quarks in QCD and we have restricted ourselves to $^3S_1$ states.
$q$ is the photon momentum, which in the rest frame of the heavy
quarkonium is $q=\left(q_{+},q_{-}, q_{\perp}\right)=(zM,0,0)$,
$q_\pm=q^0\pm q^ 3$. Different approximations to this formula are
necessary for the central ($z\sim 0.5$), lower end-point
($z\rightarrow 0$) and upper end-point ($z\rightarrow 1$) regions.

\begin{enumerate}

\item {The central region}

For $z$ away from the lower and upper end-points ($0$ and $1$
respectively), no further scale is introduced beyond those
inherent of the non-relativistic system.  The integration of the
scale $m$ in the time ordered product of currents in (\ref{gdz})
leads to local NRQCD operators with matching coefficients which
depend on $m$ and $z$. At leading order one obtains
\begin{equation}
\label{LOrate} \frac1{\Gamma_0} \frac{d\Gamma _{\rm LO}}{dz} =
\frac{2-z}{z} + \frac{z(1-z)}{(2-z)^2} + 2\frac{1-z}{z^2}\ln(1-z)
- 2\frac{(1-z)^2}{(2-z)^3} \ln(1-z),
\end{equation}
where
\begin{equation}
\Gamma_0 = \frac{32}{27}\alpha\alpha_s^2e_Q^2 \frac{\langle  V_Q
(nS)\vert {\cal O}_1(^3S_1)\vert V_Q (nS)\rangle}{m^2},
\label{gamma0}
\end{equation}
and $e_Q$ is the charge of the heavy quark. The $\alpha_s$
correction to this rate was calculated numerically in
Ref.~\cite{soto_Kramer:1999bf} for the bottomonium case. A
reasonable estimate for charmonium maybe obtained by multiplying
it by $\als (2m_c)/\als (2m_b)$. The expression corresponding to
(\ref{gamma0}) in pNRQCD is obtained at lowest order by just
making the substitution
\begin{eqnarray}
\label{singletWF}
\langle  V_Q (nS) \vert {\cal O}_1(^3S_1) \vert  V_Q (nS) \rangle &=&
 \frac{N_c}{2\pi} |R_{n0}(0)|^2,
\end{eqnarray}
where $R_{n0}(0)$ is the radial wave function at the origin. The
final result coincides with the one of the early QCD calculations
\cite{soto_Brodsky:1977du,soto_Koller:1978qg}. The NLO
contribution in the weak coupling regime reads
\cite{cyq_Bodwin:1994jh},

\begin{equation}
\label{RelCo} \frac{d\Gamma _{\rm
NLO}}{dz}=C_{\mathbf{1}}'\left(\phantom{}^3S_1\right)\frac{\langle
V_Q (nS)\vert {\cal P}_1(^3S_1)\vert V_Q (nS)\rangle}{m^4}
\end{equation}
ant it is $v^2$ suppressed with respect to (\ref{LOrate}). The new
matrix element above can be written in terms of the original one
\cite{soto_Brambilla:2002nu}

\begin{eqnarray}
\frac{\langle  V_Q (nS)\vert {\cal P}_1(^3S_1)\vert V_Q (nS)\rangle}{m^4}&=&
\left(\frac{M-2m-\mathcal{E}_1/m}{m}\right)\frac{\langle  V_Q
(nS)\vert {\cal O}_1(^3S_1)\vert V_Q (nS)\rangle}{m^2} \nonumber\\
&&\left(1+\mathcal{O}\left(v^2\right)\right)
\end{eqnarray}
In the weak coupling regime $\mathcal{E}_1/m$ is absent
\cite{soto_Gremm:1997dq}, but in the strong coupling regime it
must be kept ($\mathcal{E}_1\sim \lQ^2$ is a bound state
independent non-perturbative parameter). The matching coefficient
can be extracted from an early calculation
\cite{soto_Keung:1982jb} (see also \cite{soto_Yusuf:1996av}). It
reads

\begin{equation}
C_{\mathbf{1}}'\left(\phantom{}^3S_1\right)=-\frac{16}{27}\alpha\alpha_s^2e_Q^2\left(\left(F_B(z)+\frac{1}{2}
F_W(z)\right)\frac{1}{2}+\frac1{\Gamma_0} \frac{d\Gamma _{\rm
LO}}{dz}\right)\label{rel}
\end{equation}
where $F_B(z)$ and $F_W(z)$ are defined in Ref.
\cite{soto_Yusuf:1996av}\footnote{The last term in (\ref{rel}) was
missing in \cite{soto_GarciaiTormo:2005ch}, see footnote 4 of
\cite{soto_GarciaiTormo:2006ew}.}.

In the weak coupling regime the contributions of color octet
operators start at order $v^4$. Furthermore, away of the upper
end-point region, the lowest order color octet contribution
identically vanishes \cite{soto_Maltoni:1998nh}. Hence there is no
$1/\als$ enhancement in the central region and we can safely
neglect these contributions in this case. However, in the strong
coupling regime the color octet contributions may become order
$v^2$ and should be kept at NLO.

Then in the weak coupling regime (if we use the counting $\als
(m)\sim v^2$, $\als\left(m\als\right)\sim v$) the complete NLO
($v^2$ suppressed) contribution consists of the $\als$ correction
to (\ref{LOrate}), the relativistic corrections in (\ref{RelCo})
and the corrections to the wave function at the origin up to order
$\als^2\left(m\als\right)$
\cite{soto_Penin:1998kx,soto_Melnikov:1998ug}. Using
$m=m_c=1.6\,GeV$, $M=3.1\, GeV$, $\als(2m_c)=0.23$ and
$\als(m\als)=0.4$ we obtain the solid green curve in Fig.
\ref{direpmerg}.

\item {The lower end-point region}

For $z\rightarrow 0$, the emitted low energy photon can  only
produce transitions within the non-relativistic bound state
without destroying it. Hence the direct low energy photon emission
takes place in two steps: (i) the photon is emitted (dominantly by
dipole electric and magnetic transitions) and (ii) the remaining
(off-shell) bound state is annihilated into light hadrons. For $z$
very close to zero it has a suppression  $\sim z^ 3$ with respect
to $\Gamma_0$ (see \cite{soto_Manohar:2003xv,soto_Voloshin:2003hh}
for a recent analysis of this region in QED). Hence, at some point
the direct photon emission is overtaken by the fragmentation
contributions $\bar Q Q \rightarrow ggg \rightarrow gg\bar q q
\gamma $ \cite{soto_Catani:1994iz,soto_Maltoni:1998nh}. In
practise this is expected to happen somewhere between $0.2\lesssim
z\lesssim 0.4$, namely much before than the $z^3$ behavior of the
very low energy direct photon emission can be observed, and hence
we shall neglect the latter in the following.

\item {The upper end-point region}

In this region the standard NRQCD factorization is not applicable
\cite{soto_Rothstein:1997ac}. This is due to the fact that small
scales induced by the kinematics enter the problem and have an
interplay with the bound state dynamics. In order to study this
region, one has to take into account collinear degrees of freedom
in addition to those of NRQCD. This can be done using
Soft-Collinear Effective Theory (SCET)
\cite{soto_Bauer:2000ew,soto_Bauer:2000yr} as it has been
described in \cite{soto_Bauer:2001rh,soto_Fleming:2002sr}. This
region has only been considered in the weak coupling regime, which
we will restrict our discussion to. The color octet contributions
are only suppressed by $v^2$ or by $1-z$. Since their matching
coefficients are enhanced by $1/\als (m)$, they become as
important as the color singlet contributions if we count $\als
(m)\sim v^2\sim 1-z$. The formula one may use for the
semi-inclusive width in the end-point region, which was successful
for the bottomonium case, reads \be
\frac{d\Gamma^e}{dz}=\frac{d\Gamma^{e}_{CS}}{dz}+\frac{d\Gamma^{e}_{CO}}{dz}
\label{endp} \ee where $CS$ and $CO$ stand for color singlet and
color octet contributions respectively.

For the color singlet contribution one may use the expression with
the Sudakov resummed coefficient in Ref.
\cite{soto_Fleming:2004rk}
\[
\frac{1}{\Gamma_0}\frac{d\Gamma^{e}_{CS}}{dz} = \Theta(M-2mz) \frac{8z}9
\sum_{n \rm{\ odd}} \left\{\frac{1}{f_{5/2}^{(n)}}
\left[ \gamma_+^{(n)} r(\mu_c)^{2 \lambda^{(n)}_+ / \beta_0}  -
\gamma_-^{(n)} r(\mu_c)^{2 \lambda^{(n)}_- / \beta_0} \right]^2+\right.
\]
\begin{equation}
\label{singres}
\left.+
\frac{3 f_{3/2}^{(n)}}{8[f_{5/2}^{(n)}]^2}\frac{{\gamma^{(n)}_{gq}}^2}{\Delta^2}
\left[ r(\mu_c)^{2 \lambda^{(n)}_+ / \beta_0}  -
r(\mu_c)^{2 \lambda^{(n)}_- / \beta_0} \right]^2\right\}
\end{equation}
where the definitions for the different functions appearing in
(\ref{singres}) can be found in
\cite{soto_Fleming:2004rk,soto_GarciaiTormo:2005ch}.

For the color octet contributions we use
\begin{equation}
\frac{d\Gamma_{CO}^{e}}{dz}=\alpha_s\left(\mu_u\right)\alpha_s\left(\mu_h\right)e_Q^2\left(\frac{16M\alpha}{9m^4%
}\right)\int_z^{\frac{M}{2m}}\!\!\! C(x-z) S_{S+P}(x)dx
\end{equation}
$\mu_u\sim mv^2$ and $\mu_h\sim m$ are the ultrasoft and hard
scales respectively. $C(x-z)$ contains the Sudakov resummations of
Ref. \cite{soto_Bauer:2001rh}. The (tree level) matching
coefficients (up to a global factor) and the various shape
functions are encoded in $S_{S+P}(x)$. See
\cite{soto_GarciaiTormo:2005ch} for a precise definition of these
objects.

We would like to comment on the validity of the formulas above.
This is limited by the perturbative treatment of the collinear and
ultrasoft gluons. For the collinear gluons, entering in the jet
functions, we have $1 GeV \lesssim M\sqrt{1-z}$, which for
$J/\psi$ implies $z \lesssim 0.9$. The formalism is not reliable
beyond that point. For the ultrasoft gluons, entering in the shape
functions ($S_{S+P}(x)$), we have $1 GeV \lesssim M(1-z)$, which
implies $z \lesssim 0.7$. Hence, due to the latter, we do not
really have a reliable QCD description of the upper end-point
region for charmonium. However, for the bottomonium system, the
shape function above turns out to describe very well the data even
in the far end-point region, where it is not supposed to be
reliable either. In view of this, we believe that the formulas
above may provide a reasonable model for the description of the
region $0.7 \lesssim z \lesssim 0.9$. The outcome is the blue
dot-dashed curve of Fig. \ref{direpmerg} (the flattening of the
curve for $z > 0.85$ is an artifact, see footnote 2 of Ref.
\cite{soto_GarciaiTormo:2005ch})

\item {Merging the central and upper end-point
regions}\label{subsecmatch}

As we have seen, different approximations are necessary in the
central and upper end-point regions. It is then not obvious how
the results for the central and for the upper end-point regions
must be combined in order to get a reliable description of the
whole spectrum. When the results of the central region are used in
the upper end-point region, one misses certain Sudakov and Coulomb
resummations which are necessary because the softer scales
$M\sqrt{1-z}$ and $M(1-z)$ become relevant. Conversely, when
results for the end-point region are used in the central region,
one misses non-trivial functions of $z$, which are approximated by
their end-point ($z\sim 1$) behavior. In
\cite{soto_GarciaiTormo:2005ch} the following merging formula was
proposed, which works reasonably well for bottomonium, \be
\frac{1}{\Gamma_0}\frac{d\Gamma^{dir}}{dz}=\frac{1}{\Gamma_0}\frac{d\Gamma^{c}}{dz}+\left(\frac{1}{\Gamma_0
}\frac{d\Gamma^{e}}{dz}-\left.{\frac{1}{\Gamma_0
}\frac{d\Gamma^{e}}{dz}}\right\vert_c\right) \label{mergingNLO}
\ee $\vert_c$ means the expansion of the end-point formulas when
$z$ approaches the central region. This expansion must be carried
out at the same level of accuracy as the one we use for the
formulas in the central region.

Putting all the ingredients together in formula (\ref{mergingNLO})
we obtain the red dashed line in Fig. \ref{direpmerg} for the
direct contributions to the photon spectrum. Note that a deep is
generated for $0.8\lesssim z\lesssim 0.9$ which makes the decay
width negative. This happens in the region $0.7\lesssim z$ where
the calculation of the shape function is not reliable. A deep was
also generated in the $\Upsilon (1S)$ case, but the effect was not
so dramatic there \cite{soto_GarciaiTormo:2005ch}. We conclude
that, unlike in the $\Upsilon (1S)$ case, the limitations in the
theoretical description of the end point region make the merging
procedure deliver an unsatisfactory description of this end point
region for $J/\psi$. Clearly, further work is required to
understand better the end-point region, in particular a
description assuming $J/\psi$ in the strong coupling regime would
be desirable. For the present analysis, one should only use the
outcome of the merging procedure for $z < 0.7$, if at all. Indeed,
an alternative way to proceed would be to ignore the end-point
region and only try to describe data in the central region with
the QCD formulas for this region given above.

\begin{figure}
\centering
\includegraphics[width=7.5cm]{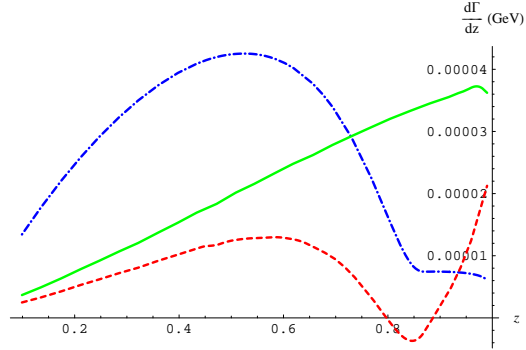}
\caption{Direct contributions in the weak coupling regime. The solid green line corresponds to the calculation for the central region at NLO, which should be reliable up to $z\lesssim  0.7$ . The blue dot-dashed line corresponds to the calculation for the upper end-point region, which is expected to provide a reasonable model for $0.7\lesssim z\lesssim 0.9$. The red dashed line is the curve obtained by merging.}
\label{direpmerg}
\end{figure}

\end{enumerate}

\subsubsection{Fragmentation contributions}\label{secfrag}

The fragmentation contributions can be written as
\begin{equation}
\frac{d\Gamma^{\rm frag}}{dz}=\sum_{a = q,\bar q, g} \int_z^1\frac{dx}{x}C_a(x)D_{a\gamma}\left(\frac{z}{x},M\right),
\end{equation}
where $C_a$ represents the partonic kernels and $D_{a\gamma}$
represents the fragmentation functions. The partonic kernels can
again be expanded in powers of $v$ \cite{soto_Maltoni:1998nh}
\begin{equation}
C_a=\sum_{\mathcal{Q}}{\hat C}_a[\mathcal{Q}]{\left< J/\psi \vert\mathcal{Q}\vert J/\psi \right>\over m^{d_\mathcal{Q}-1}}
\end{equation}
where $\mathcal{Q}$ stands for NRQCD operators, ${\hat
C}_a[\mathcal{Q}]$ for their matching coefficients, and
$d_\mathcal{Q}$ for their dimension. The leading order term in $v$
is the color singlet rate to produce three gluons
($\mathcal{Q}={\cal O}_1(^3S_1)$). The color octet contributions
have a $1/\alpha_s$ enhancement. In the weak coupling regime,
which we will assume in the following, they are $v^4$ suppressed,
but one should keep in mind that in the strong coupling regime
they may become order $v^2$. Then the color singlet fragmentation
contribution is of order $\alpha_s^3D_{g\to\gamma}$ and the color
octet fragmentation are of order $v^4\alpha_s^2D_{g\to\gamma}$
($\mathcal{Q}=O_8(\phantom{}^1S_0 )$, $O_8(\phantom{}^3P_J )$) or
$v^4\alpha_s^2D_{q\to\gamma}$ ($\mathcal{Q}=O_8(\phantom{}^3S_1
)$). We can use, as before, the counting $v^2\sim\alpha_s$ to
compare the relative importance of the different contributions.
The existing models for the fragmentation functions
\cite{soto_Aurenche:1992yc} show us that $D_{q\to\gamma}$ is much
larger than $D_{g\to\gamma}$. This causes the
$v^4\alpha_s^2D_{q\to\gamma}$ of the $O_8(\phantom{}^3S_1)$
contribution to dominate in front of the singlet
$\alpha_s^3D_{g\to\gamma}$ and the octet
$v^4\alpha_s^2D_{g\to\gamma}$ contributions. Moreover, the
$\alpha_s$ corrections to the singlet rate  will produce terms of
order $\alpha_s^4D_{q\to\gamma}$, that is of the same order as the
octet $O_8(\phantom{}^3S_1 )$ contribution, which are unknown.
This results in a large theoretical uncertainty in the
fragmentation contributions, which would be greatly reduced if the
leading order calculation of ${\hat C}_q[ O_1(\phantom{}^3S_1 ]$
(this requires a tree level four body decay calculation plus a
three body phase space integral) was known.

For the quark fragmentation function we will use the LEP
measurement \cite{soto_Buskulic:1995au} and for the gluon
fragmentation function the model \cite{soto_Owens:1986mp}. These
are the same choices as in \cite{soto_Fleming:2002sr}. For the
$O_8 (^1 S_0)$ and $O_8 (^3 P_0)$ matrix elements we will use our
estimates in \cite{soto_GarciaiTormo:2004jw}
\begin{eqnarray}
\left.\left< J/\psi \vert O_8 (^1 S_0) \vert J/\psi \right>\right|_{\mu=M} & \sim & 0.0012\, GeV^3\\
\left.\left< J/\psi \vert O_8 (^3 P_0) \vert J/\psi \right>\right|_{\mu=M} & \sim & 0.0028\, GeV^5
\end{eqnarray}
The estimate for $\left< J/\psi \vert O_8 (^1 S_0) \vert J/\psi \right>$ is compatible with the lattice results \cite{soto_Bodwin:2005gg} (nrqcd and Coulomb algorithms).
For the $O_8 (^3 S_1)$ matrix element the same lattice calculation gives
\begin{eqnarray}
 &&\left< J/\psi \vert O_8 (^3 S_1) \vert J/\psi
\right>_{\textrm{nrqcd}}=0.0005\, GeV^3 \nonumber \\
 && \left< J/\psi \vert O_8
(^3 S_1) \vert J/\psi \right>_{\textrm{coulomb}}=0.0002\, GeV^3
\end{eqnarray}
which is much smaller than using the NRQCD $v$ scaling
\begin{equation}
\left< J/\psi \vert O_8 (^3 S_1) \vert J/\psi \right>\sim v^4\left< J/\psi \vert O_1 (^3 S_1) \vert J/\psi \right>\sim 0.05\, GeV^3
\end{equation}
with $v^2\sim0.3$. With the choices above we obtain the blue
dot-dashed curves in Fig. \ref{mergfragtot} ($v$ scaling) and Fig.
\ref{mergfragtot2} (lattice, nrqcd algorithm) for the
fragmentation contributions. These curves turns out to be very
sensitive to the value assigned to $\left< J/\psi \vert O_8 (^3
S_1) \vert J/\psi \right>$. When we put together direct (red
dashed curve in Figs. \ref{mergfragtot} and  \ref{mergfragtot2})
and fragmentation contributions we obtain the solid green curves
in Figs.  \ref{mergfragtot} and  \ref{mergfragtot2}, if the
merging formula is used, and Figs.~\ref{centrfragtot} and
\ref{centrfragtot2}, if only the central region is taken into
account. The shape of this curve in Fig.  \ref{mergfragtot} is in
qualitative agreement with the early Mark II results for
$0.4\lesssim z \lesssim 0.7 $ \cite{soto_Scharre:1980yn}.

\begin{figure}
\centering
\includegraphics[width=7.5cm]{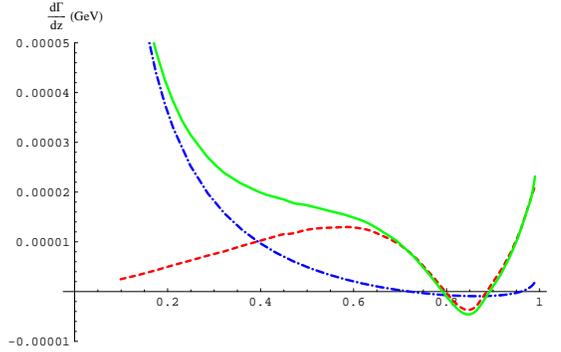}
\caption{The red dashed line corresponds to the direct contributions (merging), the blue dot-dashed line to the fragmentation contributions ($v$ scaling for $O_8(^3S_1)$) and the solid green line is the total.}
\label{mergfragtot}
\end{figure}

\begin{figure}
\centering
\includegraphics[width=7.5cm]{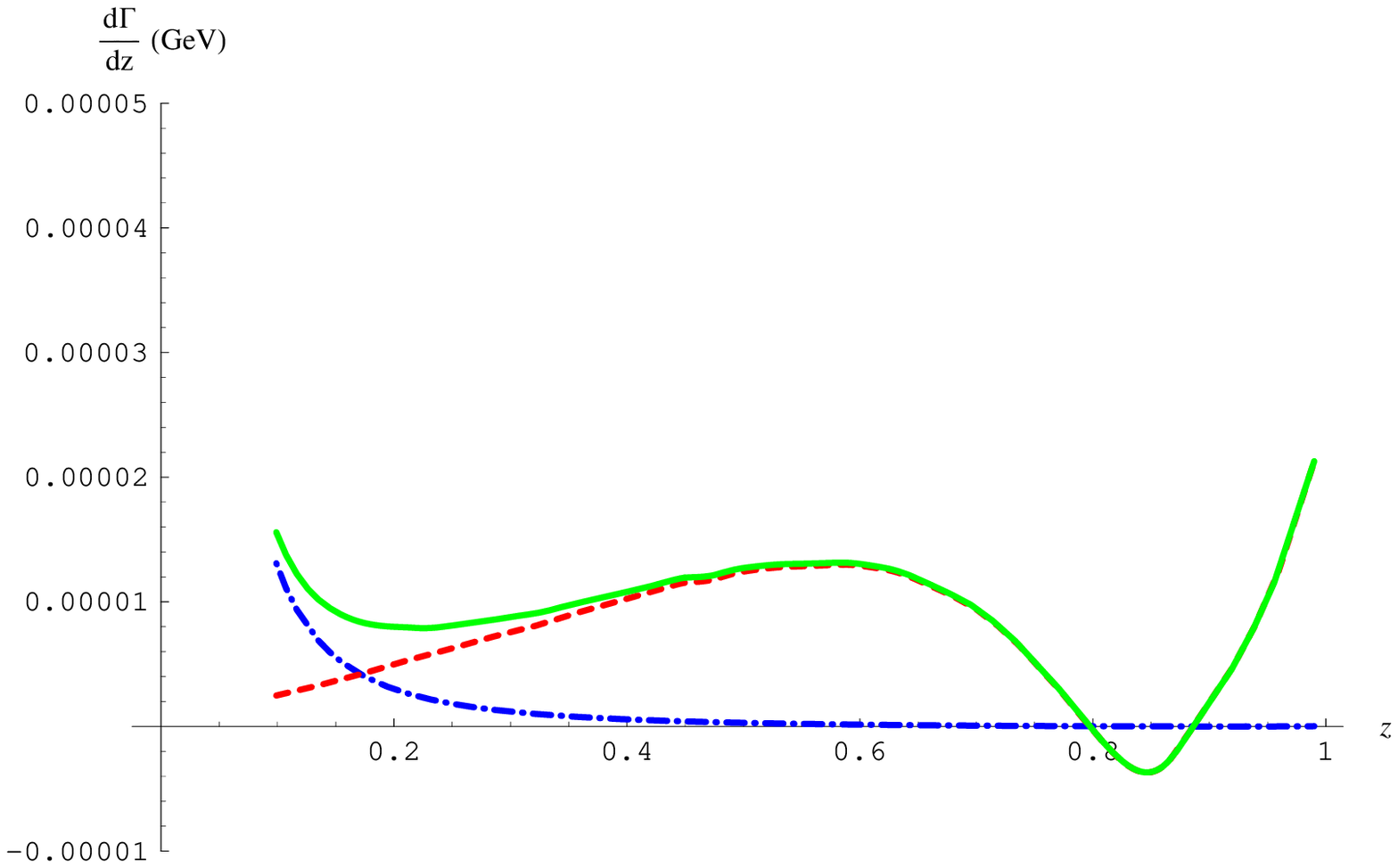}
\caption{The red dashed line corresponds to the direct contributions (merging), the blue dot-dashed line to the fragmentation contributions (lattice, nrqcd algorithm, for $O_8(^3S_1)$) and the solid green line is the total.}
\label{mergfragtot2}
\end{figure}

\begin{figure}
\centering
\includegraphics[width=7.5cm]{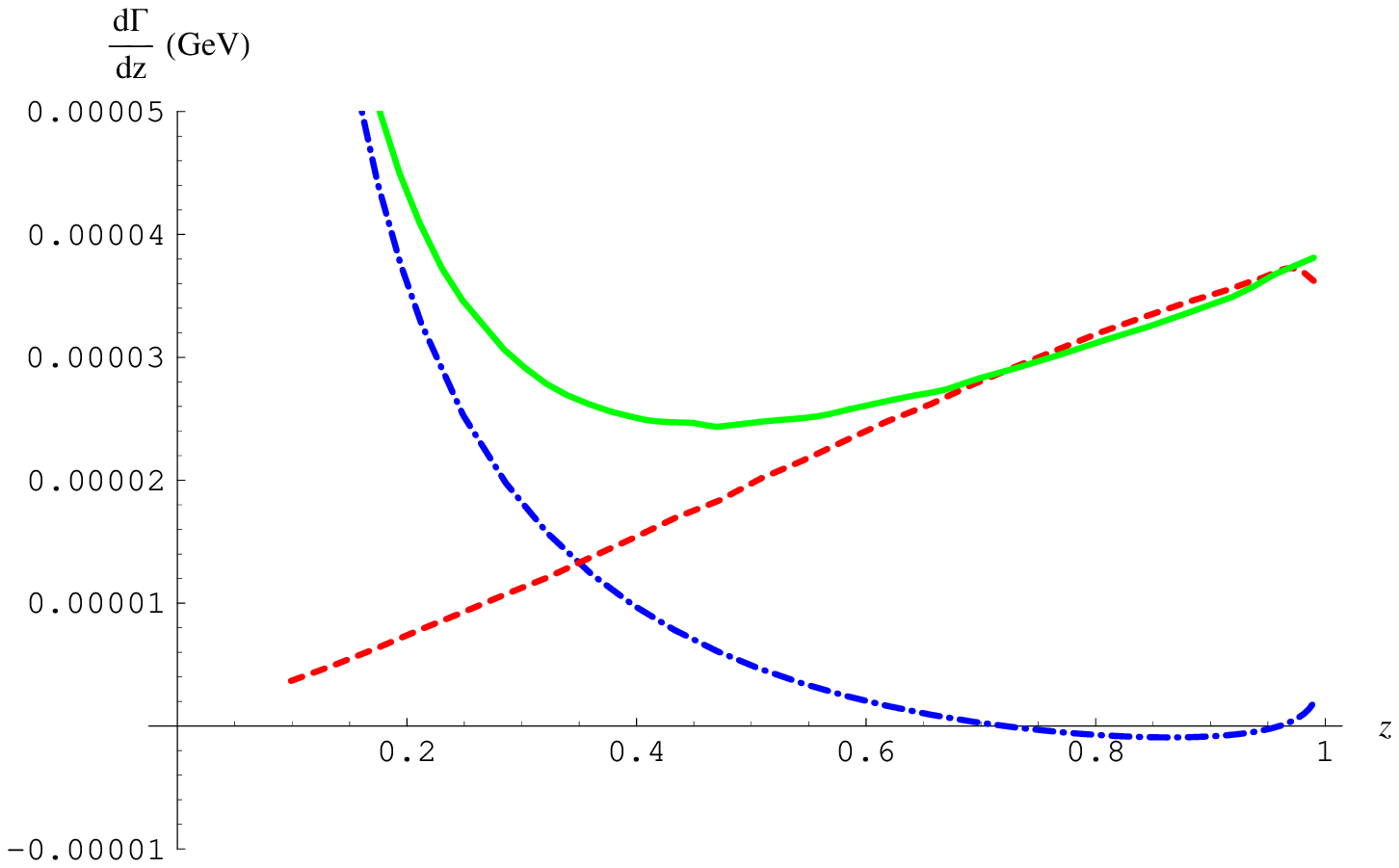}
\caption{The red dashed line corresponds to the direct contributions (central region), the blue dot-dashed line to the fragmentation contributions ($v$ scaling for $O_8(^3S_1)$) and the solid green line is the total.}
\label{centrfragtot}
\end{figure}

\begin{figure}
\centering
\includegraphics[width=7.5cm]{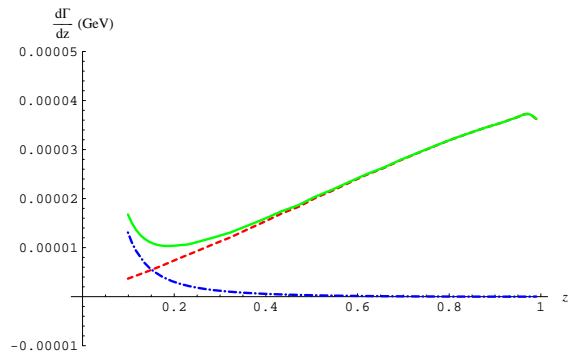}
\caption{The red dashed line corresponds to the direct
contributions (central region), the blue dot-dashed line to the
fragmentation contributions (lattice, nrqcd algorithm, for
$O_8(^3S_1)$) and the solid green line is the total.}
\label{centrfragtot2}
\end{figure}

\subsection{Extraction of $\als (M_{J/\psi})$}

As mentioned in the introduction, $\als (M_{J/\psi} )$ can in
principle be extracted from the ratio $\Gamma (J/\psi \rightarrow
\gamma_{\rm direct} X)$ over $\Gamma_{\rm strong} ( J/\psi
\rightarrow X)$, $X$ stands for light hadrons, $\gamma_{\rm
direct}$ for photons produced from the heavy quarks and
$\Gamma_{strong}$ for subtracting from $\Gamma$ the decays
mediated by a virtual photon. This maybe done in an analogous way
as it has been recently carried out for bottomonia in
\cite{soto_Besson:2005jv}. In order to obtain $\Gamma (J/\psi
\rightarrow \gamma_{\rm direct} X)$ it is important to have a good
QCD description of the photon spectrum since one should be able to
disentangle fragmentation contributions from direct ones. This may
be done by restricting the fit to data of the QCD expression for
direct contributions to the upper end-point region and the part of
the central region where fragmentation contributions are
negligible. As we have seen in the previous section, we do not
have at the moment a good QCD description of the upper end point
region for $J/\psi$ and hence a model, like the one in
\cite{soto_Field:1983cy}(see also \cite{soto_Field:2001iu}), might
be unavoidable. This expression is then used to interpolate data
to small $z$ and hence to be able to obtain the full inclusive
width for direct photons. Then $\als (M_{J/\psi} )$ may be
extracted from
 \bea R_\gamma &\equiv& \frac{\Gamma(J/\psi \to
\gamma_{\rm direct}\,
  X)}{\Gamma_{\rm strong}(J/\psi \to X)}
= \frac{36}{5}\frac{e_c^2\alpha}{\als} \left(1+{\mathcal
O}(\als)+{\mathcal O}({v^4\over \als (m)})+\cdots\right) \eea
where $e_c^2=4/9$, the ${\mathcal O}(v ^2)$ cancel in the ratio,
and $\cdots$ stand for higher order contributions. In the
extraction of $\als$ from bottomonium of
\cite{soto_Besson:2005jv}, ${\mathcal O}(\als (m) )$ corrections
were taken into account but not ${\mathcal O}({v^4\over \als})$
which are of the same order if $\als (m)\sim v^2$ and in practise
turn out to be very important. These have been included in
\cite{soto_nosaltres}.

\subsection{Learning about the nature of $J/\psi$ and $\psi (2S)$}

It has recently been shown that if two heavy quarkonium states are
in the strong coupling regime then the ratio of their total photon
spectrum in the central region is predictable from QCD at NLO
\cite{soto_GarciaiTormo:2005bs}. If the spectrum of both $J/\psi$
and $\psi (2S)$ is measured and fits well with the strong coupling
regime formula, it would indicate that both $J/\psi$ and $\psi
(2S)$ are in the strong coupling regime. Unfortunately, if it does
not, we will not be able to learn much, because it may be due to
the fact that $J/\psi$ is in the weak coupling regime or to the
fact that $\psi (2S)$ is too close to the open flavor threshold
for the strong coupling regime to hold for it (or to both).

\subsection{Conclusions}

A new measurement of the inclusive photon spectrum for radiative
$J/\psi$ decays would be of great interest since it has only been
measured before by the Mark II collaboration more than 25 years
ago. The theoretical progress which has occurred since may allow,
among other things, for a sensible extraction of $\als
(M_{J/\psi})$. The additional measurement of the photon spectrum
for $\psi (2S)$ might shed some light on the nature of these
states. No theoretical analysis are available for other states
like $\eta_c$s or $\chi_c$s. Experimental measurements would
definitively trigger them.

\section[Exclusive Radiative Decays]
{Exclusive Radiative Decays \footnote{By Changzheng Yuan}}
\label{sec:radiative}



\subsection{Introduction}
\label{intro}

At \bes3, there will be huge data samples of vector charmonium
states, such as $\jpsi$, $\psp$, and $\pspp$, and possibly not small
samples for $\psi(4040)$, $\psi(4160)$, $\psi(4415)$. The non-vector
charmonium states, including pseudoscalars, $\etac$ and $\etacp$,
and the $P$-wave states $\chicJ$, can be produced via radiative
transition of the vector states, considering the large transition
rates (except for $\psp\to \gamma \etacp$), the data samples of these 
states will
also be large. These will make detailed  studied of the radiative decays 
of charmonium into light hadrons possible.  A discussion on inclusive
radiative decays can be found in Sect.~\ref{sec:cc_inclusive_rad}.

The radiative decays of  vector charmonia have been used
extensively for the study of the light hadron spectroscopy,
especially in $\jpsi$ decays, this is reviewed in the hadron
spectroscopy part of this report. The study of the
other charmonium states is rather limited. In the following
sections, we will review the studies of radiative decays of the
states other than $\jpsi$, and the prospects for what can 
be expectd  at \bes3.

\subsection{Radiative decays of vector charmonia}
\label{Vdecays}

The dominant radiative decays of vector charmonia proceed via the
diagram shown in Fig.~\ref{feynman}. The rate of the photon radiation
from the {\it final state} quarks is expected to be very small, as has
been shown to be the case by the experimental measurements
of final-state-radiation  dominant processes like $\jpsi\to \gamma 
\piz$~\cite{ycz_QWG,ted_PDG2006}.
In pQCD, assuming the emission of hard
gluons, the inclusive radiative decay rate of 
the $\jpsi$ is around 6\%, while
that for $\psp$ decays is around 1\%~\cite{ycz_wym}.
There are no estimates for
other vector charmonia, such as the $\pspp$, $\psi(4040)$, etc.,
but it is expected that the rates are very small, since all these states
are above the open charm threshold and the dominant decay modes
are OZI-allowed.

\begin{figure*}[htbp]
\centerline{\psfig{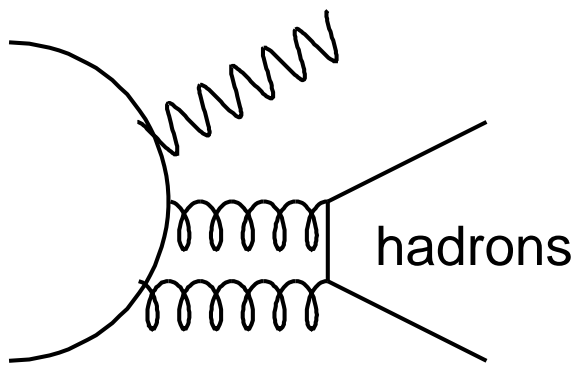}
} \caption{Radiative decays of vector charmonium state into light
hadrons.} \label{feynman}
\end{figure*}

Although radiative $\psp$ decays are expected to be about 1\%
of all decays, existing
measurements are very limited. The only observation so far is
$\psp\to \gamma \eta^\prime$~\cite{ycz_getap}, $\gamma \pi\pi$ and
$\gamma K \bar{K}$~\cite{ycz_gpp}, with a total branching
fraction about 0.05\%.  Final states with three pseudoscalars
have been measured by BESII, and no significant structures 
in the $\eta\pip\pim$ and $K\bar{K}\pi$ mass spectra
are seen; the upper limits for
the production of the pseudoscalar states $\eta(1405)$ or
$\eta(1475)$ are set at the 90\% C.L~\cite{ycz_getap2}.

The most recent study of $\psp$ radiative decays was performed by
BESII~\cite{ycz_gx}, where the total and differential branching fractions
of $\psp\to \gamma \ppb$, $\gamma 2(\pip\pim)$, $\gamma K^0_S K^+
\pim+c.c.$, $\gamma \pip\pim K^+K^-$, and $\gamma
\pip\pim \ppb$ have been measured.  For invariant masses of the
charged particle system less than 2.9~GeV/$c^2$, the total branching
fraction of all these modes is about 0.1\%, which is about 10\% of
the total radiative decays of the $\psp$. Unfortunately, due to the
low statistics,  production of intermediate states
could not be studied. The observed modes are listed in
Table~\ref{rad_exp}, together with their
measured branching fractions.

 \begin{table}
 \caption{Branching fractions of $\psp$ radiative decays, only modes
 that have been 
 observed are listed, for a complete list of all measurements,
 refer to PDG~\cite{ted_PDG2006}. \label{rad_exp}}
 \begin{center}
 \begin{tabular}{lccl}
 \hline
 Mode & Branching Fraction $(\times 10^{-5})$ & Experiment & Comment \\ \hline
 $\gamma\eta^{\prime}(958)$ & $15.4\pm 3.1\pm 2.0$ & BESI~\cite{ycz_getap} & \\
                            & $12.4\pm 2.7\pm 1.5$ & BESII~\cite{ycz_getap2} & \\
 $\gamma\eta \pip\pim$ & $36.0\pm 14.2\pm 18.3$ & BESII~\cite{ycz_getap2} & \\
 $\gamma f_2(1270)$ & $21.2\pm 1.9\pm 3.2$ & BESI~\cite{ycz_gpp} & \\
 $\gamma f_0(1710)\to \gamma \pi\pi$ & $3.01\pm 0.41\pm 1.24$ & BESI~\cite{ycz_gpp} & \\
 $\gamma f_0(1710)\to \gamma K\bar{K}$ & $6.04\pm 0.90\pm 1.32$ & BESI~\cite{ycz_gpp} & \\
 $\gamma \ppb$ & $2.9\pm 0.4\pm 0.4$ & BESII~\cite{ycz_gx} & mass $<$ 2.9~GeV/$c^2$\\
 $\gamma 2(\pip\pim)$ & $39.6\pm 2.8\pm 5.0$ & BESII~\cite{ycz_gx} & mass $<$ 2.9~GeV/$c^2$\\
 $\gamma K_S^0K^+\pim + c.c.$ & $25.6\pm 3.6\pm 3.6$ & BESII~\cite{ycz_gx} & mass $<$ 2.9~GeV/$c^2$\\
 $\gamma \pip\pim K^+K^-$ & $19.1\pm 2.7\pm 4.3$ & BESII~\cite{ycz_gx} & mass $<$ 2.9~GeV/$c^2$\\
 $\gamma \pip\pim\ppb$ & $2.8\pm 1.2\pm 0.7$ & BESII~\cite{ycz_gx} & mass $<$ 2.9~GeV/$c^2$\\
 \hline
 \end{tabular}
 \end{center}
 \end{table}

Since the dominant decay dynamics of $\psp$ or $\jpsi$ radiative decay
is the $\ccb$ emission of a real photon and subsequent annihilation into
two gluons, it is expected that glueball states, if they exist, should
be copiously produced.  This has been a widely circulated argument in
favor of careful studies of $\jpsi$ radiative decays.
The same argument also holds for $\psp$
radiative decays, in spite of the obvious disadvantage of a lower
production rate.  However, this disadvantage may no longer be
an issue at \bes3, which should have a very
large (or even huge) $\psp$ data sample, with the obvious 
advantage that the allowed mass range for investigation 
is larger than that for $\jpsi$ decays.

Lattice QCD (LQCD)~\cite{ycz_cheny} calculations
indicate that ---except for the $\jpc=0^{++}$ glueball, which 
is expected at about  1.7~GeV/$c^2$, and the
$\jpc=2^{++}$ glueball, expected  at about 2.3~GeV/$c^2$--- 
most glueball states ({\it i.e.} those 
with  other $\jpc$ values) have rather large 
masses:  above 2.5~GeV/$c^2$  for
$\jpc=0^{-+}$ and around 3~GeV/$c^2$ for $\jpc=2^{-+}$.
Since these states are espected to have broad
widths, their study in $\jpsi$ decays would be somewhat
cumbersome.   All things being equal, it woud be 
more suitable to be search  for them in $\psp$ decays.

Contrary to initial naive expectations, many $\psp$ hadronic 
decay channels are observed to be dramatically different.
Likewise, it may very well be that the notion that
$\psp$ radiative decays  have similar properties as 
those for the $\jpsi$ may be incorrect;
$\psp$ and $\jpsi$ radiative
decays may turn out to be very different.  For example, in the
decays to $\gamma \pi\pi$, the relative strengths of the 
$\pi\pi$ resonances could be very different, in which case the 
two decays would supply very different information, and
this  might provide some  helpful insights into
meson spectroscopy. Another possible
 advantage of using the $\psp$ data
is that the backgrounds are very different. For example, in $\gamma
\pip\pim$ channels, there is almost no background from $\rho\pi$ in
$\psp$ decays, while this is the dominant background for $\jpsi$
decays, since $\jpsi\to \rho\pi$ is the largest two-body hadronic
mode in $\jpsi$ decays.

While the $\jpc=0^{++}$ and $2^{++}$ states can be well
studied in their decays into a pair of pseudoscalars, such as
$\pi\pi$, $K \bar{K}$, $\eta \eta$, $\etap \etap$, or $\eta
\etap$, the $\jpc=0^{-+}$ and $2^{-+}$ states have to be studied in
their decays into three pseudoscalars, such as $\eta\pi\pi$, $K
\bar{K}\pi $, $\eta K \bar{K}$, $\etap \pi \pi$, etc. 
Final states like $\gamma \ppb$ and $\gamma \Lambda \bar{\Lambda}$
can be used to search for the glueball with any quantum numbers,
so long as the mass is above the baryon-antibaryon mass threshold.

The dominant decay modes of the glueballs are expected to be to
multi-hadron final states.  However, the spin-parity analysis of
multi-hadron final states can be very complicated, due
to the contributions of many mesons with different spin-parity,
and many different intermediate states. In the case of many
parameters,  partial wave analyses (PWA) might give
meaningless results.


\subsection{Radiative decays of pseudoscalar charmonia}
\label{Pdecays}

So far there are no measurements of  radiative decays of the $\etac$,
let alone its excited state, the $\etac(2S)$.

The decay rates of $\etac$ to a $\gamma$ and a vector meson (like
$\rho$, $\omega$ and $\phi$) is of great interest to probe the wave
function of the $\eta_c$, where it is suspected that the mixing from
$\eta$ and $\etap$ may exist to explain for example the decay rates
of $\jpsi\to \gamma \eta$ and $\gamma \etap$~\cite{ycz_QWG}.
The decays of a pure
$\ccb$ states into $\gamma V$ may be very weak, however, if the
mixing from the $\eta$ and $\etap$ exists, the decay rate may be
enhanced to a much higher level. So far there is no theoretical
estimation of the rates yet, either with or without the mixing from
light quarks.

If the photon is emitted from the $\ccb$ quark pair in the $\etac$, the 
decay will be quite similar to $\etac\to \gamma \gamma$, with one real
photon replaced by a virtual one and further coupled to a vector
meson. Since $\BR(\etac\to \gamma\gamma)=(2.8\pm 0.9)\times
10^{-4}$~\cite{ted_PDG2006}, the decay rates of $\etac\to \gamma V$ should
not be very different from that if estimated from Vector Meson Dominance
model.

\subsection{Radiative decays of P-wave charmonia}
\label{Cdecays}

There is no experimental information on the radiative decays of the
$P$-wave charmonia, including $\chicz$, $\chico$, and $\chict$.

A recent theoretical calculation of $\chicJ\to \gamma V$, $V=\rho$,
$\omega$, or $\phi$, was given in Ref.~\cite{ycz_chao} based on
nonrelativistic quantum chromodynamics, the results are shown in
Table~\ref{table_GV}. The branching fractions are at a level which
can be searched for, or be observed, using the $\psp$
data sample at \bes3.

 \begin{table}
 \caption{calculated branching fractions of $\chicJ\to \gamma V$ from
 Ref.~\cite{ycz_chao}, in unit of $\times 10^{-6}$.
 \label{table_GV}}
 \begin{center}
 \begin{tabular}{cccc}
 \hline
 Mode   &\,\,\, $\chicz$ \,\,\,&\,\,\, $\chico$\,\,\, & \,\,\,$\chict$ \,\,\,\\ \hline
 $\gamma\rho$ &   1.2    &    14    &   4.4 \\
 $\gamma\omega$ & 0.13   &   1.6     &  0.50 \\
 $\gamma\phi$ &    0.46   &  3.6    &  1.1 \\
 \hline
 \end{tabular}
 \end{center}
 \end{table}

\subsection{Discussions and conclusions}

From the above analysis, we conclude the radiative decays of the
vector charmonium states, especially of $\jpsi$ and $\psp$, should
be studied with great care for the understanding of  light hadron
spectroscopy, as well as for a better understanding of the charmonium
decay dynamics. While the decays of $\jpsi$ are much rich, the study
of the $\psp$ is rather limited, more effort should be made to find
possible new phenomena in $\psp$ decays.

The search for the decays into $\gamma$+Vector allows a probe for the
wave functions of $\etac$ and $\chicJ$ states, to test whether there
is mixture from the light quark mesons.

\chapter{Hadronic decays}
\label{sec:hadron}
\section[Light-hadron decays]
{Light hadron decay \footnote{Yu-Qi Chen and Zhi-Guo He}}
\label{sec:LightHadronDecay}

In charmonium annihilation decay,  the basic process is  the
annihilation of the $c\bar{c}$ inside the charmonium to light
quarks, gluons, or leptons.  The energy release in the process is
$2m_{c}$, which means that  space-time distance of the annihilation
process is  of order $1/(2m_{c})$. In QCD, processes happened at
such high energy scale is perturbatively calculable. In the
non-relativistic limit $v\to 0$,  the annihilation space-time region
is much smaller than the radius of charmonium at order $1/(m_{c}v)$,
where $v$ is the relative velocity between $c$ and $\bar{c}$. Thus
the annihilation decay rate  is expected to be proportional to the
probability of the $c\bar{c}$ pair appearing at the same point in
space-time and the probability in which a point-like $c\bar{c}$ pair
annihilates into light partons. The former one is a long-distance
effect depending on the dynamics of the charmonium, while the later
one is a short-distance effect. In this way, the long-distance
effects and the short-distance one are separated.

The old-fashion factorization formula to calculate  the annihilation
decay rates is the  so-called ``color singlet model"(CSM), in which
the charmonium is treated as a pure color-singlet $c\bar{c}$ pair.
The  annihilation decay amplitude is then factored into the
wavefunction at the origin of the $c\bar{c}$, which accounts for the
long-distance effects and the amplitude of the $c\bar{c}$
annihilation process, which gives the short-distance contributions.
Drawbacks of this model are obvious. When some of the annihilated
partons are soft, the whole annihilation process is no longer purely
short distance, and this naive factorization of the CSM  fails,
especially for processes with essential contributions to the decay
rate from these regions. Some typical examples appear in the P-wave
\cite{cyqhzg_Barbieri:1976fp,cyqhzg_Barbieri:1979iy,cyqhzg_Barbieri:1980yp,cyqhzg_Barbieri:1981xz}
and D-wave \cite{cyqhzg_Belanger:1987cg,cyqhzg_Bergstrom:1991dp}
decay, whose rates at $\alpha_{s}^{3}$ by CSM are infrared
sensitive.

A modern-fashion factorization scheme to calculate  the charmonium
annihilation decay rate  is the NRQCD factorization
formula\cite{cyqhzg_Bodwin:1994jh}. It improves CSM by absorbing
the contribution from the annihilated soft partons  into the
long-distance effects. The short-distance effects is that the
$c\bar{c}$ pair annihilates into hard partons. The $c\bar{c}$ pair
in the short-distance annihilation is not necessarily color-singlet
and not necessary with the same quantum number as the charmonium
state. The charmonium state is treated as superposition state of
$|Q\bar{Q}\rangle, |Q\bar{Q}g\rangle, |Q\bar{Q}gg\rangle$ and other
higher Fock states, rather than only the $|Q\bar{Q}\rangle$
color-singlet state as in CSM. The contribution of each Fock state,
organized in powers of $v^{2}$, could be written as a product of
long matrix element and short distance coefficient accordingly. The
short distance coefficients are calculated perturbatively, while the
non-perturbative long distance matrix elements could be studied in
lattice simulations, or determined phenomenologically, or estimated
through velocity scaling rules.

Below $D\bar{D}$ open flavor threshold, one expects to have ten
charmonium states by potential model calculation. They are 4
S-wave states ($\eta_{c}(nS)$ and $\psi(nS)$ with n=1,2), 4 P-wave
states ($h_{c}$ and $\chi_{cJ}$ with J=0,1,2) and 2 D-wave states
($^1D_{2}$ and $^{3}D_{2}$). Many works have been done to study
the light hadron decay of S-wave and P-wave in framework of NRQCD
factorization method while for D-wave states decay into light
hadron the numerical prediction is still given by potential model
until now. In the following, we will briefly review the knowledge
of both the short distance coefficients and long distance matrix
elements for the LH decay of S-wave and P-wave states $J/\psi$,
$\psi(2S)$, $\eta_{c}$ $\eta_{c}(2S)$, $\chi_{cJ}$ and $h_{c}$.

\subsection{$J/\psi$, $\psi'$ decay}

  The  $J/\psi$ and $\psi'$ can decay to light hadrons  via
the annihilation of the $c$ and $\bar c$.  The annihilation decay
dominates the $J/\psi$ decay rate, while it accounts for 14\% of the
$\psi'$ decay rate since hadron and the electro-magnetic transitions
govern  $\psi'$ decays. The $J/\psi$ and $\psi'$ annihilation decay
rates are small since at leading order, they can decay only through
three-gluon annihilations. Both the matrix element and the
phase-space constraint the rate.  In the NRQCD factorization
formula\cite{cyqhzg_Bodwin:1994jh}, the expressions for inclusive
decay rate of $J/\psi$ have been known at relative $v^{4}$
order:\cite{cyqhzg_Bodwin:2002hg}
\begin{eqnarray}\label{fact}
\Gamma(J/\psi\rightarrow~l.h.)=
\frac{2\mathrm{Im}f_{1}(^3S_{1})}{m_{c}^{2}}\langle\psi|\mathcal{O}_{1}(^3S_1)|\psi\rangle+
\frac{2\mathrm{Im}g_{1}(^3S_{1})}{m_{c}^{4}}\langle\psi|\mathcal{P}_{1}(^3S_1)|\psi\rangle
\nonumber\\+
\frac{2\mathrm{Im}f_{8}(^1S_{0})}{m_{c}^{2}}\langle\psi|\mathcal{O}_{8}(^1S_0)|\psi\rangle+
\frac{2\mathrm{Im}f_{8}(^3S_{1})}{m_{c}^{2}}\langle\psi|\mathcal{O}_{8}(^3S_1)|\psi\rangle
\nonumber\\+
\sum_{J=0,1,2}\frac{2\mathrm{Im}f_{8}(^3P_{J})}{m_{c}^{4}}\langle\psi|\mathcal{O}_{8}(^3P_J)|\psi\rangle+
\frac{2\mathrm{Im}h_{1}^{1}(^3S_{1})}{m_{c}^{6}}\langle\psi|\mathcal{Q}_{1}^{1}(^3S_1)|\psi\rangle
\nonumber\\+
\frac{2\mathrm{Im}h_{1}^{2}(^3S_{1})}{m_{c}^{6}}\langle\psi|\mathcal{Q}_{1}^{2}(^3S_1)|\psi\rangle.
\end{eqnarray}
where the definitions of those operators $\mathcal{O},\mathcal{P}$
and $\mathcal{Q}$ could be found in
Ref.\cite{cyqhzg_Bodwin:1994jh,cyqhzg_Bodwin:2002hg}. From the
velocity scaling rule in Ref.\cite{cyqhzg_Bodwin:1994jh}, it could
be find that the matrix element of $\mathcal{O}_{1}(^3S_1)$ is at
relative $v^{0}$ order, the matrix element of
$\mathcal{P}_{1}(^3S_1)$ is at relative $v^{2}$ order, the matrix
element of $\mathcal{O}_{8}(^1S_{0})$ is at relative $v^{3}$ order
and the matrix elements of $\mathcal{O}_{8}(^3S_1)$,
$\mathcal{O}_{8}(^3P_J)$, $\mathcal{Q}_{1}^{1}(^3S_1)$,
$\mathcal{Q}_{1}^{2}(^3S_1)$ is at relative order $v^{4}$.

The short distance coefficients have been calculated by many groups
to different order of $\alpha_{s}$. $2\mathrm{Im}f_{1}(^3S_{1})$~
were computed at $\mathcal{O}(\alpha_{s}^{3})$ and
$\mathcal{O}(\alpha_{s}^{4})$ by Mackenzie and
Lapage\cite{cyqhzg_Mackenzie:1981sf}:
\begin{equation}
2\mathrm{Im}f_{1}(^3S_{1})=\frac{20(\pi^2-9)\alpha_{s}^{3}(m_{c})}{243}\{1+[-9.46(2)C_{F}+4.13(17)C_{A}-
1.161(2)n_{f}]\frac{\alpha_{s}}{\pi}\}\footnote{We have drop the
contribution of electromagetic annihilation process
$J/\psi\rightarrow\gamma^{\ast}\rightarrow q\bar{q}$}
\end{equation}
where $C_{F}=\frac{4}{3}$, $C_{A}=3$ and $n_{f}$ is the flavors of
light quarks and equals to 3 in charmonium systems.  The coefficient
of relative order $v^{2}$ operator $\mathcal{P}(^3S_1)$ was
calculated in Ref.\cite{cyqhzg_Keung:1982jb}:
\begin{equation}
2\mathrm{Im}g_{1}(^3S_{1})=-\frac{5(19\pi^2-132)}{729}\alpha_{s}^{3},
\end{equation}
which is about 5.33 times larger than $2\mathrm{Im}f_{1}(^3S_{1})$
at $\alpha_{s}$ leading order. This indicates that the relativistic
corrections to $J/\psi$ decay into $l.h.$ is important. The
coefficients of those color octet coefficients were computed up to
$\mathcal{O}(\alpha_{s}^{3})$ in Ref.\cite{cyqhzg_Petrelli:1997ge}.
Some of they are about two orders of magnitude larger than
$2\mathrm{Im}f_{1}(^3S_{1})$ , which suggests that the higher Fock
states may play a more important role in annihilation decays than is
commonly believed. The combination
$2\mathrm{Im}h_{1}^{1}(^3S_{1})+2\mathrm{Im}h_{1}^{2}(^3S_{1})$ were
compute in Ref.\cite{cyqhzg_Bodwin:2002hg}, which is also about an
order of magnitude larger than $2\mathrm{Im}f_{1}(^3S_{1})$.

The matrix elements are expectation values in the $J/\psi$ of local
gauge-invariant NRQCD operators that measure the inclusive
probability of finding a $c \bar c$ in the $J/\psi$ at the same
point and in the color and angular-momentum state specified. They
could be studied in lattice simulations, or determined by fitting
the experiment decay data, or estimated through the operator
evolution equations. The vacuum-saturation approximation relates the
matrix element of $\mathcal{O}_{1}(^3S_1)$ in (\ref{fact}) with the
square of the non-relativistic sh$\ddot{o}$rdinger wave function at
the origin with errors up to $\mathcal{O}(v^{4})$. In pNRQCD
effective theory the matrix elements of $\mathcal{P}_{1}(^3S_1)$,
$\mathcal{O}_{8}(^1S_0)$, $\mathcal{O}_{8}(^3S_1)$, and
$\mathcal{O}_{8}(^3P_J)$ could be related with the leading order
matrix element $\langle\psi|\mathcal{O}_{1}(^3S_1)|\psi\rangle$(For
details see \cite{cyqhzg_Brambilla:2002nu} and reference therein).
And the knowledge of the the dimension 10 operators $\mathcal{Q}$ is
poor. The same expression of the decay rate hold for $\psi'$ once
the state $J/\psi$ is substituted by the $\psi'$. Since a large
number of $J/\psi$ and $\psi'$ could be directly produced at BES,
then the BES III project provide a good way to determinate those
non-perturbative parameters.

\subsection{$\eta_c$ decay}
The $\eta_c$ is the ground state of the charmonium states that could
be produced through the $\gamma\eta_{c}$ decay channel of $J/\psi$.
It can decay only via annihilation of the $c\bar{c} $. NRQCD
factorization formula for $\eta_c $ decay into light hadron at
$\mathcal{O}(v^{4})$ are:
\begin{eqnarray}
\Gamma(\eta_{c}\rightarrow~l.h.)&=&
\frac{2\mathrm{Im}f_{1}(^1S_{0})}{m_{c}^{2}}\langle\eta_{c}|\mathcal{O}_{1}(^1S_0)|\eta_{c}\rangle+
\frac{2\mathrm{Im}g_{1}(^1S_{0})}{m_{c}^{4}}\langle\eta_{c}|\mathcal{P}_{1}(^1S_0)|\eta_{c}\rangle
\nonumber\\&+&
\frac{2\mathrm{Im}f_{8}(^3S_{1})}{m_{c}^{2}}\langle\eta_{c}|\mathcal{O}_{8}(^3S_1)|\eta_{c}\rangle+
\frac{2\mathrm{Im}f_{8}(^1S_{0})}{m_{c}^{2}}\langle\eta_{c}|\mathcal{O}_{8}(^1S_0)|\eta_{c}\rangle
\nonumber\\&+&
\frac{2\mathrm{Im}f_{8}(^1P_{1})}{m_{c}^{4}}\langle\eta_{c}|\mathcal{O}_{8}(^1P_1)|\eta_{c}\rangle+
\frac{2\mathrm{Im}h_{1}^{1}(^1S_{0})}{m_{c}^{6}}\langle\eta_{c}|\mathcal{Q}_{1}^{1}(^1S_0)|\eta_{c}\rangle
\nonumber\\&+&
\frac{2\mathrm{Im}h_{1}^{2}(^1S_{0})}{m_{c}^{6}}\langle\eta_{c}|\mathcal{Q}_{1}^{2}(^1S_0)|\eta_{c}\rangle,
\end{eqnarray}
Where the definitions of those operators $\mathcal{O},\mathcal{P}$
and $\mathcal{Q}$ could also be found in
Ref.\cite{cyqhzg_Bodwin:1994jh,cyqhzg_Bodwin:2002hg}. The
coefficients $2\mathrm{Im}f_{1}(^1S_{0})$ were given up to
$\mathcal{O}(\alpha_{s}^{3})$ in
Ref.\cite{cyqhzg_Barbieri:1979be,cyqhzg_Hagiwara:1980nv}:
\begin{eqnarray}
2\mathrm{Im}f_{1}(^1S_{0})=\frac{4\pi\alpha_{s}^{2}(2m_{c})}{9}\{1+[(\frac{\pi^2}{4}-5)C_{F}
\nonumber\\+(\frac{199}{18}-\frac{13\pi^2}{24})C_{A}-\frac{8}{9}n_{f}]\frac{\alpha_{s}}{\pi}\}
\end{eqnarray}
The coefficient of the relative order $v^{2}$ operator
$\mathcal{P}(^1S_0)$ is computed in Ref.\cite{cyqhzg_Keung:1982jb}.
It is
\begin{equation}
2\mathrm{Im}g(^1S_{0})=-\frac{4C_{F}}{3N_{c}}\alpha_{s}^{2}
\end{equation}
At leading order in $\alpha_{s}$
$\frac{\mathrm{Im}g(^1S_{0})}{\mathrm{Im}f(^1S_{0})}=-\frac{4}{3}$,
which means the leading order relativistic corrections is sizable in
$\eta_{c}$ decay into light hadrons. The coefficients
$2\mathrm{Im}f_{8}(^3S_1)$ and $2\mathrm{Im}f_{8}(^1S_0)$ have been
computed in Ref.\cite{cyqhzg_Petrelli:1997ge}, which are in the same
order of magnitude as $\mathrm{Im}f(^1S_{0})$. The coefficient
$2\mathrm{Im}f_{8}(^1P_1)$ could be found in the Appendix of
Ref.\cite{cyqhzg_Bodwin:1994jh} at $\mathcal{O}(\alpha_{s}^{2})$:
\begin{equation}
2\mathrm{Im}f_{8}(^1P_1)=\frac{\pi N_{c}}{6}\alpha_{s}^{2}.
\end{equation}
The combination
$2\mathrm{Im}h_{1}^{1}(^1S_{0})+2\mathrm{Im}h_{1}^{2}(^1S_{0})$ were
compute in Ref.\cite{cyqhzg_Bodwin:2002hg}, which is also at the
same order of magnitude as $2\mathrm{Im}f_{1}(^1S_{0})$.

In heavy quark limit, the matrix elements of
$\mathcal{O}_{1}(^1S_0)$, $\mathcal{P}_{1}(^1S_0)$, $\mathcal{O}_{8}
(^3S_1)/3$, $\mathcal{O}_{8}(^1S_0)$,$\mathcal{O}_{8}(^1P_{1})/9$
$\mathcal{Q}_{1}^{1}(^1S_0)$, $\mathcal{Q}_{1}^{2}(^1S_0)$ in
$\eta_{c}$ state are equal to the matrix elements of
$\mathcal{O}_{1}(^3S_1)$, $\mathcal{P}_{1}(^3P_1)$, $\mathcal{O}_{8}
(^1S_0)$, $\mathcal{O}_{8}(^3S_1)$,$\mathcal{O}_{8}(^3P_{J})/(2J+1)$
$\mathcal{Q}_{1}^{1}(^3S_1)$, $\mathcal{Q}_{1}^{2}(^3S_1)$ in
$J/\psi$ state, respectively, with errors up to relative
$\mathcal{O}(v^{2})$. In Ref.\cite{cyqhzg_Bodwin:1996tg},
$\langle\eta_{c}|\mathcal{O}_{1}(^{1}S_{0})|\eta_{c}\rangle$ was
calculated in quenched lattice simulation, and their result in
$\overline{MS}$ scheme and at the factorization scale of $1.3$GeV is
about $0.33$GeV$^{3}$.

Comparing with the $J/\psi$, the decay width of $\eta_c$ is much
larger because $c\bar{c}$ can annihilate into two gluons and the
short distance coefficients is order $\alpha_s^2$ at the leading
order. Also in the case of $J/\psi$, there is an additional
suppression factor $\pi^2-9$ in the short distance coefficient of
$c\bar{c} \to ggg$

The $\eta_c$ have also electromagnetic decay $\eta_c \to \gamma
\gamma$. NRQCD factorization formula can also apply for $\eta_c
\to \gamma\gamma$. An interesting quantity is the ratio of the
decay rate for $\eta_c \to  {\rm light\; hadrons} $ to that for
the $\eta_c \to \gamma \gamma$, the inverse branching fraction,
denoted by $R$. It is relatively easy to be measured in
experiments. In $R$, the NRQCD matrix element cancel up to order
$v^4$, the relativistic corrections also cancel at the leading
order of $\alpha_s$. This leaves $R$ almost a pure short-distance
number. In principle, $R$ may be calculated precisely in pQCD.
However, one find that the radiative correction to $R$ is quite
large and $R$ is sensitive to the choice of the renormalization
scale $\mu$. {\it e.g.}, when $\mu$ varies from $2m_c$ to $0.5
m_c$, $ R $ varies by factor 5 for the next leading order result.
Bodwin and Chen show that the most important corrections arise
from the bubble diagram. After the bubble diagrams are resumed, a
result which is insensitive to $\mu$ is obtained, which is in
agreement with the experimental results. This quantity is expected
to be measured accurately at BES III.

\subsection{ $P$-wave $\chi_{cJ}$ decay}
    An interesting application of the NRQCD factorization
formula is the charmonium $P-$wave annihilation decay. The decay
rate was calculated by Barbieri {\it et
 al.}\cite{cyqhzg_Barbieri:1980yp,cyqhzg_Barbieri:1981xz} to the next-leading order
of $\alpha_s$ in the color-singlet model. It was found that in
$\chi_{cJ}\to ggg (J=0,2)$ processes the infrared divergence cancels
when combined with the virtual corrections, while there were
infrared divergences, which arise from the kinematic region where
the emitted gluon is soft in the process $\chi_{cJ} \to q\bar{q} g$.
This indicates that the long-distance effects enter into the
short-distance coefficients and the factorization in the
conventional color-singlet model fails. The problem was nicely
solved in the NRQCD factorization
framework\cite{cyqhzg_Petrelli:1996dp}. According to it, the soft
gluon emitting process is absorbed into the long distance matrix
elements while the short-distance process is the color-octet
$c\bar{c}$  annihilation into a light quark pair. The corresponding
long-distance matrix element is NRQCD S-wave color-octet  matrix
element. Its contribution to the annihilation decay rates of the
$P-$wave charmonia is significant since the short distance process
is the same order as the leading order color singlet process for
$\chi_{c0}$ and $\chi_{c2}$ but enhanced by order $\alpha_s(m_c)$
for $\chi_{c1}$ compared with the leading order color-single
process. The color-octet matrix elements are at the same order of
$v^{2}$ as that of the color-singlet one. In the NRQCD factorization
formula, the $P-$wave annihilation decay rates at $v^{2}$ leading
order are expressed as:
\begin{eqnarray}
\Gamma (\chi_{cJ} \to {\it l.h.} ) &=& \frac{2\mathrm{Im}f_{1}(^3P_J)}{m_{c}^{4}}
\langle\chi_{cJ}|\mathcal{O}_{1}(^3P_{J})|\chi_{cJ}\rangle\nonumber\\
&+&\frac{2\mathrm{Im}f_{8}(^3S_1)}{m_{c}^{2}}
\langle\chi_{cJ}|\mathcal{O}_{8}(^3S_{1})|\chi_{cJ}\rangle,\label{chi-cj-1}
\end{eqnarray}

A full list of those short-distance coefficients is summarized in
Ref.\cite{cyqhzg_Petrelli:1997ge}\footnote{The normalization of the
color singlet operators in Ref.\cite{cyqhzg_Petrelli:1997ge} is
different from those given in Ref.\cite{cyqhzg_Bodwin:1994jh} }.
\begin{eqnarray}
2\mathrm{Im}f_{1}(^3S_1) &=& n_{f}{\alpha_{s}^{2}\pi\over 3 } \left\{1 +
\frac{\alpha_s}{\pi}\left[
 -C_F\frac{13}{4}   \right.\right.
  +C_A \left( \frac{133}{18}+\frac{2}{3}\ln2- \frac{\pi^2}{4}  \right)
   \nonumber\\
  &&\left.\left.
      -\frac{5}{9}n_f+ 2 b_0 \log \frac{\mu}{2 m_c} \right]\right\}
+ \frac{5}{3}{\alpha_s^3}
\left(-\frac{73}{4}+\frac{67}{36}\pi^2\right)
\label{3s18}\\
2\mathrm{Im}f_{1}(^3P_0) &=& {C_F \alpha_s^2\pi  } \left\{1 +
\frac{\alpha_s}{\pi}\left[C_F \left(
 -\frac{7}{3}+\frac{\pi^2}{4}       \right) \right.\right.
  +C_A \left( \frac{427}{81}- \frac{1}{144}\pi^2  \right)
   \nonumber\\
  &&\left.\left.
      + 2 b_0 \log \frac{\mu}{2 m_c} \right]\right\}
+ n_f{\alpha_s^3}\frac{4}{27} C_F
\left( - {29\over 6} - \log{\mu_{\Lambda}\over{2m_{c}}}\right)
\label{a1} \\
2\mathrm{Im}f_{1}(^3P_1) &=& {C_F \alpha_s^3 \over 2} \left( \frac{587}{27}
-\frac{317}{144}\pi^2 \right) + n_f{\alpha_s^3}\frac{4}{27} C_F
\left( - {4\over 3}  - \log{\mu_{\Lambda} \over{2m_{c}}}\right)
 \\
2\mathrm{Im}f_{1}(^3P_2) &=&{4C_F \alpha_s^2\pi \over 15} \left\{ 1 +
\frac{\alpha_s}{\pi} \Big[ -4 \,C_F  \right.+C_A\left(
\frac{2185}{216}-\frac{337}{384}\pi^2+ \frac{5}{3} \log 2 \right)
 \nonumber\\
 && \left.\left.
+ 2 b_0 \log \frac{\mu}{2 m_{c}}\right]\right\}
 + n_f {\alpha_s^3}\frac{4}{27}C_F
\left( - {29\over 15} - \log{\mu_{\Lambda}\over{2m_c}}\right)
\nonumber \\
\end{eqnarray}

At $\mathcal{O}(v^{0})$, both the color singlet matrix elements
$\langle\chi_{cJ}|\mathcal{O}_{1}(^3P_{J})|\chi_{cJ}\rangle$ and
color octet ones
$\langle\chi_{cJ}|\mathcal{O}_{8}(^3S_{1})|\chi_{cJ}\rangle$ are the
same for different $J$, then there are only two non-perturbative
parameters in three independent processes $\chi_{cJ}\rightarrow
l.h.$. This provides a good way to test the validity of NRQCD
factorization method. In \cite{cyqhzg_Huang:1996cs}, Huang and Chao
use $\Gamma(\chi_{c1,c2}\rightarrow l.h.)$ as inputs to predict
$\Gamma(\chi_{c0}\rightarrow l.h.)$, and their result is consistent
with the experimental data.\footnote{There is an error in their
original expression, but it has only a little effect on the
numerical result.} As pointed out in chapter 4 of
Ref.\cite{cyqhzg_Brambilla:2004wf}, the ratios of the differences of
$\chi_{cJ}\rightarrow l.h.$ are not dependent on any
non-perturbative parameters, and there they give the experimental
results of the considered ratios determined from PDG 2000
\cite{cyqhzg_Groom:2000in} and PDG 2004
\cite{cyqhzg_Eidelman:2004wy} respectively, and the theory
predictions at LO and NLO are also given. Here in Table
\ref{tab:yqczhgratios}, we give the PDG 2006
\cite{cyqhzg_Yao:2006px} results of these ratios, and also list
experimental results of PDG 2004 and the theoretical LO and NLO
results in Table 4.8 of Ref.\cite{cyqhzg_Brambilla:2004wf} for
comparison. The values in Table \ref{tab:yqczhgratios} show that the
discrepancy between experimental and NLO theoretical results becomes
a little larger, but still consistent with NLO theoretical results
and the errors are reduced. In Ref.\cite{cyqhzg_Ablikim:2005yd}, the
BES Collaboration only reports their measurements of
$\Gamma(\chi_{c0})$ and $\Gamma(\chi_{c1})$ in the $\psi(2S)\to
\gamma \chi_{cJ}$ processes. With more data accumulated at BES III,
$\Gamma(\chi_{c2})$ could be measured and then the experimental
results of the considered ratios will be more accurate, and this is
helpful for our understanding of the light hadron decay of
$\chi_{cJ}$.
\begin{table}[h]
\renewcommand{\arraystretch}{1.2}
\begin{center}
\caption {Comparison of the ratios of the differences between
$\chi_{cJ}\rightarrow l.h.$. The experimental values PDG 2006 are
obtained from Ref.\cite{cyqhzg_Yao:2006px} with the inputs
$\Gamma(\chi_{c0}\rightarrow
l.h.)\approx\Gamma(\chi_{c0})=10.5\pm0.9$ MeV,
$\Gamma(\chi_{c1}\rightarrow
l.h.)\approx\Gamma(\chi_{c1})[1-\mathcal{B}(\chi_{c1}\rightarrow\gamma
J/\psi)]=0.57\pm0.04$ MeV, $\Gamma(\chi_{c2}\rightarrow
l.h.)\approx\Gamma(\chi_{c2})[1-\mathcal{B}(\chi_{c2}\rightarrow\gamma
J/\psi)]=1.56\pm0.10$ MeV. The PDG 2004 experimental values and LO
and NLO theoretical results are the same as those in Table 4.8 of
Ref.\cite{cyqhzg_Brambilla:2004wf}.}
\begin{tabular}{|c|c|c|c|c|}
\hline \textbf{Ratio}&\textbf{PDG} 2006 &\textbf{PDG} 2004 &
\textbf{LO} &  \textbf{NLO}\\
 \hline
& & & & \\
 \(\displaystyle \frac{\Gamma(\chi_{c0}\to
l.h.)-\Gamma(\chi_{c1}\to l.h.)}{\Gamma(\chi_{c2}\to
l.h.)-\Gamma(\chi_{c1}\to l.h.)}\)&
$10.0\pm0.9$& $8.9\pm1.1$& 3.75 &$\approx 7.63$ \\
& & & & \\
\hline
  & & & &\\
\(\displaystyle\frac{\Gamma(\chi_{c0}\to l.h.)-\Gamma(\chi_{c2}\to
l.h.)}{\Gamma(\chi_{c2}\to l.h.)-\Gamma(\chi_{c1}\to l.h.)}\)&
$9.0\pm0.9$ & $7.9\pm1.5$ & 2.75 &  $\approx 6.63 $\\
&  &  &  &\\
\hline
\end{tabular}
\label{tab:yqczhgratios}
\end{center}
\renewcommand{\arraystretch}{1.0}
\end{table}
Another usage of the experimental data of $\Gamma(\chi_{cJ}\to
l.h.)$ is to determine the long distance matrix elements in
$\chi_{cJ}$ production processes, since in NRQCD factorization
method the long distance matrix elements are universal and the
differences between the decay matrix elements and production matrix
elements are at relative $v^{4}$ order. With the obtained production
matrix elements from decay processes, it is certainly helpful for us
to understand the production mechanism of $\chi_{cJ}$. In
Ref.\cite{cyqhzg_Maltoni:2000km}, the color singlet matrix elements
are obtained by fitting experiment data and the best value result at
$m_{c}$=1.5GeV is
$\langle\chi_{cJ}|\mathcal{O}_{1}(^3P_{J})|\chi_{cJ}\rangle/m_{c}^{2}=3.2\pm0.4\times10^{-2}$GeV$^{3}$,
and the ratio of color singlet matrix element to color octet matrix
element is also given there. In Table \ref{tab:yqczhgmatrix} we give
the average values of
$\langle\chi_{cJ}|\mathcal{O}_{1}(^3P_{J})|\chi_{cJ}\rangle$\footnote{There
is a factor $1/(2N_{c})$ difference between our singlet operators
and those in the original paper.} and
$\langle\chi_{cJ}|\mathcal{O}_{8}(^3S_{1})|\chi_{cJ}\rangle$
obtained from PDG 2006 \cite{cyqhzg_Yao:2006px}  at different
renormalization scales.
\begin{table}[h]
\renewcommand{\arraystretch}{1.2}
\begin{center}
\caption {Results  of the long distance matrix elements
 at different renormalization scales
by fitting PDG 2006 \cite{cyqhzg_Yao:2006px}. The numerical inputs
are $\Gamma(\chi_{c0}\rightarrow
l.h.)\approx\Gamma(\chi_{c0})=10.5\pm0.9$ MeV,
$\Gamma(\chi_{c1}\rightarrow
l.h.)\approx\Gamma(\chi_{c1})[1-\mathcal{B}(\chi_{c1}\rightarrow\gamma
J/\psi)]=0.57\pm0.04$ MeV, $\Gamma(\chi_{c2}\rightarrow
l.h.)\approx\Gamma(\chi_{c2})[1-\mathcal{B}(\chi_{c2}\rightarrow\gamma
J/\psi)]=1.56\pm0.10$ MeV, $m_{c}=1.5$ GeV,
$\alpha_{s}=\alpha_{s}(\mu)$ and $\alpha_{s}(2m_{c})=0.249$.}
\begin{tabular}{|c|c|c|c|}
\hline
    & $\mu=1.0 m_{c}$& $\mu=\sqrt{2}m_{c}$& $\mu=2.0 m_{c}$\\
\hline
  $\langle\chi_{cJ}|\mathcal{O}_{1}(^3P_{J})|\chi_{cJ}\rangle$$(10^{-2}$GeV$^{5})$ & 0.97 & 1.1 & 1.4 \\
\hline
  $\langle\chi_{cJ}|\mathcal{O}_{1}(^3P_{J})|\chi_{cJ}\rangle$$(10^{-2}$GeV$^{3})$ & 0.40 & 0.38  & 0.39\\
\hline
\end{tabular}
\label{tab:yqczhgmatrix}
\end{center}
\renewcommand{\arraystretch}{1.0}
\end{table}

\section{P-wave $h_{c}$ decay} 
{P-wave $h_{c}$ decay \footnote{Yu-Qi Chen and Zhi-Guo He}}
\label{sec:LightHadronhc}
Because of C-parity conservation, $h_{c}$ can not decay into $gg$
finial state. And there is also infrared divergence
problem\cite{cyqhzg_Barbieri:1976fp} in $h_{c}\to ggg$ process in
the color-singlet model. This problem was solved in the framework of
NRQCD factorization method in Ref.\cite{cyqhzg_Huang:1996fa}, and
summarized in \cite{cyqhzg_Petrelli:1997ge}. At leading order in
$v^{2}$ the expression of $h_{c}$ decay into light hadron is:
\begin{eqnarray}
\Gamma (h_{c} \to {\it l.h.} ) &=& \frac{2\mathrm{Im}f_{1}(^1P_1)}{m_{c}^{4}}
\langle h_{c}|\mathcal{O}_{1}(^1P_{1})|h_{c}\rangle\nonumber\\
&+&\frac{2\mathrm{Im}f_{8}(^1S_0)}{m_{c}^{2}}
\langle h_{c}|\mathcal{O}_{8}(^1S_{0})|h_{c}\rangle,\label{chi-cj}
\end{eqnarray}
And at $\mathcal{O}(\alpha_{s}^{3})$ the expression of the short distance coefficients are:
\begin{eqnarray}
2\mathrm{Im}f_{1}(^1P_1)=\frac{40\alpha_{s}^{3}}{81}(-\frac{7}{3}+\frac{7\pi^2}{48}-\ln\frac{\mu_{\Lambda}}{2m_c})
\end{eqnarray}
\begin{eqnarray}
2\mathrm{Im}f_{8}(^1S_0)&=&\frac{5 C_{F}\alpha_{s}^{2}\pi}{6}\left\{1 +
\frac{\alpha_s}{\pi}\left[C_F \left(
 -5+\frac{\pi^2}{4}       \right) \right.\right.
   \nonumber\\
  &&\left.\left.
    +C_A \left( \frac{122}{9}- \frac{17}{24}\pi^2  \right)
     -\frac{8}{9}n_{f} + 2 b_0 \log \frac{\mu}{2 m_c} \right]\right\}
\end{eqnarray}

The matrix elements of both color singlet and color octet operators
are studied in quenched lattice
simulations\cite{cyqhzg_Bodwin:1996tg}. The numerical result is
$\langle
h_{c}|\mathcal{O}_{1}(^1P_{1})|h_{c}\rangle\approx8.0\times10^{-2}$GeV$^{5}$
and $\langle
h_{c}|\mathcal{O}_{8}(^1S_{0})|h_{c}\rangle\approx4.7\times10^{-3}$GeV$^{3}$
in the $\overline{MS}$ scheme with the factorization scale equaling
to $1.3$GeV. In Ref.\cite{cyqhzg_Maltoni:2000km}, the matrix
elements of $h_{c}$ are estimated through the experimental data of
$\chi_{cJ}$ owing to heavy quark spin symmetry, and the light hadron
decay width is also predicted to be $\Gamma(h_{c}\to
l.h.)=0.72\pm0.32$MeV. At present the uncertainty of $h_{c}$ decay
into light hadrons is not small on both experimental side and
theoretical side. With more $h_{c}$ are accumulated at BES III from
$\psi'\to \pi^{0}h_{c}$ process, the light hadron decay width of
$h_{c}$ will be measured more precisely.

\section[$\rho-\pi$ puzzle]
{$\rho-\pi$ puzzle\footnote{Xiaohu Mo, Ping Wang, and Changzheng Yuan}}
\label{sec:rhopi}
\renewcommand{\gb}{{\cal O}}
\newcommand{\afs}{\alpha_s}
\newcommand{\alf}{\alpha}
\renewcommand{\omegap}{\omega^{\prime}}
\newcommand{\chia}{\chi_{c0}}
\newcommand{\chib}{\chi_{c1}}
\newcommand{\chic}{\chi_{c2}}
\renewcommand{\chicj}{\chi_{cJ}}
\renewcommand{\psip}{\psi(2S)}
\renewcommand{\pspp}{\psi^{\prime \prime}}
\renewcommand{\psipp}{\psi(3770)}
\renewcommand{\DD}{D^+ D^-}
\renewcommand{\MM}{\mu^+\mu^-}
\renewcommand{\TT}{\tau^+\tau^-}
\renewcommand{\GG}{\gamma\gamma}
\renewcommand{\pp}{\pi^+\pi^-}
\renewcommand{\kk}{K^+K^-}
\renewcommand{\kskl}{K^0_SK^0_L}
\renewcommand{\kstk}{K^* \overline{K}}
\renewcommand{\KKSC}{K^{*+}K^-}
\renewcommand{\KKSN}{K^{*0}\overline{K^0}}
\renewcommand{\zz}{\bar{p} p}
\renewcommand{\rpi}{\rho\pi}
\renewcommand{\OP}{\omega\pi^0}
\renewcommand{\RP}{\rho\pi}
\renewcommand{\RET}{\rho\eta}
\renewcommand{\RETp}{\rho\eta^{\prime}}
\renewcommand{\OET}{\omega\eta}
\renewcommand{\OETp}{\omega\eta^{\prime}}
\renewcommand{\FET}{\phi\eta}
\renewcommand{\FETp}{\phi\eta^{\prime}}
\renewcommand{\ogpi}{\omega\pi^0}
\renewcommand{\omegapi}{\omega\pi^0}
\renewcommand{\rhopi}{\rho\pi}
\renewcommand{\rhopin}{\rho^0 \pi^0}
\renewcommand{\ccbar}{c\bar{c}}
\renewcommand{\ccb}{c\overline{c}}
\renewcommand{\QQb}{Q\overline{Q}}
\renewcommand{\qqb}{q\overline{q}}
\renewcommand{\nnb}{n\overline{n}}
\renewcommand{\ppbar}{p\bar{p}}
\renewcommand{\BBb}{B\overline{B}}
\renewcommand{\KKb}{K\overline{K}}
\renewcommand{\ppb}{p\overline{p}}
\renewcommand{\ddbar}{D\bar{D}}
\renewcommand{\ddb}{D\overline{D}}
\renewcommand{\DDb}{D\overline{D}}
\renewcommand{\kskp}{K^0_S K^+ \pi^- + c.c.}
\renewcommand{\kskn}{K^{*0}(892)\overline{K^0}+c.c.}
\renewcommand{\VP}{1^-0^-}
\renewcommand{\PP}{0^-0^-}
\renewcommand{\jpsipp}{J/\psi \pi^+\pi^-}
\renewcommand{\PPJP}{\pi^+\pi^- J/\psi}
\def\eref#1{(\ref{#1})}
\def\Journal#1#2#3#4{{#1} {\bf #2}, #3 (#4)}
\def\CTP{Commun. Theor. Phys.}
\def\IJMPA{Int. J. Mod. Phys. A}
\def\NIMA{Nucl. Instrum. Methods A}
\def\NPB{Nucl. Phys. B}
\def\NPA{Nucl. Phys. A}
\def\PL{Phys. Lett. }
\def\PLB{Phys. Lett. B}
\def\PLA{Phys. Lett. A}
\def\PRL{Phys. Rev. Lett.}
\def\PRD{Phys. Rev. D}
\def\PRC{Phys. Rev. C}
\def\PRP{Phys. Rep.}
\def\PTPS{Prog. Theor. Phys. Suppl.}
\def\ZPC{Z. Phys. C}
\def\EPJC{Eur. Phys. J. C}
\def\HEPNP{HEP \& NP}

\def\ctup#1{\cite{#1}}


\subsection{Introduction}

Crisply defined experimental puzzles in high-energy physics frequently
have the prospect for leading to new discoveries.
Therefore puzzles in physics often draw considerable
attention from theorists.
The ratios of hadronic decays of the $\psi(3686)$ (also know as
$\psp$) to the same decays modes of the $\jpsi$
is a puzzle that has been studied in some depth since 1983.

Ihe OZI suppressed decays of $\jpsi$ and $\psp$ to hadrons
are via three gluons or a photon.  In both cases, perturbative QCD
(pQCD) provides the relation~\cite{appelquist}
\begin{eqnarray}
Q_h &=&\frac{{\cal B}_{\psp \ra h}}{{\cal B}_{\jpsi \ra h}}
=\frac{{\cal B}_{\psp \ra \EE}}{{\cal B}_{\jpsi \ra \EE}}
\approx 12.7\%~. \label{qcdrule}
\end{eqnarray}
This relation is referred to as the ``12\% rule'' and it is expected
to hold to a reasonably good degree for both inclusive and
exclusive decays. The so-called ``$\RP$ puzzle'' is the observation that 
the Eq.~\eref{qcdrule} prediction is severely violated for the $\RP$
and several other decay channels. The first evidence for this
effect was reported by the Mark-II Collaboration in 1983~\cite{mk2}.
Since then, many theoretical explanations have been put forth
in attempts to decipher this puzzle.

With the recent BESII and CLEOc experimental results on
$\jpsi$ and $\psp$ two-body decays to vector-pseudoscalar ($VP$),
vector-tensor ($VT$), pseudoscalar-pseudoscalar ($PP$), and
baryon-antibaryon ($\BBb$) pairs, plus results on multi-body decays of
the $\jpsi$, the $\psp$ and even the $\psi(3770)$
(also known as $\pspp$)~\cite{bes3pi}-\cite{cleoklks2},
a variety of explanations have been proposed for this puzzle that can be 
tested
with higher accuracy data.
Here we survey  theoretical work on the $\RP$ puzzle
and compare them with available experimental data.
From the theoretical point of view,
since the $Q$-value for $\rpi$ is smaller than 12\%, it may be 
caused either by an enhanced $\jpsi$ or suppressed $\psp$ decay rate,
or by a combination of the two.  Thus, we classify the various
theoretical speculations into three categories:

\begin{enumerate}
\item $\jpsi$-enhancement hypothesis, which attributes the small
$Q$-value to an enhanced branching fraction for $\jpsi$ decays.
\item $\psp$-suppression hypothesis, which attributes the small $Q$-value
to a suppression of branching fractions for $\psp$ decays.
\item Other hypotheses, {\it i.e.} those not included in the
above two categories.

\end{enumerate}

\subsection{Review of theoretical work on 
the $\RP$ puzzle}\label{sect_rvrpi}

\subsubsection{$\jpsi$-enhancement models}
In the earlier days of the $\rpi$ puzzle, it was noticed that
the decay of $1^{--}$ charmonium into $\rpi$ final state violates the
Hadronic Helicity Conservation (HHC) theorem 
(this is discussed below)~\cite{brodsky81},
and, so, such decays should be suppressed. Therefore,  there 
is some speculation on possible
mechanisms that enhance  $\jpsi \to \rpi$ 
decays.
Two schemes were proposed that follow this line of reasoning.

\bnum

 \item {$\jpsi$-glueball admixture scheme}

The idea that $\jpsi$ hadronic decays proceed
 via glueball emission was first  proposed by
Freund and Nambu~\cite{freund} (FN hereafter) soon after the discovery of
the $\jpsi$ to explain the breaking of Okubo-Zweig-Iizuka (OZI)
rule~\cite{ozi}. In this mechanism, the breaking results from the mixing
of the $\omega$, $\phi$, and $\jpsi$ mesons with an $SU(4)$-singlet
vector meson $\gb$. They predicted that this $\gb$  meson would lie in
the 1.4-1.8 GeV/$c^2$ mass range with a width 
that is greater than 40 MeV/$c^2$, and that it would decay copiously into 
$\RP$ and $\kstk$, while  decays into $\KKb$, $\EE$ and $\MM$ 
final states would be severly suppressed.
These authors presented several quantitative predictions for
experimental investigation. Two of them are:
$$ R_1 =
\frac{\Gamma_{\jpsi \to \RP}}{\Gamma_{\phi \to \RP}} =0.0115-0.087~,$$
$$ R_2 =
\frac{\Gamma_{\jpsi\to\KKb}}{\Gamma_{\jpsi\to\RP}} < 8\times 10^{-5}~.$$
With currently available data and using the three pion final state as
a substitute for $\RP$ in both $\phi$~\cite{ted_PDG2006} and
$\jpsi$~\cite{bes3pi,babar3pi} decays, we obtain 
$R_1 \approx 0.003$ for the first ratio, which is almost an
order-of-magnitude smaller than the prediction. For the second 
ratio, using the
PDG~\cite{ted_PDG2006} value for $\kk$ and a new experiment result for
$\kskl$~\cite{jpsikskl}, we estimate that $\BR(\jpsi\to\KKb)
\sim 10^{-4}$, which, taken together with the results for
$\RP$~\cite{bes3pi,babar3pi}, gives $R_2 \sim 10^{-2}$ which is much
larger than the prediction.

The first attempt to explain the $\RP$ puzzle in terms of a glueball
with mass near that of
the $\jpsi$ was proposed by Hou and Soni~\cite{houws} (HS hereafter).
They attributed the enhancement of $\jpsi \to \kstk$ and $\jpsi \to \RP$
decay modes to the mixing of the $\jpsi$ with a near-by
$\jpc=1^{--}$ vector 
gluonium, also designated by $\gb$. The differences between FN's and HS's 
pictures lie in the following aspects:
\begin{itemize}
\item Based on the potential model applied to glueballs, the mass of
a low-lying three-gluon state is estimated to be 
around 2.4 GeV/$c^2$~\cite{houwsa},
rather than the 1.4 to 1.8 GeV/$c^2$ value of Ref.~\cite{freund}.
\item Mixing of $\gb$ with $\psp$ is also taken into account; this was
ignored in the previous work.
\item Since the gauge coupling constant in QCD is momentum dependent, the
mixing parameter is taken to be a function of the invariant mass $q^2$,
and, therefore, 
decreases rather sharply with increasing $q^2$. Such a propagator
effect gives rise to  suppression of the
$\psp$ decay rates to $\RP$ and $\kstk$ channels 
relative to to those of the $\jpsi$.
\end{itemize}

By virtue of these assumptions, HS suggested a search for the vector
gluonium state in certain hadronic decays of the $\psp$, 
such as $\psp \to \pi\pi +X$,
$\eta +X$, $\etap +X$, where $X$ decays into VP final states~\cite{houws}.

Based on HS's idea, Brodsky, Lepage, and Tuan~\cite{brodsky87}
(BLT) refined the glueball hypothesis for the $\RP$ puzzle. They assumed
the general validity of the pQCD theorem that the total hadron helicity is
conserved in high-momentum-transfer exclusive processes, in which case
the decays to $\rpi$ and $\kstk$ are forbidden for both the $\jpsi$ and $\psp$.
This pQCD theorem is often referred to as the rule of Hadronic Helicity
Conservation (HHC)~\cite{brodsky81}, and is based on the assumption of
the short-range ``pointlike'' interactions among the constituent quarks.
For instance, $\jpsi(\ccbar) \to 3g$ has a short range
$\simeq 1/m_c$ and a correspondingly
short time-scale for the interaction.
Nevertheless, if the three gluons subsequently resonate, thereby forming
an intermediate gluonium state $\gb$ with a large transverse size
and a correspondingly extended time scale, HHC would cease
to be valid.  In essence, the HS mechanism takes over at this latter 
stage.

Decays to final states $h$ that proceed only through the intermediate
gluonium state are expected to satisfy the ratio
\beq
Q_h = \frac{\BR(\psp \to \EE)}{\BR(\jpsi \to \EE)}
      \frac{(M_{\jpsi}- M_{\gb})^2+\Gamma^2_{\gb}/4}
           {(M_{\psp}- M_{\gb})^2+\Gamma^2_{\gb}/4}~.
\label{brodskyratio}
\eeq
The $Q_h$ value
is small when the $\gb$ mass is close to the mass of the $\jpsi$.
Experimental limits at the time~\cite{pdg86,mk2,brodsky87} implied that
the $\gb$ mass was within 80 MeV/$c^2$ of the mass of the $\jpsi$ and its
total width was less than 160 MeV/$c^2$. Brodsky {\em et al.} recommended
direct searches for the $\gb$ by scan measurements of the $\EE \to VP$
cross section across the $\jpsi$ resonance.

A related work by Chan and Hou~\cite{chan} studied
the $\jpsi$ and vector glueball ${\cal O}$ 
mixing angle $\theta_{{\cal O}\psi}$ and amplitude 
$f_{{\cal O}\psi}$ in the framework of potential models of heavy 
quarks and constituent gluons.
They obtained $|\tan \theta_{{\cal O}\psi}|=0.015$ and
$f_{{\cal O}\psi}(m^2_{{\cal O}\psi})=0.008 \mbox{ GeV}^2$.

On the experimental side, BES searched for this hypothetical particle
in a $\rpi$ scan across the $\jpsi$ mass region in $\EE$ annihilations
as well as in the decays of $\psp \to \pi\pi \gb$, $\gb \to \rpi$, and
found no evidence for its existence~\cite{besvp,xyh}. The data constrained
the mass and width of the $\gb$ to be in the 
ranges $\mid M_{\gb}-M_{\jpsi}\mid < 80$
MeV/$c^2$ and 4 $< \Gamma_{\gb} < 50 $ MeV/$c^2$~\cite{harris}.
Although the absence of any distortion to the $\jpsi \to \rpi$ line-shape
in the BES energy scan
does not rule out $M_{\gb} \simeq M_{\jpsi}$, it puts a lower bound on
$\Gamma_{\gb}$. However, as indicated in Ref.~\cite{chen}, the
experimentally constrained mass is several hundred MeV/$c^2$ lower than
the mass of the lightest vector glueball calculated in lattice simulations
of QCD without dynamical quarks~\cite{peardon}.

Recent experimental data from  BES and CLEOc are
unfavorable to this glueball hypothesis. Among these
results is the observed
large branching fraction for the isospin-violating VP mode
$\psp \to \omega\pi^{0}$~\cite{besvp1,besvp2,cleocvp}. This
contradicts the assertion that the pattern of suppression
depends on the spin-parity of the final state mesons.
In addition, according to BLT's analysis, one obtains the 
prediction~\cite{suzukia}
$$ \frac{\BR(\jpsi \to \OP)}{\BR(\jpsi \to \rho^0 \pi^0)}<0.0037~$$
which is much smaller than the PDG06 value 0.08~\cite{ted_PDG2006}.
Another experimental result that is unfavorable to this hypothesis
is the suppression of $\psp$ decays into vector-tensor (VT)
final states~\cite{bes1vt,bes2vt}.
Since hadronic VT decays, unlike the VP decays, conserve HHC, some other
mechanism(s) must be responsible for this suppression.
Furthermore, it has been argued that
the $\gb$ may also explain the decay of $J/\psi$ into $\phi f_{0}$
(previously named $S^{\ast}$) but not to $\rho a_0(980)$
(previously named $\delta$), since the $\gb$ mixes with the $\phi$ and
enhances a mode that would otherwise be suppressed~\cite{brodsky87}.
However, the decay $\psp\to\phi f_{0}$~\cite{harris} is not suppressed
experimentally, which implies the absence of an anomalous
enhancement in $J/\psi\rightarrow\phi f_{0}$, thus contradicting
this explanation. Anselmino {\em et al.} extended the idea of
$\jpsi$-$\gb$ mixing to the case of $\eta_c\rightarrow VV$ and
$p\overline{p}$~\cite{anselm}. They suggested that the enhancement of
these decays can be attributed to the presence of a tri-gluonium
pseudoscalar state with a mass close to the $\eta_c$ mass. So far
there is no experimental evidence for such a state.


 \item {Intrinsic-charm-component scheme}

 Brodsky and Karliner (BK) put forth an explanation for the puzzle
based on the hypothesized
existence of intrinsic charm $|\qqb \ccb \rangle$ Fock
components in the light vector mesons~\cite{brokar}. They noticed
that quantum fluctuations in a QCD bound state wave function
will inevitably produce Fock states containing heavy quark pairs.
The intrinsic heavy quark pairs are multiconnected to the valence quarks
of the light hadrons, and the wave functions describing these
configurations will have maximal amplitude at minimal off-shellness and
minimal invariant mass. In the case of the $\rho$ meson, 
the light-cone Fock representation becomes:
$$ \rho^+ = \psi^{\rho}_{u\overline{d}} |u\overline{d} \rangle +
\psi^{\rho}_{u\overline{d}\ccb} |u\overline{d}\ccb \rangle + \cdots~.$$
Here we expect the wave function of the $\ccb$ quarks to be in an $S$-wave
configuration with no radial nodes, in order to minimize
the kinetic energy of the charm quarks and, thus, also minimize the total
invariant mass.

The presence of such a $|u\overline{d}\ccb \rangle$ Fock state in the 
$\rho$
would allow the $\jpsi \to \rpi$ decay to proceed through rearrangement of
the incoming and outgoing quark lines; in fact, the $|u\overline{d}\ccb
\rangle$ Fock state wave function has a good overlap with the radial and
spin $|\ccb \rangle$ and $|u\overline{d} \rangle$ wave functions of the
$\jpsi$ and pion. On the other hand, the overlap with the $\psp$ wouldd be
suppressed, since the radial wave function of the $n=2$ quarkonium state
is orthogonal to the nodeless $\ccb$ in the $|u\overline{d}\ccb \rangle$
state of the $\rpi$. Similarly, the $|u\overline{s}\ccb \rangle$ Fock
component of the $K^*$ favors the $\jpsi K$ configuration,
allowing the $\jpsi \to \kstk$ decay to also proceed by quark line
rearrangement, rather than by $\ccb$ annihilation.

These authors also suggested comparing branching fractions
for the $\etac$ and $\etacp$ decays
as clues to the importance of $\eta_c$ intrinsic charm excitations in
the wavefunctions of light hadrons.

\enum

\subsubsection{$\psp$-suppressed models}

The hypothesis of the existence of a glueball to explain the $\rpi$
was brought into question soon after it was proposed.
In addition, it is also pointed out~\cite{Chernyak99} that helicity
suppression is not a strong constraint at the charmonium energy scale.
In such a case, one comes naturally to the idea that it is not
$\jpsi \to \rpi$ that is enhanced, but rather $\psp \to \rpi$
that is suppressed. Seven explanations or models along these lines
were put forth.

\bnum

 \item {Sequential-fragmentation Model}

 Karl and Roberts suggested explaining the $\rpi$ puzzle based on
the mechanism of sequential quark pair creation~\cite{karl}. The
idea is that the quark-antiquark pairs are produced sequentially,
as a result the amplitude to produce two mesons in their ground
state is an oscillatory function of the total energy of the system.
They argue that the oscillatory fragmentation probability could have
a minimum near the mass of $\psp$, which provides a mechanism for
suppressing $\psp$ decays. Even though their evaluations could
generally accommodate the data for decays of $\jpsi$ and $\psp$ to
$\rpi$ and $K^{\ast}\overline{K}$, it runs into difficulties when
it is extrapolated to $\Upsilon$ decays. According to their calculation,
the oscillations of probability amplitude are damped out in the region
of the $\Upsilon$ resonances, so the $\rpi$ channel is present in
the decay of all $\Upsilon$, $\Upsilon^{\prime}$, $\Upsilon^{\prime\prime}$,
$\cdots$ resonances with a common rate. This leads to a prediction
$\Gamma (\Upsilon \to \rpi)=0.05$ keV, or, equivalently,
$\BR(\Upsilon \to \rpi)=9.4 \times 10^{-4}$, which is above
the current upper limit $\BR(\Upsilon \to \rpi)<2 \times
10^{-4}$~\cite{ted_PDG2006}. Moreover, their model has difficulty
explaining the large branching fraction for $\phi$ decays to
$\rpi$~\cite{ted_PDG2006} due to the fact that their fragmentation
probability tends to zero as the mass of the  $\rpi$ decaying system
approaches 1~GeV/$c^2$.

 In a further analysis~\cite{karla}, Karl and Tuan pointed out that if
a suppression is observed in three-meson channels the explanation
based on sequential pair creation would be undermined. Recently such
a suppressed channel, viz. $\phi KK$, was found by CLEOc~\cite{cleolhd1}.

 \item {Exponential-form-factor model}

Guided by the suppressed ratios of $\psp$ to $\jpsi$ decays to two-body
hadronic modes, Chaichian and T\"{o}rnqvist suggested~\cite{chaichian}
that the hadronic form factors fall exponentially as described by the
overlap of wave functions within a nonrelativistic quark model. This
behavior explains the drastically suppressed two-body decay rates of
the $\psp$ compared with those of the $\jpsi$. Recently reported
observations of a number of VP channels in
$\psp$ decays~\cite{besvp1,besvp2,cleocvp}
such as $\omega \etap$, $\phi \etap$, $\rho \etap$ indicate that
the predicted decay fractions are overestimated. Moreover,
the branching fraction for $\OP$~\cite{ted_PDG2006},
is well below the prediction of this model, which is $1.04\times 10^{-4}$.


 \item {Generalized hindered M1 transition model}

  A so-called generalized hindered M1 transition model was proposed by
Pinsky as a solution for the puzzle~\cite{pinsky}. It is argued that
because $\jpsi \to \gamma\eta$ is an allowed M1 transition while
$\psp \to \gamma\etap$ is hindered (in the nonrelativistic limit), using
the vector-dominance model to relate $\psp \to \gamma\etap$ to
$\psp \to \psi\etap$ one could find the coupling $G_{\psp \psi \etac}$
is much smaller than $G_{\psi \psi \etac}$, and then, by analogy, the
coupling $G_{\omegap \rho \pi}$ would be much smaller than
$G_{\omega \rho \pi}$. Here $G_{\omega \rho \pi}$ can be extracted from
data by virtue of the analysis using the vector-dominance model and a
standard parameterization of the OZI process~\cite{pinskya}. Then, 
assuming
$\psp\to\rpi$ proceeds via $\psp$-$\omegap$ mixing, while $\jpsi\to\rpi$
proceeds
via $\jpsi$-$\omega$ mixing, one finds that $\psp\to\rpi$ is much more
severely suppressed than $\jpsi\to\rpi$. A similar estimate could be
performed for $\kstk$ and other VP final states, and one 
could expect a reduced value for $Q$:
\beq
\frac{\BR(\psp\to VP)}{\BR(\psi\to VP)}=
1.47 \frac{\Gamma_{tot}(\psi)}{\Gamma_{tot}(\psp)}
     \left( \frac{G_{V^{\prime}VP}}{G_{VVP}} \right)^2
     \frac{F_{V^{\prime}}}{F_{V}}=0.06\%~,
\label{pinskyqval}
\eeq
where $F_{V^{\prime}}/F_{V}=0.3$, $G_{\omegap\rho\pi}/G_{\omega\rho\pi}
=0.066$ according to Ref.~\cite{pinsky}. This $Q$ is much smaller than
the present experimental results~\cite{besvp1,besvp2,cleocvp}.

  Moreover, in this model, the coupling $G_{\omegap\omega f_2}$ for
$\omegap\to \omega f_2$ should not be suppressed because, by analogy, the
coupling $G_{\psp\psi \chi_{c2}}$ is not small due to the fact that the E1
transition $\psp\to\gamma \chi_{c2}$ is not hindered~\cite{chao}.
Therefore, since it proceeds
via $\psp$-$\omegap$ mixing, the $\psp \to \omegap\to \omega f_2$
decay is not expected to be suppressed, which contradicts the BES
result~\cite{bes2vt}.

 \item {Higher-Fock-state scheme}

 Chen and Braaten (CB) proposed an explanation~\cite{chen} for
the $\rho\pi$ puzzle, where they argue that the decay
$J/\psi\rightarrow\rho\pi$ is dominated by a Fock state in which
the $c\overline{c}$ pair is in a color-octet $^{3}S_{1}$ state
which decays via $c\overline{c}\rightarrow q\overline{q}$, while
the suppression of this decay mode for the $\psp$ is attributed to
a dynamical effect due to the small energy gap between the mass of
the $\psp$ and the $\ddb$ threshold. Using the BES data on the
branching fractions into $\rho\pi$ and $K^{\ast}\overline{K}$ as
input, they predicted branching fractions for many other VP
decay modes of the $\psp$, as listed in Table~\ref{cb_preds}. For
these we see that most measured values fall within the ranges of the
predictions, but we also note that for the $\omega \pi$ mode, a
deviation far
from the prediction is evident. Here it should be noticed that the
values deduced in Table~\ref{cb_preds} are calculated based on the
strength of the measured branching fractions determined from earlier
experiments; the new measurements on the branching fractions for
$\rpi$ and $K^{\ast 0} \overline{K}^{0}+c.c.$ from
BES~\cite{besvp3_rhopi,besrpi1} and CLEOc~\cite{cleocvp} may have
impact on the corresponding evaluations.

\btbl[htb]
\centering
 \caption{\label{cb_preds}Predictions and
measurements for $Q_{VP}$ in unit of 1\% for all VP final states.
The value for $\rpi$ and $K^{\ast 0} \overline{K}^{0}+c.c.$ from
Ref.~\cite{zhuys} were used as input. The theoretical parameter
$x=0.64$ is from  Ref.~\cite{bramon} and the
experimental results come from
Refs.~\cite{besvp1,besvp2,besvp3_rhopi,yrhopi}.} \vskip 0.2 cm
\center \btbu{ccc} \hline \hline
      VP        & $x=0.64$    &    Exp.            \\ \hline \hline
   $\rpi$       & $0-0.25$    &  $ 0.13 \pm 0.03$  \\
$\KKSN +c.c.$   & $1.2-3.0$   &  $ 3.2  \pm 0.08$  \\
$\KKSC +c.c.$   & $0-0.36$    &  $ 0.59^{+0.27}_{-0.36}$ \\
$\omega \eta$   & $0-1.6$     &  $ < 2.0 $         \\
$\omega \etap$  & $12-55$     &  $ 19^{+15}_{-13}$ \\
$\phi \eta$     & $0.4-3.0$   &  $ 5.1  \pm 1.9 $  \\
$\phi \etap$    & $0.5-2.2$   &  $ 9.4  \pm 4.8 $  \\
$\rho \eta$     & $14-22$     &  $ 9.2^{+3.6}_{-3.3}$    \\
$\rho \etap$    & $12-20$     &  $ 17.8^{+15.9}_{-11.1}$ \\
$\omega \pi$    & $11-17$     &  $ 4.4^{+1.9}_{-1.6}$ \\ \hline \hline
\etbu
\etbl


 \item {Survival-chamonia-amplitude explanation}

 A model put forward by G\'{e}rard and Weyers entertains the assumption
that the three-gluon annihilation amplitude and the QED amplitude add
incoherently in all channels in $\jpsi$ decays into light hadrons, while
in the case of $\psp$ decays the dominant QCD annihilation amplitude is
not into three gluons, but into a specific configuration of five
gluons~\cite{gerard}.
More precisely, they suggest that the strong annihilation
of the $\psp$ into light hadrons is a two step process: in the first
step the $\psp$ goes into two gluons in a $0^{++}$ or $0^{-+}$ state and
an off-shell $h_c(3526)$; in the second step the off-shell $h_c$ annihilate
into three gluons to produce light hadrons. Their argument implies:
(a) to leading order there is no strong decay amplitude for the processes
$\psp \to \rpi$ and $\psp \to \kstk$; (b) the 12 \% rule should hold
for hadronic processes which take place via the QED amplitude only.
As far as the second implication is concerned, the present data give
different ratios between $\psp$ and $\jpsi$ decay for $\OP$ and $\pp$ final
states, both of which are electromagnetic processes. Here even when form
factor effects are taken into account~\cite{wymfofa}, the difference
between the two types of processes is still obvious. Besides 
providing a potential explanation
for the $\rpi$ puzzle, this model predicts sizable $\psp \to
(\pi^{+}\pi^{-} \mbox{ or } \eta)~h_{1}(1170)$ branching fractions.

  In a recent paper~\cite{artoisenet}, Artoisenet, G\'{e}rard and Weyers (AGW)
update and sharpen the above idea, which leads to a somewhat 
unconventional point
of view: all non-electromagnetic hadronic decays of the $\psp$ proceed
via a transition amplitude that contains a $\ccb$ pair. AGW provide
two patterns for these two-step decays, the first is
\beq
\psp \to (2\mbox{NP}g)+(3g)~.
\label{eq_23dk}
\eeq
The physics picture is as follows: the excited $\ccb$ pair in the $\psp$
does not annihilate directly. Instead, it spits out two non-perturbative
gluons $(2\mbox{NP}g)$ and survives in a lower $\ccb$ configuration
($1^{--}$ or $1^{-+}$) which then eventually annihilate into $3g$. The
decays $\psp \to (2\pi)\jpsi$ and $\psp \to \eta\jpsi$ follow this
pattern. The second pattern is
\beq
\psp \to (3\mbox{NP}g)+(2g)~,
\label{eq_32dk}
\eeq
where the lower $\ccb$ configuration ($0^{-+}$ or $0^{++}$) annihilates
into $2g$. The only on-shell channel for this type of decay is
$\psp \to (3\pi)\etac$, whose branching fraction is estimated as 
$(1\sim2)\%$
level. A recent measurement from CLEOc~\cite{cleorep} provides
an upper limit on this decay that
is one order of magnitude below this theoretical
prediction.
Furthermore, the substitution of one photon for one gluon in
Eqs.~\eref{eq_23dk} and \eref{eq_32dk} allows
\beq
\psp \to (2\mbox{NP}g)+(2g)+\gamma~.
\label{eq_22gdk}
\eeq
This pattern corresponds to on-shell radiative decays such as
$\psp \to (\pi^+ \pi^-) \etac \gamma$ and $\psp \to \eta \etac \gamma$,
which could be larger than the observed $\psp \to \etac \gamma$  mode.

In addition to the above predications, AGW also estimate
\begin{eqnarray}
\BR(\psp \to b_1 \eta)  &=& (1.3\pm 0.3)\times 10^{-3}~, \\
\BR(\psp \to h_1 \pi^0) &=& (1.9\pm 0.4)\times 10^{-3}~, \\
\BR(\jpsi\to b_1 \eta)  &\approx &
  \BR(\psp \to b_1 \eta) \approx 1 \%~.
\end{eqnarray}
All these can be tested by experiment.

 \item {Nonvalence component explanation}

Since the $\psp$ is a highly excited state and close to the $\DDb$
threshold, it is suggested~\cite{Chernyak} that unlike the
$\jpsi$, the $\psp$ may be an admixture of large nonvalence
components in its wave function. The so-called nonvalence
components includes those with an additional gluon or a light
quark-antiquark pair (as in Ref.~\cite{Chernyak}, a $\ccb g$
component and a quasi-molecular $\DDb$ state), which make $\psp$
decays quite distinctive from those of the $\jpsi$. The
nonvalence component of the $\jpsi$ is expected to be less
significant because it is the lowest state. In a following
paper~\cite{Chernyak99}, Chernyak uses this picture to explain the
$\rpi$ puzzle. He suggests that the valence and nonvalence strong
contributions interfere destructively in $\rpi$ channel and
consequently cancel to a large extent in the total $\psp \to \rpi$
strong amplitude, while the role of nonvalence contributions is
much less significant in $\jpsi \to \rpi$. From this viewpoint,
there is no deep reason for the experimentally observed very
strong suppression of $\psp \to \rpi$, rather this is 
the result of an accidental cancellation.

Chernyak also tries to use the above idea to explain qualitatively
other decay modes, such as $VT$, $AP$, $PP$, $VV$ and $\BBb$ decay.
However, such ideas remain pure speculation,
and no concrete calculations have been performed as of yet.

 \item {$S$-$D$ wave mixing scheme}

The $\pspp$ is generally considered to be a $D$-wave charmomium state. 
Although it is primarily $c\overline{c}(1^3D_1)$, its leptonic width 
indicates some mixing with $S$-wave states, mainly the nearby 
$\psi(2^3S_1)$.
This led Rosner to propose that the small $\rhopi$ branching fraction for
the $\psp$ is due to the cancellation of the $2S$ and $1D$ matrix 
elements.
In his scheme
\beq
\begin{array}{l}
\langle\rhopi |\psp\rangle =\langle \rhopi | 2^3 S_1 \rangle \cos \theta
                  -\langle \rhopi | 1^3 D_1 \rangle \sin \theta~, \\
\langle\rhopi |\pspp\rangle=\langle \rhopi | 2^3 S_1 \rangle \sin \theta
                  +\langle \rhopi | 1^3 D_1 \rangle \cos \theta~,
\end{array}
\label{sdmix}
\eeq
where $\theta$ is the mixing angle between pure $\psi(2^3 S_1)$ and
$\psi(1^3D_1)$ states~\cite{ktchao}
and is fitted from the leptonic widths of the $\pspp$
and the $\psp$ to be $(12 \pm 2)^{\circ}$~\cite{ycz_rosner}; this is
consistent with the coupled channel estimates~\cite{eichten2,heikkila} and
with the ratio of $\psp$ and $\pspp$ partial widths to
$\jpsipp$~\cite{part4_KY90}. If the mixing and coupling of the $\psp$ and
$\pspp$ lead to complete cancellation of $\psp \to \rpi$ decay
($\langle\rhopi |\psp\rangle= 0$), the missing $\rpi$ decay mode of
the $\psp$ should show up, instead, in the decays of the $\pspp$, 
enhanced by the factor $1/\sin^2 \theta$. A concrete estimate 
gives~\cite{ycz_rosner}
\beq
{\cal B}_{\pspp\rightarrow\rhopi}=(4.1\pm1.4)\times10^{-4}~~.
\label{brphi2}
\eeq

To test this scenario with data collected at the $\pspp$ in
$\EE$ experiments, it has been pointed
out~\cite{wymrpi2_rhopi,wymfsvp} that the continuum contribution
must be considered carefully. Specifically speaking, by Rosner's
estimation, the Born-order cross section for $\pspp \to \rpi$ is
\beq \sigma_{\pspp \to \rpi}^{Born} = (4.8 \pm 1.9) \mbox{~pb}~,
\label{ressct} \eeq which is comparable in magnitude to that of
the continuum cross section, viz. \beq \sigma^{Born}_{\EE \to
\rpi}=4.4~\hbox{pb}~. \label{consct} \eeq So, what is observed is
the coherent interference of these two amplitudes. Based on the
analysis of experimental data, it has been suggested that there be
an universal phase between strong and electromagnetic amplitudes
in charmonium decays. With this assumption, the strong decay
amplitude interferes with the continuum amplitude either maximumally
destructively, e.g. for $\rhopi$, $\omega \eta$ and $\KKSC$ or
maximumally constructively, e.g. for $\KKSN$. The destructive
interference case leads to the phenomena that the measured cross
section on top of the resonance could be smaller than that off the
resonance. Recent experimental results~\cite{cleolhd2,besrpi2} on
$\rhopi$, $\omega \eta$ and $\KKSC$ have demonstrated such an
interference pattern. This provides support to the prediction of
Eq.\eref{brphi2}.  However, to determine ${\cal
B}_{\pspp\rightarrow\rhopi}$ unambiguously, currently available 
experimental data
are insufficient; the $\pspp$ resonance must be
scanned~\cite{wymmcpspp}. So, a quantitative test of Rosner's
scenario remains as a task of the future experiments at \bes3.

In a subsequent study~\cite{psippkskl}, Wang, Yuan and Mo (WYM)
extend the $S$-$D$ wave-mixing scenario to $PP$ final state, and give
a constraint for $\pspp \to \kskl$ decay,
\beq
0.12 \pm 0.07 \le 10^{5}\times
\BR(\pspp \ra \kskl) \le 3.8 \pm 1.1~,
\label{bound}
\eeq
which is compatible with both the BESII result
$\BR(\pspp \ra \kskl) < 2.1 \times 10^{-4}$ at 90\%
C.L.~\cite{bespsppkskl} and the CLEOc result
$\BR(\pspp \ra \kskl) < 1.17 \times 10^{-5}$
at 90\% C.L.~\cite{cleoklks2}.
Extrapolating these ideas to all charmless decays~\cite{wymcmless},
WYM found that this scenario could accommodate large non-$\ddb$ decay
of the $\pspp$. Although recent experimental measurements from
CLEOc favor a nearly zero non-$\ddb$
cross section at $\pspp$~\cite{cleondd},
the large errors prevent them from ruling out the existence of
non-$\ddb$ decays with a branching fraction at the 10\% level.

\enum

\subsubsection{Other Explanations}

Besides the aforementioned models, more speculations involving the
$\rpi$ puzzle are described below.

\bnum

 \item {Final-state-interaction scheme}

 Li, Bugg and Zou~\cite{lixq} (LBZ) pointed out that final state
interactions (FSI) in $\jpsi$ and $\psp$ decays give rise to effects
that are of the same order as the tree level amplitudes.  They argue
that $\jpsi \to \RP$ is strongly enhanced by the $a_2 \rho$ loop diagram,
while the direct tree-production for $\RP$ may be suppressed by the
HHC mechanism~\cite{brodsky81}. The contribution
of the $a_1 \rho$ loop diagram is much smaller than that of $a_2 \rho$
loop for the $\jpsi \to \RP$, but they have similar strength for the
$\psp \to \RP$ and may cancel each other by interfering destructively.
A similar apparent suppression for $\psp \to \kstk$ and $f_2 \omega$
may also be explained by the $K^* \overline{K}^*_{2,1}$ and $b_1 \pi$
loop, respectively. Therefore, LBZ expect that FSI might provide a
coherent explanation for all the observed suppressed modes of $\psp$
decays. However, as remarked in Ref.~\cite{tuan}, this interference model
appears to have more assumptions that predictions. The only qualitative
prediction by LBZ is that the $a_{1}\rho$ and
$K_{1}^{\ast}\overline{K}^{\ast}$ production rates should be
large for the $\psp$.
So far, no  measurements of these modes have been reported.

 \item {Large phase scheme}

Suzuki gave another FSI-based explanation for $\jpsi$
decays~\cite{suzukia}. He performed a detailed amplitude analysis
for $\jpsi \to \VP$ decay to test whether or not the
short-distance FSI dominates over the long-distance FSI in
$\jpsi$ decay. His result indicates that there is a large phase
between the three-gluon and one-photon amplitudes. Since the large
phase cannot be produced by a perturbative QCD interaction,
its source must be in the long-distance part of the strong
interaction, namely, rescattering among hadrons in their inelastic
energy region. Suzuki then performed a similar analysis for
$\jpsi \to \PP$ decay, and obtained a similar conclusion 
about the existence of a
large phase~\cite{suzukib}. His analysis also shows that the
exclusive decay ratio at the $\jpsi$ is in line with that of the
inclusive decay. This fact led him to conclude that the origin of
the relative suppression of $\psp\to\VP$ to $\jpsi\to\VP$ is not
in the $\jpsi$ but in the $\psp$.


As more $\psp$ decays were analyzed, the
experimental data at first seemed to favor a phase
close to $180^\circ$~\cite{suzukic},
contrary to the expectation that the decays of 
the $\jpsi$ and $\psp$
should not be very different.
However, it was
pointed out by Wang {\em et al.} that 
the previously published data did not take the continuum one-photon
process into account. Their reanalysis of the experimental data
showed that a phase with a value around $-90^{\circ}$ could fit
$\psp\to\VP$ data~\cite{wymfsvp} and $\pm 90^{\circ}$ could fit
$\psp\to\PP$ data~\cite{wymfspp}. The latter is confirmed by more
recent results from CLEOc~\cite{cleocpp}. Furthermore, this large
phase also shows up in the OZI-suppressed decay modes of the $\pspp$.
In many decays modes of the $\pspp$, the strong decay amplitudes
have comparable strength to the non-resonance continuum amplitude,
and a large phase around $-90^\circ$ leads to destructive or
constructive interference. In the case of destructive interference,
the observed cross sections for some modes are smaller at the peak of the
$\pspp$ than the cross section that is measured
off-resonance~\cite{wymrpi2_rhopi}. This is demonstrated by
data from CLEOc~\cite{cleolhd2}.


 \item {Mass reduction explanation}

  In a study~\cite{majp} of radiative decays of $1^{--}$ quarkonium
into $\eta$ and $\etap$, Ma presented a QCD-factorization approach,
with which he obtained theoretical predictions that are consistent with 
CLEOc measurements. The largest possible uncertainties in the analysis are 
from relativistic corrections involving the value of the charm quark mass. 
Ma argued that the effect of these uncertainties can be reduced by
using quarkonium masses instead of the quark mass. As an example
of such a reduction, he provides a modified version of the original
12\% rule
$$ Q_{\rpi} = \frac{\BR(\jpsi \to \rpi)}{\BR(\psp \to \rpi)}
            = \frac{M^8_{\jpsi}}{M^8_{\psp}}
          \frac{\BR(\jpsi \to \EE)}{\BR(\psp \to \EE)} $$
$$ = (3.6 \pm 0.6) \%~. $$
However, this reduced value is still
much larger than the experimental result
given above in Table~\ref{cb_preds}.

 \item {Vector-meson-mixing model}\label{sst_vmmm}

With the intention of providing a comprehensive description of $\jpsi$ 
two-body decays, Clavelli and Intemann (CI) proposed a 
vector-meson-mixing model in
which the vector mesons ($\rho,~\omega~,\phi~,\jpsi$) are regarded as
being admixtures of light-quark-antiquark states and 
charmed-quark-antiquark
states~\cite{clavelli}. The coupling of the $\jpsi$ to any state of light
quarks is then related to the corresponding coupling of the $\rho$, $\omega$,
and $\phi$ to the same state. With a few experiment inputs to determine
the mixing parameters, CI calculate VP, PP, and BB decay rates for the
$\jpsi$ as a function of the pseudoscalar mixing angle. Most of the
predictions agree with experimental results at the order of magnitude
level, but discrepancies are obvious for some channels, such as 
the $\kskl$ final
state~\cite{jpsikskl}. CI also extended their model to hadronic decays
of the $\psp$. However, their evaluations
for $\BR(\jpsi \to \OP)=3\times 10^{-5}$ and
$\BR(\psp \to \OP)=3\times 10^{-3}$ contradict the present
experimental values $(4.5 \pm 0.5) \times 10^{-4}$ and
$(2.1 \pm 0.6)\times 10^{-4}$~\cite{ted_PDG2006}, respectively.

  Starting from an effective Lagrangian wherein nonet-symmetry breaking
and pseudoscalar-meson mixing can be studied, Haber and Perrier
parametrized the decay modes of
$\jpsi \to PP$ (also for $\jpsi \to VV$ or $\etac \to VP$),
$\jpsi \to VP$ (also for $\jpsi \to VT$ or $\etac \to VV$),
$\jpsi \to PPP$ (also for $\jpsi \to VVP$ or $\etac \to PPV$), and
$\etac \to PPP$ (also for $\jpsi \to PPV$ or $\etac \to VVP$)~\cite{haber}.
Experimental data were used to determine their phenomenological
parameters. In a follow-on work, Seiden, Sadrozinski and Haber took
the doubly Okubo-Zweig-Iizuka suppression (DOZI) effect into consideration,
and presented a more general parameterization of amplitudes for
$\jpsi \to PP$ decays~\cite{seiden}. With this form, one could easily
derive the relative decay strengths between different final states.
However, for the most general cases of symmetry breaking, 
too many parameters are introduced for
a general analysis to be useful. In order to reduce the number of
new parameters considerably and make the analysis manageable,
only special cases have to be considered.

\btbl[htb]
 \centering
 \caption{\label{fk_preds}Comparison of predictions~\cite{feldman} and
measurements~\cite{ted_PDG2006} for the branching ratios ($10^{-5}$) for
$\psp$ decays. The upper limits are presented at 90\% C.L.}
\vskip 0.2 cm 
\btbu{ccc} \hline \hline
      VP       & Prediction  &    Measurement        \\ \hline\hline
   $\rpi$      & $1.3$       &  $ 3.2 \pm 1.2 $      \\
$\KKSN +c.c.$  & $5.1$       &  $10.9 \pm 2.0 $      \\
$\KKSC +c.c.$  & $1.3$       &  $ 1.7^{+0.8}_{-0.7}$ \\
$\omega \eta$  & $1.2$       &  $<1.1$               \\
$\omega \etap$ & $6.3$       &  $ 3.2^{+2.5}_{-2.1}$ \\
$\phi \eta$    & $1.6$       &  $ 2.8^{+1.0}_{-0.8}$ \\
$\phi \etap$   & $4.6$       &  $ 3.1 \pm 1.6 $      \\
$\rho \eta$    & $2.1$       &  $ 2.2 \pm 0.6 $      \\
$\rho \etap$   & $1.2$       &  $1.9^{+1.7}_{-1.6}$  \\
$\omega \piz$  & $3.8$       &  $ 2.1 \pm 0.6 $      \\
$\phi \piz$    & $0.01$      &  $<0.40$              \\
\hline\hline
\etbu
\etbl

  A similar parameterization of mixing-induced 
strong interaction mechanisms was proposed by Feldmann
and Kroll (FK)~\cite{feldman} for the
hadron-helicity non-conserving $\jpsi$ and $\psp$ decays, but with a
different interpretation from those put forth in
Refs.~\cite{chen,bramon,seiden,tuan}.
FK assume that, with a small probability,
charmonium states have Fock components built from light quarks only.
Through these Fock components the charmonium state decays by
a soft mechanism that is modeled by $\jpsi$-$\omega$-$\phi$ mixing and
the subsequent $\omega$ (or $\phi$) decays into the $VP$ final state.
In absence of a leading-twist perturbative QCD
contribution, the dominant mechanism is supplemented
by electromagnetic  and DOZI-violating contributions.
FK argue that this mechanism can probe the charmonium wave function
at all quark-antiquark separations and thereby experiences the
differences between the $1S$ and $2S$ radial wave functions.
The node in the latter is supposed to lead
to a strong suppression mechanism for $\psp$ decays.
With a few parameters adjusted to the experimental data,
FK obtain a numerical
description of the branching fractions for many VP decay modes of the
$\jpsi$ and $\psp$, and these agree with the measured branching fractions
at the order of magnitude level, as shown in Table~\ref{fk_preds}.
Moreover, FK have extended their mixing
approach to $\etac\to VV$ decays and obtain a reasonable description
of the branching fractions for these decays, with the $\etacp \to VV$
decays expected to be strongly suppressed.

\enum

%
%

\subsection{Summary}

Here we have presented a general review om the subject of the $\RP$ 
puzzle.
Although there is still no satisfactory explanation for all existing
experimental results, some enlightening ideas have been put forth.

As we know, physics in the charm threshold region is in the boundary
domain between perturbative and nonperturbative QCD.
Recently  observed
hadronic decays of charmonium may give new challenges to the present
theoretical understanding of the extant decay mechanisms.
Many of the new charmonium states observed by Belle and BaBar,
which have difficulty being accommodated by potential models,
have led to new theoretical
speculations about charmonium spectroscopy and novel production
mechanisms~\cite{Swanson}.

Experimentally, the expected large data sample from CLEOc in the near future,
and even larger data samples from \bes3 will open 
a new era for charmonium dynamics study, even though we may not obtain
a throughly revolutionary theory, we may acquire a more profound
understanding of the existing theory.  At the same time we can
hope for some new enlightenment on the $\rpi$ puzzle.



\section[Open-flavor decays]
{Open-flavor decays\footnote{T. Barnes}}
\label{sec:ddbar_barnes}
\def\be{\begin{equation}}
\def\ee{\end{equation}}
\def\bd{\begin{displaymath}}
\def\ed{\end{displaymath}}
\def\ba{\begin{eqnarray}}
\def\ea{\end{eqnarray}}
\def\C{\rm C}
\def\D{\rm D}
\def\F{\rm F}
\def\I{\rm I}
\def\J{\rm J}
\def\L{\rm L}
\def\M{\rm M}
\def\P{\rm P}
\def\S{\rm S}
\def\T{\rm T}
\def\X{\rm X}



It is well known from the phenomenology of even the lightest
hadrons that the dominant strong decays of mesons are those that
do not involve the annihilation of the valence $q\bar q$ pair. It
was this observation, in the context of the decays of the
$\phi(1020)$, that led Zweig to the suggestion of strange quarks.
(Specifically, he suggested that the dominance of the $\phi(1020)$
decay mode $KK$ over $\rho\pi$ could be understood if the
$\phi(1020)$ contained a valence $s\bar s$ pair that could not
easily annihilate, which is now known as the OZI rule. This
assumption would explain why the $KK$ transition ($s\bar s\to
(s\bar n)(n\bar s)$, where $n=u,d$) dominates the transition to
$\rho\pi$ ($s\bar s\to (n\bar n')(n'\bar n))$, although $KK$ has
very little phase space.

Although these open-flavor decay modes are usually the dominant
strong decay amplitudes when allowed by phase space, they remain
surprisingly poorly understood. The approach normally used by
theorists to model these decays is to assume pair production of a
$q\bar q$ pair from the vacuum with $0^{++}$ (vacuum) quantum
numbers, with a dimensionless amplitude $\gamma$ that is
independent of the initial state and the flavor of the $q\bar q$
pair. This is now known as the ${}^3$P$_0$ model, and originally
suggested by Micu in 1968 \cite{ted_Micu:1968mk} (pre-QCD), and
was subsequently developed for explicit quark model wavefunctions
by the Orsay group of LeYaouanc {\it et al.}
\cite{ted_LeYaouanc:1972ae}. The well-known flux-tube decay model
of Kokoski and Isgur \cite{ted_Kokoski:1985is} is actually a
variant of this model with some spatial modulation assumed for the
$q\bar q$ pair production amplitude. A description in terms of
Feynman diagrams and a comparison with OGE and other pair
production amplitudes was given by Ackleh {\it et al.}
\cite{ted_Ackleh:1996yt}. This ${}^3$P$_0$ model has now been
applied very extensively to most sectors of hadron Hilbert space,
including charmonia
\cite{ted_LeYaouanc:1977ux,ted_LeYaouanc:1977gm,ted_Barnes:2003vb,ted_Barnes:2005pb}.
Ref.\cite{ted_Barnes:2005pb} gives numerical predictions for the
decay amplitudes and strong branching fractions of all charmonia
expected to the mass of the $\psi(4415)$. These predictions are
especially interesting because the ${}^3$P$_0$ model is only a
simple phenomenological description, and may be inaccurate in
practice. Since the many theoretical predictions of the preferred
strong decay modes of many hadrons, including light and strange
mesons
\cite{ted_Godfrey:1985xj,ted_Barnes:1996ff,ted_Barnes:2002mu},
baryons, including suggestions for finding the ``missing" baryons
\cite{ted_Capstick:1986bm,ted_Capstick:1993kb}, and hybrid mesons
\cite{ted_Isgur:1985vy,ted_Close:1994hc} all rely on this specific
decay model, it is evidently very important to test its accuracy
in describing strong decays in an experimentally relatively clear
sector such as charmonium.

Of course one can easily construct other models for strong decay
amplitudes, and it is interesting that in their early studies of
charmonium the Cornell group \cite{ted_Eichten:1978tg} used such
an alternative model. In particular they assumed that these strong
decays were a result of the linear confining interaction, which
gave rise to $q\bar q$ pair production from the initial $c\bar c$
system with a $\gamma_0 \otimes \gamma_0$ nonlocal interaction and
a linear $br$ kernel between the vertices. (This choice of
$\gamma_0 \otimes \gamma_0$ was motivated by the belief current in
the 1970s that confinement acted as a timelike vector interaction
rather than a Lorentz scalar, which is the more usual assumption
at present.) Several recent charmonium papers also assume this
timelike vector decay model
\cite{ted_Eichten:2002qv,ted_Eichten:2004uh}. An alternative model
with an $\I \otimes \I$ linear scalar confining interaction was
developed and applied to light quarks by Ackleh {\it et al.}
\cite{ted_Ackleh:1996yt}, but has not yet been applied to
charmonium.

It will be very interesting to use charmonium decays at \bes3 to
test these strong decay models. The most sensitive tests involve
the determination of amplitude {\it ratios}, which require studies
of open-flavor decays into final states that have more than one
amplitude. These relative amplitude phases are typically very
sensitive to the assumed quantum numbers of the $q\bar q$ pair
produced during the decay, whereas simple branching fractions and
partial widths are affected by common features such as the
available phase space.

\begin{table}[htbp]
\caption{Theoretical (${}^3$P$_0$ model) open-flavor strong decay 
amplitudes and
 widths
of the four $1^{--}$ charmonium states above 3.73 GeV most easily 
accessible at \bes3
(abstracted from Ref.\cite{ted_Barnes:2005pb}). Multiamplitude decay 
channels ar
e indicated
by boldface.}
\begin{center}
\begin{tabular}{lllll}
\hline
Meson
& State \
& Mode \ \
& $\Gamma_{thy}$\ \  [$\Gamma_{expt}$] (MeV)
& Amps. (GeV$^{-1/2}$)
\\
\hline
\hline
$\psi(3770)$
& $1^3\D_1$
& DD
& 43\  \ [$ 23.0 \pm 2.7$]
& $^1$P$_1 = +0.1668$
\\
\hline
$\psi(4040)$
& $3^3\S_1$
& $\D \D$
& 0.1
& $^1$P$_1 = -0.0052$
\\
&
& $\D  \D^*$
& 33
& $^3$P$_1 = -0.0954$
\\
&
& {\bf D$^*$D$^*$}
& 33
& $^1$P$_1 = +0.0338$
\\
&
&
&
& $^5$P$_1 = -0.1510$
\\
&
&
&
& $^5$F$_1 =  0 $
\\
&
& $\D_s  \D_s$
& 7.8
& $^1$P$_1 = +0.0518$
\\
&
& {\it total}
& 74\ \ $[80 \pm 10]$
&
\\
\hline
$\psi(4160)$
& $2^3{\D}_1$
& $\D \D$
& 16
& $^1\P_1 =  -0.0522$
\\
&
& $\D \D^*$
& 0.4
& $^3\P_1 =  +0.0085$
\\
&
& {\bf D$^*$D$^*$}
& 35
& $^1\P_1 = +0.0489$
\\
&
&
&
& $^5\P_1 = -0.0219$
\\
&
&
&
& $^5\F_1 = -0.0845$
\\
&
& $\D_s \D_s$
& 8.0
& $^1\P_1 = -0.0427$
\\
&
& $\D_s \D_s^*$
& 14
& $^3\P_1 = +0.0733$
\\
&
& {\it total}
& 74\  \ [$ 103 \pm 8$]
&
\\
\hline
$\psi(4415)$
& $4^3\S_1$
& $\D \D$
& 0.4
& $^1$P$_1=+0.0066$
\\
&
& $\D \D^*$
& 2.3
&  $^3$P$_1= +0.0177$
\\
&
& {\bf D$^*$D$^*$}
& 16
& $^1$P$_1= -0.0109$
\\
&
&
&
& $^5$P$_1= +0.0487$
\\
&
&
&
& $^5$F$_1= 0$
\\
&
& {\bf DD$_1$}
& 31
& $^3$S$_1 = 0$
\\
&
&
&
& $^3$D$_1 = +0.0933$
\\
&
& {\bf DD$_1'$}
& 1.0
& $^3$S$_1 = +0.0168$
\\
&
&
&
& $^3$D$_1 = 0$
\\
&
& $\D \D_2^*$
& 23
& $^5$D$_1= -0.0881$
\\
&
& {\bf D$^*$D$_0^*$}
& 0.0
& $^3$S$_1 = -8.7\cdot 10^{-4}$
\\
&
&
& 0
& $^3$D$_1 = 0$
\\
&
& $\D_s\D_s$
& 1.3
& $^1$P$_1= -0.0135$
\\
&
& $\D_s \D_s^*$
& 2.6
& $^3$P$_1= +0.0212$
\\
&
& {\bf D$_s^*$D$_s^*$}
& 0.7
& $^1$P$_1 = +0.0027$
\\
&
&
&
& $^5$P$_1 = -0.0119$
\\
&
&
&
& $^5$F$_1 = 0$
\\
&
& {\it total}
& 78\ \ $[62 \pm 20]$
&
\\\hline
\end{tabular}
\end{center}
\label{Table_strong_Vector_decays}
\end{table}

For the decays of the most easily
accessible $1^{--}$ charmonia, studies of relative decay amplitude
phases requires final states consisting of a pair of open-charm
vectors, the ``VV modes", since the PsPs and PsV modes involve
single amplitudes (respectively ${}^1$P$_1$ and ${}^3$P$_1$). Thus
PsPs and PsV final states do not allow measurements of the
relative phases of strong decay amplitudes, whereas in the decay
$1^{--}\to VV$ there are three amplitudes present, ${}^1$P$_1$,
${}^5$P$_1$ and ${}^5$F$_1$. The ${}^3$P$_0$ model predicts very
characteristic ratios for these decay amplitudes, depending on
whether the initial $1^{--}$ $c\bar{c}$ is an S-wave (${}^3$S$_1$) or
a D-wave (${}^3$D$_1$) state.  For an S-wave one finds
${}^5$P$_1$/${}^1$P$_1  = - 2\sqrt{5}$ and ${}^5$F$_1 = 0$,
whereas for a D-wave one finds ${}^5$P$_1$/${}^1$P$_1  = -
1/\sqrt{5}$ and ${}^5$F$_1 \neq 0$ (both sets of results are
independent of the radial wavefunction). Results from a recent 
calculation~\cite{ted_Barnes:2005pb} are listed in 
Table~\ref{Table_strong_Vector_decays}. There is some evidence
that the physical states $\psi(4040)$ and $\psi(4160)$ may be
strongly mixed linear combinations of S- and D-wave basis states,
which will instead give a weighted set of decay amplitudes. 

The highest-mass $c\bar c$ state known at present is the
$1^{--}$ $\psi(4415)$, which is usually given a 4$^3$S$_1$
assignment. 
Calculations of the decay branching fractions of a
4$^3$S$_1$  $c\bar c$ $\psi(4415)$ in the $^3$P$_0$ model
\cite{ted_Barnes:2005pb} predict that the largest mode should be
the unusual DD$_1$ (the narrow D$_1$), and in pure D-wave rather
than S-wave! It would clearly be a very interesting test of strong
decay models to measure the strong decay amplitudes and branching
fractions of this state. 
There is also an ``industrial" application of the
$\psi(4415)$~\cite{ted_Barnes:2005pb,ted_Guo:2005cs}; by running
on the high mass tail of this resonance, one can expect a
relatively large branching fraction into the enigmatic
D$_{s0}(2317)$, which is otherwise difficult to produce with
useful statistics. A study of interesting decays such as the
radiative branching fraction of the D$_{s0}(2317)$ into $\gamma
$D$_s^*$ could then be carried out at \bes3; this would be valuable
in determining the relative size of the $c\bar s$ and DK
components of the D$_{s0}(2317)$.

Unfortunately, the corresponding predictions for the open-flavor
decay amplitudes of these charmonium states using other strong
decay models, such as the timelike vector model assumed by the
Cornell group \cite{ted_Eichten:1978tg}, have not yet been
published. It will be a very important exercise for theorists to
evaluate the strong decay predictions of this model for comparison
both with future \bes3 data and with the predictions of the
${}^3$P$_0$ model shown here.

An experimental study of the decays of these vector states
should begin with a simple scan in energy of the exclusive cross
sections for $e^+e^-$ into the various open-charm final states, to
establish whether the individual resonances contributions can be
clearly separated.  Such measurements of exclusive
open-charm cross sections have recently been carried out by
Belle~\cite{belle_dd:2008,belle_ddpi:2007,belle_dstrdstr:2006}.

\section[$\psi(3770)$ non-$D\overline{D}$ decays]
{$\psi(3770)$ non-$D\overline{D}$ decays\footnote{G. Rong, D.H.
Zhang, and H.L. Ma}}
\label{sec:non_ddbar}

\subsection{Introduction}

Potential Models based on QCD predict charmonium and
charmed meson properties quite well.  Until now, most
of the states predicted by the charmonium model have been
observed, and many of their decay channels have been observed and
their branching fractions have been well measured.

Based on the conventional charmonium potential model, the
$\psi(3770)$ resonance is identified as a mixture of the $1^3D_1$
and $2^3S_1$ angular momentum states, and is expected to decay
into the OZI-allowed $D\bar D$ ($D^0 \bar D^0$ and $D^+D^-$)
final states with
a branching fraction $B(\psi(3770) \rightarrow D\bar D)\ge 97\%$.
However, there is a long-standing puzzle in the understanding of
the $\psi(3770)$ production and decays.  Previously published
results~\cite{rg_rong_zhang_chen}  indicated about a $38\%$
discrepancy between the measured $D\bar
D$ production cross section $\sigma_{D\bar D}^{\rm obs}$ and the
observed $\psi(3770)$ production cross section
$\sigma_{\psi(3770)}^{\rm obs}$. Recently, the BESII Collaboration
measured the total branching fraction for ${\rm non}-D \bar D$
decays of the $\psi(3770)$ to be $B(\psi(3770) \rightarrow {\rm non}-D
\bar D)=(16.4\pm 7.3 \pm 4.2)\%$~\cite{rg_PhysRevLett_97_121801},
based on measurements of the continuum light hadron
cross section below the $D\bar D$ threshold, the observed cross
section for $D\bar D$ production and the observed cross section
for inclusive hadron production at the peak of $\psi(3770)$, and
$B(\psi(3770) \rightarrow {\rm non}-D \bar D)=(16.4\pm 7.3 \pm
4.2)\%$ ~\cite{rg_PhysLettB_141_145} obtained from an analysis of
the line-shapes of  inclusive hadron and $D\bar D$ ( $D^0\bar
D^0$ and $D^+ D^-$) production. These results indicate that either,
contrary to what is generally expected, the $\psi(3770)$ has a
substantial decay rate into non$-D\bar D$ final states, or the
measured cross sections for $D\bar D$ and $\psi(3770)$ production
suffer from large systematic shifts, even in the latest
BESII measurements. Another possibility is that there may be some other 
effect
or new phenomena that is responsible for this discrepancy.

To clarify this situation one needs to measure more precisely
$\sigma_{D\bar D}^{\rm obs}$, $\sigma_{\psi(3770)}^{\rm obs}$, the   
parameters of $\psi(3770)$ and $\psi(2S)$ resonances,
the branching fractions for $\psi(3770)\rightarrow D^0\bar
D^0,D^+D^-$, ${\rm non}-D\bar D$, and extensively search
for and study  exclusive non-$D\bar D$ decay modes of the
$\psi(3770)$
with a high statistics data sample. These
can be done with the \bes3 detector at the BEPC-II collider. In this
section we summarize the status of the experimental studies of
non-$D\bar D$ decays of $\psi(3770)$ resonance and propose an
extensive study the light hadron decays of the $\psi(3770)$ with
the \bes3 detector at the BEPC-II collider.
 
\subsection{S-D mixing and mixing angle $\theta_{\rm mix}$}

The charmonium model predicts the leptonic width of
the $1^3D_1$ $c\bar c$ state to be 
70~eV~\cite{rg_prd21_y1980_203_eichten},
while early experiments
measured a leptonic width of about 250~eV. To explain the large 
leptonic width, the $\psi(3770)$ is assumed to be a mixture of the
$1^3D_1$ and  $2^3S_1$ states plus some other
$1^{--}$ states near the $D\bar D$ thresholds, caused by some
dynamic mechanism~\cite{rg_eichten_80,rg_moxhay_rosner}. The
simplest  scheme, where the  $1^3D_1$ mixes with only the $2^3S_1$ state,
which is the dominant component of the mixed part of the $1^3D_1$ state, 
is characterized by the mixing angle $\theta_{\rm mix}$, and the
corresponding physical states $\psi(3686)$ and $\psi(3770)$ are
described as,
\begin{equation}
|\psi(3770)> = |1^3D_1>{\rm cos}\theta_{\rm mix}+|2^3S_1>{\rm
sin}\theta_{\rm mix}, \label{eq1}
\end{equation}
\begin{equation}
|\psi(3686)> =-|1^3D_1>{\rm sin}\theta_{\rm mix}+|2^3S_1>{\rm 
cos}\theta_{\rm mix}. \label{eq2}
\end{equation}
\noindent
This is enough to settle some important issues in
charmonium physics. With this simple model, the
leptonic widths of $\psi(3770)$ and $\psi(3686)$ resonances are
then calculated as a function of $\theta_{\rm mix}$
to be~\cite{ycz_rosner,
rg_ypkuang_prd65_y2002_p094024}
\begin{equation}
\Gamma(\psi(3770)\rightarrow e^+e^-) = \frac{4 \alpha^2
e_c^2}{M^2_{\psi(3770)}} \left |{\rm sin}\theta_{\rm mix}   
R_{2S}(0) + \frac{5}{2 \sqrt{2} m_c^2} {\rm cos}\theta_{\rm
mix}R^{''}_{1D}(0)\right |^2, \label{eq3}
\end{equation}
\begin{equation}
\Gamma(\psi(3686)\rightarrow e^+e^-) = \frac{4 \alpha^2
e_c^2}{M^2_{\psi(3686)}} \left |{\rm cos}\theta_{\rm mix}
R_{2S}(0) - \frac{5}{2 \sqrt{2} m_c^2} {\rm sin}\theta_{\rm  
mix}R^{''}_{1D}(0)\right |^2, \label{eq4}
\end{equation}
\noindent
where $e_c=2/3$, $R_{2S}(0)=\sqrt{4\pi}\Psi_{2S}(0)$ is the radial $2S$ 
wave
function at $r=0$, and $R^{''}_{1D}(0)$ is the second derivative of the
radial $1D$ wave function at the origin.
Experimentally, using the measured values of $\Gamma(\psi(3686)\rightarrow 
e^+e^-)$
and $\Gamma(\psi(3770)\rightarrow e^+e^-)$ one can determine the mixing 
angle
$\theta_{\rm mix}$.

Taking the ratio of  Eq.~\ref{eq3} and Eq.~\ref{eq4}, we
derive the relation
\begin{equation}
 R_{\psi(3770)/\psi(3686)}=
\frac {M^2_{\psi(3773)}\Gamma(\psi(3773)\rightarrow e^+e^-)}
{M^2_{\psi(3686)}\Gamma(\psi(3686)\rightarrow e^+e^-)}= \left
|\frac{0.734 {\rm sin}\theta_{\rm mix}+0.095 {\rm cos}\theta_{\rm
mix}} {0.734 {\rm cos}\theta_{\rm mix}-0.095 {\rm sin}\theta_{\rm
mix} } \right |^2. \label{eq5}
\end{equation}
\noindent
Using the parameters of $\psi(3770)$
and $\psi(3686)$ resonances recently measured by the   
BESII Collaboration~\cite{rg_prl97_121801_y2006}, we can determine the 
ratio of the
partial leptonic widths of the $\psi(3770)$ and the $\psi(3686)$
\begin{equation}
R^{BES}_{\psi(3770)/\psi(3686)}=0.105\pm0.015. \label{eq6}
\end{equation}
\noindent From the relations given in the Eqs.~\ref{eq5}
and~\ref{eq6}, we can extract a value for the
mixing angle, $\theta_{\rm mix}$. The relation of
$R_{\psi(3770)/\psi(3686)}$ to $\theta_{\rm mix}$ is plotted as
the parabolic curve in Fig.~\ref{xsct}, where the horizontal lines
give the measured value of $R_{\psi(3770)/\psi(3686)}$ and the
$\pm 1 \sigma$ interval of the value; the vertical lines give $\pm
1 \sigma$ intervals of the measured values of the mixing angle  
$\theta_{\rm mix}$. There are two solutions as shown in the
figure. One is $\theta_{\rm mix}=(10.6\pm 1.3)^o$ and the another
is $\theta_{\rm mix}=(-25.3\pm 1.3)^o$. However,  the solution
$\theta_{\rm mix}=(10.6\pm 1.3)^o$ is favored by the relative
decay rates for $\psi(3770) \rightarrow J/\psi \pi^+\pi^-$ and
$\psi(3686) \rightarrow J/\psi \pi^+\pi^-$. If the mixing angle is
$\theta_{\rm mix}=(-25.3\pm 1.3)^o$, the partial width for
$\psi(3770) \rightarrow J/\psi \pi^+\pi^-$ would be larger than
that for $\psi(3686) \rightarrow J/\psi \pi^+\pi^-$, which is
in conflict with measurements, as discussed below.

Measurements of $\theta_{\rm mix}$ and the rates of    
charmonium decays and transitions in experiments are essential for
testing the theoretical predictions and in the understanding of the
nature of the $\psi(3770)$.

\begin{figure}[hbt]
{\centering
\includegraphics*[width=7.5cm,height=6.5cm]
{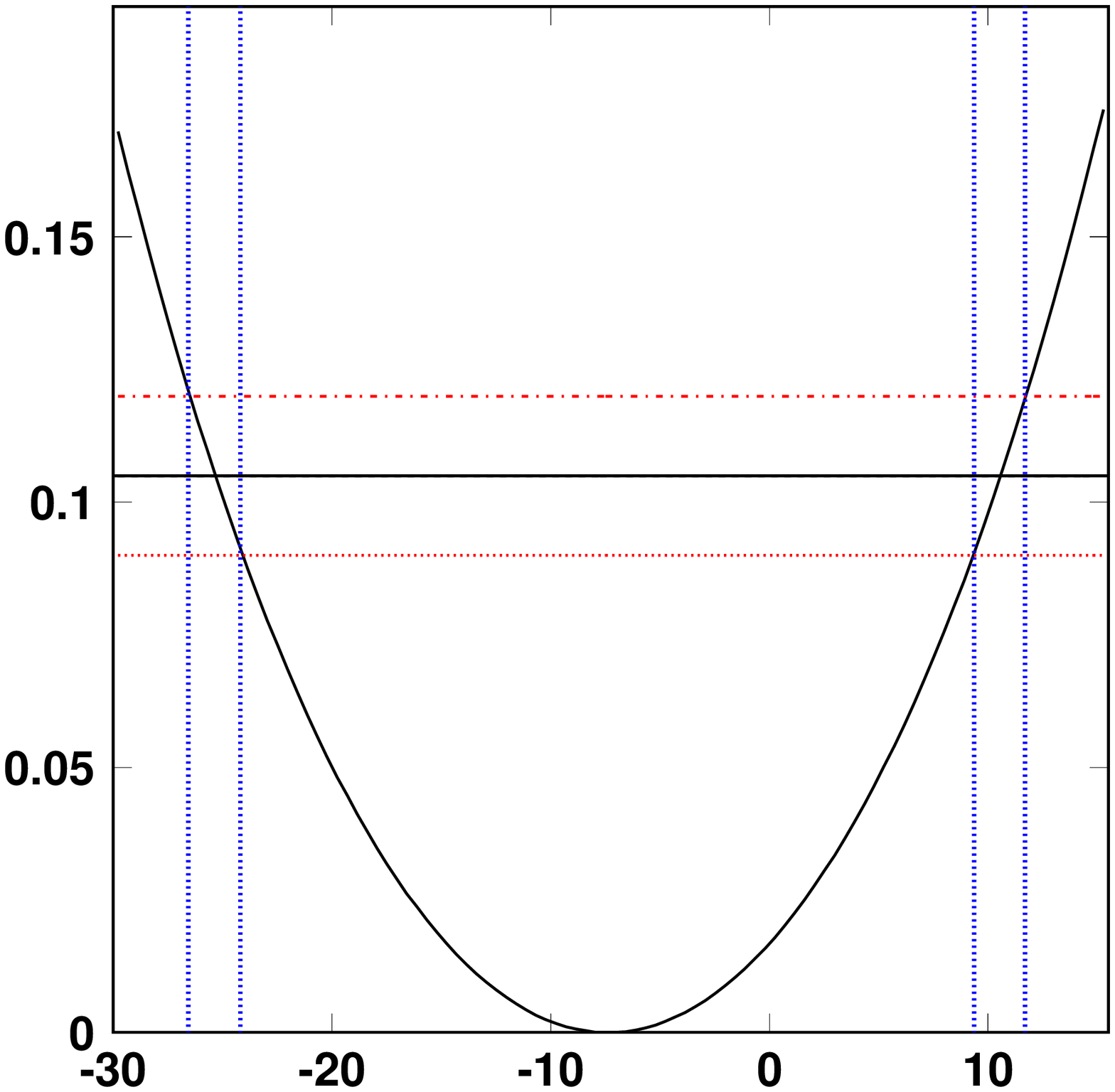}  
\put(-175,-5){\bf\large mixing angle $\theta_{\rm mix}$ [degree]}
\put(-225,65){\rotatebox{90}{\bf\large$R_{\psi(3770)/\psi(3686)}$}}
\caption{The ratio $R_{\psi(3770)/\psi(3686)}$ of the scaled  
leptonic widths as a function of mixing angle  $\theta_{\rm mix}$,
where the dashed lines show the $\pm 1 \sigma$ intervals of the 
measured quantities. } \label{xsct} }
\end{figure}

\subsection{Electromagnetic transitions}
\subsubsection{Predictions for $\psi(3770)$ electromagnetic 
transitions}

Charmonium states that are above the ground state can have 
significant decays into the states with the lower mass that proceed via
electromagnetic transitions. For the $\psi(3770)$ resonance, although
it predominantly decays to $D \bar D$ final states, there are, in
addition, electromagnetic transitions between the $\psi(3770)$ and
lower mass states. According to selection rules, the
$\psi(3770)$ can decay into $\chi_{cJ} \, (J=0,1,2)$ through
radiating an E1 photon.

The measurements of the rates of the EM transition of
$\psi(3770)$ to $\chi_{cJ} \, (J=0,1,2)$ has a special advantage.
In the $1D$ to $1P$ system the radial integrals are relatively
model-independent because of the absence of the wave function nodes in 
both states. In addition, the relativistic corrections and long distance
effects might be small. So, by directly comparing the measured
gamma transition rates from $\psi(3770)$ to $\chi_{cJ(J=0,1,2)}$  
to the non-relativistic estimates based on the potential model, one can 
get credible information about the transition mechanism. A
non-relativistic calculation for the EM transitions
of $\psi(3770) \rightarrow \gamma \chi_{cJ}\, (J=0,1,2)$ gives the
partial width~\cite{ycz_rosner,rg_Yamamoto1977}
as a function of $\theta_{mix}$:
\begin{equation}
 \Gamma(\psi(3770)\rightarrow \gamma \chi_{c0}) =
145~{\rm cos^2}\theta_{mix}(1.73+ {\rm tan}\theta_{mix})^2~{\rm
keV}, \label{eq7}
\end{equation}
\begin{equation}
 \Gamma(\psi(3770)\rightarrow \gamma \chi_{c1}) =
176~{\rm cos^2}\theta_{mix}(-0.87+ {\rm tan}\theta_{mix})^2~{\rm
keV}. \label{eq8}
\end{equation}
\begin{equation}
 \Gamma(\psi(3770)\rightarrow \gamma \chi_{c2}) =
167~{\rm cos^2}\theta_{mix}(0.17+ {\rm tan}\theta_{mix})^2~{\rm
keV}, \label{eq9}
\end{equation}
Inserting the mixing angle $\theta_{\rm mix}=(10.6\pm 1.3)^o$
determined with the parameters of the $\psi(2S)$ and $\psi(3770)$
measured by BESII~\cite{rg_prl97_121801_y2006} in the Eqs.~\ref{eq7}
-\ref{eq9} yields the partial widths
$$\Gamma(\psi(3770)\rightarrow \gamma \chi_{c0}) = 515\pm 17~~~{\rm keV}$$
$$\Gamma(\psi(3770)\rightarrow \gamma \chi_{c1}) = 79\pm 6~~~~~{\rm keV}$$
$$\Gamma(\psi(3770)\rightarrow \gamma \chi_{c2}) = 21\pm3 ~~~~~{\rm keV}.$$
These partial widths are sensitive to the mixing schemes.
Coupled-channel effects and more general mixing schemes
than the one given in the Eqs.~\ref{eq1} and~\ref{eq2} would
affect their values. A more complex mixing
scheme~\cite{rg_EEichten_prd69_y2004_p094019}, where the
$\psi(3770)$ is composed of only $52\%$ pure $c\bar c$ components, and
where, in addition to the $2^3S_1$ and $1^3D_1$ states, 
there are other $S$- and $D$-wave $1^{--}$
charmonium states included, with the
remainder of the wave function containing additional virtual or real
charmed meson pairs, predicts partial widths of:
$$\Gamma(\psi(3770)\rightarrow \gamma \chi_{c0}) =
225~~~{\rm or}~~{\it 254}~~{\rm keV}$$
$$\Gamma(\psi(3770)\rightarrow \gamma \chi_{c1}) =
59~~~{\rm or}~~{\it 183}~~{\rm keV}$$
$$\Gamma(\psi(3770)\rightarrow \gamma \chi_{c2}) =
3.9~~~{\rm or}~~{\it 3.2}~~{\rm keV},$$ where the  values in
italics are the results when the influence of the open-charm
channels is not included.  In contrast,
if the $\psi(3770)$ is a
pure $1^3D_1$ state, the partial widths would be
$$ \Gamma(\psi(3770)\rightarrow \gamma \chi_{c0}) =  434~~~ {\rm keV},$$
$$ \Gamma(\psi(3770)\rightarrow \gamma \chi_{c1}) =  133~~~ {\rm keV},$$
$$ \Gamma(\psi(3770)\rightarrow \gamma \chi_{c2}) =  4.8~~~ {\rm keV}.$$
Thus, measurements of the partial widths for the EM transitions can
provide useful information about the nature of $\psi(3770)$
resonance and potential model dynamics.

\subsubsection{Measurements at \bes3}

Measuring these branching fractions are experimentally
challenging. It requires a good electromagnetic calorimeter
to detect the low energy photons. The CLEO Collaboration measured
the partial width for $\psi(3770)\rightarrow \gamma \chi_{c1}$ to
be $\Gamma(\psi(3770)\rightarrow \gamma \chi_{c1}) = 75 \pm 18   
~~{\rm keV}$, which agrees well with most of the theoretical  
predictions~\cite{ycz_rosner,rg_EEichten_prd69_y2004_p094019,
rg_TBarnes_prd72_y2005_p054026}.  \bes3 has a good
electromagnetic calorimeter with an energy resolution of
$\sim 3\%$. With \bes3, we will be able to measure
the branching fractions for these transitions quite well.

To study how well we can measure the transition rates with the
\bes3 detector, we generated Monte Carlo events of the type
$\psi(3770)\rightarrow \gamma \chi_{cJ}\, (J=0,1,2)$, where we make
the $\chi_{cJ}\, (J=0,1,2)$ decay to $\pi^+\pi^-\pi^+\pi^-$,
$K^+K^-\pi^+\pi^-$, $p \bar p \pi^+\pi^-$ and
$\pi^+\pi^-\pi^+\pi^-\pi^+\pi^-$ final states.
Figure~\ref{chi_cj_mc} shows the distributions of the invariant
masses of the combinations for the different charged particle
final states. We find that the $\chi_{cJ}\, (J=0,1,2)$ states can
be well reconstructed with the \bes3 detector.
\begin{figure}[hbt]
{\centering
\includegraphics*[width=10.5cm,height=9.5cm]
{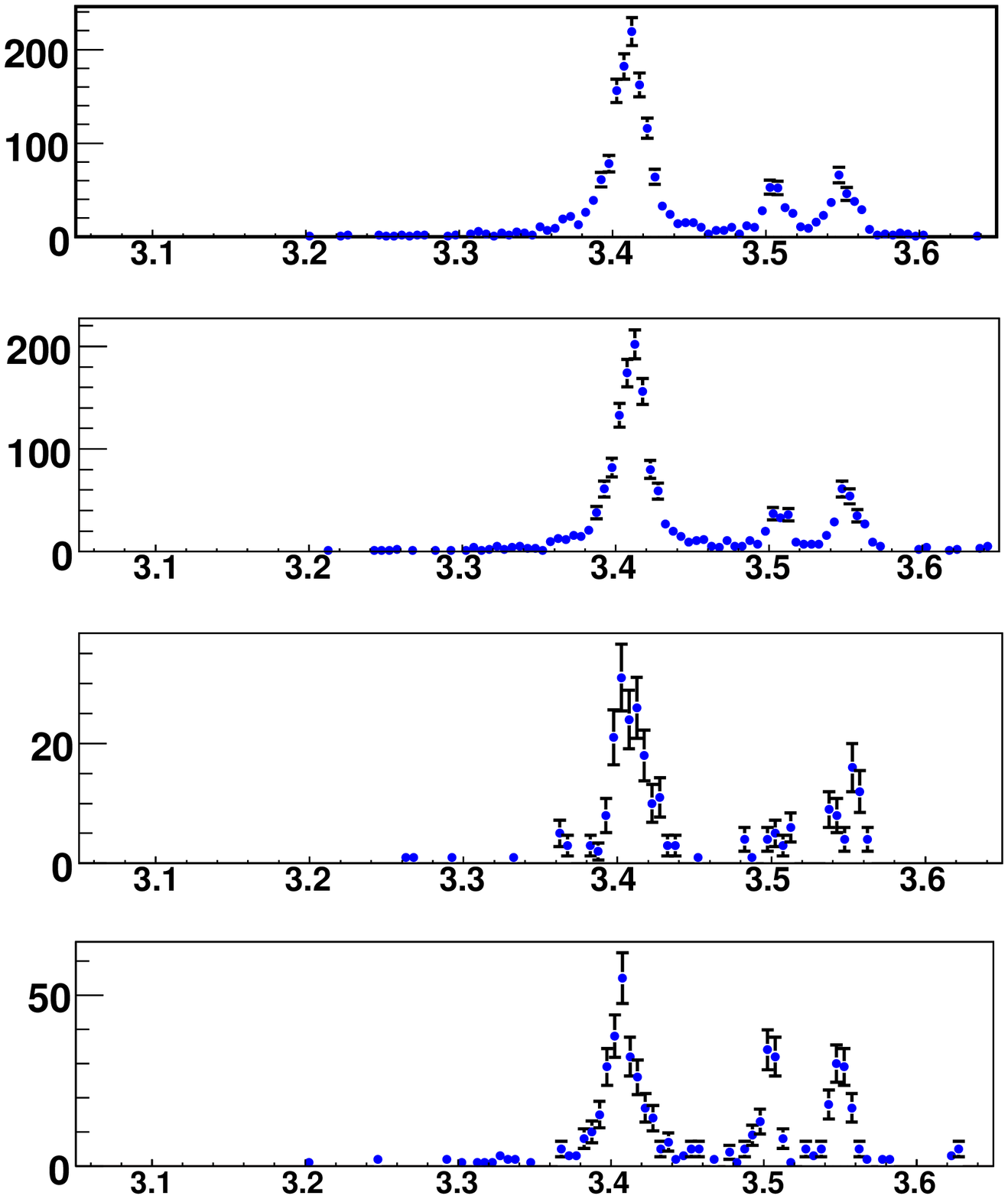} \put(-205,-10){\bf
Inv. Mass [GeV/$c^2$]} \put(-305,85){\rotatebox{90}{\bf No. of  
events}} \put(-255,240){\bf (a)} \put(-255,170){\bf (b)}
\put(-255,110){\bf (c)} \put(-255,40){\bf (d)} \put(-125,250){\bf
$\chi_{c0}$} \put(-100,235){\bf $\chi_{c1}$} \put(-78,235){\bf
$\chi_{c2}$} \caption{The distributions of the invariant masses of
(a) $\pi^+\pi^-\pi^+\pi^-$, (b) $K^+K^-\pi^+\pi^-$, (c) $p\bar p
\pi^+\pi^-$ and (d) $\pi^+\pi^-\pi^+\pi^-\pi^+\pi^-$ combinations,
where the charged particles are coming from the
$\chi_{cJ(J=0,1,2)}$ decays. } \label{chi_cj_mc} }
\end{figure}

\subsection{Hadronic transitions}

\subsubsection{Predictions for $\psi(3770)$ hadronic transition}

Measurements of the hadronic transitions of
$\psi(3770) \rightarrow J/\psi~{\rm hadron(s)}$ are important for
consideration of non-$D\bar D$ decays of the $\psi(3770)$ and to 
check the hadronic transition mechanism in the potential model.
Like the case for the $\psi(3686)$,  the emitted light hadron(s)
can be $\pi^+\pi^-$, $\pi^0$ and $\eta$ mesons that have hadronized
from radiated gluons. Other hadronic transitions that emit different
light hadrons are kinematically forbidden.
Generally, the transition rates, as  described in
Refs.~\cite{rg_moxhay,rg_billoier,part4_KY90}, are
\begin{equation}
\Gamma(2S\rightarrow 1S)=|c_1|^2G|f_{2S,1S}|^2, \label{eq10}
\end{equation}
\begin{equation}
\Gamma(1D\rightarrow 1S)=4/15|c_2|^2H|f_{1D,1S}|^2, \label{eq11}
\end{equation}
where $G$ and $H$ denote the phase space integrals ofor the two processes,
$f_{A,B}$ represents the radial matrix element between the states $A$ and $B$,
and  $c_1$ \& $c_2$ denote the strengths that appear in the soft poin matrix
elements of the gluon operators for $S$- and $D$-waves respectively.
In the $S$-$D$ mixing scheme,
the partial widths
for the hadronic $\pi^+\pi^-$ transitions of the two charmonium states
are given by ~\cite{part4_KY90}:

\begin{equation}
\Gamma(\psi(3770) \rightarrow J/\psi
\pi^+\pi^-)=|c_1|^2 \left[
   {\rm sin}^2\theta_{mix}  U
  + \frac{4}{15} \left |\frac{c_2}{c_1} \right |^2
     {\rm cos}^2\theta_{mix} V
                  \right],
                  \label{eq12}
\end{equation}
\begin{equation}
\Gamma(\psi(3686) \rightarrow J/\psi
\pi^+\pi^-)=|c_1|^2 \left[ {\rm cos}^2\theta_{mix} W
  + \frac{4}{15} \left |\frac{c_2}{c_1} \right |^2
     {\rm sin}^2\theta_{mix} X
                       \right],
                       \label{eq13}
\end{equation}
where
$U=G_{\psi(3770)}|f_{2S,1S}(M_{\psi(3770)})|^2$,
~~~~~~~~~~~~~~~~~~~~~~$V=H_{\psi(3770)} |f_{1D,1S}(M_{\psi(3770)})|^2$,
$W=G_{\psi(3686)} |f_{2S,1S}(M_{\psi(3686)})|^2$ and
$X=H_{\psi(3686)} |f_{1D,1S}(M_{\psi(3686)})|^2$.
In Eqs.~(\ref{eq12}) and (\ref{eq13}) the values for the
quantities $G$, $H$ and $f_{A,B}$ ($A=2S$ \& $1D$, and $B=1S$), can be
calculated using different models, which all give similar
results. For example, Ref.~\cite{part4_KY90}
gives
\begin{equation}
f_{2S,1S}(M_{\psi(3686)}) = 7.018~~~ {\rm GeV^{-3}},\label{eq14}
\end{equation}
\begin{equation}
f_{1D,1S}(M_{\psi(3686)}) = -8.796~~~ {\rm GeV^{-3}},\label{eq15}
\end{equation}
\begin{equation}
f_{2S,1S}(M_{\psi(3770)}) = 8.172~~~ {\rm GeV^{-3}},\label{eq16}
\end{equation}
\begin{equation}
f_{1D,1S}(M_{\psi(3770)}) = 10.266~~~ {\rm GeV^{-3}},\label{eq17}
\end{equation}
                       
\begin{equation}
G_{\psi(3686)} = 0.0353~~~ {\rm GeV^{7}},\label{eq18}
\end{equation}
\begin{equation}
H_{\psi(3686)} = 0.00291~~~ {\rm GeV^{7}},\label{eq19}
\end{equation}
\begin{equation}
G_{\psi(3770)} = 0.102~~~ {\rm GeV^{7}},\label{eq20}
\end{equation}
\begin{equation}
H_{\psi(3770)} = 0.00943~~~ {\rm GeV^{7}}.\label{eq21}
\end{equation}
However, the ratios $c_1/c_2$ are quite different for the different 
models~\cite{rg_moxhay,rg_billoier,part4_KY90}.
From the decay rates for the two hadronic transitions given in
Eqs.~\ref{eq12} and~\ref{eq13}, we have the ratio
\begin{equation}
R_{\psi(3770)/\psi(3686)}(J/\psi \pi^+\pi^-)
= \frac {\left[{\rm sin}^2\theta_{mix} U
  + \frac{4}{15} \left |\frac{c_2}{c_1} \right |^2
     {\rm cos}^2\theta_{mix} V
                       \right]}
{\left[{\rm cos}^2\theta_{mix} W
  + \frac{4}{15} \left |\frac{c_2}{c_1} \right |^2
     {\rm sin}^2\theta_{mix} X
                \right] },\label{eq22}
\end{equation}  
which can be measured experimentally. From the ratio
$R_{\psi(3770)/\psi(3686)}(J/\psi \pi^+\pi^-)$ we can extract the
ratio $c_1/c_2$.

The BESII Collaboration observed the hadronic transition process
$\psi(3770) \rightarrow J/\psi \pi^+\pi^-$ and measured the
partial width to be $\Gamma(\psi(3770) \rightarrow J/\psi
\pi^+\pi^-)=80 \pm 33 \pm 23 $ keV~\cite{rg_psipp_to_jpsipipi}.
Recently, CLEO confirmed the BESII observation for this transition
process and measured the branching fraction for $\psi(3770)
\rightarrow J/\psi \pi^+\pi^-$ to be $(0.189\pm 0.020\pm
0.020)\%$~\cite{rg_cleo_jpsipipi}. These give a weighted average
partial width of $\Gamma(\psi(3770) \rightarrow J/\psi
\pi^+\pi^-)=49.0 \pm 8.4$ keV, where the uncertainty is the combined
statistical and systematic errors. The world averaged
partial width for $\psi(3686) \rightarrow J/\psi \pi^+\pi^-$
process is $\Gamma(\psi(3686) \rightarrow J/\psi \pi^+\pi^-)=89.1
\pm 6.2$ keV. These give the ratio of partial widths
\begin{equation}
R_{\psi(3770)/\psi(3686)}(J/\psi \pi^+\pi^-)=0.55 \pm
0.10.\label{eq23}
\end{equation}
Inserting $\theta_{\rm mix}$ in Eq.~\ref{eq22} combined with the
ratio of the partial widths yields a first measurement of the
parameters of $c_1$ and $c_2$. The relation of the
$R_{\psi(3770)/\psi(3686)}(J/\psi \pi^+\pi^-)$ to the ratio
$c_2/c_1$ is plotted in Fig.~\ref{rg_fig3}, where the horizontal 
lines give the measured value of $R_{\psi(3770)/\psi(3686)}(J/\psi
\pi^+\pi^-)$ and its $\pm 1 \sigma$ interval; the
vertical lines give $\pm 1 \sigma$ interval of the measured values
of the ratio $c_2/c_1$. We find that the value of the ratio is
\begin{equation}
\frac{c_2}{c_1} = 2.08^{+0.31}_{-0.38}.\label{eq24}
\end{equation}
Inserting this solution for the ratio and the partial width  
$\Gamma(\psi(3770)\rightarrow J/\psi \pi^+\pi^-)=49.0\pm8.4$ keV
into Eq.~\ref{eq12}, we obtain
\begin{equation}
{c_1} = (7.25 \pm \Delta_{\rm err}) \times 10^{-3}. \label{eq25}
\end{equation}

\begin{figure}[hbt]
\centering
\includegraphics*[width=8.5cm,height=7.5cm]
{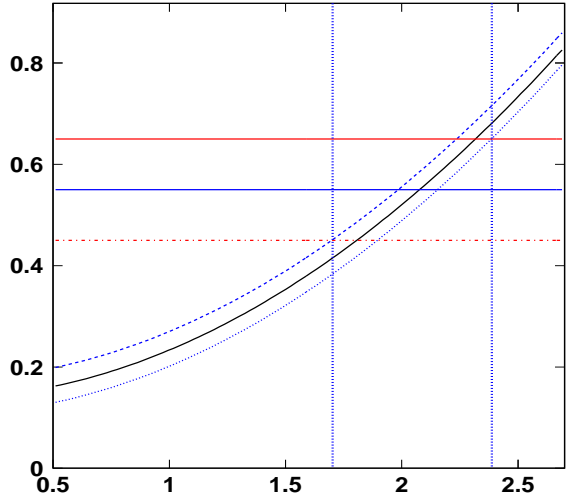}
\caption{The
ratio $R_{\psi(3770)/\psi(3686)}(J/\psi \pi^+\pi^-)$ of the
partial widths as a function of the ratio $\left
|\frac{c_2}{c_1}\right |$, where the dashed lines show the $\pm 1
\sigma$ intervals of the measured quantities. }
 \label{rg_fig3}
\end{figure}

\subsubsection{Measurements of hadronic transitions}

From an experimental point of view, since the mass difference between
$\psi(3770)$ and $J/\psi$ is 0.675 GeV, the typical momentum of 
the light hadron(s) system is low. This gives a clear kinematic
signature that can be used to cleanly separate
these decays from most other processes. The difficulty in
selecting these decays will be separating $\psi(3770)$ hadronic
transitions from the same $\psi(3686)$ hadronic
transition when the $\psi(3686)$ is produced either by ISR return
to the $\psi(3686)$ peak or is produced in its high energy tail.
To separate the $\psi(3770)$ hadronic transitions,
including $\psi(3770)\rightarrow J/\psi\pi^+\pi^-,
J/\psi\pi^0\pi^0, J/\psi\eta$, one needs a precision Monte Carlo 
event generator to simulate these processes including the effects
of ISR, and must correctly subtract the backgrounds due to the
$\psi(3686)$ hadronic transitions from the candidates for the
$\psi(3770)$ hadronic transitions observed in experiments. This
technique of subtracting the $\psi(3686)$ background in the
selected sample for $\psi(3770) \rightarrow J/\psi\pi^+\pi^-$ has
been developed for the BESII measurement of the partial
width for $\psi(3770) \rightarrow
J/\psi\pi^+\pi^-$~\cite{rg_psipp_to_jpsipipi}. The generator used
in the analysis of the decay is being developed for use in the
analysis of \bes3 data.

\subsection{Light hadron decays}

According to theory,
the branching fractions for $\psi(3770)\rightarrow {\rm light~hadrons}$
are expected to be small. However, there are some
models~\cite{rg_rosner_annl_phys_319_y2005_p1,rg_wym_prd70_114014}
that contain  mechanisms that can enhance non-$D\bar D$ decays. For
example, the re-annihilation amplitude of charmed meson pairs into
the states containing light quarks can interfere with the corresponding
amplitude for continuum light quark production of the same state. This
re-annihilation effect may result in a significant enhancement of
the production of light hadron final states at the
$\psi(3770)$~\cite{rg_h_j_lipkin_plb179_y1986_p278}. Such
re-annihilation mechanisms are also relevant for non-$K\bar K$   
decays of the $\phi$ meson~\cite{rg_h_j_lipkin_plb179_y1986_p278}.
If this re-annihilation effect is really responsible for the  
non-$D \bar D$ decays of $\psi(3770)$, the enhanced penguin  
amplitudes in $B$ decays can be explained by similar
re-annihilation effects. So the experimental study of the exclusive
non-$D\bar D$ decays of $\psi(3770)$ is important for the
understanding of a variety of puzzles in hadronic physics.

\subsubsection{Light-hadron final states}

Because of the large total width of the $\psi(3770)$, the decay amplitudes
for any exclusive channels cannot result in branching fractions as large 
as those
in the narrow resonance cases, {\it i.e.,} the J/$\psi$ and $\psi(3686)$.
As a result, the corresponding amplitudes from continuum production may 
compete
with them. On the experimental side, the most difficult item that
has to be considered is the possible interference between the
continuum amplitude and the $\psi(3770)$ resonance decay
amplitude. Because of the interference effects between these two
amplitudes, it not possible to measure the branching fractions for
the exclusive light hadron decays of $\psi(3770)$ by simply
considering the difference of yields observed far off the $\psi(3770)$
resonance and at the peak of
$\psi(3770)$~\cite{rg_wmy_ijmpa21_5163}. For example, the light
hadron decays of $\psi(3770)$ were extensively studied by both the
BES and CLEO Collaborations with data taken at
$\sqrt{s}=3.773$ GeV and at $\sqrt{s}$ around 3.66 GeV. However,  
if one only considers the observed cross sections for the light hadron
final states at the two energies, for
some decay modes the observed cross sections at $\sqrt{s}=3.773$
GeV are less than those observed at $\sqrt{s}$ around 3.66 GeV,
while, for some other decays modes, the observed cross sections at
$\sqrt{s}=3.773$ GeV are larger than those observed at $\sqrt{s}$
around 3.66 GeV. These are probably due to interference effects
between the two amplitudes. Owing to these effects, one cannot
simply determine the branching fractions of $\psi(3770)$ light
hadron decays just based on the measured cross sections at two
energy points.

To measure the branching fractions of $\psi(3770)$ light hadron
decays, one should measure the cross sections for exclusive decay
modes at different center-of-mass energies and fit these cross
sections to theoretical formulae that take interference effects
 into account. This is probably the best method to measure
the branching fractions for $\psi(3770) \rightarrow {\rm light~   
hadrons}$, as well as determine the phase difference between
the continuum light hadron production and the
$\psi(3770) \rightarrow {\rm
light~hadrons}$~\cite{rg_hekl_hepnp} amplitudes.

For families of decay modes that have some special symmetry,
for example $SU_3$ octets, the symmetry constrains
the couplings in different ways for the different channels. In this case,
one could fit the different couplings belong to various processes,
e.g. the EM coupling and the strong decay strength, using the  
observed cross sections for these channels.

\subsubsection{The ${\rm PV}$ decays of $\psi(3770)$}

Studies of the PV decay channels of $\psi(3770)$ are
interesting and very important for understanding the enhancement of
$\psi(3770)$ non-$D\bar D$ decays and the origin of the $\rho-\pi$
puzzle seen in $\psi(3686)$ decays. The most important PV
channel is the $\psi(3770)\rightarrow (0^{-+}~{\rm octet})~
(1^{--}~{\rm octet})$. CLEO has measured the production cross  
sections for $ e^+e^-\rightarrow \rho\pi$, $K^*(892) \bar K$,
$\omega\eta$, $\omega\eta'$, $\phi\eta$, $\phi\eta'$, $\rho\eta$,
$\rho\eta'$, $\omega\pi^0$ and $\phi\pi^0$~\cite{rg_cleoc_pv} at
two center-of-mass energies of 3.671 and 3.773 GeV. They claimed
that there is no evidence for significant $\psi(3770)$
decays into PV channels.
%
\vspace{0.5cm}

{\bf A. $\psi(3770) \rightarrow K^{*0}\bar K^0, K^{*\pm}K^{\mp},+{\rm c.c.}$}

From the observed decay channel $\psi(3770) \rightarrow
K^-\pi^+\pi^-\pi^+ + {\rm c.c.}$, one can analyze two correlated
modes with different charge states, i.e. $\psi(3770) \rightarrow
K^{*0}\bar K^0 + {\rm c.c.}$ with $\bar K^{*0}\rightarrow K^-
\pi^+$ and $K^0_S\rightarrow \pi^+ \pi^-$, and $\psi(3770)
\rightarrow K^{*\pm} K^{\mp}$ with $K^{*\pm}\rightarrow K^0
\pi^{\pm}$ and $K^0_S\rightarrow \pi^+ \pi^-$ by tagging the
different charge combinations of the $K^*(892)$. In this special   
example one can  determine the EM coupling and the
strong decay strength in this channel using measurements of the
two charge modes at two energy points, say at 
$\sqrt{s}=3.650$~GeV which is far away from resonance and at 
$\sqrt{s}=3.773$~GeV
which is the peak of the $\psi(3770)$ resonance, without the need from
measurements of any other PV channel at all. Unlike the
measurements of the most other PV channels, the observations can
provide us with four (not just two,  as is the case for other   
channels) observed numbers,  which
are the numbers of $K^*(892)^{\pm}K^{\mp}$ and $K^{*0}\bar
K^0+{\rm c.c.}$ observed at the two energy points. The four
observed numbers give us four equations to use to
extract three unknown parameters,  the coupling of
$K^*\bar K +{\rm c.c.}$ with $\gamma^*$, the coupling of $K^*\bar  
K +{\rm c.c}$ with $\psi(3770)$ and the phase difference between
them, and, finally, the branching fraction for
$\psi(3770) \rightarrow K^*(892)\overline{K} +{\rm c.c.}$ and the
EM coupling for PV channel in continuum production.

With this method, the BESII Collaboration measured the branching  
fraction for the decay $\psi(3770)\rightarrow K^{*0}\bar K^0+{\rm
c.c.}$ to be~\cite{rg_rongg_ichep04}
$$ B(\psi(3770)\rightarrow K^{*0}\bar K^0+{\rm c.c.}) =
(4.3^{+5.4}_{-3.4}\pm 1.3)\times 10^{-4},$$
which corresponds an upper limit on the branching fraction of      
$$B(\psi(3770)\rightarrow K^{*0}\bar K^0+{\rm c.c.})<0.12\%~~~~
{\rm at~90\%~C.L.},$$
corresponding to a limit on the partial width of
$$\Gamma(\psi(3770)\rightarrow K^{*0}\bar K^0+{\rm c.c.})
<29~~{\rm keV}~~~~~~~{\rm at~90\%~C.L.}.$$

For these two decay modes, the CLEO Collaboration observed the  
number of signal events at the two energy points ($\sqrt{s}=3.671$
GeV and $\sqrt{s}=3.773$ GeV) and measured the observed cross
sections to be~\cite{rg_cleoc_pv}
$${\bf N^{0}_{3.773 \rm GeV}=438}~~~~~~~~~~
                     \sigma^{0}_{3.773 \rm GeV}~=19.1~~~{\rm pb},$$
$${\bf N^{0}_{3.671 \rm GeV}=38}~~~~~~~~~~
                     \sigma^{0}_{3.671 \rm GeV}~=19.2~~~ {\rm pb},$$
$${\bf N^{\rm ch}_{3.773 \rm GeV}=4}~~~~~~~~~~
                     \sigma^{ch}_{3.773 \rm GeV}=0.09~~~ {\rm pb},$$
$${\bf N^{\rm ch}_{3.671 \rm GeV}=4}~~~~~~~~~~
                     \sigma^{ch}_{3.671 \rm GeV}~=1.14~~~ {\rm pb},$$
where the upper script $0$ and ${\rm ch}$ mean the neutral and
charged decay modes, respectively. When first looking at the    
numbers of the observed signal events and the values of the
observed cross sections, it seems that there is no room for the existence
of strong decays $\psi(3770)\rightarrow K^*(892)\bar K +~{\rm
c.c.}$ due to the tiny differences between the observed cross
sections at the resonance peak and off the resonance. However, the 
larger differences of the cross sections observed for the two  
different modes at each of the two energy points may indicate that
there is something else going on. This difference cannot
be explained only by the EM coupling.
\vspace{0.5cm}

{\bf B. $\psi(3770) \rightarrow \rho \pi$}

The partial width of the PV channel $\psi(3770)\rightarrow
\rho \pi$ decay might have a measureable value like that for
the $\psi(3770)\rightarrow K^*(892)\bar K$ decay mode
mentioned above.
However, BESII \cite{rg_bes_2_rhopi} did not observe
positive signals of $\psi(3770)\rightarrow \rho \pi$ or
$e^+e^-\rightarrow \rho \pi$ at either of the two energy points.
This might be due to the lack of suuficient statistics. Like  
the case for
$K^{*\pm} K^{\mp}$, there might exist a cancellation between the
amplitudes for the EM coupling and the strong decay in the decay
of $\psi(3770)\rightarrow \rho \pi$~\cite{rg_ycz_hadron05}.  
\vspace{0.5cm}

{\bf C. Fits to ${\rm PV}$ decays of the $\psi(3770) $}

Like the treatments of the ${\rm PV}$ decay channels for J$/\psi$
and $\psi(3686)$, an important task is the measurement of 
all of the ${\rm
PV}$ modes at the same time, such as $\psi(3770)\rightarrow
\rho\pi$,~$K^*(892) \bar K$,
$\omega\eta$,~$\omega\eta'$,~$\phi\eta$,~$\phi\eta'$,
$\rho\eta$,~$\rho\eta'$,
$\omega\pi^0$ and $\phi\pi^0$, etc., and to fit all of these
measured cross sections to get the information about the strong
and electro-magnetic couplings, the strength of $SU_3$ breaking,
and the mixing of $\eta-\eta'$. These types of measurements are
important for understanding the strength of $\psi(3770)$ non-$D
\bar D$ decays and can perhaps help in the understanding of the
origin of the '$\rho\pi$' puzzle. \vspace{0.5cm}

{\bf D. EM ${\rm PV}$ decays of the $\psi(3770) $}

BESII has measured the production cross sections for $e^+e^-
\rightarrow \omega \pi^0$, $\rho\eta$ and
$\rho\eta'$~\cite{rg_bes_2_ompi}. Due to isospin conservation, a
$1^{--}$ quarkonium state is only allowed to decay into
$\omega\pi^0$, $\rho\eta$ and $\rho\eta'$ through its
electro-magnetic coupling. However, the small leptonic
decay width of the $\psi(3770)$ makes the observation of the decays
$\psi(3770)\rightarrow \omega \pi^0$, $\rho\eta$ and $\rho\eta'$
very difficult (the branching fraction is at the level of $\leq
10^{-9}$), even  with \bes3 detector at the BEPC-II
collider. However, the measured cross sections for
$e^+e^-\rightarrow \omega \pi^0~{\rm etc.}$ are meaningful and
give form factor values for these modes. These indicate the common
$\gamma^*$-PV coupling of continuum hadron production in this
energy region. \vspace{0.5cm}

\subsubsection{Monte Carlo simulation}

For measurements of the branching fractions for $\psi(3770)
\rightarrow {\rm light~hadrons}$, we need to reconstruct 
$\psi(3770)$ decay final states. To understand how well we can   
do this with the \bes3 detector, we
generated the two different types of $\psi(3770)
\rightarrow {\rm light~hadrons}$ decays.  The first type
contains only stable or long-life-time charged particles
($K^{\pm}$, $\pi^{\pm}$ and $p$ or $\bar p$), and the second type
contains a promptly decaying particle, such as a $\pi^0$. To
reconstruct the final states for the first type, we 
calculate the total measured energy $E_{\rm measure}$ of the final
state containing all of the charged particles. Then we calculate
the ratio of the $E_{\rm measure}$ over the $E_{\rm cm}$. The most
probable ratio $R=E_{\rm measure}/E_{\rm cm}$ should be around 1
for the light hadron decays of $\psi(3770)$. For the second type
of decay mode, we simply reconstructed the $\pi^0$ from the  
decay mode under study. Figures~\ref{light_hadron_mc} (a), (b), (c)
and (d) show the reconstructed ratio $R$ for the final states of
$\pi^+\pi^-\pi^+\pi^-$, $K^+K^-\pi^+\pi^-$, $p \bar p\pi^+\pi^-$
and $3(\pi^+\pi^-)$, respectively; while
Figs.~\ref{light_hadron_mc}(e), (f), (g) and (h) show the
distributions of the invariant masses of the $\gamma\gamma$
combinations from the final states of $2(\pi^+\pi^-)\pi^0$,
$K^+K^-\pi^+\pi^-\pi^0$, $p \bar p\pi^+\pi^-\pi^0$ and
$3(\pi^+\pi^-)\pi^0$, respectively.

\begin{figure}[hbt]
{\centering
\includegraphics*[width=10.5cm,height=9.5cm]
{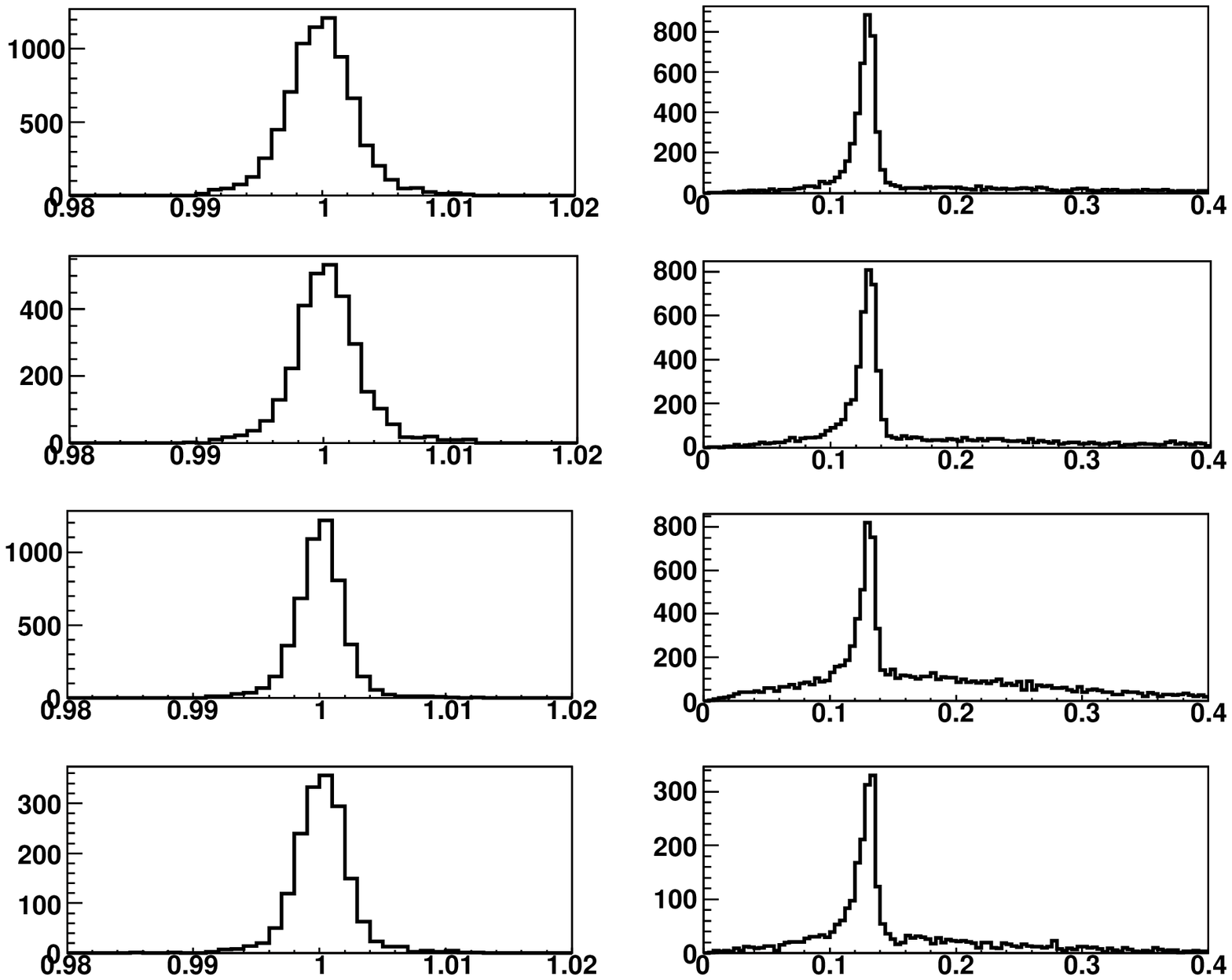} \put(-205,-10){\bf
Inv. Mass [GeV/$c^2$]} \put(-305,85){\rotatebox{90}{\bf No. of
events}} \put(-265,250){\bf (a)} \put(-265,180){\bf (b)}
\put(-265,115){\bf (c)} \put(-265,45){\bf (d)} \put(-125,250){\bf
(e)} \put(-125,180){\bf (f)} \put(-125,115){\bf (g)}
\put(-125,45){\bf (h)} \caption{The distributions of the ratio
$R=E_{\rm measure}/E_{\rm cm}$ and the invariant masses of $\gamma
\gamma$ combinations, where the (a), (b), (c) and (d) show the $R$
for the combinations of $\pi^+\pi^-\pi^+\pi^-$,
$K^+K^-\pi^+\pi^-$, $p \bar p\pi^+\pi^-$ and $3(\pi^+\pi^-)$,   
respectively; while the (e), (f), (g) and (h) show the invariant  
masses of $\gamma\gamma$ combinations for the final states of
$2(\pi^+\pi^-)\pi^0$, $K^+K^-\pi^+\pi^-\pi^0$, $p \bar
p\pi^+\pi^-\pi^0$ and $3(\pi^+\pi^-)\pi^0$, respectively. }
\label{light_hadron_mc} }
\end{figure}

Monte Carlo studies of light hadron decays and PV
decays of the $\psi(3770)$ are still in progress.

\subsection{Studies of inclusive decays}
 
\subsubsection{Measurement of branching fractions for
$\psi(3770)\rightarrow D^0 \bar D^0$, $D^+D^-$,
$D\bar D$ and $\psi(3770)\rightarrow {\rm non-}D\bar D$}

Recently, the BESII Collaboration reported measurements of the  
branching fractions for $\psi(3770) \rightarrow D^0\bar D^0,
D^+D^-$ and the branching fraction for $\psi(3770) \rightarrow   
{\rm non}-D\bar D$. Based on analysis of two different kinds of
data samples collected at $\sqrt{s}=3.773$ GeV and at $\sqrt{s}$
around 3.66 GeV, and collected in the range from 3.65 to 3.872
GeV, BESII obtained the non-$D\bar D$ branching
fraction $B(\psi(3770) \rightarrow {\rm non}-D \bar D)=(16.4\pm
7.3 \pm 4.2)\%$ ~\cite{rg_PhysRevLett_97_121801} and $B(\psi(3770)
\rightarrow {\rm non}-D \bar D)=(16.4\pm 7.3 \pm 4.2)\%$
~\cite{rg_PhysLettB_141_145}. A better way to measure the
branching fraction for $\psi(3770)\rightarrow {\rm non}-D\bar D$
is to analyze the energy dependent cross sections for
inclusive hadron, $D^0 \bar D^0$ and $D^+D^-$ event production in
the energy range covering both the $\psi(2S)$ and $\psi(3770)$
resonances, in a single scan. In this way one can also more
accurately measure the parameters of the two resonances, since
they are correlated to each other in the analysis of the cross
section scan data.

Using the same method as that used in the measurements of the
branching fractions for $\psi(3770)\rightarrow D^0 \bar D^0$,
$D^+D^-$, and non-$D\bar
D$~\cite{rg_PhysRevLett_97_121801}, we study by Monte Carlo
simulation how well we can measure the branching fractions with
the \bes3 detector at BEPC-II collider. We generated a total of
60 ${\rm pb}^{-1}$ of data at 49 energy points from 3.66 to 3.88
GeV. By analyzing these Monte Carlo events, we obtain the
branching fractions for $\psi(3770)\rightarrow D\bar D$ and
$\psi(3770)\rightarrow {\rm non-}D\bar D$.
Table~\ref{bf_psipp_to_dd} summarizes the measured
branching fractions from the Monte Carlo simulation along with the
branching fractions input into the Monte Carlo simulation. The errors
listed in the table are the statistical and systematic,
respectively. From the Monte Carlo simulation, we also obtain the
measured branching fractions for the decays $\psi(3770)
\rightarrow D^0 \bar D^0$ and $\psi(3770) \rightarrow D^+ D^-$ to
be
$$B[\psi(3770) \rightarrow D^0 \bar D^0]= (46.3\pm 1.3 \pm 1.0)\%, $$
and
$$B[\psi(3770) \rightarrow D^+ D^-]= (42.5\pm 1.2 \pm 0.9)\%, $$
which correspond to the set values of the branching fractions of
$B[\psi(3770) \rightarrow D^0 \bar D^0]= 46.8\%$ and
$B[\psi(3770) \rightarrow D^+ D^-]= 43.2\%$ in the Monte Carlo simulation,
respectively.
\begin{table}
\centering \caption{ The input and measured branching fraction for
$\psi(3770)\rightarrow {\rm non}-D\bar D$, where the "input" means
the value of the parameter set in the Monte Carlo simulation and
the "measured" means the measured value of the parameters from the
Monte Carlo simulation. }
 \label{bf_psipp_to_dd}
\begin{tabular}{ccc} \hline \hline
input/measured & $B(\psi(3770)\rightarrow D\bar D)$ [$\%$] &
$B(\psi(3770)\rightarrow {\rm non}-D\bar D)$ [$\%$] \\ \hline
input  & $90\%$ & $10\%$ \\
measured  & $88.8 \pm 2.4 \pm 2.0$ & $11.2\pm 2.4 \pm 2.0$ \\
 \hline \hline
\end{tabular}
\end{table}

Figure~\ref{psip_to_ddbar_mc} shows the observed cross sections
for inclusive hadron and $D \bar D$ production from
$\psi(3770)$ decays, where the dots with errors are the observed
cross section for the inclusive hadronic events from $\psi(3770)$
decays, the triangles with errors are the observed cross sections
for $D^+D^-$ from $\psi(3770)$ decays, and the inverted triangles
 with errors are the observed cross sections for $D^0\bar D^0$
from $\psi(3770)$ decays, and the squares with errors are the
totally observed cross sections over the neutral and the charged
$D\bar D$ production from $\psi(3770)$ decays. The lines show the
best fits to the observed cross sections.

\begin{figure}[hbt]
{\centering
\includegraphics*[width=10.5cm,height=9.5cm]
{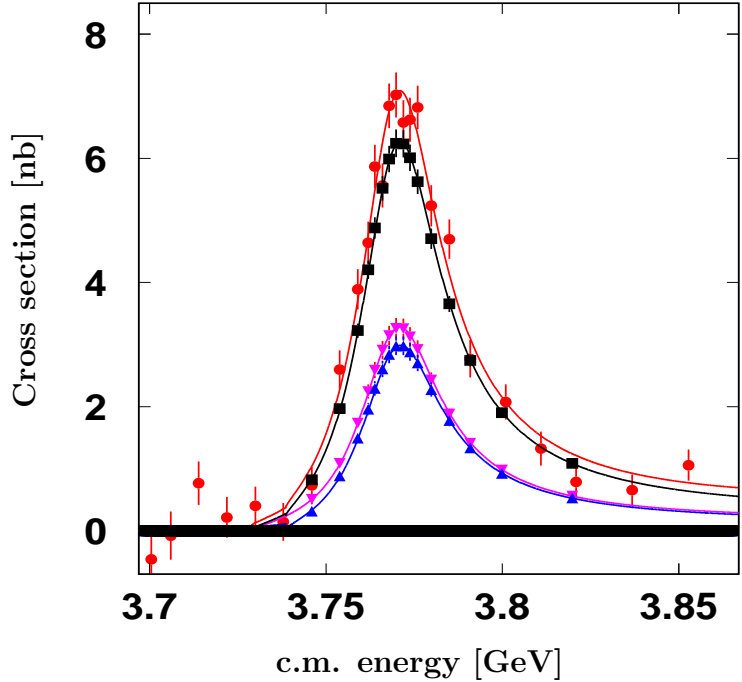} \put(-205,-10){\bf
c.m. energy [GeV]} \put(-305,90){\rotatebox{90}{\bf Cross section
[nb]}} \caption{The observed cross sections versus the nominal
c.m. energies. }
 \label{psip_to_ddbar_mc} }
\end{figure}

\subsubsection{Measurements of branching fractions for inclusive decay modes}

By fitting the cross sections for the inclusive $J/\psi$, $\eta$,
$\eta'$ ... observed at different center-of-mass energies to a
theoretical formula to describe the production of the inclusive 
final states, we can measure the branching fractions for
$\psi(3770)$ decay into these final states. These branching
fractions might prove useful for understanding of the nature of  
the $\psi(3770)$.

Monte Carlo studies of these decays are still in progress.

\subsubsection{Measurements of cross sections for
$e^+e^-\rightarrow {\rm hadrons}_{{\rm non-}D\bar D}$}

Another way to search directly  for the non-$D\bar D$ decays of
$\psi(3770)$ is to measure the cross section for
$e^+e^-\rightarrow {\rm hadrons}_{{\rm non-}D\bar D}$, where the  
${\rm hadrons}_{{\rm non-}D\bar D}$ means that the final state   
hadrons are not coming from the $D\bar D$ decays. In analysis of
the cross section scan data, if we find an enhancement in the
observed cross sections for the processes $e^+e^-\rightarrow {\rm
hadrons}_{{\rm non-}D\bar D}$ around $\psi(3770)$ resonance, we
directly observe the non-$D\bar D$ decays of the $\psi(3770)$
resonance. Actually, we can also examine the observed cross
sections for inclusive $K^0_S$, $K^{*0}$, $K^{*+}$ and $\phi$.
For these analyses, we need more data samples taken at different
center-of-mass energies.

In order to verify and calibrate the cross sections of the
non-$D\bar D$ event production in $\psi(3770)$ resonance region,
it is necessary to have data taken over the whole
energy region covered by the $\psi(3686)$ resonance.


\subsection{Summary}

In studies of the $\psi(3770)$ production in $e^+e^-$ annihilation
and its decays, there are still "puzzles" in understanding the    
physical $\psi(3770)$ state and its decays. What does the
$\psi(3770)$ consist of and how do the components of the $\psi(3770)$
effect its various non-$D\bar D$ decays, the $\gamma$-transitions,
the hadronic transitions, and the different exclusive non-$D\bar D$
decays? It is not understood how $\psi(3770)$ can decay into   
non-$D\bar D$ final states with such a large branching fraction. On
one hand, the newest BESII measurements on the branching fraction
may still suffer from a large systematic shift; on the other hand,
there may exist some other effects or new phenomena that are
responsible for the large discrepancy in the measured cross
sections $\sigma_{D\bar D}^{\rm obs}$ and
$\sigma_{\psi(3770)}^{\rm obs}$, and this results in the large   
non-$D\bar D$ branching fraction. With the \bes3 detector
at BEPC-II collider, we can measure the branching fraction
for $\psi(3770) \rightarrow {\rm non-}D\bar D$ well. With about 60
${\rm pb}^{-1}$ of data collected in the range from 3.65 to 3.88
GeV, we can measure the branching fraction with an accuracy level of
$2.5\%$. With more data taken across the $\psi(3770)$ resonance region
covering $\psi(3686)$ resonance, we can extensively study the
exclusive light hadron final states of $\psi(3770)$ decays and may
shed light on the "puzzles" mentioned above.

\section[Baryonic decays]
{Baryonic decays\footnote{Xiaohu Mo, Ronggang Ping, and Changzheng Yuan}}
\label{sec:baryon_decays}
\newcommand {\sla}[1]{ #1 \!\!\!/}
\renewcommand{\bd} {baryonic decays}
\def \nstar {N^*}
\renewcommand{\st}[1]{|#1\rangle}
\def \llb {\Lambda\bar\Lambda}
\def \xxb {\Xi^-\bar\Xi^+}
\def \ee {e^+e^-}

\newcommand{\kaz}{K^0}
\newcommand{\rop}{\rho^+}
\newcommand{\rom}{\rho^-}
\newcommand{\roz}{\rho^0}
\newcommand{\kzb}{\overline{K}^0}
\newcommand{\kszb}{\overline{K}^{*0}}
\renewcommand{\ko}{K_0}
\newcommand{\bip}{B^+}
\newcommand{\bim}{B^-}
\newcommand{\biz}{B^0}
\newcommand{\dep}{D^+}
\newcommand{\dem}{D^-}
\newcommand{\dez}{D^0}
\newcommand{\dzb}{\overline{D}^0}
\newcommand{\dszb}{\overline{D}^{*0}}
\newcommand{\ssb}{\Sigma\overline{\Sigma}}
\newcommand{\NNb}{N\overline{N}}
\renewcommand{\Dp}{D^{\prime}}
\newcommand{\Fp}{F^{\prime}}
\newcommand{\bbbar}{B\bar B}
\newcommand{\mmb}{\mu \overline{\mu}}
\newcommand{\pnb}{p\overline{n}}
\newcommand{\pbL}{\overline{p}\Lambda}
\newcommand{\SSb}{\Sigma\overline{\Sigma}}
\newcommand{\XXb}{\Xi \overline{\Xi}}
\newcommand{\SgSgbar}{\Sigma^0\bar\Sigma^0}
\newcommand{\SSbz}{\Sigma^0 \overline{\Sigma}^0}
\newcommand{\SSbp}{\Sigma^+ \overline{\Sigma}^-}
\newcommand{\SSbm}{\Sigma^- \overline{\Sigma}^+}
\newcommand{\XXbz}{\Xi^0 \overline{\Xi}^0}
\newcommand{\XXbm}{\Xi^- \overline{\Xi}^+}
\newcommand{\SzLb}{\Sigma^0 \overline{\Lambda}}
\newcommand{\SzL}{\overline{\Sigma}^0 \Lambda}
\newcommand{\SbzL}{\overline{\Sigma}^0 \Lambda}
\newcommand{\bnnb}{(n\overline{n})}
\newcommand{\bppb}{(p\overline{p})}
\newcommand{\bllb}{(\Lambda\overline{\Lambda})}
\newcommand{\bnpb}{(n\overline{p})}
\newcommand{\bnbp}{(p \overline{n})}
\newcommand{\bnlb}{(n\overline{\Lambda})}
\newcommand{\bnbl}{(\overline{n} \Lambda)}
\newcommand{\bplb}{(p\overline{\Lambda})}
\newcommand{\bpbl}{(\overline{p} \Lambda)}
\renewcommand{\VP}{1^-0^-}
\renewcommand{\PP}{0^-0^-}
\renewcommand{\fddd}{{\cal F} (\ddb \ra f)}
\renewcommand{\fndd}{\overline{\cal F} (\ddb \ra f)}
\newcommand{\rdr}{\sigma}
\newcommand{\rpdr}{\sigma^{\prime}}
\renewcommand{\M}{\frac{\sqrt{3}}{2}M}
\newcommand{\MR}{m_{\mathcal R}}

\def\eref#1{(\ref{#1})}
\def\Journal#1#2#3#4{{#1} {\bf #2}, #3 (#4)}
\def\CTP{Commun. Theor. Phys.}
\def\IJMPA{Int. J. Mod. Phys. A}
\def\NCA{Nuovo Cimento}
\def\NIM{Nucl. Instrum. Methods}
\def\NIMA{Nucl. Instrum. Methods A}
\def\NPA{Nucl. Phys. A}
\def\NPB{Nucl. Phys. B}
\def\PL{Phys. Lett. }
\def\PLB{Phys. Lett. B}
\def\PLA{Phys. Lett. A}
\def\PRL{Phys. Rev. Lett.}
\def\PRD{Phys. Rev. D}
\def\PRC{Phys. Rev. C}
\def\PRP{Phys. Rep.}
\def\PTP{Prog. Theor. Phys.}
\def\PTPS{Prog. Theor. Phys. Suppl.}
\def\RMP{Rev. Mod. Phys.}
\def\ZPC{Z. Phys. C}
\def\EPJC{Eur. Phys. J. C}
\def\HEPNP{HEP \& NP}

\def\bescoa {{BES Collaboration, J.Z. Bai, {\em et~al.}}}
\def\bescob {{BES Collaboration, M.Ablikim, {\em et~al.}}}
\def\prd#1#2#3 {{~Phys. Rev. D {#1}, #2 (#3) }}  
\def\plb#1#2#3 {{~Phys. Lett. B {#1}, #2 (#3) }}  

\newsavebox{\arrect}
\savebox{\arrect}(0,0){%
\setlength{\unitlength}{0.5cm}
\thicklines
\put(-4,1){\line(1,0){8.0}}
\put(4,1){\line(0,-1){2.0}}
\put(4,-1){\line(-1,0){8.0}}
\put(-4,-1){\line(0,1){2.0}}
\put(0,-1){\vector(0,-1){1.0}}
}

\newsavebox{\arrhomb}
\savebox{\arrhomb}(0,0){%
\setlength{\unitlength}{1cm}
\thicklines
\put(0,0.5){\line(-2,-1){1.0}}
\put(0,0.5){\line(2,-1){1.0}}
\put(0,-0.5){\line(-2,1){1.0}}
\put(0,-0.5){\line(2,1){1.0}}
\put(0,-0.5){\vector(0,-1){0.5}}
\put(-0.15,-0.7){\makebox(0,0)[r]{no}}
\put(1.0,0.0){\vector(1,0){2.0}}
\put(2.0,0.05){\makebox(0,0)[b]{yes}}
}

\newsavebox{\arrparall}
\savebox{\arrparall}(0,0){%
\setlength{\unitlength}{0.5cm}
\thicklines
\put(-3,1){\line(1,0){8.0}}
\put(5,1){\line(-1,-1){2.0}}
\put(3,-1){\line(-1,0){8.0}}
\put(-5,-1){\line(1,1){2.0}}
\put(0,-1){\vector(0,-1){1.0}}
}
\newsavebox{\arrparalla}
\savebox{\arrparalla}(0,0){%
\setlength{\unitlength}{0.5cm}
\thicklines
\put(-3,1){\line(1,0){8.0}}
\put(5,1){\line(-1,-1){2.0}}
\put(3,-1){\line(-1,0){8.0}}
\put(-5,-1){\line(1,1){2.0}}
}

\subsection{Introduction}\label{xct_int}
The discovery of the $\jpsi$ particle~\cite{jpsidiskv} opened a new
epoch of hadronic physics. Many theoretical concepts and tools have
been put forth to describe features of charmonium decays (see
Ref.~\cite{sBianco03} for a general review). However, since the
structure of baryons is comparatively more complicated than that for
mesons, theoretical research in the area of baryonic decays are at 
present relatively limited, and scattered in a variety of
references. Here we collect various
discussions and descriptions that are pertinent to baryonic decays,
so that we can identify what is known about  mechanisms for
baryonic decays, what remains unclear, and what needs further exploration.

\subsection{Theoretical Framework}\label{xct_fwk}

\subsubsection{Fock expansion}\label{sxct_fockepn}
Exclusive $\jpsi$ decays into baryon anti-baryon ($\bbbar$) pairs have
been investigated by a number of authors in the framework of perturbative
QCD (pQCD)~\cite{exdec:dun80,exdec:bro81,exdec:che82,exdec:bai85};
for a recent review see Ref.~\cite{HeavyQuackiniuPhysics}. The
dominant dynamical mechanism is $\ccbar$ annihilation into the
minimum number of gluons allowed by symmetries and the subsequent
creation of light quark-antiquark pairs that form the final state
baryons.
The decay amplitude is expressed as a convolution of a hard scattering
amplitude and a factor that involves the initial-state
charmonium wave function and the final-state baryonic wave function. 
As shown in Ref.~\cite{exdec:BKS}, the charmonium wave function
can be organized into a hierarchy of scales associated with powers of
the velocity of the $c$ quark in the charmonium meson. The Fock 
expansions for the charmonium states start as:
 \begin{eqnarray}
|\jpsi\rangle & = & \underbrace{|\ccbar_{1}({}^3S_1)\rangle} +
\underbrace{|\ccbar_{8}({}^3P_J)\, g \rangle} +
\underbrace{|\ccbar_{8}({}^3S_1)\, gg \rangle} + \dots , \nonumber\\
&& \hspace{5mm}{{\cal O}(1)}\hspace{15mm}{{\cal O}({\it
v})}\hspace{15mm}{{\cal O}({\it v}^2)}
\nonumber\\[0.2em]
|\;\;\eta_{c}\;\rangle & = &
\underbrace{|\ccbar_{1}({}^1S_0)\rangle} +
\underbrace{|\ccbar_{8}({}^1P_1)\, g\rangle} +
\underbrace{|\ccbar_{8}({}^1S_0)\, gg \rangle} + \dots,
\nonumber\\
&& \hspace{5mm}{{\cal O}(1)}\hspace{15mm}{{\cal O}({\it
v})}\hspace{15mm}{{\cal O}({\it v}^2)}
\nonumber \\[0.2em]
  |\;\chi_{c J}\rangle & = &
\underbrace{|\ccbar_{1}({}^3P_J)\rangle} +
\underbrace{|\ccbar_{8}({}^3S_1)\, g \rangle} + \dots,
\label{eq:exdec-Fockexpansion}
\\
&& \hspace{5mm}{{\cal O}(1)}\hspace{15mm}{{\cal O}({\it v})}
\nonumber
\end{eqnarray}
where the subscripts to the $\ccbar$ pair specify whether it is in a
colour-singlet ($1$) or colour-octet ($8$) state; ${\cal O}(1)$,
${\cal O}(v)$ and ${\cal O}(v^2)$ are the orders to which the
corresponding Fock states contribute, once evaluated in a matrix
element.

As shown in Ref.~\cite{exdec:wong}, decays of $P$-wave charmonium  
into baryon anti-baryon pairs are suppressed by a factor of 
$1/M$ relative to those for $S$-wave charmonium. For charmonium 
decays into
$\bbbar$, the decay amplitude can be expressed as:
\begin{equation}{\label{ampli}}
\mathcal{M} \sim f_c\phi_c(x)\otimes f_N\phi_N(x)\otimes f_{\bar
N}\phi_{\bar N}\otimes T_H(x),
\end{equation}
where $T_H$ is the hard perturbative part, and $f_i$ and $\phi_i$
are the decay constant and the hadronic wave function for charmonium
and baryon/anti-baryon, respectively. It is easy to use power
counting in Eq.(\ref{ampli}) to compare $S$-wave and
$P$-wave charmonium, as well as the color-singlet and octet
contributions to the decay width. For vector charmonia decays into
$\bbbar$, the decay amplitude dependence on  $M$ goes
as:
\begin{eqnarray}
\mathcal{M}^{(1)}_{S}\sim M{f^{(1)}_c\over M}\left({f_N\over M^2}\right)^2\sim
{1\over
M^4},\nonumber \\
\mathcal{M}^{(8)}_{S}\sim M{f^{(8)}_c\over M^2}\left({f_N\over
M^2}\right)^2\sim {1\over M^5}.
\end{eqnarray}
For example, in the case of $\jpsi$ decays to $\bbbar$, the color-octet
contribution is suppressed by the energy scale $1/M$. Therefore, the
color octet contribution can be neglected in $S$-wave decays.
However, for $P$-wave charmonium decays, the color-octet contribution
cannot be neglected. For example, the amplitudes of
$\chi_{cJ}\to\bbbar$ depend on the energy scale as:
\begin{eqnarray}
\mathcal{M}^{(1)}_P\sim M{f^{(1)}_c\over M^2}\left({f_N\over M^2}\right)^2\sim
{1\over
M^5},\nonumber \\
\mathcal{M}^{(8)}_P\sim M{f^{(8)}_c\over M^2}\left({f_N\over M^2}\right)^2\sim
{1\over M^5}.
\end{eqnarray}
To evaluate the decay widths, information on decay
constants of charmonium can be determined from the leptonic decay
width. For example:
\begin{equation}
\Gamma(\jpsi\to e^+e^-)={4\pi\over
3}{e_c^2\alpha_{em}^2f^2_{\jpsi}\over M_{\jpsi}}.
\end{equation}
One gets $f_{\jpsi}=409$MeV,$f_{\psip}=282$MeV. The other soft
physics information required is the leading-twist baryon
distribution amplitude~\cite{exdec:bolz97,exdec:bolz95}.

\subsubsection{Hadronic helicity conservation}
The leading-twist formation for the light hadrons in the final state
has implications for their helicity configurations. As a consequence
of the vector nature of QCD (and QED), time-like virtual gluons (or
photons) create light, (almost) massless quarks and antiquarks in
opposite helicity states. To leading-twist accuracy, such partons
form the valence quarks of the light hadrons and transfer their
helicities to them. Hence, the total hadronic helicity is
zero
\begin{equation}
\lambda_{1} + \lambda_{2} = 0~. \label{eq:exdec-hsr}
\end{equation}
The conservation of hadronic helicities is a dynamical consequence
of QCD (and QED) that holds to leading-twist order. The violation
of helicity conservation in a decay process signals the presence of
higher-twist, higher Fock states and/or soft, non-factorizable
contributions.

We note that hadronic helicity conservation does not hold either in
$\eta_c$ or $\chi_{c0}$ decays into baryon--antibaryon pairs where,
in the charmonium rest frame, angular momentum conservation requires
$\lambda_{B\phantom{\overline{B}}}\hspace*{-0.4cm} =
\lambda_{\overline{B}}$. A systematic investigation of higher-twist
contributions to these processes is still lacking despite some
attempts at estimating them. For a review see Ref.~\cite{exdec:CZ}.
Recent progress in classifying higher-twist distribution amplitudes
and understanding their properties~\cite{exdec:braun90,exdec:braun}
now permits such analyses. The most important question to be
answered is whether or not factorization holds for these decays to
higher-twist order. It goes without saying that besides higher-twist
effects, the leading-twist forbidden channels might be under control
of other dynamical mechanisms such as higher Fock state
contributions or soft power corrections.

There is no reliable theoretical interpretation of these decays as
yet. One proposition \cite{exdec:anselmino} is the use of a diquark
model, a variant of the leading-twist approach in which baryons are
viewed as being composed of quarks and quasi-elementary diquarks.
With vector diquarks as constituents, one may overcome the helicity
sum rule (\ref{eq:exdec-hsr}). The diquark model in its present
form, however, has some difficulties. Large momentum transfer
data on the Pauli form factor of the proton, as well as a helicity
correlation parameter for Compton scattering off protons are in
severe conflict with predictions from the diquark model. Other
phenomenological models argue that $\chi_{cJ}$ helicity
violation decays might proceed via uncommon mechanisms, such as
quark pair creation in $\chi_{cJ}\to B\bar B$ decays
\cite{exdec:3p0} and the exchange of an intermediate state in
$\chi_{cJ}$ decays \cite{exdec:zhouhq}.

\subsubsection{Analysis based on $SU(3)$ symmetry}\label{sxct_anasu3}
In $SU(3)_{flavour}$ representations, the baryons are arranged in 
singlet, octet,
and decuplets:
$$ {\mathbf 3} \otimes {\mathbf 3} \otimes {\mathbf 3}  = {\mathbf 1}_A
\oplus {\mathbf 8}_{M_1} \oplus {\mathbf 8}_{M_2} \oplus {\mathbf 10}_S~.$$
The subscripts indicate antisymmetric, mixed-symmetric or symmetric
multiplets under interchange of flavor labels of any two quarks. Each
multiplet corresponds to a unique baryon number, spin, parity and its
members are classified by $I,~I_3$, and $S$. The lowest lying singlet,
octet, and decuplet states, denoted $B_1$, $B_8$, and $B_{10}$,
correspond to $J^P = \frac{1}{2}^-, \frac{1}{2}^+,\frac{3}{2}^+$,
respectively.

In an $SU(3)$-symmetric world, only the decays into final states
$B_1 \overline{B}_1$, $B_8 \overline{B}_8$, and $B_{10} \overline{B}_{10}$
are allowed, with the same decay amplitudes within a given decay family if
electromagnetic contributions are neglected. Nevertheless, $SU(3)$ symmetry
can be broken in several ways~\cite{kopke}, so in phenomenological
analyses both symmetric and symmetry-breaking terms have to be included.

As a specific example, the parameterization forms for octet-baryon-pair
final state are worked out and presented in Table~\ref{octetbynform}.

\begin{table}[hbt]
\caption{\label{octetbynform}
Amplitude parameterization forms~\cite{kopke,Kowalski:1976mc}
for decays of the $\jpsi$ into a pair
of octet baryons (phase space is not included).
General expressions in terms of singlet $A$,
as well as symmetric and antisymmetric charge-breaking ($D,F$) and
mass-breaking terms ($\Dp,\Fp$) are given..}
\center
\begin{tabular}{ll}\hline \hline
  Final state    & Amplitude parameterization form  \\ \hline
  $\ppb$         & $A+D+F-\Dp+\Fp$    \\
  $\nnb$         & $A-2D-\Dp+\Fp$     \\
  $\SSbp$        & $A+D+F+2\Dp$       \\
  $\SSbz$        & $A+D+2\Dp$         \\
  $\SSbm$        & $A+D-F+2\Dp$       \\
  $\XXbz$        & $A-2D-\Dp-\Fp$     \\
  $\XXbm$        & $A+D-F-\Dp-\Fp$    \\
  $\LLb$         & $A-D-2\Dp$         \\
  $\SzLb+\SbzL$  & $\sqrt{3}D$        \\
\hline \hline
\end{tabular}
\end{table}

Here, we add a remark concerning the treatment of charge-conjugated 
final states. Applying the operator for charge conjugation
to a baryon-antibaryon system,
\beq
C|{B}_n \overline{B}_m \rangle =| \overline{B}_n {B}_m \rangle \left\{
\begin{array}{ll}
   = |{B}_n \overline{B}_m \rangle & \mbox{  for } n=m \\
\neq |{B}_n \overline{B}_m \rangle & \mbox{  for } n \neq m
\end{array} \right. ~,
\label{eq_cpbbnm}
\eeq
generally leads to a different state. Charge conjugated states will
nevertheless be produced with the {\bf same} branching ratio if {\bf 
isospin}
is conserved in the decay of the final state particles. We therefore
adopt the convention that charge conjugated states are implicitly
included in the measurement of branching ratios. In fact, the
parameterization in Table~\ref{octetbynform} has followed such
a convention.

\subsubsection{12\% rule and decay ratio}
The ratio derived based on the perturbative QCD:
\beq
Q_h =\frac{{\cal B}_{\psp \ra h}}{{\cal B}_{\jpsi \ra h}}
=\frac{{\cal B}_{\psp \ra \LL}}{{\cal B}_{\jpsi \ra \LL}}
\approx (12.4\pm 0.4) \%~, \label{qhvalue}
\eeq
appears to be valid for a number of hadronic final states, but is
violated severely in the $\RP$ and several other decay channels. This
is the so-called ``$\RP$ puzzle.'' However, the naive prediction for $Q_h$
suffers from several apparently simplistic  approximations.
More detailed and quantitative analyses are needed to refine the 
estimate.
One such refinement was put forth by Chernyak~\cite{Chernyak:1999cj}.
Based on the investigation of the asymptotic behaviour of hadronic 
exclusive processes within the framework of QCD~\cite{exdec:CZ}, 
Chernyak suggested the adoption of
the ratio of ``reduced'' decay amplitudes instead of branching fractions
to describe the relation between the $\jpsi$ and $\psp$ decays:
\beq
\kappa_h \equiv \left| \frac{A^{\prime}}{A} \right|=
\left\{ \frac{1}{f_{\kappa}} \cdot \frac{\BR_{\psp \to h}}{\BR_{\jpsi \to h}}
\right\}^{1/2}~,
\label{defkappa}
\eeq
where $A^{\prime}$ and $A$ are the ``reduced'' decay amplitudes
of the $\jpsi$ and $\psp \to h$ decays and $f_{\kappa}$ is a scaling 
factor:
\beq
f_{\kappa} = \frac{\BR_{\psp \to\LL}}{\BR_{\jpsi \to \LL}} \cdot
\left( \frac{M_{\jpsi}}{M_{\psp}} \right)^{n_{eff}} \cdot
\left( \frac{\xi^{\prime}}{\xi} \right)^{k} ~.
\label{fkappa}
\eeq
Here $n_{eff}$ is the effective index obtained by asymptotic dynamic analysis
of component quarks and contains the corresponding helicity
information~\cite{Chernyak:1999cj}, $k$ is a kinetic index for $\xi$
and $\xi^{\prime}$, which are phase space factors for the $\jpsi$ 
and
$\psp$ decays. For a two-body decay, $\xi$ or $\xi^{\prime}$ can be
written explicitly as
$$
\xi = \frac{|\mathbf{p}|}{M}
= \frac{\sqrt{[M^2-(m_1+m_2)^2][M^2-(m_1-m_2)^2]}}{2M^2}~,
$$
where $\mathbf{p}$ is the center-of-mass momentum of one of decay 
particles;
 $M$ and $(m_1,m_2)$ correspond to the masses of the parent and two
decay particles, respectively. For some special decay modes, the values of
$n_{eff}$ and $k$ are given in Ref.~\cite{Chernyak:1999cj}
(see Table~\ref{kappaval}).

\begin{table}[bth]
\caption{\label{kappaval}The values of $n_{eff}$ and $k$ for
some kinds of the $\jpsi$ and $\psp$ decay modes .}
\center
{
\begin{tabular}{ccccccc} \hline \hline
  Mode    & $VP$ & $VT,~AP$ & $\BBb$ & $PP,~VV$ & $\gamma T$ & $\gamma P$ \\ \hline
$n_{eff}$ & 6    & 6        & 8      & 4        & 2          & 4$\dagger$ \\
  $k$     & 3    & 1        & 1      & 2        & 1          & 3          \\ \hline\hline
\end{tabular} \\
$\dagger$: the value only for $\gamma \eta$ and $\gamma \etap$.
}
\end{table}

Although from a theoretical point of view $\kappa_h$ seems more
reasonable that $Q_h$, as discussed in Ref.~\cite{Mo:2006cy},
from an experimental point of view the estimation of $Q_h$ value 
affords us
some clues concerning the exploration of charmonium decay dynamics.
Therefore, in the following discussions $Q_h$ and/or $\kappa_h$
will be used.

\subsection{Two body decays}\label{xct_stopt}
In this section,  experimental data for baryonic decays are collected
as much as possible and compared with various kinds of theoretical predictions
and suggestions. For the $\chi_{cJ}$, $\jpsi$, and $\psp$, 
only the results on
two-body decays are presented and discussed while for the $\eta_c$, 
$\eta_c^{\prime}$
$\pspp$, and other higher-charmonium states, all results involving
baryonic pairs are mentioned since available measurements are rather 
limited.

\subsubsection{$\eta_c$ and $\eta_c^{\prime}$}
All experimental results up to now relevant to $\etac$ decays into
final-states containing baryons are summarized in Table~\ref{tab_expetac};
results involving baryon-pair for $\etacp$ decays are still not
available. Inclusive decays of the $\eta_c$ and
$\eta_c^{\prime}$ are discussed in section~\ref{sbxct_inkdk}.
For the $\eta_c$ and $\eta_c^{\prime}$ decays, more work is needed
both theoretically and experimentally.

\begin{table}[bth]
\caption{\label{tab_expetac}Experimental results for $\etac$ decays into
baryon-containing final states. The upper limits are at the 90\% 
confidence
level. $\BR_{\etac\to X}=\BR_{\jpsi,\psp\to \gamma\etac, ~\etac \to X}/
\BR_{\jpsi,\psp\to \gamma\etac}$ and the values of $\BR_{\jpsi,\psp\to \gamma\etac}$
are from PDG06~\cite{PDG2006}; the values with $\star$ are for $\psp\to \gamma\etac$
while the others are for $\jpsi\to \gamma\etac$.}
\center
{
\begin{tabular}{llllcc} \hline \hline
   Mode   & Number           & $\BR_{\jpsi,\psp\to \gamma\etac, ~\etac \to X}$
          &$\BR_{\etac\to X}$& $\BR_{\jpsi,\psp\to \gamma\etac}$ & Reference \\
      & of events        & ($\times 10^{-4}$)
      &($\times 10^{-3}$)& ($\times 10^{-2}$)         &      \\   \hline
 $\ppb$   & $213\pm 33$      & $0.19 \pm 0.03\pm 0.03$
          & $1.5\pm 0.6$     & $1.27\pm 0.36$             &~\cite{Bai:2003tr} \\
          & $18\pm 6$        & $0.13 \pm 0.04\pm 0.03$
      &$1.0\pm0.3\pm0.4$ & $1.27\pm 0.36$             &~\cite{Bisello:1990re} \\
      & $23\pm 1$        & $0.14 \pm 0.07$
      & $1.1\pm 0.6$     & $1.27\pm 0.36$             &~\cite{Baltrusaitis:1985mr}\\
      &                  & $0.08^{+0.08}_{-0.04}$
        &$2.9^{+2.9}_{-1.5}$ & $0.28\pm 0.06\star$        &~\cite{Himel:1980dj} \\
$[$average$]$&                  &
          & $1.3\pm 0.4$     &                            &~\cite{PDG2006} \\
 $\LLb$   & $<6$             & $<0.25$
          & $<2$             & $1.27\pm 0.36$             &~\cite{Bisello:1990re} \\
          &                  &
          & $0.87^{+0.38}_{-0.37}$ &                      &~\cite{Wu:2006vx} \\
$\ppb\pp$ &                  & $<0.05$
          & $<12$            & $0.28\pm 0.06\star$        &~\cite{Himel:1980dj} \\
\hline\hline
\end{tabular} \\
}
\end{table}

\subsubsection{$\chi_{cJ}$}

The colour-singlet contribution to the decays $\chi_{cJ}\to
p\bar{p}$ ($J=1,2$) has been investigated by a number of authors
\cite{exdec:che82,exdec:CZ,exdec:ste94,exdec:dam85}. Employing the
proton distribution amplitude (Ref.~\cite{exdec:bolz95}) or a similar
one, one again finds results that are clearly below experiment,
which again signals the importance of colour-octet contributions. An
analysis of  $\chi_{c1(2)}$ decays into the octet and decuplet
baryons along the same lines as for the pseudoscalar meson channels
\cite{exdec:BKS} has been carried out by Wong \cite{exdec:wong}.
The branching ratios have been evaluated from the baryon wave
functions and the same colour-octet $\chi_{cJ}$ wave function as in
Ref.~\cite{exdec:BKS}. Some of the results obtained in Ref.~\cite{exdec:wong}
are shown and compared to experiment in Table~\ref{tab:exdec-chi}.
As can be seen from the table, the results for the $p\bar{p}$
channels are in good agreement with experiment while the
branching ratios for $\Lambda\overline{\Lambda}$ channels are much
smaller than experiment \cite{exdec:BES-LLbar} although the errors
are large. A peculiar fact should be noted: the experimental
$\Lambda\overline{\Lambda}$ branching ratios are larger than the
proton--antiproton ones although the difference is only at the two
standard deviation level.

\begin{table}[tbh]
\caption[Comparison of theoretical and experimental branching ratios
         for various $\chi_{cJ}$ decays into pairs of light hadrons]
        {Comparison of theoretical and experimental branching ratios
         for various $\chi_{cJ}$ decays into pairs of light
         hadrons. The theoretical values have been computed within the
         modified perturbative approach, colour-singlet and -octet
         contributions are taken into account.
         The branching ratios are quoted in units 
         $10^{-5}$. The data are
         taken from ef.~\cite{PDG2006}. The values listed for the
         $\ppbar$ branching ratios do not include the most recent
         values $\left(27.4^{+4.2}_{-4.0}\pm 4.5\right)\, \cdot \,
         10^{-5}$, $\left(5.7^{+1.7}_{-1.5}\pm 0.9\right)\, \cdot \,
         10^{-5}$ and $\left(6.9^{+2.5}_{-2.2}\pm 1.1\right)\, \cdot
         \, 10^{-5}$ measured by BES \cite{exdec:BES-ppbar} for
         $\chi_{c0}$, $\chi_{c1}$ and $\chi_{c2}$ respectively.}
\label{tab:exdec-chi}
\renewcommand{\arraystretch}{1.2}
\begin{center}
\begin{tabular}{|c  || c |c |}
\hline
process & theory & experiment \\
\hline \hline ${\cal B}(\chi_{c0}\to\, p\; \bar{p}\,)$ & $-$  &
$22.4\pm 2.7$
\\\hline ${\cal B}(\chi_{c1}\to\, p\; \bar{p}\,)$ & $6.4\,$
\cite{exdec:wong} &
                                                         $7.2\pm1.3$\\ \hline
${\cal B}(\chi_{c2}\to\, p\; \bar{p}\,)$ & $7.7\,$ \cite{exdec:wong}
&
                                                         $6.8\pm0.7$\\ \hline
${\cal B}(\chi_{c0}\to\, \Lambda\, \overline{\Lambda}\,)$ & $-$ &
                                             $47\pm 16$  \\ \hline
${\cal B}(\chi_{c1}\to\, \Lambda\, \overline{\Lambda}\,)$ & $3.8\,$
\cite{exdec:wong}
                                          & $26\pm12$ \\ \hline
${\cal B}(\chi_{c2}\to\, \Lambda\, \overline{\Lambda}\,)$ & $3.5\,$
\cite{exdec:wong}
                                          & $34\pm17$ \\ \hline
\end{tabular}
\end{center}
\renewcommand{\arraystretch}{1.0}
\end{table}

The present analyses of baryonic $\chi_{cJ}$ decays suffer from the rough
treatment of the colour-octet charmonium wave function. As
mentioned above, a reanalysis of the decays into the $PP$ and
$B\overline{B}$ channels, as well as an extension to $VV$ decays is
required. Our knowledge of the colour-octet wave function has been
improved recently due to the intense analyses of inclusive processes
involving charmonia (see Ref.~\cite{exdec:inc}). This new information
may be used to ameliorate the analysis of the $\chi_{cJ}\to PP,
B\overline{B}$ decays and, perhaps, to reach a satisfactory
quantitative understanding of these processes.  We finally 
remark that the colour-octet contribution not only plays an
important role in the $\chi_{cJ}$ decays into $PP$ and
$B\overline{B}$ pairs but is potentially important
for their two-photon decays as 
well~\cite{exdec:man95,exdec:ma02,Ma:2002ev}.

The leading-twist-forbidden $\chi_{c0}\to B\overline{B}$ decays have
sizeable experimental branching ratios; see
Table~\ref{tab:exdec-chi}.  As of now, there is no reliable theoretical
interpretation for these decays. One phenomenological model
argues that the quark-pair creation mechanism makes a large
contribution to these helicity violating decays \cite{exdec:3p0}. It
is assumed that the $c\bar c$ quarks annihilate into two gluons, and
then the two gluons materialize into two quark-antiquark
pairs. Due to the quark-gluon coupling, another quark pair is
allowed to be created from the QCD vacuum with quantum numbers
$J^{PC} = 0^{++}$, thereafter the three (anti)quarks hadronize into
the outgoing (anti)baryon. It is found that the quark pair creation
mechanism, together with the $SU(3)_f$ symmetry breaking effect, give
calculated branching fractions for $\chi_{cJ}\to\Lambda\bar\Lambda$
that are comparable with the measured values.

Experimentally, 
tests for the colour-octet contribution involve comparisons between
reliable theoretical width calculations, or other observables, 
such as helicity amplitudes and angular distributions etc.,
with measurements. More experiments  on
charmonium decays to baryon-antibaryon pair will be useful for
determining the color-octet wave function.

\subsubsection{$\jpsi$ and $\psp$}
The experimental data for two-body baryonic decays
of the $\jpsi$ and $\psp$ are summarized in Table~\ref{brydkdata}.
Values of $Q_h$ and $\kappa_h$, as defined in Eqns.~\eref{qhvalue} and 
\eref{defkappa},
are also presented. From Table~\ref{brydkdata}, we see 
that for baryonic decays,
$Q_h$ is fairly consistent with the 12\% rule, within experimental 
errors.
The scaling factor defined in Eq.~\eref{fkappa}, 
ranges from 3.3 to 4.6, and the corresponding 
$\kappa_h$ value is almost a constant around 2. 
All this indicates that the dynamics of baryonic decays
of the $\jpsi$ and $\psp$ are fairly well 
described by pQCD.
This point is further confirmed by the amplitude analyses
discussed below.

\begin{table}[hbtp]
\caption{\label{brydkdata}
Experiemental data~\cite{PDG2006} on two-body baryonic decays
of the $\jpsi$ and $\psp$. The $Q_h$ and $\kappa_h$ defined in
Eq.~\eref{qhvalue} and \eref{defkappa} respectively, are also listed. The
upper limits are given at 90\% C.L. }
\center
\begin{tabular}{lllll}\hline \hline
 Mode        & $\BR_{\jpsi\to\BBb}$ & $\BR_{\psp\to\BBb}$ & $Q_h$  &$\kappa_h$ \\
             &  ($10^{-3}$)    &  ($10^{-4}$)    & (\%)            & (\%)      \\ \hline
$\ppb$       & $2.17\pm 0.08$  &  $2.65\pm 0.22$ & $12.2\pm 1.1 $  & $1.91\pm 0.09$  \\ 
$\nnb$       & $2.2 \pm 0.4 $  &  $ - $          & $ - $           & $ -  $          \\ 
$\SSbp$      & $ -$            &  $2.6 \pm 0.8 $ & $ -$            & $ -$            \\ 
$\SSbz$      & $1.31\pm 0.10$  &  $2.1 \pm 0.7 $ & $16.0\pm 5.5 $  & $2.09\pm 0.36$  \\ 
$\SSbm$      & $ -$            &  $ -$           & $ -$            & $ -$            \\ 
$\XXbz$      & $ -$            &  $2.8 \pm 0.9 $ & $ -$            & $ -$            \\ 
$\XXbm$      & $0.9 \pm 0.2 $  &  $1.5 \pm 0.7 $ & $16.7\pm 8.2 $  & $2.01\pm 0.52$  \\ 
$\LLb$       & $1.54\pm 0.19$  &  $2.5 \pm 0.7 $ & $16.2\pm 5.0 $  & $2.14\pm 0.33$  \\ 
$\SzLb+\SbzL$& $<0.15$         &  $- $           & $ - $           & $ -  $          \\ 
$\Delta(1232)^{++}\overline{\Delta}(1232)^{--}$
             & $1.10\pm 0.29$  &  $1.28\pm 0.35$ & $11.6\pm 4.5 $  & $1.76\pm 0.34$  \\ 
$\Delta(1232)^{+}\overline{p}$
             & $<0.1$          &  $ -$           & $ -$            & $ - $           \\ 
$\Sigma(1385)^{-}\overline{\Sigma}(1385)^{+}$ (or c.c.)
             & $1.03\pm 0.13$  &  $1.1 \pm 0.4 $ & $10.7\pm 4.2 $  & $1.53\pm 0.30$  \\ 
$\Sigma(1385)^{-}\overline{\Sigma}^{+}$ (or c.c.)
             & $0.31\pm 0.05$  &  $ -$           & $ -$            & $ - $           \\ 
$\Sigma(1385)^{0}\overline{\Lambda}$
             & $< 0.2$         &  $-$            & $ - $           & $ -  $          \\ 
$\Xi(1530)^{0}\overline{\Xi}(1530)^{0}$
             & $ -$            &  $<0.81 $       & $ - $           & $ -  $          \\ 
$\Xi(1530)^{0}\overline{\Xi}^{0}$
             & $0.32\pm 0.14$  &  $-$            & $ - $           & $ -  $          \\ 
$\Xi(1530)^{-}\overline{\Xi}^{+}$
             & $0.59\pm 0.15$  &  $-$            & $ - $           & $ -  $          \\ 
$\Omega^{-}\overline{\Omega}^{+}$
             &                 &  $<0.73 $       &                 &                 \\ 

\hline \hline
\end{tabular}
\caption{\label{rdbrpznfm}
Reduced branching ratios $\tilde{\BR}$
for two-body octet baryonic final states
 compared with the parameterization of the squared amplitude 
described in the text.
The numbers in the parentheses are the calculated results with the fitted 
parameters
given in Table~\ref{bdkfit2} for $\phi=-90^{\circ}$.}
\center
\newcommand{\f}{e^{i\phi}}
\begin{tabular}{llll}\hline \hline
 Final state   & Parameterization        &$\tilde{\BR}_{\jpsi}$
                                                     &$\tilde{\BR}_{\psp}(10^{-1})$ \\ \hline
 $\ppb$        & $|A+\f(D+F)-\Dp+\Fp|^2$ & $1.29\pm 0.05(1.30)$  & $  1.46\pm 0.13(1.47)$  \\
 $\nnb$        & $|A-\f(2D)-\Dp+\Fp|^2$  & $1.31\pm 0.24(1.31)$  & $  ~~~~~-~~~~~~(1.11)$  \\
 $\SSbp$       & $|A+\f(D+F)+2\Dp|^2$    & $~~~~~-~~~~~~(0.96)$  & $  1.61\pm 0.50(1.62)$  \\
 $\SSbz$       & $|A+\f(D)+2\Dp|^2$      & $0.97\pm 0.07(0.97)$  & $  1.31\pm 0.46(1.25)$  \\
 $\SSbm$       & $|A+\f(D-F)+2\Dp|^2$    & $~~~~~-~~~~~~(0.99)$  & $  ~~~~~-~~~~~~(1.04)$  \\
 $\XXbz$       & $|A-\f(2D)-\Dp-\Fp|^2$  & $~~~~~-~~~~~~(0.79)$  & $  1.89\pm 0.61(1.58)$  \\
 $\XXbm$       & $|A+\f(D-F)-\Dp-\Fp|^2$ & $0.82\pm 0.18(0.81)$  & $  1.02\pm 0.48(1.34)$  \\
 $\LLb$        & $|A-\f(D)-2\Dp|^2$      & $1.05\pm 0.13(1.06)$  & $  1.49\pm 0.42(1.36)$  \\
 $\SzLb+\SbzL$ & $|\sqrt{3}D|^2$         & $~~~~~-~~~~~~(0.00)$  & $  ~~~~~-~~~~~~(0.00)$  \\
\hline \hline
\end{tabular}
\end{table}

As discussed in section~\ref{sxct_anasu3}, two-body octet baryon
decays can be parameterized by five quantities:
$A$ (the strong amplitude in the limit of exact $SU(3)$), $D,F$ (the 
symmetric and antisymmetric charge-breaking amplitudes),and
$\Dp,\Fp$ (the symmetric and antisymmetric mass-breaking amplitudes).
Specific espressions for the different decay modes are listed
in Table~\ref{octetbynform}.
These five amplitudes could, in principle, be complex,
each with a magnitude and a phase, giving a total of
ten parameters, which are too many for the nine baryonic decay modes.
Thus, we adopt the following assumptions to make our fit practical:
\begin{enumerate}
\item The isospin-violating decay $\jpsi \to \SzLb+\SbzL$ is proportional to 
$D$. As indicated in Table~\ref{brydkdata}, the branching fraction for this
channel is at least one order-of-magnitude smaller than the others. Accordingly,
we set $D=0$.\footnote{As a result, it is 
important to measure the $\SzLb+\SbzL$ final state; even a determination
of an upper limit would be very useful for phenomenological or theoretical 
analysis.}
\item There are no relative phases between the amplitudes $A$, $\Dp$, and $\Fp$.
\end{enumerate}

In the context of these assumptions, the parameterization forms are given in
Table~\ref{rdbrpznfm}, where $A$, $D$, $F$, $\Dp$, and $\Fp$ are now real 
numbers.
In addition, another point should be noted. The parameterization forms
in Table~\ref{octetbynform} and \ref{rdbrpznfm} do no include phase space
effects. This could be corrected for by the  use of so-called ``reduced'' 
branching 
fractions, defined as
$$ \tilde{\BR} = \BR/ PSF_2~,$$
where $PSF_2$ is the phase space factor for two-body decay. 
These 
are tabulated in
Table~\ref{rdbrpznfm}. With this information and the corresponding 
parameterization
forms, we do the fit and present the results in the Table~\ref{bdkfit2}.
\begin{table}[htb]
\caption{\label{bdkfit2}
Fit results for two-body baryonic decay. The parameterization forms
and data are given in Table~\ref{rdbrpznfm}. The $\star$ indicates the
phase value is fixed in the fit.}
\center
\begin{tabular}{ccccc} \hline\hline
Paramter & \multicolumn{2}{c}{$\jpsi$ decay} & \multicolumn{2}{c}{$\psp$ decay} \\ \hline
  $A$    & $ 1.008\pm 0.029$ & $ 1.008\pm 0.029$ & $ 1.132\pm 0.096$ & $ 1.143\pm 0.105$ \\
  $\Dp$  & $-0.010\pm 0.016$ & $-0.010\pm 0.016$ & $-0.023\pm 0.053$ & $-0.013\pm 0.051$ \\
  $\Fp$  & $ 0.127\pm 0.110$ & $ 0.127\pm 0.114$ & $-0.167\pm 0.181$ & $-0.103\pm 0.157$ \\
  $F$    & $ 0.008\pm 0.107$ & $ 0.020\pm 0.249$ & $-0.221\pm 0.207$ & $-0.290\pm 0.316$ \\
  $\phi$ & $-91.12\pm 3.675$ & $-90 \star~~~~~~$ & $-91.11\pm 2.505$ & $-90 \star~~~~~~$ \\
\hline
$\chi^2$ &$4.3\time 10^{-3}$ &$4.0\time 10^{-3}$ &$0.4 $             &$0.4 $             \\
\hline\hline
\end{tabular}
\end{table}

From Table~\ref{bdkfit2}, we notice the following points
\begin{itemize}
\item The absolute values of $\Dp$, $\Fp$, and $F$ are much smaller than that for
$A$, which means the $SU(3)$-breaking effects,  both charge-breaking 
($D,F$) and
mass-breaking ($\Dp,\Fp$), are weak for $\jpsi$ and 
$\psp$ baryonic decays.
In addition, the antisymmetric breaking term is about one order of
magnitude stronger than the symmetric term.
\item The phase between the strong and electromagnetic amplitudes is almost 
$-90^{\circ}$ from the free parameter fit for both the $\jpsi$ and $\psp$ decays. 
This large phase is consistent with a previous analysis~\cite{NNmode}, and
also consistent with results from 
the analyses for two-body mesonic decays, such $1^+0^-$($90^\circ$)~\cite{suzuki01},
$1^-0^-$ $(106 \pm 10)^\circ$~\cite{VPmode,castro},
$1^-1^-$ $(138 \pm 37)^\circ$~\cite{kopke,castro,suzuki99},
$0^-0^-$ $(89.6 \pm 9.9)^\circ$~\cite{kopke,castro,suzuki99} and so on.
Moreover, the fit results also support 
the idea of a universal phase proposed in Ref.~\cite{wymrpi2}.
\item Electromagnetic-breaking (due to value $F$) is much stronger for
the $\psp$ than for the $\jpsi$. This phenomenon has been
noticed in an analysis of $\psp \to K^*(892)\overline{K}+c.c.$ 
decays~\cite{besvp3}.
\end{itemize}
Using the fit parameters, the reduced branching fractions are calculated and
results are given in Table~\ref{rdbrpznfm}. Since there are many 
parameters  
to adjust, the calculated branching fractions agree with experiment
fairly well, as might be expected. Moreover, some branching fractions
that have not be measured are also calculated to provide predictions for
further experimental work.

\begin{table}[hbtp]
\caption{\label{tab:exdec-jpsi}
Comparison between theoretical evaluation~\cite{exdec:bolz97} and
experiemental data~\cite{PDG2006} (or refer to Table~\ref{brydkdata}).
Here the branching fractions are transformed into decay widths by
using $\Gamma_{\jpsi}=(93.4\pm 2.1)$ keV~\cite{PDG2006}.
${\cal R} \equiv \Gamma_{\bbbar}/\Gamma_{\ppb}$.}
\center
\begin{tabular}{lccccc}\hline \hline
 Mode        & \multicolumn{2}{c}{$\Gamma_{the.}$ (eV)} & $\Gamma_{exp.}$ (eV)
                         & ${\cal R}_{the.}$ & ${\cal R}_{exp.}$  \\
             & $m_s=150$ MeV & $m_s=350$ MeV &                 &      \\ \hline
$\ppb$       & 174  &  174  & $203\pm ~9$  & 1     & 1                \\ 
$\SSbz$      & 128  &  113  & $122\pm 10$  & 0.649 & $0.604\pm 0.052$ \\ 
$\LLb$       & 133  &  117  & $144\pm 19$  & 0.672 & $0.710\pm 0.092$ \\ 
$\XXbm$      & 92.8 &  62.5 & $~84\pm 19$  & 0.359 & $0.415\pm 0.094$ \\ 
$\Delta^{++}\overline{\Delta}^{--}$
             &      &  105  & $103\pm 28$  & 0.603 & $0.507\pm 0.135$ \\ 
$\Sigma^{*-}\overline{\Sigma}^{*+}$
             &      &  66.1 & $~96\pm 13$  & 0.380 & $0.475\pm 0.065$ \\ 

\hline \hline
\end{tabular}
\end{table}

Now we turn to another aspect of comparison between theorectical predictions and
experimental measurements. As mentioned in section~\ref{sxct_fockepn}, the decay
width of the $\jpsi$ (or $\psp$) can be calculated with information about 
hadronic distribution amplitudes. In Ref.~\cite{exdec:bolz97}, a modified 
perturbative
approach is adopted, where quark transverse momenta are retained and Sudakov
suppressions, comprising those gluonic radiative corrections not included in
the evolution of the distribution amplitude, are taken into account. The advantage
of this modified perturbative approach is the strong suppression of the soft
end-point regions where the pQCD can not be applied.

An evaluation of $\jpsi\to \bbbar$ via the hard process
$c\bar c\to 3g^*\to 3(q\bar q)$ is described in Ref.~\cite{exdec:bolz97} and
listed in Table \ref{tab:exdec-jpsi}. The theoretical values are
in good agreement with the measured values within errors.\footnote{In
Ref.~\cite{exdec:bolz97}, the decay widths for $\psp\to \bbbar$
are also presented, which are actually scaled results
assuming the  12\% rule. Thus, the consistency between
theoretical envaluations and experimental measurements for $\psp$
decay depends only on those of the $\jpsi$ and the validity of 12\% rule.}
Predictions for the absolute value of a decay width are subject
to many uncertainties~\cite{exdec:bolz97} while the ratios of any two
$\bbbar$ decay widths are robust since many uncertainties cancel
to a large extent. For this reason the ratios 
of branching fractions (normalized to the
$\ppb$ final state) are also presented in Table~\ref{tab:exdec-jpsi}.
Although good agreement can be seen for both absolute decay widths and
relative ratios, there still exist a 10\% to 20\% difference between
the theoretical and experimental values. Here the results from the 
amplitude
analysis should be noted, particularly electromagnetic breaking effects
which are at the level of 10\% to 20\%. Moreover, the relative phase between
the strong and electromagnetic interaction is around $-90^{\circ}$ and is
a crucial piece of information for the improvement of the calculation in
Ref.~\cite{exdec:bolz97}.

\begin{table}[htb]
\center
\caption{\label{expang}
Angular distribution parameter $\alpha$ for $\jpsi\to\bbbar$
decays. They are assumed to be the form of $dN/d\cos\theta\propto
1+\alpha\cos^2\theta$.}
\begin{tabular}{lrcc}
\hline\hline
& & \multicolumn{2}{c}{Calculated value of $\alpha$}\\
Decay mode & Measured value of $\alpha$ & Ref. \cite{exdec:bro81} & Ref.
\cite{carimalo}\\\hline
$\jpsi\to\ppbar$& $0.68\pm 0.06$\cite{plb591} &1 &0.69 \\
$\jpsi\to\llb$& $0.65\pm 0.11$\cite{plb632} & 1&0.51 \\
$\jpsi\to\ssb$& $-0.24\pm 0.20$\cite{plb632} &1 &0.43 \\
$\jpsi\to\xxb$& $-0.13\pm 0.59$\cite{prd29} &1 & 0.27\\
$\psi'\to\ppbar$&$0.67\pm0.16$\cite{plb610} &1 &0.80 \\
\hline\hline
\end{tabular}
\end{table}

Last, but not least, are measurements of the angular distributions for
the processes $e^+e^-\to (\jpsi \textrm{ or } \psp )\to B_8\overline{B}_8$.
This has the form : 
\begin{equation}
{d\Gamma\over d\cos\theta}\propto 1+\alpha\cos^2\theta,
\end{equation}
where $\theta$ is the angle between the out-going baryon and the
$e^+e^-$ beam. Table \ref{expang} summarizes the measured angular
distribution parameters and provides
comparisons with theoretical predictions.
In the limit of the helicity conservation, $\alpha=+1$.
This is prominently violated for the $\jpsi\to\ssb$ and
$\xxb$ modes, where not only the magnitude is smaller then expected,
but also the sign contradicts the prediction.
This violation has been attributed to a constituent
quark~\cite{carimalo,exdec:mur95} and/or hadron mass
effect~\cite{exdec:clau82} or final state interactions~\cite{pingrg},
both of these effects are part of the $\mathcal O(v^2)$ and
high-twist/power corrections. Also electromagnetic effects on the
value of $\alpha$ have been investigated.

\subsubsection{$\pspp$ and other higher charmonium 
states}\label{sxct_psppahi}
Charmless decays of the $\pspp$ can shed light on $S$-$D$ mixing,
missing $\psp$ decays such as $\psp \to\rho\pi$,
the discrepancy between the total and $\DDb$ cross section at the
$\pspp$, and rescattering effects contributing to an enhanced
$b \to s$ penguin amplitude in $B$ meson decays~\cite{Rosner:2004mi}.
A phenomenological analysis~\cite{Wang:2004kf} also indicates
the possibility of a large charmless branching fraction for the $\pspp$.
Stimulated by theorectical suggestions and based on observations with
data samples of 55.8~pb$^{-1}$ on the peak of the $\pspp$ and 20.70~pb$^{-1}$
at the continuum ($\sqrt{s}$=3.67~GeV) (which is used for background
subtraction), CLEOc reported results of searches for a wide variety of
non-$\DDb$ modes~\cite{Huang:2005fx}, among which the results involving
baryonic pairs are reproduced in Table~\ref{psppbdkdt}.
However, no signals are reported and only the upper limits are given,
which indicates that the charmless fraction is rather small. This seems also 
consistent with CLEOc's inclusive results~\cite{Besson:2005hm}.

\begin{table}[htb]
\center
\caption{\label{psppbdkdt}
Experiment results for exclusive baryonic decays of 
the $\pspp$~\cite{Huang:2005fx}.
Listed are cross section upper limits that include systematic errors (90\% 
C.L.),
and the branching ratio upper limit (90\% C.L.).}
\newcommand{\PPPiPi}   {$p \bar{p} \pi^+ \pi^-$}
\newcommand{\PPTriPi}  {$p \bar{p} \pi^+ \pi^-\pi^0$}
\newcommand{\EtaPP}    {$\eta p \bar{p}$}
\newcommand{\OmegaPP}  {$\omega p \bar{p}$}
\newcommand{\PPKK}     {$p \bar{p} K^+ K^-$}
\newcommand{\PhiPP}    {$\phi p \bar{p}$}
\newcommand{\LamLam}   {$\Lambda \bar\Lambda$}
\newcommand{\LLPiPi}   {$\Lambda\bar\Lambda\pi^+\pi^-$}
\newcommand{\LamPK}    {$\Lambda \bar{p} K^+$}
\newcommand{\LamPKPiPi}{$\Lambda\bar{p}K^+\pi^+\pi^-$}
\begin{tabular}{lcc}\hline\hline
mode & $\sigma$ U.L.  & $\cal B$ U.L. \\
     & (pb)           & ($10^{-4}$)   \\ \hline
\PPPiPi   &    4.5 &  5.8 \\
\PPTriPi  &   14.4 & 18.5 \\
\EtaPP    &    4.2 &  5.4 \\
\OmegaPP  &    2.2 &  2.9 \\
\PPKK     &    2.5 &  3.2 \\
\PhiPP    &    1.1 &  1.3 \\
\LamLam   &    1.0 &  1.2 \\
\LLPiPi   &    2.0 &  2.5 \\
\LamPK    &    2.2 &  2.8 \\
\LamPKPiPi&    4.9 &  6.3 \\
\hline\hline
\end{tabular}
\end{table}

Since the masses of higher charmonium states ({\i.e.} the 
$\psi$-family members with mass above $\pspp$)
are above the threshold for open charm production, the $\bbbar$ branching
fractions are likely to be every small and it is unlikely that they 
will be measured.
Therefore, up to now, no attempts have been made nor have reports been 
forthcoming
for higher charmonium state decays into baryon-pair-containing final
states.\footnote{The only exception is for $Y(4260)$, the upper
limit at 90\% C.L. is reported: $\BR(Y\to \ppb)/\BR(Y\to \PPJP) <13\%$. }

\subsection{Three body decays}\label{xct_sun}
Compared with two-body decays, the theoretical analysis relevant
to three-body decays is more difficult and only special final
states have been discussed phenomenologically~\cite{Sinha:1984qn}.
Recently,  theoretical interest in baryon-containing final states
has been revived thanks to recent experimental discoveries, especially 
the phenomena of baryon-anitbaryon invariant mass enhancements
near thresholds.

\subsubsection{Baryon-antibaryon enhancements}
In charmonium decays, the BESII collaboration observed
 $\ppb$ and $p\overline{\Lambda}$ enhancements 
in $\jpsi \ra \gamma \ppb$~\cite{besgppb},
$\psp \ra \piz \ppb,~\eta \ppb$~\cite{bespaeppb}, and $\jpsi \ra p
\overline{\Lambda} K^- +c.c.$~\cite{bespkl}, as well as in $\psp
\ra p \overline{\Lambda} K^- +c.c.$ decays~\cite{bespkl}. 
In $B$ decays, many baryon-antibayron-pair-containing
final states have been measured by the CLEO, Belle, 
and BABAR collaborations,
such as $B^{\pm} \ra \ppb K^{\pm}$~\cite{belbppk},
$\biz \ra p \overline{\Lambda} \pi^-$~\cite{belbplpi},
$\bip \ra \LLb \kap$~\cite{belbllk},
$\biz \ra D^{*-} p \overline{n}$~\cite{cleobdcy},
$\overline{\biz} \ra D^{*0} \ppb,~D^{0} \ppb$~\cite{belbdpp},
$\bip \ra \ppb \pi^+,~\ppb K^{*+}$, $\biz \ra \ppb \kaz$~\cite{belbpppk},
$\biz \ra \overline{D}^{*0} \ppb,~\overline{D}^{0} \ppb$~\cite{bababdcy},
and so on, with enhancements observed in 
$\ppb$, $p\overline{\Lambda}$,
and $\LLb$ mass spectra. Other than the enhancement in $\jpsi \ra
\gamma \ppb$, which is claimed to be very narrow and below the
$\ppb$ mass threshold, all the other enhancements are slightly above the
baryon antibaryon mass threshold and have widths that range from a few 
tens of MeV/$c^2$ to less than 200~MeV/$c^2$.

To interpret these various enhancements in a uniform framework,
Yuan, Mo and Wang (YMW),
motivated by the Fermi-Yang-Sakata (FYS) model~\cite{fermi_yang,sakata},
suggested a nonet scheme to accommodate
the baryon-antibaryon enhancements observed in charmonium
and $B$ decays~\cite{wymbb9}. In the YMW approach,
the baryon and antibaryon are presumed to be bound together
by residual forces that originate from the strong interaction
between the quarks and gluons inside the baryon or antibaryon.
On one hand, the masses of the three-quark systems
(the baryon and the antibaryon) increase by a small amount
due to the residual forces required to form the bound state;
on the other hand, the binding energy between the two three-quark
systems reduces the mass of the baryon-antibaryon system to lower
than the sum of the masses of the three-quark systems, but very close to
the baryon-antibaryon mass threshold. This supplies a phenomenological
surmise, based on which the further analysis suggests many
experimental consequences.

\begin{table}[bthp]
\caption{\label{tab_cch}Decay modes containing
baryon-antibaryon nonets in charmonium decays. The first $J^P$ is
for the accompanying particle while the second for the
baryon-antibaryon resonance.}
\center
{\footnotesize
\begin{tabular}{clc} \hline \hline
             &~~~~~~~~decay mode                     & Note \\
                                                            \hline

$1^-\&0^{-}$  & $\roz\bppb,~\rop\bnpb,~\rom\bnbp$   &      \\
              & $K^{*+}\bpbl,~K^{*-}\bplb$          &$\ast$\\
              & $K^{*0}\bnbl,~\overline{K}^{*0}\bnlb$&$\ast$\\
              & $\omega\bppb$                       &      \\
              & $\phi\bllb$                         &$\ast$\\
 $1^+\&0^{-}$ & $b_1^0(1235)\bppb$                  &$\ast$\\
              & $b_1^+(1235)\bnpb,~b_1^-(1235)\bnbp$&$\ast$\\
              & $h_1(1170)\bppb,~h_1(1170)\bllb$    &$\ast$\\
              & $K_1^{+}(1270)\bpbl,~K_1^{-}(1270)\bplb$
                                                    &$\ast$\\
              & $K_1^{0}(1270)\bnbl,~
                 \overline{K}_1^{0}(1270)\bnlb$     &$\ast$\\
              & $K_1^{+}(1400)\bpbl,~K_1^{-}(1400)\bplb$
                                                    &$\ast$\\
              & $K_1^{0}(1400)\bnbl,~
                 \overline{K}_1^{0}(1400)\bnlb$     &$\ast$\\
 \hline
 $0^-\&1^{-}$ & $\piz\bppb,~\pip\bnpb,~\pim\bnbp$   &      \\
              & $\kap\bpbl,~\kaz\bnbl$              &      \\
              & $\kam\bplb,~\kzb\bnlb$              &      \\
              & $\eta\bppb,~\eta\bllb$              &      \\
              & $\etap\bppb,~\etap\bllb$            &$\ast$\\
 $0^+\&1^{-}$ & $a_0^0(980)\bppb$                   &      \\
              & $a_0^+(980)\bnpb,~a_0^-(980)\bnbp$  &      \\
              & $a_0^0(1450)\bppb$                  &$\ast$\\
              & $a_0^+(1450)\bnpb,~a_0^-(1450)\bnbp$&$\ast$\\
              & $f_0(980)\bppb,~f_0(980)\bllb$      &      \\
              & $f_0(1370)\bppb,~f_0(1370)\bllb$    &$\ast$\\
              & $\ko^{*+}(1430)\bpbl,~\ko^{*-}(1430)\bplb$
                                                    &$\ast$\\
              & $\ko^{*0}(1430)\bnbl,~
                 \overline{\ko}^{*0}(1430)\bnlb$    &$\ast$\\
 $1^+\&1^{-}$ & $a_1^0(1260)\bppb$                  &$\ast$\\
              & $a_1^+(1260)\bnpb,~a_1^-(1260)\bnbp$&$\ast$\\
              & $f_1(1285)\bppb,~f_1(1420)\bllb$    &$\ast$\\
              & $K_1^{+}(1270)\bpbl,~K_1^{-}(1270)\bplb$
                                                    &$\ast$\\
              & $K_1^{0}(1270)\bnbl,~
                 \overline{K}_1^{0}(1270)\bnlb$     &$\ast$\\
              & $K_1^{+}(1400)\bpbl,~K_1^{-}(1400)\bplb$
                                                    &$\ast$\\
              & $K_1^{0}(1400)\bnbl,~
                 \overline{K}_1^{0}(1400)\bnlb$     &$\ast$\\
 $2^+\&1^{-}$ & $a_2^0(1320)\bppb$                  &$\ast$\\
              & $a_2^+(1320)\bnpb,~a_2^-(1320)\bnbp$&$\ast$\\
              & $f_2(1270)\bppb$                    &$\ast$\\
              & $K_2^{*+}(1430)\bpbl,~K_2^{*-}(1430)\bplb$
                                                    &$\ast$\\
              & $K_2^{*0}(1430)\bnbl,~
                 \overline{K}_2^{*0}(1430)\bnlb$    &$\ast$\\
\hline \hline
\end{tabular} \\
$\ast$: not allowed in $\jpsi$ decays.} 
\end{table}

First, because of the large phase space, $B$-meson
decays play important roles in the
study of the baryon-antibaryon resonances. A detailed discussion
can be found in Ref.~\cite{wymbb9}.

Second, charmonium provides another domain to study baryon-antibaryon 
states.
Unlike $B$ decays, conservation laws limit the  possible decay modes
Some possible modes
are listed in Table~\ref{tab_cch}.
The production of a $0^-$ baryon-antibaryon bound states in
$\jpsi$ (or $\psp$) decays can be accompanied by a vector meson. For
the iso-vector bound states, one may look at $\rho N \overline{N}$
(nucleon antinucleon) final states,
including $\rho^+ n \overline{p}$, $\rho^0 p \overline{p}$ and
$\rho^- p \overline{n}$; for the iso-scalar bound state, one may look 
at the $\omega p \overline{p}$ final state; while for the strange
states, one may look at $K^{*+} \Lambda \overline{p} +
c.c.$ and $K^{*0} \Lambda \overline{n} + c.c.$ final states. The neutron or
anti-neutron, which is not detected, may be 
reconstructed by a kinematic fit to the rest of the tracks in the event. 
An $SU(3)$ singlet state can be searched for in the $\phi \LLb$
final state. The measurement of the $0^-$ baryon-antibaryon bound states
produced
together with an axial-vector meson is less promising since almost all 
of the axial-vector mesons are resonances.

The production of $1^-$ baryon-antibaryon bound states can be
accompanied by a pseudoscalar ($\pi$, $\eta$, $\eta'$, $K$),
scalar, tensor or axial-vector meson. The most promising way
to look for them is in the decays with a pseudoscalar meson:
{\it i.e.} analyze $\pi N \overline{N}$ for the iso-vector bound states;
$\eta p \overline{p}$ for iso-scalar bound state;
$K^{+} \Lambda \overline{p} + c.c.$ and $K^{0} \Lambda
\overline{n} + c.c.$ for the strange bound states. An $SU(3)$ singlet
bound state can be searched for in the $\eta' \LLb$ channel.

It should be noted that neutral non-strange $0^-$ baryon-antibaryon
bound states can also be produced via radiative decays of the $\jpsi$ (or
$\psp$); $1^-$ baryon-antibaryon bound states can not be
produced this way because of spin-parity conservation.

Although among various charmonium decays, the 
$\jpsi$ provides a good source of the
baryon-antibaryon bound states because of the large data samples,
it has some disadvantages: the phase space is small and there are
many $N^*$'s near the nucleon-meson mass threshold, and these affect the
identification of the states~\cite{Nstar}.
$\psp$ decays have larger phase space,
however the data samples are smaller, and there is a large fraction of
charmonium transitions. CLEOc and \bes3 will sudy these decays with
higher statistics, and partial wave analyses will be required to account 
for the $N^*$ contributions correctly.

At last, it worthwile to mention that another meticulous and mathematical
study of baryon-antibaryon enhancements
is described in Ref.~\cite{Ding:2005tr}, where the group theory
technique employed
in the Jaffe's studies on the $\rm{q}^2\overline{\rm{q}}^2$ system~\cite{Jaffe:2005md}
is extended to the $\rm{q}^3\overline{\rm{q}}^3$ system.

\subsubsection{Experiment results}

Some experimental results on three body decays involving baryon and
antibaryon pairs are summarized in Table~\ref{tab_3bdbary}. In fact, the
available results for three body baryonic decay are at present rather
limited. For $\jpsi$ decays, all quoted results are experimental
measurements before 1990; for $\psp$ decays, the results are mainly
due to BES~\cite{besmppb2,besmttat2} and CLEOc~\cite{cleomthd2}; for
$\pspp$ and $\chicj$ decays, the results are solely from
 CLEOc~\cite{cleochicj}. As to the energy region above the $\pspp$,
results are still unavailable as  mentioned above
in section~\ref{sxct_psppahi}.

However, even with these limited results, we
can see some interesting
phenomena. One interesting fact, as pointed out in Ref.~\cite{besmppb2},
is that for the $\piz\ppb$ final state, $Q_{\piz\ppb}= 12.2\pm 1.9$,
in agreement with the 12\% rule. In contrast,
for the $\eta \ppb$ final state, $Q_{\eta\ppb}=2.8\pm 0.6$, which
seems to be suppressed significantly. Another suppression seems in
effect for the $\omega \ppb$ final state, while for $\kap\pbL$ final
state, there is no suppression. Here more accurate data are needed for
further study.

\begin{table}[hbt]
\caption{\label{tab_3bdbary}
Branching ratios for three-body decays involving baryons for charmonium states.
Most results are from PDG06~\cite{PDG2006}, except for $\chicj$ decays, which
are from CLEOc~\cite{cleochicj}. The upper limits are given at 90\% C.L. 
and
$Q_h = \BR_{\psp\to h}/\BR_{\jpsi\to h }$ is also given for possible final states.
The numbers in parentheses are the ``reduced'' branching fractions, defined as
$ \tilde{\BR} = \BR/ PSF_3$, where $PSF_3$ is the phase space factor for the
three-body decay 
The units of $PSF_3$ are $10^1$ and $10^{-1}$ for the $\jpsi$ and $\psp$
decays respectively.
}
\center
\begin{tabular}{lllll}\hline \hline
 Final      &$\jpsi$             &$\psp$               & $Q_h$          &$\chicj~(J=0,1,2)$ \\
 State      &($10^{-4}$)         &($10^{-5}$)          &(\%)            & ($10^{-3}$) \\ \hline
$\piz\ppb$  &$10.9\pm 0.9(1.84)$ &$13.3\pm 1.7(6.83) $ &$12.2\pm 1.9$   & $0.59\pm 0.13$  \\
            &                    &                     &                & $0.12\pm 0.06$  \\
            &                    &                     &                & $0.44\pm 0.10$  \\
$\pim\pnb$  &$20.0\pm 1.0(3.40)$ &                     &                &                 \\
$\eta\ppb$  &$20.9\pm 1.0(7.23)$ &$~6.0\pm 1.2(4.48) $ &$~2.8\pm 0.6$   & $0.39\pm 0.12$  \\
            &                    &                     &                & $ <0.16 $       \\
            &                    &                     &                & $0.19\pm 0.08$  \\
$\roz\ppb$  &$~<3.7~~~~~~~~~~~~$ &$~5.0\pm 2.2(5.24) $ &$~>5.9~~~$      &                 \\
$\omega\ppb$&$13.0\pm 2.5(9.91)$ &$~6.9\pm 2.1(7.31) $ &$~5.3\pm 1.9$   &                 \\
$\phi \ppb$ &$~4.5\pm 1.5(15.6)$ &$~<2.4~~~~~~~~~~~ $  &$~<9.3~~~~~$    &                 \\
$\piz\LLb$  &$~2.2\pm 0.6(0.84)$ &                     &                &                 \\
$\kap\pbL$  &$~8.9\pm 1.6(4.65)$ &$10.0\pm 1.4(9.18) $ &$11.2\pm 2.5$   & $1.07\pm 0.21$  \\
            &                    &                     &                & $0.33\pm 0.10$  \\
            &                    &                     &                & $0.85\pm 0.18$  \\
$\kam p \overline{\Sigma}^0$
            &$~2.9\pm 0.8(2.05)$ &                     &               \\
$\kam p \overline{\Sigma}(1385)^0$
            &$~5.1\pm 3.2(10.2)$ &                     &               \\
$\pim \overline{p} \Delta^{++}(1231)$
            &$~1.6\pm 0.5(0.52)$ &                     &               \\
$\etap(958) \ppb$
            &$~9~~\pm~~4~(18.4)$ &               &               \\
\hline \hline
\end{tabular}
\end{table}

\subsection{Multi-body and Semi-inclusive decay}\label{xct_mtbdk}

\subsubsection{Multi-body decay}\label{sbxct_mtpwa}

By multi-body decays, we mean charmonium decays into a state with
at least four particles, including one pair of baryons. Since
baryon pair masses are around 2~GeV/$c^2$ or larger, the number
of allowed
decay modes is not large, especially for the low lying charmonium
states like the $\eta_c$ and $\jpsi$. Since the kaon mass is around
0.5~GeV/$c^2$, modes with kaons are further suppressed.

So far, only the $\pip\pim\ppb$ mode has been measured in the decays of
most of the charmonium states below open charm threshold, with
observations for $\jpsi$, $\psp$, and $\chicj$ decay; $\pip\pim\piz
\ppb$ is observed in $\jpsi$ and $\psp$ decays; while $\jpsi\to
\pip\pim \NNb$ was measured with 100\% uncertainty.

For final states with strangeness, $\pip\pim \LLb$ has only been
observed in $\psp$ decays.  Although it has been searched for in
$\chicj$ decays, no signal has been seen. An observed five-body decay
is $\psp\to \bar{p} K^+ \Lambda \pip \pim$. $K_S^0 K_S^0
\ppb$ has been searched for in $\chicj$ and $\psp$ decays, but
with no signal.  As a byproduct of a search for the penta-quark
state candidate ($\Theta(1540)$),  upper limits for $\jpsi$ and
$\psp\to K_S^0 p K^- \bar{n}$ branching fractions have been
determined.

Generally speaking, except for the fact that
the selection efficiency will not be
high because of the high multiplicity and low track momentum, the
identification of final states with four or more final 
particles including a baryon pair is quite easy and the purity is
relatively high. This makes it possible to study the interaction
between baryon and mesons, such as $N\pi\pi$, $N\pi\pi\pi$, $NK$,
$NK\pi$, $N\rho$, $N\omega$, those with the nucleon replaced by a
$\Lambda$ etc. However, as these excited Baryon states are
generally broad, and the mass difference between states is small,
the successful identifications of the states or the decay mode
require partial wave analyses, as has been performed for $\jpsi\to
p \bar{n} \pim + c.c.$ As one can imagine, when there are more
final states particles, the PWA will be correspondingly 
more difficult, more model-dependent, and, sometimes, more
unreliable~\cite{ycz_hadron05}.

\subsubsection{Semi-inclusive decays}\label{sbxct_inkdk}

\bnum

 \item Semi-inclusive $\eta_c$ and $\eta_c^{\prime}$ decays

In an earlier study of $\eta_c$ and $\eta_c^{\prime}$ hadronic decays,
Lee, Quigg, and Rosner advanced~\cite{bwLee} a statistical model appropriate
to particle decay according to which the mean mulitplicity
of decay products is
\beq
\begin{array}{rcl}
\langle n \rangle & =& n_0+
{\displaystyle \left( \frac{4}{\pi} \right)^{1/4}
\frac{\zeta(3)}{[3\zeta(4)]^{3/4}}
\left(\frac{E}{E_0} \right)^{3/4} } \\
   &=& n_0+0.528 (E/E_0)^{3/4}~.
\end{array}
\label{eq:avgnmb}
\eeq
Here $\zeta(x) \equiv \sum_{i=1}^{\infty} 1/i^x$, is the 
Riemann Zeta function and
$E$ is the energy availiable in excess of the rest 
mass of the lowest-multiplicity
($n_0$) decay channel. For the decay like 
$\eta_c \to \NNb+(m \mbox{ pions })^0$,
$E=(M_{\eta_c}-M_{N}-M_{\overline{N}})c^2$ and $n_0=2$. 
The scale $E_0$ is given
by the hadronic radius $R_0$:
\beq
E_0 \equiv \hbar c/R_0~.
\label{eq:ezrz}
\eeq
For a radius of 1 fm (typical for bag models of hadrons~\cite{aChodos}),
$E_0 \simeq 0.2$ GeV. As for the $\etac$ and $\etacp$ decay calculation, the
value $E_0 \simeq 0.17$ GeV is suggested in Ref.~\cite{cQuigg77}.
A Poisson distribution is asssumed for the variable ($\langle n \rangle-n_0$)
with neutral decay particles incorporated by means of statistical isospin 
weights. The calculated results for $\etac,\etacp \to \NNb + (m \mbox{ pions })$
can be found in Ref.~\cite{cQuigg77,mkGaillard75}, where
it is noted that the $\ppb$ mode accounts for about
5\% of all decays $\etac\to \NNb+\mbox{ pions }$. In the statistical
model~\cite{cQuigg77}, the semi-inclusive rates as fractions of the total
baryonic decay rate of $\etac$ are estimated to be
\[\begin{array}{lr}
\Gamma(\NNb+\mbox{pions})  & 44\% \\
\Gamma(\LLb+\mbox{pions})  &  7\% \\
\Gamma(\Lambda \overline{\Sigma} \mbox{ or } \overline{\Lambda} \Sigma +\mbox{pions})
                           & 24\% \\
\Gamma(\SSb+\mbox{pions})  & 20\% \\
\Gamma(\XXb+\mbox{pions})  &  6\% ~.
\end{array}\]
On this basis, the $\ppb$ mode is approximately 2\% of all the baryonic decays.
For $\jpsi$, the $\ppb$ mode makes up about 0.2\% of the direct hadronic
decays~\cite{PDG2006}. This suggests that the semi-inclusive baryonic decays
of $\etac$ makes up about 10\% of its hadronic width.

\item {The Lund model and general inclusive decays}

J{\scriptsize ETSET} and
P{\scriptsize YTHIA} are well-known Monte Carlo 
packages used for simulating inclusive (multihadron) final states at 
fairly high 
energies ({\it i.e.} around or above the $Z_0$ mass). 
The generation of inclusive final states in
J{\scriptsize ETSET} or P{\scriptsize YTHIA} is based on the famous Lund
model~\cite{Anderssonbk1998}, which adopts and synthesizes many physics 
models,
such as the massless relativistic string model, the dipole cascade model, the
linked dipole chain model, the Feynman parton model, 
the string fragmention model, etc.
The simulated results work fairly well at high energies.

Recently, by virtue of the basic Lund-model ``area law,'' a theoretical 
formalism has been
developed for simulation of inclusive hadronic final state at 
comparatively
low energies, including the charmonium 
region~\cite{Andersson:1999ui}.
This formalism has been used to generate hadronic final state for 
$R$-value measurement studies and the simulated results agree 
fairly well with data~\cite{huhm2001}.

We will not go into further discussion about the Lund model, 
the area law, and the details
involving Monte Carlo simulation here.  It is only necessary to
note that the Lund model is a fairly sucessful model for the generation
of  general inclusive decays.  Moreover, it provides an intuitive, 
visualized and  calculable decay model for hadronization processes 
and, in fact, provides us with many ideas for dynamics exploration. 
However, studies of inclusive final states in the charmonium region are
still rather limited, due to a dearth of data; much is left
to be done with \bes3.

\enum

\subsection{Summary and discussion}\label{xct_sum}

Herein we have collected and sumarized some theoretical ideas
concerning the charmonium dynamics and some phenomenological
models commonly used in high energy physics.

Experimental measurements are used to the extent possible to
test these theoretical predictions and provide clues for further
advances in understanding charmonium decay dynamics.

From the previous review, we notice that on one hand most of
the theoretical
explanations are still qualitative and on the other hand the experiment data
are rather limited and with fairly large uncertainties. Anyway,
with the unprecedentedly large data samples  forthcoming from
\bes3, we could make  systematic experimental investigations of
hadronic decays of charmonium, from the $\etac$ to 
to the $\pspp$ or even higher, both
on resonances and off resonances, for strong interactions and
electromagnetic interactions, covering phenomenology and dynamics, and
hopefully finding some empirical rules and establishing a universal 
model, or, even better, come across an unexpected discovery.

\section[$\psi(3770)$ and $\psi(4415)$ decays to $p\bar{p}$]
{$\psi(3770)$ and $\psi(4415)$ decays to $p\bar{p}$\footnote{T. Barnes}}
\label{sec:baryon_barnes}
\def\be{\begin{equation}}
\def\ee{\end{equation}}
\def\bd{\begin{displaymath}}
\def\ed{\end{displaymath}}
\def\ba{\begin{eqnarray}}
\def\ea{\end{eqnarray}}
\def\C{\rm C}
\def\D{\rm D}
\def\F{\rm F}
\def\I{\rm I}
\def\J{\rm J}
\def\L{\rm L}
\def\M{\rm M}
\def\P{\rm P}
\def\S{\rm S}
\def\T{\rm T}
\def\X{\rm X}


Another very interesting question that can be addressed using the
$\psi(3770)$ is the strength of the coupling of orbitally excited
$c\bar c$ states to $p\bar p$. This is a very important question
for the future PANDA experiment \cite{ted_GSI_report} at GSI,
which plans to use $p\bar p$ annihilation to produce excited
$c\bar c$ and charmonium hybrids. At present we have the
intriguing experimental observation that the L=1 $c\bar c$
$\chi_J$ states couple much more strongly to $p\bar p$ than the
L=0 $J/\psi$, but whether this trend continues to L=2 is an open
question. BES can easily answer this question through a
high-statistics search for $\psi(3770) \to p\bar p$. Charmonium
decays to final states including a $p\bar p$ pair, such as $\Psi
\to p\bar p$ and $\Psi \to p\bar p m $ (where $\Psi$ is a generic
charmonium or charmonium hybrid resonance and $m$ is a light
meson) are also very interesting in this regard, and can be used
to estimate the associated production cross section for $p \bar p
\to m \Psi $ (see
Refs.\cite{ted_Lundborg:2005am,ted_Barnes:2006ck}); this reaction
in particular will be used by PANDA to search for
J$^{\P\C}$-exotic charmonium hybrids.

\chapter[Rare and forbidden charmonium decays]
{Rare and forbidden charmonium decays\footnote{By Haibo Li,
Jianping Ma and Xinmin Zhang}}
\label{chpt:rare}
\indent At present, two general trends can be distinguished in
accelerator particle physics. On one hand,  very high energy
accelerators, for example the LHC,  provide the ability
to explore physics at high energy frontier.
On the other hand, smaller experiments at lower
energies but with very high intensities and low
backgrounds, for example the $B$ factories and \bes3 ,
provide capabilities for
performing precise tests and accurate determinations of many
parameters of the Standard Model (SM).  Moreover, the close scrutiny of
rare
processes may illuminate new physics in a complementary fashion to
high-energy colliders.

\indent With huge $J/\psi$ and $\psi(2S)$ data samples, the \bes3
experiment will approach a precision level where rare $\psi$
decays can used to provide important tests of the Standard Model, with
the accompanying possibilty for uncovering new-physics induced deviations.


\section{Weak Decays of Charmonium}
\label{sec:weak_cc_decays}

\indent  The low-lying charmonium states, {\it i.e.} those below the
open-charm threshold, usually decay through intermediate photons or
gluons produced by the annihilation of
the parent $c\bar{c}$ quark pair.
These OZI-violating but flavor-conserving decays result in narrow
natural widths of the $J/\psi$ and $\psi(2S)$ states. In the SM
framework, flavor-changing weak decays of these states are also possible,
although they are expected to have rather low branching fractions.
The huge $J/\psi$ data samples at \bes3 will provide opportunities to
search for such rare decay processes, which in some cases
may be detectable, even at SM levels.
The observation of an anomalous production rate for single charmed mesons
in $J/\psi$ or $\psi(2S)$ decays at \bes3 would be a hint of
possible new physics, either in underlying continuum processes via
flavor-changing-neutral-currents~\cite{lihb_neutral} or in the
decays of the $\psi$ resonances due to
unexpected effects of quark dynamics~\cite{lihb_weak}.

\subsection{ Semileptonic Decays of Charmonium}

\indent  The inclusive branching fraction for $J/\psi$ weak decays
via a single quark (either the $c$ or the $\bar{c}$) has been
estimated to be $(2\sim4) \times 10^{-8}$ 
by simply  
using the $D^0$ lifetime~\cite{lihb_sl_psi}. Such a
small branching fraction  makes the observation of weak decays
of the $J/\psi$ or $\psi(2S)$ quite challenging, despite the expected
cleanliness of the events. However, BEPC-II, running at design
luminosity, will produce of order  $10^{10}$
$J/\psi$ events per year of data taking,
leading to $ \cong 400$ weak decays for the
predicted SM branching fraction.   The semi-leptonic decay of
a $c\bar{c}~(1^{--})$ vector charmonium state below
the open-charm threshold is induced by
the weak quark-level transition $c \rightarrow q W^*$, where $W^*$
is a virtual intermediate boson. Hence, the
accessible exclusive semi-leptonic channels are:
\begin{eqnarray}
\psi(nS) \rightarrow D_q l \nu,
\label{eq:semi_c_allowed_1}
\end{eqnarray}
\begin{eqnarray}
\psi(nS) \rightarrow D_q^* l \nu,
\label{eq:semi_c_allowed_2}
\end{eqnarray}
where $n = 1$~or~$2$, and $q$ can be either a $d$- or $s$-quark, which
corresponds to a $D^{\pm}$(Cabibbo-suppressed mode) or $D_s$   
(Cabibbo-allowed mode) meson.
 Semi-leptonic weak decays of the
$J/\psi$ will offer several advantages over the purely hadronic ones
from both the experimental and theoretical points
of view: the prompt charged lepton $l = e, \mu$ can be used to tag the
events,
removing a large fraction of conventional $\psi(nS)$ hadronic decays.
In addition, the missing energy due to the escaping neutrino can be also
exploited to remove backgrounds.  The identification of the charm meson in
the final state would provide an unambiguous signature
of the semi-leptonic weak decays of $\psi(nS)$.
Meanwhile, decays of the excited
mesons $D^*_s$ and $D^{*\pm}$ produced in
reaction (\ref{eq:semi_c_allowed_2}) would
provide useful additional experimental handles.  In the lab system,   
the detectable photons from the $D^{*\pm}_s \rightarrow D^{\pm}_s \gamma$,
radiative transition are  in the 90$\sim$200~MeV energy interval. These,
and
the soft pion produced from
$D^{*\pm}$ in $D^{*\pm} \rightarrow D^{0} \pi^{\pm}$ decay,
can provide powerful constraints to help identify a $D_s$ or
$D^0$ meson produced in the weak decay of a charmonium state.

A specific calculation based on Heavy Quark Spin Symmetry
(HQSS)~\cite{lihb_hqss_1, lihb_hqss_2},
gives branching fractions for the Cabibbo-allowed
mode of~\cite{lihb_sl_psi}:  
\begin{eqnarray}
BR(J/\psi \rightarrow D_s l \nu) &\cong& 0.26 \times 10^{-8},  \nonumber 
\\
BR(J/\psi \rightarrow D^*_s l \nu) &\cong& 0.42 \times 10^{-8}.
\label{eq:semi_c_expect}
\end{eqnarray}
Summing over both modes, one gets a total $BR \cong 0.7 \times 10^{-8}$,
which  is about 20\% of the expected total rate for weak decays ($4 \times
10^{-8}$) estimated in
Ref.~\cite{lihb_sl_psi}. Taking into account the overall theoretical
uncertainty ($\cong 40\%$ ), the expected branching ratios are
within the range $(0.4 \sim 1.0) \times 10^{-8}$.  For the
Cabibbo-suppressed
decay modes, one can obtain the following ratio:
\begin{eqnarray}
 \frac{BR(J/\psi \rightarrow D_s l \nu)}{BR(J/\psi \rightarrow D^{\pm}
l^{\mp} \nu)} =
 \frac{BR(J/\psi \rightarrow D^*_s l \nu)}{BR(J/\psi \rightarrow
D^{*\pm} l^{\mp} \nu)}
\cong  \frac{|V_{cs}|^2}{|V_{cd}|^2} \cong 20,
\label{eq:semi_c_ratio}
\end{eqnarray}
where $V_{cs}(\cong 1.0)$ and $V_{cd}(\cong 0.22)$
denote the relevant Cabibbo-Kobayashi-Mashkawa (CKM) mixing matrix
elements.

\subsection{Two-body Weak Hadronic Decays of Charmonium}

Non-leptonic,  two-body weak hadronic decays of the
$J/\psi$ and $\psi(2S)$  are addressed
in the context of the factorization scheme
for both the Cabibbo-allowed ($c \rightarrow s$) and
Cabibbo-suppressed ($c \rightarrow d$) quark-level transitions
in Ref.~\cite{lihb_twobodys_weak}.  There, expressions for branching
fractions for $\psi \rightarrow PP/PV$ decays
(where $P$ and $V$ represent pseudoscalar mesons and vector
mesons, respectively) are given. Using the decay rate formula (5) in
Ref.~\cite{lihb_twobodys_weak},  the
$\psi \rightarrow PP$ branching ratio values
listed in Table~\ref{tab:psi_pp} are computed.
Among the Cabibbo-allowed decays, one
finds that the dominant  mode is $\psi \rightarrow D^+_s \pi^-$,
with a branching ratio
\begin{eqnarray}
BR(\psi \rightarrow D^+_s \pi^-) = 0.87 \times 10^{-9};
\label{eq:twobodys_weak_large}
\end{eqnarray}
the next strongest is $\psi \rightarrow D^0 K^0$, with a branching
ratio
\begin{eqnarray}
BR(\psi \rightarrow D^0 K^0) = 0.28 \times 10^{-9}.
\label{eq:twobodys_weak_large2}
\end{eqnarray}
The branching ratios of $\psi \rightarrow PV$ decays in
Cabibbo-allowed and Cabibbo-suppressed channels are listed in
Table~\ref{tab:psi_pv}.
For the color enhanced modes of the Cabibbo-allowed channels, one finds
\begin{eqnarray}
BR(\psi \rightarrow D^+_s \rho^-) = 0.36 \times 10^{-8},
\label{eq:twobodys_weak_large_3}
\end{eqnarray}
which is higher than the $\psi \rightarrow D^+_s \pi^-$
branching ratio.
The following
relative ratio
is used in the discussions below~\cite{lihb_twobodys_weak}
\begin{eqnarray}
\frac{BR(\psi \rightarrow D^+_s
\rho^-)}{BR(\psi \rightarrow D^+_s \pi^-)} = 4.2.
\label{eq:twobodys_weak_ratio}
\end{eqnarray}
The $\psi \rightarrow D^+_s \rho^-$ mode may be
measureable at the SM level at \bes3.
\begin{table}[htbp]
   \caption{ SM branching fractions for $\psi \rightarrow PP$
   from Ref.~\cite{lihb_twobodys_weak}. The
transition mode, $\Delta C = \Delta S = +1$,
    corresponds to Cabibbo-allowed decay modes, while $\Delta C = +1$,
$\Delta S = 0$ corresponds to
    Cabibbo-suppressed decay modes.}
  \label{tab:psi_pp}
  \centering
  \begin{tabular}{c|c|c}
\hline
Transition Mode & Decay Modes & Branching ratio($\times 10^{-10}$)
\\ \hline
  $\Delta C = \Delta S = +1$ & & \\
                             & $\psi \rightarrow D^+_s\pi^-$ & 8.74 \\
                             & $\psi \rightarrow D^0 K^0$ & 2.80 \\
  \hline
  $\Delta C = +1$, $\Delta S = 0$ & &  \\
                                  & $\psi \rightarrow D^+_s K^-$ & 0.55 \\
                                  & $\psi \rightarrow D^+ \pi^-$ & 0.55 \\
                                  & $\psi \rightarrow D^0 \eta$ & 0.016 \\
                                 & $\psi \rightarrow D^0 \eta^{\prime} $ &
0.003 \\
                                  & $\psi \rightarrow D^0 \pi^0$ & 0.055 
\\
 \hline
  \end{tabular}
\end{table}
   
\begin{table}[htbp]
   \caption{ SM branching fractions for $\psi \rightarrow PV$
            from Ref.~\cite{lihb_twobodys_weak}. The
transition mode, $\Delta C = 
           \Delta S = +1$, corresponds to Cabibbo-allowed decay
            modes, while $\Delta C = +1$, $\Delta S = 0$ corresponds
            to Cabibbo-suppressed decay modes.}
  \label{tab:psi_pv}
  \centering
  \begin{tabular}{c|c|c}
\hline   
Transition Mode & Decay Modes & Branching ratio($\times 10^{-10}$)
\\ \hline
  $\Delta C = \Delta S = +1$ & & \\
                             & $\psi \rightarrow D^+_s\rho^-$ & 36.30 \\
                               & $\psi \rightarrow D^0 K^{*0} $ & 10.27 \\
  \hline
  $\Delta C = +1$, $\Delta S = 0$ & &  \\
                                  & $\psi \rightarrow D^+_s K^{*-}$ &2.12 
\\
                                  & $\psi \rightarrow D^+ \rho^- $ &2.20 
\\
                                 & $\psi \rightarrow D^0 \rho^0$ & 0.22 \\
                                  & $\psi \rightarrow D^0 \omega$ & 0.18 
\\
                                  & $\psi \rightarrow D^0 \phi $ &0.65 \\
 \hline
  \end{tabular}
\end{table}

The $\psi(nS)$ semi-leptonic decay modes can be related to   
the two-body
hadron decay modes by applying both  spin symmetry and the non-recoil
approximation to the semi-leptonic decay
rates~\cite{lihb_sl_psi}. For $J/\psi
\rightarrow
D^+_s( D^{*+}_s) \pi^-$ decay modes, $q^2 = (p_{\psi} - p_{D})^2 =
m_{\pi}^2$ (here $p_{\psi}$ and $p_{D}$ are the four momenta
of the initial and final state heavy mesons) and, assuming
factorization as suggested
by Bjorken~\cite{lihb_bjorken} for $B$ decays, and in the non-recoil
approximation for the hadronic transition amplitudes~\cite{lihb_voloshin},
equations (7) and (8) in Ref.~\cite{lihb_sl_psi} give
the relation between relative branching ratios:
\begin{eqnarray}
r=\frac{BR(J/\psi \rightarrow D^{*+}_s
\pi^-)}{BR(J/\psi \rightarrow D^+_s \pi^-)} \cong
\left[\frac{d\Gamma(J/\psi \rightarrow D^{*+}_s
l^- \nu )/dq^2}{d\Gamma(J/\psi \rightarrow D^+_s l^- \nu)/dq^2}\right
]_{q^2 = m^2_{\pi}} \cong 3.5.
\label{eq:twobodys_weak_ratio_2}
\end{eqnarray}
If a $\rho$ is substituted for the $\pi$ one gets $r \cong 1.4$.
In this way, the
estimated  
branching ratios in Table~\ref{tab:psi_pp} for $\psi(nS) \rightarrow PP$
channels can
be related to  $\psi(nS) \rightarrow VP$ channels with the pseudoscalar
charm mesons replaced by vector charm mesons.

All of the above estimates show an overall enhancement of final-state
vector charm mesons with respect to the pseudoscalar ones.  This suggests
the use of $D^*_s$ or $D^{*\pm}$ as signals in searches for
weak decays of the $\psi(nS)$ in non-leptonic decay channels as well.

\subsection{Searches at \bes3}

At \bes3, assuming a $10^{10}$ $J/\psi$ event sample,
the central value $BR \cong 0.7 \times 10^8$
translates into about 70 semi-leptonic decay events of the type $J/\psi
\rightarrow D_s (D^*_s) l \nu$. The following event selection
criteria would be useful for
searches for such exclusive semi-leptonic channels :
\begin{itemize}
 \item The prompt charged lepton can be used to tag the weak decay:  
in order to suppress cascade decay backgrounds from $J/\psi$
strong decays, the tagging lepton momentum could be required
to be between 0.5 GeV and 1.0 GeV, close
to the upper kinematic limit for the decay under consideration.  High
quality lepton discrimination from
charged pions or kaons is needed for the measurement.
 \item The missing mass of the reconstructed candidates must be consistent
with the (nearly) zero mass of the undetected neutrino.
 \item The reconstruction of a $D_s$ or $D^{\pm}$ meson would provide  
an unambiguous signature for a weak decay
of a below-open-charm threshold $\psi(nS)$.  Good invariant mass
resolution of the ${D_s}$ decay products will be important for reducing
combinatorial backgrounds.
 \item Soft photons in the energy interval (90$\sim$200)~MeV
from the $D^*_s\rightarrow \gamma D_s$ transition and
soft charged pions from $D^{*\pm}\rightarrow \pi^{\pm}D$ decay can
provide further suppression of combinatorial backgrounds: the
additional
constraint of an intermediate $D^*_s$ state would reconfirm the $D_s$
signal.
\end{itemize}

 In general, exclusive hadronic decays are probably too
tiny to look for in any specific fully reconstructed
decay channel. Therefore, it
seems that an inclusive search for $J/\psi \rightarrow D^*_s + X$    
at \bes3 may be more fruitful. The $\gamma$ from the
decay of a $D^*_s$ meson should
be useful as a kinematic constraint to clean up
any $D_s$ meson signal, as discussed in~\cite{lihb_sl_psi}.

 Finally, in the Standard Model, weak flavor-changing-neutral-current
(FCNC) $J/\psi$ decays are predicted to be unobservably
small~\cite{lihb_weak} and, thus, any observation of such would
provide a signal for new physics.
In Ref.~\cite{lihb_weak}
the predictions of various models,
such as TopColor models, minimal supersymmetric standard
model (MSSM) with R-parity violation and a general two-Higgs-doublet 
model, are discussed. These authors find that the
branching fraction for
$J/\psi \rightarrow D/\bar{D} X_u$, which is mediated by
the weak $c \rightarrow u$ transition, could be as large as
$10^{-6}\sim 10^{-5}$ in some new physics scenarios.

At \bes3  it will be difficult to isolate pure,
$c \rightarrow u$ mediated, hadronic $J/\psi \rightarrow D/\bar{D}
X_u$ decays. On the other hand the decays $J/\psi
\rightarrow D^0/\bar{D^0} l^+ l^-$ ($l= e$, $\mu$) and $J/\psi
\rightarrow D^0/\bar{D^0} \gamma $ decays, which are
also dominated by FCNC processes, would be quite distinct.

\section{Search for the invisible decays of Quarkonium}
\label{sec:invisible_cc_decays}

Invisible decays of quarkonium states such as the $J/\psi$ and the
$\Upsilon$ etc., offer a window into what may lie beyond the Standard
Model~\cite{lihb_Part4_Fayet:1979qi,lihb_Part4_Fayet:2006sp}.
The reason is that other than
neutrinos, the SM includes no other invisible final
particles that these states can decay into.  BESII explored such a window
by establishing the first experimental
limits on invisible decays of the $\eta$ and $\eta'$, which complemented
the limit of $2.7 \ 10^{-7}$ that was previously established
for invisible  decays of
the $\,\pi^\circ$~\cite{lihb_Part4_Artamonov:2005cu}.

Some theories of beyond the SM physics predict new
particles with masses that are accessible at \bes3, such as the
light dark matter (LDM) particles
discussed in Ref.~\cite{lihb_Part4_Boehm:2003hm}. These
can have the right relic abundance to constitute the nonbaryonic dark
matter of the Universe, if they are coupled to the SM via a new
light gauge boson $U$~\cite{lihb_Part4_Fayet:1980ad}, or exchanges of 
heavy
fermions.  A light neutralino with
a coupling to the SM that is mediated by a light scalar singlet in the
next-to-minimal supersymmetric standard model
has also been considered~\cite{lihb_Part4_Ellis:1988er}.

These considerations have received a boost in interest by the
recent observation of a bright~511 keV $\gamma$-ray line from the
galactic bulge reported by the SPI spectrometer on the
INTEGRAL satellite~\cite{lihb_Part4_Jean:2003ci}.  The corresponding 
galactic
positron flux, as well as the smooth symmetric morphology of the
511~keV $\gamma$ emission, could be interpreted as originating from the
annihilation
of LDM particles into $e^+e^-$
pairs ~\cite{lihb_Part4_Boehm:2003hm,lihb_Part4_Beacom:2004pe}.
It would be very interesting to see evidence for
such light invisible particles in collider
experiments.  CLEO gave an upper bound on $\Upsilon(1S) \rightarrow
\gamma + \mbox{invisible}$, which is sensitive to dark matter
candidates lighter than about 3 GeV/$c^2$~\cite{lihb_Part4_Balest:1994ch}, 
and
also provided an upper limit on the axial coupling of any new $U$
boson to the $b$ quark.  It is important, in addition, to search for the
invisible decays of othr light quarkonium states
($q\bar{q}$, $q= u$,$d$, or $s$ quark), since these can be used to
constrain the masses of LDM particles and the couplings of a
$U$ boson to the light quarks~\cite{lihb_Part4_Fayet:2006sp}.

 It has been shown
that measurements of the $J/\psi$ invisible decay widths can
be used to constrain new physics
models~\cite{lihb_Part4_ng_invisible}.
It is straightforward for one to calculate the SM ratio of
branching fractions for $J/\psi$  invisible decays and its
measured branching fraction for
decays into electron-positron pairs~\cite{lihb_Part4_ng_invisible}.
Within the SM, the invisible mode consists solely of
decays into the three types of neutrino-antineutrino pairs. Neglecting
polarization effects and taking into account $e^+e^-$ production
through a photon only, one gets~\cite{lihb_Part4_ng_invisible}:
\begin{eqnarray}
\frac{\Gamma(J/\psi \rightarrow
\nu\bar{\nu})}{\Gamma(J/\psi \rightarrow e^+e^-)} =   
\frac{27 G^2 M^4_{J/\psi}}{256
\pi^2\alpha^2} \left(1-\frac{8}{3}sin^2(\theta_W \right)^2 = 4.54 \times
10^{-7},
\label{eq:invisible_ratio}
\end{eqnarray}
where $G$ and $\alpha$ are the Fermi and fine-structure constants, 
respectively, and $M_{J/\psi}$ is the $J/\psi$ mass. The
uncertainty
of the above relation is about 2$\sim$3\% and comes mainly from
corrections to the $J/\psi$ wave function, $e^+e-$ production
via the $Z$ boson and electroweak radiative
corrections~\cite{lihb_Part4_ng_invisible}.

At \bes3, one can tag charmonium states that decay invisibly
by looking for a particular cacscade transition, such as $\psi(2S)
\rightarrow \pi^+\pi^- J/\psi$, $\psi(2S) \rightarrow \gamma \chi_c$ and
so on, where the soft $\pi^+\pi^-$ pairs or the monoenergetic radiative
$\gamma$ serves as a tag for the invisibly decaying $J/\psi$ or
$\chi_c$ state.
A list of potentially useful decay chains is provided in
Table~\ref{tab:invisible_mode_cc}.
\begin{table}[htbp]
  \caption{ $\psi(2S)$ and $J/\psi$ decay modes that can be used
to search for invisible
           decays of the $J/\psi$, $\chi_{c0}$, $\chi_{c1}$,
$\chi_{c2}$, $\eta_c(1S)$ and $\eta_c(2S)$.
            The branching fractions are taken from the
PDG~\cite{lihb_Part4_PDG_2004}. For
each mode, a
            ``tagging topology" is given, which is the set of visible
particles that are seen within the detector's acceptance. In
            each case the tagging topology has well defined kinematics.
           The number of events are the expected event yield in a 3
billion $\psi(2S)$ (10 billion $J/\psi$) data set, in which we did not 
consider the decay probabilities of
            the tagging particles.  }
  \label{tab:invisible_mode_cc}
  \centering
  \small
  \begin{tabular}{c|c|c|c|c}\hline
\hline
$\psi(2S)$ & Branching  &Number of events  & Invisible & Tagging  
 \\
    decay mode & fraction ($10^{-2}$) & /3 billion $\psi(2S)$s
 & decay mode & topology  \\\hline

   $\psi(2S) \rightarrow \pi^+ \pi^- J/\psi$ & $31.7 \pm 1.1$ & $9.3\times
10^8$ & $J/\psi \rightarrow \mbox{invisible}$
    & $\pi^+ \pi^-$  \\ \hline
  $\psi(2S) \rightarrow \pi^0 \pi^0 J/\psi$ & $18.6 \pm 0.8$& $5.6\times
10^8$ & $J/\psi \rightarrow  \mbox{invisible}$
    & $\pi^0 \pi^0$  \\ \hline
   $\psi(2S) \rightarrow \eta J/\psi$ & $3.08 \pm 0.17$ &$9.3
\times 10^7$ & $J/\psi \rightarrow  \mbox{invisible}$ 
    & $\eta$ \\ \hline
  $\psi(2S) \rightarrow \pi^0 J/\psi$ & $0.123 \pm 0.018$ &$3.7
\times 10^6$ & $J/\psi \rightarrow  \mbox{invisible}$
    & $\pi^0$  \\ \hline
  $\psi(2S) \rightarrow \gamma \chi_{c0}$ & $9.0 \pm
0.4$ &$2.7\times 10^8$ & $\chi_{c0}  \rightarrow  \mbox{invisible}$
    & $\gamma$ \\ \hline
  $\psi(2S) \rightarrow \gamma \chi_{c1}$ & $8.7 \pm 0.5$ &$2.6
\times 10^8$ & $\chi_{c1}  \rightarrow  \mbox{invisible}$
    & $\gamma$  \\ \hline
 $\psi(2S) \rightarrow \gamma \chi_{c2}$ & $8.2 \pm 0.3$ &$2.5
\times 10^8$ & $\chi_{c2} \rightarrow  \mbox{invisible}$
    & $\gamma$  \\ \hline
 $\psi(2S) \rightarrow
\gamma \eta_c(1S)$ & $0.26 \pm 0.04$ &$7.8\times 10^6$ & $\eta_c(1S)
\rightarrow  \mbox{invisible}$
    & $\gamma$  \\ \hline
 $J/\psi \rightarrow \gamma \eta_c(1S)$ & $1.3 \pm 0.4$ & $1.3
\times 10^8$ & $\eta_c(1S) \rightarrow  \mbox{invisible}$
    & $\gamma$ \\ \hline
    
  \end{tabular}
\end{table}
\begin{figure}[htbp]
  \centering
  \includegraphics[width=0.5\textwidth,
height=0.2\textheight]{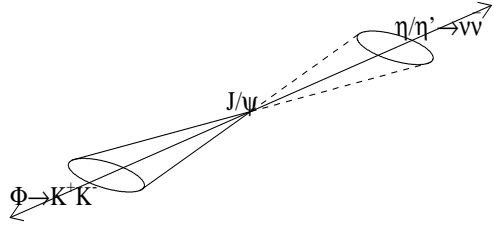}
  \caption{ The demonstration of  J/$\psi \rightarrow \phi\eta$ or
           $\phi\eta^{\prime}$, the $\phi$,  which is reconstructed in the
           $K^+K^-$ decay mode, and can be used to tag the invisible 
decays
           of the missing particles. }
  \label{fig:tag}
\end{figure}

It is also interesting to search for invisible decays of the $\eta$,
$\eta^{\prime}$,  $\rho$, $\omega$ and $\phi$ light mesons,   
using  two-body decay modes of the $J/\psi$.  For example, the
two-body decay modes $J/\psi \rightarrow \phi\eta$ or
$\phi\eta^{\prime}$ can be selected using only the very clean and
distinct $\phi\rightarrow K^+K^-$ decays, which then tag the
presence of an $\eta$ or $\eta^{\prime}$ meson
that has decayed into an invisible final state,
as illustrated in Fig.~\ref{fig:tag}.    Since both the $\phi$ and
$\eta$ ($\eta^{\prime}$) have natural widths that are negligible
compared with the detector resolution, the shape of the momentum
distribution of the $\phi$ is approximately Gaussian.  The mean value
of the $\phi$ momentum distribution is 1.320 GeV/$c$ for 
$J/\psi\to\phi\eta$
and 1.192 GeV/$c$ for $J/\psi\to\phi\eta^\prime$.  The missing
momentum, ${P}_{miss} = |\vec{P}_{miss}|$, is a powerful
discriminating variable to separate signal events from possible
backgrounds in which the missing side is not from an $\eta$
($\eta^{\prime}$). Here, $\vec{P}_{miss} = -
\vec{P}_{\phi}$. In addition, the regions of the detector
where the $\eta$ and $\eta^{\prime}$ decay products
are expected to go are easily defined thanks to the strong
boost from the $J/\psi$ decay, as illustrated in
Fig.~\ref{fig:tag}.
\begin{figure}[htbp]
  \centering
  \begin{minipage}{1.0\linewidth}
\includegraphics[width=\textwidth,height=0.25\textheight]{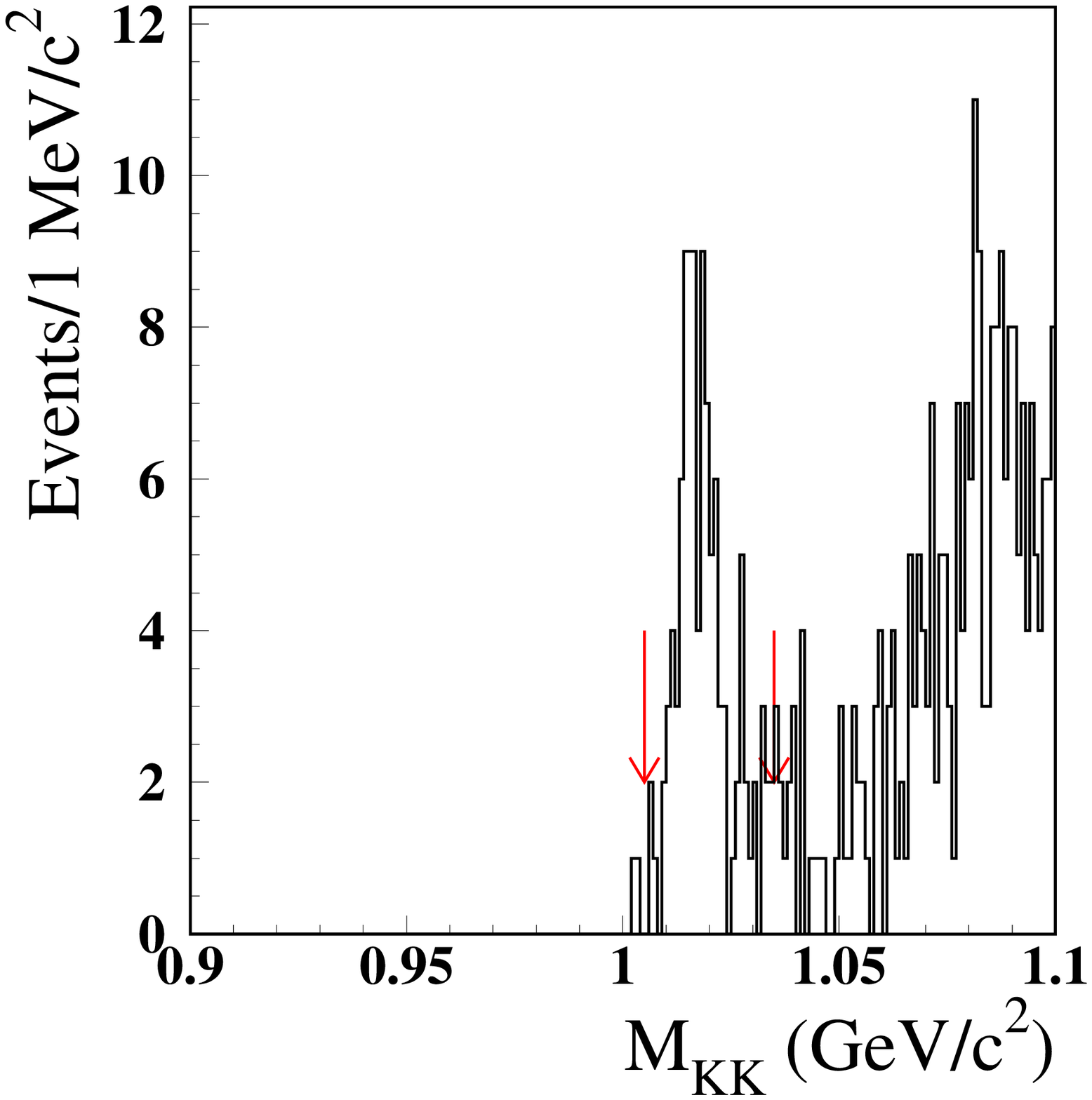}
  \put(-60,140){(a)}
  \end{minipage}
  \begin{minipage}{1.0\linewidth}
\includegraphics[width=\textwidth,height=0.25\textheight]{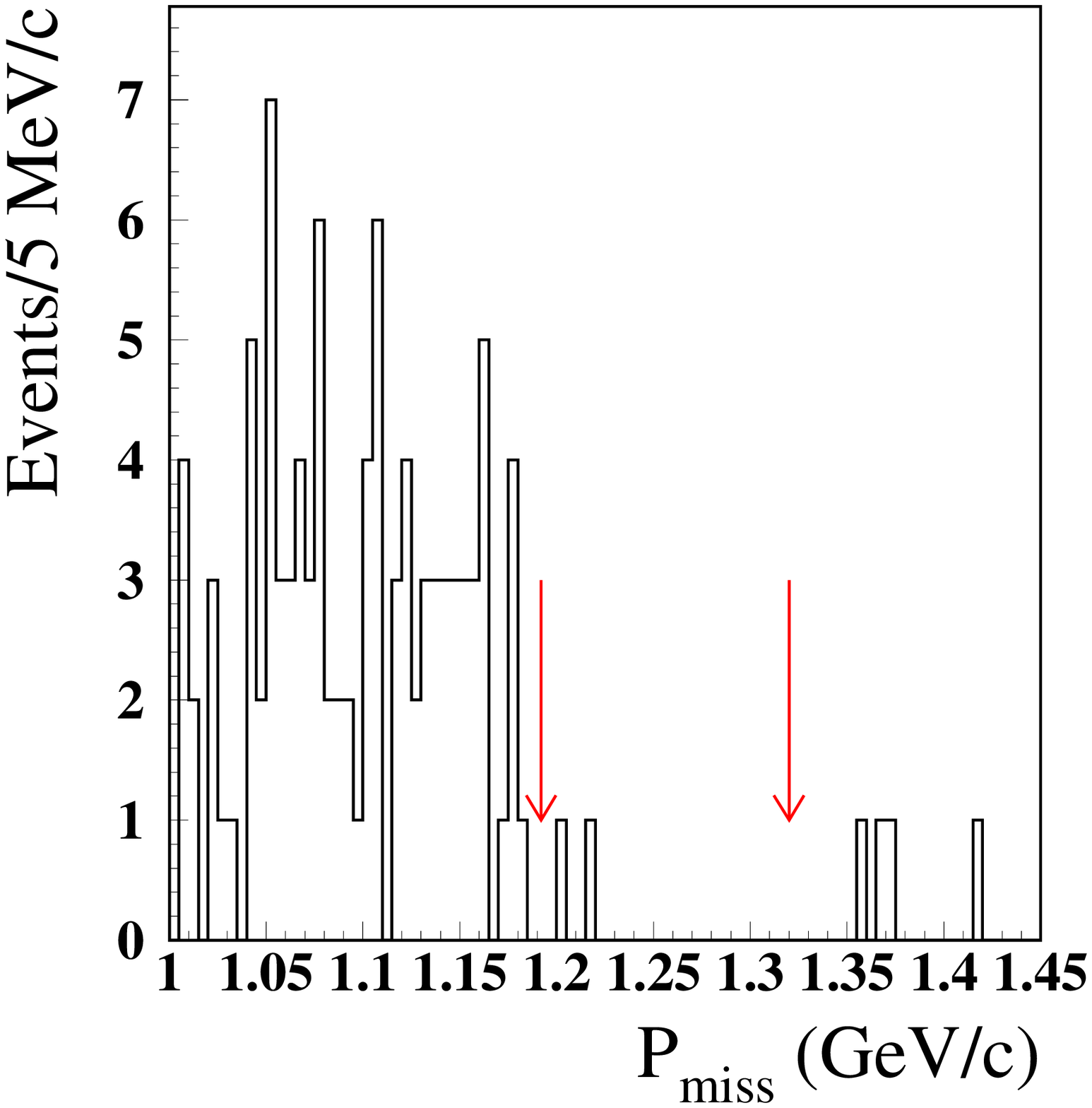}
  \put(-60,140){(b)}
  \end{minipage}
  \caption{(a) The $m_{KK}$ distribution for candidate events.
    The arrows on the plot indicate the signal region of $\phi$   
    candidates.  (b) $P_{miss}$ distribution for the events with
    $1.005<m_{KK}<1.035$ GeV/$c^2$ in (a).  The mean vaues of the missing
    momenta  for $J/\psi\to\phi\eta$ and
    $J/\psi\to\phi\eta^\prime$ are located around 1.32 and 1.20
    GeV/$c$, respectively, as indicated by the two arrows.}   
  \label{fig:data}
\end{figure}

Using $\phi\rightarrow K^+K^-$-tagged  $J/\psi \rightarrow
\phi \eta$ and $\phi \eta^{\prime}$ decays
in a 58 million $J/\psi$ event sample, BES II  searched for
invisible decays of the
$\eta$ and  $\eta^{\prime}$ mesons~\cite{lihb_bes-ii-invisible}.
Candidate events are those where the only charged tracks seen
in the detector are the $K^+K^-$ pair from the $\phi$ decay.
In addition, events with any
neutral energy custers within a $30^0$ cone around the
axis of the charged tracks (see Fig.~\ref{fig:tag}) are rejected.
The $M(K^+K^-)$ distribution for selected events is shown
in Fig.~\ref{fig:data}(a), where a $\phi\rightarrow K^+K^-$
peak is evident at the $\phi$ mass (1.02~GeV/$c^2$).  The
$P_{miss}$ distribution for events in the $\phi$ signal
region (indicated by the arrows in Fig.~\ref{fig:data}(a))
is shown in Fig.~\ref{fig:data}(b).   No evidence is seen
for either $\eta$ or $\eta^{\prime}$ invisible decays.
    
An unbinned extended maximum likelihood (ML) fit was used to extract
the event yield for J/$\psi \rightarrow \phi\eta (\eta^{\prime})$
[$\phi \rightarrow K^+K^-$ and $\eta(\eta^{\prime}) \rightarrow
{\rm invisible}$].  In the ML fit, $P_{miss}$
was used as the discriminating
variable and the signal region was defined as  $1.00
<P_{miss}<1.45$ GeV/$c$, shown in Fig.~\ref{fig:data}(b),
over which the background shape was well understood.
Probability density
functions (PDFs) for the $P_{miss}$ distributions for
($\mathcal{F}^{\eta}_{sig}$ and $\mathcal{F}^{\eta^{\prime}}_{sig}$)
signals and background ($\mathcal{F}_{bkgd}$)
were constructed using detailed MC
simulations of signal and background.  The PDFs for the signals were
parameterized by double Gaussian distributions with common means, one
relative fraction and two distinct widths, which are all fixed to the
MC simulation. The PDF for the background was a bifurcated Gaussian plus a
first-order polynomial ($P_1$).  All parameters related to the   
background shape were floated in the fit to data. The PDFs for signals
and background were combined in a likelihood function $\mathcal{L}$,
defined as a function of the free parameters $N^{\eta}_{sig}$,
$N^{\eta^{\prime}}_{sig}$, and $N_{bkgd}$,
\begin{eqnarray}
  \mathcal{L}(N^{\eta}_{sig},N^{\eta^{\prime}}_{sig}, N_{bkgd}) =
  \frac{ e^{-(N^{\eta}_{sig} +
      N^{\eta^{\prime}}_{sig} + N_{bkgd} )} }{N!}
  \times \prod^N_{i=1}[N^{\eta}_{sig}\mathcal{F}^{\eta}_{sig}(P^i_{miss})+
  \nonumber \\
   N^{\eta^{\prime}}_{sig}\mathcal{F}^{\eta^{\prime}}_{sig}(P^i_{miss}) +
  N_{bkgd}\mathcal{F}_{bkgd}(P^i_{miss})],   
 \label{eq:likelihood}
\end{eqnarray}
where $N^{\eta}_{sig}$ and $N^{\eta^{\prime}}_{sig}$ are the number of
$J/\psi \rightarrow \phi(\rightarrow K^+K^-)\eta(\rightarrow
{\rm invisible})$ and $J/\psi \rightarrow \phi(\rightarrow
K^+K^-)\eta^\prime (\rightarrow {\rm invisible})$ signal events;
$N_{bkgd}$ is the number of background events.  The fixed
parameter $N$ is the total number of selected events in the fit
region, and $P^i_{miss}$ is the value of $P_{miss}$ for the $i$th
event.  The negative log-likelihood ($- \ln \mathcal{L}$) was
minimized with respect to $N^{\eta}_{sig}$, $N^{\eta^{\prime}}_{sig}$,
and $N_{bkgd}$ in the data sample. A total of 105 events were used in the
fit, and the resulting fitted values of $N^{\eta}_{sig}$,
$N^{\eta^{\prime}}_{sig}$, and $N_{bkgd}$ were $-2.8\pm 1.4$, $2.2
\pm 3.4$, and $106\pm 11$, where the errors are statistical.
Figure~\ref{fig:table_eta_etap} shows the $P_{miss}$ distribution and
fitted result superimposed.  No significant signal is observed for the
invisible decay of either the $\eta$ or the $\eta^\prime$. Upper
limits were obtained
by integrating the normalized likelihood distribution over the   
positive values of the number of signal events. The upper limits at the
90\% confidence level were 3.56 events for the $\eta$  and 5.72 events
for the $\eta^\prime$, respectively.
\begin{figure}[htbp]
  \centering
\includegraphics[width=0.4\textwidth,height=0.3\textheight]{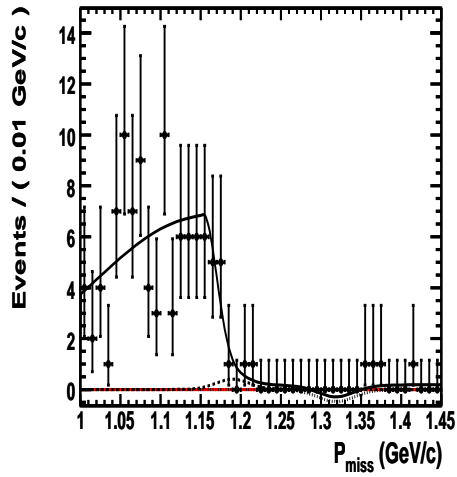}
  \caption{The $P_{miss}$ distribution for candidate events. The data
    (black crosses) are compared to the total fit results. The dotted
    curve is the projection of $\eta$ signal component, and the dashed
    curve is the projection of $\eta^{\prime}$ signal component,
    and the solid curve is the total likelihood fit result.}
\label{fig:table_eta_etap}
\end{figure}

The branching fraction of $\eta (\eta^{\prime})\rightarrow \gamma
\gamma$ was also determined in $J/\psi \rightarrow \phi \eta 
(\eta^{\prime})$ decays, in order to provide the ratio
$\mathcal{B}(\eta(\eta^\prime) \rightarrow {\rm invisible})$ to
$\mathcal{B}(\eta(\eta^\prime) \rightarrow \gamma \gamma)$.  The
advantage of measuring
$\displaystyle\frac{\mathcal{B}(\eta(\eta^\prime) \rightarrow
  {\rm invisible})}{\mathcal{B}(\eta(\eta^\prime) \rightarrow \gamma 
  \gamma)}$ is that the uncertainties due to the total number of
$J/\psi$ events, tracking efficiency, PID, the number of the charged 
tracks,
the cut on $M(KK)$, and residual noise in the BSC all cancel.

For these $\eta (\eta^{\prime}\rightarrow\gamma\gamma$ measurements, two
good photons with energy greater than 50~MeV were required to be within
the  $30^0$ cone.  A 4C kinematic fit was applied to the
selected $K^+K^-\gamma\gamma$ combination and the
$\eta (\eta^{\prime})\gamma\gamma$ signal yields of
$\eta$  and  $\eta^{\prime}$ events were determined
from fits to  peaks seen in the $\gamma\gamma$ invariant mass
distribution.

The upper limit on the ratio of the $\mathcal{B}(\eta\to
{\rm invisible})$ to $\mathcal{B}(\eta\to\gamma\gamma)$ was
calculated with
\begin{eqnarray}
  \frac{\mathcal{B}(\eta\to
{\rm invisible})}{\mathcal{B}(\eta\to\gamma\gamma)} <
\frac{n^{\eta}_{UL}/\epsilon_\eta}{n^{\eta}_{\gamma\gamma}/\epsilon^\eta_{\gamma\gamma}}\cdot
  \frac{1}{(1-\sigma_\eta)},
  \label{eq:upper}
\end{eqnarray}
where $n^{\eta}_{UL}$ is the 90\% upper limit of the number of
observed events for $J/\psi\to\phi\eta$, $\phi\to K^+K^-$, $\eta\to  
{\rm invisible}$ decay, $\epsilon_{\eta}$ is the MC determined  
efficiency for the signal channel, $n^\eta_{\gamma\gamma}$ is the
number of events for the $J/\psi\to\phi\eta$, $\phi\to K^+K^-$,
$\eta\to\gamma\gamma$ decay, $\epsilon^\eta_{\gamma\gamma}$ is the MC
determined efficiency for the decay mode, and $\sigma_{\eta}$ is
$\sqrt{(\sigma^{sys}_{\eta})^2+(\sigma^{stat}_{\eta})^2} = 8.1\%$,
where $\sigma^{sys}_{\eta}$ and $\sigma^{stat}_{\eta}$ are the total
relative systematical error for the $\eta$ case from
Table~\ref{tab:br} and the relative statistical error of
$n^\eta_{\gamma\gamma}$, respectively. For $\eta^\prime$,
$\sigma_{\eta^\prime}$ is
$\sqrt{(\sigma^{sys}_{\eta^\prime})^2+(\sigma^{stat}_{\eta^\prime})^2}=21.6\%$.
The relative statistical error of the fitted yield for $J/\psi
\rightarrow \phi \eta(\eta^\prime)$,
$\eta(\eta^\prime)\to\gamma\gamma$, is 2.8\% (18.5\%) according to the
results from the fit to the invariant mass of $\gamma \gamma$.
The upper limit on the ratio of the
$\mathcal{B}(\eta^\prime \to {\rm invisible})$ to
$\mathcal{B}(\eta^\prime\to\gamma\gamma)$ was determined by replacing
$\eta$ with
$\eta^\prime$ in Eq.~(\ref{eq:upper}).
\begin{table}[htbp]
  \centering
  \caption{ The numbers used in the calculations of the ratios in
    Eq.~(\ref{eq:upper}), where $n^\eta_{UL}~(n^{\eta^\prime}_{UL})$ is
    the upper limit of the signal events at the 90\% confidence level,
    $\epsilon_\eta~(\epsilon_{\eta^\prime})$ is the selection
    efficiency,
    $n^\eta_{\gamma\gamma}~(n^{\eta^\prime}_{\gamma\gamma})$ is the   
    number of the events of $J/\psi\to\phi\eta(\eta^\prime)$, $\phi\to
    K^+ K^-$, $\eta(\eta^\prime)\to\gamma\gamma$,
    $\epsilon^\eta_{\gamma\gamma}~(\epsilon^{\etap}_{\gamma\gamma})$ is
    its selection efficiency.  $\sigma_\eta^{stat}$
    ($\sigma_{\eta^\prime}^{stat}$) is the relative statistical error
    and  $\sigma_\eta^{sys}$ ($\sigma_{\eta^\prime}^{sys}$) is the 
    relative statistical error
    of $n^\eta_{\gamma\gamma}~(n^{\eta^\prime}_{\gamma\gamma})$,
    $\sigma_\eta~(\sigma_{\eta^\prime})$ is the total relative error.}
  \begin{tabular}{l|cc}
    \hline \hline  
    quantity & \multicolumn{2}{|c}{value} \\
    & $\eta$ & $\etap$ \\\hline
    $n^{\eta}_{UL}$ ($n^{\eta^\prime}_{UL}$) & 3.56 & 5.72 \\
    $\epsilon_\eta$ ($\epsilon_{\eta^\prime}$) & $23.5$\% & $23.2$\% \\
    $n^{\eta}_{\gamma\gamma}$ ($n^{\eta^\prime}_{\gamma\gamma}$) & $1760.2
\pm 49.3$ & $71.6 \pm 13.2$ \\
    $\epsilon^\eta_{\gamma\gamma}$ ($\epsilon^{\etap}_{\gamma\gamma}$) &
$17.6$\% & $15.2$\% \\
    $\sigma_\eta^{stat}$ ($\sigma_{\eta^\prime}^{stat}$) & 2.8\% & 18.5\% 
\\
   $\sigma_\eta^{sys}$ ($\sigma_{\eta^\prime}^{sys}$) & 7.7 &11.1 \\
    $\sigma_\eta$ ($\sigma_{\eta^\prime}$) & 8.1\% & 21.6\% \\   
    \hline\hline
  \end{tabular}
  \label{tab:br}
\end{table}
    
Using the numbers in Table~\ref{tab:br}, the upper limit on the ratio 
of ${\mathcal B} (\eta(\etap) \rightarrow {\rm invisible})$ and
${\mathcal B}(\eta(\etap) \rightarrow \gamma \gamma)$ was obtained at the
90\% confidence level of $1.65\times 10^{-3}$ ($6.69\times 10^{-2}$).
    
Table~\ref{tab:invisible_lighter} lists possible two-body
$J/\psi$ decay modes that can be used to study the invisible decays
of the $\eta$, $\eta^{\prime}$, $\rho$, $\omega$ and $\phi$
mesons at \bes3.
    
\begin{table}[htbp]
  \caption{ $J/\psi$ decay modes that can be used to study invisible
decays of $\eta$, $\eta^{\prime}$, $\rho$, $\omega$ and $\phi$ mesons.
 The branching fractions are from the PDG~\cite{lihb_Part4_PDG_2004}. For
each mode, a
``tagging topology"
is given, which is the set of visible tracks
in the detector's acceptance. In each case the tagging topology has
well defined kinematics.  The produced number of events are the expected
events in 10 billion $J/\psi$ event data set at \bes3, with
the decay probabilites of the tagging particles included. }
  \label{tab:invisible_lighter}
  \centering   
  \small
  \begin{tabular}{c|c|c|c|c}\hline
    $J/\psi$ & Branching & Invisible & Tagging & Number of events \\
    decay mode & fraction ($10^{-4}$)  & decay mode & topology & /10  
billion $J/\psi$s \\\hline
   $J/\psi \rightarrow \phi \eta $ & $6.5 \pm 0.7$ &  $\eta \rightarrow  
     \mbox{invisible}$
    & $\phi \rightarrow K^+ K^-$ & $(31.4 \pm 3.4)\times 10^{5}$\\
      & $6.5 \pm 0.7$ & $\phi \rightarrow  \mbox{invisible}$
    & $\eta \rightarrow \gamma \gamma $ &  $(25.7 \pm 2.8 )\times 10^{5}$
\\\hline
      $J/\psi \rightarrow \phi \eta^{\prime} $ & $3.3 \pm 0.4$ &
$\eta^{\prime} \rightarrow  \mbox{invisible}$
    & $\phi \rightarrow K^+ K^- $ &  $(16.2 \pm 1.9 )\times 10^{5}$ \\
         & $3.3 \pm 0.4$ & $\phi \rightarrow  \mbox{invisible}$
    & $\eta^{\prime} \rightarrow \gamma \rho^0 $ &
$(9.6\pm 1.2 )\times 10^{5}$ \\\hline

      $J/\psi \rightarrow \omega \eta $ & $15.8 \pm
1.6$ & $\eta \rightarrow  \mbox{invisible}$ 
    & $\omega \rightarrow \pi^+\pi^-\pi^0 $ &
$(13.9 \pm 1.4 )\times 10^{6}$ \\
       & $15.8 \pm 1.6$ & $\omega \rightarrow  \mbox{invisible}$
    & $\eta \rightarrow \gamma \gamma $ &
$(6.2 \pm 0.6 )\times 10^{6}$\\ \hline
  
      $J/\psi \rightarrow \omega \eta^{\prime} $ & $1.67 \pm 0.25$ &
$\eta^{\prime} \rightarrow  \mbox{invisible}$
    & $\omega \rightarrow  \pi^+\pi^-\pi^0 $ & $(1.5 \pm 0.2)\times 
10^{6}$\\
        & $1.67 \pm 0.25$ & $\omega \rightarrow  \mbox{invisible}$
    & $\eta^{\prime} \rightarrow \gamma \rho^0 $ & $(0.7 \pm 0.1 )\times 
10^{6}$\\ \hline
    
      $J/\psi \rightarrow \rho^0
\eta $ & $1.93 \pm 0.23$ & $\eta \rightarrow  \mbox{invisible}$
    & $\rho^0 \rightarrow \pi^+\pi^- $ &  $(1.9 \pm 0.2 )\times 10^{6}$\\
      & $1.93 \pm 0.23$ & $\rho^0 \rightarrow  \mbox{invisible}$
    & $\eta \rightarrow \gamma \gamma $ & $(0.8 \pm 0.09 )\times 10^{6}$ 
\\
\hline
         
      $J/\psi \rightarrow \rho^0 \pi^0 $ & $56 \pm
7$ & $\rho^0 \rightarrow  \mbox{invisible}$
    & $\pi^0 \rightarrow \gamma\gamma $ &   $(55.3 \pm 5.8 )\times 
10^{6}$\\
\hline

  \end{tabular}
\end{table}
    
\section{Search for $C$ or $P$ violating processes in $J/\psi$ decays}
\label{sec:c_cc_decays}
        
With its huge $J/\psi$ and $\psi(2S)$ data samples, the \bes3 experiment 
will be approaching the statistics regime
where studies of rare $\psi$ decays can
provide important tests of the SM and possibly
uncover deviations. Among the interesting examples are $C$,  $P$ or
$CP$ violating processes in $J/\psi$ decays.  An example of such
modes would be $\psi(nS) \rightarrow V^0 V^0$, where $V^0$ is used to
denote $J^{PC} = 1^{--}$ vector mesons ($\phi$, $\omega$, $\rho^0$ and
$\gamma$).  A distinct signal for this class of event
would be $\psi(nS) \rightarrow \phi \phi$  detected
in $\psi(nS) \rightarrow K^+K^- K^+K^-$ final states.  Because of the $C$
violation,  $\psi(nS)\rightarrow V^0 V^0$ decays can only occur
in the SM via $c\bar{c}$ annihilation via a $Z^0$ or $W$-exchange
decays as discussed in Ref.~\cite{lihb_double_phi}. The rate for this
type of weak decay can provide a measurement of the charmonium
wave function at the origin~\cite{lihb_double_phi}.

In order to
make a rough estimate for the rate, we first consider just the rate due
to the $W$-exchange contribution, which
is straightforward to compute~\cite{lihb_goggi_cp}
\begin{eqnarray}
\frac{\Gamma(J/\psi \rightarrow s\bar{s})^{weak}} {\Gamma (J/\psi
\rightarrow e^+e^-)} \cong \frac{1}{2} \left(
\frac{m_{J/\psi}}{m_W} \right)^4,
\label{eq:cp_vv}
\end{eqnarray}
where $m_{J/\psi}$ and $m_{W}$ are the masses of
$J/\psi$ and $W$ boson,  respectively.  This leads to $BR(J/\psi
\rightarrow
s\bar{s})^{weak} \cong 10^{-7}$ for this weak contribution.
To form the $\phi\phi$ final state, another $s\bar{s}$ pair must be
produced from
the vacuum and these $s$-quarks have to
bind with the outgoing $s\bar{s}$ from the $c\bar{c}$ decay
to produce the $ \phi \phi$
final state~\cite{lihb_double_phi}.
When this is considered, it seems that one can expect that the
SM exclusive $BR(\psi(nS) \rightarrow \phi \phi)$ rate should be below the
level of $10^{-8}$ and probably out of reach of the \bes3 experiment.

Experimentally, there are some possible backgrounds that will dilute the
signal for $J/\psi \rightarrow \phi\phi$ decays. One major
background is $J/\psi \rightarrow \gamma \phi \phi$, which is mainly from
$J/\psi \rightarrow \gamma \eta_c(1S)$, $\eta_c(1S) \rightarrow \phi
\phi$. This
background can be  removed by doing a constrained kinematic
fit. A detailed calculation had been done to estimate
the background from
$J/\psi \rightarrow \gamma \phi \phi$~\cite{lihb_double_phi}.
Another background appears if one studies only  
$2(K\bar{K})$ invariant pair mass distributions. It arises from the
$C$ and $P$-conserving reaction $J/\psi \rightarrow \phi
(K\bar{K})_{S-wave}$,
due to the fact that the $\phi$ mass is only two $S$-wave-widths away
from the
$K\bar{K}$~$S$-wave resonance mass, for example, $f_0(980) \rightarrow
K\bar{K}$.
Although it may be difficult to subtract in a small statistical sample, 
one
can, in principle, remove this kind of background by
either a spin-parity analysis of the $ K\bar{K}$ pairs in a narrow
window about $\phi$ mass, or by a subtraction normalized to an
observed $S$-wave mass peak.  To avoid the
$S$-wave contribution, one can reconstruct one $\phi$ from
$K^+K^-$ and another $\phi$ from the $K_S K_L$ mode, which is not allowed 
to form an $S$-wave. It will be easy to look for the missing mass of one
$\phi$ reconstructed from $K^+K^-$, to see if there is any peak under the
$\phi$ mass region by also requiring $K_S$ and $K_L$ information
in the final states.


 It is noted that there is possible continuum background   
produced via a two-photon annihilation process. It is a peaking
background that cannot be removed without considering detail
angular distributions in a high statistics sample.  It is very
hard to deal with this kind of peaking source with
a small sample of signal events.
One way is to use off-peak data which are taken below the
$J/\psi$ mass peak. The $e^+e^- \rightarrow \gamma \gamma$ process
has been investigated before~\cite{lihb_two_photons}, and it has a   
unique production angle ($\theta^*$) distribution, which is
defined as the angle between $\phi$ and $e^-$ beam direction in
the Center-of-Mass (CM) frame. The production angle distribution for
the two real photon annihilation process has the form of
\begin{eqnarray}
 \sigma (cos\theta^*)_{e^+e^- \rightarrow \gamma \gamma} =
\frac{1+cos\theta^{*2}}{1-cos\theta^{*2}},
 \label{eq:angular_twophoton}
\end{eqnarray}
while, in the process of two virtual photon into $V^0V^0$ pairs, the
distribution is (to first order)~\cite{lihb_stan}:
\begin{eqnarray}
 \sigma (cos\theta^*)_{e^+e^- \rightarrow \gamma^*
\gamma^* \rightarrow V^0V^0} =
\frac{1+cos\theta^{*2}}{k^2-cos\theta^{*2}},
 \label{eq:angular_twophoton_1}
\end{eqnarray}
where factor $k$ is:
\begin{eqnarray}
 k= \frac{2m^2_{V^0}-S}{\sqrt{S^2 - 4Sm^2_{V^0} }},
 \label{eq:angular_twophoton_2}
\end{eqnarray}
where $S$ is the square of
CM energy. In principle, by using an  angular analysis, one can
remove the peaking background with high statistic data sample.
To avoid the peaking background from the
continuum, $\psi(2S) \rightarrow \pi \pi J/\psi$ could be used
to study this kind of rare $J/\psi$ decays with 3 billion
$\psi(2S)$ sample, but the statistics will be substantially reduced.

\section{Lepton flavor violating processes in decays of $J/\psi$ }
\label{sec:lfv_cc_decays}

Standard Model lepton-flavor-violating (LFV) processes are suppressed
by powers of the very small neutrino
masses~\cite{lpv_zhang}. Therefore, such decays can be used as a
probe for possible new physics. At present, there are many stringent
bounds for $\mu$, $\tau$ and $Z$ boson decays, such as $BR(\mu
\rightarrow 3e) \le 10^{-12}$, $BR(\mu \rightarrow e \gamma
\gamma) \le 10^{-10}$ and somewhat weaker $O(10^{-6}$ bounds on
LFV $\tau$ decays~\cite{ted_PDG2006}. There have been a lot of studies,
both theoretically and experimentally, on testing the lepton flavor  
conservation law~\cite{lpv_zhang,lpv_yue}. With a huge $J/\psi$
event sample, the \bes3 experiment will be able to make high sensitivity
experimental searches for lepton flavor violating processes of the type
$J/\psi \rightarrow l l^{\prime}$ ($l$ and $l^{\prime} = \tau,
\mu, e$, $l \ne l^{\prime}$).

 To estimate the branching ratio for the lepton flavor violating $J/\psi$
decays that are allowed by the current experimental data, Peccei, Wang
and Zhang used a model-independent approach to new physics and
introduced an effective four-fermion contact
interaction~\cite{lpv_zhang,lpv_zhang_2,lpv_zhang_3}: 
\begin{eqnarray}
\frac{4\pi}{\Lambda^2} \bar{c} \gamma^{\mu}c \bar{l}
\gamma^{\mu}l^{\prime},
\label{eq:lpv_four_fermion}
\end{eqnarray}
where $\Lambda$ is the new physics cutoff. This effective operator
is forbidden in the standard model, however, it will be generated
in theories where lepton flavor is not conserved, such as the MSSM
with and/or without R parity, and models with large extra
dimension~\cite{lpv_laregN}.  Any observed signal would be
direct evidence for non-standard physics and improve our
understanding of flavor dynamics, especially in the lepton sector.

There is no direct experimental limit on the $\Lambda$ cutoff in   
eqn.~\ref{eq:lpv_four_fermion}. However, at the one-loop level, attaching
a neutral gauge boson $Z$ to the charm quark loop generates an 
effective coupling of $Z$ to $\bar{l} l^{\prime}$. From the
limits given in the PDG on $BR(Z \rightarrow
\bar{l} l^{\prime})$~\cite{ted_PDG2006}, one obtains upper
bounds on the branching fractions of $J/\psi$ decay into
leptons~\cite{lpv_zhang_3}:
\begin{eqnarray}
BR(J/\psi \rightarrow \tau^+ e^-) < 2.7 \times 10^{-5}; \\
BR(J/\psi \rightarrow \tau^+ \mu^-) < 4.9 \times 10^{-5}; \\  
BR(J/\psi \rightarrow \mu^+ e^-) < 8.3 \times 10^{-6}.
\label{eq:lpv_psi_predict}
\end{eqnarray}  
Recently, Nussinov, Peccei and Zhang~\cite{lpv_zhang} have
also examined ``unitarity inspired" relations between two- and
three-body lepton flavor violating decays and found that the
existing strong bounds on $\mu \rightarrow 3e$ and $\mu
\rightarrow e \gamma \gamma$ servery constrain two-body lepton
flavor violating decays of vector mesons, such as the $J/\psi$,  
$\Upsilon(1S)$ and $\phi$, into $\mu^{\pm} e^{\mp}$ final states. 
In Ref.~\cite{lpv_zhang}, using $BR(\mu \rightarrow 3e)\le
10^{-12}$ and data pertaining to the $e^+e^-$ widths of the
$J/\psi$, the bound on the branching ratio for the two-body LFV
decay $J/\psi\rightarrow \mu^{\pm}e^{\mp}$ decay is
\begin{eqnarray}
BR(J/\psi \rightarrow \mu^{\pm} e^{\mp}) < 4 \times 10^{-13}.      
\label{eq:lpv_psi_predict_2}
\end{eqnarray}
Likewise, the generic upper bounds on LPV $\tau$ decays
$BR(\tau \rightarrow ll^{\prime} \bar{l^{\prime}}) \le 10^{-6}$ yields
\begin{eqnarray}
BR(J/\psi \rightarrow \tau^{\pm} l^{\mp}) < 6 \times 10^{-7}.
\label{eq:lpv_psi_predict_3}
\end{eqnarray}  
with $l/l^{\prime} = e/\mu$.  These inferred bounds are unlikely
to be improved by future experimental data on two-body decays,
especially at \bes3. However, all the bounds derived in
Ref.~\cite{lpv_zhang,lpv_zhang_3} can be evaded if there is
a kinematical suppression or some
cancellations~\cite{lpv_zhang}. Searching for lepton flavor
violating decays of $J/\psi$ with a huge sample a \bes3 remains a
worthwhile experimental challenge~\cite{lpv_zhang_3}.

With a 58 M $J/\psi$ event sample at BESII, the following upper limits
have been established~\cite{lpv_besii}:
\begin{eqnarray}
BR(J/\psi \rightarrow \tau^{\pm} e^{\mp}) < 8.3 \times 10^{-6}; \\
BR(J/\psi \rightarrow \tau^{\pm} \mu^{\mp}) < 2.0 \times 10^{-6}; \\
BR(J/\psi \rightarrow \mu^{\pm} e^{\mp}) < 1.1 \times 10^{-6}. 
\label{eq:lpv_psi_besii}
\end{eqnarray}  
The limits on the two-body  lepton
flavor violating decays of the $J/\psi$ could be
reduced to the $10^{-8} \sim 10^{-9}$ level
at \bes3 with a one year full-luminosity
run at the $J/\psi$ peak. This would be
a significant improvement.

\chapter{Miscellaneous}
\label{sec:mis}
\section[Bell inequalities in high energy physics]
{Bell inequalities in high energy physics\footnote{Junli Li and Cong-Feng 
Qiao}}
\label{sec:bell}


In 1935, Einstein, Podolsky, and Rosen (EPR) \cite{qiao_epr}
demonstrated that quantum mechanics (QM) could not provide a
complete description of the ``physical reality'' for two spatially
separated but quantum mechanically correlated particle systems.
Alternatively, local hidden variable theories (LHVTs) have been
developed to restore the completeness of QM. In 1964, Bell
\cite{qiao_bell} showed that in realistic LHVTs two-particle
correlation functions satisfy a set of Bell inequalities (BI),
whereas the corresponding QM predictions can violate such
inequalities in some region of parameter space. This leads to the
possibility of experimental testing of the validity of LHVTs in
comparison with QM.

The definition of correlations for LHVTs and QM read respectively as:
\begin{eqnarray}
E(\textbf{a},\textbf{b}) & = & \int \mathrm{d}\lambda
\rho(\lambda)
A(\textbf{a},\lambda)B(\textbf{b},\lambda)\; , \\
E(\textbf{a},\textbf{b}) & = & \langle \psi| \sigma \cdot
\textbf{a} \otimes \sigma \cdot \textbf{b} |\psi \rangle =
-\textbf{a} \cdot \textbf{b}\; . \label{bohm}
\end{eqnarray}
Here, $\rho(\lambda)$ is the distribution of a hidden variable
regardless of whether $\lambda$ is a single variable or a set, or
even a set of functions. These variables can be either discrete or
continuous. $\textbf{a}$ and $\textbf{b}$ indicate spin
directions. One of the Bell inequalities, the CHSH inequality,
takes the  form \cite{qiao_CHSH}
\begin{eqnarray}
S = |E(\textbf{a},\textbf{b}) - E(\textbf{a},\textbf{b}')| +
|E(\textbf{a}',\textbf{b}) + E(\textbf{a}',\textbf{b}')|
 \leq 2\; . \label{bell}
\end{eqnarray}
The correlation function $E$ above can be calculated by
\begin{eqnarray}
E(\textbf{a},\textbf{b}) = P_{++}(\textbf{a},\textbf{b}) -
P_{+-}(\textbf{a},\textbf{b}) - P_{-+}(\textbf{a},\textbf{b}) +
P_{--}(\textbf{a},\textbf{b})\; ,
\end{eqnarray}
where $P_{\pm\pm} = N_{\pm\pm}(\textbf{a},\textbf{b})/N$, $N$ is the
total number of particle pairs, and $N_{++(+-)}$ means that two
particle has the same (opposite) spin directions.

In 1992, Hardy proved without using
inequalities, a kind of definite conflict that can occur for any
non-maximally entangled state composed of two two-level
subsystems~\cite{qiao_hardy}. Later, Hardy's argument was expanded on by 
Jordan
\cite{qiao_jordan}, who demonstrated that there exist four
projection operators satisfying
\begin{eqnarray}
\langle FG \rangle = 0\; , & & \langle D(1-G) \rangle = 0\;
,\label{jordan0}\\ \langle (1-F)E \rangle = 0\; , & & \  \ \ \langle
DE \rangle
> 0\; , \label{jordan}
\end{eqnarray}
which are in contradiction with LHVTs. Jordan also demonstrated in
a converse way that for any choice of four
different measurements, there exists a state satisfying Hardy's
argument~\cite{qiao_jordan}. 
In 1995, Garuccio~\cite{qiao_garuccio} found that the
contradiction between QM and LHVT can be embedded in Clauser-Horne
(CH) inequalities~\cite{qiao_Clauser-Horne}, {\it i.e.},
\begin{eqnarray}
\langle DE \rangle \leq \langle FG \rangle + \langle D(1-G)
\rangle + \langle (1-F)E \rangle\; . \label{eber}
\end{eqnarray}

Many experiments in regard of the Bell inequalities have been
carried out by using the entangled photons. In the optical
experiment the correlation of polarizers in orientations
\textbf{a} and \textbf{b} is defined as follows:
\begin{eqnarray}
E(\textbf{a}, \textbf{b}) = \frac{N_{++}(\textbf{a}, \textbf{b}) +
N_{--}(\textbf{a}, \textbf{b}) - N_{+-}(\textbf{a}, \textbf{b}) -
N_{-+}(\textbf{a}, \textbf{b}) }{N_{++}(\textbf{a}, \textbf{b}) +
N_{+-}(\textbf{a}, \textbf{b}) + N_{-+}(\textbf{a}, \textbf{b}) +
N_{--}(\textbf{a}, \textbf{b})}\; ,
\end{eqnarray}
where $N_{+-}$ is the coincidence rate of photon polarizations;
$+$ for parallel and $-$ for perpendicular to the chosen
direction. Of the various optical experiments, one of the
important ones was carried out by Aspect \textit{et al.}
\cite{qiao_aspect1}. Their measurement gave
\begin{eqnarray}
S_{exp} = 2.697\pm 0.015\; .
\end{eqnarray}
This result is in excellent agreement with the predictions of
quantum mechanics, which, for their polarizer efficiencies and lens
apertures, gives $S_{QM} = 2.7\pm 0.05$.

It has been noted that the experiments testing the
completeness of QM are mainly limited to the electromagnetic
interaction regime, {\it i.e.}, by employing entangled photons, no
matter whether the photons are generated from atomic cascades or the
PDC method. Considering the fundamental importance of the questions
involved,  experimenatl tests of LHVT with massive
quanta and with other kinds of interactions are 
necessary~\cite{qiao_abel}.

As early as the 1960s, the EPR-like features of the $K^0\bar{K^0}$ pair
in decays of the $J^{PC}=1^{--}$ vector particles was noticed by
Lipkin \cite{qiao_lipkin}. Other early attempts at testing LHVTs
through the Bell inequality in high energy physics focused on
exploiting the nature of particle spin correlations
\cite{qiao_abel,qiao_tqvist,qiao_privitera}. In Ref.
\cite{qiao_tqvist} T\"ornqvist suggested  measuring the BI via the
\begin{eqnarray}
e^+e^-\rightarrow\Lambda\bar{\Lambda}\rightarrow\pi^-\; p\; \pi^+\;
\bar{p}
\end{eqnarray}
process. A similar process was suggested by Privitera
\cite{qiao_privitera}, {\it i.e.,}
\begin{eqnarray}
e^+e^- \rightarrow \tau^+ \tau^- \rightarrow \pi^+ \bar{\nu}_{\tau}
\pi^-\nu_{\tau}\; .
\end{eqnarray}
The DM2 Collaboration \cite{qiao_dm2} observed $7.7 \times 10^6\
J/\psi$ events with about $10^3$ of them identified as being from
the process
$J/\psi \rightarrow \Lambda \bar{\Lambda} \rightarrow
\pi^-p\pi^+\bar{p}$. The experimental measurement unfortunately
does not give a very significant result \cite{qiao_tqvistbook} due
to insufficient statistics.

However, it was subsequently realized that these proposals have
controversial assumptions \cite{qiao_book}. It was found that
for testing the LHVTs in high energy physics, using the
``quasi-spin" to mimic the photon polarization in the construction
of entangled states is a more practical approach. A typical process that 
produces
an entangled  $K^0 \bar{K}^0$ state is  $ e^+e^-
\rightarrow \phi \rightarrow K^0 \bar{K}^0$. The wave function of
the $J^{PC} = 1^{--}$ particles, like the $\phi$ which decays into
$K^0 \bar{K}^0$, can be formally configured as
\cite{qiao_cptviolation}:
\begin{eqnarray}
|\phi\rangle = \frac{1}{\sqrt{2}}\{ |K^0\rangle |\bar{K}^0\rangle
- |\bar{K}^0\rangle |K^0\rangle \}\; . \label{kaonentangle}
\end{eqnarray}
Similar expressions apply to $\Upsilon(4S) \rightarrow B^0
\bar{B}^0$, $\Upsilon(5S) \rightarrow B_{s}^0 \bar{B}_{s}^0$, and
$\psi(3770)\rightarrow D^0 \bar{D}^0$ cases. There are two different
methods in this ``quasi-spin" scheme. In the first method, one fixes
the quasi-spin, but leaves the time free. The second method is to 
leave the  quasi-spin free but  fix the time.

Consider the first case: we can choose different times to measure the
final states, the kaons, on each side. We choose the quantum number
of strangeness as the quasi-spin in our consideration, but neglect
$CP$ violation effects, which in some sense is a good
approximation. With the time evolution, the initial entangled state,
like in (\ref{kaonentangle}), becomes:
\begin{eqnarray}
|\Psi(t_l, t_r)\rangle = \frac{1}{\sqrt{2}} \{ e^{-i(m_Lt_l +
m_St_r)}e^{-\frac{\Gamma_L}{2}t_l-\frac{\Gamma_S}{2}t_r}|K_L\rangle
|K_S\rangle \nonumber \\ - e^{-i(m_St_l +
m_Lt_r)}e^{-\frac{\Gamma_S}{2}t_l-\frac{\Gamma_L}{2}t_r}|K_S\rangle
|K_L\rangle\}\; .
\end{eqnarray}
In the above, the small letters $l$ and $r$ denote left side and
right side; we name the two entangled particles to be left
and right without lose generality. Choosing different measurement
times for the two sides, the expectation value of correlation is
\cite{qiao_quant-ph/0501069}:
\begin{eqnarray}
E(t_l,t_r) = - \cos(\Delta m\Delta t)
e^{-\frac{\Gamma_L+\Gamma_S}{2}(t_l+t_s)}\; .
\label{kaoncorrelation}
\end{eqnarray}
Inserting this correlation directly into the CHSH inequality, one
finds that the violation of the inequality depends on the
ratio of $x = \Delta m/\Gamma$  \cite{qiao_Bertlmann}, where
$\Delta m$ characterizes the strangeness oscillation and 
$\Gamma$ characterizes the weak decay lifetime. Among the known neutral
mesons, the $B^0_S \bar{B}^0_S$ system has the largest
value of $x$, and hence the violation of inequalities
might be found there.

The EPR-type strangeness correlation in the process
$p\bar{p}\rightarrow K^0\bar{K}^0$ has been tested at the CPLEAR
detector \cite{qiao_cplear} at CERN. In the experiment the
$K^0\bar{K}^0$ pairs were created in $J^{PC} = 1^{--}$
configuration. The wave function at proper time $t_l = t_r = 0$ is
\begin{eqnarray}
|\Psi(0, 0)\rangle = \frac{1}{\sqrt{2}} [|K^0\rangle_l
|\bar{K}^0\rangle_r - |\bar{K}^0\rangle_l  |K^0\rangle_r]\; .
\end{eqnarray}
\begin{figure}[t,m,u]
\begin{center}
\includegraphics[width=7cm,height=5cm]{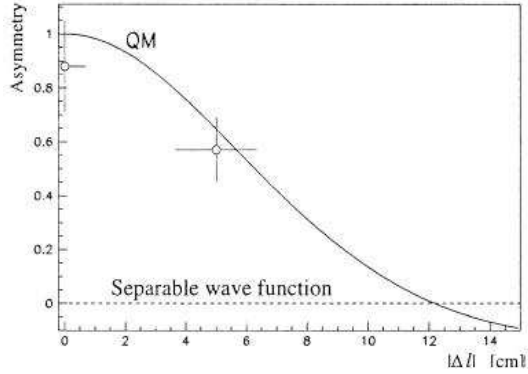}
\end{center}
\caption{\label{cplear} \small The best fit to the experimental
measurement \cite{qiao_cplear}. The two points with error bars
correspond to time difference $\Delta t = 0$ and $\Delta t = 1.2
\tau_s$. The solid line represents the QM prediction.}
\end{figure}
The strangeness was tagged via strong interaction with absorbers
away from the creation point. From Fig.\ref{cplear} one notices that
the non-separability hypothesis of QM is strongly favoured by
experiment.

The $B^0\bar{B^0}$ entangled system produced at the $\Upsilon(4S)$
resonance has also been measured in the KEKB 
$B$-factory~\cite{qiao_b0}.
The wave function for $\Upsilon(4S) \rightarrow B^0\bar{B^0}$ has
the same form as the spin singlet:
\begin{eqnarray}
|\Upsilon\rangle = \frac{1}{\sqrt{2}}\{ |B^0\rangle_l
|\bar{B^0}\rangle_r - |\bar{B^0}\rangle_l |B^0\rangle_r \}\; .
\end{eqnarray}
\begin{figure}
\begin{center}
\includegraphics[width=7cm,height=5cm]{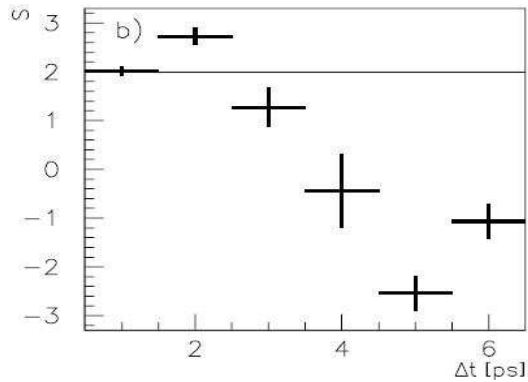}
\end{center}
\caption{\label{bb} \small The experimental result on the violation
of the inequality~\cite{qiao_b0}. The horizontal axis refers to
$\Delta t$ and the vertical axis to the S. The LHVTs limit of 2 is
shown by the solid line.}
\end{figure}
Here, the $b$-flavor quantum number plays the role of spin
polarization in the spin correlation system. The experiment, which
is based on a data sample of $80\times10^6$ $\Upsilon(4S)
\rightarrow B\bar{B}$ decays at Belle detector at the KEKB
asymmetric collider in Japan, gives $S=2.725\pm 0.167_{stat}$. It
is obviously a result that violate the Bell inequality, as shown in
Fig.\ref{bb}. However, debates on whether this was genuine test of
LHVTs or not is still ongoing~\cite{qiao_verified}.

Recently, expanding on Hardy's approach, Bramon and Garbarino
proposed a new scheme to test the
local realism by virtue of entangled neutral 
kaons~\cite{qiao_prl88,qiao_prl89}.  After
neglecting the small $CP$-violation effect, the initial $K_S K_L$
pair from $\phi$ decay, or proton-antiproton annihilation has the form:
\begin{eqnarray}
|\phi(T=0)\rangle=\frac{1}{\sqrt{2}}[K_SK_L-K_LK_S]\ ,\label{phi0}
\end{eqnarray}
where $K_S=(K^0+\bar{K^0})/\sqrt{2}$ and
$K_L=(K^0-\bar{K^0})/\sqrt{2}$ are the mass eigenstates of the $K$
mesons. One of the key points of the scheme is to test the LHVTs 
by generating a nonmaximally entangled (asymmetric) state. That is
\begin{eqnarray}
|\phi(T)\rangle=\frac{1}{\sqrt{2+|R|^{\,2}}}\,
[K_SK_L-K_LK_S-re^{-i(m_L-m_S)\,T+[(\Gamma_S-\Gamma_L)/2]
T}K_LK_L]\ .\label{jpsiT}
\end{eqnarray}
Here, $r$ is the regeneration parameter of the order of
magnitude $10^{-3}$~\cite{qiao_prl89}, $\Gamma_L$ and $\Gamma_S$
are the $K_L$ and $K_S$ decay widths, respectively, and $T$ is the
evolution time of kaons after their production. Technically, this
asymmetric state can be achieved by placing a thin regenerator
close to the $\phi$ decay point.

Four specific transition probabilities for joint measurements from
QM take the following forms:
\begin{eqnarray}
P_{QM}(K^0,\bar{K^0}) & \equiv & |\langle
K^0\bar{K^0}|\phi(T) \rangle|^{\ 2}\nonumber \\
 & = & \frac{|2+\mathrm{R}e^{i\varphi}|^{\,2}}
 {4(2+|R|^2)}\ ,
 \label{tp1} \\
P_{QM}(K^0,K_L) & \equiv & |\langle K^0 K_L|\phi
(T)\rangle|^{\ 2}\nonumber \\
& = & \frac{|1 + \mathrm{R}e^{i\varphi}|^{\,2}}{2(2 + |R|^2)}\ ,
 \label{tp2}\\
P_{QM}(K_L,\bar{K^0}) & \equiv & |
\langle K_L\bar{K_0}|\phi(T)\rangle|^{\ 2}\nonumber \\
& = & \frac{|1+\mathrm{R}e^{i\varphi}|^{\,2}}{2(2+|R|^2)}\ ,
 \label{tp3} \\
P_{QM}(K_SK_S)& \equiv & |\langle K_SK_S|\phi(T)\rangle|^{\ 2}= 0
\ ,
 \label{tp4}
\end{eqnarray}
where ${\rm R}=-|{\rm R}|=-|r|e^{[(\Gamma_S-\Gamma_L)/2]T}$ and
$\varphi$ is the phase of R. In Ref. \cite{qiao_prl89} the
special case of $R=-1$ was considered, in which
\begin{eqnarray}
P_{QM}(K^0,\bar{K^0}) & = & 1/12\; , \label{4.1}\\ P_{QM}(K^0,K_L) &
= & 0\; , \label{4.2}\\ P_{QM}(K_L,\bar{K^0}) & = & 0\; , \label{4.3} \\
P_{QM}(K_S,K_S) & = & 0\; . \label{4.4}
\end{eqnarray}
These equations has the same form as (\ref{jordan}), and are
in conflict with QM.

For simplicity we consider an ideal case, in which the detection
efficiency for the kaon decays is 100 percent. Then the Eberhard
inequality (EI) for the kaon system takes the similar form as
Eq.(\ref{eber}) \cite{qiao_Eberhard,qiao_garuccio}. It reads
\begin{eqnarray}
P_{LR}(K^0,\bar{K^0}) & \leq &
P_{LR}(K^0,K_L)+P_{LR}(K_S,K_S)+P_{LR}(K_L,\bar{K^0})\ .\label{ei}
\end{eqnarray}
\begin{figure}[t,m,u]
\begin{center}
\includegraphics[width=7cm,height=6cm]{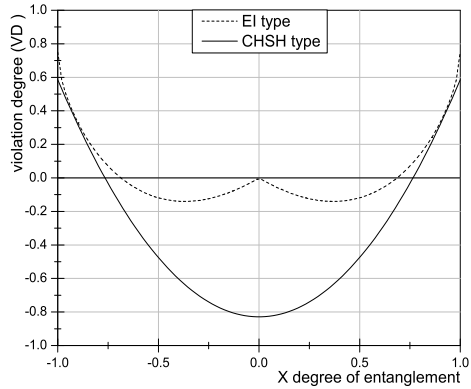}
\end{center}
\caption{\label{breaking1}\small The violation degree of the Bell
inequalities (the dashed line for EI type and the solid line for
CHSH type) in terms of the entanglement. Here, for the sake of
transparency, we make the coordinate transformation $C = 1-x^2$. 
Magnitudes of $V\!\!D$ less than zero indicate the breaking of the BIs. }
\end{figure}
For the case of QM, substituting equations (\ref{tp1}) -
(\ref{tp4}) into the inequality (\ref{ei}) and assuming $\varphi =
0$, we have
\begin{eqnarray}
\frac{(2+\mathrm{R})^2}{4(2+\mathrm{R}^2)} & \leq &
\frac{(1+\mathrm{R})^2}{2(2+\mathrm{R}^2)}\; +\; 0\; + \;
\frac{(1+\mathrm{R})^2}{2(2+\mathrm{R}^2)}\ . \label{ei1}
\end{eqnarray}
The above inequality is apparently violated by QM while
$\mathrm{R} = -1$. In Ref. \cite{qiao_ourself} we generalized the
method used in \cite{qiao_prl89} to heavy quarkonium. This
straightforward generalization however leads to some new
observations of a nonlocal property. Upon further analyzing the
R value when it produces a violation of Eq.(\ref{ei1}), we find out
that there exists a period of time during which the violation
becomes larger via its time evolution. In quantum information
theory the entanglement property of two pure state qubits is well
understood and can be characterized by the concurrence $C$
\cite{qiao_concurrence}. We can also see how the degree of
entanglement evolves with time. According to the definition of
concurrence we have:
\begin{eqnarray}
C(J/\psi)=|\langle
J/\psi|\widetilde{J/\psi}\rangle|=\frac{2}{2+|R|^{\,2}}=
\frac{2}{2+|r|^{\,2}e^{(\Gamma_S-\Gamma_L)\,T}}\ , \label{ev1}
\end{eqnarray}
where $|\widetilde{J/\psi}\rangle = \sigma^1_y \sigma^2_y
|(J/\psi)^{*}\rangle\ $ and $\sigma^{1,\ 2}$ are Pauli matrices.
$C$ ranges from null to unity for no entanglement and full
entanglement. Equation~(\ref{ev1}) shows that the state becomes less
entangled as time evolves. So, considering  Ref.~\ref{ei1}
we realize that the violation  does not decrease monotonically
with the degree of entanglement. To clarify this phenomenon we
express the degree of violation ($V\!\!D$) of the inequalities 
(left side minus the right side) in terms of $C$ and compare it with the
usual CHSH inequality~\cite{qiao_CHSH}. In Fig.~\ref{breaking1}, the
differing  $V\!\!D$ behavior of CHSH's and EI's inequalities 
are presented. For the CHSH case, the $V\!\!D_{\rm CHSH}$ is obtained by
the same condition since the maximal violation happens at full
entanglement, {\it i.e.} $C = 1$. We have:
\begin{eqnarray}
V\!\!D_{CHSH} = \sqrt{2}(1+C)-2\ .
\end{eqnarray}
In fact, the above $V\!\!D_{\rm CHSH}$ can be deduced from the
results given in Refs.~\cite{qiao_gisin,qiao_kar,qiao_degree of
two}. For the EI case,
\begin{eqnarray}
V\!\!D_{EI}=\frac{-3(1-C)+2\sqrt{2}\sqrt{C-C^2}}{4}\ .
\end{eqnarray}
Here,  in EI the counterintuitive quantum effect shows up, {\it i.e.}
less entanglement corresponds to a larger $V\!\!D$ in some
regions (see Fig.\ref{breaking1}). It is worth noting that with
the time evolution, when $R$ becomes less than $-\frac{4}{3}$, both
QM and LHVTs satisfy the inequality (\ref{ei}). Thus given a
certain asymmetrically entangled state, the Hardy state
\cite{qiao_hardy}, in the region of $R\in[-4/3,0)$  QM and
LHVTs can be distinguished from  EI.

In a recent work \cite{qiao_brachingratio}, an improved
measurement of branching ratio $B(J/\psi \rightarrow K_S^0K_L^0) =
(1.80 \pm 0.04 \pm 0.13)\times 10^{-4}$ is reported, which is
significant larger than previous ones. Entangled kaon pairs from
heavy quarkonium decays can be easily space-like separated.
Thus, a little evolution time $T$ will guarantee the locality
condition \cite{qiao_ourself}, and  enable us to test the
full range of $R$ and the peculiar quantum effects. It is
promising and worthwhile to implement such tests in future
tau-charm factories, because of both the experimental feasibility
and their theoretical importance.

\section[Special topics in $\bbbar$ final states]
{Special topics in $\bbbar$ final states\footnote{Xiaohu Mo,
Ronggang Ping and Changzheng Yuan}}
\label{sec:others}

In this section we discuss the study of $SU(3)$ flavor
symmetry breaking effects, searches for $CP$ violation, exotic
states, bound states, tests of the Bell inequality and other special
topics in hadronic decays of charmonium states.

\subsection{SU(3) flavor symmetry breaking effects}

At the level of $SU(3)$ symmetry, only $\jpsi$ baryonic decays
of the type 
$$\jpsi\to B_1\bar B_1,B_8\bar B_8,B_{10}\bar B_{10}$$
are allowed, with a single decay amplitude for each given decay
family if electromagnetic contributions are neglected. However, the
MarkII collaboration first published experimental results showing that
a large SU(3) flavor symmetry breaking takes place in the $\jpsi$
decays into baryon pairs~\cite{Mark2:84}, especially, into
octet-decuplet baryon pairs. This was confirmed by
the DM2~\cite{DM2:87} and MarkIII~\cite{Mark3:86} collaborations. 
Table~\ref{flavorBreaking}
summarizes the DM2 results on the $\jpsi$ SU(2) or SU(3) forbidden
decays. For the $\jpsi\to \Lambda\bar\Lambda \pi^0$, it seems that
a large contamination from $\jpsi\to \Sigma\bar\Lambda\pi^0$
would lead to a small branching fraction. These SU(3) flavor
symmetry decays will be studied at \bes3.

\begin{table}[htbp]
\caption{Summary of $\jpsi$ SU(2) and SU(3) forbidden decay mode
measurements.\label{flavorBreaking}}
\begin{center}
\begin{tabular}{lcc}
\hline \hline \\
Decay Mode & Number of events & Branching fraction ($\times
10^{-4}$)\\\hline
 \multicolumn{3}{c}{SU(3) forbidden decay
modes}\\\hline
$\jpsi\to \Sigma(1385)^-\bar\Sigma^+$&$74\pm8$&$3.0\pm0.3\pm0.8$\\
$\jpsi\to \Sigma(1385)^+\bar\Sigma^-$&$77\pm9$&$3.4\pm0.4\pm0.8$\\
$\jpsi\to \Xi(1530)^-\bar\Xi^+$&$80\pm9$&$5.9\pm0.7\pm1.5$\\
$\jpsi\to \Xi(1530)^0\bar\Xi^0$&$24\pm5$&$3.2\pm0.7\pm1.5$\\\hline
 \multicolumn{3}{c}{SU(2) forbidden decay
modes}\\\hline
$\jpsi\to \Lambda\bar\Lambda\pi^0$&$19\pm4$&$2.2\pm0.5\pm0.5$\\
$\jpsi\to \Sigma(1385)^0\bar\Lambda$&13&$<2.0(90\%CL)$\\
$\jpsi\to \Sigma^0\bar\Lambda$&$11$&$<0.9(90\%CL)$\\
$\jpsi\to \Delta^+\bar p$&$50$&$<1.0(90\%CL)$\\\hline\hline
\end{tabular}
\end{center}
\end{table}
As discussed in the literature, $SU(3)$-flavor symmetry can be broken
in several ways:
\begin{itemize}
\item One photon processes, i.e. $c\bar c\to \gamma \to B_{10}\bar
B_8$. Because the direct product $8\otimes\bar{10}$ contains an
octet contribution, this is possible via the octet component of the
photon. It also occurs via the processes $c\bar c\to gg\gamma\to
B_{10}\bar B_8$, which represent a direct electromagnetic decay.
As calculated in the framework of pQCD, the ratio $R$ for this decay
amplitude to that for the three-gluon decay is a few percent,
$R_{QCD}=-4\alpha/(5\alpha_s)$ \cite{korner:87}; 
in the framework of vector meson dominance,
$R_{VMD}=24\alpha/(5\alpha_s)$.

\item A second SU(3) breaking mechanism arises from the mass
difference of light and strange quarks. The decay chain $c\bar
c\to (u\bar u+d\bar d+s\bar s)_1\to \alpha(u\bar u+d\bar
d)_{1\oplus 8}+\beta (s\bar s)_{1\oplus 8}\to B_{10}\bar B_8$ can
occur if the coupling $\alpha$ and $\beta$ differ. The mass
breaking can equivalently be described by an octet \cite{haber:99}
or 27-plet representation to the $\jpsi$ wave function
\cite{DM2:87}.

\item The third pssible mechanism can come from intermediate
states. As pointed by Genz {\it et al.}~\cite{genz:85},  an
intermediate $q\bar q$ state could lead to an apparent $SU(3)$
violation. A generalization to multi-quark intermediate states
would also make the contribution of a 27-plet possible. If the
decay amplitudes are decomposed into the contributions from
one-photon ($D$), octet SU(3) breaking ($D'$), and 27-plet terms
($D^"$), the ratios of branching fractions for $\jpsi$ decays into
octet-decuplet baryon pairs are given by
\begin{eqnarray}\label{}
R_1&=&{B(\Xi(1530)^0\bar\Xi^0)\over
B(\Sigma(1385)^+\bar\Sigma^-)}\propto \left |{2D+D'+{3\over
2}D"\over
2D+D'-D"}\right |^2,\nonumber \\
R_2&=&{B(\Xi(1530)^-\bar\Xi^+)\over
B(\Sigma(1385)^-\bar\Sigma^+)}\propto \left |{D'+{3\over 2}D"\over
D'-D"}\right |^2,\\
R_3&=&{B(\Sigma(1385)^+\bar\Sigma^-)-B(\Sigma(1385)^-\bar\Sigma^+)\over
B(\Xi(1530)^0\bar\Xi^0)-B(\Xi(1530)^-\bar\Xi^+)}\propto {|
{2D+D'-D"}|^2-|D'-D"|^2\over |2D+D'+{3\over 2}D"|^2-|D'+{3\over
2}D"|^2}\nonumber.
\end{eqnarray}
Octet dominance $(D'>>D^{''})$  predicts that
$R_1=R_2=R_3=1$, in contradiction with the measured values
$R_1=1.3\pm 0.6, R_2=2.8\pm 1.0$ and $R_3=-0.1\pm 0.3$. The more
sophisticated model of K\"{o}rner \cite{korner:96}, which allows
for strong mass breaking effects and final state dependent
electromagnetic amplitudes but neglects a 27-plet contribution,
runs into similar problems. While a model allowing electromagnetic
contributions is ruled out, some electromagnetic component seems
to be required, since $B(\jpsi\to\Xi^0(1530)\bar\Xi^0)\neq
B(\jpsi\to\Xi(1530)^-\bar\Xi^+)$. In the framework of the given
model, the data can be well described if both electromagnetic and
strong isospin breaking effects are taken into account.
\end{itemize}

\subsection{$CP$ violation}

The decays of $\jpsi$ (or $\psip$) $\to B_8B_8$ ($B_8$: octet
baryon) can be used to search for an electric dipole momentum
(EDM) of baryons. A non-zero value of EDM would indicate that $CP$
symmetry is violated. As shown in Ref.~\cite{hexg:93}, for
$\jpsi\to B(p_1)\bar B(p_2)$ the decay amplitudes can be
parameterized as
\begin{equation}
\mathcal{M}=\epsilon^\mu\bar
u(p_1)[\gamma_\mu(a+b\gamma_5)+(p_{1\mu}-p_{2\mu})(c+id\gamma_5)]v(p_2)\equiv
\epsilon^\mu A_\mu ,
\end{equation}
where $\epsilon^{\mu}$ is the polarization of the $\jpsi$. If $CP$ is
violated, $d \neq 0$.

From an experimental point of view, the decay $\jpsi\to
\Lambda\bar\Lambda$ is a good laboratory to search for an EDM of
the $\Lambda$, since this channel has a large branching fraction
($B(\jpsi\to \Lambda\bar\Lambda)=(1.54\pm0.19)\times 10^{-3}$) and
can be well reconstructed with almost no background. The
polarization of the $\Lambda~(\bar\Lambda)$ particles are measured by
analyzing the subsequent $\Lambda( \textbf{s}_1)\to p(
\textbf{q}_1)\pi^-$, $\bar \Lambda(\textbf{s}_2)\to \bar p(
\textbf{q}_2)\pi^+$ decays with density matrices
$\rho_\Lambda=1+\alpha_+\textbf{s}_1\cdot
\textbf{q}_1/|\textbf{q}_1|$ and $\rho_{\bar
\Lambda}=1-\alpha_-\textbf{s}_2\cdot \textbf{q}_2/|\textbf{q}_2|$.
Experimental observables $O$ can be constructed from
$\textbf{p},\textbf{q}_i$ and the electron beam direction
\textbf{k}. Their expectation values are given by
\begin{equation}
\langle O \rangle ={\sqrt{1-4m^2/M^2}\over 2M\Gamma(\jpsi\to \Lambda\bar
\Lambda)8\pi} {1\over (4\pi)^3}\int d\Omega_p d\Omega_{q_1}
d\Omega_{q_2} OTr\{R_{ij}\rho_{ji}\rho_\Lambda\rho_{\bar \Lambda}\},
\end{equation}
where $R_{ij}=A_iA_j^*$ and $\rho_{ij}$ are the density matrices for
$\jpsi$ decays into $\Lambda\bar \Lambda$ and $\jpsi$ production
from $e^+e^-$, respectively. The $CP$-odd observable $A$ and
$CPT$-even observable $B$ are constructed as:
\begin{eqnarray}
A&=&\theta(\hat{p}\cdot (\hat q_1\times \hat q_2))-\theta(-\hat
p\cdot (\hat q_1\times \hat q_2)) \nonumber \\
B&=&\hat p\cdot (\hat q_1\times \hat q_2)~,
\end{eqnarray}
where $\theta(x)$ is 1 if $x>0$ and 0 if $x<0$. The
expectation values can be expressed as:
\begin{eqnarray}
\langle A \rangle &=& -{\alpha^2_-\beta^2\over 96M\Gamma(\jpsi\to\Lambda\bar
\Lambda)}M^2[2mRe(da^*)+(M^2-4m^2)Re(dc^*)] \nonumber \\
\langle B \rangle&=&-{48\over 27\pi} \langle A \rangle~.
\end{eqnarray}

The quantity $ \langle A \rangle$ is equal to
\begin{equation}
\langle A \rangle ={N^+-N^-\over N^++N^-},
\end{equation}
where $N^\pm$ indicate events with sgn[$\bf p\cdot (q_1\times
q_2)]=\pm$, respectively.

The EDM $d_\Lambda$ of he$\Lambda$ is related to the quantity $\langle A 
\rangle $ by
the Lagrangian:
$$L_{\textrm{dipole}}=i{d_\Lambda\over 2}\bar \Lambda \sigma_{\mu\nu}\gamma_5\Lambda
F^{\mu\nu},$$
where $F^{\mu\nu}$ is the field strength of the electromagnetic
field. Exchanging a photon between the $\Lambda$ and a $c$ quark, the
$CP$ violating $c-\Lambda$ interaction is expressed by
$$L_{c-\Lambda}=-{2\over 3M^2}ed_\Lambda(p_1^\mu-p_2^\mu)
\bar c\gamma_\mu c\bar \Lambda i\gamma_5 \Lambda$$
From these relations one has $d=-{2.5\over 3M^2}ed_\Lambda$, and
\begin{equation}
|\langle A \rangle |=\left\{
 \begin{array}{cc}
  5.6\times 10^{-3}d_\Lambda/(10^{-16}e\mbox{ cm}),& \textrm{if the $a$ term dominates}\\
  1.25\times 10^{-2}d_\Lambda/(10^{-16}e \mbox{ cm}),& \textrm{if the $c$ 
term dominates}.
 \end{array}
 \right.
\end{equation}
The current experimental upper bound on $d_\Lambda$ is $1.5\times
10^{-16}e$ cm~\cite{PDG2006}. If $d_\Lambda$ indeed has a value
close to its experimental upper bound, the asymmetry $|\langle A
\rangle |$ can be as large as $10^{-2}$. So $\langle A \rangle $
can be used to improve the bound on $d_\Lambda$. If $10^{10}$ 
$\jpsi$ events are produced, one can improve the upper
bound on $d_\Lambda$ by more than an order of magnitude. The same
analysis can be easily extended to $\jpsi$ to $\Sigma$, $\Xi$, etc.

Quantitative predictions for $CP$ violation in hyperon decays
indicate that $A={\displaystyle
\frac{\alpha_\Lambda+\alpha_\Lambda}
{\alpha_\Lambda-\alpha_\Lambda} }$ should be in the range
$(-2\times 10^{-5}\sim -1\times 10^{-4})$. Present experimental
results dose not have sufficient sensitivity to observe such a
small effect, but \bes3 will have an opportunity to measure this
quality. As shown by DM2 collaboration, the decay of
$\jpsi\to\Lambda\bar\Lambda$ can be used to look for $CP$ violation
by studies of the correlations between the $p$ and $\bar p$ momenta in 
the mother system frame~\cite{DM2:88}. The differential cross-section
for the $\jpsi\to\Lambda\bar\Lambda\to p\pi^-\bar p\pi^+$ 
decay chain can be expressed as:
\begin{eqnarray}
{d\Gamma\over d \cos\theta d\Omega^{\prime}d\Omega^{\prime\prime}}& \propto
{\displaystyle 2\left| {A_{++}\over A_{--}}\right | } \sin^2\theta
[1-\alpha_\Lambda\alpha_{\bar\Lambda}(\cos\theta^{\prime} \cos\theta^{\prime\prime}
-\sin\theta^{\prime} \sin\theta^{\prime\prime}
\cos(\phi^{\prime}-\phi^{\prime\prime})]\nonumber\\
&+(1+\cos^2\theta)(1+\alpha_\Lambda\alpha_{\bar\Lambda}
\cos\theta^{\prime}\cos\theta^{\prime\prime},
\end{eqnarray}
where $\alpha_\Lambda~(\alpha_{\bar\Lambda})$ is the
$\Lambda~{(\bar\Lambda)}$ decay constant and $A_{\lambda_1\lambda_2}$
and $\theta$ are the helicity amplitude and polar angle of the
out-going $\Lambda$ for the $\jpsi\to\Lambda\bar\Lambda$,
respectively. The angular variables $\Omega'$ and $\Omega"$ are
defined as shown in Fig.~\ref{llpp}. Using this equation, the
quantity $\alpha_\Lambda\alpha_{\bar\Lambda}$ can be obtained
experimentally.

\begin{figure}[hbt]
\begin{minipage}{7cm}
\centerline{\psfig{file=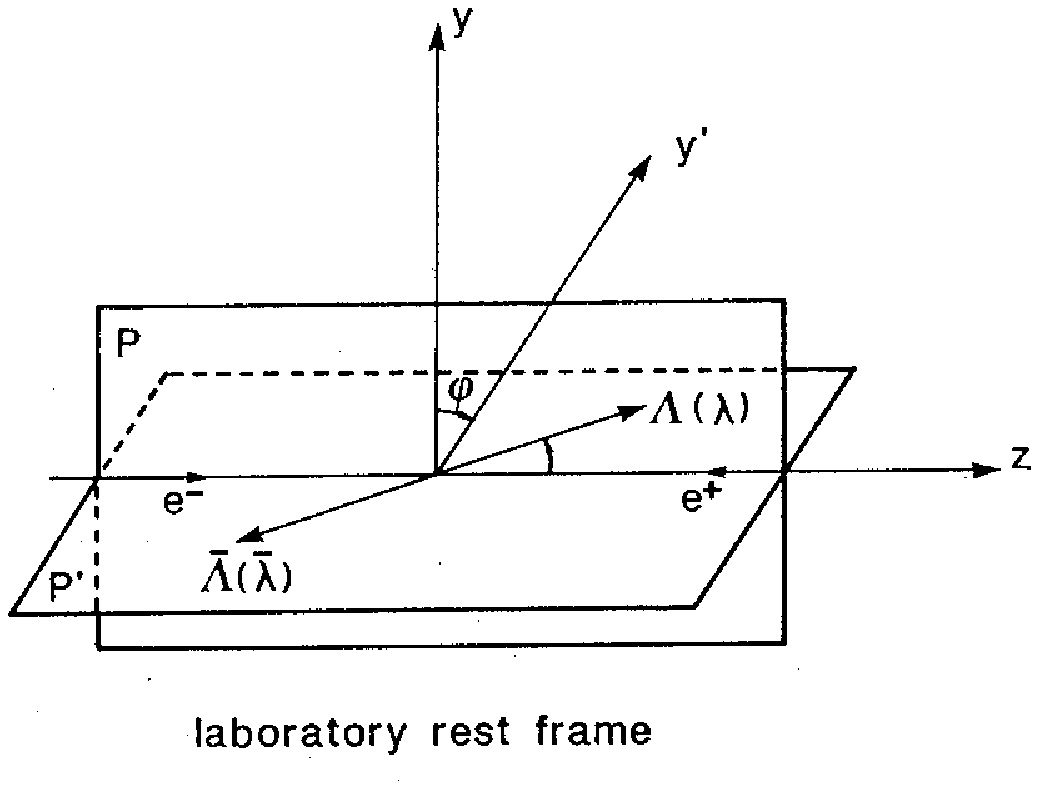,height=5.0cm
,width=7.cm}}
\end{minipage}
\begin{minipage}{3.5cm}
\centerline{\psfig{file=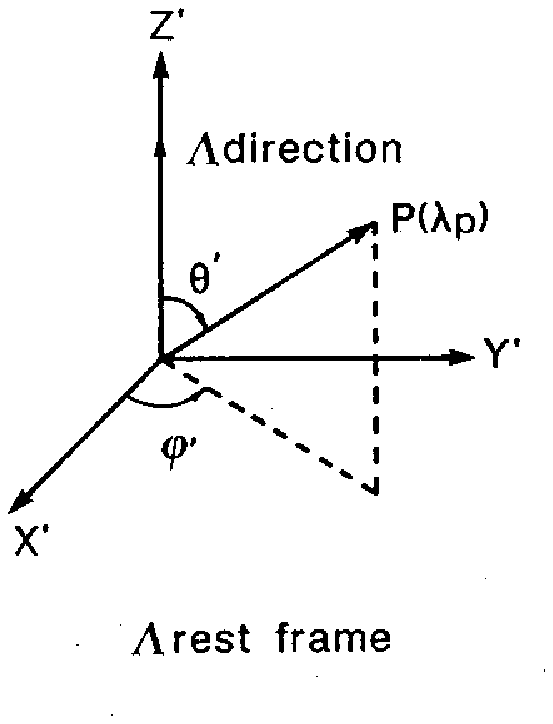,height=5.0cm
,width=3.5cm}}
\end{minipage}
\begin{minipage}{3.5cm}
\centerline{\psfig{file=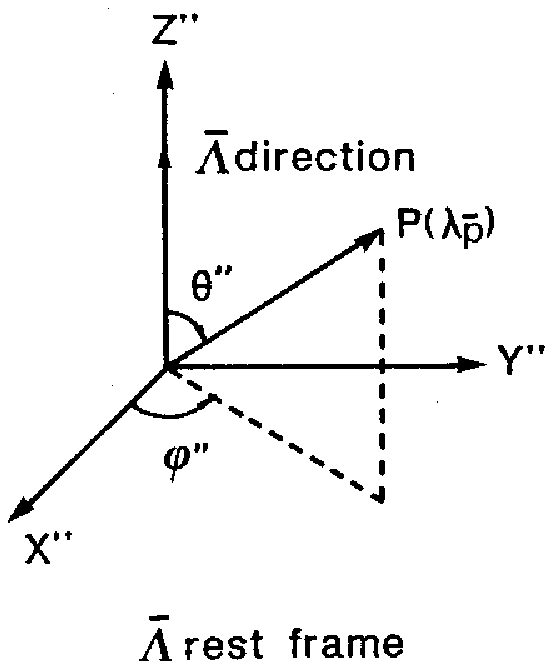,height=5.0cm
,width=3.5cm}}
\end{minipage}
\caption{\label{llpp}The definition of helicity system for $J/\psi\to
\Lambda\bar\Lambda,~\Lambda(\bar\Lambda)\to p\pi-(\bar p\pi+)$.}
\end{figure}

As T\"ornqvist \cite{Tornqvist:81} first demonstrated, the decays 
$\eta_c$ and $\jpsi\to\Lambda\bar\Lambda$ are experimental
realizations of Bell's conceptual proposition for testing Quantum
Mechanics versus local hidden variable theories. The initial state
is well known and due to parity symmetry breaking, the decay of 
the $\Lambda$ serves 
as a spin analyser. The proton direction plays the
same part that the direction of external polarimeter in classical
experiments with photons~\cite{asp:82}. 
The important quantity is the
scalar product of $p$ and $\bar p$ 3-momentum vectors
in the $\Lambda$ and
$\bar\Lambda$ rest frames. The differential cross-section of the
$\eta_c\to\Lambda\bar\Lambda$ decay is directly proportional to
$\vec a\cdot \vec b$. So it is the most sensitive test of Quantum
Mechanics since this scalar product can be compared to Bell's
inequality. The decay $\eta_c\to \Lambda \bar{\Lambda}$ was
recently observed by the Belle Collaboration~\cite{etacLL_belle} with
a branching fraction of $(8.7\pm 3.7)\times 10^{-4}$. This would
correspond to
a $\eta_c\to \Lambda \bar{\Lambda}$ sample of about
100,000 events in a 10~billion $\jpsi$ data sample, it will be a very
suitable sample for this study. For the $\jpsi\to\Lambda\bar\Lambda$
case, Tornqvist reformulated the differential cross-section as:
\begin{equation}
{d\Gamma\over d \cos\theta d\Omega'd\Omega^{"}}\propto
2(1-{p_\Lambda^2\over
E_\Lambda^2}\sin^2\theta)(1-\alpha_\Lambda^2a_nb_n)+{p_\Lambda^2\over
E_\Lambda^2}\sin^2\theta[1-\alpha_\Lambda^2(\vec a\cdot \vec
b-2a_xb_x)],
\end{equation}
where $\vec a$ and $\vec b$ are the proton and antiproton
momentum, respectively in the $\Lambda~(\bar\Lambda)$ rest frame,
$x$ is the direction orthogonal to the $\Lambda\bar\Lambda$
direction and to the $e^+e^-$ beam axis and $\vec n$ is an axis
defined to take into account the suppression of 0-spin projection
in the $\jpsi$ decays. The terms containing $a_nb_n$ or $a_xb_x$
only reduce the sensitivity of the test since they do not depend
on the nature of the theory, and they play the same role as hidden
parameters \cite{Tornqvist:81}. The contribution of the $\vec
a\cdot \vec b$ term is important for the test of Quantum
Mechanics. Unfortunately $p_\Lambda^2/E_\Lambda^2$ from
$\jpsi$ decay is
only equal to 0.48 and $\alpha_\Lambda^2$ is 0.412; this reduces
the contribution of the $\vec a\cdot\vec b$ term to the experimental
measurement.

The DM2 collaboration~\cite{DM2:88} measured
the $\Lambda$-$\bar{\Lambda}$ decay parameter asymmetry to be:
$$A={\alpha_\Lambda+\alpha_{\bar\Lambda}\over \alpha_\Lambda-\alpha_{\bar\Lambda} }
=0.01\pm 0.10$$ 
with 1077 observed $\jpsi\to \Lambda \bar \Lambda$ events. 
The precision of this measurement does not
permit one to conclude anything about $CP$ violation; with $10^{10}$
$\jpsi$ events, the sensitivity is expected to improve to $8\times
10^{-4}$.

The measurement of the correlation between the proton and
antiproton in $\jpsi\to\llb$ decay is associated with the test of
Bell's inequality. For example, in the
$\eta_c\to\llb\to\ppbar\pi^+\pi^-$ decay, the spin correction
between the two nucleon predicted by Quantum Mechanics can be
expressed by \cite{Tornqvist:81}:
\begin{equation}
I(\vec a,\vec b)\propto 1+\alpha^2 \vec a\cdot \vec b,
\end{equation}
while a hidden measurement of $\Lambda$ polarization before the
decay would reduce the slope to $\alpha^2/3$, {\it i.e.}
\begin{equation}
I(\vec a,\vec b)\propto 1+{\alpha^2\over 3} \vec a\cdot \vec b.
\end{equation}
Using invariance under rotations and reflections, one can derive
a special bound for Bell's inequality: $$|E(\theta)|\le 1-{2\over
\pi}\theta,~0\le\theta\le\pi.$$ Figure~\ref{bellEquality} shows
the distribution in the angle $\theta$ between two pions as
predicted by Quantum Mechanics and the area bounded by Bell's
inequality.

\begin{figure}[hbt]
\centering
\centerline{\psfig{file=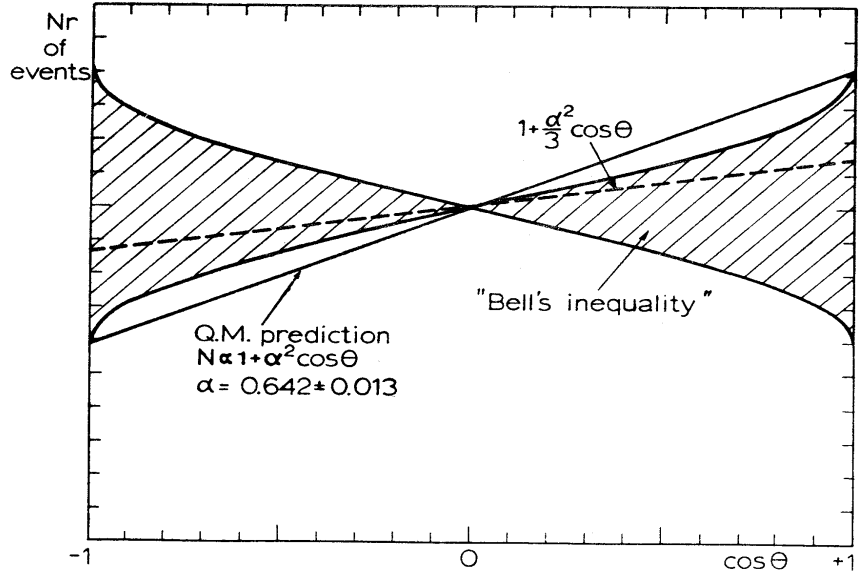,height=12.0cm,width=8.0cm,angle=-90}}
\caption{\label{bellEquality} The distribution in the angle
$\theta$ between the $\pi^+$ and the $\pi^-$ as predicted by
Quantum Mechanics (solid line), and if a "hidden" $\Lambda$
polarization measurement is done before the decay (dashed line).
The shaded area gives the domain where the inequality is
satisfied.}
\end{figure}

\subsection{Exotic states}

As is well known, in addition to conventional 
$q\bar{q}$ mesons and $qqq$ baryons, QCD theory also
predicts the existence of multiquark states, hybrid states and
other exotic states. Searching for such exotic states has been
attempted for a long time, but none are established experimentally.
One of the difficulties in identifying exotic states is the need
to find signature properties that distinguish them from the common
states or get information about their mixing. For this it is important
for experimentalists to collaborate closely with theoretical
physicists. 
Hadronic spectroscopy will continue to be a key tool to search for
$N^*$ states (see the section of "Baryon spectroscopy") and exotic
states. Various models and methods have been used to predict the
spectrum of hybrid mesons/baryons, such as the bag model, QCD sum
rule, the flux tube model, and so on. Though each model assumes a
particular description of exited glue, fortunately they often
reach similar conclusions regarding the quantum numbers and
approximate masses of these states. For instance, the predictions
on the light hybrid mesons are in good agreement with each other,
with the so-called exotic number $J^{PC}=1^{-+}$ states having masses
about 1.5-2.0 GeV. In the baryon sector, the Roper resonance, {\it 
i.e.} the $N^*(1440)$ has been suggested as a potential candidate for a
hybrid baryon for a long time.

As shown in Ref.~\cite{BES:2006nstar}, $J/\psi$ decays provide an
excellent place for studying the Roper resonance. Using $58\times
10^{6}$ $\jpsi$ decays, the $N^*(1440)$ is clearly seen with a
statistical significance of 11$\sigma$. For the identification of 
the Roper
resonance as a hybrid state, information on transition
amplitudes from partial wave analyses will play an important role
at \bes3. As demonstrated in Ref.~\cite{prg:2004}, if the Roper
resonance is assigned as a pure hybrid state, numerical results
show that the ratios $ \Gamma(\jpsi^{(\Lambda)}\to\bar p\nstar)/
\Gamma(\jpsi^{(\Lambda)}\to\bar pp)  <2\%, \textrm{and }
\Gamma(\jpsi^{(\Lambda)}\to\bar \nstar\nstar)/
\Gamma(\jpsi^{(\Lambda)}\to\bar pp)  <0.2\%$, and their angular
distribution parameters are $\alpha_*=0.42\sim
 0.57$ and $\alpha_{**}=(-0.1)-(-0.9)$, respectively.
However, when the Roper resonance is assumed to be a common $2S$
state, the results are quite different, with $
\Gamma(\jpsi^{(\Lambda)}\to\bar p\nstar)/
\Gamma(\jpsi^{(\Lambda)}\to \bar pp)  = 2.0\sim 4.5, \textrm{and }
\Gamma(\jpsi^{(\Lambda)}\to\bar \nstar\nstar)/
\Gamma(\jpsi^{(\Lambda)}\to\bar pp)  = 3.2\sim 22.0 $, and with the
angular distribution parameter $\alpha_*=0.22\sim 0.70
,\alpha_{**}=0.06\sim 0.08$. This implies
 that, not only the dynamics of three gluons and created
quarks , but also the structure of the final cluster state , {\it i.e.}
$\st{qqq}$ or $\st{qqqg}$, play important roles in the evaluation
of the amplitudes in these decay processes. So it is suggestive
that an accurate measurement of the decay widths and angular
distributions of these channels may provide a novel tool to
probe the structure of the Roper resonance. If the Roper resonance
is assumed to be a mixture of a pure quark state $\st{qqq}$ with a
hybrid state $\st{qqqg}$ with a mixing parameter $\delta$,
the results show that the hybrid constituent makes a large
contribution to the decay width of the $\jpsi$ into $ \bar
pN^*(1440)$ and $\bar N^*(1440)N^*(1440)$.

In a search for the pentaquark state
$\Theta(1540)^+$ in $\jpsi$ and $\psip$ decays into $p\bar
nK_S^0K^-$ and $\bar p nK_S^0K^+$ with 14 million $\psip$ and 58
million $\jpsi$ events accumulated at the BESII detector, no
$\Theta(1540)$ signal events are observed, and upper limits are set for
$\mathcal B(\psip\to\Theta\bar\Theta\to K_S^0pK^-\bar n+K_S^0\bar
pK^+n)<0.84\times 10^{-5}$ and $\mathcal
B(\jpsi\to\Theta\bar\Theta\to K_S^0pK^-\bar n+K_S^0\bar
pK^+n)<1.1\times 10^{-5}$ at the 90\% confidence level
\cite{bes:2004theta}. So far there have been a number of other
high statistics experiments, none of which have found any evidence
for the $\Theta^+$; all attempts to confirm the two other
claimed pentaquark states have led to negative results. As
reviewed by Particle Data Group 2006, {\it "The conclusion
that Pentaquarks in general,
and the $\Theta^+$, in particular, do not exist, appears compelling."}

\subsection{Meson-nucleon bound states}

Searching for nucleon-nucleon and meson-nucleon bound states is an
intriguing problem in studies of the nuclear interaction. The recent
discoveries of near-threshold baryon-anitbaryon enhancements,
as discussed above, motivates 
searches for other possible bound states.
In this section, we first introduce some ideas on such possible
bound states from the point view of nuclear physics, then turn to
ways to search for them at \bes3.

The idea of explaining the interaction between nucleons at the quark
level was put forth three decades ago~\cite{daLiberman77ao}.
Resorting to the quark and gluon theory, the nucleon interaction
at short distances which had been credited to vector-meson
exchanges could then be interpreted as a manifestation of
diquark-diquark and quark-quark interactions. 
Along these lines, a QCD
van der Waals force was introduced to describe a special kind of
bound states -- charmonium-nucleon bound
states~\cite{sjBrodsky90}, say an $\etac$-$N$ or a $\jpsi$-$N$ bound
state. In the QCD picture, the nuclear forces are identified with the
residual strong color interactions due to quark interchange and
multiple-gluon exchange. The peculiar feature of
charmonium-nucleon bound state lies in the fact that because of the
distinct flavors of the quarks involved in the charmonium-nucleon
interaction there is no quark exchange to first order in the
elastic processes and, therefore, no single-meson-exchange potential
from which to build a usual nuclear potential. The nuclear
interaction in this case is purely gluonic and, thus, of a
distinctive nature from the ordinary nuclear forces. The
production of a charmonium-nucleon bound state would be the first
realization of a hadronic nucleus with exotic components bound by a
purely gluonic potential. In Ref.~\cite{sjBrodsky90}, a
nonrelativistic Yukawa-type attractive potential $V_{(\QQb)A}=
-\alpha e^{-\mu r}/r$ was utilized to characterize the QCD van der
Waals interaction. Estimates indicated that the QCD van der
Waals interaction due to multiple-gluon exchange can provide a
kind of attractive nuclear force capable of binding heavy
quarkonia to nuclei.

Following the above idea, the possibility of a
$\phi$-$N$ bound state was studied
in Ref.~\cite{hGao01}. Similar to the charmonium state, here the
$\phi$ meson is almost a pure $s\overline{s}$ state, so one could
expect that the attractive QCD van der Waals force dominates the
$\phi$-$N$ interaction. Using a variational method and following
Ref.~\cite{sjBrodsky90} to assume $V_{(\QQb)N}= -\alpha e^{-\mu
r}/r$, it was found~\cite{hGao01} that a $\phi$-$N$ bound state
is possible, with a binding energy of around 1.8~MeV.

Recently, a $\phi$-$N$ bound state structure with spin-parity
$J^P=3/2^-$ and $J^P=1/2^-$ were dynamically studied in both the
chiral $SU(3)$ quark model and the extended chiral $SU(3)$ quark
model~\cite{fHuang06}. In the chiral $SU(3)$ quark model, the
quark-quark interaction containing confinement, one gluon exchange
(OGE) and boson exchange stemming from scalar and pseudoscalar
nonets, and short range quark-quark interactions  provided by OGE
and quark exchange effects are included.  It remains a controversial
problem for low-energy hadron physics whether gluons or Goldstone
bosons are the proper effective degree of freedom in addition to the
constituent quarks. Thus the $SU(3)$ quark model has been extended
to include the coupling of the quark and vector chiral fields, and
the role of OGE in the short range quark-quark interaction is then
nearly replaced by the vector meson exchanges. This so-called
extended chiral SU(3) quark model can successfully reproduce as many
physics features as does of the chiral SU(3) quark model.
Therefore, both models are adopted to study $\phi$-$N$ bound
state, where $N$ and $\phi$ are treated as two clusters and the
corresponding resonating group method (RGM) equation are solved.
The calculation results indicate the mechanisms of the quark-quark
short-range interaction are quite different for the two chiral models.
Moreover, one result from the extended chiral SU(3) quark model
indicates the existence of a $\phi$-$N$ bound state.

Next we turn to experiment aspects of searching for a $\phi$-$N$ bound
state in \bes3. The calculation~\cite{zhangzy} gives
$M_{\phi N} \simeq 1950 \sim 1957$~MeV and $\Gamma_{\phi N} \simeq
4.4$~MeV, the intrinsic width of the $\phi$ meson. The
invariant mass of the $\phi$ reconstructed from two kaons will be less
than that of free $\phi$ due to the existence of some bounding energy.
This mass deficit can be understood as a result of the fact
that the $\phi$ decays off its energy shell when bound to the
nucleus.

Since the total mass of the $\phi N$ system and a $\bar N$ is close
to 3.0~GeV, the production of such state in $\jpsi$ decays will be
suppressed due to small phase space, however, it could be
produced in $\psp$ decays, or from $\chicj$ decays with much
larger phase space, although with fewer statistics.

The decays of the $\phi N$ bound state could be $K \bar{K} N$,
with a $K \bar{K}$ mass that is below the $\phi$ resonance.
Experimentally,
one expects to see a shoulder on the low-mass side of the $\phi$ meson
peak or another narrow peak below the $\phi$ resonance. In principle,
there could be interference between $\phi\to K \bar{K}$ and the
$K\bar K$ from $\phi N$ bound state decays, this may make the
identification of the bound state difficult.

The annihilation of the $s\bar{s}$ quark inside the $\phi$ meson
is not small, so another way of searching for the $\phi N$ bound
state is through its decay into $\pi^+\pi^-\pi^0+N$, considering
the large branching fraction of $\phi\to \pi^+\pi^-\pi^0$, with
even smaller phase space to $K \bar{K} N$, the decay rate for the
$\phi N$ bound state to $\pi^+\pi^-\pi^0+N$ could even be larger.
However, the background may be more serious in the
$\pi^+\pi^-\pi^0+N$ mode.

\part[Charm Physics]{Charm Physics \\
\vspace*{2cm}
 {\centering  \Large Conveners \\
 Hai-Bo Li, Zhi-Zhong Xing}\\
 \vspace*{1cm}
\Large Contributors \\
 D.~M.~Asner, I.~I.~Bigi, J.~Charles, J.~C.~Chen, H.~Y.~Cheng, S.~
Descotes-Genon, K.~L.~He, H.~B.~Li, J. Liu, H. H. Liu, H.~L.~Ma, G.~Rong, L.~Roos, S.
S. Sun, S.~T$^\prime$Jampens, Y.~L.~Wu,
Z.~Z.~Xing, M. Yang, M.~Z.~Yang, Y.~D.~Yang, D.~Y.~Zhang, M.~Zhong and Jia-Heng Zou}
\label{part:five}
\def\bes3{\mbox{\slshape B\kern-0.1em{E}\kern-0.1em S-III}}
\def\qqbar {\ensuremath{q\overline q}\xspace}
\def\ccbar {\ensuremath{c\overline c}\xspace}
\def\ssbar {\ensuremath{s\overline s}\xspace}
\def\Kbar    {\kern 0.18em\overline{\kern -0.18em K}{}\xspace}
\def\Kb      {\ensuremath{\Kbar}\xspace}
\def\KK      {\ensuremath{K\Kbar}\xspace}
\def\Kz      {\ensuremath{K^0}\xspace}
\def\Kzb     {\ensuremath{\Kbar^0}\xspace}
\def\KzKzb   {\ensuremath{\Kz {\kern -0.16em \Kzb}}\xspace}
\def\Ku      {\ensuremath{K^+}\xspace}
\def\Kub     {\ensuremath{K^-}\xspace}
\def\Kp      {\ensuremath{\Ku}\xspace}
\def\Km      {\ensuremath{\Kub}\xspace}
\def\Kpm     {\ensuremath{K^\pm}\xspace}
\def\Kmp     {\ensuremath{K^\mp}\xspace}
\def\Ks     {\ensuremath{K_S}\xspace}
\def\Kl     {\ensuremath{K_L}\xspace}
\def\KsKs   {\ensuremath{\Ks {\kern -0.16em \Ks}}\xspace}
\def\KlKl   {\ensuremath{\Kl {\kern -0.16em \Kl}}\xspace}
\def\KsKl   {\ensuremath{\Ks {\kern -0.16em \Kl}}\xspace}
\def\KlKs   {\ensuremath{\Kl {\kern -0.16em \Ks}}\xspace}
\def\Dbar    {\kern 0.18em\overline{\kern -0.18em D}{}\xspace}
\def\DD    {\ensuremath{D\Dbar}\xspace}
\def\Dz      {\ensuremath{D^0}\xspace}
\def\Dzb     {\ensuremath{\Dbar^0}\xspace}
\def\DzDzb   {\ensuremath{\Dz {\kern -0.16em \Dzb}}\xspace}
\def\DsP      {\ensuremath{D_S^+}\xspace}
\def\Ds      {\ensuremath{D^+_s}\xspace}
\def\Dp      {\ensuremath{D^+}\xspace}

\def\DsM     {\ensuremath{D_S^-}\xspace}
\def\DspDsm    {\ensuremath{\DsP {\kern -0.16em \DsM}}\xspace}
\def\Dp      {\ensuremath{D^+}\xspace}
\def\Dm     {\ensuremath{D^-}\xspace}
\newcommand{\DpDm}{\ensuremath{\Dp {\kern -0.16em \Dm}}\xspace}

\def\DpBm    {\ensuremath{\Dp {\kern -0.16em \Dm}}\xspace}
\def\Bbar    {\kern 0.18em\overline{\kern -0.18em B}{}\xspace}
\def\Bb      {\ensuremath{\Bbar}\xspace}
\def\BB      {\ensuremath{B\Bbar}\xspace}
\def\Bz      {\ensuremath{B^0}\xspace}
\def\Bzb     {\ensuremath{\Bbar^0}\xspace}
\def\BzBzb   {\ensuremath{\Bz {\kern -0.16em \Bzb}}\xspace}
\def\Bu      {\ensuremath{B^+}\xspace}
\def\Bub     {\ensuremath{B^-}\xspace}
\def\Bp      {\ensuremath{\Bu}\xspace}
\def\Bm      {\ensuremath{\Bub}\xspace}
\def\Bpm     {\ensuremath{B^\pm}\xspace}
\def\Bmp     {\ensuremath{B^\mp}\xspace}
\def\BpBm    {\ensuremath{\Bu {\kern -0.16em \Bub}}\xspace}
\def\Bs      {\ensuremath{B_s}\xspace}
\def\Bsb     {\ensuremath{\Bbar_s}\xspace}

\def\op{{\bf P}}
\def\oc{{\bf C}}
\def\ot{{\bf T}}
\def\cp{{\bf CP}}
\def\cpt{{\bf CPT}}

Nobody doubts that the discovery of hadrons with charm of the hidden and
open variety -- {\it i.e.} 
charmonium as well as $D$ mesons -- was instrumental in the acceptance of
the Standard Model (SM) in general, and 
of quarks as {\em physical} degrees of freedom, rather than objects of 
mere {\em mathematical}  
convenience, in particular. Yet charm is all too often viewed as 
a quantum
number with a great past and with no particularly interesting future. 
This is due to SM predictions of a rather dull electroweak phenomenology 
of  CKM parameters, a low frequency for
$D^0-\bar{D^0}$ oscillations, tiny (at best) {\it CP} asymmetries,
and extremely rare flavor changing neutral currents which, in any case,
are swamped by huge backgrounds
from long-distance dynamics. 

However, this pessimistic view focuses merely on the surface. 
A more  profound perspective starts from the observation 
that charmed quark dynamics is full of challenges and promises. 
There is actually a three-fold  motivation for {\em further 
dedicated and comprehensive} studies of charmed-quark physics: 
\begin{enumerate}
\item 
Because the SM weak forces are well known, charmed particle decays provide 
an excellent laboratory for studying 
the impact of nonperturbative-QCD dynamics and for testing the 
validity of theoretical methods for dealing with them. 
\item 
This, in turn, provides a calibration of the theoretical tools 
that are available for dealing with $B$ decays. 
\item 
Charm decays provide a novel window on New Physics in the weak sector,
precisely because there is so 
little SM Background, in particular in {\it CP} violating processes. 
\end{enumerate}
None of the novel successes that the SM has scored since the turn of the
millenium in the heavy flavour sector invalidates at all the case for 
the incompleteness of the SM. However, they strongly suggest that we 
{\em cannot} count on a numerically massive impact of New Physics 
on heavy flavour transitions. 
Thus, high accuracy and reliability -- on the experimental as well as
theoretical side -- are
{\em essential} for the exploitation of the indirect probes of New 
Physics that are accessible in heavy flavor decays, 
as is also the need  to search in unusual places. 
Items (2) and (3) above can be expanded as follows: 
\begin{itemize}
\item 
In order to 
saturate the discovery potential provided by
$B$ and $D$ meson decays, we have to
bring them under as precise theoretical 
control as possible.  This requires much more than 
simply obtaining ``engineering''
input such as absolute
branching ratios or the spectrum and quantum numbers of charmed 
hadrons. In addition we have to map out and understand in detail
the Dalitz plots of many 
three-body $D^0$, $D^+$ and $D_s^+$ decay channels, and extend 
such analyses to four-body final states. 
\item 
As mentioned above, the SM predicts at best small {\it CP}-violating 
asymmetries in charmed particle decays
(basically in singly Cabibbo suppressed modes only) -- a 
domain we are just now starting to probe in a quantitatively 
meaning way.   While  
we cannot count on New Physics inducing large {\it CP} violation in 
charmed particle decays, the SM Background is either absent or very 
small.  In this sense, the NP
signal-to-noise ratio is favorable for
{\it CP} studies in charm transitions.  On the other hand,
large data samples are required,
and complex final states 
have to be analyzed with good control over the systematics. 

\end{itemize}
In the following, these issues are addressed in considerable detail.
While there are already several very good   
reviews in the recent literature~\cite{CHARMREV},  we 
have striven to make this
exposition as self-contained as reasonably possible. 

\chapter[Charm production and $D$ tagging]{Charm production and $D$ tagging \footnote{By Kang-Lin He, Sheng-Sen Sun, Ming Yang and Da-Yong Zhang}}
\label{chapter:charm_prod_tagging}

\section{Charmed particle production cross sections}
\label{sec:cross_section}

\subsection{Introduction}

Evidence for the onset of charmed particle production is clearly seen in 
the energy-dependence of the $R$ values for
$e^+e^- \rightarrow {\mbox hadrons}$ shown in Fig.~\ref{fig:rcharm}. 
Below the open charm $D\bar{D}$ threshold, the strikingly narrow 
$J/\psi$ and $\psi^{\prime}$ peaks have
been assigned to the $1S$ and $2S$ $c\bar{c}$ bound states
predicted by potential models that incorporate a color
Coulomb term at short distances and a linear scalar confining term
at large distances.  Above the $D\bar{D}$ threshold, there are 
several broad resonance peaks that decay predominantly into 
pairs of open-flavor charmed meson final states and, thus, 
have the potential of serving as ``factories'' for the production of 
charmed mesons.  In the strong decays of these above-threshold
resonances,  the initial $c\bar{c}$ meson decays
via the production of a light $q\bar{q}$ quark-antiquark pair
($q=u,d,s$), forming  $c\bar{q}$-$\bar{c}q$ systems that 
subsequently  separate into two charmed mesons. The
mechanism of this open-charm decay process of the charmonium
resonances is still poorly understood. 
In quark-model calculations, the process
is modeled by a simple phenomenological $q\bar{q}$ pair
production amplitude, where the $q\bar{q}$ pair is usually assumed
to be produced with vacuum ($0^{++}$) quantum numbers; 
variants of this decay model make different assumptions regarding
the spatial dependence of the pair production amplitude relative
to the initial $c\bar{c}$ pair.

\begin{figure}\centering
\includegraphics[width=6in,height=3in]{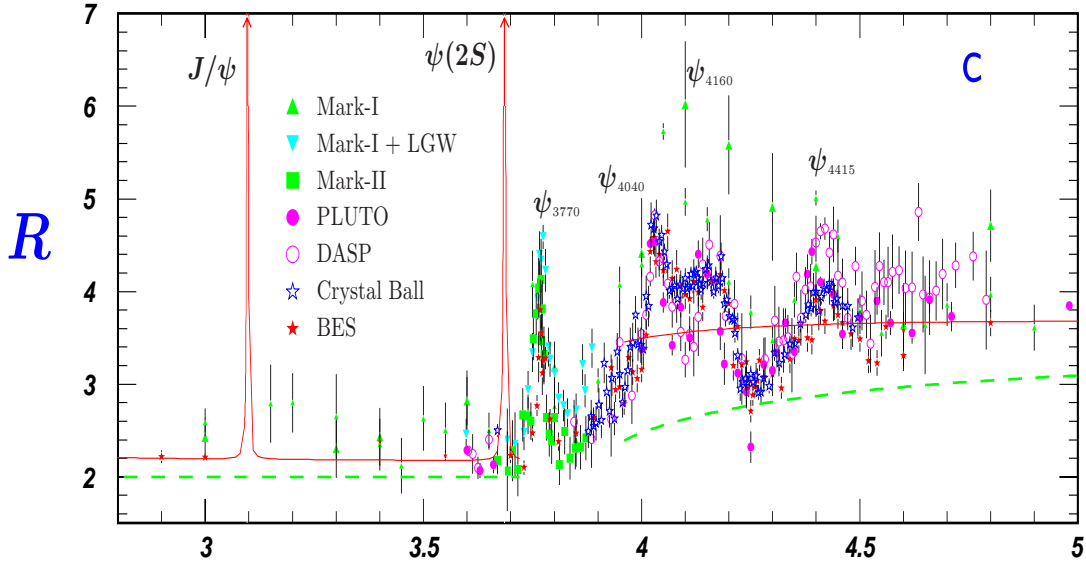}
\caption{Measured $R$ values in the region of the open charmed
particle threshold.}
\label{fig:rcharm}
\end{figure}

\par

A detailed study of the charmed particle
production cross section above the 
$D\bar{D}$ mass
threshold may provide a wealth of information about the strong
dynamics of heavy and light quarks. Experimentally,  charmed
cross sections ($\sigma_\textrm{charm}$) are determined from:
\begin{equation}
\displaystyle\sigma_{\textrm{charm}} = \frac{N_{\textrm{charm}}}{\mathscr{L}}
\end{equation}
where $\mathscr{L}$ is the integrated luminosity, and
$N_{\textrm{charm}}$ is the number of produced charmed meson pairs.
$N_{\textrm{charm}}$ can be obtained using a tagging technique that
measures  the number of charmed mesons that decay
via a certain ``tag'' decay  mode, and relating this to
$N_{\textrm{charm}}$ via
$${\mathcal S}=N_{\textrm{charm}}\cdot\epsilon\cdot B,$$
where ${\mathcal S}$ is the number of observed tags, 
$\epsilon$ is the detection efficiency, 
and $B$ is the branching fraction for the tag mode.  Accurate 
measurements of charm production
cross sections require both a large amount of 
integrated luminosity and high efficiency
$(\epsilon\cdot B)$.  The high statistics charm data obtained 
at \bes3 will provide precision
charm production cross section measurements above the $D\bar{D}$
threshold.
\begin{figure}
\centering
\includegraphics[width=4in,height=3in]{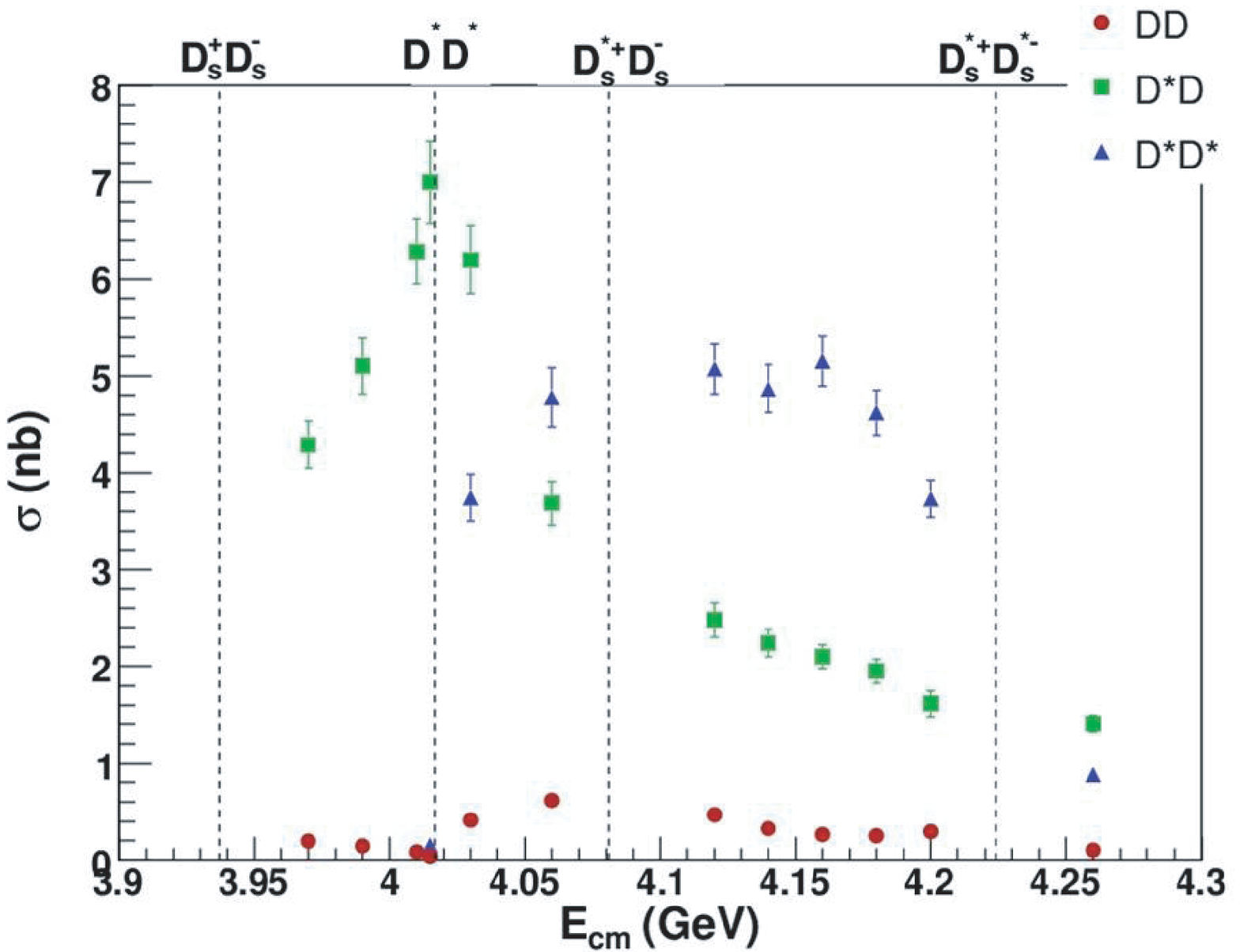}
\includegraphics[width=4in,height=3in]{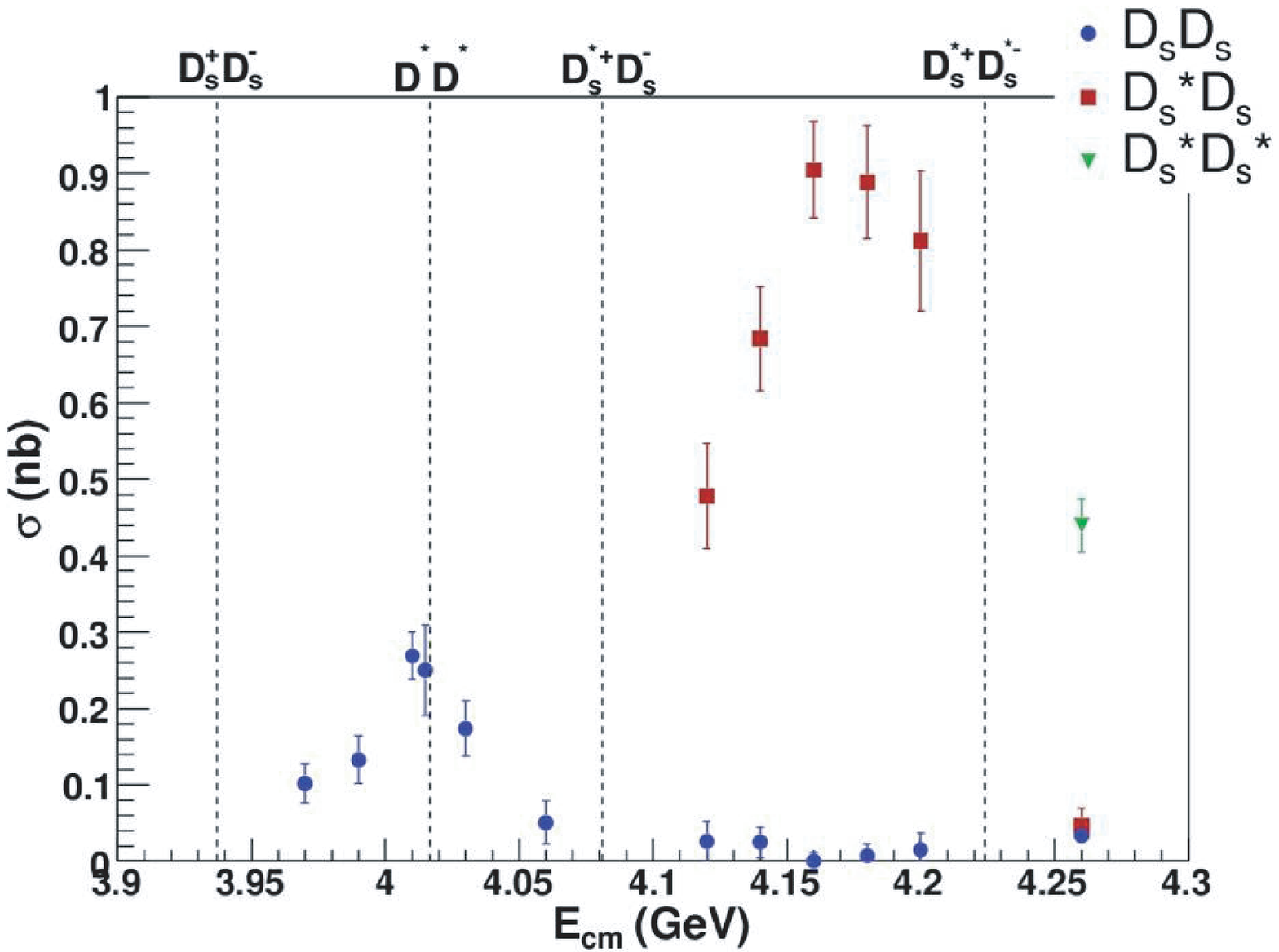}
\caption{Cross sections for $D\bar{D}$,
$D^{*}\bar{D}$, $D^{*}\bar{D^{*}}$,
$D_{s}\bar{D_{s}}$, $D_{s}^{*}\bar{D_{s}}$ and
$D_{s}^{*\pm}D_{s}^{*\mp}$ from CLEO-c.}
 \label{fig:cleo_dsscan}
\end{figure}
\par
In the early fall of 2005 and in the summer of 2006, 
the CLEO-c experiment spent two months of data-taking time 
scanning the  3.97$-$4.26GeV center-of-mass (c.m.) energy range
in order to determine
the optimal c.m. energy value for studying $D_{s}$ decays.
They measured cross sections at twelve c.m. energies
with a total luminosity of about 60~pb$^{-1}$. 
At each energy point, three,
five and eight tag decay modes were used for $D^{0}$, $D^{+}$ and
$D_{s}^{+}$ meson production measurements, respectively. 
The different production channels can
be distinguished based on the kinematics of the reconstructed tags
using the kinematic variables $M_{\textrm{inv}}$ and
$p_{D}$ ({\it i.e.} the invariant mass and c.m. momentum of the tagged
$D$'s).  
Preliminary cross section
results (with partially
evaluated systematic uncertainties, no correction for multi-body
contributions and not radiatively corrected)~\cite{part5:ref:cleo_dsscan}
are shown in~\figurename~\ref{fig:cleo_dsscan}. 
The CLEO-c results agree well with previous
 measurements~\cite{part5:ref:bes_4030_Ds,part5:ref:bes_4030_D,
part5:ref:mark2_4160,part5:ref:mark3_4160_D,part5:ref:mark3_4160_Ds}.
\par

As shown in \figurename~\ref{fig:cleo_dsscan}, there is very
little $D\bar{D}$ production at any energy covered by the scan. 
Instead there is a sharply peaked $D^{*}\bar{D}$ structure near 
the $D^{*}\bar{D^{*}}$ threshold and a broad $D^{*}\bar{D^{*}}$ 
peak or plateau that sets in just
above threshold.  The total charmed-particle production
cross section throughout this region is considerable, and
comparable to that for $D\bar{D}$ production at the 
$\psi(3770)$ ($\sim 6$nb). There is a
visible, but disappointingly small, peak in $D_{s}\bar{D_{s}}$
production near the $D^*\bar{D^*}$ threshold ($\sim0.3$nb), 
but a more  impressive broad peak near 4.17~GeV, where there
is about 1~nb of $D_{s}^{*}\bar{D_{s}}$ production.
\par
The CLEO-c scan data suggest the existence of ``multi-body''
production such as $e^{+}e^{-}\rightarrow D^{*}\bar{D}\pi$,
which is reflected in the fact that the sum of
the two-body modes that are measured does not account for all 
of charmed-meson production as determined from inclusive measurements. 
This is a very interesting possibility
that can be studied at \bes3.

\subsection{The $1^{--}$ resonances around 4.0GeV}

The four known $1^{--}$~ $c\bar{c}$ states above the $D\bar{D}$
threshold, {\it i.e.} the $\psi(3770)$, $\psia$, $\psi(4160)$ 
and $\psic$, are of
special interest because they are easily produced at 
an $e^{+}e^{-}$ collider. The cross section in the region 
around 4.0~GeV can be understood as the successive onset of 
specific charmed meson channels: 
$D\bar{D}$, $D^{*}\bar{D}$,
$D_{s}\bar{D_{s}}$, etc. \figurename~\ref{fig:charm} shows
the $c\bar{c}$ spectrum above the
$D\bar{D}$ threshold, with the thresholds of the lowest-lying 
charmed-meson decay channels indicated at the right. 
\tablename~\ref{tab:charm_decay_width} 
lists Ref.~\cite{part5:ref:high_charmonium}'s
predicted decay widths for these four
$1^{--}$ $c\bar{c}$ states.

\begin{figure}\centering
\includegraphics[width=5in, height=3in]{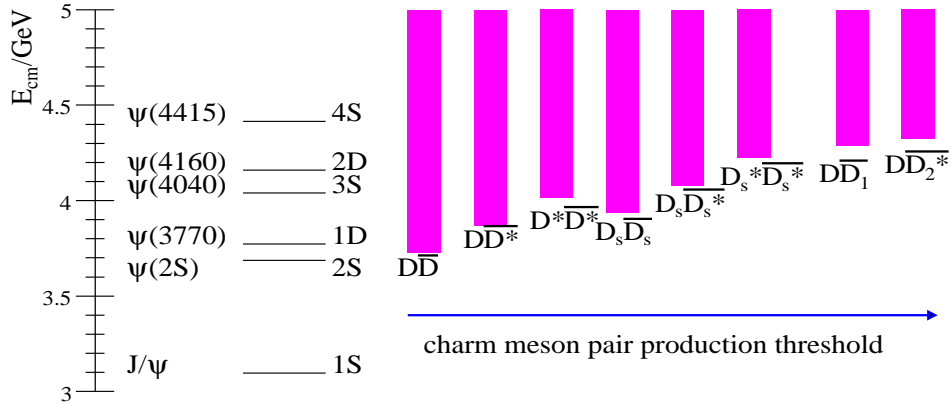}
\caption{The $c\bar{c}$ spectrum above the
$D\bar{D}$ threshold, and the thresholds
for the lowest-lying charmed-meson
decay channels. A comparison of the masses with  potential model
predictions indicate $\psi(3770)$, $\psia$, $\psib$ and $\psic$ 
assignments as the $1^{3}D_{1}$, $3^{3}S_{1}$, $2D^{3}D_{1}$ and
$4^{3}S_{1}$ $c\bar{c}$ states, respectively.  } \label{fig:charm}
\end{figure}

\begin{table}
\centering
\begin{tabular}{|c|c|c|c|c|} \hline
Mode  & $\psi(3770)$ &$\psia$ &$\psib$ &$\psic$ \\ \hline %
 $D\bar{D}$ &43 &0.1 &16 &0.4 \\ %
$D^{*}\bar{D}$ & &33 &0.4 &23 \\   %
$D^{*}\bar{D^{*}}$ &  &33 &35 &16 \\ %
$D_{s}\bar{D_{s}}$   &  &7.8  &8.0 &1.3 \\  %
$D_{s}^{*}\bar{D_{s}}$  &  &    &14   &2.6  \\  %
$D_{s}^{*}\bar{D_{s}^{*}}$ &  &   &     &0.7  \\  %
$D_{1}\bar{D}$ &  &  &  &32 \\  %
$D_{2}^{*}\bar{D}$ &  &  &  &23 \\ \hline %
total &43 &74 & 74 &78 \\  %
experiment\cite{Part5_pdg06} &($23.6\pm2.7$) &($52\pm10$) &($78\pm20$) &($43\pm15$) \\ \hline %
\end{tabular}
\caption{ Open-charm  strong decays widths for the $\psi(3770)$,
$\psia$, $\psib$ and $\psic$, as predicted 
by Ref.~\cite{part5:ref:high_charmonium} (in MeV).}
 \label{tab:charm_decay_width}
\end{table}

\subsubsection{{\Large $\psi(3770)$}} 

The $\psi(3770)$ lies below the
$D\bar{D^{*}}$ threshold and decays predominately into
$D\bar{D}$ pairs, making it an ideal ``$D$-meson factory.'' The
$\psi(3770)$ is generally considered to be the $1^{3}D_{1}$~
$c\bar{c}$ state. Some theoretical models predict
$\Gamma(\psi(3770)\rightarrow
D\bar{D})=43$~MeV~\cite{part5:ref:high_charmonium} for a pure
$^{3}D_{1}$ state, which is much wider than its measured
value of $23.6 \pm 2.7$~MeV\cite{Part5_pdg06}. 
This can be explained by the
influence of an admixture of a $2^{3}S_{1}$ component:
\begin{equation}
  \left|\psi(3770)\right>=\cos\theta\left|1^{3}D_{1}\right>+
\sin\theta\left|2^{3}S_{1}\right> ,
\end{equation}
in this case the experimental $\psi(3770)$ width 
can be accommodated with  a mixing
angle of $\theta=-17.4^{\circ}\pm 2.5^{\circ}$.

\par

On the basis of isospin conservation and phase space 
considerations alone, one expects
\begin{equation}
\displaystyle \frac{\sigma(\psi(3770)\rightarrow
D^{0}\bar{D^{0}})}{\sigma(\psi(3770)\rightarrow D^{+}D^{-})}=
\left(\frac{p_{D^{0}}}{p_{D^{+}}}\right)^{3}=1.45 .
\end{equation}
However, in the calculation of Ref.~\cite{part5:ref:couple_channel}, 
the ratio of $D^{0}$ to $D^{+}$ produced at 
$\psi(3770)$ peak is predicted to be lower, namely 1.36, 
because of the influence of
a $^{3}D_{1}$ form-factor suppression.  Precise determinatons of
the $D^{0}$ and $D^{+}$ cross sections near $\psi(3770)$ 
will make it possible to measure
this momentum-dependent form factor.

\par

Charmed meson cross sections at the $\psi(3770)$ have been  measured 
by Mark-III~\cite{part5:ref:mark3_3770}, BES-II~\cite{part5:ref:bes_3770} 
and CLEO-c~\cite{part5:ref:cleo_3770}. The results are listed in
\tablename~\ref{tab:xsec_3770}.

\begin{table}\centering
   \begin{tabular}{|c|c|c|c|} \hline
        &$\sqrt{s}$ &$\sigma(D^{0}\bar{D^{0}})$
        &$\sigma(D^{+}D^{-})$ \\ \hline
        Mark-III\cite{part5:ref:mark3_3770} &3.768 &$2.9\pm0.25\pm0.2$ &$2.1\pm0.3\pm0.15$ \\
        BES-II\cite{part5:ref:bes_3770} &3.773 &$3.36\pm0.29\pm0.18$ &$2.34\pm0.28\pm0.12$ \\
        CLEO-c\cite{part5:ref:cleo_3770}&3.773 &$3.60\pm0.07^{+0.07}_{-0.04}$
        &$2.79\pm0.07^{+0.10}_{-0.05}$ \\ \hline
   \end{tabular}
   \caption{\label{tab:xsec_3770} Charmed meson cross sections 
at the $\psi(3770)$ (in nb).}
\end{table}

\subsubsection{$\psia$} 

The $\psia$ is generally considered to be the $3^{3}S_{1}$
charmonium state.
Studies of its strong decays are interesting 
because there are four kinematically allowed open-charm modes:
$D\bar{D}$, $D^{*}\bar{D}$, $D^{*}\bar{D^{*}}$ and
$D_{s}\bar{D_{s}}$.
Experimental results for the
three non-strange modes as well as the $D_{s}\bar{D_{s}}$ mode from
BES-I and CLEO-c, are listed in \tablename~\ref{tab:xsec_4040_4160}.
Here $\sigma_{D\bar{D}}$ is pretty low, about ~0.3~nb, while the cross
sections for $D^{*}\bar{D}$ and $D^{*}\bar{D^{*}}$ are
approximately equal, even though $D^{*}\bar{D^{*}}$ has very
little phase space. The reported  relative branching fractions
(scaled by $p^{-3}$, where $p$ is the c.m. frame momentum) 
show a very strong preference for $D^{*}$
final states: $D^{*0}\bar{D^{*0}}\gg
D^{*0}\bar{D^{0}}\gg D^{0}\bar{D^{0}}$. This motivated
suggestions that the $\psia$ might be a $D^{*}\bar{D^{*}}$
molecule.
\par
The cross sections reported by BES-I
and CLEO-c for $D_{s}\bar{D_{s}}$ production
at $E_{\rm cm}$ near 4.04~GeV
are both around 0.3~nb. This corresponds to a $\psia$ branching
fraction of about $4\%$, which is lower than the predicted
value of $\sim11\%$. This branching fraction is of special interest
because it determines the event rates available for studies of
$D_{s}$ decays.
\par
The $D^{*}\bar{D^{*}}$ mode has three independent decay
amplitudes: $^{1}P_{1}$, $^{5}P_{1}$ and $^{5}F_{1}$.
Since the c.m. energy is very near the $D^{*}\bar{D^{*}}$
threshold, the $^{5}F_{1}$ amplitude can 
probably be ignored,  in which case the
amplitude ratio is predicted to be
$^{5}P_{1}/^{1}P_{1}=-2/\sqrt{5}$.

\begin{table}
\footnotesize
\centering
\begin{tabular}{|c|c|c|c|c|c|c|} \hline
\multicolumn{2}{|c|}{}  &\multicolumn{2}{|c|}{$\psia$} &\multicolumn{3}{|c|}{$\psib$} \\ \hline
\multicolumn{2}{|c|}{Modes}
&BES-I\cite{part5:ref:bes_4030_Ds,part5:ref:bes_4030_D}
&CLEO-c\cite{part5:ref:cleo_dsscan} &Mark-II\cite{part5:ref:mark2_4160}
&Mark-III\cite{part5:ref:mark3_4160_D, part5:ref:mark3_4160_Ds} &CLEO-c\cite{part5:ref:cleo_dsscan} \\
\hline

  &$D^{0}\bar{D^{0}}$ &$0.19\pm0.05$  &   &    &    & \\
\rb{$D\bar{D}$} &$D^{+}D^{-}$ &$0.13\pm0.04$ &\rb{$\sim0.2$}
&\rb{$0.4\pm0.3$} &\rb{$0.23\pm0.04\pm0.05$}   &\rb{$\sim0.2$} \\
\hline
  &$D^{*0}\bar{D^{0}}$ &$2.46\pm0.60$   &    &   &  &  \\
\rb{$D^{*}\bar{D}$}  &$D^{*\pm}D^{\mp}$ &$2.31\pm0.70$
&\rb{$\sim7$} &\rb{$2.0\pm0.5$} &\rb{$1.5\pm0.1\pm0.3$}
&\rb{$\sim2$}  \\ \hline
  &$D^{*0}\bar{D^{*0}}$ &$2.07\pm0.40$  &  &  &  & \\
\rb{$D^{*}\bar{D^{*}}$} &$D^{*\pm}D^{*\mp}$  &$0.87\pm0.30$
&\rb{$\sim3.6$} &\rb{$4.4\pm0.7$} &\rb{$3.6\pm0.2\pm0.6$}
&\rb{$\sim5$}  \\ \hline
\multicolumn{2}{|c|}{$D^{+}_{s}D^{-}_{s}$} &$0.320\pm0.056\pm0.081$ 
&$\sim0.3$ & & &$\sim0.02$ \\
\hline \multicolumn{2}{|c|}{$D^{*\pm}_{s}D^{\mp}_{s}$}
&\multicolumn{2}{|c|}{} & &$0.83\pm0.17\pm0.31$ & $\sim0.9$ \\
\hline
\end{tabular}
\caption{Charmed meson production  cross sections  at
the $\psia$ and  $\psib$ resonance peaks (in nb).}
\label{tab:xsec_4040_4160}
\end{table}

\subsubsection{{\Large $\psib$}}

The generally preferred charmonium assignment for the
$\psib$ is the $2^{3}D_1$ state.
Like the $\psi(3770)$,  it may also have a significant $S$-wave
$c\bar{c}$ component, since it has a $e^{+}e^{-}$ width 
that is larger than expected for a pure $D$-wave
$c\bar{c}$ state.  There are five open-charm decay modes
available for the $\psib$: $D\bar{D}$, $D^{*}\bar{D}$,
$D^{*}\bar{D^{*}}$, $D_{s}\bar{D_{s}}$ and
$D_{s}^{*}\bar{D_{s}}$.  Experimental 
cross section results from Mark-II,
Mark-III and CLEO-c are listed in
\tablename~\ref{tab:xsec_4040_4160}. The leading mode is
$D^{*}\bar{D^{*}}$, with a branching fraction 
that is greater than
50\%, followed by substantial $D^{*}\bar{D}$ and
$D_{s}^{*}\bar{D_{s}}$ modes, plus a weak
$D\bar{D}$ mode.   The $D_{s}\bar{D_{s}}$ cross section is very
small.
\par
The $\psib\to D^{*}\bar{D^{*}}$ decay mode is 
interesting because of the three decay amplitudes allowed for this
final state. For a pure $D$-wave $c\bar{c}$ assignment for the $\psib$,
the ratio of the two $D^{*}\bar{D^{*}}$ $P$-wave amplitudes,
$^{5}P_{1}/^{1}P_{1}=-1/\sqrt{5}$, is independent of 
the radial wave function. 
The $^{5}F_{1}$ amplitude is
predicted to be the largest, whereas it is zero for an $S$-wave
$c\bar{c}$ assignment.   A determination of these
$D^{*}\bar{D^{*}}$ decay amplitude ratios would provide an
interesting test of the charmonium assignment 
for the $\psib$.

\subsubsection{{\Large $\psic$}}

The $\psic$ is usually considered to be the $4^{3}S_{1}$ $c\bar{c}$
state.   There are more than ten kinematically allowed
open-charm strong decay modes, seven with 
$c\bar{n}$ meson states ($n=u,d$), and three
with $c\bar{s}$.  To date, no experimental results on 
exclusive charmed hadronic decay modes of the $\psic$
have been reported. 
Some of the theoretical predictions are: 
\begin{itemize}
\item[1)]The largest exclusive mode is predicted to be the 
$D_{1}\bar{D}$,  $S+P$-wave meson
combination.  The $D_{1}$ is the $1^{++}$
axial mesons near 2.425~GeV. Since it is rather narrow
($\Gamma\approx20-30$ MeV) and decays dominantly to $D^{*}\pi$,
the experimental signal for $\psic \to DD^{*}\pi$ should be 
quite distinct. 
\item[2)] The second-largest decay mode is predicted to
be another $S+P$-wave meson mode, namely $D_{2}^{*}\bar{D}$. 
The $D_{2}^{*}$ is moderately narrow
and has significant branching fractions  to
both $D^{*}\pi$ and $D\pi$; the $D_{2}^{*}\bar{D}$ mode of
the $\psic$ should be observable in both of these channels.
\item[3)] The $D^{*}\bar{D^{*}}$ mode is predicted to have comparable
strength to that for $D_{2}^{*}\bar{D}$. If the $\psic$ is indeed 
predominantly an $S$-wave $c\bar{c}$ state,  
the expected amplitude  ratio would be
$^{5}P_{1}/^{1}P_{1}=-2/\sqrt{5}$, with 
a zero $^{5}F_{1}$ amplitude.
 \item[4)] It is interesting to note that
the high mass tail of $\psic$
decays may provide access to the recently discovered
$D_{s0}(2317)$, even though the channel
$D^{*}_{s}D_{s0}(2317)$ has a threshold of 4429~MeV, 
which is 14~MeV above the nominal $\psic$ mass. Since
the decay $\psic\rightarrow D^{*}_{s}D_{s0}(2317)$ 
would be purely $S$-wave, with no centrifugal barrier,  
$D_{s0}(2317)$ production just above
threshold, near $E_{\textrm{cm}}=4435$~MeV, might be significant.
\end{itemize}

\subsubsection{Angular distributions and correlations}

The strong decays of the $\psia$ and $\psib$ vector states to
$D^{*}\bar{D^{*}}$ are of special interest, because 
in each case this is
their only multi-amplitude decay mode. The decay to
$D\bar{D}$ and $D^{*}\bar{D}$ are single amplitude
modes, $^{1}P_{1}$ and $^{3}P_{1}$, respectively, and not much
is learned from angular distribution measurements for these 
channels.  In contrast,
decays to $D^{*}\bar{D^{*}}$, have
three allowed amplitudes, $^{1}P_{1}$, $^{5}P_{1}$ and
$^{5}F_{1}$, and an experimental determination of the ratios of
these amplitudes can provide important tests of the decay
model, in particular about the quantum numbers of the light
$q\bar{q}$ pair produced in the decay.  In this section we
give a brief description of the angular distribution in
non-relativistic language.
\par
$D\bar{D}$ production is purely $P$-wave with an amplitude
\begin{equation}
\mathscr{A}_{D\bar{D}}\propto \vec{\eta}\cdot\vec{p},
\end{equation}
where $\vec{\eta}$ is the $\gamma^{*}$ polarization vector and
$\vec{p}$ is the $D$ meson three-momentum  in the c.m.
system. For unpolarized beams, the angular distribution is
simply:
\begin{equation}
\amplitd{D\bar{D}}^{2}\propto 1-\cos^{2}\theta=\sin^{2}\theta .
\end{equation}

\par
For $D^{*}\bar{D}$ (or $\bar{D^{*}}D$) production, $S=1$.
To have the correct parity, $L$ has to be odd, and, since $J=1$,
it cannot be greater than $L=2$; thus $L=1$. 
The  $D^{*}\bar{D}$ decay amplitude has the form
\begin{equation}
\mathscr{A}_{D^{*}\bar{D}} \propto
\vec{\eta}\cdot(\vec{p}\times\epsilon).
\end{equation}
The $D^{*}$ decays either into $D\pi$ or $D\gamma$, each with a
single amplitude:
\begin{equation}
\begin{matrix}
\mathscr{A}_{D^{*}\rightarrow D\pi} \propto \vec{\epsilon}\cdot\vec{q},\\
\mathscr{A}_{D^{*}\rightarrow D\gamma}\propto
\vec{\epsilon}\cdot(\hat{\bfk}\times\hat{E})\propto\vec{\epsilon}\cdot\hat{B},
\end{matrix}
\end{equation}
where $\vec{\epsilon}$ is the $D^{*}$ polarization vector, $\vec{q}$
the pion momentum, $\hat{\bfk}$ the $\gamma$ direction, and
$\hat{E}$ and $\hat{B}$ represent the photon's electric and magnetic
polarization vectors, where
\begin{equation}
\sum_{\textrm{pol}}{\hat{E_{i}}\hat{E_{j}}}=
\sum_{\textrm{pol}}{\hat{B_{i}}\hat{B_{j}}}=
\delta_{ij}-\hat{\bf k_{i}}\hat{\bf k_{j}}.
\end{equation}
 The measureable angular distributions have the form
\begin{equation}
\begin{matrix}
\amplitd{D^{*}\bar{D}}^{2}\propto 1 + \cos^{2}\theta, \\
\amplitd{D^{*}\bar{D},D^{*}\rightarrow D\pi}^{2} \propto 
1+\cos^{2}\theta_{\pi}\propto 1-\cos^{2}\theta_{D\pi}, \\
\displaystyle \amplitd{D^{*}\bar{D},D^{*}\rightarrow
D\gamma}^{2} \propto 1 - \frac{1}{3}\cos^{2}\theta_{\gamma}.
\end{matrix}
\end{equation}

\par
For $D^{*}\bar{D^{*}}$ production, $S=0$,~1~and~2 are possible.
However, since $P=(-1)^{L}$ and $C=(-1)^{L+S}$, $L$ has to be odd 
and $S$ has to be even. If $S=2$, then $L=1$~or~3. For a purely 
$^{3}S_{1}$
$c\bar{c}$ parent state, the amplitude for $L=3$ ($F$-wave) is zero. 
In any case, if the c.m. energy is near the
$D^{*}\bar{D^{*}}$ threshold, the $F$-wave term can be ignored
because of the centrifugal barrier.  Using $\mathscr{A}_{0}$ and
$\mathscr{A}_{2}$ to denote the production amplitudes for $S=0$ and
$S=2$ states, we have
\begin{equation}
\displaystyle \mathscr{A}_{D^{*}\bar{D^{*}}} =
\mathscr{A}_{0}
(\vec{\epsilon}\cdot\vec{\bar{\epsilon}})(\vec{p}\cdot\vec{\eta})+\mathscr{A}_{2}\left[
\frac{1}{2}(\vec{\epsilon}\cdot\vec{p})(\vec{\bar{\epsilon}}\cdot\vec{\eta})+
\frac{1}{2}(\vec{\epsilon}\cdot\vec{\eta})(\vec{\bar{\epsilon}}\cdot\vec{p})-
\frac{1}{3}(\vec{\epsilon}\cdot\vec{\bar{\epsilon}})(\vec{p}\cdot\vec{\eta})
\right],
\end{equation}
where $\vec{\epsilon}$ and $\vec{\bar{\epsilon}}$ are the
$D^{*}$ and $\bar{D^{*}}$  polarization
vectors. The amplitudes are
normalized in such a way that the total cross section is given by
$\displaystyle
\amplitd{D^{*}\bar{D^{*}}}^{2}\propto\amaz^{2}+\frac{5}{9}\amatwo^{2}$.
Then the production angular distribution has the 
form
\begin{equation}
\displaystyle \amplitd{D^{*}\bar{D^{*}}}^{2} \propto 1 -
\frac{\amatwo^{2}+18\amaz^{2}}{7\amatwo^{2}+18\amaz^{2}}\cos^{2}\theta .
\end{equation}
The angular distributions for pions and photons produced in the
process $e^{+}e^{-}\rightarrow
D^{*}\bar{D^{*}}\rightarrow(D\pi_{1})(\bar{D}\pi_{2})$,
$(D\pi)(D\gamma)$, and $(D\gamma_{1})(\bar{D}\gamma_{2})$,
and their correlations are as
follows~\cite{part5:ref:angular_dis}:
\begin{equation}
\begin{matrix}
  \displaystyle \amplitd{D^{*}\bar{D^{*}}}^{2} \propto
  \left(1+\frac{1}{3}\cos^{2}\theta_{\pi\pi}\right)\amatwo^{2}+6\cos^{2}\theta_{\pi\pi}\amaz^{2},\\
  \\
  \displaystyle \amplitd{D^{*}\bar{D^{*}}}^{2} \propto
  \left(1-\frac{1}{7}\cos^{2}\theta_{\gamma\pi}\right)\amatwo^{2}+
  \frac{18}{7}\left(1-\cos^{2}\theta_{\gamma\pi}\right)\amaz^{2}, \\
  \\
  \displaystyle \amplitd{D^{*}\bar{D^{*}}}^{2} \propto
  \left(1+\frac{1}{13}\cos^{2}\theta_{\gamma\gamma}\right)\amatwo^{2}+
  \frac{18}{13}\left(1+\cos^{2}\theta_{\gamma\gamma}\right)\amaz^{2},\\
    \\
  \displaystyle \amplitd{D^{*}\bar{D^{*}}}^{2} \propto
  \left(1-\frac{21}{47}\cos^{2}\theta_{\pi}\right)\amatwo^{2}+\frac{12}{47}\amaz\amatwo\cos\varphi\left(1-3\cos^{2}\theta_{\pi}\right)+
  \frac{72}{47}\amaz^{2}, \\
  \\
  \displaystyle \amplitd{D^{*}\bar{D^{*}}}^{2}\propto
  \left(1+\frac{21}{73}\cos^{2}\theta_{\gamma}\right)\amatwo^{2}+\frac{\amaz\amatwo}{73}\cos\varphi\left(3\cos^{2}\theta_{\gamma}-1\right)+
  \frac{144}{73}\amaz^{2},
\end{matrix}
\end{equation}
where $\varphi$ is the relative phase between the amplitudes
$\mathscr{A}_{0}$ and $\mathscr{A}_{2}$.

\par

The two principal models currently used by theorists to study 
$c\bar{c}$ decays to open charm are the $^{3}P_{0}$
model~\cite{part5:ref:3p0model} and the Cornell~(timelike vector)
model~\cite{part5:ref:cnlmodel}. They give different predictions for the
relative $D^{*}\bar{D^{*}}$ decay amplitudes, which have not
been tested experimentally.  At \bes3, we can measure these
amplitude ratios, which will guide the
formulation of more accurate models for $c\bar{c}$ strong
decays and  improve our general understanding of
QCD strong-decay processes.

\subsection{Charmed meson cross sections below the $D^{*}\bar{D}$ 
threshold}

Near the open charm threshold, charmed mesons
are produced in pairs, {\it i.e.} 
$D^{0}\bar{D^{0}}$,
$D^{+}D^{-}$ and $D_{s}^{+}D_{s}^{-}$, that are
nearly at rest.  Here the
double-tag method can be applied to obtain 
decay-mode-independent cross section measurements. 
For specific hadronic 
decay modes, $i$ and $j$, 
the number of produced $D\bar{D}$ pairs can
be expressed as the number of single and double tags:
\begin{equation} \label{eq:nddb}
\displaystyle N_{D\bar{D}}= \left\{
   \begin{array}{cc}
      \displaystyle \frac{1}{2}\times\frac{{\mathcal S}_{i}\times
{\mathcal S}_{j}}{{\mathcal D}_{ij}}\times 
\frac{\epsilon_{ij}}{\epsilon_{i}\times\epsilon_{j}}
&
i\ne j \\
\displaystyle
\frac{1}{4}\times\frac{{\mathcal S}_{i}^{2}}{{\mathcal D}_{ii}}
\times\frac{\epsilon_{ii}}{\epsilon_{i}^{2}}
& i=j,
    \end{array}\right.
\end{equation}
where $N_{D\bar{D}}$ is the total number of produced
$D\bar{D}$ pairs, summed over all decay modes,
${\mathcal S}_{i}$ and ${\mathcal D}_{ij}$ 
are the number of single and double tags, and
$\epsilon_{i}$ and $\epsilon_{ij}$, $\epsilon_{ii}$ are the detection
efficiencies for the single and double tag decays.
\par
Many systematic uncertainties cancel in the double-tag
measurement.  Applying the (reasonably good) approximations
$\displaystyle\frac{\epsilon_{ij}}{\epsilon_{i}\epsilon_{j}}\approx
1$, $\displaystyle\frac{\epsilon_{ii}}{\epsilon_{i}^{2}}\approx 1$
to Eq.~\eqref{eq:nddb}, and ignoring the error on the
number of single tags, the precision on the number of
produced charm meson pairs is estimated to be
\begin{equation} \label{eq:prec_nddb}
\displaystyle \frac{\Delta N} {N} \approx 
\frac{1}{\sqrt{\sum {\mathcal D}_{ij}}}=\frac {1}{\sqrt{\mathcal D}},
\end{equation}
where ${\mathcal D}$ presents the total number of double-tags.
\par
At \bes3, about 400,000 and 200,000 $D^{0}$ and $D^{+}$ double
tags are expected to be reconstructed in a 15~fb$^{-1}$ 
data sample accumulated at
$\sqrt{s}=3.773$ GeV.  Thus, the statistical error on the number of 
produced
$D\bar{D}$ pairs can be ignored. Ultimately, the dominant systematic
uncertainty on the cross section measurement will be 
that from the luminosity 
($\mathscr{L}$) measurement,
which is expected to be at the 1\% level. For a 3~fb$^{-1}$ data
sample taken at 4.03~GeV or 4.17~GeV, about 750 and 2,200 $D_{s}$ 
double-tag events will be detected, and the
corresponding statistic errors will be 2.0\% or
1.2\%.  Statistical and systematic errors 
will contribute roughly equally to the
$D_{s}$ cross section measurement errors.

\subsection{Charmed cross sections above the $D^{*}\bar{D}$ threshold}
Both the $\psia$ and $\psib$ are above the $D^{*}$ threshold and
decay mainly into $D^{*}$ final states. Since the hadronic decay 
modes of the $D^{*}$ 
meson have a low $Q$ value, 
the charmed meson momentum spectra can be used to distinguish
the different production channels and measure 
their cross sections.

\par
The $D$-meson  momentum distributions are
monochromatic for $D\bar{D}$ production and 
for the ``bachelor'' $D$ meson produced 
directly at the $\psia\to D\bar{D^*}$ production point, but 
different for the ``daughter'' $D$ mesons from
$D^{*}$ decays. Daughter $D$ mesons from $D^{*}\to\pi^{0}D$ 
or $\pi^{+}D^{0}$ decays have a narrower momentum distribution
than those for $D^{*}$ decays to $\gamma D$. 
Figure~\ref{fig:pd_403} shows the momentum
distributions for $D^{0}$ and $D^+$ mesons that are  produced 
in $\psia\to D\bar{D^{*}}$ (or $\bar{D}D^*$) decays, where 
the contributions from the
bachelor $D$ mesons and the $D^*\to\pi D$
and $D^*\to\gamma D$ daughters are indicated.

\subsubsection{The momentum distribution of $D$ mesons}

The $p_{D}$  distribution for $D$ nesons produced by the decay of
a $D^{*}$ depends on the momentum of the parent $D^{*}$ in the lab
frame and the angle of emission of the $D$ in the $D^{*}$ rest
frame. In the case of $D^{*}\bar{D}$, the angular distribution
of the $D$ in the $D^{*}$ frame can be uniquely predicted, while, in the
case of $D^{*}\bar{D^{*}}$, it is much more complicated. Initial
state radiation further distorts the shape of the $p_{D}$
distribution by reducing the effective center-of-mass energy.
\par
The shape of the distribution for $p_{D}$ can be simulated 
with a Monte Carlo event
generator~\cite{part5:ref:KKMC,part5:ref:EvtGen}, with
main contributions  from the following
channels:
\begin{equation}\label{eq:pd}
  \begin{matrix}
     p_{D^{+}}\rightarrow\begin{pmatrix}
                         D^{*+}D^{*-} & D^{*\pm}\rightarrow\pi^{0}D^{\pm} \\
                         D^{*+}D^{*-} & D^{*\pm}\rightarrow\gamma D^{\pm} \\
                         D^{*\pm}D^{\mp} & D^{*\pm}\rightarrow\pi^{0}D^{\pm} \\
                         D^{*\pm}D^{\mp} & \textrm{Direct }D^{\mp} \\
                         D^{*\pm}D^{\mp} & D^{*\pm}\rightarrow\gamma D^{\pm} \\
                         D^{+}D^{-}      & \textrm{Direct }D^{\pm}
                      \end{pmatrix}
   &p_{D^{0}}\rightarrow\begin{pmatrix}
                        D^{*+}D^{*-} & D^{*+}\rightarrow\pi^{+}D^{0} \\
                        D^{*0}\bar{D^{*0}} & D^{*0}\rightarrow\pi^{0}D^{0} 
\\
                        D^{*0}\bar{D^{*0}} & D^{*0}\rightarrow\gamma D^{0} 
\\
                        D^{*\pm}D^{\mp} & D^{*\pm}\rightarrow\pi^{\pm}D^{0} \\
                        D^{*0}\bar{D^{0}} & D^{*0}\rightarrow\pi^{0}D^{0} 
\\
                        D^{*0}\bar{D^{0}} & D^{*0}\rightarrow\gamma D^{0} 
\\
                        D^{*0}\bar{D^{0}} & \textrm{Direct } D^{0} \\
                        D^{0}\bar{D^{0}}  & \textrm{Direct } D^{0}
                      \end{pmatrix}
  \end{matrix}
\end{equation}
Charmed meson production cross sections and $D^*$ decay branching can be 
determined by simultaneous fits to the $p_{D^0}$ and $p_{D^+}$ 
distributions.

\begin{figure}\centering
\includegraphics[width=4in,height=4in]{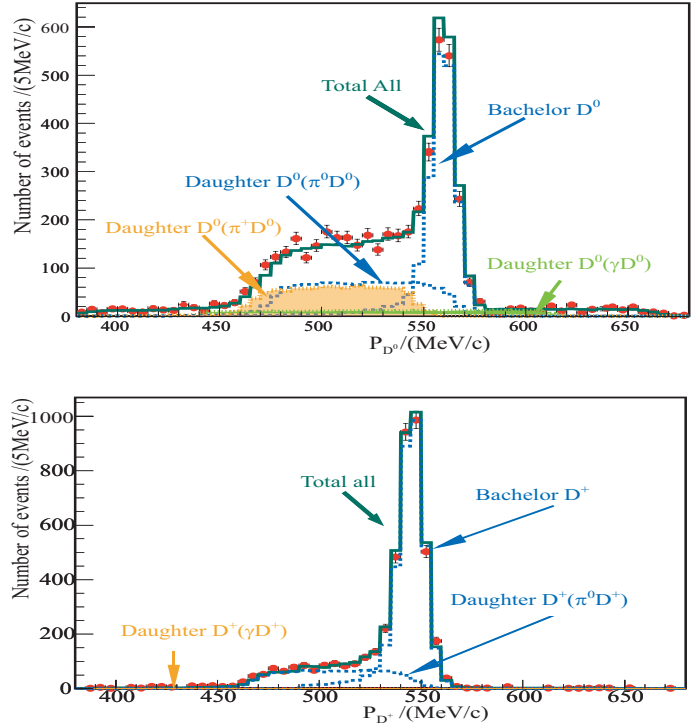}
\caption{\label{fig:pd_403} 
Simulated $p_{D^{0}}$ and $p_{D^{+}}$ distribution
from $D^{*}\bar{D}$ production at $\sqrt{s}=4.03$GeV.}
\end{figure}



The $p_{D}$ distributions are distorted by the effects of ISR  and 
the $\psi(3770)$, $\psia$ and $\psib$ line shapes.  Systematic
uncertainties can be estimated by varying
the resonance parameters of the $\psi$'s in the MC generator. As shown in
Table~\ref{tab:charm_decay_width}, the current values of the
$\psia$ and $\psib$ widths have large uncertainties. In addition, 
the partial decay widths of the $\psi$'s to different
charm meson pair channels have an energy dependence given by
the relation
\begin{equation}
\displaystyle
\Gamma_{i}=\Gamma^{0}_{i}\left(\frac{q}{q_{0}}\right)^{2L+1}\frac{m_{0}}{m},
\end{equation}
where $q$ and $q_{0}$ are the c.m. frame decay momenta in for 
$E_{cm}=mc^2$ and $m_{0}c^2$, respectively, with the subscript 0 
denoting 
the $q$ and $m$ values at the resonance peak,
and the $L$ is the orbital angular momentum.

The mass peaks of the three higher $\psi$
resonances are close to each other and,  
since they have large total
widths,  interference between them (as well as with the continuum)
can distort their lineshapes.  Fine scan measurements can be used to
investigate the effects of interference in each decay channel.

\subsubsection{Full reconstruction of $D\bar{D}$ events }

In the discussion of $D$ meson momentum spectra above
the $D^{*}$ production threshold, we have assumed that
events containing charmed $D$ mesons arise  
only from $D\bar{D^*}$ and
$D^{*}\bar{D^*}$ production ($D\bar{D}$  production is
simple), so that the momentum of
single detected $D$ in an event
can be used to infer that the recoiling system is 
either a monochromatic bachelor $\bar{D}$, a
$D^{*}$ (from $D\bar{D^*}$ production), 
or a $D^*\pi$ (or $D^*\gamma$) combination 
for $D^*\bar{D^*}$ production.  
Note that $D^{0}$  tags can
come from neutral $D^{(*)}$ pair production or charged
$D^{(*)}$ pair production, while $D^{+}$ tags come solely 
from  charged $D^{(*)}$ pair production, since neutral 
$D^{*}$'s are kinematically forbidden 
to decay to the charged $D^{\pm}$ mesons.
We assumed that the neutral $D^{*}$ mesons decay to
$D^{0}$ mesons with a 100$\%$ branching fraction, and the 
charged $D^{*}$ mesons
decay to $D^{0}$ meson
with a decay fraction Br$(D^{*+}\rightarrow\pi^{+}D^{0})$ and the
correspondingly produced  $D^{+}$ mesons
with a decay fraction $1-\mathrm{Br}(D^{*+}\rightarrow\pi^{+}D^{0})$.
Table~\ref{tab:stdt_gtthe}
lists the observed probability of single- and double-tags for different
charmed $D$ meson production channels.

\par
Events containing either one or two reconstructed
$D$ mesons are selected.  The momentum distributions of
reconstructed $D$ tags provide additional information to identify the
$D\bar{D}$,$D\bar{D^*}$,$D^{*}\bar{D^*}$ production channels. 
By comparing the observed number of single-tag $D$ meson events
with the number of partially reconstructed double-tag $D$ meson events,
we determined the different charmed $D$ meson production
rates using a $\chi^{2}$ minimization fit. 

\par
To determine the individual branching ratios ($B_{i}$) and the number of
produced $D\bar{D^{*}}$, $D^{*}\bar{D^{*}}$ pairs 
($N=\sigma {\mathscr L}$),
the observed number of single-tags($S_{i}$) and double-tags($D_{ij}$) is
 expected to be (see Table ~\ref{tab:stdt_gtthe})
\begin{equation}\label{eq:onestar}
\quad\left. \begin{array}{ccc}
S^{0}_{i} & = &
2N^{0}\epsilon_{i}B_{i}+N^{+}\epsilon_{i}B_{i}B^{0}_{+} \\
S^{+}_{i} & = &
N^{+}\epsilon_{i}B_{i}+N^{+}\epsilon_{i}B_{i}(1-B^{0}_{+})  \\
D^{00}_{ij} & = &
\delta_{ij}N^{0}\epsilon_{ij}B_{i}B_{j}  \\
D^{++}_{ij} & = &
\delta_{ij}N^{+}\epsilon_{ij}B_{i}B_{j}(1-B^{0}_{+})  \\
D^{0+}_{ij} & = &
N^{+}\epsilon_{ij}B_{i}B_{j}B^{0}_{+}
\end{array}
\quad\right\}\quad\mathrm{for}\quad D\bar{D^{*}}\quad\mathrm{production}
\end{equation}

and

\begin{equation}\label{eq:twostar}
\quad\left. \begin{array}{ccc}
S^{0}_{i} & = &
2N^{0}\epsilon_{i}B_{i}+2N^{+}\epsilon_{i}B_{i}B^{0}_{+} \\
S^{+}_{i} & = &
2N^{+}\epsilon_{i}B_{i}(1-B^{0}_{+})  \\
D^{00}_{ij} & = &
\delta_{ij}(N^{0}\epsilon_{ij}B_{i}B_{j}+N^{+}\epsilon_{ij}B_{i}B_{j}
(B^{+}_{0})^{2})\\
D^{++}_{ij} & = &
\delta_{ij}N^{+}\epsilon_{ij}B_{i}B_{j}(1-B^{0}_{+})^{2}  \\
D^{0+}_{ij} & = &
2N^{+}\epsilon_{ij}B_{i}B_{j}B^{0}_{+}(1-B^{0}_{+})
\end{array}
\quad\right\}\quad\mathrm{for}\quad
D^{*}\bar{D^{*}}\quad\mathrm{production},
\end{equation}
where $N^{0}=\sigma_{D^{0}\bar{D^{*0}}}\times {\mathscr L}$ \&
$N^{+}=\sigma_{D^{\pm}D^{*\mp}}\times {\mathscr L}$ 
in Eq.~\eqref{eq:onestar},
$N^{0}=\sigma_{D^{*0}\bar{D^{*0}}}\times {\mathscr L}$ \&
$N^{+}=\sigma_{D^{*\pm}D^{*\mp}}\times L$ in Eq.~\eqref{eq:twostar},
$B_{i,j}$ are the individual branching fractions for
$D$ decay modes $\{i,j\}$, $\epsilon_{i}$ 
the efficiency for reconstructing a single-tag in the $i^{\mathrm{th}}$
$D$ decay mode, $\epsilon_{ij}$ the reconstruction efficiency
for $D\bar{D}$(from $D\bar{D^{*}}$ and $D^{*}\bar{D^{*}}$)
decay mode $\{i,j\}$,
and $B^{0}_{+}=\mathrm{Br}(D^{*+}\rightarrow\pi^{+}D^{0})$.
The efficiencies are determined from a detailed Monte
Carlo simulation of $D\bar{D^{*}}$, $D^{*}\bar{D^{*}}$ production
and decay, including a full simulation of the
detector response. Finally, we form a $\chi^{2}$
expression:
\begin{equation}\label{eq:nbrfit}
\chi^{2}=\sum_{i}{\frac
{(S^{i}_{\mathrm{measure}}-S^{i}_{\mathrm{predict}})^{2}}
{\sigma^{2}_{S^{i}_{\mathrm{measure}}}}}+\sum_{ij}{\frac
{(D^{ij}_{\mathrm{measure}}-D^{ij}_{\mathrm{predict}})^{2}}
{\sigma^{2}_{D^{ij}_{\mathrm{measure}}}}},
\end{equation}
\noindent
where the indices $\{$measure,predict$\}$ represent for the number of 
tags obtained from the $\{$measurement, prediction$\}$ and $\sigma$ are 
the measurement errors.
In the fit, one can extract the individual branching fractions
($B_{i}$), and the number of
produced $D\bar{D^{*}}$ and $D^{*}\bar{D^{*}}$ pairs 
($N=\sigma {\mathscr L}$).

\begin{table}
\begin{tabular}{|c|c|c|c|c|c|c|} \hline
\multicolumn{2}{|c|}{ } &\multicolumn{2}{c|}{Single Tags}
 &\multicolumn{3}{c|}{Double Tags}
\\ \hline
\multicolumn{2}{|c|}{Modes} &$D^{0}$ &$D^{+}$ &$D^{0}$ vs
$\bar{D^{0}}$ &$D^{+}$ vs $D^{-}$ &$D^{0}$ vs $D^{-}$ \\ \hline
&0 &$2\epsilon_{i}B_{i}$ &0 &$\delta_{ij}\epsilon_{ij}B_{i}B_{j}$ &0
&0
\\ 
\rb{$D\bar{D^{*}}$} &$\pm$ &$\epsilon_{i}B_{i}B^{+}_{0}$
&$\epsilon_{i}B_{i}(1+B^{+}_{+})$ & 0
&$\delta_{ij}\epsilon_{ij}B_{i}B_{j}B^{+}_{+}$
&$\epsilon_{ij}B_{i}B_{j}B^{+}_{0}$ \\ \hline &0
&$2\epsilon_{i}B_{i}$ &0 &$\delta_{ij}\epsilon_{ij}B_{i}B_{j}$ &0 &0
\\ 
\rb{$D^{*}\bar{D^{*}}$} &$\pm$ &$2\epsilon_{i}B_{i}B^{+}_{0}$
&$2\epsilon_{i}B_{i}B^{+}_{+}$
&$\delta_{ij}\epsilon_{ij}B_{i}B_{j}(B^{+}_{0})^{2}$
&$\delta_{ij}\epsilon_{ij}B_{i}B_{j}(B^{+}_{+})^{2}$
&$2\epsilon_{ij}B_{i}B_{j}B^{+}_{0}B^{+}_{+}$  \\ \hline &0
&$2\epsilon_{i}B_{i}$ &0 &$\delta_{ij}\epsilon_{ij}B_{i}B_{j}$ &0 &0
\\ 
\rb{$D\bar{D}$} &$\pm$ &0
&$2\epsilon_{i}B_{i}$ &0 &$\delta_{ij}\epsilon_{ij}B_{i}B_{j}$ &0 \\
\hline
\end{tabular}
\caption{\label{tab:stdt_gtthe}The observation probability for
single-tags and double-tags above charm threshold. Here,
$B_{i},B_{j}$ are the decay fractions for $D\rightarrow(i^{th},
j^{th})$ tag channels, $\epsilon_{i}$ is the 
single-tag and $\epsilon_{ij}$ the double-tag acceptance,
$B^{0}_{+}=$Br$(D^{*+}\rightarrow\pi^{+}D^{0})$,
$B^{+}_{+}=1-B^{0}_{+}$, and $\delta_{ij}=\left\{^{1
(i=j)}_{2 (i\neq j)}\right.$. The acceptance $\epsilon$ is weighted
by the detection efficiencies for $D$ mesons originating from different 
$D^{*}$ decay modes.}
\end{table}


\subsection{Corrections to the observed cross sections}

The Born cross section for charmed mesons are obtained by correcting
the observed cross section for the effects of initial state radiation
(ISR).   A detailed discussion of ISR corrections can be found in the
Section~\ref{sec:radiativeinee}.   For completeness, we provide here 
a brief description of the correction procedure. 

The ISR correction is dependent on the cross section at all
energies lower than the nominal c.m. energy. Kuraev and
Fadin~\cite{part5:ref:kf1,part5:ref:kf2,part5:ref:kf3} give the observed
cross section $\sigma$ as an integral over the idealized 
radiatively corrected
Born cross section $\bar{\sigma}$: 
\begin{displaymath}
\sigma(s)=\int\bar{\sigma}(s(1-x))F_{KF}(x,s),
\end{displaymath}
where $s=W^{2}$ and $x=(W^{2}-W^{2}_{\textrm{eff}})/W^{2}$, $W$ is
the nominal c.m. energy, $W_{\textrm{eff}}$ is the effective
energy after ISR and $F_{KF}(x,s)$ is the Kuraev-Fadin 
kernel function given by
\begin{equation}
\begin{array}{c}
\displaystyle F_{KF}(x,s)=
tx^{t-1}\left[1+\frac{3}{4}t+\frac{\alpha}{\pi}\left(\frac{\pi^{2}}{3}-\frac{1}{2}\right)
+t^{2}\left(\frac{9}{32}-\frac{\pi^{2}}{12}\right)\right]
-t\left(1-\frac{x}{2}\right) \\
\displaystyle +\frac{t^{2}}{8}\left[4\left(2-x\right)\ln\frac{1}{x}
-\frac{1+3(1-x)^{2}}{x}\ln(1-x)-6+x\right],
\end{array}
\end{equation}
where $\displaystyle
t=2\frac{\alpha}{\pi}\left(\ln\frac{W^{2}}{m^{2}_{e}}-1\right)$.
Kuraev and Fadin claim a 0.1\% accuracy for
$F_{KF}(x,s)$~\cite{part5:ref:kf3}.
The idealized Born cross section $\bar{\sigma}$ for the
charm region and below is sketched in Figure~\ref{fig_xsec_born}.

\begin{figure}\centering
\includegraphics[totalheight=6cm,width=12cm]{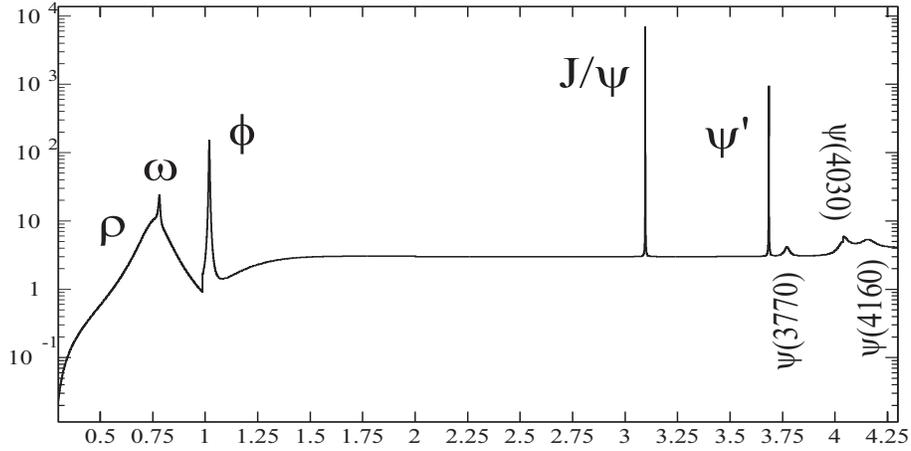}
\caption{The Born cross section for $e^{+}e^{-}\rightarrow
\mathrm{hadrons}$ from 0.3~GeV to 4.3~GeV (in units of $R$).}
\label{fig_xsec_born}
\end{figure}

For a resonance, the Breit-Wigner amplitude and the cross section 
are given by
\begin{equation}
\begin{matrix}
\displaystyle{\cal{A}}(W)=\frac{\sqrt{\kappa}}{W-M+i\Gamma/2} \\
\displaystyle\sigma(W)
=\left|{\cal{A}}\right|^{2}=\frac{\kappa}{(W-M)^{2}+\Gamma^{2}/4},
\end{matrix}
\label{eq_bw_cs}
\end{equation}
where
\begin{equation}
\kappa=3\pi\Gamma_{ee}\Gamma/M^{2}.
\end{equation}
Note that the amplitude ${\cal{A}}$ is complex, with a phase
$\phi_{\textrm{res}}=-\tan^{-1}\left[(\Gamma/2)/(W-M)\right]$ that
starts near $0^{\circ}$ at low $W$, passes through $90^{\circ}$ at
$W=M$, and approaches $180^{\circ}$ at large $W$. If 
interference occurs between different 
overlapping $\psi$ resonances with 
relative phases
$\alpha$'s,  the total amplitude and the cross section
is given by:
\begin{equation}
{\cal{A}}(W)=\sum_{i}{{\cal{A}}_{i}\exp(i\alpha_{i})},
\quad\sigma(W)=\left|{\cal{A}}\right|^{2} .
\end{equation}

In BEPCII,
synchrotron radiation and the replacement of the radiated energy by
the RF cavities generates an energy spread in each beam, resulting in an 
essentially Gaussian distribution
for the c.m. energy $W^{\prime}$ centered on the nominal $W$ value
\begin{equation}
G(W,W^{\prime})=\frac{1}{\sqrt{2\pi}\Delta}
\exp\left[-\frac{(W-W^{\prime})^2}{2\Delta^{2}}\right],
\end{equation}
where $\Delta$ is the rms c.m. energy spread. The final convolution has to 
be done numerically in order to include effects of 
the Breit-Wigner, the ISR tail, and the c.m. energy spread.

The observed partial width, $\Gamma_{ee}$, has to be corrected for all
effects that contribute to  $1^{--}\rightarrow
e^+e^-$,  including $1^{--}\rightarrow e^+e^-\gamma$ decays.  
This correction will depend on the minimum detectable
photon energy and, thus, the total scan width will not be equal to
$\Gamma_{tot}$.   
In $1^{--}\rightarrow e^+e^-(\gamma)$ decay, the
divergences at the soft limit for photon emission
and in the vertex correction cancel~\cite{part5:ref:soft_cancel}. 
But in $e^{+}e^{-}\rightarrow 1^{--}$ production there is no photon in
the initial state and, thus, no cancelation. The correct
procedure is to use the final state definition of
$\Gamma_{ee}=\Gamma(1^{--}\rightarrow e^+e^-(\gamma))$, and to
correct the lowest-order prediction for the $e^+e^-\rightarrow
1^{--}$ process to account for the radiative effects that are
included in the definition of $\Gamma_{ee}$:
\begin{equation}
\displaystyle \Gamma_{ee} =\Gamma_{ee}^{0}(1+\delta_{\textrm{vac}}),
\end{equation}
where $\Gamma_{ee}$ is the experimental width, $\Gamma_{ee}^{0}$ is
the lowest-order width, and $(1+\delta_{\textrm{vac}})$ is
the vacuum polarization factor that includes both leptonic and hadronic
terms. Its variation from charm threshold to 4.14~GeV is less than
$\pm2\%$. To a reasonable approximation it can be
treated as a constant with the value
\begin{equation}
(1+\delta_{\textrm{vac}}) = 1.047\pm 0.024.
\end{equation}

\section{$D$-meson tagging }
\label{sec:charm_tagging}

BEPCII will run near the charm threshold and provide unprecedented
opportunities to study the physics of weak decay of charmed mesons.
At the $\psi(3770)$ resonance,
$D^{0}\bar{D^{0}}$ and $D^{+}D^{-}$ pairs are produced with no 
other accompanying particles, thereby providing
extremely clean and pure charmed-meson signals. 
The large production cross section at the peak of the
$\psi(3770)$, its large decay fractions to charmed meson pairs,
and high tagging efficiencies of \bes3 will allow us 
to accumulate large samples of events where both $D$ mesons
are reconstructed.  By applying the double-tag method to these event 
samples, 
many systematic uncertainties will cancel.
The quantum coherence of the two produced $D^0$ mesons from
 $\psi(3770)\to D^0\bar{D^0}$ decays will provide opportunities
to measure $D^{0}-\bar{D^{0}}$ mixing parameters, 
determine strong phases, and search for $CP$ violation.

  The general technique used for charm physics studies at
the $\psi(3770)$ is referred as tagging. 
Since the $D\bar{D}$ are pair produced just above 
threshold,   the identification of one $D$ meson from a subset 
of tracks in an event 
guarantees that the remaining tracks have originated from the
decay of the recoiling $\bar{D}$. The reconstructed $D$ is referred 
to as the tagged $D$ or simply the tag, while everything not associated
with the tag is referred to as the recoil. Once a tag is found,
the recoil tracks can be examined for the decay mode of interest.
Usually Cabibbo-favored hadronic decay modes with 
large branching ratios, such as
$D^0\rightarrow K^{-}\pi^{+}$($\sim4\%$), $D^0 \rightarrow
K^{-}\pi^{+}\pi^{+}\pi^{-}$($\sim8\%$), $D^0 \rightarrow
K^{-}\pi^{+}\pi^{0}$($\sim14\%$), $D^{+}\rightarrow
K^{-}\pi^{+}\pi^{+}$($\sim9\%$), $D^{+}\rightarrow \bar{K}^{0}
\pi^{+}$($\sim3\%$), $D^{+}\rightarrow \bar{K}^{0}
\pi^{+}\pi^{0}$($\sim14\%$),$D^{+}\rightarrow \bar{K}^{0}
\pi^{+}\pi^{+}\pi^{-}$($\sim6\%$), $D^{+}\rightarrow
K^{-}\pi^{+}\pi^{+}\pi^{0}$($\sim6\%$) are
used for reconstructing tags.
At the $\psia$ or $\psib$ energies, $D_s^+$ mesons 
can be tagged using the $D_s^+ \rightarrow
K^{-}K^{+}\pi^{+}$($\sim5\%$) and 
$D_s^+ \rightarrow \bar{K}^{0}
K^{+}$($\sim4\%$) modes, and some 
$D_s^+$ decays to final states containing an $\eta$
or an $\eta^{\prime}$'.

\subsection{Tag reconstruction}
\label{part5:sec:tagrec} 

Tag reconstruction begins with the charged track
selection. All charged tracks must have a good helix fit, and are
required to be measured in the fiducial region of the drift chamber.
Their parameters must be corrected for energy loss and multiple
scattering according to the assigned mass hypotheses. Tracks not
associated with $K_S$ reconstruction are required to have originated
from the interaction point. Kaons and pions are identified by the PID
algorithm to reduce the combinatoric background.

Neutral kaon candidates are mainly reconstructed in the $K_S
\rightarrow\pi^{+}\pi^{-}$ decay mode. The decay vertex formed by the
$\pi^{+}\pi^{-}$ pair is required to be separated from the interaction
point, and the momentum vector of $\pi^{+}\pi^{-}$ pair must be aligned
with the position vector from the interaction point
to the decay vertex.  The
$\pi^{+}\pi^{-}$ invariant mass is required to be consistent with the
$K_S$ mass. The track parameters and the error matrices are
recalculated at the secondary vertex.   $K_L$ reconstruction
will be useful in some double tag analyses. In this case, 
the recoil mass of all detected particles should be
within the range of the $K_L$ mass, and an interaction 
should be observed in a restricted angular region of
the EMC or muon counters.

Neutral $\pi^0$ mesons are reconstructed from
$\pi^{0}\rightarrow\gamma\gamma$ decays using photons observed
in the barrel and endcap regions of the EMC. 
In the identification of neutral tracks in the EMC, 
one has to distiguish genuine photons from a
number of processes that can produce spurious
showers. The major source of these ``fake photons" arise from
interactions of $\pi$  or $K$ mesons with the material
before or in the EMC crystals; the secondary particles from these
interactions can ``split-off" and create energy clusters that are not 
associated to the original track's shower by the pattern
recognition algorithm.  Other sources of
fake photons are particle decays, back splash from particle interactions
behind the EMC, 
beam-associated
backgrounds and electronic noise. A shower is selected as an isolated
photon by requiring a minimum energy deposit, e.g.
$E_{\gamma}>40$MeV, and a spatial separation from the nearest
charged tracks. In addition, $\eta/\eta^{\prime}$ meson
candidates can be reconstructed
in the $\eta\rightarrow\gamma\gamma$,
$\eta\rightarrow\pi^{+}\pi^{-}\pi^{0}$,
$\eta^{\prime}\rightarrow\gamma\rho^{0}$ and
$\eta^{\prime}\rightarrow\eta\pi^{+}\pi^{-}$ decay modes. 
For all these modes, $3\sigma$ consistency with the 
$\pi^{0}/\eta/\eta^{\prime}$ mass is
required, followed by a kinematic mass constraint.

Finally, for $D$-meson reconstruction, all tracks with consistent
mass hypotheses and, if appropriate, reconstructed $K_S$'s,
$\pi^{0}$'s and $\eta/\eta^{\prime}$'s are permuted to form the
candidate combinations.  To be accepted as a $D$ tag, the candidate
combination of final particles with a reconstructed total energy
$E_{D}$, and total momentum $p_{D}$ must fulfill the 
requirement that the energy difference, 
$\Delta E=E_{D}-E_{\textrm{beam}}$, is consistent with zero.

\subsection{Beam constrained mass}
\label{part5:sec:mbc} 

The conventional method for observing resonant
signals in particle physics analyses is by selecting a set of tracks
and studying the invariant mass ($M_{\textrm{inv}}$):
\begin{equation}
\displaystyle M_{\textrm{inv}} =
\sqrt{\left(\sum_{i}{E_{i}}\right)^{2}-\left(\sum_{i}{\vec{p_{i}}}\right)^{2}},
\label{eq_minv}
\end{equation}
where $(E, \vec{p})_{i}$ are the energy and three-momentum of the track
$i$. For reconstructing $D$ decays at the $\psi(3770)$ peak, the
resolution of the invariant mass at \bes3 is typically 6~to 8~MeV for
modes containing only charged tracks and about 12~MeV for
modes containing a single $\pi^{0}$. 
Improvement can be obtained by exploiting the kinematics of pair production
of $D$ mesons  near threshold. Since the $D$'s are pair
produced, each has
an energy equal to that of the beam in the c.m. frame. Another 
quantity, known as the beam-constrained mass ($M_{\textrm{bc}}$), 
can be constructed by
replacing the energy of the $D$ ($E_{D}$ in Eq.~\ref{eq_minv})
with the energy of the beam in the c.m. frame ($E_{\textrm{beam}}$):
\begin{equation}
\displaystyle M_{\textrm{bc}} =
\sqrt{E_{\textrm{beam}}^{2}-\left(\sum_{i}{\vec{p_{i}}}\right)^{2}}
=\sqrt{E_{\textrm{beam}}^{2}-p^{2}_{D}}.
 \label{eq_mbc}
\end{equation}
\noindent
This quantity is simply a function of the total momentum of the
decay products, $p_{D}=\sum_{i}{\vec{p_{i}}}$. The resolution of
$M_{\textrm{bc}}$ can be computed from
\begin{equation}
\displaystyle \delta  M_{\textrm{bc}} = \frac{E_{\textrm{beam}}}{
M_{\textrm{bc}}}\delta
E_{\textrm{beam}}\oplus\frac{p_{D}}{ M_{\textrm{bc}}}\delta p_{D}.
\end{equation}
The energy spread of the BEPCII beam is small
($\delta E_{\textrm{beam}} \sim0.9$ MeV) as is  
the $D$ meson momentum (at the $\psi(3770)$, 
$p_{D}$($\simeq 270$MeV for $D^{0}$ 
and $\simeq 242$~MeV for $D^{+}$) and its measurement 
($\delta p_{D}\simeq 5$~MeV). As a result, the
$M_{\textrm{bc}}$ resolution is of
order  1.2$-$2~MeV.

Another advantage of having this second independent mass variable is
in the reduction of misidentification background that it provides. 
The total
energy (or the invariant mass) is sensitive to the mass hypotheses
of the decay products, while the beam-constrained mass only relies
on the track momentum (the measured momentum
is dependent on the mass hypotheses only to the extent of a
small correction for $dE/dx$).  Imposing a requirement on
$\Delta E$, and leaving the $ M_{\textrm{bc}}$ value
to be examined for a ``signal peak,'' or fitting both the $\Delta E$ 
and $ M_{\textrm{bc}}$ distributions simultaneously
can improve the signal-to-background ratio.

\subsection{Multiple Counting}
\label{part5:sec:dblcnt} 

Sometimes one event could have two different track combinations
that satisfy the tag candidate requirements.  Such multiple counting 
can  occur in two ways:
\begin{itemize}
\item[1)] Two or more different tag channels can be reconstructed
for a given event, for example, when both $D$'s in an
event are reconstructed.
\item[2)] Two or more possible combinations of tracks can yield a
consistent tag for a given channel; tag modes with higher
multiplicity, such as $D^{+}\rightarrow 
\bar{K^{0}}\pi^{+}\pi^{+}\pi^{-}$, and
those containing $\pi^{0}$'s tend to be more susceptible to this
problem.
\end{itemize}
To count the actual number of tagged events in an unbiased manner,
the following criteria are  used to select only one tag combination
per event:
\begin{itemize}
\item[1)] If more than one tag channel can be reconstructed, select
the channel with the largest signal-to-noise ratio.
\item[2)] For all tags containing charged tracks only, if more than
one combination of tracks form a tag, choose the combination whose
lowest momentum track has a momentum higher than other combinations.
\item[3)] For all tags containing neutrals, if more than one
combination of a photon pair can form a tag, choose the
combination with the lowest $\chi^{2}$ from the $\pi^{0}$ fit, or
choose the combination whose lowest energy track has an energy higher 
than that of the other combinations.
\end{itemize}
Items 2) and 3) are based on the fact that the tracking efficiency
is higher for higher momentum/energy tracks, and the measurement is
more reliable.

\subsubsection*{Mass plot fitting}
\label{part5:sec:massfit}

\begin{figure}
\centering
\includegraphics[width=7.5cm,height=5.0cm]{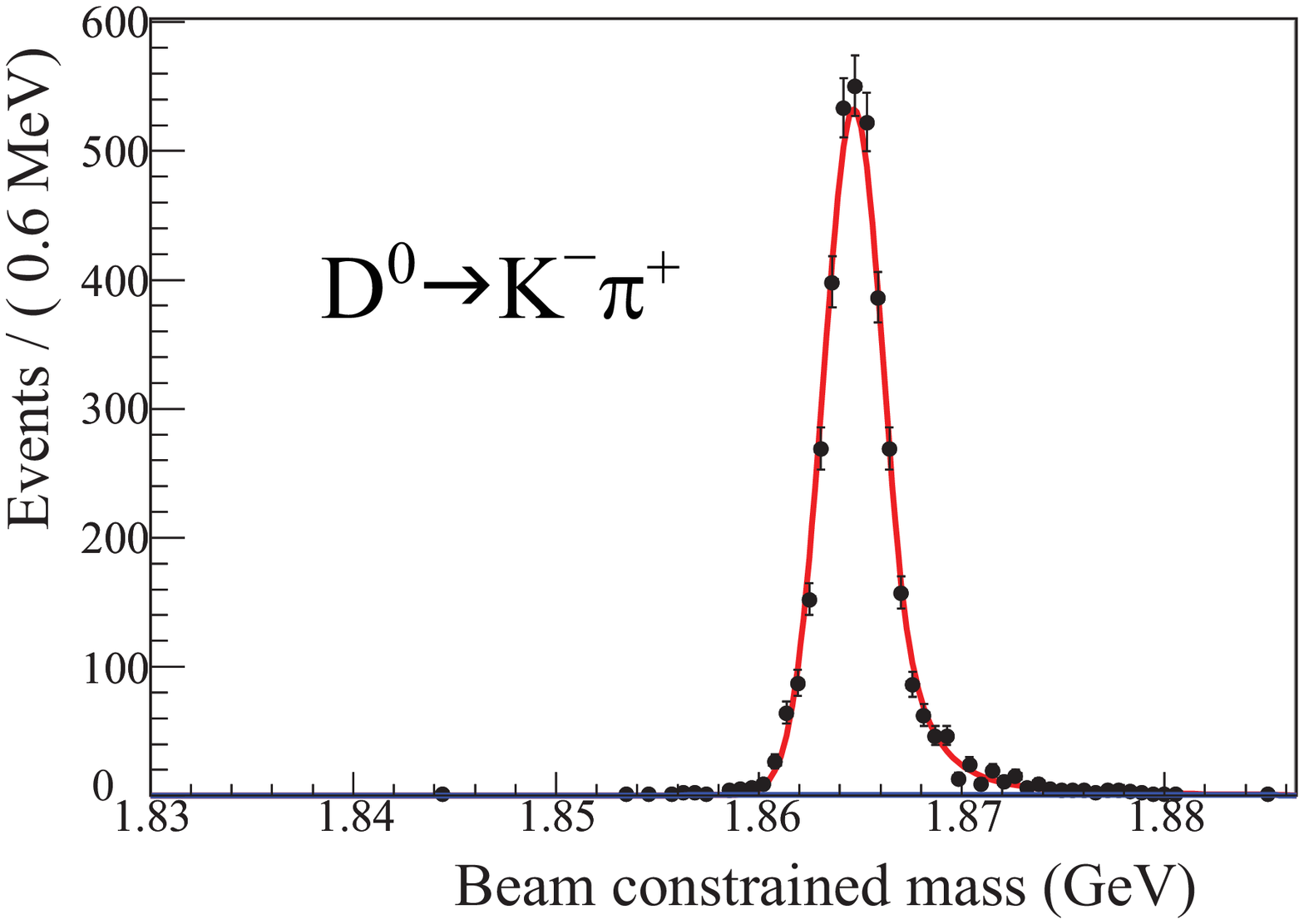}\quad
\includegraphics[width=7.5cm,height=5.0cm]{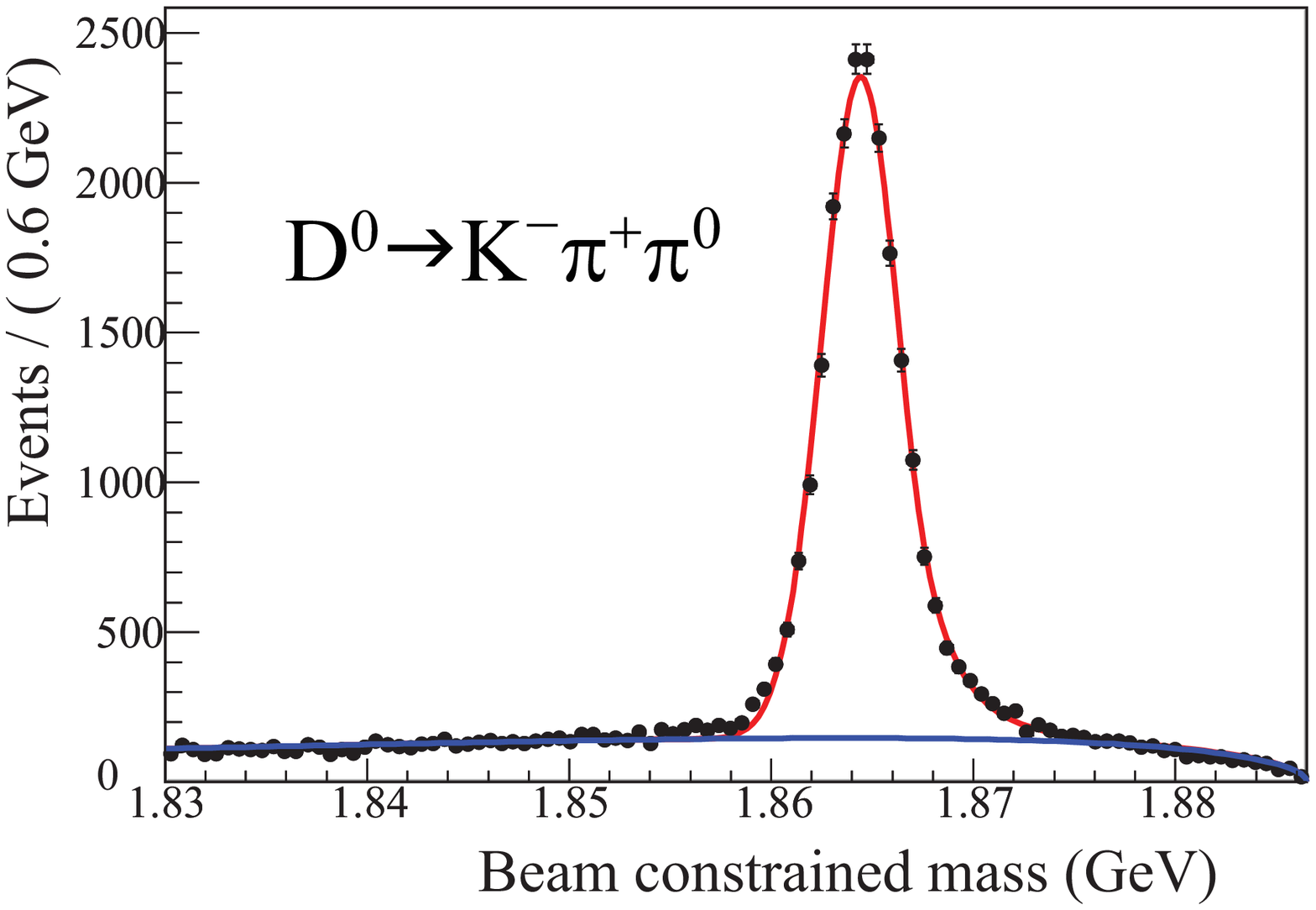}
\caption{Beam constrained mass distributions for
$D^0 \rightarrow K^{-}\pi^{+}$ and $D^0 \rightarrow
K^{-}\pi^{+}\pi^{0}$ tags. The red curves show the sum of signal and
background functions. The blue curves indicate the background fits.}
 \label{fig_mbc_d0}
\end{figure}

The distribution of beam-constrained masses for tag-$D$ candidates 
that survive the above-described  selection procedure   
is shown in Fig.~\ref{fig_mbc_d0}. The
number of tags is determined by an extended unbinned maximum
likelihood fit to a signal function on top of a background. The
background shape is represented by the well known ARGUS
function~\cite{part5:ref:argusbg} which is an empirical formula to model
the phase space of multi-body decays near threshold and is
frequently used in $B$ physics. The ARGUS function has the
form:
\begin{equation}
\displaystyle \mathcal
{B}(M_{\textrm{bc}})=
M_{\textrm{bc}}\left[1-\left(\frac{M_{\textrm{bc}}}{E_{\textrm{beam}}}\right)^{2}\right]^{p}
\cdot\exp\left(c\left[1-\left(\frac{M_{\textrm{bc}}}{E_{\textrm{beam}}}\right)^{2}\right]\right),
\label{part5:eq_argus}
\end{equation}
where $p$ is the power term (usually taken to be 0.5) and $c$
is a scale factor for the exponential term. The shape parameters are
usually determined by fitting $
M_{\textrm{bc}}$ distributions extracted from 
$\Delta E$ sidebands.

There is a tail towards the high-mass end of 
the $M_{\textrm{bc}}$  signal
distribution that is caused by  initial state radiation (ISR).
The effects of ISR, the $\psi(3770)$ resonance parameters and  
line-shape,
the beam-energy spread, and the detector
resolution can all contribute to the signal shape,
which is usually taken to have the
form~\cite{part5:ref:cbline}:
\begin{equation}
\mathcal{S}(M_{\textrm{bc}})=\left\{ \begin{matrix} \displaystyle A\cdot
\exp\left[-\frac{1}{2}\left(\frac{ M_{\textrm{bc}}
-M_{D}}{\sigma_{M_{\textrm{bc}}}}\right)^{2}\right]
&& \textrm{  for } M_{\textrm{bc}} <
M_{D}-\alpha\cdot\sigma_{M_{\textrm{bc}}} \\
&& \\
 \displaystyle
A\cdot\frac{\left(\frac{n}{\alpha}\right)^{n}\exp\left(-\frac{1}{2}\alpha^{2}\right)}
{\left(\frac{M_{\textrm{bc}}-M_{D}}{\sigma_{M_{\textrm{bc}}}}+\frac{n}{\alpha}-\alpha\right)^{n}}
&& \textrm{  for } M_{\textrm{bc}} >
M_{D}-\alpha\cdot\sigma_{M_{\textrm{bc}}},
\end{matrix}\right.
\label{part5:eq_cbline}
\end{equation}
which is similar to the form used to extract photon signals from
electromagnetic calorimeters. In Eq.~\ref{part5:eq_cbline}, the
normalization constant $A$ is related to the other parameters by
\begin{equation}
\displaystyle
A^{-1}=\sigma_{M_{\textrm{bc}}}\cdot\left[\frac{n}{\alpha}\cdot\frac{1}{n-1}\exp\left(-\frac{1}{2}\alpha^{2}\right)+
\sqrt{\frac{\pi}{2}}\left(1+\textrm{erf}\left(\frac{\alpha}{\sqrt{2}}\right)\right)\right],
\label{part5:a}
\end{equation}
where $M_{D}$ is the "true" (or most likely) mass, $\sigma_{M_{\textrm{bc}}}$
is the mass resolution, and $n$ and $\alpha$ are parameters
governing the shape of the high mass tail.

\chapter[Leptonic, semileptonic $D(D_S)$ decays and CKM matrix]{Leptonic, semileptonic $D(D_S)$ decays and CKM
matrix elements}
\label{sec:charm_lepton}
\section{Leptonic Decays}
\subsection[Theoretical Review]{Theoretical Review\footnote{By Ya-Dong Yang}}
\label{sec:review_lepton}

Purely leptonic decays are the simplest and the cleanest 
decay modes of the  pseudoscalar charged $D^+$ meson. The 
hadronic dynamics is simply factorized into the decay constant
$f_{D}$, which is defined as
\begin{equation}
\label{fd_1}
\langle
0|\bar{d}\gamma_{\mu}\gamma_{5}c|D^{+}(p)\rangle=if_{D^{+}}p_{\mu}.
\end{equation}
The decays $D^{+}\to \ell^+ \nu_{l}$ proceed via the 
mutual annihilation of the $c$- and
$\bar{d}$-quarks into a virtual $W^+$ boson with a decay
width given in the Standard Model (SM) by
\begin{equation}\label{rate_part5}
\Gamma(D^+ \to \ell^+ \nu)=\frac{G_F^2}{8\pi}f^2_{D^+}m^2_l
M_{D^+} \left( 1-\frac{m^2_l}{M^2_{D^+}}\right)|V_{cd}|^2 ,
\end{equation}
where $M_{D^+}$ is the $D^+$ mass,  $m_l$ is the mass of the 
final-state lepton, $V_{cd}$
is the CKM matrix element,  $f_{D^+}$ is 
the decay constant  and  $G_{F}$ is the Fermi coupling constant.
Figure~\ref{part5:fy:pure-lep} shows the Feynman diagram for 
the $D^+ \to l^+ \nu_{\ell}$ process. 
Because the $D$ meson is a pseudoscalar, the decay is helicity 
suppressed; the decay rate is proportional to $m^2_l$.  The dynamics are 
the same as that for
$\pi^- \to \mu \bar{\nu_{\mu}}$,~\&~$e \bar{\nu_{e}} $.  
Because of this
helicity suppression, $D^{+}\to e^+ \nu$ is very small:  the 
decay widths for $D^+ \rightarrow \tau^{+}\nu$,~$\mu^+ \nu$, 
and $e^+\nu$ 
are expected to have relative
values of $2.65:1:2.3\times 10^{-5}$, respectively.
Standard Model expectations for the branching fractions
for the different leptonic decay modes of the $D^+$ and $D_s^+$
charmed mesons are listed in
Table~\ref{bf}~\cite{richman95}.
\begin{table}
\begin{center}
\caption{Predicted SM branching fractions for
$D^+$ and $D_s$ purely leptonic decays
assuming $f_D=f_{D_s}=200$ MeV, $|V_{cd}|=0.21$ and $|V_{cs}|=0.97$.}
\begin{tabular}{|c|c|}
\hline
Decay Mode & Branching fraction \\
\hline
$D^+\rightarrow e^+ \nu_e$  & $7.5\times 10^{-9}$ \\
$D^+\rightarrow \mu^+ \nu_{\mu}$  & $3.2\times 10^{-4}$ \\
$D^+\rightarrow \tau^+ \nu_{\tau}$   & $7.2\times 10^{-4}$\\
$D_s^+\rightarrow e^+ \nu_e$  & $7.5\times 10^{-8}$ \\
$D_s^+\rightarrow \mu^+ \nu_{\mu}$ & $3.2\times 10^{-3}$ \\
$D_s^+\rightarrow \tau^+ \nu_{\tau}$  & $2.9\times 10^{-2}$ \\
\hline
\end{tabular}
\label{bf}
\end{center}
\end{table}

Although, $D^+ \to \mu^{+}\nu$ has a smaller rate than $D^+\to \tau^+
\nu$, it is the most favorable mode for experimental measurement
because of the complications caused by the additional neutrino(s) 
produced in $\tau$ decays. 
\begin{figure}[htbp]
\begin{center}
\scalebox{0.9}{\epsfig{file=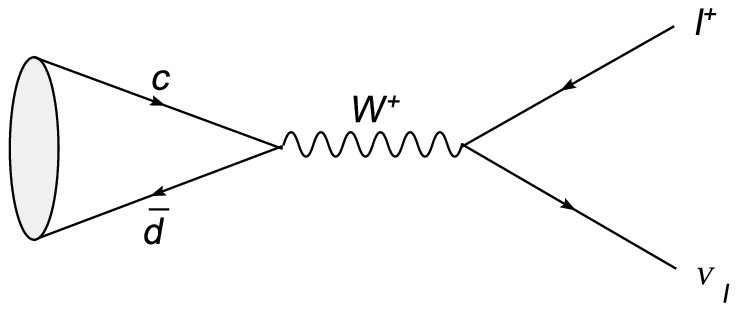}}
\end{center}
 \caption{The Feynman diagram for $D^+ \to l^+ \nu_{\ell}$  }
\label{part5:fy:pure-lep}
\end{figure}
From Eq.~\ref{rate_part5}, in the context of the SM,
a  measurement of $D^+ \to \mu^+ \nu$ 
determines $f_{D^+}|V_{cd}|$.  Thus, 
the value of  $|V_{cd}|$ is needed in order
to extract $f_{D^+}$.

The magnitude of $|V_{cd}|$ has been deduced from  neutrino and
anti-neutrino production of charmed particles on valence $d$ quarks in 
fixed
target experiments~\cite{ep}. The current world-average value
is~\cite{Part5_pdg06}
\begin{equation}
|V_{cd}|=0.224\pm 0.012.
\end{equation}
In the Wolfenstein parameterization of the CKM-matrix, 
$V_{cd}=-V_{us}$ up to  $\mathcal{O}(\lambda^4)$, where 
$\lambda=|V_{us}|$.  With a $281$pb$^{-1}$ data sample
taken at the  $\psi(3770)$ resonance, the
CLEO-c group  determined~\cite{CLEO05}
\begin{equation}
\mathcal{B}(D^+ \to \mu^+ \nu)=(4.40\pm0.66^{+0.09}_{-0.12})\times
10^{-4}.
 \end{equation}
Using  $|V_{cd}|=|V_{us}|=0.2238\pm0.0029$, this translates
into~\cite{CLEO05}
\begin{equation}
f_{D^+}=(222.6\pm16.7^{+2.8}_{-3.4})MeV.
\end{equation}

In the literature, $f_{D^+}$ has been extensively studied in
a variety of theoretical approaches~\cite{fdlat, fdth};  it has been 
measured by the MARKIII~\cite{fdmark3},
BES~\cite{fdbes} and CLEO-c~\cite{CLEO05, fdexp1} collaborations.
The situation  is summarized  in
Fig.~\ref{part5:lqcd:exp:comp}.
\begin{figure}[htbp]
\begin{center}
\scalebox{0.5}{\epsfig{file=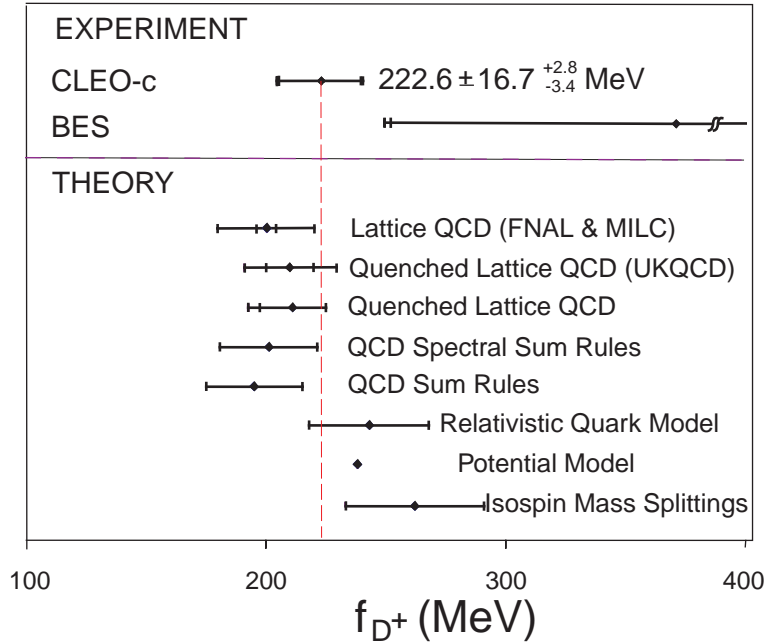}}
\end{center}
 \caption{Summary of theoretical predictions and experimental results for 
$f_{D^+}$ }
\label{part5:lqcd:exp:comp}
\end{figure}
From this figure, we can see that the CLEO-c measurement agrees with
the Lattice QCD  calculations by the Fermilab Lattice, MILC and HPQCD
Collaborations~\cite{lat05}, which find
$f_{D^+}=201\pm3\pm17$ MeV.  Clearly, both the 
experimental measurements and theoretical predictions still lack
sufficient precision to put SM predictions to a 
very stringent test.


The decay constant is a very fundamental parameter that can be
used to test our knowledge of hadronic dynamics.  For example,
$f_{B}$ is hard to determine experimentally because
purely leptonic $B^{\pm}$ decays are suppressed by the very
small value of $V_{ub}=(3.6\pm0.7)\times 10^{-3}$.  Thus,  
theoretical calculations on the Lattice, or with QCD Sum Rules, etc., 
are needed.  However, a self-consistent calculation should include 
a prediction for $f_D$.  If $f_D$ is known with good precision, 
one can validate theoretical calculations of $f_B$, which is
a parameter of critical importance for $B$ physics. 

The purely  leptonic decay $D^{+}_s \to \ell^+ \nu$  is very
similar to $D^{+} \to \ell^+ \nu$. However, it involves the larger
CKM element $V_{cs}$ and, thus, purely leptonic  $D^{+}_s$
decay rates are substantially larger than those for the
corresponding $D^{+} \to \ell^+ \nu$ processes.

The value of $|V_{cs}|$ can be obtained from $W^+$ boson 
charm-tagged hadronic decays. In the SM, the branching 
fractions of the $W^+$ boson
depend on the six CKM elements $V_{ij}$({\it i={u,c}, j={d, s, b}}). In
terms of these six CKM elements, the leptonic branching ratio
$\mathcal{B}(W\to \ell\bar{\nu_{ell}})$ is given by~\cite{Part5_pdg06}
 \begin{equation}
 \frac{1}{\mathcal{B}(W\to \ell\bar{\nu}_{\ell})}=
 3\left\{
 1+\left[
 1+\frac{\alpha_{s}(M^2_W)}{\pi}
   \right]\sum_{i,j}|V_{ij}|^2
   \right\}.
 \end{equation}
Taking $\alpha_{s}(M^2_W)=0.121\pm0.002$, 
and $\mathcal{B}(W\to
\ell\bar{\nu_{ell}})=(10.69\pm0.06({\rm stat.})\pm0.07({\rm syst.}))\%$ 
from LEP measurements~\cite{Part5_pdg06}, we have
 \begin{equation}
\sum_{i,j}|V_{ij}|^2 =2.039\pm0.025\pm0.001.
\end{equation}
Using  experimental results from the PDG~\cite{Part5_pdg06}, 
one can get $|V_{ud}|^2
+|V_{us}|^2 +|V_{ub}|^2 +|V_{cd}|^2 +|V_{cb}|^2 =1.0477\pm0.0074$\cite{Part5_pdg06}.
Then the LEP result can be converted  into a measurement of
$|V_{cs}|$
\begin{equation}
|V_{cs}|=0.996\pm0.013.
\end{equation}

With this value for $|V_{cs}|$,  measurements of $D^{+}_s \to \ell^+ 
\nu$ decays can be used to
determine  $f_{D_{s}}$.  At present, there are eight
published measurements of $D^{+}_s$ decaying to $\mu^+ \nu_{\mu}$
and/or $\tau^+ \nu_{\tau}$: WA75~\cite{AOKI93}, BES~\cite{BAI95},
E653~\cite{KODAMA96}, L3~\cite{ACCIARRI97F},
CLEO~\cite{CHADA98},
BEATRICE~\cite{ALEXANDROV00}, OPAL~\cite{ABBIENDI01L},
and ALEPH~\cite{HEISTER02I}. All of these experiments
other than BES either 
explicitly or
implicitly measure the leptonic branching ratio relative to the
nonleptonic mode $\mathcal{B}(D^+ \to \phi \pi^+ )$ or relative
to semileptonic $D^+_s$  decays.  The PDG weighted average
for the leptonic branching fraction from all experiments
other than BES is~\cite{Part5_pdg06}
\begin{equation}
\mathcal{B}(D^+ \to \mu^+ \nu_{\mu})=0.00547\pm0.00067\pm0.00132.
\end{equation}
Since this average is $1.5\sigma$ below the BES result of
$\mathcal{B}(D^+ \to \mu^+
\nu_{\mu})=0.015^{+0.013+0.003}_{-0.006-0.002}$, the PDG used the
negative uncertainties of the BES measurement to recalculate the
weighted average for all experiments, including BES, to be:
\begin{equation}
\mathcal{B}(D^+ \to \mu^+ \nu_{\mu})=0.00596\pm0.00144.
\end{equation}
Using this value, and including the relatively minor uncertainties
on other parameters relevant to the decay, the world average
$D^+_s$ decay constant is extracted to be~\cite{Part5_pdg06}
\begin{equation}
f_{D_s}=(267\pm33) \mbox{MeV}.
\end{equation}

\subsection[Decay Constants]{Decay Constants\footnote{By Ya-Dong Yang}}

\subsubsection*{The ratio of  $f_{D^+}$ and $f_{D^+_s}$}
\label{fd_ratio}

The $D^0$ decay constant is very important for interpreting measurements
of $D^{0}-\bar{D}^0$ mixing, which is a good testing ground for the SM.
Because of the smallness of the \textit{u} and \textit{d} quark masses, 
chiral symmetry gives
\begin{equation}
f_{D^0}=f_{D^{\pm}}
\end{equation}
It should be noted that in the SM, the decay 
$D^0_u \to \mu^+ \mu^+$ is strongly
suppressed by the GIM mechanism; the expected branching
fraction is less than $10^{-12}$.  
Any observation of this decay at present or near-future colliders will 
be a signal of new physics beyond the SM.

The ratio $f_{D_s}/f_{D}$ is a very important quantity in flavor
physics. It characterizes $SU(3)_V$ breaking. If we take
$m_s ,$ \& $m_{u,d} \to 0$, flavor $SU(3)_V$ is an exact symmetry and
$f_{D_s}/f_{D}=1$. However, in nature $m_s \gg m_{u,d}$, and
this ratio deviates from unity. To the one-loop level in Chiral
Perturbation Theory, Grinstein \textit{et al}.,~\cite{Grinstein} find
\begin{equation}
f_{D_s}/f_{D_d}=1.2,
\end{equation}
which is very nearly equal to $f_{K}/f_{\pi}=1.25$. 
The relative values of the $f_{D_s}/f_{D^{\pm}}$  
and $f_{B_s}/f_{B}$ ratios are characterized by the 
quantity $R_{1}$~\cite{Oakes}:
\begin{equation}
R_{1}=\frac{f_{B_s}/f_{B_d}}{f_{D_s}/f_{D_d}},
\end{equation}
and it has been shown that $R_1$ deviates  from unity by very
small correction factors~\cite{HuangT}.

It is known that the $f_{B_s}/f_{B_d}$ ratio is a measurement of
the relative strength of $B_{s}-\bar{B}_s$ and $B_{d}-\bar{B}_d$
mixing, which is parameterized by the ratio $R_2$
\begin{equation}
R_2 =\frac{(\triangle M/\Gamma)_{B_s}}{(\triangle
M/\Gamma)_{B_d}}.
\end{equation}
Many uncertainties  cancel out in the SM calculation of $R_2$:
\begin{equation}
R_2 =\frac{|V_{ts}|^2}{|V_{td}|^2}
\frac{f^2_{B_s}}{f^2_{B_d}}\frac{B_{B_s}}{B_{B_d}}.
\end{equation}
For example, the ratio of vacuum insertion  parameters
$B_{B_s}/B_{B_d}$ is unity  with very small corrections. Therefore, to
a good approximation
\begin{equation}
R_2 =\frac{|V_{ts}|^2}{|V_{td}|^2} \frac{f^2_{D_s}}{f^2_{D}}.
\end{equation}

It should be noted that leptonic decays of the $B^0_s$,  $B^0_d$ and
$B^{\pm}$ are sensitive to New Physics that does not effect $D$ mesons.  
Therefore, measurements of $B^{0}-\bar{B}^{0}$
 and $B^{0}_{s}-\bar{B}^{0}_{s}$ mixing, in conjunction  with
precise values of $f_{D_{s}}$ and $f_{D}$, will be 
very helpful in clarifying whether or not there is New Physics in   
$B^{0}-\bar{B}^{0}$ and $B^{0}_{s}-\bar{B}^{0}_{s}$ mixing.

\subsubsection{Theoretical extrapolations from $f_D$ to $f_B$ }
Heavy meson decay constants are among the simplest quantities
that can be computed with  Heavy Quark Effective Theory (HQET)~\cite{HQET}.

In HQET, one can derive the relation
\begin{equation}
f_{M}\sqrt{m_{M}}=\hat{C}_{M_Q}F_{ren}+\mathcal{O}(1/m_Q),
\end{equation}
where
\begin{equation}
\hat{C}_{P}(M_Q)=[\alpha_{s}(m_Q)]^{2/\beta_0}\left\{
1+\frac{\alpha_{s}(m_Q)}{\pi}(Z_{hl}+\frac{2}{3})\right\},
\end{equation}
\begin{equation}
\hat{C}_{V}(M_Q)=[\alpha_{s}(m_Q)]^{2/\beta_0}\left\{
1+\frac{\alpha_{s}(m_Q)}{\pi}Z_{hl} \right\},
\end{equation}
\begin{equation}
Z_{hl}=3\frac{153-19n_{f}}{(33-2n_{f})^2}-\frac{381+28\pi^2 -30n_f
}{36(33-2n_{f})}-\frac{4}{3}.
\end{equation}
This expresses the well known asymptotic-scaling law, 
rigorously derived in HQET, that 
$f_{M}\sqrt{m_{M}}$ approaches a constant as $m_M \to\infty$.
To  leading order
in $\mathcal{O}(1/m_Q )$, one finds that
\begin{eqnarray}
r_1 &=&\frac{f_B}{f_D}=\sqrt{\frac{m_D}{m_B}} \left(
 \frac{\alpha_{s}(m_c)}{\alpha_{s}(m_b)}\right)^{6/25}
 \left\{1+0.894\frac{\alpha_{s}(m_c)-\alpha_{s}(m_b)}{\pi}
\right\},\\
r_2 &=&\frac{f_V}{f_P}=1-\frac{\alpha_{s}(m_Q)}{3\pi}.
\end{eqnarray}
At subleading order, one must include $1/m_{Q}$ and the
$SU(3)_V$ corrections.  To the order
of $1/m_Q$, one obtains
\begin{equation}
f_{M}\sqrt{m_{M}}=\hat{C}_{M_Q}F_{ren}\left\{
1+\frac{G_{1}(m_Q)}{m_Q}+\frac{2d_{M}}{m_Q} \left[ \hat{G}_{2}(m_Q
)-\frac{\bar{\Lambda}}{12}\right]\right\},
\end{equation}
where $d_P =3$ and $d_V =-1$. The non-trivial hadronic parameters
$\hat{G}_{i}(m_Q )$ need to be computed using non-perturbative
techniques. Using QCD sum rules, Ball~\cite{ball} has estimated
\begin{equation}
\hat{G}_{1}(m_b )=-0.81\pm0.15GeV, ~~~~~~~~ \hat{G}_{1}(m_c
)=-0.72\pm0.15GeV,
\end{equation}
and Neubert~\cite{neubert} finds
\begin{equation}
\hat{G}_{2}(m_b )=-26\pm4 MeV, ~~~~~~~~ \hat{G}_{2}(m_c
)=-0.44\pm7MeV.
\end{equation}
\noindent
The $SU(3)_V$-violating correction to $f_{B_s}/f_{f_{B}}$ is
estimated to be
\begin{equation}
\frac{f_{B_s}}{f_{B}}=\sqrt{\frac{1-(\frac{7}{3}+3g^2
)\frac{m^2_{K}}{16\pi^2 f^2 } \ln\frac{m^2_K }{\mu^2}    }
{1-\frac{2}{3}(1-3g^2 )\frac{m^2_{K}}{16\pi^2 f^2 } \ln\frac{m^2_K
}{\mu^2} }}\approx 1.138,
\end{equation}
for $\mu=1$GeV, $f=f_K$, $g^2 \approx0.4$, and $f_K =1.25
f_{\pi}$.

\subsubsection*{Previous measurements of purely leptonic $D$
and $D_s$ decays}

{\bf (a) Measurements of 
${\mathcal B}(D^+\rightarrow \mu\nu_{\mu}$) and $f_{D^+}$}

With data collected in the BES-I detector, BES measured
$f_D=(300^{+180+80}_{-150-40})$~MeV by fully reconstructing
$D^{*+}D^-$ events at $\sqrt{s}$=4.03 GeV.  They searched for
$D^- \rightarrow \mu^- \bar{\nu}_{\mu}$ events recoiling from
a sample of tagged $D^{*+}\rightarrow \pi^+D^0$ with
$D^0\rightarrow K^-\pi^+$ decays~\cite{besfd}. Subsequently,
with the upgraded BESII detector, BES measured $f_{D}$ again,
in this case by  reconstructing both $D$ mesons
in $D^+D^-$ events produced
close to threshold with data taken around $\sqrt{s}$=3.773 GeV,
and obtained the branching fraction value
${\mathcal B}(D^+ \rightarrow \mu^+\nu)
=(0.122^{+0.111}_{-0.053}\pm 0.010)\%$,
and a value for the decay constant 
of $f_{D^+}=(371^{+129}_{-119}\pm 25)$~MeV~\cite{bes2fd}.

The MARK-III group determined an upper limit of 
$f_D\leq 290$~MeV with data taken at
3.770 GeV at the SPEAR $e^+e^-$ collider.
The $D^+ \rightarrow \mu^+\nu$ events were selected
based on missing mass and momentum signatures~\cite{mark3}. Recently,
the CLEO-c group made two measurements of the branching fraction
for $D^+ \rightarrow \mu^+\nu_{\mu}$. In the first one, they 
used a 60~pb$^{-1}$ data sample taken at $\sqrt{s}$=3.770~GeV 
and found 8 events (including 1 background event).
They selected events based on the measured missing mass 
$M_{miss}^2=(E_{beam}-E_{\mu})^2-(-p_{D^-} - p_{\mu})^2$ 
and had a total systematic error of $16.4\%$~\cite{cleofd1}. The
branching fraction was measured to be
${\mathcal B}(D^+\rightarrow \mu^+\nu_{\mu})=(0.035\pm0.014\pm0.006)\%$
and $f_{D^+}$ was determined to be $(202\pm41 \pm 17)$~MeV.
Their second measurement was based on a 
281~pb$^{-1}$ data set, in which they found 
50 $D^+\rightarrow \mu^+\nu_{\mu}$ events
(including $2.81\pm0.30\pm0.27$ background events) 
on the recoil side of $158,354\pm496$ $D^-$ single-tags.
The measured  branching
fraction is
${\mathcal B}(D^+\rightarrow \mu^+\nu_{\mu})=
(4.40\pm0.66^{+0.09}_{-0.12})\times 10^{-4}$ from which they
extract $f_{D^+}=(222.6\pm 16.7^{+2.8}_{-3.4})$~MeV. In this second
measurement, they reduced the systematic error to $(^{+2.1}_{-2.5})\%$. 
Most of this 
improvement came  from a reduction in  the 
uncertainty of the background contamination
(from
$15.4\%$ to $(^{+0.6}_{-1.7})\%)$~\cite{cleofd2}.
Measured $f_D$ values are summarized in Table~\ref{fd}.
\begin{table}
\begin{center}
\caption{Measured values of $f_D$.}
\footnotesize
\begin{tabular}{|c|c|c|c|c|c|}
\hline\hline
Experiment & $E_{cm}$&Lum&Events&$B(D^+ \rightarrow \mu
\nu_{\mu})($\%$)$ &$f_D(MeV)$\\ \hline
BESI&4.03 GeV &22.3$pb^{-1}$&1&$0.08^{+0.16+0.05}_{-0.05-0.02}$
&$300^{+180+80}_{-150-40}$ \\
BESII&3.773 GeV &33&$2.67\pm1.74$&$0.122^{+0.111}_{-0.053}\pm{0.10}$
&$371^{+129}_{-119}\pm{25}$ \\
MARKIII&3.773 GeV &9.3&0&$\leq 0.072 (C.L. 90\%)$ &$\leq 290$ (C.L. $90\%$)
\\
CLEOC&3.770 GeV &60&$7.0\pm2.8$&$0.035\pm 0.014\pm 0.006$
&$202\pm{41}\pm{17}$ \\
CLEOC&3.770 GeV &281&$47.2\pm7.1^{+0.3}_{-0.8}$&$0.044\pm
0.0066^{+0.0009}_{-0.0012}$
&$222.6\pm{16.7}^{+2.8}_{-3.4}$ \\ \hline
PDG06~\cite{Part5_pdg06}&  & &&$0.044\pm 0.007$ & \\
\hline \hline
\end{tabular}
\label{fd}
\end{center}
\end{table}

{\bf (b) Pioneering measurements of ${\mathcal B}(D_s \rightarrow 
\mu\nu_{\mu}$)  and $f_{D_s}$}

The first measurement of leptonic $D_s$ decays
was the BES result ${\mathcal  B}(D_s \rightarrow
\mu\nu)=(1.5^{+1.3+0.3}_{-0.6-0.2})\%$ from which they determined
$f_{D_s}=(430^{+150}_{-130}\pm 40)$~MeV.
The measurement was done by fully reconstructing $D_s^+D_s^-$
events with the  BESI detector operating at $\sqrt{s}$=4.03~GeV. The
$D_s$ were tagged using the 
$D_s \rightarrow \phi \pi, K^{*0}K, K^{0}K, ...$ modes
and the recoil systems
were examined for $D_s \rightarrow l \nu_l$ signals~\cite{besds}.

WA75 reported $f_{D_s}=(232\pm45\pm20\pm48)$~MeV using muons from $D_s^+$
leptonic decay seen in emulsion~\cite{wa75}.
 E653 determined a value of $(194\pm 35\pm 20\pm 14)$~MeV from
one prong muon decays  seen in an emulsion~\cite{e653}. BEATRICE
published a value of
${\mathcal B}(D_s\rightarrow \mu\nu)
=(0.83\pm0.23_{stat}\pm0.06_{sys}\pm 0.18_
{{\mathcal B}(\phi \pi)})\%$ using $\pi^-$ interactions
on copper and tungsten targets. The
resulting value for the $D_s$ decay constant was 
$f_{D_s}=323\pm 44\pm 12\pm 34$~MeV~\cite{beatrice}. 
CLEO measured $f_{D_s}$ twice,  their first result was
$f_{D_s}=344\pm 37\pm 52\pm 42$~MeV and determined from
the measured value of the ratio $\Gamma(D_s^+
\rightarrow \mu^+\nu)/\Gamma(D_s^+ \rightarrow \phi \pi)$ from a
2.13~fb$^{-1}$ data sample taken at the $\Upsilon(4S)$~\cite{cleofds1}; 
later they
presented a improved determination of $f_{D_s}$ using a 4.75~fb$^{-1}$
data sample. Their later value is 
$f_{D_s}=280\pm 19\pm 28\pm 34$~MeV~\cite{cleofds2}.

Measured $f_{D_s}$ values from these pioneering
measurements are listed in  Table~\ref{fds}.  More
recent results are discussed in the following section.

\begin{table}
\footnotesize
\begin{center}
\caption{Measured values of $f_{D_s}$.}
\begin{tabular}{|c|c|c|c|c|}
\hline \hline
Experiment & $E_{cm}$&Lum&$B(D_s \rightarrow \mu
\nu_{\mu})(\%)$&$f_{D_s} (MeV)$\\ \hline
BESI&4.03 GeV &22.3$pb^{-1}$&$1.5^{+1.3+0.3}_{-0.6-0.2}$
&$430^{+150+40}_{-130-40}$\\
WA75&350 GeV & &$0.4^{+0.18+0.08}_{-0.14-0.06}\pm 1.7$
&$232\pm 45 \pm 20 \pm 48$\\
E653&600 GeV &  &$0.30\pm 0.12 \pm 0.06 \pm 0.05$&$194\pm 35 \pm 20 \pm 14$
\\
BEATRICE&350 GeV &  &$0.83\pm 0.23 \pm 0.06 \pm 0.18$
&$323\pm 44 \pm 12 \pm 34$ \\
CLEO&$\Upsilon(4S)$ &2.13 $fb^{-1}$&
&$344\pm 37 \pm 52 \pm 42$ \\
CLEO&$\Upsilon(4S)$ &4.79 $fb^{-1}$&
&$280\pm 19 \pm 28 \pm 34$ \\\hline
PDG06\cite{Part5_pdg06}&  & & $0.61\pm 0.19$ & \\
\hline \hline
\end{tabular}
\label{fds}
\end{center}
\end{table}

\subsection[Probing for new physics in leptonic $D_{(s)}$ decays]{Probing 
for new physics in leptonic $D_{(s)}$ decays\footnote{By Hai-Bo Li and 
Jia-Heng Zou}}
\label{sec:charm_lepton_new_physics}

Purely leptonic decays of heavy mesons are of great interest both
theoretically and experimentally. Measurements of the
decays $B^+ \rightarrow l^+ \nu$, $\Ds \rightarrow l^+\nu$ and
$\Dp \rightarrow l^+\nu$,  provide an experimental determination of
the product of CKM elements and decay constants.  If the CKM
element is measured from other reactions, the leptonic decays 
can access the decay constants, which can be
used to test lattice QCD predictions for heavy quark
systems.

In the Standard Model the purely leptonic decays $B^+
\rightarrow  l^+ \nu$, $\Dp \rightarrow l^+ \nu$ and 
$\Ds \rightarrow l^+ \nu$ proceed via
annihilation of the meson's constituent
quarks into virtual $W^{\pm}$ boson. Akeroyd
 and Chen~\cite{part5:akeroyd2007} point out that leptonic decay widths are
 modified by new physics. The charged Higgs bosons in the
two $SU(2)_L \times U(1)_Y$ Higgs doublets 
with  hypercharge $Y=1$ model (2HDM) would modify the SM predictions for 
$D^+$ and $D_s^+$ leptonic  decays~\cite{part5:akeroyd2007}. The
Feynman diagram for  $\Ds \rightarrow l^+\nu$ is shown 
in Fig.~\ref{tree}.
 The tree-level partial width in the 2HDM is given 
by~\cite{part5:akeroyd2007}
\begin{eqnarray}
 \Gamma(\Ds \rightarrow l^+ \nu) &= & \frac{G^2_F m_{\Ds} m^2_l
 f^2_{\Ds}}{8\pi} |V_{cs}|^2 \left(1-\frac{m_l^2}{m^2_{\Ds}}
 \right)^2 \times r_s,
 \label{eq:lepton-amp}
\end{eqnarray}
where $G_F= 1.16639 \times 10^{-5} \, \mbox{GeV}^{-2}$ is the
Fermi constant, $m_l$ is the mass of the lepton, $m_{\Ds}$ is the
mass of the $\Ds$ meson, $V_{cs}$ is the Cabibbo-Kobayashi-Maskawa
(CKM) matrix element, and $f_{\Ds}$ is the decay constant. In the
2HDM (with model-II type Yukawa couplings), 
the process is modified at tree level by the scaling
factor $r_s$, given by~\cite{part5:akeroyd2007}
\begin{eqnarray}
 r_s &=& \left[ 1- m^2_{\Ds} \frac{tan^2\beta}{m^2_{H^{\pm}}}
 \left(\frac{m_s}{m_c+m_S}\right) \right]^2 \nonumber \\
 &=& \left[ 1- m^2_{\Ds}R^2 \left(\frac{m_s}{m_c+m_s}\right)
 \right]^2,
 \label{eq:define}
\end{eqnarray}
where $m_{H^\pm}$ is the charged Higgs mass, $m_c$ is the charm
quark mass, $m_s$ is the strange quark mass (for $\Dp$ decays, it
is replaced by the $d$-quark mass, $m_d$), $\tan\beta$ is the ratio 
of the
vacuum expectation values of the two Higgs doublets,  and the
$H^\pm$ contribution to the decay rate depends on $R =
\frac{\tan\beta}{m_{H^{\pm}}}$. The contribution from the $H^{\pm}$
interferes destructively with the $W^{\pm}$-mediated SM diagram.
As discussed in Ref.~\cite{part5:rosner2008}, the recent
experimental measurements of ${\cal B}(B^\pm \rightarrow \tau^\pm
\nu_{\tau})$~\cite{part5:belleB2006,part5:babar2007} provide an upper limit of
$R<0.29$ GeV$^{-1}$ at 90\% C.L.. For values of $R$ in the interval 
$0.20<R<0.30$ GeV$^{-1}$,  the charged Higgs contribution could have a 
sizable effect on  the $\Ds$ leptonic decay 
rate~\cite{part5:akeroyd2007,part5:rosner2008}.
\begin{figure}[h]\vspace{-3.5cm}
 \centerline{\psfig{file=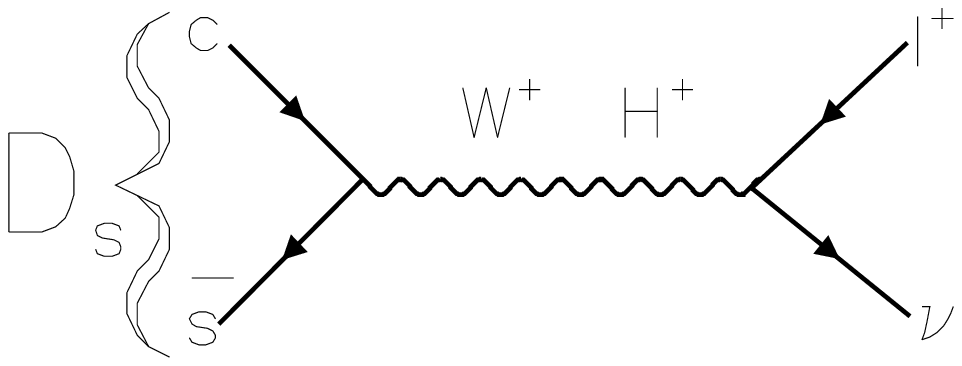,width=11cm,height=13cm}}\vspace{-6.5cm}
 \caption{The tree-level annihilation diagram for pure $\Ds$ leptonic 
decays.} 
\label{tree}
\end{figure}

For $\Dp$ leptonic decays, $m_d << m_c$, the modification is 
negligible and the scaling factor $r_d \approx 1$. However, 
in the case of the $\Ds$, the scaling factor $r_s$ may 
sizable due to the non-negligible value of $m_s /m_c$.  
Although the contribution from new
physics to the rate is small in comparison to the SM rate for 
$\Ds \rightarrow l^+\nu$ decays, measureable effects may be
accessible since the decay rate for $\Ds \rightarrow
\mu^+\nu_{\mu}$ is much larger than that for $B$ leptonic
decays, and can be measured with good precision.

\begin{table*}
  \centering
  \caption{Recent experimental results for ${\cal B}(\Ds \rightarrow \mu
  \nu_{\mu})$, ${\cal B}(\Ds \rightarrow \tau \nu_{\tau})$ and
  $f_{\Dp}$. The decay constants are extracted from the measured
  branching ratios using $|V_{cs}| = 0.9737$, and a $\Ds$ lifetime
  of 0.50 ps~\cite{Part5_pdg06} (from Ref.~\cite{part5:rosner2008}). }
  \begin{tabular}{cccc} \hline\hline
   Experiments & Decay mode  & ${\cal B}$  & $f_{\Ds}$ \\\hline
   CLEO-c~\cite{part5:cleoc2007} & $\mu \nu_{\mu}$ & 
 $(5.94\pm0.66\pm0.31)\times 10^{-3}$ & $264\pm 15 \pm7$ \\
   CLEO-c~\cite{part5:cleoc2007} & $\tau \nu_{\tau}$ & $(8.0\pm 1.3 
\pm0.4)\times 10^{-2}$ & $310\pm 25 \pm8$ \\
   CLEO-c~\cite{part5:cleoc2008} & $\tau \nu_{\tau}$ & $(6.17\pm 0.71 
\pm0.36)\times 10^{-2}$ & $273\pm 16 \pm8$ \\
   CLEO-c combination~\cite{part5:rosner2008} & & & $274 \pm 10 \pm 5$\\
   Belle~\cite{part5:belle2007} &$\mu \nu_{\mu}$ & 
$(6.44\pm0.76\pm0.52)\times 10^{-3}$ & $275\pm 16 \pm12$ \\
   BaBar~\cite{part5:babar2007:d}&$\mu \nu_{\mu}$ & 
 $(6.74\pm0.83\pm0.26\pm 0.66)\times 10^{-3}$ & $283\pm 17 \pm 7 \pm 14$
   \\\hline
   Our average & & & $276 \pm 9$ \\
  \hline \hline
 \end{tabular}
  \label{tab:fds}
\end{table*}

\begin{table*}
  \centering
  \caption{Recent theoretical predictions for $f_{\Dp}$,
  $f_{\Ds}$ and $f_{\Ds}/f_{\Dp}$ from lattice QCD calculations. The most 
precise calculation is from
  HPQCD+UKQCD~\cite{part5:hpqcd2008}, which determines $\Dp$ and $\Ds$ 
decay constants with 2\% errors, four times better than
experimental and previous theoretical results (from 
Ref.~\cite{part5:rosner2008}).}
  \begin{tabular}{cccc} \hline\hline
   Physical Model & $f_{\Dp}$  & $f_{\Ds}$ & $f_{\Ds}/f_{\Dp}$ \\\hline
   Lattice (HPQCD+UKQCD)~\cite{part5:hpqcd2008} & $208\pm 4$ & $241\pm 3$ & $1.164 \pm 0.011$ \\
   Lattice (FNAL+MILC+HPQCD)~\cite{part5:fnal2005} & $201\pm 3\pm 17$ & $249\pm 3 \pm 16$ & $1.24 \pm 0.01 \pm 0.07$ \\
   Quenched lattice (QCDSF)~\cite{part5:qcdsf}  & $206\pm 6 \pm 3 \pm 22$ & $220\pm 6\pm 5\pm 11$ & $1.07 \pm 0.02 \pm 0.02$ \\
   Quenched lattice (Taiwan)~\cite{part5:taiwan2005} & $235\pm 8\pm14 $ & $266\pm 10\pm 18$ & $1.13 \pm 0.02\pm 0.05$ \\
   Quenched Lattice~\cite{part5:ukqcd2001} & $210\pm 10^{+17}_{-16}$ & $236\pm 8^{+17}_{-14}$ & $1.13 \pm 0.02^{+0.04}_{0.02}$ \\
   Experiment (world averages) & $223\pm 17$~\cite{part5:rosner2008} & $276 \pm 9$ & $1.23\pm 0.10$~\cite{part5:rosner2008} \\
  \hline \hline
 \end{tabular}
  \label{tab:theo}
\end{table*}

Experimental measurements of the branching fraction for
$\Dp \rightarrow \mu \nu_{\mu}$ are summarized
in Table~\ref{fd} of the previous section. 
The most precise result is the CLEO-c 
measurement that is based on
a 281$~pb^{-1}$ data sample taken at the $\psi(3770)$ peak. The measured
decay rate of the $\Dp \rightarrow \mu \nu_{\mu}$ is $(4.40 \pm
0.66^{+0.09}_{-0.12})\times 10^{-4}$~\cite{part5:cleocfd}. 
In the context of the SM, using the
well measured $\Dp$ lifetime of $1.040 \pm 0.007$ ps and assuming
$|V_{cd}| =|V_{us}| = 0.2238(29)$, they
determine~\cite{part5:cleocfd}
\begin{eqnarray}
(f_{\Dp})_{\mbox{CLEO-c}} = (222.6 \pm 16.7^{+2.8}_{-3.4})
\mbox{~MeV}.
 \label{eq:dp}
\end{eqnarray}

Recently, measurements 
of $\Ds \rightarrow l^+\nu$ decays with 
precision levels comparable to that for
$\Dp \rightarrow \mu^+\nu$ decays
have been reported by
CLEOc~\cite{part5:cleoc2007,part5:cleoc2008},
BaBar~\cite{part5:babar2007} and Belle~\cite{part5:belle2007};
these are summarized in Table~\ref{tab:fds}.  For the $\Ds
\rightarrow \mu \nu_{\mu}$ decay mode, the combined decay rate
from the CLEO-c, Belle and BaBar experiments is
$(6.26\pm0.43\pm0.25)\times 10^{-3}$. For the $\Ds \rightarrow \tau^+
\nu_{\tau}$ decay mode, combining the two $\tau$ decay channels
($\tau^+ \rightarrow \pi^- \bar{\nu}_{\tau}$ and $e^+\nu_{e}
\bar{\nu}_{\tau}$) from CLEO-c~\cite{part5:cleoc2008}, one obtains
${\cal B}(\Ds \rightarrow \tau^+ \nu_{\tau})=(6.47\pm0.61\pm
0.26)\%$. Using the $\Ds$ lifetime of 0.50 ps and
$|V_{cs}| =0.9737$~\cite{Part5_pdg06} in the SM relation, 
one determines the decay constant $f_{\Ds}$ from
the $\Ds \rightarrow \mu^+ \nu_{\mu}$ mode to be
\begin{eqnarray}
(f_{\Ds})^{\mu}_{exp} = (272 \pm 11) \, \mbox{~MeV},
 \label{eq:fds_exp_mu}
\end{eqnarray}
and that from the $\Ds \rightarrow \tau^+ \nu_{\tau}$ decay mode to be
\begin{eqnarray}
(f_{\Ds})^{\tau}_{exp} = (285 \pm 15) \, \mbox{~MeV}.
 \label{eq:fds_exp_tau}
\end{eqnarray}
The average of the $\tau \nu_{\tau}$ and $\mu \nu_{\mu}$ values is
\begin{eqnarray}
(f_{\Ds})_{exp} = (276 \pm 9)\, \mbox{~MeV}.
 \label{eq:fds_exp_ave}
\end{eqnarray}
Table~\ref{tab:theo} summarizes recent lattice QCD predictions
for the decay constants. The HPQCD+UKQCD collaboration
claims better than 2\% precision for their unquenched calculations:  
\begin{eqnarray}
(f_{\Dp})_{QCD} &= &(208 \pm 4) \mbox{MeV}, \nonumber \\
(f_{\Ds})_{QCD} &= &(241\pm 3)\mbox{MeV}.
 \label{eq:fds_theo}
\end{eqnarray}
As pointed out in
Ref.~\cite{part5:bogdan2008},  
there is a 15\% ($3.8\sigma$) discrepancy
between the experimental and lattice QCD values of  $f_{\Ds}$ 
(Eqs.~\ref{eq:fds_exp_ave} and~\ref{eq:fds_theo}). The
discrepancy is seen in both the $\tau\nu_{\tau}$ mode, where
it is 18\% (2.9 $\sigma$), and the  $\mu\nu_{\mu}$ 
where it is 13\% (2.7 $\sigma$).  

Equation~\ref{eq:lepton-amp} shows that the charged
Higgs would lower the $\Ds$ decay rate relative to the 
SM prediction.  However, the LQCD predicted value
(Eq.~\ref{eq:fds_theo}) is {\em below} the measured value
by more than 3$\sigma$.   This indicates  that there is
no value of $m_{H^+}$ in the 2HDM that can accommodate 
the measured $f_{D_s}$ value.  If we take the discrepancy
seriously, there must be new physics that  {\em enhances}
the predicted leptonic decay rate.  

Measurements of $f_{\Ds}$ and its world average
are shown in Fig.~\ref{fig:exp:fds}
together with the LQCD prediction.
With 20 fb$^{-1}$ at $E_{CM} = 4170$ MeV, the \bes3
sensitivity for the measurement of the
leptonic $\Ds$ decay branching fraction would be
about 2\%~\cite{part5:li2006}, which
corresponds to a 1.0\% uncertainty level for $f_{\Ds}$, as indicated
in Fig.~\ref{fig:exp:fds}.  Assuming that the central
value for the combined experimental $f_{\Ds}$ result persists, 
the discrepancy between the SM prediction and a \bes3 
measurement would be more than 8$\sigma$, and a
signal for new physics beyond the SM.

\begin{figure}[h]
 \centerline{\psfig{file=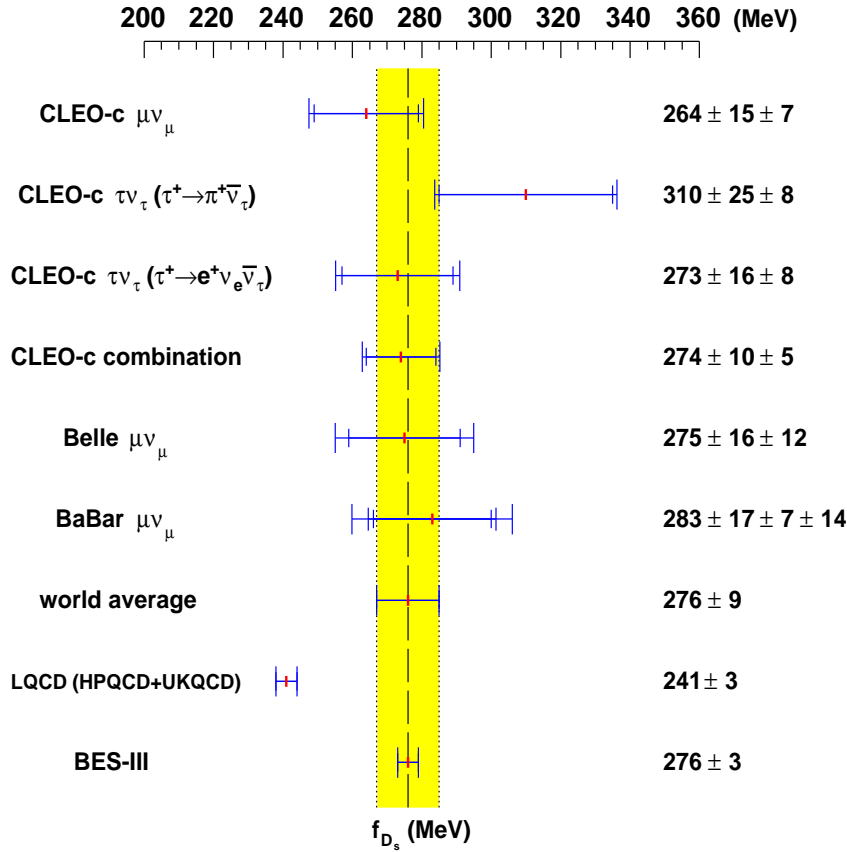,width=12cm,height=12cm}}\vspace{0.cm}
 \caption{Values of $f_{\Ds}$ extracted from different
 experiments in the context of the SM.  The world average is $f_{\Ds} = 
276 \pm 9$ MeV, with an uncertainty of about 3.3\%.
 The 1\% \bes3 sensitivity to $f_{\Ds}$   is 
indicated with the assumption that the current world average central value
persists.} \label{fig:exp:fds}
\end{figure}
\begin{figure}[h]
 \centerline{\psfig{file=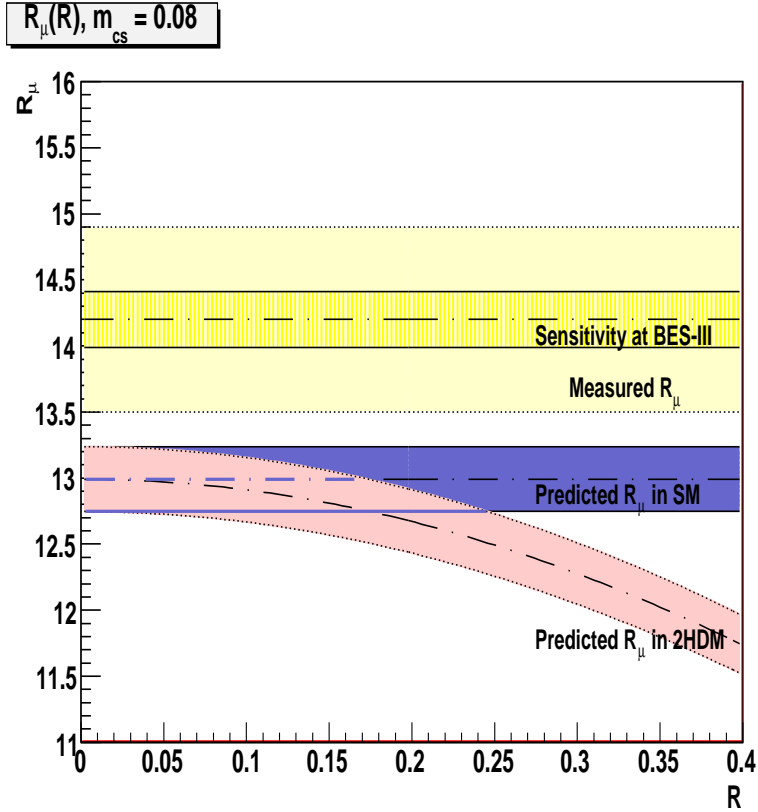,width=11cm,height=11cm}}\vspace{0.cm}
 \caption{ ${\cal R}_{\mu}$ as a function of $R = \mbox{tan}\beta/m_{H^\pm}$
  for $m_{sc} = m_s/(m_s + m_c)=0.08$ and $f_{\Ds}/f_{\Dp}=
  1.164 \pm 0.011$ from LQCD calculations.  The uncertainty on the 
  theoretical
  prediction of ${\cal R}_{\mu}$ is shown as the gray band.
  The expected $\pm 1\sigma$ \bes3 uncertainty experimental range of 
 ${\cal  R}_{\mu}$ is indiated by the yellow band.
  The sensitivity for the measurement of the 
  ratio ${\cal R}$ at BES-III is about 1.5\% level
  is also shown with the assumption that the current central value for 
  ${\cal R}_{\mu}$ persists.} \label{fig:exp:rmur}
\end{figure}


Another, more conservative
approach, is to use the LQCD prediction for the ratio $f_{\Ds}/f_{\Dp}$,
which is inherently more precise than those for the
individual $f_D$ values ({\it c.f.} the discussion in
Sect.~\ref{fd_ratio}). 
A significant deviation of this
ratio from the SM prediction
would be a very robust sign of new physics beyond the SM.

Experimentally, the ratio $f_{\Ds}/f_{\Dp}$ can be extracted from
the measured ratio ${\cal R}_{\mu}$ of the leptonic decay rates of
the $\Ds$ and the $\Dp$. In the SM, one has~\cite{part5:akeroyd2007}:
\begin{eqnarray}
{\cal R}_{\mu} &\equiv& \frac{{\cal BR}(\Ds \to \mu^+ \nu)}{{\cal
BR}(\Dp  \to \mu^+ \nu)}= \left|\frac{f_{\Ds}}{f_{\Dp}}\right|^2
\left|\frac{V_{cs}}{V_{cd}}\right|^2\frac{m_{\Ds}}{m_{\Dp}}\times \nonumber \\
& &
\left(\frac{1-m^2_{\mu}/m^2_{\Ds}}{1-m^2_{\mu}/m^2_{\Dp}}\right)\times
\frac{\tau_{\Ds}}{\tau_{\Dp}}.
 \label{eq:ratio:rate}
\end{eqnarray}
In the case of the 2HDM, new physics only modifies the
$\Ds$ terms, and the ratio ${\cal R}_{\mu}$ in
Eq.~\ref{eq:ratio:rate} is corrected by a factor $r_s$ defined in
Eq.~\ref{eq:define}.

Using only CLEOc measurents and the SM relation, 
the experimental value  for the 
$f_{\Ds}/f_{\Dp}$ ratio is~\cite{part5:rosner2008}
\begin{eqnarray}
r_{\Ds/\Dp} \equiv \frac{f_{\Ds}}{f_{\Dp}} = 1.23 \pm 0.10.
 \label{eq:ratio_exp_cleoc}
\end{eqnarray}
The most precise LQCD prediction~\cite{part5:hpqcd2008} 
is $f_{\Ds}/f_{\Dp} = 1.164 \pm 0.011$, which has a
claimed precision that is better than 1\%, and an
order of magnitude better than the existing experimental 
measurement.  

In Fig~\ref{fig:exp:rmur}, ${\cal R}_{\mu}$ is plotted
as a function of $R=\mbox{tan}\beta/m_{H^\pm}$
for the case of the 2HDM, 
using $m_{sc} = m_s/(m_s + m_c)=0.08$ and $f_{\Ds}/f_{\Dp}= 1.164
\pm 0.011$ from the
LQCD calculation.   The SM prediction is
${\cal R}_{\mu}$ is $(12.99 \pm 0.25)$, where the error is
from the uncertainty on the LCQCD prediction for $f_{\Ds}/f_{\Dp}$.  
Compared to the measured value ${\cal R}_{\mu} = 14.2 \pm
0.7$, we see that the SM prediction is almost 2 standard
deviations low.  If the LQCD calculation
is reliable, this indicates that we need a
modification to the SM that has
constructive interference to accommodate the
discrepancy~\cite{part5:bogdan2008}. It
may be concluded that the 2HDM discussed in
Ref.~\cite{part5:akeroyd2007} is disfavored by the current data.
It would be very interesting  if the experimental precision on
the $f_{\Ds}/f_{\Dp}$ ratio 
could be improved to match the one percent level
of the theoretical errors  in the near future.

As discussed in the following Section, a MC study of the
\bes3 sensitivity for the ${\cal R}_{\mu}$ ratio measurement 
indicates a precision about 5\% for 20 fb$^{-1}$ data samples taken at 
$E_{CM} = 3773$ and $4030$~MeV (Sect.~\ref{sec:charm_lepton_exp}).
As mentioned in Sect.~\ref{sec:cross_section},
the cross section for $e^+e^-\rightarrow D_s^*\bar{D_s}$
at 4.17~GeV has been measured by CLEOc to be four times the
cross section  for $e^+e^-\rightarrow D_s\bar{D_s}$ 
at 4.03~GeV~\cite{part5:ref:cleo_dsscan}.
If techniques that are currently
being developed to tag $D_s$ mesons produced
in $D_s^*\bar{D_s}$ final states are successful, and
measurements from the $D_s^+\rightarrow \tau^+ \nu_{\tau}$   
are included, another factor of two precision
will be gained by running at 4.17~GeV.
Branching fraction errors of $\sim 2\%$ translate into a 
$\sim 1.5\%$ precision level on the
$f_{\Ds}/f_{\Dp}$ ratio.

\subsection[Measurements of Leptonic Decays at \bes3]{Measurements of Leptonic Decays at
\bes3\footnote{By Jiang-Chuan Chen, Jian Liu and Gang Rong}}
\label{sec:charm_lepton_exp}

The \bes3 detector has very good $\mu$ identification capabilities:
the solid angle coverage of the Muon Chamber system
is $89\%\times 4\pi$. Current \bes3 plans include the accumulation 
of a total of 20~fb$^{-1}$ $\psi(3770)$
data plus a smaller $D_s^+D_s^-$ data sample at 4.03~GeV or 
a $D_s^+D_s^{*-}$ sample at 4.17~GeV. 
By measuring the leptonic decays
$D^+\rightarrow \mu^+\nu_{\mu}$ and $D_s^+\rightarrow \mu^+\nu_{\mu}$
in these data sets, 
\bes3 will measure $f_D$ and $f_{D_s}$ with high precision. 

\subsubsection{Monte Carlo samples and event selection}

The decay constant, $f_D$, can be measured using data 
taken at the peak of the $\psi(3770)$ where 
$D\bar{D}$ mesons are produced in pairs. If we reconstruct a $D^-$ from the
$D^+D^-$ pair (it is called a single-tag $D^-$), the remaining tracks
in the event must come from the ``recoil  $D^+$.''
Single-tag $D^-$ mesons can be reconstructed via 
hadronic decays $D^-\rightarrow mKn\pi$, and
$D^+\rightarrow \mu^+\nu_{\mu}$ decay events
can be isolated among the accompanying recoils.

According  to the Eichten coupled-channel charmed meson
production model~\cite{Eichten}, the peak 
of $D_s^+D_s^-$ production cross section ($R_{D_s^+D_s^-}\sim 0.1$) 
is around $\sqrt{s}=4.03$ GeV, which is just 
above the $e^+e^-\rightarrow D_s^+D_s^-$ threshold, and below the threshold of 
$D_s^+D_s^{*-}$ production.  If data are taken around 4.03~GeV,  
leptonic $D_s^+\rightarrow \mu^+\nu_{\mu}$ events can be selected from
among the recoil systems accompanying  single-tag $D_s^-$ mesons.  
The tagged $D_s^-$ can be reconstructed 
via $D_s^-$ hadronic decays  such as $D_s\to KK\pi$($\phi \pi$), 
$\bar{K}^{*0}K$($\bar{K}^{*0}\rightarrow K^-\pi^+$),  
$\bar{K}^0K$($K_s^0\rightarrow \pi^+\pi^-$), $KK\pi\pi$, etc.

To study the \bes3 capabilities for
 leptonic decay $D^+\rightarrow \mu^+\nu_{\mu}$ measurements, 
Monte Carlo $e^+e^- \to D\bar D$ events are generated at
3.773 GeV, where the $D$ and $\bar D$ mesons are allowed to decay 
into all possible final states using branching fractions taken from the
PDG~\cite{Part5_pdg06}. These are simulated and reconstructed
using BOSS version 6.02.
A total of 5~million $D\bar D$ Monte Carlo events were generated,
which  corresponds
to a data sample with an integrated luminosity of about 
800~pb$^{-1}$. 

For the  study of decay $D_s^+\rightarrow \mu^+\nu_{\mu}$, 
a $D_s^+D_s^-$ Monte Carlo sample was generated 
at a c.m. energy of 4.03~GeV.
A total of $1.2\times 10^{5}$ 
($D_s^+\rightarrow X$ versus $ D_s^-\rightarrow \phi \pi^-$) 
events were generated, which  corresponds
to a data sample with an integrated luminosity of
about 10~fb$^{-1}$.

To select good events,  charged tracks are required
to satisfy a fiducial cut $|\rm{cos\theta}|<0.93$
and originate from the 
interaction region. For pion and kaon identification, a combined
confidence level, calculated using the $dE/dx$ and TOF measurements,
is used.  Neutral $\pi^0$ and $K_S^0$ mesons are reconstructed
in their $\pi^0 \to \gamma\gamma$ and 
$K^0_S \to \pi^+\pi^-$ decay modes.

\subsubsection{Single-tag $D^-$ and $D_s^-$ meson reconstruction}

Single-tag $D^-$ mesons are reconstructed in the four hadronic decay modes 
$K^+\pi^-\pi^-$, $K^+K^-\pi^-$, $K^0\pi^-$ and $K^+\pi^-\pi^-\pi^0$.
In order to reduce the background and improve the momentum resolution, 
a beam energy constraint is imposed on each $mKn\pi$ combination of the 
$D^-$ tag modes. The beam-constrained masses of the $mKn\pi$ combinations
are calculated using 
\begin{equation}
\label{beammass}
M_{tag}=\sqrt{E^2_{beam}-p^2_{mKn\pi}},
\end{equation}
where $E_{beam}$ is the beam energy in the c.m. frame 
and $p_{mKn\pi}$  is the total momentum
of the  $mKn\pi$ combination. 
To suppress backgrounds from misidentified particles and fake 
photons, 
an $M_{tag}$ signal region
of $\pm 3\sigma_{\Delta E}$ around the $m_{D^-}$ is defined, 
where $\Delta E=E_{mKn\pi}-E_{beam}$ and 
$E_{mKn\pi}$ is the total energy of the $mKn\pi$ combination.

Figure~\ref{dntag} shows the beam-constrained mass distributions for the 
four $D^-$ tag  modes $K^+\pi^-\pi^-$, $K^+K^-\pi^-$, $K^0\pi^-$ and 
$K^+\pi^-\pi^-\pi^0$. 
The observed number of single-tag $D^-$ mesons can be obtained from a 
maximum likelihood fit to the mass spectrum with a Gaussian
function for the $D$ signal and a special
function~\cite{besd0semi,besdpsemi} to describe the background
shape for each tag mode. The total number of
single-tag $D^-$ mesons is found to be  $N_{D^-}=232,803.0\pm802$. 
\begin{figure}[htb]
\begin{center}
  \includegraphics[width=12cm,height=10cm]
{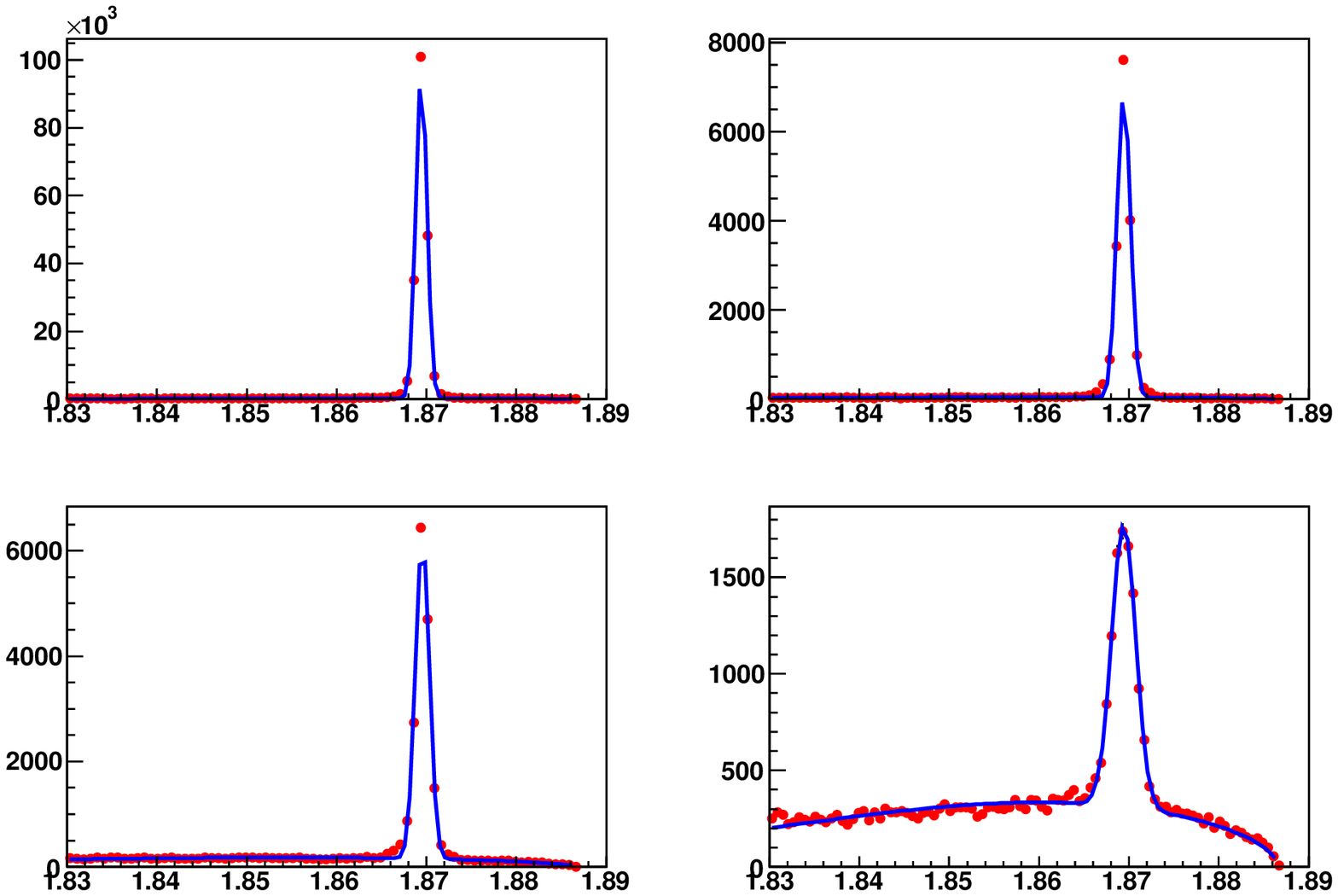}
      \put(-300,230){\Large \bf (a)}
      \put(-130,230){\Large \bf (b)}
      \put(-300,107){\Large \bf (c)}
      \put(-130,107){\Large \bf (d)}
      \put(-285,-15){\bf \large Beam Constrained Mass (GeV/$c^2$)}
      \put(-360,100){\rotatebox{90}{\large Events}}
   \caption{\normalsize The beam constrained mass
distributions for single 
tag $\bar D^-$ candidates in the modes: 
(a) $K^+\pi^-\pi^-$, (b) $K^0\pi^-$, (c) $K^-K^+\pi^-$ and
(d) $K^+\pi^-\pi^-\pi^0$.
\label{dntag}}
\end{center}
\end{figure}

Figure~\ref{dstag} shows the  beam
constrained mass distributions for the  $D_s^-\to \phi \pi^-$ 
tag mode from the MC $D_s^+D_s^-$ sample. 
A maximum likelihood fit to the mass spectrum with a Gaussian
function for the $D_s^-$ signal yields an observed number of 
single-tag $D_s^-$ mesons of $N_{D_s^-} = 26,793\pm 318$.
\begin{figure}[htb]
\begin{center}
  \includegraphics[width=8cm,height=7cm]
{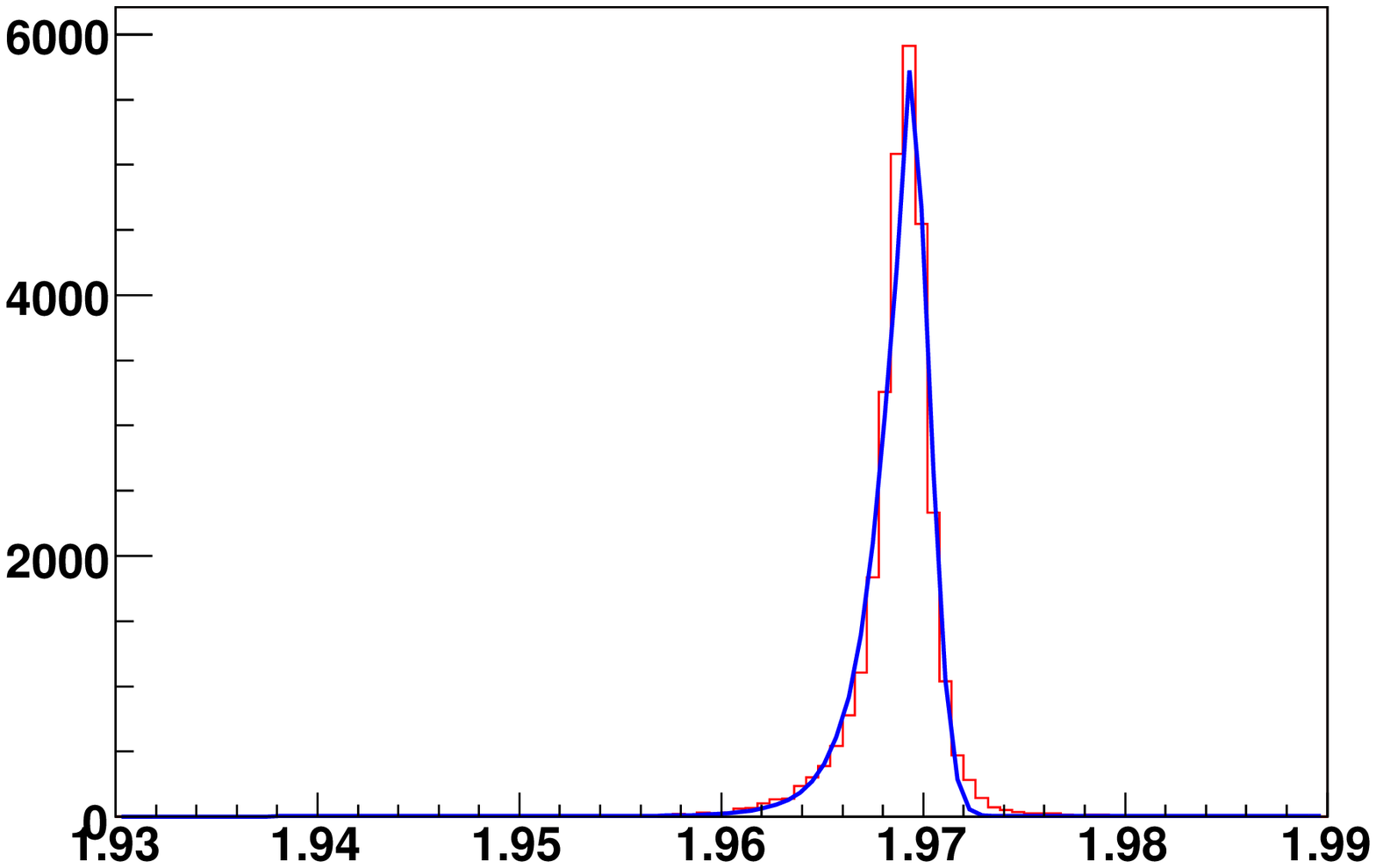}
      \put(-250,100){\rotatebox{90}{\large Events}}
      \put(-210,-15){\bf \large Beam Constrained Mass (GeV/$c^2$)}
      \caption{\normalsize The fitted beam constrained masses for 
single-tag $D_s^-$ candidates.
\label{dstag}
}
\end{center}
\end{figure}

\subsubsection{Leptonic event selection}

To select leptonic decay events from the recoils  to the
single-tag $D^-(D_s^-)$ mesons, we require that the event satisfy
the following criteria:
\begin{enumerate}
\item the event should not contain any isolated photons that are not used in the 
reconstruction of the single tag $D^-(D_s^{-})$ meson;
\item only one well reconstructed charged track is seen n the recoil 
side, and it should have a charge opposite that of the tag meson. 
\end{enumerate}

In the leptonic decay  $D^{+}(D_s^{+})\rightarrow \mu^+\nu_{\mu}$, the 
neutrino is undetected.  Thus, the missing mass
$$M^2_{\rm miss}=E^2_{\rm miss}-p^2_{\rm miss},$$ 
where $E_{\rm miss}=E_{cm}-\sum\sqrt{\vec{p_i}^2+m_i^2}$ and 
$p_{\rm miss}=-|\sum{\vec{p_i}}|$, should be that of the 
undetected neutrino.  Since the neutrino mass is (nearly) zero,
the $M^2_{\rm miss}$ distribution should peak around zero.
A further criterion requires that the momentum of the
candidate  muon track
should be within the region (0.78, 1.08)~GeV/$c$ for $D^+ \to
\mu^+\nu_{\mu}$, or (0.78, 1.22)~GeV/$c$ for $D_s^{+}\to
\mu^+\nu_{\mu}$, as shown in Fig.~\ref{pmuon}.
\begin{figure}[htb]
\begin{center}
  \includegraphics[width=8cm,height=6cm]
{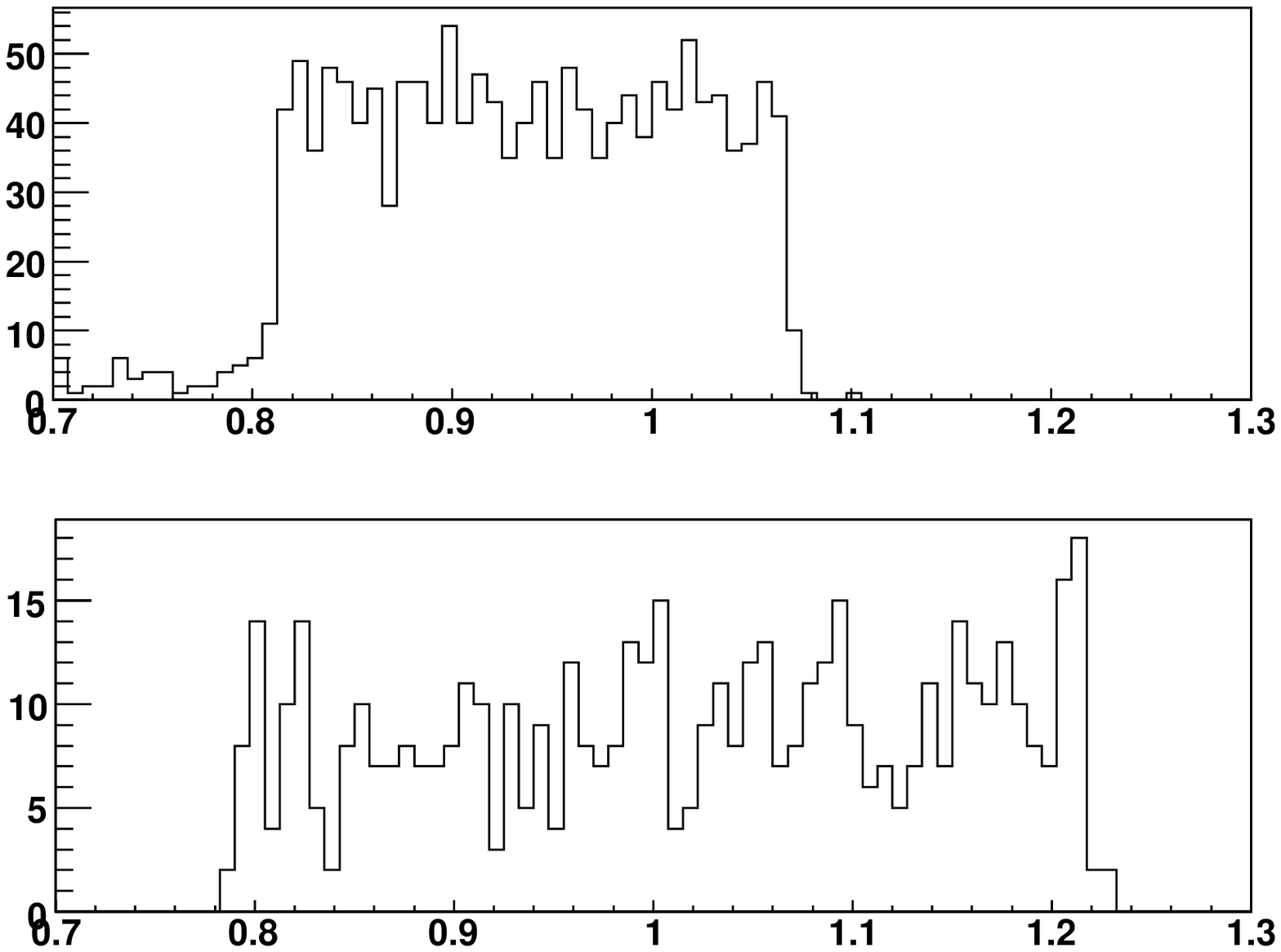}
      \put(-195,140){\large (a)}
      \put(-195,60){\large (b)}
      \put(-150,-15){\large $p_{\mu}$ (GeV/$c^2$)}
      \put(-230,80){\rotatebox{90}{\large Events}}
      \caption{\normalsize The candidate muon momentum 
distributions for (a) $D^+ 
\rightarrow
\mu^+\nu_{\mu}$ and (b) $D_s^{+}\rightarrow \mu^+\nu_{\mu}$.
  \label{pmuon}}
\end{center}
\end{figure}

Figure~\ref{umissD} shows the $M^2_{\rm miss}$ distribution 
for the $D^+\rightarrow \mu^+\nu_{\mu}$ candidates on the 
recoil side of the four $D^-$ tag modes;
Fig.~\ref{umissDs} shows the $M^2_{\rm miss}$ distribution
for the $D_s^+\rightarrow \mu^+\nu_{\mu}$ candidates on the recoil side of 
the $D_s^-\to\phi \pi^-$ tags. 
The number of signal events can be obtained from the $M^2_{\rm miss}$ 
distributions. A window within $\pm3\sigma_{M^2_{\rm miss}}$ around zero is taken
as the signal region,
where $\sigma_{M^2}$ is the standard deviation of the $M^2_{\rm miss}$ 
for  the $D^+$ ($D_s^+$) sample. 
The numbers of the $D^+\rightarrow \mu^+\nu_{\mu}$ and 
$D_s^+\rightarrow \mu^+\nu_{\mu}$
candidates can be extracted from the events in the signal regions, where the 
large
peaks near 0.25~GeV$^2$ for $D^+$ and 0.3~GeV$^2$ for $D_s^+$ in
Figs.~\ref{umissD} and~\ref{umissDs} are from the main background 
sources 
$D^+\rightarrow K_L\pi^+$ and $D_s^+\rightarrow K_L\pi^+$. The total number 
of $D^+\rightarrow \mu^+\nu_{\mu}$ and  $D_s^+\rightarrow\mu^+\nu_{\mu}$
candidates are determined to be $91 \pm 9.5$ and $63 \pm 7.9$, respectively.

\begin{figure}[htb]
\begin{center}
 \includegraphics[width=8cm,height=6cm]
{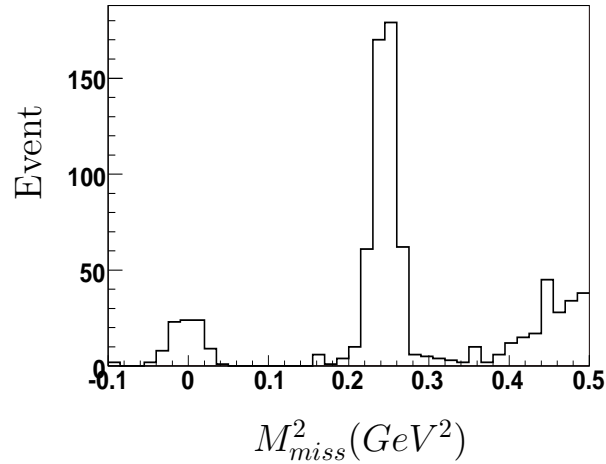}
      \put(-240,80){\rotatebox{90}{\large Event}}
      \put(-150,-15){\large $M^2_{miss}(GeV^2)$}
\caption{
The $M^2_{\rm miss}$ distribution of the muon candidates on the recoil-side of 
the four $D^-$ tag modes $K^+\pi^-\pi^-$ , $K^+K^-\pi^-$, $K_S^0\pi^-$ 
and $K^+\pi^-\pi^-\pi^0$ (for a 800pb$^{-1}$ MC data sample).
\label{umissD}
}
\end{center}
\end{figure}

\vspace{1.0cm}
\begin{figure}[htb]
\begin{center}
  \includegraphics[width=8cm,height=6cm]
{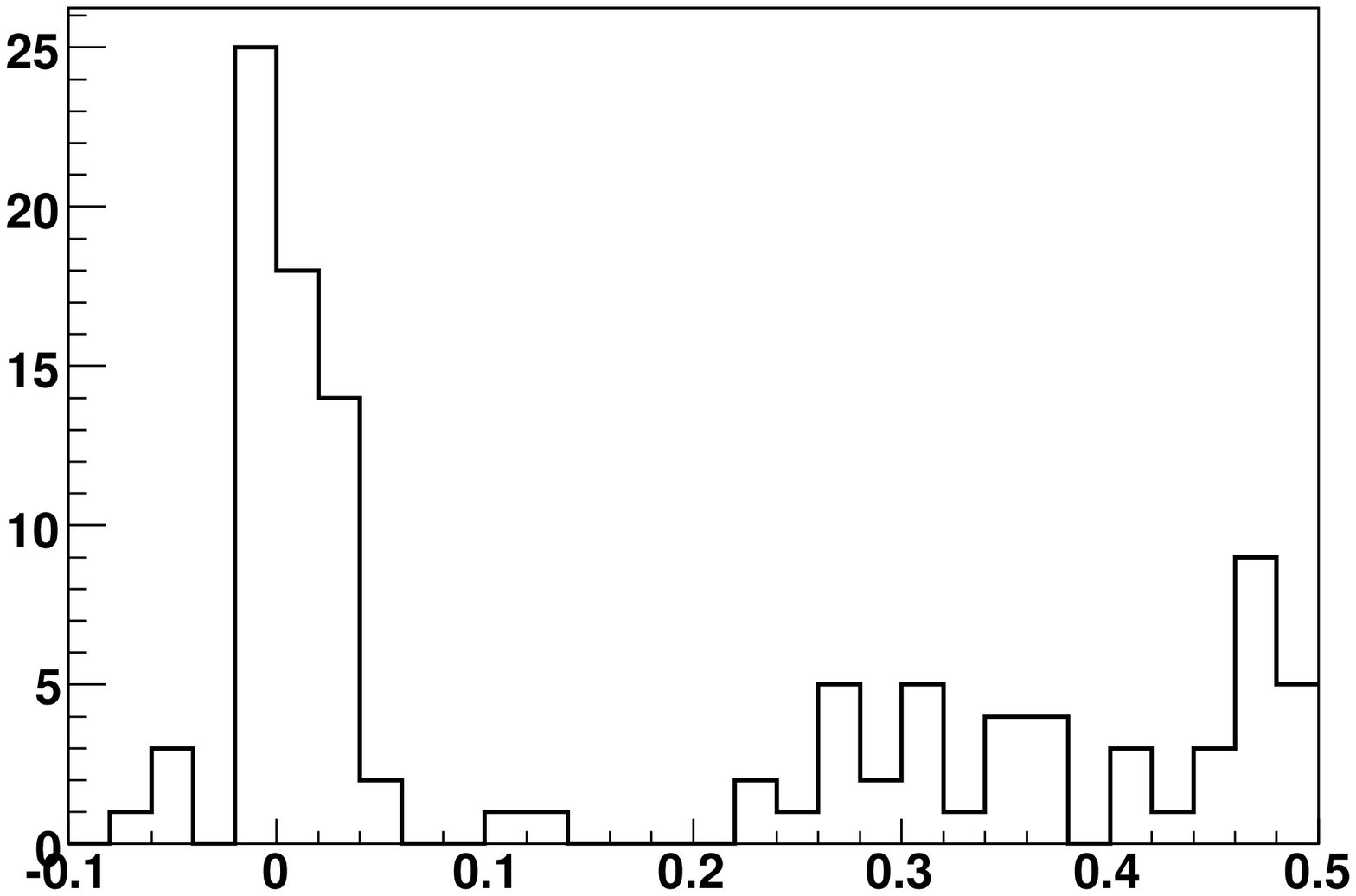}
      \put(-230,80){\rotatebox{90}{\large Event}}
      \put(-150,-15){\large $M^2_{miss}(GeV^2)$}
\vspace{0.5cm}
\caption{
The $M^2_{\rm miss}$ distribution of the muon candidates on the recoil-side of 
the $D_s^- \to \phi \pi^-$ tags (for a 10~fb$^{-1}$ MC data sample).
\label{umissDs}
}
\end{center}
\end{figure}

\subsubsection{Branching fractions for $D^+$ 
and $D_s^+$ leptonic decays}

The detection efficiency for  $D^+\rightarrow 
\mu^+\nu_{\mu}$  and $D_s^+\rightarrow\mu^+\nu_{\mu}$ leptonic decays
can be estimated using MC events ($D^+\rightarrow 
\mu^+\nu_{\mu}~vs~D^-_{\rm tag}$) and ($D_s^+\rightarrow
\mu^+\nu_{\mu}~vs~D_s^-(\rm tag)$).  They are $53\pm 1\%$ for 
$D^+\rightarrow \mu^+\nu_{\mu}$
 and $52\pm 1\%$  for $D_s^+ \to \mu^+\nu_{\mu}$, respectively.

The branching fractions for $D^+\rightarrow \mu^+\nu_{\mu}$ and 
$D_s^+\rightarrow \mu^+\nu_{\mu}$  are determined using the relation: 
\begin{equation}
{\mathcal B}(D_{(s)}^+\rightarrow 
\mu^+\nu_{\mu})=\frac{N_{\mu^+}}{N_{D_{(s)\rm tag}^-}\times 
\epsilon_{D_{(s)}^+\rightarrow \mu^+\nu_{\mu}}},
\label{brdmu} 
\end{equation}
where $N_{\mu^+}$ is the observed numbers of $D^+$ or $D_{s}^+\rightarrow
\mu^+\nu_{\mu}$ events,
$N_{D_{(s)\rm tag}^-}$ are numbers of single-tag $D^-$ or $D_s^-$ mesons and
$\epsilon_{D_{(s)}^+\rightarrow \mu^+\nu_{\mu}}$ are the efficiencies. 
For the MC experment, the branching fractions for the leptonic decays $D^+ \rightarrow
\mu^+\mu_{\nu}$ and $D_s^+ \rightarrow \mu^+\nu_{\mu}$ are determined to be
$${\mathcal B}(D^+ \to \mu^+\nu_{\mu}) = (0.074\pm 0.008) \% ~({\rm input} 
 = 0.08\%)$$
and $${\mathcal B}(D_s^+ \to \mu^+\nu_{\mu}) = (0.452\pm 0.057) \% ~({\rm input} = 
 0.40\%)$$

The statistical error on ${\mathcal B}(D^+ \to \mu^+\nu_{\mu})$ is 
about 10.5\%  for an 
integrated luminosity of 800~pb$^{-1}$ at the $\psi(3770)$ and four tag 
modes. This extrapolates to an error of $\sim 2\% $ 
for a 20~fb$^{-1}$ data sample and six tag modes $K^+\pi^-\pi^-$ , 
$K^+K^-\pi^-$, $K_S^0\pi^-$, $K^+\pi^-\pi^-\pi^0$, $K_S^0\pi^-\pi^0$ and 
$K_S^0\pi^-\pi^-\pi^+$ (the PDG-2006~\cite{Part5_pdg06} 
uncertainty is about 16\%). For ${\mathcal B}(D_s^+ \to \mu^+\nu_{\mu})$, the 
statistical error is about 12.5\% for an integerated luminosity of 
10~fb$^{-1}$ 
of $D_s^+D_s^-$ data  for one tag mode. This extrapolates to be about 5\%  
for  a 20~fb$^{-1}$ sample and six tag modes $\phi\pi^-$, 
$K^{*0}K^-$, $K_S K^-$, $K_S^0\pi^-$,  
$K_S^0K^+\pi^-\pi^-$ and $f_0\pi^-$ (the 
PDG-2006~\cite{Part5_pdg06} uncertainty is about 31.1\%).

As mentioned in Sect.~\ref{sec:cross_section}, 
CLEOc has found that the
cross section for $e^+e^-\rightarrow D_s^*\bar{D_s}$
at 4.17~GeV is a factor of four times the
the  $e^+e^-\rightarrow D_s\bar{D_s}$ cross section 
at 4.03~GeV~\cite{part5:ref:cleo_dsscan}.  
If techniques  being developed to 
tag $D_s$ mesons produced in $D_s^*\bar{D_s}$ 
final states are successful, another factor two in precision 
can be obtained, and branching ratio measurements for 
 $D^+ \to \mu^+\nu_{\mu}$ and  $D_s^+ \to \mu^+\nu_{\mu}$
at the $\sim 2\%$ precision levels can be expected.

\subsubsection{Determination of $f_D$  and $f_{D_s}$}

The decay constant $f_D$ ($f_{D_s}$) can be 
obtained by inserting the measured leptonic branching fractions, the mass of the 
muon, the mass of the $D^+$ ($D_s^+$) meson, the 
CKM matrix element $|V_{cd}|$ ($|V_{cs}|$), the Fermi coupling constant $G_F$, 
the lifetime of the $D^+$ ($D_s^+$) into Eq.~\ref{rate_part5}. The 
systematic error on $f_{D^+}$ ($f_{D_s^+}$) are mainly due to uncertainties  
of the $D^+$ ($D_s^+$) lifetime,  
$|V_{cd}|$ ($|V_{cs}|$), and the measured branching 
fraction. The latter is due to the uncertainties in the track-finding efficiency, particle 
identification, photon selection, background estimation and the number of the 
single-tag $D^-$ ($D_s^-$) mesons. 
The expected errors on $f_{D^+}$ and $f_{D_s^+}$  are estimated using 
the relation 
\begin{equation}
\frac{\delta 
f_{D_q}}{f_{D_q}}=\sqrt{(\frac{\Delta\tau_q}{2\tau_q})^2+(\frac{\Delta 
B}{2B})^2+(\frac{\Delta |V_{cq}|}{|V_{cq}})^2},
\label{dfdq}
\end{equation}
where $D_q=D+(D_s^+)$, $\tau_q=\tau_{D^+}(\tau_{D_s^+})$ and
$V_{cq}=V_{cd}(V_{cs})$.  Table~\ref{fderr} lists the expected 
errors of $f_{D^+}$ and $f_{D_s^+}$ with 20~fb$^{-1}$ data samples at \bes3. 

\begin{table}[htb]
\footnotesize
\begin{center}
\caption{The expected errors on $f_{D^+}$ and $f_{D_s^+}$ from \bes3 
measurements with 20~fb$^{-1}$ data samples.\label{fderr} }
\begin{tabular}{|c|c|c|c|c|c|}
\hline \hline
Decays &Decay Const.&$\frac{\delta 
B}{B}$&$\frac{\delta\tau}{\tau}$~\cite{Part5_pdg06} &$\frac{\delta
|V_{cq}|}{|V_{cq}}$~\cite{dvcq}&$\frac{\delta f_{D_q}}{f_{D_q}}$\\ \hline
$D^+ \rightarrow \mu^+\nu_{\mu}$& $f_{D^+}$& 2\%&0.6\%&1.1\%&1.5\%\\
$D_s^+ \rightarrow \mu^+\nu_{\mu}$& $f_{D_s^+}$& 2\%&1.0\%&0.06\%&1.3\%\\
\hline \hline
\end{tabular}
\end{center}
\end{table}

\subsubsection{Conclusion}
The decay constants $f_D$ and $f_{D_s}$ have been measured by many experiments
but the precision of the measured values is still not sufficient for stringent
tests of the SM.
\bes3 is expected to measure  $f_{D}$ and $f_{D_s}$ with significantly  
improved precision. From a Monte Carlo study, we estimate 
that the statistical error for ${\mathcal B}(D^+ \to \mu^+\nu_{\mu})$ 
would be about 
$\sim 2\%$ for a 20~fb$^{-1}$ sample with six tag modes $K^+\pi^-\pi^-$ , 
$K^+K^-\pi^-$, $K_S^0\pi^-$, $K^+\pi^-\pi^-\pi^0$, $K_S^0\pi^-\pi^0$ and 
$K_S^0\pi^-\pi^-\pi^+$ (the PDG-2006~\cite{Part5_pdg06} 
uncertainty is about 16\%); for ${\mathcal B}(D_s^+ \to \mu^+\nu_{\mu})$, the 
statistical error is expected to be about $\sim 5\%$  
for a 20~fb$^{-1}$ data sample at 4.03~GeV
with six tag modes $\phi\pi^-$, 
$K^{*0}K^-$, $K_S^0\pi^-$, $K_S^0K^-$, 
$K_S^0K^+\pi^-\pi^-$ and $f_0\pi^-$ (the 
PDG-2006~\cite{Part5_pdg06} uncertainty is about 31.1\%). 
respectively.  If techniques being developed to tag 
$D_s$ mesons produced in $D_s^*\bar{D_s}$ final states
prove successful, the higher cross section for 
$e^+e^-\rightarrow D_s^*\bar{D_s}$ at the
4.17~GeV cms energy can be exploited
to gain as much as a factor of two improvement in 
statistical precision.  Ultimate errors
on $f_{D}$ and $f_{D_s}$ in the
 $1\sim 2$\% range may be attainable.

\section{Semileptonic Decays}
\subsection[Theoretical Review]{Theoretical Review\footnote{By Mao-Zhi Yang}}
\label{part5:section:review:semi}

Charm mesons can decay into other hadrons by emitting a 
$\ell^+\nu$ lepton pair via the weak interactions. At the 
quark level, this
process is induced by the semileptonic charm quark decay: $c\to
q\ell^+\nu$, where $q=d,~s$. The light $d$ or $s$ daughter quark is
bound to the initial light quark of the charm meson by the strong
interaction to form a new hadron $X$, as depicted
by the Feynman diagram of Fig.~\ref{Feynman}.
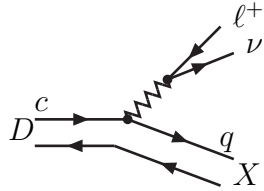
\begin{figure}[h]
\begin{center}
\begin{picture}(65,65) (55,-20)
\SetWidth{1.0}
\ArrowLine(45,-5)(80,-5) \ArrowLine(75,-15)(45,-15)
\Vertex(80,-5){1.83} \ZigZag(80,-5)(95,10){2.5}{4}
\ArrowLine(95,10)(120,20) \ArrowLine(115,30)(95,10)
\Vertex(95,10){1.83}
\ArrowLine(115,-30)(75,-15) \ArrowLine(80,-5)(120,-20)
\put(45,-2){$c$}\put(115,-15){$q$}
\put(35,-13){$D$}\put(120,-30){$X$}
\put(125,20){$\nu$}\put(120,30){$\ell^+$}
\end{picture}
\end{center}
\caption{Feynman diagram for semileptonic D decay} \label{Feynman}
\end{figure}

In semileptonic decays.
the two leptons do not feel the strong interaction, 
and are thus free of strong binding effects. 
Therefore, they can be factored out of the hadronic matrix
element in the amplitude of the semileptonic decay process
\begin{equation}
A=\frac{G_F}{\sqrt{2}}V_{cq}^*
\bar{\nu}\gamma_\mu(1-\gamma_5)l\langle X|\bar{q}\gamma^\mu
(1-\gamma_5)c|D\rangle ,
\end{equation}
where all strong interactions are included in the hadronic matrix
element $\langle X|\bar{q}\gamma^\mu (1-\gamma_5)c|D\rangle $. The
amplitude of the semileptonic decay process depends both on the
hadronic matrix element and the quark-mixing parameter
$V_{cq}$---the Cabibbo-Kobayashi-Maskawa (CKM) matrix element. Thus,
the semileptonic charm meson decay process is a good laboratory for both
studying the quark-mixing mechanism and testing theoretical
techniques developed for calculating the hadronic matrix element.

The hadronic matrix element can be decomposed into several form
factors according to its Lorentz structure. The form factors are
generally controlled by non-perturbative dynamics,  since
perturbative QCD can not be applied directly. Several 
theoretical methods can
be used to calculate these transition form factors, including:
Lattice QCD \cite{LQCD1,LQCD2}, QCD sum rules
\cite{SVZ,sum,BBD,cd,dly}, light-cone sum rules \cite{LCSR}, quark
model \cite{BSW,ISGW,ISGW2,MS}, light-front approach
\cite{LightFront}, and large-energy and heavy-quark-effective
theory \cite{fk}.  Among these, LQCD, QCD sum rules and LCSR are
modeled on QCD.

Depending on the total quantum numbers of the daughter-quark system,
the final hadron produced in the semileptonic decay process can be
a pseudoscalar, vector, or scalar meson, etc. In the following, the
charm meson decay to pseudoscalar, vector, or
scalar meson processes are denoted as $D\to P\ell^+\nu$, $D\to
V\ell^+\nu$, and $D\to S\ell^+\nu$, respectively.

\begin{center}{\bf \normalsize{(1) Transitions to pseudoscalar mesons $D\to
P\ell^+\nu$}}
\end{center}

According to its Lorentz structure, the hadronic matrix element of
$D\to P$ transition process can be decomposed as
\begin{equation}
\langle P|\bar{q}\gamma^\mu (1-\gamma_5)c|D\rangle = (p_1+p_2)_\mu
F_+(q^2)+(p_1-p_2)_\mu F_-(q^2),
\end{equation}
where $p_1$ and $p_2$ are the momenta of the initial $D$ and final
pseudoscalar mesons, respectively, $q=p_1-p_2$, and $F_+ (q^2)$
and $F_-(q^2)$ are the form factors. Equivalently, the
decomposition can be expressed in the form
\begin{equation}
\langle P|\bar{q}\gamma^\mu (1-\gamma_5)c|D\rangle =
\left(p_1+p_2-\frac{m_D^2-m_P^2}{q^2}q\right)_\mu
F_+(q^2)+\frac{m_D^2-m_P^2}{q^2}q_\mu F_0 (q^2)
\label{part5:eq:semi:rev}
\end{equation}
where $F_0(q^2)$ and $F_+(q^2)$ are the longitudinal
and transverse form factors, respectively.

In general, the form factors in Eq.~\ref{part5:eq:semi:rev}
can depend on all of the Lorentz scalars that can be 
formed from the two momenta $p_1$ and $p_2$, i.e., $p_1^2$, $p_2^2$
and $p_1\cdot p_2$. However, $p_1^2$ and $p_2^2$ are not
variables, they are the on-shell masses of the initial- and final-state
particles. Therefore, the form factors can only depend on the
Lorentz scalar $p_1\cdot p_2$, which can equivalently be
represented by  $q^2$, according to the relation
$q^2=p_1^2-2p_1\cdot p_2+p_2^2$.

For the $q^2$ dependence of the form factors, one usually
assumes nearest-pole dominance~\cite{BSW}:
\begin{equation}
F_{\pm}(q^2)=\frac{F_{\pm}(0)}{1-q^2/m_{\tiny\mbox{pole}}^2}
\end{equation}
One can choose the value of $m_{\tiny\mbox{pole}}$ as the mass of
the nearest charm resonance with the same $J^P$ as the hadronic
weak current that induces the $c\to q$ transition. This is only an
assumption. In practice, one can also fit the data to obtain the
effective pole mass. Any deviation from the mass of the nearest
charm resonance will indicate the presence of contributions of the higher
resonances. In this regard, one can also fit the data with the
modified $q^2$ distribution~\cite{BK,114002}
\begin{equation}\label{double-pole}
F_{\pm}(q^2)=\frac{F_{\pm}(0)}{(1-q^2/m_{D^*_{(s)}}^2)(1-\alpha
q^2/m_{D^*_{(s)}}^2)},
\end{equation}
where the parameter $\alpha$ describes the deviation from the
single resonance contribution.

In addition to the pole dominance ansatz discussed above, there is
another model for the $q^2$ dependence of the form factors 
that is commonly used in the
literature~\cite{ISGW,ISGW2}:
\begin{equation}
F_{\pm}(q^2)=F_{\pm}(0)e^{\alpha q^2}.
\end{equation}
Although this exponential form is quite different from the
pole-dominance ansatz, it is difficult in practice 
to see the difference in cases where the final meson is 
heavy because of the short range of $q^2$ that is kinematically 
accessible.   Experimentally, it is possible to distinguish 
between the $q^2$ dependence of different models
in decays to light final-state  meson
production, such as the $D\to\pi$ transition. 

Neglecting the lepton mass, the differential decay 
width of $D\to P\ell^+\nu$ is given by
\begin{equation}
\frac{d\Gamma}{dq^2}(D\to P\ell^+\nu )=\frac{G_F^2}{192\pi^3
m_D^3} |V_{cq}|^2[(m_D^2+m_P^2-q^2)-4m_D^2m_P^2]^{3/2}|F_+(q^2)|^2,
\end{equation}
where only the $F_+(q^2)$ form factor contributes. The
contribution of $F_-(q^2)$ is proportional to the squared lepton
mass $m_\ell^2$, and is, therefore, neglected. The $q^2$ 
distribution covers the range $0\le q^2\le (m_D-m_P)^2$. 
The branching ratio for the
semileptonic decay can be obtained by integrating the differential
decay width over the entire physical $q^2$ range
\begin{equation}
Br(D\to P\ell^+\nu)=\tau_D\int_0^{(m_D-m_P)^2}
dq^2\frac{\Gamma}{dq^2},
\end{equation}
where $\tau_D$ is the mean life time of the $D$ meson.

The semileptonic decays $D\to\pi\ell^+\nu$ and $D\to K
\ell^+\nu$ have been investigated both theoretically and 
experimentally.
Theoretically the $F_+^{D\pi}$ and $F_+^{DK}$ form factors have
been calculated in the quark model
and with QCD sum rules, QCD light-cone sum
rules (LCSR), lattice QCD (LQCD), etc. Some numerical results are
summarized in Table~\ref{t1}.
\begin{table}[h]
\caption{The $F_+^{D\pi}(0)$ and $F_+^{DK}(0)$ form factors.}
\begin{center}
\begin{tabular}{c|c|c}\hline
     &$F_+^{D\pi}(0)$ &$F_+^{DK}(0)$\\
 \hline
 LQCD1\cite{LQCD1}& $0.57\pm 0.06^{+0.01}_{-0.00}$ & $ 0.66\pm 0.04^{+0.01}_{-0.00}$
                    \\\cline{2-2}
                  & $0.57\pm 0.06^{+0.02}_{-0.00}$  &  \\ \hline
 LQCD2\cite{LQCD2}& $0.64\pm 0.03\pm 0.06$ & $0.73\pm 0.03\pm 0.07$ \\   \hline
 QCD SR\cite{BBD} &$0.5\pm 0.1$ & $0.6^{+0.15}_{-0.10}$\\   \hline
 LCSR\cite{LCSR}  &$0.65\pm 0.11$ & $0.785\pm 0.11$\\   \hline
 LCSR\cite{LCSR2}  &$0.67\pm 0.19$ & $0.67\pm 0.20$\\   \hline
 Quark Model\cite{MS}& 0.69 & 0.78\\   \hline
 Light-Front\cite{LightFront} &0.67& \\   \hline\hline
 BES\cite{BES}(Exp.)    & $0.73\pm 0.14\pm 0.06$ & $0.78\pm 0.04\pm
           0.03$\\   \hline
                &\multicolumn{2}{c}{$\frac{|F_+^{D\pi}(0)|^2|V_{cd}|^2}
                {|F_+^{DK}(0)|^2|V_{cs}|^2}=0.038^{+0.007+0.005}_{-0.007-0.003}$}\\
                \cline{2-3}
 CLEO\cite{Part5_CLEO_semi}(Exp.)  & \multicolumn{2}{c}{Input $|V_{cd}|^2/|V_{cs}|^2=0.052\pm
                0.001$}\cite{Part5_pdg06}\\ 
                &\multicolumn{2}{c}{$\Rightarrow \frac{|F_+^{D\pi}(0)|^2}{|F_+^{DK}(0)|^2}=0.86
                \pm 0.07^{+0.06}_{-0.04}\pm 0.01$}\\ \hline\hline
\end{tabular}\label{t1}
 \end{center}
\end{table}
Existing BES collaboration data~\cite{BES} are not precise enough
to challenge these theoretical prediction. The precision will be
highly improved by \bes3.

For the $q^2$ dependence of the form factors, QCD sum rules
confirm the pole dominance behavior for $F_+^{D\pi}(q^2)$ and
$F_+^{DK}(q^2)$. The fitted pole masses from QCD sum rules are:
$m_{\tiny\mbox{pole}}^{D\to\pi}=(1.95\pm 0.10)~\mbox{GeV}$ and
$m_{\tiny\mbox{pole}}^{D\to K }=(1.81\pm 0.10)~\mbox{GeV}$
\cite{BBD}, which are compatible with the experimental values
$m_{\tiny\mbox{pole}}^{D\to\pi}=1.86^{+0.10+0.07}_{-0.06-0.03}~\mbox{GeV}$,
$m_{\tiny\mbox{pole}}^{D\to K}=1.89\pm
0.05^{+0.04}_{-0.03}~\mbox{GeV}$~\cite{Part5_CLEO_semi}.
The $m_{\tiny\mbox{pole}}^{D\to\pi}$ value is consistent with the 
mass of the $D^*$, while $m_{\tiny\mbox{pole}}^{D\to K}$ is definitely
distinct from the mass of $D^*_{(s)}$. This indicates that the single
pole behavior of $F_+^{DK}(q^2)$ is the mean effect of a set
of resonances in the $c\bar{s}$ channel.

Using the modified pole form of Eq.~\ref{double-pole}, the value
of $\alpha$ for both $D\to \pi$ and $D\to K$ from LCSR is
consistent with zero~\cite{LCSR}, which implies the strong
dominance of the $D_{(s)}^*$ pole. Experiment gives $\alpha^{D\to
\pi}=0.37^{+0.20}_{-0.31}\pm 0.15$ and $\alpha^{D\to K}=0.36\pm
0.10 ^{+0.03}_{-0.07}\pm 0.15$~\cite{Part5_CLEO_semi}. 
The non-zero value of
$\alpha^{D\to K}$ suggests the existence of contributions beyond
the pure $D_s^*$ pole to $F_+^{DK}(q^2)$.

Different theoretical techniques predict slightly different
$q^2$ dependencies of the form factors. High precision experimental
measurements of the partial decay width over different ranges of
$q^2$ will distinguish which method correctly describes the 
non-perturbative dynamics of QCD.

\begin{table}[h]
\caption{Branching fractions for $D^0\to \pi^-\ell^+\nu$ and $D^0\to
K^-\ell^+\nu$.}
\begin{center}
\begin{tabular}{c|c|c}\hline
     &$Br(D^0\to \pi^-\ell^+\nu)$(\%) &$Br(D^0\to K^-\ell^+\nu)$(\%)\\
 \hline
 LQCD1\cite{LQCD1}& $0.23\pm 0.06$ & $2.83\pm 0.45$
                    \\\cline{2-3}
                  & $0.24\pm 0.06$  & $2.99\pm 0.45$ \\ \hline
 LQCD2\cite{LQCD2}& $0.32\pm 0.02\pm 0.06\pm0.03$ & $3.77\pm 0.29\pm 0.74\pm 0.08$ \\   \hline
 QCD SR\cite{BBD} &$0.16\pm 0.03$ & $2.7\pm 0.6$\\   \hline
 LCSR\cite{LCSR}  &$0.27\pm 0.10$ & $3.6\pm 1.4$\\   \hline
 LCSR\cite{LCSR2}  &$0.30\pm 0.09$ & $3.9\pm 1.2$\\   \hline\hline
 BES\cite{BES}(Exp.) & $0.33\pm 0.13\pm 0.03$ & $3.82\pm 0.40\pm
          0.27$\\   \hline
 CLEO\cite{CLEO2}(Exp.)    & $0.262\pm 0.025\pm 0.008$ & $3.44\pm 0.10\pm
          0.10$\\   \hline\hline
\end{tabular}\label{t2}
 \end{center}
\end{table}

Theoretical calculations of the branching fractions for $D^0\to
\pi^-\ell^+\nu$ and $D^0\to K^-\ell^+\nu$ are compared with
current experimental data in Table~\ref{t2}. Most of the
theoretical predictions are consistent with experimental data
at the present level of precision, except for the results from 
QCD sum rules~\cite{BBD}, which are somewhat lower than 
the experimental values.

Isospin invariance has been tested in $D$ meson semileptonic
decays~\cite{CLEO3}. Absolute branching fractions for $D^+\to\pi^0
e^+ \nu$ and $D^+\to \bar{K}^0 e^+\nu$ are measured to be:
$Br(D^+\to\pi^0 e^+ \nu )=(0.44\pm 0.06\pm 0.03)\%$, $Br(D^+\to
\bar{K}^0 e^+\nu )=(8.71\pm 0.38\pm 0.37)\% $. The ratio of the
decay widths $D^0\to K^- e^+\nu $ and $D^+\to \bar{K}^0 e^+\nu
$ is $\frac{\Gamma (D^0\to K^- e^+\nu)}{\Gamma (D^+\to \bar{K}^0
e^+\nu )}=1.00\pm 0.05\pm 0.04 $ and is consistent with 
the isospin invariance expectation for 
this ratio, which is unity.  Thus, the experimental data 
confirms the applicability of isospin
symmetry to $D\to K\ell^+\nu$ semileptonic decays.

The ratio of $\frac{\Gamma (D^0\to \pi^- e^+\nu)}{2\Gamma (D^+\to
\pi^0 e^+\nu )} $ is measured to be $0.75^{+0.14}_{-0.11}\pm 0.04$
\cite{CLEO3}. Isospin invariance says that this ratio should also 
be unity. The CLEO central values imply the presence of a
large isospin breaking effect in the semileptonic decay of
$D\to \pi e^+\nu $. However, the present experimental error is
too large to be conclusive.  The precision will be 
improved by future \bes3 measurements.

\begin{table}[h]
\caption{Branching fractions for $D$ and $D_s^+$ to $\eta$, $\eta'$,
$K$ transitions.}
\begin{center} 
\begin{tabular}{c|c|c}\hline
   channel  & $Br(\%)$ & Ref.\\
 \hline
$D^+\to \eta\ell^+\nu$   & 0.10 & Ref.\cite{fk} (double pole)\\
                             & 0.15 & Ref.\cite{fk} (single pole)\\
                             & $<0.5$ & Exp.\cite{Part5_pdg06} \\ \hline
$D^+\to \eta'\ell^+\nu$  & 0.016 & Ref.\cite{fk} (double pole)\\
                             & 0.019 & Ref.\cite{fk} (single pole)\\
                             & $<1.1$ & Exp.\cite{Part5_pdg06} \\ \hline
$D_s^+\to \eta\ell^+\nu$ & 1.7 & Ref.\cite{fk} (double pole)\\
                             & 2.5 & Ref.\cite{fk} (single pole)\\
                             &$2.3\pm 0.4$& Ref.\cite{cd}\\
                            & $2.5\pm 0.7$ & Exp.\cite{Part5_pdg06} \\ \hline
$D_s^+\to \eta'\ell^+\nu$ & 0.61 & Ref.\cite{fk} (double pole)\\
                             & 0.74 & Ref.\cite{fk} (single pole)\\
                             &$1.0\pm 0.2$& Ref.\cite{cd}\\
                            & $0.89\pm 0.33$ & Exp.\cite{Part5_pdg06} \\ \hline
$D_s^+\to K^0\ell^+\nu$ & 0.20& Ref.\cite{fk} (double pole)\\
                             & 0.32 & Ref.\cite{fk} (single pole)\\ \hline

 \end{tabular}\label{t3} 
 \end{center}
\end{table}

There are other $D_{(s)}$ to pseudoscalar semileptonic decay
channels that deserve further studies. The present theoretical
calculations and experimental measurements are listed in
Table~\ref{t3}.


\begin{center}{\bf\normalsize{(2) Transitions to vector mesons $D\to
V\ell^+\nu$}}\end{center}

The Lorentz decomposition of the transition matrix element of \
$D\to V$ is
\begin{eqnarray}
&&\langle V(\varepsilon,p_2)|\bar{q}\gamma_\mu (1-\gamma_5)c|
D(p_1)\rangle
=\varepsilon_{\mu\nu\alpha\beta}\varepsilon^{*\nu}p_1^\alpha
p_2^\beta
\frac{2V(q^2)}{m_{D}+m_{V}} \nonumber \\
&&-i(\varepsilon^*_\mu-\frac{\varepsilon^*\cdot q}{q^2}
q_\mu)(m_{D}+m_{V})A_1(q^2)   +i[(p_1+p_2)_\mu
-\frac{m_{D}^2-m_{V}^2}{q^2} q_\mu ]\nonumber\\
&&\times \varepsilon^*\cdot q \frac{A_2(q^2)}{m_{D}+m_{V}}
-i\frac{2m_V \varepsilon^*\cdot q}{q^2}q_\mu A_0(q^2),
\label{part5:eq:semi:rev:vector}
\end{eqnarray}
where the form factor $V(q^2)$ receives contributions from 
the vector current $\bar{q}\gamma_\mu c$, and the form factors
$A_{0,1,2}(q^2)$ from the axial-vector current $\bar{q}\gamma_\mu
\gamma_5 c$.

The differential and total $D\to V\ell^+\nu$ decay rates
can be calculated from the above decomposition of the hadronic
matrix element. There are three polarization states for
the $V$ meson: one longitudinal state and two transverse 
states (right-handed and left-handed). The differential 
decay rate to a
longitudinally polarized $V$ meson is given by
\begin{eqnarray}
\frac{d\Gamma_L}{d
q^2}&=&\displaystyle\frac{G_F^2|V_{cq}|^2}{192\pi ^3
   m_{D}^3} \sqrt{\lambda
   (m_{D}^2,m_{V}^2,q^2)}\left|\frac{1}{2m_{V}}
   \left[(m_{D}^2-m_{V}^2-q^2)\right.\right.
   \nonumber\\[4mm]
    &&\left.\left.\times (m_{D}+m_{V})A_1(q^2)-
    \displaystyle\frac{\lambda (m_{D}^2,m_{V}^2,q^2)}{m_{D}+m_{V}}
    A_2(q^2)\right]\right|^2 ,
\end{eqnarray}
where $\lambda (m_{D}^2,m_{V}^2,q^2)\equiv
(m_{D}^2+m_{V}^2-q^2)^2-4m_{D}^2m_{V}^2$.

The differential decay rates to the transverse states is given by
\begin{equation}
\frac{d\Gamma_T^\pm}{d
q^2}=\displaystyle\frac{G_F^2|V_{cq}|^2}{192\pi ^3
   m_{D}^3}q^2 \lambda
   (m_{D}^2,m_{V}^2,q^2)^{3/2}\left|\frac{V(q^2)}{m_{D}+m_{V}}
   \mp \frac{(m_{D}+m_{V})A_1(q^2)}{\sqrt{\lambda
   (m_{D}^2,m_{V}^2,q^2)}}\right|^2 ,
\end{equation}
where $+$ and $-$ correspond to the right- and left-handed
states, respectively. Finally, the combined transverse and total
differential decay rates are given by
\begin{equation}
\frac{d\Gamma_T}{d q^2}=\frac{d}{dq^2}(\Gamma_T^+ +\Gamma_T^-),
~~~~\frac{d\Gamma}{d q^2}=\frac{d}{dq^2}(\Gamma_L +\Gamma_T).
\end{equation}

The vector meson is tagged by its decay products.  For example, for
$D\to K^*\ell^+\nu$, the $K^*$ is tagged according to its decay
process $K^*\to K\pi$. The angular distribution for $D\to
K^*\ell^+\nu$ $(K^*\to K\pi)$ is
\begin{eqnarray}
\displaystyle\frac{d\Gamma}{dq^2d\cos\theta_Kd\cos\theta_\ell d
\chi}&=&\displaystyle\frac{3G_F^2|V_{cs}|^2}{8(4\pi)^4}
\frac{p_{K^*}q^2}{m_D^2}[(1+\cos\theta_\ell)^2\sin^2\theta_K|H_+(q^2)|^2\nonumber\\
&&+(1-\cos\theta_\ell)^2\sin^2\theta_K|H_-(q^2)|^2
-4\sin^2\theta_\ell \cos^2\theta_K|H_0(q^2)|^2\nonumber\\
&&+4\sin\theta_\ell(1+\cos\theta_\ell)\sin\theta_K\cos\theta_K\cos\chi
H_+(q^2)H_0(q^2)\nonumber\\
&&-4\sin\theta_\ell(1-\cos\theta_\ell)\sin\theta_K\cos\theta_K\cos\chi
H_-(q^2)H_0(q^2)\nonumber\\
&&-2\sin^2\theta_\ell \sin^2\theta_K\cos2\chi
H_+(q^2)H_-(q^2)]\nonumber\\
&&\times Br(K^*\to K\pi),
\end{eqnarray}
where $\theta_\ell$ is the polar angle of the lepton in the rest
frame of the $\ell^+\nu$ lepton pair, $\theta_K$ the polar angle
of the kaon in the $K^*$ rest frame, and $\chi$ is the relative
angle between the $D\to K^*\ell^*\nu$ and $K^*\to K\pi$ decay
planes.  The helicity functions $H_+(q^2)$, $H_-(q^2)$ and
$H_0(q^2)$ are
\begin{eqnarray}
H_\pm (q^2)&=& (m_D+m_{K\pi})A_1(q^2)\mp
\displaystyle\frac{\sqrt{\lambda (m_{D}^2,m_{K\pi}^2,q^2)}}{m_D+m_{K\pi}}\nonumber\\
H_0(q^2)&=&\frac{1}{2m_{K\pi}\sqrt{q^2}}
   \left[(m_{D}^2-m_{K\pi}^2-q^2)(m_{D}+m_{K\pi})A_1(q^2)\right.\nonumber\\
&&\left.-\displaystyle\frac{\lambda(m_{D}^2,m_{K\pi}^2,q^2)}{m_{D}+m_{K\pi}}
    A_2(q^2)\right].
\end{eqnarray}

 The $D$ to vector meson transition form factors can be
calculated by a variety of methods, such as the quark model (QM), QCD
sum rules (SR), light-cone sum rules (LCSR), light-front approach
(LF), etc. Some numerical results are collected in Table~\ref{t4}.

\begin{table}[h]
\caption{Form factors for $D\to V$ transitions.}
\begin{center}
\begin{tabular}{c|cccc|c}\hline
   mode  & $V(0)$ & $A_0(0)$ & $A_1(0)$& $A_2(0)$& Ref.\\
 \hline \hline
$D\to K^*$&0.82         &0.47&0.57&0.75& QM \cite{gh}\\
          &1.03         &    &0.66&0.49& QM \cite{MS}\\
          &$1.1\pm 0.25$&    &$0.50\pm 0.15$&$0.60\pm  0.15$&SR\cite{BBD}\\
          &$0.8\pm 0.10$&    &$0.59\pm 0.10$&$0.55\pm  0.08$&LCSR\cite{LCSR2}\\
          &0.99         &1.12&0.62&0.31& Ref.\cite{fk}\\
          \hline
$D\to \rho$&0.65        &0.35&0.41&0.50& QM \cite{gh} \\
           &0.90        &    &0.59&0.49& QM \cite{MS}\\
           &$1.0\pm 0.2$&    &$0.5\pm 0.2$&$0.4\pm 0.1$&SR\cite{BBD}\\
          &$0.72\pm 0.10$&    &$0.57\pm 0.08$&$0.52\pm  0.07$&LCSR\cite{LCSR2}\\
           &0.86        &0.64&0.58&0.48&LF \cite{LightFront}\\
           &1.05        &1.32&0.61&0.31& Ref.\cite{fk}\\
           \hline
$D\to \omega$&1.05        &1.32&0.61&0.31& Ref.\cite{fk}\\
           \hline
$D_s\to \phi$&1.10        &1.02&0.61&0.32& Ref.\cite{fk}\\
          &$1.21\pm 0.33$& $0.42\pm 0.12$&$0.55\pm 0.15$&$0.59\pm 0.17$&SR\cite{dly}\\
           \hline
$D_s\to K^*$&1.16        &1.19&0.60&0.33& Ref.\cite{fk}\\
           \hline\hline
 \end{tabular}\label{t4}
 \end{center}
\end{table}

For some modes, different theoretical methods give consistent
results, for others, different methods give different results.
These different predictions provide opportunities to test the
theoretical methods. However, it is difficult to measure the
absolute values of $D\to V$ transition form factors directly;
usually experiments measure ratios of form factors. 
The ratios of the $D\to V$ transition form factors are
defined as
\begin{equation}
r_V\equiv \frac{V(0)}{A_1(0)},~~r_2\equiv \frac{A_2(0)}{A_1(0)}.
\end{equation}
Measured values of $r_V$ and $r_2$ for each decay mode can be
compared with theoretical calculations to test the theoretical
techniques (see Table~\ref{t5}).

\begin{table}[h]
\caption{Comparisons of $r_V$ and $r_2$ with experimental data.}
\begin{center}
\begin{tabular}{c|cc|c}\hline
   mode  & $r_V$ & $r_2$ & Ref.\\
 \hline \hline
$D\to K^*$&1.44         &1.32& QM \cite{gh}\\
          &2.10         &1.35& QM \cite{MS}\\
          &$2.2\pm 0.2$ &$1.2\pm 0.2$&SR\cite{BBD}\\
          &$1.36\pm 0.39$ &$0.93\pm 0.29$&LCSR\cite{LCSR2}\\
          &3.19         &2.00& Ref.\cite{fk}\\ \cline{2-4}
          &$1.62\pm 0.08$&$0.83\pm 0.05$& Exp.\cite{Part5_pdg06}\\
          \hline
$D\to \rho$&1.3        &0.82& QM \cite{gh} \\
           &1.84        &1.20& QM \cite{MS}\\
           &1.79        &1.21&LF \cite{LightFront}\\
           &3.39        &1.97& Ref.\cite{fk}\\ \cline{2-4}
           &-           & - & Exp.\\
           \hline
$D\to \omega$&3.39        &1.97 & Ref.\cite{fk}\\ \cline{2-4}
            &-           & -  & Exp.\\
           \hline
$D_s\to \phi$&3.44        &1.91& Ref.\cite{fk}\\
          &1.569&0.865&LF \cite{LightFront}\\
          &$2.20\pm 0.85$& $1.07\pm 0.43$&SR\cite{dly}\\ \cline{2-4}
          &$1.92\pm 0.32$&$1.60\pm 0.24$& Exp.\cite{Part5_pdg06}\\
           \hline
$D_s\to K^*$&3.52        &1.82& Ref.\cite{fk}\\ \cline{2-4}
           &-           & - & Exp.\\
           \hline\hline
 \end{tabular}\label{t5}
 \end{center}
\end{table}

The $q^2$ dependence of the form factors can be calculated by
theory. The QCD sum rule result shows that the behavior of
$V(q^2)$ and $A_0(q^2)$ is compatible with the pole model
\cite{BBD,dly}
\begin{equation}
V(0)=\frac{V(0)}{1-q^2/m^V_{\tiny\mbox{pole}}},~~
A_0(0)=\frac{A_0(0)}{1-q^2/m^{A_0}_{\tiny\mbox{pole}}}.
\end{equation}
The fitted pole mass $m^V_{\tiny\mbox{pole}}$ is consistent with
the low-lying $J^P=1^-$ charmed meson resonance. While
the $q^2$ dependence of $A_1(q^2)$ and $A_2(q^2)$ is weak.
This behavior implies that the pole-dominance assumption for
$A_1(q^2)$ and $A_2(q^2)$ is inadequate.

\begin{table}[thb]
\caption{Ratios of the polarized decay widths and total branching
fractions for $D\to V \ell^+\nu$ decays.}
\begin{center}
\begin{tabular}{c|cc|c}\hline
   mode  & $\Gamma_L/\Gamma_T$ & $Br(\%)$ & Ref.\\
 \hline \hline
$D^0\to K^{*-}\ell^+\nu$ &        &4.0& QM \cite{gh}\\
                             &1.28    &2.46& QM \cite{MS}\\
                             &$1.15\pm 0.10$ &$2.0\pm 0.5$&LCSR\cite{LCSR2}\\
                             &1.149         &2.2& Ref.\cite{fk}\\
                             &        &$2.15\pm 0.35$&  Exp.\cite{Part5_pdg06}\\
  \hline
$D^+\to \bar{K}^{*0}\ell^+\nu$ &$0.86\pm 0.06$ &$4.0\pm 1.6$&SR\cite{BBD}\\
                                   &1.13     &5.6& Ref.\cite{fk}\\
                             &        &$5.73\pm 0.35$& Exp.\cite{Part5_pdg06}\\
                                \hline
$D^0\to \rho^-\ell^+\nu$&        &0.29& QM \cite{gh} \\
                            &1.16    &0.17& QM \cite{MS}\\
                            &$1.17\pm 0.09$ &$0.14\pm 0.035$&LCSR\cite{LCSR2}\\
                            &1.10     &0.20& Ref.\cite{fk}\\
                            &    -   &$0.194\pm 0.039\pm 0.013$& Exp.\cite{CLEO2}\\
  \hline
$D^+\to \rho^0\ell^+\nu$&1.10        &0.25& Ref.\cite{fk}\\
                    &-    & $0.21\pm 0.04\pm 0.01$ & Exp.\cite{CLEO3}\\
           \hline
$D^+\to \omega\ell^+\nu$&1.10        &0.25 & Ref.\cite{fk}\\
            &-           & $0.16^{+0.07}_{-0.06}\pm 0.01$& Exp.\cite{CLEO3}\\
           \hline
$D_s^+\to \phi\ell^+\nu$&       &2.5& QM \cite{MS}\\
             &1.08        &2.4& Ref.\cite{fk}\\
          &$0.99\pm 0.43$& $1.8\pm 0.5$&SR\cite{dly}\\
          &$0.72\pm 0.18$&$2.0\pm 0.5$& Exp.\cite{Part5_pdg06}\\
           \hline
$D_s^+\to K^{*0}\ell^+\nu$ &1.21        &0.19& QM \cite{MS}\\
             &1.03        &0.22& Ref.\cite{fk}\\
           &-           & - & Exp.\\
           \hline\hline
 \end{tabular}\label{t6}
 \end{center}
\end{table}

Ratios of the polarized decay widths and total branching ratios
of $D\to V \ell^+\nu$ decays are shown in Table~\ref{t6}.

Isospin symmetry in $D\to V \ell^+\nu$ decays can be tested by
measuring $\frac{\Gamma (D^0\to K^{*-}\ell^+\nu
)}{\Gamma (D^+\to \bar{K}^{*0}\ell^+\nu )}$ and $\frac{\Gamma
(D^0\to \rho^{-}\ell^+\nu )}{2\Gamma (D^+\to \rho^{0}\ell^+\nu
)}$, which isospin invariance says should be unity.
Deviations  from unity provide  measures of the degree of
isospin breaking. CLEO measures:
$\frac{\Gamma (D^0\to K^{*-}e^+\nu )}{\Gamma
(D^+\to \bar{K}^{*0}e^+\nu )}=0.98\pm 0.08\pm 0.04$
and $\frac{\Gamma
(D^0\to \rho^{-}e^+\nu )}{2\Gamma (D^+\to \rho^{0}e^+\nu
)}=1.2^{+0.4}_{-0.3}\pm 0.1$ \cite{CLEO3}, both of which are
consistent with unity. The former ratio implies that isospin
symmetry in $D\to K^*$ channel is a good symmetry at the few
percentage level, while the latter needs significant
improvement in precision to match that of the former.

\begin{center}{\bf\normalsize{(3) Transitions to scalar mesons $D\to
S\ell^+\nu$}}\end{center}

A large number of scalar mesons have been found 
experimentally~\cite{Part5_pdg06}, including the:  
$\sigma$ [or $f_0(600)$], $f_0(980)$,
$f_0(1370)$, $f_0(1500)$, $f_0(1710)$, $a_0(980)$, $a_0(1450)$,
$\kappa$, $K_0^*(1430)$, etc. Their structure is still not well
established theoretically. Suggestions about the composition
of the scalar mesons include
$q\bar{q}$, $q\bar{q}q\bar{q}$ and meson-meson bound states. To
investigate the structure of the scalar mesons, a large amount of
experimental data and theoretical studies are necessary.

Semileptonic $D$ meson transitions to scalar mesons are important
processes for studying the nature of scalars, because of the
cleanliness of semileptonic decays as compared to hadronic decays.

The Lorentz decomposition of the hadronic matrix element of $D\to
S$ transition is
\begin{equation}
 \langle S(p_2)|\bar{q}\gamma^\mu (1-\gamma_5) c|D\rangle
 =i[ (p_1+p_2)^\mu F_+(q^2) +(p_1-p_2)^\mu F_-(q^2)].
\end{equation}
The form factor $F_-(q^2)$ does not contribute to the
semileptonic decay amplitude in the limit of zero lepton mass.
The differential decay width for $D\to S\ell^+\nu$ in
this limit is
\begin{equation}
\frac{d\Gamma}{d q^2}=\displaystyle\frac{G_F^2|V_{Qq}|^2}{192
\pi^3m_{D}^3}F_+(q^2)^2[(m_{D}^2+m_{S}^2-q^2)^2-4m_D^2m_{S}^2]^{3/2}.
\end{equation}
If the form factor is known,  one can predict the decay width
theoretically. The form factor not only depends on the dynamics of
the strong binding effects, but also depends on the constituents of
the scalar mesons. The $D\to S$ transition form factors have been
studied with QCD sum rules by treating the scalars as quark-antiquark
bound states~\cite{dfnn,yang}; the results are listed in 
Table~\ref{t7}.
\begin{table}[thb]
\caption{Form factors for $D_{(s)}\to S \ell^+\nu$ decays.}
\begin{center}
\begin{tabular}{c|c|c}\hline
   mode      & $F_+(0)$ &  Ref.\\
 \hline \hline
$D\to \sigma$& $0.50\pm 0.07$ & SR \cite{dfnn}\\
$D\to \kappa$& $0.52\pm 0.03$ & SR \cite{dfnn}\\
$D\to K_0^*(1430)$& $0.57\pm 0.19$ &SR \cite{yang}\\
$D_s\to K_0^*(1430)$& $0.51\pm 0.20$ &SR \cite{yang}\\
           \hline\hline
 \end{tabular}\label{t7}
 \end{center}
\end{table}

The $q^2$ dependence of $D\to S$ transition form factors
calculated with QCD sum rules is consistent with the pole-model
\begin{equation}
 F_+(q^2)=\frac{F_+(0)}{1-q^2/m_{\tiny\mbox{pole}}^2}.
 \end{equation}
The fitted pole masses for each mode are~\cite{dfnn,yang}
\begin{eqnarray}
m^{D\kappa}_{\tiny\mbox{pole}}~=& 2.05\pm 0.15~\mbox{GeV},\nonumber\\
m^{DK_0^*}_{\tiny\mbox{pole}}~=& 2.9\pm 0.3~\mbox{GeV},\\
m^{D_sK_0^*}_{\tiny\mbox{pole}}~=& 1.96\pm
0.12~\mbox{GeV}.\nonumber
\end{eqnarray}
The total decay widths for each mode are
\begin{eqnarray}
&\Gamma (D\to \sigma\ell^+\nu)=(8.0\pm 2.5)\times
10^{-16}~\mbox{GeV}~\cite{dfnn},\nonumber\\
&\Gamma (D\to \kappa\ell^+\nu)=(5.5\pm 1.0)\times
10^{-15}~\mbox{GeV}~\cite{dfnn},\nonumber\\
&\Gamma (D\to
  K_0^*(1430)\ell^+\nu)=(2.9^{+2.3}_{-1.6})\times
  10^{-16}~\mbox{GeV}~\cite{yang},\nonumber\\
&\Gamma (D_s^+\to
  K_0^*(1430)^0\ell^+\nu)=(3.2^{+3.0}_{-2.0})\times
  10^{-17}~\mbox{GeV}~\cite{yang},
\end{eqnarray}
and the corresponding branching fractions are
\begin{eqnarray}
&Br(D^+\to \sigma\ell^+\nu)=(1.26\pm 0.40)\times
10^{-3},\nonumber\\
&Br(D^{0}\to \kappa^-\ell^+\nu)=(3.43\pm 0.62)\times
10^{-3},\\
&Br(D^{+}\to \bar{\kappa}^0\ell^+\nu)=(8.7\pm 1.6)\times
10^{-3},\nonumber\\
&Br(D^{0}\to K_0^*(1430)^-\ell^+\nu)=(1.8^{+1.5}_{-1.0})\times
10^{-4},\\
&Br(D^{+}\to
\bar{K}_0^*(1430)^0\ell^+\nu)=(4.6^{+3.7}_{-2.6})\times
10^{-4},\nonumber\\
&Br(D_s^+\to K_0^*(1430)^0\ell^+\nu)=(2.4^{+2.2}_{-1.5})\times
10^{-5}.\nonumber
\end{eqnarray}
\noindent
These numerical results indicate that many of the scalar semileptonic
decay modes are rare. Measuring them experimentally will require large
data sample.  In return, however, these measurements will provide valuable 
information on the nature of the light scalar mesons. Such 
measurements are highly desired.

Other decay modes that are worth-while  to measure are $D\to
f_0(980)\ell^+\nu$, $D\to a_0(980)\ell^+\nu$, etc.

\begin{center}{\bf\normalsize{(4) CKM matrix elements $V_{cd}$ and
$V_{cs}$ in semileptonic $D$ decays}}\end{center}

The CKM matrix elements are fundamental SM parameters 
that describe the  mixing of quark fields due to the weak
interaction. These parameters cannot be predicted from the basic
SM theory, they must be measured by experiment. Measurements of the
CKM matrix elements are important for understanding the dynamics of
quark mixing and the source of $CP$ violation. In theories with more
than three generations of quarks, the CKM matrix provides
phase parameters that generate non-SM $CP$ violations.

Semileptonic $D$ decays are sensitive to $V_{cd}$ and $V_{cs}$. 
Precise branching fraction measurements
for semileptonic $D$ meson 
decays can improve the precision of
$V_{cd}$ and $V_{cs}$. The Particle Data Group (2004)
gives magnitudes for $V_{cd}$ and $V_{cs}$ of~\cite{Part5_pdg06}
\begin{equation}
|V_{cd}|=0.224\pm 0.012,~~~~~~~|V_{cs}|=0.996\pm 0.13.
\end{equation}
The magnitude of $|V_{cd}|$ is deduced from neutrino and
antineutrino production of charm from valence $d$ quarks. The
present error on $|V_{cd}|$ is about 5.4\%. Values for $|V_{cs}|$
obtained from neutrino production of charm have errors that exceed
10\% mainly because of theoretical uncertainties. 
The error of the present 
direct measurement value, given above, is about 13\%.
With the 
application of requirements for
unitarity of the three-generation CKM matrix, the precision for
$|V_{cs}|$ can be greatly improved. 

In \bes3, the absolute branching fractions for the semileptonic decay
modes $D\to \pi\ell^+\nu$, $D\to K\ell^+\nu$, $D\to \eta
(\eta')\ell^+\nu$, $D\to \rho\ell^+\nu$, $D\to K^*\ell^+\nu$,
$D_s\to \phi\ell^+\nu$, etc., will be measured with precisions
at the 1\% level. The form factor  $q^2$-distribution slope will
be as precise as 1.5\%~\cite{lihb}. The differential decay width
is proportional to squared product of $|V_{cd(s)}|$ and the relevant
semileptonic transition form factor $F(q^2)$:
\begin{equation}
\displaystyle\frac{d\Gamma (D\to X_{d,s}\ell^+\nu)}{dq^2}\propto
|V_{cd(s)}|^2|F(q^2)|^2.
\end{equation}
High-precision values of $|V_{cd(s)}|^2|F(q^2)|^2$ can be extracted
from precisely measured absolute branching fractions. This will
lead to determinations of  $V_{cd}$ and
$V_{cs}$ with 1\% precision if the form factors can be
theoretically calculated at the 1.5\% precision level. It 
is hoped that Lattice QCD can reach this level in next few
years.

\subsection[Exclusive Semileptonic Decays]{Exclusive Semileptonic Decays
\footnote{By Hui-Hui Liu, Hai-Long Ma and Gang Rong}}

In exclusive semileptonic decays of $D$ mesons, 
the effects of weak and strong interactions can be be 
well separated theoretically. Therefore,  these channels provide a
good laboratory both for studying the quark-mixing mechanism 
and for testing theoretical  techniques developed to calculate
hadronic matrix elements.

At \bes3, we will collect a $\psi(3770)$ data sample with an integrated
luminosity of about $\mathcal L=$ 20 fb$^{-1}$, which will be
enough to allow for systematic characterizations of the features
of exclusive semileptonic $D$ meson decays. 
The absolute branching fractions for many exclusive semileptonic decays of
$D$ mesons will be precisely measured, including:
$D^0\to K^-\ell^+\nu_\ell(\ell=e,\hspace{0.1cm}{\rm or}\hspace{0.1cm}\mu)$;
$D^0\to\pi^-\ell^+\nu_\ell$; $D^0 \to K^{*-}\ell^+\nu_\ell$, $D^0 \to \rho^-
\ell^+\nu_\ell$; $D^+ \to \overline K^0\ell^+\nu_\ell$; $D^+ \to \pi^0\ell^+
\nu_\ell$; $D^+ \to \overline K^{*0}\ell^+\nu_\ell$; $D^+ \to \rho^0\ell^+
\nu_\ell$; $D^+ \to \omega\ell^+\nu_\ell$; etc. The 
magnitude of the CKM matrix element
$|V_{cs(d)}|$ can then be extracted with  high precision.
Here we describe a MC simulation of a measurement of the simplest
pseudoscalar exclusive semileptonic decays: $D^0 \to K^-e^+\nu_e$ and $D^0
\to \pi^-e^+\nu_e$, to demonstrate the experimental capabilities of
\bes3 in this area.

A Monte Carlo sample corresponding to an integrated luminosity
of about 800~pb$^{-1}$ at 3.773~GeV is used to simulate a
measurement of the branching  fractions
for $D^0 \to K^-e^+\nu_e$ and
$D^0 \to \pi^-e^+\nu_e$. We first reconstruct single-tag
$\bar D^0$ mesons from the sample, and then 
select candidates for the
semileptonic decays $D^0 \to K^-e^+\nu_e$ and $D^0 \to \pi^-e^+\nu_e$
in the system recoiling against the tag $\bar D^0$ meson.  We
measure the branching fractions using a  method previously 
used by BESII~\cite{besd0semi,besdpsemi}. The form factor $|f^{K(\pi)}_+(0)|$,
the CKM matrix element $|V_{cs(d)}|$, and the ratio $|V_{cd}|/|V_{cs}|$ are
extracted. Based on the results of this simulation, we estimate the expected
precision of these measurements for a 20 fb$^{-1}$ $\psi(3770)$ data set.

\subsubsection{\bf 1. Overview of the study of the decays
$D^0 \to K^-e^+\nu_e$ and $D^0 \to \pi^-e^+\nu_e$}
The exclusive semileptonic $D^0 \to K^-e^+\nu_e$ and $D^0 \to \pi^-e^+\nu_e$
decay final states contain only two easily reconstructed charged particles
and have been studied by many different
experiments~\cite{besd0semi,e691,Part5_cleo_ma,cleo2,mark3,
Part5_cleoc,cleo95,e687}.   The
measured branching fractions for $D^0 \to K^-e^+\nu_e$ and $D^0 \to \pi^-
e^+\nu_e$ are summarized in Tables~\ref{brd0kev} and~\ref{brd0piev},
respectively, including the number of signal events associated with
each entry. In some experiments,
the $D^0 \to K^-e^+\nu_e$ (or $D^0 \to\pi^-e^+\nu_e$) branching fractions
are measured relative to the topologically similar mode $D^0 \to K^-\pi^+$
(or $D^0 \to K^-e^+\nu_e$).
Absolute branching fractions for $D^0 \to K^-e^+\nu_e$ and $D^0\to \pi^-e^+\nu_e$
can be extracted by multiplying by the branching fraction of ${\mathcal
B}(D^0\to K^-\pi^+)$ or ${\mathcal B} (D^0 \to K^-e^+\nu_e)$ \cite{Part5_pdg06}.
In the
tables, the first error is statistical, the second is systematic, and the
third arises from the uncertainty in ${\mathcal B}(D^0 \to K^-\pi^+)$ or
${\mathcal B}(D^0\to K^-e^+\nu_e)$.
The relative errors of these branching fractions are 3.1\% for
${\mathcal B}(D^0 \to K^-e^+\nu_e)$ 
and 6.8\% for ${\mathcal B} (D^0 \to \pi^-e^+\nu_e)$,
with the most precise measurements 
coming from the CLEO Collaboration~\cite{Part5_cleoc}. 
\bes3 should be able to improve these errors significantly.

\begin{table}[h]
\caption{\normalsize Summary of measurements of the
$D^0 \to K^-e^+\nu_e$ branching fraction
from different experiments; here `absolute' means
that the branching fraction is made directly.}
\begin{center}
\resizebox{!}{2.0cm}{
\begin{tabular}{cccccc} \hline \hline
Experiment & Number    & Normalization & Ratio of   &
${\mathcal B}(D^0 \to K^-e^+\nu_e)$ \\
           & of Events  &   Mode      & branching fraction& (\%)       \\ \hline
E691\cite{e691}&250 &$\frac{D^0 \to K^-e^+\nu_e}{D^0\to K^-\pi^+}$
&$0.91\pm0.07\pm0.11$&$3.46\pm0.27\pm0.42\pm0.08$\\
CLEO~\cite{Part5_cleo_ma}&584&$\frac{D^0 \to K^-e^+\nu_e}{D^0\to K^-\pi^+}$&$0.90\pm0.06\pm0.06$&$3.42\pm0.23\pm0.23\pm0.08$\\
CLEOII~\cite{cleo2}&2510&$\frac{D^0 \to K^-e^+\nu_e}{D^0\to K^-\pi^+}$&$0.978\pm0.027\pm0.044$&$3.72\pm0.10\pm0.17\pm0.09$\\
MARKIII~\cite{mark3}&55 &absolute     & &$3.4\pm0.5\pm0.4$\\
BESII~\cite{besd0semi}&104&absolute   & &$3.82\pm0.40\pm0.27$\\
CLEOc~\cite{Part5_cleoc}    &1311&absolute   & &$3.44\pm0.10\pm0.10$\\ \hline
PDG average~\cite{Part5_pdg06}    &    &           & &$3.51\pm0.11$       \\ \hline
\end{tabular}
}
\label{brd0kev}
\end{center}
\end{table}

\begin{table}[h]
\caption{\normalsize Summary of  measurements of  the
$D^0 \to \pi^-e^+\nu_e$ branching fraction from different experiments; hhere `absolute' means
that the branching fraction is made directly.}
\begin{center}
\resizebox{!}{2.0cm}{
\begin{tabular}{cccccc} \hline \hline
Experiment & Number    & Normalization & Ratio of   &
${\mathcal B} (D^0 \to \pi^-e^+\nu_e)$ \\
           & of Events  &   Mode      & branching fraction& (\%) \\ \hline
CLEO~\cite{cleo95}&87    &$\frac{D^0 \to \pi^-e^+\nu_e}{D^0 \to K^-e^+\nu_e}$
                            &$0.103\pm0.039\pm0.013$&$0.37\pm0.14\pm0.05\pm0.02$\\
E687~\cite{e687}& 91($e$ and $\mu$)&$\frac{D^0 \to \pi^-l^+\nu_l}{D^0 \to K^-l^+\nu_l}$
                                      &$0.101\pm0.020\pm0.003$&$0.36\pm0.07\pm0.01\pm0.02$\\
MARKIII~\cite{mark3}&7 &absolute     & &$0.39^{+0.23}_{-0.11}\pm0.04$\\
BESII~\cite{besd0semi}&9&absolute   & &$0.33\pm0.13\pm0.03$\\
CLEOc~\cite{Part5_cleoc}    &117&absolute   & &$0.262\pm0.025\pm0.008$\\ \hline
PDG average~\cite{Part5_pdg06}    &    &           & &$0.281\pm0.019$       \\ \hline
\end{tabular}
}
\label{brd0piev}
\end{center}
\end{table}

\subsubsection{\bf 2. Event simulation and selection}
Monte Carlo $e^+e^- \to D\bar D$ events are generated at 3.773~GeV,
with the $D$ and $\bar D$ mesons allowed to decay into all possible final
states with branching fractions taken from the PDG tables~\cite{Part5_pdg06}.
A total of 4.9 million $D\bar D$ Monte Carlo events are generated,
corresponding to an integrated luminosity of about 800 pb$^{-1}$.
Events are fully simulated using the GEANT4 package and reconstruced using
the \bes3 software version BOSS~6.0.2.

To select good candidate events, we required at least two charged tracks
are well reconstructed in the MDC. All charged tracks are required to satisfy
a geometrical requirement $|\rm{cos\theta}|<0.93$, where $\theta$ is the polar angle
with respect to the beam axis. Each track must originate from the
interaction region, which is defined as $V_{xy}<1.0$ cm and $|V_z|<5.0$ cm,
where $V_{xy}$ and $|V_z|$ are the distances of closest approach of the 
charged track in the $xy$-plane and $z$ direction.
Pions and kaons are identified using the $dE/dx$ and TOF measurements,
while neutral kaons are reconstructed through the decay $K^0_S \to\pi^+\pi^-$.
Neutral pions are reconstructed via their $\pi^0 \to \gamma\gamma$ decay mode.

Single-tag $\bar D^0$ mesons are reconstructed in four hadronic decay
modes:  $K^+\pi^-$, $K^+\pi^-\pi^-\pi^+$, $K^+\pi^-\pi^0$ and $K^0\pi^+\pi^-$.
The method used is similar to that described in Section 24.1.3.2.
Figure~\ref{bes3d0tags} shows the resulting beam-constrained
mass distributions for
$Kn\pi(n=1,2,3)$ modes for the single-tag $\bar D^0$ mesons.
A maximum likelihood fit to the mass spectrum with a Gaussian function 
representing
the $\bar D^0$ signal and a special background function~\cite{besd0semi,besdpsemi}
yields the number of the single-tag $\bar D^0$ mesons found in
each mode. Their sum,  $N_{\bar D^0_{\rm tag}}$, is the total number of
reconstructed single-tag $\bar D^0$ mesons. In each mass distribution, events
within  $\pm 3\sigma_{M_{\bar D^0_i}}$ of the fitted $\bar
D^0$ mass $M_{\bar D^0_i}$ value are defined as single-tag $\bar D^0$ candidates,
where $\sigma_{M_{\bar D^0_i}}$ is the standard deviation of the mass
spectrum for the $i^{\rm th}$ tag mode. The region outside of 
$\pm 4\sigma_{M_{\bar D^0_i}}$
window around the fitted $\bar D^0$ mass are used as a $\bar D^0$
sideband sample for
estimating the background in the $\bar D^0$ signal region.

\begin{figure}[htbp]
\begin{center}
  \includegraphics[width=12cm,height=9cm]
{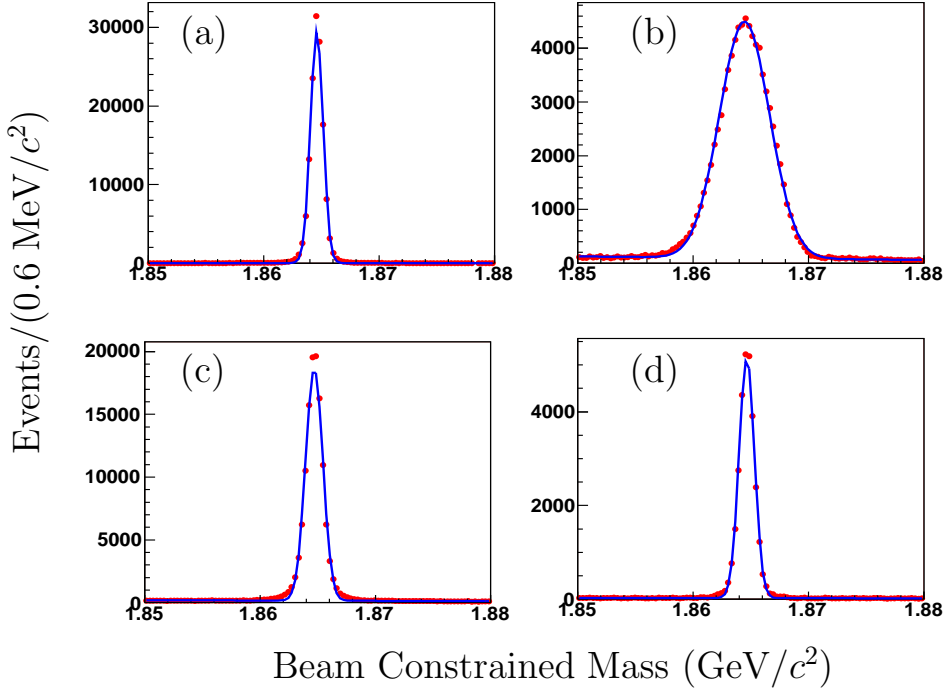}
      \put(-300,225){\large (a)}
      \put(-130,225){\large (b)}
      \put(-300,97){\large (c)}
      \put(-130,97){\large (d)}
      \put(-265,-15){\large Beam Constrained Mass (GeV/$c^2$)}
      \put(-365,70){\rotatebox{90}{\large Events/(0.6 MeV/$c^2$)}}
  \caption{\normalsize The beam constrained masses of the $Kn\pi$ 
single-tag $\bar D^0$ decay modes:
(a) $K^+\pi^-$,
(b) $K^+\pi^-\pi ^0$,
(c) $K^+\pi^-\pi^-\pi^+$ and
(d) $K^0\pi^+\pi^-$.}
  \label{bes3d0tags}
\end{center}
\end{figure}

Candidate $D^0 \to K^-e^+\nu_e$ and
$D^0 \to \pi^-e^+\nu_e$ decays are selected from the surviving
tracks in the system recoiling against the tag $\bar D^0$ mesons.
We require that there are only two oppositely charged tracks, one of
which is identified as an electron and the other as a kaon or pion.
For modes other than $K^0\pi^+\pi^-$, the electron's charge is 
required to be opposite to the charm of the tag $\bar D^0$. 
In order to reduce background events from decays such as 
$D^0 \to K^-\pi^0e^+(\mu^+)\nu_{e(\mu)}$ and 
$D^0 \to K^-\pi^+\pi^0$, we require that there
are no extra charged tracks or isolated photons that have not been
used in the reconstruction of the tag $\bar D^0$ meson. There are
still possible backgrounds for each semileptonic decay due to 
misidentified pions faking an electron.  For example, the decay 
$D^0\to K^-\pi^+$ may be misidentified
as $D^0 \to K^-e^+\nu_e$. These events are suppressed by requiring that
the invariant mass of the $K^-e^+$ combination is less 
than 1.8~GeV.

In semileptonic decays, there is one undetected  massless neutrino.
This can be reconstructed  using the kinematic
quantity
$$U_{miss} \equiv E_{miss} - p_{miss},$$
where $E_{miss}$ and $p_{miss}$ are the total energy and momentum of all the
missing particles. The value of $U_{miss}$ should be close to zero for
correctly reconstructed signal events.
Figure~\ref{kev} shows the $U_{miss}$ distribution 
for tagged $D^0 \to K^-e^+\nu_e$ candidates.
Figure~\ref{piev} shows the
$U_{miss}$ distribution for $D^0 \to \pi^-e^+\nu_e$ candidates
where background from $K^-$ misidentified as $\pi^-$ shows up as peak B.
The signals around zero are the signals for
$D^0 \to K^-e^+\nu_e$ and $D^0 \to \pi^-e^+\nu_e$.
Fitting each $U_{miss}$ distribution with a Gaussian signal function 
and a polynomial background gives
$N_{D^0 \to K^-e^+\nu_e}=5528\pm75$ and
$N_{D^0 \to \pi^-e^+\nu_e}=707\pm27$.  
There is no peak around zero observed in the $U_{miss}$
distributions for selected events with tags in the $\bar D^0$ sideband
regions.

Monte Carlo studies show that the dominant background for $D^0 \to
K^-e^+\nu_e$ is from $D^0 \to K^-\mu^+\nu_{\mu}$, and the main background
for $D^0 \to \pi^-e^+\nu_e$ is from $D^0 \to \pi^-\mu^+\nu_{\mu}$,
$D^0 \to K^-e^+\nu_e$ and $D^0 \to \rho^-e^+\nu_e$
where the $\mu^+(K^-)$ is misidentified as a $e^+(\pi^-)$ and the
$\pi^0$ is missed.  In addition to the tail of
$D^0 \to K^-e^+\nu_e$,  events in peak C are
mainly from $D^0 \to K^{*-}\pi^+, K^{*-} \to K^-\pi^0$ and $D^0 \to
K^-\rho^+, \rho^+ \to \pi^+\pi^0$ where the $K^-\pi^+$ pair is misidentified
as a $\pi^-e^+$ pair the $\pi^0$ is missed.

\begin{figure}[htbp]
\begin{center}
 \begin{minipage}[t]{7cm}
  \includegraphics[height=6.0cm,width=8.0cm]{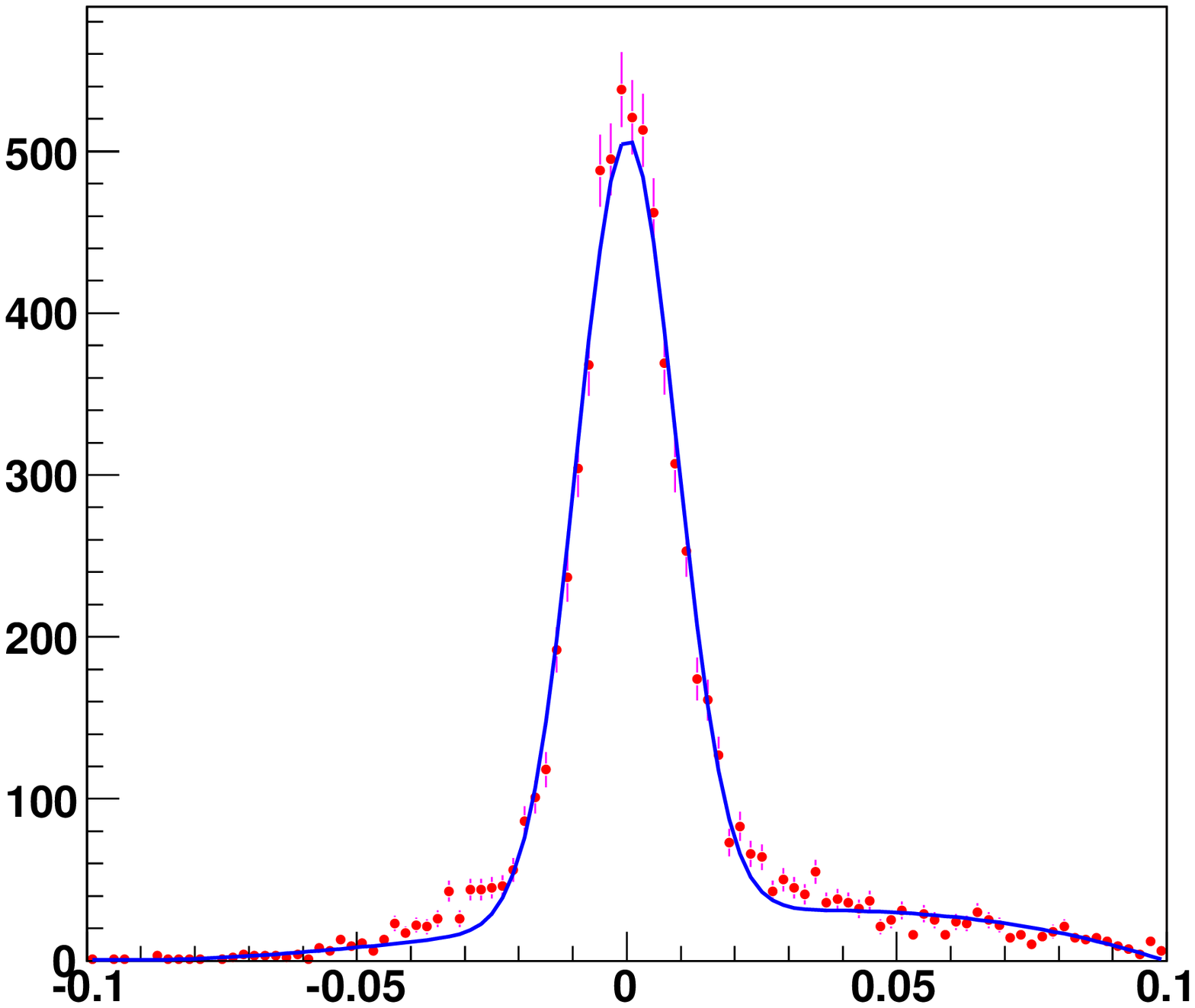}
      \put(-150,-15){\large $U_{miss}$ (GeV)}
      \put(-235,20){\rotatebox{90}{\large Events/(0.002 GeV)}}
  \caption{\normalsize The $U_{miss}$ distribution for
$D^0 \to K^-e^+\nu_e$ candidate events.}
  \label{kev}
  \end{minipage}
\hspace{0.25cm}
  \begin{minipage}[t]{7cm}
  \includegraphics[height=6.0cm,width=8.0cm]{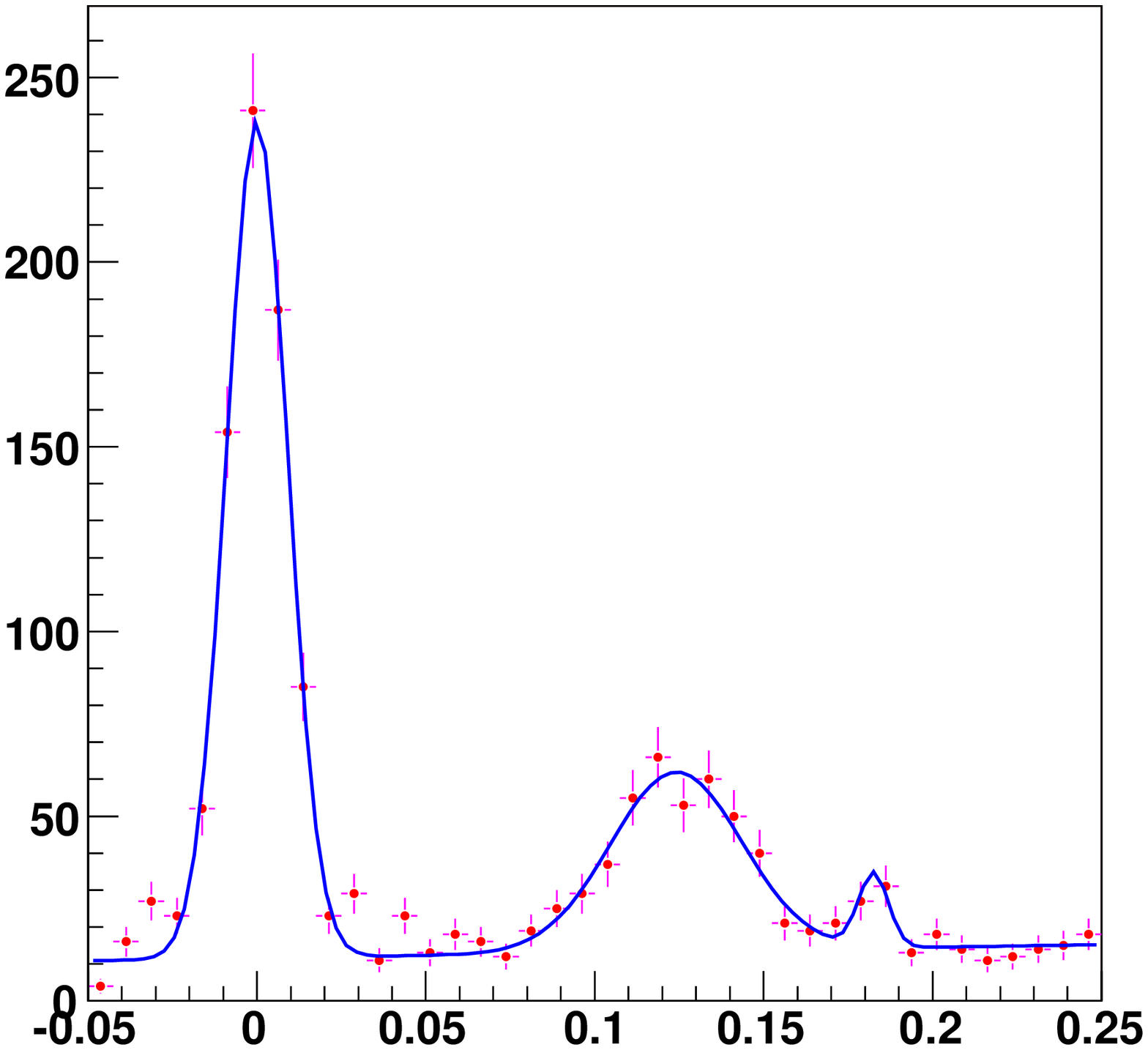}
      \put(-150,-15){\large $U_{miss}$ (GeV)}
      \put(-235,20){\rotatebox{90}{\large Events/(0.0075 GeV)}}
      \put(-175,140){\large A}
      \put(-100,50){\large B}
      \put(-70,50){\large C}
  \caption{\normalsize The $U_{miss}$ distribution for
$D^0 \to \pi^-e^+\nu_e$ candidate events.}
\label{piev}
\end{minipage}
\end{center}
\end{figure}

\subsubsection{\bf 3. Branching fraction determinations}
The  $D^0 \to K^-(\pi^-)e^+ \nu_e$ branching fraction
is determined from the relation
\begin{equation}
\mathcal {B}(D^0 \to K^-(\pi^-)e^+\nu_e)=\frac{N_{D^0 \to K^-(\pi^-)e^+\nu_e}}
{N_{\bar D^0_{\rm tag}}\times\epsilon_{D^0 \to K^-(\pi^-)e^+\nu_e}},
\label{brsemi}
\end{equation}
where $N_{D^0 \to K^-(\pi^-)e^+\nu_e}$ is the number of signal events,
$N_{\bar D^0_{\rm tag}}$ is the total
number of the single-tag $\bar D^0$ mesons and $\epsilon_{D^0 \to
K^-(\pi^-)e^+\nu_e}$ is the MC-determined detection efficiency. 

Inserting numbers into Eqn.~\ref{brsemi}, we obtain the branching fractions
$$\mathcal{B}(D^0 \to K^-e^+\nu_e)=(3.41\pm0.05\pm0.03)\%,$$
$$\mathcal{B}(D^0 \to \pi^-e^+\nu_e)=(0.409\pm0.015\pm0.04)\%,$$
where the first error is statistical and the second systematic.
The systematic errors 
mainly arise from uncertainties in tracking simulation, particle
identification, photon selection, fitting the $U_{miss}$ distribution
and the number of  single-tag $\bar D^0$ mesons.
The total systematic error  expected at \bes3 is 1.0\%.

By extrapolating the simulation results to 20~fb$^{-1}$, 
the expected statistical errors for \bes3 measurements of
$\mathcal{B}(D^0 \to K^-e^+\nu_e)$ and $\mathcal{B}(D^0\to \pi^-e^+\nu_e)$
are estimated to be
0.3\% and 0.7\%, respectively, as shown in Table~\ref{exbr}.

\begin{table}[h]
\caption{\normalsize Expected statistical errors for
$\mathcal{B}(D^0 \to K^-e^+\nu_e)$ and $\mathcal{B}(D^0\to \pi^-e^+\nu_e)$
at \bes3.}
\begin{center}
\begin{tabular}{c|c|c} \hline \hline
$\mathcal L$ & 4 fb$^{-1}$ & 20 fb$^{-1}$ \\ \hline
$\frac{\Delta \mathcal{B}(D^0\to K^-e^+\nu_e)}
{\mathcal{B}(D^0\to K^-e^+\nu_e)}_{\rm stat.}$   & 0.6\% & 0.3\%\\
$\frac{\Delta \mathcal{B}(D^0\to \pi^-e^+\nu_e)}
{\mathcal{B}(D^0\to \pi^-e^+\nu_e)}_{\rm stat.}$ & 1.6\% & 0.7\%\\
 \hline
\end{tabular}
\label{exbr}
\end{center}
\end{table}

\subsubsection{\bf 4. Form factors and CKM matrix elements}
The semileptonic decay width for $D^0 \to K^-(\pi^-)e^+\nu_e$ is
related to the form factor $|f^{K(\pi)}_+(0)|$ and the CKM matrix element
$|V_{cs(d)}|$ via the relations~\cite{besd0semi,morrison,ablikim}
\begin{equation}
\Gamma(D^0\to K^-e^+\nu_e)=\frac{\mathcal{B}(D^0\to
K^-e^+\nu_e)}{\tau_{D^0}}=1.53|V_{cs}|^2|f^K_+(0)|^2\times 10^{11} s^{-1}
\label{vcs}
\end{equation}
\begin{equation}
\Gamma(D^0\to \pi^-e^+\nu_e)=\frac{\mathcal{B}(D^0\to
\pi^-e^+\nu_e)}{\tau_{D^0}}=3.01|V_{cs}|^2|f^{\pi}_+(0)|^2\times 10^{11} {\rm s}^{-1}
\label{vcd}
\end{equation}
\noindent
The form factors $|f^K_+(0)|$ and $|f^{\pi}_+(0)|$ can be extracted from
the measured branching fractions and the lifetime of the $D^0$ meson.
Inserting the values $|V_{cs}|=0.996\pm 0.013$ and $|V_{cd}|=0.224\pm
0.012$~\cite{Part5_pdg06}, the lifetime $\tau_{D^0}=(410.1\pm1.5)\times
10^{-15}$s and the branching fractions into 
Eqs.~\ref{vcs}~and~\ref{vcd}, the form factors are determined to be
$$|f^K_+(0)|=0.74\pm0.01\pm0.01$$
\noindent
and
$$|f^{\pi}_+(0)|=0.81\pm0.04\pm0.04,$$
where the first error is statistical and the second systematic.
In the form factor determination, the systematic error arises mainly
from the uncertainties in the measured branching fractions ($\Delta
\mathcal{B}/\mathcal{B}$ is taken to be 1.0\%), the lifetime
$\tau_{D^0}$ (0.4\%), and the values of the CKM matrix elements $|V_{cs(d)}|$
(1.3\% for $|V_{cs}|$ and 5.4\% for $|V_{cd}|$),
\begin{equation}
\frac{\Delta f^{K(\pi)}_+(0)}{f^{K(\pi)}_+(0)} =
\sqrt{
\left (\frac{\Delta \mathcal{B}}{2\mathcal{B}} \right )^2 +
\left (\frac{\Delta \tau_{D^0}}{2\tau_{D^0}} \right )^2 +
\left (\frac{\Delta V_{cs(d)}}{V_{cs(d)}} \right )^2}.
\label{errorform}
\end{equation}
From this, the total systematic errors in the determination of
$f^{K(\pi)}_+(0)$ are estimated to be
$$\frac{\Delta f^{K}_+(0)}{f^{K}_+(0)} = 1.4\%,$$
\noindent
and
$$\frac{\Delta f^{\pi}_+(0)}{f^{\pi}_+(0)} = 5.4\%,$$
and are dominated by the uncertainties in $|V_{cs(d)}|$,
see Table \ref{formfactor}.

\begin{table}[h]
\caption{\normalsize The errors in the determination of
the $|f^{K(\pi)}_+(0)|$ form factors with a 20~fb$^{-1}$ $\psi(3770)$ data
at \bes3.}
\begin{center}
\begin{tabular}{c|c|c|c|c|c} \hline \hline
&$\frac{\Delta \mathcal{B}}{2\mathcal{B}}_{\rm stat.}$
&$\frac{\Delta \mathcal{B}}{2\mathcal{B}}_{\rm sys.}$
&$\frac{\Delta \tau_{D^0}}{2\tau_{D^0}}$
&$|V_{cs(d)}|$& Total error \\ \hline
$|f^{K}_+(0)|$  & 0.15\% & 0.5\% & 0.2\% & 1.3\% & 1.4\% \\
$|f^{\pi}_+(0)|$& 0.35\% & 0.5\% & 0.2\% & 5.4\% & 5.4\% \\
 \hline
\end{tabular}
\label{formfactor}
\end{center}
\end{table}

Using the measured branching fractions for $D^0 \to K^-e^+\nu_e$
and $D^0 \to \pi^-e^+\nu_e$, and the predicted form factors
$f^{K}_+(0)=0.66\pm0.04^{+0.01}_{-0.00}$ and
$f^{\pi}_+(0)=0.57\pm0.06^{+0.01}_{-0.00}$ \cite{lqcd} from  lattice~QCD,
we can extract the CKM matrix elements $|V_{cs}|$ and $|V_{cd}|$ from 
Eqs.~\ref{vcs} and \ref{vcd}:
$$|V_{cs}|=1.12\pm0.06\pm0.02,$$
$$|V_{cd}|=0.32\pm0.02\pm0.01,$$
where the first error is statistical and the second systematic.
In the determinations of the CKM elements, the systematic error
can be written as
\begin{equation}
\frac{\Delta V_{cs(d)}}{V_{cs(d)}} =
\sqrt{
\left (\frac{\Delta \mathcal{B}}{2\mathcal{B}} \right )^2 + 
\left (\frac{\Delta \tau_{D^0}}{2\tau_{D^0}} \right )^2 +
\left (\frac{\Delta f^{K(\pi)}_+(0)}{f^{K(\pi)}_+(0)}\right )^2}.
\label{error}
\end{equation}
In the next few years, the form factor uncertainties 
from lattice-QCD are expected to be about 1.5\%, in which
case the total systematic error in the determination of
the CKM matrix element is estimated to be
$$\frac{\Delta V_{cs(d)}}{V_{cs(d)}} = 1.6\%,$$
and dominated by the uncertainty in $f^{K(\pi)}_+(0)$
(see Table~\ref{vcsvcd}).

\begin{table}[h]
\caption{\normalsize The errors of in the determination of
$|V_{cs(d)}|$ with a 20~fb$^{-1}$ $\psi(3770)$ data sample at \bes3.}
\begin{center}
\begin{tabular}{c|c|c|c|c|c} \hline \hline
&$\frac{\Delta \mathcal{B}}{2\mathcal{B}}_{\rm stat.}$
&$\frac{\Delta \mathcal{B}}{2\mathcal{B}}_{\rm sys.}$
&$\frac{\Delta \tau_{D^0}}{2\tau_{D^0}}$
&$|f^{K(\pi)}_+(0)|$ & Total error \\ \hline
$|V_{cs}|$      & 0.15\% & 0.5\% & 0.2\% & 1.5\% & 1.6\% \\
$|V_{cd}|$      & 0.35\% & 0.5\% & 0.2\% & 1.5\% & 1.6\% \\
 \hline
\end{tabular}
\label{vcsvcd}
\end{center}
\end{table}

Combining Eqs.~\ref{vcs}~and~\ref{vcd}, the ratio of the 
$D^0 \to K^-e^+\nu_e$ and $D^0 \to\pi^-e^+\nu_e$ decay widths
is related
to the form factor $|f^{K(\pi)}_+(0)|$ and the CKM matrix element
$|V_{cs(d)}|$ as
\begin{equation}
\frac{\mathcal{B}(D^0 \to \pi^-e^+\nu_e)}{\mathcal{B}(D^0\to K^-e^+\nu_e)} =
\frac{\Gamma(D^0\to \pi^-e^+\nu_e)}{\Gamma(D^0 \to K^-e^+\nu_e)}
=1.967 \times \frac{|f^K_+(0)|^2}{|f^{\pi}_+(0)|^2}\frac{|V_{cd}|^2}{|V_{cs}|^2}.
\label{ratio}
\end{equation}
\noindent
Using the measured branching fractions for $D^0 \to K^-e^+\nu_e$ and $D^0
\to\pi^-e^+\nu_e$, the ratio $\frac{|f^K_+(0)|^2|V_{cd}|^2}{|f^{\pi}_+(0)|^2
|V_{cs}|^2}$ is determined to be $0.0615 \pm 0.0031 \pm 0.0006$, where the
first error is the statistical error, and the second systematic.
In this ratio, some of the systematic uncertainties in the branching fraction
measurements, such as the
electron tracking simulation, particle identification,
photon selection, and the number of the single-tag $\bar D^0$ mesons
cancel completely. In addition, the ratio is
independent of the lifetime $\tau_{D^0}$. The 
remaining systematic error arises
mainly from uncanceled uncertainties in the branching fraction measurements,
including the $K/\pi$ tracking simulation, particle
identification, and fitting the $U_{miss}$ distributions. The
statistical error can be neglected in the case of 20~fb$^{-1}$
$\psi(3770)$ data sample, 
while the total systematic error is estimated to be~1\%.

If the error on the lattice QCD ratio of
the form factors can reach the 1\% level, 
the ratio $|V_{cd}|/|V_{cs}|$ can
be measured with a precision of 1.1\% with a 20~fb$^{-1}$ $\psi(3770)$ 
data sample. The  $|V_{cd}|/|V_{cs}|$ ratio measurement will
be more precise than the individual $|V_{cs}|$ and 
$|V_{cd}|$ determinations.

\subsection[Inclusive Semileptonic Decays]{Inclusive Semileptonic Decays
\footnote{By Hai-Long Ma, Hui-Hui Liu and Gang Rong}}

With a Monte Carlo sample corresponding to an intergrated luminosity
of about 800~pb$^{-1}$  at
3.773 GeV, we developed analysis methods to  measure branching
fractions for inclusive semileptonic $D$-meson decays.

\subsubsection{\bf 1. Overview of studies of  inclusive
semileptonic $D$ decays}
The branching fractions for  inclusive $D$ semileptonic decays
have been measured by many experiments~\cite{argus_ex,cleo2_ex,mark3_ex,
hybr_ex,mark2_ex,cleoc_ex,dlco_ex}. The measured results for
$D^0 \to Xe^+\nu_e$, $D^0 \to X\mu^+\nu_\mu$ and $D^+ \to Xe^+\nu_e$ are
summarized in Tables~\ref{d0ex},~\ref{d0mux}~and~\ref{dpex},
respectively, along with the numbers of signal events associated
with each entry.

The relative errors are 4.1\% for $D^0 \to Xe^+\nu_e$,
12.3\% for $D^0 \to X\mu^+\nu_\mu$, and 11.0\% for $D^+ \to Xe^+\nu_e$.
Recently, the CLEO Collaboration reported a precision measurement
of ${\mathcal B}(D \to Xe^+\nu_e)$~\cite{cleoc_ex}.  However,
there have only been a few ${\mathcal B}(D \to X\mu^+\nu_\mu)$ 
measurements reported during the thirty years that have
passed since the discovery of the $D$ mesons.
Because of advantages provided by the unique capabilities of the 
\bes3 $\mu$ detection system, significant improvements of the
${\mathcal B}(D \to X\mu^+\nu_\mu)$ measurement can be 
expected.

\begin{table}[h]
\caption{\normalsize Summary of measurements of the branching fractions for $D^0
\to Xe^+\nu_e$ from different experiments;
here the superscript $^N$ indicates measurements that are not used in the 
PDG average.}
\begin{center}
\begin{tabular}{ccccc} \hline \hline
Experiments            &Number&Comment                 &${\mathcal B}(D^0\to Xe^+\nu_e)$\\
                    &of Events&                        &(\%)                       \\ \hline
ARGUS~\cite{argus_ex}  &1670&$e^+e^-\sim10$ GeV      &$6.9\pm0.3\pm0.5$          \\
CLEOII~\cite{cleo2_ex}    &4609&$e^+e^-\sim\Upsilon(4S)$&$6.64\pm0.18\pm0.29$       \\
MARKIII~\cite{mark3_ex}   &137 &$e^+e^-$ 3.77 GeV       &$7.5\pm1.1\pm0.4$          \\
HYBR$^{N}$~\cite{hybr_ex} &  &$\pi p$, $pp$ 360, 400 GeV&$15\pm5$                   \\
MARKII$^N$~\cite{mark2_ex}&12  &$e^+e^-$ 3.771 GeV      &$5.5\pm3.7$                \\
CLEOc$^N$~\cite{cleoc_ex}&2246&$e^+e^-$ 3.773 GeV       &$6.46\pm0.17\pm0.13$       \\ 
\hline
PDG average~\cite{Part5_pdg06}        &    &                        &$6.87\pm0.28$              
\\ 
\hline
\end{tabular}
\label{d0ex}
\end{center}
\end{table}

\begin{table}[h]
\caption{\normalsize Summary of measurements of the branching fraction for $D^+
\to Xe^+\nu_e$ from different experiments;
here $^N$ indicates that the measurement is not used in the PDG average.}
\begin{center}
\begin{tabular}{ccccc} \hline \hline
Experiments            &Number&Comment                 &${\mathcal B}(D^+\to Xe^+\nu_e)$\\
                    &of Events&                        &(\%)              \\ \hline
HYBR~\cite{hybr_ex}    &       &$\pi p$, $pp$ 360, 400 GeV&$20^{+9}_{-7}$    \\
MARKIII~\cite{mark3_ex}&158    &$e^+e^-$ 3.77 GeV       &$17.0\pm1.9\pm0.7$\\
MARKII~\cite{mark2_ex}&23     &$e^+e^-$ 3.771 GeV      &$16.8\pm6.4$      \\
DELCO$^N$~\cite{dlco_ex}&       &$e^+e^-$ 3.77 GeV       &$22.0^{+4.4}_{-2.2}$\\
CLEOc$^N$~\cite{cleoc_ex}&8798 &$e^+e^-$ 3.773 GeV      &$16.13\pm0.20\pm0.33$\\ \hline
PDG average~\cite{Part5_pdg06}     &       &                        &$17.2\pm1.9$   \\ 
\hline
\end{tabular}
\label{dpex}
\end{center}
\end{table}

\begin{table}[h]
\caption{\normalsize Summary of measurements of the branching fraction for $D^0
\to X\mu^+\nu_\mu$ from different experiments.}
\begin{center}
\begin{tabular}{ccccc} \hline \hline
Experiments            &Number&Comment          &${\mathcal B}(D^0\to X\mu^+\nu_\mu)$\\
                    &of Events&                 &(\%)\\ \hline
ARGUS~\cite{argus_ex}    &310&$e^+e^-\sim10$ GeV&$6.0\pm0.9\pm1.2$ \\ \hline
PDG average~\cite{Part5_pdg06}        &    &                 &$6.5\pm0.8$ \\ \hline
\end{tabular}
\label{d0mux}
\end{center}
\end{table}

\subsubsection{\bf 2. Analysis method for inclusive semileptonic $D$ decays}
For $D^0 \to Xe^+\nu_e$, only three hadronic decays $\bar D^0 \to K^+\pi^-$,
$\bar D^0 \to K^+\pi^-\pi^0$ and $\bar D^0 \to K^+\pi^-\pi^-\pi^+$ are used
to reconstruct single-tag $\bar D^0$ mesons, since the mode $\bar D^0 \to K^0
\pi^+\pi^-$ does not determine the charm of the $D^0$ meson.
For $D^+ \to Xe^+\nu_e$, only the hadronic decay $D^- \to K^+\pi^-\pi^-$ is
used to reconstruct single-tag $D^-$ mesons, since
this mode has a large branching fractiona and low background.

The resulting  beam-constrained mass distributions for the
$Kn\pi(n=1,2,3)$ single-tag $\bar D$ modes are
shown in Fig.~\ref{bes3d4tags}. A maximum likelihood fit to the mass spectrum with
a Gaussian function representing the $\bar D$ signal and a special 
background function~\cite{besd0semi,besdpsemi} yields the number of 
single-tag $\bar D$ mesons for each mode.

\begin{figure}[h]
\begin{center}
  \includegraphics[width=12cm,height=10cm]
{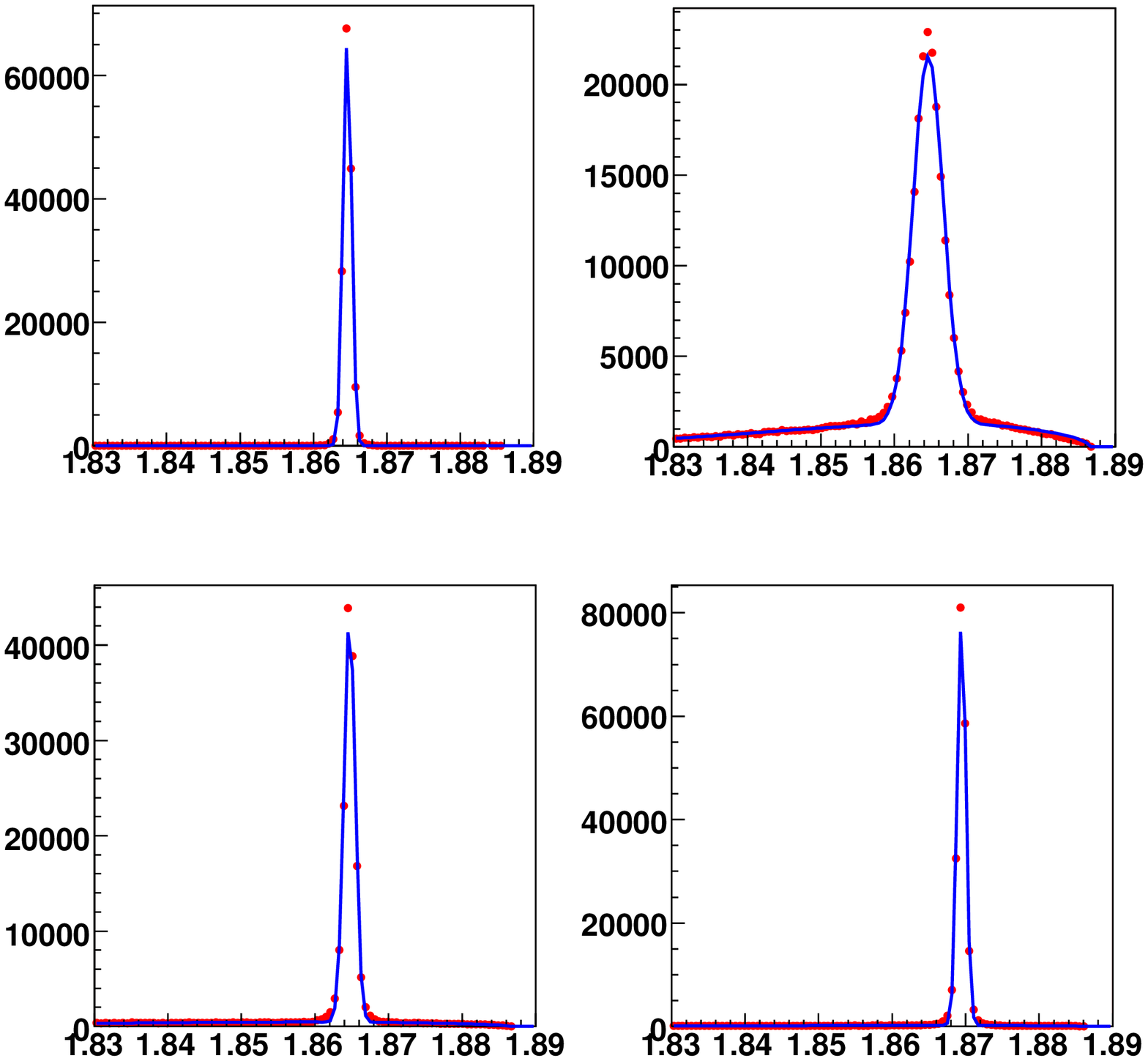}
      \put(-300,250){\large (a)}
      \put(-130,250){\large (b)}
      \put(-300,107){\large (c)}
      \put(-130,107){\large (d)}
      \put(-265,-15){\large Beam Constrained Mass (GeV/$c^2$)}
      \put(-365,70){\rotatebox{90}{\large Events/(0.6 MeV/$c^2$)}}
  \caption{\normalsize The beam constrained mass
distributions for the $Kn\pi$ 
single-tag $\bar D$ modes:
(a) $\bar D^0 \to K^+\pi^-$,
(b) $\bar D^0 \to K^+\pi^-\pi ^0$,
(c) $\bar D^0 \to K^+\pi^-\pi^-\pi^+$ and
(d) $D^- \to K^+\pi^-\pi^-$.}
  \label{bes3d4tags}
\end{center}
\end{figure}

In the system recoiling against the single-tag $\bar D$ mesons, electrons,
kaons and pions are selected from the surviving tracks.
Electrons, kaons and pions are classified into right-sign and wrong-sign
samples according to their charge-correlation relative to the flavor tag.
Since the wrong-sign electrons are
all from the background, they can be used to estimate the
background level. The wrong-sign unfolded yield of electrons accounts
for the the charge-symmetric background, which is mostly produced by
$\pi^0\to\gamma e^+e^-$ decays and $\gamma$ conversions.

Figures~\ref{bes3dexrec}(a) to (d') show the fitted beam constrained mass
distributions for the $Kn\pi$ single-tag
combinations for events with one right-sign ((a)-(d))
or wrong-sign ((a')-(d')) electron
observed in the system recoiling against the tag $\bar D$ mesons.
Because the detection efficiency for a particle and the probability of
misidentifying a particle depends on momentum, 
we divide the momentum into $n$ bins.
The yield $N_e^{{\rm obs},i}$ of electrons in the $i^{\rm th}$ momentum bin is
obtained by fitting to the corresponding mass spectrum.
Similar analyses give the yields $N_K^{{\rm obs},i}$ and
$N_{\pi}^{{\rm obs},i}$ of kaons and pions in each momentum range.

\begin{figure}[h]
\begin{center}
  \includegraphics[width=12cm,height=10cm]
{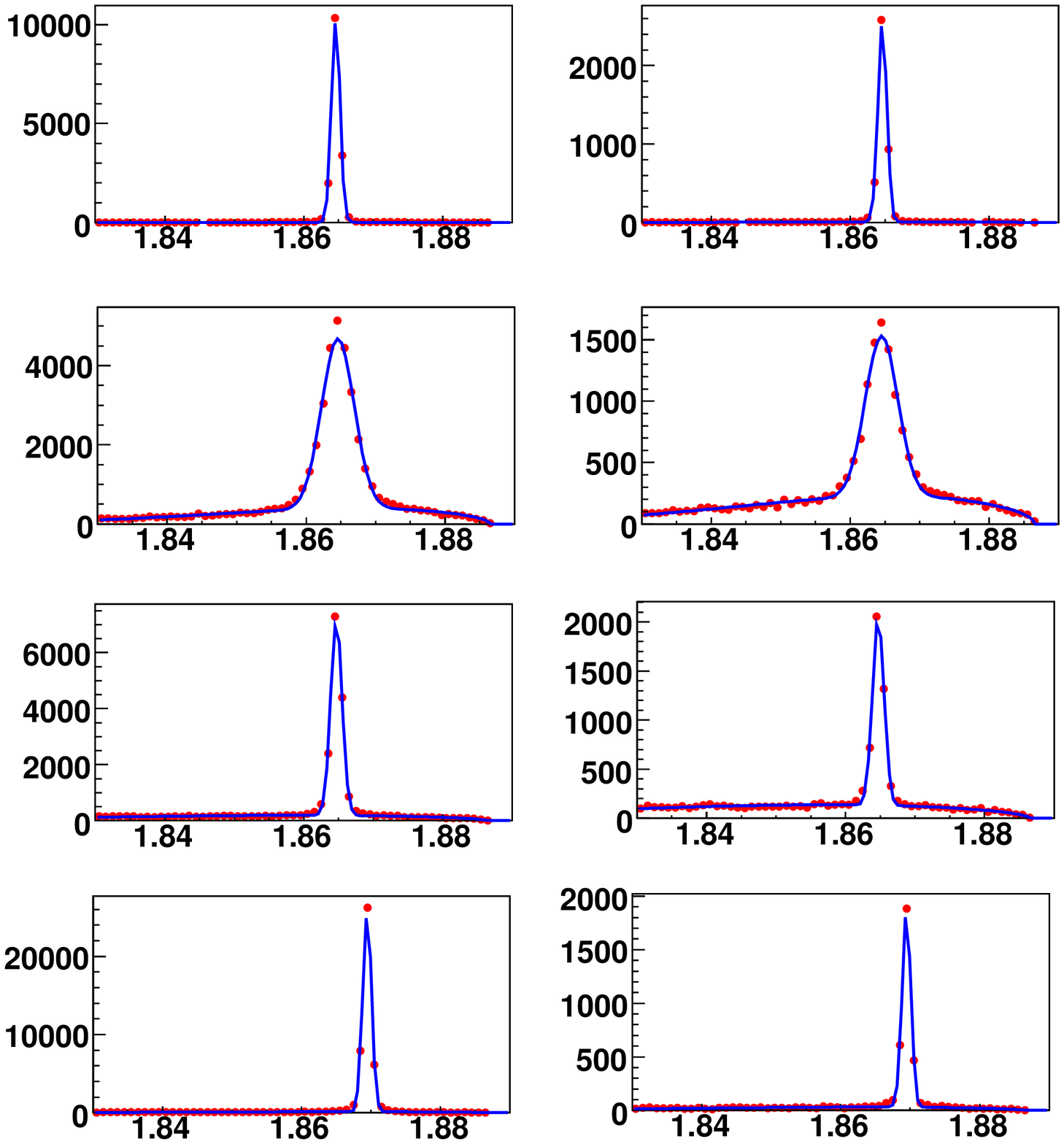}
      \put(-300,260){\large (a)}
      \put(-130,260){\large (a')}
      \put(-300,190){\large (b)}
      \put(-130,190){\large (b')}
      \put(-300,120){\large (c)}
      \put(-130,120){\large (c')}
      \put(-300,50) {\large (d)}
      \put(-130,50) {\large (d')}
      \put(-265,-15){\large Beam Constrained Mass (GeV/$c^2$)}
      \put(-365,70){\rotatebox{90}{\large Events/(1 MeV/$c^2$)}}
  \caption{\normalsize The fitted beam constrained mass
distributions for $Kn\pi$ tag $D$ mesons with
one electron  observed in the recoil system:
(a), (a') $\bar D^0 \to K^+\pi^-$;
(b), (b') $\bar D^0 \to K^+\pi^-\pi^0$;
(c), (c') $\bar D^0 \to K^+\pi^-\pi^-\pi^+$ and
(d), (d') $D^- \to K^+\pi^-\pi^-$;
here (a), (b), (c) and (d) are the right-sign events,
while (a'), (b'), (c') and (d') are the wrong-sign events.}
  \label{bes3dexrec}
\end{center}
\end{figure}

In the $i^{\rm th}$ momentum bin, the true number of electrons with the right-sign
and wrong-sign samples are obtained through an unfolding procedure, using
the matrix:
\[
\left(
\begin{array}{c}
N_e^{{\rm obs},i} \\
N_K^{{\rm obs},i} \\
N_{\pi}^{{\rm obs},i} \\
\end{array} \right )
=
\left(
\begin{array}{lcr}
\epsilon^i_e & f^i_{\pi \rightarrow e} & f^i_{K \rightarrow e} \\
f^i_{e \rightarrow K} & \epsilon^i_K & f^i_{\pi \rightarrow K}\\
f^i_{e \rightarrow \pi} & f^i_{K \rightarrow \pi} & \epsilon^i_{\pi} \\
\end{array} \right )
\left(
\begin{array}{c}
N_e^{{\rm real},i} \\
N_K^{{\rm real},i} \\
N_{\pi}^{{\rm real},i} \\
\end{array} \right ),
\]
where $N_e^{{\rm obs},i}$, $N_K^{{\rm obs},i}$, $N_{\pi}^{{\rm obs},i}$ represent
the numbers of the electrons, kaons and pions observed in the system recoiling
against the tag $\bar D$ meson; $N_e^{{\rm real},i}$, $N_K^{{\rm
real},i}$, $N_{\pi}^{{\rm real},i}$ denote their corresponding true numbers; the
efficiencies $\epsilon^i_e$, $\epsilon^i_K$, $\epsilon^i_{\pi}$ account for
the losses due to track finding, track selection criteria, and particle
identification; the off-diagonal element $f^i_{a \to b}$ is the
probability of misidentifying particle type $a$ as $b$.
In analyses of real data, an electron sample selected from
the radiative Bhabha scattering events, and  kaon and
pion samples selected from $J/\psi \to \phi K^+K^-$ and $J/\psi
\to \omega\pi^+\pi^-$ events can be used to measure  $f^i_{a \to b}$. 
In the simulation analysis,
the efficiencies and the misidentification probabilities determined by
Monte Carlo simulation for each kind of track is used. A detailed analysis of these
particle samples give $\epsilon^i_e$, $\epsilon^i_K$, $\epsilon^i_{\pi}$ and
$f^i_{a \to b}$ for each momentum bin.

Subtracting the number $N_e^{{\rm real},i}(R)$ of right-sign electrons by
the number $N_e^{{\rm real},i}(W)$ of wrong-sign electrons, we obtain the
net number of electrons in $i^{\rm th}$ momentum range,
$$N_e^{{\rm real},i}(net)=N_e^{{\rm real},i}(R)-N_e^{{\rm real},i}(W).$$
Adding the net number of electrons in each momentum range yields the total
net number $N^{\rm net}_{D \to X\ell^+\nu_\ell}$ of electrons,
$$N^{\rm net}_{D \to X\ell^+\nu_\ell}=\sum_i^n N_e^{{\rm real},i}(net).$$

\subsubsection{\bf 3. Branching fractions}
The branching fraction for $D \to X\ell^+\nu_\ell$ is
determined from the relation
\begin{equation}
{\mathcal B}(D \to X\ell^+\nu_\ell)=\frac{N^{\rm net}_{D \to X\ell^+\nu_\ell}}
{N_{\bar D_{\rm tag}}\times \epsilon_{D \to X\ell^+\nu_\ell}},
\label{brinsemi}
\end{equation}
where $N^{\rm net}_{D \to X\ell^+\nu_\ell}$ is the net number of
$D \to X\ell^+\nu_\ell$ events, $N_{\bar D_{\rm tag}}$ is the total number of the
single-tag $D^-$ or $\bar D^0$ mesons, and $\epsilon_{D \to X\ell^+\nu_\ell}$
is the detection efficiency.  The detection
efficiency is determined by Monte Carlo simulations for each single-tag $\bar D$
mode, where the recoil $D \to X\ell^+\nu_\ell$ includes all exclusive semileptonic
channels.  A MC analysis procedure similar to that described above gives the detection
efficiency.
Inserting these numbers into Eqn.~\ref{brinsemi}, we can obtain the branching
fraction for the inclusive semileptonic decay.

In the $D \to X\ell^+\nu_\ell$ branching fraction measurement, 
the dominant systematic errors are due to uncertainties in the electron
tracking simulation and particle identification, the efficiencies and
misidentification probabilities, and the number of
the single-tag $\bar D$ mesons. As previously discussed,
in the case of a 20 fb$^{-1}$ $\psi(3770)$ data sample, 
the statistical error can be neglected relative to
the systematic error, which is expected to be at the 1.0\% level.

\subsubsection{\bf 4. Determination of the ratio $\frac{\Gamma^{\rm SL}_{D^+}}
{\Gamma^{\rm SL}_{D^0}}$}
Using the measured branching fractions for the inclusive $D$ semileptonic
decays ${\mathcal B}(D^+ \to X\ell^+\nu_\ell)$ and ${\mathcal B}(D^0 \to
X\ell^+\nu_\ell)$ in conjunction with 
the well measured lifetimes of the $D^+$ and $D^0$
mesons, $\tau_{D^+}=(1040\pm7)\times 10^{-15}$ s and $\tau_{D^0}=(410.1
\pm1.5)\times 10^{-15}$ s, the ratio of the partial widths for the
two inclusive semileptonic decays can be determined:
\begin{equation}
R=\frac{\Gamma^{\rm SL}_{D^+}}{\Gamma^{\rm SL}_{D^0}}
 =\frac{{\mathcal B}(D^+ \to X\ell^+\nu_\ell)/\tau_{D^+}}
       {{\mathcal B}(D^0 \to X\ell^+\nu_\ell)/\tau_{D^0}}.
\end{equation}
In the determination of this ratio with a 20~fb$^{-1}$ 
$\psi(3770)$ data sample, the statistical errors
on ${\mathcal B}(D \to X\ell^+\nu_\ell)$ 
can be neglected. The systematic error can be written as
\begin{equation}
\frac{\Delta R}{R}=
\sqrt{
\left (\frac{\Delta \tau_{D^+}}{\tau_{D^+}} \right )^2 +
\left (\frac{\Delta \tau_{D^0}}{\tau_{D^0}} \right )^2 +
\left (\frac{\Delta \mathcal{B}(D^+ \to X\ell^+\nu_\ell)}{\mathcal{B}(D^+
\to X\ell^+\nu_\ell)} \right )^2 +
\left (\frac{\Delta \mathcal{B}(D^0 \to X\ell^+\nu_\ell)}{\mathcal{B}(D^0
\to X\ell^+\nu_\ell)} \right )^2},
\end{equation}
where the uncertainties in the lifetimes of $D$ mesons are $\Delta
\tau_{D^+}/\tau_{D^+}=0.7\%$ and $\Delta \tau_{D^0}/\tau_{D^0}=0.3\%$,
the uncanceled uncertainty in each measured branching fraction is
conservatively taken to be $\Delta \mathcal{B}(D \to X\ell^+\nu_\ell)
/\mathcal{B}(D \to X\ell^+\nu_\ell)=1.0\%$. The resulting estimate
of the total systematic error is about 1.6\%.


\def\ie{{i.e.}}

\def\eg{{e.g.}}

\def\ea{{\em et al.}}

%


%

\def\rhobar {\ensuremath{\overline \rho}\xspace}

\def\etabar {\ensuremath{\overline \eta}\xspace}

\newcommand\rhoeta{(\rhobar,\etabar)}

\newcommand\ckmfitter{{CKMfitter}\xspace}

\newcommand\rfit{{\em R}fit}

\newcommand\tho{{\rm theo}}

\newcommand\experi{{\rm exp}}

\newcommand\modckm{{\rm mod}}

\newcommand\QCD{{\rm QCD}}

\newcommand\xexp{x_{\experi}}

\newcommand\xthe{x_{\tho}}

\newcommand\ymod{y_{\modckm}}

\newcommand\ythe{y_{\tho}}

\newcommand\yQCD{y_{\QCD}}

\newcommand\Lik{{\cal L}}

\newcommand\Likexp{{\cal L}_{\experi}}

\newcommand\Likthe{{\cal L}_{\tho}}

\newcommand\mini{{\rm min}}

\newcommand\ChiMinGlob{\chi^2_{\mini ;\ymod}}

\newcommand\Mu{\mu}

\section[Impact on CKM Measurements]{Impact on CKM
Measurements\footnote{J. Charles, S. Descotes-Genon, H.
Lacker, H. B. Li, L. Roos, S. T$^\prime$Jampens, Z. Z. Xing}}
 
\subsection{The Role of Charm in Precision CKM Physics}

In the Standard Model (SM), quark-flavor mixing
is described by the $3\times 3$ Cabibbo-Kobayashi-Maskawa (CKM)
matrix $V$~\cite{part5:CKM},
\begin{equation}
V = \left ( \begin{matrix} V^{}_{ud} & V^{}_{us} & V^{}_{ub} \cr
V^{}_{cd} & V^{}_{cs} & V^{}_{cb} \cr V^{}_{td} & V^{}_{ts} &
V^{}_{tb} \end{matrix} \right ) \; .
\end{equation}
Unitarity is the only, albeit powerful, constraint
on $V$. Without loss of generality, $V$ can be
parameterized in terms of three mixing angles and one phase
\cite{Part5_pdg06}:
\begin{equation}
V = \left( \begin{matrix} c^{}_{12}c^{}_{13} & s^{}_{12}c ^{}_{13} &
s^{}_{13} e^{-i\delta} \cr -s^{}_{12}c^{}_{23}
-c^{}_{12}s^{}_{23}s^{}_{13} e^{i\delta} & c^{}_{12}c^{}_{23}
-s^{}_{12}s^{}_{23}s^{}_{13} e^{i\delta} & s^{}_{23}c^{}_{13} \cr
s^{}_{12}s^{}_{23} -c^{}_{12}c^{}_{23}s^{}_{13} e^{i\delta} &
-c^{}_{12}s^{}_{23} -s^{}_{12}c^{}_{23}s^{}_{13} e^{i\delta} &
c^{}_{23}c^{}_{13} \end{matrix} \right) \; ,
\end{equation}
where $c^{}_{ij} \equiv \cos\theta_{ij}$ and $s^{}_{ij} \equiv
\sin\theta_{ij}$ (for $ij=12,23$ and $13$). The irremovable phase
$\delta$ is the unique source of CP violation in quark
flavor-changing processes within the SM.

The goal of precision CKM physics is threefold: (a) to
measure the mixing and CP-violating parameters of $V$ as
accurately as possible; (b) to test the self-consistency of the
CKM picture for quark mixing and CP violation; (c) to search for
possible new physics beyond the CKM mechanism. 
It is, therefore, important to measure very precisely the various
entries of the CKM matrix. A test of the self-consistency of the 
CKM picture is provided by unitarity conditions.

Charm physics may impact the study of the CKM matrix in various ways:
\begin{enumerate}
\item  The current status of the first row (from direct measurements
only) is:
 $|V^{}_{ud}|^2 + |V^{}_{us}|^2 +
|V^{}_{ub}|^2 = 0.9992 \pm 0.0011$ \cite{Part5_pdg06}; {\it i.e.}, 
unitarity
holds at the $10^{-3}$ level. 
As for the second row, we have (from direct measurements
only):
$|V^{}_{cd}|^2 + |V^{}_{cs}|^2 + |V^{}_{cb}|^2 = 0.968 \pm
0.181$~\cite{Part5_pdg06},
where the error is dominated by the theoretical
uncertainty in $|V^{}_{cs}|$. 

\item  The unitarity of $V$ implies that $|V^{}_{us}| -
|V^{}_{cd}| =O(\lambda^5)$~\cite{part5:Xing96}, where
$\lambda \equiv \sin\theta^{}_{12} \approx 0.22$ is the
well-known Wolfenstein parameter \cite{part5:Wolfenstein83}. 
At present, the best direct determination of $|V^{}_{cd}|$ is based on
deep inelastic scattering of neutrinos and 
antineutrinos \cite{Part5_pdg06}: $|V^{}_{cd}|
= 0.230 \pm 0.011$, which has an error that is larger than that
for $|V^{}_{us}|$ extracted from kaon semileptonic decays
($|V^{}_{us}| = 0.2257 \pm 0.0021$ \cite{Part5_pdg06}). 
Recent results
on semileptonic $K_{\ell 3}$ decays from NA38 are likely to
change the value of $|V_{us}|$~\cite{part5:NA48,part5:Vus}, 
which will make more accurate determinations of
$|V^{}_{cd}|$ and $|V^{}_{cs}|$ all the more interesting. 
Inconsistencies in the upper-left square of the CKM
matrix
would imply the existence of new physics.

\item  If the elements in the first and second rows of $V$ are all
determined to a sufficiently high degree of precision, it is then
possible to establish a ``charming" unitarity triangle based on
the orthogonality condition $V^{}_{ud}V_{cd}^* + V^{}_{us}V_{cs}^*
+ V^{}_{ub}V_{cb}^* =0$. That will be another (CP-conserving) way
to cross-check the CKM mechanism for quark flavor mixing.

\item Reliable information on
hadronic $D$ and $D^{}_s$ decays (such as $D^0\rightarrow K\pi$
and $D^{}_s\rightarrow \phi\pi$) can be used to normalize
results in $B$ physics. It can also improve the understanding of
$B$-decays into final states containing a charmed meson. For instance,
comparing $B^-\to D^0K^-$ and $B^-\to \bar{D}^0K^-$
where $D^0$ and $\bar{D}^0$ both decay into $K_s\pi^+\pi^-$ provides 
a determination of $\gamma$ that relies on models for the $D$ 
decay~\cite{part5:Gronau07}. The
uncertainty on $\gamma$ could be reduced to $3^\circ$ with
\bes3 measurements of the relevant Dalitz plots in conjunction with 
measurements at $B$-machines.

\item Ratios of quantities related to charm and beauty mesons
can be determined with a fairly good accuracy, in particular through
lattice simulations. This allows for an interplay of charm and bottom 
physics in the era of high-precision heavy flavor physics, as exemplified 
in the next section.
\end{enumerate}

\subsection{Impact of BESIII measurements}

\subsubsection{The \ckmfitter package} \label{sec:ckmfitter}

The \ckmfitter package is a comprehensive tool for CKM matrix analysis. It allows the user to:

\begin{itemize}
\item quantify the agreement between theory (Standard Model or beyond) and
experimental measurements;
\item obtain the best estimate of a given set of theoretical parameters
within a given theory (\eg, CKM parameters in the Standard Model).
\end{itemize}
\noindent
In either case, a  major issue for the fit of the CKM matrix is how to deal
with the theoretical uncertainties. For most of the measured observables that
enter into the fit, these uncertainties are non-negligible, or 
even dominate over the experimental ones. 
While experimental uncertainties can usually be considered as Gaussian, the
meaning of the theoretical uncertainties is often not well defined (an
exception consists in unquenched lattice QCD predictions).
In the \ckmfitter package, this issue is addressed by allowing  theoretical
parameters to vary only within the range defined by their uncertainties.

\subsubsection{The statistical approach}

The \ckmfitter package is based on the frequentist approach \rfit,
described in \cite{part5:ckmfitter1} and \cite{part5:ckmfitter2} and recalled here.
The likelihood function $\Lik$ is defined as the product of two components
$\Likexp$ and $\Likthe$: $$
\label{eq:likFunction}
   \Lik(\ymod) = \Likexp(\xexp-\xthe(\ymod))\cdot \Likthe(\yQCD)~.
$$
where $\xexp$ is a set of $N_{exp}$ experimental measurements, and $\xthe$
the $N_{exp}$  corresponding theoretical predictions. $\xthe$ depends on
$N_{mod}$ parameters $\ymod$, which are either free parameters of the
theory (\eg, the CKM Matrix parameters) or,  approximately known QCD
related quantities (denoted $\yQCD$). Each individual measurement entering
into the $\Likexp$ component is, in general, considered as
Gaussian,\footnote{In the case of a non-Gaussian experimental errors,
the exact description of the associated likelihood is directly used in the
fit.} and correlations between variables, if known, are taken into 
account.
The experimental systematics are added in quadrature to the statistical
errors, whereas the theoretical systematics are dealt with through the
theoretical component $\Likthe$. The uncertainties on the theoretical
parameters $\yQCD$ define the allowed range of values for each parameter.
In other words, each individual likelihood component  $\Likthe(\yQCD(i))$
is one within the allowed range and zero outside.
The fit is performed on all the parameters $\ymod$ by minimizing 
$\chi^2(\ymod)\equiv-2\ln(\Lik(\ymod))$. The minimum value is denoted $\ChiMinGlob$. One quantifies the agreement between theory and data by the probability to observe $\chi^2$ values greater or equal to $\ChiMinGlob$.
In the present study, we focus on a subset of the $\ymod$ parameters, namely $\rhoeta$, defined as:
$$\rhobar + i\etabar \equiv-\frac{V^{}_{ud}V^{*}_{ub}}{V^{}_{cd}V^{*}_{cb}}.$$ 

Let us denote $a=\rhoeta$ and $\mu$ the remaining parameters, such that $\ymod=(a,\mu)$.
The minimum  value  $\chi^2_{\mini ;\,\Mu}(a)$ is computed for a set of fixed
value $a$, while varying $\mu$. In the following graphics, we represent
the Confidence Level obtained from
$\chi2$ difference $\Delta\chi^2(a)=\chi^2_{\mini ;\,\Mu}(a)-\ChiMinGlob$.

\subsubsection{The Global CKM Fit}~\label{sec:globalfit}

\begin{figure}[htb]

\hspace{-0.7cm}

\includegraphics[scale=0.8]{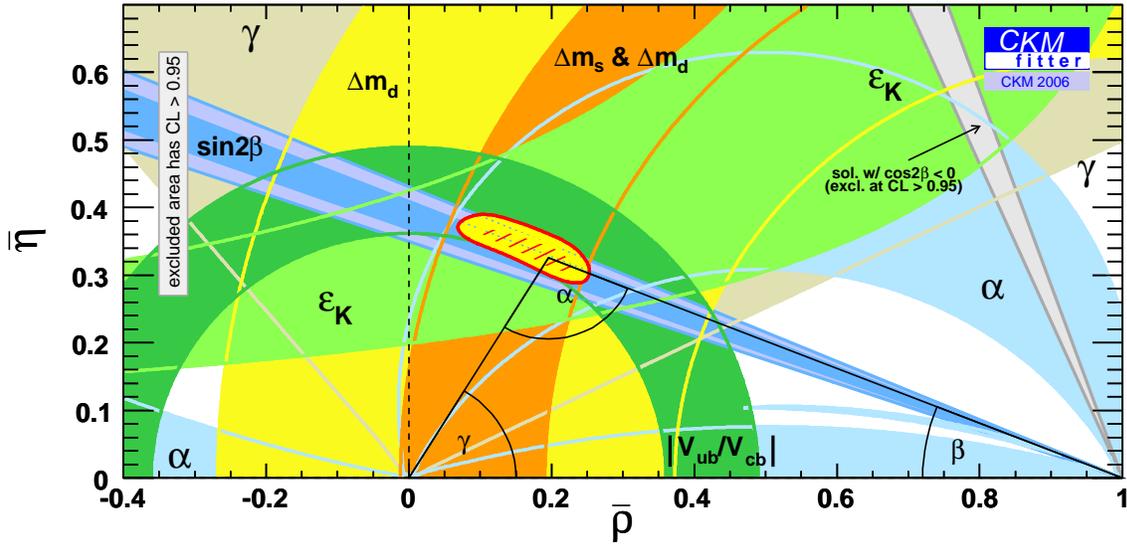}

\vspace{-0.5cm}

\caption{Individual constraints and the global CKM fit on the
($\overline{\rho},\overline{\eta}$) plane (as of Winter 2007). The shaded
areas have 95\% CL.}\label{ckm06:fig_global}
\end{figure}

The inputs to the global fit are observables where the theoretical
uncertainties are quantitatively under control in order to test the
Standard Model: $|V_{ud}|$, $|V_{us}|$, $|V_{cb}|$ (to fix the length
scale of the UT and the constraints on $A$ and $\lambda$), and the
following quantities that are particularly sensitive to
$(\overline{\rho},\overline{\eta})$, {\it i.e.}, $|V_{ub}|$, ${\cal 
B}(B\to \tau
\nu)$, $\varepsilon_K$, $\Delta m_d$, $\Delta m_d \, \& \, \Delta m_s$,
$\sin2\beta$, $\cos2\beta$, $\alpha$ and $\gamma$. $\lambda$ is determined
from $|V_{ud}|$ (superallowed nuclear transitions) and $|V_{us}|$
(semileptonic kaon decays) to a combined precision of 0.5\%. $A$ is
determined from $|V_{cb}|$ (inclusive and exclusive semileptonic $B$
decays) to a combined precision of about 1.7\%. While $\lambda$ and $A$ are
well-known, the parameters $\overline{\rho}$ and $\overline{\eta}$ are
much more uncertain (about 20\% for $\overline{\rho}$ and 7\% for
$\overline{\eta}$). 

The main goal of CP-violation experiments is to
over-constrain these parameters by measuring both the three angles and the
sides of the $B_d$ unitarity triangle and, possibly, to find
inconsistencies suggesting the existence of physics beyond the SM. What is
important is, thus, the capability of the CKM mechanism to describe flavor
dynamics of many constraints from vastly different scales and not the
measurement of the CKM phase's value \textit{per se}.
The unitarity triangle checks for the consistency of the information
obtained from mixing with the information obtained from decay.  This is
only one of many tests of the CKM matrix. The strategy of placing all CKM
constraints on the ($\overline{\rho},\overline{\eta}$) plane is a
convenient way to compare the overconstraining measurements and a way to
search for New Physics by looking for inconsistencies. The 95\% CL of the
individual constraints and the result of the global fit are displayed in
Fig.~\ref{ckm06:fig_global}. For the detailed inputs, see
Ref.~\cite{part5:ckmfitter2}. 

The $\gamma$ and $\alpha$ measurements together with
$|V_{ub}/V_{cb}|$ determine \rhobar and \etabar from (effectively)
tree-level processes, independently of mixing, and agree with the other
loop-induced constraints. Present CKM fits provide a consistency check of
the Standard Model hypothesis. The rather large allowed regions provided
by each constraint individually (other than the angles) are mainly due to
theoretical uncertainties from lattice QCD. 

\subsubsection{The 2012 Prospective scenario}

After the many  measurements done by the $B$-factory and 
Tevatron
experiments, the near future will be devoted to precision flavor physics in
order to uncover physics beyond the Standard Model and to probe the flavour
structure of new physics  
that may be discovered elsewhere. A crucial
ingredient for precision measurements is to enhance or validate the
theoretical control over QCD to reach the per cent level accuracy. The most
promising tool is the simulation of QCD on the lattice, which needs to be
confronted to precision measurements. The charm sector offers the possibility
to validate forthcoming lattice QCD calculations at the few percent
level. These can then be used to make precise measurements of CKM elements,
$|V_{cd}|$, $|V_{cs}|$, $|V_{ub}|$, $|V_{cb}|$ and $|V_{ts}|$.

Charm physics opens an interesting window on the strong and weak sectors of the Standard Model. 
At the theoretical level, lattice simulations of QCD are a particularly relevant (and almost unique) tool 
to tackle charm dynamics, since the natural scales of these simulations lie between the strange and 
the charm quark masses. Therefore, lattice simulations can simulate almost directly charm dynamics, whereas they have to 
rely on extrapolations in the case of $b$-physics. In addition, they can help to reduce uncertainties
 in the computation of $b$-physics quantities~\cite{part5:Asner}, since many
long-distance effects are similar 
in $B$- and $D$-observables, 
as can be checked explicitly using effective field theories 
({\it e.g.}, Heavy Meson Chiral Perturbation Theory). 
In particular, there are ratios of quantities that are useful
for precision physics in the charm and beauty sectors: these ratios are
 experimentally more accurately determined than absolute values, their
estimate on the lattice does not suffer from large uncertainties related
to the determination of an absolute scale, and they are less affected by
 the systematics from the extrapolation in quark masses.
We, however, stress that the study of charm physics does not alleviate all of the difficulties encountered 
in lattice simulations ({\it e.g.}, finite volume effects, renormalisation issues).

For the prospective part, it is difficult to anticipate the progress in
lattice simulations over the next five 
years~\cite{part5:lattice}. 
For the present exercise, we  take very rough estimates for theory
uncertainties in 2012, and assume that the following 
accuracy can be obtained from $B$-machines (super-$B$ factories and LHCb)
\begin{equation} \label{eq:accur}
\sin(2\beta) \to 0.011 \qquad
\alpha \to 5^\circ \qquad 
\gamma \to 3^\circ \qquad
|V_{ub}| \to 4\% \qquad |V_{cb}| \to 1.5\% .
\end{equation}
$|V_{ub}|$ can be determined through either inclusive or exclusive processes.
For the exclusive
determination from $Br(B \to \pi \ell \nu)$, we take an uncertainty of 4\% on
the experimentally measured branching ratio 
and 4\% for the lattice determination of $F_{B \to \pi}$.
We assume that the inclusive extraction from $Br(B \to X_u \ell \nu)$,
will provide a second determination of $|V_{ub}|$ at 5\%. The error on 
$|V_{ub}|$ in Eq.~(\ref{eq:accur}) corresponds to an average
of the two determinations (inclusive and exclusive).
For the other relevant observables, the projected situation 
in 2012 for lattice and experiment
is summarised in  Table~\ref{part5:tab:ckm1}.

\begin{table}[htbp]
\begin{center}
\caption{The projected precision for Lattice and experiments in 2012.}
\begin{tabular}{|c|c||c|c||c|c||}
\hline
Observable & CKM                 &  Had. param & Lattice error 
& Exp. error \\
\hline
$Br(B \to \tau \nu)$ & $|V_{ub}|$ & $f_{B}$    &  4\%   &
10\% \\
$\Delta m_s$         & $|V_{ts}V_{tb}|$ & $f_{Bs}\sqrt{B_{Bs}}$    &  3\%  &
0.7\% \\
$\frac{\Delta m_s}{\Delta m_d}$ & $\left|\frac{V_{ts}}{V_{td}}\right|$   
              & $\xi$       & 1.5\%   & For $\Delta m_d$ : 0.8\%\\
$\varepsilon_K$ & $V_{qs}V_{qd}^*$ & $B_K$ & 2\% & 0.4\%\\
\hline
\end{tabular}
\label{part5:tab:ckm1}
\end{center}
\end{table}

For the problem at hand, we propose to represent the combination of the
CKM constraints in the plane that is relevant to the $D$ meson Unitarity
Triangle (DUT). In analogy with the exact and rephasing-invariant
expression of $(\overline\rho,\overline\eta)$~\cite{part5:ckmfitter2} we define
the coordinates of the apex of the DUT
\begin{equation}
\overline\rho_D +i\,\overline\eta_D \equiv
-\frac{V_{ud}V_{cd}^*}{V_{us}V_{cs}^*}
\end{equation}
where $\overline\rho_D=1+\mathcal{O}(\lambda^4)$ and
$\overline\eta_D=\mathcal{O}(\lambda^4)$. 
One can see that this triangle has two sides with length very close to 1
and a small side of order $O(\lambda^4)$, with angles 
$\alpha_D=-\gamma$, $\beta_D=\gamma +\pi +O(\lambda^4)$ and
$\gamma_D=O(\lambda^4)$.
The individual constraints as well as the combination from the usual
observables are shown for 2012 in the $(\overline\rho_D,\overline\eta_D)$
plane in Fig.~\ref{DUT2012}.
\begin{figure}[htb]

\hspace{-0.7cm}

\includegraphics[scale=0.8]{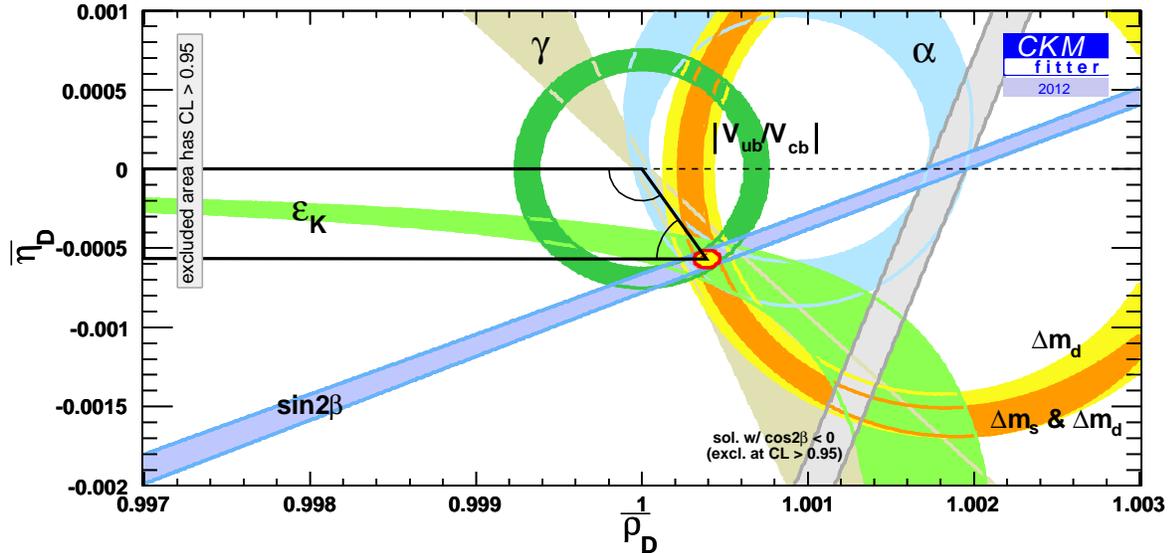}

\vspace{-0.5cm}

\caption{Individual constraints and the global CKM fit on the
($\overline{\rho}_D,\overline{\eta}_D$) plane (as of our prospective in
2012). The shaded areas have 95\% CL. Only a part of the $D$ Unitarity 
Triangle is visible 
(in black solid lines) : the two apices associated with large angles are shown, whereas
the missing apex is situated at the origin, far away on the left.}\label{DUT2012}
\end{figure}

For this prospective exercise, we  only consider the impact of \bes3
measurements for charm-related CKM matrix elements, which is less
difficult to quantify
than the more indirect one for $B$-physics quantities. 
For the charm-related CKM matrix elements, several observables can be of
interest to determine $|V^{}_{cd}|$ and $|V^{}_{cs}|$ :
\begin{itemize}
\item Precision measurements of leptonic $D$ and $D^{}_s$
decays can help determine the CKM matrix elements 
$|V^{}_{cd}|$ and $|V^{}_{cs}|$ provided that $f^{}_D$ and
$f^{}_{D^{}_s}$ are theoretically known to a good degree of accuracy.

\item Precision measurements of semileptonic $D$ and
$D^{}_s$ decays can help determine the CKM matrix elements
$|V^{}_{cd}|$ and $|V^{}_{cs}|$ to a good degree of accuracy, 
provided the relevant form factors are well
predicted from lattice QCD and other theoretical models. 
\end{itemize}

Assuming a 20~${\rm fb}^{-1}$ data sample at the 
$\psi(3770)$, the uncertainties for \bes3
experimental measurements can be estimated
by considering both statistical and systematic errors (tracking, PID,
and neutral particle reconstruction). 
We quote here the corresponding  systematic uncertainties
and perform a projection for the lattice errors in 2012. The expected
situation is summarized in  Table~\ref{part5:tab:ckm2}.
\begin{table}[htbp]
\begin{center}
\caption{Summary of lattice and experimental sensitivity in 2012.}
\begin{tabular}{|c|c||c|c||c|c|}
\hline
Observable & CKM                 &  Had. param & Lattice error & Exp. measure
& Exp. error \\
\hline
$Br(D  \to \ell \nu)$  & $|V_{cd}|$            & $f_D$        &  2\%   & $f_D|V_{cd}|$  & 1.1\%  \\
$Br(D_s \to \ell \nu)$ & $|V_{cs}|$            & $f_{Ds}$     &  1.5\% & $f_{Ds}|V_{cs}|$ & 0.7\%  \\
$\frac{Br(D_s \to \ell \nu)}{Br(D  \to \ell \nu)}$ &
  $\left|\frac{V_{cs}}{V_{cd}}\right|$     & $\frac{f_{Ds}}{f_D}$ &  1\% & $\left|\frac{V_{cs}f_{Ds}}{V_{cd}f_D}\right|$ & 0.8\% \\
$d\Gamma(D^0\to \pi^-)/ds$    & $|V_{cd}|$ & $F_{D\to \pi}(0)$     & 4\%  & $|V_{cd}|F_{D\to \pi}(0)$ & 0.6\% \\ 
$d\Gamma(D^0\to K^-)/ds$      & $|V_{cs}|$ & $F_{D\to K}(0)$       & 3\%  & $|V_{cs}|F_{D\to K}(0)$ & 0.5\%\\
$d\Gamma(D_s\to K)/ds$    & $|V_{cd}|$ & $F_{D_s\to K}(0)$     & 2\%  &
$|V_{cd}|F_{D_s\to K}(0)$& 1.2\% \\
$d\Gamma(D_s\to \phi)/ds$ & $|V_{cs}|$ & $F_{D_s\to \phi}(0)$  & 1\%  &
$|V_{cs}|F_{D_s\to \phi}(0)$& 0.8\% \\
\hline
\end{tabular}
\label{part5:tab:ckm2}
\end{center}
\end{table}

The most striking outcome of \bes3 will be accurate measurements of
quantities related to $|V_{cd}|$, $|V_{cs}|$ and their ratio. Currently, the
best direct determination comes from dimuon production in deep-inelastic
scattering of neutrinos and anti-neutrinos on nucleons for $|V_{cd}|$ and from
charm-tagged W decays for $|V_{cs}|$~\cite{Part5_pdg06,part5:PDG04}
\begin{eqnarray}
|V_{cd}| = 0.230\pm 0.011              &\qquad& \sigma(|V_{cd}|)/|V_{cd}|= 5\%\\
|V_{cs}| =   0.97\pm 0.09\pm 0.07   &\qquad& \sigma(|V_{cs}|)/|V_{cs}|= 12\% ,
\end{eqnarray}
with little hope to improve on the accuracy in the case of $|V_{cd}|$. 
An alternative determination, which is superseding the previous one, comes from semileptonic 
$D\to K\ell\nu$ and $D\to\pi\ell\nu$ decays combined with lattice inputs for the form factors. 
The current CLEO-c data provide~\cite{part5:HockerLigeti}:
\begin{eqnarray}
|V_{cd}| &=& 0.213\pm 0.008 \pm 0.021  \qquad \sigma(|V_{cd}|)/|V_{cd}|= 11\% \\
|V_{cs}| &=& 0.957\pm 0.017\pm 0.093    \qquad \sigma(|V_{cs}|)/|V_{cs}|=  10\% .
\end{eqnarray}

In 2012, with the inputs quoted above, the global fit from CKMfitter is
expected to determine $|V_{cd}|$ and $|V_{cs}|$ with an accuracy of
\begin{eqnarray}
\sigma(|V_{cd}|)/|V_{cd}|&=& 0.4\% 
\qquad \sigma(|V_{cs}|)/|V_{cs}|= 0.02\%\\
\frac{\sigma(|V_{cd}|/|V_{cs}|}{|V_{cd}|/|V_{cs}|)} &= & 0.4\% .
\end{eqnarray}
We stress that these values are  obtained assuming that the CKM
mechanism is the only source for CP violation and including only the
experimental inputs described in Sec.~\ref{sec:globalfit}. In particular,
these predictions are made without any direct experimental input on
$|V_{cd}|$ and $|V_{cs}|$. The main constraint comes from the accurate
determination of $|V_{ud}|$, which, thanks to unitarity,
fixes $\lambda$
in the Wolfenstein parametrization and, thus, the two CKM matrix elements of
interest. 

This accuracy in the indirect determination from the global fit can be
compared with the expected 
accuracy from \bes3 and lattice QCD on the same CKM matrix elements from
leptonic decays:
\begin{eqnarray}
\sigma(|V_{cd}|)/|V_{cd}| &=& 2.3\%  \qquad
\sigma(|V_{cs}|)/|V_{cs}| = 1.7\% \\
\frac{\sigma(|V_{cd}|/|V_{cs}|)}{|V_{cd}|/|V_{cs}|}&=& 1.3\% ,
\end{eqnarray}
and from semileptonic decays (respectively $D_s\to K$ and $D_s\to\phi$):
\begin{equation}
\sigma(|V_{cd}|)/|V_{cd}| = 2.4\% \qquad
\sigma(|V_{cs}|)/|V_{cs}|= 1.3\% ,
\end{equation}
which means that the accuracy of leptonic and semileptonic measurements
will allow a meaningful and detailed comparison 
with the CKMfitter predictions in a corner of the CKM
matrix that has been tested only with a limited precision. This is
particularly interesting in view of the recent hint of an anomaly in the
$s\to u$  weak current~\cite{part5:NA48}:
a  disagreement
between the predictions of the global fit and \bes3 data supplemented
by lattice results could provide a very valuable indication of physics
beyond the Standard Model in the first two quark generations. This short
and presumably very naive  prospective exercise exemplifies how indirect
methods (CKMfitter) and direct measurements (\bes3) can help each other
to test the Standard Model in its less well-known aspects and which
improvement can be expected from such combined analysis in the coming
years.

\chapter[Hadronic $D(D_S)$ decays]{Hadronic $D(D_S)$ decays\footnote{By
Yue-Liang Wu and Ming Zhong}}
\label{sec:charm_hadron}
\newcommand {\s}{\vspace{.25in}}

\newcommand{\tto}{\!\to\!}

\newcommand{\gsim}{\lower.7ex\hbox{$
\;\stackrel{\textstyle>}{\sim}\;$}}
\newcommand{\lsim}{\lower.7ex\hbox{$
\;\stackrel{\textstyle<}{\sim}\;$}}
\renewcommand{\Im}{{\rm Im}\,}

\newcommand{\bibit}[1]{\bibitem{#1} \marginpar{\vspace*{.4cm}~~\tiny[#1]}}

\newcommand{\aver}[1]{\langle #1\rangle}

\newcommand{\La}{\overline{\Lambda}}
\newcommand{\Si}{\overline{\Sigma}}
\newcommand{\Lam}{\Lambda_{\rm QCD}}
\newcommand{\mhad}{\mu_{\rm hadr}}

\newcommand{\sigp}{\vec\sigma \vec\pi}

\def\op{{\bf P}}
\def\oc{{\bf C}}
\def\ot{{\bf T}}
\def\cp{{\bf CP}}
\def\cpt{{\bf CPT}}

\newcommand{\ga}{\gamma}
\newcommand{\de}{\delta}
\newcommand{\la}{\lambda}
\newcommand{\as}{\alpha_s}
\newcommand{\GeV}{\,\mbox{GeV}}
\newcommand{\MeV}{\,\mbox{MeV}}
\newcommand{\matel}[3]{\langle #1|#2|#3\rangle}
\newcommand{\state}[1]{|#1\rangle}
\newcommand{\ve}[1]{\vec{\bf #1}}

\newcommand{\gsl}{\Gamma_{\rm sl}(b\!\to\!c)}
\newcommand{\asMS}{\alpha_s^{\overline{\rm MS}}}

\newcommand{\eod}{\end{document}}

\newcommand{\msp}[1]{\mbox{\hspace*{#1mm}~}}
\newcommand{\note}[1]{\marginpar{\hspace*{-2mm}\tiny #1}}
\newcommand{\od}{\marginpar{{\large **}}}


\section{Present Status and Implication for QCD}\label{status}

The SM's electroweak phenomenology of charm-changing transitions
appears dull, with the  CKM parameters well known due to
three-family unitarity constraints, the very slow $D^0 - \bar D^0$ oscillation
frequency,  CP~violating asymmetries that are small at best, and 
with loop-driven
decays  that are extremely rare and swamped by huge backgrounds due to long distance
dynamics. Yet this very dullness can be utilized to gain new
insights into nonperturbative dynamics, make progress in
establishing theoretical control over them and calibrate the
theoretical tools for $B$ studies.

The issue at stake here is {\em not} whether QCD is the theory of
the strong forces -- there is no alternative -- but our ability to
perform calculations. Here charm hadrons can act as a bridge
between the worlds of light flavours -- as carried by $u$, $d$ and
$s$ quarks with masses lighter or at most comparable to $\Lambda
_{QCD}$ and described by chiral perturbation theory -- and that of
the {\it bona-fide} heavy $b$ quark, with $\Lambda_{QCD} \ll m_b$ and
treatable by heavy quark theory. Only lattice QCD (LQCD) carries
the promise for a truly quantitative treatment of charm hadrons
that can be improved {\em systematically}. Furthermore LQCD  is
the only framework available that allows one to approach charm from
lower as well as higher mass scales, which involves different
aspects of nonperturbative dynamics and thus -- if successful --
would provide impressive validation.

At present, such a program can be carried  most explicitly for
exclusive semileptonic decays of charm hadrons, as described in
detail in Sect.~26.2, especially since lattice QCD (LQCD) is reaching a
stage where it can make rather accurate predictions for such
modes. The theoretical challenges posed by nonleptonic decays are
obviously more formidable. The complexities increase considerably
for exclusive nonleptonic transitions, in particular because of the
importance of final state interactions (FSI), which are much
harder to bring under theoretical control even by using
state-of-the-art LQCD.

Yet there are some strong motivations for obtaining a reliable
description of exclusive nonleptonic charm decays:
\begin{itemize}
\item Their dynamics is largely determined by the transition
region from the perturbative to the nonperturbative domain. Thus,
we can gain novel insights there. One should also not give up hope
for a future theoretical breakthrough in LQCD (or the advent of
another similarly powerful theoretical technology) allowing us to
extract numerically reliable lessons. \item The most sensitive
probes for New Physics are CP~asymmetries in nonleptonic
channels. Search strategies and subsequent interpretations
depend on hadronic matrix elements, FSI and their phases. As
already indicated, we do not know how to compute them, yet one can
profit here from a pragmatic exercise in `theoretical
engineering:' providing a phenomenological, yet comprehensive
framework for a host of charm modes allows one to extract quantitative
information on hadroniic matrix elements and FSI phases and
evaluate their reliability through overconstraints. The huge
datasets already obtained by the $B$ factories, CLEO-c and BESII
and to be further expanded including also future \bes3 studies
will be of essential help here. \item Analogous decays of $B$ 
mesons are
being studied also as a means to extract the complex phase of
$V_{ub}$. One could hope that $D$ decays might serve as a
validation analysis. \item Careful analysis of branching fractions
can teach us novel lessons on light-flavour hadron spectroscopy,
like on characteristics of resonances such as the scalar mesons,
or on $\eta$-$\eta^{\prime}$ mixing and possible
non-$\bar qq$ components inside them.
\end{itemize}
\section{Theoretical Review}
\label{ThRev}
\subsection{The Effective Weak Hamiltonian}
\label{EWH}

The theoretical description starts from constructing an effective
$\Delta C \neq 0$ Hamiltonian through an operator product
expansion (OPE) in terms of {\em local} operators $O_i$ and their
coefficients $c_i$:
\begin{equation}
\langle f |{\cal H}_{eff} |D\rangle =\frac{G_F}{\sqrt {2}} V_{CKM}
\sum \limits_i c_i(\mu)\langle f | O_i|D\rangle {(\mu)}
\end{equation}
The {\em auxiliary} scale $\mu$ has been introduced -- and this is
a central element of the Wilsonian prescription for the OPE -- to
separate contributions from long- and short-distance dynamics: long
distance $> 1/\mu >$ short distance. Degrees of freedom with
 mass scales above $\mu$ are integrated out into the
coefficients $c_i$ typically using perturbation theory, while
degrees of freedom with scales below $\mu$ remain dynamical and
are contained in the operators $O_i$.  Nonperturbative dynamics
enters through their hadronic expectation values.

Observables, of course, cannot depend at all on the choice of $\mu$,
{\it i.e.}, the $\mu$ dependence of the coefficients has to cancel
against that of the matrix elements when one does a complete
calculation. However, in practice one has to keep the following in
mind:
\begin{itemize}
\item The perturbatively treated coefficients also contain the
strong coupling $\alpha_S$. To keep it in the perturbative domain,
one needs \begin{equation} \mu \gg \Lambda_{QCD} \end{equation}
\item Yet, at the same time, one does not want to choose too high a
value for $\mu$, since it also provides the momentum cut-off in
the hadronic wave function with which the matrix element is
evaluated.

\end{itemize}
These two contravening requirements can be met by $\mu \sim 1 -
1.5$GeV, which happens to be close to the charmed quark mass. Thus
$\mu = m_c$ provides a reasonable ansatz. In practice one has to
rely on additional approximations of various kinds, and these cause
the computed rates to contain some sensitivity at least to $\mu$,
which can, in turn, provide a gauge for the reliability of the result.

We do not know yet how to calculate these hadronic matrix elements
from QCD's first principles in a numerically accurate way,
although several different `second generation' theoretical
technologies have been brought to bear on them: $1/N_C$
expansions, QCD sum rules and lattice QCD. While there is
reasonable hope that the latter will be validated in
(semi)leptonic $D$ decays, exclusive nonleptonic transitions
provide qualitatively new challenges.


While in the SM the weak decays are driven by charged currents,
the intervention of QCD affects the strength of the charged
current product and induces a product of effective neutral currents
in a way that depends on $\mu$. For Cabibbo-allowed transitions,
one can write down the effective weak Lagragian
 \begin{equation} {\cal L}_{eff}^{\Delta C=1}(\mu = m_c) =
-\frac{G_F}{\sqrt{2}} V_{ud}V^*_{cs} \cdot [c_- O_- + c_+ O_+] \ ,
\label{EFFWEAKLAG} \end{equation} \begin{equation}
O_{\pm}=\frac{1}{2}\left[ (\bar s_L\gamma_{\nu}c_L)(\bar
u_L\gamma_{\nu}d_L)] \pm (\bar u_L\gamma_{\nu}c_L)(\bar
s_L\gamma_{\nu}d_L)\right]  \; , \end{equation} which is
conveniently rewritten  as:
$$
{\cal L}_{eff}^{\Delta C=1}(\mu = m_c) = -\frac{G_F}{\sqrt{2}}
V_{ud}V^*_{cs} \cdot [c_1O_1 + c_2O_2] \ ,
$$
\begin{equation} O_1 = (\bar s_L\gamma_{\nu}c_L)(\bar u_L\gamma_{\nu}d_L) \; ,
\; \ \ O_2 =  (\bar u_L\gamma_{\nu}c_L)(\bar s_L\gamma_{\nu}d_L)\
, \label{EFFWEAKLAG2} \end{equation} with \begin{equation} c_1 =
\frac{1}{2}(c_+ + c_-) \; , \; \ \ \ c_2 = \frac{1}{2}(c_+ - c_-)
\ .\end{equation} Using different schemes, one typically gets
\cite{Part5_ming_bbl}:
\begin{equation} c_1(m_c)=1.25 \pm 0.03 \; , \; \ \ \  c_2(m_c)= - 0.48 \pm
0.05. \end{equation}

\subsection{Factorization and First Generation Theoretical
Techniques} \label{sec:factor}

All decay amplitudes can then be expressed as linear combinations
of two terms: \begin{equation} {\cal A}(D \to f) \propto
a_1\langle f|J_{\mu}^{(ch)}J^{\prime(ch)\mu}|D\rangle + a_2\langle
f|J_{\mu}^{(neut)}J^{\prime(neut)\mu}|D\rangle \ ,
\end{equation} with
\begin{equation} a_1 = c_1 + \xi c_2 \; , \; \ \ \ a_2 = c_2 + \xi c_1 \ .
\label{A1A2} \end{equation} It should be noted that the quantities
$c_1$ and $c_2$ on one hand and $\xi$ on the other are of {\em
completely different} origin despite their common appearance in
$a_1$ and $a_2$: while $c_{1,2}$ are determined by short-distance
dynamics and $\xi$ parametrizes the impact of long distance dynamics
on the size of matrix elements including effects due to FSI. 
Equation~\ref{A1A2} contains two very
important implicit assumptions, namely that the value of $\xi$ is
the same in the expressions for $a_1$ and $a_2$ and that it does
not depend on the final state.

A very convenient ansatz is to write the nonleptonic transition
matrix element as a product of two simpler matrix elements
\cite{Part5_ming_naiveF}
\begin{eqnarray}
 \langle f |J_{\mu}J^{\prime\mu}|D\rangle \equiv \langle f_1f_2 |
J_{\mu}J^{\prime\mu}|D\rangle \simeq \langle f_1 |J_{\mu}|0\rangle
\langle f_2 |J^{\prime\mu}|D\rangle \ , \label{FACT}
\end{eqnarray}
where $f_1$ and $f_2$ are "{\em effective particles}" that can
contain any number of final state particles. The basic assumption
here is that the color flow mediated by gluon exchanges {\em
between} the two `clusters' $0 \to f_1$ and $D\to f_2$ can be
ignored and all the strong interaction effects lumped into two
simpler transition amplitudes. Clearly this factorization ansatz
can be only an approximation rather than an identity. One should
also note that Eq.~\ref{FACT} is $\mu$ dependent; i.e. changing
the value of $\mu$ will transform factorized contributions into
non-factorized ones and {\it vice versa}. The best chance for this
ansatz to represent a decent approximation is for the separation
scale $\mu$ to be around ordinary hadronic scales of about 1 GeV.
This value happens to be close to $m_c$, yet that is a
coincidence, since heavy quark masses are extraneous to QCD.

Besides these two types of diagrams, which are usually referred as
color favored and color suppressed diagrams, other types of
considerations are the weak annihilation (WA) contributions
including annihilation and exchange diagrams where the matrix
element is approximately written as
\begin{eqnarray}
\label{WA} \langle f_1f_2 | J_{\mu}J^{\prime\mu}|D\rangle \simeq
\langle f_1f_2 |J_{\mu}|0\rangle \langle 0 |J^{\prime\mu}|D\rangle
\ . \label{FACT_2}
\end{eqnarray}

 Having assumed factorization, we have greatly restricted the
number of free parameters. In principle at least,
the amplitudes  $\langle f_1
|J_{i\mu}|0\rangle$ and $\langle f_2 |J^{\prime\mu}_i|D\rangle $
can be taken from  (semi)leptonic $D$ decay data,
although in practice that information is augmented by some
theoretical arguments. The two quantities $a_{1,2}$ are then
treated as free parameters fitted from experiment, although, in
practice again, some theoretical judgment has to be applied
concerning if and to what degree WA diagrams are included in
addition to the spectator diagrams and corrections for FSI have to
be applied.

Such an analysis was first carried out by Bauer, Stech and Wirbel (BSW)
for charmed meson two-body decays, yielding \cite{Part5_ming_bsw}
\begin{equation} a_1|_{exp} \simeq 1.2 \pm 0.1 \; , \; \ \
a_2|_{exp} \simeq - 0.5 \pm 0.1 , \label{A1A2EXP} \end{equation} to
be compared with the theoretical expectations \begin{equation}
a_1|_{QCD} \simeq 1.25 - 0.48 \xi \; , \; \ \ a_2|_{QCD} \simeq -
0.48 + 1.25 \xi \ .\label{A1A2NAIVE} \end{equation} It is
remarkable that with just two fit parameters one can get a decent
description of a host of nonleptonic rates. However one might say
that those parameters have the wrong values: naively just counting
colors one expects $\xi \simeq 1/N_C = 1/3$ and thus $a_1|_{QCD}
\simeq 1.09$ and $a_2|_{QCD} \simeq - 0.06$; for $a_2$ this is
inconsistent with the experimental fit value. $\xi \simeq 0$ would
reconcile Eqs.~\ref{A1A2EXP} and \ref{A1A2NAIVE}.


\subsection{The $1/N_C$ ansatz} \label{BURAS}
%
The fit result $\xi \simeq 0$ leads to an intriguing speculation
that these weak two-body decays can be described more rigorously
using $1/N_C$ expansions~\cite{Part5_ming_GERARD}. These are invoked to
calculate hadronic matrix elements. The procedure is as
follows: One employs the effective weak transition operator
${\cal L}_{eff}(\Delta C=1)$ given explicitly in
Eq.~\ref{EFFWEAKLAG}; since it describes short distance dynamics,
one has kept $N_C=3$ there. Then one expands the matrix element
for a certain transition driven by these operators in $1/N_C$
\begin{equation} {\cal A}(D\to f) = \langle f|{\cal
L}_{eff}(\Delta C=1)|D\rangle = \sqrt{N_C}\left( b_0 +
\frac{b_1}{N_C} + {\cal O}(1/N_C^2) \right) \ . \end{equation}

%

Using the rules for $1/N_C$ expansions, it is easy to show that
the following simplifying properties hold for the
leading $1/N_C$ contributions:
\begin{itemize}
\item
one has to consider {\em valence} quark wave functions only;
\item
{\em factorization} holds;
\item
{\em WA} has to be ignored as have FSI.
\end{itemize}
To leading order in $1/N_C$, only the term $b_0$ is retained; then one
has effectively $\xi =0$ since $\xi \simeq 1/N_C$ represents a higher
order contribution.  However, the next-to-leading term $b_1$ is, in
general, beyond theoretical  control. $1/N_C$ expansions therefore do not
enable us to decrease  the uncertainties {\em systematically}.

The $N_C \to \infty$ prescription is certainly a very compact one with
transparent rules, and it provides a not-bad first approximation --
but no more. One can ignore neither FSI nor WA completely.

\subsection{Treatment with QCD sum rules} \label{BLOK}
%
A treatment of $D \to PP$ and $D\to PV$ decays based on a
judicious application of QCD sum rules was developed in a series
of papers~\cite{Part5_ming_BLOKSHIFMAN}. The authors analyzed four-point
correlation functions between the weak Lagrangian ${\cal L}(\Delta
C=1)$ and three currents. As usual, an OPE is applied to the
correlation function in the Euclidean region; nonperturbative
dynamics is incorporated through condensates $\langle0|m\bar
qq|0\rangle$, $\langle0|G\cdot G|0\rangle$, etc., the  numerical
values of which are extracted from other {\em light}-quark
systems. They extrapolated their results to the Minkowskian domain
through a (double) dispersion relation and succeeded in finding a
stability range for matching it with phenomenological hadronic
expressions.
%

 The analysis has some nice features:

\noindent $\oplus$ It has a clear basis in QCD, and includes,
 in principle at least, nonperturbative dynamics in a well defined way.

\noindent $\oplus$ It incorporates different quark-level processes
-- external and internal $W$ emission, WA and Pauli interference
-- in a natural manner.

\noindent $\oplus$ It allows one to include nonfactorizable contributions
systematically.

In practice, however, it suffers at the same time from some shortcomings:

\noindent $\ominus$ The charm scale is not sufficiently high for one
to have full confidence in the various extrapolations undertaken.

\noindent $\ominus$ To make these very lengthy calculations at all
manageable, some simplifying assumptions had to be made, like
$m_u=m_d=m_s=0$ and $SU(3)_{Fl}$ breaking beyond $m_K > m_{\pi}$
had to be ignored; in particular $\langle 0|\bar ss|0\rangle$ =
$\langle 0|\bar dd|0\rangle$ = $\langle 0|\bar uu|0\rangle$ was
used. Thus, for example, one cannot expect $SU(3)_{Fl}$ breaking to be
reproduced correctly.

 \noindent $\ominus$ {\em Prominent} FSI that vary rapidly with the
energy scale -- like
effects due to narrow  resonances - cannot be described in
this treatment; the extrapolation from the Euclidean to the
Minkowskian domain amounts to some averaging or `smearing' over
energies.

A statement that the predictions did not provide an excellent fit
to the data on about twenty-odd $D^0$, $D^+$ and $D_s^+$ modes --
while correct on the surface, especially when $SU(3)_{Fl}$
breaking is involved -- misses the main point:
\begin{itemize}
\item
No {\it a priori} model assumption like factorization had to be made.
\item
In principle,
the theoretical description does not contain any free parameters,
although in practice there is some leeway in the size of some
decay constants.
\end{itemize}

\subsection{Modern Developments} \label{MODERN}
%
As the data improved, the BSW prescription became inadequate,
however most subsequent attempts to describe nonleptonic decays in
the $D$ system -- except for the sum rules approach sketched above
-- use the assumption of naive factorization as a starting point.

Improvements and generalizations of the BSW description have been
made in three areas:
\begin{enumerate}
\item Different parameterizations for the $q^2$ dependence of the
form factors are used and different evaluations of their
normalization are made. This is similar to what was addressed in
our discussion of exclusive semileptonic decays. One appealing
suggestion is to use only those expressions for
form factors that asymptotically -- {\it i.e.} for $m_c$, $m_s \to
\infty$ -- exhibit heavy quark symmetry. \item Attempts have been
made to incorporate FSI more reliably. Non-factorized
contributions in general have been considered.  \item
Contributions due to WA and Penguin operators have been included.
\end{enumerate}

Two frameworks that are more firmly based on QCD than quark models
have been developed to treat two-body decays of $B$ mesons, namely
`QCD factorization' \cite{Part5_ming_bbns} and `pQCD' \cite{Part5_ming_lixn}. While
there is little reason to expect the more aggressive pQCD approach
to work for charm decays, a treatment based on QCD
factorization is worth a try despite the fact that the charm mass barely
exceeds ordinary hadronic scales. To illustrate the present
status, the branching fractions for $D\to \pi\pi$ decays  inferred
from naive factorization and QCD factorization approaches are
listed and compared with experimental data \cite{Part5_ming_wuyin}:
\begin{eqnarray}
&&BR(D^0\to \pi^+\pi^-)=\left\{
\begin{array}{ll} 1.86\times 10^{-3} \ , & (\rm Naive \ \ Factorization)\\
1.69\times 10^{-3} \ , & (\rm QCD \ \ Factorization)\\
(1.364\pm 0.032)\times 10^{-3} \ , & (\rm PDG06 \cite{Part5_pdg06})
\end{array} \right.\nonumber\\
&&BR(D^+\to \pi^+\pi^0)=\left\{
\begin{array}{ll} 1.68\times 10^{-3} \ , & \ \ \ (\rm Naive \ \ 
Factorization)\\
1.94\times 10^{-3} \ , & \ \ \ (\rm QCD \ \ Factorization)\\
(1.28\pm 0.09)\times 10^{-3} \ , & \ \ \ (\rm PDG06 \cite{Part5_pdg06})
\end{array} \right.\nonumber\\
&&BR(D^0\to \pi^0\pi^0)=\left\{
\begin{array}{ll} 2.44\times 10^{-5} \ , & \ \ \ \ (\rm Naive \ \ 
Factorization)\\
2.06\times 10^{-5} \ , & \ \ \ \ (\rm QCD \ \ Factorization)\\
(7.9\pm 0.8)\times 10^{-4} \ . & \ \ \ \ (\rm PDG06 \cite{Part5_pdg06}).
\end{array}
\right. \nonumber
\end{eqnarray}
Both the naive factorization and the QCD factorization predictions
for $D^0\to \pi^+\pi^-$ and $D^+\to \pi^+\pi^0$ are of the same 
order as the
experimental results, while the predictions for
$D^0\to \pi^0\pi^0$ are about forty times smaller than the
experiment. For proper perspective one should note that
the modes $D^0\to \pi^+\pi^-$ and $D^+\to \pi^+\pi^0$ are
described by color-{\em favored} tree diagram $T$, whereas $D^0\to
\pi^0\pi^0$ is dominated by a color-{\em suppressed} tree diagram
$C$. Non-factorizable corrections are found to be larger for the
latter and, at present, are beyond theoretical control.

 In summary, a theoretical description of exclusive
nonleptonic decays of charmed mesons based on general principles
is not yet possible. Even though the short distance contributions
can be calculated and the effective weak Hamiltonian has been
constructed at next-to-leading order, the evaluation of its matrix
elements requires nonperturbative techniques. Some decent
phenomenological descriptions have been achieved, but
realistically few frameworks provide opportunities for
systematic improvements, especially when they are applied to
multi-body channels.


\subsection{Symmetry Analysis} \label{su3}

\subsubsection{Isospin SU(2) Symmetry}
Symmetry-based arguments are a powerful weapon in our theoretical
arsenal. Isospin invariance should hold at the ${\cal O}(1\% )$
level, and no evidence to the contrary has been found. Taking
two-body decays as an example, it leads to triangle relations
among the decay amplitudes:
\begin{eqnarray}
&&{\cal A}(D^0\to
\pi^+\pi^-)+\sqrt{2}{\cal A}(D^0\to \pi^0\pi^0)-\sqrt{2}{\cal A}(D^+\to \pi^+\pi^0)=0 \ , \\
&&{\cal A}(D^0\to
K^-\pi^+)+\sqrt{2}{\cal A}(D^0\to \bar{K}^0\pi^0)-{\cal A}(D^+\to \bar{K}^0\pi^+)=0 \ , \\
&&{\cal A}(D^0\to \pi^+K^{*-})+\sqrt{2}{\cal A}(D^0\to \pi^0\bar
{K}^{*0})-{\cal A}(D^+\to \pi^+\bar
{K}^{*0})=0 \ , \\
&&{\cal A}(D^0\to \rho^+K^-)+\sqrt{2}{\cal A}(D^0\to \rho^0\bar
{K}^0)-{\cal A}(D^+\to \bar {K}^0\rho^+)=0 \ .
\end{eqnarray}
The measured rates tell us that these amplitudes possess large
relative phases indicating strong FSI. Considering the three $D\to
\pi\pi$ modes, the transition amplitudes can be decomposed into
\begin{eqnarray}
&{\cal A}(D^0\to
\pi^+\pi^-)=\sqrt{\frac{2}{3}}{\cal A}_0+\sqrt{\frac{1}{3}}{\cal A}_2 \ , \\
&{\cal A}(D^0\to
\pi^0\pi^0)=\sqrt{\frac{1}{3}}{\cal A}_0-\sqrt{\frac{2}{3}}{\cal A}_2 \ , \\
&{\cal A}(D^+\to \pi^+\pi^0)=\sqrt{\frac{3}{2}}{\cal A}_2 \ .
\end{eqnarray}
The subscripts $0$ and $2$ of the ${\cal A}$ describe the isospin
$I=0$ and $2$ components of the $\pi\pi$ system. From the experimental
data, the amplitude ratio $|A_2/A_0|$ and the relative phase
$\delta=\delta_2-\delta_0$ are determined to be~\cite{Part5_ming_prl71}
\begin{eqnarray}
|A_2/A_0|=0.72\pm 0.13\pm 0.11 \ , \ \ \cos \delta=0.14\pm 0.13\pm
0.09 \ .
\end{eqnarray}
\subsubsection{Flavor $SU(3)$ Symmetry and Its Breaking} It would
seem tempting to argue that $SU(3)$-flavor symmetry holds to within,
say, 20 - 30 \%. This, however, does not seem to be the case, at
least not for exclusive channels, as can be read off most
dramatically from the difference of the ratio \begin{equation}
\frac{{\cal A} (D^0 \to K^+K^-)}{{\cal A} (D^0 \to \pi^+\pi^-)}
\simeq 1.8 \end{equation} from unity. This significant $SU(3)$
symmetry violation may come from the finite strange quark mass,
FSI and resonances \cite{Part5_ming_bre}. There are indications, however,
that for inclusive rates, $SU(3)$-flavor breaking does not exceed the
20\% level \cite{Part5_ming_ex0309021}.
\section{Two-Body Decays}
\label{2B}
Two-body modes in charmed meson nonleptonic decays have drawn
much attention since the 1980s because they have a number of
advantages, in comparison with multi-body ones, including:
\begin{itemize}
\item Nonleptonic charm meson decays have been observed to
proceed mainly via two-body channels, where a resonance is
considered as a single body.  A large accumulation of precise 
experimental data on
two-body decays, including branching fractions for about 60 decay
modes, is contained in the PDG tables
\cite{Part5_pdg06}. \item The phase space is trivial and the number of
form factors are quite limited. \item There are fewer color
sources in the form of quarks and antiquarks, and fewer different
combinations of color flux tubes that can form.\item Quite a number of
two-body modes allow for sizeable momentum transfers thereby,
hopefully, reducing the predominance of long-distance dynamics.
\item This is the one class of nonleptonic decays where one can
harbor reasonable hopes of some success. It is not utopian to
expect lattice QCD to treat these transitions some day in full
generality. Such results will, however, only be reliable if
obtained with the incorporation of fully dynamical fermions -- i.e.
without "quenching" and without reliance on a $1/m_c$ expansion.

\end{itemize}
\subsection{Kinematics and Topologies of Amplitudes}
In the center-of-mass frame, the differential decay rate for
$n$-body charmed meson decay is
\begin{equation}
\label{n-bparwid}
d\Gamma=\frac{1}{2m_D}(\prod_{i=1}^{n}\frac{d^3p_i}{(2\pi)^3}\frac{1}{2E_i}
)|{\cal A}(m_D\to
\{p_1,p_2,\cdots,p_n\})|^2(2\pi)^4\delta^4(p_D-\sum_{i=1}^{n}p_i).
\end{equation}
If the number of final state particles is set to two, one can
easily perform the integral over phase space to obtain the
decay rate
\begin{equation}
\Gamma(D \to f_1f_2) = \frac{p}{8 \pi M_D^2}|{\cal A}|^2 \ ,
\end{equation}
where
\begin{displaymath}
p=\frac{\sqrt{(M_D^2-(m_1+m_2)^2)(M_D^2-(m_1-m_2)^2)}}{2M_D}
\end{displaymath}
denotes the center-of-mass 3-momentum of each final-state particle. The
branching fraction for the $D \to f_1f_2$ transition is
the ratio of this partial decay width to the full width of the $D$ meson:
\begin{equation}
{\it B}(D \to f_1f_2)=\frac{\Gamma(D \to f_1f_2)}{\Gamma(D)}\ .
\end{equation}

 The amplitude ${\cal
A}$ can be decomposed into six distinct quark-graph topologies
\cite{Part5_ming_cc}: (1) color-favored tree amplitude $T$, (2)
color-suppressed tree amplitude $C$, (3) $W$-exchange amplitude $E$,
(4) $W$-annihilation amplitude $A$, (5) horizontal $W$-loop amplitude
$P$ and (6) vertical $W$-loop amplitude $D$. The penguin diagrams
$P$ and $D$ play little role in practice because the relation of
the CKM matrix elements $V^*_{cs}V_{us}\approx -V^*_{cd}V_{ud}$
results in cancellations among them.

\subsection{$D\to PP$, $D\to PV$ and $D\to VV$ Decays}
Among the experimental data on charm nonleptonic two-body
decays, charmed mesons decaying to two pseudoscalar mesons ($D\to
PP$), to one pseudoscalar and one vector meson ($D\to PV$) and to
two vector mesons ($D\to VV$) have the best precision. The light
pseudoscalar and vector mesons are two classes of particles that
are distinct because of their well established basic properties 
such as mass,
lifetime, width, quark component and decay rate. The form factors
for charmed mesons transforming to light pseudoscalar and vector
mesons have been calculated in a variety of theoretical models.
Based on the resulting form factors, most predictions on charmed
meson semileptonic decays are consistent with experimental data,
as discussed in Sect.~\ref{part5:section:review:semi}.  
As a result, $PP$, $PV$ and $VV$ decays are  an ideal place to
test the factorization assumption and develop an
understanding of the mysteries of
FSI and unfactorizable contributions.

Using the form factor definitions given in
Eqs.~\ref{part5:eq:semi:rev} and~\ref{part5:eq:semi:rev:vector}, the four 
relevant amplitudes for
$D\to P_1P_2$ in the factorization approach are:
\begin{eqnarray}
\label{ft}
T&=&i\frac{G_F}{\sqrt{2}}V_{q_1q_2}V^*_{cq_3}a_1f_{_{P_1}}
(m^2_{_D}-m^2_{_2})F^{D\rightarrow P_2}_0(m^2_{_1}), \\
\label{fc}
C&=&i\frac{G_F}{\sqrt{2}}V_{q_1q_2}V^*_{cq_3}a_2f_{_{P_1}}
(m^2_{_D}-m^2_{_2})F^{D\rightarrow P_2}_0(m^2_{_1}), \\
\label{fe} E&=&i\frac{G_F}{\sqrt{2}}V_{q_1q_2}V^*_{cq_3}a_2f_{_D}
(m^2_{_1}-m^2_{_2})F^{P_1P_2}_0(m^2_{_D}), \\
\label{fa} A&=&i\frac{G_F}{\sqrt{2}}V_{q_1q_2}V^*_{cq_3}a_1f_{_D}
(m^2_{_1}-m^2_{_2})F^{P_1P_2}_0(m^2_{_D}).
\end{eqnarray}

The amplitudes for $D\to PV$ are a little more complicated than
$D\to PP$, since one has to distinguish terms where the spectator
quark ends up in the pseudoscalar or vector particle in the final state.
Using the
subscripts $P$ and $V$ to distinguish 
between the cases where the spectator quark is in
the pseudoscalar or vector final-state meson, respectively, one can read 
off the
amplitudes:
\begin{eqnarray}
\label{ftv}
T_V&=&2\frac{G_F}{\sqrt{2}}V_{q_1q_2}V^*_{cq_3}a_{1}f_{_P}m_{_V}(\varepsilon^*\cdot p_{_D})A^{D\rightarrow V}_0(m^2_{_P}), \\
\label{ftp}
T_P&=&2\frac{G_F}{\sqrt{2}}V_{q_1q_2}V^*_{cq_3}a_{1}f_{_V}m_{_V}(\varepsilon^*\cdot p_{_D})F^{D\rightarrow P}_+(m^2_{_V}), \\
\label{fcv}
C_V&=&2\frac{G_F}{\sqrt{2}}V_{q_1q_2}V^*_{cq_3}a_{2}f_{_P}m_{_V}(\varepsilon^*\cdot p_{_D})A^{D\rightarrow V}_0(m^2_{_P}), \\
\label{fcp}
C_P&=&2\frac{G_F}{\sqrt{2}}V_{q_1q_2}V^*_{cq_3}a_{2}f_{_V}m_{_V}(\varepsilon^*\cdot p_{_D})F^{D\rightarrow P}_+(m^2_{_V}), \\
\label{fev}
E&=&2\frac{G_F}{\sqrt{2}}V_{q_1q_2}V^*_{cq_3}a_{2}f_{_D}m_{_V}(\varepsilon^*\cdot p_{_D})A^{PV}_0(m^2_{_D}), \\
\label{fap}
A&=&2\frac{G_F}{\sqrt{2}}V_{q_1q_2}V^*_{cq_3}a_{1}f_{_D}m_{_V}(\varepsilon^*\cdot
p_{_D})A^{PV}_0(m^2_{_D}).
\end{eqnarray}

Charm decays to two vector meson final states have a
richer structure than those with at least one pseudoscalar in
the final state.  Here
\begin{eqnarray} T(D\to
V_1V_2)&=&
{G_F\over\sqrt{2}}V_{q_{_1}q_{_2}}V_{cq_{_3}}^*a_1[if_{_{V_1}}
m_{_1}(m_{_D}+m_{_2})A_1^{D\to
V_2}(m_{_1}^2)\varepsilon^{*}_{_1}\cdot \varepsilon^{*}_{_2}
\nonumber\\
&&-i{1\over m_{_D}+m_{_2}}f_{_{V_1}} m_{_1} A_2^{D\to
V_2}(m^2_{_1})\varepsilon^{*}_{_1}\cdot(p_{_D}+p_{_2})\varepsilon^{*}_{_2}\cdot(p_{_D}-p_{_2})
\nonumber\\
&&-{2\over m_{_D}+m_{_2}}f_{_{V_1}} m_{_1} V^{D\to
V_2}(m^2_{_1})\epsilon_{
\mu\nu\alpha\beta}\varepsilon^{*\mu}_{_1}\varepsilon^{*\nu}_{_2}p^\alpha_{_D}p_{_2}^\beta]
\ , \\
C(D\to
V_1V_2)&=&{G_F\over\sqrt{2}}V_{q_{_1}q_{_2}}V_{cq_{_3}}^*a_2[if_{_{V_1}}
m_{_1}(m_{_D}+m_{_2})A_1^{D\to
V_2}(m_{_1}^2)\varepsilon^{*}_{_1}\cdot \varepsilon^{*}_{_2}
\nonumber\\
&&-i{1\over m_{_D}+m_{_2}}f_{_{V_1}} m_{_1} A_2^{D\to
V_2}(m^2_{_1})\varepsilon^{*}_{_1}\cdot(p_{_D}+p_{_2})\varepsilon^{*}_{_2}\cdot(p_{_D}-p_{_2})
\nonumber\\
&&-{2\over m_{_D}+m_{_2}}f_{_{V_1}} m_{_1} V^{D\to
V_2}(m^2_{_1})\epsilon_{
\mu\nu\alpha\beta}\varepsilon^{*\mu}_{_1}\varepsilon^{*\nu}_{_2}p^\alpha_{_D}p_{_2}^\beta]
\ .
\end{eqnarray}
The terms proportional to $A_1$, $A_2$ and $V$ represent $S$,
longitudinal $D$ and $P$ waves respectively. The omitted
expressions for exchange and annihilation topologies contain the
vector-to-vector form factor.

From a long history of phenomenological analyses, we can draw some
general conclusions on the factorization formalism that
serve as guides for studies of the other charmed meson decay
modes like those to two-body final states containing scalar ($S$),
axial-vector ($A$) and tensor ($T$) particles as well as 
multi-body final states.
\begin{itemize}
    \item Nonfactorizable corrections
that result from spectator interactions, FSI, resonance effects, etc.,
are known to be significant~\cite{Part5_ming_fsi}. Some
phenomenological models based on the one-particle-exchange 
method~\cite{Part5_ming_lidu}, resonance 
formation~\cite{Part5_ming_blmps}, and the combination of
heavy quark effective theory and chiral perturbation 
theory~\cite{Part5_ming_Bajc} have been developed to try
to get some insights into these corrections.
Effects of  $q\bar{q}$ resonance formation are probably 
most important for hadronic charm decays, owing to the existence
of an abundant spectrum of resonances that are known to exist at energies
close to the mass of charmed mesons. Most of the resonance
properties conform to unitarity and the effects of
resonance-induced nonfactorizable contributions can be described
in a model-independent manner in terms of the masses and decay
widths of the contributing resonances~\cite{Part5_ming_zenc}.

    \item The parameters $a_1$ and $a_2$ were found to be non-universal 
and, instead, process or class dependent. For illustration purposes, 
we consider some examples
\begin{eqnarray}
\label{chi1}
&&a_1(\mu)=c_1(\mu)+(\frac{1}{N_c}+\chi_1(\mu))c_2(\mu) \ ,  \\
\label{chi2}
&&a_2(\mu)=c_2(\mu)+(\frac{1}{N_c}+\chi_2(\mu))c_1(\mu) \ ,
\end{eqnarray}
with $\chi_1(\mu)$ and $\chi_2(\mu)$ partially denoting
nonfactorizable effects in the case of $N_c=3$. $\chi_2(\mu)$ has
been determined to be  \cite{Part5_ming_cheng94}
\begin{eqnarray}
&&\chi_2(D\to \bar{K}\pi)\simeq -0.33 \ , \nonumber\\
&& \chi_2(D\to \bar{K}^*\pi)\simeq -(0.45\sim 0.55) \ , \nonumber\\
&&\chi_2(D\to \bar{K}^*\rho)\simeq -(0.6\sim 0.65) \ .
\end{eqnarray}
    \item The light-meson to light-meson form factors 
involved in the  above formulae
are believed to be negligibly small. Thus, the factorization
of the exchange and annihilation diagrams has no effect
on the overall amplitudes. The main contributions from these
diagrams may arise from the nonfactorizable parts. Through
intermediate states, they relate to the tree diagram $T$ and
color-suppressed diagram $C$ \cite{Part5_ming_zenc,Part5_ming_chengEPJC}. As a
consequence, they have sizable magnitudes that
can be comparable to the $T$
and $C$ amplitudes and large strong phases relative to the $T$
amplitude, as was demonstrated in a $SU(3)$-flavor symmetry 
analysis~\cite{Part5_ming_rosner,Part5_ming_chiang}.  
This is especially true in the 
case of decay mode $D^0\to
K^0\bar{K}^0$, which proceeds completely via the $E-E$ diagram
representation, and the factorizable contribution is too trivial to be
consistent with the experimentally measured branching fraction 
${\it B}=(7.1\pm 1.9)\times 10^{-4}$. 
Many studies have been performed on 
this 
decay~\cite{Part5_ming_DKK} and they find that a 
nonfactorizable correction of the same order as 
that of the $E$ diagram can 
account for the experimental results~\end{itemize}.

Results from a variety of calculations for $D \to PP$
decays are presented in Table~\ref{tab:predict1}; results for
$D\to PV$ decays are in Table~\ref{tab:predict2}.
Note that all of these calculations have
introduced some number of free parameters to describe the
nonfactorizable contributions, and these are determined from fits to
the experimental data.  Results for $D\to VV$ decays can be found in
Refs.~\cite{Part5_ming_Bajc,Part5_ming_KaVeSi,Part5_ming_BeDaMa,Part5_ming_HiKa}.
\begin{table}[hbp]
\caption{Predictions for $D\to PP$ branching fractions 
($\times 10^{-2}$).  
Most decay modes involving a neutral K meson are given
as $K^0_S$ in PDG06 and $\bar {K}^0$ in PDG04, which
are presented as well.  \label{tab:predict1}} \vspace{8pt}
\begin{small}
\begin{center}
\begin{tabular}{|l|c|c|c|l|}
\hline  Decay Modes& Buccella~{\it et al.} \cite{Part5_ming_lp2002}
     &Du~{\it et al.}\cite{Part5_ming_lidu}& Wu~{\it et al.} \cite{Part5_ming_wzz2005}& PDG~06/04  \\

    \hline
 $D^0\to K^- \pi^+$ &3.847
     & 3.72 &3.79\ ;\ 3.80 &$3.80 \pm 0.07$\\
 $\quad~\to\bar{K}^0 \pi^0$ &1.310
     &2.09 &2.27\ ;\ 2.24 & $2.28 \pm 0.22$~({\footnotesize PDG04})\\
     &  & & &$1.14 \pm 0.12$\\
 $\quad~\to\bar{K}^0 \eta$ &
     &  &0.80\ ;\ 0.81 & $0.76 \pm 0.11$~({\footnotesize PDG04})\\
      & & & &$0.38 \pm 0.06$\\
 $\quad~\to\bar{K}^0 \eta\,'$ &
     &  & 1.85\ ;\ 1.88& $1.87 \pm 0.28$~({\footnotesize PDG04})\\
      & & & &$0.91 \pm 0.14$ \\
 $\quad~\to\pi^+ \pi^-$ &0.151
     & 0.149 &0.144\ ;\  0.144 & $0.1364 \pm 0.0032$\\
 $\quad~\to\pi^0 \pi^0$ &0.115
     & 0.106&0.078\ ;\  0.097 &$0.079 \pm 0.008$ \\
 $\quad~\to K^+ K^-$ &0.424
     & 0.40 &0.413\ ;\ 0.413 &$0.384 \pm 0.010$  \\
 $\quad~\to K^0 \bar{K}^0$ &0.130
    & 0.0573 &0.069\ ;\ 0.062 &$0.071 \pm 0.019$~({\footnotesize PDG04})\\
      & & & &$0.037 \pm 0.007$ \\
 $\quad~\to K^+\pi^-$& 0.033 & 0.0141 &0.0150\ ;\ 0.0151 & $0.0143 \pm 0.0004$ \\
  $\quad~\to\eta \pi^0$ &
     &  &0.069\ ;\ 0.068  &$0.056\pm 0.014$ \\
 $\quad~\to\eta\,' \pi^0$ &
     &  &0.088\ ;\  0.091& --- \\
  $\quad~\to\eta \eta$ &
     &  &0.011\ ;\ 0.016& --- \\
 $\quad~\to\eta \eta\,'$ &
     &  &0.026\ ;\ 0.030 & --- \\
  $\quad~\to K^0\pi^0$&0.008 & 0.0284 &0.002\ ;\ 0.005 &--- \\
 $\quad~\to K^0 \eta$ &
     & &0.001\ ;\ 0.002 &--- \\
 $\quad~\to K^0 \eta\,'$ &
     &  &0.0\ ;\ 0.0&---\\
\hline  $D^+\to\bar{K}^0 \pi^+$ &2.939
     &  &2.76\ ;\ 2.76& $2.77 \pm 0.18$~({\footnotesize PDG04})\\
     &  & & &$1.47\pm 0.06$ \\
  $\quad~\to\pi^+ \pi^0$  &0.185 & 0.18 &0.25
     \ ;\ 0.19& $0.128 \pm
  0.009$ \\
  $\quad~\to\eta\pi^+$&  & &0.34\ ;\ 0.37&$0.35 \pm 0.032$\\
  $\quad~\to\eta'\pi^+$&  &  &0.45\ ;\ 0.42&$0.53 \pm 0.11$
  \\
   $\quad~\to K^+ \bar{K}^0$ &0.764
     & 0.64  &0.62 \ ;\ 0.62 & $0.58 \pm 0.06$~({\footnotesize PDG04})\\
      & & & &$0.296 \pm 0.019$ \\
  $\quad~\to K^0\pi^+$&0.053  & 0.0756 &0.012\ ;\ 0.026 & --- \\
  $\quad~\to K^+\pi^0$& 0.055 & 0.0296 &0.021\ ;\ 0.023& $<0.042$ \\
  $\quad~\to K^+\eta$&   &  &0.011\ ;\ 0.012
  & ---\\
  $\quad~\to K^+\eta'$&   &
  &0.005\ ;\ 0.006&--- \\
\hline
 $D^+_s\to\bar{K}^0 K^+$ &4.623
     &   &3.06\ ;\ 3.13&$4.4 \pm 0.9$ \\
 $\quad~\to\pi^+\eta$&  1.131 &  &1.05\ ;\ 1.09& $2.11 \pm 0.35$ \\
 $\quad~\to\pi^+\eta\,'$&   &  &4.19\ ;\ 4.43& $4.7 \pm 0.7$\\
  $\quad~\to\pi^+ K^0$ &0.373
     &  &0.24\ ;\ 0.26&  $<0.9$ \\
  $\quad~\to\pi^0 K^+$ &0.146
     &  &0.047\ ;\ 0.090&---  \\
  $\quad~\to\eta K^+$& 0.300  & &0.055\ ;\ 0.040& ---\\
  $\quad~\to\eta\,' K^+$&   &
   &0.090\ ;\ 0.102&---\\
  $\quad~\to K^+K^0$&  0.012  & &0.014\ ;\ 0.010 &---  \\
\hline
\end{tabular}
\end{center}
\end{small}
\end{table}
\begin{table}[hbp]
\caption{Predictions for $D\to PV$ branching fractions for $D\to PV$ 
($\times 10^{-2}$). Most decay modes involving a neutral K meson are 
given as $K^0_S$ in PDG06 and $\bar {K}^0$ in PDG04, which
are presented as well.
 \label{tab:predict2}} \vspace{8pt}
\begin{center}
\begin{tabular}{|l|c|c|c|l|}
\hline Decay Modes& Buccella~{\it et al.} \cite{Part5_ming_lp2002}
     & Du~{\it et al.}\cite{Part5_ming_lidu}&Wu~{\it et al.} \cite{Part5_ming_wzz2005}& PDG~06/04\\
  \hline
 $D^0\to K^{*-} \pi^+$ &4.656
     & 5.22  &5.93 \ ;\ 5.97  & $5.9 \pm 0.4$~({\footnotesize PDG04})\\
   $\quad~\to K^- \rho^+$ &11.201
     &  11.1 &9.99 \ ;\ 9.90 &$10.1 \pm 0.8$~({\footnotesize PDG04}) \\
   $\quad~\to\bar{K}^{*0} \pi^0$ &3.208
     & 2.72  &2.72\ ;\ 2.81&$2.8 \pm 0.4$~({\footnotesize PDG04}) \\
   $\quad~\to\bar{K}^0 \rho^0$ &0.759
     & 1.25&1.49\ ;\ 1.25& $1.55^{+0.12}_{-0.16}$ ~({\footnotesize PDG04}) 
\\
   $\quad~\to\bar{K}^{*0} \eta$ &
     &    &1.50\ ;\ 1.94 &$1.8 \pm 0.4$~({\footnotesize PDG04}) \\
  $\quad~\to\bar{K}^0 \omega$ &1.855
     &  &2.11\ ;\ 1.80& $2.3 \pm 0.4$~({\footnotesize PDG04})\\
 & & & &$1.1 \pm 0.2$\\
  $\quad~\to \bar{K}^0 \phi$ &
     &  &0.95 \ ;\ 0.90 & $0.94 \pm 0.11$~({\footnotesize PDG04})\\
  $\quad~\to K^+ K^{*-}$ &0.290
     &  &0.25\ ;\ 0.25&$0.20\pm0.11$ \\
  $\quad~\to K^- K^{*+}$ &0.431
     &  &0.43\ ;\ 0.43&$0.37\pm0.08$ \\
   $\quad~\to K^0 \bar{K}^{*0}$ &0.052
    &  &0.08\ ;\ 0.16& $<0.17$~({\footnotesize PDG04}) \\
     & & & & $<0.08$\\
  $\quad~\to\bar{K}^0 K^{*0}$ &0.062
     &    &0.08\ ;\ 0.16& $<0.09$~({\footnotesize PDG04}) \\
    &  & & & $<0.04$\\
  $\quad~\to\pi^0 \phi$ &0.105
     &  &0.12\ ;\  0.12&$0.074\pm 0.005$  \\
  $\quad~\to\bar{K}^{*0} \eta\,'$ &
  &  &0.004\ ;\ 0.003 & $< 0.10$~({\footnotesize PDG04}) \\
  $\quad~\to\eta \phi$ &
     &  &0.035\ ;\ 0.034&$0.014\pm0.004$ \\
  $\quad~\to\pi^+ \rho^-$ &0.485
     &  0.36 &0.34\ ;\  0.35& $0.45\pm 0.04$  \\
   $\quad~\to\pi^- \rho^+$ &0.706
     & 0.73  &0.62\ ;\ 0.61 &  $1.0\pm0.06$ \\
  $\quad~\to\pi^0 \rho^0$ &0.216
     & 0.11  &0.19\ ;\  0.16& $0.32\pm0.04$   \\
  $\quad~\to\pi^0 \omega$ &0.013
     &   &0.020\ ;\ 0.003 & $<0.026$  \\
  $\quad~\to\eta \omega$ &
     &   &0.13\ ;\ 0.10&  \\
   $\quad~\to\eta\,' \omega$ &
     &  &0.0007\ ;\ 0.0003&  \\
  $\quad~\to\eta \rho^0$ &
     &  &0.0039 \ ;\ 0.0015& \\
   $\quad~\to\eta\,' \rho^0$ &0.039
     &  &0.012\ ;\  0.009&  \\
  $\quad~\to K^{*+}\pi^-$&0.025 & &0.029\ ;\ 0.029& \\
  $\quad~\to K^+\rho^-$&0.004 & &0.016\ ;\ 0.016&  \\
  $\quad~\to K^{*0}\pi^0$& 0.008& &0.0052\ ;\ 0.0064&  \\
  $\quad~\to K^0\rho^0$& & &0.0069\ ;\ 0.0059&  \\
  $\quad~\to K^{*0}\eta$& &
   &0.0030\ ;\ 0.0041&  \\
  $\quad~\to K^{*0}\eta'$& &
   &0.0\ ;\ 0.0& \\
   $\quad~\to K^0\omega$&0.002 & &0.0076\ ;\ 0.0056&  \\
   $\quad~\to K^0\phi$&  & &0.0\ ;\ 0.0006& \\
\hline
\end{tabular}
\end{center}
\end{table}
\addtocounter{table}{-1}
\begin{table}[hbp]
\caption{({\it continued}) Predictions for $D\to PV$ branching fractions 
($\times 10^{-2}$).  Most decay modes involving a neutral K meson
are given as $K^0_S$ in PDG06 and $\bar {K}^0$ in
PDG04, which are presented as well.} \vspace{8pt}
\begin{center}
\begin{tabular}{|l|c|c|c|l|}
\hline Decay Modes& Buccella~{\it et al.} \cite{Part5_ming_lp2002}
     & Du~{\it et al.}\cite{Part5_ming_lidu}  &Wu~{\it et al.} \cite{Part5_ming_wzz2005}& PDG~06/04\\
  \hline
 $D^+\to\bar{K}^{*0} \pi^+$ &1.996
     &  1.93 &1.96\ ;\ 1.96&$1.95 \pm 0.19$~({\footnotesize PDG04}) \\
  $\quad~\to\pi^+ \phi$  & 0.619 &
  &0.64\ ;\ 0.62 & $0.65\pm 0.07$\\
 $\quad~\to\bar{K}^0 \rho^+$ &12.198
     & 7.01 &7.56 \ ;\ 8.43&$6.6 \pm 2.5$ ~({\footnotesize PDG04}) \\
  $\quad~\to\pi^+ \rho^0$ &0.104
     & 0.13  &0.088\ ;\ 0.088& $0.107\pm0.011$ \\
  $\quad~\to K^+ \bar{K}^{*0}$ &0.436
     &   &0.44\ ;\ 0.44& $0.43\pm0.06$~({\footnotesize PDG04}) \\
   $\quad~\to\bar{K}^0 K^{*+}$ &1.515
     &  &1.43 \ ;\ 1.25& $3.1\pm1.4$ ~({\footnotesize PDG04}) \\
      & & & &$1.6\pm 0.7$  \\
  $\quad~\to K^+\rho^0$&  0.029 & &0.030\ ;\ 0.025&$0.025 \pm 0.007$
  \\
   $\quad~\to K^{*0}\pi^+$&0.027 &  &0.024\ ;\ 0.022 &$0.030 \pm 0.006$  \\
    $\quad~\to K^+\phi$&   & &0.0066\ ;\ 0.0067 &$<0.013$~({\footnotesize 
PDG04})\\
   $\quad~\to \pi^+\omega$&   &
   &0.57\ ;\ 0.58& $<0.034$\\
   $\quad~\to \eta\rho^+$&   & &0.24\ ;\ 0.43& $<0.7$\\
   $\quad~\to\eta'\rho^+$&   & &0.15\ ;\ 0.15&$<0.6$
  \\
  $\quad~\to\pi^0 \rho^+$ &0.451
     & 0.31 &0.28\ ;\ 0.35&  \\
  $\quad~\to K^0\rho^+$& 0.042  & &0.025\ ;\ 0.022&   \\
   $\quad~\to\pi^0K^{*+}$&0.057 & &0.037\ ;\ 0.036&  \\
   $\quad~\to K^+\omega$&    &  &0.012\ ;\ 0.011& \\
  $\quad~\to K^{*+}\eta$&    &  &0.015\ ;\ 0.015
  & \\
   $\quad~\to K^{*+}\eta'$&    &
  &0.00014\ ;\ 0.00016 & \\
\hline  $D^+_s\to\bar{K}^{*0} K^+$ &4.812
     &  &3.34 \ ;\ 3.42&$3.3 \pm 0.9$ ~({\footnotesize PDG04})\\
  $\quad~\to\bar{K}^0 K^{*+}$ &2.467
     &  &4.98 \ ;\ 4.66& $5.3 \pm 1.3$ \\
  $\quad~\to\pi^+ \rho^0$ &
     & &0.06 \ ;\ 0.06 &$<0.07$~({\footnotesize PDG04})\\
  $\quad~\to\pi^+ \phi$ &4.552
     &   &3.08\ ;\  2.93& $4.4 \pm 0.6$ \\
   $\quad~\to\pi^+ K^{*0}$ &0.445
     &  &0.33\ ;\ 0.35 & $0.65\pm0.28$~({\footnotesize PDG04})  \\
  $\quad~\to K^+ \rho^0$ &0.198
     &   &0.12\ ;\  0.12& $0.26\pm0.07$  \\
   $\quad~\to K^+\phi$& 0.008  &  &0.032\ ;\  0.033& $<0.06$ \\
  $\quad~\to K^+ \omega$ &0.178
     &  &0.40\ ;\ 0.39 &  \\
   $\quad~\to K^0 \rho^+$ &1.288
     &   &0.91\ ;\ 0.77&  \\
   $\quad~\to\pi^0 K^{*+}$ &0.076
     &  &0.13 \ ;\  0.13&    \\
   $\quad~\to \eta K^{*+}$& 0.146 & &0.038\ ;\ 0.047& \\
   $\quad~\to\eta\,' K^{*+}$&  &
  &0.068\ ;\ 0.059&\\
  $\quad~\to K^{*0}K^+$& 0.006 & &0.0015\ ;\ 0.0015&  \\
   $\quad~\to K^{*+}K^0$&0.018  & &0.0076\ ;\ 0.0085&  \\
\hline
\end{tabular}
\end{center}
\end{table}

\subsection{$D\to SP$ Decays}
\label{SP}
 Scalar meson production measurements in charm
decays are now available from
ARGUS \cite{Part5_ming_ARGUS}, CLEO \cite{Part5_ming_CLEO}, E687 \cite{Part5_ming_E687}, E691
\cite{Part5_ming_E691}, E791 \cite{Part5_ming_E791}, FOCUS \cite{Part5_ming_FOCUS}, and BaBar
\cite{Part5_ming_BaBar}. Specifically, the decays $D\to f_0\pi(K)$, $D\to
a_0\pi(K)$, $D\to \bar K^*_0\pi$ and $D^+\to \sigma\pi^+$
have been observed in Dalitz plot analysis of three-body
decays. The results of various experiments are summarized in 
Table~\ref{ScaExp}, where the products of ${\cal B}(D\to SP_3)$ and
${\cal B}(S\to P_1P_2)$ are listed. In order to extract the
branching fractions for $D\to f_0 P$, one should use the the result
from a recent analysis \cite{Part5_ming_ANS}: $\Gamma(f_0\to
\pi\pi)=(64\pm8)$~MeV, $\Gamma(f_0\to K\bar K)=(12\pm 1)$~MeV
and $\Gamma_{f_0 ~total}=(80\pm 10)$~MeV. In this case one has
 \begin{eqnarray}
 {\cal B}(f_0(980)\to K^+K^-)=0.08\pm 0.01\,, \qquad {\cal B}(f_0(980)\to\pi^+\pi^-)=0.53\pm 0.09\,.
 \end{eqnarray}
For $D\to a_0P$, one can apply the PDG average $ \Gamma(a_0\to
K\bar K)/\Gamma(a_0\to \pi\eta)=0.183\pm 0.024 $ \cite{Part5_pdg06}
to obtain
 \begin{eqnarray} \label{aKK}
 {\cal B}(a_0^0(980)\to \pi^0\eta) &=& 0.845\pm 0.017 ,\,
 \nonumber\\
 {\cal B}(a_0^+(980)\to K^+\bar K^0) &=& {\cal B}(a_0^-(980)\to K^- 
K^0)=0.155\pm 0.017\,,
 \nonumber \\  {\cal B}(a_0^0(980)\to K^+K^-) &=& 0.078\pm 0.009\,.
 \end{eqnarray}
 For $D\to K^*_0(1430)P$, the branching fraction ${\cal B}(K^*_0(1430)
 \to K^+\pi^-)=\frac{2}{3}(0.93\pm 0.10)$ should be used~\cite{Part5_pdg06}. 
Some scalar meson decays are not listed in the
PDG tables. Precise measurements of these branching fractions would 
of great value for understanding charmed meson to scalar meson
transitions.

\begin{table}[hbp]
\caption{Experimental branching fractions of various $D\to SP$
decays measured by ARGUS, E687, E691, E791, CLEO, FOCUS and BaBar.
For simplicity and convenience, mass identifications for the
$f_0(980)$, $a_0(980)$ and $K^*_0(1430)$ have been dropped.
\label{ScaExp} }
 \begin{scriptsize}
\begin{center}
\begin{tabular}{l l l l   }
\hline
Collaboration & ${\cal B}(D\to SP)\times {\cal B}(S\to P_1P_2)$  & ${\cal B}(D\to SP)$ \\
 \hline
  PDG06& ${\cal B}(D^+\to f_0\pi^+){\cal B}(f_0\to\pi^+\pi^-)=(2.1\pm 0.5)\times 10^{-4}$ &
 ${\cal B}(D^+\to f_0\pi^+)=(4.0\pm 1.2)\times 10^{-4}$ \\
 E791 & ${\cal B}(D^+\to f_0\pi^+){\cal B}(f_0\to\pi^+\pi^-)=(1.9\pm 0.5)\times 10^{-4}$ &
 ${\cal B}(D^+\to f_0\pi^+)=(3.6\pm 1.1)\times 10^{-4}$ \\
 FOCUS & ${\cal B}(D^+\to f_0K^+){\cal B}(f_0\to K^+K^-)=(3.84\pm 0.92)\times 10^{-5}$ &
 ${\cal B}(D^+\to f_0K^+)=(4.8\pm 1.3)\times
 10^{-4}$ \\
  PDG06 & ${\cal B}(D^+\to f_0K^+){\cal B}(f_0\to \pi^+\pi^-)=(5.7\pm 3.5)\times 10^{-5}$ &
 ${\cal B}(D^+\to f_0K^+)=(1.1\pm 0.7)\times 10^{-4}$ \\
 FOCUS & ${\cal B}(D^+\to f_0K^+){\cal B}(f_0\to \pi^+\pi^-)=(6.12\pm 3.65)\times 10^{-5}$ &
 ${\cal B}(D^+\to f_0K^+)=(1.2\pm 0.7)\times 10^{-4}$ \\
 FOCUS & ${\cal B}(D^+\to a_0^0\pi^+){\cal B}(a_0^0\to
 K^+K^-)=(2.38\pm0.47)\times 10^{-3}$ & ${\cal B}(D^+\to
 a_0^0\pi^+)=(3.1\pm 0.7)\%$ \\
 E791 & ${\cal B}(D^+\to\sigma\pi^+){\cal B}(\sigma\to\pi^+\pi^-)=(1.4\pm0.3)\times 10^{-3}$ &
 ${\cal B}(D^+\to\sigma\pi^+)=(2.1\pm0.5)\times 10^{-3}$ \\
 E791 & ${\cal B}(D^+\to\kappa\pi^+){\cal B}(\kappa\to K^-\pi^+)=(4.4\pm1.2)\%$ &
 ${\cal B}(D^+\to\kappa\pi^+)=(6.5\pm1.9)\%$ \\
 E691,E687 & ${\cal B}(D^+\to\bar K_0^{*0}\pi^+){\cal B}(\bar 
K_0^{*0}\to K^-\pi^+)=(2.3\pm 0.3)\%$  &
 ${\cal B}(D^+\to\bar K_0^{*0}\pi^+)=(3.7\pm0.6)\%$ \\
 PDG06 & ${\cal B}(D^+\to\bar K_0^{*0}\pi^+){\cal B}(\bar K_0^{*0}\to 
K^-\pi^+)=(2.41\pm 0.24)\%$  &
 ${\cal B}(D^+\to\bar K_0^{*0}\pi^+)=(3.9\pm0.6)\%$ \\
 E791 & ${\cal B}(D^+\to\bar K_0^{*0}\pi^+){\cal B}(\bar K_0^{*0}\to 
K^-\pi^+)=(1.14\pm 0.16)\%$  &
 ${\cal B}(D^+\to\bar K_0^{*0}\pi^+)=(1.8\pm0.3)\%$ \\
  PDG06 & ${\cal B}(D^+\to\bar K_0^{*0}K^+){\cal B}(\bar K_0^{*0}\to 
K^-\pi^+)=(3.7\pm 0.4)\times 10^{-3}$  &
 ${\cal B}(D^+\to\bar K_0^{*0}K^+)=(6.0\pm0.9)\times 10^{-3}$ \\
PDG06 &${\cal B}(D^+\to f_0(1370)\pi^+){\cal B}(f_0(1370)\to
\pi^+\pi^-) =
 (8\pm 6)\times 10^{-5}$&  \\
 FOCUS &${\cal B}(D^+\to f_0(1370)\pi^+){\cal B}(f_0(1370)\to K^+K^-) =
 (6.2\pm 1.1)\times 10^{-4}$&  \\
  \hline
  PDG06 & ${\cal B}(D^0\to \sigma\pi^0){\cal B}(f_0\to \pi^+\pi^-)<2.7\times 10^{-5}$  &
 ${\cal B}(D^0\to \sigma\pi^0)<4.1\times 10^{-5}$ \\
  PDG06 & ${\cal B}(D^0\to f_0\pi^0){\cal B}(f_0\to \pi^+\pi^-)<3.4\times 10^{-6}$  &
 ${\cal B}(D^0\to f_0\pi^0)<6.4\times 10^{-6}$ \\
  PDG06 & ${\cal B}(D^0\to f_0K_s^0){\cal B}(f_0\to \pi^+\pi^-)=(1.36^{+0.30}_{-0.22})\times 10^{-3}$  &
 ${\cal B}(D^0\to f_0K_s^0)=(2.6^{+0.7}_{-0.6})\times 10^{-3}$ \\
 ARGUS,E687 & ${\cal B}(D^0\to f_0\bar K^0){\cal B}(f_0\to 
\pi^+\pi^-)=(3.2\pm 0.9)\times 10^{-3}$  &
 ${\cal B}(D^0\to f_0\bar K^0)=(6.0\pm  2.0)\times 10^{-3}$ \\
 CLEO & ${\cal B}(D^0\to f_0\bar K^0){\cal B}(f_0\to 
\pi^+\pi^-)=(2.5^{+0.8}_{-0.5})\times 10^{-3}$  &
 ${\cal B}(D^0\to f_0\bar K^0)=(4.7^{+1.7}_{-1.2})\times 10^{-3}$ \\
  PDG06 & ${\cal B}(D^0\to f_0 K_s^0){\cal B}(f_0\to K^+K^-)<1.0\times 10^{-4}$  &
 ${\cal B}(D^0\to f_0 K_s^0)<1.3 \times 10^{-3}$ \\
 BaBar & ${\cal B}(D^0\to f_0\bar K^0){\cal B}(f_0\to K^+K^-)=(1.2\pm 
0.9)\times 10^{-3}$  &
 ${\cal B}(D^0\to f_0\bar K^0)=(1.5\pm  1.1)\%$ \\
 PDG06 & ${\cal B}(D^0\to a_0^+K^-){\cal B}(a_0^+\to K^+ K_s^0)=(6.1\pm 1.8)\times 10^{-4}$ &
 ${\cal B}(D^0\to a_0^+K^-)=(7.9\pm2.5)\time 10^{-3}$ \\
 BaBar & ${\cal B}(D^0\to a_0^+K^-){\cal B}(a_0^+\to K^+\bar K^0)=(3.3\pm 
0.8)\times 10^{-3}$ &
 ${\cal B}(D^0\to a_0^+K^-)=(2.1\pm0.6)\%$ \\
 PDG06 & ${\cal B}(D^0\to a_0^-K^+){\cal B}(a_0^-\to K^- K_s^0)<1.1\times 10^{-4}$ &
 ${\cal B}(D^0\to a_0^-K^+)<1.4\times 10^{-3}$ \\
 BaBar & ${\cal B}(D^0\to a_0^-K^+){\cal B}(a_0^-\to K^-\bar 
K^0)=(3.1\pm1.9)\times 10^{-4}$ &
 ${\cal B}(D^0\to a_0^-K^+)=(2.0\pm1.2)\times 10^{-3}$ \\
 PDG06 & ${\cal B}(D^0\to a_0^0 K_s^0){\cal B}(a_0^0\to K^+K^-)=(3.0\pm 0.4)\times 10^{-3}$ &
 ${\cal B}(D^0\to a_0^0 K_s^0)=(3.8\pm0.7)\%$ \\
 BaBar & ${\cal B}(D^0\to a_0^0\bar K^0){\cal B}(a_0^0\to K^+K^-)=(5.9\pm 
1.3)\times 10^{-3}$ &
 ${\cal B}(D^0\to a_0^0\bar K^0)=(7.6\pm1.9)\%$ \\
 BaBar & ${\cal B}(D^0\to a_0^+\pi^-){\cal B}(a_0^+\to K^+\bar 
K^0)=(5.1\pm4.2)\times 10^{-4}$ &
 ${\cal B}(D^0\to a_0^+\pi^-)=(3.3\pm2.7)\times 10^{-3}$  \\
 BaBar & ${\cal B}(D^0\to a_0^-\pi^+){\cal B}(a_0^-\to K^-K^0)=(1.43\pm1.19)\times 10^{-4}$ &
 ${\cal B}(D^0\to a_0^-\pi^+)=(9.2\pm 7.7)\times 10^{-4}$ \\
  PDG06 & ${\cal B}(D^0\to K_0^{*-}\pi^+){\cal B}(K_0^{*-}\to K_s^0\pi^-)=(2.8^{+0.6}_{-0.4})\times 10^{-3}$ &
 ${\cal B}(D^0\to K_0^{*-}\pi^+)=(9.0^{+2.3}_{-1.7})\time 10^{-3}$ \\
 ARGUS,E687 & ${\cal B}(D^0\to K_0^{*-}\pi^+){\cal B}(K_0^{*-}\to \bar 
K^0\pi^-)=(7.3\pm 1.6)\times 10^{-3}$ &
 ${\cal B}(D^0\to K_0^{*-}\pi^+)=(1.2\pm0.3)\%$ \\
 CLEO & ${\cal B}(D^0\to K_0^{*-}\pi^+){\cal B}(K_0^{*-}\to \bar 
K^0\pi^-)=(4.3^{+1.9}_{-0.8})\times 10^{-3}$ &
 ${\cal B}(D^0\to K_0^{*-}\pi^+)=(7.0^{+3.2}_{-1.5})\times 10^{-3}$ \\
 PDG06 & ${\cal B}(D^0\to K_0^{*-}\pi^+){\cal B}(K_0^{*-}\to K^-\pi^0)=(4.6\pm 2.2)\times 10^{-3}$ &
 ${\cal B}(D^0\to K_0^{*-}\pi^+)=(1.5\pm0.7)\%$ \\
 CLEO & ${\cal B}(D^0\to K_0^{*-}\pi^+){\cal B}(K_0^{*-}\to K^-\pi^0)=(3.6\pm 0.8)\times 10^{-3}$ &
 ${\cal B}(D^0\to K_0^{*-}\pi^+)=(1.2\pm0.3)\%$ \\
 PDG06 & ${\cal B}(D^0\to \bar K_0^{*0}\pi^0){\cal B}(\bar 
K_0^{*0}\to K^-\pi^+)=
 (5.8^{+4.6}_{-1.5})\times 10^{-3}$ &
 ${\cal B}(D^0\to \bar K_0^{*0}\pi^0)=(9.4^{+7.5}_{-2.6})\times 10^{-3}$  
\\
 CLEO & ${\cal B}(D^0\to \bar K_0^{*0}\pi^0){\cal B}(\bar K_0^{*0}\to 
K^-\pi^+)=
 (5.3^{+4.2}_{-1.4})\times 10^{-3}$ &
 ${\cal B}(D^0\to \bar K_0^{*0}\pi^0)=(8.5^{+6.8}_{-2.5})\times 10^{-3}$  
\\
PDG06 & ${\cal B}(D^0\to f_0(1370)K_s^0){\cal
B}(f_0(1370)\to\pi^+\pi^-)=
 (2.5\pm0.6)\times 10^{-3}$ & \\
 PDG06 & ${\cal B}(D^0\to a_0K_s^0){\cal B}(a_0\to\eta\pi^0)=
 (6.2\pm2.0)\times 10^{-3}$ & \\
 ARGUS,E687 & ${\cal B}(D^0\to f_0(1370)\bar K^0){\cal 
B}(f_0(1370)\to\pi^+\pi^-)=
 (4.7\pm1.4)\times 10^{-3}$ & \\
 CLEO & ${\cal B}(D^0\to f_0(1370)\bar K^0){\cal 
B}(f_0(1370)\to\pi^+\pi^-)=
 (5.9^{+1.8}_{-2.7})\times 10^{-3}$ & \\
 PDG06 & ${\cal B}(D^0\to f_0(1400) K_s^0){\cal B}(f_0(1400)\to K^+K^-)=
 (1.7\pm1.1)\times 10^{-4}$ & \\
  \hline
  PDG06 & ${\cal B}(D_s^+\to f_0\pi^+){\cal B}(f_0\to K^+K^-)=(5.7\pm 2.5)\times 10^{-3}$ &
 ${\cal B}(D_s^+\to f_0\pi^+)=(7.1\pm 3.2)\%$ \\
 E687 & ${\cal B}(D_s^+\to f_0\pi^+){\cal B}(f_0\to K^+K^-)=(4.9\pm 2.3)\times 10^{-3}$ &
 ${\cal B}(D_s^+\to f_0\pi^+)=(6.1\pm 3.0)\%$ \\
 E791 & ${\cal B}(D_s^+\to f_0\pi^+){\cal B}(f_0\to \pi^+\pi^-)=(5.7\pm 1.7)\times 10^{-3}$ &
 ${\cal B}(D_s^+\to f_0\pi^+)=(1.1\pm 0.4)\%$ \\
 FOCUS & ${\cal B}(D_s^+\to f_0\pi^+){\cal B}(f_0\to \pi^+\pi^-)=(9.5\pm 2.7)\times 10^{-3}$ &
 ${\cal B}(D_s^+\to f_0\pi^+)=(1.8\pm 0.6)\%$ \\
 FOCUS & ${\cal B}(D_s^+\to f_0\pi^+){\cal B}(f_0\to K^+K^-)=(7.0\pm1.9)\times 10^{-3}$ &
 ${\cal B}(D_s^+\to f_0\pi^+)=(8.8\pm 2.6)\%$ \\
 FOCUS & ${\cal B}(D_s^+\to f_0K^+){\cal B}(f_0\to K^+K^-)=(2.8\pm1.3)\times 10^{-4}$ &
 ${\cal B}(D_s^+\to f_0K^+)=(3.5\pm1.7)\times 10^{-3}$ \\
 PDG06 & ${\cal B}(D_s^+\to \bar K_0^{*0}K^+){\cal B}(\bar 
K_0^{*0}\to K^-\pi^+)=(4.8\pm2.5)\times 10^{-3}$ &
 ${\cal B}(D_s^+\to \bar K_0^{*0}K^+)=(7.7\pm4.1)\times 10^{-3}$  \\
 E687 & ${\cal B}(D_s^+\to \bar K_0^{*0}K^+){\cal B}(\bar K_0^{*0}\to 
K^-\pi^+)=(4.3\pm2.5)\times 10^{-3}$ &
 ${\cal B}(D_s^+\to \bar K_0^{*0}K^+)=(6.9\pm4.1)\times 10^{-3}$  \\
 FOCUS & ${\cal B}(D_s^+\to K_0^{*0}\pi^+){\cal B}(K_0^{*0}\to K^+\pi^-)=(1.4\pm 0.8)\times 10^{-3}$ &
 ${\cal B}(D_s^+\to K_0^{*0}\pi^+)=(2.3\pm1.3)\times 10^{-3}$  \\
 E791 &${\cal B}(D^+_s\to f_0(1370)\pi^+){\cal B}(f_0(1370)\to\pi^+\pi^-) =
 (3.3\pm 1.2)\times 10^{-3}$&  \\
 \hline
\end{tabular}
\end{center}
\end{scriptsize}
\end{table}

The theoretical approaches to $D\to SP$ are very similar to $D\to PP$
except for the fact that the quark structure of the scalar mesons,
especially the $f_0(980)$ and $a_0(980)$, is still not 
well established. (For a recent review, see Ref.~\cite{Part5_pdg06} and 
references
therein). Thus, one faces a 'Scylla and Charybdis' dilemma 
with limited theoretical control on
{\em both} the factorizable
and nonfactorizable contributions. Still, there is no
doubt that the study of charmed meson decays will provide a new
avenue to the understanding of  light scalar meson
spectroscopy. One might resort to the knowledge gained from the
$D\to PP$ and $D\to PV$ modes and try to gain some new understanding
of some old puzzles related to the internal structure and parameters,
{\it e.g.} the masses and widths, of light scalar mesons through the
study of $D\to 
SP$~\cite{Part5_ming_0012120,Part5_ming_0109158,Part5_ming_0212117}. 
Or {\it vice versa}:  one could 
start with an assumed structure for the scalar mesons and make
predictions for the 
decays~\cite{Part5_ming_9601343,Part5_ming_0007207,Part5_ming_0304193}.
Some of the theoretical results in the literature are given in
Table~\ref{scalarBr}. In either case, the following factorization
formulae are useful:

\begin{eqnarray}
\label{fts_2}
T_S&=&-\frac{G_F}{\sqrt{2}}V_{q_1q_2}V^*_{cq_3}a_{1}f_{_P}(m^2_{_D}-m^2_{_S})F_0^{D\rightarrow S}(m^2_{_P}), \\
\label{ftp_2}
T_P&=&-\frac{G_F}{\sqrt{2}}V_{q_1q_2}V^*_{cq_3}a_{1}f_{_S}(m^2_{_D}-m^2_{_P})F_0^{D\rightarrow P}(m^2_{_S}), \\
\label{fcv_2}
C_S&=&-\frac{G_F}{\sqrt{2}}V_{q_1q_2}V^*_{cq_3}a_{2}f_{_P}(m^2_{_D}-m^2_{_S})F_0^{D\rightarrow S}(m^2_{_P}), \\
\label{fcp_2}
C_P&=&-\frac{G_F}{\sqrt{2}}V_{q_1q_2}V^*_{cq_3}a_{2}f_{_S}(m^2_{_D}-m^2_{_P})F_0^{D\rightarrow
P}(m^2_{_S}).
\end{eqnarray}

\begin{table}[hbp]
\caption{Branching fractions for various $D\to SP$ decay modes.
Experimental results are taken from Table~\ref{ScaExp}.
\label{scalarBr}
 }
\begin{center}
\begin{tabular}{|l| l| l| l |  }
\hline
Decay & Buccella~{\it et al.}\cite{Part5_ming_9601343} & Cheng~\cite{Part5_ming_0212117} & Experiment \\
 \hline
 $D^+\to f_0\pi^+$ &$2.8\times 10^{-4}$
  & $3.5\times 10^{-4}$ & $(3.6\pm 1.1)\times 10^{-4}$ \\
 \quad~ $\to f_0K^+$ &  $2.3\times 10^{-5}$ &
 $2.2\times 10^{-5}$ & $ \sim10^{-4}$ \\
 \quad~ $\to a_0^+\bar K^0$ &  $6.4\times 10^{-3}$ &
 $1.7\times 10^{-2}$ &  \\
 \quad~ $\to a_0^0\pi^+$ &
 $5.9\times 10^{-4}$ & $1.7\times 10^{-3}$ & $(3.1\pm 0.7)\%$ \\
 \quad~ $\to \sigma\pi^+$ &   & input
  & $(2.1\pm0.5)\times 10^{-3}$ \\
 \quad~ $\to\kappa\pi^+$ & & input  &
 $(6.5\pm1.9)\%$ \\
  \quad~ $\to \bar K_0^{*0}\pi^+$ &  & input
   & $(1.8\pm0.3)\%$ \\
\quad~ $\to a_0^{0}\pi^+$ & $5.9\times 10^{-4}$ &
   & \\
   \quad~ $\to a_0^{+}\pi^0$ & $1.2\times 10^{-4}$ &
   &  \\
   \quad~ $\to a_0^{+}\eta$ & $7.4\times 10^{-4}$ &
   &  \\
   \quad~ $\to a_0^{0}K^+$ & $6.2\times 10^{-5}$ &
   &  \\
   \quad~ $\to f_0K^+$ & $2.3\times 10^{-5}$ &
   &  \\
  \hline
  $D^0\to f_0\bar K^0$ &  $7.4\times 10^{-4}$
  & input   & $\sim10^{-3}- 10^{-2}$ \\
  \quad~ $\to a_0^+K^-$ &  $7.8\times 10^{-4}$ &
  $1.1\times 10^{-3}$ & $(2.1\pm0.6)\%$ \\
 \quad~ $\to a_0^0\bar K^0$ &  $2.2\times 10^{-3}$
 & $3.6\times 10^{-3}$ & $(7.6\pm1.9)\%$ \\
  \quad~ $\to a_0^-K^+$ &  $4.0\times 10^{-5}$ & $7.9\times 10^{-5}$ &
  $(2.0\pm1.2)\times 10^{-3}$ \\
  \quad~ $\to a_0^+\pi^-$ &  $3.0\times 10^{-5}$ &
  $6.5\times 10^{-5}$ & $(3.3\pm 2.7)\times 10^{-3}$ \\
  \quad~ $\to a_0^-\pi^+$ &  $7.0\times 10^{-4}$ &
  $1.3\times 10^{-3}$ & $(9.2\pm 7.7)\times 10^{-4}$ \\
  \quad~ $\to K_0^{*-}\pi^+$ &   &
  $1.1\times 10^{-2}$ &  $\sim10^{-3}- 10^{-2}$ \\
  \quad~ $\to \bar K_0^{*0}\pi^0$ &
  & $3.7\times 10^{-3}$ & $(8.5^{+6.8}_{-2.5})\times 10^{-3}$  \\
   \quad~ $\to f_0\pi^0$ &  $6.0\times 10^{-6}$ &
 &  \\
  \quad~ $\to f_0\eta$ &  $4.0\times 10^{-5}$ &
   & \\
    \quad~ $\to a_0^0\pi^0$ &  $1.1\times 10^{-4}$ &
   &  \\
    \quad~ $\to a_0^0\eta$ &  $1.5\times 10^{-4}$ &
  &  \\
  \hline
  $D_s^+\to f_0\pi^+$ &$1.1\%$  & input
    & $(1.8\pm0.6)\%$ \\
  \quad~ $\to f_0K^+$ &  $6.9\times 10^{-4}$ &
   $1.2\times 10^{-3}$ &  $(3.5\pm1.7)\times 10^{-3}$ \\
  \quad~ $\to \bar K_0^{*0}K^+$ &
  & $1.5\times 10^{-3}$ & $(6.9\pm4.1)\times 10^{-3}$ \\
  \quad~ $\to K_0^{*0}\pi^+$ &
   & $1.1\times 10^{-3}$ & $(2.3\pm1.3)\times 10^{-3}$  \\
   \quad~ $\to a^+_0K^0$ &$3.0\times 10^{-5}$  &
    &  \\
\quad~ $\to a^0_0K^+$ &$7.0\times 10^{-5}$  &
    &  \\
 \quad~    $\to a^+_0\eta$ &$7.0\times 10^{-5}$  &
    &  \\
  \hline
\end{tabular}
\end{center}
\end{table}

\subsection{$D\to AP$ Decays}
\label{AP}
 There are two different types of axial vector
mesons: $^3P_1$ and $^1P_1$, with quantum numbers
$J^{\rm PC}=1^{++}$ and $1^{+-}$, respectively. The isovector
non-strange axial vector mesons $a_1(1260)$ and $b_1(1235)$, which
correspond to $^3P_1$ and $^1P_1$, respectively, cannot 
mix because of their opposite $G$-parities. However, the
isodoublet strange mesons $K_{1}(1270)$ and $K_{1}(1400)$ are a
mixture of $^3P_1$ and $^1P_1$ states due to the strange and
non-strange light-quark mass difference. One usually expresses these as
\begin{eqnarray}
K_1(1270)&=&K_{1A}\sin\theta+K_{1B}\cos\theta \ , \nonumber\\
K_1(1400)&=&K_{1A}\cos\theta-K_{1B}\sin\theta \ ,
\end{eqnarray}
where $K_{1A}$ and $K_{1B}$ are the strange partners of the
$a_1(1260)$ and $b_1(1235)$, respectively.

Two-body hadronic $D\to AP$ decays have been studied in
\cite{Part5_ming_Kamal,Part5_ming_XYPham,Part5_ming_TNPham,Part5_ming_Kamal94,Part5_ming_Katoch,Part5_ming_Lipkin,Part5_ming_0301198}. 
In the factorization approximation, the decay amplitudes are
\begin{eqnarray}
\label{afta}
T_A&=&2\frac{G_F}{\sqrt{2}}V_{q_1q_2}V^*_{cq_3}a_{1}f_{_P}m_{_A}(\varepsilon^*\cdot p_{_D})A^{D\rightarrow A}_0(m^2_{_P}), \\
\label{aftp}
T_P&=&2\frac{G_F}{\sqrt{2}}V_{q_1q_2}V^*_{cq_3}a_{1}f_{_A}m_{_A}(\varepsilon^*\cdot p_{_D})F^{D\rightarrow P}_+(m^2_{_A}), \\
\label{afcv}
C_A&=&2\frac{G_F}{\sqrt{2}}V_{q_1q_2}V^*_{cq_3}a_{2}f_{_P}m_{_A}(\varepsilon^*\cdot p_{_D})A^{D\rightarrow A}_0(m^2_{_P}), \\
\label{afcp}
C_P&=&2\frac{G_F}{\sqrt{2}}V_{q_1q_2}V^*_{cq_3}a_{2}f_{_A}m_{_A}(\varepsilon^*\cdot
p_{_D})F^{D\rightarrow P}_+(m^2_{_A}).
\end{eqnarray}
When compared with experimental data,
the predicted branching fractions derived from the factorizable 
contributions for
$D^0\to K^-a_1^+$ and $D^0\to K_1^-(1270)\pi^+$ and those for
$D^0\to\bar K^0 a_1^+$ and $D^+\to \bar K_1^0(1400)\pi^+$ in 
Refs.~\cite{Part5_ming_Kamal,Part5_ming_XYPham,Part5_ming_TNPham,Part5_ming_Kamal94,Part5_ming_Katoch} 
are found to be too small by roughly factors of 5 and 2, respectively.
One explanation is that the factorization approach
may only be suitable for energetic two-body decays such as $D\to PP$
and $D\to PV$; for $D\to AP$ there is very little energy release and the
approximation is questionable, since the nonperturbative
contributions are large. A recent analysis~\cite{Part5_ming_0301198}
that considers the sizable FSI effects indicated that the predictions,
which are presented in Table~\ref{ap1} and Table~\ref{ap2}, are
improved greatly.

\begin{table}[hbp]
\caption{Branching fractions for $D\to Ka_1(1260)$ and $D\to
Kb_1(1235)$. Decay modes involving a neutral K meson are
given as $K^0_S$ in PDG06 and $\bar {K}^0$ in PDG04
which are presented as well. \label{ap1}
 }
\begin{center}
\begin{tabular}{l c c c}
\hline &  \multicolumn{2}{c}{Theory \cite{Part5_ming_0301198}}  \\
\cline{2-3}
 \raisebox{1.5ex}[0cm][0cm]{Decay}
& without FSI & with FSI &
 \raisebox{1.5ex}[0cm][0cm]{Experiment \cite{Part5_pdg06}} \\
\hline
 $D^+\to \bar K^0a_1^+(1260)$ & 12.1\% & 12.1\% &  $(3.6\pm0.6)\%$   \\
 & & & $(8.2\pm1.7)\%$~({\footnotesize PDG04})  \\
 $D^0\to K^-a_1^+(1260)$ & 3.8\% & 6.2\% & $(7.5\pm 1.1)\%$ \\
 $D^0\to \bar K^0a_1^0(1260)$ & $3.3\times 10^{-4}$ & $5.6\times 10^{-4}$ 
& $<1.9\%$ \\
 \hline
 $D^+\to \bar K^0b_1^+(1235)$ & $1.7\times 10^{-3}$ & $1.7\times 10^{-3}$ 
&  \\
 $D^0\to K^-b_1^+(1235)$ & $3.7\times 10^{-6}$ & $5.9\times 10^{-6}$ & \\
 $D^0\to \bar K^0b_1^0(1235)$ & $3.9\times 10^{-4}$ & $6.7\times 10^{-4}$ 
&  \\
 \hline
\end{tabular}
\end{center}
\end{table}

\begin{table}[hbp]
\caption{Branching fractions for $D\to K_1(1270)\pi$ and $D\to
K_1(1400)\pi$ calculated for various $K_{1A}-K_{1B}$ mixing
angles. \label{ap2}
 }
\begin{center}
\begin{tabular}{lcccccc}
\hline &  \multicolumn{4}{c}{Theory \cite{Part5_ming_0301198}}  \\
\cline{2-5}
 \raisebox{1.5ex}[0cm][0cm]{Decay}
& $-37^\circ$ & $-58^\circ$ & $37^\circ$ & $58^\circ$ &
 \raisebox{1.5ex}[0cm][0cm]{Experiment \cite{Part5_pdg06}} \\
 \hline
 $D^+\to \bar K_1^0(1270)\pi^+$ & $6.4\times 10^{-3}$ & $7.8\times 
10^{-3}$
 & $2.9\%$ & $4.7\%$ & $<7\times 10^{-3}$ \\
 $D^+\to \bar K_1^0(1400)\pi^+$ & $2.9\%$ & $4.0\%$
 & $6.6\%$  & $6.6\%$ & $(4.3\pm1.5)\%$ \\
 $D^0\to K^-_1(1270)\pi^+$ & $6.3\times 10^{-3}$ & $5.5\times 10^{-3}$
 & $4.9\times 10^{-4}$  & $4.4\times 10^{-5}$ & $(1.12\pm 0.31)\%$ \\
 $D^0\to K^-_1(1400)\pi^+$ & $3.7\times 10^{-8}$ & $4.2\times 10^{-4}$
 & $3.0\times 10^{-3}$ & $3.2\times 10^{-3}$ & $<1.2\%$ \\
 $D^0\to \bar K^0_1(1270)\pi^0$ & $8.4\times 10^{-3}$ & $8.4\times 
10^{-3}$
 & $8.4\times 10^{-3}$ & $8.4\times 10^{-3}$ &  \\
 $D^0\to \bar K^0_1(1400)\pi^0$ & $5.7\times 10^{-3}$ & $5.5\times 
10^{-3}$
 & $5.7\times 10^{-3}$ & $5.5\times 10^{-3}$ & $<3.7\%$ \\
 \hline
\end{tabular}
\end{center}
\end{table}
\subsection{$D\to TP$ Decays}
\label{TP}

 The $J^P=2^+$ tensor mesons $f_2(1270)$,
$f_2'(1525)$, $a_2(1320)$ and $K_2^*(1430)$ form a $SU(3)$
$1\,^3P_2$ nonet with quark content $q\bar q$. Hadronic charm
decays to a pseudoscalar meson and a tensor meson $f_2(1270)$,
$a_2(1320)$ or $K^*_2(1430)$ were found in early experiments by
ARGUS~\cite{Part5_ming_ARGUS} and E687~\cite{Part5_ming_E687}, and
more recently by E791~\cite{Part5_ming_E791}, CLEO~\cite{Part5_ming_CLEO}, 
FOCUS~\cite{Part5_ming_FOCUS} and BaBar~\cite{Part5_ming_BaBar}, 
although  some of these measurements do not have
a compelling statistical significance. The results from various
experiments are summarized in Table~\ref{TenExp}, where the
products of ${\cal B}(D\to TP_3)$ and ${\cal B}(T\to P_1P_2)$ are
shown. It is evident that most of the $D\to TP$ decays
that are listed have
branching fractions of order $10^{-3}$, even though some of them are
Cabibbo-suppressed. In order to extract the branching ratios for
$D\to T P$ decays, one must use the branching fractions for the
strong decays of the tensor mesons~\cite{Part5_pdg06}:
 \begin{eqnarray}
  &&{\cal B}(f_2(1270)\to\pi\pi) =
  (84.7^{+2.5}_{-1.2})\%, \qquad~~
 {\cal B}(f_2(1270)\to K\bar K) =(4.6\pm 0.4)\%,
 \nonumber   \\ && {\cal B}(a_2(1320)\to K\bar K)=(4.9\pm 0.8)\%, \qquad
 {\cal B}(K_2^*(1430)\to K\pi)=(49.9\pm 1.2)\%.\\
 \end{eqnarray}

Theoretical calculations based on the factorization 
hypothesis~\cite{Part5_ming_KatVer,Part5_ming_MuRoLo,Part5_ming_0303195} 
are listed in Table~\ref{preTP}, where one sees that most of the 
theoretical
predictions are not consistent with experimental data. At
first glance, some decays like $D\to \bar K_2^*(1430)K$ and
$D^0\to f_2'(1525)\bar K^0$ etc., appear to be kinematically
not allowed since the total mass of the final state particles lies
outside of the phase space for the decay. Nevertheless, they
are possible because the tensor mesons have widths of
order several hundred MeV~\cite{Part5_pdg06}.

\begin{table}[hbp]
\caption{Experimental branching fractions for various $D\to TP$
decays measured by BaBar, CLEO, E791, FOCUS and PDG06. For
simplicity and convenience, we have dropped mass
identifications for the $f_2(1270)$, $a_2(1320)$ and $K^*_2(1430)$.
\label{TenExp}
 }
 \begin{scriptsize}
\begin{center}
\begin{tabular}{l l l l   }
\hline
Collaboration & ${\cal B}(D\to TP)\times {\cal B}(T\to P_1P_2)$  & ${\cal B}(D\to TP)$ \\
 \hline
 PDG06&${\cal B}(D^+\to f_2\pi^+){\cal B}(f_2\to\pi^+\pi^-)=(4.8\pm 1.3)\times 10^{-4}$ &
 ${\cal B}(D^+\to f_2\pi^+)=(8.5\pm 2.3)\times 10^{-4}$ \\
 E791 & ${\cal B}(D^+\to f_2\pi^+){\cal B}(f_2\to\pi^+\pi^-)=(6.0\pm 1.1)\times 10^{-4}$ &
 ${\cal B}(D^+\to f_2\pi^+)=(1.1\pm 0.2)\times 10^{-3}$ \\
 FOCUS & ${\cal B}(D^+\to f_2\pi^+){\cal B}(f_2\to\pi^+\pi^-)=(3.8\pm 0.8)\times 10^{-5}$ &
 ${\cal B}(D^+\to f_2\pi^+)=(6.8\pm 1.4)\times 10^{-4}$ \\
 FOCUS & ${\cal B}(D^+\to f_2\pi^+){\cal B}(f_2\to K^+K^-)=(7.0\pm 1.9)\times 10^{-5}$ &
 ${\cal B}(D^+\to f_2\pi^+)=(3.1\pm 0.9)\times 10^{-3}$ \\
 PDG06&${\cal B}(D^+\to K^{*0}_2\pi^+){\cal B}(K_2^{*0}\to K^+\pi^-)=(5.2\pm 3.5)\times 10^{-5}$ &
 ${\cal B}(D^+\to K^{*0}_2\pi^+)=(1.6\pm 1.1)\times 10^{-4}$ \\
 E791 & ${\cal B}(D^+\to \bar K^{*0}_2\pi^+){\cal B}(\bar K_2^{*0}\to 
K^-\pi^+)=(4.6\pm 2.0)\times 10^{-4}$ &
 ${\cal B}(D^+\to \bar K^{*0}_2\pi^+)=(1.4\pm 0.6)\times 10^{-3}$ \\
 PDG06 &    &${\cal B}(D^+\to a_2^+K_S^{0})<1.5\times 10^{-3}$ \\
 \hline
 PDG06 & ${\cal B}(D^0\to f_2 K_S^0){\cal B}(f_2\to\pi^+\pi^-)=(1.3^{+1.1}_{-0.7})\times 10^{-4}$ &
 ${\cal B}(D^0\to f_2K_S^0)=(2.3^{+2.0}_{-1.3})\times 10^{-4}$ \\
 CLEO & ${\cal B}(D^0\to f_2\bar K^0){\cal 
B}(f_2\to\pi^+\pi^-)=(1.6^{+2.4}_{-1.3})\times 10^{-4}$ &
 ${\cal B}(D^0\to f_2\bar K^0)=(2.8^{+4.3}_{-2.3})\times 10^{-3}$ \\
 BaBar & ${\cal B}(D^0\to a_2^-\pi^+){\cal B}(a_2^-\to K^0K^-)=(3.5\pm2.1)\times 10^{-5}$ &
 ${\cal B}(D^0\to a_2^-\pi^+)=(7.0\pm 4.3)\times 10^{-4}$ \\
 PDG06 & ${\cal B}(D^0\to K^{*-}_2\pi^+){\cal B}(K_2^{*-}\to K_S^0\pi^-)=(3.2^{+2.1}_{-1.1})\times 10^{-4}$ &
 ${\cal B}(D^0\to K^{*-}_2\pi^+)=(2.0^{+1.3}_{-0.7})\times 10^{-3}$ \\
 CLEO & ${\cal B}(D^0\to K^{*-}_2\pi^+){\cal B}(K_2^{*-}\to 
\bar K^0\pi^-)=(6.5^{+4.2}_{-2.2})\times 10^{-4}$ &
 ${\cal B}(D^0\to K^{*-}_2\pi^+)=(2.0^{+1.3}_{-0.7})\times 10^{-3}$ \\
 BaBar & ${\cal B}(D^0\to K^{*+}_2 K^-){\cal B}(K_2^{*+}\to K^0\pi^+)=(6.8\pm 4.2)\times 10^{-4}$ &
 ${\cal B}(D^0\to K^{*+}_2 K^-)=(2.0\pm 1.3)\times 10^{-3}$ \\
 BaBar & ${\cal B}(D^0\to \bar K^{*0}_2 K^0){\cal B}(\bar K_2^{*0}\to 
K^-\pi^+)=(6.6\pm 2.7)\times 10^{-4}$ &
 ${\cal B}(D^0\to \bar K^{*0}_2 K^0)=(2.0\pm 0.8)\times 10^{-3}$ \\
 PDG06 &    &${\cal B}(D^0\to a_2^+K^{-})<2\times 10^{-3}$ \\
 \hline
 PDG06 & ${\cal B}(D_s^+\to f_2\pi^+){\cal B}(f_2\to\pi^+\pi^-)=(1.2\pm 0.7)\times 10^{-3}$ &
 ${\cal B}(D_s^+\to f_2\pi^+)=(2.1\pm 1.3)\times 10^{-3}$ \\
  E791 & ${\cal B}(D_s^+\to f_2\pi^+){\cal B}(f_2\to\pi^+\pi^-)=(2.0\pm 0.7)\times 10^{-3}$ &
 ${\cal B}(D_s^+\to f_2\pi^+)=(3.5\pm 1.2)\times 10^{-3}$ \\
 FOCUS & ${\cal B}(D_s^+\to f_2\pi^+){\cal B}(f_2\to\pi^+\pi^-)=(1.0\pm 0.3)\times 10^{-3}$ &
 ${\cal B}(D_s^+\to f_2\pi^+)=(1.8\pm 0.5)\times 10^{-3}$ \\
 FOCUS & ${\cal B}(D_s^+\to f_2 K^+){\cal B}(f_2\to\pi^+\pi^-)=(2.0\pm 1.3)\times 10^{-4}$ &
 ${\cal B}(D_s^+\to f_2 K^+)=(3.5\pm 2.3)\times 10^{-4}$ \\
 \hline
\end{tabular}
\end{center}
\end{scriptsize}
\end{table}

\begin{table}[hbp]
\caption{Branching fractions for various $D\to TP$ decays.
Experimental results are taken from Table~\ref{TenExp}.
\label{preTP}
 }
 \begin{scriptsize}
\begin{center}
\begin{tabular}{l c c c c c}
\hline
 \raisebox{-1.5ex}[0cm][0cm]
{Decay} & \raisebox{-1.5ex}[0cm][0cm]{Katoch~{\it et
al.}\cite{Part5_ming_KatVer}}&\raisebox{-1.5ex}[0cm][0cm]{Mu$\tilde 
{\rm n}$oz~{\it
et al.}\cite{Part5_ming_MuRoLo}}
& \multicolumn{2}{c}{Cheng \cite{Part5_ming_0303195}}& \raisebox{-1.5ex}[0cm][0cm]{${\rm Experiment}$} \\
\cline{4-5}
 &    &    &  ${\rm without~FSIs}$ & ${\rm with~FSIs}$ &   \\ \hline
 $D^+\to f_2(1270)\pi^+$ & &$7.97\times 10^{-6}$
 & $2.9\times 10^{-5}$ & $2.2\times 10^{-4}$ & $(0.9\pm 0.1)\times 10^{-3}$ \\
 $D^0\to f_2(1270)\pi^0$ & &$2.47\times 10^{-7}$ &
&  &  \\
 $D^0\to f_2(1270)\bar K^0$ & $9.0\times 10^{-5}$ & &
 $1.0\times 10^{-4}$ & $2.5\times 10^{-4}$ & $(4.5\pm 1.7)\times 10^{-3}$ \\
 $D_s^+\to f_2(1270)\pi^+$ & $3.6\times 10^{-4}$ & &
 $6.6\times 10^{-5}$ & $2.1\times 10^{-3}$ & $(2.1\pm 0.5)\times 10^{-3}$ \\
 $\quad~\to f_2(1270)K^+$ &  &
 & $5.2\times 10^{-6}$ & $4.9\times 10^{-5}$
 & $(3.5\pm 2.3)\times 10^{-4}$ \\
 \hline
 $D^+\to f_2'(1525)\pi^+$ & &$7.18\times 10^{-9}$
 & $1.4\times 10^{-6}$ & $3.7\times 10^{-6}$ &  \\
  $D^0\to f_2'(1525)\pi^0$ & &$2.18\times 10^{-10}$
 & &  &  \\
 $D^0\to f_2'(1525)\bar K^0$ & & &
 $2.5\times 10^{-7}$ & $6.0\times 10^{-7}$ &  \\
 $D_s^+\to f_2'(1525)\pi^+$ & $1.3\times 10^{-2}$ & &
 $1.6\times 10^{-4}$ & $1.5\times 10^{-4}$ &  \\
 $\quad~\to f_2'(1525)K^+$ &  &
 & $4.9\times 10^{-6}$ & $7.5\times 10^{-6}$
 &  \\
 \hline
 $D^+\to a_2^+(1320)\pi^0$ & &$9.05\times 10^{-7}$ &  &
  &  \\
  $D^+\to a_2^0(1320)\pi^+$ & &$5.55\times 10^{-6}$ &  &
  &  \\
 $D^+\to a_2^+(1320)\bar K^0$ & $1.1\times 10^{-4}$ & & $1.3\times 
10^{-6}$ &
 $1.3\times 10^{-6}$ & $<3\times 10^{-3}$ \\
 $D^0\to a_2^-(1320)\pi^+$ & & $4.21\times 10^{-6}$& $5.7\times 10^{-6}$ & $6.1\times 10^{-6}$ &
 $(7.0\pm 4.3)\times 10^{-4}$  \\
$\quad~\to a_2^0(1320)\pi^0$ & & $1.72\times 10^{-7}$&  & &
   \\
 $\quad~\to a_2^+(1320)K^-$ & $0$ & &0 & $8.9\times 10^{-8}$ & $<2\times 10^{-3}$ \\
 $\quad~\to a_2^0(1320)\bar K^0$ & $1.7\times 10^{-5}$ & & &  & \\
 \hline
 $D^+\to \bar K_2^{*0}(1430)\pi^+$ & $9.9\times 10^{-3}$ & & $2.6\times 
10^{-4}$ &
 $2.6\times 10^{-4}$ & $(1.4\pm 0.6)\times 10^{-3}$ \\
 $D^0\to K_2^{*-}(1430)\pi^+$ & $4.1\times 10^{-3}$ & &$1.0\times 10^{-4}$ & $1.1\times 10^{-4}$
 & $(2.0^{+1.3}_{-0.7})\times 10^{-3}$ \\
 $\quad~\to \bar K_2^{*0}(1430)\pi^0$ & 0& & 0 & $1.3\times 10^{-5}$
 & $<3.4\times 10^{-3}$ \\
 $\quad~\to K_2^{*+}(1430)K^-$ & & & 0 & $1.3\times 10^{-6}$
 & $(2.0\pm 1.3)\times 10^{-3}$   \\
 $\quad~\to \bar K_2^{*0}(1430)K^0$ & &
 & 0 & $\sim 10^{-8}$ & $(2.0\pm 0.8)\times 10^{-3}$ \\
 $D_s^+\to \bar K_2^{*0}(1430)K^+$ & 0& &  &
 &  \\
 $\quad~\to \bar K_2^{*+}(1430)\bar K^0$ &$4.2\times 10^{-5}$ &
 & & & \\
 \hline
\end{tabular}
\end{center}
\end{scriptsize}
\end{table}

\subsection{Other Decay Modes}
Measurements of other nonleptonic two-body modes, such as $D\to
AV$ etc., have been reported. The PDG table lists  two
fairly strong decay modes~\cite{Part5_pdg06}: ${\cal B}(D^+\to
\bar{K}^{*0}a_1(1260)^+)=(9.4\pm1.9)\times 10^{-3}$ and
${\cal B}(D_s^+\to \phi a_1(1260)^+)=(2.9\pm0.7)\%$, even though the
total mass of the final state mesons exceeds the available phase
spaces. Given the relevant form factors, the branching fractions
can be worked out in the factorization approach. However, one
doubts the reliability of factorization for
these modes because the nonfactorized corrections
may be quite large at such small momentum transfers.
\section{Three-Body Decays}
\label{mb}
\subsection{Kinematics and Dalitz Plot}

Starting from Eq.~\ref{n-bparwid} and integrating over the
solid angles, the decay rate for $D\to M_1M_2M_3$ can be obtained
\begin{eqnarray}
d\Gamma=\frac{1}{(2\pi)^3}\frac{1}{32m_{D}^3}|{\cal
A}|^2dm^2_{{12}}dm^2_{{23}} \ ,
\end{eqnarray}
where $m_{_{ij}}$ is the invariant mass of particles $i$ and $j$.
For a given value of $m^2_{12}$ in the range $(m_1+m_2)^2\leq
m^2_{12}\leq (m_D-m_3)^2$, the upper and lower bounds of
$m^2_{23}$ are determined
\begin{eqnarray}
(m^2_{{23}})_{max}&=&(E^*_2+E^*_3)^2-(\sqrt{E^{*2}_2-m^2_2}-\sqrt{E^{*2}_3-m^2_3})^2
\ , \\
(m^2_{{23}})_{max}&=&(E^*_2+E^*_3)^2-(\sqrt{E^{*2}_2-m^2_2}+\sqrt{E^{*2}_3-m^2_3})^2
\ .
\end{eqnarray}
Here $E^*_2$ and $E^*_3$ are the respective
energies of final state mesons $M_2$
and $M_3$ in the rest frames of $M_1$ and $M_2$:,
\begin{eqnarray}
E^*_2&=&\frac{m^2_{12}-m^2_1+m^2_2}{2m_{12}} \ , \\
E^*_3&=&\frac{m^2_{D}-m^2_{12}-m^2_2}{2m_{12}} \ .
\end{eqnarray}
The scatter plot of $m^2_{12}$ versus $m^2_{23}$ is called the
Dalitz plot. For a detailed introduction of Dalitz plot techniques,
please refer to Chapter 4 and Ref.~\cite{Part5_ming_Dalitz}. The amplitude
$|{\cal A}|^2$ of a nonresonant decay is parameterized as a constant
without variation in magnitude or phase across the Dalitz plot,
in which case the allowed region of the plot is uniformly populated
with events. A nonuniformity with bands near the mass of the
resonance in the plot will reflect the influence of a resonance
contribution. One can find a review of Dalitz plot applications 
specific to charm decays in Ref.~\cite{Part5_ming_CharmDali}.

\subsection{Resonant Three-Body Decays}
Charmed meson three-body decays proceed dominantly via
quasi-two-body decays containing an intermediate resonance state
that subsequently decays into two particles. The analysis of these 
resonant decays using Dalitz plot techniques enables one to study the
dynamical properties of various resonances. In theoretical
studies, resonant decays are often divided into the product of two
subprocesses: ${\cal B}(D\to RM_3)\times {\cal B}(R\to M_1M_2)$, just
as we have shown in Section \ref{2B}. In this case we reduce the
multi-body decay into a pair of two-body decays.

\subsection{Nonresonant Three-Body Decays}

The nonresonant contribution is usually a small fraction of the
total three-body decay rate. Experimentally, they are hard to
measure since the interference between nonresonant and
quasi-two-body amplitudes makes it difficult to disentangle these
two distinct contributions and, then, extract the nonresonant one.
Theoretically, the matrix element for $D$ decaying into three mesons
in general has two different formalisms in the factorization
approximation that differ on how the three final mesons
are distributed into two ``clusters".

For one type with a ``cluster" where $D$ transits to a light meson,
one has
 \begin{eqnarray}
 \langle M_1M_2M_3|J_{i\mu}J^{\prime\mu}_i|D\rangle \sim \langle
 M_1M_2|J_{i\mu}|0\rangle\langle
 M_3|J^{\prime\mu}_i|D\rangle .
 \end{eqnarray}
 Here it is evident that its contribution is negligibly small since 
the matrix element $\langle M_1M_2|J_{i\mu}|0\rangle$, which also appears 
in the factorizable contributions
 of weak annihilation in two-body decays,  vanishes in the chiral limit.

 For the other type of ``clustering" where $D$ transits
to two light mesons, the factorized formula is
 \begin{eqnarray}
 \langle M_1M_2M_3|J_{i\mu}J^{\prime\mu}_i|D\rangle \sim \langle
 M_1|J_{i\mu}|0\rangle\langle
 M_2M_3|J^{\prime\mu}_i|D\rangle ,
 \end{eqnarray}
where a matrix element $\langle
M_2M_3|J^{\prime\mu}_i|D\rangle$ is introduced. This has the general
form~\cite{Part5_ming_Lee}
\begin{eqnarray}
 \langle M_2(p_{_2})M_3(p_{_3})|J^{\prime\mu}_i|D(p_{_D})\rangle &=&
 ir(p_{_D}-p_{_2}-p_{_3})^\mu+i\omega_+(p_{_2}+p_{_3})^\mu + i\omega_-(p_{_3}-p_{_2})^\mu
\nonumber \\
 &&+h\epsilon^{\mu\nu\alpha\beta}p_{_D\nu} (p_{_2}+p_{_3})_\alpha
 (p_{_3}-p_{_2})_\beta \ ,
 \end{eqnarray}
where $r$, $\omega_\pm$ and $h$ are form factors. In general these
receive two distinct contributions: one from the point-like weak
transition and the other from the pole diagrams that involve
four-point strong vertices. Models based on chiral symmetry and
heavy quark effective theory have been developed to make some
estimates of
them~\cite{Part5_ming_Lee,Part5_ming_9309301,Part5_ming_9710422}.

Charmed meson to three pseudoscalar nonresonant decays have been
studied in the approach of an effective $SU(4)_L\times SU(4)_R$
chiral Lagrangian
\cite{Part5_ming_Singer,Part5_ming_KP,Part5_ming_Cheng86,Part5_ming_CC90,Part5_ming_Botella}. 
For these, the predictions of the branching ratios are in general
smaller than experimental measiurements.  With the advent of heavy meson
chiral perturbation theory (HMChPT)
\cite{Part5_ming_Yan,Part5_ming_Wise,Part5_ming_Burdman},
nonresonant $D$ decays can be studied reliably at least in the
kinematical region where the final pseudoscalar mesons are soft
\cite{Part5_ming_Zhang,Part5_ming_Ivanov,Part5_ming_cheng02}. 
Some theoretical results are collected in Table~\ref{3BNon}.

 \vskip 0.4cm
\begin{table}[hbp]
\caption{Branching fractions (in \%) for nonresonant three-body $D$
decays from various models. Most decay modes involving a neutral K
meson are given as $K^0_S$ in PDG06 and $\bar {K}^0$
in PDG04, which are presented as well. \label{3BNon}}
\begin{footnotesize}
\begin{center}
\begin{tabular}{|l|c|c|c|l|}
\hline
 Decay mode & Chau~{\it et al.} \cite{Part5_ming_CC90} &
Botella~{\it et al.} \cite{Part5_ming_Botella} & Cheng~{\it et
al.}\cite{Part5_ming_cheng02} & PDG~06/04
\\ \hline
 $D^0\to\bar K^0\pi^+\pi^-$  & 0.13&0.19
 &0.03\ ;\ 0.17 & $0.026^{+0.059}_{-0.016}$   \\
  & & & &$0.054^{+0.120}_{-0.034}$~({\footnotesize $pdg04$})   \\
 $\quad~\to K^-\pi^+\pi^0$ &0.18
  & 0.76&0.61\ ;\ 0.28 &$1.13^{+0.54}_{-0.20}$
  \\
  $\quad~\to \bar K^0K^+K^-$  & 0.02 &0.006 &0.16\ ;\
  0.01 &   \\
  $\quad~\to \pi^+\pi^-\pi^0$ &0.04 &0.11&  &  \\
  $\quad~\to K^+K^-\pi^0$ & &0.013&  &  \\
  $\quad~\to K^0K^-\pi^+$ & &0.007&  & $0.11\pm 0.11$ \\
   & & & &$0.23\pm 0.23$~({\footnotesize $pdg04$})   \\
  $\quad~\to \bar K^0K^+\pi^-$  & &0.013&  & $0.19^{+0.11}_{-0.08}$ \\
   & & & &$0.38^{+0.23}_{-0.19}$~~~~({\footnotesize $pdg04$})   \\
  $\quad~\to\bar K^0\pi^0\pi^0$  & &
 & & $0.42\pm 0.11$   \\
  & & & &$0.85\pm 0.22$~({\footnotesize $pdg04$})   \\
  \hline
  $D^+\to \bar K^0\pi^+\pi^0$  &0.76
   & 1.9&1.5;0.7 & $0.9\pm 0.7$ \\
    & & & &$1.3\pm 1.1$~~~({\footnotesize $pdg04$})   \\
  $\quad~\to K^-\pi^+\pi^+$  &1.71  & 0.95&6.5\ ;\ 1.6 &
  $9.0\pm0.7$ \\
  $\quad~\to\pi^+\pi^+\pi^-$  &0.15
   &0.19 &0.50\ ;\ 0.067 &
   \\
  $\quad~\to K^-K^+\pi^+$  &0.02  &
   0.016& 0.48\ ;\ 0.004 &
   \\
  $\quad~\to K^+\pi^+\pi^-$   &
   &0.0032 & & \\
   $\quad~\to K^+K^+K^-$   &
   &$1.58\times10^{-5}$ & & \\
$\quad~\to\pi^+\eta\eta$ &
   &0.016\ ;\ 0.032 & &  \\
   $\quad~\to\pi^+\eta\eta'$ &
   &0.032 & &  \\
   \hline
  $D_s^+\to K^-K^+\pi^+$  & 0.42 & 0.32  &1.0\ ;\ 0.69 &
   \\
  $\quad~\to\pi^+\pi^+\pi^-$  & $5\times 10^{-5}$ & $4.7\times10^{-4}$& &
   \\
  $\quad~\to \pi^+\pi^0\eta$ &
   &1.1\ ;\ 0.95 & & $< 5$\\
 $\quad~\to \pi^+\pi^0\eta'$ &
   &0.158 & & $<1.8$\\
 $\quad~\to K^+\pi^+\pi^-$ &
   &0.047 & &$0.1\pm 0.04$ \\
   \hline
\end{tabular}
\end{center}
\end{footnotesize}
\end{table}

\subsection{Beyond Three-Body Decays}
Some multi-body charm meson decays, up to seven-body, have
been experimentally measured~\cite{Part5_pdg06}. 
However, the available
theoretical tools lose much of their power when applied to genuine
multi-body transitions. The kinematic structure and strong
dynamics becomes more-and-more complicated and ultimately gets out of
control when the number of final-state particles increases.

\chapter[Charm baryon production and decays]{Charm baryon production and
decays\footnote{By Hai-Yang Cheng}}
\label{sec:charm_baryon}

\section{Introduction}

In the past years many new excited charmed baryon states have been
discovered by BaBar, Belle and CLEO. In particular, $B$ factories
have provided a very rich source of charmed baryons both from $B$
decays and from the continuum $e^+e^-\to c\bar c$. A new chapter
for the charmed baryon spectroscopy is opened by the rich mass
spectrum and the relatively narrow widths of the excited states.
Experimentally and theoretically, it is important to identify the
quantum numbers of these new states and understand their
properties. Since the pseudoscalar mesons involved in the strong
decays of charmed baryons are soft, the charmed baryon system
offers an excellent ground for testing the ideas and predictions
of heavy quark symmetry of the heavy quark and chiral symmetry of
the light quarks.

The observation of the lifetime differences among the charmed
mesons $D^+,~D^0$ and charmed baryons is very interesting since it
was realized very early that the naive parton model gives the same
lifetimes for all heavy particles containing a heavy quark $Q$,
while experimentally, the lifetimes of $\Xi_c^+$ and $\Omega_c^0$
differ by a factor of six ! This implies the importance of the
underlying mechanisms such as $W$-exchange and Pauli interference
due to the identical quarks produced in the heavy quark decay and
in the wavefunction of the charmed baryons. With the advent of
heavy quark effective theory, it was recognized in early nineties
that nonperturbative corrections to the parton picture can be
systematically expanded in powers of $1/m_Q$.  Within the
QCD-based heavy quark expansion framework, some phenomenological
assumptions can be turned into some coherent and quantitative
statements and nonperturbative effects can be systematically
studied.

Contrary to the significant progress made over the last 20 years
or so in the studies of the heavy meson weak decay, advancement in
the arena of heavy baryons is relatively slow. Nevertheless, the
experimental measurements of the charmed baryon hadronic weak
decays have been pushed to the Cabibbo-suppressed level. Many new
data emerged can be used to test a handful of phenomenological
models available in the literature. Apart from the complication
due to the presence of three quarks in the baryon, a major
disparity between charmed baryon and charmed meson decays is that
while the latter is usually dominated by factorizable amplitudes,
the former receives sizable nonfactorizable contributions from
$W$-exchange diagrams which are not subject to color and helicity
suppression. Besides the dynamical models, there are also some
considerations based on the symmetry argument and the quark
diagram scheme.

The exclusive semileptonic decays of charmed baryons like
$\Lambda_c^+\to\Lambda e^+(\mu^+)\nu_e$, $\Xi_c^+\to \Xi^0
e^+\nu_e$ and $\Xi_c^0\to \Xi^-e^+\nu_e$ have been observed
experimentally. Their rates depend on the heavy baryon to the
light baryon transition form factors. Experimentally, the only
information available so far is the form-factor ratio measured in
the semileptonic decay $\Lambda_c\to\Lambda e\bar{\nu}$.

Although radiative decays are well measured in the charmed meson
sector, e.g. $D^*\to D\gamma$ and $D_s^+\to D_s^+\gamma$, only
three of the radiative modes in the charmed baryon sector have
been observed, namely, $\Xi'^0_c\to\Xi_c^0\gamma$, $\Xi'^+_c\to
\Xi^+_c\gamma$ and $\Omega_c^{*0}\to\Omega_c^0\gamma$. Charm
flavor is conserved in these electromagnetic charmed baryon
decays.  However, it will be difficult to measure the rates of
these decays because these states are too narrow to be
experimentally resolvable.  There are also charm-flavor-conserving
weak radiative decays such as $\Xi_c \to \Lambda_c \gamma$ and
$\Omega_c \to \Xi_c \gamma$. In these decays, weak radiative
transitions arise from the diquark sector of the heavy baryon
whereas the heavy quark behaves as a ``spectator". The
charm-flavor-violating weak radiative decays, e.g.,
$\Lambda_c^+\to\Sigma^+\gamma$ and $\Xi_c^0\to\Xi^0\gamma$, arise
from the $W$-exchange diagram accompanied by a photon emission
from the external quark.

Two excellent review articles on charmed baryons can be found in
Refs.~\cite{Korner94,Bigireview}.

\section{Production of charmed baryons at \bes3}
Production and decays of the charmed baryons can be studied at
\bes3 once its center-of-mass energy $\sqrt{s}$  is upgraded to
the level above 4.6 GeV. In order to estimate the number of
charmed baryon events produced at \bes3, it is necessary to know
its luminosity, the cross section $\sigma(e^+e^-\to c\bar c)$ at
the energies of interest and the fragmentation function of the $c$
quark into the charmed baryon. 

\section{Spectroscopy}

Charmed baryon spectroscopy provides an ideal place for studying
the dynamics of the light quarks in the environment of a heavy
quark. The charmed baryon of interest contains a charmed quark and
two light quarks, which we will often refer to as a diquark. Each
light quark is a triplet of the flavor SU(3). Since ${\bf 3}\times
{\bf 3}={\bf \bar 3}+{\bf 6}$, there are two different SU(3)
multiplets of charmed baryons: a symmetric sextet {\bf 6} and an
antisymmetric antitriplet ${\bf \bar 3}$. For the ground-state
$s$-wave baryons in the quark model, the symmetries in the flavor
and spin of the diquarks are correlated. Consequently, the diquark
in the flavor-symmetric sextet has spin 1, while the diquark in
the flavor-antisymmetric antitriplet has spin 0. When the diquark
combines with the charmed quark, the sextet contains both spin
$1/2$ and spin 3/2 charmed baryons. However, the antitriplet
contains only spin 1/2 ones. More specifically, the $\Lambda_c^+$,
$\Xi_c^+$ and $\Xi_c^0$ form a ${\bf \bar 3}$ representation and
they all decay weakly. The $\Omega_c^0$, $\Xi'^+_c$, $\Xi'^0_c$
and $\Sigma_c^{++,+,0}$ form a {\bf 6} representation; among them,
only $\Omega_c^0$ decays weakly. Note that we follow the Particle
Data Group (PDG)~\cite{Part5_pdg06} to use a prime to distinguish the
$\Xi_c$ in the {\bf 6} from the one in the ${\bf \bar 3}$.

The lowest-lying orbital excited baryon states are the $p$-wave
charmed baryons with their quantum numbers listed in Table
\ref{tab:pwave}. Although the separate spin angular momentum
$S_\ell$ and orbital angular momentum $L_\ell$ of the light
degrees of freedom are not well defined, they are included for
guidance from the quark model. In the heavy quark limit, the spin
of the charmed quark $S_c$ and the total angular momentum of the
two light quarks $J_\ell=S_\ell+L_\ell$ are separately conserved.
It is convenient to use them to enumerate the spectrum of states.
There are two types of $L_\ell=1$ orbital excited charmed baryon
states: states with  the unit of orbital angular momentum between
the diquark and the charmed quark, and states with the unit of
orbital angular momentum between the two light quarks. The orbital
wave function of the former (latter) is symmetric (antisymmetric)
under the exchange of two light quarks. To see this, one can
define two independent relative momenta ${\bf k}={1\over 2}({\bf
p}_1-{\bf p}_2)$ and ${\bf K}={1\over 2}({\bf p}_1-{\bf p}_2-2{\bf
p}_c)$ from the two light quark momenta ${\bf p}_1$, ${\bf p}_2$
and the heavy quark momentum ${\bf p}_c$. (In the heavy quark
limit, ${\bf p}_c$ can be set to zero.) Denoting the quantum
numbers $L_k$ and $L_K$ as the eigenvalues of ${\bf L}_k^2$ and
${\bf L}_K^2$, the $k$-orbital momentum $L_k$ describes relative
orbital excitations of the two light quarks, and the $K$-orbital
momentum $L_K$ describes orbital excitations of the center of the
mass of the two light quarks relative to the heavy quark
\cite{Korner94}. The $p$-wave heavy baryon can be either in the
$(L_k=0,L_K=1)$ $K$-state or the $(L_k=1,L_K=0)$ $k$-state. It is
obvious that the orbital $K$-state ($k$-state) is symmetric
(antisymmetric) under the interchange of ${\bf p}_1$ and ${\bf
p}_2$.

\begin{table}[h]
\caption{The $p$-wave charmed baryons and their quantum numbers,
where $S_\ell$ ($J_\ell$) is the total spin (angular momentum) of
the two light quarks. The quantum number in the subscript labels
$J_\ell$. The quantum number in parentheses is referred to the
spin of the baryon. In the quark model, the upper (lower) four
multiplets have even (odd) orbital wave functions under the
permutation of the two light quarks. That is, $L_\ell$ for the
former is referred to the orbital angular momentum between the
diquark and the charmed quark, while $L_\ell$ for the latter is
the orbital angular momentum between the two light quarks. The
explicit quark model wave functions for $p$-wave charmed baryons
can be found in \cite{Pirjol}.} \label{tab:pwave}
\begin{center}
\begin{tabular}{|c|cccc||c|cccc|} \hline
~~~~~State~~~~~ & SU(3) & ~~$S_\ell$~~ & ~~$L_\ell$~~&
~~$J_\ell^{P_\ell}$~~ & ~~~~~State~~~~~ & SU(3) & ~~$S_\ell$~~ &
~~$L_\ell$~~&
~~$J_\ell^{P_\ell}$~ \\
 \hline
 $\Lambda_{c1}({1\over 2},{3\over 2})$ & ${\bf \bar 3}$ & 0 & 1 &
 $1^-$ & $\Xi_{c1}({1\over 2},{3\over 2})$ & ${\bf \bar 3}$ & 0 & 1 & $1^-$ \\
 $\Sigma_{c0}({1\over 2})$ & ${\bf 6}$ & 1 & 1& $0^-$ &
 $\Xi'_{c0}({1\over 2})$ & ${\bf 6}$ & 1 & 1& $0^-$ \\
 $\Sigma_{c1}({1\over 2},{3\over 2})$ & ${\bf 6}$ & 1 & 1 & $1^-$
 &  $\Xi'_{c1}({1\over 2},{3\over 2})$ & ${\bf 6}$ & 1 & 1 &
 $1^-$\\
 $\Sigma_{c2}({3\over 2},{5\over 2})$ & ${\bf 6}$ & 1 & 1 & $2^-$
 &  $\Xi'_{c2}({3\over 2},{5\over 2})$ & ${\bf 6}$ & 1 & 1 &
 $2^-$\\  \hline
 $\tilde \Sigma_{c1}({1\over 2},{3\over 2})$ & ${\bf 6}$ & 0 & 1 & $1^-$
 &  $\tilde\Xi'_{c1}({1\over 2},{3\over 2})$ & ${\bf 6}$ & 0 & 1 & $1^-$
 \\
 $\tilde\Lambda_{c0}({1\over 2})$ & ${\bf \bar 3}$ & 1 & 1 & $0^-$ &
 $\tilde\Xi_{c0}({1\over 2})$ & ${\bf \bar 3}$ & 1 & 1 & $0^-$  \\
 $\tilde\Lambda_{c1}({1\over 2},{3\over 2})$ & ${\bf \bar 3}$ & 1 & 1 & $1^-$
 &  $\tilde\Xi_{c1}({1\over 2},{3\over 2})$ & ${\bf \bar 3}$ & 1 & 1 &
 $1^-$ \\
 $\tilde\Lambda_{c2}({3\over 2},{5\over 2})$ & ${\bf \bar 3}$ & 1 & 1 & $2^-$
 &  $\tilde\Xi_{c2}({3\over 2},{5\over 2})$ & ${\bf \bar 3}$ & 1 & 1 & $2^-$ \\
 \hline
\end{tabular}
\end{center}
\end{table}

\begin{figure}[t]
\centerline{\psfig{file=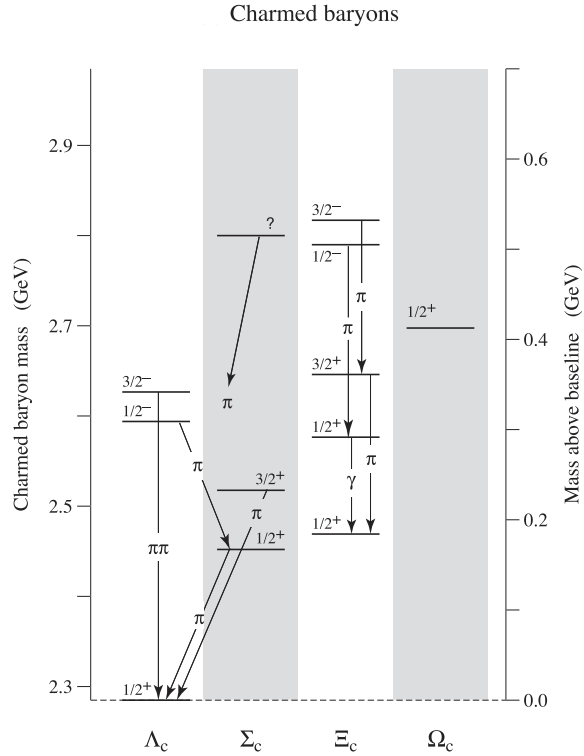,width=3.0in}}
\caption{Charmed baryons and some of their orbital excitations
\cite{Part5_pdg06}.} \label{fig:charmspect}
\end{figure}

The observed mass spectra and decay widths of charmed baryons are
summarized in Table \ref{tab:spectrum} (see also Fig.
\ref{fig:charmspect}).  $B$ factories have provided a very rich
source of charmed baryons both from $B$ decays and from the
continuum $e^+e^-\to c\bar c$.  For example, several new excited
charmed baryon states  such as
$\Lambda_c(2765)^+,\Lambda_c(2880)^+,\Lambda_c(2940)^+$,
$\Xi_c(2815),\Xi_c(2980)$ and $\Xi_c(3077)$  have been measured
recently and they are not still not in the list of 2006 Particle
Data Group~\cite{Part5_pdg06}. By now, the $J^P={1\over 2}^+$ and ${1\over
2}^-$ ${\bf \bar 3}$ states: ($\Lambda_c^+$, $\Xi_c^+,\Xi_c^0)$,
($\Lambda_c(2593)^+$, $\Xi_c(2790)^+,\Xi_c(2790)^0)$, and
$J^P={1\over 2}^+$ and ${3\over 2}^+$ ${\bf 6}$ states:
($\Omega_c,\Sigma_c,\Xi'_c$), ($\Omega_c^*,\Sigma_c^*,\Xi'^*_c$)
are established. Notice that except for the parity of the lightest
$\Lambda_c^+$, none of the other $J^P$ quantum numbers given in
Table \ref{tab:spectrum} has been measured. One has to rely on the
quark model to determine the $J^P$ assignments.

\begin{table}[!]
\caption{Mass spectra and decay widths (in units of MeV) of
charmed baryons. Experimental values are taken from the Particle
Data Group~\cite{Part5_pdg06} except $\Lambda_c(2880)$, $\Lambda_c(2940)$,
$\Xi_c(2980)^{+,0}$, $\Xi_c(3077)^{+,0}$ and $\Omega_c(2768)$ for
which we use the most recent available BaBar and Belle
measurements.} \label{tab:spectrum}
\begin{center}
\begin{tabular}{|c|c|c|c|c|} \hline \hline
~~State~~ & ~~Quark content~~ & ~~$J^P$~~ & ~~~~~~~~~Mass~~~~~~~~~ & ~~~~Width~~~~ \\
\hline
 $\Lambda_c^+$ & $udc$ & ${1\over 2}^+$ & $2286.46\pm0.14$ & \\
 \hline
 $\Lambda_c(2593)^+$ & $udc$ & ${1\over 2}^-$ & $2595.4\pm0.6$ & $3.6^{+2.0}_{-1.3}$ \\
 \hline
 $\Lambda_c(2625)^+$ & $udc$ & ${3\over 2}^-$ & $2628.1\pm0.6$ & $<1.9$ \\
 \hline
 $\Lambda_c(2765)^+$ & $udc$ & $?^?$ & $2766.6\pm2.4$ & $50$ \\
 \hline
 $\Lambda_c(2880)^+$ & $udc$ & ${5\over 2}^+$ & $2881.5\pm0.3$ & $5.5\pm0.6$ \\
 \hline
 $\Lambda_c(2940)^+$ & $udc$ & $?^?$ & $2938.8\pm1.1$ & $13.0\pm5.0$ \\ \hline
 $\Sigma_c(2455)^{++}$ & $uuc$ & ${1\over 2}^+$ & $2454.02\pm0.18$ & $2.23\pm0.30$ \\
 \hline
 $\Sigma_c(2455)^{+}$ & $udc$ & ${1\over 2}^+$ & $2452.9\pm0.4$ & $<4.6$ \\
 \hline
 $\Sigma_c(2455)^{0}$ & $ddc$ & ${1\over 2}^+$ & $2453.76\pm0.18$ & $2.2\pm0.4$ \\
 \hline
 $\Sigma_c(2520)^{++}$ & $uuc$ & ${3\over 2}^+$ & $2518.4\pm0.6$ & $14.9\pm1.9$ \\
 \hline
 $\Sigma_c(2520)^{+}$ & $udc$ & ${3\over 2}^+$ & $2517.5\pm2.3$ & $<17$ \\
 \hline
 $\Sigma_c(2520)^{0}$ & $ddc$ & ${3\over 2}^+$ & $2518.0\pm0.5$ & $16.1\pm2.1$ \\
 \hline
 $\Sigma_c(2800)^{++}$ & $uuc$ & $?^?$ & $2801^{+4}_{-6}$ & $75^{+22}_{-17}$ \\
 \hline
 $\Sigma_c(2800)^{+}$ & $udc$ & $?^?$ & $2792^{+14}_{-5}$ & $62^{+60}_{-40}$ \\
 \hline
 $\Sigma_c(2800)^{0}$ & $ddc$ & $?^?$ & $2802^{+4}_{-7}$ & $61^{+28}_{-18}$ \\
 \hline
 $\Xi_c^+$ & $usc$ & ${1\over 2}^+$ & $2467.9\pm0.4$ & \\ \hline
 $\Xi_c^0$ & $dsc$ & ${1\over 2}^+$ & $2471.0\pm0.4$ & \\ \hline
 $\Xi'^+_c$ & $usc$ & ${1\over 2}^+$ & $2575.7\pm3.1$ & \\ \hline
 $\Xi'^0_c$ & $dsc$ & ${1\over 2}^+$ & $2578.0\pm2.9$ & \\ \hline
 $\Xi_c(2645)^+$ & $usc$ & ${3\over 2}^+$ & $2646.6\pm1.4$ & $<3.1$ \\
 \hline
 $\Xi_c(2645)^0$ & $dsc$ & ${3\over 2}^+$ & $2646.1\pm1.2$ & $<5.5$ \\
 \hline
 $\Xi_c(2790)^+$ & $usc$ & ${1\over 2}^-$ & $2789.2\pm3.2$ & $<15$ \\
 \hline
 $\Xi_c(2790)^0$ & $dsc$ & ${1\over 2}^-$ & $2791.9\pm3.3$ & $<12$ \\
 \hline
 $\Xi_c(2815)^+$ & $usc$ & ${3\over 2}^-$ & $2816.5\pm1.2$ & $<3.5$ \\
 \hline
 $\Xi_c(2815)^0$ & $dsc$ & ${3\over 2}^-$ & $2818.2\pm2.1$ & $<6.5$ \\
 \hline
 $\Xi_c(2980)^+$ & $usc$ & $?^?$ & $2971.1\pm1.7$ & $25.2\pm3.0$ \\
 \hline
 $\Xi_c(2980)^0$ & $dsc$ & $?^?$ & $2977.1\pm9.5$ & $43.5$ \\
 \hline
 $\Xi_c(3077)^+$ & $usc$ & $?^?$ & $3076.5\pm0.6$ & $6.2\pm1.1$ \\
 \hline
 $\Xi_c(3077)^0$ & $dsc$ & $?^?$ & $3082.8\pm2.3$ & $5.2\pm3.6$ \\
 \hline
 $\Omega_c^0$ & $ssc$ & ${1\over 2}^+$ & $2697.5\pm2.6$ & \\
 \hline
 $\Omega_c(2768)^0$ & $ssc$ & ${3\over 2}^+$ & $2768.3\pm3.0$ & \\
 \hline \hline
\end{tabular}
\end{center}
\end{table}

In the following we discuss some of the new excited charmed baryon
states:
 \begin{itemize}
 \item
The highest $\Lambda_c(2940)^+$ was first discovered by BaBar in
the $D^0p$ decay mode \cite{BaBar:Lamc2940} and  confirmed by
Belle in the decays $\Sigma_c^0\pi^+,\Sigma_c^{++}\pi^-$ which
subsequently decay into $\Lambda_c^+\pi^+\pi^-$
\cite{Belle:Lamc2880,Mizuk}. The state $\Lambda_c(2880)^+$ first
observed by CLEO \cite{CLEO:Lamc2880} in $\Lambda_c^+\pi^+\pi^-$
was also seen by BaBar in the $D^0p$ spectrum
\cite{BaBar:Lamc2940}. It was originally conjectured that, based
on its narrow width, $\Lambda_c(2880)^+$ might be a
$\tilde\Lambda^+_{c0}({1\over 2})$ state \cite{CLEO:Lamc2880}.
Recently, Belle has studied the experimental constraint on the
$J^P$ quantum numbers of $\Lambda_c(2880)^+$
\cite{Belle:Lamc2880}. The angular analysis of
$\Lambda_c(2880)^+\to\Sigma_c^{0,++}\pi^\pm$ indicates that
$J=5/2$ is favored over $J=1/2$ or $3/2$, while the study of the
resonant structure of $\Lambda_c(2880)^+\to\Lambda_c^+\pi^+\pi^-$
implies the existence of the $\Sigma_c^*\pi$ intermediate states
and
$\Gamma(\Sigma_c^*\pi^\pm)/\Gamma(\Sigma_c\pi^\pm)=(24.1\pm6.4^{+1.1}_{-4.5})\%$.
This value is in agreement with heavy quark symmetry predictions
\cite{IW} and favors the $5/2^+$ over the $5/2^-$
assignment.\footnote{Strictly speaking, the argument in favor of
the $5/2^+$ assignment is reached in \cite{Belle:Lamc2880} by
considering only the $F$-wave contribution and neglecting the
$P$-wave contribution to $\Lambda_c(2880)^+\to\Sigma_c^*\pi$ (see
\cite{Cheng06} for more discussions).}
Therefore, it is not a $\tilde\Lambda^+_{c2}({5\over 2})$ state
either. Since $J_\ell=2,S_\ell=0,L=2$ for the diquark system of
$\Lambda_c(2880)^+$, this is the first observation of a $d$-wave
charmed baryon. It is interesting to notice that, based on the
diquark idea, the assignment $J^P=5/2^+$ has already been
predicted in \cite{Selem} for the state $\Lambda_c(2880)$ before
the Belle experiment.
As for $\Lambda_c(2980)^+$, it was recently argued that it is an
exotic molecular state of $D^{*0}$ and $p$ \cite{He}.

\item The new charmed strange baryons $\Xi_c(2980)^+$ and
$\Xi_c(3077)^+$ that decay into $\Lambda_c^+K^-\pi^+$ were first
observed by Belle \cite{Belle:Xic2980} and confirmed by BaBar
\cite{BaBar:Xic2980}. In the recent BaBar measurement
\cite{BaBar:Xic2980}, the $\Xi_{c}(2980)^+$ is found to decay
resonantly through the intermediate state $\Sigma_c(2455)^{++}K^-$
with $4.9\,\sigma$ significance and non-resonantly to
$\Lambda_c^+K^-\pi^+$ with $4.1\,\sigma$ significance. With
$5.8\,\sigma$ significance, the $\Xi_c(3077)^+$ is found to decay
resonantly through $\Sigma_c(2455)^{++}K^-$, and with
$4.6\,\sigma$ significance, it is found to decay through
$\Sigma_c(2520)^{++}K^-$. The significance of the signal for the
non-resonant decay $\Xi_c(3077)^+\rightarrow\Lambda_c^+K^-\pi^+$
is $1.4\,\sigma$.

\item The highest isotriplet charmed baryons
$\Sigma_c(2800)^{++,+,0}$ decaying into $\Lambda_c^+\pi$ were
first measured by Belle \cite{Belle:Sigc2800}.  They are most
likely to be the $J^P=3/2^-$ $\Sigma_{c2}$ states because the
$\Sigma_{c2}({3\over 2})$ baryon decays principally into the
$\Lambda_c\pi$ system in a $D$-wave, while $\Sigma_{c1}({3\over
2})$ decays mainly into the two pion system $\Lambda_c\pi\pi$. The
state $\Sigma_{c0}({1\over 2})$ can decay into $\Lambda_c\pi$ in
an $S$-wave, but it is very broad with width of order 406 MeV
\cite{Cheng06}. Experimentally, it will be very difficult to
observe it.

\item The new $3/2^+$ $\Omega_c(2768)$ was recently observed by
BaBar in the electromagnetic decay
$\Omega_c(2768)\to\Omega_c\gamma$ \cite{BaBar:Omegacst}. With this
new observation, the $3/2^+$ sextet is finally completed.

\item Evidence of double charm states has been reported by SELEX
in $\Xi_{cc}(3519)^+\to\Lambda_c^+K^-\pi^+$ \cite{Selex02}.
Further observations of $\Xi_{cc}^{++}\to\Lambda_c^+K^-\pi^+\pi^+$
and $\Xi_{cc}^+\to pD^+K^-$ were also announced by SELEX
\cite{Selex04}. However, none of the double charm states
discovered by SELEX has been confirmed by FOCUS, BaBar
\cite{BaBar:dc} and Belle \cite{Mizuk} despite the $10^6$
$\Lambda_c$ events produced in $B$ factories versus 1630
$\Lambda_c$ events observed at SELEX.

\end{itemize}

Charmed baryon spectroscopy has been studied extensively in
various models. The interested readers are referred to
Ref.~\cite{theorycharmspect} for further references. In heavy quark
effective theory, the mass splittings between spin-${3\over 2}$
and spin-${1\over 2}$ sextet charmed baryon multiplets are
governed by the chromomagnetic interactions so that
 \begin{eqnarray}
 m_{\Sigma_c^*}-m_{\Sigma_c}=m_{\Xi'^*_c}-m_{\Xi'_c}=m_{\Omega_c^*}-m_{\Omega_c},
 \end{eqnarray}
 up to corrections of $1/m_c$. This relation is borne out by experiment:
 $m_{\Sigma_c^{*+}}-m_{\Sigma_c^+}=64.6\pm2.3$ MeV, $m_{\Xi'^{*+}_c}-m_{\Xi'^+_c}=70.9\pm3.4$ MeV and
 $m_{\Omega_c^*}-m_{\Omega_c}=70.8\pm1.5$ MeV.

\section{Strong decays}

Due to the rich mass spectrum and the relatively narrow widths of
the excited states, the charmed baryon system offers an excellent
ground for testing the ideas and predictions of heavy quark
symmetry and light flavor SU(3) symmetry. The pseudoscalar mesons
involved in the strong decays of charmed baryons such as
$\Sigma_c\to\Lambda_c\pi$ are soft. Therefore, heavy quark
symmetry of the heavy quark and chiral symmetry of the light
quarks will have interesting implications for the low-energy
dynamics of heavy baryons interacting with the Goldstone bosons.

The strong decays of charmed baryons are most conveniently
described by the heavy hadron chiral Lagrangians in which heavy
quark symmetry and chiral symmetry are incorporated
\cite{Part5_ming_Yan,Part5_ming_Wise}. The Lagrangian involves two coupling constants
$g_1$ and $g_2$ for $P$-wave transitions between $s$-wave and
$s$-wave baryons \cite{Part5_ming_Yan}, six couplings $h_{2}-h_7$ for the
$S$-wave transitions between $s$-wave and $p$-wave baryons, and
eight couplings $h_{8}-h_{15}$ for the $D$-wave transitions
between $s$-wave and $p$-wave baryons \cite{Pirjol}. The general
chiral Lagrangian for heavy baryons coupling to the pseudoscalar
mesons can be expressed compactly in terms of superfields. We will
not write down the relevant Lagrangians here; instead the reader
is referred to Eqs. (3.1) and (3.3) of Ref.~\cite{Pirjol}.
Nevertheless, we list some of the partial widths derived from the
Lagrangian \cite{Pirjol}:
 \begin{eqnarray} \label{eq:swavecoupling}
 \Gamma(\Sigma_c^*\to \Sigma_c\pi)={g_1^2\over 2\pi
 f_\pi^2}\,{m_{\Sigma_c}\over m_{\Sigma_c^*}}p_\pi^3, &&
 \Gamma(\Sigma_c\to \Lambda_c\pi)={g_2^2\over 2\pi
 f_\pi^2}\,{m_{\Lambda_c}\over m_{\Sigma_c}}p_\pi^3, \nonumber \\
 \Gamma(\Lambda_{c1}(1/2)\to \Sigma_c\pi)={h_2^2\over 2\pi
 f_\pi^2}\,{m_{\Sigma_c}\over m_{\Lambda_{c1}}}E_c^2\pi_\pi, &&
 \Gamma(\Sigma_{c0}(1/2)\to \Lambda_c\pi)={h_3^2\over 2\pi
 f_\pi^2}\,{m_{\Lambda_c}\over m_{\Sigma_{c0}}}E_\pi^2p_\pi,
 \nonumber \\
 \Gamma(\Sigma_{c1}(1/2)\to \Sigma_c\pi)={h_4^2\over 4\pi
 f_\pi^2}\,{m_{\Sigma_c}\over m_{\Sigma_{c1}}}E_\pi^2p_\pi, &&
 \Gamma(\tilde\Sigma_{c1}(1/2)\to \Sigma_c\pi)={h_5^2\over 4\pi
 f_\pi^2}\,{m_{\Sigma_c}\over m_{\tilde\Sigma_{c1}}}E_\pi^2p_\pi, \nonumber \\
 \Gamma(\tilde\Xi_{c0}(1/2)\to \Xi_c\pi)={h_6^2\over 2\pi
 f_\pi^2}\,{m_{\Xi_c}\over m_{\tilde\Xi_{c0}}}E_\pi^2p_\pi, &&
 \Gamma(\tilde\Lambda_{c1}(1/2)\to \Sigma_c\pi)={h_7^2\over 2\pi
 f_\pi^2}\,{m_{\Sigma_c}\over m_{\tilde\Lambda_{c1}}}E_\pi^2p_\pi,\nonumber \\
 \end{eqnarray}
where $p_\pi$ is the pion's momentum and $f_\pi=132$ MeV.
Unfortunately, the decay $\Sigma_c^*\to\Sigma_c\pi$ is
kinematically prohibited since the mass difference between
$\Sigma_c^*$ and $\Sigma_c$ is only of order 65 MeV. Consequently,
the coupling $g_1$ cannot be extracted directly from the strong
decays of heavy baryons.

\subsection{Strong decays of $s$-wave charmed baryons}
In the framework of heavy hadron chiral pertrubation theory
(HHChPT), one can use some measurements as input to fix the
coupling $g_2$ which, in turn, can be used to predict the rates of
other strong decays. We shall use $\Sigma_c\to\Lambda_c\pi$ as
input~\cite{Part5_pdg06}
 \begin{eqnarray}
\Gamma(\Sigma_c^{++})=\Gamma(\Sigma_c^{++}\to\Lambda_c^+\pi^+) &=&
2.23\pm0.30\,{\rm MeV}.
 \end{eqnarray}
>From which we obtain
 \begin{eqnarray} \label{eq:g2}
 |g_2|=0.605^{+0.039}_{-0.043}\,,
 \end{eqnarray}
where we have neglected the tiny contributions from
electromagnetic decays. Note that $|g_2|$ obtained from
$\Sigma_c^0\to\Lambda_c^+\pi^-$ has the same central value as Eq.
(\ref{eq:g2}) except that the errors are slightly large. If
$\Sigma_c^*\to\Lambda_c\pi$ decays are employed as input, we will
obtain $|g_2|=0.57\pm0.04$ from
$\Sigma_c^{*++}\to\Lambda_c^+\pi^+$ and $0.60\pm0.04$ from
$\Sigma_c^{*0}\to\Lambda_c^+\pi^-$. Hence, it is preferable to use
the measurement of $\Sigma_c^{++}\to\Lambda_c^+\pi^+$ to fix
$|g_2|$.\footnote{For previous efforts of extracting $g_2$ from
experiment using HHChPT, see \cite{Cheng97,Pirjol}. }

As pointed out in \cite{Part5_ming_Yan}, within in the framework of the
non-relativistic quark model, the couplings $g_1$ and $g_2$ can be
related to $g_A^q$, the axial-vector coupling in a single quark
transition of $u\to d$, via
 \begin{eqnarray}
 g_1={4\over 3}g_A^q, \qquad\qquad g_2=\sqrt{2\over 3}g_A^q.
 \end{eqnarray}
Using $g_A^q=0.75$ which is required to reproduce the correct
value of $g_A^N=1.25$, we obtain
  \begin{eqnarray}
 g_1=1, \qquad\qquad g_2=0.61\,.
 \end{eqnarray}
Hence, the quark model prediction is in good agreement with
experiment, but deviates $2\sigma$ from the large-$N_c$ argument:
$|g_2|=g_A^N/\sqrt{2}=0.88\,$ \cite{Guralnik}. Applying
(\ref{eq:g2}) leads to (see also Table \ref{tab:strongdecayS})
 \begin{eqnarray}
\Gamma(\Xi_c^{'*+})=\Gamma(\Xi_c^{'*+}\to\Xi_c^+\pi^0,\Xi_c^0\pi^+)
&=& {g_2^2\over 4\pi
 f_\pi^2}\left({1\over 2}{m_{\Xi_c^+}\over m_{\Xi'^+_c}}p_\pi^3+
 {m_{\Xi_c^0}\over m_{\Xi'^+_c}}p_\pi^3\right)=
(2.8\pm 0.4)\,{\rm MeV},   \nonumber \\
\Gamma(\Xi_c^{'*0})=\Gamma(\Xi_c^{'*0}\to\Xi_c^+\pi^-,\Xi_c^0\pi^0)
&=& {g_2^2\over 4\pi
 f_\pi^2}\left({m_{\Xi_c^+}\over m_{\Xi'^0_c}}p_\pi^3+{1\over 2}
 {m_{\Xi_c^0}\over m_{\Xi'^0_c}}p_\pi^3\right)=
(2.9\pm 0.4)\,{\rm MeV}.\nonumber \\
 \end{eqnarray}
Note that we have neglected the effect of $\Xi_c-\Xi'_c$ mixing in
calculations (for recent considerations, see \cite{Boyd,Ito}).
Therefore, the predicted total width of $\Xi_c^{'*+}$ is in the
vicinity of the current limit $\Gamma(\Xi_c^{'*+})<3.1$ MeV
\cite{CLEOb}.

\begin{table}[t]
\caption{Decay widths (in units of MeV) of $s$-wave charmed
baryons. Theoretical predictions of \cite{Tawfiq} are taken from
Table IV of \cite{Ivanov}.} \label{tab:strongdecayS}
\begin{scriptsize}
\begin{center}
\begin{tabular}{|c|c|c|c|c|c|c|} \hline \hline
~~~~~~~Decay~~~~~~~ & HHChPT & Tawfiq & Ivanov & Huang & Albertus
&
Expt.  \\
& & et al. \cite{Tawfiq} &
 et al. \cite{Ivanov} &  et al. \cite{Huang95} & ~~et al. 
\cite{Albertus}~~ &~~\cite{Part5_pdg06}~~ \\
\hline
 $\Sigma_c^{++}\to\Lambda_c^+\pi^+$ & input & $1.51\pm0.17$ & $2.85\pm0.19$
 &  2.5  & $2.41\pm0.07$ & $2.23\pm0.30$ \\ \hline
 $\Sigma_c^{+}\to\Lambda_c^+\pi^0$ & $2.6\pm0.4$ & $1.56\pm0.17$ & $3.63\pm0.27$ &
 3.2 & $2.79\pm0.08$ & $<4.6$ \\ \hline
 $\Sigma_c^{0}\to\Lambda_c^+\pi^-$ & $2.2\pm0.3$  & $1.44\pm0.16$ & $2.65\pm0.19$ &
 2.4 & $2.37\pm0.07$ & $2.2\pm0.4$ \\ \hline
 $\Sigma_c(2520)^{++}\to\Lambda_c^+\pi^+$ & $16.7\pm2.3$ & $11.77\pm1.27$ & $21.99\pm0.87$ &
 8.2 & $17.52\pm0.75$ & $14.9\pm1.9$  \\  \hline
 $\Sigma_c(2520)^{+}\to\Lambda_c^+\pi^0$ & $17.4\pm2.3$ & $$ & $$
 &8.6
  & $17.31\pm0.74$ & $<17$  \\  \hline
 $\Sigma_c(2520)^{0}\to\Lambda_c^+\pi^-$ & $16.6\pm2.2$ & $11.37\pm1.22$ & $21.21\pm0.81$ &
 8.2 & $16.90\pm0.72$ & $16.1\pm2.1$  \\  \hline
 $\Xi_c(2645)^+\to\Xi_c^{0,+}\pi^{+,0}$ & $2.8\pm0.4$ & $1.76\pm0.14$ & $3.04\pm0.37$ &
 & $3.18\pm0.10$ & $<3.1$  \\  \hline
 $\Xi_c(2645)^0\to\Xi_c^{+,0}\pi^{-,0}$ & $2.9\pm0.4$ & $1.83\pm0.06$ & $3.12\pm0.33$ &
 & $3.03\pm0.10$ & $<5.5$  \\ \hline \hline
\end{tabular}
\end{center}
\end{scriptsize}
\end{table}

It is clear from Table \ref{tab:strongdecayS} that the predicted
widths of $\Sigma_c^{++}$ and $\Sigma_c^0$ by HHChPT are in good
agreement with experiment. The strong decay width of $\Sigma_c$ is
smaller than that of $\Sigma_c^*$ by a factor of $\sim 7$,
although they will become the same in the limit of heavy quark
symmetry. This is ascribed to the fact that the pion's momentum is
around 90 MeV in the decay $\Sigma_c\to\Lambda_c\pi$ while it is
two times bigger in $\Sigma_c^*\to\Lambda_c\pi$. Since $\Sigma_c$
states are significantly narrower than their spin-3/2
counterparts, this explains why the measurement of their widths
came out much later. Instead of using the data to fix the coupling
constants in a model-independent manner, there exist some
calculations of couplings in various models such as the
relativistic light-front model \cite{Tawfiq}, the relativistic
three-quark model \cite{Ivanov} and light-cone sum rules
\cite{Huang95,Zhu}. The results are summarized in Table
\ref{tab:strongdecayS}.

It is worth remarking that although the coupling $g_1$ cannot be
determined directly from the strong decay such as
$\Sigma_c^*\to\Sigma_c\pi$, some information of $g_1$ can be
learned from the radiative decay $\Xi_c^{'*0}\to\Xi_c^0 \gamma$,
which is prohibited at tree level by SU(3) symmetry but can be
induced by chiral loops. A measurement of
$\Gamma(\Xi_c^{'*0}\to\Xi_c^0\gamma)$ will yield two possible
solutions for $g_1$.  Assuming the validity of the quark model
relations among different coupling constants, the experimental
value of $g_2$ implies $|g_1|=0.93\pm 0.16$~\cite{Cheng97}. 

\subsection{Strong decays of $p$-wave charmed baryons}
Some of the $S$-wave and $D$-wave couplings of $p$-wave baryons to
$s$-wave baryons can be determined. In principle, the coupling
$h_2$ is readily extracted from $\Lambda_c(2593)^+\to
\Sigma_c^0\pi^+$ with $\Lambda_c(2593)^+$ identified as
$\Lambda_{c1}({1\over 2})^+$. However, since
$\Lambda_c(2593)^+\to\Sigma_c\pi$ is kinematically barely allowed,
the finite width effects of the intermediate resonant states  will
become important \cite{Falk03}.

\begin{table}[t]
\caption{Same as Table \ref{tab:strongdecayS} except for $p$-wave
charmed baryons \cite{Cheng06}.} \label{tab:strongdecayP}
\begin{scriptsize}
\begin{center}
\begin{tabular}{|c|c|c|c|c|c|c|} \hline \hline
~~~~~~~Decay~~~~~~~ & ~HHChPT~ & Tawfiq & Ivanov & Huang & Zhu &
Expt.  \\
& & ~~et al. \cite{Tawfiq}~~ &
 ~~~et al. \cite{Ivanov}~~~ &  ~~et al. \cite{Huang95}~~ & ~~\cite{Zhu}~~ 
&~~\cite{Part5_pdg06}~~ \\
\hline
 $\Lambda_c(2593)^+\to (\Sigma_c^{+}\pi\pi)_R$ & input & $$
 & $$ & $2.5$ & & $2.63^{+1.56}_{-1.09}$ \\ \hline
 $\Lambda_c(2593)^+\to \Sigma_c^{++}\pi^-$ & $0.62^{+0.37}_{-0.26}$ & $1.47\pm0.57$ &
 $0.79\pm0.09$ & $0.55^{+1.3}_{-0.55}$ & 0.64 & $0.65^{+0.41}_{-0.30}$ \\ \hline
 $\Lambda_c(2593)^+\to \Sigma_c^{0}\pi^+$ & $0.67^{+0.40}_{-0.28}$ & $1.78\pm0.70$
 & $0.83\pm0.09$ & $0.89\pm0.86$ & 0.86 & $0.67^{+0.41}_{-0.30}$ \\ \hline
 $\Lambda_c(2593)^+\to \Sigma_c^{+}\pi^0$ & $1.34^{+0.79}_{-0.55}$ & $1.18\pm0.46$
 & $0.98\pm0.12$ & $1.7\pm0.49$ & 1.2 & $$ \\ \hline
 $\Lambda_c(2625)^+\to \Sigma_c^{++}\pi^-$ & $\leq 0.011$ & $0.44\pm0.23$ & $0.076\pm0.009$ &
 $0.013$ & 0.011 & $<0.10$ \\ \hline
 $\Lambda_c(2625)^+\to \Sigma_c^{0}\pi^+$ & $\leq 0.015$ & $0.47\pm0.25$ & $0.080\pm0.009$
 & 0.013 & 0.011 & $<0.09$ \\ \hline
 $\Lambda_c(2625)^+\to \Sigma_c^{+}\pi^0$ & $\leq 0.011$ & $0.42\pm0.22$ & $0.095\pm0.012$
 & 0.013 & 0.011 & $$ \\ \hline
 $\Lambda_c(2625)^+\to \Lambda_c^+\pi\pi$ & $\leq 0.21$ & $$ & $$
 & 0.11 &  & $<1.9$ \\ \hline
 $\Sigma_c(2800)^{++}\to\Lambda_c\pi,\Sigma_c^{(*)}\pi$ & input &
 & & & & $75^{+22}_{-17}$ \\ \hline
 $\Sigma_c(2800)^{+}\to\Lambda_c\pi,\Sigma_c^{(*)}\pi$ & input &
 & & & & $62^{+60}_{-40}$ \\ \hline
 $\Sigma_c(2800)^0\to\Lambda_c\pi,\Sigma_c^{(*)}\pi$ & input &
 & & & & $61^{+28}_{-18}$ \\ \hline
 $\Xi_c(2790)^+\to\Xi'^{0,+}_c\pi^{+,0}$ & $7.7^{+4.5}_{-3.2}$ &
 $$ & $$ & & & $<15$ \\ \hline
 $\Xi_c(2790)^0\to\Xi'^{+,0}_c\pi^{-,0}$ & $8.1^{+4.8}_{-3.4}$ &
 $$ & $$ & & & $<12$ \\ \hline
 $\Xi_c(2815)^+\to\Xi'^{*+,0}_c\pi^{0,+}$ & $3.2^{+1.9}_{-1.3}$ &
 $2.35\pm0.93$ & $0.70\pm0.04$ & & & $<3.5$ \\ \hline
 $\Xi_c(2815)^0\to\Xi'^{*+,0}_c\pi^{-,0}$ & $3.5^{+2.0}_{-1.4}$ &
 $$ & $$ & & & $<6.5$ \\ \hline \hline
\end{tabular}
\end{center}
\end{scriptsize}
\end{table}

Pole contributions to the decays
$\Lambda_c(2593)^+,\Lambda_c(2625)^+\to \Lambda_c^+\pi\pi$ have
been considered in \cite{Cho,Huang95,Pirjol} with the finite width
effects included. The intermediate states of interest are
$\Sigma_c$ and $\Sigma_c^*$ poles. The resonant contribution
arises from the $\Sigma_c$ pole, while the non-resonant term
receives a contribution from the $\Sigma_c^*$ pole. (Since
$\Lambda_c(2593)^+,\Lambda_c(2625)^+\to\Lambda_c^*\pi$ are not
kinematically allowed, the $\Sigma_c^*$ pole is not a resonant
contribution.) The decay rates thus depend on two coupling
constants $h_2$ and $h_8$. The decay rate for the process
$\Lambda_{c_1}^+(2593)\to \Lambda_c^+\pi^+\pi^-$ can be calculated
in the framework of heavy hadron chiral perturbation theory to be
\cite{Cheng06}
 \begin{eqnarray}
 \Gamma(\Lambda_c(2593)\to\Lambda_c^+\pi\pi)&=& 14.48h_2^2+27.54h_8^2-3.11h_2h_8,
 \nonumber
 \\
 \Gamma(\Lambda_c(2625)\to\Lambda_c^+\pi\pi)&=& 0.648h_2^2+0.143\times
 10^6h_8^2-28.6h_2h_8.
 \end{eqnarray}
It is clear that the limit on $\Gamma(\Lambda_c(2625))$ gives an
upper bound on $h_8$ of order $10^{-3}$ (in units of MeV$^{-1}$),
whereas the decay width of $\Lambda_c(2593)$ is entirely governed
by the coupling $h_2$. This indicates that the direct non-resonant
$\Lambda_c^+\pi\pi$ cannot be described by the $\Sigma_c^*$ pole
alone. Identifying the calculated
$\Gamma(\Lambda_c(2593)\to\Lambda_c^+\pi\pi)$ with the resonant
one, we find
 \begin{eqnarray} \label{eq:h2fw}
 |h_2|=0.427^{+0.111}_{-0.100}\,, \qquad\quad |h_8|\leq 3.57\times
 10^{-3}\,.
 \end{eqnarray}

Assuming that the total decay width of the $\Lambda_c(2593)$ is
saturated by the resonant $\Lambda_c^+\pi\pi$ 3-body decays,
Pirjol and Yan obtained $|h_2|=0.572^{+0.322}_{-0.197}$ and
$|h_8|\leq (3.50-3.68)\times 10^{-3}\,{\rm MeV}^{-1}$
\cite{Pirjol}. Using the updated hadron masses and
$\Gamma(\Lambda_c(2593)\to\Lambda_c^+\pi\pi)$,\footnote{The CLEO
result $\Gamma(\Lambda_c(2593))=3.9^{+2.4}_{-1.6}$ MeV \cite{CLEO}
is used in \cite{Pirjol} to fix $h_2$.}
we find $|h_2|=0.499^{+0.134}_{-0.100}$. Taking into account the
fact that the $\Sigma_c$ and $\Sigma_c^*$ poles only describe the
resonant contributions to the total width of $\Lambda_c(2593)$, we
finally reach the $h_2$ value given in (\ref{eq:h2fw}).  Taking
into account the threshold (or finite width) effect in the strong
decay $\Lambda_c(2593)^+\to \Lambda_c\pi\pi$, a slightly small
coupling $h^2_2=0.24^{+0.23}_{-0.11}$ is obtained in
\cite{Falk03}. For the spin-${3\over 2}$ state $\Lambda_c(2625)$,
its decay is dominated by the three-body channel
$\Lambda_c^+\pi\pi$ as the major two-body decay $\Sigma_c\pi$ is a
$D$-wave one.

Some information on the coupling $h_{10}$ cane be inferred from
the strong decays of $\Lambda_c(2800)$. As noticed in passing, the
states $\Sigma_c(2800)^{++,+,0}$ are most likely to be
$\Sigma_{c2}({3\over 2})$. Assuming their widths are dominated by
the two-body modes $\Lambda_c\pi$, $\Sigma_c\pi$ and
$\Lambda_c^*\pi$, we have \cite{Pirjol}
 \begin{eqnarray}
\Gamma\left(\Sigma_{c2}({3\over 2})^{++}\right) &\approx&
\Gamma\left(\Sigma_{c2}({3\over
2})^{++}\to\Lambda_c^+\pi^+\right)+
\Gamma\left(\Sigma_{c2}({3\over 2})^{++}\to\Sigma_c^+\pi^+\right)+ \nonumber
\\
&& \Gamma\left(\Sigma_{c2}({3\over
2})^{++}\to\Sigma_c^{*+}\pi^+\right)
\nonumber \\
&=& {4h_{10}^2\over 15\pi f_\pi^2}\,{m_{\Lambda_c}\over
m_{\Sigma_{c2}}}p_c^5+{h_{11}^2\over 10\pi
f_\pi^2}\,{m_{\Sigma_c}\over m_{\Sigma_{c2}}}p_c^5+ {h_{11}^2\over
10\pi f_\pi^2}\,{m_{\Sigma_c^*}\over m_{\Sigma_{c2}}}p_c^5,
 \end{eqnarray}
and similar expressions for $\Sigma_c(2800)^+$ and
$\Sigma_c(2800)^0$. Using the quark model relation
$h_{11}^2=2h_{10}^2$ [see also Eq. (\ref{eq:QMh8})] and the
measured widths of $\Sigma_c(2800)^{++,+,0}$ (Table
\ref{tab:spectrum}), we obtain
 \begin{eqnarray}
|h_{10}|=(0.85^{+0.11}_{-0.08})\times 10^{-3}\,{\rm MeV}^{-1}\,.
 \end{eqnarray}
Since the state $\Lambda_{c1}({3\over 2})$ is broader, even a
small mixing of $\Lambda_{c2}({3\over 2})$ with
$\Lambda_{c1}({3\over 2})$ could enhance the decay width of the
former \cite{Pirjol}. In this case, the above value for $h_{10}$
should be regarded as an upper limit of $|h_{10}|$. Using the
quark model relation $|h_8|=|h_{10}|$ (see Eq. (\ref{eq:QMh8})
below), the calculated partial widths of $\Lambda_c(2625)^+$ are
shown in Table \ref{tab:strongdecayP}.

The $\Xi_c(2790)$ and $\Xi_c(2815)$ baryons form a doublet
$\Xi_{c1}({1\over 2},{3\over 2})$. $\Xi_c(2790)$ decays to
$\Xi'_c\pi$, while $\Xi_c(2815)$ decays to $\Xi_c\pi\pi$,
resonating through $\Xi_c^*$, i.e. $\Xi_c(2645)$. Using the
coupling $h_2$ obtained (\ref{eq:h2fw}) and the experimental
observation that the $\Xi_c\pi\pi$ mode in $\Xi_c(2815)$ decays is
consistent with being entirely via $\Xi_c(2645)\pi$, the predicted
$\Xi_c(2790)$ and $\Xi_c(2815)$ widths are shown in Table
\ref{tab:strongdecayP} and they are consistent with the current
experimental limits.

Couplings other than $h_2$ and $h_{10}$ can be related to each
other via the quark model. The $S$-wave couplings between the
$s$-wave and the $p$-wave baryons are related by \cite{Pirjol}
 \begin{eqnarray}
 {|h_3|\over |h_4|}={\sqrt{3}\over 2}, \quad {|h_2|\over |h_4|}={1\over 2},
 \quad {|h_5|\over |h_6|}={2\over \sqrt{3}},\quad {|h_5|\over
 |h_7|}=1\,.
 \end{eqnarray}
The $D$-wave couplings satisfy the relations
 \begin{eqnarray} \label{eq:QMh8}
 |h_8|=|h_9|=|h_{10}|, \quad {|h_{11}|\over |h_{10}|}={|h_{15}|\over |h_{14}|}=\sqrt{2},
 \quad {|h_{12}|\over |h_{13}|}=2, \quad {|h_{14}|\over
 |h_{13}|}=1\,.
 \end{eqnarray}
The reader is referred to Ref.~\cite{Pirjol} for further details.

\section{Lifetimes}
 The lifetime differences among the charmed mesons $D^+,~D^0$ and charmed
baryons have been studied extensively both experimentally and
theoretically since the late 1970s. It was realized very early that
the naive parton model gives the same lifetimes for all heavy
particles containing a heavy quark $Q$ and that the underlying
mechanism for the decay width differences and the lifetime
hierarchy of heavy hadrons comes mainly from the spectator effects
like $W$-exchange and Pauli interference due to the identical
quarks produced in the heavy quark decay and in the charmed
baryons (for a review, see \cite{Bigi,Bigireview,BS94}). The
spectator effects were expressed in 1980s in terms of local
four-quark operators by relating the total widths to the imaginary
part of certain forward scattering amplitudes
\cite{Bilic,Guberina,SV}. (The spectator effects for charmed
baryons were first studied in \cite{Ruckl}.) With the advent of
heavy quark effective theory (HQET), it was recognized in early
1990s that nonperturbative corrections to the parton picture can
be systematically expanded in powers of $1/m_Q$
\cite{Bigi92,BS93}. Subsequently, it was demonstrated that this
$1/m_Q$ expansion is applicable not only to global quantities such
as lifetimes, but also to local quantities, e.g. the lepton
spectrum in the semileptonic decays of heavy hadrons
\cite{Bigi93}. Therefore, the above-mentioned phenomenological
work in 1980s acquired a firm theoretical footing in 1990s, namely
the heavy quark expansion (HQE), which is a generalization of the
operator product expansion (OPE) in $1/m_Q$. Within this QCD-based
framework, some phenomenological assumptions can be turned into
some coherent and quantitative statements and nonperturbative
effects can be systematically studied.

Based on the OPE approach for the analysis of inclusive weak
decays, the inclusive rate of the charmed baryon is schematically
represented by
 \begin{eqnarray}
 \Gamma({\cal B}_c\to f) = {G_F^2m_c^5\over
192\pi^3}V_{\rm CKM}\left(A_0+{A_2\over m_c^2}+{A_3\over
m_c^3}+{\cal O}({1\over m_c^4})\right).
 \end{eqnarray}
The $A_0$ term comes from the $c$ quark decay and is common to all
charmed hadrons. There is no linear $1/m_Q$ corrections to the
inclusive decay rate due to the lack of gauge-invariant
dimension-four operators \cite{Chay,Bigi92}, a consequence known
as Luke's theorem \cite{Luke}. Nonperturbative corrections start
at order $1/m_Q^2$ and they are model independent. Spectator
effects in inclusive decays due to the Pauli interference and
$W$-exchange contributions account for $1/m_c^3$ corrections and
they have two eminent features: First, the estimate of spectator
effects is model dependent; the hadronic four-quark matrix
elements are usually evaluated by assuming the factorization
approximation for mesons and the quark model for baryons. Second,
there is a two-body phase-space enhancement factor of $16\pi^2$
for spectator effects relative to the three-body phase space for
heavy quark decay. This implies that spectator effects, being of
order $1/m_c^3$, are comparable to and even exceed the $1/m_c^2$
terms.

The lifetimes of charmed baryons are measured to be~\cite{Part5_pdg06}
  \begin{eqnarray} \label{eq:exptlifetime}
&& \tau(\Lambda^+_c)= (200\pm6)\times 10^{-15}s, \qquad
\tau(\Xi^+_c)= (442\pm26)\times 10^{-15}s,
  \nonumber \\
&& \tau(\Xi^0_c)= (112^{+13}_{-10})\times 10^{-15}s, \qquad~~
\tau(\Omega^0_c)= (69\pm12)\times 10^{-15}s.
  \end{eqnarray}
As we shall see below, the lifetime hierarchy
$\tau(\Xi_c^+)>\tau(\Lambda_c^+)>\tau(\Xi_c^0)
>\tau(\Omega_c^0)$ is qualitatively understandable in the OPE approach but not
quantitatively.

\begin{figure}[t]
\centerline{\psfig{file=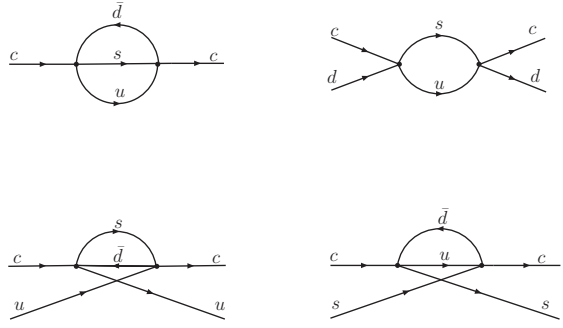,width=3.0in}}
\caption{Contributions to nonleptonic decay rates of charmed
baryons from four-quark operators: (a) $c$-quark decay, (b)
$W$-exchange, (c) destructive Pauli interference and (d)
constructive interference.} \label{fig:fourquarkNL}
\end{figure}

In general, the total width of the charmed baryon ${\cal B}_c$
receives contributions from inclusive nonleptonic and semileptonic
decays: $\Gamma({\cal B}_c)=\Gamma_{\rm NL}({\cal
B}_c)+\Gamma_{\rm SL}({\cal B}_c)$. The nonleptonic contribution
can be decomposed into
 \begin{eqnarray}
 \Gamma_{\rm NL}({\cal B}_c)=\Gamma^{\rm dec}({\cal B}_c)+\Gamma^{\rm ann}
 ({\cal B}_c)+\Gamma^{\rm
 int}_-({\cal B}_c)+\Gamma^{\rm int}_+({\cal B}_c),
 \end{eqnarray}
corresponding to the $c$-quark decay, the $W$-exchange
contribution, destructive and constructive Pauli interferences. It
is known that the inclusive decay rate is governed by the
imaginary part of an effective nonlocal forward transition
operator $T$. Therefore, $\Gamma^{\rm dec}$ corresponds to the
imaginary part of Fig. \ref{fig:fourquarkNL}(a) sandwiched between
the same ${\cal B}_c$ states. At the Cabibbo-allowed level,
$\Gamma^{\rm dec}$ represents the decay rate of $c\to su\bar d$,
and $\Gamma^{\rm ann}$ denotes the contribution due to the
$W$-exchange diagram $cd\to us$. The interference $\Gamma^{\rm
int}_-$ ($\Gamma^{\rm int}_+$) arises from the destructive
(constructive) interference between the $u$ ($s$) quark produced
in the $c$-quark decay and the spectator $u$ ($s$) quark in the
charmed baryon ${\cal B}_c$. Notice that the constructive Pauli
interference is unique to the charmed baryon sector as it does not
occur in the bottom baryon sector. From the quark content of the
charmed baryons (see Table \ref{tab:spectrum}), it is clear that
at the Cabibbo-allowed level, the destructive interference occurs
in $\Lambda_c^+$ and $\Xi_c^+$ decays, while $\Xi_c^+,\Xi_c^0$ and
$\Omega_c^0$ can have $\Gamma^{\rm int}_+$. Since $\Omega_c^0$
contains two $s$ quarks, it is natural to expect that $\Gamma^{\rm
int}_+(\Omega_c^0)\gg \Gamma^{\rm int}_+(\Xi_c)$. $W$-exchange
occurs  only for $\Xi_c^0$ and $\Lambda_c^+$ at the same
Cabibbo-allowed level. In the heavy quark expansion approach, the
above-mentioned spectator effects can be described in terms of the
matrix elements of local four-quark operators.

Within this QCD-based heavy quark expansion approach, some
phenomenological assumptions can be turned into some coherent and
quantitative statements and nonperturbative effects can be
systematically studied. To begin with, we write down the general
expressions for the inclusive decay widths of charmed hadrons.
Under the heavy quark expansion, the inclusive nonleptonic decay
rate of a charmed baryon ${\cal B}_c$ is given by
\cite{Bigi92,BS93}
 \begin{eqnarray}\label{eq:nl}
 \Gamma_{\rm NL}({\cal B}_c) &=& {G_F^2m_c^5\over
192\pi^3}N_c\,V_{\rm CKM}\, {1\over 2m_{{\cal B}_c}} \Bigg\{
\left(c_1^2+c_2^2+{2c_1c_2\over N_c}\right)
 \Big[I_0(x,0,0)\langle {\cal B}_c|\bar cc|{\cal B}_c\rangle   \nonumber \\
&-& {1\over m_c^2}I_1(x,0,0) \langle {\cal B}_c|\bar c\sigma \cdot
G c|{\cal B}_c\rangle \Big]   -{4\over m_c^2}\,{2c_1c_2\over
N_c}\,I_2(x,0,0) \langle {\cal B}_c|\bar c\sigma\cdot G c|{\cal
B}_c\rangle\Bigg\} \nonumber \\
&+& {1\over 2m_{{\cal B}_c}}\langle {\cal B}_c|{\cal L}_{\rm
spec}|{\cal B}_c\rangle+{\cal O}\left({1\over m_c^4}\right),
 \end{eqnarray}
where $\sigma\!\cdot\! G=\sigma_{\mu\nu}G^{\mu\nu}$,
$x=(m_s/m_c)^2$, $N_c$ is the number of colors, $c_1,~c_2$ are
Wilson coefficient functions, $N_c=3$ is the number of color and
$V_{\rm CKM}$ takes care of  the relevant CKM matrix elements. In
the above equation, $I_{0,1,2}$ are phase-space factors
 \begin{eqnarray}
I_0(x,0,0) &=& (1-x^2)(1-8x+x^2)-12x^2\ln x,   \nonumber \\
I_1(x,0,0) &=& {1\over 2}(2-x{d\over dx})I_0(x,0,0)=(1-x)^4,   \nonumber \\
I_2(x,0,0) &=& (1-x)^3,
 \end{eqnarray}
for $c\to su\bar d$ transition.

In heavy quark effective theory, the two-body matrix element
$\langle {\cal B}_c|\bar cc|{\cal B}_c\rangle$  in Eq.
(\ref{eq:nl}) can be recast to
 \begin{eqnarray} \label{eq:2bodyme}
{\langle {\cal B}_c|\bar cc|{\cal B}_c\rangle\over 2m_{{\cal
B}_c}}=1-{K_H\over 2m_c^2}+{G_H\over 2m_c^2},
 \end{eqnarray}
with
 \begin{eqnarray}
 && K_H\equiv -{1\over 2m_{{\cal B}_c}}\langle {\cal B}_c|\bar
c(i{D_\perp})^2c|{\cal B}_c\rangle=
-\lambda_1,   \nonumber \\
&& G_H\equiv{1\over 2m_{{\cal B}_c}}\langle {\cal B}_c|\bar
c{1\over 2}\sigma\cdot Gc|{\cal B}_c\rangle=d_H\lambda_2,
 \end{eqnarray}
where $d_H=0$ for the antitriplet baryon and $d_H=4$ for the
spin-${1\over 2}$ sextet baryon. It should be stressed that the
expression (\ref{eq:2bodyme}) is model independent and it contains
nonperturbative kinetic and chromomagnetic effects which are
usually absent in the quark model calculations. The
nonperturbative HQET parameters $\lambda_1$ and $\lambda_2$ are
independent of the heavy quark mass. Numerically, we shall use
$\lambda_1^{\rm baryon}=-(0.4\pm0.2)\,{\rm GeV}^2$ \cite{Neubert}
and $\lambda_2^{\rm baryon}=0.055\,{\rm GeV}^2$ for charmed
baryons \cite{Chenglife}. Spectator effects in inclusive decays of
charmed hadrons are described by the dimension-six four-quark
operators ${\cal L}_{\rm spec}$ in Eq. (\ref{eq:nl}) at order
$1/m_c^3$. Its complete expression can be found in, for example,
Eq. (2.4) of \cite{Chenglife}.

For inclusive semileptonic decays, there is an additional
spectator effect in charmed-baryon semileptonic decay originating
from the Pauli interference of the $s$ quark for charmed baryons
$\Xi_c$ and $\Omega_c$ \cite{Voloshin}.  The general expression of
the inclusive semileptonic widths is given by
 \begin{eqnarray} \label{eq:sl}
 \Gamma_{\rm SL}({\cal B}_c) &=& {G_F^2m_c^5\over
192\pi^3}V_{\rm CKM}\,{ \eta(x,x_\ell,0)\over 2m_{{\cal B}_c}}
\Big[ I_0(x,0,0)\langle {\cal B}_c|\bar cc|{\cal
B}_c\rangle-{1\over m_c^2}\,I_1(x,0,0) \langle {\cal B}_c|\bar
c\sigma\cdot G c|{\cal B}_c\rangle \Big] \nonumber \\
&-& {G_F^2m_c^2\over 6\pi}|V_{cs}|^2{1\over 2m_{{\cal
B}_c}}(1-x)^2\Big[(1+{x\over 2}) (\bar c s)(\bar s c)-(1+2x)\bar
c(1-\gamma_5)s\bar s(1+\gamma_5)c\Big],\nonumber \\
 \end{eqnarray}
where $\eta(x,x_\ell,0)$ with $x_\ell=(m_\ell/m_Q)^2$ is the QCD
radiative correction to the semileptonic decay rate and its
general analytic expression is given in \cite{Hokim}. Since both
nonleptonic and semileptonic decay widths scale with the fifth
power of the charmed quark mass, it is very important to fix the
value of $m_c$. It is found that the experimental values for $D^+$
and $D^0$ semileptonic widths~\cite{Part5_pdg06} can be fitted by the
quark pole mass $m_c=1.6$ GeV. Taking $m_s=170$ MeV, we obtain the
charmed-baryon semileptonic decay rates
 \begin{eqnarray} \label{eq:slrate}
 && \Gamma(\Lambda_c\to Xe\bar\nu)=\Gamma(\Xi_c\to Xe\bar\nu)=
\,1.533\times 10^{-13}{\rm GeV},   \nonumber \\
&& \Gamma(\Omega_c\to Xe\bar\nu)=\,1.308\times 10^{-13}{\rm GeV}.
 \end{eqnarray}
The prediction  (\ref{eq:slrate}) for the $\Lambda_c$ baryon is in
good agreement with experiment~\cite{Part5_pdg06}
 \begin{eqnarray}
\Gamma(\Lambda_c\to Xe\bar\nu)_{\rm expt}=\,(1.480\pm 0.559)\times
10^{-13}{\rm GeV}.
 \end{eqnarray}
We shall see beolw that the Pauli interference effect in the
semileptonic decays of $\Xi_c$ and $\Omega_c$ can be very
significant, in particular for the latter.

The baryon matrix element of the four-quark operator $\langle
{\cal B}_c|(\bar c q_1)(\bar q_2q_3)|{\cal B}_c\rangle$ with
$(\bar q_1q_2)=\bar q_1\gamma_\mu(1-\gamma_5)q_2$ is customarily
evaluated using the quark model. In the non-relativistic quark
model (for early related studies, see \cite{Bilic,Guberina}), the
matrix element is governed by the charmed baryon wave function at
origin, $|\psi^{{\cal B}_c}_{cq}(0)|^2$, which can be related to
the charmed meson wave function $|\psi^D_{cq}(0)|^2$. For example,
the hyperfine splittings between $\Sigma_c^*$ and $\Sigma_c$, and
between $D^*$ and $D$ separately yield \cite{Rosner}
 \begin{eqnarray}
|\psi^{\Lambda_c}_{cq}(0)|^2=|\psi^{\Sigma_c}_{cq}(0)|^2=\,{4\over
3}\,{m _{\Sigma^*_c}-m_{\Sigma_c}\over
m_{D^*}-m_D}\,|\psi^D_{c\bar q}(0)|^2.
 \end{eqnarray}
This relation is supposed to be robust as $|\psi_{cq}(0)|^2$
determined in this manner does not depend on the strong coupling
$\alpha_s$ and the light quark mass $m_q$ directly. Defining
 \begin{eqnarray}
 |\psi^{{\cal B}_c}_{cq}(0)|^2=r_{{\cal B}_c}|\psi^D_{cq}(0)|^2,
 \end{eqnarray}
we have
 \begin{eqnarray}
r_{\Lambda_c}=\,{4\over 3}\,{m_{\Sigma_c^*}-m_{\Sigma_c}\over
m_{D^*}-m_D}, \quad r_{\Xi_c}=\,{4\over
3}\,{m_{\Xi_c^*}-m_{\Xi'_c}\over m_{D^*}-m_D},\quad
r_{\Omega_c}=\,{4\over 3}\,{m_{\Omega_c^*}-m_{\Omega_c}\over
m_{D^*}-m_D}.
  \end{eqnarray}
In terms of the parameter $r_{{\cal B}_c}|\psi^D_{cq}(0)|^2$ we
have \cite{Chenglife}
 \begin{eqnarray} \label{eq:nspect}
 \Gamma^{\rm ann}(\Lambda_c) &=& {G_F^2m_c^2\over
\pi}\,r_{\Lambda_c}(1-x)^2\Big
(\eta(c_1^2+c_2^2)-2c_1c_2\Big)|\psi^D(0)|^2,   \nonumber \\
 \Gamma^{\rm int}_-(\Lambda_c) &=& -{G_F^2m_c^2\over
4\pi}\,r_{\Lambda_c} (1-x)^2(1+x)
\Big(\eta c_1^2-2c_1c_2-N_cc_2^2\Big)|\psi^D(0)|^2,   \nonumber \\
 \Gamma^{\rm ann}(\Xi_c)/r_{\Xi_c} &=& \Gamma^{\rm
ann}(\Lambda_c)/r_{\Lambda_c},~~~~~~~\Gamma^{\rm
int}_-(\Xi_c^+)/r_{\Xi_c}=\,\Gamma^{\rm int}_-(\Lambda_c)/r_{\Lambda_c},   \nonumber \\
\Gamma^{\rm int}_+(\Xi_c) &=& -{G_F^2m_c^2\over 4\pi}\,r_{
\Xi_c}(1-x^2)(1+x)\Big(\eta c_2^2-2c_1c_2-N_cc_1^2\Big)|\psi^D(0)|^2,   \nonumber \\
\Gamma^{\rm int}_+(\Omega_c) &=& -{G_F^2m_c^2\over 6\pi}\,r_{
\Omega_c}(1-x^2)(5+x)\Big(\eta
c_2^2-2c_1c_2-N_cc_1^2\Big)|\psi^D(0)|^2, \nonumber \\
 \Gamma^{\rm ann}(\Omega_c) &=&
6{G_F^2m_c^2\over \pi}\,r_{\Omega_c}(1-x^2)\Big(\eta
(c_1^2+c_2^2)-2c_1c_2\Big)|\psi^D(0)|^2, \nonumber \\
 \Gamma^{\rm int}(\Xi_c\to Xe\bar\nu) &=& {G_F^2m_c^2\over 4\pi}
\,r_{\Xi_c}(1-x^2)(1+x)|\psi^D(0)|^2,   \nonumber \\
\Gamma^{\rm int}(\Omega_c\to Xe\bar\nu) &=& {G_F^2m_c^2\over
6\pi}\,r_{\Omega_c}(1-x^2)(5+x)|\psi^D(0)|^2,
 \end{eqnarray}
where the parameter $\eta$ is introduced via
 \begin{eqnarray}
 \langle {\cal B}_c|(\bar cc)(\bar qq)|{\cal B}_c\rangle=-\eta
 \langle {\cal B}_c|(\bar cq)(\bar qc)|{\cal B}_c\rangle,
 \end{eqnarray}
so that $\eta=1$ in the valence quark approximation. In the zero
light quark mass limit ($x=0$) and in the valence quark
approximation, the reader can check that results of
(\ref{eq:nspect}) are in agreement with those obtained in
Refs.~\cite{Bilic,Guberina,Guberina98} except  the Cabibbo-suppressed
$W$-exchange contribution to $\Omega_c^0$, $\Gamma^{\rm
ann}(\Omega_c)$. We have a coefficient of 6 arising from the
matrix element $\langle \Omega_c|(\bar cs)(\bar
sc)|\Omega_c\rangle=-6|\psi^{\Omega_c}_{cs}(0)|^2 (2m_{\Omega_c})$
\cite{Chenglife}, while the coefficient is claimed to be ${10\over
3}$ in \cite{Guberina98}.

Neglecting  the small difference between $r_{\Lambda_c}$,
$r_{\Xi_c}$ and $r_{\Omega_c}$ and setting $x=0$, the inclusive
nonleptonic rates of charmed baryons in the valence quark
approximation have the expressions:
 \begin{eqnarray} \label{eq:lifetimes}
 \Gamma_{\rm NL}(\Lambda_c^+) &=& \Gamma^{\rm
 dec}(\Lambda_c^+)+\cos_C^2\Gamma^{\rm ann}+\Gamma^{\rm
 int}_-+\sin_C^2\Gamma^{\rm int}_+,  \nonumber \\
 \Gamma_{\rm NL}(\Xi_c^+) &=& \Gamma^{\rm
 dec}(\Xi_c^+)+\sin_C^2\Gamma^{\rm ann}+\Gamma^{\rm
 int}_-+\cos_C^2\Gamma^{\rm int}_+,  \nonumber \\
 \Gamma_{\rm NL}(\Xi_c^0) &=& \Gamma^{\rm
 dec}(\Xi_c^0)+\Gamma^{\rm ann}+\Gamma^{\rm
 int}_-+\Gamma^{\rm int}_+,  \nonumber \\
 \Gamma_{\rm NL}(\Omega_c^0) &=& \Gamma^{\rm
 dec}(\Omega_c^0)+6\sin_C^2\Gamma^{\rm ann}+{10\over 3}\cos_C^2\Gamma^{\rm int}_+,
 \end{eqnarray}
with $\theta_C$ being the Cabibbo angle.

Assuming the  $D$ meson wavefunction at the origin squared
$|\psi^D_{c\bar q}(0)|^2$ being given by ${1\over 12}f_D^2m_D$, we
obtain $|\psi^{\Lambda_c}(0)|^2=7.5\times 10^{-3}{\rm GeV}^3$ for
$f_D=220$ MeV.\footnote{The recent CLEO measurement of
$D^+\to\mu^+\nu$ yields $f_{D^+}=222.6\pm16.7^{+2.8}_{-3.4}$ MeV
\cite{CLEOfD}.}
To proceed to the numerical calculations, we use the Wilson
coefficients $c_1(\mu)=\,1.35$ and $c_2(\mu)=-0.64$ evaluated at
the scale $\mu=1.25$ GeV. Since $\eta=1$ in the valence-quark
approximation and since the wavefunction squared ratio $r$ is
evaluated using the quark model, it is reasonable to assume that
the NQM and the valence-quark approximation are most reliable when
the baryon matrix elements are evaluated at a typical hadronic
scale $\mu_{\rm had}$. As shown in \cite{Neubert97}, the
parameters $\eta$ and $r$ renormalized at two different scales are
related via the renormalization group equation, from which we
obtain  $\eta(\mu)\simeq 0.74\eta(\mu_{\rm had})\simeq 0.74$ and
$r(\mu) \simeq 1.36\,r(\mu_{\rm had})$ \cite{Chenglife}.

The results of calculations are summarized in Table
\ref{tab:lifetime}. It is clear that the lifetime pattern
 \begin{eqnarray} \label{eq:lifepattern}
\tau(\Xi_c^+)>\tau(\Lambda_c^+)>\tau(\Xi_c^0)>\tau(\Omega_c^0)
 \end{eqnarray}
is in accordance with experiment.  This lifetime hierarchy is
qualitatively understandable. The $\Xi_c^+$ baryon is
longest-lived among charmed baryons because of the smallness of
$W$-exchange and partial cancellation between constructive and
destructive Pauli interferences, while $\Omega_c$ is
shortest-lived due to the presence of two $s$ quarks in the
$\Omega_c$ that renders the contribution of $\Gamma^{\rm int}_+$
largely enhanced. From Eq. (\ref{eq:nspect}) we also see that
$\Gamma^{\rm int}_+$ is always positive, $\Gamma^{\rm int}_-$ is
negative and that the constructive interference is larger than the
magnitude of the destructive one. This explains why
$\tau(\Xi_c^+)>\tau(\Lambda_c^+)$. It is also clear from Table
\ref{tab:lifetime} that, although the qualitative feature of the
lifetime pattern is comprehensive, the quantitative estimates of
charmed baryon lifetimes and their ratios are still rather poor.

\begin{table}
\caption{Various contributions to the decay rates (in units of
$10^{-12}$ GeV) of charmed baryons. The charmed meson wavefunction
at the origin squared $|\psi^D(0)|^2$ is taken to be ${1\over
12}f_D^2m_D$. Experimental values are taken from~\cite{Part5_pdg06}.}
\label{tab:lifetime}
\begin{center}
\begin{tabular}{|c|c c c c l l  l  l|} \hline \hline
 & $\Gamma^{\rm dec}$ & $\Gamma^{\rm ann}$ & $\Gamma^{\rm int}_-$ &
$\Gamma^{\rm int}_+$ & ~ $\Gamma_{\rm SL}$ & ~$\Gamma^{\rm tot}$ &
~$\tau(10^{-13}s)$~ & ~ $\tau_{\rm expt}(10^{-13}s)$~ \\
\hline
 $\Lambda_c^+$~ & ~1.006~ & ~1.342~ & ~$-0.196$~ & & ~0.323~ &
~2.492~ & ~  2.64~ &  ~$2.00\pm 0.06$~   \\
 $\Xi_c^+$ & 1.006 & 0.071 & $-0.203$ & 0.364 & ~0.547 &
~1.785~ & ~ 3.68~ & ~$4.42\pm0.26$  \\
 $\Xi_c^0$ & 1.006 & 1.466 & & 0.385 & ~0.547 &
~3.404 & ~ 1.93 & ~$1.12^{+0.13}_{-0.10}$ \\
 $\Omega_c^0$ & 1.132 & 0.439 & & 1.241 & ~1.039 &
~3.851 & ~ 1.71  & ~$0.69\pm 0.12$  \\
\hline \hline
\end{tabular}
\end{center}
\end{table}
\vskip 0.4cm

In \cite{Guberina98}, a  much larger charmed baryon wave function
at the origin is employed.  This is based on the argument originally
advocated in \cite{BS94}. The physical charmed meson decay
constant $f_D$ is related to the asymptotic static value $F_D$ via
 \begin{eqnarray}
 f_D=F_D\left(1-{|\mu|\over m_c}+{\cal O}({1\over m_c^2})\right).
 \end{eqnarray}
It was argued in \cite{BS94} that one should not use the physical
value of $f_D$ when relating $|\psi^{{\cal B}_c}(0)|^2$ to
$|\psi^D(0)|^2$ for reason of consistency since the widths have
been calculated through order $1/m_c^3$ only. Hence, the part of
$f_D$ which is not suppressed by $1/m_c$ should not be taken into
account. However, if we use $F_D\sim 2f_D$ for the wave function
$|\psi^D(0)|^2$, we find that the predicted lifetimes of charmed
baryons become too short compared to experiment except
$\Omega_c^0$. By contrast, using
$|\psi^{\Lambda_c}(0)|^2=2.62\times 10^{-2}{\rm GeV}^3$ and the
so-called hybrid renormalization, lifetimes
$\tau(\Lambda_c^+)=2.39$, $\tau(\Xi_c^+)=2.51$,
$\tau(\Xi_c^0)=0.96$ and $\tau(\Omega_c^0)=0.61$ in units of
$10^{-13}s$ are obtained in \cite{Guberina98}. They are in better
agreement with the data except $\Xi_c^+$. The predicted ratio
$\tau(\Xi_c^+)/\tau(\Lambda_c^+)=1.05$ is too small compared to
the experimental value of $2.21\pm0.15$. By inspecting Eq.
(\ref{eq:lifetimes}), it seems to be very difficult to enhance the
ratio by a factor of 2.

In short, when the lifetimes of charmed baryons are analyzed
within the framework of the heavy quark expansion, the qualitative
feature of the lifetime pattern is understandable, but a
quantitative description of charmed baryon lifetimes is still
lack. This may be ascribed to the following possibilities:
 \begin{enumerate}
\item Unlike the semileptonic decays, the heavy quark expansion in
inclusive nonleptonic decays cannot be justified by analytic
continuation into the complex plane and local duality has to be
assumed in order to apply the OPE directly in the physical region.
The may suggest a significant violation of quark-hadron local
duality in the charm sector.
 \item Since the $c$ quark is not heavy enough, it casts doubts
on the validity of heavy quark expansion for inclusive charm
decays. This point can be illustrated by the following example. It
is well known that the observed lifetime difference between the
$D^+$ and $D^0$ is ascribed to the destructive interference in
$D^+$ decays and/or the constructive $W$-exchange contribution to
$D^0$ decays. However, there is a serious problem with the
evaluation of the destructive Pauli interference $\Gamma^{\rm
int}(D^+)$ in $D^+$. A direct calculation analogous to
$\Gamma^{\rm int}_-({\cal B}_c)$ in the charmed baryon sector
indicates that $\Gamma^{\rm int}(D^+)$ overcomes the $c$ quark
decay rate so that the resulting nonleptonic decay width of $D^+$
becomes negative \cite{BS94,Chernyak}. This certainly does not
make sense. This example clearly indicates that the $1/m_c$
expansion in charm decay is not well convergent and sensible, to
say the least. It is not clear if the situation is improved even
after higher dimension terms are included.
 \item To overcome the aforementioned difficulty with $\Gamma^{\rm
int}(D^+)$, it has been conjectured in \cite{BS94} that
higher-dimension corrections amount to replacing $m_c$ by $m_D$ in
the expansion parameter $f_D^2m_D/m_c^3$, so that it becomes
$f_D^2/m_D^2$. As a consequence, the destructive Pauli
interference will be reduced by a factor of $(m_c/m_D)^3$. By the
same token, the Pauli interference in charmed baryon decay may
also be subject to the same effect. Another way of alleviating the
problem is to realize that the usual local four-quark operators
are derived in the heavy quark limit so that the effect of
spectator light quarks can be neglected. Since the charmed quark
is not heavy enough, it is very important, as stressed by Chernyak
\cite{part5:Chernyak}, to take into account the nonzero momentum of
spectator quarks in charm decay. In the framework of heavy quark
expansion, this spectator effect can be regarded as higher order
$1/m_c$ corrections.
 \item One of the major theoretical uncertainties comes from
the evaluation of the four-quark matrix elements. One can hope
that lattice QCD will provide a better handle on those quantities.
 \end{enumerate}

\section{Hadronic weak decays}

Contrary to the significant progress made over the last 20 years
or so in the studies of the heavy meson weak decay, advancement in
the arena of heavy baryons, both theoretical and experimental, has
been relatively slow. This is partly due to the smaller baryon
production cross section and the shorter lifetimes of heavy
baryons. From the theoretical point of view, baryons being made
out of three quarks, in contrast to two quarks for mesons, bring
along several essential complications. First of all, the
factorization approximation that the hadronic matrix element is
factorized into the product of two matrix elements of single
currents and that the nonfactorizable term such as the
$W$-exchange contribution is negligible relative to the
factorizable one is known empirically to be working reasonably
well for describing the nonleptonic weak decays of heavy mesons.
However, this approximation is {\it a priori} not directly
applicable to the charmed baryon case as $W$-exchange there,
manifested as pole diagrams, is no longer subject to helicity and
color suppression. This is different from the naive color
suppression of internal $W$-emission. It is known in the heavy
meson case that nonfactorizable contributions will render the
color suppression of internal $W$-emission ineffective. However,
the $W$-exchange in baryon decays is not subject to color
suppression even in the absence of nonfactorizable terms. A simple
way to see this is to consider the large-$N_c$ limit. Although the
$W$-exchange diagram is down by a factor of $1/N_c$ relative to
the external $W$-emission one, it is compensated by the fact that
the baryon contains $N_c$ quarks in the limit of large $N_c$, thus
allowing $N_c$ different possibilities for $W$ exchange between
heavy and light quarks \cite{Korner}. That is, the pole
contribution can be as important as the factorizable one. The
experimental measurement of the decay modes
$\Lambda_c^+\to\Sigma^0\pi^+,~ \Sigma^+\pi^0$ and
$\Lambda^+_c\to\Xi^0K^+$, which do not receive any factorizable
contributions, indicates that $W$-exchange indeed plays an
essential role in charmed baryon decays. Second, there are more
possibilities in drawing the quark daigram amplitudes as depicted
in Fig. \ref{fig:QDS}; in general there exist two distinct
internal $W$-emissions and several different $W$-exchange diagrams
which will be discussed in more detail shortly.

Historically, the two-body nonleptonic weak decays of charmed
baryons were first studied by utilizing the same technique of
current algebra as in the case of hyperon decays \cite{Korner79}.
However, the use of the soft-meson theorem makes sense only if the
emitted meson is of the pseudoscalar type and its momentum is soft
enough. Obviously, the pseudoscalar-meson final state in charmed
bayon decay is far from being ``soft''. Therefore, it is not
appropriate to make the soft meson limit. It is no longer
justified to apply current algebra to heavy-baryon weak decays,
especially for $s$-wave amplitudes. Thus one has to go back to the
original pole model, which is nevertheless reduced to current
algebra in the soft pseudoscalar-meson limit, to deal with
nonfactorizable contributions. The merit of the pole model is
obvious: Its use is very general and is not limited to the soft
meson limit and to the pseudoscalar-meson final state. Of course,
the price we have to pay is that it requires the knowledge of the
negative-parity baryon poles for the parity-violating transition.
This also explains why the theoretical study of nonleptonic decays
of heavy baryons is much more difficult than the hyperon and heavy
meson decays.

   The nonfactorizable pole contributions to hadronic weak decays
of charmed baryons have been studied in the literature
\cite{CT92,CT93,XK92}. In general, nonfactorizable $s$- and
$p$-wave amplitudes for ${1\over 2}^+\to {1\over 2}^++P(V)$ decays
($P$: pseudoscalar meson, $V$: vector meson), for example, are
dominated by ${1\over 2}^-$ low-lying baryon resonances and
${1\over 2}^+$ ground-state baryon poles, respectively. However,
the estimation of pole amplitudes is a difficult and nontrivial
task since it involves weak baryon matrix elements and strong
coupling constants of ${1\over 2}^+$ and ${1\over 2}^-$ baryon
states. This is the case in particular for $s$-wave terms as we
know very little about the ${1\over 2}^-$ states. As a
consequence, the evaluation of pole diagrams is far more uncertain
than the factorizable terms.  In short, $W$-exchange plays a
dramatic role in the charmed baryon case and it even dominates
over the spectator contribution in hadronic decays of
$\Lambda_c^+$ and $\Xi_c^0$ \cite{Cheng92}.

Since the light quarks of the charmed baryon can undergo weak
transitions, one can also have charm-flavor-conserving weak
decays, e.g., $\Xi_c\to\Lambda_c\pi$ and $\Omega_c\to\Xi_c\pi$,
where the charm quark behaves as a spectator. This special class
of weak decays usually can be calculated more reliably than the
conventional charmed baryon weak decays.

\subsection{Quark-diagram scheme}
Besides dynamical model calculations, it is useful to study the
nonleptonic weak decays in a way which is as model independent as
possible. The two-body nonleptonic decays of charmed baryons have
been analyzed in terms of SU(3)-irreducible-representation
amplitudes \cite{Savage,Verma}. However, the quark-diagram scheme
(i.e., analyzing the decays in terms of quark-diagram amplitudes)
 has the advantage that it is more intuitive and easier for implementing
model calculations. It has been successfully applied to the
hadronic weak decays of charmed and bottom mesons
\cite{Chau,CC86}. It has provided a framework with which we not
only can do the least-model-dependent data analysis and give
predictions but also make evaluations of theoretical model
calculations.

\begin{figure}[t]
\centerline{\psfig{file=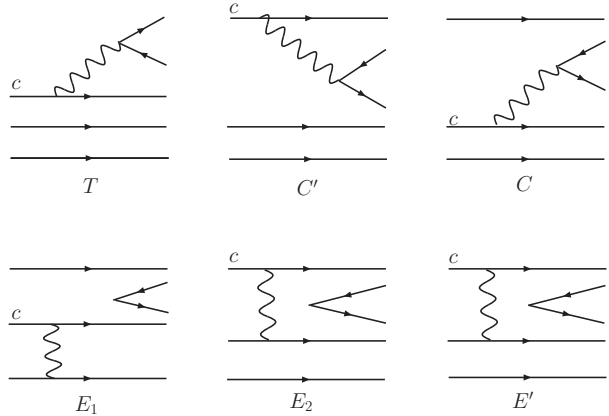,width=3.2in}}
 \caption{Quark
diagrams for charmed baryon decays} \label{fig:QDS}
\end{figure}

A general formulation of the quark-diagram scheme for the
nonleptonic weak decays of charmed baryons has been given in
\cite{CCT} (see also \cite{Kohara}). The general quark diagrams
shown in Fig. \ref{fig:QDS} are: the external $W$-emission tree
diagram $T$, internal $W$-emission diagrams $C$ and $C'$,
$W$-exchange diagrams $E_1,~E_2$ and $E'$ (see Fig.~2 of
\cite{CCT} for notation and for details). There are also
penguin-type quark diagrams which are presumably negligible in
charm decays due to GIM cancellation. The quark diagram amplitudes
$T,~C,~C'\cdots$ etc. in each type of hadronic decays are in
general not the same. For octet baryons in the final state, each
of the $W$-exchange amplitudes has two more independent types: the
symmetric and the antisymmetric, for example, $E_{1A}$, $E_{2A}$,
$E_{2S}$, $E'_A$ and $E'_S$ \cite{CCT}. The antiquark produced
from the charmed quark decay $c\to q_1q_2\bar q_3$ in diagram $C'$
can combine with $q_1$ or $q_2$ to form an outgoing meson.
Consequently, diagram $C'$ contains factorizable contributions but
$C$ does not. It should be stressed that all quark graphs used in
this approach are topological with all the strong interactions
included, i.e. gluon lines are included in all possible ways.
Hence, they are {\it not} Feynman graphs. Moreover, final-state
interactions are also classified in the same manner. A good
example is the reaction $D^0\to\bar K^0\phi$, which can be
produced via final-state rescattering even in the absence of the
$W$-exchange diagram. Then it was shown in \cite{CC86} that this
rescattering diagram belongs to the generic $W$-exchange topology.

Since the two spectator light quarks in the heavy baryon are
antisymmetrized in the antitriplet charmed baryon ${\cal B}_c({\bf
\bar 3})$ and  the wave function of the decuplet baryon ${\cal
B}({\bf 10})$ is totally symmetric, it is clear that factorizable
amplitudes $T$ and $C'$ cannot contribute to the decays of type
${\cal B}_c({\bf \bar 3})\to{\cal B}({\bf 10})+M({\bf 8})$; it
receives contributions only from the $W$-exchange and penguin-type
diagrams (see Fig.~1 of \cite{CCT}). Examples are
$\Lambda_c^+\to\Delta^{++}K^-,\Sigma^{*+}\rho^0,\Sigma^{*+}\eta,\Xi^{*0}K^+$
and $\Xi_c^0\to\Sigma^{*+}\bar K^0$. They can only proceed via
$W$-exchange. Hence, the experimental observation of them implies
that the $W$-exchange mechanism plays a significant role in
charmed baryon decays. The quark diagram amplitudes for all
two-body decays of (Cabibbo-allowed, singly suppressed and doubly
suppressed) $\Lambda_c^+,\Xi_c^{+,0}$ and $\Omega_c^0$ are listed
in \cite{CCT}. In the SU(3) limit, there exist many relations
among various charmed baryon decay amplitudes, see \cite{CCT} for
detail. For charmed baryon decays, there are only a few decay
modes which proceed through factorizable external or internal
$W$-emission diagram, namely, Cabibbo-allowed
$\Omega_c^0\to\Omega^-\pi^+(\rho^+),~\Xi^{*0}\bar{K}^0
(\bar{K}^{*0})$ and Cabibbo-suppressed $\Lambda_c^+\to p\phi$.

\subsection{Dynamical model calculation}

To proceed we first consider the Cabibbo-allowed decays ${\cal
B}_c( {1\over 2}^+)\to {\cal B}({1\over 2}^+)+P(V)$. The general
amplitudes are
 \begin{eqnarray} \label{eq:spwave}
 M[{\cal B}_i(1/2^+)\to {\cal B}_f(1/2^+)+P] &=& i\bar{u}_f(p_f)(A+B
\gamma_5)u_i(p_i),    \\
 M[{\cal B}_i(1/2^+)\to {\cal B}_f(1/2^+)+V] &=&
\bar{u}_f(p_f)\varepsilon^{*\mu}[A_1
\gamma_\mu\gamma_5+A_2(p_f)_\mu\gamma_5+B_1\gamma_\mu+B_2(p_f)_\mu]
u_i(p_i), \nonumber
 \end{eqnarray}
where $\varepsilon_\mu$ is the polarization vector of the vector
meson, $A$, $(B,B_1,B_2)$ and $A_2$ are $s$-wave, $p$-wave and
$d$-wave amplitudes, respectively, and $A_1$ consists of both
$s$-wave and $d$-wave ones. The QCD-corrected weak Hamiltonian
responsible for Cabibbo-allowed hadronic decays of charmed baryons
reads
 \begin{eqnarray}
{\cal H}_W=\,{G_F\over\sqrt{2}}\,V_{cs}V_{ud}^*(c_1O_1+c_2O_2),
 \end{eqnarray}
where $O_1=(\bar{s}c)(\bar{u}d)$ and $O_2=(\bar{s}d)(\bar{u}c)$
with $(\bar{q}_1q_2)\equiv\bar{q}_1\gamma_\mu(1-\gamma_5)q_2$.
>From the expression of $O_{1,2}$, it is clear that factorization
occurs if the final-state meson is $\pi^+(\rho^+)$ or $\bar
K^0(\bar K^{*0})$. Explicitly,
 \begin{eqnarray}
A^{\rm fac}({\cal B}_i\to {\cal B}_f+\pi^+) &=& \lambda\,a_1f_P(m_i-m_f)f_1(m^2_\pi),   \nonumber \\
B^{\rm fac}({\cal B}_i\to {\cal B}_f+\pi^+) &=& \lambda\,
a_1f_P(m_i+m_f)g_1(m^2_\pi), \nonumber \\
A^{\rm fac}({\cal B}_i\to {\cal B}_f+\bar K^0) &=& \lambda\, a_2f_P(m_i-m_f)f_1(m^2_K), \nonumber \\
B^{\rm fac}({\cal B}_i\to {\cal B}_f+\bar K^0) &=& \lambda\,
a_2f_P(m_i+m_f)g_1(m^2_K),
 \end{eqnarray}
and
 \begin{eqnarray}
A_1^{\rm fac}({\cal B}_i\to {\cal B}_f+\rho^+) &=&
-\lambda\,a_1f_\rho m_\rho
[g_1(m_\rho^2)+g_2(m^2_\rho)(m_i-m_f)],   \nonumber \\
A_2^{\rm fac}({\cal B}_i\to {\cal B}_f+\rho^+) &=& -2\lambda
\,a_1f_\rho m_\rho g_2(m^2_\rho),   \nonumber \\
B_1^{\rm fac}({\cal B}_i\to {\cal B}_f+\rho^+) &=& \lambda\,
a_1f_\rho m_\rho[f_1(m_\rho^2)-f_2(m^2_\rho)(m_i+m_f)],  \nonumber \\
B_2^{\rm fac}({\cal B}_i\to {\cal B}_f+\rho^+) &=& 2\lambda\,
a_1f_\rho m_\rho f_2(m^2_\rho),  \nonumber
 \end{eqnarray}
and similar expressions for ${\cal B}_i\to {\cal B}_f+\bar
K^{*0}$, where $\lambda=G_F V_{cs}V_{ud}^*/\sqrt{2}$,  $f_i$ and
$g_i$ are the form factors defined by ($q=p_i-p_f$)
 \begin{eqnarray} \label{eq:FF}
 \langle {\cal B}_f(p_f)|V_\mu-A_\mu|{\cal B}_i(p_i)\rangle &=& \bar{u}_f(p_f)
[f_1(q^2)\gamma_\mu+if_2(q^2)\sigma_{\mu\nu}q^\nu+f_3(q^2)q_\mu   \nonumber \\
&&
-(g_1(q^2)\gamma_\mu+ig_2(q^2)\sigma_{\mu\nu}q^\nu+g_3(q^2)q_\mu)\gamma_5]
u_i(p_i).\nonumber \\
 \end{eqnarray}

In the naive factorization approach, the coefficients $a_1$ for
the external $W$-emission amplitude and $a_2$ for internal
$W$-emission are given by $(c_1+{c_2\over N_c})$ and
$(c_2+{c_1\over N_c})$, respectively. However, we have learned
from charmed meson decays that the naive factorization approach
never works for the decay rate of color-suppressed decay modes,
though it usually operates for color-allowed decays. Empirically,
it was learned in the 1980s that if the Fierz-transformed terms
characterized by $1/N_c$ are dropped, the discrepancy between
theory and experiment is greatly improved \cite{Fuk}. This leads
to the so-called large-$N_c$ approach for describing hadronic $D$
decays \cite{Buras}. Theoretically, explicit calculations based on
the QCD sum-rule analysis \cite{BS} indicate that the Fierz terms
are indeed largely compensated by the nonfactorizable corrections.

As the discrepancy between theory and experiment for charmed meson
decays gets much improved in the $1/N_c$ expansion method, it is
natural to ask if this scenario also works in the baryon sector?
This issue can be settled down by the experimental measurement of
the Cabibbo-suppressed mode $\Lambda_c^+\to p\phi$, which receives
contributions only from the factorizable diagrams. As pointed out
in \cite{CT92}, the large-$N_c$ predicted rate is in good
agreement with the measured value. By contrast, its decay rate
prdicted by the naive factorization approximation is too small by
a factor of 15. Therefore, the $1/N_c$ approach also works for the
factorizable amplitude of charmed baryon decays. This also implies
that the inclusion of nonfactorizable contributions is inevitable
and necessary. If nonfactorizable effects amount to a redefinition
of the effective parameters $a_1$, $a_2$ and are universal (i.e.,
channel-independent) in charm decays, then we still have a new
factorization scheme with the universal parameters $a_1,~a_2$ to
be determined from experiment. Throughout this paper, we will thus
treat $a_1$ and $a_2$ as free effective parameters.

\begin{figure}[t]
\centerline{\psfig{file=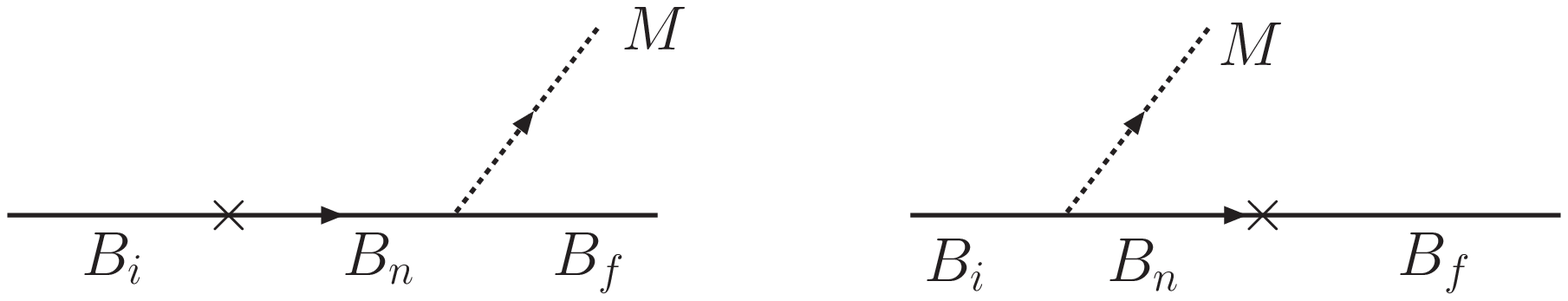,width=3.2in}}
 \caption{Pole
diagrams for charmed baryon decay ${\cal B}_i\to{\cal B}_f+M$. }
\label{fig:pole}
\end{figure}

At the hadronic level, the decay amplitudes for quark diagrams $T$
and $C'$are conventionally evaluated using the factorization
approximation.  How do we tackle with the remaining
nonfactorizable diagrams $C,~E_1,~E_2,~E'$ ? One popular approach
is to consider the contributions from all possible intermediate
states. Among all possible pole contributions, including
resonances and continuum states, one usually focuses on the most
important poles such as the low-lying ${1\over 2}^+,{1\over 2}^-$
states, known as pole approximation. More specifically,  the
$s$-wave amplitude is dominated by the low-lying $1/2^-$
resonances and the $p$-wave one governed by the ground-state
$1/2^+$ poles (see Fig. \ref{fig:pole}):
 \begin{eqnarray} \label{eq:poleamp}
 A^{\rm nf} &=& -\sum_{{\cal B}^*_n(1/2^-)}\left({g_{_{{\cal B}_f{\cal B}_{n^*}
P}}b_{_{n^*i}}\over m_i-m_{n^*}}+{b_{_{fn^*}}g_{_{{\cal
B}_{_{n^*}}{\cal B}_iP}}\over m_f-m_ {n^*}}\right)+\cdots,
\nonumber \\
 B^{\rm nf} &=&-\sum_{{\cal B}_n}\left({g_{_{{\cal B}_f{\cal B}_nP}}
 a_{_{ni}}\over m_i-m_n}+{a_{_{fn}}g_{_
{{\cal B}_n{\cal B}_iP}}\over m_f-m_n}\right)+\cdots,
 \end{eqnarray}
where $A^{\rm nf}$ and $B^{\rm nf}$ are the nonfactorizable $s$-
and $p$-wave amplitudes of ${\cal B}_c\to{\cal B}P$, respectively,
ellipses in Eq.(\ref{eq:poleamp}) denote other pole contributions
which are negligible for our purposes, and $a_{ij}$ as well as
$b_{i^*j}$ are the baryon-baryon matrix elements defined by
 \begin{eqnarray}
\langle {\cal B}_i|{\cal H}_{_W}|{\cal B}_j\rangle =
\,\bar{u}_i(a_{ij}-b_{ij} \gamma_5 )u_j,  \qquad \langle{\cal
B}^*_i(1/2^{^-})|{\cal H}^{\rm pv}_{_W}|{\cal B}_j\rangle =
\,ib_{i^*j}\, \bar{u}_iu_j,
 \end{eqnarray}
with $b_{ji^*}=-b_{i^*j}$. Evidently, the calculation of $s$-wave
amplitudes is more difficult than the $p$-wave owing to the
troublesome negative-parity baryon resonances which are not well
understood in the quark model.  In \cite{CT92,CT93}, the form
factors appearing in factorizable amplitudes and the strong
coupling constants and baryon transition matrix elements relevant
to nonfactorizable contributions are evaluated using the MIT bag
model \cite{MIT}. Two of the pole model calculations for branching
ratios \cite{CT93,XK92} are displayed in Table \ref{tab:BRs}. The
study of charmed baryon hadronic decays in \cite{Verma98} is
similar to \cite{CT93,XK92} except that the effect of $W$-exchange
is parametrized in terms of the baryon wave function at origin.
Sharma and Verma \cite{Verma98} defined a parameter
$r=|\psi^{{\cal B}_c}(0)|^2/|\psi^{\cal B}(0)|^2$ and argued that
its value is close to 1.4\,. A variant of the pole model has been
considered in \cite{Zen} in which the effects of
pole-model-induced SU(4) symmetry breaking in parity-conserving
and parity-violating amplitudes are studied.

\begin{table}
\caption{Branching ratios of Cabibbo-allowed ${\cal B}_c\to{\cal
B}+P$ decays in various models. Results of \cite{Korner,XK92,CT93}
have been normalized using the current world averages of charmed
baryon lifetimes~\cite{Part5_pdg06}. Branching ratios cited from
\cite{Verma98} are for $\phi_{\eta-\eta'}=-23^\circ$ and
$r=1.4$\,. The experimental branching fraction for $\Xi_c^+\to 
\Xi^0\pi^+$ is relative to $\Xi_c^+\to \Xi^-\pi^+\pi^+$.} \label{tab:BRs}
\begin{footnotesize}
\begin{center}
\begin{tabular}{|l|c|c|c|c|c|c|c|}
\hline\hline ~~~~Decay & K\"{o}rner,             & Xu,
& Cheng,             & \,\,\, Ivanov  \, & \.Zenczy- & Sharma & Expt. \\
        & Kr\"{a}mer \cite{Korner}& Kamal \cite{XK92}
& Tseng \cite{CT93} & et al. \cite{Ivanov98} & kowski\cite{Zen} &  
\cite{Verma98}& \cite{Part5_pdg06} \\
\hline
$\Lambda^+_c\to \Lambda \pi^+$ & input & 1.62 & 0.88 & 0.79 & 0.54 & 1.12 & 0.90$\pm$ 0.28\\
\hline
$\Lambda^+_c\to \Sigma^0 \pi^+$ & 0.32 & 0.34 & 0.72 & 0.88 & 0.41 & 1.34 & 0.99$\pm$ 0.32\\
\hline
$\Lambda^+_c\to \Sigma^+ \pi^0$ & 0.32 & 0.34 & 0.72 & 0.88 & 0.41 & 1.34 & 1.00$\pm$ 0.34\\
\hline
$\Lambda^+_c\to \Sigma^+ \eta$ & 0.16 & & & 0.11 & 0.94 & 0.57 & 0.48$\pm$ 0.17\\
\hline
$\Lambda^+_c\to \Sigma^+ \eta^\prime$ & 1.28 & & & 0.12& 0.12 & 0.10 &  \\
\hline
$\Lambda^+_c\to p \bar K^0$ & input & 1.20 & 1.26 & 2.06 & 1.79 & 1.64 & 2.3$\pm$ 0.6\\
\hline
$\Lambda^+_c\to \Xi^0 K^+$ & 0.26 & 0.10 & & 0.31 & 0.36 & 0.13 & 0.39$\pm$ 0.14\\
\hline
$\Xi^+_c\to \Sigma^+ \bar K^0$ & 6.45 & 0.44 & 0.84 & 3.08 & 1.56 & 0.04 &  \\
\hline
$\Xi^+_c\to \Xi^0 \pi^+$ & 3.54 & 3.36 & 3.93 & 4.40 & 1.59 & 0.53 &  $0.55\pm0.16$ \\
\hline
$\Xi^0_c\to \Lambda \bar K^0$ & 0.12 & 0.37 & 0.27 & 0.42 & 0.35 & 0.54 &  seen \\
\hline
$\Xi^0_c\to \Sigma^0 \bar K^0$ & 1.18 & 0.11 & 0.13 & 0.20 & 0.11 & 0.07 & \\
\hline
$\Xi^0_c\to \Sigma^+ K^-$ & 0.12 & 0.12 &  & 0.27 & 0.36 & 0.12 & \\
\hline
$\Xi^0_c\to \Xi^0 \pi^0$ & 0.03 & 0.56 & 0.28 & 0.04 & 0.69 & 0.87 &  \\
\hline
$\Xi^0_c\to \Xi^0 \eta$ & 0.24 & & & 0.28 & 0.01 & 0.22 &  \\
\hline
$\Xi^0_c\to \Xi^0 \eta^\prime$ & 0.85 & & & 0.31 & 0.09 & 0.06 & \\
\hline
$\Xi^0_c\to \Xi^- \pi^+$ & 1.04 & 1.74 & 1.25 & 1.22 & 0.61 & 2.46 & seen \\
\hline
$\Omega^0_c\to \Xi^0 \bar K^0$ & 1.21 & & 0.09 & 0.02 & & & \\
\hline \hline
\end{tabular}
\footnotetext[1]{Branching ratio relative to
$\Xi_c^+\to\Xi^-\pi^+\pi^+$.}
\end{center}
\end{footnotesize}
\end{table}

\begin{table}
\caption{The predicted asymmetry parameter $\alpha$ for
Cabibbo-allowed ${\cal B}_c\to{\cal B}+P$ decays in various
models. Results cited from \cite{Verma98} are for
$\phi_{\eta-\eta'}=-23^\circ$ and $r=1.4$\,.} \label{tab:decasy}
\begin{footnotesize}
\begin{center}
\begin{tabular}{|l|c|c|c|c|c|c|c|}
\hline\hline ~~~~Decay & K\"{o}rner,             & Xu,
& Cheng,             & \,\,\, Ivanov  \, & \.Zenczy- & Sharma, & Expt. \\
        & Kr\"{a}mer \cite{Korner}& Kamal \cite{XK92}
& Tseng \cite{CT93} & et al. \cite{Ivanov98} & kowski\cite{Zen} &  
\cite{Verma98}& \cite{Part5_pdg06} \\
\hline $\Lambda^+_c\to \Lambda \pi^+$ & $-$0.70 & $-$0.67 &
$-$0.95 & $-$0.95 & $-$0.99 & $-$0.99 & $-$0.91 $\pm$ 0.15\\
\hline
$\Lambda^+_c\to \Sigma^0 \pi^+$ & 0.70 & 0.92 & 0.78 & 0.43 & 0.39 & $-$0.31 &  \\
\hline $\Lambda^+_c\to \Sigma^+ \pi^0$ & 0.71 & 0.92 & 0.78 & 0.43
& 0.39 & $-$0.31 & $-$0.45$\pm$ 0.32  \\
\hline
$\Lambda^+_c\to \Sigma^+ \eta$ & 0.33 & & & 0.55 & 0 & $-$0.91 & \\
\hline
$\Lambda^+_c\to \Sigma^+ \eta^\prime$ & $-$0.45 & & & $-$0.05 & $-$0.91 & 0.78 &  \\
\hline
$\Lambda^+_c\to p \bar K^0$ & $-$1.0 & 0.51 & $-$0.49 & $-$0.97 & $-$0.66 & $-$0.99 &  \\
\hline
$\Lambda^+_c\to \Xi^0 K^+$ & 0 & 0 & & 0 & 0 & 0 & \\
\hline
$\Xi^+_c\to \Sigma^+ \bar K^0$ & $-$1.0 & 0.24 & $-$0.09 & $-$0.99 & 1.00 & 0.54 & \\
\hline
$\Xi^+_c\to \Xi^0 \pi^+$ & $-$0.78 & $-$0.81 & $-$0.77 & $-$1.0 & 1.00 & $-$0.27 &  \\
\hline
$\Xi^0_c\to \Lambda \bar K^0$ & $-$0.76 & 1.0 & $-$0.73 & $-$0.75 & $-$0.29 & $-$0.79 & \\
\hline
$\Xi^0_c\to \Sigma^0 \bar K^0$ & $-$0.96 & $-$0.99 & $-$0.59 & $-$0.55 & $-$0.50 & 0.48 & \\
\hline
$\Xi^0_c\to \Sigma^+ K^-$ & 0 & 0 &  & 0 & 0 & 0 & \\
\hline
$\Xi^0_c\to \Xi^0 \pi^0$ & 0.92 & 0.92 & $-$0.54 & 0.94 & 0.21 & $-$0.80 & \\
\hline
$\Xi^0_c\to \Xi^0 \eta$ & $-$0.92 & & & $-$1.0 & $-$0.04 & 0.21 & \\
\hline
$\Xi^0_c\to \Xi^0 \eta^\prime$ & $-$0.38 & & & $-$0.32 & $-$1.00 & 0.80 &  \\
\hline
$\Xi^0_c\to \Xi^- \pi^+$ & $-$0.38 & $-$0.38 & $-$0.99 & $-$0.84 & $-$0.79 & $-$0.97 & $-0.6\pm0.4$ \\
\hline
$\Omega^0_c\to \Xi^0 \bar K^0$   & 0.51 &  & $-$0.93 & $-$0.81 & & &  \\
\hline \hline
\end{tabular}
\end{center}
\end{footnotesize}
\end{table}

\begin{table}
\caption{Branching ratios of Cabibbo-allowed ${\cal B}_c\to{\cal
B}+V$ decays in various models. The experimental value denoted by
the superscript $*$ is the branching ratio relative to
$\Xi_c^+\to\Xi^-\pi^+\pi^+$.} \label{tab:BRsBV}
\begin{center}
\begin{tabular}{|l|c|c|c|c|}
\hline\hline ~~~~Decay & K\"{o}rner,
 & \.Zenczykowski & Cheng,  & ~~Experiment~~ \\
        &  Kr\"{a}mer \cite{Korner}  & \cite{Zen} & Tseng \cite{CT92} & 
\cite{Part5_pdg06} \\
\hline
$\Lambda^+_c\to \Lambda \rho^+$ & 19.08  & 1.80 & 2.6 & $<5$ \\
\hline
$\Lambda^+_c\to \Sigma^0 \rho^+$  & 3.14  & 1.56 & 0.19 & 0.99$\pm$ 0.32\\
\hline
$\Lambda^+_c\to \Sigma^+ \rho^0$  & 3.12  & 1.56 & 0.19 & $<1.4$ \\
\hline
$\Lambda^+_c\to \Sigma^+ \omega$  & 4.02 &  1.10 & & 2.7$\pm$ 1.0\\
\hline
$\Lambda^+_c\to \Sigma^+ \phi$ & 0.26  & 0.11 & & 0.32$\pm$ 0.10\\
\hline
$\Lambda^+_c\to p \bar K^{*0}$  & 3.08  & 5.03 & 3.3 & 1.6$\pm$ 0.5\\
\hline
$\Lambda^+_c\to \Xi^0 K^{*+}$ & 0.12  & 0.11 & & 0.39$\pm$ 0.14\\
\hline
$\Xi^+_c\to \Sigma^+ \bar K^{*0}$  & 2.34  & 7.38 & & $0.81\pm0.15^*$ \\
\hline
$\Xi^+_c\to \Xi^0 \rho^+$  & 95.83  & 5.48& & $$ \\
\hline
$\Xi^0_c\to \Lambda \bar K^{*0}$ & 1.12  & 1.15 & & $$ \\
\hline
$\Xi^0_c\to \Sigma^0 \bar K^{*0}$ & 0.62  & 0.77 & & \\
\hline
$\Xi^0_c\to \Sigma^+ K^{*-}$ & 0.39  & 0.37 & & \\
\hline
$\Xi^0_c\to \Xi^0 \rho^0$ & 1.71  & 1.22 & & \\
\hline
$\Xi^0_c\to \Xi^0 \omega$ & 2.33  & 0.15 & & \\
\hline
$\Xi^0_c\to \Xi^0 \phi$ & 0.18 & 0.10 & & \\
\hline
$\Xi^0_c\to \Xi^- \rho^+$  & 12.29  & 1.50 & &  \\
\hline
$\Omega^0_c\to \Xi^0 \bar K^{*0}$ & 0.59  & & & \\
\hline \hline
\end{tabular}
\end{center}
\end{table}

Instead of decomposing the decay amplitude into products of strong
couplings and two-body weak transitions, K\"orner and Kr\"amer
\cite{Korner} have analyzed the nonleptonic weak processes using
the spin-flavor structure of the effective Hamiltonian and the
wave functions of baryons and mesons described by the covariant
quark model. The nonfactorizable amplitudes are then obtained in
terms of two wave function overlap parameters $H_2$ and $H_3$,
which are in turn determined by fitting to the experimental data
of $\Lambda_c^+\to p\bar{K}^0$ and $\Lambda_c^+\to\Lambda\pi^+$,
respectively. Despite the absence of first-principles calculation
of the parameters $H_2$ and $H_3$, this quark model approach has
fruitful predictions for not only ${\cal B}_c\to {\cal B}+P$, but
also ${\cal B}_c\to {\cal B}+V,~{\cal B}^*(3/2^+)+P$ and ${\cal
B}^*(3/2^+)+V$ decays. Another advantage of this analysis is that
each amplitude has one-to-one quark-diagram interpretation. While
the overlap integrals are treated as phenomenological parameters
to be determined from a fit to the data, Ivanov {\it et al.}
\cite{Ivanov98} developed a microscopic approach to the overlap
integrals by specifying the form of the hadron-quark transition
vertex including the explicit momentum dependence of the Lorentz
scalar part of this vertex.

\begin{table}
\caption{Branching ratios and decay asymmetries (in parentheses)
of Cabibbo-allowed ${\cal B}_c\to{\cal B}(3/2^+)+P(V)$ decays in
various models. Experimental values denoted by the superscript $*$
are the branching ratios relative to $\Xi_c^+\to\Xi^-\pi^+\pi^+$.
The model calculations of Xu and Kamal are done in two different
schemes \cite{XK92b}. } \label{tab:BRs3/2}
\begin{footnotesize}
\begin{center}
\begin{tabular}{|l|c|c|c|c|}
\hline\hline ~~~~Decay & K\"{o}rner,
 & Xu \& Kamal & Cheng &~~Experiment~~ \\
        &  Kr\"{a}mer \cite{Korner}  & \cite{XK92b} & \cite{Cheng97b} & 
\cite{Part5_pdg06} \\
\hline
$\Lambda^+_c\to \Delta^{++}K^-$  & 2.70  & 1.00(0.00);~1.04(0.43) & & $0.86\pm0.30$ \\
\hline
$\Lambda^+_c\to \Delta^+\bar K^0$ & 0.90  & 0.34(0.00);~0.34(0.43) & & 0.99$\pm$ 0.32\\
\hline
$\Lambda^+_c\to \Xi^{*0}K^+$  & 0.50  & 0.08(0.00);~0.08(0.25) & & $0.26\pm0.10$ \\
\hline
$\Lambda^+_c\to \Sigma^{*+}\pi^0$  & 0.50  & 0.22(0.00);~0.24(0.40) & &   \\
\hline
$\Lambda^+_c\to \Sigma^{*0}\pi^+$ & 0.50 & 0.22(0.00);~0.24(0.40) & &  \\
\hline
$\Lambda^+_c\to \Sigma^{*+}\eta$ & 0.54 & & & $0.85\pm0.33$ \\
\hline
$\Xi^+_c\to \Sigma^{*+} \bar K^{0}$  & 0  & 0 & & $1.0\pm0.5^*$  \\
\hline
$\Xi^+_c\to \Xi^{*0} \pi^+$ & 0  & 0 & & $<0.1^*$ \\
\hline
$\Xi^0_c\to \Omega^-K^+$  & 0.34  & 0.15(0.00);~0.16(0.27) & & seen \\
\hline
$\Xi^0_c\to \Sigma^{*0} \bar K^{0}$ & 0.25 & 0.09(0.00);~0.10(0.43) & & \\
\hline
$\Xi^0_c\to \Sigma^{*+} K^{-}$  & 0.49  & 0.18(0.00);~0.19(0.43) & & \\
\hline
$\Xi^0_c\to \Xi^{*0} \pi^0$  & 0.28  & 0.12(0.00);~0.13(0.40) & & \\
\hline
$\Xi^0_c\to \Xi^{*-} \pi^+$  & 0.56  & 0.25(0.00);~0.27(0.40) & &  \\
\hline
 $\Omega_c^0\to\Omega^-\pi^+$ & 0.35$a_1^2$ & 1.47$a_1^2(0)$;
 $1.44a_1^2(0)$ &  0.92$a_1^2(0.17)$ & seen \\ \hline
 $\Omega_c^0\to\Xi^{*0}\bar K^0$  & 0.40$a_2^2$ & 0.69$a_2^2(0)$;
 $0.61a_2^2(0)$ & 1.06$a_2^2(0.35)$ & \\ \hline
 $\Omega_c^0\to\Omega^-\rho^+$  & 2.02$a_1^2$ & 8.02$a_1^2(-0.08)$;
 $7.82a_2^2(-0.21)$ & 3.23$a_1^2(0.43)$ & \\ \hline
 $\Omega_c^0\to\Xi^{*0}\bar K^{*0}$  & 2.28$a_2^2$ & 3.15$a_2^2(-0.09)$;
 $1.13a_2^2(-0.27)$ & 1.60$a_2^2(0.28)$ & \\ \hline
\hline
\end{tabular}
\end{center}
\end{footnotesize}
\end{table}

\subsection{Discussions}

Various model predictions of the branching ratios and decay
asymmetries for Cabibbo-allowed ${\cal B}_c\to{\cal B}+P(V)$
decays are summarized in Tables \ref{tab:BRs}-\ref{tab:BRs3/2}. In
the following we shall first discuss the decay asymmetry parameter
$\alpha$ and then turn to the decay rates.

\subsubsection{Decay asymmetry}
A very  useful information is provided by the study of the
polarization of the daughter baryon ${\cal B}'$ in the decay
${\cal B}\to{\cal B}'\pi$. Its general expression is given by
 \begin{eqnarray}
 {\bf P_{\cal B'}}={(\alpha_{\cal
 B}+{\bf P_{\cal B}}\cdot {\bf n}){\bf n}+\beta_{\cal
B}({\bf n}\times {\bf P_{\cal B}})+\gamma_{\cal
 B}{\bf n}\times({\bf n}\times {\bf P_{\cal
 B}})\over 1+\alpha_{\cal B} {\bf P_{\cal
 B}}\cdot {\bf n}},
 \end{eqnarray}
where ${\bf P_{\cal B}}$ is the parent baryon polarization,
$\alpha_{\cal B}$, $\beta_{\cal B}$ and $\gamma_{\cal B}$ are the
parent baryon asymmetry parameters and ${\bf n}$ is a unit vector
along the daughter baryon ${\cal B'}$ in the parent baryon frame.
If the parent baryon is unpolarized, the above equation reduces to
${\bf P_{\cal B'}}=\alpha_{\cal B}{\bf n}$, which implies that the
baryon ${\cal B'}$ obtained from the decay of the unpolarized
baryon ${\cal B}$ is longitudinally polarized by the amount of
$\alpha_{\cal B}$.  The transverse polarization components are
measured by the parameters $\beta_{\cal B}$ and $\gamma_{\cal B}$.
In terms of the $s$- and $p$-wave amplitudes in Eq.
(\ref{eq:spwave}), the baryon parameters have the expressions
 \begin{eqnarray}
 \alpha={2{\rm Re}(S^*P)\over |S|^2+|P|^2}, \qquad
 \beta={2{\rm Im}(S^*P)\over |S|^2+|P|^2}, \qquad
 \gamma={|S|^2-|P|^2\over |S|^2+|P|^2}, \qquad
 \end{eqnarray}
where
 \begin{eqnarray}
 S=\sqrt{2m_{\cal B'}(E'+m_{\cal B'})}\,A, \qquad
 P=\sqrt{2m_{\cal B'}(E'-m_{\cal B'})}\,B.
 \end{eqnarray}
When $CP$ is conserved and final-state interactions are
negligible, $\beta$ vanishes. Since the sign of $\alpha_{\cal B}$
depends on the relative sign between $s$- and $p$-wave amplitudes,
the measurement of $\alpha$ can be used to discriminate between
different models.

The model predictions for the decay asymmetry $\alpha$ in
$\Lambda_c^+\to\Lambda\pi^+$ range from $-0.67$ to $-0.99$ (see
Table \ref{tab:decasy}). The current world average of $\alpha$ is
$-0.91\pm0.15$~\cite{Part5_pdg06}, while the most recent measurement is
$-0.78\pm0.16\pm0.19$ by FOCUS \cite{FOCUS:alpha}. The agreement
between theory and experiment implies the $V-A$ structure of the
decay process $\Lambda_c^+\to\Lambda\pi^+$.

It is evident from Table \ref{tab:decasy} that all the models
except one model in \cite{Verma98} predict a positive decay
asymmetry for the decay $\Lambda_c^+\to\Sigma^+\pi^0$. Therefore,
the measurement of $\alpha=-0.45\pm0.31\pm0.06$ by CLEO
\cite{CLEO:alpha} is a big surprise. If the negative sign of
$\alpha$ is confirmed in the future, this will imply an opposite
sign between $s$-wave and $p$-wave amplitudes for this decay,
contrary to the model expectation. The implication of this has
been discussed in detail in \cite{CT92}. Since the error of the
previous CLEO measurement is very large, it is crucial to have
more accurate measurements of the decay asymmetry for
$\Lambda_c^+\to\Sigma^+\pi^0$.

The decays $\Lambda_c^+\to\Xi^0K^+$ and $\Xi_c^0\to\Sigma^+K^-$
share some common features that they can proceed via $W$-exchange
\cite{CCT} and that their $s$-wave amplitudes are very small. As a
consequence, their decay asymmetries are expected to be very tiny.
Indeed, all the existing models predict vanishing $s$-wave
amplitude and hence $\alpha=0$ (cf. Table \ref{tab:decasy}).

\subsubsection{$\Lambda_c^+$ decays}
Experimentally, nearly all the branching ratios of the
$\Lambda_c^+$ are measured relative to the $pK^-\pi^+$ mode. Some
Cabibbo-suppressed modes such as $\Lambda_c^+\to\Lambda K^+$ and
$\Lambda_c^+\to\Sigma^0K^+$ have been recently measured by BaBar
\cite{BaBar:LamcCS}. Theoretically, only one model \cite{Uppal}
gives predictions for the Cabibbo-suppressed decays.

The first measured Cabibbo-suppressed mode $\Lambda_c^+\to p\phi$
is of particular interest because it receives contributions only
from the factorizable diagram and is expected to be color
suppressed in the naive factorization approach. An updated
calculation in \cite{CT96} yields
 \begin{eqnarray}
 {\cal B}(\Lambda_c^+\to p\phi)=2.26\times 10^{-3}a_2^2, \qquad
 \alpha(\Lambda_c^+\to p\phi)=-0.10\,.
 \end{eqnarray}
>From the experimental measurement ${\cal B}(\Lambda_c^+\to
p\phi)=(8.2\pm2.7)\times 10^{-4}$~\cite{Part5_pdg06}, it follows that
 \begin{eqnarray} \label{eq:a2}
 |a_2|_{\rm expt}=0.60\pm0.10\,.
 \end{eqnarray}
This is in excellent agreement with the $1/N_c$ approach where
$a_2=c_2(m_c)=-0.59$\,.

\subsubsection{$\Xi_c^+$ decays}
No absolute branching ratios have been measured. The branching
ratios listed in Tables \ref{tab:BRs} and \ref{tab:BRsBV} are the
ones relative to $\Xi_c^+\to\Xi^-\pi^+\pi^+$. Several
Cabibbo-suppressed decay modes such as $p\bar K^{*0}$,
$\Sigma^+\phi$ and $\Xi(1690)K^+$ have been observed.

The Cabibbo-allowed decays $\Xi_c^+\to {\cal B}(3/2^+)+P$ have
been studied and they are believed to be forbidden as they do not
receive factorizable and $1/2^\pm$ pole contributions
\cite{XK92b,Korner}. However, the $\Sigma^{*+}\bar K^0$ mode was
seen by FOCUS before \cite{FOCUS:SigK} and this may indicate the
importance of pole contributions beyond low-lying $1/2^\pm$
intermediate states.

\subsubsection{$\Xi_c^0$ decays}
No absolute branching ratios have been measured so far. However,
there are several measurements of ratios of branching fractions,
for example~\cite{Part5_pdg06},
 \begin{eqnarray}
 R_1={\Gamma(\Xi_c^0\to\Lambda
K_S^0)\over\Gamma(\Xi_c^0\to\Xi^-\pi^+)}=0.21\pm0.02\pm0.02, \quad
R_2=
  {\Gamma(\Xi_c^0\to\Omega^-
K^+)\over\Gamma(\Xi_c^0\to\Xi^-\pi^+)}=0.297\pm0.024\,. \nonumber \\
 \end{eqnarray}
Most models predict a ratio of $R_1$ smaller than 0.18 except the
model of \.Zenczykowski \cite{Zen} which yields $R_1=0.29$ due to
the small predicted rate of $\Xi_c^0\to\Xi^-\pi^+$ (see Table
\ref{tab:BRs}). The model of K\"orner and Kr\"amer \cite{Korner}
predicts $R_2=0.33$ (Table \ref{tab:BRs3/2}), in agreement with
experiment, but its prediction $R_1=0.06$ is too small compared to
the data.

\subsubsection{$\Omega_c^0$ decays}
One of the unique features of the $\Omega_c^0$ decay is that the
decay $\Omega_c^0\to\Omega^-\pi^+$ proceeds only via external
$W$-emission, while $\Omega_c^0\to \Xi^{*0}\bar K^0$ proceeds via
the factorizable internal $W$-emission diagram $C'$. The general
amplitudes for ${\cal B}_c\to B^*({3\over 2}^+)+P(V)$ are:
 \begin{eqnarray} \label{eq:BstP}
M[{\cal B}_i\to {\cal B}_f^*(3/2^+)+P] &=&
iq_\mu\bar{u}^\mu_f(p_f)(C+D
\gamma_5)u_i(p_i),   \nonumber \\
M[{\cal B}_i\to {\cal B}_f^*(3/2^+)+V] &=&
\bar{u}_f^\nu(p_f)\varepsilon^{*\mu}[
g_{\nu\mu}(C_1+D_1\gamma_5)  \nonumber \\
&& +p_{1\nu}\gamma_\mu(C_2+D_2\gamma_5)+p_{1\nu}
p_{2\mu}(C_3+D_3\gamma_5)]u_i(p_i), \nonumber \\
 \end{eqnarray}
with $u^\mu$ being the Rarita-Schwinger vector spinor for a
spin-${3\over 2}$ particle. Various model predictions of
Cabibbo-allowed $\Omega_c^0\to {\cal B}(3/2^+)+P(V)$ are displayed
in Table \ref{tab:BRs3/2} with the unknown parameters $a_1$ and
$a_2$. From the decay $\Lambda_c^+\to p\phi$ we learn that
$|a_2|=0.60\pm0.10$. Notice a sign difference of $\alpha$ for
$\Omega_c\to {3\over 2}^++V$ in \cite{XK92b} and \cite{Cheng97b}.
It seems to us that the sign of $A_i$ and $B_i$ in Eq.~(58) of
\cite{XK92b} should be flipped. A consequence of this sign change
will render $\alpha$ positive in $\Omega_c\to {3\over 2}^++V$
decay. In the model of Xu and Kamal \cite{XK92b}, the $D$-wave
amplitude in Eq. (\ref{eq:BstP}) and hence the parameter $\alpha$
vanishes in the decay $\Omega_c\to {3\over 2}^++P$ due to the fact
that the vector current is conserved at all $q^2$ in their scheme
1 and at $q^2=0$ in scheme 2.

\subsection{Charm-flavor-conserving weak decays}
There is a special class of weak decays of charmed baryons which
can be studied in a reliable way, namely, heavy-flavor-conserving
nonleptonic decays. Some examples are the singly
Cabibbo-suppressed decays $\Xi_c\to\Lambda_c\pi$ and
$\Omega_c\to\Xi'_c\pi$. The idea is simple: In these decays only
the light quarks inside the heavy baryon will participate in weak
interactions; that is, while the two light quarks undergo weak
transitions, the heavy quark behaves as a ``spectator". As the
emitted light mesons are soft, the $\Delta S=1$ weak interactions
among light quarks can be handled by the well known short-distance
effective Hamiltonian. The synthesis of heavy-quark and chiral
symmetries provides a natural setting for investigating these
reactions \cite{ChengHFC}. The weak decays $\Xi_Q\to\Lambda_Q\pi$
with $Q=c,b$  were also studied in \cite{Voloshin00}.

The combined symmetries of heavy and light quarks severely
restricts the weak interactions allowed. In the symmetry limit, it
is found that there cannot be ${\cal B}_{\bar 3}-{\cal B}_6$ and
${\cal B}^*_6-{\cal B}_6$ nonleptonic weak transitions. Symmetries
alone permit three types of transitions:  ${\cal B}_{\bar 3}-{\cal
B}_{\bar 3}$,  ${\cal B}_{6}-{\cal B}_6$ and ${\cal B}^*_6-{\cal
B}_6$ transitions. However, in both the MIT bag and diquark
models, only ${\cal B}_{\bar 3}-{\cal B}_{\bar 3}$ transitions
have nonzero amplitudes.

The predicted rates are \cite{ChengHFC}
 \begin{eqnarray}
 \Gamma(\Xi_c^0\to\Lambda_c^+\pi^-) &=& 1.7\times 10^{-15}\,{\rm GeV},
 \qquad
 \Gamma(\Xi_c^+\to\Lambda_c^+\pi^0) = 1.0\times 10^{-15}\,{\rm GeV},
 \nonumber \\
 && \Gamma(\Omega_c^0\to\Xi'^+_c\pi^-) = 4.3\times 10^{-17}\,{\rm
 GeV},
 \end{eqnarray}
and the corresponding branching ratios are
 \begin{eqnarray}
 {\cal B}(\Xi_c^0\to\Lambda_c^+\pi^-) &=& 2.9\times 10^{-4},
 \qquad
 {\cal B}(\Xi_c^+\to\Lambda_c^+\pi^0) = 6.7\times 10^{-4},
 \nonumber \\
 &&  {\cal B}(\Omega_c^0\to\Xi'^+_c\pi^-) = 4.5\times 10^{-6}.
 \end{eqnarray}
As stated above, the ${\cal B}_{6}-{\cal B}_6$ transition
$\Omega_c^0\to\Xi'^+_c\pi^-$ vanishes in the chiral limit. It
receives a finite factorizable contribution as a result of
symmetry-breaking effect. At any rate, the predicted branching
ratios for the charm-flavor-conserving decays
$\Xi_c^0\to\Lambda_c^+\pi^-$ and $\Xi_c^+\to\Lambda_c^+\pi^0$ are
of order $10^{-3}\sim 10^{-4}$ and  should be readily accessible
in the near future.

\section{Semileptonic decays}

{\tiny
\begin{table}
\caption{Predicted semileptonic decay rates (in units of
$10^{10}s^{-1}$) and decay asymmetries (second entry) in various
models. Dipole $q^2$ dependence for form factors is assumed
whenever the form-factor momentum dependence is not calculated in
the model. Predictions of \cite{Marcial} are obtained in the
non-relativistic quark model and the MIT bag model (in
parentheses).  } \label{tab:SL}
\begin{scriptsize}
\begin{center}
\begin{tabular}{|c|c|c|c|c|c|c|c|c|} \hline \hline
 Process & P\'erez-Marcial & Singleton & Cheng, & ~~Ivanov~~ & ~~Luo~~ & Marques de &
 Huang, & Expt.  \\
 & et al. \cite{Marcial} & \cite{Singleton} & Tseng \cite{CT96} & \cite{Ivanov96}
 & \cite{Luo} & Carvalho \cite{Carvalho} & Wang \cite{Huang} & 
\cite{Part5_pdg06} \\ \hline
 $\Lambda_c^+\to\Lambda^0 e^+\nu_e$  & 11.2 (7.7)  & 9.8 & 7.1 &
 7.22 & 7.0 &  $13.2\pm1.8$ & $10.9\pm3.0$ & $10.5\pm 3.0$ \\
 & & & & $-0.812$ & & $-1$ & $-0.88\pm0.03$ & $-0.86\pm0.04$ \\ \hline
 $\Xi_c^0\to\Xi^- e^+\nu_e$  & 18.1 (12.5) &
8.5 & 7.4 & 8.16 & 9.7 & & & seen \\ \hline
 $\Xi_c^+\to\Xi^0 e^+\nu_e$  & 18.4~(12.7)
& 8.5 & 7.4 & 8.16 & 9.7 & & & seen  \\ \hline \hline
\end{tabular}
\end{center}
\end{scriptsize}
\end{table}}

The exclusive semileptonic decays of charmed baryons:
$\Lambda_c^+\to\Lambda e^+(\mu^+)\nu_e$, $\Xi_c^+\to \Xi^0
e^+\nu_e$ and $\Xi_c^0\to \Xi^-e^+\nu_e$ have been observed
experimentally. Their rates depend on the ${\cal B}_c\to{\cal B}$
form factors $f_i(q^2)$ and $g_i(q^2)$ ($i=1,2,3$) defined in Eq.
(\ref{eq:FF}). These form factors have been evaluated in the
non-relativistic quark model \cite{Marcial,Singleton,CT96}, the
MIT bag model \cite{Marcial}, the relativistic quark model
\cite{Ivanov96}, the light-front quark model \cite{Luo} and QCD
sum rules \cite{Carvalho,Huang}. Experimentally, the only
information available so far is the form-factor ratio measured in
the semileptonic decay $\Lambda_c\to\Lambda e\bar{\nu}$. In the
heavy quark limit, the six $\Lambda_c\to\Lambda$ form factors are
reduced to two:
 \begin{eqnarray}
\langle
\Lambda(p)|\bar{s}\gamma_\mu(1-\gamma_5)c|\Lambda_c(v)\rangle=\,\bar{u}
_{_\Lambda}\left(F_1^{\Lambda_c\Lambda}(v\cdot p)+v\!\!\!/
F_2^{\Lambda_c \Lambda}(v\cdot
p)\right)\gamma_\mu(1-\gamma_5)u_{_{\Lambda_c}}.
 \end{eqnarray}
Assuming a dipole $q^2$ behavior for form factors, the ratio
$R=\tilde{F}_2^{\Lambda_c\Lambda}/\tilde{F}_1^{\Lambda_c\Lambda}$
is measured by CLEO to be \cite{CLEO:sl}
 \begin{eqnarray} R=-0.31\pm 0.05\pm 0.04\,.
 \end{eqnarray}

Various model predictions of the charmed baryon semileptonic decay
rates and decay asymmetries are shown in Table \ref{tab:SL}.
Dipole $q^2$ dependence for form factors is assumed whenever the
form factor momentum dependence is not available in the model. The
predicted rates cited from \cite{Marcial} include QCD corrections.
However, as stressed in \cite{Korner94}, it seems that QCD effects
computed in \cite{Marcial} are unrealistically too large.
Moreover, the calculated heavy-heavy baryon form factors in
\cite{Marcial} at zero recoil do not satisfy the constraints
imposed by heavy quark symmetry \cite{CT96}. From Table
\ref{tab:SL} we see that the computed branching ratios of
$\Lambda_c^+\to\Lambda e^+\nu$ lie in the range $1.4\%\sim 2.6\%$,
in agreement with experiment, $(2.1\pm0.6)\%$~\cite{Part5_pdg06}.
Branching ratios of $\Xi_c^0\to\Xi^-e^+\nu$ and $\Xi_c^+\to\Xi^0
e^+\nu$ are predicted to fall into the ranges $(0.8\sim 2.0)\%$
and $(3.3\sim 8.1)\%$, resepctively. Experimentally, only the
ratios of the branching fractions are available so far~\cite{Part5_pdg06}
 \begin{eqnarray}
 {\Gamma(\Xi_c^+\to\Xi^0
 e^+\nu)\over \Gamma(\Xi_c^+\to\Xi^-\pi^+\pi^+)}= 2.3\pm0.6^{+0.3}_{-0.6},
 \qquad
 {\Gamma(\Xi_c^0\to\Xi^-
 e^+\nu)\over\Gamma(\Xi_c^0\to\Xi^-\pi^+)} =
 3.1\pm1.0^{+0.3}_{-0.5}\,.
 \end{eqnarray}

\section{Electromagnetic and Weak Radiative decays}

Although radiative decays are well measured in the charmed meson
sector, e.g. $D^*\to D\gamma$ and $D_s^+\to D_s^+\gamma$, only
three of the radiative modes in the charmed baryon sector have
been seen, namely, $\Xi'^0_c\to\Xi_c^0\gamma$, $\Xi'^+_c\to
\Xi^+_c\gamma$ and $\Omega_c^{*0}\to\Omega_c^0\gamma$. This is
understandable  because $m_{\Xi'_c}-m_{\Xi_c}\approx 107$ MeV and
$m_{\Omega_c^*}-m_{\Omega_c}\approx 71$ MeV. Hence, $\Xi'_c$ and
$\Omega_c^*$ are governed by the electromagnetic decays. However,
it will be difficult to measure the rates of these decays because
these states are too narrow to be experimentally resolvable.
Nevertheless, we shall systematically study the two-body
electromagnetic decays of charmed baryons and also weak radiative
decays.

\subsection{Electromagnetic decays}

In the baryon sector, the following two-body electromagnetic
decays are of interest:
 \begin{eqnarray}
B_6 \to B_{\overline{3}} + \gamma & : & \Sigma_c \rightarrow
\Lambda_c + \gamma,  \quad \Xi^\prime_c \rightarrow \Xi_c + \gamma , \nonumber \\
 B^\ast_6 \rightarrow B_{\overline{3}} + \gamma & : &
\Sigma^\ast_c \rightarrow \Lambda_c + \gamma ,  \quad \Xi^{\prime
\ast}_c \rightarrow
\Xi_c + \gamma ,  \nonumber \\
B^\ast_6 \rightarrow B_6 + \gamma & : & \Sigma^\ast_c \rightarrow
\Sigma_c + \gamma , \quad \Xi^{\prime \ast}_c \rightarrow
\Xi^\prime_c + \gamma ,  \quad \Omega^\ast_c \rightarrow \Omega_c
+ \gamma ,
 \end{eqnarray}
where we have denoted the spin $\frac{1}{2}$ baryons as $B_6$ and
$B_{\overline{3}}$ for a symmetric sextet {\bf 6}  and
antisymmetric antitriplet {\bf \={3}}, respectively, and the spin
$\frac{3}{2}$ baryon by $B^\ast_6$.

An ideal theoretical framework for studying the above-mentioned
electromagnetic decays is provided by the formalism in which the
heavy quark symmetry and the chiral symmetry of light quarks are
combined \cite{Part5_ming_Yan,Part5_ming_Wise}. When supplemented by the nonrelativistic
quark model, the formalism determines completely the low energy
dynamics of heavy hadrons. The electromagnetic interactions of
heavy hadrons consist of two distinct contributions: one from
gauging electromagnetically the chirally invariant strong
interaction Lagrangians for heavy mesons and baryons given in
\cite{Part5_ming_Yan,Part5_ming_Wise}, and the other from the anomalous magnetic moment
couplings of the heavy particles.  The heavy quark symmetry
reduces the number of free parameters needed to describe the
magnetic couplings to the photon.   There are two undetermined
parameters for the ground state heavy baryons. All these
parameters are related simply to the magnetic moments of the light
quarks in the nonrelativistic quark model.  However, the charmed
quark is not particularly heavy ($m_c \simeq 1.6$ GeV), and it
carries a charge of $\frac{2}{3} e$.  Consequently, the
contribution from its magnetic moment cannot be neglected. The
chiral and electromagnetic gauge-invariant Lagrangian for heavy
baryons can be found in Eqs. (3.8) and (3.9) of \cite{Cheng93}.

The amplitudes of electromagnetic decays are given by
\cite{Cheng93}
 \begin{eqnarray}
 A(B_6\to B_{\bar 3}+\gamma) &=& i\eta_1\bar u_{\bar
3}\sigma_{\mu\nu}k^\mu \varepsilon^\nu u_6,   \nonumber \\
A(B^*_6\to B_{\bar 3}+\gamma) &=&
i\eta_2\epsilon_{\mu\nu\alpha\beta}
\bar u_{\bar 3}\gamma^\nu k^\alpha\varepsilon^\beta u^\mu, \nonumber \\
A(B^*_6\to B_6+\gamma) &=& i\eta_3\epsilon_{\mu\nu\alpha\beta}
\bar u_6\gamma^\nu k^\alpha\varepsilon^\beta u^\mu,
 \end{eqnarray}
where $k_\mu$ is the photon 4-momentum and $\varepsilon_\mu$ is
the polarization 4-vector. The corresponding decay rates are
\cite{Cheng93}
 \begin{eqnarray}
\Gamma(B_6\to B_{\bar 3}+\gamma) &=& \eta_1^2\,{k^3\over \pi},  \nonumber \\
\Gamma(B_6^*\to B_{\bar 3}+\gamma) &=& \eta_2^2\,{k^3\over
3\pi}\,{ 3m_i^2+m_f^2\over 4m_i^2},   \nonumber \\
\Gamma(B_6^*\to B_6+\gamma) &=& \eta_3^2\,{k^3\over 3\pi}\,{3m_i^2
+m_f^2\over 4m_i^2},
 \end{eqnarray}
where $m_i$ ($m_f$) is the mass of the parent (daughter) baryon.

\begin{figure}[t]
\centerline{\psfig{file=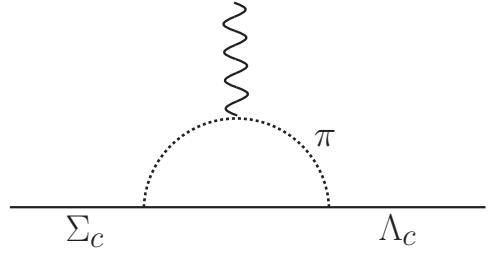,width=2.6in}}
 \caption{Chiral loop contribution to the E2 amplitude of $\Sigma_c^*\to\Lambda_c\gamma$.}
\label{fig:E2}
\end{figure}

The coupling constants $\eta_i$ can be calculated using the quark
model \cite{Cheng93}; some of them are
 \begin{eqnarray} \label{eq:eta}
\eta_1(\Sigma_c^+\to\Lambda_c^+)=\,{e\over 6\sqrt{3}}\left({2\over
M_u}+{1 \over M_d}\right),&&
\eta_2(\Sigma_c^{*+}\to\Lambda_c^+)=\,{e\over 3\sqrt{6}}
\left({2\over M_u}+{1\over M_d}\right),   \nonumber  \\
\eta_3(\Sigma_c^{*++}\to\Sigma_c^{++})=\,{2\sqrt{2}e\over
9}\left({1\over M_u} -{1\over
M_c}\right),&&\eta_3(\Sigma_c^{*0}\to\Sigma_c^0)=\,{2\sqrt{2}e\over
9}\left(-{1\over 2M_d}-{1\over M_c}\right),   \nonumber  \\
\eta_3(\Sigma_c^{*+}\to\Sigma_c^+)=\,{\sqrt{2}e\over
9}\left({1\over M_u} -{1\over 2M_d}-{2\over M_c}\right),
&&\eta_3(\Xi_c^{'*+}\to\Xi_c^+)=\,{e\over 3\sqrt{6}}
\left({2\over M_u}+{1\over M_s}\right),  \nonumber  \\
\eta_3(\Xi_c^{'*0}\to\Xi_c^0)=\,{e\over 3\sqrt{6}}\left(-{1\over
M_d}+{1\over M_s}\right), &&
\eta_3(\Omega_c^{*0}\to\Omega_c^0)=\,{2\sqrt{2}e\over
9}\left(-{1\over 2M_s}-{1\over M_c}\right).\nonumber \\
 \end{eqnarray}
Using the constituent quark masses, $M_u=338$ MeV, $M_d=322$ MeV
and $M_s=510$ MeV~\cite{Part5_pdg06}, the calculated results are
summarized in the second column of Table \ref{tab:em}. A similar
procedure is followed in \cite{Tawfiq01} where the heavy quark
symmetry is supplemented with light-diquark symmetries to
calculate the widths of $\Sigma_c^+\to\Lambda_c^+\gamma$ and
$\Sigma_c^*\to\Sigma_c\gamma$. The authors of \cite{Ivanov96}
apply the relativistic quark model to predict various
electromagnetic decays of charmed baryons. Besides the magnetic
dipole (M1) transition, the author of \cite{Savage95} also
considered and estimated the electric quadrupole (E2) amplitude
for $\Sigma_c^{*+}\to\Lambda_c^+\gamma$ arising from the chiral
loop correction (Fig. \ref{fig:E2}). A detailed analysis of the E2
contributions was presented in \cite{Pich}. The E2 amplitudes
appear at different higher orders for the three kinds of decays:
${\cal O}(1/\Lambda_\chi^2)$ for $B^*_6\to B_6+\gamma$, ${\cal
O}(1/m_Q\Lambda_\chi^2)$ for $B^*_6\to B_{\bar 3}+\gamma$ and
${\cal O}(1/m_Q^3\Lambda_\chi^2)$ for $B_6\to B_{\bar 3}+\gamma$.
Therefore, the E2 contribution to $B_6\to B_{\bar 3}+\gamma$ is
completely negligible.

\begin{table}[t]
\caption{Electromagnetic decay rates (in units of keV) of charmed
baryons.} \label{tab:em}
\begin{center}
\begin{tabular}{|l|c|c|c|c|c|}
\hline\hline ~~~~Decay & ~HHChPT~ & Ivanov
 & Ba\~nuls & Tawfiq & ~~Experiment~~ \\
 & +QM \cite{Cheng97,Cheng93} & ~~et al. \cite{Ivanov96}~~  & ~~et al. \cite{Pich}~~
 & et al. \cite{Tawfiq01} & \cite{Part5_pdg06} \\
\hline
 $\Sigma^+_c\to \Lambda_c^+\gamma$ & 88 & $60.7\pm1.5$  & & 87 &  \\ \hline
 $\Sigma_c^{**+}\to\Sigma_c^{++}\gamma$ & 1.4 & & & 3.04 & \\ \hline
 $\Sigma_c^{*+}\to\Sigma_c^+\gamma$ & 0.002 & $0.14\pm0.004$ & &
 0.19 & \\ \hline
 $\Sigma_c^{*+}\to\Lambda_c^+\gamma$ & 147 & $151\pm4$ & & &
 \\ \hline
 $\Sigma_c^{*0}\to\Lambda_c^0\gamma$ & 1.2 & $$ & & 0.76 & \\ \hline
 $\Xi'^+_c\to\Xi_c^+\gamma$ & 16 & $12.7\pm1.5$ & & & seen \\ \hline
 $\Xi'^0_c\to\Xi_c^0\gamma$ & 0.3 & $0.17\pm0.02$ & $1.2\pm0.7$ & & seen \\ \hline
 $\Xi'^{*+}_c\to\Xi_c^+\gamma$ & 54 & $54\pm3$ & & & \\ \hline
 $\Xi'^{*0}_c\to\Xi_c^0\gamma$ & 1.1 & $0.68\pm0.04$ & $5.1\pm2.7$ & & \\ \hline
 $\Omega_c^{*0}\to\Omega_c^0\gamma$ & 0.9 & & & & seen \\ \hline
 \hline
\end{tabular}
\end{center}
\end{table}

Chiral-loop corrections to the M1 electromagnetic decays and to
the strong decays of heavy baryons have been computed at the one
loop order in \cite{ChengSU3}. The leading chiral-loop effects we
found are nonanalytic in the forms of $m/ \Lambda_\chi$ and
$(m^2/\Lambda^2_\chi)\ln(\Lambda^2/ m^2)$ (or $m_q^{1/2}$ and
$m_q\ln m_q$, with $m_q$ being the light quark mass). Some results
are \cite{ChengSU3}
 \begin{eqnarray}
\Gamma(\Sigma_c^+\to\Lambda_c^+\gamma)= \,112\,{\rm keV}, \quad
\Gamma({\Xi'}_c^+\to\Xi_c^+\gamma)= \,29\,{\rm keV}, \quad
\Gamma({\Xi'}_c^0\to\Xi_c^0\gamma)= \,0.15\,{\rm keV},\nonumber \\
 \end{eqnarray}
which should be compared with the corresponding quark-model
results: 88 keV, 16 keV and 0.3 keV (Table \ref{tab:em}).

The electromagnetic decay $\Xi'^{*0}_c\to\Xi_c^0\gamma$ is of
special interest. It has been advocated in \cite{Lu} that a
measurement of its branching ratio will determine one of the
coupling constants in HHChPT, namely, $g_1$. The radiative decay
$\Xi_c^{'*0}\to\Xi_c^0\gamma$ is forbidden at tree level in SU(3)
limit [see Eq. (\ref{eq:eta})]. In heavy baryon chiral
perturbation theory, this radiative decay is induced via chiral
loops where SU(3) symmetry is broken by the light current quark
masses. By identifying the chiral loop contribution to
$\Xi'^{*0}_c\to\Xi_c^0\gamma$ with the quark model prediction
given in Eq. (\ref{eq:eta}), it was found in \cite{Cheng97} that
one of the two possible solutions is in accord with the quark
model expectation for $g_1$.

For the electromagnetic deacys of $p$-wave charmed baryons, the
search of $\Lambda_c(2593)^+\to\Lambda_c^+\gamma$ and
$\Lambda_c^+(2625)^+\to\Lambda_c^+\gamma$ has been failed so far.
On the theoretical side, the interested reader is referred to
Refs.~\cite{Tawfiq01,Ivanov96,Lu,Zhu,Cho,Cohen} for more details.

The electromagnetic decays considered so far do not test
critically the heavy quark symmetry nor the chiral symmetry.  The
results follow simply from the quark model.  There are examples in
which both the heavy quark symmetry and the chiral symmetry enter
in a crucial way. These are the radiative decays of heavy baryons
involving an emitted pion. Some examples which are kinematically
allowed are
 \begin{eqnarray}
\Sigma_c \rightarrow \Lambda_c \pi \gamma,~~ \Sigma^\ast_c
\rightarrow \Lambda_c \pi \gamma,~~ \Sigma^\ast_c \rightarrow
\Sigma_c \pi \gamma,~~  \Xi^\ast_c \rightarrow \Xi_c \pi \gamma.
 \end{eqnarray}
For an analysis of the decay $\Sigma_c\to\Lambda_c\pi\gamma$, see
\cite{Cheng93}.

\subsection{Weak radiative decays}

\begin{figure}[t]
\centerline{\psfig{file=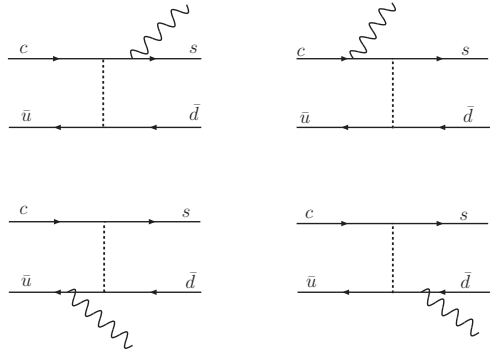,width=2.6in}}
 \caption{$W$-exchange diagrams contributing to the quark-quark
bremsstrahlung process $c+\bar{u}\to s+\bar{d}+\gamma$. The
$W$-annihilation type diagrams are not shown here.}
\label{fig:rad}
\end{figure}

At the quark level, there are three different types of processes
which can contribute to the weak radiative decays of heavy
hadrons, namely, single-, two- and three-quark transitions
\cite{Kamal82}. The single-quark transition mechanism comes from
the so-called electromagnetic penguin diagram. Unfortunately, the
penguin process $c\to u\gamma$ is very suppressed and hence it
plays no role in charmed hadron radiative decays. There are two
contributions from the two-quark transitions: one from the
$W$-exchange diagram accompanied by a photon emission from the
external quark, and the other from the same $W$-exchange diagram
but with a photon radiated from the $W$ boson. The latter is
typically suppressed by a factor of $m_qk/M_W^2$ ($k$ being the
photon energy) as compared to the former bremsstrahlung process
\cite{Kamal83}. For charmed baryons, the Cabibbo-allowed decay
modes via $c\bar{u}\to s\bar{d}\gamma$ (Fig. \ref{fig:rad}) or
$cd\to us\gamma$ are
 \begin{eqnarray} \label{eq:rad}
  \Lambda_c^+\to\Sigma^+\gamma,\qquad \Xi_c^0\to\Xi^0\gamma.
 \end{eqnarray}
Finally, the three-quark transition involving $W$-exchange between
two quarks and a photon emission by the third quark is quite
suppressed because of very small probability of finding three
quarks in adequate kinematic matching with the baryons
\cite{Kamal82,Hua}.

The general amplitude of the weak radiative baryon decay reads
 \begin{eqnarray}
 A({\cal B}_i\to {\cal B}_f\gamma)=\,i\bar{u}_f(a+b\gamma_5)\sigma_{\mu\nu}\varepsilon^\mu
k^ \nu u_i,
 \end{eqnarray}
where $a$ and $b$ are parity-conserving and -violating amplitudes,
respectively. The corresponding decay rate is
 \begin{eqnarray}
 \Gamma({\cal B}_i\to {\cal B}_f\gamma)=\,{1\over 8\pi}\left( {m^2_i-m^2_f\over
m_i}\right) ^3(|a|^2+|b|^2).
 \end{eqnarray}

Nonpenguin weak radiative decays of charmed baryons such as those
in (\ref{eq:rad}) are characterized by emission of a hard photon
and the presence of a highly virtual intermediate quark between
the electromagnetic and weak vertices. It has been shown in
\cite{Chengweakrad} that these features should make possible to
analyze these processes by perturbative QCD; that is, these
processes are describable by an effective local and gauge
invariant Lagrangian:
 \begin{eqnarray}
 {\cal H}_{\rm eff}(c\bar{u}\to s\bar{d}\gamma)=\,{G_F\over
2\sqrt{2}}V_{cs} V_{ud}^*(c_+{O}^F_++c_-{O}^F_-),
 \end{eqnarray}
with
 \begin{eqnarray}
{O}^F_\pm(c\bar{u}\to s\bar{d}\gamma)=\,{e\over m_i^2-m_f^2}
&\Bigg\{& \left( e_s{m_f\over m_s}+e_u{m_i\over
m_u}\right)\left(\tilde{F}_{\mu\nu}+iF_{\mu\nu}
\right)O_\pm^{\mu\nu}   \\
&-&\left(e_d{m_f\over m_d}+e_c{m_i\over
m_c}\right)\left(\tilde{F}_{\mu\nu}
-iF_{\mu\nu}\right)O_\mp^{\mu\nu}\Bigg\},
 \end{eqnarray}
where $m_i=m_c+m_u$, $m_f=m_s+m_d$, $\tilde{F}_{\mu\nu}\equiv
{1\over 2}\epsilon_{\mu\nu\alpha\beta}F^{ \alpha\beta}$ and
  \begin{eqnarray}
 O_\pm^{\mu\nu}=\,\bar{s}\gamma^\mu(1-\gamma_5)c\bar{u}\gamma^\nu(1-\gamma_5)d
\pm \bar{s}\gamma^\mu(1-\gamma_5)d\bar{u}\gamma^\nu(1-\gamma_5)c.
 \end{eqnarray}

For the charmed baryon radiative decays, one needs to evaluate the
matrix element $\langle {\cal B}_f|O_\pm^{\mu\nu}|{\cal
B}_i\rangle$.  Since the quark-model wave functions best resemble
the hadronic states in the frame where both baryons are static,
the static MIT bag model \cite{MIT} was thus adopted in
\cite{Chengweakrad} for the calculation. The predictions are
 \begin{eqnarray}
 && {\cal B}(\Lambda_c^+\to
\Sigma^+\gamma)=\,4.9\times 10^{-5},\qquad \alpha
(\Lambda_c^+\to\Sigma^+\gamma)=-0.86\,,   \nonumber \\
&& {\cal B}(\Xi_c^0\to\Xi^0\gamma)=3.6\times 10^{-5}, \qquad
\alpha(\Xi_c^0 \to\Xi^0\gamma)=-0.86\,.
 \end{eqnarray}
A different analysis of the same decays was carried out in
\cite{Uppalrad} with the results
 \begin{eqnarray}
&& {\cal B}(\Lambda_c^+\to \Sigma^+\gamma)=\,2.8\times
10^{-4},~~~~\alpha (\Lambda_c^+\to\Sigma^+\gamma)=0.02\,,
\nonumber \\ && {\cal B}(\Xi_c^0\to\Xi^0\gamma)=1.5\times
10^{-4},~~~~\alpha(\Xi_c^0 \to\Xi^0\gamma)=-0.01\,.
 \end{eqnarray}
Evidently, these predictions (especially the decay asymmetry) are
very different from the ones obtained in \cite{Chengweakrad}.

Finally, it is worth remarking that, in analog to the
heavy-flavor-conserving nonleptonic weak decays as discussed 
above, there is a special class of weak radiative decays in
which heavy flavor is conserved. Some examples are $\Xi_c \to
\Lambda_c \gamma$ and $\Omega_c \to \Xi_c \gamma$.  In these
decays, weak radiative transitions arise from the diquark sector
of the heavy baryon whereas the heavy quark behaves as a
spectator. However, the dynamics of these radiative decays is more
complicated than their counterpart in nonleptonic weak decays,
e.g., $\Xi_c \to \Lambda_c \pi$. In any event, it deserves a
detailed study.

\chapter[$\Dz - \Dzb$ Mixing]{$\Dz - \Dzb$ Mixing}
\label{sec:charm_mixing}

Meson-antimeson oscillations have been of central importance in the
evolution of the Standard Model (SM) and in searches for new physics. 
This is based on two
seminal features~\cite{CHARMREV}:
(a) In such oscillations quantum mechanical effects will build up over
macroscopic distances, which
makes tiny mass differences measurable. (b) Oscillations and the
rare decay reactions  $s \rightarrow d l^+l^-$, $s \rightarrow d
\nu\overline{\nu}$,
$b \rightarrow s \gamma$, and $b \rightarrow sl^+l^-$ are driven by flavor
changing neutral currents
(FCNC), which are highly suppressed in the SM, where they are given by
1-loop processes.


A program of searches for new physics in charm is complementary to the
corresponding programs in bottom or strange systems, since the situation
is fundamentally different for neutral $D$ mesons built from the {\em up}-type quark
$c$. For example,  $D^0 - \bar D^0$ mixing or FCNC decays are
sensitive to the dynamics of ultra-heavy down-type particles. 
Although, the $D^0 - \bar D^0$ oscillations have to be rather slow in the SM, 
large statistics usually available in charm physics experiment makes
it possible to probe small effects that might be generated by the presence
of new physics particles and interactions. 
 
On the other hand, $D^0 - \bar D^0$ oscillations can be generated
at second order in $SU(3)_{Fl}$ breaking from long-distance contributions. 
As discussed below, within the
SM $D^0 - \bar D^0$ oscillations are driven by long distance dynamics, over
which our theoretical control
is rather limited. It is quite difficult to make the statement that these
oscillations are slow.
Nevertheless, it is mandatory to probe $D^0 - \bar D^0$ oscillations as
sensitively as possible;
while their observation by themselves might not provide conclusive proof for
the intervention of New Physics, it is an essential element in searches for
$CP$ violation, where such conclusive proof can be obtained.

\section[Theoretical Review]{Theoretical Review\footnote{by Edited by I.~I.~Bigi and H.~B.~Li}}

\subsection{Oscillation Formalism: the Phenomenology}
\label{OSCFORM}

It has become customary to use the terms `oscillations' and
`mixing' in an interchangeable way. This is unfortunate. For while
the two terms describe phenomena that are related, they are of a
different nature with `mixing' denoting the more general concept
and `oscillations' the more specific one.

{\em Mixing} means that classically  distinct states are not
necessarily so in quantum mechanics and therefore can {\em
interfere}. For example in atomic physics wave functions are said
to be {\em mixtures} of `right' and `wrong' components whose
interference generates parity odd observables; it is the weak
neutral current that induces such wrong parity components. Mass
eigenstates of quarks (and leptons) contain components of
different flavors giving rise to the non-diagonal CKM (or PMNS)
matrix described below. This is usually referred to as quark
mixing. Such mixing creates a plethora of observable effects.

The most intriguing mixings arise when the violation of a certain quantum
number -- like strangeness --
leads to stationary or mass eigenstates that are not eigenstates of 
that quantum number. This
induces {\em oscillations} like matter-antimatter oscillations discussed
above or neutron-antineutron
oscillations or neutrino oscillations. {\em Oscillations} thus require {\em
mixing}, but go beyond it in the
sense that they generate transitions with a very peculiar time evolution,
namely an oscillatory one rather than the usual exponentially damped one.

The time evolution of the $D^0 -\overline{D}^0$ system is described by the
Schr\"odinger-like equation as
\begin{equation}
i \frac{\partial}{\partial t}\,\left(
\begin{array}{c}
D^0(t)\\
\bar{D^0}(t)
\end{array}\right)
=\left({\bf M} - \frac{i}{2}{\bf \Gamma} \right)\,\left(
\begin{array}{c}
D^0(t)\\
\bar{D^0}(t)
\end{array}\right),
\label{eq:Schrodinger}
\end{equation}
where the ${\bf M}$ and ${\bf \Gamma}$ matrices are Hermitian, and are
defined as
\begin{equation}
\left({\bf M} - \frac{i}{2}{\bf \Gamma} \right) = \left(
\begin{array}{cc}
M_{11} - \frac{i}{2}\Gamma_{11} & M_{12} - \frac{i}{2}\Gamma_{12}\\
M^*_{12} - \frac{i}{2}\Gamma^*_{12} & M_{22} - \frac{i}{2}\Gamma_{22}\\
\end{array}\right).
\label{eq:operator}
\end{equation}
$CPT$ invariance imposes
\begin{equation}
M_{11} = M_{22}  \equiv M \, , \, \Gamma_{11} = \Gamma_{22}  \equiv \Gamma.
\label{eq:mgamma}
\end{equation}
The off-diagonal elements of these matrices describe the dispersive and
absorptive parts of $D^0 -\overline{D}^0$ mixing (for details
see Ref.~\cite{bigi_sanda}).
The two eigenstates $D_1$ and $D_2$ of the effective Hamiltonian matrix
$({\bf M} -\frac{i}{2}{\bf \Gamma})$ are given by
\begin{equation}
|D_1\rangle = \frac{1}{\sqrt{|p|^2+|q|^2}}(p|D^0\rangle + q |\overline{D}^0
\rangle )\, , \,
|D_2\rangle = \frac{1}{\sqrt{|p|^2+|q|^2}}(p|D^0\rangle - q |\overline{D}^0
\rangle ).
\label{eq:eigenstates}
\end{equation}
The corresponding eigenvalues are
\begin{equation}
\lambda_{D_1} \equiv m_{1} - \frac{i}{2} \Gamma_{1} = \left( M - \frac{i}{2}
\Gamma \right) + \frac{q}{p} \left( M_{12} - \frac{i}{2}\Gamma_{12} \right
),
\label{eq:eigenvalue1}
\end{equation}
\begin{equation}
\lambda_{D_2} \equiv m_{2} - \frac{i}{2} \Gamma_{2} = \left( M -
\frac{i}{2}
\Gamma \right) - \frac{q}{p} \left( M_{12} - \frac{i}{2}\Gamma_{12} \right
),
\label{eq:eigenvalue2}
\end{equation}
where $m_1$($m_2$) and $\Gamma_1$($\Gamma_2$) are the mass and width of
$D_1$ ($D_2$), respectively, and
\begin{equation}
\frac{q}{p} = \left( \frac{M^*_{12} - \frac{i}{2}
\Gamma^*_{12}}{M_{12} - \frac{i}{2} \Gamma_{12}}\right)^{1/2}.
\label{eq:cp_mixing}
\end{equation}

From Eqs.~\ref{eq:eigenvalue1} and \ref{eq:eigenvalue2}, one can get
the differences in mass and width:
\begin{equation}
\Delta m \equiv m_2 - m_1 = -2 {\mbox Re}\left[\frac{q}{p}(M_{12} -
\frac{i}{2}\Gamma_{12})\right],
\label{eq:diff_mass}
\end{equation}
\begin{equation}
\Delta \Gamma \equiv \Gamma_2 - \Gamma_1 = -2 {\mbox Im}\left[\frac{q}{p}(M_{12} -
\frac{i}{2}\Gamma_{12})\right].
\label{eq:diff_width}
\end{equation}
The subscripts  are mere labels at this point. We have chosen the
definitions of
$\Delta M$ and $\Delta \Gamma$ such tha when applied to the kaon sector 
with
$K_S =K_1$ and $K_L = K_2$, we get both differences positive.

A pure $D^0$ state generated at $t=0$ could decay to $K^+ \pi^-$ state
either by $D^0 -\overline{D}^0$ mixing or by DCSD, and the two amplitudes
may interfere. The time evolutions of $|D_1  \rangle$ and  $|D_2  \rangle$
are given by
\begin{equation}
|D_i  \rangle = e_i |D_i  \rangle \, , \, e_i = e^{-im_i t - \frac{1}{2}
\Gamma_i
t} \, , \, (i = 1, 2).
\label{eq:time_evolution1}
\end{equation}
Under the phase convention $CP|D^0 \rangle = |\overline{D}^0
\rangle$, a state that is purely $|D^0 \rangle$ ($|\overline{D}^0\rangle$)
prepared by the strong interaction at $t=0$ will evolve to
$|D^0_{\mbox{phys}}(t)
\rangle$
($|\overline{D}^0_{\mbox{phys}}(t)\rangle$):
\begin{equation}
|D^0_{\mbox{phys}}(t) \rangle = g_+(t) |D^0\rangle +
\frac{q}{p}g_-(t) |\overline{D}^0 \rangle \, ,
\label{eq:mixing_evolve1}
\end{equation}
\begin{equation}
|\overline{D}^0_{\mbox{phys}}(t) \rangle = \frac{p}{q} g_- (t) |D^0\rangle +
g_+(t) |\overline{D}^0 \rangle \, ,
\label{eq:mixing_evolve2}
\end{equation}
where
\begin{equation}
g_{\pm}(t) = \frac{1}{2}(e_1 \pm e_2).
\end{equation}
The probability to find a $\overline{D}^0$ at time $t$ in an
initially pure $D^0$ beam is given by
\begin{equation}
|\langle \overline{D}^0|D^0_{\mbox{phys}}(t) \rangle = \frac{1}{4} \left|
\frac{q}{p}\right|^2 e^{-\Gamma_2 t} \left(1+e^{-\Delta \Gamma t} - 2
e^{-\frac{1}{2} \Delta \Gamma t}\mbox{cos}\Delta m t \right).
\label{eq:mixing_D_in_Dphys}
\end{equation}
While the flavour of the initial meson is tagged by its production, the
flavour of the final meson is inferred from its decay.

There are two dimensionless ratios describing the interplay between
oscillations and decays:
\begin{equation}
x \equiv \frac{\Delta m}{\Gamma}\, , \, y \equiv \frac{\Delta
\Gamma}{2\Gamma} ,
\label{eq:mixing_x_and_y}
\end{equation}
where $\Gamma \equiv (\Gamma_1 + \Gamma_2)/2$ is the averaged $D^0$ width.

\subsection{Time-dependent Rate for Incoherent $D$ Decays}
\label{subsec:incoherent}

Searches for mixing attempt to identify the process $|D^0 \rangle
\rightarrow |\overline{D}^0\rangle$ ($|\overline{D}^0\rangle \rightarrow
|D^0\rangle$) by analyzing the decay products of a particle known to be
created as a $|D^0\rangle$ ($|\overline{D}^0\rangle$). In practice, this
means reconstructing the state $|f\rangle$ in an attempt to
observe
\begin{equation}
|D^0 \rangle \rightarrow |\overline{D}^0\rangle \rightarrow
|f \rangle
\label{eq:mix_amplitude}
\end{equation}
The difficulty comes from the fact that for hadronic systems, the decay
\begin{equation}
|D^0 \rangle \rightarrow |f \rangle\, 
\label{eq:mix_DCSD}
\end{equation}
can occur directly, without any mixing at all. Distinguishing
process~(\ref{eq:mix_amplitude}) from~(\ref{eq:mix_DCSD}) is the primary 
goal of
$D$ mixing searches, and it relies on the fact that the decay-time
distribution of the final state $|f \rangle$ is different for the
two processes. The most sensitivity to mixing will occur when the
amplitude for process~(\ref{eq:mix_DCSD}) is as small as possible and,
therefore, doubly Cabibbo-suppressed (DCS) decays are chosen for this type
of analysis.

Let us define $A_f \equiv \langle f|{\cal H}|D^0\rangle$, $\overline{A}_f
\equiv \langle f||{\cal H}|\overline{D}^0\rangle$ with $\rho_f \equiv
\frac{\overline{A}_f}{A_f}$; and $A_{\overline{f}} \equiv \langle
\overline{f}|{\cal H}|D^0\rangle$, $\overline{A}_{\overline{f}}
\equiv \langle \overline{f}||{\cal H}|\overline{D}^0\rangle$ with
$\overline{\rho}_{\overline{f}} \equiv
\frac{A_{\overline{f}}}{\overline{A}_{\overline{f}}}$. Now the time-dependent
{\it wrong sign} decay amplitude for states of initially pure $|D^0\rangle$
($|\overline{D}^0\rangle$) to decay to $|f\rangle$ ($|\overline{f}\rangle$)
is given by (with $f = K^+ \pi^-$ and $|\overline{f}\rangle \equiv
CP|f\rangle$)
\begin{equation}
A(|D^0_{\rm phys}(t) \rightarrow f ) = \frac{q}{p}\overline{A}_f \left[
\lambda^{-1}_f g_+(t) + g_-(t) \right],
\label{eq:mix_wrong_sign_D0}
\end{equation}
and
\begin{equation}
A(|\overline{D}^0_{\rm phys}(t) \rightarrow \overline{f}) =
\frac{p}{q} A_{\overline{f}}
\left[
\lambda_{\overline{f}} g_-(t) + g_+(t) \right],
\label{eq:mix_wrong_sign_antiD0}
\end{equation}
where
\begin{equation}
\lambda_f \equiv \frac{q}{p}\frac{\overline{A}_f}{A_f} \equiv \frac{q}{p}
\rho_f \, , \, \lambda_{\overline{f}} \equiv \frac{q}{p}
\frac{\overline{A}_{\overline{f}}}{A_{\overline{f}}}
\equiv  \frac{q}{p}
\overline{\rho}_{\overline{f}}.
\label{eq:mix_lambda_f}
\end{equation}
and
\begin{equation}
\frac{q}{p} = (1+A_M) e^{i\phi}.
\end{equation}
In order to describe the three types of $CP$ violation in a convenient way,
one can also parameterize $\lambda_f$ ($\lambda_{\overline{f}}$)
as~\cite{Part5_Nir,Part5_CLEO_GODANG00}
\begin{equation}
\lambda_f \equiv \frac{q}{p}\frac{\overline{A}_f}{A_f}  =
\frac{(1+A_M)}{\sqrt{R_D}(1+A_D)} e^{i(\delta + \phi)}
 \label{eq:mix_lambda_f_cpphase},
\end{equation}
\begin{equation}
\lambda_{\overline{f}} \equiv  \frac{q}{p}
\frac{\overline{A}_{\overline{f}}}{A_{\overline{f}}}
 = \frac{\sqrt{R_D}(1+A_M)}{(1+A_D)} e^{-i(\delta -\phi)}
\label{eq:mix_lambda_f_cpphase_2},
\end{equation}
where $|q|/|p| = (1+A_M)$ and $|\overline{A}_f|/|A_f| =
\sqrt{R_D}(1+A_D)$, and $\delta$ is the strong phase difference
between $\overline{A}_f$ and $A_f$. Here $\phi$ represents the
convention-independent weak phase difference between the ratio of
decay amplitudes and the mixing matrix. In
Chapter~\ref{sec:charm_cpv}, we will use these definitions to
describe the three types of $CP$ violation in detail. In the limit
of $CP$ conservation, $A_M$, $A_D$ and $\phi$ are all zero.

It is usual to normalize the {\it wrong sign} decay distributions to the
integrated rate of {\it right sign} decays and to express time in units of
the precisely measured $D^0$ mean lifetime, $\overline{\tau}_{D^0} =
1/\Gamma = 2/(\Gamma_1 +\Gamma_2)$. Therefore the time-dependent rates of
production of the {\it wrong sign} final states relative to the integrated
{\it right sign} states are:
\begin{equation}
r(t) = \left|\frac{q}{p}\right|^2
\left| \lambda^{-1}_f g_+(t) + g_-(t) \right|^2,
\label{eq:mix_rate_D0}
\end{equation}
and
\begin{equation}
\overline{r}(t) = \left|\frac{p}{q}\right|^2\left|
\lambda_{\overline{f}} g_-(t) + g_+(t) \right|^2.
\label{eq:mix_rate_antiD0}
\end{equation}
We will expand $r(t)$ and $\overline{r}(t)$ to second order in time for the
modes where the ratio of the decay amplitudes $R_D = |A_f/\overline{A}_f|^2$
is very small (it is the ratio of double-cabibbo-favored (DCS) decay rate and cabibbo-favored (CF) decay rate).

\subsection{Standard Model Predictions for Oscillation Parameters}
Oscillations arise from $\Delta C = 2$ interactions that
generate off-diagonal terms in mass matrix as in Eq.\ref{eq:operator} for
$D^0$ and $\overline{D}^0$
mesons. The expansion of the off-diagonal terms in the neutral $D$ mass
matrix to second order in perturbation theory is
\begin{equation}
M_{12} = \langle \overline{D}^0|{\cal H}_w^{\Delta C=2} | D^0 \rangle + P
\sum_n {\langle \overline{D}^0 | {\cal H}_w^{\Delta C=1} | n
 \rangle\, \langle n | {\cal H}_w^{\Delta C=1} | D^0 \rangle
\over m^2_D-E^2_n}\, ,
\label{eq:theory_off_m12}
\end{equation}
\begin{equation}
\Gamma_{12} = \sum_n { \rho_n \langle \overline{D}^0 | {\cal H}_w^{\Delta
C=1} |
n
 \rangle\, \langle n | {\cal H}_w^{\Delta C=1} | D^0 \rangle }\, ,
\label{eq:theory_off_G12}
\end{equation}
where the sum runs over all relevant intermediate states (virtual ones for
$M_{12}$ and real ones for
$\Gamma_{12}$), $P$ denotes the principle
value, and $\rho_n$ is the density of the states $n$.

The operator in the first term -- ${\cal H}_w^{\Delta C=2}$ -- is a {\em
local} one at scale $\mu \sim m_D$, so it contributes to the $M_{12}$
(but not to the $\Gamma_{12}$) part of the generalized mass matrix.
New Physics could induce such a contribution of potentially significant
size. The SM generates such a term, namely from the quark box diagram with $b$
quarks as internal quarks, yet it is truly tiny due to its highly suppressed CKM
parameters.
The second term in Eq.~\ref{eq:theory_off_m12} comes from a
double insertion of $\Delta C=1$ operators;
it contributes to both $M_{12}$ and $\Gamma_{12}$.  The dominant SM
contribution comes from here as described below; here New Physics could 
make a significant contribution to $M_{12}$.

%
\begin{figure}[htb]
\begin{center}
\scalebox{0.8}{\epsfig{file=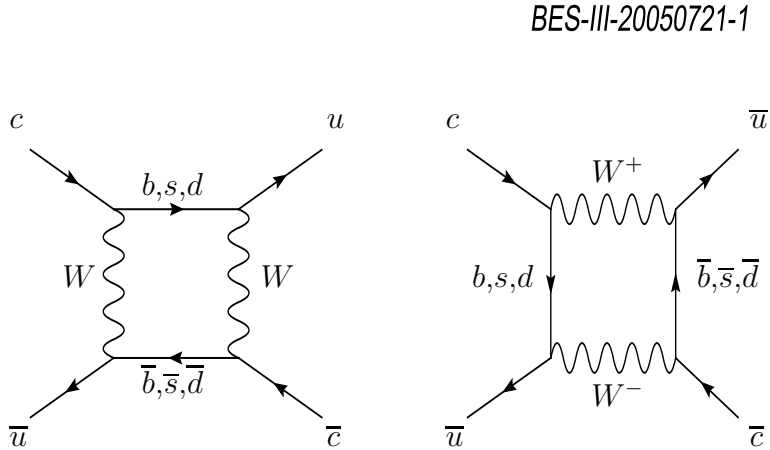,height=7cm,width=12cm}}
\put(-280,110){$c$}\put(-230,85){$b$,$s$,$d$}\put(-230,10){$\overline{b}$,$\overline{s}$,$\overline{d}$}
\put(-280,-10){$\overline{u}$}\put(-260,50){$W$}\put(-185,50){$W$}\put(-160,110){$u$}\put(-160,-10){$\overline{c}$}
\put(-115,110){$c$}\put(-115,-10){$\overline{u}$}\put(-60,90){$W^+$}\put(-60,5){$W^-$}
\put(-105,50){$b$,$s$,$d$}\put(-20,50){$\overline{b}$,$\overline{s}$,$\overline{d}$}
\put(0,110){$\overline{u}$}\put(0,-10){$\overline{c}$}
\caption{Standard Model box diagrams of flavor-changing neutral currents
contributing to $D^0 -\overline{D}^0$ mixing at the quark level.}
\label{fig:box}
\end{center}
\end{figure}


\subsubsection*{Short-Distance Contribution to $x$ and $y$}

In the SM there is a {\em bona fide} short-distance $\Delta C =2$ 
operator, which
is inferred from the quark box diagram with $b$ quarks as the intermediate quarks, see
Fig.~\ref{fig:box}.
The effects due to
intermediate $b$ quarks are evaluated in a straightforward way since they
are far off-shell~\cite{bigi_review}:
\begin{equation}
\Delta m^{b\overline{b}} \simeq - \frac{G_F^2 m_b^2}{8\pi ^2}
\left| V^*_{cb}V_{ub}\right| ^2
\frac{\langle D^0|(\overline{u} \gamma_{\mu} (1 \!-\! \gamma _5)c)
(\overline{u} \gamma_{\mu} (1 \!- \!\gamma _5)c) |\overline{D}^0
\rangle}{2M_D}\; ;
\label{eq:theory_box_b}
\end{equation}
however they are highly suppressed by
the tiny CKM parameters. Using factorization to
estimate the  matrix element one finds $x^{b\overline{b}} \sim 10^{-6}$.
Loops with one $b$ and one light quark likewise are suppressed.

For the light intermediate quarks -- $d,s$ -- the momentum scale is set by
the {\em external} mass $m_c$. However, it is highly GIM suppressed
\begin{eqnarray}
\Delta m^{(s,d)} &&\simeq
- \frac{G_F^2 m_c^2}{8\pi ^2}
\left| V^*_{cs}V_{us}\right| ^2
\frac{\left( m_s^2 \!-\! m_d^2\right) ^2}{m_c^4} \times \nonumber\\
&& \frac{\langle D^0 | (\overline{ u} \gamma_{\mu} (1 \!-\! \gamma _5)c)
(\overline{u} \gamma_{\mu} (1 \!-\! \gamma _5)c) +
(\overline{u} (1 \!+ \!\gamma _5)c)
(\overline{u}  (1 \!+ \!\gamma _5)c)|\overline{D}^0 \rangle}{2M_D}
\;.
\label{eq:theory_box_sd}
\end{eqnarray}
In contrast to $K^0 -\overline{K}^0$ and $B^0-\overline{B}^0$ mixing,
the internal quarks in the box diagrams
here are down-type quarks. The $b$-quark contribution, which would give, 
in principle, the largest GIM violation, is suppressed by small CKM mixing
factors $V^*_{cb}V_{ub}$. The leading contribution, as shown in
Eq.~\ref{eq:theory_box_sd},
is given by the strange quark and therefore results in a very effective GIM
suppression.

The contribution to $\Delta \Gamma$ from the bare quark box is
greatly suppressed by a factor $m_s^6$. The GIM mass insertions yield
a factor $m_s^4$. Contrary to a claim in Ref.~\cite{part5_Falk}, the 
additional factor of $m_s^2$ is {\em not} due to helicity suppression -- 
the GIM factors already take care of that effect; it is of an accidental
nature: it arises because the weak currents are purely $V-A$ and only
in four dimensions. Including radiative QCD corrections to the
box diagram yields contributions $\propto m_s^4\alpha_S/\pi$.
Numerically one finds:
\begin{eqnarray}
y^{box} \ll x^{box} \simeq (10^{-6}) - (10^{-5}).
\label{eq:theory_box_x_y}
\end{eqnarray}

\subsubsection*{Long-distance Contribution to $x$ and $y$}
\begin{figure}[htb]
\begin{center}
\scalebox{0.8}{\epsfig{file=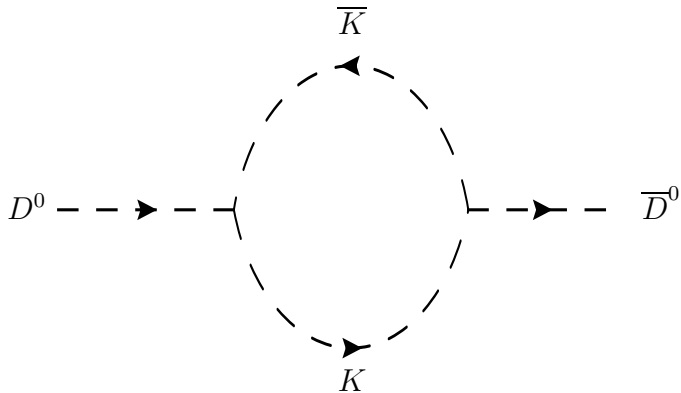,height=7cm,width=12cm}}
\put(-290,50){$D^0$}\put(-165,120){$\overline{K}$}
\put(-50,50){$\overline{D}^0$}\put(-165,-15){$K$}
\caption{A hadron-level diagram of a long-distance physics contribution to
$D^0 -\overline{D}^0$ mixing.}
\label{fig:long_mixing}
\end{center}
\end{figure}
The long-distance contributions to $D^0 - \overline{D}^0$ oscillations are
inherently non-perturbative, and we have not yet learned how to 
calculate them from first principles. It is, however, extremely important 
to estimate their size in
order to understand the origin of a possible experimental observation.
These contributions come from transitions to final states $|f\rangle$ that
are accessible
to both $|D^0\rangle$ and $|\overline{D}^0\rangle$. For example,
Fig.~\ref{fig:long_mixing} illustrates a contribution to mixing from
transitions to two pseudoscalars. GIM cancellations are such that they
become complete in the $SU(3)_{Fl}$ limit; i.e., no oscillations can occur then.

%
\begin{figure}[htb]
\begin{center}
\epsfig{file=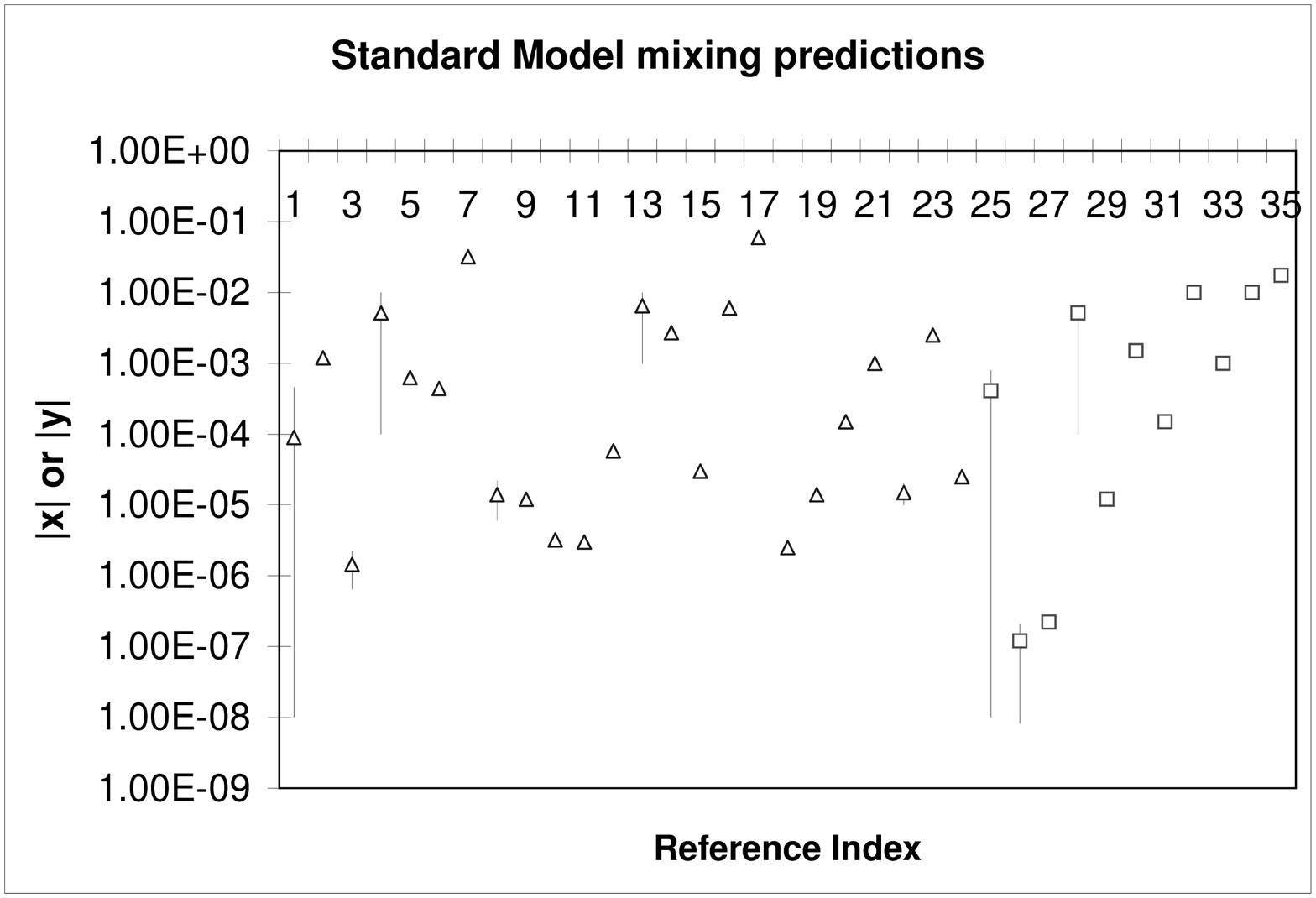,height=10cm,width=12cm}
\caption{Standard Model predictions for $|x|$ (open triangles)
and $|y|$ (open squares). Horizontal line references are tabulated in
Table 4 in Ref.~\cite{Part5_petrov:2001hx}.}
\label{mixing_fig1}
\end{center}
\end{figure}
\begin{figure}[htb]
\begin{center}
\epsfig{file=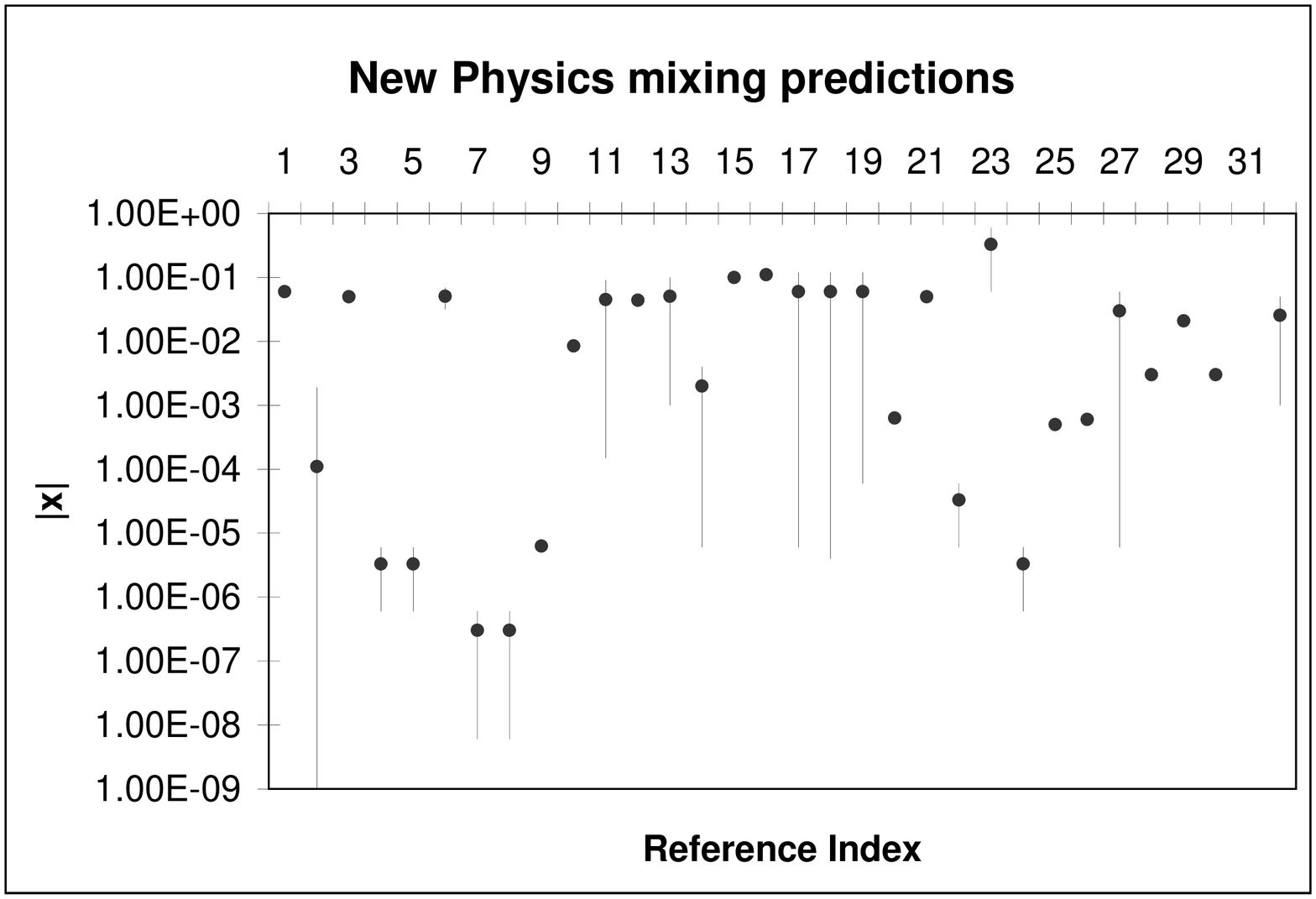,height=10cm,width=12cm}
\caption{New Physics predictions for $|x|$. Horizontal line references are
tabulated in Table 5 in Ref.~\cite{Part5_petrov:2001hx}.}
\label{mixing_fig2}
\end{center}
\end{figure}
%

Within the SM $x$ and $y$ are generated only at second order in
SU(3)$_F$ breaking,

\begin{equation}
x\,,\, y \sim \sin^2\theta_C \times [SU(3) \mbox{ breaking}]^2\,,
\end{equation}
where $\theta_C$ is the Cabibbo angle.  The SM predictions for $x$ and $y$
thus depend crucially on estimating the
size of SU(3)$_F$ breaking.  Although $y$ is expected to be determined
by SM processes, its value nevertheless affects
significantly the sensitivity to new physics of experimental analyses of $D$
mixing~\cite{Part5_Bergmann:2000id}. This circumstance would lead to the naive
estimate
\begin{equation}
x\,,\, y \sim \sin^2\theta_C \times
\left(\frac{m_s}{\Lambda_{hadron}}\right)^2 \, \leq O(10^{-3}),
\label{eq:theory_long_su3}
\end{equation}
with $\Lambda_{hadron} \sim O(1)$ GeV a typical hadronic scale.
Beyond this simple estimate, there are two main approaches to estimating the
long-distance contributions to mixing: an ${\it inclusive}$ approach using
an operator product expansion (OPE), and an ${\it exclusive}$ approach that
sums over intermediate hadronic states using experimental data. Neither
approach yields accurate predictions.

The {\em inclusive} approach applies Heavy-Quark Expansions (HQE) to
calculate contributions to $D$ oscillations, an approach first taken by
Georgi~\cite{Part5_h_gorgi} and later extended by others~\cite{Part5_Ohl,
Part5_Bigi_HQET}.
There are two main assumptions. The first is that the mass of the $c$-quark
is large, $m_c \gg \Lambda_{hadr.}$. The second is that one can construct
local quark-level operators that can be applied to hadron-level processes,
i.e. that quark-hadron duality is applicable already at the charm scale. 
Then $x$ and $y$ are evaluated through the OPE as an expansion in powers 
of ($\Lambda_{hadr.}/m_c$).
The result of this type of approach is~\cite{Part5_Bigi_HQET}
\begin{equation}
x \sim y \sim O(10^{-3}).
\label{eq:theory_long_x_y_hqet}
\end{equation}

The {\em exclusive} approach takes all of the known hadronic states common
to $|D^0\rangle$ and $|\overline{D}^0\rangle$, and groups them both according
to their respective $SU(3)_F$ multiplets and to the number of particles in the
final state. An example of such a set would be ($\pi^+\pi^-$, $\pi^+K^-$,
$K^+K^-$, $K^+\pi^-$). In the limit of a perfect $SU(3)_F$ symmetry, the
individual contributions within each of these groups would cancel, and there
would be no mixing. If one knows the relative amplitudes and strong phases
for these states, calculations of $x$ and $y$ can be done for each
multiplet. For the example set above, this calculation gives a small
contribution due to cancellations, a reasonable result since all of the
states in the set are far from threshold and not affected as much by phase
space considerations. Contributions to $x$ are not required to be on-shell,
so in this case there is no symmetry breaking caused by limited phase space.
If one assumes that all of the sets contribute incoherently in roughly the
same amount, one concludes that~\cite{bigi_review}
\begin{equation}
x  \leq O(10^{-3}).
\label{eq:theory_long_x_y_exclusive}
\end{equation}

By contrast, contributions to $y$ are due to on-shell states, so phase space
is a significant source of symmetry breaking. Considering phase space as the
only source of symmetry breaking, one can calculate the contribution to $y$
of each of the final state multiplets for which there is data using the
measured masses of the final-particles~\cite{Part5_petrov:2001hx}. The largest
calculable contribution comes from the final-state multiplet comprising four
pseudoscalars, whose elements are either near the production threshold with
relatively large branching fractions, or are above threshold and entirely
absent. This ansatz leads to ~\cite{Part5_petrov:2001hx}
\begin{equation}
y  \leq O(10^{-2}).
\label{eq:theory_long_x_y_exclusive_2}
\end{equation}

The results from these two methods should not be seen as inconsistent.
Rather they indicate
the range of uncertainty, which can be phrased as follows: While the best a
priori SM estimate
yields $x$, $y$ $\sim O(10^{-3})$, we cannot conclude that values as `high'
as $10^{-2}$ would
necessarily establish the intervention of New Physics. To be more specific:
while the presence of New Physics can enhance $x$ to its present upper bound of few
$\times 10^{-2}$, it
should not affect $y$ in a significant way, since $\Delta \Gamma$ is
generated from
{\em on-}shell transitions. The minimal requirement for any claim of New
Physics thus is $x|_{\mbox{experim}} \gg y|_{\mbox{experim}}$. This appears as a rather iffy scenario at
present. Nevertheless, it is mandatory to probe for oscillations with as 
much
sensitivity as possible for three
main reasons: (i) With oscillations being an intriguing quantum mechanical
phenomenon their
observation carries intrinsic intellectual value. (ii) We might be only one
breakthrough in our computational control over nonperturbative dynamics away
from making precise predictions.
(iii) Last and most importantly: $CP$ asymmetries that involve oscillations
-- see Sect.~\ref{sec:cp_mixing_decay} below --
would conclusively establish the existence of New Physics. Having an
independent measurement of those oscillations  would provide a most powerful validation of such
asymmetries.

In summary: to the best of our present knowledge even values for $x$ and
$y$ as `high' as
0.01 could be due entirely to SM dynamics of otherwise little interest. It
is likewise possible that a
large or even dominant part of $x \sim 0.01$ in particular is due to New
Physics. While one should
never rule out a theoretical breakthrough, we are less than confident that
even the usual panacea, namely
lattice QCD, can provide a sufficiently fine instrument in the foreseeable
future.

Yet despite this lack of an unequivocal statement from theory,
one wants to probe
these oscillations as accurately as possible, even in the absence of the
aforementioned breakthrough,
since they represent an intriguing quantum mechanical phenomenon and -- on
the more
practical side -- constitute an important ingredient for $CP$~asymmetries
arising in $D^0$ decays
due to New Physics as explained in the next Chapter.

\section[Experimental Review]{Experimental Review\footnote{by Edited by D.~M.~Asner and H.~B.~Li}}

 $D^0 -\overline{D}^0$ mixing and $CP$ violation in charm sector have been
 searched for by various experimental facilities with different techniques.
The principal production processes are $e^+e^- \rightarrow c\overline{c}$
at center of mass energy from threshold up
to $Z^0$ boson peak, hadroproduction at both
fixed-target experiments and the Fermilab Tevatron, photoproduction and so
on.  The cross sections vary from a few nb to microbarns for
photoproduction, and to order of a millibarn at the Tevatron. However, the
ratio of signal to background cross section range from 1:1 in
$e^+e^-$ annihilation to 1:500 at the Tevatron as listed in
Table~\ref{tab:experiments}.
\begin{table}[htbp]
  \centering
  \caption{Summary of recent charm experiments including:
techniques, luminosity, charm production cross section and signal
to noise ratio. }
  \begin{tabular}{c|c|c|c|c|c} \hline
 Experiment & Beam & Lumin. & Cross-Section & \#events $c\overline{c}$ & $\sigma(c\overline{c})/\sigma_{Total}$  \\
            &      & (cm$^{-2}$s$^{-1}$)        &  & Per year\\ \hline
BaBar      & $e^+e^-$ ($\Upsilon(4S)$)    & $3\times 10^{33}$  & $1.3$ nb
 &   $40\times 10^6$        & $\sim 1/5$ \\
Belle      & $e^+e^-$ ($\Upsilon(4S)$)    & $3\times 10^{33}$  & $1.3$ nb
 &  $40\times 10^6$        & $\sim 1/5$ \\
CLEO-c & $e^+e^-$ ($\psi(3770)$) & $2\times 10^{32}$  & 6.4 nb  & 6.4$\times 10^{6}$ & $\sim 1$ \\
BES-III & $e^+e^-$ ($\psi(3770)$) & $1\times 10^{33}$  & 6.4 nb  & 32$\times 10^{6}$ & $\sim 1$ \\
LHC-b   & $pp$($\sqrt{s}=14$ TeV) & $2\times 10^{32}$ & 1.0 mb & 1$\times 10^{11}$ & $\sim 1/100$ \\
\hline
 \end{tabular}
  \label{tab:experiments}
\end{table}

The techniques that can be used to search for mixing  can be
roughly divided into four classes: mixing in semi-leptonic decays,
time-dependent measurements in wrong-sign decays to hadronic
non-$CP$ eigenstates, decays to $CP$ eigenstates and mixing
measurements via quantum coherence at threshold.

\subsection{Semileptonic Decays}

The manifestation of $D^0$ mixing in semileptonic decays is relatively
simple, since such transitions are flavor specific in the standard model
or some of its extensions. Because of the flavor specificity of $D^0
\rightarrow l^+ X^-$ and $\overline{D}^0 \rightarrow l^- X^+$, it is not
necessary to study the time-dependent $D$ decay modes.

In semileptonic $D$ decays, the wrong-sign decay amplitudes 
$A_f = \overline{A}_{\overline{f}} = 0$. 
Then in the limit of weak mixing, where $|ix+y| \ll 1$, from
Eq.~\ref{eq:mix_rate_D0}, $r(t)$ is given by
\begin{equation}
r(t) = \left|\frac{q}{p}\right|^2
\left| g_-(t) \right|^2 \sim \frac{e^t}{4}(x^2+y^2) t^2
\left|\frac{q}{p}\right|^2,
\label{eq:mix_rate_D0_semi}
\end{equation}
In the limit of $CP$ conservation, $r(t)=\overline{r}(t)$, and integrating
Eq.~\ref{eq:mix_rate_D0_semi} over all time gives
\begin{equation}
R_M = \int^{\infty}_0 r(t) dt = \left|\frac{q}{p}\right|^2
\frac{x^2+y^2}{2+x^2+y^2} \simeq \frac{x^2+y^2}{2},
\label{eq:mix_rate_semi_RM}
\end{equation}

The traditional method of looking for {\em like-sign}
$\mu^\pm \mu^\pm$ pairs is an example at fixed target
experiments~\cite{Part5_Aubert_fix,Part5_Benvenuti_fix,Part5_Louis_fix}. 
However, in $e^+e^-$ experiments, such as those at a $B$ factory, there 
are enough kinematic constraints to infer the
neutrino momentum. Specifically, momentum conservation prescribes $P_{\nu} =
P_{CM} - P_{\pi_s K l} -P_{\rm rest}$~\cite{Part5_Belle_Bitenc}, where $P_{CM}$ is
the four-momentum of the $e^+e^-$ center-of-mass (CM) system, $\pi_s$, $K$,
and $l$ are daughters from decay $D^*+ \rightarrow D^0 \pi_s \rightarrow
\pi_s Kl\nu$, and $P_{\rm rest}$ is the four-momentum of the remaining
particles in the event. In  $B$-factory experiments, the magnitude of 
$|P_{\rm rest}|$ is
rescaled to satisfy $(P_{CM} - P_{\rm rest})^2 = m^2_{D^*}$ and, after 
this rescaling, the direction of ${\bf \overrightarrow{p_{\rm rest}}}$ is
adjusted to satisfy $P^2_{\nu}(=m^2_{\nu}) =  0$.  
Table~\ref{tab:time_d_mixing_semi} gives a summary of the experimental 
status
of $R_M$ measurements in semileptonic decays.
\begin{table}[htbp]
  \centering
 \caption{Results for $R_M$ in $D^0$ semileptonic decays from
HFAG~\cite{part5:ref:hfag}.}
  \begin{tabular}{c|c|c|c} \hline
 Year & Experiment & Final states & $R_M$ (\%) \\ \hline
1996 & E791~\cite{Part5_e791_semi} & $K^{+} e^- \overline{\nu}$ & $0.11^{+0.30+0.00}_{-0.27-0.014}$ \\ \hline
2005 &  CLEO-II.V~\cite{Part5_CAW_cleo} &  $K^{(*)+} e^- \overline{\nu}$ & $0.16\pm 0.29\pm 0.29$ \\ \hline
2004 & BaBar~\cite{Part5_Semi_Ba} &  $K^{(*)+} e^- \overline{\nu}$  & $0.23\pm 0.12\pm 0.04$ \\ \hline
2005 & Belle~\cite{Part5_Belle_Bitenc} & $K^{(*)+} e^- \overline{\nu}$ & $0.02 \pm 0.047 \pm 0.014$ \\ \hline
2007 & BaBar~\cite{Part5_Semi_Ba2} &  $K^{(*)+} e^- \overline{\nu}$ &
$0.004^{+0.070}_{-0.060}$ \\ \hline 
 & Average from HFAG~\cite{part5:ref:hfag} &    & $0.0173 \pm 0.0387$ (CL=0.965)  \\ \hline 
 \end{tabular}
 \label{tab:time_d_mixing_semi}
\end{table}

The best places to use the semileptonic method are probably at \bes3 and
CLEO-c operating near the charm threshold. The idea is to search for 
$e^+e^- \rightarrow \psi(3770) \rightarrow D^0 \overline{D}^0 \rightarrow
(K^-l^+\nu)(K^-l^+\nu)$ or $e^+e^- \rightarrow  D^- \overline{D}^{*+}
\rightarrow (K^+ \pi^-\pi^-)(K^+l^-\nu)+\pi_s$. The latter is probably the
only process where the semileptonic method does not suffer from a large
background, since there is only one
neutrino missing in the entire event, threshold kinematic constraints should
provide a clean signal.

However, it has been pointed out that one can not claim a $D^0$
mixing signal based on the semileptonic decay alone (unless accompanied by
information on the decay time of $D^0$ which is possible at a $B$ 
factory).
Bigi~\cite{Part5_bigi_semi} pointed out that an observation of
a signal on $D^0 \rightarrow l^-X^+$ establishes only that a
certain selection rule is violated in the processes where the
charm quantum number is changed, namely, the rule $\Delta C =
-\Delta Q_l$ where $Q_l$ denotes leptonic charge. This violation
can occur either through $D^0 -\overline{D}^0$ mixing (with the
unique attribute of the decay time-dependence of mixing), or
through new physics beyond the SM (which could be independent of
time). Nevertheless, one can always use this method to set an upper
limit for mixing.

\subsection{Hadronic Final States}
\label{subsecc:hadron_final}

{\it Wrong-sign} hadronic decay modes can occur either through $D^0 -
\overline{D}^0$ mixing or through DCSD as illustrated in
Eqs.~\ref{eq:mix_amplitude} and~\ref{eq:mix_DCSD}. The major
complication for this method is the need to distinguish between DCSD and
mixing. In principle, there are at least three ways to distinguish between
DCSD and mixing candidates experimentally: (1) use the difference in the
decay time-dependence~\cite{Part5_Ba_Kpi, Part5_belle_Kpi}; (2) use the possible difference in the resonance
substructure between DCSD and mixing events on the Dalitz plot in three-body
$D^0 \rightarrow K^+ \pi^- \pi^0$
decay~\cite{Part5_babar_dp}, or multi-body decays, like $D^0
\rightarrow K^+ \pi^- \pi^+ \pi^-$, etc.; (3) use the quantum correlations 
between  the production and decay processes at the $\psi(3770)$
peak~\cite{Part5_bigi_semi,Part5_cor_asner}.  Method (1)
is popular at $B$ factory since the $D^0$ is highly boosted,  so its
decay time information can be used.  Method (2) requires knowledge of the
structures of the DCSD decay on the Dalitz plots, which can be done at both
a $B$ factory and a $\tau$-charm factory.  Method (3) can be done by
\bes3 at charm threshold region.   In this subsection, we only discuss
method (1), specifically. Method (2) and (3) are discussed later. 

According to Eq.~\ref{eq:mix_rate_D0}, one has:
\begin{equation}
r(t) = \left|\frac{q}{p}\right|^2 \left( \lambda^{-2}_f g^2_+(t) +
\lambda^{-1}_f g_+(t)g^*_-(t) + (\lambda^{-1}_f)^* g^*_+(t)g_-(t) + g^2_-(t)
\right),
\label{eq:mix_hadronic_decay_WS}
\end{equation}
We can simplify Eq.~\ref{eq:mix_hadronic_decay_WS} under the
assumption of small mixing, $|ix+y| \ll 1$, and express
\begin{eqnarray}
 \lambda^{-2}_f g^2_+(t) & = & |\lambda^{-1}_f|^2 \frac{e^{-t}}{2}\left[{\rm
cosh}(y t) +{\rm cos}(xt) \right] \nonumber \\
& \simeq &  |\lambda^{-1}_f|^2 e^{-t},
\label{eq:mix_dcsd_term}
\end{eqnarray}
\begin{eqnarray}
g^2_-(t) & = & \frac{e^{-t}}{2}\left[{\rm
cosh}(y t) - {\rm cos}(xt) \right] \nonumber \\
 & \simeq &  e^{-t} \left(\frac{x^2+y^2}{4} \right) t^2,
\label{eq:mix_mixing_term}
\end{eqnarray}
and
\begin{eqnarray}
&& \lambda^{-1}_f g_+(t)g^*_-(t) + (\lambda^{-1}_f)^* g^*_+(t)g_-(t) \nonumber\\
&&  =  \frac{e^{-t}}{2} \left(e^{-i(\delta+\phi)} ({\rm sinh}(yt)-i{\rm sin}(xt))+
e^{i(\delta+\phi)}({\rm sinh}(yt)+i{\rm sin}(xt)) \right) \nonumber \\
&& \simeq |\lambda^{-1}_f| e^{-t}(y{\rm cos}(\delta + \phi) - x{\rm
sin}(\delta + \phi)) t.
\label{eq:mix_interfer_term}
\end{eqnarray}
If we define
\begin{eqnarray}
y^{\prime}_{\pm} \equiv y^{\prime}{\rm cos}\phi \pm x^{\prime}{\rm
sin}\phi = y{\rm cos}(\delta \mp \phi) - x{\rm sin}(\delta \mp \phi),
\end{eqnarray}
where
\begin{eqnarray}
y^{\prime} \equiv y{\rm cos}\delta - x{\rm sin}\delta\, , \, x^{\prime} \equiv x{\rm
cos}\delta + y{\rm sin}\delta,
\end{eqnarray}
and combine Eqs.~\ref{eq:mix_lambda_f_cpphase},
\ref{eq:mix_dcsd_term}, \ref{eq:mix_mixing_term} and
\ref{eq:mix_interfer_term} in the limit of $CP$ conservation
($A_D =0$, $A_M = 0$ and $\phi = 0$), we obtain the standard form
for the time-dependent decay rate, including $D$ mixing:
\begin{eqnarray}
r(t) = \overline{r}(t) = e^{-t} \left( R_D + \sqrt{R_D} y^{\prime} t +
\frac{1}{2} R_M t^2 \right)
\label{eq:mix_standard_rate}
\end{eqnarray}
The above wrong-sign decay rate includes three components: one from the
DCSD, another from mixing, and one form the interference between
DCSD and mixing.  The time-integrated wrong-sign rate relative to the integrated
right-sign rate is
\begin{eqnarray}
R_{WS} = \int^{\infty}_0 r(t)dt =  R_D + \sqrt{R_D} y^{\prime} +
\frac{1}{2} R_M \, .
\label{eq:mix_standard_rate_inte}
\end{eqnarray}
As shown in Eq.~\ref{eq:mix_standard_rate}, $D$ mixing is
characterized in the decay rate by a small deviation  from a
pure exponential. In order to have the most sensitivity to
$(x^2+y^2)$, a decay channel for which $R_D$ is relatively small
is desireable. The analysis technique benefits from the ability to
compare the signal distribution, given by
Eq.~\ref{eq:mix_standard_rate}, to the CF decay distribution,
which may be treated as pure exponential. In this way, systematic
bias is significantly limited.

The ratios $R_{WS}$ and $R_D$ are the most readily accessible experimental
quantities.
Table~\ref{tab:mixing_inte_R_WS} gives recent measurements of $R_{WS}$ and
$R_D$ ifor $D^0 \rightarrow K^+ \pi^-$ decays. 
\begin{table}[htbp]
  \centering
 \caption{ Results for $R_{WS}$ and $R_D$ in $D^0 \rightarrow K^+
\pi^-$, where it is assumed that there is no mixing and no $CP$ violation 
in DCSD decays.}
  \begin{tabular}{c|c|c|c} \\ \hline \hline
  Experiments & comments &  $R_{WS}$ (\%) & $R_D$  (\%) \\ \hline
  E791~\cite{Part5_e791_kpi_rd} & 500 GeV $\pi^-$N interactions & $0.68^{+0.34}_{-0.33} \pm 0.07$ & -- \\ \hline  
  CLEO~\cite{Part5_CLEO_GODANG00} & 9.0 fb$^{-1}$ near $\Upsilon(4S)$  & $0.332^{+0.063}_{-0.065} \pm 0.040$ & $0.47^{+0.11}_{0.12}\pm 0.040$  \\ \hline
  FOCUS~\cite{Part5_FOCUS_KPi}  &  $\gamma$ BeO &  $0.429\pm 0.063\pm 0.028$ & $0.381^{+0.167}_{-0.163}\pm 0.092$ \\ \hline
  Belle~\cite{Part5_belle_Kpi} & 400 fb$^{-1}$ near $\Upsilon(4S)$  & $0.377\pm 0.008 \pm 0.005$ & $0.364\pm 0.017$ \\ \hline
  BaBar~\cite{Part5_Ba_Kpi} & 384 fb$^{-1}$ near $\Upsilon(4S)$ &--  & $0.303 \pm 0.016\pm 0.010$  \\ \hline
  CDF~\cite{Part5_CDF_2006} & 0.35 fb$^{-1}$  at $\sqrt{s} = 1.96$ TeV  &$0.405 \pm 0.021 \pm 0.011$ & -- \\ \hline
 \hline
 \end{tabular}
 \label{tab:mixing_inte_R_WS}
\end{table}
\begin{table}[htbp]
  \centering
 \caption{ $D^0 \rightarrow K^+ \pi^-$ results for $x^{\prime 2}$ and
$y^{\prime}$ from HFAG~\cite{part5:ref:hfag}. These results assume no $CPV$.}
  \begin{tabular}{c|c|c|c} \\ \hline \hline
 Experiment &  comments & $x^{\prime 2}$ (\%)  & $y^{\prime}$ (\%)  \\ \hline
 CLEO~\cite{Part5_CLEO_GODANG00} & 9.0 fb$^{-1}$ near $\Upsilon(4S)$ & 
 $0.0 \pm 1.5 \pm 0.2 $ & $-2.3^{+1.3}_{-1.4} \pm 0.3$  \\ \hline 
 Belle~\cite{Part5_belle_Kpi} & 400 fb$^{-1}$ near $\Upsilon(4S)$  &
$0.018^{+0.021}_{-0.023}$ & $0.06^{+0.40}_{-0.39}$ \\ \hline 
 BaBar~\cite{Part5_Ba_Kpi} & 384 fb$^{-1}$ near $\Upsilon(4S)$ & $-0.022\pm
0.030\pm 0.021$ & $0.97\pm 0.44\pm 0.31$  \\ \hline
 \hline
 \end{tabular}
  \label{tab:time_d_mixing}
\end{table}
%
\begin{figure}[htb]
\begin{center}
\epsfig{file=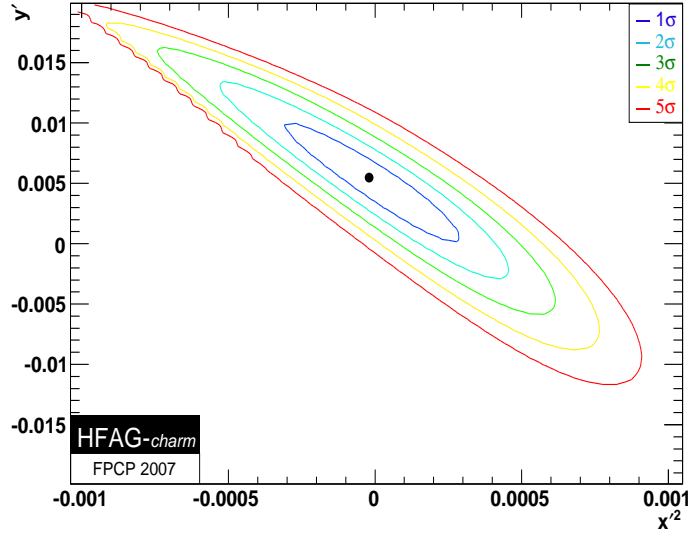,height=7cm,width=9cm}
\caption{Contours (1 through 5$\sigma$) of the allowed region in the 
$x^{\prime 2}$ vs $y^{\prime}$ plane are shown for  $D^0 \rightarrow
K^+\pi^-$ decay by combining BaBar and Belle's results. The significance of
the oscillation effect exceeds 4$\sigma$.}
\label{fig:kpi_combi_ba_belle_mixng}
\end{center}
\end{figure}

The interference causes the measured $x$ and $y$ to be rotated
through an angle $\delta$, the phase difference between the DCS
and CF decay processes. By measuring the time dependence of the
decay rate it is possible to sort out the mixing from the DCS
decay. At $B$ factory and the CLEO experiments, wrong-sign candidate
events of the types $D^0 \rightarrow K^+ \pi^-$ and
$\overline{D}^0 \rightarrow K^- \pi^+$ are selected by requiring
the soft $\pi_s$ from the $D^*$ decay and the daughter $K$ of
the $D^0$ to have identical charge (wrong-sign tag). In order to
determine the wrong-sign and right-sign yields, two powerful
distinguishing variables are used by Belle,
$M_{K\pi}$ and the released energy $Q \equiv M^* - M_{K\pi} -
m_{\pi_s}$, where $M^*$ is the reconstructed mass of the $K^+\pi^-
\pi^+_{s}$ system, $M_{K\pi}$ is the reconstructed mass of
$K^+\pi^-$ system, and $m_{\pi_s}$ is the charged pion mass.
In comparison, the BaBar and CLEO experiments used the mass difference
$\delta m = M^* - M_{K\pi}$. 
To detect a deviation from an
exponential decay in wrong-sign events, a likelihood fit to the
distribution of the reconstructed proper decay time $t$ is performed
for each experiment. The likelihood fit includes a signal and a
background component and models each as the convolution of a
decay-time distribution and a resolution function. 

Results for $x^{\prime 2}$ and $y^{\prime}$ from several experiments are
listed in Table~\ref{tab:time_d_mixing}.
While no experiment claims an effect, it is interesting that the Belle result is
consistent with no mixing only at the 3.9\% C.L.~\cite{Part5_belle_Kpi}. 
Figure~\ref{fig:kpi_combi_ba_belle_mixng} shows the 95\% C.L. region in 
the $x^{\prime 2}$-$y^{\prime}$ plane from HFAG~\cite{part5:ref:hfag}.  
The significance of  the oscillation effect exceeds 4$\sigma$.

Mixing has also been searched for in the wrong-sign multibody final
states $K^+\pi^- \pi^0$ and 
$K^+ \pi^- \pi^+\pi^-$~\cite{Part5_e791_multi,Part5_cleo_2001,Part5_belle-multi-2005}.
Using 281~fb$^{-1}$ of data,
Belle has done a time-integrated analysis, and measured the
time-independent ratio of wrong-sign to right-sign decays. The results are
$R_{WS} = (0.229 \pm 0.015^{+0.013}_{-0.009})\%$ for the $D^0 \rightarrow
K^+\pi^- \pi^0$ and $R_{WS} = (0.320\pm0.018^{+0.018}_{-0.013})\%$ for the
$D^0 \rightarrow K^+ \pi^- \pi^+\pi^-$.
\begin{figure}[htb]
\begin{center}
\epsfig{file=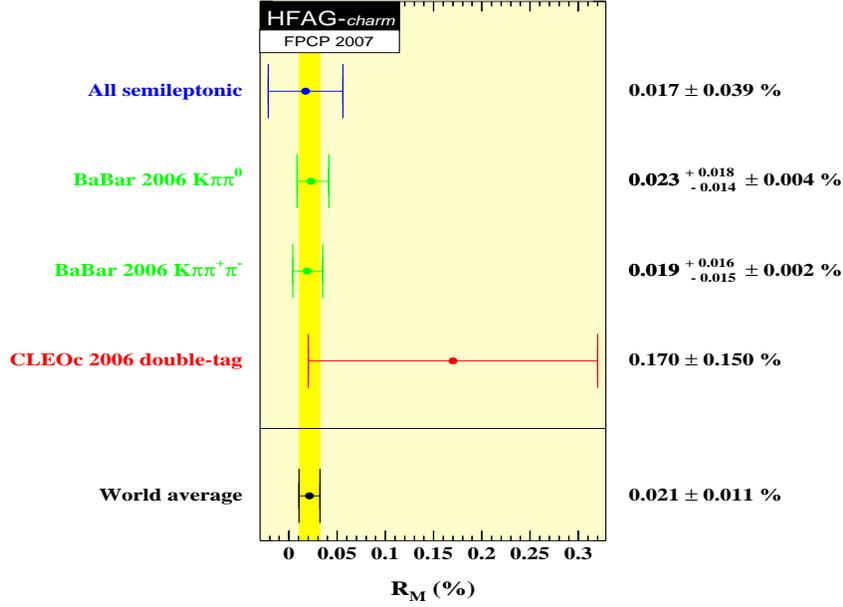,height=8cm,width=11cm}
\caption{$R_M$ combination:  the world averaged
$R_M$ by combining the results of semileptonic $D$
decays and these from the wrong-sign multibody final states.}
\label{fig:part5:rm:all}
\end{center}
\end{figure}

BaBar reported a measurement of the mixing
rate $R_M$ in the decay  $D^0 \rightarrow K^+ \pi^- \pi^0$ based on
a time-dependent analysis~\cite{Part5_babar_dp}. There are two key
motivations to select $D^0 \rightarrow K^+ \pi^- \pi^0$: (1) one
expects the Dalitz-plot structure of $DCS$ decay to differ from that
of CF decay.  We note that DCS decays proceed primarily through the
resonance $D^0 \rightarrow K^{*+} \pi^-$, $K^{*0} \pi^0$, while CF
decays proceed primarily through the resonance $D^0 \rightarrow
K^- \rho^+$. The measurement sensitivity to $R_M$ is increased
by selecting regions of the Dalitz plot where $CF$ decays
contribute with a large amplitude relative to the corresponding
DCS decays. (2) The time-integrated mixing rate $R_M =
(x^2+y^2)/2$ is independent of the decay mode and is expected to
be consistent between differemt mixing measurement methods.
The sensitive regions of the
phase space (i.e., the Dalitz plot) are selected by removing the $K^*$
resonance.  In the limit of $CP$ conservation, the time-dependent ratio of
wrong-sign to right-sign is expressed approximately as:
\begin{eqnarray}
R_{WS} =   \tilde{R}_D &+& \alpha \sqrt{\tilde{R}_D} \tilde{y}^{\prime}\, t +
\frac{1}{2} R_M t^2 \\
&& 0 \leq \alpha \leq 1 \nonumber \, ,
\label{eq:mix_standard_rate_Babar}
\end{eqnarray}
where the tilde indicates quantities that have been integrated
over any choice of phase space regions. Here $\tilde{R}_D$ is the
integrated DCS ratio of DCS decays to CF decays;
$\tilde{y}^{\prime} = y{\rm cos}\tilde{\delta} - x {\rm
sin}\tilde{\delta}$, $\tilde{\delta}$ is an unknown integrated
strong-phase difference between the CF and the DCS decay
amplitudes; $\alpha$ is a suppression factor that accounts for
possible strong-phase variations over the regions. The time-integrated
mixing rate $R_M$ is independent of decay mode.  By fitting the
proper time distribution,  they get the results $R_M =
(0.023^{+0.018}_{-0.014} \pm 0.004)\%$ and $\tilde{R}_D =
(0.164^{+0.026}_{-0.022} \pm 0.012)\%$ with the assumption of $CP$
invariance. An upper limit is established as $R_M < 0.054\%$ at the 95\%
confidence level. They conclude that the observed data are
consistent with no mixing at the 4.5\% confidence level.  Using
a similar method and idea, BaBar also measured the
time-integrated mixing rate $R_M$ in $D^0 \rightarrow K^+ \pi^-
\pi^+\pi^+$ decay mode. Assuming $CP$ conservation, they get $R_M
= (0.019^{+0.016}_{-0.015}\pm 0.002)\%$, and $R_M < 0.048\%$ at
95\% C.L.. Furthermore, they combined results from both decay
modes, and found that $R_M = (0.020^{+0.011}_{-0.010})\%$ and $R_M
< 0.042\%$ at 95\% C.L. which is the best limit with current data.
The combined data sets are consistent with the no-mixing
hypothesis with 2.1\% confidence.

By combining the results of semileptonic $D$ decays and these from 
the wrong-sign multibody final states, one can obtain the world averaged 
$R_M$ shown in Fig.~\ref{fig:part5:rm:all}. The averaged mixing rate is $R_Ms
= (0.021\pm 0.011 )\%$ with $CL = 0.795$.

\subsection{$CP$ Eigenstates}

$D^0$ mixing parameters can be measured by comparing the lifetimes extracted
from the analysis of $D$ decays into the $CP$-even and $CP$-odd final
states.  When the final state is a $CP$ eigenstate $f$ (i.e.,
$|\overline{f}\rangle \equiv CP | f\rangle = \pm | f \rangle$), such as
$K^+K^-$, $\pi^+\pi^-$ and $K_S \pi^0$, there is no strong phase difference
 between $\overline{A}_f$ and $A_f$. Assuming $|\overline{A}_f| =|A_f|$ (no
direct $CP$ violation), $\lambda_f = -|q|/|p| e^{i\phi}$ and
$\lambda_{\bar{f}} = -|p|/|q| e^{-i\phi}$, where $\phi$ is the weak phase
difference. Inserting these terms into Eqs.~\ref{eq:mix_rate_D0} and
\ref{eq:mix_rate_antiD0} gives
\begin{eqnarray}
R(D^0 \rightarrow K^+K^-) \simeq |A_{K^+K^-}|^2 e^{-t} e^{-|q/p| (y {\rm
cos} \phi - x {\rm sin}\phi)t},
\label{eq:mix_y_CP}
\end{eqnarray}
and
\begin{eqnarray}
R(\overline{D}^0 \rightarrow K^+K^-) \simeq |A_{K^+K^-}|^2 e^{-t} e^{-|p/q| (y {\rm
cos} \phi + x {\rm sin}\phi)t}.
\label{eq:mix_y_CP_antiD0}
\end{eqnarray}
Equations~\ref{eq:mix_y_CP} and~\ref{eq:mix_y_CP_antiD0} imply that 
the
measured $D^0$ and $\overline{D}^0$ inverse lifetimes are slightly different
each other. We define
\begin{eqnarray}
y_{CP} = \frac{\tau_{D^0 \rightarrow K^- \pi^+}}{\tau_{D^0 \rightarrow K^+K^-}} -1 = 
\frac{|q|}{|p|} (y {\rm
cos} \phi - x {\rm sin}\phi),
\label{eq:define_y_CP}
\end{eqnarray}
for $D^0$ decays, and for $\overline{D}^0$ decays, it is
\begin{eqnarray}
y_{CP} = \frac{\tau_{\overline{D}^0 \rightarrow K^+
\pi^-}}{\tau_{\overline{D}^0 \rightarrow K^+K^-}} -1 = \frac{|p|}{|q|} (y {\rm
cos} \phi + x {\rm sin}\phi).
\label{eq:define_y_CP_antiD0}
\end{eqnarray}
For $|q/p| = 1$, i.e., no $CP$ violation in mixing, $y_{CP} = y {\rm
cos}\phi$ for samples with equal numbers of $D^0$ and $\overline{D}^0$ 
decays. If, in addition,
$\phi = 0$ (no $CP$ violation), $y_{CP} = y$.

One can also combine the two $D \rightarrow K^+K^-$ modes. To
understand the consequences of such an analysis, one has to consider the
relative weight of $D^0$ and $\overline{D}^0$ in the sample. We define
 $A_{\rm prod}$ as the production asymmetry of $D^0$ and $\overline{D}^0$:
\begin{eqnarray}
A_{\rm prod} = \frac{N(D^0) - N(\overline{D}^0)}{N(D^0) + N(\overline{D}^0)}
\label{eq:define_A_prod}.
\end{eqnarray}
Then
\begin{eqnarray}
y_{CP} & =&   \frac{\tau_{D^0 \rightarrow K^- \pi^+}}{\tau_{D \rightarrow K^+K^-}} -1 \nonumber \\
& = & y{\rm cos} \phi \left[\frac{1}{2}\left(\left|\frac{p}{q}\right| +
\left|\frac{q}{p}\right|\right) + \frac{A_{\rm prod}}{2}
\left(\left|\frac{p}{q}\right| - \left|\frac{q}{p}\right|\right)
\right] \nonumber \\
& & - x {\rm sin}\phi \left[\frac{1}{2}\left(\left|\frac{p}{q}\right| -
\left|\frac{q}{p}\right|\right) + \frac{A_{\rm prod}}{2}
\left(\left|\frac{p}{q}\right| + \left|\frac{q}{p}\right|\right)
\right] ,
\label{eq:define_y_CP_bothD0}
\end{eqnarray}
\begin{table}[htbp]
  \centering
  \caption{Summary of $y_{CP}$results.}
  \begin{tabular}{c|c|c|c} \hline
Year &  Experiment & comments & $y_{CP}$(\%)  \\ \hline
2000 &  FOCUS~\cite{Part5_44_focus} & $\gamma$N interactions & $3.4 \pm 1.4 \pm  0.7$ \\ \hline
2002 &  CLEO~\cite{Part5_43_cleo} & 9.0 fb$^{-1}$ near $\Upsilon(4S)$ & $-1.2 \pm 2.5 \pm 1.4$  \\ \hline
2002 & Belle~\cite{Part5_45_Belle} &  23.4 fb$^{-1}$ near $\Upsilon(4S)$; no $D^*$ tag & $-0.5 \pm 1.0 \pm 0.8$ \\ \hline
2007 & Belle~\cite{Part5_46_07_Belle} & 540 fb$^{-1}$ near $\Upsilon(4S)$ &$1.31\pm 0.32 \pm 0.25$  \\ \hline
2007 &  BaBar~\cite{Part5_47_07_Ba} & 384 fb$^{-1}$ near $\Upsilon(4S)$  & $1.03 \pm 0.33 \pm 0.19 $ \\ \hline
    & Average & & $1.132 \pm 0.266$ \\ \hline 
 \end{tabular}
 \label{tab:time_d_mixing_ycp}
\end{table}

The value of $y_{CP}$ is determined from the difference in slopes of the 
decay-time distributions 
for the $D^0 \rightarrow K^- \pi^+$ sample, which is an equal mixture of
$CP$-even and $CP$-odd final states, and the $D^0 \rightarrow K^+K^-$ or
$\pi^+\pi^-$ samples, which are $CP$-even final states. An unbinned 
maximum likelihood
fit to the distribution of the reconstructed proper time $t$ of the $D^0$
candidates is performed. To date, five experiments have measured 
$y_{CP}$, as listed in Table~\ref{tab:time_d_mixing_ycp}. 

Since it requires a fit to the proper-time distribution, this method can 
not be applied at the \bes3 and CLEO-c experiments, which operate near 
the  open charm threshold.

\subsection{Mixing parameters from a Dalitz Plot analysis}

The first time-dependent Dalitz plot analysis was done by CLEO for the 
$D^0 \rightarrow K_S \pi^+\pi^-$ decay mode~\cite{Part5_dalitz3_cleo}.  
In 2007, Belle  also presented results with a 540~fb$^{-1}$ data 
sample at the $\Upsilon(4S)$ peak~\cite{part5:belle:2007:kspp}. 
They use a self-conjugate final state that is not a $CP$ eigenstate 
as reflected by substructures with either
$L= 0$ ($CP$-even) or $L=1$ ($CP$-odd) in the three body decay.
The decay rate to $K_S \pi^+\pi^-$ with ($m^2_{K_S \pi^-}$,
$m^2_{\pi^+\pi^-}$) at time $t$ of a particle tagged as $| D^0 \rangle$ at $
t= 0$ is
\begin{equation}
d\Gamma(m^2_{K\pi}, m^2_{\pi\pi}, t) = \frac{1}{256\pi^3 M^3}|{\cal M}|^2
dm^2_{K\pi} dm^2_{\pi\pi},
\label{eq:rate_dalitz_diff}
\end{equation}
where ${\cal M} = \langle f|{\it H}| D^0(t)
\rangle$,  and $\langle f | = \langle K_S \pi^+\pi^-
(m^2_{K_S\pi^-},m^2_{\pi^+\pi^-}) |$. An expression for $| D^0(t)\rangle$ 
is given in Eq.~\ref{eq:mixing_evolve1}.

The decay channels can be collected into those that are $CP$-even or 
$CP$-odd (with amplitudes ${\rm Amp_+}$ or ${\rm Amp_-}$) and to those 
that are $D^0$ or $\overline{D}^0$ flavor eigenstates (with amplitudes 
${\rm Amp_f}$ or $\overline{\rm Amp}_f$):
\begin{equation}
\langle f |{\cal H} | D_{+,-}\rangle = \sum a_{j} e^{i\delta_j} {\cal
A}^j_{+,-} = p {\rm Amp}_{+,-} \, ;
\label{eq:dalitz_amp_cp}
\end{equation}
\begin{equation}
\langle \overline{f} |{\cal H} | D_{+,-}\rangle = \sum a_{j} e^{i\delta_j} {\cal
\overline{A}}^j_{+,-} = q \overline{\rm Amp}_{+,-}\, ;
\label{eq:dalitz_amp_cp_2}
\end{equation}
\begin{equation}
\langle f |{\cal H} | D^0 \rangle = \sum a_{j} e^{i\delta_j} {\cal A}^j = {\rm Amp}_f \, ;
\label{eq:dalitz_amp_flavor}
\end{equation}
\begin{equation}
\langle \overline{f} |{\cal H} | \overline{D}^0\rangle = \sum a_{j} e^{i\delta_j}
{\cal \overline{A}}^j = p \overline{\rm Amp}_{\overline{f}}\, .
\label{eq:dalitz_amp_flavor_2}
\end{equation}
Here $a_j$ and $\delta_j$ are the explicitly $CP$ conserving amplitudes 
and relative strong phases (Ref.~\cite{Part5_cleo-2004} fixes 
$a_{\rho} = 1$ and $\delta_{\rho}=0$;
all other strong phases are relative to the $\rho$ amplitude) for the 
$j^{\rm th}$ quasi-two-body state.
${\cal A}^j_{+,-} = {\cal A}^j_{+,-} (m^2_{K_S\pi^-}, m^2_{\pi^+\pi^-})$ is
the Breit-Wigner amplitude for resonance $j$ with $D$ decay to $CP = +$ or $CP = -$
quasi-two-body contributions. In Ref.~\cite{Part5_cleo-2004}, CLEO 
considered the following ten modes: $K^{*-} \pi^+$, $K^*_0(1430)^-
\pi^+$, $K^*_2(1430)^-\pi^+$, $K^*(1680)^- \pi^+$, $K_S \rho$ $K_S \omega$,
$K_S f_0(980)$, $K_S f_2(1270)$, $K_S f_0(1370)$, and the "wrong-sign"
$K^{*+} \pi^-$ plus a small non-resonant component.
Collecting terms with similar time dependence and combining
Eqs.~\ref{eq:mixing_evolve1} and~\ref{eq:mixing_evolve2},  one finds
\begin{eqnarray}
{\cal M} &=& \langle f |{\cal H} |D^0(t)\rangle  =  \frac{1}{2p} ( \langle f |{\cal H} |D_1(t)\rangle + \langle f |{\cal H} |D_2 \rangle ) \nonumber \\
& = & \frac{1}{2p}( \langle f |{\cal H} |(pD^0 + q \overline{D}^0) \rangle
 e_1(t) +  \langle f |{\cal H} |(pD^0 - q \overline{D}^0) \rangle e_2(t)) \nonumber \\
& = &  \frac{1}{2p} ([p({\rm Amp}_f + {\rm Amp}_+) + q (\overline{\rm Amp}_f +
\overline{\rm Amp}_+)] e_1(t) \nonumber \\
& &+ [p({\rm Amp}_f + {\rm Amp}_-) - q(\overline{\rm Amp}_f + \overline{\rm Amp}_-)] e_2(t)) \nonumber \\
& = & \frac{1}{2} ([(1+\chi_f){\rm Amp}_f + (1+\chi_+){\rm Amp}_+] e_1(t)
\nonumber \\
&&+[(1 - \chi_f){\rm Amp}_f + (1- \chi_-){\rm Amp}_-] e_2(t)) \nonumber \\
& \equiv & e_1(t)A_1 + e_2 A_2 \, ,
\label{eq:ampli_dalitz_1}
\end{eqnarray}
and
\begin{eqnarray}
 \overline{\cal M}  &=& \langle \overline{f} |{\cal H} |\overline{D}^0(t)\rangle  =
 \frac{1}{2q} ( \langle \overline{f} |{\cal H} |D_1(t)\rangle - \langle
\overline{f} |{\cal H} |D_2 \rangle ) \nonumber \\
&=& \frac{1}{2} ([(\chi^{-1}_{\overline{f}} + 1)\overline{\rm
Amp}_{\overline{f}} + (\chi^{-1}_+ + 1)\overline{\rm Amp}_+] e_1(t)
\nonumber \\
&& - [(\chi^{-1}_{\overline{f}} -1)\overline{\rm Amp}_{\overline{f}} + (
\chi^{-1}_- -1)\overline{\rm Amp}_-] e_2(t)) \nonumber \\
& \equiv & e_1(t) \overline{A}_1 + e_2 \overline{A}_2 \, ,
\label{eq:ampli_dalitz_2}
\end{eqnarray}
for $D^0$ and $\overline{D}^0$, respectively. The CLEO experiment
defines
\begin{equation}
\chi_f = \frac{q}{p} \frac{\overline{\rm Amp}_f}{{\rm Amp}_f} = \left |
\frac{\overline{\rm Amp}_f}{{\rm Amp}_f} \right | (1+A_M) e^{i(\delta + \phi)}
\label{eq:cp_cleo_dalitz}
\end{equation}
\begin{equation}
\chi_{\overline{f}} = \frac{q}{p} \frac{\overline{\rm Amp}_{\overline{f}}}{{\rm
Amp}_{\overline{f}}} = \left |
\frac{\overline{\rm Amp}_{\overline{f}}}{{\rm
Amp}_{\overline{f}}}  \right | (1+A_M) e^{-i(\delta - \phi)}
\label{eq:cp_cleo_dalitz_2}
\end{equation}
\begin{equation}
\chi_{\pm} = \pm \frac{q}{p} \frac{\overline{\rm Amp}_{\pm}}{{\rm
Amp}_{\pm}} = \pm \left |
 \frac{\overline{\rm Amp}_{\pm}}{{\rm
Amp}_{\pm}}  \right | (1+A_M) e^{\pm i \phi)} \, ,
\label{eq:cp_cleo_dalitz_3}
\end{equation}
where $\delta$ is the relative strong phase between $D^0$ and
$\overline{D}^0$ to $K_S \pi^+\pi^-$ and, in the limit of $CP$ 
conservation,
the real $CP$-violating parameters, $A_D$ and $\phi$, are zero. Squaring the
amplitude and factoring out the time-dependent functions yields
\begin{equation}
|{\cal M}|^2 = |e_1 (t)|^2 |A_1|^2 + |e_2(t)|^2 |A_2|^2 + 2 \mbox{Re}[e_1(t)
e^*_2(t) A_1 A^*_2]\, .
\label{eq:cleo_dalitz_mag_1}
\end{equation}
\begin{equation}
|\overline{\cal M}|^2 = |e_1 (t)|^2 |\overline{A}_1|^2 + |e_2(t)|^2
|\overline{A}_2|^2 + 2 \mbox{Re}[e_1(t)
e^*_2(t) \overline{A}_1 \overline{A}^*_2]\, .
\label{eq:cleo_dalitz_mag_2}
\end{equation}
The time-dependent terms are given explicitly by
\begin{equation}
|e_1(t)|^2 = e^{\Gamma_1 t} = e^{- \Gamma(1- y)t}\, ,
\label{eq:cleo_dalitz_time_1}
\end{equation}
\begin{equation}
|e_2(t)|^2 = e^{\Gamma_2 t} = e^{- \Gamma(1+y)t} ,
\label{eq:cleo_dalitz_time_2}
\end{equation}
\begin{eqnarray}
e_1(t)e_2^*(t) &=&  e^{-\lambda_{D_1} t} e^{+\lambda_{D_2} t} = e^{- \Gamma(1+
ix)t}, \nonumber \\
& =& e^{\Gamma t} ({\rm cos}(\Delta m t) - i {\rm sin}(\Delta m t)).
\label{eq:cleo_dalitz_time_3}
\end{eqnarray}

Experimentally, $y$ modifies the lifetime of certain contributions to the
Dalitz plot while $x$ introduces a sinusoidal rate variation. Then, $|{\cal
M}|^2$ can be expressed as
\begin{eqnarray}
|{\cal M}|^2 & = & |A_1|^2  e^{- \Gamma(1- y)t} + |A_2|^2 e^{- \Gamma(1+y)t} \nonumber \\
& &+ 2 e^{\Gamma t} (\mbox{Re}{A_1 A^*_2}{\rm cos}(\Delta m t) + \mbox{Im}{A_1 A^*_2}{\rm sin}(\Delta m
t))\, .
\label{eq:cleo_dalitz_time_mag_final}
\end{eqnarray}

\begin{table}[t]
\centering
\renewcommand{\arraystretch}{1.2}
\caption{Results for mixing parameters from 
$t$-dependent fits to the $D^0\ra K^0_S\,\pi^+\pi^-$ Dalitz plot. For 
CLEO's
results, the errors
are statistical, experimental systematic,
and decay model systematic, respectively.} 
\label{tab:cleo_dalitz_time} \vskip 0.1 in
\begin{tabular}{|c|ccc|}
\hline
{\bf Year/Exp.} & {\bf Param.} & {\bf Result\,(\%)} &
{\bf {\boldmath Comment}} \\
\hline
\hline
2005/CLEO II.V~\cite{Part5_dalitz3_cleo} 
&
\begin{tabular}{c}
$x$ \\ $y$ \\
\end{tabular}
&
\begin{tabular}{c}
$1.9\,^{+3.2}_{-3.3}\,\pm 0.4\,\pm 0.4$  \\
$-1.4\,\pm 2.4 \,\pm 0.8\,\pm 0.4$ \\
\end{tabular}
&
\begin{tabular}{c}
9.0 fb$^{-1}$ near $\Upsilon(4S)$\\
\end{tabular}
\\
\hline\hskip-0.20in
\begin{tabular}{c}
2007/Belle~\cite{part5:belle:2007:kspp}
\end{tabular}
&
\begin{tabular}{c}
$x$ \\ $y$ \\
\end{tabular}
&
\begin{tabular}{c}
$0.80\,\pm 0.29 \,\pm 0.17$  \\
$0.33\,\pm 0.24\,\pm 0.15$ \\
\end{tabular}
&
\begin{tabular}{c}
540 fb$^{-1}$ near $\Upsilon(4S)$ \\
\end{tabular}
\\
\hline\hskip-0.20in
\begin{tabular}{c}
Average \\
\end{tabular}    
&
\begin{tabular}{c}
$x$ \\ $y$ \\
\end{tabular}
&
\begin{tabular}{c}
$0.811\,\pm 0.334$  \\
$0.309\,\pm 0.281$ \\
\end{tabular}
&
\begin{tabular}{c}
CL = 0.74 \\
CL = 0.50 \\
\end{tabular}
\\
\hline
\end{tabular}
\end{table}

Using the above probability density function, CLEO does an unbinned 
maximum
likelihood fit to the Dalitz plot and the time $t$ distribution to determine
$a_j$, $\delta_j$, $x$ and $y$. There is a systematic uncertainty arising 
from the decay model, {\it i.e.}, one must decide which intermediate 
states to 
include in the fit. Equation~\ref{eq:cleo_dalitz_time_mag_final} depends 
linearly on $x$ ($x<1$) and is, therefore, sensitive to the sign of $x$. 
The 
fit results are listed in Table~\ref{tab:cleo_dalitz_time}.

The first significant results in $D^0 \rightarrow K_S \pi^+\pi^-$ 
are from Belle~\cite{part5:belle:2007:kspp} as 
listed in  Table~\ref{tab:cleo_dalitz_time}. This is 2.7$\sigma$ 
from the non-mixing hypothesis. The 95\% CL intervals are $0< x < 0.016$ 
and $-0.0035<y<0.010$.  

The analysis relies on the amplitude throughout the Dalitz plot, but its
modelling has only been tested so far with rates. In the region 
of the
Dalitz plot corresponding to large $K^{**}$ masses ($K^{**}$ denotes heavy
kaon states which decay to $K_S \pi$) the ratio of the DCS and CF rates is
significantly enhanced in the Belle model~\cite{part5:belle:2007:kspp}
compared to that for $D \rightarrow K\pi$. While this is possible
theoretically, it is less pronounced in the BaBar
model~\cite{part5:babar:2006,part5:ligeti:2007}. Data on $CP$-tagged $D
\rightarrow K_S \pi^+\pi^-$ decays expected soon from CLEO-c and \bes3
could help reduce the uncertainties. With more data, one can also attempt 
a model-independent analysis, as in the extraction of the CKM angle
$\gamma$~\cite{part5:gamma:d}. 

In summary, combining all experimental results obtained without allowing for
$CP$ violation, HFAG finds a 5.7$\sigma$ signal for $D^0 -\bar{D}^0$ mixing,
with the projections~\cite{part5:ref:hfag}
\begin{eqnarray}
 x = (8.7^{+3.0}_{-3.4}) \times 10^{-3}, \,\, y = (6.6 \pm 2.1)\times
 10^{-3}. 
 \label{part5:x:y:combine:all}
\end{eqnarray}
\begin{figure}[htb]
\begin{center}
\epsfig{file=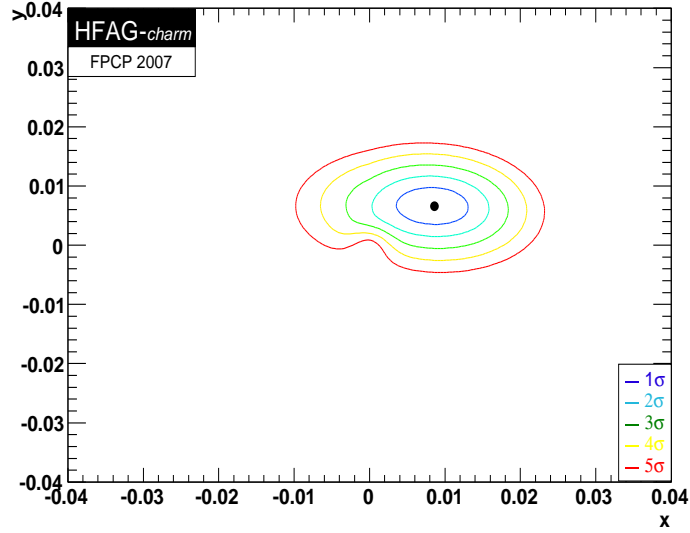,height=7cm,width=9cm}
\caption{Contours (1 through 5$\sigma$) of the allowed region in the
$x$ vs $y$ plane are shown from HFAG by combining all results. The significance of
the oscillation effect is about 5.7$\sigma$.}
\label{fig:part5:fig:combined:all}
\end{center}
\end{figure}
\noindent
In Fig.~\ref{fig:part5:fig:combined:all}, 
the overall likelihood function with
contours (1 through 5$\sigma$) for the allowed region in the
$x$ vs $y$ plane from HFAG is shown~\cite{part5:ref:hfag}.


\newpage
\section[Measurements at the $\psi(3770)$ peak]{Measurements at the 
$\psi(3770)$
peak\footnote{Edited by H.~B.~Li}}

At \bes3,  $\psi(3770)$ decays will provide another opportunity to
search for $D^0-\overline{D}^0$ mixing and to investigate sources of $CP$
violation in the charm system. The amplitude for $\psi(3770)$ decaying to
$D^0\overline{D}^0$ is $\langle D^0 \overline{D}^0 |{\cal H}|\psi(3770)\rangle$,
and the $D^0\overline{D}^0$ final-state system has 
charge parity
$C=-1$, which can be expressed as
\begin{eqnarray}
|D^0 \overline{D}^0\rangle^{C=-1} = \frac{1}{\sqrt{2}} [
|D^0\rangle|\overline{D}^0 \rangle -| \overline{D}^0\rangle |D^0 \rangle ].
\label{eq:cleo_dalitz_time_4}
\end{eqnarray}
Even though there is a weak current contribution to $\psi(3770)
\rightarrow D^0 \overline{D}^0$ decay that may not conserve
charge-parity, the $D^0 \overline{D}^0$ pair can still not be in a
$C=+1$ state.  The reason is that the relative orbital angular
momentum of the $D^0 \overline{D}^0$ pair must be $L=1$ because
of angular momentum conservation, and a boson-pair with $L=1$ must 
be in an anti-symmetric state; the anti-symmetric state of a
particle-anti-particle pair must have $C=-1$. The
$D^0$ and $\overline{D}^0$ mesons will, therefore, be entangled
with the same quantum numbers as the parent resonance.

In general, as shown in Ref.~\cite{Part5_Rosner}, a $D^0\overline{D}^0$ pair
produced via a virtual photon in the reaction $e^+e^- \rightarrow D^0
\overline{D}^0 + m \gamma +n \pi^0$ is in a $C=(-1)^{m+1}$ state. Thus, at
the $\psi(3770)$, where no additional fragmentation particles are produced,
there can only be $C=-1$, while at energies above $D^* 
\overline{D}$ threshold, both $C=-1$ and $C=+1$ eigenstates
can be accessed, such as $e^+ e^-
\rightarrow \psi(4140) \rightarrow \gamma (D^0 \overline{D}^0)_{C=+1}$
or $\pi^0 (D^0 \overline{D}^0)_{C=-1}$.

We now consider decays of these correlated systems into various
final states, and look in particular for interference effects
that depend on $\delta$.  In all cases, we integrate with respect to
the proper time, since the vertex separation at a symmetric $e^+e^-$
``charm factory" is likely to be problematic.
Xing~\cite{Part5_xing} and Gronau {\it et
al.}~\cite{Part5_Grossman} have considered time-integrated decays
into correlated pairs of states, including some effects of a
non-zero final state phase difference.  In
Ref.~\cite{Part5_Grossman}, Gronau {\it et al.} derived general
expressions for time-integrated decay rates into a pair of final
states $f_1$ and $f_2$, from $C=-1$ and $C=+1$ $D^0\overline{D}^0$
states:
\begin{eqnarray}
 \Gamma^{C=-1} (f_1, f_2) &=& \frac{1}{2} |a^-|^2 \left[\frac{1}{1-y^2} +
\frac{1}{1+x^2} \right]  +\frac{1}{2}|b^-|^2 \left[\frac{1}{1-y^2} -
\frac{1}{1+x^2}\right], \nonumber \\
\label{eq:besiii_master_1}
\end{eqnarray}
\begin{eqnarray}
 \Gamma^{C=+1} (f_1, f_2)  & = & \frac{1}{2} |a^+|^2
\left[\frac{1+y^2}{(1-y^2)^2} +
\frac{1-x^2}{(1+x^2)^2}\right] \nonumber \\
& +& \frac{1}{2}|b^-|^2\left[\frac{1+y^2}{(1-y^2)^2} - \frac{1}{(1-x^2)^2} \right] \nonumber
\\
&+&  2  {\cal R}e \left\{ a^{+*} b^+ \left[\frac{y}{(1-y^2)^2} + \frac{ix}{(1+x^2)^2}
\right]\right\},
\label{eq:besiii_master_2}
\end{eqnarray}
where
\begin{eqnarray}
a^{\pm} = \langle f_1 |D^0 \rangle \langle f_2| \overline{D}^0 \rangle \pm
\langle f_1 | \overline{D}^0 \rangle \langle f_2|D^0 \rangle ,
\label{eq:part5_amp_1}
\end{eqnarray}
\begin{eqnarray}
b^{\pm} = \langle f_1 |D^0 \rangle \langle f_2| D^0 \rangle \pm
\langle f_1 | D^0 \rangle \langle f_2|D^0 \rangle .
\label{eq:part5_amp_2}
\end{eqnarray}
These can be easily generalized to allow for $CP$
violation~\cite{Part5_Grossman}:
\begin{eqnarray}
a^{\pm} = \frac{p}{q} A_1 A_2 (\lambda_2 \pm \lambda_1),
\label{eq:part5_amp_cp1}
\end{eqnarray}
\begin{eqnarray}
b^{\pm} = \frac{p}{q} A_1 A_2 (1\pm \lambda_1 \lambda_2),
\label{eq:part5_amp_cp2}
\end{eqnarray}
where
\begin{eqnarray}
A_i \equiv \langle f_i | D^0 \rangle, \,\,\,\, \overline{A}_i \equiv \langle
f_i | \overline{D}^0 \rangle, \,\,\,\, \lambda_i \equiv \frac{q}{p}
\frac{\overline{A}_i}{A_i}.
\label{eq:part5_cp}
\end{eqnarray}
The quantities in the above equation are discussed in detail in
Sect.~\ref{subsec:incoherent}.

Following Refs.~\cite{Part5_xing,Part5_Grossman,Part5_petrov,Part5_asner}, 
we consider the
following categories of $D^0$ and $\overline{D}^0$ final states:
\begin{itemize}
\item Hadronic final states, $f$ or $\overline{f}$,  but not $CP$ eigenstates, such as $K^- \pi^+$,
which is produced via CF $D^0$ transitions or DCS $\overline{D}^0$
transitions;
\item Semileptonic or pure leptonic final states, $l^+$ or $l^-$, which, in
the absence of mixing, tag unambiguously the flavor of the parent $D$;
\item $CP$-even ($S_+$) and $CP$-odd ($S_-$) eigenstates, respectively.
\end{itemize}
We discuss different final states in the context of the assumption
of $CP$ conservation.  Taking into account  $x$, $y\ll 1$, keeping
terms up to order $x^2$, $y^2$ and $R_D$ ($R_D$ is the ratio between DCS and
CF decay rate as defined in Eq.~\ref{eq:mix_lambda_f_cpphase_2}), and 
neglecting $CP$ violation
in decay and mixing,  one gets the following results for various
cases~\cite{Part5_Grossman}\\

$C= -1$ $D^0 \overline{D}^0$ states:
\begin{itemize}
\item $(K^- \pi^+)(K^-\pi^+)$:
 \begin{eqnarray}
      \Gamma^{C=-1} (K^- \pi^+)(K^-\pi^+) = \frac{1}{2} A^4 |1-R_D e^{-2 i
\delta}|^2 (x^2 + y^2) \approx \frac{1}{2} A^4 (x^2+y^2), \nonumber \\
 \label{eq:correlation_1}
\end{eqnarray}
where $A = |\langle K^- \pi^+ | D^0 \rangle |$ is the real-valued decay
amplitude. This process serves to measure mixing effects.
\item $(K^- \pi^+)( K^+ \pi^-)$:
\begin{eqnarray}
\Gamma^{C=-1} (K^- \pi^+)(K^+\pi^+) =  A^4 \left[1- 2R_D
\mbox{cos}2\delta - \frac{1}{2}(x^2 -y^2) \right].
 \label{eq:correlation_2}
\end{eqnarray}
This process is not sensitive to the mixing measurements since $x$ and
$y$ are small. It can be used as a normalization for the previous case.
\item $(K^-\pi^+)( S_{\chi})$:
\begin{eqnarray}
\Gamma^{C=-1} (K^- \pi^+)(S_{\chi}) =  A^2 A^2_{S_{\chi}} | 1+ \chi
\sqrt{R_D} e^{-i \delta}|^2 (1+y^2) \approx A^2 A^2_{S_{\chi}}
 (1+2\chi \sqrt{R_D} \mbox{cos} \delta) , \nonumber \\
\label{eq:correlation_3}
\end{eqnarray}
where $S_{\chi}~\chi=\pm 1$ denotes $CP$ eigenstate decays.  
In the $SU(3)_{fl}$ limit $R_D = \mbox{tan}^4 \theta_C \approx 0.0025$. By
comparing the rate for $\chi = +1$ final states, such as $K^+K^-$, and 
$\chi = -1$ final states, such as $K_S$ ($\pi^0$ or $\omega$, $\phi$), one 
can measure  cos$\delta$.
\item $(K^-\pi^+)(l^- X)$:

      At the $\psi(3770)$ peak, using a leptonic $\overline{D}^0$ flavor 
tag and
defining $A_{l^-} = \langle l^- X| \overline{D}^0 \rangle$, one
finds~\cite{Part5_Grossman,Part5_xing}
\begin{eqnarray}
\Gamma^{C=-1} (K^- \pi^+)(l^- X) =  A^2 A^2_{l^-}\left[ 1- \frac{1}{2}(x^2
- y^2)\right].
\label{eq:correlation_4}
\end{eqnarray}
This process is not sensitive to mixing parameters and serves as a
normalization for the next process.
\item $(K^-\pi^+)(l^+ X)$:
\begin{eqnarray}
\Gamma^{C=-1} (K^- \pi^+)(l^+ X) =  A^2 A^2_{l^+}\left[ R_D+\frac{1}{2}(x^2
+ y^2)\right].
\label{eq:correlation_5}
\end{eqnarray}
where $A_{l^+} = \langle l^+X | D^0 \rangle = A_{l^-}$.  This process is 
interesting  if $x^2$, $y^2$ and $R_D$ are of comparable
in size. But, as mentioned above, it is  likely that 
$x^2$, $y^2 \ll R_D$ in which
case this process can be used to measure $R_D$.
\item $(S_{\chi})(l^+X)$:
 \begin{eqnarray}
\Gamma^{C=-1} (S_{\chi})(l^+ X) =  A^2_{S_{\chi}} A^2_{l^+}(1+y^2).
\label{eq:correlation_6}
\end{eqnarray}
Here $A^2_{S_{\chi}}$ is already of order $\sqrt{R_D}$. 
This process serves as a normalization for the others.
\end{itemize}

$C= +1$ $D^0 \overline{D}^0$ states: \\
\begin{itemize}
\item $(K^- \pi^+)(K^-\pi^+)$.
 \begin{eqnarray}
      \Gamma^{C=+1} (K^- \pi^+)(K^-\pi^+) = 4 A^4 \left[ R_D + \sqrt{R_D}
y^\prime + \frac{3}{8} (x^2 + y^2) \right].
 \label{eq:correlation_1_p}
\end{eqnarray}
In this case, the three terms ($R_D$, $y^\prime$ and $R_M$) may be
measurable. However, in most SM estimates  $x$ and $y$ 
are considerably smaller than a
percent~\cite{Part5_nelson}. If this is the case, the last term in
Eq.~\ref{eq:correlation_1_p} may be inaccessible even though
evidence exists for a non-zero $\sqrt{R_D} y^\prime$ term. Moreover,
we need an independent determination of $\delta$, which can be
obtained from Eq.~\ref{eq:correlation_3}. 
\item $(K^- \pi^+)( K^+ \pi^-)$:
\begin{eqnarray}
\Gamma^{C=+1} (K^- \pi^+)(K^+\pi^+) =  A^4 \left[1+ 2R_D
\mbox{cos}2\delta + 4 \sqrt{R_D}(y \mbox{cos}\delta + x \mbox{sin}\delta) -
\frac{3}{2} (x^2 -y^2) \right].  \nonumber \\
 \label{eq:correlation_2_p}
\end{eqnarray}
Here the correction terms are probably unmeasurable and, so, this process 
can serve as a normalization for comparison with the previous one.
\item $(K^-\pi^+)(S_{\chi})$:
 \begin{eqnarray}
\Gamma^{C=+1} (K^- \pi^+)(S_{\chi}) =
\approx  A^2 A^2_{S_{\chi}} (1-2 \chi \sqrt{R_D} \mbox{cos}\delta)(1-2\chi y).
 \label{eq:correlation_3_p}
\end{eqnarray}
This process provides information that constrains $\mbox{cos}\delta$ if
$R_D$ and $y$ are known.
\item $(K^-\pi^+)(l^- X)$:
  \begin{eqnarray}
\Gamma^{C=+1} (K^- \pi^+)(l^- X) =  A^2 A^2_{l^-}
\left[1+2 \sqrt{R_D}(y \mbox{cos}\delta + x \mbox{sin}\delta)- \frac{3}{2}(x^2
-y^2) \right].\nonumber \\
 \label{eq:correlation_4_p}
\end{eqnarray}
\item $(K^- \pi^+)(l^+ X)$:
  \begin{eqnarray}
\Gamma^{C=+1} (K^- \pi^+)(l^+ X) =  A^2 A^2_{l^+}
\left[R_D +2 \sqrt{R_D}y^\prime+ \frac{3}{2} (x^2 + y^2) \right].
 \label{eq:correlation_5_p}
\end{eqnarray}

\end{itemize}

\subsection{The mixing rate: $R_M$}
\label{subsec:mixing_RM_3770}

The \bes3 experiment at BEPCII will search for $D^0 - \overline{D}^0$ 
mixing at the $\psi(3770)$ 
by observing semileptonic modes of $D^0$'s:
\begin{eqnarray}
\frac{N(l^{\pm} l^\pm)}{N(l^\pm l^\mp)} = \frac{x^2+y^2}{2} = R_M,
\label{eq:besiii_rm_measure}
\end{eqnarray}
hadronic decay modes:
\begin{eqnarray}
\frac{N[(K^- \pi^+ )(K^-\pi^+)]}{N[(K^-\pi^+)(K^+\pi^-)]} \approx \frac{x^2+y^2}{2} =
R_M.
\label{eq:besiii_rm_measure_hadron}
\end{eqnarray}
and for cases where one final state is hadronic and the other 
semileptonic:
\begin{eqnarray}
\frac{N[(l^+ )(K^-\pi^+)]}{N[(l^+)(K^+\pi^-)]} \approx
\frac{x^2+y^2}{2}+R_D ,
\label{eq:besiii_rm_measure_hadron_semi}
\end{eqnarray}
where $R_D$ is defined in Eq.~\ref{eq:mix_lambda_f_cpphase_2}.

The measurement of $R_M$ can be performed unambiguously using the
reactions:
\begin{eqnarray}
(i)\,\, e^+ e^- \rightarrow \psi(3770) \rightarrow D^0 \overline{D}^0
\rightarrow
(K^\pm \pi^\mp)( K^\pm \pi^\mp), \nonumber \\
(ii)\, \, e^+ e^- \rightarrow \psi(3770) \rightarrow D^0 \overline{D}^0
\rightarrow
(K^- e^+ \nu )(K^- e^+ \nu),  \nonumber \\
(iii) \, \, e^+e^- \rightarrow D^- D^{*+} \rightarrow (K^+ \pi^- \pi^-) (
\pi^+_{soft} [K^+ e^- \nu]).
\label{eq:besiii_rm_corr_1}
\end{eqnarray}
\noindent
The observation of reaction $(i)$ would be definite evidence for
the existence of $D^0 -\overline{D}^0$ mixing since the final
state $(K^\pm \pi^\mp)( K^\pm \pi^\mp)$ can not be produced from
DCS decay due to quantum
statistics~\cite{bigi_sanda,Part5_bigi_semi}. In particular, the
initial $D^0\overline{D}^0$ pair is in an odd eigenstate of $C$,
which precludes, in the absence of mixing between the $D^0$
and $\overline{D}^0$ over time, the formation of a symmetric
state, which is required by Bose statistics for decays  to the same
final state.  This final state is also very appealing
experimentally, because it involves a two-body decay of both charm
mesons, with energetic charged particles in the final state that
form an overconstrained system. Particle identification is crucial
in this measurement: if both the kaon and pion are
misidentified in one of the two $D$-meson decays in the event, it
is impossible to discern whether or not mixing has occurred. At
\bes3, with an expectd integrated luminosity of 20~fb$^{-1}$
at the $\psi(3770)$ peak,  the sensitivity will be 
$\sqrt{R_M} \simeq 0.4\%$, but only if the particle identification
capabilities are adequate. If it were possible to obtain 500
fb$^{-1}$ at the $\psi(3770)$, the sensitivity would be  $\sqrt{R_M}
\simeq 0.08\%$~\cite{CHARMREV}.

Reactions $(ii)$ and $(iii)$
offer unambiguous evidence for mixing in that the mixing would be seen in
semileptonic decays for which there is no DCS decay contamination. 
Since
the time-evolution is not measured, the observation of Reactions $(ii)$ 
and $(iii)$ would actually indicate a violation of the SM selection 
rule relating the change in charm to the change in leptonic 
charge~\cite{Part5_bigi_semi}.

In Table~\ref{tab:mixing_r_m}, the sensitivity for $R_M$ measurements in
different decay modes are estimated for a four year run at 
BEPCII ({\it i.e.}, 20~fb$^{-1}$).
\begin{table}[htbp]
  \centering
\caption{The sensitivity for $R_M$ measurements at BESIII with different
decay modes for a four year run at BESPCII}
  \begin{tabular}{c|c|c} \hline
 \multicolumn{3}{c}{$D^0\overline{D}^0$ Mixing} \\ \hline

 Reaction & Events  & Sensitivity of $R_M$  \\
         & Right Sign &                  \\ \hline
 $\psi(3770) \rightarrow (K^-\pi^+)(K^-\pi^+)$ & 10,3600  & $ 1\times 10^{-4}$\\ \hline
 $\psi(3770) \rightarrow (K^- e^+ \nu)(K^- e^+ \nu )$ & 8,8705  &   \\
 $\psi(3770) \rightarrow (K^- e^+ \nu)(K^- \mu^+ \nu )$ & 8,0617  & $ 3.7 \times 10^{-4}$\\
 $\psi(3770) \rightarrow (K^- \mu^+ \nu)(K^- \mu^+ \nu )$ & 7,3268  &  \\ \hline

$D^{*+} D^- \rightarrow [\pi^+_s(K^+ e^- \overline{\nu})(K^+  \pi^- \pi^- )]$ & 76000  &  \\
$D^{*+} D^- \rightarrow [\pi^+_s(K^+ \mu^- \overline{\nu})(K^+ \pi^- \pi^-)]$ & 60000  &  \\
$D^{*+} D^- \rightarrow [\pi^+_s(K^+ e^- \overline{\nu})$(other $D^-$ tag)] & 60000  &  $4.7 \times 10^{-5}$ \\
$D^{*+} D^- \rightarrow [\pi^+_s(K^+ \mu^- \overline{\nu})$(other $D^-$ tag)] & 60000  & \\ \hline
\label{tab:mixing_r_m}
\end{tabular}
\end{table}

\begin{table}[htbp]
  \centering
\caption{The expected mixing signal for
$N_{sig}=N(K^\pm\pi^\mp)(K^\pm\pi^\mp)$, background $N_{bkg}$,
and the Poisson probability $P(n)$ for 10 fb$^{-1}$ and 20
fb$^{-1}$  \bes3 data samples at the $\psi(3770)$ peak. 
Here, we assume a mixing rate of $R_M=1.18\times 10^{-4}$. }
  \begin{tabular}{c|c|c} \hline \hline
   & 10 fb$^{-1}$ ($\psi(3770)$)  &   20 fb$^{-1}$ ($\psi(3770)$) \\
   & 36 million $D^0\Dzb$ &72 million
$D^0\Dzb$ \\ \hline
  $N_{\rm sig}$  & 1.5 & 3.0      \\
  $N_{\rm bkg}$  & 0.3 & 0.6      \\ \hline
  $P(n=0)$  & 15.7\% & 2.5\%  \\
 $P(n=1)$   & 29.1\% & 9.1\%  \\
 $P(n=2)$  & 26.9\% & 16.9\%  \\
 $P(n=3)$   & 16.6\% & 20.9\%  \\
 $P(n=4)$  & 7.7\% & 19.3\%  \\
 $P(n=5)$   & 2.8\% & 14.3\%  \\
  $P(n=6)$  & 0.9\% & 8.8\%  \\
 $P(n=7)$   & 0.2\% & 4.7\%  \\
  $P(n=8)$  & 0.1\% & 2.2\%  \\
 $P(n=9)$   & 0.01\% & 0.9\%  \\ \hline \hline
\end{tabular}
\label{part5:new:tab:possion}
\end{table}

In the limit of $CP$ conservation, the combined measurements of
$x$ from $D^0 \rightarrow K_S \pi\pi$ and $y_{CP}$ from
Belle~\cite{part5:new:li:yang},  gives
$R_M =(1.18\pm 0.6)\times 10^{-4}$.  With 20~fb$^{-1}$ of
data at \bes3, about 12 events for the precess $D^0\Dzb
\rightarrow (K^\pm \pi^\mp) (K^\pm \pi^\mp)$ will be produced. Only
3.0 events are expected to be observed, on average, because the 
\bes3 detection efficiency for a four-charged-particle 
final state is about 25\%.  The background contamination 
due to double particle misidentification in the same sample
is estimated to be about 0.6 events. 
Table~\ref{part5:new:tab:possion} lists the
expected mixing signal for
$N_{sig}=N(K^\pm\pi^\mp)(K^\pm\pi^\mp)$, background $N_{bkg}$ ,
and the Poisson probability $P(n)$, where $n$ is the possible
number of events that are observed  in the experiment. In
Table~\ref{part5:new:tab:possion}, we assume $R_M =1.18\times 10^{-4}$,
and the expected numbers of mixing signal events are estimated for
10fb$^{-1}$ and 20fb$^{-1}$ data samples.

\subsection{Lifetime differences and the strong phase $\delta_{K\pi}$ }

Doubly Cabibbo suppressed (DCS) decays of the $D^0$ mesons, and
$D^0$ mixing, give rise to identical final states. The two
processes can only be distinguished by their different time
dependence or, at the $\psi(3770)$ peak, by taking advantage of
effects due to quantum statistics as discussed in
Sect.~\ref{subsec:mixing_RM_3770}.  In
Eq.~\ref{eq:mix_standard_rate_inte} of
Sect.~\ref{subsecc:hadron_final}, the wrong-sign decay rate
relative to the right-sign rate is defined as
\begin{eqnarray}
R_{WS}  =  R_D + \sqrt{R_D} y^{\prime} +
\frac{1}{2} R_M \, .
\label{eq:dbcs_rate_3770}
\end{eqnarray}
In absence of mixing, $R_{WS}  =  R_D = |A_f/\overline{A}_f|^2$.
In general, the ratio of DCS decay rate relative to CF decay rate is
$R_D \sim \mbox{tan}^4 \theta_C \sim 0.25\%$, where $\theta_C$ is the Cabibbo
angle. However, as pointed out in Ref.~\cite{bigi_sanda}, $ \mbox{tan}^4 \theta_C$
is not the only suppression factor.  Final state interactions can
cause  the  ratio to be differerent for each
final state, such as $R_D \sim 2.1  \times \mbox{tan}^4 \theta_C$ for $D^0 \rightarrow K^+
\pi^-$, while $R_D \sim 0.45 \times \mbox{tan}^4 \theta_C$ for $D^0 \rightarrow K^+
\rho^-$.

One can also measure $R_D$ in the multibody channels $D^0 \rightarrow K^+
\pi^- \pi^0$ and $D^0 \rightarrow K^+ \pi^- \pi^+\pi^+$ as discussed in
Sect.~\ref{subsecc:hadron_final}.

At the $\psi(3770)$ peak, semileptonic decays can be used to tag
hadronic decays on the recoil side.  Using
Eq.~\ref{eq:besiii_rm_measure_hadron_semi}
and neglecting mixing effects, one
has~\cite{Part5_Grossman}
\begin{eqnarray}
\frac{N[(l^+ )(K^-\pi^+)]}{N[(l^+)(K^+\pi^-)]} \sim R_D.
\label{eq:besiii_rm_measure_hadron_semi_again}
\end{eqnarray}
Since it is likely that $x^2$, $y^2 \ll R_D$, this process can
be used to measure $R_D$ directly.

In the limit of $SU(3)_{fl}$ symmetry, $A_{K^+ \pi^-}=\langle K^+ \pi^- 
|{\cal
H}|D^0 \rangle$ and $\overline{A}_{K^+ \pi^-} =  \langle K^+ \pi^- |{\cal H}|\Dzb
\rangle$ ($A_{K^- \pi^+} = \langle K^- \pi^+ |{\cal H}|D^0
\rangle$ and $\overline{A}_{K^-
\pi^+} =  \langle K^- \pi^+ |{\cal H}|\Dzb \rangle$) 
are simply related by CKM factors, $A_{K^+ \pi^-}
=(V_{cd}V^*_{us}/V_{cs}V^*_{ud})\overline{A}_{K^+
\pi^-}$~\cite{Part5_Grossman}. In particular, $A_{K^+ \pi^-}$ and
$\overline{A}_{K^+ \pi^-}$ have the same strong phase.  But
$SU(3)_{fl}$ symmetry is broken; according to the recent
measurements from the $B$ factories, the ratio~\cite{Part5_Nir}:
\begin{eqnarray}
{\cal R} = \frac{{\cal BR}(D^0 \rightarrow K^+\pi^-)}{{\cal BR}
(\Dzb \rightarrow K^+\pi^-)} \left| \frac{V_{ud}V^*_{cs}}{V_{us}
V^*_{cd}} \right|^2,
 \label{part5:new:eq:rt_su30}
\end{eqnarray}
which is unity in the $SU(3)_{fl}$ symmetry limit, is measured to 
be~\cite{Part5_pdg06}
\begin{eqnarray}
{\cal R}_{exp} = 1.21\pm 0.03.
 \label{part5:new:eq:rt_su3}
\end{eqnarray}
Since $SU(3)_{fl}$ is apparently broken in $D
\rightarrow K\pi$ decays at the level of 20\%, the
strong phase $\delta$ should be non-zero. Recently, a 
time-dependent analysis of $D \rightarrow K\pi$ has been performed
based on a 384 fb$^{-1}$ data sample at the 
$\Upsilon(4S)$~\cite{part5:new:1}. With the assumption of
$CP$ conservation, the following neutral
$D$ mixing results are obtained~\cite{part5:new:1}:

\begin{eqnarray}
R_D &=& (3.03 \pm 0.16 \pm 0.10)\times 10^{-3} ; \nonumber \\
x^{\prime^2}& =& (-0.22 \pm 0.30 \pm 0.21) \times 10^{-3};
\nonumber \\
y^\prime &=& (9.7 \pm 4.4 \pm 3.1)\times 10^{-3}.
 \label{part5:new:eq:kp_babar}
\end{eqnarray}

\begin{table}[htbp]
  \centering
  \caption{Current experimental results. The quoted errors are
the quadrature sum of the statistical and systematic contributions.}
  \begin{tabular}{c|c|c|c} \hline\hline
  Parameter & BaBar ($\times 10^{-3}$)  & Belle($\times 10^{-3}$) &
Technique \\ \hline
  $x^{\prime^2}$ & -$0.22\pm 0.37$~\cite{part5:new:1}
&$0.18^{+0.21}_{-0.23}$~\cite{part5:new:belle_kp_06} & $K\pi$ \\
  $y^{\prime}$ & $9.7\pm 5.4$~\cite{part5:new:1} &
$0.6^{+4.0}_{-3.9}$~\cite{part5:new:belle_kp_06} & $K\pi$ \\
  $R_D$ & $3.03\pm 0.19$~\cite{part5:new:1} & $3.64\pm
0.17$~\cite{part5:new:belle_kp_06} &
$K\pi$ \\
  $y_{CP}$ & - & $13.1\pm 4.1$~\cite{part5:new:2} & $K^+K^-$, $\pi^+\pi^-$ \\
  $x$ & -& $8.0\pm3.4$~\cite{part5:new:marko_belle_07} & $K_S \pi^+\pi^-$ \\
  $y$ & -& $3.3\pm 2.8$~\cite{part5:new:marko_belle_07} & $K_S \pi^+\pi^-$ \\
  \hline \hline
 \end{tabular}
  \label{part5:new:tab:experiments}
\end{table}

The results are inconsistent with the non-mixing hypothesis with a
significance of 3.9 standard deviations.  The BaBar
and Belle results for the $y^\prime$ measurement using $D \rightarrow
K\pi$ agree within 2 standard deviations (see
Table~\ref{part5:new:tab:experiments}).  As indicated in
Eq.~\ref{part5:new:eq:rt_su3}, the strong phase $\delta$ should be
non-zero as a result of $SU(3)_{fl}$ violation.   
In order to extract  the direct mixing parameters, $x$ and $y$,
the strong phase difference has to be known. 
However, it is hard to do this at the $B$ factories in a
model-independent way~\cite{Part5_Grossman,CHARMREV}. To
extract the strong phase $\delta$ we need 
$CP$ tagged $D^0$ decays near the
$D\overline{D}$ threshold, as discussed in
Ref.~\cite{Part5_Grossman}.  Here, we investigate existing
information on the strong phase $\delta$ from
recent $B$ factory results for different decay
modes.  This can give us some idea about 
\bes3's sensitivity for measuring the strong phase.

In Ref~\cite{part5:new:2}, Belle reported results for
$y_{CP}=\frac{\tau(D^0\rightarrow K^+\pi^-)}{\tau(D^0 \rightarrow
f_{CP})} -1$,  where $f_{CP} = K^+K^-$ and $\pi^+\pi^-$,
\begin{eqnarray}
 y_{CP} = (13.1 \pm 3.2 \pm 2.5) \times 10^{-3}.
 \label{part5:new:eq:kk_belle}
\end{eqnarray}
This result is about a 3.2$\sigma$ significant deviation from zero
(non-mixing). In the limit of $CP$ symmetry, $y_{CP} =
y$~\cite{part5:new:nir_2000,part5:new:petrov_2005}. 
In the decay of $D^0 \rightarrow
K_S \pi^+\pi^-$, Belle did a Dalitz plot (DP)
analysis~\cite{part5:new:marko_belle_07} and obtained the direct mixing
parameters $x$ and $y$ as
\begin{eqnarray}
 x = (8.0 \pm 3.4) \times 10^{-3}, \,\, y = (3.3 \pm 2.8)\times
 10^{-3},
 \label{part5:new:eq:kspp_belle}
\end{eqnarray}
where the errors include both statistic and systematic
uncertainties. Since the parameterizations  of the resonances 
contributing to the Dalitz plot (DP)
are model-dependent,  the results suffer from large
systematic uncertainties. In their analysis, they see a
2.4$\sigma$ significant deviation from non-mixing. Here, we
use the value of $x$ measured in the DP analysis for further
discussion. Once $y$,
$y^\prime$ and $x$ are known, it is straightforward to extract the
strong phase difference between the DCS and CF 
$D^0 \rightarrow K\pi$ decay amplitudes. 
Using the measured central values of
$x$, $y_{CP}(\approx y)$ , and $y^\prime$ as input parameters, we
find a two-fold ambiguous  solution for $\mbox{tan}\delta$:
\begin{eqnarray}
 \mbox{tan}\delta = 0.35\pm 0.63, \,\, \mbox{or} \, \, -7.14 \pm 29.13,
 \label{part5:new:eq:phase_solution}
\end{eqnarray}
corresponding to $(19 \pm 32)^0$ and $(-82^0 \pm 30)^0$, respectively.


In order to extract the mixing parameter $y$
at the $\psi(3770)$ peak, one can
make use of rates for exclusive $D^0\Dzb$ combinations, where both
the $D^0$ final states are specified (known as double tags or DT),  
as well as inclusive rates, where either the $D^0$ or $\Dzb$ is
identified and the other $D^0$ decays generically (known as single
tags or ST).  With the DT 
technique~\cite{part5:new:markiii_1,part5:new:markiii_2}, 
one can fully exploit the quantum correlations in the 
$C=-1$ $D^0\Dzb$ pairs produced from $\psi(3770$ 
decays~\cite{Part5_bigi_semi,bigi_sanda,part5:new:asner_2005}.\footnote{The 
DT technique can also be used at energies above the $\psi(3770)$  
to exploit quantum correlations in $C=-1$ and $C=+1$ 
$D^0\Dzb$ pairs produced, respectively, via the
reactions $e^+e^- \rightarrow D^0 \Dzb(n\pi^0)$ and $e^+e^-
\rightarrow D^0 \Dzb \gamma (n\pi^0)$.}

For the ST sample, in the limit of $CP$ conservation, the rate of $D^0$
decays into a $CP$ eigenstate is given as~\cite{part5:new:asner_2005}:
\begin{eqnarray}
 \Gamma_{f_\eta}\equiv\Gamma(D^0 \rightarrow f_{\eta}) = 2A_{f_{\eta}}^2
\left[1-\eta  y \right],
 \label{part5:new:eq:st_cp}
\end{eqnarray}
where $f_{\eta}$ is a $CP$ eigenstate with eigenvalue $\eta = \pm
1$, and $A_{f_{\eta}}= |\langle f_\eta | {\cal H}|D^0 \rangle|$ is
the real-valued decay amplitude.

For the DT case,
Xing~\cite{Part5_xing} and
Gronau {\it et. al.}~\cite{Part5_Grossman}
have considered time-integrated decays into
correlated pairs of states, including the effects of non-zero
final state phase difference. As discussed in
Ref.~\cite{Part5_Grossman}, the rate of ($D^0 \Dzb)^{C=-1}
\rightarrow (l^\pm X)(f_\eta)$ is given
as~\cite{Part5_Grossman}:
\begin{eqnarray}
\Gamma_{l;f_\eta}\equiv \Gamma[(l^\pm X)(f_{\eta})] &= & A_{l^\pm
X}^2A_{f_\eta}^2 (1+ y^2) \nonumber \\
&\approx&  A_{l^\pm X}^2 A_{f_\eta}^2, \label{part5:new:eq:dt_cp}
\end{eqnarray}
where $A_{l^\pm X} = | \langle l^\pm X|{\cal H}|D^0\rangle|$ is
the real-valued amplitude for semileptonic decays; here, we have 
neglected the $y^2$ term since $y\ll1$.

For $C=-1$ initial $D^0\Dzb$ states,  $y$ can be expressed in terms
of ratios of DT rates and double ratios of ST rates to DT
rates~\cite{part5:new:asner_2005}:
\begin{eqnarray}
y = \frac{1}{4 } \left( \frac{\Gamma_{l; f_+}
\Gamma_{f_-}}{\Gamma_{l;f_-}\Gamma_{f_+}} -\frac{\Gamma_{l; f_-}
\Gamma_{f_+}}{\Gamma_{l;f_+}\Gamma_{f_-}} \right ).
\label{part5:new:eq:dt_cp_y}
\end{eqnarray} 
For a small $y$, its error, $\Delta (y)$, is approximately
$1/\sqrt{N_{l^\pm X}}$, where $N_{l^\pm X}$ is the total number of
$(l^\pm X)$ events tagged with $CP$-even and $CP$-odd eigenstates.
The number $N_{l^\pm X}$ of $CP$ tagged events is related to the
total number of $D^0 \Dzb$ pairs $N(D^0 \Dzb)$ through $N_{l^\pm
X} \approx N(D^0 \Dzb)[ {\cal BR}(D^0 \rightarrow l^\pm +X)\times
{\cal BR}(D^0 \rightarrow f_{\pm})\times \epsilon_{tag}] \approx
1.5\times 10^{-3} N(D^0 \Dzb)$, here we take the branching
ratio-times-efficiency factor (${\cal BR}(D^0 \rightarrow
f_{\pm})\times \epsilon_{tag}$) for tagging $CP$ eigenstates 
to be 1.1\% (the total branching ratio into $CP$ eigenstates is
larger than about 5\%~\cite{Part5_pdg06}). We find
\begin{eqnarray}
\Delta(y) = \frac{\pm 26}{\sqrt{N(D^0\Dzb)}} = \pm 0.003.
 \label{part5:new:eq:dt_cp_dy}
\end{eqnarray}
If we take the central value of $y$ from the Belle measurement of
$y_{CP}$~\cite{part5:new:2}, we infer that at the
\bes3 experiment~\cite{part5:bepcii}, with a
20$fb^{-1}$ data sample taken at the $\psi(3770)$
peak, the significance of the $y$ measurement 
would be around 4.3$\sigma$.

 We can also take advantage of the coherence of the $D^0$
mesons produced at the $\psi(3770)$ peak to extract the strong
phase difference $\delta$ between DCS and CF decay amplitudes~\cite{Part5_Grossman,part5:new:asner_2005}. Because the
$CP$ properties of the final states produced in the decay of the
$\psi(3770)$ are anti-correlated~\cite{Part5_bigi_semi,bigi_sanda}, one
$D^0$ state decaying into a final state with definite $CP$
properties immediately identifies or tags the $CP$ properties of
the other side.  As discussed in Ref.~\cite{Part5_Grossman}, the
process of one $D^0$ decaying to $K^-\pi^+$, while the other $D^0$
decaying to a $CP$ eigenstate $f_{\eta}$  can be described as
\begin{eqnarray}
\Gamma_{K\pi;f_\eta}\equiv \Gamma[(K^- \pi^+)(f_{\eta})] &\approx
& A^2A^2_{f_{\eta}}|1+ \eta
\sqrt{R_D} e^{-i \delta} |^2  \nonumber \\
&\approx&  A^2A^2_{f_{\eta}}(1+2 \eta \sqrt{R_D}
\mbox{cos}\delta),\nonumber \\
 \label{part5:new:eq:besiii_delta_rD}
\end{eqnarray}
where $A = |\langle K^- \pi^+ |{\cal H}| D^0 \rangle |$ and
$A_{f_{\eta}} = |\langle f_{\eta} |{\cal H}| D^0 \rangle |$ are
the real-valued decay amplitudes, and we have neglected the $y^2$
terms in Eq.~\ref{part5:new:eq:besiii_delta_rD}. In order to estimate the
total sample of events needed to perform a useful measurement of
$\delta$,  we use the asymmetry~\cite{CHARMREV,Part5_Grossman} 
\begin{eqnarray}
{\cal A} \equiv \frac{\Gamma_{K\pi;f_+} - \Gamma_{K\pi;
f_-}}{\Gamma_{K\pi;f_+} +\Gamma_{K\pi; f_-}},
\label{part5:new:eq:besiii_delta_rD_a}
\end{eqnarray}
where $\Gamma_{K\pi;f_\pm}$, defined in
Eq.~\ref{part5:new:eq:besiii_delta_rD},  is the rate for the
$\psi(3770) \rightarrow D^0 \Dzb$ configuration to decay into
a $K\pi$ flavor eigenstate and a $CP$-eigenstate $f_\pm $.
Equation~\ref{part5:new:eq:besiii_delta_rD} 
implies a small asymmetry: ${\cal
A} = 2 \sqrt{R_D} \mbox{cos} \delta$. 
In general, a small asymmetry, 
has an error $\Delta {\cal A}$ that is
approximately $1/\sqrt{N_{K^-\pi^+}}$, where $N_{K^-\pi^+}$ is the
total number of events tagged with $CP$-even and $CP$-odd
eigenstates. Thus one obtains
\begin{eqnarray}
\Delta (\mbox{cos} \delta) \approx \frac{1}{2\sqrt{R_D}
\sqrt{N_{K^-\pi^+}}}. \label{part5:new:eq:besiii_delta_rD_est}
\end{eqnarray}
\begin{figure}[htb]
\centering
 \epsfig{file=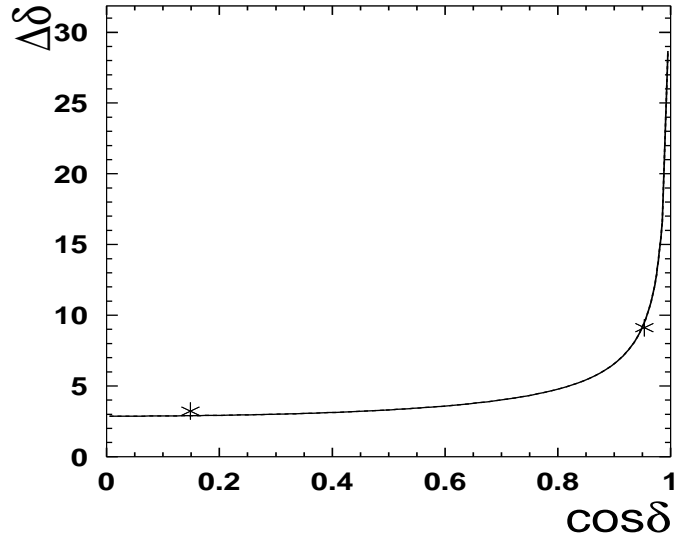,width=10cm,height=8cm}
\caption{ An illustrative plot of the expected error ($\Delta \delta$)
on the strong phase angle, in degrees, for various central values of
$\mbox{cos}\delta$.  The expected \bes3 error on $\mbox{cos}\delta$ is
0.04 for a 20~fb$^{-1}$ data sample taken at the $\psi(3770)$ peak.
The asterisks correspond to $\delta = 19^0$ and $-82^0$.} 
\label{part5:new:figure1}
\end{figure}
The expected number of $CP$-tagged events, $N_{K^-\pi^+}$, can be 
connected to the total number of $D^0 \Dzb$ pairs $N(D^0 \Dzb)$
through $N_{K^-\pi^+} \approx N(D^0 \Dzb){\cal BR}(D^0 \rightarrow
K^- \pi^+) \times {\cal BR}(D^0 \rightarrow f_{\pm})\times
\epsilon_{tag} \approx 4.2 \times 10^{-4} N(D^0
\Dzb)$~\cite{Part5_Grossman}, where, as in Ref~\cite{Part5_Grossman},
we take the branching ratio-times-efficiency factor to be ${\cal BR}(D^0
\rightarrow f_{\pm})\times \epsilon_{tag}=1.1\%$.  Using the
measured value of
$R_{D} = (3.03\pm 0.19)\times 10^{-3}$ and ${\cal BR}(D^0
\rightarrow K^- \pi^+)=3.8\%$~\cite{Part5_pdg06}, one
finds~\cite{Part5_Grossman}
\begin{eqnarray}
\Delta (\mbox{cos}\delta) \approx \frac{\pm 444}{\sqrt{N(D^0
\Dzb)}}. \label{part5:new:eq:besiii_delta_rD_est_num}
\end{eqnarray}
At \bes3,  about $72 \times 10^6$ $D^0 \Dzb$ pairs will be
collected in a four year run at the $\psi(3700)$. 
Considering both $K^- \pi^+$
and $K^+\pi^-$ final states,  we estimate that 
an accuracy level of about 0.04 for cos$\delta$ can be reached.
Figure~\ref{part5:new:figure1} shows the expected error 
of the strong phase
$\delta$ for various central values of $\mbox{cos}\delta$. With
the expected error of $\Delta(\mbox{cos}\delta)= \pm 0.04$,  the
sensitivity to the strong phase {\em angle} varies with the physical 
value of
$\mbox{cos}\delta$. For $\delta = 19^0$ and $-82^0$, the expected
errors would be $\Delta(\delta) = \pm 8.7^0$ and $\pm 2.9^0$,
respectively.

\chapter[CP and T Violation]{CP and T Violation\footnote{By Ikaros Bigi and Hai-Bo Li}}
\label{sec:charm_cpv}
The violation of the $CP$ symmetry, where $C$ and $P$ are the
charge-conjugation and parity-transformation operators, respectively, is one
of the fundamental and most exciting phenomena in particle physics. Although
weak interactions are not invariant under $P$ (and $C$) transformations, as
discovered in 1957, it was believed for several years that the product $CP$
was preserved. However, in 1964, it was discovered through the observation
of $K_L \rightarrow \pi^+\pi^-$ decays that weak interactions are not
invariant under $CP$ transformations~\cite{Part5_Cronin_1964}. After this
discovery, many observations show us that $CP$ violation has been
established in both $K$ and $B$ systems~\cite{Part5_Nir_2005_10}. All these measurements are
consistent with the Kobayashi-Maskawa (KM) picture of $CP$ violation.

However, people still believe  that there must be new sources of $CP$
violation beyond the SM prediction for, at least, the following two 
reasons:
\begin{itemize}
\item {\it The baryon asymmetry of the Universe}: \\
      Baryogenesis is a consequence of $CP$ violating
processes~\cite{Part5_Sakharov_1967}. Therefore, the present baryon 
number,
which is accurately deduced from Big Bang Nucleosynthesis (BBN) and Cosmic
Microwave Background Radiation (CMBR) constraints,
\begin{eqnarray}
Y_B \equiv \frac{n_B - n_{\overline{B}}}{s} \approx 9\times 10^{-11},
\label{eq:Part5_Sakharov}
\end{eqnarray}
is essentially a $CP$ violating observable. The surprising point is that the
KM mechanism for $CP$ violation fails to account for
it~\cite{Part5_Nir_2005_10}.
\item {\it Non-vanishing neutrino masses:} \\
 It is also interesting to note that the evidence for non-vanishing neutrino
masses that has emerged over the last few years points towards an origin
beyond the SM~\cite{Part5_Altarelli_2003}, raising the question of having $CP$ violation in the
neutrino sector, which could be studied, in the more distant future, at a
dedicated neutrino factory~\cite{Part5_Rujula_1999}.
\end{itemize}

It would be very interesting to look for $CP$ violation in the $D$ system;
most factors favour dedicated searches for $CP$ violation in Charm
transitions:
\begin{itemize}
\item New physics might be just around the corner, since
baryogenesis implies the existence of New Physics (NP) in $CP$
violation dynamics. It will be of interest to undertake dedicated
searches for $CP$ asymmetries in Charm decays, where the SM
predicts very small effects,
{\it i.e.} smaller than ${\cal O}(10^{-3})$, and they 
can arise only in {\it singly Cabibbo-suppressed} (SCS)
transitions. Significantly larger values would signal NP. Any
asymmetry in CF and DCS decays requires the intervention of NP
(except for $D^\pm \rightarrow K_S \pi^\pm$~\cite{CHARMREV}, where
the $CP$ impurity in $K_S$ induces an asymmetry of $3.3 \times
10^{-3}$).

\item Secondly, the neutral $D$ system is the only one where the external
up-sector quarks are involved. Thus it probes models in which the up-sector
plays a special role, such as supersymmetric models with
alignment~\cite{Part5_Nir_1993,Part5_Leurer_1994} and, more
generally, models in which CKM mixing is generated in the up sector.

\item Third, SCS decays are sensitive to new-physics contributions to
penguin and dipole operators. As far as this point is concerned, among all 
hadronic $D$
decays, the SCS decays ($c \rightarrow u q\overline{q}$) are uniquely 
sensitive to new contributions to 
$\Delta C =1$ QCD penguin and chromomagnetic dipole
operators~\cite{Part5_Grossman_2006,Part5_Du_2006}. In particular, such 
contributions would
affect neither CF ($c\rightarrow s \overline{d}u$) nor the DCS ($c
\rightarrow d \overline{s} u$) decays.

\item There is a rich assortment of light resonances in 
the $D$ mass region and, thus, final state interactions 
involving these resonances are expected to be important 
sources of strong phase shifts that are necessary
for producing direct $CP$ violation.

\item Decays to final states of more than two pseudoscalar or one
pseudoscalar and one vector mesons contain more dynamical information than
that given by their partial widths.  Dalitz plot 
analyses can exhibit $CP$ asymmetries
that might be considerably larger than those from
decay rates only~\cite{Part5_Bigi_charm2006}.

\end{itemize}

 $CP$ asymmetries in integrated partial widths depend on hadronic matrix
elements and (strong)
phase shifts, neither of which can be predicted accurately. However the
craft of theoretical
engineering can be practised with profit here. One makes an ansatz for the
general form of the matrix
elements and phase shifts that are included in the description of 
$D\to PP,~PV,~VV$~etc.
channels, where $P$ and $V$ denote pseudoscalar and vector mesons, and fits
them to the
measured branching ratios on the Cabibbo-allowed, once- and 
twice-Cabibbo-forbidden levels.
If one has sufficiently accurate and comprehensive data,
one can use these fitted values of the hadronic parameters to predict
$CP$ asymmetries.
Such analyses have been undertaken in the past \cite{Part5_LUSIG},
but the data base was not as broad and precise as one would
like~\cite{Part5_Bigi_charm2006}.
CLEO-c and \bes3 measurements will certainly lift such studies to a new 
level of reliability.

\section{Formalism and Review}

In order to discuss the $CP$ violation in neutral $D$ system, we
use the following notations as described in the previous sections:
\begin{eqnarray}
&& \tau \equiv \Gamma_D t, \, \, \Gamma_D \equiv \frac{\Gamma_{D_H} +
\Gamma_{D_L} }{2}, \nonumber \\
&& A_f \equiv A(D^0 \rightarrow f), \, \, \overline{A}_f \equiv
A(\overline{D}^0 \rightarrow f), \,\, \nonumber \\
&&A_{\overline{f}} \equiv A(D^0 \rightarrow \overline{f}), \, \,
\overline{A}_{\overline{f}} \equiv
A(\overline{D}^0 \rightarrow \overline{f}), \,\, \nonumber \\
&&x \equiv \frac{\Delta m_D}{\Gamma_D} \equiv \frac{m_{D_H}
-m_{D_L}}{\Gamma_D}, \, \, y \equiv \frac{\Delta \Gamma_D}{2 \Gamma_D}
\equiv \frac{\Gamma_{D_H} - \Gamma_{D_L}}{2\Gamma_D}, \nonumber \\
&& \lambda_f \equiv \frac{q}{p}\frac{\overline{A}_f}{A_f}, \, \, A_M \equiv
\left| \frac{q}{p} \right| -1 , \,\, R_f \equiv  \left|
\frac{\overline{A}_f}{A_f}\right|,
\label{eq:Part5_define}
\end{eqnarray}
where $D_H$ and $D_L$ stand for the heavy and light mass
eigenstates, and $q$ and $p$ are defined  in 
Eq.~\ref{eq:eigenstates}. We distinguish three types of
$CP$-violation effects in neutral $D$ meson decays:
\begin{itemize}
\item $CP$ violation in decay is defined by
\begin{eqnarray}
\left| \frac{\overline{A}_{\overline{f}}}{A_f} \right| \neq 1 .
\label{eq:Part5_define_direct}
\end{eqnarray}
In the charged $D$ decays without mixing effects, it is the only possible
source of $CP$ asymmetries:
\item $CP$ violation in mixing is defined as
\begin{eqnarray}
\left| \frac{q}{p} \right| \neq 1.
\label{eq:Part5_define_cp_mixing}
\end{eqnarray}
In charged-current semileptonic decays of neutral 
$D$, $|A_{l^+X}| = |\overline{A}_{l^-X}|$ and 
$A_{l^-X} = \overline{A}_{l^+X} = 0$ in the
SM and most of the reasonable extensions of SM, where
$|p|/|q|\neq 1$ would be the only source of
$CP$ violation. This can be measured in the asymmetry 
of ``wrong-sign" decays
induced by oscillations.
\item $CP$-violation in the interference between a decay without mixing,
$D^0 \rightarrow f$, and a decay with mixing, $D^0 \rightarrow
\overline{D}^0 \rightarrow f$, and is defined by
\begin{eqnarray}
{\cal I}m \left( \frac{q}{p} \frac{\overline{A}_{f}}{A_f}
\right) \neq  0.
\label{eq:Part5_define_cp_mixing_decay}
\end{eqnarray}
This form of $CP$ violation can be observed in the asymmetry of $D^0$ and
$\overline{D}^0$ decays into common final states, such as $CP$ eigenstates
$f_{CP}$.
\end{itemize}
\noindent
Example of these three 
different types of $CP$ violation are given in the 
following sections.

There are several ways to study $CP$ violation in
charm decays~\cite{Part5_patrov_2005}:  we can look for direct $CP$
violation, even in charged decays; we can look for $CP$ violation via
mixing; $T$ violation can be examined in 4-body $D$ meson decays, assuming
$CPT$ conservation, by measuring triple-product
correlations~\cite{Part5_Bigi_KAON2001};  the quantum coherence
present in correlated $D^0\overline{D}^0$ decays of the $\psi(3770)$ can be
exploited~\cite{Part5_Cinabro_2006}. 

Most existing $CP$ limits are for direct $CP$ violation. A few results 
from CLEO,
BaBar and Belle experiments consider $CP$ violation in mixing. 
Tables~\ref{tab:Part5_CP_neutral_D} and \ref{tab:Part5_CP_charge_D} are
summaries of measurements of $CP$ violations in neutral $D$ and 
charged
$D$ decays, respectively. No evidence for $CP$ violation is observed and all
results are consistent with the SM expectations. 

Large samples of $D$ mesons will be available at the
\bes3 experiment. One year's running  at \bes3 will provide an
intrinsic statistical precision of $<1.0\%$. For this purpose, one has to
pay great attention to systematic biases. Initial-state asymmetries and
detector asymmetries will be the main concerns. 

Finally, in Table~\ref{tab:Part5_CP_D_life}, we list results
of $CP$ violation measurements done by looking at 
$D^0$ lifetime asymmetries for different decay modes.

\begin{table}[htbp]
  \centering
  \begin{tabular}{c|c|c|c} \hline
Year & Experiment & Decay Mode & $A_{CP}$(\%)  \\ \hline
2007 & BABAR~\cite{part5:cp:babar:kk} & $D^0 \rightarrow K^+K^-$ & $+0.00 \pm 0.34 \pm 0.13$ \\ 
2005 & CDF~\cite{Part5_cp_cdf1} & $D^0 \rightarrow K^+K^-$ & $+2.0 \pm 1.2 \pm 0.6$  \\ 
2002 & CLEO~\cite{part5:cp:cleo2002:kk} &$D^0 \rightarrow K^+K^-$ & $+0.0 \pm 2.2 \pm 0.8$  \\ 
2000 & FOCUS~\cite{part5:cp:focus2000:kk} & $D^0 \rightarrow K^+K^-$ & $-0.1 \pm 2.2 \pm 1.5$ \\ 
1998 & E791~\cite{part5:cp:e791:1998:kk} & $D^0 \rightarrow K^+K^-$ & $-1.0 \pm 4.9 \pm 1.2$ \\ 
1995 & CLEO~\cite{part5:cp:cleo:1995:kk} & $D^0 \rightarrow K^+K^-$ & $+8.0 \pm 6.1$ \\ 
1994 & E687~\cite{part5:cp:e687:1994:kk} & $D^0 \rightarrow K^+K^-$ & $+2.4 \pm 8.4$ \\ 
     & Average                           &  $D^0 \rightarrow K^+K^-$ & $+0.15 \pm 0.34$ \\ \hline \hline
2007 & BABAR~\cite{part5:cp:babar:kk} & $D^0 \rightarrow \pi^+\pi^-$ & $-0.24 \pm 0.52 \pm 0.22$  \\
2005 &  CDF~\cite{Part5_cp_cdf1} & $D^0 \rightarrow \pi^+\pi^-$ & $1.0 \pm 1.3 \pm 0.6$ \\
2002 & CLEO~\cite{part5:cp:cleo:2002:pipi} &  $D^0 \rightarrow \pi^+\pi^-$ & $+1.9 \pm 3.2 \pm 0.8$ \\ 
2000 & FOCUS~\cite{part5:cp:focus2000:kk} &  $D^0 \rightarrow \pi^+\pi^-$ & $+4.8 \pm 3.9 \pm 2.5$\\
1998 & E791~~\cite{part5:cp:e791:1998:kk} &  $D^0 \rightarrow \pi^+\pi^-$ & $-4.9 \pm 7.8 \pm 3.0$ \\
     & Average &  $D^0 \rightarrow \pi^+\pi^-$ & $+0.02 \pm 0.51$ \\ \hline \hline 
2001 & CLEO~\cite{part5:cp:cleo:2001:ksks} &  $D^0 \rightarrow K_S K_S$ & $-23 \pm 19$ \\ \hline\hline
2005 & CLEO~\cite{Part5_cp_cleo} & $D^0 \rightarrow \pi^+ \pi^- \pi^0$ & $1^{+9}_{-7} \pm 8$ \\ \hline \hline
2001 & CLEO~\cite{part5:cp:cleo:2001:ksks} &  $D^0 \rightarrow K_S \pi^0$ & $+0.1 \pm 1.3$ \\ \hline \hline  
2007 & CLEO-c~\cite{part5:cp:cleoc:2007:kpp0} &  $D^0 \rightarrow K^- \pi^+ \pi^0$ & $+0.2 \pm 0.4 \pm 0.8 $ \\ 
2001 & CLEO~\cite{Part5_cp_cleo1} & $D^0 \rightarrow K^- \pi^+ \pi^0$ & $-3.1 \pm 8.6$  \\ 
    & Average &  $D^0 \rightarrow K^- \pi^+ \pi^0$ &  $+0.16 \pm 0.89$ \\ \hline \hline
2005 & Belle~\cite{Part5_Belle_CP} &  $D^0 \rightarrow K^+ \pi^- \pi^0$ &  
$-0.6 \pm 5.3$ \\ 
2001 & CLEO~\cite{part5:cp:cleo:2001:kppmp0} &  $D^0 \rightarrow K^+ \pi^- \pi^0$ & $+9.0^{+25}_{-22}$ \\ 
     &  Average &  $D^0 \rightarrow K^+ \pi^- \pi^0$ & $-0.14 \pm 5.17$ \\ \hline \hline
2004 & CLEO~\cite{Part5_cp_cleo2} & $D^0 \rightarrow K_S \pi^+\pi^-$ & $-0.9 \pm 2.1^{+1.0+1.3}_{-4.3-3.7}$   \\ \hline \hline
2005 & Belle~\cite{Part5_Belle_CP} & $D^0 \rightarrow K^+ \pi^- 
\pi^+\pi^-$ & $-1.8 \pm 4.4$  \\ \hline \hline
2005 & FOCUS~\cite{Part5_cp_focus1} & $D^0 \rightarrow K^+ K^- \pi^+\pi^- $ & $-8.2 \pm 5.6 \pm 4.7$ \\ \hline \hline 
  \end{tabular}
  \caption{Measurements of $CP$ violating asymmetries in neutral 
$D$ decays in different modes. The averaged results are from HFAG. } 
 \label{tab:Part5_CP_neutral_D}
\end{table}

\begin{table}[htbp]
  \centering
  \begin{tabular}{c|c|c|c} \hline
Year & Experiment & Decay Mode & $A_{CP}$(\%)  \\ \hline
2007 & CLEO-c~\cite{part5:cp:cleoc:2007:kpp0} & $D^+ \rightarrow K_S \pi^+$ & $-0.6 \pm 1.0 \pm 0.3 $ \\  
2002 & FOCUS~\cite{part5:cp:focus:2002:kspip} &  $D^+ \rightarrow K_S \pi^+$ & $-1.6\pm 1.5 \pm 0.9$ \\ 
     & Average & $D^+ \rightarrow K_S \pi^+$ & $-0.86 \pm 0.90$ \\ \hline \hline 
2002 & FOCUS~\cite{part5:cp:focus:2002:kspip} & $D^+ \rightarrow K_S K^+$  & $+7.1 \pm 6.1 \pm 1.2$ \\ \hline \hline  
1997 & E791~\cite{part5:cp:e791:3pi} &  $D^+ \rightarrow \pi^+\pi^-\pi^+$ & $-1.7 \pm 4.2$ \\ \hline \hline  
2007 & CLEO-c~\cite{part5:cp:cleoc:2007:kpp0} & $D^+ \rightarrow K^- \pi^+\pi^+$ & $-0.5 \pm 0.4\pm 0.9$  \\ \hline \hline  
2007 &  CLEO-c~\cite{part5:cp:cleoc:2007:kpp0} & $D^+ \rightarrow K_S \pi^+\pi^0$ & $+0.3 \pm 0.9 \pm 0.3$ \\ \hline \hline 
2007 &  CLEO-c~\cite{part5:cp:cleoc:2007:kpp0} & $D^+ \rightarrow K^+K^- \pi^+$ & $-0.1\pm 1.5 \pm 0.8$ \\
2005 & BaBar~\cite{Part5_cp_babar} & $D^+ \rightarrow K^+K^- \pi^+$  & 
$1.4 \pm 1.0 \pm 0.8$ \\
2000 & FOCUS~\cite{part5:cp:focus2000:kk} & $D^+ \rightarrow K^+K^- \pi^+$ & $+0.6 \pm 1.1 \pm 0.5$ \\ 
1997 & E791~\cite{part5:cp:e791:3pi} & $D^+ \rightarrow K^+K^- \pi^+$ & $-1.4 \pm 2.9$ \\ 
1994 & E687~\cite{part5:cp:e687:1994:kk} & $D^0 \rightarrow  K^+K^- \pi^+$ & $-3.1 \pm 6.8$ \\ 
     & Average &  $D^0 \rightarrow  K^+K^- \pi^+$  $+0.59 \pm 0.75$ \\ \hline \hline  
2007 & CLEO-c~\cite{part5:cp:cleoc:2007:kpp0} & $D^+ \rightarrow K^- \pi^+\pi^-\pi^0$ & $+1.0 \pm 0.9 \pm 0.9$ \\ \hline \hline 
2007 & CLEO-c~\cite{part5:cp:cleoc:2007:kpp0} & $D^+ \rightarrow K_S \pi^+\pi^+ \pi^-$ & $+0.1 \pm 1.1 \pm 0.6$  \\ \hline \hline
2005 &  FOCUS~\cite{Part5_cp_focus1} & $D^0 \rightarrow K_S K^+\pi^+ \pi^-$ & $-4.2 \pm 6.4 \pm 2.2$ \\ \hline \hline 
2005 & BaBar~\cite{Part5_cp_babar} & $D^+ \rightarrow \phi \pi^+$ & $0.2 
\pm 1.5 \pm 0.6$ \\ \hline \hline 
2005 & BaBar~\cite{Part5_cp_babar} & $D^+ \rightarrow \overline{K}^{*0} 
K^+$ & $0.9 \pm 1.7 \pm 0.7$\\ \hline \hline
  \end{tabular}
  \caption{Measurements of $CP$ violating asymmetries in charged $D$
decays in different modes. The averaged results are from HFAG. } 
 \label{tab:Part5_CP_charge_D}
\end{table}

\begin{table}[htbp]
  \centering
  \begin{tabular}{c|c|c|c} \hline
Year & Experiment & Decay Mode & $(\tau_{\bar{D}^0} -\tau_{D^0})/(\tau_{\bar{D}^0} + \tau_{D^0})$(\%)  \\ \hline
2007 & Belle~\cite{part5:cp:belle:life:2007} & $D^0 \rightarrow K^+K^-$ 
and $\pi^+\pi^-$ & $+.010 \pm 0.300 \pm 0.150$ \\
2007 & Babar~\cite{part5:cp:babar:life:2007} & $D^0 \rightarrow K^+K^-$ 
and
$\pi^+\pi^-$ & $+0.260 \pm 0.360 \pm 0.080$  \\
   & Average & $D^0 \rightarrow K^+K^-$ and $\pi^+\pi^-$ &$ +0.123 \pm 0.248$ \\ \hline    
  \end{tabular}
  \caption{$CP$ lifetime asymmetries in $D^0$ decay modes.} 
 \label{tab:Part5_CP_D_life}
\end{table}

\section{$CP$ Violation in $\Dz - \Dzb$ Mixing}

A mixing $CP$ asymmetry can be best isolated in semileptonic 
decays of neutral $D$ mesons, as discussed in Ref.~\cite{Part5_Buros_2005}. 
In the case of the $D^0$ meson, this can be measured as an  asymmetry 
of  ``wrong-sign" decays ($A_{SL}$):
\begin{eqnarray}
A_{SL} & \equiv &  \frac{\Gamma(\overline{D}^0(t)_{\mbox{phys}} \rightarrow l^+ X) -
\Gamma(D^0(t)_{\mbox{phys}} \rightarrow l^-
X)}{\Gamma(\overline{D}^0(t)_{\mbox{phys}}
\rightarrow l^+ X) + \Gamma(D^0(t)_{\mbox{phys}} \rightarrow l^- X)} \nonumber \\
&=& \frac{1-|q/p|^4}{1+|q/p|^4}.
\label{eq:Part5_define_cp_mixing_semi}
\end{eqnarray}
Here $\overline{D}^0(0)_{\mbox{phys}} =\overline{D}^0$ and
$D^0(0)_{\mbox{phys}}
= D^0$. Note that all the final states in 
Eq.~\ref{eq:Part5_define_cp_mixing_semi} contain ``wrong-sign"
leptons and can only be reached via $D^0 -
\overline{D}^0$ oscillations.  This asymmetry represents the
difference between the rates for $\overline{D}^0 \rightarrow D^0
\rightarrow l^+X$ and $D^0 \rightarrow \overline{D}^0 \rightarrow
l^- X$. If the phases in the $\overline{D}^0
\rightarrow D^0$ and $D^0 \rightarrow \overline{D}^0$ 
transition amplitudes differ from each
other, a non-vanishing $CP$ violation follows. This asymmetry is
expected to be tiny, both in the SM and many of its extensions.
The corresponding observable has been studied in semileptonic decays of
neutral $K$ and $B$ mesons.
Since $A_{SL}$ is controlled by $(\Delta \Gamma/\Delta M){\rm
sin}\phi_{weak}$,
it is predicted to be small in both cases, albeit for different reasons:    
(i) while $(\Delta \Gamma_K/\Delta M_K) \sim 1$, one has sin$\phi_{weak}^K
\ll 1$ leading to
$A_{SL}^K = \delta _l  \simeq (3.32 \pm 0.06)\cdot 10^{-3}$ as observed;
(ii) for $B^0$ mesons, one has
$(\Delta \Gamma_B/\Delta M_B)\ll 1$ leading to $A_{SL}^B < 10^{-3}$.

For $D^0$, on the other hand, both $\Delta M_D$ and $\Delta \Gamma_D$ are
small, but
$\Delta \Gamma_D/\Delta M_D$ is not: present data indicate it is about unity
or even larger.
$A_{SL}$ is given by the smaller of $\Delta \Gamma_D/\Delta M_D$ or its
inverse multiplied by
sin$\phi_{weak}^D$, which might not be that small: {\it i.e.}, 
while the rate for `wrong-sign'
leptons is certainly small in semileptonic decays of neutral
$D$ mesons, their $CP$ asymmetry might not be if New Physics intervenes to
induce a non-zero value for $\phi_{weak}^D$.  

At the $\psi(3770)$ peak, this kind of $CP$-violating signal can
manifest itself in like-sign dilepton events of
$(D^0\overline{D}^0)$ pairs:
\begin{eqnarray}
A_{SL} \equiv \frac{R(l_1^+X, l_2^+ X) - R(l_1^-X, l_2^- X) }{R(l_1^+X,
l_2^+ X) + R(l_1^-X, l_2^- X)}  =  \frac{1-|q/p|^4}{1+|q/p|^4},
\label{eq:Part5_define_cp_mixing_semi_3770}
\end{eqnarray}
where $R(l_1^+X, l_2^+ X)$ and $ R(l_1^-X, l_2^- X)$ are the production
rates for the like-sign dileptons ot the $\psi(3770)$, as defined in
Ref.~\cite{Part5_Li_Yang_2006}. Note that this asymmetry is not only
independent of the time distributions, but also independent of the
charge-conjugation parity $C$ of the $(D^0\overline{D}^0)$ pair. 
Thus, it can
be measured using time-integrated dilepton events at either
the $\psi(3770)$ or $\psi(4170)$ resonances.

\section{$CP$ Violation in Decay}

Decay $CP$ violation (also called ``direct $CP$ violation") occurs
when the absolute value of the decay amplitude $A_f$ for $D$ decaying to a
final state $f$ is different from the one for the corresponding 
$CP$-conjugated
amplitude, i.e. $|A_f| \neq |\overline{A}_{\overline{f}}|$. This kind of $CP$
violation would be induced by $\Delta C =1$ effective operators, and 
could produce asymmetries in both charged and neutral $D$ decays.
For charged $D$ meson decays, where
mixing effects are absent, this is the only possible observable $CP$
asymmetry;
\begin{eqnarray}
A^{CP}_{f^\pm} \equiv \frac{\Gamma(D^- \rightarrow f^-) - \Gamma(D^+ \rightarrow
f^+)}{\Gamma(D^- \rightarrow f^-) - \Gamma(D^+ \rightarrow
f^+)} = \frac{|\overline{A}_{f^-}/A_{f^+}|^2 -1
}{|\overline{A}_{f^-}/A_{f^+}|^2 +1 },
\label{eq:Part5_define_cp_decay}
\end{eqnarray}
where $\Gamma_{f^\pm}$ represents the $D^{\pm} \rightarrow f^\pm$ decay
rate. A two-component decay amplitude with weak and strong phase differences
is required for this type of $CP$ violation. If, for example, there are two
such contributions, $A_f = a_1 + a_2$, we have
\begin{eqnarray}
A_f = |a_1|e^{+i\phi_T} [1+ r_f e^{i(\Delta_f + \theta_f)}], \nonumber \\
\overline{A}_{\overline{f}} = |a_1| e^{-i\phi_T}[ 1+
r_f e^{i(\Delta_f - \theta_f)}],
\label{eq:Part5_define_cp_decay_amp}
\end{eqnarray}
where $\Delta_f$ corresponds to the strong phase difference and
$\theta_f$ corresponds to the weak phase difference between the
$CP$-conserving ($a_1$ from tree level contribution in the SM) and
$CP$-violating parts of the decay amplitude and $r_f$ represents
the small ratio, $r_f = |a_2|/|a_1|$;  $\phi_T$ is the weak phase
from the SM tree-level contribution.

It is straightforward to evaluate the $CP$ asymmetry for charged $D$
decays:
\begin{eqnarray}
A^{CP}_{f^\pm} = - \frac{2|r_f|\mbox{sin}(\Delta_f)\mbox{sin}(\theta_f)}{1+|r_f|^2 + 2|r_f|
\mbox{cos}(\Delta_f)\mbox{cos}(\theta_f)}.
\label{eq:Part5_define_cp_decay_amp_1}
\end{eqnarray}
No reliable model-independent predictions exist for
$\Delta_f$; it is believed that it could be quite large due to the
abundance of light-quark resonances in the vicinity of the
$D$-meson mass that can induce large final-state interaction (FSI)
phases. The quantity of most interest to theory is the weak phase
difference $\theta_f$. Its extraction from the asymmetry requires,
however, that the amplitude ratio $r_f$ and the strong phase
difference $\Delta_f$ are known. Both quantities are difficult to
calculate due to non-perturbative hadronic parameters.  In the SM,
relative weak phases can  obtain in SCS decays via, for
instance, interference between spectator and penguin
amplitudes. Since the most optimistic model-dependent estimates put
the SM predictions for the asymmetry $A^{CP} <
0.1\%$~\cite{Part5_Buccella_1996}, an observation of any
$CP$-violating signal in the current round of experiments would be a
sign of new physics.

Specific model calculations~\cite{Part5_Buccella_1993} for $D \rightarrow K\overline{K}$,
$\pi\pi$, $K^* \overline{K}$, three-body modes, etc. yield this 
order-of-magnitude effect. New physics could enter, for example, via
large phases in the penguin diagram. These could produce asymmetries of 
the order of $1\%$ or larger. On the other hand, CF decays do not have two 
amplitudes with different weak phases and, therefore, their $CP$ 
asymmetries are zero in the SM. Some new physics scenarios may provide 
extra phases that could give asymmetries as large as 1\%.

\section{$CP$ Violation in the interference between decays with and without
mixing}
\label{sec:cp_mixing_decay}

This type of $CP$ violation is possible for common final states to which
both $D^0$ and $\overline{D}^0$ can decay. It is usually associated with the
relative phase between mixing and decay contributions as described in
Eq.~\ref{eq:Part5_define_cp_mixing_decay}. It can be studied with both
time-dependent and time-integrated asymmetries. \\

{\centering \bf (1) $CP$ eigenstate} \\
In general, for a $CP$ even (odd) eigenstate, the decay amplitudes can be
written as
\begin{eqnarray}
A_f&=& |a_1|e^{+i\phi_T} [1+ r_f e^{i(\Delta_f + \theta_f)}], \nonumber \\
\overline{A}_{f} &=& \eta^{CP}_f \times \overline{A}_{\overline{f}}
 = \eta^{CP}_f |a_1| e^{-i\phi_T}[ 1+
r_f e^{i(\Delta_f - \theta_f)}],
\label{eq:Part5_define_cp_decay_amp_2}
\end{eqnarray}
where $\eta^{CP}_f = + (-)$ for a $CP$ even (odd) state and we have used
$CP|D^0\rangle = - |\overline{D}^0 \rangle$. Neglecting $r_f$ in
Eq.~\ref{eq:Part5_define_cp_decay_amp_2}, $\lambda_f$ can be written as
\begin{eqnarray}
\lambda_f \equiv - \eta^{CP}_{f} R_{m}  e^{i\phi},  \nonumber \\
\overline{\lambda}_{\overline{f}} \equiv - \eta^{CP}_f R^{-1}_m e^{-i\phi},
\label{eq:Part5_define_2}
\end{eqnarray}
where $\phi$ is the relative weak phase between the mixing
amplitude and the decay amplitude, and $R_m = |q/p|$.  The time-integrated $CP$
asymmetry for a final $CP$ eigenstate $f$ is defined as:
\begin{eqnarray}
 A^{CP}_f \equiv \frac{\Gamma(D^0 \rightarrow f) - \Gamma(\overline{D}^0 \rightarrow
f)}{\Gamma(D^0 \rightarrow f) + \Gamma(\overline{D}^0 \rightarrow
f)}.
\label{eq:Part5_define_CP_mixing_decay_inter}
\end{eqnarray}
Given experimental constraints, one can take $x$, $y$, $r_f \ll 1$ and expand to
the leading order in these parameters, and get~\cite{Part5_Grossman_2006}:
\begin{eqnarray}
 A^{CP}_f = 2 r_f \mbox{sin}\Delta_f \mbox{sin}\theta_f - \eta^{CP}_f
\frac{y}{2} (R_m -R^{-1}_m) \mbox{cos} \phi + \eta^{CP}_f \frac{x}{2}(R_m  +
R^{-1}_m)\mbox{sin}\phi,
\label{eq:Part5_define_CP_mixing_decay_inter_1}
\end{eqnarray}
where the first term represents $CP$ violation in the decay, the second term
is related to $CP$ violation in mixing, and the third term is for $CP$
violation in the  interference between mixing and decay amplitudes.
If $CP$ violation in mixing and decay is neglected, one has:
\begin{eqnarray}
 A^{CP}_f \approx  \eta^{CP}_f x \mbox{sin}\phi.
\label{eq:Part5_define_CP_mixing_decay_inter_2}
\end{eqnarray}
The above discussion is only for incoherent $D^0 \overline{D}^0$ decays. In
the case of $D^0 \overline{D}^0$ produced coherently at \bes3,
the $D^0 \overline{D}^0$ pair system is in
a state with charge parity $C= \eta$,
which can be defined
as~\cite{Part5_Du_2006,Part5_Li_Yang_2006}
\begin{equation}
| D^0 \overline{D}^0 \rangle^{C=\eta} = \frac{1}{\sqrt{2}} \left [ |D^0\rangle
|\overline{D}^0\rangle + \eta |\overline{D}^0\rangle |D^0 \rangle\right ],
\label{eq:Part5_d0d0}
\end{equation}
where $\eta$ is the charge conjugation
parity or orbital angular momentum of the  $D^0 \overline{D}^0$ pair. Thus,
it is easy to see that $D^0 \overline{D}^0$ occur in a $P$-wave ($L=1$) 
in the reactions
\begin{eqnarray}
& & e^+ e^- \rightarrow \gamma^* \rightarrow D^0 \overline{D}^0, \nonumber \\
& & e^+ e^- \rightarrow \gamma^* \rightarrow D^0 \overline{D}^{*0},\, D^{*0}
\overline{D}^0 \rightarrow  D^0 \overline{D}^0 \pi^0, \nonumber \\
&& e^+ e^- \rightarrow \gamma^* \rightarrow D^{*0} \overline{D}^{*0} \rightarrow D^0
\overline{D}^0\pi^0 \pi^0,
\label{eq:Part5_d0d0_reaction_odd}
\end{eqnarray}
and in an $S$-wave ($L=0$) in the reactions
\begin{eqnarray}
& & e^+ e^- \rightarrow \gamma^* \rightarrow D^0 \overline{D}^{*0},\, D^{*0}
\overline{D}^0 \rightarrow  D^0 \overline{D}^0 \gamma, \nonumber  \\
&& e^+ e^- \rightarrow \gamma^* \rightarrow D^{*0} \overline{D}^{*0}
\rightarrow D^0
\overline{D}^0 \gamma \pi^0.
\label{eq:Part5_d0d0_reaction_even}
\end{eqnarray}
One can use the semileptonic decay of one $D$ meson to tag the other $D$
decaying to a $CP$ eigenstate $f$. We define the leptonic-tagged $CP$
asymmetry $A^{CP}_{fl^\pm}$ as
\begin{equation}
A^{CP}_{fl^\pm} = \frac{R(l^-X, f) -R(l^+ X, f)}{R(l^-X, f) + R(l^+ X, f)} =
\frac{N(l^-X, f) -N(l^+ X, f)}{N(l^-X, f) + N(l^+ X, f)},
\label{eq:Part5_d0d0_asymmetry}
\end{equation}
where $R(l^-X, f)$ and $R(l^+ X, f)$ are the time integrated decay rates of $| D^0
\overline{D}^0 \rangle^{C=\eta}$ into $(l^-X, f)$ and $(l^+ X, f)$ final
states, respectively, and are defined as~\cite{Part5_Du_2006}:
\begin{eqnarray}
R(l^- X, f) = \int^{\infty}_0 dt_1 dt_2 | \langle (l^-X, f)| {\cal H} |  D^0
\overline{D}^0 \rangle^{C=\eta}|^2,
\label{eq:Part5_d0d0_rate_1}
\end{eqnarray}
\begin{eqnarray}
R(l^+ X, f) = \int^{\infty}_0 dt_1 dt_2 | \langle (l^+X, f)| {\cal H} |  D^0
\overline{D}^0 \rangle^{C=\eta}|^2,
\label{eq:Part5_d0d0_rate_2}
\end{eqnarray}
and are proportional to the numbers of 
lepton-tagged $D^0 (\overline{D}^0)\to f$ events, 
$N(l^-X, f)$ and  $N(l^+X, f)$.
After a complicated calculation, Du
finds~\cite{Part5_Du_2006}:
\begin{equation}
A^{CP}_{fl^\pm} = (1+ \eta) \eta^{CP}_f \left[ -\frac{y}{2} (R_m -R^{-1}_m)
\mbox{cos}\phi + \frac{x}{2}(R_m +R^{-1}_m) \mbox{sin} \phi \right ].
\label{eq:Part5_d0d0_asymmetry_final}
\end{equation}
Comparing Eqs. \ref{eq:Part5_d0d0_asymmetry_final} and
\ref{eq:Part5_define_CP_mixing_decay_inter_1}, and neglecting $CP$
violation in the decay (namely, the first term in
Eq.~\ref{eq:Part5_define_CP_mixing_decay_inter_1} is zero), Du finds that
$A^{CP}_{fl^\pm}$  is just twice as large as  $A^{CP}_f$ when charge
conjugation parity (or the orbital angular momentum $L$) is even.
In Table~\ref{tab:Part5_CP_eigenstate}, we present an estimate 
of the number of all lepton-tagged events of the type 
$K (\pi) e\nu$  and $K(\pi) \mu \nu$ --- where the neutrino 
is the only missing particle --- with $CP$-eigenstates that can be
collected in a  year of running at $\sqrt{s}$ = 4.17
GeV with luminosity of ${\cal L} = 10^{33}$ cm$^{-2}$ sec$^{-1}$. The chain
of the reactions considered is $ e^+ e^- \rightarrow \gamma^* \rightarrow
 D^0 \overline{D}^{*0},\, D^{*0}
\overline{D}^0 \rightarrow  D^0 \overline{D}^0 \gamma$ and $(D^0
\overline{D}^0) \rightarrow
(l^+X, f)$. An estimated production cross-section of
$\sigma( e^+ e^- \rightarrow D^0 \overline{D}^{*0}) = 2.6 $~nb
is used and the branching ratio
for the decay $D^{*0} \rightarrow D^0 \gamma$ is taken to be 38\%. 
The $CP$-eigenstate branching ratios are taken from 
PDG2006~\cite{Part5_pdg06}.  The efficiencies are estimated
based on solid angle and PID criteria. All the numbers are normalized via the
branching ratios and efficiencies for the above decay chain.
As the results in Table~\ref{tab:Part5_CP_eigenstate} indicate, there will
be a total of about 10K events and 
the observed $CP$ asymmetry, $A^{CP}_{fl^\pm}$, could
be measured with an accuracy of 1.2\%.
 \begin{table}[htbp]
  \centering
  \begin{tabular}{c|c|c|c|c} \hline
 Eigenstate & $\eta^{CP}_f$ & Branching Ratio(\%) & Efficiency  & $D^0
\overline{D}^0 \gamma$ Events\\ \hline
$K^+K^-$ & +1 & ($0.38\pm 0.01$) & 0.50 &  1040 \\ \hline
$\pi^+\pi^-$ & +1 & ($0.14\pm 0.003$) & 0.80 &  460 \\ \hline
$K_S K_S$ & +1 & ($0.037\pm 0.001$) & 0.26 &  30 \\ \hline
$\rho^0 \pi^0$ & +1 & ($0.32 \pm 0.04$) & 0.70 & 3140 \\ \hline
$\pi^0 \pi^0$ & +1 & ($0.079 \pm 0.008$) & 0.50 & 500 \\ \hline
$K_S \pi^0$ & -1 & ($1.14 \pm 0.12$) & 0.26 & 770 \\ \hline
$K_S \eta$ & -1 & ($0.38 \pm 0.06$) & 0.12 & 290 \\ \hline
$K_S \rho^0$ & -1 & ($0.75 \pm 0.07$) & 0.42 & 460 \\ \hline
$K_S \phi$ & -1 & ($0.42\pm 0.032$) & 0.05 & 60 \\ \hline
$K_S \omega$ & -1 & ($1.10\pm 0.20$) & 0.06 &320 \\ \hline
   \end{tabular}
  \caption{Estimates of the numbers of fully reconstructed 
lepton-tagged $CP$-eigenstate decays in a one year at  \bes3.}
 \label{tab:Part5_CP_eigenstate}
\end{table}
\\

{\centering \bf (2) Non-$CP$ eigenstate} \\

Another promising channel for probing $CP$ symmetry is $D^0(t) \to 
K^+\pi^-$.  Since this mode is doubly Cabibbo suppressed, it should 
{\it a priori} exhibit a higher sensitivity to a New Physics amplitude.  
Furthermore it cannot exhibit a direct $CP$ violation
in the SM. 
\begin{equation}
\frac{\Gamma (D^0(t) \to K^+\pi^-)}{\Gamma (D^0(t) \to K^-\pi^+)} =
\left| \frac{T(D^0 \to K^+\pi^-)}{T(D^0 \to K^-\pi^+)}\right| ^2 \times
\label{part5:eq:DKPI}
\end{equation}
\begin{equation}
\nonumber
\left[ 1 + \left( \frac{t}{\tau_D}\right)^2\left(\frac{x^2 + y^2}{4{\rm
tg}\theta_C^4}\right)
\left|\frac{q}{p}\right|^2
\left| \hat \rho_{K\pi}\right|^2 + \left(\frac{t}{\tau_D}\right)
\left|\frac{q}{p}\right| \left| \hat \rho_{K\pi}\right|
\left( \frac{y^{\prime}{\rm cos}\phi_{K\pi} +
x^{\prime}{\rm sin}\phi_{K\pi}}{{\rm tg}\theta_C^2}\right),
\right]
\end{equation}
\begin{equation}
\frac{\Gamma (\bar D^0(t) \to K^-\pi^+)}{\Gamma (\bar D^0(t) \to K^+\pi^-)}
=
\left| \frac{T(\bar D^0 \to K^-\pi^+)}{T(\bar D^0 \to K^+\pi^-)}\right| ^2
\times
\label{BARDKPI}
\end{equation}
\begin{equation}
\nonumber
\left[ 1 + \left( \frac{t}{\tau_D}\right)^2\left(\frac{x^2 + y^2}{4{\rm
tg}\theta_C^4}\right) \left|\frac{p}{q}\right|^2
\left| \hat \rho_{K\pi}\right|^2 + \left(\frac{t}{\tau_D}\right)
\left|\frac{p}{q}\right|  \left| \hat \rho_{K\pi}\right|
\left( \frac{y^{\prime}{\rm cos}\phi_{K\pi} -
x^{\prime}{\rm sin}\phi_{K\pi}}{{\rm tg}\theta_C^2}\right),
\right]
\end{equation}
with
\begin{eqnarray}
\nonumber
\frac{q}{p} \frac{T(D^0 \to K^+\pi^-)}{T(D^0 \to K^-\pi^+)} &\equiv &
 - \frac{1}{{\rm tg}^2\theta_C} (1+ A_M) |\hat \rho
_{K\pi}|e^{-i(\delta - \phi_{K\pi})}, \\
 \frac{q}{p} \frac{T(\bar D^0 \to K^-\pi^+)}{T(\bar D^0 \to K^+\pi^-)}
&\equiv &
 - \frac{1}{{\rm tg}^2\theta_C} \frac{1}{1+A_M} |\hat \rho
_{K\pi}|e^{-i(\delta + \phi_{K\pi})},
\end{eqnarray}
yielding an asymmetry
$$
\frac{\Gamma (D^0(t) \to K^+\pi^-) - \Gamma (\bar{D}^0(t) \to K^-\pi^+)}
{\Gamma (D^0(t) \to K^+\pi^-) + \Gamma (\bar{D}^0(t) \to K^-\pi^+)} \simeq
$$
\begin{equation}
 \left(\frac{t}{\tau_D}\right) \left| \hat \rho_{K\pi}\right|
 \left( \frac{y^{\prime}{\rm cos}\phi_{K\pi}A_M +
x^{\prime}{\rm sin}\phi_{K\pi}}{{\rm tg}\theta_C^2}\right) -
 \left(\frac{t}{\tau_D}\right)^2 \left| \hat \rho_{K\pi}\right|^2
\frac{A_M(x^2 + y^2)}
 {2{\rm tg}\theta_C^4} .
\end{equation}
 Here we have again assumed for simplicity $|A_M| \ll 1$ ($|q|/|p| =
1+A_M$),  $ \rho_{K\pi} = \frac{A(D^0 \to K^+\pi^-)}{\bar{D}^0(t) \to
K^-\pi^+)}$ and {\em no direct} $CP$ violation.

BaBar has searched for a time dependent $CP$ asymmetry in $D^0 \to
K^+\pi^-$ vs.
$\bar D^0(t) \to K^- \pi^+$, but has not found any evidence for it
at about the 1\% level~\cite{part5_new_BABAROSC}. 
However, with $x^{\prime}$ and $y^{\prime}$
capped at
about 1\%, no nontrivial bound can be placed on the weak phase $\phi_{K\pi}$
that could be induced by New Physics. 
On the other hand, a further increase in
experimental sensitivity might reveal a signal.

\section{Rate of the $CP$ Violation in the Coherent $D^0 \overline{D}^0$
Pair}

Let us consider the reaction:
\begin{eqnarray}
 e^+ e^- \rightarrow \psi(3770) \rightarrow D^0 \overline{D}^0 \rightarrow
f_a f_b,
\label{eq:Part5_d0d0_reaction_peak}
\end{eqnarray}
where $f_a$ and $f_b$ represent $CP$ eigenstates with the same $CP$ 
parity,
i.e.
\begin{eqnarray}
&& CP| f_a \rangle = \eta_a |f_a \rangle, \nonumber \\
&& CP| f_b \rangle = \eta_b |f_b \rangle, \nonumber \\
&&  \eta_a \eta_b = +1 .
\label{eq:Part5_cp_value_d0d0_peak}
\end{eqnarray}
The process in Eq.~\ref{eq:Part5_d0d0_reaction_peak} can proceed only in
the presence of $CP$ violation, because:
\begin{eqnarray}
CP| \psi(3770) \rangle = + | \psi(3770) \rangle,
\label{eq:Part5_cp_value_d0d0}
\end{eqnarray}
whereas
\begin{eqnarray}
CP| f_a f_b \rangle = \eta_a \eta_b (-1)^{l=1} | f_a f_b \rangle = - | f_a
f_b \rangle.
\label{eq:Part5_cp_value_d0d0_2}
\end{eqnarray}
Thus $CP(\mbox{initial}) \neq CP(\mbox{final})$, and $CP$ invariance is broken.
More explicitly, for $x \ll 1$ one has:
$$ 
{\rm BR}(\psi (3770) \to D^0\bar D^0 \to f_a f_b \simeq
{\rm BR}(D \to f_a) {\rm BR}(D \to f_b) \cdot 
$$
\begin{equation} 
\left[ (2+x^2)\left| \frac{q}{p}  \right|^2\left|\bar \rho(f_a) -
\bar \rho(f_b) \right|^2
+ x^2 \left| 1 - \frac{q}{p}\bar \rho(f_a)\frac{q}{p}\bar
\rho(f_b)  \right|^2\right],
\label{CPEX}
\end{equation}
where $\bar \rho(f)$ is defined in Eq.~\ref{eq:mix_lambda_f} of
Sect.~\ref{subsec:incoherent}.
The second contribution in the square brackets can occur only via
oscillations, but includes $f_a = f_b$; moreover it is heavily suppressed
by $x^2 \leq 10^{-4}$
making it practically unobservable. The first term arises even with $x =
0$, yet
requires $f_a \neq f_b$. It is possible that the 
$\left|\bar \rho(f_a) -  \bar \rho(f_b) \right|^2$ term provides
a larger signal for
$CP$ violation than that from either 
the $\left|1-  |\rho(f_a)|^2\right|$ or
$\left|1-  |\rho(f_b)|^2\right|$ terms. 
Equation~\ref{CPEX}  
also holds when the final states are not $CP$ eigenstates, as
long as the modes are
common  to $D^0$ and $\bar D^0$. 
Consider, for example, $e^+e^- \to D^0 \bar D^0 \to f_a f_b$
with $f_a = K^+K^-$, $f_b = K^{\pm}\pi^{\mp}$. Measurements
of these rates would
yield unique information on the strong phase shifts.

Note that all these arguments cannot be applied for $D$ decays into vector
states $D \rightarrow V_1V_2$. These decay modes, in contrast to $D
\rightarrow PP$ and $PV$, are described by more than one independent 
amplitude. Therefore, the process 
$e^+ e^- \rightarrow D^0 \overline{D}^0 \rightarrow
(K^* \rho )(K^* \rho)$ can occur even in the absence of mixing and $CP$
violation, since the two decays could be  described by different 
combinations of decay amplitudes. Analyses of these
decays could yield useful
information on final state interactions in these decay modes.

Table~\ref{tab:CP_D_threshold} summarizes 
the sensitivities of measurements 
of $CP$ asymmetries in coherent $D^0
\overline{D}^0$ decays into $CP$ eigenstates at \bes3.
\begin{table}[htbp]
  \centering
  \begin{tabular}{c|c|c} \hline
 \multicolumn{3}{c}{CP Violation} \\ \hline
 Reaction  & Event  &  Comment \\\hline
 $D^{*0} \overline{D^0} \rightarrow [\gamma_s$(semileptonic)($CP$ eigenstates)] & 26280 &
 Measure mixing-dependent \\
 & &  $CP$ violation \\
 &        &  Asymmetry determined to 1\% \\\hline
$\psi(3770) \rightarrow$ (semileptonic)($CP$ eigenstates) &
136000 & Measure magnitude  \\
 &  & of $CP$ violating \\
 &   & amplitude to 0.5 \% \\\hline
$\psi(3770) \rightarrow$ ($CP\pm$ eigenstates)($CP\pm$ eigenstates) &
16000 & Sensitive to phase  \\
 &  & of direct $CP$ violating \\
 &  & amplitude (1.0\%) \\\hline
 \end{tabular}
  \caption{ Summary of $CP$ violation measurements 
using coherent $D^0\overline{D}^0$ 
decays to $CP$ eigenstates at \bes3}
 \label{tab:CP_D_threshold}
\end{table}

%
%
%
%
%
%
\section{T Violation}
\label{sec:part5:t}

\subsection{Triple-Product Correlation in $D$ Decays}
\label{sec:part5:cp:t}

There are other types of $CP$-violating effects that can also reveal the
presence of new physics, namely Triple-product (TP)
correlations~\cite{Part5_Valencia_1989}. These take the form
$\vec{v}_1\cdot(\vec{v}_2 \times \vec{v}_3)$, where each $\vec{v}_i$ is
a spin or momentum vector. 
Since $T (\vec{v}_i) = - \vec{v}_i$, these TP's are odd 
under time reversal ($T$) and, hence, by the $CPT$ theorem, also 
constitute potential signals for $CP$
violation. One can construct the non-zero TP by measuring the non-zero value
of the asymmetry:
\begin{eqnarray}
A_{T} \equiv \frac{\Gamma(\vec{v}_1\cdot(\vec{v}_2 \times \vec{v}_3)>0) -
\Gamma(\vec{v}_1\cdot(\vec{v}_2 \times
\vec{v}_3)<0)}{\Gamma(\vec{v}_1\cdot(\vec{v}_2 \times \vec{v}_3)>0) +
\Gamma(\vec{v}_1\cdot(\vec{v}_2 \times \vec{v}_3)<0)},
\label{eq:Part5_T_odd_asym}
\end{eqnarray}
where $\Gamma$ is the decay rate for a given process. However, a
non-vanishing value for $A_{T}$ is not necessarily due to the $T$
transformation. This is because, in addition to reversing spins and momenta,
the time reversal symmetry $T$ also exchanges the initial and final states.
In particular, non-zero TP correlations can be due to final state
interactions, even if there is no $CP$ violation.  One typically
finds 
\begin{eqnarray}
A_{T} \propto \mbox{sin}(\phi+\delta),
\label{eq:Part5_T_odd_asym_prop}
\end{eqnarray}
where $\phi$ is a weak, $CP$-violating phase and $\delta$ is a strong phase.
From this we see that if $\delta \neq 0$, a TP correlation will appear, even
without $CP$ violation.  To perform a stringent test of $CP$ invariance, one
has to also measure
$\overline{A}_T$ for the $CP$-conjugate decay process.  One then gets a
$T$-violating asymmetry:
\begin{eqnarray}
{\cal A}^{CP}_{T} \equiv \frac{1}{2} (A_T + \overline{A}_T).
\label{eq:Part5_T_vio_asym}
\end{eqnarray}
This is a true $T$-violating signal that is non-zero only if $\phi
\neq 0$.

In fact, TP asymmetries are similar to direct $CP$ asymmetries in two
ways~\cite{Part5_Datta_2004}: (i) they are both obtained by comparing a
signal in a given decay with the corresponding signal in the
$CP$-transformed process, and (ii) they both need the interference between
two different decay amplitudes.  However, there is one important difference
between the direct $CP$ and TP asymmetries. The direct $CP$ asymmetry can be
written
\begin{eqnarray}
A^{CP}_{dir} \propto \mbox{sin}\phi \mbox{sin}\delta,
\label{eq:Part5_CP_direct}
\end{eqnarray}
while, for the true $T$-violating asymmetry is given by
\begin{eqnarray}
{\cal A}^{CP}_{T} \propto \mbox{sin}\phi \mbox{cos}\delta.
\label{eq:Part5_T_vio_prop}
\end{eqnarray}
The key point here is that one can produce a direct $CP$ asymmetry only if
there is a non-zero strong-phase difference between the two decay
amplitudes. However, TP asymmetries are maximal when the strong phase
difference is zero. Thus, it may be more promising to look for TP
asymmetries than direct $CP$ asymmetries in $D$ and $B$ decays.

Consider the weak decay of a $D$ meson to two vector mesons, $D
\rightarrow V_1(k_1,\epsilon_1)V_2(k_2,\epsilon_2)$, where $k_1$ and
$\epsilon_1$
 ($k_2$ and $\epsilon_2$) denote the polarization and momentum of $V_1$
($V_2$).  For example, in the decay of $D^+ \rightarrow
\overline{K}^{*0}(k_1, \epsilon_1)
K^{*+}(k_2, \epsilon_2)$, one can study the triple correlation
\begin{eqnarray}
A_T = \frac{N(k_1 \cdot (\epsilon_1 \times \epsilon_2)>0) - N(k_1 \cdot
(\epsilon_1 \times \epsilon_2)<0)}{N(k_1 \cdot (\epsilon_1 \times
\epsilon_2)>0) + N(k_1 \cdot
(\epsilon_1 \times \epsilon_2)<0)}
\label{eq:Part5_T_Dkstat}
\end{eqnarray}
and the true $T$ asymmetry, ${\cal A}^{CP}_{T} \equiv \frac{1}{2}
(A_T + \overline{A}_T)$, can be studied  by considering the $CP$
conjugate decay mode $D^- \rightarrow K^{*0}K^{*-}$. $CP$ symmetry
is violated if ${\cal A}^{CP}_{T} \neq 0$.

The only reported experimental search for $T$-odd 
asymmetries is from FOCUS in the  $D^0
\rightarrow K^+ K^- \pi^+\pi^- $ and $D_S \rightarrow \overline{K}^0 K^-
\pi^+ $  decay modes, as listed in Table~\ref{tab:Part5-CP-T-odd}. No 
evidence
for a $T$-odd asymmetry is observed. The large \bes3 
data samples are expected to provide enhanced sensitivity to
possible $T$-violating asymmetries.     
\begin{table}[htbp]
  \centering
  \begin{tabular}{c|c|c|c} \hline
Year & Experiment & Decay Mode & $A^{CP}_T$(\%)  \\ \hline
2005 & FOCUS~\cite{Part5_cp_focus1} & $D^0 \rightarrow K^+ K^- \pi^+\pi^- $ & $1.0 \pm 5.7 \pm 3.7$ \\ \hline
2005 &  FOCUS~\cite{Part5_cp_focus1} & $D^+ \rightarrow \overline{K}^0 K^- \pi^+\pi^- $ & $2.3 \pm 6.2 \pm 2.2$ \\ \hline
2005 & FOCUS~\cite{Part5_cp_focus1} & $D_S \rightarrow \overline{K}^0 K^- \pi^+ $ & $-3.6 \pm 6.7 \pm 2.3$ \\ \hline
  \end{tabular}
  \caption{ $T$-violating asymmetries in $D$ meson decays from the FOCUS 
experiment.}
 \label{tab:Part5-CP-T-odd}
\end{table}

\subsection[Searches for $CP$ Violation via $T$-odd moments in $D$ 
decays]{Searches for $CP$ Violation via $T$-odd moments in $D$
decays\footnote{By Ikaros Bigi}}
\label{sec:part5:cp:bigi}

Decays to final states of {\em more than} two pseudoscalar or one pseudoscalar and one vector
mesons contain 
more dynamical information than given by their partial widths; their distributions as described by Dalitz plots 
or $T$-odd moments can exhibit $CP$~asymmetries that can be 
considerably larger than those for the integrated partial 
width. Final state interactions, while not necessary for the emergence of 
such effects, can fake a signal; however, these
can be disentangled by comparing $T$-odd moments for $CP$~conjugate modes, 
as explained below.

All $CP$~asymmetries observed to date in $K_L$ and $B_d$ decays concern 
partial widths -- 
$\Gamma (P \to f) \neq \Gamma (\bar P \to \bar f)$ -- except for one. 
The notable exception, 
namely the $T$-odd moment found in the rare mode $K_L \to \pi^+\pi^- 
e^+e^-$,  can teach 
us important lessons for future searches in charm decays. Define 
$\phi$ as the angle between the planes 
spanned by 
the two pions and the two leptons in the $K_L$ 
restframe:  
\begin{equation}    
\phi \equiv \angle ( \vec n_l, \vec n_{\pi})\; , \;   
\vec n_l = \vec p_{e ^+}\times \vec p_{e ^-}/
|\vec p_{e ^+}\times \vec p_{e ^-}| \; , \;  
\vec n_{\pi} = \vec p_{\pi ^+}\times \vec p_{\pi ^-}/ 
|\vec p_{\pi ^+}\times \vec p_{\pi ^-}| \; . 
\label{PHISEHGAL}
\end{equation}    
One analyzes 
the decay rate as a function of $\phi$: 
\begin{equation} 
\frac{d\Gamma}{d\phi} = \Gamma _1 {\rm cos}^2\phi + 
\Gamma _2 {\rm sin}^2\phi + 
\Gamma _3 {\rm cos}\phi \, {\rm sin} \phi .
\end{equation} 
Since  
\begin{equation} 
{\rm cos}\phi \, {\rm sin} \phi = 
(\vec n_l \times \vec n_{\pi}) \cdot 
(\vec p_{\pi ^+} + \vec p_{\pi ^-}) 
(\vec n_l \cdot \vec n_{\pi})/
|\vec p_{\pi ^+} + \vec p_{\pi ^-}| ,
\end{equation}
one notes that 
\begin{equation} 
{\rm cos}\phi \, {\rm sin} \phi \; \; \; 
\stackrel{{T},{CP}}{\longrightarrow} \; \; \; 
- \; {\rm cos}\phi \, {\rm sin} \phi 
\end{equation}    
under both $T$~ and $CP$~transformations; i.e. the observable  
$\Gamma _3$ represents a $T$- and $CP$-odd correlation. 
It can be projected out by comparing the $\phi$ 
distribution integrated over two quadrants: 
\begin{equation} 
A = 
\frac{\int _0^{\pi/2} d\phi \frac{d\Gamma}{d\phi} - 
\int _{\pi /2}^{\pi} d\phi \frac{d\Gamma}{d\phi}}
{\int _0^{\pi} d\phi \frac{d\Gamma}{d\phi}} = 
\frac{2\Gamma _3}{\pi (\Gamma _1 + \Gamma _2)} .
\end{equation}
It was first measured by KTEV and then confirmed by NA48 \cite{Part5_pdg06}:  
\begin{equation} 
A = (13.7 \pm 1.5)\% \, .
\label{KTEVSEHGAL2}
\end{equation} 
$A\neq 0$ is induced by $\epsilon_K$, the $CP$~violation in the $K^0 - 
\bar K^0$ mass matrix, 
leading to the prediction \cite{part5:SEGHALKL}
\begin{equation} 
A = (14.3 \pm 1.3)\% \, .
\end{equation}
The observed value for the $T$~odd moment $A$ is fully consistent with 
$T$~violation. 

It is actually easy to see how this sizable forward-backward asymmetry is generated from the tiny quantity 
$|\eta_{+-}| \simeq 0.0023$. For 
$K_L \to \pi ^+ \pi ^- e^+ e^-$ is driven by the two sub-processes 
\begin{eqnarray} 
K_L &\stackrel{\not {CP} \&\Delta S =1}{\longrightarrow} \pi^+\pi^- 
\stackrel{E1}{\longrightarrow} \pi^+\pi^- \gamma ^* \to \pi ^+ \pi ^- e^+ e^- 
\\
K_L &\stackrel{M1\& \Delta S =1}{\longrightarrow} \pi^+\pi^- \gamma ^* \to \pi ^+ \pi ^- e^+ e^- \; , 
\end{eqnarray}
where the first reaction is suppressed, since it requires $CP$~violation 
in  $K_L \to 2\pi$, as is the second one, since it involves an $M1$ 
transition.  Those two {\em a priori} very 
different suppression mechanisms happen to yield comparable amplitudes, 
which thus generate a 
sizable interference. The price one pays is the small branching ratio, namely 
BR$(K_L \ra \pi ^+ \pi ^- e^+ e^-) = 
(3.32 \pm 0.14 \pm 0.28 ) \cdot 10^{-7}$.  
{\it I.e.}, one has `traded away' rate for a much larger asymmetry.  

$D$ decays can be treated in an analogous way \cite{part5:BEIJING06}.  
Consider the Cabibbo-suppressed channel 
\footnote{This mode can exhibit direct $CP$~violation even within the SM.}
\begin{equation} 
\stackrel{(-)}D \to K \bar K \pi^+\pi^-
\end{equation}
and now define $\phi$ as the angle between the $K \bar K$ and 
$\pi^+\pi^-$ planes. Then 
one has 
\begin{eqnarray} 
\frac{d\Gamma}{d\phi}(D \to K \bar K\pi^+\pi^-) &=& \Gamma_1 {\rm cos}^2 \phi + 
\Gamma_2 {\rm sin}^2 \phi + \Gamma_3 {\rm cos} \phi {\rm sin}\phi , \\
\frac{d\Gamma}{d\phi}(\bar D \to K \bar K\pi^+\pi^-) &=& \bar \Gamma_1 {\rm cos}^2 \phi + 
\bar \Gamma_2 {\rm sin}^2 \phi + \bar \Gamma_3 {\rm cos} \phi {\rm 
sin}\phi . 
\end{eqnarray}
The partial width for $D[\bar D] \to K\bar K \pi^+\pi^-$ is given by 
$\Gamma_{1,2}~[\bar \Gamma_{1,2}]$; $\Gamma_1 \neq \bar \Gamma_1$ or 
$\Gamma_2 \neq \bar \Gamma_2$ represents direct $CP$~violation in the 
partial width. 
$\Gamma_3~\&~\bar \Gamma_3$ constitute $T$~odd correlations. By 
themselves,  they do not necessarily 
indicate $CP$~violation, since they can be induced by strong final state 
interactions. However 
\begin{equation} 
\Gamma_3 \neq \bar \Gamma_3 \; \; \Longrightarrow CP~{\rm violation!}
\end{equation} 
It is quite possible, or even likely, that a difference 
in $\Gamma_3$ {\it vs.}  $\bar \Gamma_3$ 
is significantly larger than in $\Gamma_1$ {\it vs.} $\bar \Gamma_1$ or 
$\Gamma_2$ {\it vs.} $\bar \Gamma_2$. 
Furthermore, one can expect that differences in detection 
efficiencies can be handled by comparing $\Gamma_3$ with $\Gamma_{1,2}$ and 
$\bar \Gamma_3$ with $\bar \Gamma_{1,2}$. A pioneering search for such an effect has been 
undertaken by FOCUS \cite{part5:PEDRINI}. 

The recent evidence observed for $D^0 - \bar D^0$ 
oscillations even suggests that the 
$T$~odd moments $\Gamma_3$ and $\bar \Gamma_3$ would show a 
{\em time dependence} \cite{part5:TAIWAN}. 

\chapter[Rare and Forbidden Charm Decays]{Rare and Forbidden Charm
Decays\footnote{By Hai-Bo Li and Ya-Dong Yang}}
\label{sec:charm_rare}

\section{Introduction}
 The remarkably successful Standard Model of particle physics describes 
matter's  most basic elements and the forces through which they interact. 
Physicists have
 tested its predictions to better than $1\%$   precision.
 Yet despite its many successes, we know the SM  is not the whole story.
 The questions of  so-called fine-tuning,   mass  hierarchy, etc. 
 remain unexplained.   Testing the SM is one of the most  important missions
 in particle physics.    One important feature of the SM is the absence of
 FCNC at tree level.  Generally, FCNC processes begin at one-loop level 
and are GIM suppressed. They have proven to be powerful tools for 
probing the structure  of electro-weak  interactions.   
Examples are rare $K$ and $B$ decays and 
oscillations  that induced by penguin or box Feynman diagrams. 
Because of the large  top quark mass,  
the GIM suppression is mild for down-type quarks.   
For charm quark FCNC processes, however, 
only down type quarks can propagate in the loop, 
as a result, the GIM suppression is very 
strong and  SM predictions for FCNC
are very tiny, and beyond the sensitivities of 
any running or planned experiment. Thus,
searches for these decays would  be sensitive 
probes of new, beyond the SM physics.  However, long distance 
contributions,  which
are not always reliably calculable, could be 
considerably  larger than  SM short distance effects.

Potentially interesting  decays include: 
(i) $D\to V\gamma (V = \rho, \omega,  \phi,  ... )$ ,  
(ii)$ D\to X \ell^{+} \ell^{-}~(X=\pi, K, \eta, \rho, \omega, \phi, ... ) $,  
(iii)    $D\to X \nu_\ell {\bar \nu}_\ell \ (X = \pi, K, \eta,
\dots)$, (iv) $D\to \gamma \gamma$, and
(v) $D\to \ell^+ \ell^-$.

There have been many attempts
to calculate both short- and long-distance contributions to these
decays as reliably as possible in both the SM
and possible new physics scenarios that
potentially have large effects. In a number of cases, 
extensions of the SM  can give contributions
that are sometimes orders
of magnitude larger than those of the SM~\cite{bghp2}.    
In order to establish the existence
 of a clean  window for observation of new physics in a given
 observable in rare charm decays,  SM contributions first should be made 
clear.  In this regard, it is very important to include long-distance
contributions due to the propagation of light quarks, although
they are hard to calculate with analytical methods. In the
following,  we review these decays in the  
context of the SM and
then look at them in new physics  models.

\section{Rare charm decays in the SM }

In principle, the task of producing SM predictions for FCNC $D$
meson decays is straightforward.  There are two components to the
analysis: short-distance (SD) and long-distance (LD), which must
be separately calculated. It is known that the SD contribution can
be calculated in a well defined framework, but phenomenological
methods have to be resorted to for LD contributions.

\subsection{Radiative charm decays}
\begin{figure}[t]
\begin{center}
\scalebox{0.6}{\epsfig{file=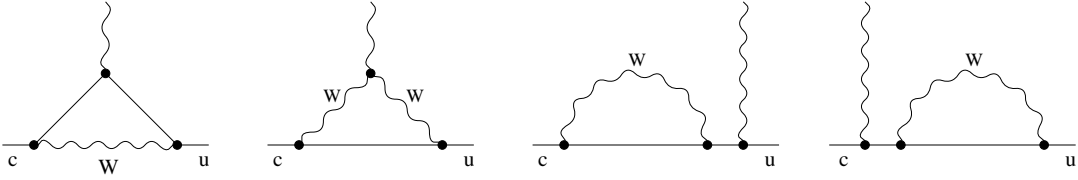}}
\caption{One-loop diagrams inducing the $c\to u \gamma$ }
\end{center}
\label{part5:rare:fig1}
\end{figure}

Radiative  decay modes  with one photon include $D
\rightarrow V \gamma_, \ D_s \to V \gamma$ and $D\to X_{u}\gamma$.
The short distance effective Hamiltonian for $\bar{c} \rightarrow \bar{u} +
\gamma$ starts from the one-loop diagrams shown in 
Fig.~\ref{part5:rare:fig1} and gives
\begin{equation} \label{Lint}
L_{int} = -\frac{4 G_{\scs F}}{\sqrt{2}} \; A \; \frac{e}{16 \pi^2} \; m_c
                (\bar{u} \sigma_{\mu\nu} P_{\scs R} c) F^{\mu\nu},
\end{equation}
with
\begin{equation} \label{1loop}
\Delta A_{\rm 1\;loop} \simeq -\frac{5}{24}
                \sum_{q=d,s,b} V^*_{cq} V_{uq} \left( \frac{m_q}{M_W} \right)^2.
\end{equation}
The CKM factors in the above equation have very different orders of
magnitude:
\begin{equation}
|V^*_{cd} V_{ud}| \simeq |V^*_{cs} V_{us}| \simeq 0.22
\hspace{1cm} \mbox{and} \hspace{1cm}
|V^*_{cb} V_{ub}| \simeq (1.3 \pm 0.4) \times 10^{-4}.
\end{equation}
Consequently, $|\Delta A_{\scs 1\;loop}| \sim 2 \times 10^{-7}$.
The extraordinary smallness of this number is due to the
tiny factors $(m_q/M_W)^2$ for the light quarks and the small CKM
elements governing the $b$-quark contributions.

Since the important suppression factors are independent of gauge
couplings, it is possible that higher orders in perturbation
theory give dominant  contributions to the radiative amplitude
considered above because they may not suffer from the same dramatic
suppression factors and are reduced only by powers of the gauge couplings.

With RGE,  one can re-sum short distance  QCD corrections in the
leading logarithmic approximation which results in the effective
Hamiltonian
\begin{figure}[t]
\begin{center}
\scalebox{0.6}{\epsfig{file=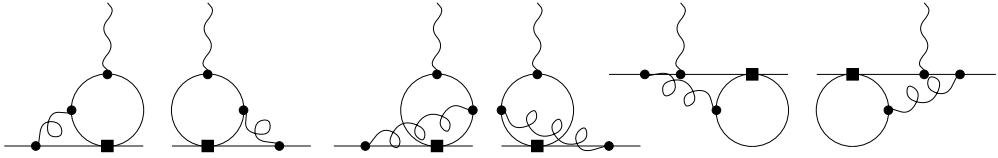}}
\caption{ Two-loop diagrams  for the $c\to u \gamma$ }
\end{center}
\label{part5:rare:fig2}
\end{figure}

\begin{equation}
H_{eff}(m_b > \mu > m_c) =\frac{4 G_{\scs F}}{\sqrt{2}} \sum_{q=d,s}
V^*_{cq} V_{uq} [ C_1(\mu) O^q_1 + C_2(\mu) O^q_2 +
                                 \sum_{i=3}^8 C_i(\mu) O_i ] \label{H2},
\end{equation}
where
\begin{eqnarray}
O^q_1 &=& (\bar{u}_{\alpha} \gamma_{\mu} P_{\scs L} q_{\beta})
          (\bar{q}_{\alpha} \gamma^{\mu} P_{\scs L} c_{\beta}),
                                                        \hspace{1cm} q=d,s,b\\
O^q_2 &=& (\bar{u}_{\alpha} \gamma_{\mu} P_{\scs L} q_{\alpha})
          (\bar{q}_{\beta}  \gamma^{\mu} P_{\scs L} c_{\beta}),
                                                        \hspace{1cm} q=d,s,b\\
O_7   &=&  \frac{e}{16 \pi^2} \; m_c
      (\bar{u}_{\alpha} \sigma_{\mu\nu} P_{\scs R} c_{\alpha}) F^{\mu\nu}.
\end{eqnarray}
The Wilson co-efficients $C_{i}$ and remaining operators $O_i$
are given explicitly in Ref.~\cite{bghp1}. It is found that the
leading logarithmic QCD correction will enhance the $c\to u\gamma$
by more than one order-of-magnitude:
\begin{equation} \label{numLLA}
|\Delta A_{\scs LLA}| =
[ 0.001 C_1(m_b) + 0.055 C_2(m_b)]\;|V^*_{cb} V_{ub}| =
0.060 \; |V^*_{cb} V_{ub}| \simeq (8 \pm 3) \times 10^{-6}.
\end{equation}

Moreover, when the two loop contributions 
depicted in Fig. \ref{part5:rare:fig2} are included,
one finds that the strength of   $c\to u\gamma$
is further  enhanced by more than two orders-of-magnitude
\begin{equation} \label{num2loop}
|A| = |V^*_{cs} V_{us}| \frac{\al(m_c)}{4 \pi} \; (0.86 \pm 0.19)
   = (4.7 \pm 1.0) \times 10^{-3}.
\end{equation}
After these  surprising enhancements,  the predicted
branching ratio $({\cal B})$ for $c \rightarrow u \gamma$ is found
to be $\sim 10^{-8}$. However, even this is very small compared to
long-distance contributions from the $s$ and $u$ channel poles
from nearby states and VMD contributions from $\rho,~\omega$,
and $\phi$. These estimates suffer from uncertainties from several
sources. The mass $m_c$ is too large for the chiral symmetry
approximation and it is too small for the HQET approximation.
Thus, one has to resort to inspired guesswork~\cite{bghp1,greab,bajc}. 
In any case, the end results seem
reasonable and agree with  recent measurements.  The branching fractions
are in the range of $10^{-5}$ to $10^{-7}$.  The recent
measurement of ${\cal B} (D^0 \rightarrow \phi^0 \gamma) \sim
2.6\times 10^{-5}$ by Belle~\cite{tajima} is well within the
expected range. However, it would be extremely hard to extract the
small short-distance contributions.

\subsection{GIM-suppressed decays}

\begin{figure}[t]
\begin{center}
\scalebox{0.4}{\epsfig{file=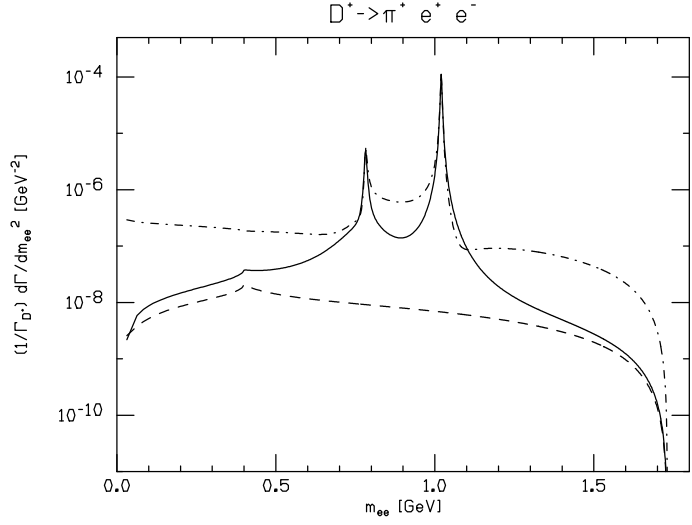, angle=90}}
\caption{Predicted dilepton mass distributions for $D^+\to
\pi^+ e^+ e^-$.  The dashed (solid) line is the short-distance
(total) SM contribution.  The dot-dashed line is the
$R$-parity violating SUSY contribution.  }\label{part5:rare:fig:pill}
\end{center}
\end{figure}

\subsubsection{The short-distance component}\label{subsec:sd}

Short-distance amplitudes are concerned with
the QCD degrees of freedom (quarks, gluons) and any relevant
additional fields (leptons, photons).  Thus, the short-distance
part of the $D \to X_u \ell^+ \ell^-$ amplitude involves the
quark process $c \to u \ell^+ \ell^-$.  It is usually
most natural to employ an effective description in which the weak
Hamiltonian is expressed in terms of local multiquark operators
and Wilson coefficients~\cite{bbl}.  For example, the effective
Hamiltonian for $c\to u\ell^+\ell^-$ with renormalization scale
$\mu$ in the range $m_b \ge \mu \ge m_c$ is~\footnote{Quantities
with primes have had the explicit $b$-quark contributions
integrated out}
\begin{equation}
{\cal H}_{\rm eff}^{c\to u\ell^+\ell^-} = - {4 G_F \over \sqrt{2}}
\left[ \sum_{i=1}^2~\left(\sum_{q=d,s} ~C_i^{(q)}(\mu)
{\cal O}_i^{(q)}(\mu) \right) +
\sum_{i=3}^{10} ~C_i^{'}(\mu) {\cal O}_i^{'}(\mu) \right] \ \ .
\label{weakham}
\end{equation}
In the above expression, ${\cal O}_{1,2}^{(q)}$ are four-quark current-current
operators, ${\cal O}^{'}_{3-6}$ are the QCD penguin operators,
${\cal O}_7$ (${\cal O}_8$) is the electromagnetic
(chromomagnetic) dipole operator and ${\cal O}_{9,10}$
explicitly couple quark and lepton currents.  For example, we
have
\begin{equation}
O_7^{'} = \frac{e}{16\pi^2}m_c(\bar{u}_L\sigma_{\mu\nu}c_R)F^{\mu\nu}\ ,
\qquad O_9^{'} = \frac{e^2}{16\pi^2} (\bar{u}_L\gamma_\mu c_L)
( \bar{\ell}\gamma^\mu \ell) \ \ .
\label{examples}
\end{equation}
The famous Inami-Lim functions~\cite{il} contribute to the Wilson
coefficients $C_{7-10}$ at scale $\mu = M_{\rm W}$.

Figure~\ref{part5:rare:fig:pill} displays the predicted dilepton
mass spectrum for $D^+ \to \pi^+ \ell^+\ell^-$.
Several distinct kinds of contributions are included.
The short-distance SM component corresponds to the dashed
line, which is seen to lie beneath the other two curves.
For reference, we note the predicted `short distance'
{\em inclusive} branching ratio,
\begin{equation}
{\cal B}r_{D^+\to X_u^+ e^+e^-}^{\rm (sd)} \simeq 2\times10^{-8} \ \ .
\label{sdbr}
\end{equation}

\subsubsection{The long-distance component}\label{subsec:ld}

\begin{figure}
\begin{center}
\scalebox{0.8}{\epsfig{file=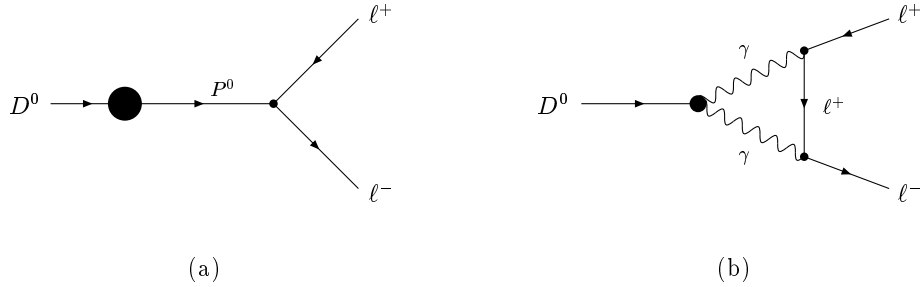}} \caption{Long distance
contributions to $D^0 \to \ell^+\ell^-$.}
\end{center}
\label{part5:rare:fig:ll}
\end{figure}

\begin{figure}
\begin{center}
\scalebox{0.8}{\epsfig{file=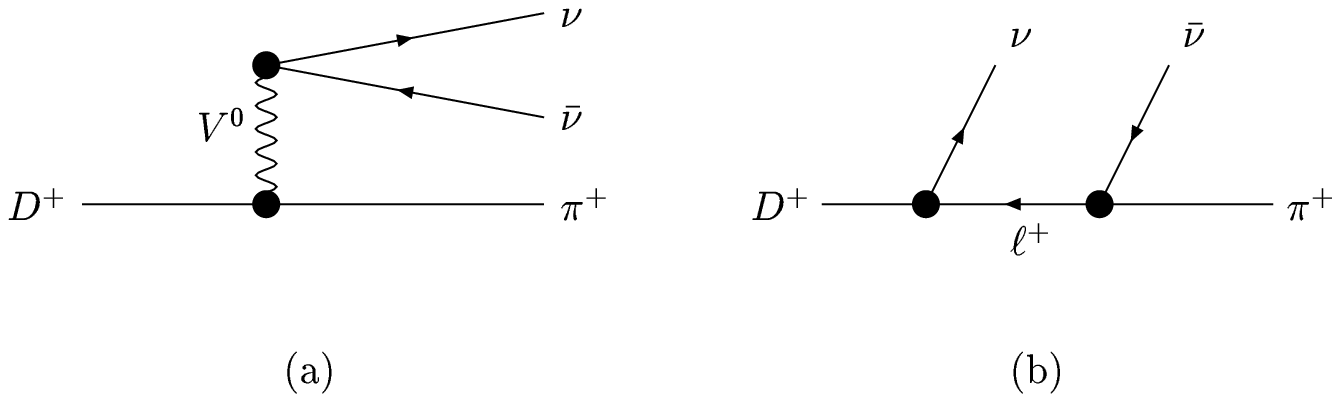}} \caption{Long distance
contributions to $D^+ \to \pi^+\ell^+\ell^-$.}
\end{center}
\label{part5:rare:fig:pll}
\end{figure}

The long-distance component to a transition amplitude is often
cast in terms of hadronic entities rather than the underlying
quark and gluonic degrees of freedom. For charm decays, the
long-distance amplitudes are typically important but difficult to
determine with any rigor. There are generally several
long-distance mechanisms for a given transition, {\it e.g.} as
indicated for $D^0 \to \ell^+ \ell^-$ in Fig.~\ref{part5:rare:fig:ll} and for $D^+ \to
\pi^+ {\bar\nu} \nu$ in Fig.~\ref{part5:rare:fig:pll}.

We return to the case of $D^+ \to \pi^+ \ell^+\ell^-$ depicted
in Fig.~\ref{part5:rare:fig:pill}.  The solid curve represents the {\it
total} SM signal, summed over both SD and LD contributions.  In
this case the LD component dominates, and from studying the
dilepton mass distribution we can see what is happening.  The
peaks in the solid curve must correspond to intermediate
resonances ($\phi$, etc.).  The corresponding Feynman graph
would be analogous to that in Fig. \ref{part5:rare:fig:pll} (b) in which the final state
neutrino pair is replaced by a charged lepton pair. One finds
numerically that \begin{equation} {\cal B}r_{D^+\to \pi^+ e^+e^-}^{\rm (SM)}
\simeq {\cal B}r_{D^+\to \pi^+ e^+e^-}^{\rm (\ell d)} \simeq
2\times10^{-6} \ \ . \label{ldbr} \end{equation}

\subsubsection{The Standard Model Predictions}\label{subsec:SM}

In an  analysis based in part on existing literature~\cite{lit},
 both SD and LD amplitudes for a number
of FCNC $D$-meson transitions have been calculated~\cite{bghp2}. Results
are collected in Table~\ref{tab:br_rare}.  As stated earlier, the
current database for processes appearing in Table~\ref{tab:br_rare}
consists entirely of upper bounds (or in the case of $D^0 \to
\gamma \gamma$ no data entry at all).  In all cases, existing
experimental bounds are much larger than the SM predictions, so
there is no conflict between the two. For some cases ({\it e.g.} $D \to
\pi \ell^+ \ell^-$) the gap between SM theory and experiment is
not so large and there is hope for detection in the near future.
In other cases ({\it e.g.} $D^0 \to \ell^+ \ell^-$) the gap is
enormous, leaving plenty of opportunities for signals from New Physics
to appear. This point is sometimes not fully appreciated and, thus,
warrants some emphasis. It is why, for example, attempts to detect
$\Delta M_{\rm D}$ via $D^0$-${\bar D}^0$ mixing experiments are
so important.
\begin{table}[ht]
\caption{ Standard Model predictions and current experimental
limits for the branching fractions due to
short and long distance contributions for various rare $D$ meson
decays.\label{tab:br_rare}}
\vspace{0.4cm}
\begin{center}
\begin{tabular}{|l|c|c|c|} \hline
Decay Mode & Experimental Limit & ${\cal B}r_{S.D.}$ & ${\cal B}r_{L.D.}$
\\ \hline
$D^+\to X_u^+ e^+e^-$ & & $2\times 10^{-8}$ & \\
$D^+\to\pi^+e^+e^-$ & $<4.5\times 10^{-5}$ & &$2\times10^{-6}$ \\
$D^+\to\pi^+\mu^+\mu^-$ & $<1.5\times 10^{-5}$ & &$1.9\times10^{-6}$ \\
$D^+\to\rho^+e^+e^- $ & $<1.0\times 10^{-4}$ & &$4.5\times10^{-6}$ \\
$D^0\to X_u^0+e^+e^-$ & & $0.8\times 10^{-8}$ & \\
$D^0\to\pi^0e^+e^-$ & $<6.6\times 10^{-5}$ &  &
$0.8\times10^{-6}$ \\
$D^0\to\rho^0e^+e^-$ & $<5.8\times 10^{-4}$ & & $1.8\times10^{-6} $ \\
$D^0\to\rho^0\mu^+\mu^-$ & $<2.3\times 10^{-4}$ & & $1.8\times10^{-6} $ \\
\hline
$D^+\to X_u^+\nu\bar\nu$ & & $1.2\times 10^{-15}$ & \\
$D^+\to\pi^+\nu\bar\nu$ & &  & $5\times 10^{-16}$ \\
$D^0\to\bar K^0\nu\bar\nu$ &  & & $2.4\times10^{-16}$ \\
$D_s\to \pi^+\nu\bar\nu$ & & & $8\times10^{-15}$ \\ \hline
$D^0\to\gamma\gamma$ & & $4\times10^{-10}$ & few~$\times 10^{-8}$ \\ \hline
$D^0\to\mu^+\mu^-$ & $<3.3\times 10^{-6}$ & $1.3\times 10^{-19}$ &
${\rm few}~\times 10^{-13}$ \\
$D^0\to e^+e^-$ & $<1.3\times 10^{-5}$ & $(2.3-4.7)\times 10^{-24}$ & \\
$D^0\to\mu^\pm e^\mp$ & $<8.1\times 10^{-6}$ & $0$ & $0$ \\
$D^+\to\pi^+\mu^\pm e^\mp$ & $<3.4\times 10^{-5}$ & $0$ & $0$ \\
$D^0\to\rho^0\mu^\pm e^\mp$ & $<4.9\times 10^{-5}$ & $0$ & $0$ \\ \hline
\end{tabular}
\end{center}
\end{table}

\section{New Physics Analysis}

There is a wide collection of possible
New Physics models leading to FCNC $D$-meson transitions.  Among
those considered in Ref.~\cite{bghp2} are
(i) Supersymmetry (SUSY): $R$-parity conserving, $R$-parity violating,
(ii) Extra Degrees of Freedom: Higgs bosons, Gauge bosons, Fermions,
Spatial dimensions, (iii) Strong Dynamics:
Extended technicolor, Top-condensation.

We  restrict most of our attention to the case of
supersymmetry and add a few remarks
on the topic of large extra dimensions.
The SUSY discussion divides naturally according
to how the $R$-parity $R_{\rm P} $ is treated, where
\begin{equation}
R_{\rm P} \ = \ (-)^{3(B-L)+2S} \ = \ \left\{
\begin{array}{cc}
+1 & {\rm (particle)} \\
-1 & {\rm (sparticle)} \ \ .
\end{array}
\right.
\end{equation}

\subsection{$R$-parity conserving SUSY}

$R$-parity conserving SUSY will contribute to charm FCNC amplitudes
via loops. To calculate $R$-parity conserving SUSY contributions,
the so-called {\it mass insertion approximation} is always
employed~\cite{lhall};  this is oriented towards phenomenological
studies and is also model independent.  Let us first describe what
is actually done and then provide a brief explanation of the
underlying rationale.

\begin{figure}
\begin{center}
\scalebox{0.5}{\epsfig{file=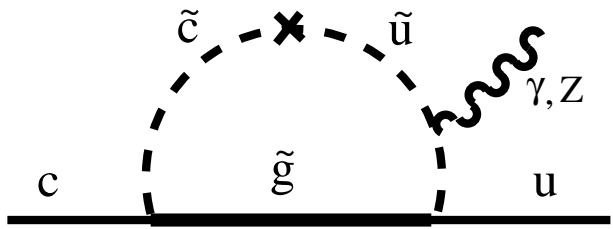, angle=-90}}
\caption{A typical contribution to $c\to u$ FCNC transitions in
the MSSM. The cross denotes one  mass insertion
$(\delta^u_{12})_{\lambda\lambda'}$ and $\lambda,\lambda'$
are helicity labels.}
\end{center}
\label{part5:rare:fig:mssm}
\end{figure}

In this approach, a squark propagator becomes modified by a mass
insertion ({\it e.g.} the `$\times$' 
in Fig.~\ref{part5:rare:fig:mssm} that changes the
squark flavor)~\cite{lhall,susyfcnc}. For convenience, one expands
the squark propagator in powers of the dimensionless quantity
$(\delta^u_{ij})_{\lambda\lambda'}$,
\begin{equation}
(\delta^u_{ij})_{\lambda\lambda'}={(M^u_{ij})^2_{\lambda\lambda'}
\over M^2_{\tilde q}}\ \ ,
\label{delta}
\end{equation}
where $i\neq j$ are generation indices, $\lambda,\lambda'$ denote
the chirality, $(M^u_{ij})^2$ are the off-diagonal elements of the
up-type squark mass matrix and $M_{\tilde q}$ represents the
average squark mass.  The exchange of squarks in loops thus leads
to FCNC through diagrams such as the one in Fig.~\ref{part5:rare:fig:mssm}. The role of
experiment is either to detect the predicted (SUSY-induced) FCNC
signal or to constrain the contributing
$(\delta^u_{ij})_{\lambda\lambda'}$.

This topic is actually part of the super-CKM problem. If one works
in a basis that diagonalizes the fermion mass matrices, then
sfermion mass matrices (and thus sfermion propagators) will
generally be nondiagonal.  As a result, flavor-changing processes
can occur.  One can use phenomenology to restrict these FCNC
phenomena.  The $Q=-1/3$ sector has yielded fairly strong
constraints but thus far only $D^0$-${\bar D}^0$ mixing has been
used to limit the $Q=+2/3$ sector.  In the analysis of Ref.~\cite{bghp2},
charm FCNCs have been taken to be as large as allowed by the
$D$-mixing upper bounds.

For the decays $D \to X_u \ell^+ \ell^-$ discussed
above in Sect.~\ref{subsec:sd}, the gluino contributions will
occur additively relative to those from the SM and, so, we can
write for the Wilson coefficients,
\begin{equation}
C_i = C_i^{\rm (SM)} + C_i^{\rm \tilde g} \ \ .
\label{wilson}
\end{equation}
To get some feeling for the dependence on the
$(\delta^u_{12})_{\lambda\lambda'}$ parameters, we display
the examples
\begin{equation}
C_7^{\rm \tilde g} \ \propto \ (\delta^u_{12})_{\rm LL}\ {\rm and} \
(\delta^u_{12})_{\rm LR}\ ,
\qquad
C_9^{\rm \tilde g} \ \propto \ (\delta^u_{12})_{\rm LL}\ \ ,
\label{right}
\end{equation}
whereas for quark helicities opposite~\footnote{We use the
notation ${\hat C}$ for the associated Wilson coefficients.}
to those in the operators of Eq.~(\ref{examples}), one finds
\begin{equation}
{\hat C}_7^{\rm \tilde g} \ \propto \ (\delta^u_{12})_{\rm RR}\
{\rm and} \ (\delta^u_{12})_{\rm LR}\ ,
\qquad
{\hat C}_9^{\rm \tilde g} \ \propto \ (\delta^u_{12})_{\rm RR}\ \ .
\label{wrong}
\end{equation}
Moreover, the
term in ${\hat C}_7^{\rm \tilde g}$ that contains
$(\delta^u_{12})_{\rm LR}$ experiences the enhancement factor
$M_{\tilde g}/m_c$.

\begin{figure}
\scalebox{0.5}{\psfig{figure=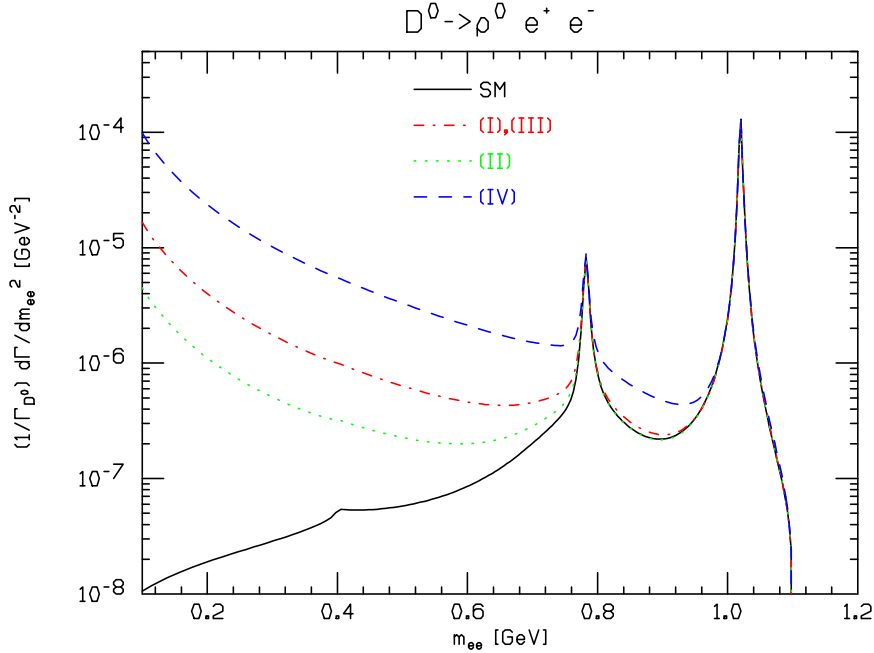,angle=90}}
\caption{Dilepton mass distributions for $D^0 \to \rho^0 e^+ e^-$
in the mass insertion approximation of MSSM.  The SM prediction
(solid curve) is provided for reference and the MSSM curves
refer to (i) $M_{\tilde g} = M_{\tilde q} = 250~{\rm GeV}$,
(ii) $M_{\tilde g} = 2 M_{\tilde q} = 500~{\rm GeV}$,
(iii) $M_{\tilde g} = M_{\tilde q} = 1000~{\rm GeV}$ and
(iv) $M_{\tilde g} = M_{\tilde q}/2 = 250~{\rm GeV}$.}
\label{part5:rare:fig:rhollmssm}
\end{figure}

The effects in $c\to u\ell^+\ell^-$ are studied numericaly for
the range of masses: (I) $M_{\tilde g}=M_{\tilde q}=250$~GeV, (II)
$M_{\tilde g}=2\,M_{\tilde q}=500$~GeV, (III) $M_{\tilde g}=
M_{\tilde q}=1000$~GeV and (IV) $M_{\tilde g}=(1/2)\,M_{\tilde
q}=250$~GeV. For some $D \to X_u \ell^+ \ell^-$ modes, the
effect of the squark-gluino contributions can be large relative to
the SM component, both in the total branching ratio and for
certain kinematic regions of the dilepton mass.  The mode $D^0 \to
\rho^0 e^+ e^-$ is given in Fig.~\ref{part5:rare:fig:rhollmssm}. 
This figure demonstrates the importance of measuring the low
$m_{\ell^+\ell^-}$ part of the dilepton mass spectrum.

\subsection{$R$-parity violating SUSY}
The effect of assuming that $R$-parity can be violated is to
allow additional interactions between particles
and sparticles.  Ignoring bilinear terms that are
not relevant to our discussion of FCNC effects, we introduce
the $R$-parity violating (RPV) super-potential of trilinear couplings,
\begin{equation}
{\cal W}_{{\not R}_p} =\epsilon_{ab}\left[
\frac{1}{2}\lambda_{ijk}L^a_iL^b_j\bar{E}_k
+\lambda'_{ijk}L_i^aQ^b_j\bar{D}_k
+\frac{1}{2}\epsilon_{\alpha\beta\gamma}\lambda^{''}_{ijk}\bar{U}^\alpha_i
\bar{D}^\beta_j\bar{D}^\gamma_k \right]\ \ ,
\label{rpv1}
\end{equation}
where $L$, $Q$, $\bar E$, $\bar U$ and $\bar D$ are the standard
chiral super-fields of the MSSM and $i,j,k$ are generation indices.
The quantities $\lambda_{ijk}$, $\lambda_{ijk}'$ and
$\lambda^{''}_{ijk}$ are {\it a priori} arbitrary couplings
with a total of $9+27+9=45$ unknown parameters in the theory.

The presence of RPV means that {\it tree-level} amplitudes become
possible in which a virtual sparticle propagates from one of the
trilinear vertices in Eq.~\ref{rpv1} to another. In order to
avoid significant FCNC signals (which would be in contradiction
with current experimental limits), bounds must be placed on the
(unknown) coupling parameters. As experimental probes become more
sensitive, the bounds become ever tighter. In particular, the FCNC
sector probed by charm decays involves the $\{ \lambda_{ijk}'\}$.
Introducing matrices ${\cal U}_{\rm L}$, ${\cal D}_{\rm R}$ to
rotate left-handed up-quark fields and right-handed down-quark
fields to the mass basis, we obtain for the relevant part of the
superpotential
\begin{equation} {\cal W}_{\lambda'} = {\tilde \lambda}_{ijk}' \left[
-\tilde{e}_L^i\bar{d}_R^k u_L^j - \tilde{u}_L^j\bar{d}_R^ke_L^i
-(\tilde{d}_R^k)^*(\bar{e_L^i})^cu_L^j + \dots \right] \ \ ,
\label{rpv2}
\end{equation}
where neutrino interactions are not shown, and we
define
 \begin{equation} \tilde{\lambda'}_{ijk}\equiv \lambda'_{irs} {\cal
U}^L_{rj} {\cal D}^{*R}_{sk} \ \ . \label{rpv3}
 \end{equation}
 Some bounds on
the $\{ \tilde{\lambda'}_{ijk} \}$ are already available from data
on such diverse sources as charged-current universality, the ratio
$\Gamma_{\pi\to e\nu_e}/\Gamma_{\pi\to \mu \nu_\mu}$, the
semileptonic decay $D\to K \ell \nu_\ell$, {\it etc}.~\cite{addr}.
The  additional experimental implications of the preceding
formalism are considered:

(i) For the decay $D^+\to \pi^+ e^+ e^-$, the effect of RPV  is
displayed as the dot-dash line in Fig.~\ref{part5:rare:fig:pill}. Here the
effect is proportional to ${\tilde\lambda}'_{11k}\cdot
{\tilde\lambda}'_{12k}$ and we have employed existing limits on
these couplings. Although the effect on the branching ratio is not
large, the dilepton spectrum away from resonance poles is seen
to be sensitive to the RPV contributions. This case is not optimal
because the current experimental limit on ${\cal B}r_{D^+\to \pi^+
e^+e^-}$ is well above the dot-dash curve.

(ii) For $D^+\to \pi^+ \mu^+ \mu^-$, the current experimental
limit on ${\cal B}r_{D^+\to \pi^+ \mu^+ \mu^-}$ actually
provides the new bound
\begin{equation}
{\tilde\lambda}'_{11k}\cdot {\tilde\lambda}'_{12k} \le 0.004 \ \ .
\label{rpv4}
\end{equation}

(iii) Another interesting mode is $D^0 \to \mu^+\mu^-$.
Upon using the bound of Eq.~(\ref{rpv4}) we obtain
\begin{equation}
{\cal B}r^{\not R_p}_{D^0\to\mu^+\mu^-} < 3.5\times 10^{-6}\;
\left(\frac{\tilde{\lambda}'_{12k}}{0.04}\right)^2
\,\left(\frac{\tilde{\lambda}'_{11k}}
{0.02}\right)^2 \ \ .
\label{d2mu_bound}
\end{equation}
A modest improvement of the existing limit on
${\cal B}r_{D^0\to\mu^+\mu^-}$ will yield a new bound on
the product ${\tilde\lambda}'_{11k}\cdot {\tilde\lambda}'_{12k}$.

(iv) Lepton flavor violating processes are allowed by the
RPV Lagrangian.  One example is the mode $D^0 \to e^+ \mu^-$,
for which existing parameter bounds predict
\begin{equation}
{\cal B}r^{\not R_p}_{D^0\to\mu^+e^-} <0.5 \times10^{-6}\times
\left[ \left(\frac{\tilde{\lambda}'_{11k}}{0.02}\right)
\left(\frac{\tilde{\lambda}'_{22k}}{0.21}\right)
 +
\left(\frac{\tilde{\lambda}'_{21k}}{0.06}\right)
\left(\frac{\tilde{\lambda}'_{12k}}{0.04}\right)
\right] \ \ .
\label{d2mue_2body}
\end{equation}
An order-of-magnitude improvement in
${\cal B}r^{\not R_p}_{D^0\to\mu^+e^-}$ will provide a new
bound on the above combination of RPV couplings.

\subsection{Large Extra Dimensions}

For several years, the study of large extra dimensions
(`large' means much greater than the Planck scale) has
been an area of intense study.  This approach might hold the
solution of the hierarchy problem while having verifiable
consequences at the TeV scale or less.  Regarding the subject
of rare charm decays, one's reaction might be to ask
{\it How could extra
dimensions possibly affect the decays of ordinary hadrons?}.
We provide a few examples in the following.

Suppose the spacetime of our world amounts to a $3+1$ brane
which together with a manifold of additional dimensions
(the bulk) is part of some higher-dimensional space.
A field $\Theta$ that can propagate in a large extra
dimension will exhibit a Kaluza-Klein (KK) tower of
states $\{ \Theta_n \}$, detection of which would signal existence
of the extra dimension.  Given our ignorance regarding properties
of the bulk or of which fields are allowed to propagate
in it, one naturally considers a variety of different models.

Assume, for example, the existence of an extra dimension of
scale $1/R \sim 10^{-4}$~eV such that the gravitational field
(denote it simply as $G$) alone can propagate in the extra
dimension~\cite{add}.  There are then bulk-graviton KK states
$\{ G_n\}$ which couple to matter.  In principle there will be
FCNC transitions $c \to u ~G_n$ and, since the $\{ G_n\}$
remain undetected, there will be apparent missing energy.
However this mechanism leads to too small a rate to be observable.

Another possibility that has been studied is that the scale
of the extra dimension is $1/R \sim 1$~TeV and that SM gauge
fields propagate in the bulk~\cite{ant}.  However, precision
electroweak data constrain the mass of the first gauge KK excitation to
be in excess of 4 TeV~\cite{tgr} and, thus, their contributions to rare
decays are small~\cite{deshxd}.

More elaborate constructions, such as allowing fermion fields to
propagate in the five-dimensional bulk of the Randall-Sundrum
localized-gravity model~\cite{rs}, are currently being actively
explored~\cite{frank}.  Interesting issues remain and a good deal
more study deserves to be done.

\section{Summary}

FCNC charm decays are very rare in the SM  due to the strong
GIM suppression. In sharp contrast to $B\to X_{s} \gamma,
K^{*}\gamma $,  the SD contributions in radiative charm decays are
much smaller than the LD contributions in the SM, 
which makes the
extraction SD contribution from future measurements of $D\to
X_{u}\gamma$ and $D\to V\gamma $ more complicated. 
Although the LD contributions
still dominate the rates,  just as in the radiative decays,  there
are decays modes such as  $D\to \pi\ell^{+}\ell^{-}$ and  $D\to
\rho\ell^{+}\ell^{-}$  where it is possible to access the  SD
physics away from the resonance contributions in the low di-lepton
invariant mass region. As illustrated in Figs.~\ref{part5:rare:fig:pill}
and~\ref{part5:rare:fig:rhollmssm},  for low
di-lepton mass  the sum of long and short distance effects leaves
a large window where the physics beyond the SM can be observed.

 In summary,  the FCNC modes are most sensitive to the effects of
 some new physics scenarios.  If the sensitivity of experiment could
 reach below $10^{-6}$, these new physics effects could be tightly 
constrained. Such an experimental sensitivity, of course, makes many 
radiative $D$ decays accessible. However, these may not illuminate short 
distance physics.

Past searches have set upper
limits for the dielectron and dimuon decay modes~\cite{Part5_pdg06}. 
In Table~\ref{tab:rare} and Table~\ref{tab:rare_d0}, the current
limits and expected sensitivities at \bes3 are
summarized for $D^+$ and $D^0$, respectively.  Detailed
descriptions of rare charm decays can be found in Refs~\cite{Burdman,ian}.
Charmed meson radiative decays are also very important to
understand final state interaction that may enhance the decay rates. 
In Refs.~\cite{Burdman,ian}, the decay rates of $D \rightarrow V \gamma$ 
($V$ can be $\phi$, $\omega$, $\rho$ and $K^*$ ) are
estimated to be in the $10^{-5} - 10^{-6}$ range, which can be reached at 
\bes3. 

\begin{table}[htbp]
 \centering
\begin{tabular}{@{}llll}
\hline
                & Reference     & Best Upper  &  \bes3  \\
Mode            & Experiment              & limits($10^{-6}$)      & ($\times 10^{-8}$)  \\
\hline
$ \pi^+ e^+e^-$     & CLEO-c~\cite{cleo-c-rare} & 7.4 & 5.6  \\
$ \pi^+ \mu^+\mu^-$ & FOCUS~\cite{focus}  & 8.8 & 8.7  \\
$ \pi^+ \mu^+ e^-$ & E791~\cite{e791}  & 34 & 5.9  \\      
$ \pi^- e^+ e^+$ & CLEO-c~\cite{cleo-c-rare}  & 3.6 & 5.6  \\      
$ \pi^- \mu^+\mu^+$ & FOCUS~\cite{focus}  & 4.8 & 8.7  \\
$ \pi^- \mu^+ e^+$ & E791~\cite{e791}  & 50 & 5.9  \\      
$ K^+ e^+e^- $ & CLEO-c~\cite{cleo-c-rare}  & 6.2 & 6.7  \\    
$ K^+ \mu^+\mu^- $ & FOCUS~\cite{focus}  & 9.2 & 10.5  \\     
$ K^+ \mu^+ e^- $ & E791~\cite{e791}  & 68 & 8.3  \\    
$ K^- e^+ e^+ $ & CLEO-c~\cite{cleo-c-rare}  & 4.5 & 6.7  \\
$ K^- \mu^+ \mu^+ $ & FOCUS~\cite{focus}  & 13 & 10.4  \\
$ K^- \mu^+ e^+ $ & E687~\cite{e687_rare}  & 130 & 8.3  \\
\hline
\end{tabular}
\caption{ Current and projected 90\%-CL upper limits on
rare $D^+$ decay modes at \bes3 with a 20 fb$^{-1}$ data 
sample taken at the $\psi(3770)$ peak.}
\label{tab:rare}
\end{table}

\begin{table}[htbp]
 \centering
\begin{tabular}{@{}llll}
\hline
                & Reference     & Best Upper  &  \bes3  \\
Mode            & Experiment              & limits($10^{-6}$)      & ($\times 10^{-8}$)  \\
\hline
$\gamma \gamma $  & CLEO~\cite{cleo-d0} & 28 & 5.0  \\
$\mu^+\mu^-$ & D0~\cite{d0-coll}  & 2.4 & 17.0  \\
$\mu^+ e^-$ & E791~\cite{e791}  & 8.1 & 4.3  \\      
$e^+ e^-$ & E791~\cite{e791}  & 6.2 & 2.4  \\      
$\pi^0 \mu^+\mu^-$ & E653~\cite{e653_rare}  & 180 & 12.3  \\
$\pi^0 \mu^+ e^+$ & CLEO~\cite{cleo-d01}  & 86 & 9.7  \\      
$\pi^0 e^+e^- $ & CLEO~\cite{cleo-d01}  & 45 & 7.9  \\    
$K_S \mu^+\mu^- $ & E653~\cite{e653_rare}  & 260 & 10.6  \\     
$K_S \mu^+ e^- $ & CLEO~\cite{cleo-d01}  & 100 & 9.6 \\    
$K_S e^+ e^- $ & CLEO~\cite{cleo-d01}  & 110 & 7.5  \\
$\eta \mu^+ \mu^- $ & CLEO~\cite{cleo-d01}  & 530 & 15.0  \\
$\eta \mu^+ e^- $ & CLEO~\cite{cleo-d01}  & 100 & 12.0  \\
$\eta e^+ e^- $ & CLEO~\cite{cleo-d01}  & 110 & 10.0  \\
\hline
\end{tabular}
\caption{Current and projected 90\%-CL upper limits on
rare $D^0$ decay modes at \bes3 with a 20 fb$^{-1}$ data 
sample taken at the $\psi(3770)$ peak.}
\label{tab:rare_d0}
\end{table}

\part[$\tau$ Physics]{$\tau$ Physics\\
\vspace*{2cm}
 {\centering \Large Conveners \\
  Antonio Pich, Changzheng Yuan}\\
\vspace*{1cm}
\Large Contributors \\
I.~I.~Bigi, I.~R.~Boyko, D.~Dedovich, C.~D.~Fu, 
X.~H.~Mo, A.~Pich, A.~Stahl,
Y.~K.~Wang, and~C.~Z.~Yuan}
\label{part:six}
\renewcommand{\ee}   {\ensuremath{\mathrm{e}^+\mathrm{e}^-}}
\newcommand{\tauK} {\ensuremath{\tau^-\rightarrow \mathrm{K}^-\:\nu_\tau}}
\newcommand{\taupi}{\ensuremath{\tau^-\rightarrow \pi^-\:\nu_\tau}}
\newcommand{\tauKpo}{\ensuremath{\tau^-\rightarrow \mathrm{K}^-\:\pi^0\:\nu_\tau}}

\chapter[Tau Physics near Threshold]{Tau Physics near Threshold\footnote{By Achim Stahl}}
\label{chp:achim}

\section{Introduction}	
\label{sec:introduction}

The tau-charm factory currently under construction at the Institute of
High Energy Physics in Beijing will give us excellent opportunities for
interesting physics with tau leptons. Such an {\ee} collider, running in the
energy regime of the tau and charm thresholds, will produce large samples of
tau leptons. It could produce up to 50 million tau pairs per year, but this
is not the real advantage of the tau-charm factory.
Today the $B$-factories have tau samples of several hundred million tau 
pairs 
and the LHC will produce {$10^{12}$} tau pairs per
year even at low luminosity.
The advantages  of the tau-charm factory are the excellent experimental
conditions that will allow the experiments to analyze many aspects of
the tau decays with low systematics. Aspects that are hardly accessible
at other machines.

The cross section for the production of tau pairs rises above threshold like
\begin{equation}
\sigma(s) = \frac{4\,\pi\,\alpha^2}{3\,s}\:\beta\:
            \frac{3\,-\,\beta^2}{2}\, ,
\label{eqn:crosssection}
\end{equation}
where {$s$} is the center-of-mass energy and {$\beta$} the velocity of one of
the tau leptons in the overall rest frame.
The formula is the result of a first-order calculation.
More detailed calculations
are available \cite{achim_Ruiz-Femenia:2002wm}. The main backgrounds are the
production of light quark pairs ($u$-, $d$-, and $s$-quarks) with a cross 
section of
approximately {$173.6\;\mathrm{nbarn} / s$} ($s$ in {$\mathrm{GeV}^2$}) 
and
charm production. For tau physics there are four interesting running points
in this region:
\begin{itemize}
 \item[1)] {$\sqrt s = 3.50\:\mathrm{GeV}$}\\
       This energy is just below the production threshold for tau pairs.
       It allows one to measure and investigate the light quark 
       background.
       The experimentally determined background at this point can
       be extrapolated to the other running points with the help of a Monte
       Carlo program. The background varies only slightly with the
       center-of-mass energy and no significant systematic error is expected
       from the extrapolation.
 \item[2)] {$\sqrt s = 3.55\:\mathrm{GeV}$}\\
       This is right at the production threshold and the tau leptons are
       produced at rest. The cross section is
       different from zero due to the Coulomb attraction between the
       {$\tau^+$} and {$\tau^-$}. The cross section at threshold is
       0.1 nbarn for ideal beams. For a realistic number the energy spread
       of the machine must be taken into account. This is the ideal energy
       for the study of tau decays as explained in the following section.
 \item[3)] {$\sqrt s = 3.69\:\mathrm{GeV}$}\\
       This is an energy above the {$\Psi(2S)$}, but still below the production
       threshold for open charm. The tau cross section has risen to 2.4 nbarn.
       The background situation is as good as at point 2, but the tau leptons
       are no longer produced at rest. The kinematic identification
       described in the following section will not work here.
 \item[4)] {$\sqrt s = 4.25\:\mathrm{GeV}$}\\
       At this energy the cross section for tau production is at its maximum
       (3.5 nbarn). But here there is substantial charm background on top
       of the light quark background.
\end{itemize}

\begin{figure}[pb]
   \centerline{\psfig{file=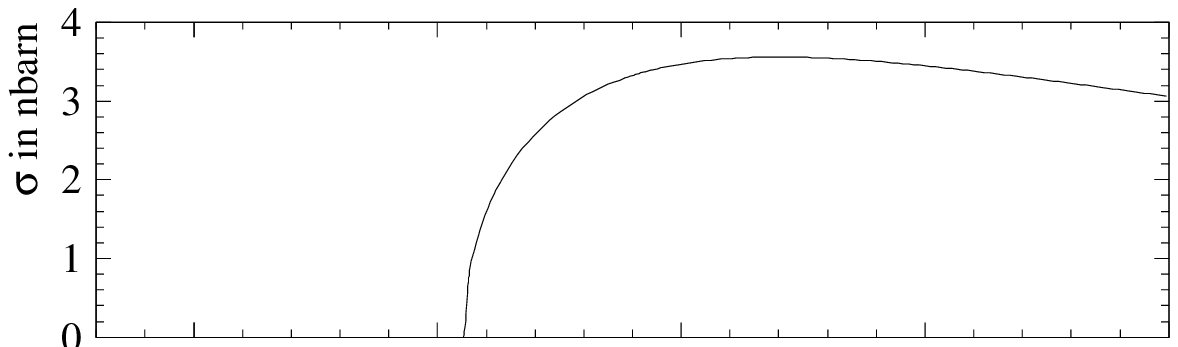,width=12.0cm}}
   \vskip .4cm
   \centerline{\psfig{file=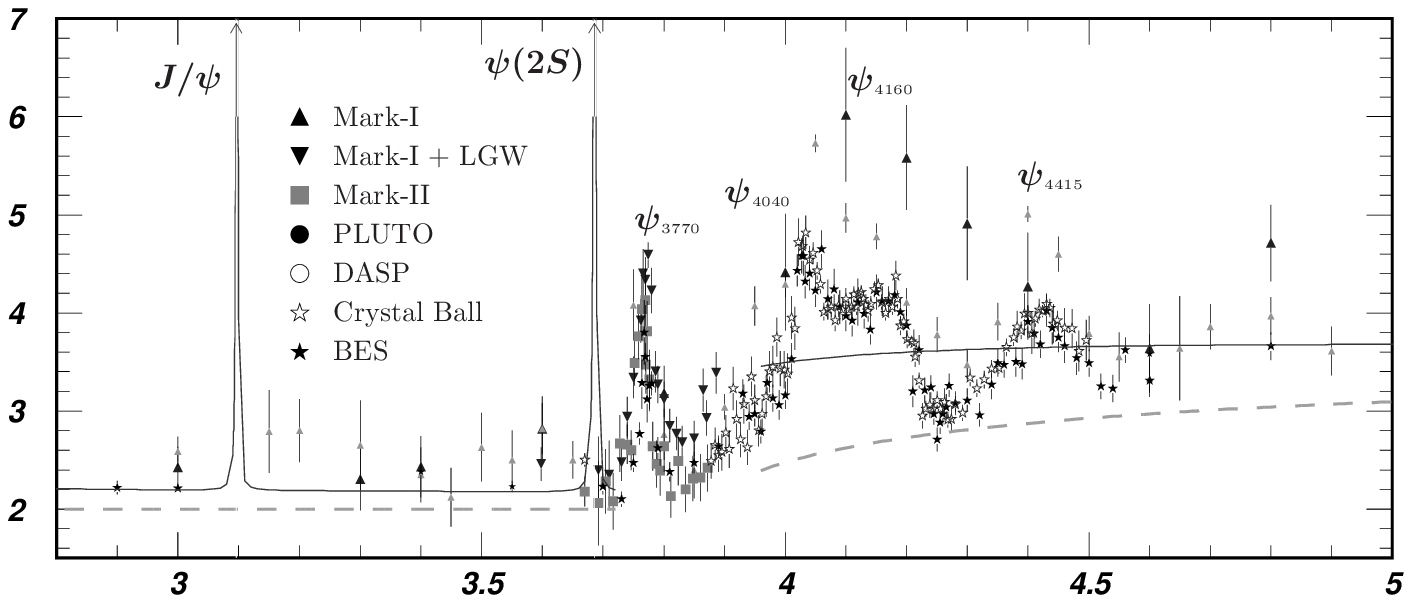,width=12.0cm}}
\vspace*{8pt}
\caption{Top: Production cross section for tau pairs. Bottom: The 
$R$-ratio
showing the expected background from quark production (from PDG). Cross
section and $R$-ratio are shown versus the center-of-mass energy in GeV on
identical scales.
\label{fig:crosssection}}
\end{figure}

\section{Tau Production at Threshold}	
Data taking right at the tau production threshold, where the tau leptons are
produced at rest, is the most favored situation. 
In addition to its two leptonic
decay modes, the tau lepton decays with a branching fraction of
{$64.8\:\%$}~\cite{pich_PDG} into a tau neutrino plus hadrons.
The decay can be summarized as a two-body decay {$\tau^- \rightarrow 
\nu_\tau \mathrm{had}^-$} where the hadronic system {$\mathrm{had}^-$} 
might fragment
into several mesons.\footnote{All arguments apply similarly to the 
{$\tau^+$}}
This description is kinematically correct even if no
identifiable intermediate resonance is present.
As a consequence, the hadronic system has a fixed energy and momentum in 
the
tau rest frame:
\begin{equation}
  E_{\mathrm{had}}^\star =\frac{m_\tau^2 + m^2_\mathrm{had}}{2\,m_\tau},
  \hspace{1cm}
  p_{\mathrm{had}}^\star =\frac{m_\tau^2 - m^2_\mathrm{had}}{2\,m_\tau},
  \label{eqn:Ehad}
\end{equation}
where {$m_\tau = 1776.99\:\mathrm{MeV}$} is the mass of the tau lepton and
{$m_\mathrm{had}$} the invariant mass of the hadronic system in this
particular decay. A mass of the tau neutrino in the sub-eV range is negligible
in this context. At the production threshold, the tau leptons are at rest 
in the center-of-mass system and the hadron energy in this system is 
directly given by {$E_{\mathrm{had}}^\star$}. Above threshold the energy 
of the hadronic system is given by
\begin{equation}
  E_{\mathrm{had}} = \gamma\,E_{\mathrm{had}}^\star + \beta\,\gamma\,
                     p_{\mathrm{had}}^\star\,\cos\theta^\star ,
\end{equation}
with the Lorentz factors {$\beta$} and {$\gamma$} and the unknown decay angle
{$\theta^\star$}.
The monoenergetic spectrum at threshold broadens under the influence of the
angle {$\theta^\star$}.
For example, for a {$\rho$} meson with mass of 770 MeV, the 
the $E_{\mathrm{had}}^\star$=1055~MeV value at threshold
is smeared out to have a {$\pm 24\:\mathrm{MeV}$} width for 
production just 1~MeV above threshold. The kinematic constraint rapidly 
vanishes with increasing center-of-mass energy.

\begin{figure}[pb]
\centerline{\psfig{file=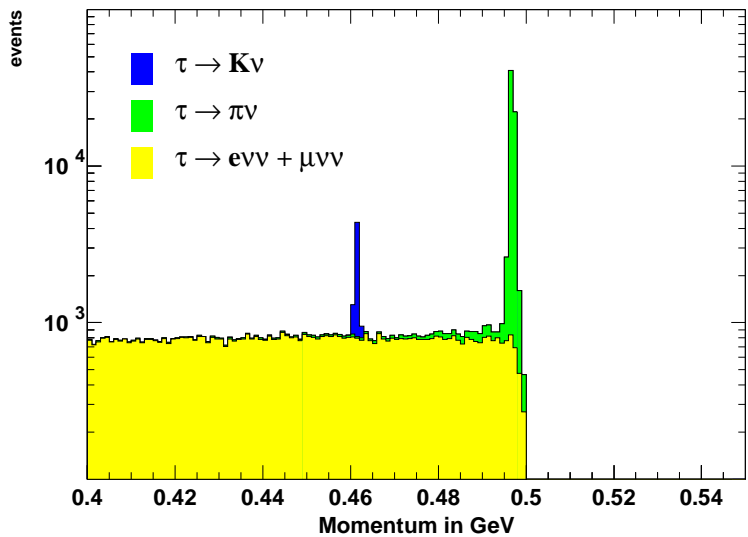,width=10.0cm}}
\vspace*{8pt}
\caption{Momentum spectrum of charged tracks from tau leptons produced at
rest at the production threshold. No particle identification applied.}
\label{fig:kaon}
\end{figure}

Two-body kinematics at threshold provide powerful possibilities to 
distinguish different hadronic
decay channels from kinematics only. In simple decays 
such as, for example, 
{\tauK} (branching ratio {$0.691\:\%$}) the kaons
must have a fixed energy or momentum in the center-of-mass system.
Figure \ref{fig:kaon} shows the expected spectrum
derived from a fast simulation of 300.000 events generated with
TAUOLA \cite{achim_TAUOLA}.
A momentum resolution of {$\Delta p/p = 0.32\:\%\times p \oplus 0.37\:\%$}
with the two terms added in quadrature is used, similar to the expected
resolution of the \bes3 detector. No particle identification is applied.
All negative tracks in the events are considered. For clarity, the plot shows
only the background from the leptonic decays (branching ratio of {$35.3\:\%$})
and {\taupi} (branching ratio {$10.90\:\%$}).
The others are small. The peak for the kaon decay is very well separated from
the decay to the pion.

\begin{figure}[pb]
\centerline{\psfig{file=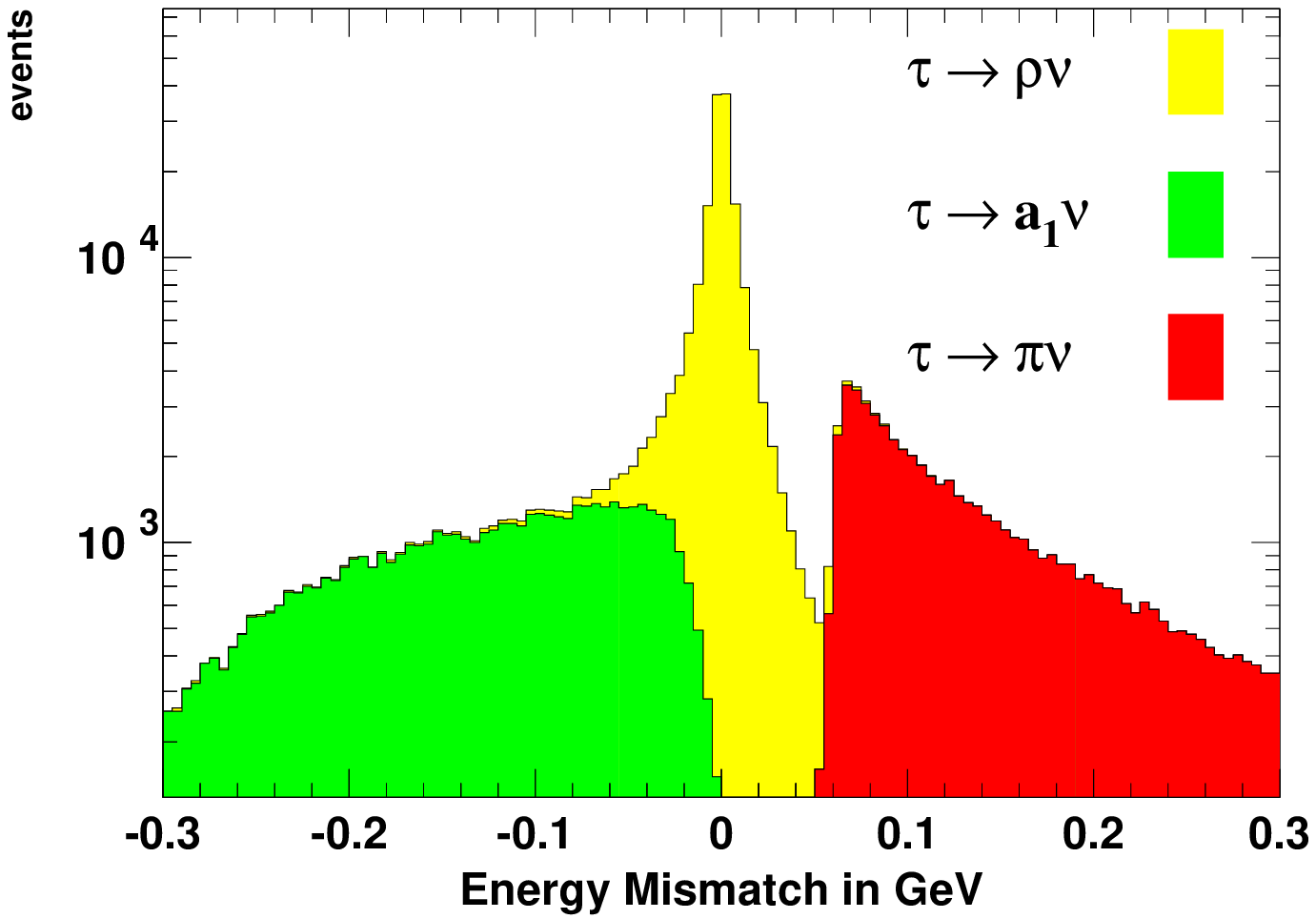,width=10.0cm}}
\centerline{\psfig{file=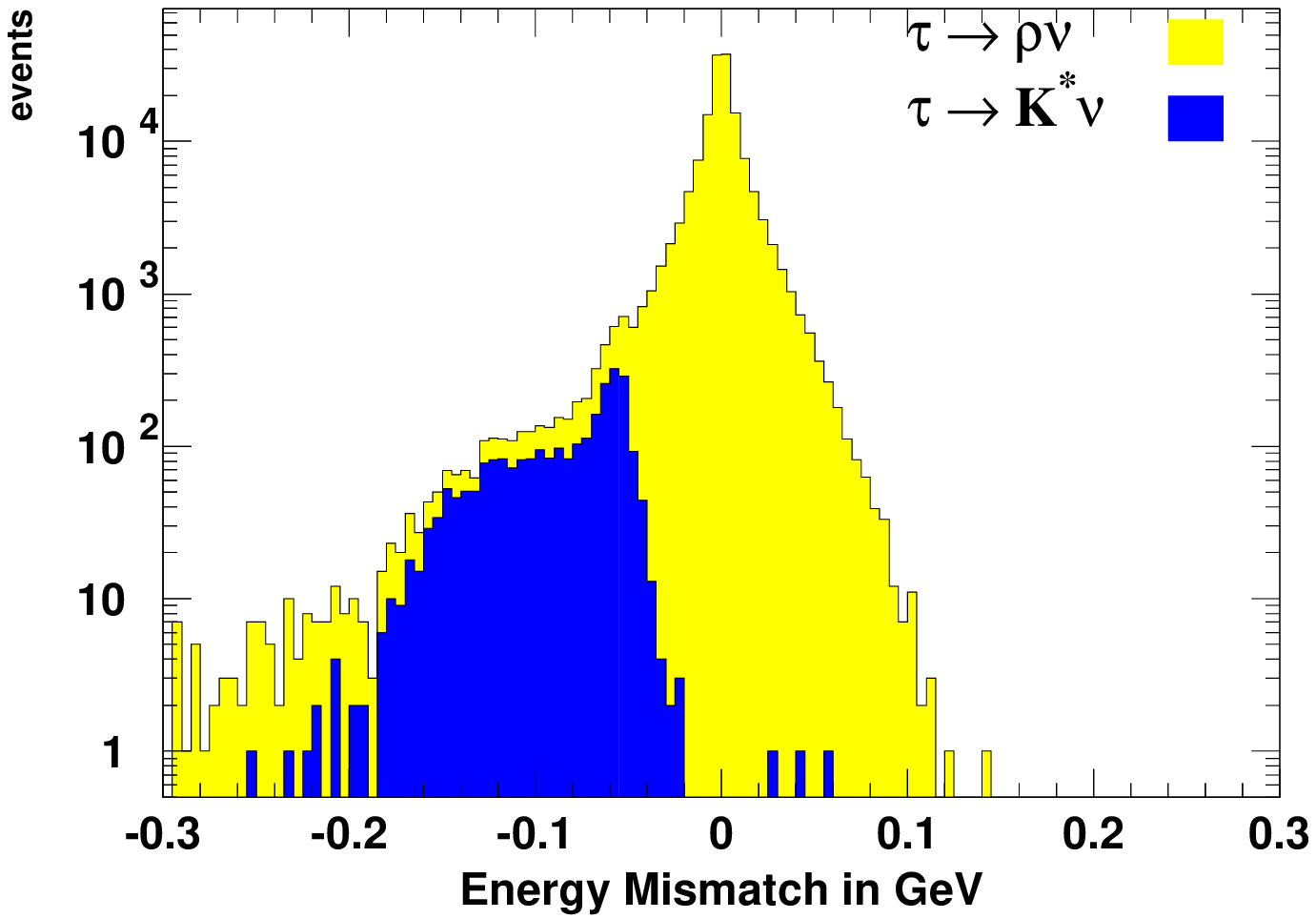,width=10.0cm}}
\vspace*{8pt}
\caption{Measured energy minus predicted energy for a {$\tau^- \rightarrow
\pi^-\:\pi^0\:\nu_\tau$} hypothesis for negative charged tracks and the
most energetic {$\pi^0$} in the events. The two plots show the signal together
with the main backgrounds. No particle identification has been applied.}
\label{fig:rho}
\end{figure}

In more complicated decays, the detector records certain particles from
the decays of the two tau leptons and measures their momenta.
To identify the decay of the {$\tau^-$} a hypothesis is formed about which
particles belong to its decay and what their identity is (whether they are
pions or kaons). The invariant mass is calculated under this hypothesis and
then {$E_{\mathrm{had}}$} can be predicted from Eq.~\ref{eqn:Ehad}.
The hypothesis is correct only if the predicted energy of the hadronic system
matches the measured one. Figure \ref{fig:rho} illustrates the identification.
It shows the difference of the measured and the predicted energy. Again
300,000 events generated with TAUOLA are processed through a fast 
simulation
with the same momentum resolution for charged tracks as above and an energy
resolution for neutral pions of {$\Delta E / E = 2\:\% / \sqrt[4]{E} \oplus
2.5\:\%$}. Only the most energetic neutral pion in an event is considered;
the others are assumed to be lost. A simple model for the creation of fake
{$\pi^0$}'s from charged hadrons is also included in the simulation. 
Again, 
the signal can be well separated from the background without any particle
identification.

Because of these kinematic properties, running directly at threshold (point 2
in section~\ref{sec:introduction}) is the preferred configuration 
for for a number of tau physics measurements.

\section{Physics Opportunities}

The tau physics opportunities at a tau-charm factory are numerous, even 
compared to the high statistics of the $B$-factories. For many 
measurements, statistics are not the problem; 
efficiency, background and resolution are the main issues.
For these measurements, the tau-charm factory is in a better situation.
A list of measurements that a tau-charm factory should be
able to improve significantly is given below.\\
\begin{itemize}
 \item Measurement of the non-strange spectral function with low background
       and much improved mass resolution.
 \item First measurement of the strange spectral function with good statistics.
       Separation into vector and axial-vector parts through a detailed
       analysis of all decay modes.
       This improvement, together with the first point, will reduce the
       uncertainties of the measurement of the strong coupling constant
       {$\alpha_s$} from tau decays.
 \item Identification and measurement of second-class currents and 
       Wess-Zumino anomalies.
 \item Tests of calculations from chiral perturbation theory at low 
       {$Q^2$} and
       a determination of their range of validity.
 \item Many branching ratios can be improved, especially those with charged
       kaons.
 \item Some exclusive channels provide good opportunities for spectroscopy
       of light mesons, for example the measurement of the mass and width of
       the charged and neutral $\rho$ mesons.

 \item It would be interesting to improve the knowledge on the tau mass which
       is currently dominated by the BES-II 
       measurement~\cite{achim_Bai:1995hf}.
       This would imply the reduction of all systematics including the
       knowledge of the energy spread of the machine and its stability.
 \item The measurement of the Michel parameters can be improved for the
       leptonic decays and the hadronic decays due to higher statistics
       combined with lower background. This is especially interesting as the
       LHC might tell us for what kind of new currents we are looking for.
 \item Other more exotic topics might be the search for a deviation of the
       g-factor of the tau lepton from 2 or the search for $CP$ violation 
       in tau decays.
\end{itemize}
Measurements that cannot be improved at a tau-charm factory are the tau
lifetime, the neutral current couplings, the mass of the tau neutrino,
a search for very rare or forbidden decays, or 
a search for $CP$-violation in the {\em production} of tau leptons.

\section{Running Strategy}

The tau-charm factory is a machine that can do very interesting physics in
several areas and a running strategy must balance the requests of the 
different communities of users. 
Two months of running for tau physics should be
sufficient during the initial operation of the machine and below threshold.

Running on threshold would produce 100,000 very clean tau pairs. From
kinematically tagged decays one should be able to study the particle
identification (time-of-flight, $dE/dx$, calorimeter, etc.), to understand
the efficiencies, purities, and resolution, and to tune the detector
simulation. A few physics analysis can already be improved from this data set,
for example the {\tauK} branching ratio or the {\tauKpo} mass spectrum.

Running below threshold is necessary to understand the background from
the production of light quarks and to tune the corresponding Monte Carlo
generators. Once this has been done, the Monte Carlo can be used to subtract
the background at the other running points.

The understanding of the identification of the tau leptons and their decay
modes from the initial running at threshold will enable the collaboration
to use luminosity collected at a later stage at higher energies for charm
physics to do some of the tau measurements.
The experience gained during the initial running at threshold will be a good
basis to decide how much more luminosity one would like to accumulate 
at the tau threshold at some later stage.

%

\chapter[Tau Decays]{Tau Decays}
\label{sec:tau}

\section[Hadronic $\tau$ Decays]{Hadronic $\tau$ Decays}
\label{sec:hadronic_tau}
The pure leptonic or semileptonic character of $\tau$ decays
provides a clean laboratory for performing very precise tests of the
electroweak gauge structure at the 0.1\% to 1\% level. Moreover, 
hadronic $\tau$ decays turn out to be a beautiful laboratory for
studying strong interaction effects at low energies
\cite{pich_taurev98,pich_taurev06a,pich_taurev06b,pich_Stahl00}. 
Accurate determinations
of the QCD coupling and the strange quark mass have been obtained
with $\tau$-decay data. More recently, a very competitive estimate of
the quark mixing $|V_{us}|$ has been also extracted from Cabibbo
suppressed $\tau$ decays.

The excellent experimental conditions offered by the tau-charm
factory will allow for further analyses of many aspects of $\tau$ physics
with low systematics. The BEPCII collider could produce large
statistical samples as many as
50 million $\tau^+\tau^-$ pairs per year,  providing the opportunity
for an extensive programme of high-precision measurements
with $\tau$ leptons. The $B$-factories have already produced  much
larger data samples, which will be further increased at LHC and, if
approved, at future Super-$B$ factories. However, the threshold
region makes possible a much better control of backgrounds and
systematic errors for a number of measurements. 
Thus, the tau-charm factory combines the optimum
conditions to perform a number of very accurate measurements.

\subsection[A Laboratory for QCD]{A Laboratory for QCD\footnote{By Antonio Pich}}
\label{sec:part6:qcd}

The inclusive character of the total $\tau$ hadronic width renders
possible an accurate calculation of the ratio [$(\gamma)$ represents
additional photons or lepton pairs]
\begin{equation}\label{eq:r_tau_def}
R_\tau \equiv { \Gamma [\tau^- \to \nu_\tau \,\mathrm{hadrons}\,
(\gamma)] \over \Gamma [\tau^- \to \nu_\tau e^- {\bar \nu}_e
(\gamma)] }\, = \, R_{\tau,V} + R_{\tau,A} + R_{\tau,S}\, ,
\end{equation}
using analyticity constraints and the Operator Product Expansion
\cite{pich_BR:88,pich_NP:88,pich_BNP:92,pich_LDP:92a,pich_QCD:94}. One can separately compute
the contributions associated with specific quark currents.
$R_{\tau,V}$ and $R_{\tau,A}$ correspond to the Cabibbo--allowed
decays through the vector and axial-vector currents, while
$R_{\tau,S}$ contains the remaining Cabibbo--suppressed
contributions.

The theoretical prediction for $R_{\tau,V+A}$ can be expressed as
\cite{pich_BNP:92}
\begin{equation} R_{\tau,V+A} = N_C\, |V_{ud}|^2\, S_{\mathrm{EW}} \left\{ 1 +
\delta_{\mathrm{P}} + \delta_{\mathrm{NP}} \right\} , \end{equation}
where $N_C=3$ denotes the number of quark colours and
$S_{\mathrm{EW}}=1.0201\pm 0.0003$ contains the electroweak
radiative corrections \cite{pich_MS:88}. The dominant correction ($\sim
20\%$) is the perturbative QCD contribution $\delta_{\mathrm{P}}$,
which is fully known to $O(\alpha_s^3)$ \cite{pich_BNP:92} and includes a
resummation of the most important higher-order effects
\cite{pich_LDP:92a}.

Non-perturbative contributions are suppressed by six powers of the
$\tau$ mass \cite{pich_BNP:92} and, therefore, are very small. Their
numerical size has been determined from the invariant--mass
distribution of the final hadrons in $\tau$ decay, through the study
of weighted integrals that can be calculated theoretically in the
same way as $R_{\tau}$ \cite{pich_LDP:92b}:
\begin{equation}\label{eq:klMoments}
R_{\tau}^{kl} \equiv \int_0^{m_\tau^2} ds\, \left(1 - {s\over
m_\tau^2}\right)^k\, \left({s\over m_\tau^2}\right)^l\, {d
R_{\tau}\over ds} \, .
\end{equation}
The predicted suppression \cite{pich_BNP:92} of the non-perturbative
corrections has been confirmed by ALEPH \cite{pich_ALEPH:05}, CLEO
\cite{pich_CLEO:95} and OPAL \cite{pich_OPAL:98}. The most recent analysis
\cite{pich_ALEPH:05} gives
\begin{equation}\label{eq:del_np}
\delta_{\mathrm{NP}} \, =\, -0.0043\pm 0.0019 \, .
\end{equation}

The QCD prediction for $R_{\tau,V+A}$ is then completely dominated
by the perturbative contribution; non-perturbative effects being
smaller than the perturbative uncertainties from uncalculated
higher-order corrections. The result turns out to be very sensitive
to the value of $\alpha_s(m_\tau^2)$, 
thereby allowing for an accurate
determination of the fundamental QCD coupling \cite{pich_NP:88,pich_BNP:92}.
The experimental measurement $R_{\tau,V+A}= 3.471\pm0.011$ implies
\cite{pich_DHZ:05}
%
\begin{equation}\label{eq:alpha}
\alpha_s(m_\tau^2)  =  0.345\pm
0.004_{\mathrm{exp}}\pm 0.009_{\mathrm{th}} \, .
\end{equation}
%

%
\begin{figure}[t]
\label{fig:alpha_s} \centering
\includegraphics[width=10cm]{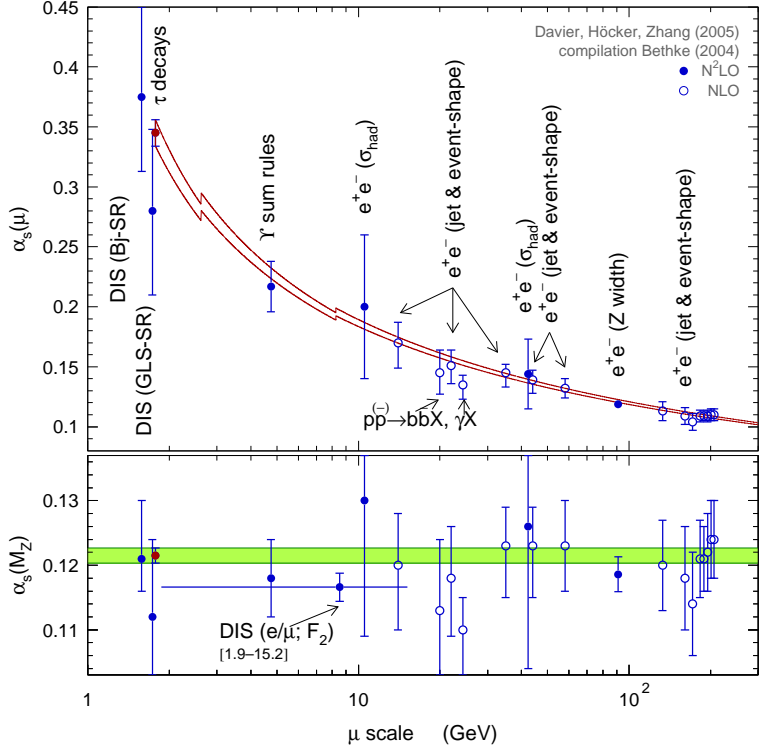}
\caption{Measured values of $\alpha_s$ at different scales. The
curves show the energy dependence predicted by QCD, using
$\alpha_s(m_\tau^2)$ as input. The corresponding extrapolated
$\alpha_s(M_Z^2)$ values are shown at the bottom, where the shaded
band displays the $\tau$ decay result \cite{pich_DHZ:05}.}
\end{figure}

The strong coupling measured at the $\tau$ mass scale is
significantly larger than the values obtained at higher energies.
From the hadronic decays of the $Z$, one gets $\alpha_s(M_Z^2) =
0.1186\pm 0.0027$ \cite{pich_LEPEWWG}, which differs from the $\tau$
decay measurement by more than twenty standard deviations. After
evolution up to the scale $M_Z$ \cite{pich_Rodrigo:1998zd}, the strong
coupling constant in Eq.~\ref{eq:alpha} 
decreases to \cite{pich_DHZ:05}
\begin{equation}\label{eq:alpha_z}
\alpha_s(M_Z^2)  =  0.1215\pm 0.0012 \, ,
\end{equation}
in excellent agreement with the direct measurements at the $Z$ peak
and with a similar accuracy. The comparison of these two
determinations of $\alpha_s$ in two extreme energy regimes, $m_\tau$
and $M_Z$, provides a beautiful test of the predicted running of the
QCD coupling; {\it i.e.}, a very significant 
experimental verification of
{\it asymptotic freedom}.

With $\alpha_s(m_\tau^2)$ fixed to the value in
Eq.~\ref{eq:alpha}, the same theoretical framework gives definite
predictions for the semi-inclusive $\tau$ decay widths $R_{\tau,V}$,
$R_{\tau,A}$ and $R_{\tau,S}$, in good agreement with the
experimental measurements. Moreover, using the measured invariant
mass distributions, one can study (for each separate $V$, $A$ and
$S$ component) 
the integrated moments defined in Eq.~\ref{eq:klMoments},
with arbitrary weight functions and/or varying the upper end of
integration in the range $s_0\le m_\tau^2$. This allows one to
investigate many non-perturbative aspects of the strong interactions
\cite{pich_PI:89}. For instance, $R_{\tau,V}-R_{\tau,A}$ is a pure
non-perturbative quantity; basic QCD properties force the associated
mass distribution to obey a series of chiral sum rules
\cite{pich_PI:89,pich_WE:67}, which relate the $\tau$ measurements with
low-energy non-perturbative observables such as the pion decay
constant $f_\pi$ or the electromagnetic pion mass difference
$m_{\pi^\pm}-m_{\pi^0}$. One can also extract the non-perturbative
contributions to the OPE of the QCD vector and axial-vector current
correlators. The determination of these effects is needed to perform
many theoretical predictions of other important observables, such as, for
instance, the kaon $CP$-violating ratio $\varepsilon'/\varepsilon$. The
measured vector spectral distribution can also be used to estimate
the hadronic vacuum polarization contribution to $\alpha(M_Z)$ and
to the muon anomalous magnetic moment.

\subsection[Determinations of $m_s$ and $V_{us}$ in Hadronic $\tau$ Decays]
{Determinations of $m_s$ and $V_{us}$ in Hadronic $\tau$ Decays\footnote{By
A. Pich}}
\label{sec:vus_tau}

Separate measurements of the $|\Delta S|=0$ and $|\Delta S|=1$ \
$\tau$ decay widths will allow us to pin down the $SU(3)_{fl}$ breaking 
effects induced by the strange quark 
mass~\cite{pich_Davier,pich_PP:99,pich_ChDGHPP:01,pich_ChKP:98,pich_MW:06,pich_GJPPS:05,pich_BChK:05},
through the differences \cite{pich_PP:99}
\begin{equation}
\delta R_\tau^{kl} \;\equiv\;
  {R_{\tau,V+A}^{kl}\over |V_{ud}|^2} - {R_{\tau,S}^{kl}\over |V_{us}|^2}
  \;\approx\;  24\, {m_s^2(m_\tau^2)\over m_\tau^2} \, \Delta_{kl}(\alpha_s)
  - 48\pi^2\, {\delta O_4\over m_\tau^4} \, Q_{kl}(\alpha_s)\, .
\end{equation}
The perturbative QCD corrections $\Delta_{kl}(\alpha_s)$ and
$Q_{kl}(\alpha_s)$ are known to $O(\alpha_s^3)$ and $O(\alpha_s^2)$,
respectively \cite{pich_PP:99,pich_BChK:05}.
Since the longitudinal contribution to $\Delta_{kl}(\alpha_s)$ does
not converge well, the $J=0$ QCD expression is replaced by its
corresponding phenomenological hadronic parametrization
\cite{pich_GJPPS:05}, which is much more precise because it is 
strongly dominated
by the well known kaon pole. The small non-perturbative
contribution, $\delta O_4 \equiv\langle 0| m_s \bar s s - m_d \bar d
d |0\rangle
 = -(1.5\pm 0.4)\times 10^{-3}\;\mbox{\rm GeV}^4$,
has been estimated with Chiral Perturbation Theory techniques
\cite{pich_PP:99}.

%
%
\begin{figure}[t]
\label{fig:DS=1} \centering
\includegraphics[width=10cm]{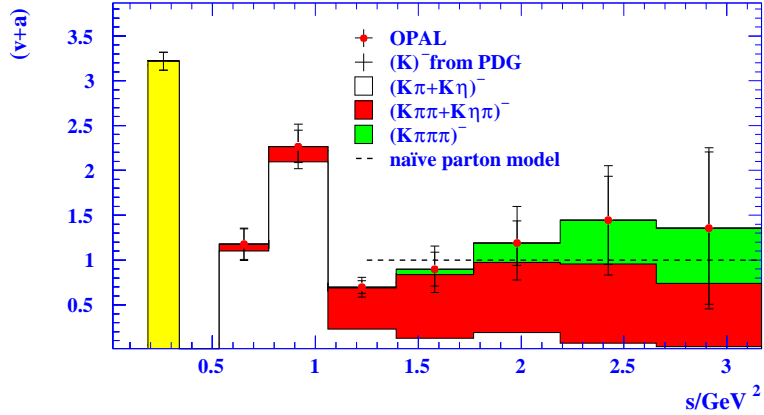}
\caption{The OPAL measurement of the spectral distribution in $|\Delta
S|=1$ $\tau$ decays~\cite{pich_OPALms}.}
\end{figure}

From the measured moments $\delta R_\tau^{k0}$ ($k=0,1,2,3,4$)
\cite{pich_ALEPHms,pich_OPALms}, it is possible to determine the strange quark
mass; however, the extracted value depends sensitively on the
modulus of the Cabibbo--Kobayashi--Maskawa matrix element
$|V_{us}|$. It appears, then, more natural to turn things around and, with
an input for $m_s$ obtained from other sources, 
determine $|V_{us}|$~\cite{pich_GJPPS:05}. The most sensitive moment is
$\delta R_\tau^{00}$:
\begin{equation}\label{eq:Vus_formula} |V_{us}|^2 =
\frac{R^{(0,0)}_{\tau,S}}{\frac{R^{(0,0)}_{\tau,V+A}}{|V_{ud}|^2}-\delta
R^{(0,0)}_{\tau,{\mathrm{th}}}} \, .
\end{equation}
Using $m_s(2~\mathrm{GeV})= (94\pm 6)~\mathrm{MeV}$, which includes
the most recent determinations of $m_s$ from lattice and QCD Sum
Rules \cite{pich_JOP:06}, one obtains $\delta R^{00}_{\tau,{\mathrm{th}}}
= 0.240 \pm 0.032$ \cite{pich_GJPPS:05}. This prediction is much smaller
than $R^{(0,0)}_{\tau,V+A}/|V_{ud}|^2$, making the theoretical
uncertainty in Eq.~\ref{eq:Vus_formula} negligible in comparison with
the experimental inputs $R^{(0,0)}_{\tau,V+A}=3.471\pm 0.011$ and
$R^{(0,0)}_{\tau,S}=0.1686\pm 0.0047$ \cite{pich_DHZ:05}. Taking
$|V_{ud}|=0.97377\pm 0.00027$ \cite{pich_PDG}, one gets~\cite{pich_GJPPS:05}
\begin{equation}\label{eq:Vus_value}
 |V_{us}| =0.2220 \pm 0.0031_{\mathrm{exp}} \pm
0.0011_{\mathrm{th}}
\, .
\end{equation}
This result is competitive with the standard $K_{e3}$ determination,
$|V_{us}| =0.2234 \pm 0.0024$ \cite{pich_SMEW:07}. The precision should be
considerably improved in the near future because the error is
dominated by the experimental uncertainty, which can be reduced with
the much better data samples from BaBar, Belle and \bes3.
Thus,  $\tau$ data has the potential to provide the best
determination of $|V_{us}|$.

With future high-precision $\tau$ data, a simultaneous fit of $m_s$
and $|V_{us}|$ should also become possible. A better understanding
of the perturbative QCD corrections $\Delta_{kl}(\alpha_s)$ would be
very helpful to improve the precision on 
$m_s$~\cite{pich_MW:06,pich_GJPPS:05}.


\subsection[Tau Hadronic Spectral Functions]{Tau Hadronic Spectral
Functions\footnote{By I.~Boyko and D.~Dedovich}}
\label{part6:sec:tau_spectral}

Hadronic tau decays give unique possibilities for performing detailed
investigations of  hadronic production from the QCD vacuum through the 
determination of 
{\it spectral functions}. Spectral functions play an important role in the 
understanding of hadron dynamics  in the intermediate
 energy range and provide
the basic input for QCD studies and for calculation of the low-energy 
contributions  from the hadronic vacuum polarization.

\subsubsection{Definition}
The spectral function $ v_1(a_1,a_0) $, where subscript refers to 
the spin of hadronic system, is defined for a non-strange (${\Delta}S=0 $) 
or strange ( ${\Delta}S=1 $) vector (axial-vector)
 tau decay channel $ V^{-}\nu_\tau $
($A^- \nu_\tau$), and is obtained by dividing
the invariant mass-squared distribution $(1/N_{V/A})(dN_{V/A}/ds) $
by the appropriate kinematic factor and is normalized by the ratio
 of vector/axial vector branching fraction  
${\cal B}(\tau \rightarrow V^-/A^-\nu_\tau)$ to the 
branching fraction of  the decay to a massless lepton (electron) :.

\begin{equation}
\label{eq:spectral_v1/a1}
v_1/a_1=\frac{m^2_{\tau}}{6|V_{CKM}|^2S_{EW}}
~\frac{{\cal B}(\tau \rightarrow V^-/A^-\nu_\tau)}{{\cal B}(\tau \rightarrow e^-\nu_e\nu_\tau)}
~\frac{dN_{V/A}}{N_{V/A}ds} 
{\left [{\left ( 1-\frac{s}{m^2_\tau}\right ) }^2 \left 
(1+\frac{2s}{m^2_\tau} \right ) \right ]}^{-1} ,
\end{equation}

\begin{equation}
\label{eq:spectral_a0}
a_0=\frac{m^2_{\tau}}{6|V_{CKM}|^2S_{EW}}
~\frac{{\cal B}(\tau \rightarrow \pi^-\nu_\tau)}{{\cal B}(\tau \rightarrow e^-\nu_e\nu_\tau)}
~\frac{dN_{A}}{N_{A}ds}
 {\left (1-\frac{s}{m^2_\tau}\right )}^{-2} .
\end{equation}

\noindent Here $ S_{EW} $ is the electroweak radiative correction factor
that is  introduced in the previous section.
Due to the conserved vector current (CVC), there is no $ J=0 $  
contribution 
to the vector spectral function;
the only contribution to $ a_0 $ is  assumed to be 
from the pion pole, with $ dN_A/N_Ads=\delta (s-m^2_\pi) $ .

Using unitarity and analyticity, the spectral functions are connected 
to the imaginary parts of the two-point correlation (hadronic vacuum 
polarization) functions~\cite{part6:vacpol1,part6:vacpol2}

\begin{eqnarray}
\label{eq:vac_pol}
\Pi^\mu_{i,j,U} (q^2) & \equiv & \int{d^4xe^{iqx}\langle 0 | T(U^\mu_{ij}(x)U^\mu_{ij}(0)^+)|0\rangle} \nonumber\\
& = & (-g^{\mu\nu}q^2 + q^\mu q^\nu)\Pi^{(1)}_{ij,U}(q^2)+q^\mu 
q^\nu\Pi^{(0)}_{ij,U}(q^2) ,
\end{eqnarray}

\noindent of vector( $ U^{\mu\nu}_{ij} = V^\nu_{ij} = \overline{q}_j \gamma^\mu q_i $ ) or
 axial-vector ( $ U^\mu_{ij} = V^\mu_{ij} = \overline{q}_j \gamma^\mu 
\gamma_5 q_i $ ) quark currents for 
time-like momenta-squared $ q^2>0 $.
The polarization functions $ \Pi^{\mu\nu}_{i,j,U} (s) $ have a branch cut along the real axis in the complex $ s=q^2 $ plane. Their 
imaginary parts give the spectral functions defined 
in Eq.~\ref{eq:spectral_v1/a1}. For non-strange currents:

\begin{eqnarray}
\label{eq:pol-spectral}
Im \Pi^{(1)}_{\overline{u}d,V/A}(s) = \frac{1}{2\pi}v_1/a_1(s) , \nonumber\\
Im\Pi^{(0)}_{\overline{u}d,A}(s) = \frac{1}{2\pi}v_1/a_0(s) .
\end{eqnarray}

Analytic  functions $\Pi^{(J)}{ij,U} (q^2)$ obey the dispersion relation

\begin{equation}
\label{eq:pol_dispers}
\Pi^{(J)}{ij,U} (q^2)=\frac{1}{\pi} 
\int^\infty_0{ds\frac{Im\Pi^{(J)}_{ij,U} (s)}{s-q^2-i\epsilon}} .
\end{equation}
\noindent
This dispersion relation allows one to connect the experimentally 
accessible spectral functions
to the correlation functions $ \Pi^{(J)}{ij,U} (q^2)$, which can be derived from QCD theory and 
are used for theoretical calculations of total cross sections and decay widths.

\subsubsection{Tau spectral functions and electron-positron annihilation data}
\begin{figure}[h]
  \begin{center} \mbox{\epsfig{file=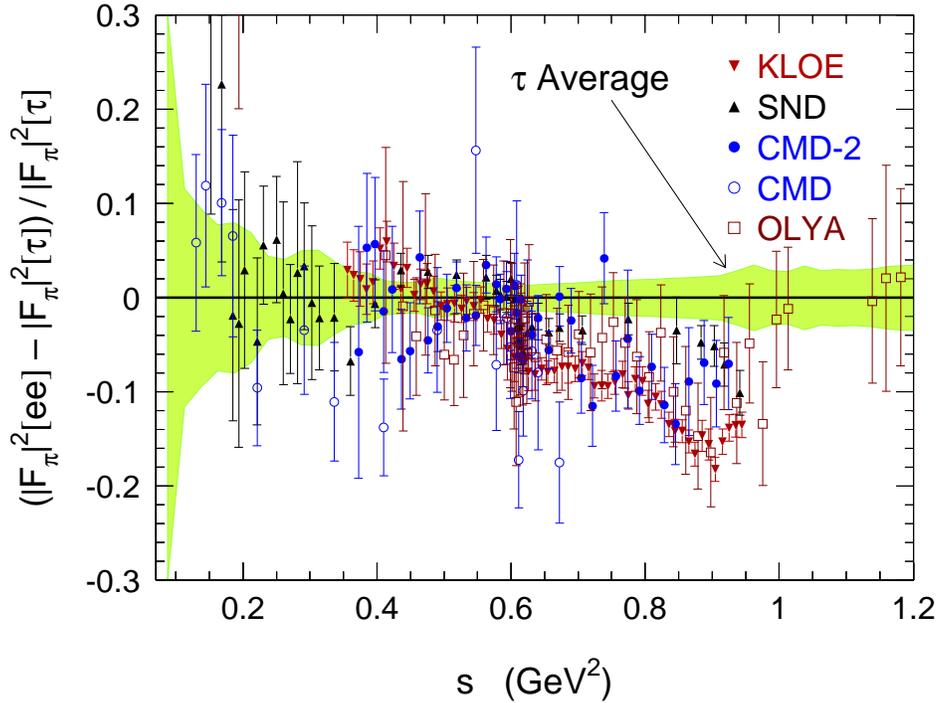,height=10cm}}
  \end{center}
  \caption[.]{Relative comparison of the isospin-breaking-corrected $\tau$
    data (world average) and $\pi^+\pi^-$ spectral function\ from $ e^+e^- $ annihilation,
    expressed as a ratio to the $\tau$ spectral function. The shaded band gives the 
    uncertainty on the $\tau$ spectral function. 
  }
  \label{part6:fig:sf_dif}
\end{figure}

In the limit of isospin invariance, the vector current is conserved
(CVC), so that the spectral function  for a vector $\tau$ decay mode 
$X^-\nu_\tau$
in a given isospin state for the hadronic system is related 
to the $ e^+e^- $  annihilation cross section of the corresponding 
isovector final state $X^0$ :

\begin{equation}
\label{eq:cvc}
  \sigma_{e^+e^-\to X^0}^{I=1}(s) \:=\:
         \frac{4\pi\alpha^2}{s}\,v_{1,\,X^-}(s)~,
\end{equation}

\noindent where $\alpha$ is the electromagnetic fine 
structure constant. In reality, 
isospin symmetry is broken, particularly due to electromagnetic effects,
and corrections must be applied to compare (and combine) $ \tau $-decay 
and 
$ e^+e^- $ data. A more complete review and further references about
sources and value of the symmetry breaking can be found in 
Refs.~\cite{part6:aleph_05,part6:Davier&Zhang_06}. 
A comparison of two-pion spectral functions obtained from
tau decays and $ e^+e^- $ data is shown 
in Fig.~\ref{part6:fig:sf_dif}, which is taken from
Ref.~\cite{part6:Davier&Zhang_06}.  Here
it is evident  that, although the absolute difference
 is relatively small, there is a  clear 
discrepancy between data, especially above the $\rho $ peak.  
\noindent Another, more ``quantitative'' way to compare spectral functions  is to use 
$ e^+e^- $ data to calculate hadronic tau 
branching fractions. For the most studied
decay, $\tau \rightarrow \nu_\tau\pi\pi^0 $,  different authors 
\cite{part6:aleph_05,part6:Davier&Zhang_06,part6:tau06} 
give $ \tau  -  e^+e^- $  discrepancies in the range $2.9 \sigma-4.5  
\sigma $ with differences 
depending mainly on which $ e^+e^- $ data are included in the analysis, 
while in all cases the $ \tau $-decay data are dominated
by the ALEPH results. 
In a recent work~\cite{part6:muon07}, a new calculation 
of  isospin breaking corrections
is presented, and this reduces the discrepancies mentioned above to $ 2.6 
\sigma $ and 
substantially improves the agreement between
the shapes of spectral functions  obtained from
$ \tau $-decay and $ e^+e^- $ annihilation. 
Nevertheless, the situation is still far from the
``good agreement'' state. In addition, the predicted value 
for the anomalous muon magnetic moment $ g_\mu$-2
 calculated using combined the $ \tau $ and  $ e^+e^- $ data is 
now $ 3.6 \sigma $  away from the most precise  measurement.
Thus, we now face a very intriguing situation: differences between 
predictions and  measurements are substantial on one side, but still too 
small to be a sign of new physics on the other side;  new 
reliable measurements will be very useful.

\subsubsection{Measurement of  Tau Hadronic Spectral Functions at \bes3}
 Experimental tau physics data can be divided into two main 
groups: Data
produced at LEP  have low background and high selection efficiency, 
but are statistically limited. Data from $B$-factories have almost 
unlimited statistics, but much worse background conditions due to 
relatively low multiplicity of hadronic events and the high ratio of 
the inclusive hadron and the tau-pair production  cross-sections. The 
\bes3 experiment has no advantage over the $B$-factories when working at 
the charmonium resonances. On the other hand,
a dedicated ``tau'' run at an energy  slightly below the
$ \psi(2S) $  resonance would allow a combination of 
high statistics and excellent background conditions. 
Using leptonic $ \tau $-decay tagging
in combination with the usual  kinematic selection criteria 
(high missing momenta, acollinearity and acomplanarity, broken $P_t$ balance ) 
 will make it is possible
to select an extremely clean sample of $\tau$ decays,
with backgrounds well below 1\%.  Another obvious advantage of using 
lepton-tagged events is that that the selection (in)efficiency causes no 
bias in the measured values, and only affects the available statistics. 
Thus, if more strict selection criteria are used, 
we can say that ``we are buying low systematics with statistics.''
The table  below gives a comparison of \bes3 and ALEPH 
experimental conditions
for $ \tau $ hadronic branching measurements and spectral 
functions determinations.  The \bes3 values
are computed  assuming a three-month dedicated tau run and 
an $ 80\% $ tagging efficiency.

\hspace{1.2cm}

\begin{tabular}{|l|l|l|}
\hline
             &{\bf ALEPH}&{\bf  \bes3}  \\
\hline
 $\tau $-decays selected & $ \sim $ 327000       &   $ 1.6\times10^6 $          \\
\hline
$\tau \rightarrow \pi\pi_0\nu_\tau $ 
decays selected           &  $ \sim $ 81000       &    $ \sim $ 280 000  \\
\hline
external background & $1.2\% $           &  $ <1\%$\\
\hline
hadronic mass reconstruction accuracy & $ \sim 80 $ MeV &  $ < 50$ MeV  \\
\hline 
\end{tabular}

\hspace{1.2cm}
~

Another very important 
consideration is the neutral hadron ($\pi^0$ most of all)  
identification efficiency.  A preliminary 
simulation shows that the single  $ \pi^0 $ registration efficiency for
the \bes3 detector will be about 95\% for decays that are within 
the acceptance of the calorimeter, 
which is at least as good as that for ALEPH.
It should be noted that the tau leptons in LEP events are highly boosted
and this results in a large number
of merged or overlapped clusters in electromagnetic calorimeters, 
which require a very 
complicated (and, therefore, vulnerable to error) analysis. 
At \bes3, the $\tau $ decay products are distributed  almost 
uniformly throughout the deector.   The main contribution
to the neutral pion reconstruction inefficiency 
will be due to  the geometrical 
acceptance (95\%), which can
be calculated with high accuracy and 
should not result in a substantial systematic error.     
Another attractive possibility is related to 
the determination of strange spectral 
functions. At LEP experiments, kaons were identified only on a statistical 
basis. At \bes3, kaons produced in tau decay will have momenta below 0.8 
GeV/c, {\it i.e.} in the momentum range 
where they can be selected  with high 
purity  by the TOF and $dE/dX$ measurements.  

Thus,  a three-month dedicated tau run at \bes3 
has a good chance of providing the most accurate 
measurements of the hadronic 
spectral functions.



\section[Leptonic Tau decays]{Leptonic Tau decays}
\label{sec:lepton_tau}

The precise measurement of the different exclusive $\tau$ decays
provides very valuable information to test the Standard Model, both
in the electroweak and strong sectors. The threshold region, with
its kinematical advantages and low backgrounds, makes 
accurate studies of the lowest-multiplicity decay modes possible.

\subsection[Leptonic Decays and Universality Tests]{Leptonic Decays and Universality
Tests\footnote{A. Pich}}

The leptonic decays $\tau^-\to l^-\bar\nu_l\nu_\tau$ ($l=e,\mu$) are
theoretically understood at the level of the electroweak radiative
corrections. Within the Standard Model
\begin{equation}
\label{eq:leptonic} \Gamma_{\tau\to l} \, \equiv \, \Gamma (\tau^-
\rightarrow \nu_{\tau} l^- \bar{\nu}_l)  \, = \,
  {G_F^2 m_{\tau}^5 \over 192 \pi^3} \,
  f\!\left({m_l^2 \over m_{\tau}^2}\right) \, r_{EW},
\end{equation}
where $f(x) = 1 - 8 x + 8 x^3 - x^4 - 12 x^2 \log{x}$. The factor
$r_{EW}$ takes into account radiative corrections not included in
the Fermi coupling constant $G_F$, and the non-local structure of
the $W$ propagator \cite{pich_MS:88}; these effects are quite small
[$\alpha(m_\tau) = 1 / 133.3 $]:
\begin{equation}\label{eq:e_EW}
 r_{EW} \, = \, \left[1 + {\alpha(m_\tau) \over 2 \pi}
 \left({25 \over 4}  - \pi^2 \right) \right] \, \left[ 1 +
 { 3\over 5 } {m_\tau^2 \over M_W^2} - 2 {m_l^2 \over M_W^2} \right]
 \, = \,  0.9960 \, .
\end{equation}
\noindent
Using the value of $G_F$ measured in $\mu$ decay in
Eq.~(\ref{eq:leptonic}), yields a relation between the $\tau$
lifetime and the leptonic branching ratios:
\begin{equation}\label{relation}
 B_{\tau\to e} \,\equiv\,\mathrm{Br}(\tau^-\to e^-\bar\nu_e\nu_\tau)\,
 = \, {B_{\tau\to\mu} \over 0.972564\pm 0.000010} \, = {
 \tau_{\tau} \over (1632.1 \pm 1.4) \times 10^{-15}\, {\rm s} } \, .
\end{equation}
Here the quoted errors reflect the present uncertainty of $0.3$ MeV in
the value of $m_\tau$.

\begin{figure}[htb]\centering
\includegraphics[width=10cm]{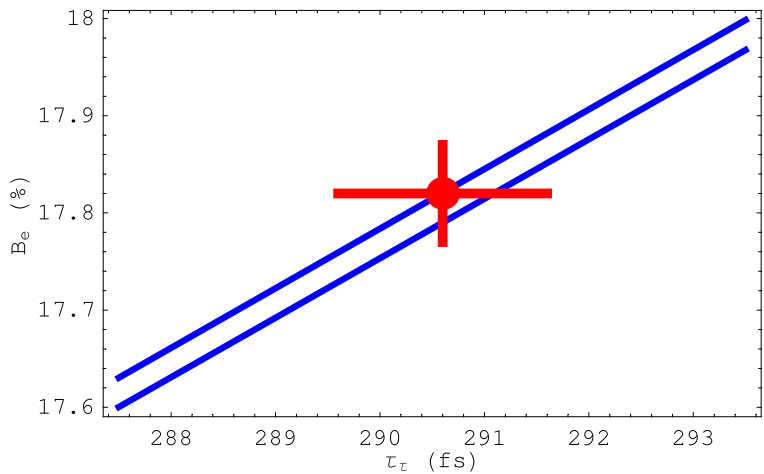}
\caption{The relation between $B_{\tau\to e}$ and $\tau_\tau$. The
diagonal band corresponds to Eq.~(\protect\ref{relation}).}
\label{fig:BeLife}
\end{figure}

The predicted value of $B_{\tau\to\mu}/B_{\tau\to e}$ is in excellent
agreement with the measured ratio $B_{\tau\to\mu}/B_{\tau\to e} =
0.9725 \pm 0.0039$. As shown in Fig.~\ref{fig:BeLife}, the
relation between $B_{\tau\to e}$ and $\tau_\tau$ is also well
satisfied by the present data. Note that this relation is very
sensitive to the value of the $\tau$ mass [$\Gamma_{\tau\to
l}\propto m_\tau^5$].

These measurements can be used to test the universality of the $W$
couplings to the leptonic charged currents, 
{\it i.e.}  \ $g_e = g_\mu
= g_\tau \equiv g\, $. The $B_{\tau\to\mu}/B_{\tau\to e}$ ratio
constraints $|g_\mu/g_e|$, while the $B_{\tau\to e}/\tau_\tau$
relation provides information on $|g_\tau/g_\mu|$. As shown in
Table~\ref{tab:ccuniv}, the present data verify the universality of
the leptonic charged-current couplings to the 0.2\% level.\footnote{
$\mathrm{Br}(W\to\nu_\tau\tau)$ is $2.1\,\sigma/2.7\,\sigma$ larger
than $\mathrm{Br}(W\to \nu_e e / \nu_\mu\mu)$. The stringent limits
on $|g_\tau/g_{e,\mu}|$ from $W$-mediated decays makes it unlikely that
this is a real physical effect.}

\begin{table}[t] 
\centering \caption{Present constraints on $|g_l/g_{l'}|$
\cite{pich_taurev06b,pich_SMEW:07}.} \label{tab:ccuniv} \vspace{0.2cm}
\renewcommand{\tabcolsep}{0.8pc} 
\renewcommand{\arraystretch}{1.3} 
\begin{tabular}{ccccc}
\hline\hline &
 $B_{\tau\to e}\,\tau_\mu/\tau_\tau$ & $\Gamma_{\tau\to\pi}/\Gamma_{\pi\to\mu}$ &
 $\Gamma_{\tau\to K}/\Gamma_{K\to\mu}$ &
 $B_{W\to\tau}/B_{W\to\mu}$ \\
 $|g_\tau/g_\mu|$  & $1.0004\pm 0.0022$ & $0.996\pm 0.005$ &
 $0.979\pm 0.017$ & $1.039\pm 0.013$
 \\ \hline\hline &
 $B_{\tau\to\mu}/B_{\tau\to e}$ & $B_{\pi\to \mu}/B_{\pi\to e}$ &
 $B_{K\to \mu}/B_{K\to e}$ & $B_{K\to\pi\mu}/B_{K\to\pi e}$ \\
 $|g_\mu/g_e|$ & $1.0000\pm 0.0020$ & $1.0017\pm 0.0015$ &
 $1.012\pm 0.009$ & $1.0002\pm 0.0026$
 \\ \hline\hline &
 $B_{W\to\mu}/B_{W\to e}$ &\multicolumn{1}{||c}{} &
 $B_{\tau\to\mu}\,\tau_\mu/\tau_\tau$ & $B_{W\to\tau}/B_{W\to e}$\\
 $|g_\mu/g_e|$ & $0.997\pm 0.010$ &
 \multicolumn{1}{||c}{$|g_\tau/g_e|$} &
 $1.0004\pm 0.0023$ & $1.036\pm 0.014$
 \\ \hline\hline
\end{tabular}
\end{table}
%

The $\tau$ leptonic branching fractions and the $\tau$ lifetime are
known with a precision of $0.3\%$. Slightly improved lifetime
measurements are expected from BaBar and Belle. For comparison,
the $\mu$ lifetime is known with an accuracy of $10^{-5}$, which
should be further improved to $10^{-6}$ by the MuLan experiment at
PSI \cite{pich_LY06}.

Universality tests also  require a precise determination of
$m_\tau^5$, which is only known to the $0.08\%$ level. Two new
preliminary measurements of the $\tau$ mass have been presented
recently:
\begin{equation}
 m_\tau =\left\{ \begin{array}{lr} 1776.71\pm 0.13\pm 0.35~\mathrm{MeV}
 &\qquad [\mathrm{Belle}],\\
 1776.80\, {}^{+\, 0.25}_{-\, 0.23} \pm 0.15~\mathrm{MeV} &\qquad
 [\mathrm{KEDR}]. \end{array}\right.
\end{equation}
The Belle value~\cite{pich_Shapkin} 
is based on a pseudomass analysis of
$\tau\to\nu_\tau 3\pi$ decays, while
the KEDR result~\cite{pich_KEDR} comes from a measurement of
$\tau^+\tau^-$ threshold production, taking advantage of a precise
energy calibration through the resonance depolarization method. In
both cases the achieved precision is getting close to the present
BES-I dominated value, 
$m_\tau = 1776.99\, {}^{+\, 0.29}_{-\, 0.26}$
\cite{pich_PDG}. KEDR aims to obtain a final accuracy of 0.15 MeV. A
precision of better than 0.1 MeV should be easily achieved at 
\bes3~\cite{pich_MO}, through a detailed analysis of
$\sigma(e^+e^-\to\tau^+\tau^-)$ at 
threshold~\cite{achim_Ruiz-Femenia:2002wm,pich_Pedro,pich_Voloshin}, as 
discussed in detail in Chapt.~\ref{part6:sec:tau_mass}.

\subsection[Lorentz Structure]{Lorentz Structure\footnote{By A. Pich}}


With high statistics, the leptonic $\tau$ decay modes 
provide opportnities to
investigate the Lorentz structure of the decay amplitude, through
the analysis of the energy and angular distribution of the final
charged lepton. The most general, local, derivative-free,
lepton-number conserving, four-lepton interaction Hamiltonian,
consistent with locality and Lorentz invariance
\cite{pich_MI:50,pich_SCH:83,pich_FGJ:86,pich_PS:95},
%
\begin{equation}\label{eq:hamiltonian}
 {\cal H} \, =\,  4\, \frac{G_{l'l}}{\sqrt{2}}\,
 \sum_{n,\epsilon,\omega} g^n_{\epsilon\omega} \left[
 \overline{l'_\epsilon} \Gamma^n {(\nu_{l'})}_\sigma \right]\, \left[
 \overline{({\nu_l})_\lambda} \Gamma_n l_\omega \right]\, ,
\end{equation}
contains ten complex coupling constants or, since a common phase is
arbitrary, nineteen independent real parameters that could be
different for each leptonic decay. The sub-indices $\epsilon , \omega
, \sigma, \lambda$ label the chiralities (left-handed, right-handed)
of the  corresponding  fermions, and $n$ the type of interaction:
scalar ($I$), vector ($\gamma^\mu$) and tensor
($\sigma^{\mu\nu}/\sqrt{2}$). For given $n, \epsilon , \omega $, the
neutrino chiralities $\sigma $ and $\lambda$ are uniquely
determined.

Taking out a common factor $G_{l'l}$, which is determined by the
total decay rate, the coupling constants $g^n_{\epsilon\omega}$ are
normalized to \cite{pich_FGJ:86}
\begin{eqnarray}\label{eq:normalization}
 1 &\!\!\! = &\!\!\! {1\over 4}\,\left( |g^S_{RR}|^2 + |g^S_{RL}|^2
 + |g^S_{LR}|^2 + |g^S_{LL}|^2 \right)
 \, + \, 3 \,\left( |g^T_{RL}|^2 + |g^T_{LR}|^2 \right)
 \nonumber \\ & &\!\!\!\! \mbox{} + \left(
 |g^V_{RR}|^2 + |g^V_{RL}|^2 + |g^V_{LR}|^2 + |g^V_{LL}|^2 \right) \, .
\end{eqnarray}
In the Standard Model, $g^V_{LL} = 1$  and all the other
$g^n_{\epsilon\omega} = 0 $. The sums of all contributions in
Eq.~\ref{eq:normalization} with identical initial and final
chiralities, ${\cal Q}_{\epsilon\omega}$, can be interpreted as the
probabilities for the decay of an $\omega$-handed $l^-$ into an
$\epsilon$-handed daughter lepton. Upper bounds on any of these
(positive-semidefinite) probabilities translate into corresponding
limits for all couplings with the given chiralities. The measurement
of the $\tau$ polarization is possible due to the fact that the
spins of the $\tau^+\tau^-$ pair produced in $e^+e^-$ annihilation
are strongly correlated. 
Table~\ref{tab:Michel} shows the present
90\% C.L. experimental bounds on the $g^n_{\epsilon\omega}$
couplings.


\begin{table}[tbh]\centering
\caption{90\% C.L. experimental bounds
\protect\cite{pich_PDG} for the normalized $\tau$-decay couplings
$g'^n_{\epsilon\omega }\equiv g^n_{\epsilon\omega }/ N^n$, where
$N^n \equiv \protect\mbox{\rm max}(|g^n_{\epsilon\omega }|) =2$, 1,
$1/\protect\sqrt{3} $ for $n =$ S, V, T.} \label{tab:Michel}
 \vspace{0.2cm}
 \renewcommand{\tabcolsep}{1.1pc} 
 \renewcommand{\arraystretch}{1.3} 
 \begin{tabular}{cccc}
 \hline\hline $\tau^-\to e^-\bar\nu_e\nu_\tau$\\ \hline
 $|g_{RR}^S| < 0.70$ & $|g_{LR}^S| < 0.99$ & $|g_{RL}^S| < 2.01$ 
 & $|g_{LL}^S| < 2.01$ \\
 $|g_{RR}^V| < 0.17$ & $|g_{LR}^V| < 0.13$ & $|g_{RL}^V| < 0.52$
 & $|g_{LL}^V| < 1.01$ \\
 $|g_{RR}^T| \equiv 0$ & $|g_{LR}^T| < 0.08$ & $|g_{RL}^T| < 0.51$
 & $|g_{LL}^T|\equiv 0$ \\
 \hline\hline $\tau^-\to \mu^-\bar\nu_\mu\nu_\tau$\\ \hline
 $|g_{RR}^S| < 0.72$ & $|g_{LR}^S| < 0.95$ & $|g_{RL}^S| < 2.01$
 & $|g_{LL}^S| < 2.01$ \\
 $|g_{RR}^V| < 0.18$ & $|g_{LR}^V| < 0.12$ & $|g_{RL}^V| < 0.52$
 & $|g_{LL}^V| < 1.01$ \\
 $|g_{RR}^T| \equiv 0$ & $|g_{LR}^T| < 0.08$ & $|g_{RL}^T| < 0.51$
 & $|g_{LL}^T|\equiv 0$ \\
 \hline\hline
 \end{tabular}
\end{table}


\subsection[Study of the Lorentz structure at \bes3]{Study of
the Lorentz structure at \bes3 \footnote{By Igor R. Boyko.
and Dedovich Dima}}

\subsubsection{Michel parameters}
\label{part6:mich}

%
%
%

\begin{figure}[htb]
\begin{center} \mbox{\epsfig{file=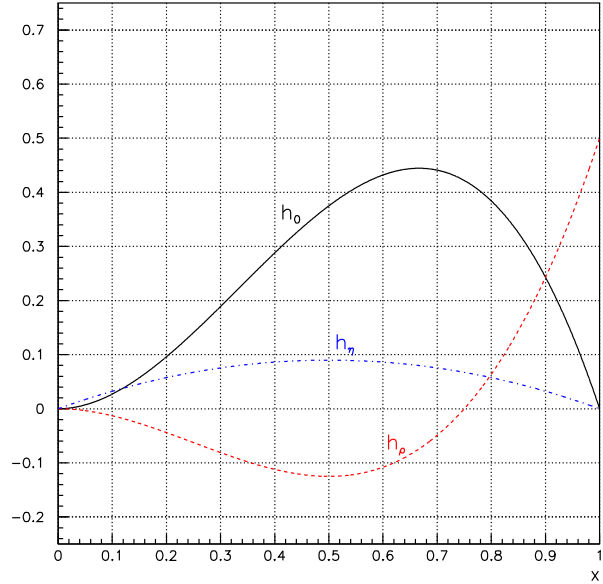,width=9cm}}
\end{center}
\caption{The shapes of polynomial functions $h_i$ for the Michel
parametrisation of tau decays.}
\label{part6:fig:poly}
\end{figure}

The coupling constants, $g^n_{\epsilon\omega}$ in Eq.~\ref{eq:hamiltonian}, can be experimentally accessed via
the energy spectra of the daughter leptons from  tau decays.
The polarization of the daughter leptons usually cannot be measured;
however the polarization of the tau lepton $P_{\tau}$ in principle
can be measured through its decay spectra. Under these
assumptions the spectrum of the tau decays predicted by 
Eq.~\ref{eq:hamiltonian} can be parametrized at 
the Born level by the following sum of polynomials $h_i$:

\begin{equation}
\label{eq:poly}
\frac 1 \Gamma \frac {d \Gamma} {d x_\ell}  =
h_0(x_\ell) + { \bf \eta} h_{\eta}(x_\ell)
+ { \bf \rho} h_{\rho}(x_\ell) -
P_{\tau}  \left[
{ \bf \xi} h_{\xi}(x_\ell) + { \bf \xi \delta} h_{\xi\delta}
(x_\ell) \right],
\end{equation}

\noindent
where $P_{\tau}$ is the average tau polarization and
$x_\ell = E_\ell/E_{max}$ is the ``reduced energy'' of the
daughter lepton, or the ratio of its energy to the maximum possible energy.
Examples of the polynomials $h_i$ 
are illustrated in Fig.~\ref{part6:fig:poly}.

The coefficients $\eta$, $\rho$, $\xi$, $\xi\delta$, known as the Michel
parameters \cite{michel}, are bilinear combinations of the coupling 
constants in Eq.~\ref{eq:hamiltonian}. The  four Michel 
parameters carry the full information on the coupling constants
that can be extracted from the decay spectra of the tau leptons
(without the measurement of the polarisation of the final state leptons).

In the Standard Model, the Michel parameters are predicted to be: 
$\rho = 0.75$; ~ $\eta = 0$; ~ $\xi = 1$; ~ $\xi\delta = 0.75$.
An observation of different values of the Michel
parameters would indicate a violation of the Standard Model.

\subsubsection{Anomalous tensor coupling}

The parametrization presented in Eq.~\ref{eq:hamiltonian}
is based on certain assumptions; namely it assumes
the Hamiltonian to be
 lepton-number conserving, derivative-free, local,
Lorentz invariant, and a 4-fermion point interaction.
While most of these assumptions are quite natural,
there is no fundamental reason to assume that the
interaction Lagrangian does not include derivatives.

An anomalous interaction involving derivatives
(which can only be a tensor interaction)
can be represented in the following form \cite{chijov}:

\begin{equation}
\label{eq:lagr}
{\cal L} = \frac{g}{\sqrt{2}}~W^{\alpha}~
\left\{\overline{\tau} \gamma_{\alpha}
\frac{1-\gamma^{5}}{2}
\nu +
\frac{{ \kappa^W_\tau}}{2m_{\tau}}
\partial_{\beta} \left(
\overline{\tau}\sigma_{\alpha\beta}
\frac{1-\gamma^{5}}{2}\nu \right) \right\} + {\mathrm h.c.},
\end{equation}

\noindent
where $\kappa^W_\tau$ is the strength of the anomalous coupling.
Such an anomalous interaction can be studied through the possible
distortions of the energy spectra of tau decays.
Since the Lagrangian of Eq.~\ref{eq:lagr} explicitly contains 
derivatives,
the distortions of the energy spectra can not be described
in terms of the known Michel parameters.

The matrix element for the purely leptonic tau decays 
then takes the form:

\begin{equation}
\label{eq:matten}
{\cal M}  =
\frac{4 G}{\sqrt{2}}
\langle    \overline{v}_{l} \left| \; \gamma_\alpha \right| v_{\overline{\nu}_l}
 \rangle
\left(
\langle \overline{v}_{\nu_\tau} \left| \; \gamma_\alpha \right| u_{\tau_L} \rangle
- i \frac{\kappa^W_\tau}{2m_\tau} q_\beta
\langle \overline{v}_{\nu_\tau} \left| \; \sigma_{\alpha\beta} \right| u_{\tau_R
} \rangle
  \right)  ,
\end{equation}

\noindent
where $q$ is the 
four-momentum of the $W$. The first summand in Eq.~\ref{eq:matten}
is the Standard Model prediction, while the second one is the contribution
of the anomalous coupling. (In the framework of the Standard Model
$\kappa^W_\tau = 0$.) In principle, both the anomalous coupling 
$\kappa^W_\tau$
and the ``standard'' couplings $g^\gamma_{ij}$ can take non-Standard Model
values simultaneously. In this case the first summand in (\ref{eq:matten})
has to be replaced by the full parametrisation of
Eq.~\ref{eq:hamiltonian}.

Like the case of Michel parameters, the contribution of the anomalous
tensor coupling can be also parametrized in terms of the polynomials.
The approximate shape of the energy spectrum is

\begin{equation}
\label{eq:tenz}
\frac 1 \Gamma \frac {d \Gamma} {d x_\ell}  \sim
x_\ell^2 \cdot \left ( -2x_\ell + 3 + 2\cdot\kappa^W_\tau\cdot x_\ell 
\right ).
\end{equation}

\begin{figure}[htb]
\begin{center} \mbox{\epsfig{file=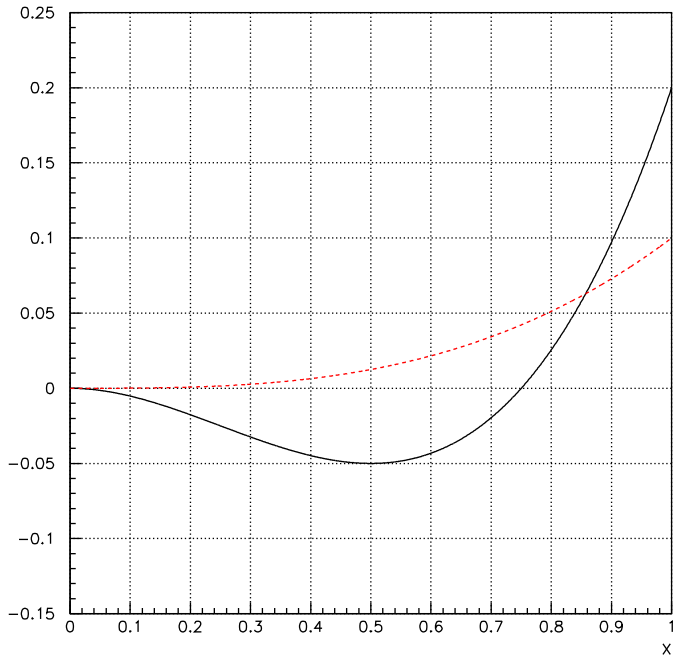,width=10cm}}
\end{center}
\caption{The deviation of the spectrum of leptons from tau decays
from the Standard Model prediction. Dashed line: for the non-zero value of
the anomalous tensor coupling. Solid line: for $\rho \ne 0.75$.}
\label{part6:fig:tenz}
\end{figure}

\noindent Figure~\ref{part6:fig:tenz} compares the deviations of the decay 
spectrum from the Standard Model prediction for non-Standard Model
values of $\kappa^W_\tau$ and the Michel parameter $\rho$.
One can see that the change in shape is significantly different
for the two cases, which makes it possible to measure
the two parameters simultaneously.

\subsubsection{Measurement of Michel parameters and search for an
anomalous tensor coupling}

The Michel parameters have been extensively measured in tau lepton
decays by many experiments. The current experimental uncertainty
on the parameter $\rho$ is about 0.008 and for $\eta$ the uncertainty
is 0.07 . To date, the anomalous tensor coupling 
has only been studied by  the DELPHI experiment. The coupling
constant $\kappa^W_\tau$ was measured with a precision of 0.04 and was
found to be consistent with zero.

The \bes3 experiment will have an abundant sample of tau lepton decays
and an accurately measured energy spectrum. This
will provide an excellent
possibility to improve significantly the current knowledge
on the Lorentz structure in tau decays. This section presents
a Monte-Carlo estimate of the possible reach of \bes3 for measurements of
the Michel parameters and in the search for the anomalous tensor coupling.
For simplicity, only two Michel parameters $\rho$  
and $\eta$ are considered.

The simulation was based on version 5.1 of the \bes3 software.
The events were generated at a c.m. energy of 3.69~GeV,
where background from hadronic events is minimal.
The following samples were simulated: 100K Bhabha scattering events
$ee \rightarrow ee$; 20K dimuon events $ee \rightarrow \mu\mu$;
 100K hadronic events $ee \rightarrow qq$. The total number of simulated
tau pair events was 100K. The simulated signal was limited to the
decay channel 
$ee \rightarrow \tau\tau \rightarrow e \mu \ (+\,\mathrm{neutrinos})$.
This channel represents only about 3\% of the total tau pairs,
but these are rather easy events to select. The possible 
inclusion of hadronic tau decays can increase the available statistics 
significantly, therefore the results presented in this section
can be considered as a conservative estimate of the \bes3 reach.

\begin{figure}[htb]
\begin{center}
\mbox{\epsfig{file=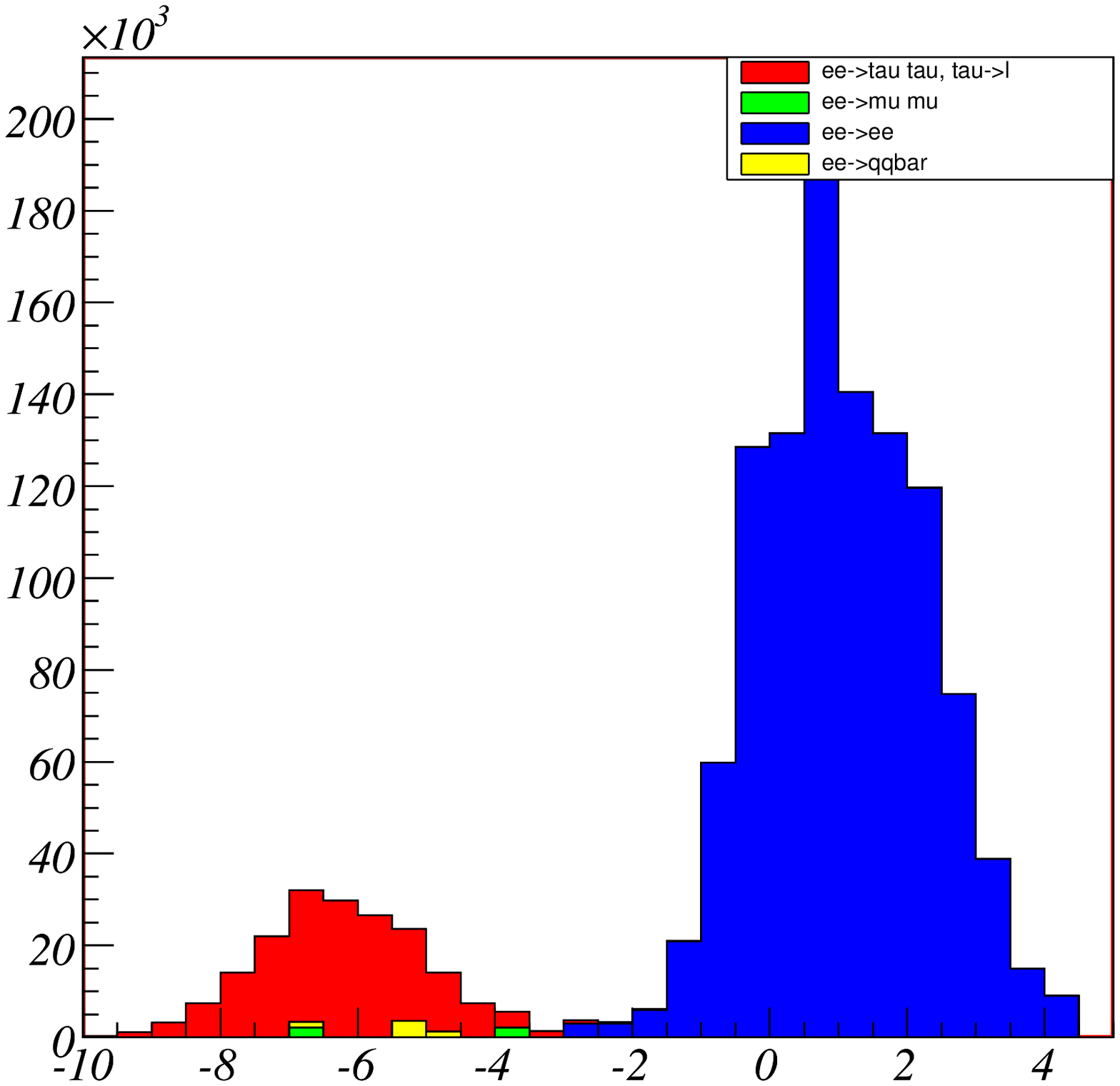,width=6cm}\epsfig{file=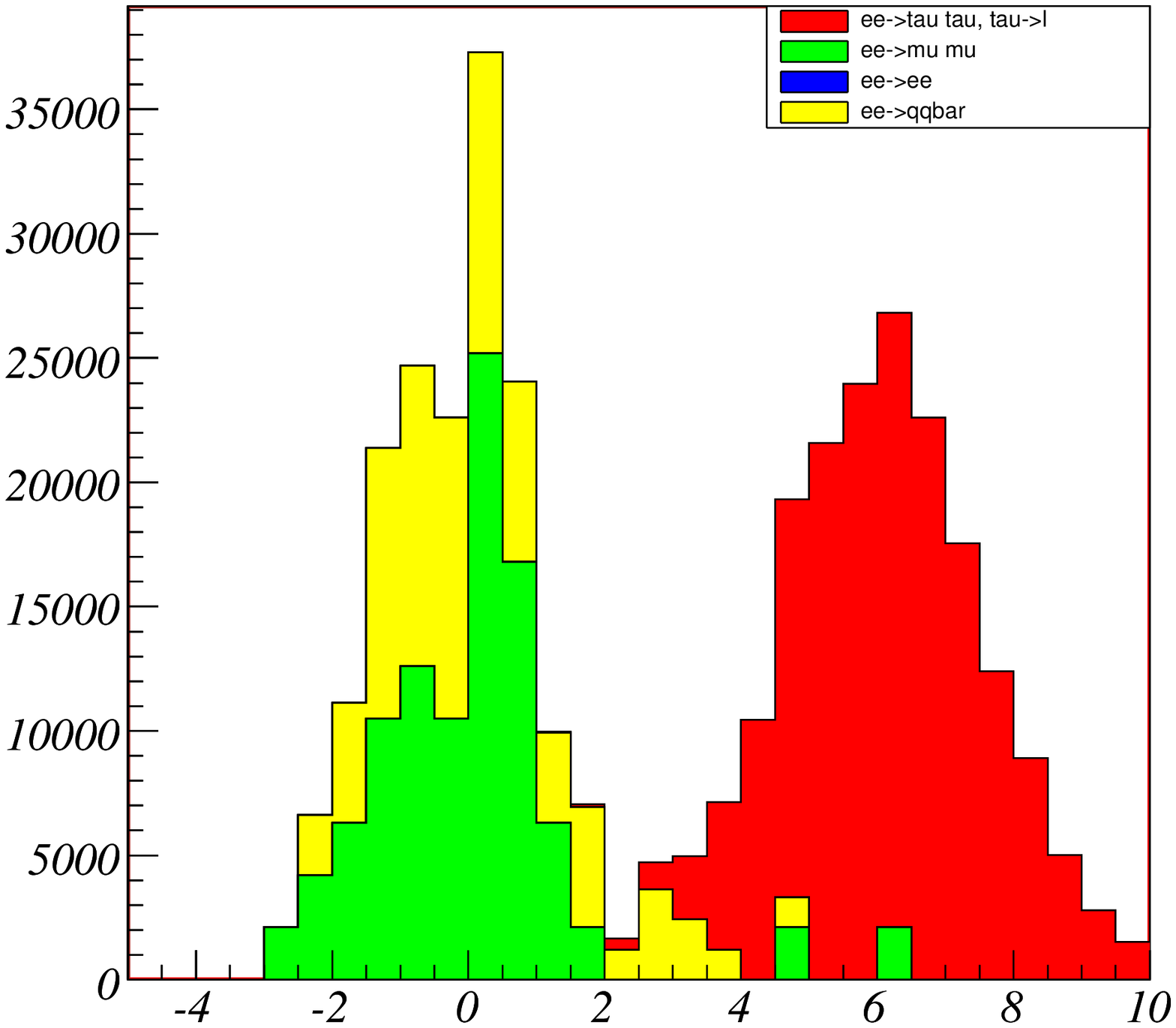,width=6cm}}
\end{center}
\caption{Left: the muon hypothesis pull for the electron candidate.
Right: the electron hypothesis pull for the muon candidate.
The selection cuts were +3 for the electron candidate and
-3 for the muon candidate. }
\label{part6:fig:dedx_pull}
\end{figure}

The analysis was restricted to the angular region $|\cos{\theta}|<0.83$.
The event selection was based on the particle identification.
Exactly two reconstructed charged particles were required in the 
event. One of them was required to be identified as an electron 
and the other had to be a muon. 
The main criterion for particle identification was 
based on the $dE/dx$ pull variable, {\it i.e.} 
the deviation of the measured $dE/dx$ value
from the expected one, expressed in units  of the $dE/dx$ uncertainty.
Figure~\ref{part6:fig:dedx_pull} shows the pulls for
the muon and electron
 hypotheses for the electron and muon candidates.
One can see that the dimuon and Bhabha events are rejected
very efficiently.

\begin{figure}[htb]
\begin{center}
\mbox{\epsfig{file=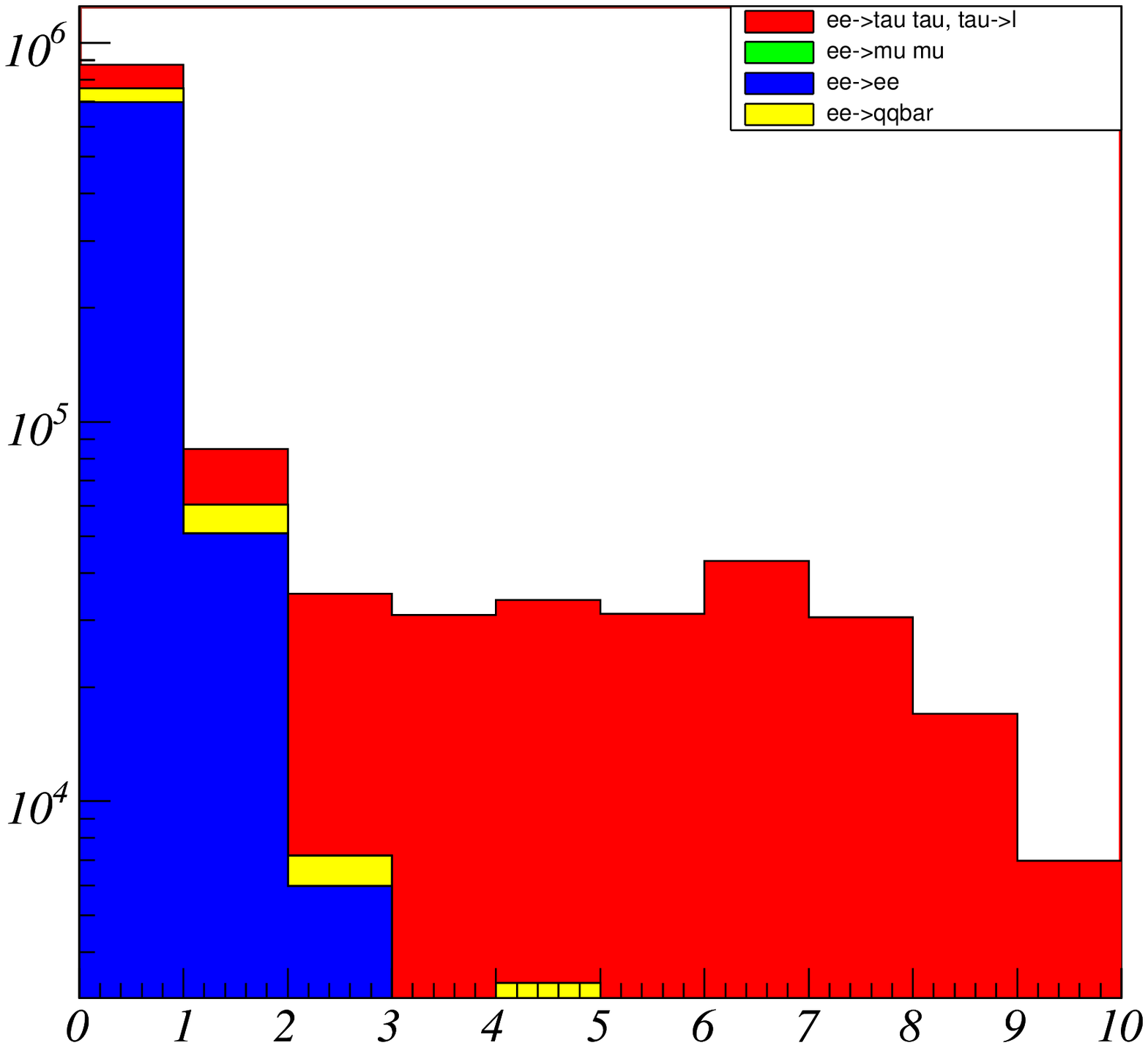,width=6cm}\epsfig{file=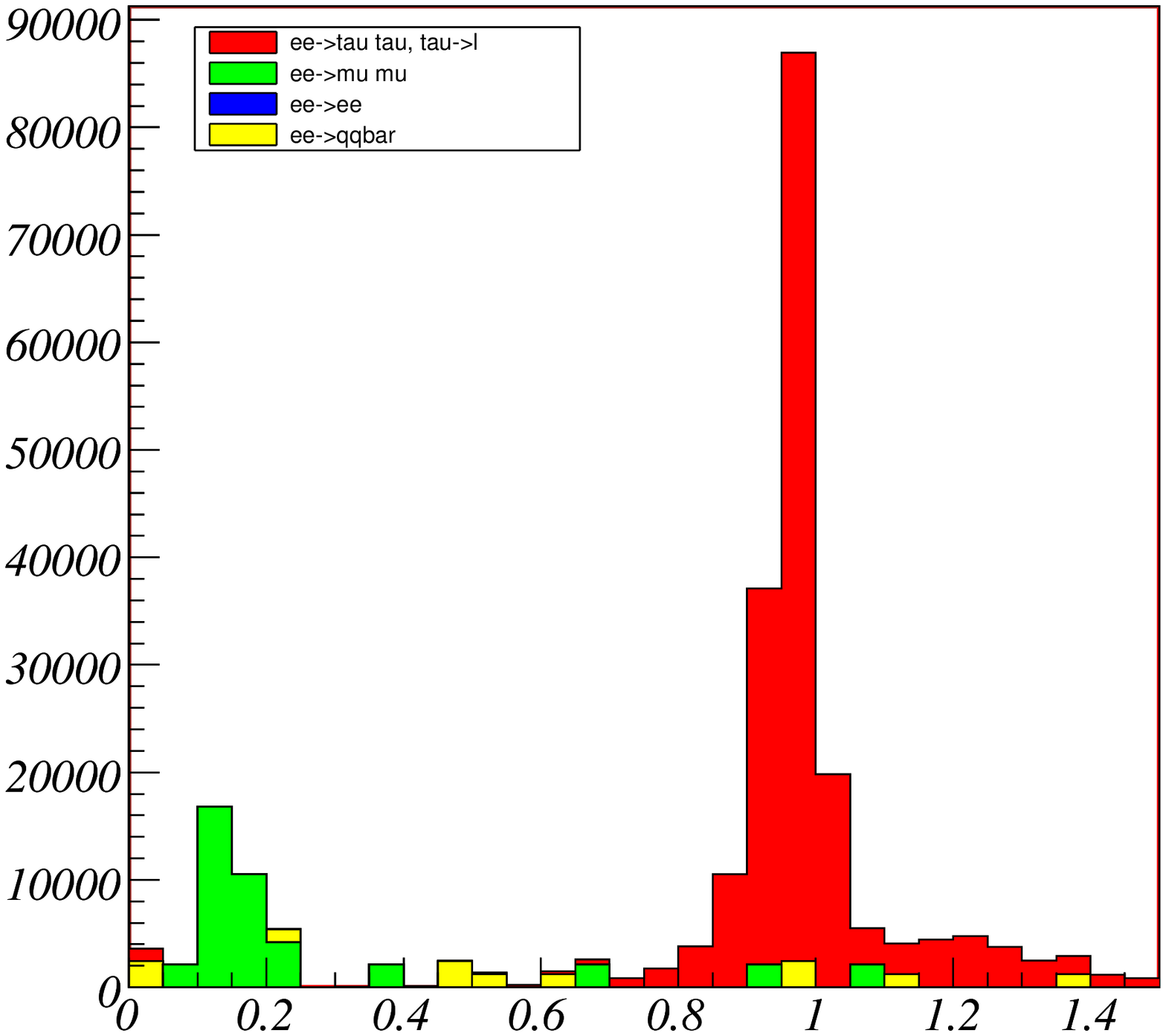,width=6cm}}
\end{center}
\caption{Left: the number of hits in muon chambers associated
with the muon candidates. Right: the relative electromagnetic
energy deposition E/P in the EMC for the electron candidates.
The selection cuts were: $N_{HIT}\ge 3$ for muons
and $E/P>0.8$ for electrons.}
\label{part6:fig:id}
\end{figure}

Several additional cuts were applied for further
background suppression. The electron and muon candidates
were required to be identified as such by 
the electromagnetic calorimeter (EMC) 
and muon chambers, respectively.
Figure~\ref{part6:fig:id} illustrates the selection criteria.
In addition, TOF information was used to reject
protons and kaons from hadronic events in the momentum
regions where $dE/dx$ of these particles is close to
that of electrons.

\begin{figure}[htb]
\begin{center}
\mbox{\epsfig{file=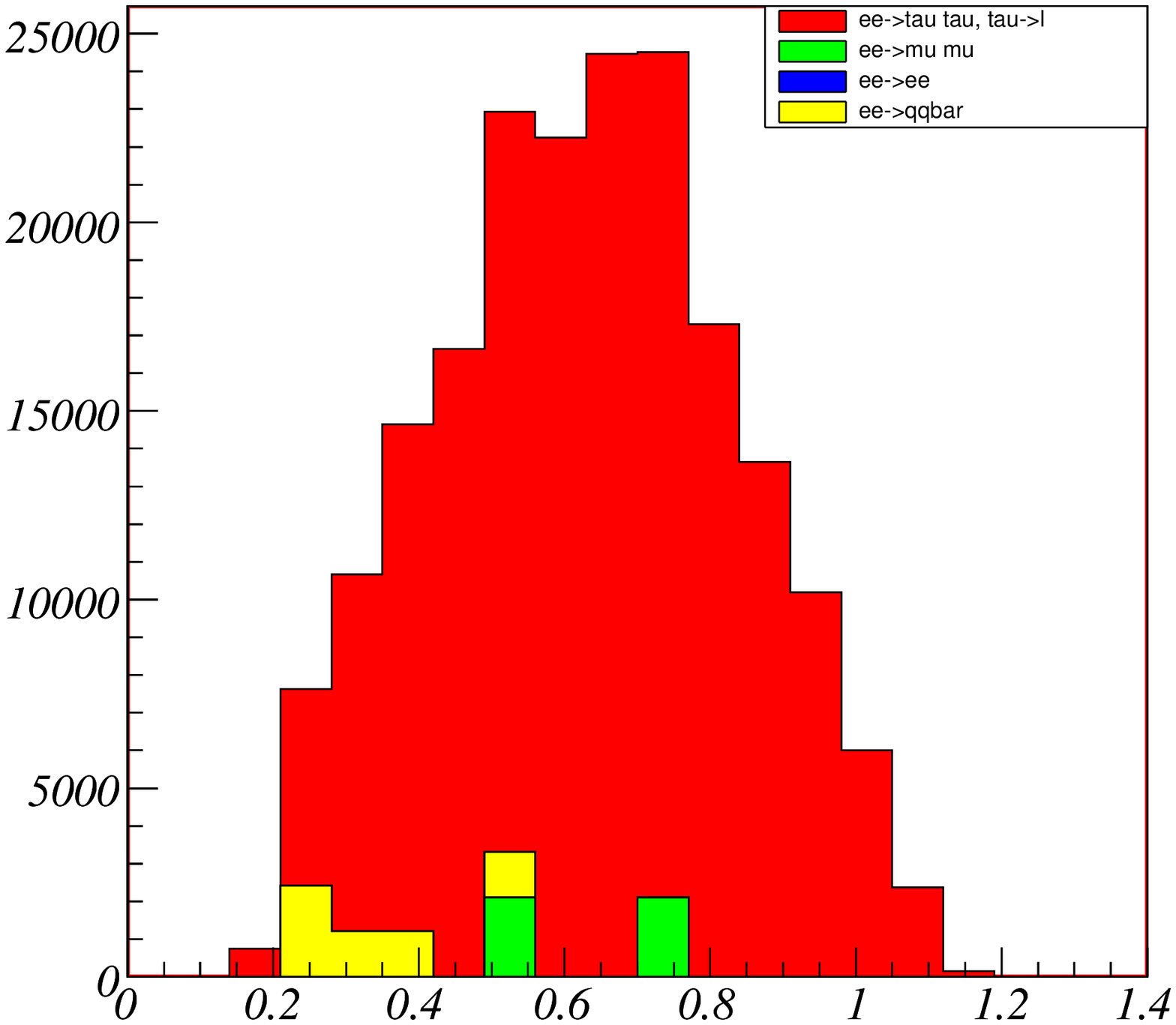,width=6cm}\epsfig{file=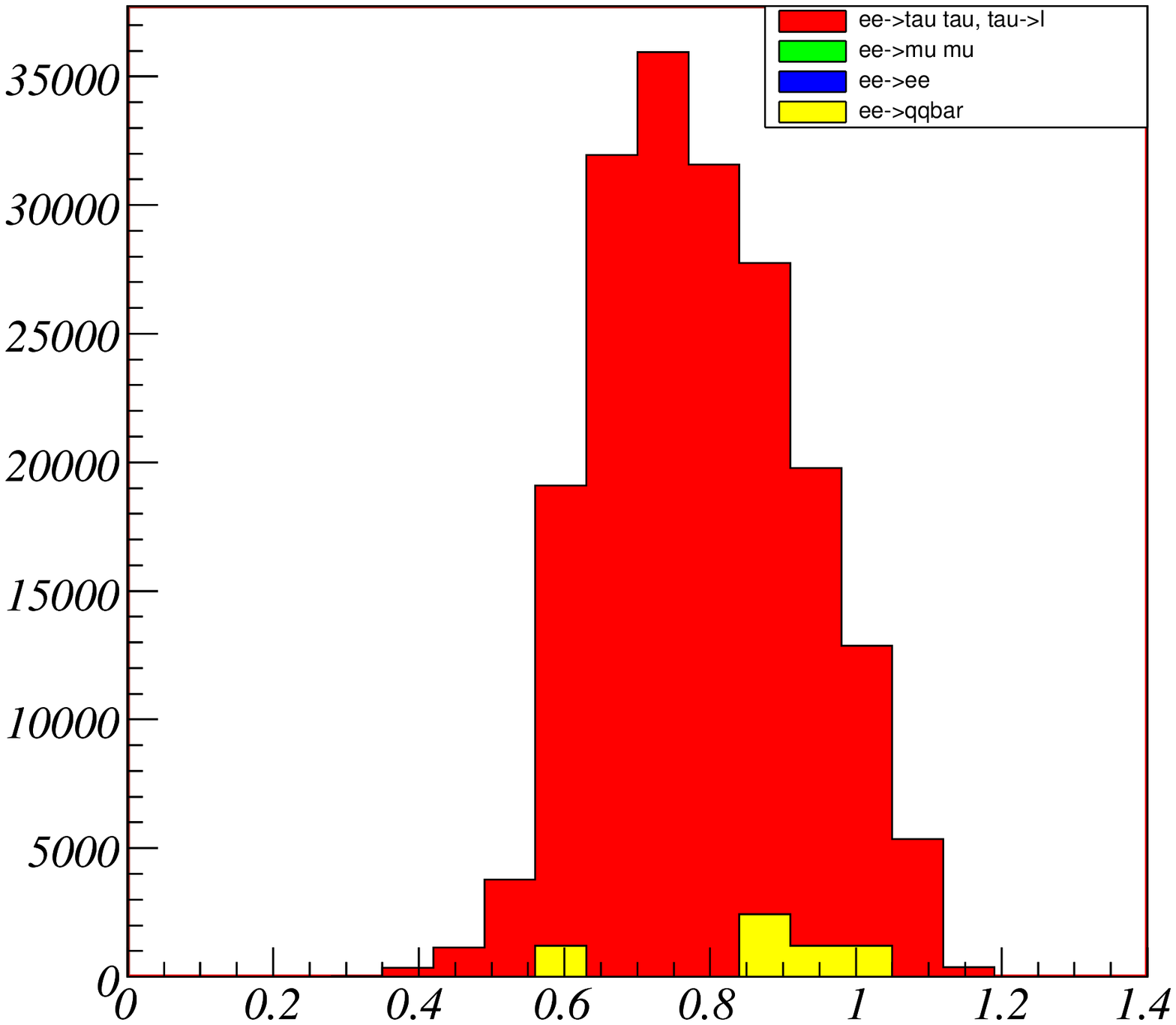,width=6cm}}
\end{center}
\caption{The momentum spectra of the selected muon (left)
and electron (right) candidates. The statistics corresponds to
a 5~fb$^{-1}$ data sample.}
\label{part6:fig:spectrum}
\end{figure}

The signal selection efficiency was found to be about 30\%
(with respect to the full solid angle).
For an integrated luminosity of 5~fb$^{-1}$, this
efficiency corresponds to 180K selected signal events.
The residual background is 6\%, about half of which are hadronic events
$ee \rightarrow qq$. The momentum spectra of the selected
electron and muon candidates are presented in 
Fig.~\ref{part6:fig:spectrum}

The spectra of reconstructed momenta of electrons and muons
from tau decays were fitted to the expectations for different
values of the Michel parameter $\rho$ and the anomalous coupling
constant $\kappa^W_\tau$.
The statistical uncertainties of the fit parameters are:

$$ \sigma(\rho) = 0.003 $$
$$ \sigma(\eta) = 0.02 $$
$$ \sigma(\kappa^W_\tau) = 0.002 $$
\noindent
The statistical errors correspond to a 5 fb$^{-1}$ 
data sample collected
at a c.m. energy of \mbox{3.69 GeV.}

\subsubsection{Summary}

The \bes3 experiment will provide excellent opportunities
to study the Lorentz structure of tau decays, including the measurements
of the Michel parameters and a search for an anomalous tensor coupling.
Conservative estimates based on Monte-Carlo simulation
of leptonic tau decays
suggest that the current precision on the Michel parameters
can be improved by factors of 2-4, and the limits on the
anomalous coupling constant $\kappa^W_\tau$ can be improved by at least
a factor of 10 with a 5~fb$^{-1}$ data sample collected at 
$E_{cm}=3.69$~GeV.
The inclusion of hadronic tau decays into analysis
would significantly improve the precision.

\newpage
\section[Semileptonic Decays]{Semileptonic Decays\footnote{By A. Pich}}
\label{sec:part6:semilep}

\subsection{Two-body Semileptonic Decays}

The $\tau$ is the only known lepton massive enough to decay into
hadrons. Its semileptonic decays are, thus, ideally suited for
studying the hadronic weak currents in very clean conditions.
The 
decay $\tau^-\to\nu_\tau H^-$ probes the matrix element of the
left--handed charged current between the vacuum and the final
hadronic state $H^-$.

For the decay modes with lowest multiplicity,
$\tau^-\to\nu_\tau\pi^-$ and $\tau^-\to\nu_\tau K^-$, the relevant
matrix elements (the so-called decay constants $f_{\pi,K}$) are
already known from the measured decays $\pi^-\to\mu^-\bar\nu_\mu$
and  $K^-\to\mu^-\bar\nu_\mu$. The corresponding $\tau$ decay widths
can then be accurately predicted:
\begin{eqnarray}\label{eq:R_tp}
 R_{\tau/\pi} & \!\!\!\equiv &\!\!\!
 {\Gamma(\tau^-\to\nu_\tau\pi^-) \over
 \Gamma(\pi^-\to \mu^-\bar\nu_\mu)} =
\Big\vert {g_\tau\over g_\mu}\Big\vert^2 {m_\tau^3\over 2 m_\pi
m_\mu^2} {(1-m_\pi^2/ m_\tau^2)^2\over
 (1-m_\mu^2/ m_\pi^2)^2}
\left( 1 + \delta R_{\tau/\pi}\right) , \qquad
\\ \label{eq:R_tk}
R_{\tau/K} &\!\!\! \equiv &\!\!\! {\Gamma(\tau^-\to\nu_\tau K^-)
\over
 \Gamma(K^-\to \mu^-\bar\nu_\mu)} =
\Big\vert {g_\tau\over g_\mu}\Big\vert^2 {m_\tau^3\over 2 m_K
m_\mu^2} {(1-m_K^2/m_\tau^2)^2\over (1-m_\mu^2/ m_K^2)^2} \left( 1 +
\delta R_{\tau/K}\right) . \qquad
\end{eqnarray}
Owing to the different energy scales involved, the radiative
corrections to the $\tau^-\to\nu_\tau\pi^-/K^-$ amplitudes are,
however, not the same as the corresponding effects in
$\pi^-/K^-\to\mu^-\bar\nu_\mu$. The relative corrections
have been estimated \cite{pich_MS:93,pich_DF:94} to be:
\begin{equation}\label{eq:dR_tp_tk}
 \delta R_{\tau/\pi} = (0.16\pm 0.14)\% \ , \qquad\qquad
 \delta R_{\tau/K} = (0.90\pm 0.22)\%  \ .
\end{equation}
Using these numbers, the measured $\tau^-\to\pi^-\nu_\tau$ and
$\tau^-\to K^-\nu_\tau$ decay rates imply the $|g_\tau/g_\mu|$
ratios given in Table~\ref{tab:ccuniv}.

Assuming universality in the $W^\pm$ quark couplings, these decay
modes determine the ratio \cite{pich_PDG,pich_JOP:06}
\begin{equation}
 \frac{|V_{us}|\, f_K}{|V_{ud}|\, f_\pi} \, = \,\left\{
 \begin{array}{cc}
 0.27618\pm 0.00048 & \;\: [\Gamma_{K/\pi\to\nu_\mu\mu} ],
 \\[5pt]
 0.267\pm 0.005 & \;\: [\Gamma_{\tau\to\nu_\tau K/\pi}].
 \end{array}\right.
\end{equation}
The very different accuracies reflect the present poor precision on
$\Gamma(\tau^-\to\nu_\tau K^-)$. \bes3 could considerably improve
the measurements of the $\tau^-\to\nu_\tau K^-$ and
$\tau^-\to\nu_\tau \pi^-$ branching ratios. The monochromatic
kinematics of the final hadron at threshold will make possible a
clean separation of each decay mode and, therefore, 
excellent accuracy.

\subsection{Decays into Two Hadrons}

\begin{figure}[tbh]\centering
\begin{minipage}{0.45\textwidth}
\includegraphics[angle=-90,width=7.3cm,clip]{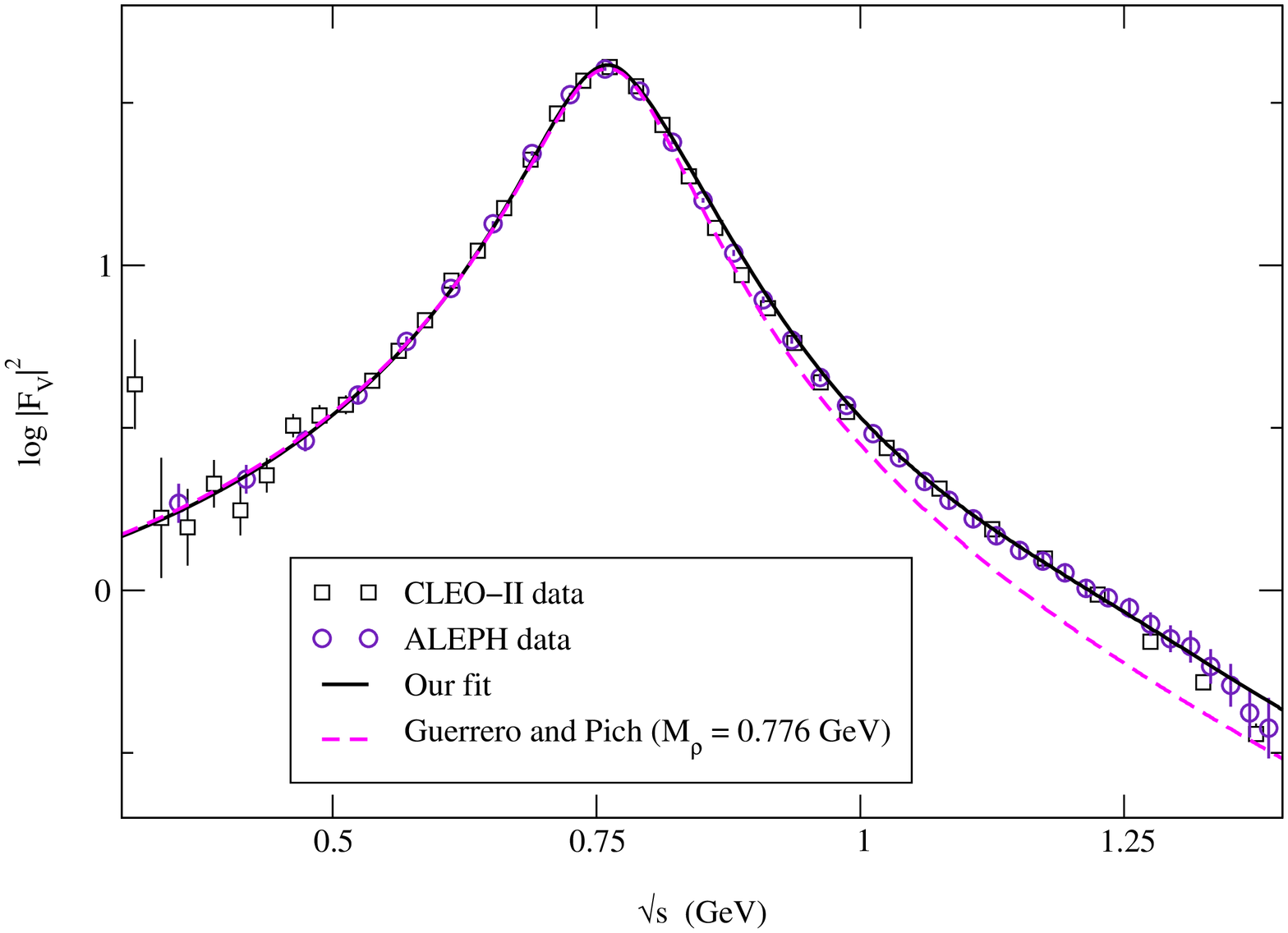}
\end{minipage}
\hskip 1cm
\begin{minipage}{0.45\textwidth}
\includegraphics[angle=-90,width=7cm,clip]{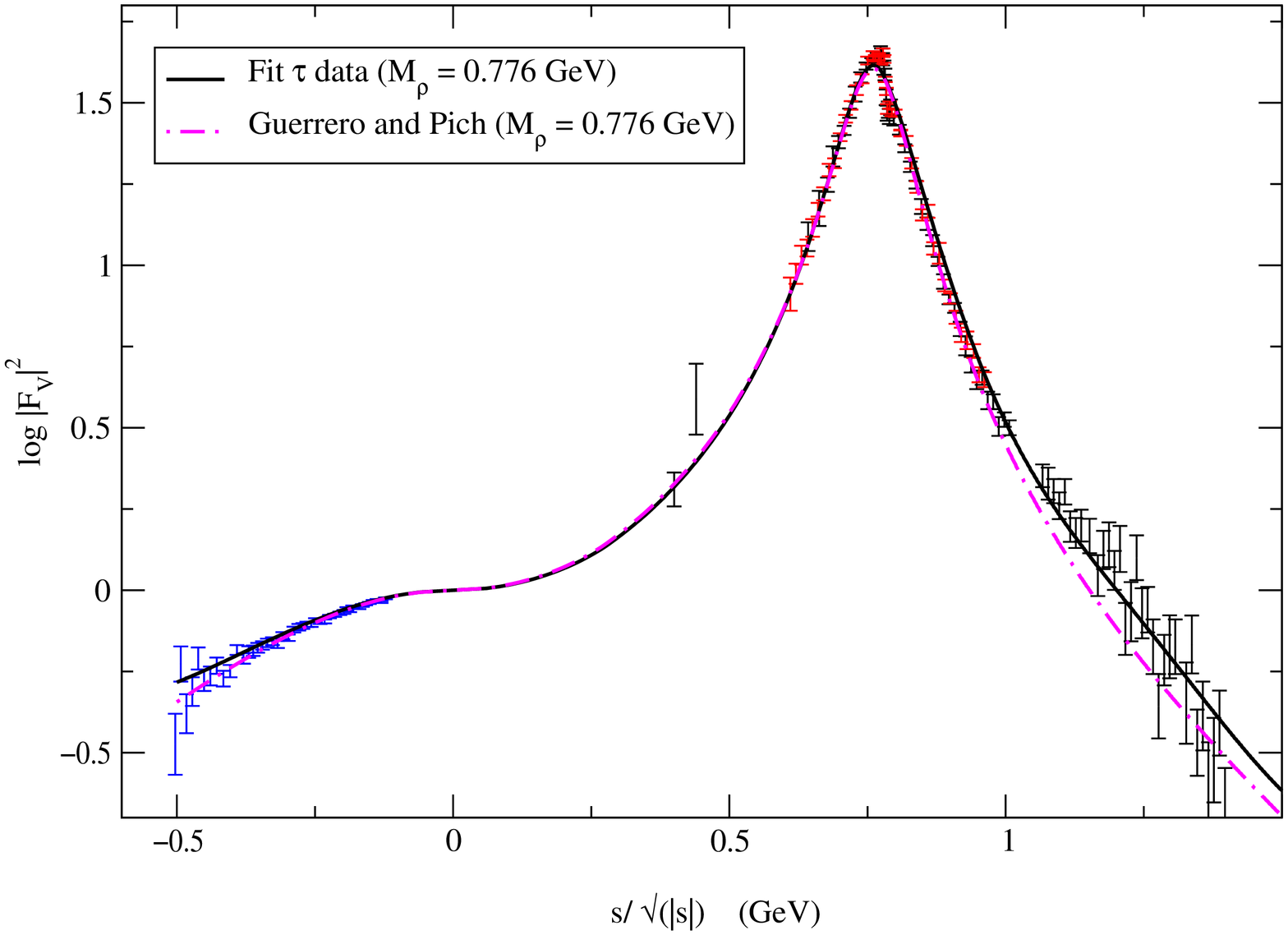}
\end{minipage}
\vspace*{8pt}\caption{The pion form factor from $\tau$ data
\protect\cite{pich_ALEPHpiff,pich_CLEOpiff} (left) and $e^+e^-$ data
\protect\cite{pich_BA85,pich_AM86} (right), compared with theoretical
predictions \protect\cite{pich_Portoles,pich_GP:97}. The dashed lines
correspond to the expression in Eq.~\ref{eq:PFF_GP}.}
\label{fig:pionth}
\end{figure}

For the two--pion final state, the hadronic matrix element is
parameterized in terms of the so-called pion form factor \ [$s\equiv
(p_{\pi^-}\! + p_{\pi^0})^2$]:
\begin{equation}\label{eq:Had_matrix}
 \langle \pi^-\pi^0| \bar d \gamma^\mu  u | 0\rangle \equiv
 \sqrt{2}\, F_\pi(s)\, \left( p_{\pi^-}-p_{\pi^0}\right)^\mu \, .
\end{equation}
A dynamical understanding of the pion form factor can be achieved
\cite{pich_Portoles,pich_GP:97,pich_DPP:00,pich_Juanjo}, using analyticity, unitarity
and some general properties of QCD, such as chiral symmetry
\cite{pich_GL:85} and the short-distance asymptotic behavior
\cite{pich_EGPR:89,pich_Tempe}. Putting all these fundamental ingredients
together, one gets the result \cite{pich_GP:97}
\begin{equation}\label{eq:PFF_GP}
F_\pi(s) = {M_\rho^2\over M_\rho^2 - s - i M_\rho \Gamma_\rho(s)}
\exp{\left\{-{s \,\mbox{\rm Re} A(s)           
\over 96\pi^2f_\pi^2} \right\}} ,
\end{equation}
where
\begin{equation}
 A(s) \equiv \log{\left({m_\pi^2\over
 M_\rho^2}\right)} + 8 {m_\pi^2 \over s} - {5\over 3} + \sigma_\pi^3
 \log{\left({\sigma_\pi+1\over \sigma_\pi-1}\right)}
\end{equation}
contains the one-loop chiral logarithms,
$\sigma_\pi\equiv\sqrt{1-4m_\pi^2/s}$ and the off-shell $\rho$ width
\cite{pich_GP:97,pich_DPP:00} is given by $\Gamma_\rho(s)\, =\,
\theta(s-4m_\pi^2)\,\sigma_\pi^3\, M_\rho\, s/(96\pi f_\pi^2)$.
This prediction, which only depends on $M_\rho$, $m_\pi$ and the
pion decay constant $f_\pi$, is compared with the data in
Fig.~\ref{fig:pionth}. The agreement is rather impressive and
extends to negative $s$ values, where the $e^-\pi$ elastic data
applies.

\begin{figure}[t]\centering
\centering
\includegraphics[width=7.5cm,clip]{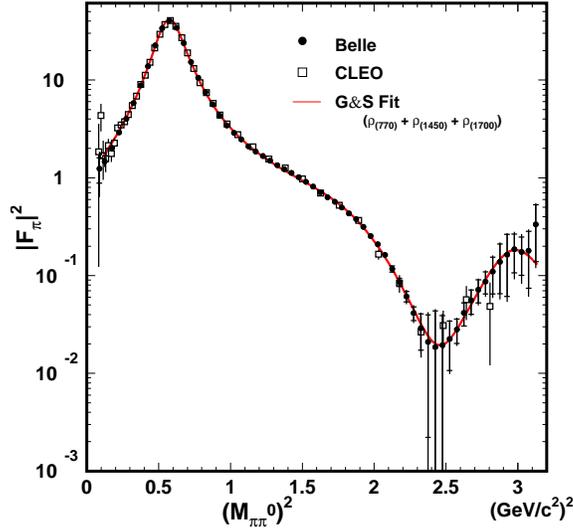}
\vspace{-.5cm} \caption{Preliminary Belle measurements of the pion
form factor from $\tau^-\to\nu_\tau \pi^-\pi^0$ decays
\protect\cite{pich_Fujikawa}.} \label{fig:BellePFF}
\end{figure}

The small effects of heavier $\rho$ resonance contributions and
additional next-to-leading  order $1/N_C$ corrections can be easily
included, at the price of having some free parameters that decrease
the predictive power~\cite{pich_Portoles,pich_Juanjo}. This gives a better
description of the $\rho'$ shoulder around 1.2 GeV (continuous lines
in Fig.~\ref{fig:pionth}).
A clear signal for the $\rho''(1700)$ resonance in
$\tau^-\to\nu_\tau \pi^-\pi^0$ events 
has been reported by Belle (see Fig.~\ref{fig:BellePFF}),
with a data sample 20 times larger than that of previous 
experiments~\cite{pich_Fujikawa}.

The $\tau^-\to\nu_\tau\pi^-\pi^0$ decay amplitude can be related
through an isospin rotation with the isovector piece of
$e^+e^-\to\pi^+\pi^-$. Thus, for $s< m_\tau^2$, $F_\pi(s)$ can be
obtained from the two sets of data. At present, there exists a
serious discrepancy between $e^+e^-$ and $\tau$ data. From $e^+e^-$
data one predicts $\mathrm{Br}(\tau\to\nu_\tau 2\pi) = (24.48\pm
0.18)\% $, which is $4.5\,\sigma$ smaller than the direct $\tau$
measurement $(25.40\pm 0.10)\% $~\cite{pich_DEHZ:03}. This discrepancy
translates into two different estimates of the hadronic vacuum
polarization to the anomalous magnetic moment of the muon; while the
$e ^+e^-$ data leads to a theoretical prediction for $(g-2)_\mu$
which is $3.3\,\sigma$ below the BNL-E821 measurement,  the
prediction obtained from the $\tau$ data is in much better agreement
($0.9\,\sigma$) with the experimental value~\cite{pich_taurev06b}.

Clearly, new precise $e^+e^-$ and $\tau$ data sets are needed. The
present experimental situation is very unsatisfactory, showing
internal inconsistencies among different $e^+e^-$ and $\tau$
measurements. The KLOE $e^+e^-$ invariant-mass distribution does not
agree with CMD2 and SND, while the most recent Belle measurement of
the $\tau$ decay spectrum~\cite{pich_Fujikawa} slightly disagrees with
ALEPH and CLEO \cite{pich_DEHZ:03}. The accurate measurement of
$F_\pi(s)$ at \bes3 could clarify this important issue.

\begin{figure}[tbh]\centering
\begin{minipage}{0.55\textwidth}\centering
 \includegraphics[width=8.5cm,clip]{./Part6/Toni/Tau/psfigs/dGTau.eps}
 \caption{Predicted $\tau\to\nu_\tau K\pi$ 
 distribution, together with the separate contributions from the
 $K^*(892)$ and $K^*(1410)$ vector mesons as well as the scalar
 component residing in 
 $F_0^{K\pi}(s)$ \protect\cite{pich_JPP:06}.}
 \label{fig:KpSpectrum}\end{minipage}
 \hfill
\begin{minipage}{0.4\textwidth}\centering
 \includegraphics[width=5.5cm,clip]{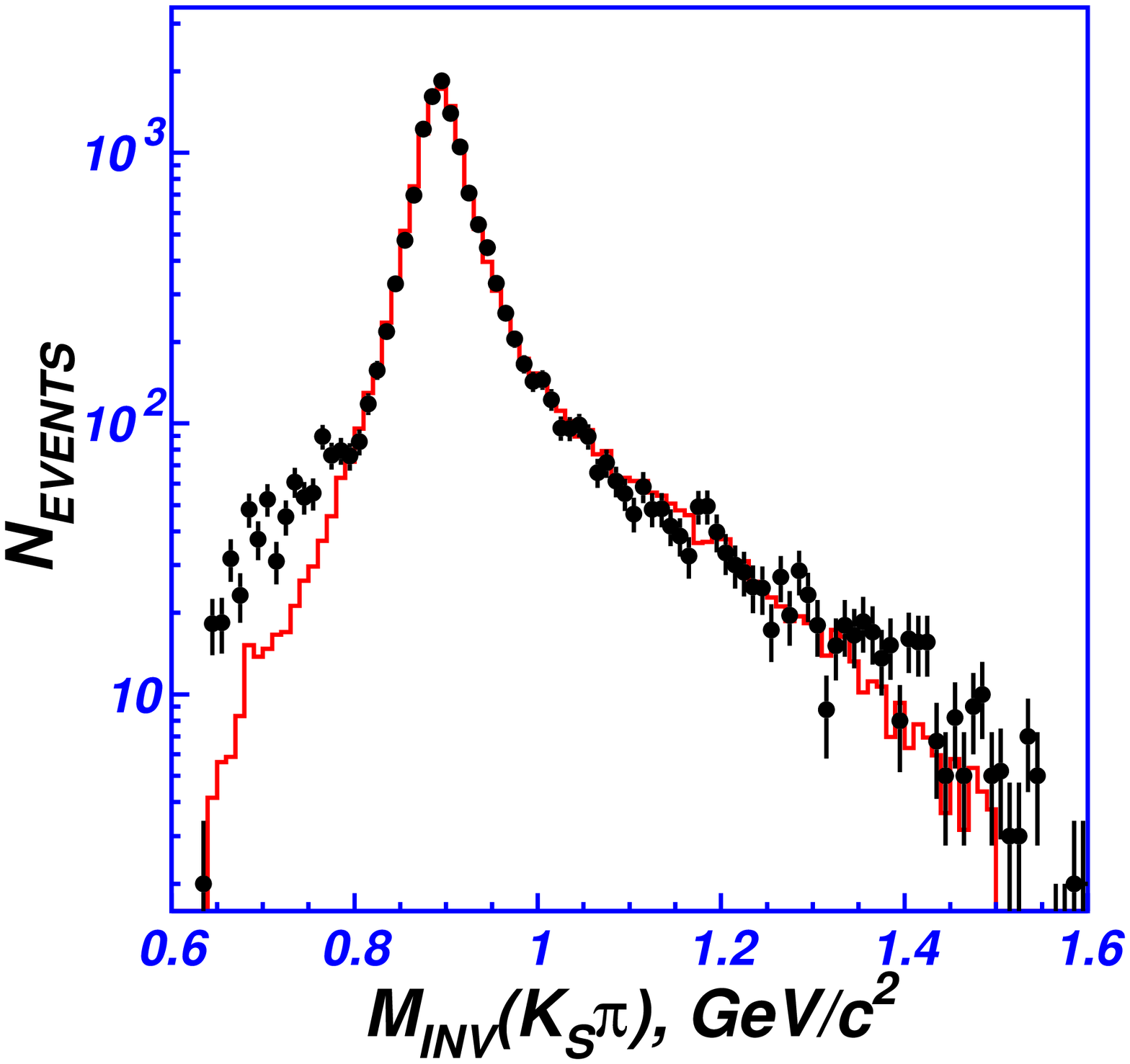}
 \caption{$K_S\pi$ invariant-mass distribution from
 BELLE $\tau\to\nu_\tau K_S\pi$ events. The histogram shows the
 expected $K^*(892)$ contribution \protect\cite{pich_Shwartz}.}
 \label{fig:Shwartz}\end{minipage}
\end{figure}

More recently, the decay $\tau\to\nu_\tau K\pi$ has been studied in
Ref.~\cite{pich_JPP:06}. The hadronic spectrum, shown in
Fig.~\ref{fig:KpSpectrum}, is characterized by two form factors,
\begin{equation}\label{dGtau2kpi}
 \frac{d\Gamma_{K\pi}}{d\sqrt{s}} =
\frac{G_F^2|V_{us}|^2 m_\tau^3}{32\pi^3s}
\biggl(1-\frac{s}{m_\tau^2}\biggr)^{\! 2}\Biggl[
\biggl(1+2\,\frac{s}{m_\tau^2} \biggr) q_{K\pi}^3 |F_+^{K\pi}(s)|^2
+ \frac{3\Delta_{K\pi}^2}{4s}\,q_{K\pi}^{\phantom{3}}
 |F_0^{K\pi}(s)|^2 \Biggr]\, ,
\end{equation}
where
$q_{K\pi}^{\phantom{3}}=\frac{1}{2\sqrt{s}}\,\lambda^{1/2}(s,m_K^2,m_\pi^2)$
and $\Delta_{K\pi}= m_K^2-m_\pi^2$. The vector form factor
$F_+^{K\pi}(s)$ has been described in an analogous way to
$F_\pi(s)$, while the scalar component $F_0^{K\pi}(s)$ takes
into account additional information from $K\pi$ scattering data
through dispersion relations \cite{pich_JOP:06,pich_JOP}. The decay width is
dominated by the $K^*(892)$ contribution, with a predicted branching
ratio Br$[\tau\to\nu_\tau K^*] = (1.253 \pm 0.078)\%$, while the
scalar component is found to be Br$[\tau\to\nu_\tau(K\pi)_{\rm
S-wave}]=(3.88\pm 0.19)\cdot 10^{-4}$.

Preliminary measurements of the $\tau^-\to\nu_\tau K_S\pi^-$
(Belle~\cite{pich_Shwartz})
and $\tau^-\to\nu_\tau K^-\pi^0$ (BaBar~\cite{pich_Nugent}) 
distributions show clear evidence for the 
scalar contribution at low invariant masses and a $K^*(1410)$ vector
component at large $s$ (see Fig.~\ref{fig:Shwartz}).

The dynamical structure of other hadronic final states can be
investigated in a similar way. The $\tau\to\nu_\tau 3\pi$ decay mode
was studied in Ref.~\cite{pich_DPP:01}, where a theoretical description
of the measured structure functions \cite{pich_CLEO3pi,pich_OPAL3pi,pich_ALEPH:05}
was provided. A detailed analysis of other $\tau$ decay modes into
three final pseudoscalar mesons is in progress \cite{pich_RPP:06}. The
more complicated $\tau\to\nu_\tau 4\pi$ and $e^+e^-\to 4\pi$
transitions have also been studied~\cite{pich_EU:02}. Accurate
experimental measurements of the hadronic decay distributions would
provide a very valuable data set to perform important tests of QCD
in the non-perturbative regime.


Violation in $\tau$

\section[Search for $CP$ Violation in $\tau$ decays]{Search for $CP$ Violation in $\tau$
decays\footnote{By I.~I.~Bigi}}
\label{sec:tau_cp}

There are two powerful motivations for probing {\it CP}~symmetry in lepton 
decays:
\begin{itemize}
\item
The discovery of {\it CP}~asymmetries in $B$ decays that are close to 100 
\% in a sense `de-mystifies' {\it CP}~violation, in that it established 
that  complex {\it CP}~phases are not intrinsically small and 
can even be close to 90 degrees.  This de-mystification would be 
completed, if {\it CP}~violation were found in the decays of leptons as
well.
\item
We know that CKM dynamics, which is so successful in describing quark flavour transitions, is utterly irrelevant to baryogenesis. There are actually intriguing arguments
for baryogenesis being merely a secondary effect driven by primary leptogenesis \cite{part6:Buchmuller:2005eh}. To make the latter less speculative, one has to find
{\it CP}~violation in leptodynamics.
\end{itemize}
The strength of these motivations has been well recognized in 
the community, as can be seen from the planned experiments to measure 
{\it CP}~violation in neutrino
oscillations and the ongoing heroic efforts to find 
an electron EDM. Yet there are other avenues to this 
goal as well that certainly are at least as challenging,
namely to probe {\it CP}~symmetry in $\tau$ decays. 
There is also a less orthodox
probe, namely attempting to extract an 
EDM for $\tau$ leptons from $e^+e^- \to \tau^+\tau^-$.  
It is understood that the Standard Model does not produce an observable
effect here. One should also note that one is searching for a 
{\it CP}-odd effect in an {\em electromagnetic} production process unlike 
in $\tau$ decays, which are controlled by the weak force.

The betting line is that $\tau$ decays -- next to the 
electron EDM and $\nu$ oscillations -- provide the best 
stage to search for manifestations of {\it CP}~breaking
leptodynamics. There exists a considerable literature on 
the subject started by discussions on a tau-charm factory 
more than a decade 
ago~\cite{part6:Tsai:1989ez,part6:Tsai:1994rc,part6:Kuhn:1992nz,
part6:Kuhn:1996dv}, which has recently
attracted renewed interest 
recently~\cite{part6:Bigi:2005ts,part6:Datta:2006kd,
part6:Bernabeu:2006wf,part6:Delepine:2006fv}
especially stressing the following points:
\begin{itemize}
\item
There are many more decay channels for tau lepons
than for muons, making 
the  constraints imposed by {\it CPT}  symmetry much less restrictive.
\item
The $\tau$ lepton has sizable rates into multibody final states. 
Due to their nontrivial kinematics, asymmetries can emerge in the 
final-state distributions, where they are likely to be significantly 
larger than in the integrated widths. The channel
$$
K_L \to \pi^+\pi^-e^+e^-
$$
illustrates this point.  It commands only the tiny branching ratio of 
$3\cdot 10^{-7}$.  The
forward-backward asymmetry $\langle A \rangle$
in the angle between the $\pi^+\pi^-$ and $e^+e^-$ planes constitutes
a {\it CP}-{odd} observable. It has been measured by KTeV and NA48 to 
be truly large, namely
about 13 \%, although it is driven by the small value of 
$|\epsilon _K| \sim 0.002$. One can, thus, trade
branching ratio for the size of a {\it CP} asymmetry.

\item
New Physics in the form of multi-Higgs models can 
contribute on the tree-level, such as the SM W exchange.
\item
Some of the channels could exhibit enhanced sensitivity to New Physics.
\item
Having polarized $\tau$ leptons provides a powerful 
handle on {\it CP}~asymmetries as well as control over systematics.
\end{itemize}
These features will be explained in more detail below. 
It seems clear that such measurements can be performed only 
in $e^+e^-$ annihilation, {\it i.e.} at \bes3, the 
existing $B$ factories,
or better still at a Super-Flavour factory. 
There one has the added advantage that one can realistically 
obtain highly polarized $\tau$ leptons: This can be
achieved directly by having the electron beam longitudinally 
polarized or more indirectly even with unpolarized beams by 
using the spin alignment of the produced
$\tau$ pair to `tag' the spin of the $\tau$ under study 
by the decay of the other $\tau$ like $\tau \to \nu \rho$.

\subsection{$\tau \to \nu K \pi$}\label{TAUCPDET}

The most promising channels for exhibiting 
{\it CP}~asymmetries are $\tau^- \to \nu K_S \pi^-$, $\nu K^-\pi^0$  
\cite{part6:Kuhn:1996dv}:
\begin{itemize}
\item
Due to the heaviness of the lepton and quark 
flavours they are most sensitive to nonminimal Higgs 
dynamics while being Cabibbo suppressed in the SM.
\item
They can show asymmetries in the final state distributions.
\end{itemize}
The SM does generate a {\it CP}~asymmetry in $\tau$ decays that should be 
observable. Based on known physics one can reliably predict a
{\it CP}~asymmetry \cite{part6:Bigi:2005ts}:
\begin{equation}
\frac{\Gamma(\tau^+\to K_S \pi^+ \overline \nu)-\Gamma(\tau^-\to 
K_S \pi^- \nu)}{\Gamma(\tau^+\to K_S \pi^+ \overline \nu)+
\Gamma(\tau^-\to K_S \pi^- \nu)}= (3.27 \pm 0.12)\times 10^{-3}
\label{CPKS}
\end{equation}
due to $K_S$'s preference for antimatter over matter. 
Strictly speaking, this prediction is more general than 
the SM: no matter what produces the {\it CP}~impurity in
the $K_S$ wave function, the effect underlying 
Eq.~\ref{CPKS} has to be present, while of course not affecting 
$\tau^{\mp} \to \nu K^{\mp}\pi^0$.

To generate a {\it CP}~asymmetry, one needs two different amplitudes 
that contribute coherently. This requirement is satisfied, 
since the $K\pi$ system can be produced from
the (QCD) vacuum in a vector and scalar configuration with 
form factors $F_V$ and $F_S$, respectively. Both are present in 
the data,  with the vector component (mainly
in the form of the $K^*$) dominant as 
expected~\cite{pich_taurev06a}. 
Within the SM, there is no weak phase between them 
at any observable level, yet it can
readily be provided by a charged Higgs 
exchange in non-minimal Higgs models, which contribute to $F_S$.

A few general remarks on the phenomenology might be helpful to 
set the stage. For a {\it CP}~violation in the underlying weak dynamics to 
generate an observable asymmetry in partial widths or energy 
distributions, 
one also needs a relative strong phase between the two amplitudes:
$$
\Gamma (\tau ^- \to \nu K^- \pi^0) - \Gamma (\tau ^+ \to \bar \nu K^+ \pi^0) ,
\frac{d}{dE_K} \Gamma (\tau ^- \to \nu K^- \pi^0) -
\frac{d}{dE_K} \Gamma (\tau ^+ \to \bar \nu K^+ \pi^0) \propto
$$
\begin{equation}
{\rm Im}( F_HF_V^*) {\rm Im}g_Hg_W^* \; ,
\end{equation}
where $F_H$ denotes the Higgs contribution to $F_S$ and $g_H$ 
its weak coupling. This should not represent a serious restriction, 
since the $K\pi$ system is produced
in a mass range with several resonances. If, on the other hand, one is 
searching for a {\it T}-odd correlation such as
\begin{equation}
O_T \equiv \langle \vec \sigma _{\tau} \cdot 
(\vec p_K \times \vec p_{\pi})\rangle \; ,
\end{equation}
then {\it CP}~violation can surface even with{\em out} a relative strong 
phase
\begin{equation}
O_T \propto {\rm Re}( F_HF_V^*) {\rm Im}g_Hg_W^* \; .
\label{OT1}
\end{equation}
However, there is a caveat: final state interactions can 
generate {\it T}-odd moments even from {\it T}-invariant dynamics, where
one has
\begin{equation}
O_T \propto {\rm Im}( F_HF_V^*) {\rm Re}g_Hg_W^* \; .
\label{OT2}
\end{equation}
Fortunately one can differentiate between the two scenarios 
of Eqs.~\ref{OT1} and \ref{OT2} at \bes3
by comparing directly the
{\it T}-odd moments for the {\it CP}-conjugate 
pair $\tau^+$ and $\tau ^-$: 
\begin{equation}
O_T(\tau^+) \neq O_T(\tau^-)  \; \; \; 
\Longrightarrow \; \; \; {\rm {\it CP}~violation!}
\end{equation}

A few numerical scenarios might illuminate the situation: 
a Higgs amplitude 1\% or 0.1\% the strength of the SM $W$-exchange 
amplitude -- the former [latter] contributing
[mainly] to $F_S$ [$F_V$] -- is safely in the 
`noise' of present measurements of partial widths; 
yet it could conceivably create a {\it CP}~asymmetry as large 1\% or
0.1\%, respectively.  More generally a {\it CP}-odd observable in a SM 
allowed process is merely {\em linear} in a New Physics amplitude, 
since the SM provides the other
amplitude. On the other hand SM-forbidden transitions -- say 
lepton-flavour violation as in $\tau \to \mu \gamma$ --  have 
to be {\em quadratic} in the New Physics amplitude.
\begin{equation}
{\it CP}-{\rm odd} \propto |T^*_{SM}T_{NP}|   \; \; {\rm vs.} \; \; {\rm 
LFV} \propto |T_{NP}|^2
\end{equation}
Probing {\it CP}~symmetry on the 0.1\% level in $\tau \to \nu K\pi$ thus 
has roughly the same sensitivity for a New Physics amplitude as searching for
BR$(\tau \to \mu \gamma)$ on the $10^{-8}$ level.

CLEO has undertaken a pioneering search for 
a {\it CP}~asymmetry in the angular distribution of $\tau \to \nu K_S \pi$ 
placing an upper bound of a few percent
\cite{part6:Anderson:1998ke}.

\subsection{Other $\tau$ decay modes}

It appears unlikely that analogous asymmetries could be 
observed in the Cabibbo allowed channel 
$\tau \to \nu \pi \pi$, yet detailed studies of 
$\tau \nu 3\pi/4\pi$ look promising, also because the more complex final 
state allows one to form {\it T}-odd correlations with unpolarized $\tau$ 
leptons; on the other hand, decays of polarized $\tau$ leptons might
exhibit much larger {\it CP}~asymmetries~\cite{part6:Datta:2006kd}.

Particular attention should be paid to $\tau \to \nu K 2\pi$, 
which has potentially very significant additional advantages:

\noindent
$\oplus$: One can interfere {\em vector} 
with {\em axial vector} $K2\pi$ configurations.

\noindent
$\oplus$: The larger number of kinematical variables 
and of specific channels should provide more
internal cross checks of systematic uncertainties such as 
detection efficiencies for positive vs. negative particles.

\chapter[$\tau$ mass near threshold]{$\tau$ mass near
threshold\footnote{By Xiao-Hu Mo, Y.~K. Wang, C.~Z.~Yuan and C.~D.~Fu}}
\label{part6:sec:tau_mass}

\newsavebox{\talupright}
\savebox{\talupright}(0,0){%
\setlength{\unitlength}{0.5mm}
\linethickness{0.2mm}
\put(0,0){\line(1,0){10.0}}
\put(0,0){\line(0,1){10.0}}
\put(0,10){\line(1,-1){10.0}}
}
\newsavebox{\talupleft}
\savebox{\talupleft}(0,0){%
\setlength{\unitlength}{0.5mm}
\linethickness{0.2mm}
\put(0,0){\line(1,0){10.0}}
\put(10,0){\line(0,1){10.0}}
\put(0,0){\line(1,1){10.0}}
}
\newsavebox{\taldwright}
\savebox{\taldwright}(0,0){%
\setlength{\unitlength}{0.5mm}
\linethickness{0.2mm}
\put(0,10){\line(1,0){10.0}}
\put(0,0){\line(0,1){10.0}}
\put(0,0){\line(1,1){10.0}}
}
\newsavebox{\taldwleft}
\savebox{\taldwleft}(0,0){%
\setlength{\unitlength}{0.5mm}
\linethickness{0.2mm}
\put(0,10){\line(1,0){10.0}}
\put(10,0){\line(0,1){10.0}}
\put(10,0){\line(-1,1){10.0}}
}

\section{Introduction}
\label{part6:xct_int}
\begin{figure}[tbh]
\begin{center}
\includegraphics[height=7cm,width=10.cm]{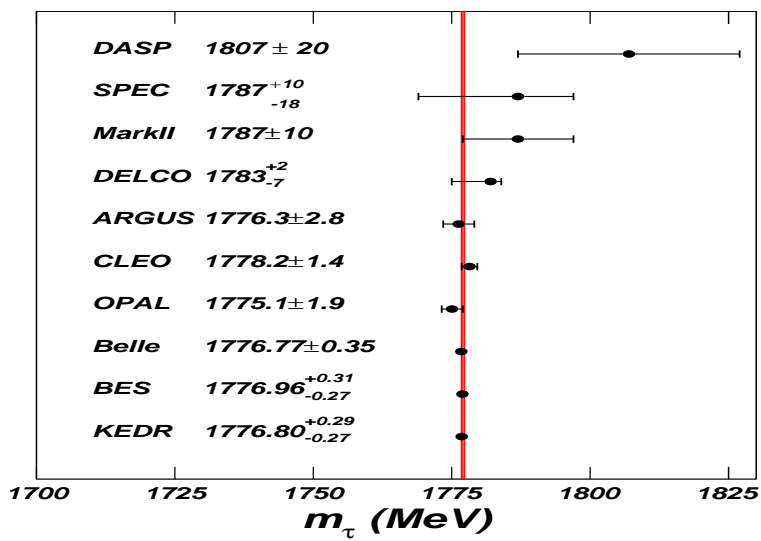}
\caption{\label{taumm}A comparison of  different measurements of
the $\tau$ mass. The vertical line indicates the current world average value:
$1776.99^{+0.29}_{-0.26}$ MeV~\cite{pich_PDG}, which is the averaged result
of the measurements from
Refs.~\cite{part6_taudelco,part6_tauargus,part6_taubes,part6_taucleo,part6_tauopal}.}
\end{center}
\end{figure}

The mass of the $\tau$ lepton is a fundamental parameter of the Standard
Model; many 
experiments~\cite{part6_taudasp,part6_tauspec,part6_taumak2,part6_taudelco,
part6_tauargus,part6_taubes,part6_taucleo,part6_tauopal,part6_taubelle,part6_taukedr}
have measured it as shown in Fig.~\ref{taumm}. Experimentally, 
the depolarization technique  developed by the KEDR collaboration 
has been used to realize a highly accurate beam energy 
calibration --- at the level of one part in $10^{6}$ for c.m. 
energies near the
$\tau$ mass threshold~\cite{part6_Bogomyagkov}--- while theoretically,
accurate calculations have claimed precisions
at the level of one part in $10^{4}$ 
for the near-threshold $\tau$-pair production 
cross section~\cite{part6_voloshin,part6_smith,part6_pich,part6_ruizf}. 
Large $\tau$-pair  data samples are expected at \bes3 and,
therefore, it is of great interest to understand how
accurate a $\tau$ mass measurement we can look forward 
to having in the near future.

%

Usually, either pseudomass and threshold-scan methods are employed to
measure the $\tau$ mass. The scan method adopted by the BES-I 
collaboration achieved the most accurate single
measurement of the $\tau$ mass~\cite{part6_taubes}: 
\beq m_{\tau} =
1776.96^{+0.18+0.25}_{-0.21-0.17} \mbox{ MeV }. \label{bestama} \eeq
Note that the relative statistical ($1.6\times 10^{-4}$) and
systematic ($1.7\times 10^{-4}$) uncertainties are comparable in
magnitude, improvements are needed in both categories.

\section{Statistical Optimization}
\label{part6:xct_stopt}
\subsection{Methodology}\label{sect_two}
We need to develop a scheme that provides the most precise $\tau$ mass 
measurement one can acheive given a specific period of data-taking time 
or, equivalently, a given amount of integrated luminosity.
A sampling technique 
is utilized to simulate various data taking possibilities, among which 
the optimal one is to be found. The likelihood function for a given scheme
is constructed as~\cite{part6_taubes,part6_taufirst,part6_tauxcold}:
\begin{equation}
LF(m_{\tau})=\prod\limits_i^n
\frac{\mu_i^{N_i} e^{-\mu_i} }{N_i!}~ ,
\label{lklihd}
\end{equation}
where $N_i$ is the observed number of $\TT$ events detected
in the  $e \mu$ final 
state\footnote{For simplicity only the $e \mu$ channel is considered 
at this time; the statistical significance will be improved if more 
channels are taken into account, see Sect.~\ref{sect_dis} for 
a detailed discussion.} at 
the c.m. energy scan point $i$ and $\mu_i$ is 
the expected number of events given by
\begin{equation}
 \mu_i(m_{\tau})=[\epsilon \cdot B_{e\mu} \cdot 
\sigma_{exp} (m_{\tau},E^i_{cm})+\sigma_{BG}] \cdot {\cal L}_i~~.
\label{mudef}
\end{equation}
In Eq.~\eref{mudef}, ${\cal L}_i$ is the integrated luminosity at 
$i^{\rm th}$ point, $\epsilon$ is the overall detection 
efficiency for the $e\mu$ final
states (including the trigger and event selection efficiencies), 
$B_{e\mu}$ is the combined branching ratio for 
decays $\tau^+ \rightarrow e^+ {\nu}_e \overline{\nu}_{\tau}$ and
$\tau^- \rightarrow \mu^- \overline{\nu}_{\mu} \nu_{\tau}$ (or
the corresponding charge conjugate mode) and $\sigma_{exp}$ is the
experimentally observed cross section, which has the form~\cite{part6_rad}
\begin{equation}
\sigma_{exp}
(s,m_{\tau},\Delta)=\int_{0}^{\infty}
d\sqrt{s'} G(\sqrt{s'},\sqrt{s})
\int_{0}^{ 1-\frac{ 4m_{\tau}^2 }{ s' } } dx
F(x,s') \frac{ \bar{\sigma}(s'(1-x),m_{\tau}) }{|1-\Pi (s'(1-x))|^2}~.
\label{expsec:1}
\end{equation}
Here $\sqrt{s}$ is the c.m. energy; $F(x,s)$ is the 
initial-state radiation factor~\cite{part6_rad}, $\Pi$ is the vacuum 
polarization factor~\cite{part6_ruizf,part6_berends,part6_rodrigo}, 
and $G(\sqrt{s'},\sqrt{s})$, which is usually treated as a
Gaussian distribution~\cite{part6_vitual}, depicts the energy 
spread of the $e^+e^-$ collider. The production cross section 
$\bar{\sigma}$ can be expressed as\footnote{
Here the Voloshin's formula in Ref.~\cite{part6_voloshin} is adopted. 
This takes  into account: (a) radiation from 
the initial electron and positron; (b) vacuum polarization of 
the time-like photon; (c) corrections to the special density 
of the electromagnetic current of the tau leptons; and (d)
the interference between the effects (a)-(c) which start 
from the relative order $\alpha^2$.} 
\begin{equation}
\bar{\sigma} (v) = \frac{2\pi\alpha^2}{3s}
 v(3-v^2)F_c(v)\left(1+\frac{\alpha}{\pi}S(v)-
\frac{\pi \alpha}{2v}+h(v)\right),
\label{prdsig}
\end{equation}
Here $v=\sqrt{1-4m_{\tau}^2/s}$ is the velocity of either of the 
$\tau$ leptons in the c.m. frame and $F_c(v)$ is the so-called Coulomb 
factor, which is defined as
\beq
F_c(v)=\frac{\pi\alpha/v}{1-\exp(-\pi\alpha/v)}~~.
\label{fcdef}
\eeq
A description of the correction function $S(v)$ can be found in 
Schwinger's textbook~\cite{part6_schwinger}:
\beq
\begin{array}{ll}
 S(v)=& {\displaystyle \frac{1}{v} \left\{ (1+v^2) \left[ \frac{\pi^2}{6}+
     \ln{ \left( \frac{1+v}{2} \right) }
     \ln{ \left( \frac{1+v}{1-v} \right) }
   +2\mbox{Li}_2 \left( \frac{1-v}{1+v} \right)
   +2\mbox{Li}_2 \left( \frac{1+v}{2} \right) \right. \right. }\\ \\
      & {\displaystyle \left. -2\mbox{Li}_2 \left( \frac{1-v}{2} \right)
       -4\mbox{Li}_2(v)+ \mbox{Li}_2(v^2) \right] }\\ \\
      & {\displaystyle  +\left[ \frac{11}{8}(1+v^2)-3v
       +\frac{1}{2}\frac{v^4}{(3-v^2)} \right]
       \ln{ \left( \frac{1+v}{1-v} \right) } }\\ \\
      & {\displaystyle \left. + 6v\ln \left( \frac{1+v}{2} \right)-4v\ln {v}
       +\frac{3}{4}v\frac{(5-3v^2)}{(3-v^2)} \right\} }~,
      \end{array}
\label{svdef}
\eeq
with
$$
 \mbox{Li}_2(x)=-\int^x_0\ln(1-t)dt/t=\sum \limits_{n=1}^\infty
   x^n/n^2.
$$
The correction function $h(v)$ is expressed in terms of a
double integral~\cite{part6_voloshin,part6_smith}:
\beq  h(v)=\frac{2\alpha}{3\pi} \left[
-2\lambda Im \int^{\infty}_{0}dt\int^{\infty}_{1}dx
  (\frac{1+t}{t})^{i\lambda}\frac{(t+izxv^{-1})^{i\lambda-1}}
   {(t+1+izxv^{-1})^{i\lambda+1}}(1+\frac{1}{2x^2})
   \frac{\sqrt{x^2-1}}{x^2} \right]~~,
\label{hvdef}
\eeq
with
$$ z=m_e/m_{\tau} \ \ , \ \ \lambda=\frac{\alpha}{2v}~.$$
The function $h(v)$ contains corrections from two sources:  
the so-called hard correction due to a finite radiative effect in 
the $\tau$ electromagnetic vertex at the threshold, 
and from the modification of the Coulomb interaction
due to running of the coupling $\alpha$, which is described by the
Uehling-Serber radiative correction to the potential~\cite{part6_uehling}. 

In the following study, we take $\epsilon=14.2\%$~\cite{part6_tauxc}, 
the c.m. energy spread\footnote{The c.m. energy 
spread is calculated from
the empircal formula:
$\Delta= (0.16203 E^2_{cm}/4+0.89638)\times 10^{-3}~\mbox{ GeV}$,
which gives $\Delta$ = 1.4 MeV at 
$E_{cm}$ = 1.77699 GeV~\cite{part6_tauxc}.} $\Delta = 1.4$~ MeV and
$B_{e\mu} = 0.06194$~\cite{pich_PDG}. We neglect the corresponding 
uncertainties, which are left for the systematic study. As for 
$\sigma_{BG}$, previous experience~\cite{part6_taufirst} indicates 
that $\sigma_{BG} \approx 0.024 \mbox{ pb}^{-1}$, which is small 
compared to the $\TT$ production cross section near threshold of
($\simeq0.1~{\rm nb}^{-1}$).  In any case, a large data sample
 can be taken below the threshold to determine $\sigma_{BG}$ accurately. 
For simplicity, we set 
$\sigma_{BG}$ to be zero, which means it is background free.
In fact, if $\sigma_{BG}$ is a constant, it has no effect on the 
optimization of the data taking strategy.  

Since we want to optimize the accuracy of 
the $m_{\tau}$ measurement, the value of $\tau$  mass itself is 
assumed to be known. In fact, the optimal number of points and the 
luminosity distribution among these points are correlated.
To resolve this dilemma, we use an iterative procedure:
we start from a simple distribution and
look for the optimal number of points; we then look for 
an optimal distribution; with such a distribution, we then reoptimize
the number of points.

\subsection{First Optimization}\label{sect_opone} 
As a starting point, we study an energy interval that
is evenly divided, {\it viz.}
\begin{equation}
E_i=E_0+i\times \delta E,\ \ \ \  (i=1,2,...,N_{pt}).
\label{engdiv}
\end{equation}
Here the initial point is $E_0=3.545~\mbox{GeV}$,
the final point is $E_f=3.595~\mbox{GeV}$ and the fixed step is
$\delta E=(E_f-E_0)/N_{pt}$, where $N_{pt}$ is the number of energy
points. The total luminosity is distributed equally 
(${\cal L}_i = {\cal L}_{tot}/N_{pt}$) at each point,\footnote{
Another scheme is to apportion the total number of events evenly at 
each point, the luminosity at each point is determined by
relation ${\cal L}_i = {\cal L}_{tot} / (\sigma_i \cdot \sum 1/\sigma_i)$
with $\sigma_i$ denotes the cross section at $i$-th point. 
This scheme leads to the same final conclusion of this study.} 
and the fitted $m_{\tau}$ and corresponding uncertainty 
$S_{m_\tau}^2(m_\tau) $ using Eq.~\eref{lklihd} is averaged 
over all the samplings ($N_{samp}$) for each $N_{pt}$ value
in order to suppress or reduce the statistical 
fluctuation~\cite{part6_dataanalysis}:
\begin{equation}
\overline{m}^i_\tau=\frac{1}{N_{samp}} \sum\limits_{j=1}^{N_{samp}}
m^i_{\tau j}\  , \label{mbar}
\end{equation}
\begin{equation}
S_{m_\tau}^2(m^i_\tau)=\frac{1}{N_{samp}-1}\sum\limits_{j=1}^{N_{samp}}
(m^i_{\tau j}-\overline{m}^i_{\tau })^2 \ . \label{smtau}
\end{equation}
Here $i$ indicates the  scheme being tested while $j$ indicates 
the sampling times, which is 500 in this study.

\begin{figure}[htbp]
\begin{minipage}{6cm}
\includegraphics[height=5cm,width=6.cm]{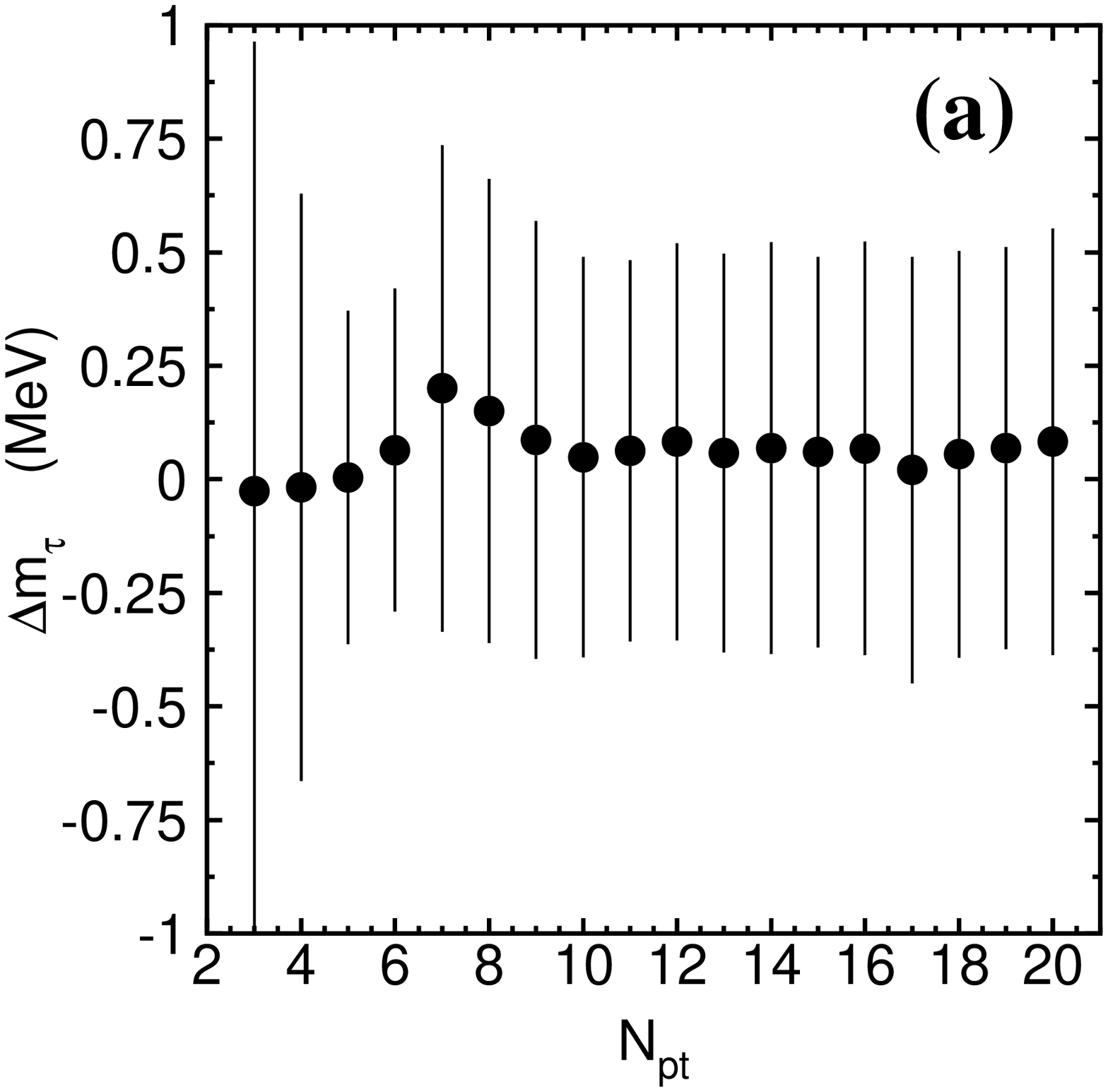}
\end{minipage}
\hskip 2cm
\begin{minipage}{6cm}
\includegraphics[height=5cm,width=6.cm]{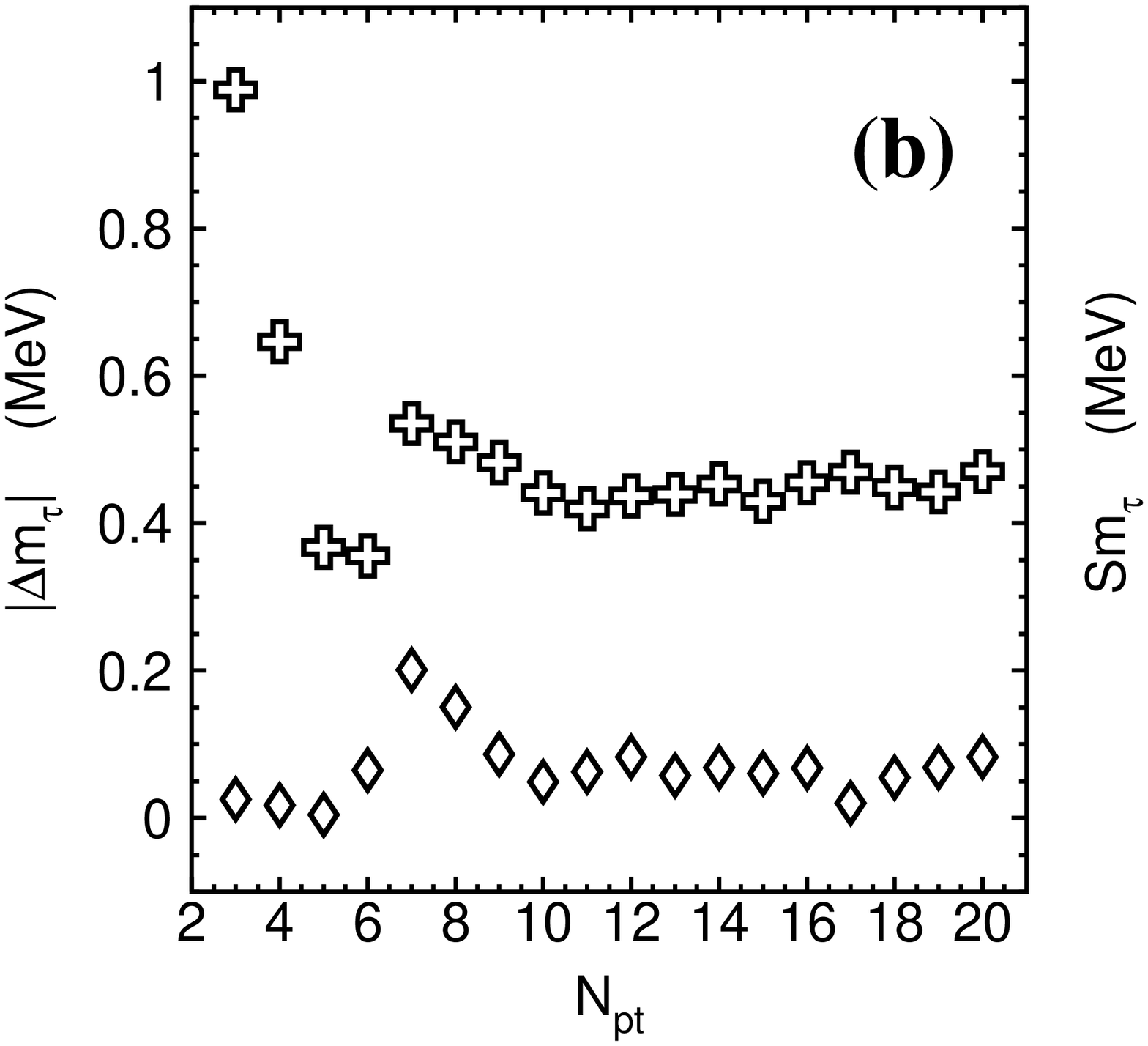}
\end{minipage}
\caption{\label{lmnpt}The variation of $\Delta m_\tau$ 
(and $|\Delta m_\tau|$) 
and $S_{m_\tau}$ {\it versus} the number of 
measurement points ${N}_{pt}$. 
In (a) the dots and bars represent $\Delta m_\tau$ and 
$S_{m_\tau}$, respectively.
In (b) the diamonds denote $|\Delta m_\tau |$ and
the crosses $S_{m_\tau}$.}
\end{figure}

Using the experiment parameters $\epsilon$, $\Delta$, and $B_{e\mu}$
given in Sect.~\ref{sect_two}
and setting ${\cal L}_{tot}=30\mbox{ pb}^{-1}$,
we obtain the fitted results for
$N_{pt}$ ranging from 3 to 20 
shown in  Figs.~\ref{lmnpt}(a) and (b), 
where $\Delta m_\tau = \overline{m}_\tau - m^0_\tau$, the difference between
the average fitted $m_{\tau}$ and the input value ($m^0_\tau=1776.99$ MeV 
according to PDG06~\cite{pich_PDG}), and $S_{m_{\tau}}$ is the 
corresponding $rms$ uncertainty.

The $\Delta m_\tau$ values, shown as the dots in Fig.~\ref{lmnpt}(a), 
indicate the fit bias due to the limited number of 
events.  With increased luminosity (and increased 
number  of events), the bias tends to zero.  
This point is demonstrated more clearly in 
a separate study 
of the variation of  $\Delta m_\tau$ with luminosity,
discussed below (in reference to Table~\ref{lumauty}). 

For the fit uncertainties, $S_{m_{\tau}}$,  two points should be 
noted: first, $S_{m_{\tau}}$ is much larger than absolute value of
the bias $|\Delta m_\tau|$. Thus,  from the point view of accuracy, 
the optimization of the former is much more crucial 
than that of the latter.
Therefore, in the following study $S_{m_{\tau}}$ is used as the 
goodness-of-fit quantity. 
Second, it is evident in  Fig.~\ref{lmnpt}(b)
that taking very few data  points yields
a large uncertainty, while having many points makes no contribution
to the improvement of the accuracy.  From the figure one can
see that $N_{pt}=5$ is near the optimal number of
measurement points for the evenly-divided-distribution scheme.

\begin{figure}
\begin{center}
\includegraphics[height=6cm,width=7.cm]
{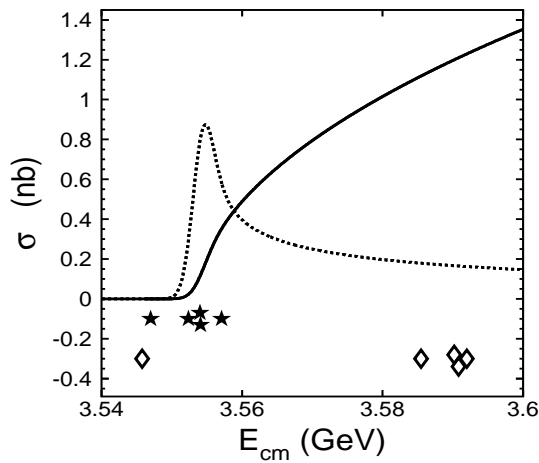}
\caption{\label{rdskn}The distributions of the data taking points
for the schemes with the smallest (denoted by stars) and largest 
(denoted by diamonds) values of $S_{m_\tau}$. 
The solid curve is the calculated observed cross section,
and the dashed line the corresponding derivative of the cross section
with respect to energy (with a scale factor of $10^{-2}$).}
\end{center}  
\end{figure}

With five points, we search further for ways to minimize the
fit uncertainty. Without any theoretical considerations, a sampling 
technique is employed and the energy points 
are distributed randomly over the chosen interval. 
In a total of 200 samplings, the one with the smallest uncertainty  
has $S_{m_\tau}=0.152$ MeV and the one with the greatest 
uncertainty has  $S_{m_\tau}=1.516$ MeV.
The two extreme distributions are indicated in Fig.~\ref{rdskn}. 
It is apparent that the small uncertainty measurement has scan
points that are crowded near the threshold,
while the scheme with the large uncertainty has
scan points that are far from the threshold.\footnote{
For both schemes, the lowest energy point in 
Fig.~\ref{rdskn} corresponds 
to an expected number of events of zero, since they are below 
threshold.} 
Intuitively, one expects that the smallest uncertainty occurs
at energies near the point where the derivative of the cross section 
is largest.  This must be, therefore, close to the optimal position for 
data taking, as discussed in the next section.

\subsection{Second Optimization}\label{sect_optwo} 
Based on the previous study, we embark on studies to determine 
(a) the energy region most sensitive to the fit uncertainty, 
(b) the optimal number of points that should be taken in that region, 
and  (c) the locations of the optimal points. 

\subsubsection{Optimal region}
\begin{figure}[htbp]
\begin{center}
\begin{minipage}{6cm}
\includegraphics[height=5cm,width=6.cm]{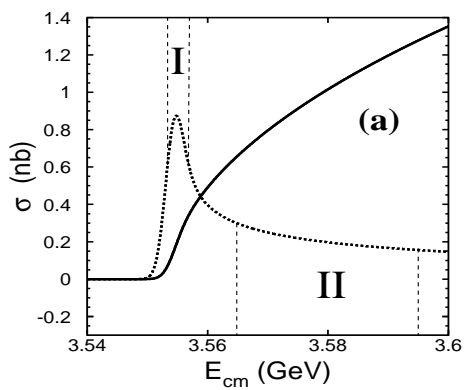}
(a) Energy regions $I$ and $II$
\end{minipage}
\hskip 2cm
\begin{minipage}{6cm}
\includegraphics[height=5cm,width=6.cm]{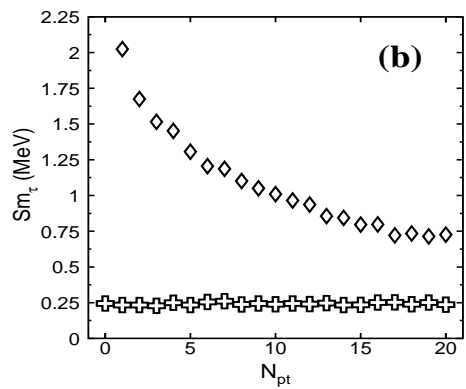}
(b) results for different schemes
\end{minipage}
\caption{\label{regdy}(a) The locations of the
energy subregions $I$ and $II$, 
with different derivative features. Here the solid line denotes the 
observed cross section and the dashed line the corresponding derivative
value (with a scale factor of $10^{-2}$). 
(b) The fit uncertainties for the two measurement schemes, the crosses and 
the  diamonds denote, respectively, the results for the first and second 
schemes as described in the text.}
\end{center}
\end{figure}

To hunt for the most sensitive energy region, two regions are defined 
as shown in Fig.~\ref{regdy}(a):
region $I$ ($E_{cm} \subset (3.553,~3.558)~\mbox{GeV}$ ), where
the derivative is greater than 75\% of its maxinum value and
region $II$ ($E_{cm} \subset (3.565,~3.595)~\mbox{GeV}$ ), where
the variation of the derivative is smoother than it is in region $I$.

To confirm the afore-mentioned speculation, 
two schemes are designed. In the first scheme, two points are taken in
region $I$, one at 3.55398 GeV as the threshold point and
the other at 3.5548 GeV corresponding to the energy point where
the derivative of the cross section is largest.  In addition
points are taken in
the region $II$, with the number of points $N_{pt}$ is
varied from 1 to 20, with each point having a 
luminosity of 5~pb$^{-1}$.  The fit
results for this scheme, displayed as crosses in 
Fig.~\ref{regdy}(b), show no improvement in precision
resulting from an increased number of data points 
tin region $II$
($S_{m_{\tau}}$ stays very near 0.25~MeV 
for any number of points).  To further examine 
this point, we investigated a second scheme, 
which only uses energy points in  region 
$II$, again with $N_{pt}$ varied from 1 to 20.
The fit results for this scheme are displayed 
as diamonds in Fig.~\ref{regdy}(b).
As expected, with the increasing number of points, $S_{m_{\tau}}$
decreases, but even with 20 points spread over region
$II$, the value of $S_{m_{\tau}}=0.7256$~MeV is still much larger than
that of $S_{m_{\tau}}$ with only two points in region $I$.
From this we conclude that the data taken
at energy points within region $I$ are much more
useful for optimal data taking.

\subsubsection{Optimal position}
\begin{figure}[htbp]
\center
\includegraphics[height=5cm,width=5.cm]{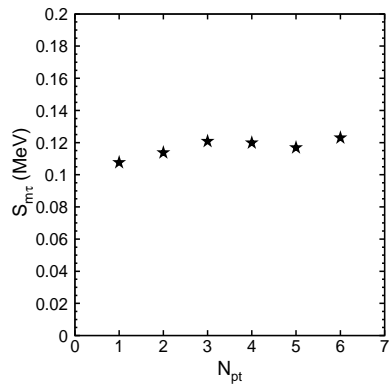}
\caption{\label{sper}The relation between $S_{m_\tau}$
and the number of measurement points within the energy region 
between 3.553~GeV  and 3.555~GeV.}
\end{figure}

In this subsection, we investigate the number of energy points that
are optimal in the large derivative region ($I$). 
Using a procedure similar to that described 
in Sect.~\ref{sect_opone}, the total luminosity
${\cal L}_{tot}=45$ pb$^{-1}$ is evenly distributed into $N_{pt}$
points ($N_{pt}=1,2, \cdots, 6$) inside the energy region 
between 3.553~GeV and 3.557~GeV. 
The results for  $S_{m_\tau}$
are shown in Fig.~\ref{sper}, where it is seen
that the number of points has a weak effect on the final uncertainty.
In other words, within the large derivative region, a single point 
is sufficient to give a small uncertainty. This is easy to understand 
since  there is only one free parameter ($m_{\tau}$) to be fit in the 
$\TT$ production cross section, even one measurement will fix
the normalization of the curve. The larger the derivative, the
more sensitivity to the mass of $\tau$ lepton.

\begin{figure}[htbp]
\begin{center}
\begin{minipage}{6cm}
\includegraphics[height=5cm,width=6.cm]{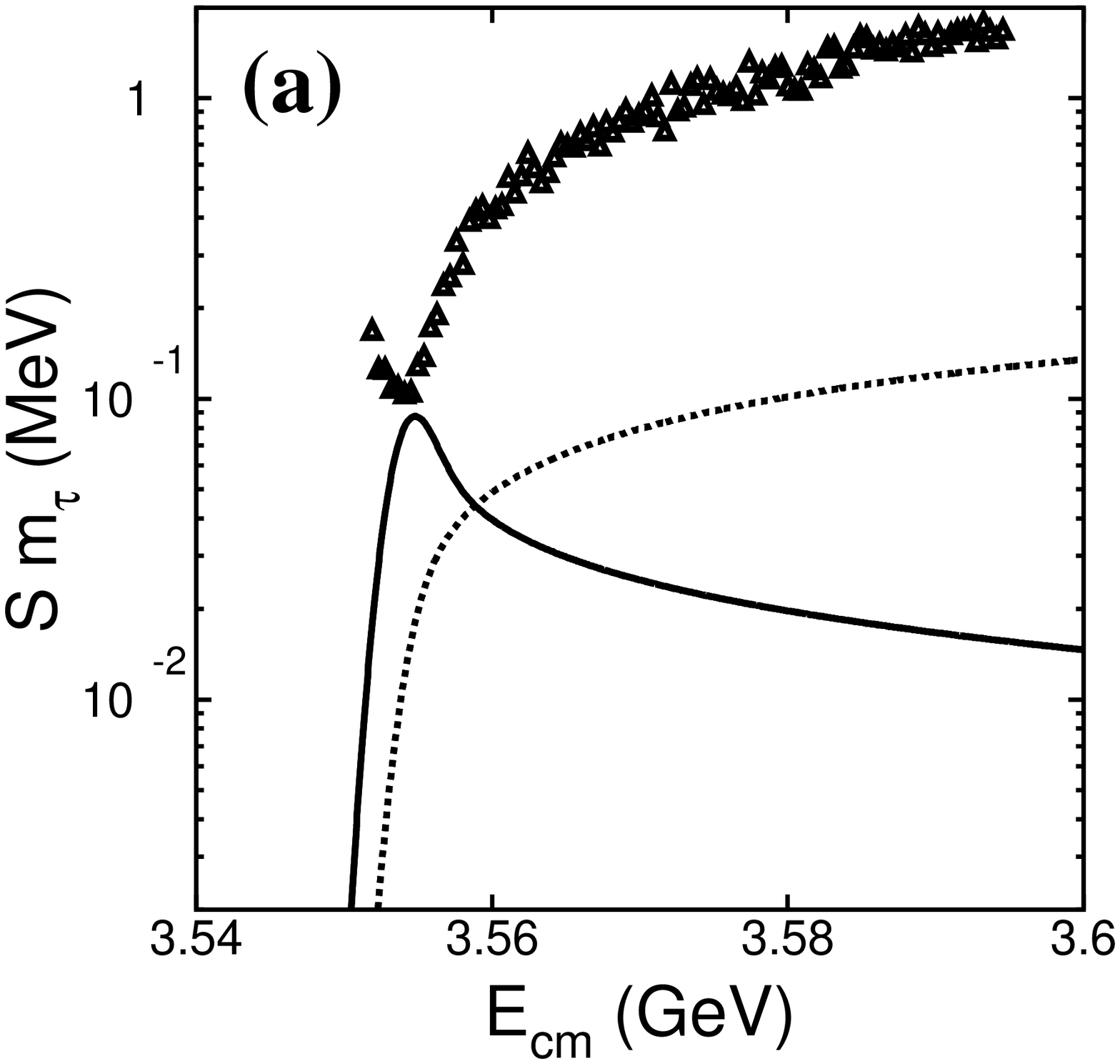}
(a) scan region from 3.551 to 3.595~GeV
\end{minipage}
\hskip 2cm
\begin{minipage}{6cm}
\includegraphics[height=5cm,width=6.cm]{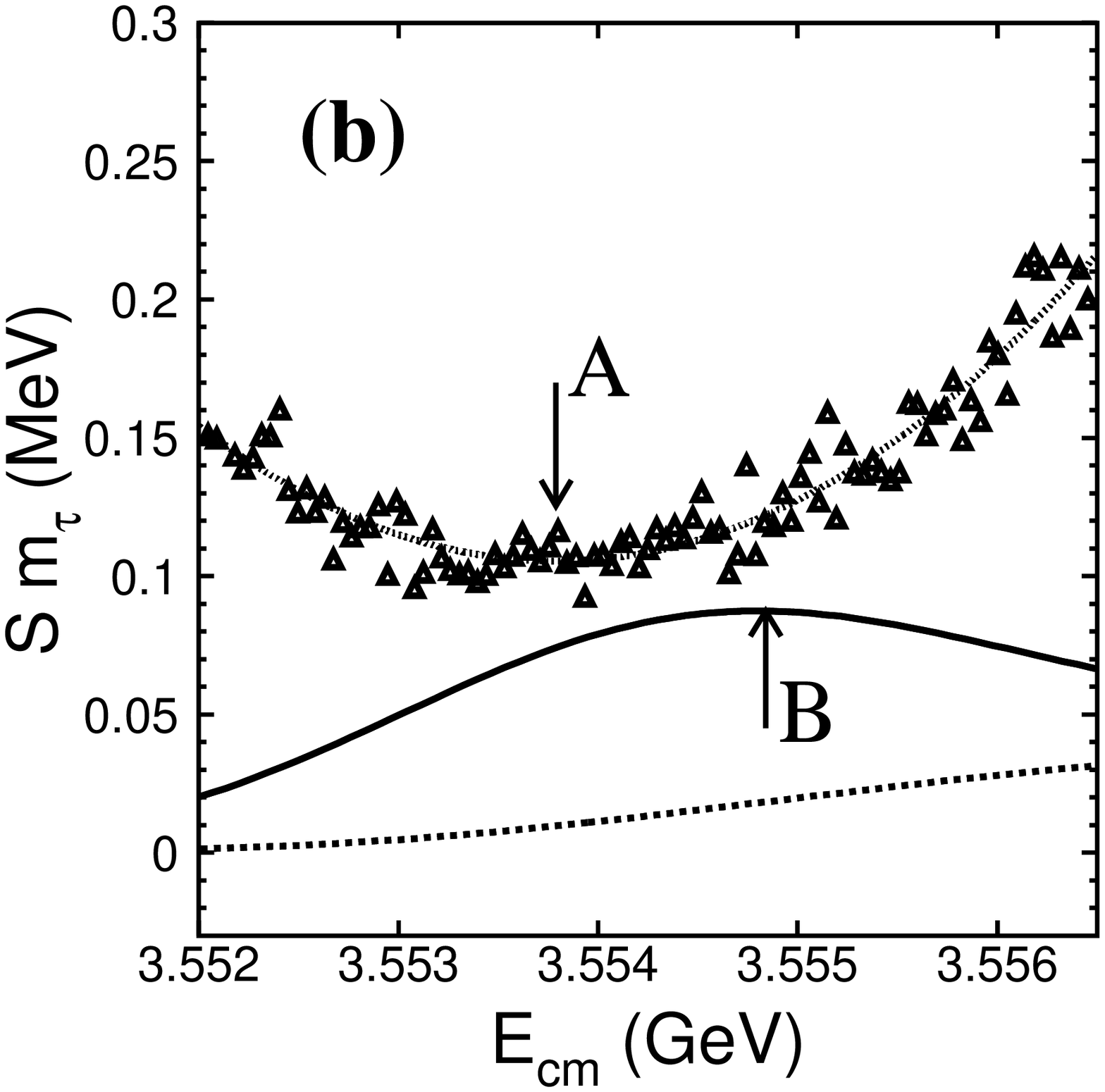}
(b) 3.552~GeV to 3.5565~GeV scan region
\end{minipage}
\caption{\label{dvskn}The variation of $S_{m_\tau}$ {\it versus} 
energy from a one-point measurement with ${\cal L}_{tot}=45$ pb$^{-1}$.
(a) over the scan region from 3.551~GeV to 3.595~GeV 
(b) over the scan region from 3.5533~GeV to 3.55694~GeV. 
The solid line denotes the cross section derivative with a scale factor
of $10^{-3}$ and the dashed line the observed cross section
with a scale factor of $10^{-1}$.}
\end{center}
\end{figure}

If one point is enough, an immediate question is where is the optimal 
point located? To answer this, a one-point scan with the luminosity
${\cal L}_{tot}=45$ pb$^{-1}$ was made and the results are shown in 
Figs.~\ref{dvskn}(a) and (b). 
As indicated in previous study, the small uncertainty is near 
the peak of the derivative. Actually, the most precise result, 
$S_{m_\tau} =0.105$~MeV,
is obtained  at the $m_{\tau}$ threshold, at 3.55398~MeV;   
the measurement taken right
at the peak of the cross section derivative, 
$S_{m_\tau}$ is slightly worse.
In addition, the study indicates that within a 2~MeV region the 
variation of $S_{m_\tau}$ is fairly small 
(from 0.105~MeV to 0.127~MeV), which is
very good for actual data taking.

\subsubsection{luminosity and uncertainty}

The last question we investigate is the relation between
the uncertainty and the luminosity. For the fit with one 
point in the large-derivative region, results are listed in 
Table~\ref{lumauty}.
The second and third column present the results at
$E_{cm}=3.55398$~GeV, corresponding to the threshold while the
last two columns give the results at $3.55484$~GeV,
corresponding to the point with the largest 
cross section derivative. From the results
in Table~\ref{lumauty}, we see that the precision is
inversely proportional to the luminosity and a luminosity of
63 pb$^{-1}$ is sufficient to provide an accuracy 
of less than 0.1~MeV.

\begin{table}[htb]
\caption{\label{lumauty}The relation between luminosity and
uncertainty.}
\center
\begin{tabular}{ccccc} \hline \hline
$L_{tot}$ & \multicolumn{2}{c}{$E_{cm}=3.55398$ GeV}
                           & \multicolumn{2}{c}{$E_{cm}=3.5548$ GeV} \\
          & $Sm_\tau$  & $\Delta m_\tau$ & $Sm_\tau$ & $\Delta m_\tau$ \\
(pb$^{-1}$) 
          & (MeV)      &  (MeV)          & (MeV)     & (MeV)\\ \hline\hline
      9  & 0.24874  & 0.02931  & 0.29240  &$ 0.02114$ \\
     18  & 0.16926  & 0.01550  & 0.19635  &$ 0.00756$ \\
     27  & 0.14024  & 0.01234  & 0.15670  &$ 0.00475$ \\
     36  & 0.12130  & 0.00812  & 0.14384  &$ 0.00504$ \\
     45  & 0.10653  & 0.00824  & 0.12717  &$ 0.00292$ \\
     54  & 0.09783  & 0.00717  & 0.10714  &$-0.00037$ \\
     63  & 0.09035  & 0.00726  & 0.09923  &$-0.00003$ \\
     72  & 0.08424  & 0.00520  & 0.09297  &$ 0.00008$ \\
    100  & 0.06781  & 0.00129  & 0.07876  &$-0.00002$ \\
   1000  & 0.02146  & 0.00016  & 0.02515  &$ 0      $ \\
  10000  & 0.00684  & 0        & 0.00805  &$ 0      $ \\ \hline \hline
\end{tabular}
\end{table}

\subsection{Discussion}\label{sect_dis}
At \bes3, the design peak luminosity 
is around 1 nb$^{-1}{\rm s}^{-1}$. If the average luminosity 
is taken to be 50\% of the peak value, two-days of data taking 
will give to 
a statistical uncertainty of less than 0.1~MeV. 
Notice that this evaluation 
is solely for $e\mu$-tagged event, other channels, 
such as $ee$, $e\mu$, $e h$, $\mu\mu$, $\mu h$ and $h h$ 
will have at least five times the number
of $e\mu$-tagged events~\cite{part6_taufirst,part6_taubes},
and these can significantly improve the uncertainty.
Therefore at \bes3, a one-week data-taking run will lead to a 
statistical uncertainty of order 0.017~MeV.

\section{Systematic uncertainty}
\label{part6:xct_sun}

In this section systematic uncertainties on $m_{\tau}$ measurements
will be examined, including the theoretical accuracy, energy spread,
energy scale, luminosity, efficiency, backgrounds, etc. 

\subsection{Theoretical accuracy}
\begin{figure}[bth]
\begin{center}
\includegraphics[height=8.cm,width=9.cm]{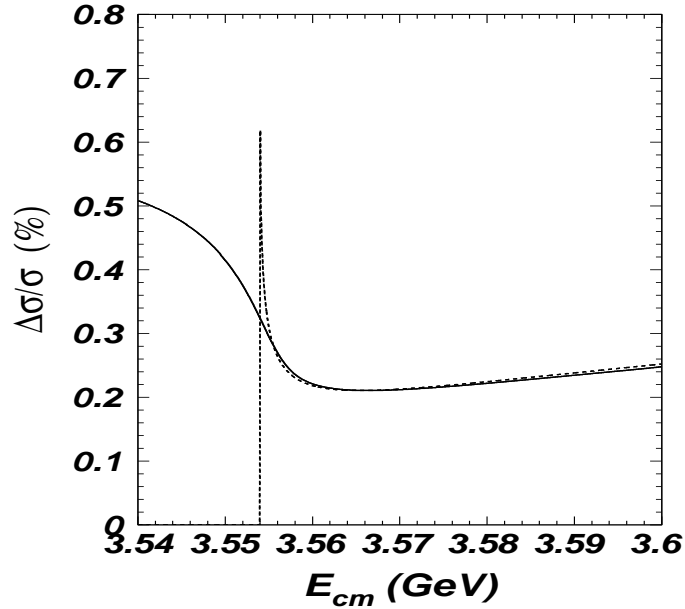}
\caption{\label{cmpfmu}The relative differences of the 
latest and previous cross section
calculations from Voloshin~\cite{part6_tauxc}.
The result that includes effects of the beaam energy spread 
($\sigma_{exp}$) are shown as the solid line. 
Results without beam energy effects ($\bar{\sigma}$)
are shown as the dashed line.}
\end{center}
\end{figure}

The experimentally observed cross section has the form~\cite{part6_rad}
$$\sigma_{exp} (m_{\tau},s,\Delta)=\int_{0}^{\infty}
d\sqrt{s'} G(\sqrt{s'},\sqrt{s}) $$
\beq
\times \int_{0}^{ 1-\frac{ 4m_{\tau}^2 }{ s' } } dx F(x,s')
\frac{ \bar{\sigma}(s'(1-x),m_{\tau}) }{|1-\Pi (s'(1-x))|^2}~,
\label{expsec:2}
\eeq
where $s=E^2_{cm}$, $F(x,s)$ is the initial state radiation
factor~\cite{part6_rad}, $\Pi$ is the vacuum polarization
factor~\cite{part6_ruizf,part6_berends,part6_rodrigo}, and
$G(\sqrt{s'},\sqrt{s})$, describes the energy-spread
of the $e^+e^-$ collider, which is usually treated as a
Gaussian function~\cite{part6_vitual}, 
As mentioned in Sect.~\ref{part6:xct_int}, the high-accuracy 
calculations of the production cross section ($\bar{\sigma}$) 
have  only recently become available.
Voloshin's improved formulae~\cite{part6_voloshin}
are used here; the production cross section 
(denoted as $\bar{\sigma}^*$) used in in the BES-I $m_{\tau}$
fit was based on Voloshin's earlier results~\cite{part6_tauxcold}.
The relative differences in the cross sections calculated with
the two sets of Voloshin's formulae are shown in Fig.~\ref{cmpfmu}.

To estimate the effect due to theoretical calculation accuracy,
the fitted $\tau$ masses by using the two formulae are compared.
The comparison shows the uncertainty due to this effect is at the
level of $10^{-3}$~MeV.

\subsection{Energy spread}
As indicated in Eq.~\eref{expsec:2}, the experimentally measured cross
section $\sigma_{exp}$ depends on the energy spread ($\Delta$); the
effect on the $m_{\tau}$ measurement is considered here.

In fact, the value of $\Delta$ in the $\tau$ 
threshold region is usually interpolated from 
the energy spreads measured at the $\jpsi$ ($\Delta_{\jpsi}$) and
$\psp$ ($\Delta_{\psp}$) regions, assuming the relation
\beq
\frac{\Delta - \Delta_{\jpsi}}{\Delta_{\psp} - \Delta_{\jpsi}}=
\frac{f(E) - f(E_{\jpsi})}{f(E_{\psp}) - f(E_{\jpsi})}~.
\label{eq_espd}
\eeq
It is assumed that $\Delta \propto f(E)$, where $f(E)$ denotes the
dependence of $\Delta$ on the beam energy $E_{cm}$.
Since a rigorous form for $f(E)$ is not available,
a generic form is assumed:
\beq
f(E)=a\cdot E + b\cdot E^2 + c\cdot E^3~,
\label{eq_fenerg}
\eeq
and linear, quadradiic, cubic or a mixed-type of energy-dependence
is used. 
At BES-I, the fit results give $\delta m_{\tau} < 1.5 \times 10^{-3}$ MeV.
Even if $\Delta$ is artificially changed to $3\Delta$ the fit indicates that
$\delta m_{\tau} < 6 \times 10^{-3}$ MeV.

\subsection{Energy scale}
 In the BES-I measurement, precisely known mass values of the $\jpsi$ 
($M_{\jpsi}$) and the $\psp$ ($M_{\psp}$) were used as the scale to 
calibrate  the energies in $\tau$ ($E_{\tau}$) 
threshold region,
\beq
\frac{E_{\tau} - E_{\jpsi}}{E_{\psp} - E_{\jpsi}}=
\frac{E_s - M_{\jpsi}}{M_{\psp} - M_{\jpsi}}~,
\eeq
where $E_s$ is the scaled energy value. If the energy and
mass peak have only a small relative shift $\delta$,
which in actuality are
at the level of $10^{-4}$, the relation can be written as
\beq
\frac{E_s - M_{\jpsi}}{M_{\psp} - M_{\jpsi}}=
\frac{\delta_s - \delta_{\jpsi}}{\delta_{\psp} - \delta_{\jpsi}}~.
\label{eq_escale}
\eeq
Similar to the energy spread, if we could assume
$\delta \propto f(E)$, the uncertainty due to the energy scale
would only be about $8 \times 10^{-3}$ MeV.

However, the relation $\delta \propto f(E)$ is merely a speculation and
probably not a safe assumption. 
At present, there are two approaches to determine the absolute energy 
scale directly, one is depolarization and the other is Compton back-scattering, 
both were developed by the KEDR group~\cite{part6_taukedr}.

\begin{figure}[bth]
\begin{center}
\includegraphics[height=5.cm,width=6.cm]{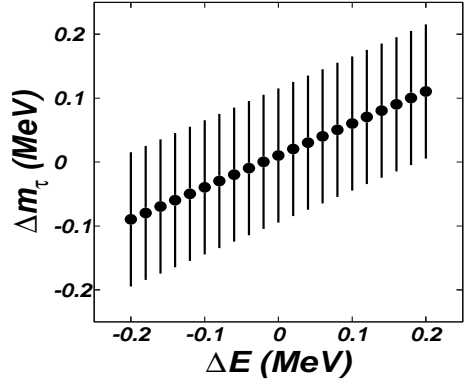}
\caption{\label{eskeff}The effects of uncertainty in the absolute
energy calibration on $m_{\tau}$ measurement.}
\end{center}
\end{figure}

It is proposed to adopt the Compton back-scattering technique to
measure the BEPCII beam energies. The relative precision of this technique 
is expected to be at the $5 \times 10^{-5}$ level, which directly 
translates to a systematic uncertainty on $m_{\tau}$ of
$5 \times 10^{-5}$ (relative error) or 
$0.09$ MeV (absolute error).  This error is an
 order-of-magnitude  larger than the 
combined error from other sources,
which is $8 \times 10^{-3}$~MeV.
Since the uncertainty of the energy scale will transfer 
directly to the final 
$m_{\tau}$ measurement as displayed in Fig.~\ref{eskeff},
the absolute calibration of energy scale is the
\underline{bottleneck} for the $m_{\tau}$ measurement.

\subsection{Other factors}
The systematic uncertainties due to other experimental factors are quantified
by reasonable variations of the corresponding quantities, as listed
in Table~\ref{sytuty}. The total uncertainty is estimated to be
at the level of 0.1 MeV. 

\begin{table}[htb]
\caption{\label{sytuty}Systematic uncertainties for $m_{\tau}$
measurement.} \center
\begin{tabular}{ccc} \hline \hline
Source & $\delta m_\tau$           & $\delta m_\tau/m_\tau$  \\
       & ( $10^{-3}$ MeV)          &  ( $10^{-6}$ )         \\ \hline\hline
Luminosity (2\%)           &   14.0     &    7.9   \\
Efficiency (2\%)           &   14.0     &    7.9   \\
Branching Fraction (0.5\%) &   3.5      &    2.0   \\
Background (10\%)          &   1.7      &    1.0   \\
Energy spread (30\%)       &   3.0      &    1.7   \\
Theoretical accuracy       &   3.0      &    1.7   \\
Energy scale               & 100        &   56.3   \\ \hline 
Summation                  & 102 & 57.5  \\  \hline\hline
\end{tabular}
\end{table}
%

\section{Energy measurement at \bes3}
\label{part6:xct_egskbes3}

As we mentioned above, the absolute energy calibration plays a crucial
role in the $\tau$ mass measurement. A technique based on the Compton 
back-scattering principle is proposed to measure measure the 
BEPCII beam  energies with high precision.

\begin{figure}[bthp]
\center
\includegraphics[height=8.cm,width=10.cm]{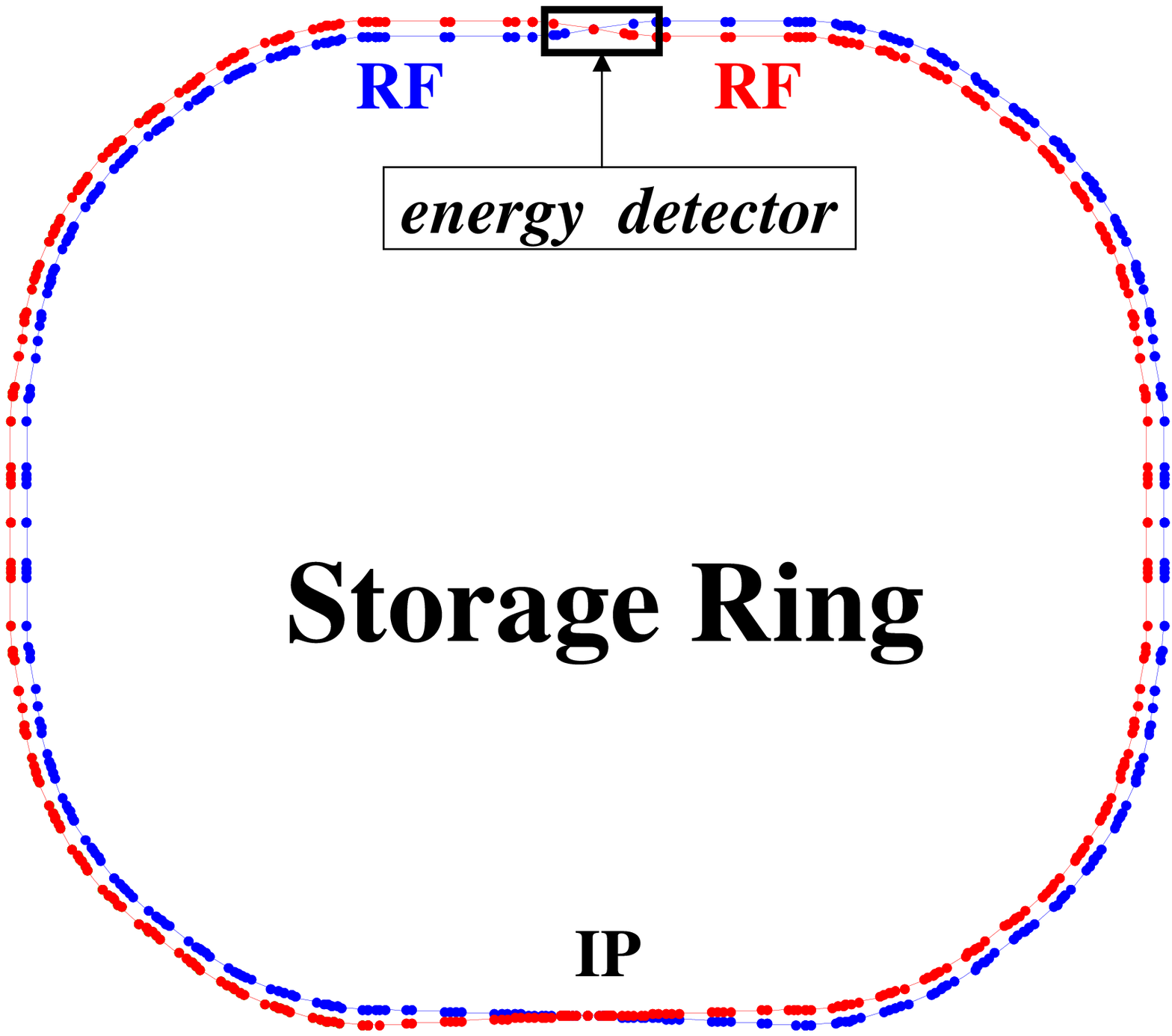}
\caption{\label{psnofdet}The location of the energy measurement
system at in BEPCII.}
\includegraphics[height=8.cm,width=10.cm]{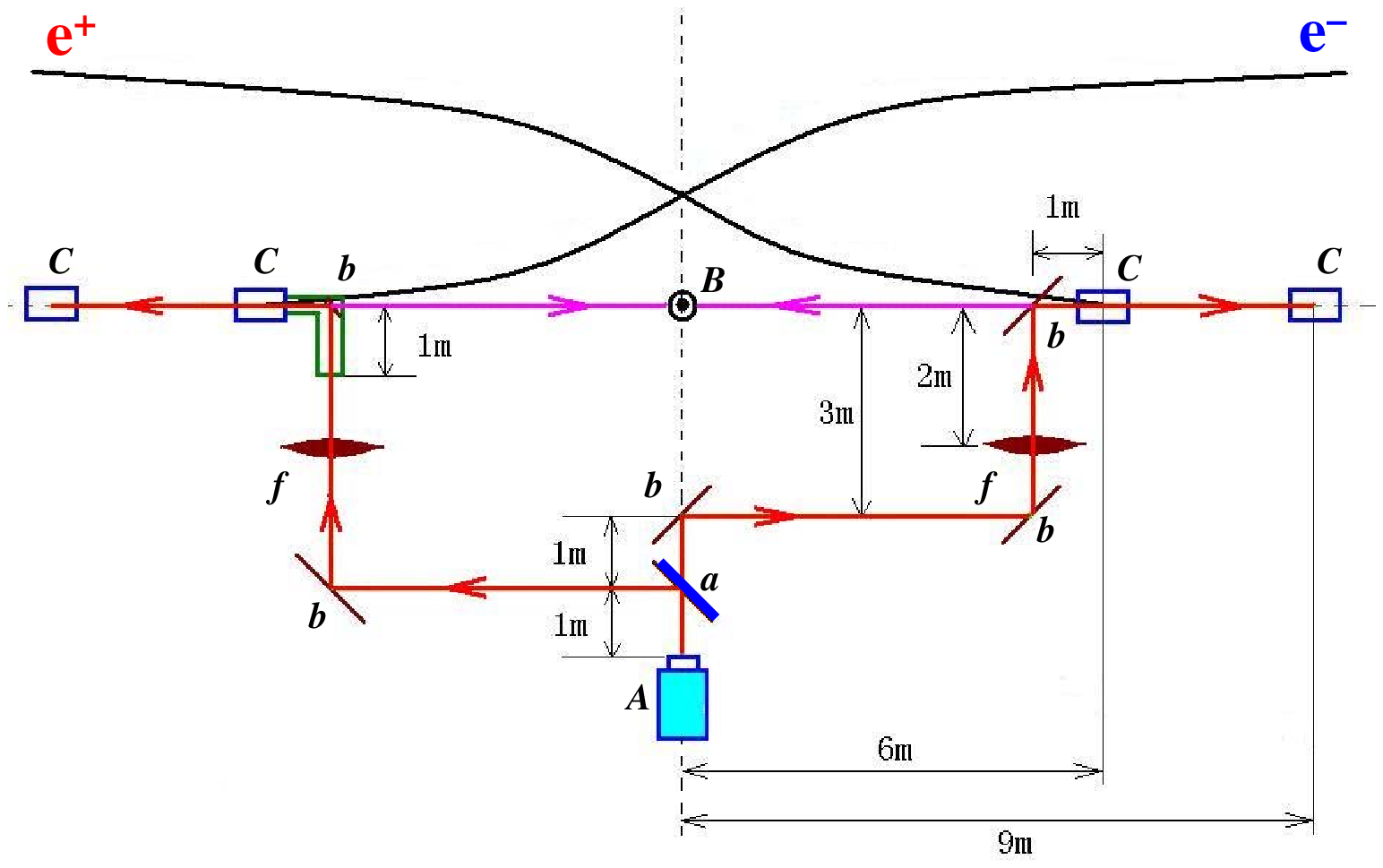}
\caption{\label{fig:engmsys}A schematic diagram of
the layout for the energy measurement
system. The two black cured lines denote the positron and electron beams.
The bottle-like box ($A$) indicates the laser source; the circled dot ($B$)
indicates the HPGe detector; rectangles ($C$) denote 
BEPCII bending magnets; $a$ indicates
a half-transmissive/half-reflective lens; $b$'s indicate 
reflective lenes and $f$'s indicate focusing lenses.}
\end{figure}

The detection system would be located at the north interaction 
point (IP) of the BEPCII storage ring as shown in 
Figs.~\ref{psnofdet}~and~\ref{fig:engmsys}.

The proposed beam energy measurement system (denoted as energy detector 
in Fig.~\ref{psnofdet}) is comprised of the following parts: 
\begin{enumerate}
\item the laser source and optics system;
\item the interaction regions where the laser beams collides with the 
electron or
positron beam;
\item a high purity Geranium detector (HPGe) to measure back-scattering
high energy $\gamma$-rays or $X$-rays.
\end{enumerate}

In the following subsections, we will describe each subsystem.
In operation, the energy measured at the north IP has to be 
corrected for synchrontron radiation produced by
the beams when the travel to the collision point
in the south intersection point.  This is at the level 
of 200 keV, with a corresponding uncertainty that is less 
than 10\%, or, equivalently, 20 keV.

\subsection{Laser source and optics system}
A simple and extensively used expression for a
laser beam amplitude is~\cite{part6_bk:wscChang}
\beq
\psi(x,y,z)=C \cdot \exp \left[ -\frac{x^2+y^2}{r^2(z)}\right] \cdot
\exp \left[ i \theta(x,y,z) \right]~.
\label{gstem00}
\eeq
Here $\psi(x,y,z)$ describes a TEM$_{00}$ mode Gaussian beam, where TEM is the 
abbreviation for transverse electric and magnetic wave. 
In Eq.~\eref{gstem00}, $\theta(x,y,z)$ is a phase factor and $r(z)$ is a 
function of $z$ (the propagation distance):
\beq
r(z)= r_0 \sqrt{1+\left( \frac{z}{z_0} \right)^2},
~~z_0=\frac{\pi r_0}{\lambda}~,
\label{xpn_raz}
\eeq
where $r_0$ is the waist width of laser beam given at $z=0$ and $\lambda$ is 
the wavelength of the laser beam. Using Eq.~\eref{gstem00}, we  
obtain the photon density in the laser beam to be
\beq
\rho_{\gamma} = |\psi(x,y,z)|^2 =
\rho_0 \cdot f_{\gamma}(x,y,z), ~~
f_{\gamma}(x,y,z) = \exp \left[ -\frac{2(x^2+y^2)}{r^2(z)}\right]~. 
\label{rholaser}
\eeq
To determine the constant $\rho_0~(=C^2)$ in Eq.~\eref{rholaser}, we 
consider
the differential relation between the laser power ($P$) and 
the photon density:
\beq
dP = \omega_{\gamma} \cdot \rho_{\gamma} \cdot \Delta S \cdot c~,
\label{prrelation}
\eeq
where $\omega_{\gamma}$ is the energy of laser beam, $c$ the velocity of light, 
and $\Delta S$ is a cross-sectional area element of the laser beam.
Integrating the above relation:
$$P=\int dP =
\omega_{\gamma} \cdot \rho_0 \cdot c \int \int f_{\gamma}(x,y,z) dxdy~,$$
we find
$$P=\rho_0 \cdot ( \omega_{\gamma} \cdot c \cdot \pi \cdot r^2(z))~,$$
or
\beq
\rho_0 = \frac{P}{\omega_{\gamma} \cdot c \cdot \pi \cdot r^2(z)}~.
\label{fatrho0}
\eeq
From dimensional analysis, we notice that $\rho_0$ has dimension of inverse
volume, so the photon density $\rho_{\gamma}$ is actually a volume 
density distribution.
 
\subsection{Electron beam} 
Similarly, the density function for
the electron (positive or negative) beam
can  be expressed as
\beq
\rho_{e} = \rho^{\prime}_0 \cdot f_{e}(x,y,z), ~~
f_{e}(x,y,z) = \exp \left[ - \left( \frac{x^2}{\sigma^2_x} 
 + \frac{y^2}{\sigma^2_y} \right) \right]~, 
\label{rhoelek}
\eeq
where $\sigma_x$ and $\sigma_y$ are standard deviations of electron beam
in $x$ and $y$ directions, respectively; $\rho^{\prime}_0$ is the normalization 
factor which is determined by the differential relation between the 
current intensity ($I$) and electron density:
\beq
dI = e \cdot \rho_{e} \cdot \Delta S \cdot u_e~,
\label{perelation}
\eeq
where $e$ is the electron charge, $u_e$ is the velocity of electron beam, 
and $\Delta S$ is a cross-sectional area element of the electron beam. 
From the integration of the above relation:
$$I=\int dI =
e \cdot \rho^{\prime}_0 \cdot u_e \int \int f_{e}(x,y,z) dxdy~,$$
we get
$$P=\rho^{\prime}_0 \cdot ( u_e \cdot e \cdot \pi \cdot
\sigma_x \sigma_y)~,$$
or
\beq
\rho^{\prime}_0 = \frac{I}{u_e \cdot e \cdot \pi \cdot \sigma_x \sigma_y}~.
\label{fatrhop0}
\eeq
From dimensional analysis, we notice that $\rho^{\prime}_0$ has dimension of 
inverse volume, so the electron density $\rho_{e}$ is actually a volume 
density 
distribution.

\subsection{Compton back-scattering principle}
The interaction of the electron and laser beams is described by 
the Compton back-scattering 
principle~\cite{part6_bk:pRullhusenetal,part6_bk:ldLandau}. 
The energy of the back-scattered photon ($\omega_2$) is
\beq
\omega_2 = \frac{\omega_1 (1- \beta \cos \phi_1)}
{1- \beta \cos \phi_2+
{\displaystyle \frac{\omega_1}{\gamma m}} (1- \cos [\phi_1 - \phi_2])}~~.
\label{eq:omega2}
\eeq
When unpolarized light is scattered by unpolarized electrons and
neither the spin of the residual electron nor the polarization of the
final photon are observed, the differential cross section 
can be expressed in terms of 
relativistic invariants as 
\beq
\frac{d\sigma}{dt} = 2 \pi r^2_0 \frac{1}{(m x_1)^2}
\left\{4y(1+y) -\frac{x_1}{x_2}-\frac{x_2}{x_1} \right\}~,
\label{eq:dsigdt}
\eeq
where $r_0$ is the classical electron radius and 
\begin{eqnarray}
x_1&=& 2 \gamma {\displaystyle\frac{\omega_1}{m}}(1- \beta \cos \phi_1)~, 
\label{eqn:x1} \\ 
x_2&=& -2 \gamma {\displaystyle\frac{\omega_2}{m}}(1- \beta \cos \phi_2)~,
\label{eqn:x2} \\
y&=& {\displaystyle \frac{1}{x_1} + \frac{1}{x_2} } ~.
\label{eqn:y}
\end{eqnarray}
In a scattering experiment, $s$ (and therefore $x_1$) is fixed by the
energies of the initial electron and photon. The total cross section 
is obtained from Eq.~\eref{eq:dsigdt} using $t=-m^2(x_1+x_2)$ and
integrating over $x_2$ at a fixed value of $x_1$ 
\beq
\sigma = 2 \pi r^2_0 \frac{1}{x_1} 
\left\{ \left(1-\frac{4}{x_1}-\frac{8}{x^2_1} \right) \ln (1+x_1) 
+\frac{1}{2}+\frac{8}{x_1}-\frac{1}{2(1+x_1)^2} \right\}~,
\label{eq:intsig}
\eeq
When observed in the laboratory, however, the Lorentz 
transformation concentrates the photon flux into a small 
cone with half-angle $\phi_2$ of
the order of $1/\gamma$, which
increases  the photon flux at backward scattering
angles.  In Ref.~\cite{part6_bk:jmRauch}, the differential cross 
section per unit solid angle is given by
\beq
\frac{d\sigma}{d\Omega} = 2 r^2_0 
\left( \frac{\omega_2}{m x_1} \right)^2
\left\{4y(1+y) -\frac{x_1}{x_2}-\frac{x_2}{x_1} \right\}~.
\label{eq:dsigdmg}
\eeq
The formulae in Eqs.~\ref{eq:intsig} and~\ref{eq:dsigdmg} are used 
below to calculate the interaction rate between the laser beam and
the high energy electron beam.

\subsection{A simple simulation}\label{sbxct_sisisy}

Based on the principles introduced above, we establish a simplified
system to simulate the procedure of the absolute energy measurement. 
We evaluate the intensity of the back-scattered photons
and simulate the detected photon energy distribution. Finally, the 
precison of the beam energy measurement is obtained. 

The intensity of back-scattered photons can be calcuated by the following 
formula:
\beq
N_{\gamma} = u_e \sigma_T \cdot
\int  \int \int~ \rho_{\gamma} \cdot \rho_e~ dxdydz
\eeq
where $\rho_{\gamma}$ denotes the volume density of the laser beam
described by Eqs.~\ref{rholaser} and~\ref{fatrho0}; 
$\rho_e$ is the volume density of the electron beam described by
Eqs.~\ref{rhoelek} and~\ref{fatrhop0};
$\sigma_T$ is the total cross section given in Eq.~\ref{eq:dsigdmg} 
(or approximately by Eq.~\ref{eq:intsig}).  
We can rewrite the above equation as:
\beq
N_{\gamma} = \frac{PI \sigma_T}{\omega_{\gamma} \cdot c \cdot e \cdot \pi^2}
\int  \int \int~ \frac{1}{r^2(z)} f_{\gamma} \cdot f_{e}~ dxdydz~.
\label{ngamma}
\eeq

Inserting the parameters provided in Table~\ref{parmtlsrelk} into
Eq.~\ref{ngamma}, we find
$$ N_{\gamma} = 2.7 \times 10^8 s^{-1}~.$$

\begin{table}[htb]
\caption{\label{parmtlsrelk}Some input parameters for the laser and 
electron beam.}
\center
\begin{tabular}{ll} \hline \hline
     laser beam                               & electron  beam       \\ \hline 
  power P= 50 W                               & $I$ = 9. 8 mA        \\
  wave-length $\lambda$ = 10. 59 $\mu$ m      & $\sigma_x$ = 1.6 mm  \\
  laser energy $\omega_{\gamma}$ =0.117 eV    & $\sigma_y$ = 0.16 mm \\
  waist radius $r_0$ = 2mm                    & $\sigma_z$ = 15 mm   \\
\hline \hline
\end{tabular}
\end{table}

\begin{figure}[hbt]
\begin{minipage}{7.5cm}
\centerline{\psfig{file=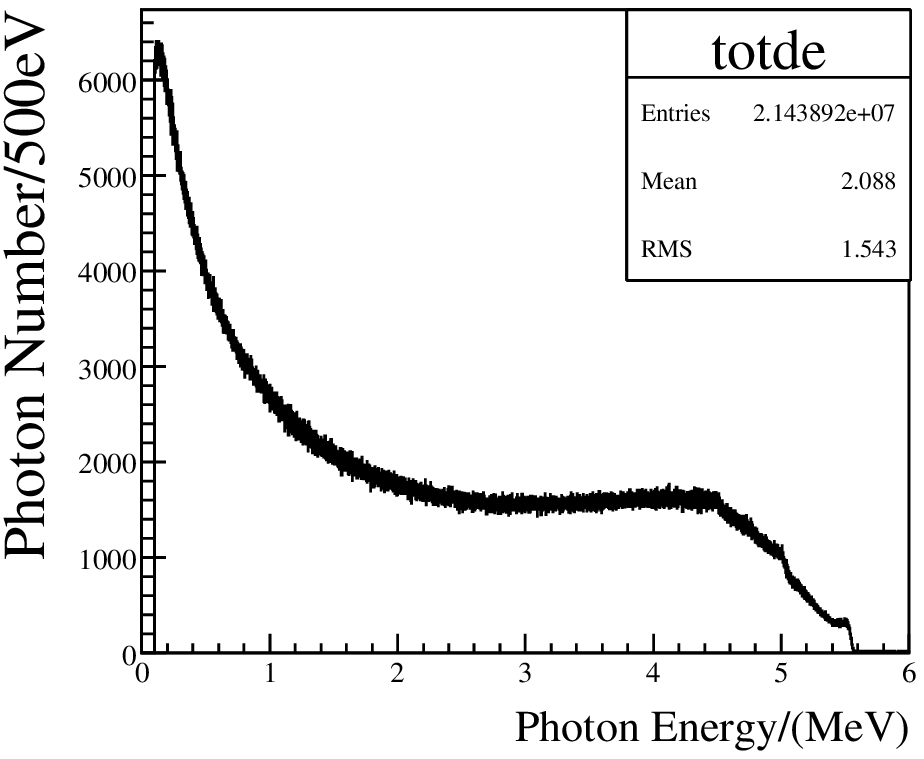,height=6.0cm ,width=7.5cm}}
(a) The detected photon energy spectrum between 0 and 6~MeV.
\end{minipage}
\hskip 1.0 cm
\begin{minipage}{7.5cm}
\centerline{\psfig{file=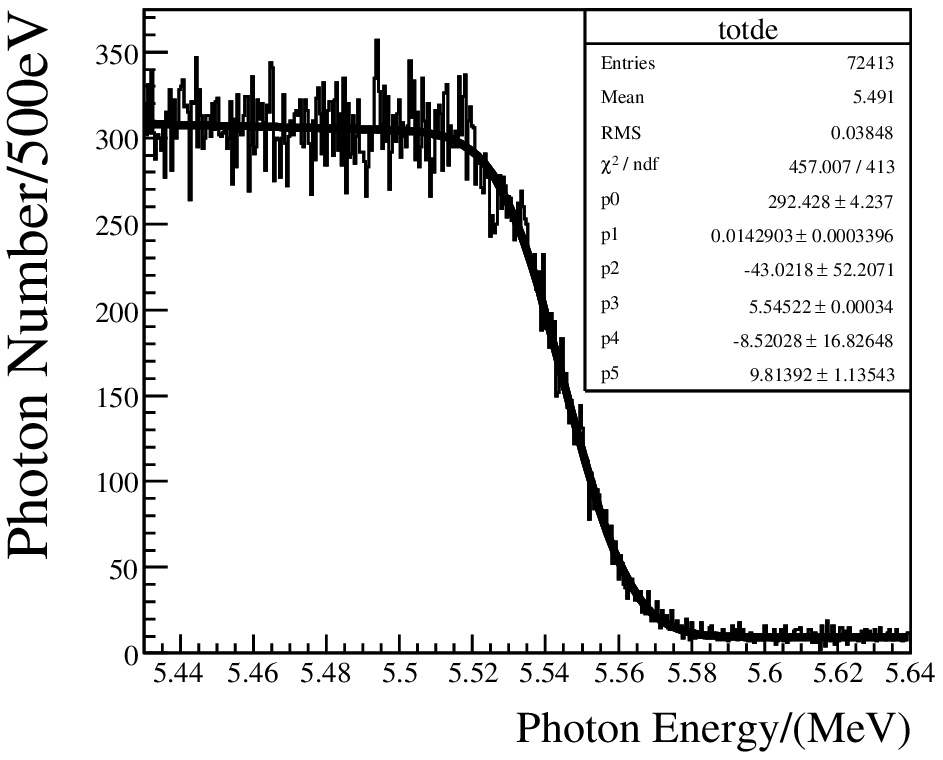,height=6.0cm ,width=7.5cm}}
(b) the detected photon energy spectrum between 5.43~MeV and 5.64~MeV
\end{minipage}
\caption{\label{smhpge} The simulated photon energy spectrum measured by 
the HPGe detector.}
\end{figure}

The detection of the back-scattered photons 
in the HPGe is simulated using the GEANT4 
package~\cite{part6_fucda}. Figure~\ref{smhpge} shows the photon 
energy spectrum, where a clear Compton edge is evident;
this edge is used  the measure the absolute beam energy.

The edge of the Compton spectrum is fitted with the six parameter
function~\cite{part6_nMuchnoi}:
\beq
g(x,\vec{p})=\frac{1}{2}(p_2(x-p_0)+p_3)\cdot
\mbox{erfc} \left[ \frac{x-p_0}{\sqrt{2} p_1}\right]-
\frac{p_1 p_2}{\sqrt{2\pi}} \cdot 
\exp \left[-\frac{(x-p_0)^2}{2 p^2_1}\right]+
p_4(x-p_0)+p_5~, 
\label{cmpedgft}
\eeq
where $p_0$ is the edge position; 
$p_1$ is the edge width; $p_2$ is the left slope; $p_3$ is the edge 
amplitude;
$p_4$ is the right slope; and $p_5$ is the background. 
The $p_0$ parameter
gives the information about the average electron beam energy 
during the data acquisiton period, while $p_1$ is mostly coupled 
to the electron beam energy spread. 

In addition to the simple simulation described above, 
a number of factors have to
be taken into account in a real measurement, inluding: the effect
of the focusing system on the laser beam,  the dynamics and trajectories
of the electron beam, the calibration of the HPGe detector, etc., 
in order to estimate the accuracy of the beam energy measurement. 
%
%
%

\section{Summary}
\label{part6:xct_sum}

The statistical and systematic uncertainties of  a $m_{\tau}$
measurement at \bes3 are studied based on
previous experience and recent theoretical calculations. Monte Carlo
simulation and sampling techniques are employed to obtain an optimal
data taking strategy. It is found that:
\begin{enumerate}
\item the optimal measurement energy is located near the point with 
the largest derivative of the cross section with respect to energy;
\item one measurement point in this region is sufficient to provide the 
smallest statistical error for a given amount of integrated luminosity;
\item an integrated  luminosity of 63 pb$^{-1}$ will produce  a 
statistical accuracy that is better than 0.1 MeV.
\end{enumerate}

In addition, many factors have been taken into account to estimate possible
systematic uncertainties. The expected total relative error is at the 
level of $5.8 \times 10^{-5}$. The absolute calibration of energy scale 
will be crucial for further improvement of the accuracy of the $m_{\tau}$ 
measurement.
If the laser back-scattering  technique is applied at \bes3, an
ultimate systematic uncertainty of around 0.09 MeV could be achieved.




\clearpage
\section*{Acknowledgements}
\addcontentsline{toc}{section}{Acknowledgements}

We acknowledge the encouragement and strong support from IHEP director 
He-sheng Chen and  former IHEP directors Zhi-peng Zheng and Ming-han Ye. 
We also benefited from excellent advice and many useful suggestions from
BES-II spokespersons Wei-guo Li and Fred Harris, and former CLEO-c spokesperson 
Ian Shipsey.  Ying-hua Jia provided valuable assistance to all phases 
of the preparation of this document. 
This work is supported in part by the National Natural Science Foundation of China under
contracts Nos. 10491300, 10225524, 10225525, 10425523, 10575108,10521003,
10735080, 10675131,10605031 (IHEP),  the Chinese Academy of Sciences under contract Nos. KJ95T-03
and KJCX3-SYW-N2, the 100 Talents
Program of CAS under Contract Nos. U-11, U-24, U-25, and the
Knowledge Innovation Project of CAS under Contract Nos. U-602, the Research
and Development Project of Important Scientific Equipment 
of CAS under contracts No.  H7292330S7,  the National Natural Science
Foundation of China under Contract Nos. 10775179, 10491306 (GUCAS),  
the National Natural Science Foundation of China
under Contract Nos. 10625521, 10721063 (Peking University), the National Natural Science Foundation of China
under Contract Nos. 10225522,10775077, 10635030 (Tsinghua University), 
the Department of Energy under Contract No.DE-FG02-04ER41291 (U Hawaii),
U.S. National Science Foundation through grant NSF-PHY-0244786 at the
University of Tennessee, U.S. Department of Energy under contract
DE-AC05-00OR22725 at Oak Ridge National Laboratory, the NSF under grant
number PHY-0355098 (Univ. of Notre Dame), 
the European Network Flavianet under contract MRTN-CT-2006-035482,
the network Flavianet MRTN-CT-2006-035482,  the MEC (Spain) under grant  
Nos. FPA2007-60275, FPA2007-60323,  the Spanish Consolider-Ingenio 2010 
Programme CPAN (CSD2007-00042), the catalan grant SGR2005-00916, 
the ANR contract ANR-06-JCJC-0056, the EU Contract No. MRTN-CT-2006-035482,
\lq\lq FLAVIAnet'',  the National Science Council of Taiwan,  the RFBR-NSFC
grant No. 06-02-39009,  and the grant of President of Russian Federation MK-2952.2006.2.

\appendix
\chapter[Statistics in HEP data analysis]{Statistics in HEP data analysis\footnote{By Yong-Sheng Zhu}}
\label{appen_a_stat}

This Appendix introduces an overview of two aspects of statistical
methods used in High Energy Physics (HEP)- parameter estimation and
hypothesis testing. For the detail, it is recommended to refer to
the relevant textbooks~\cite{Eadie,Frodesen,Zhuys} and
literatures~\cite{PDG}, and references quoted in this Appendix. In
an experiment of HEP, an observable $x$ usually is a random
variable, its $pdf$ is expressed as $f(x,\theta)$ with parameter
$\theta$, what the experiment obtained is an sample of
$x$:$\overrightarrow{x}=(x_{1},...,x_{n})^{T}$. The task of the
statistical inference is based on the sample of data to determine
the value and error, or a confidence interval at given confidence
level for the parameter $\theta$, or infer observable's $pdf$
$f(x,\theta)$.
\section{Parameter estimation}
Parameter $\theta$ is estimated with a function of sample of
observed data: $\hat{\vartheta}(x_{1},...,x_{n})$, which is called
an estimator of $\theta$. The sample of observed data
$\overrightarrow{x}=(x_{1},...,x_{n})^{T}$ is also a random
variable, the value of the estimator to a specific measurement of
$(x_{1},...,x_{n})^{T}$ is called an estimate. Throughout this
Appendix, we will use same notation to denote estimate and
estimator. An good estimator should have properties of consistency,
unbiasedness and high efficiency.

The consistency means when the size of the data sample
$(x_{1},...,x_{n})$ goes to infinity, the estimator $\hat{\theta}$
converges to the true value of parameter $\theta$.

The bias of an estimator is defined as the difference of the
expectation of the estimator and the true value $\theta$:
$E(\hat{\theta)}=\theta+b(\theta)$. The unbiasedness is a property
of an estimator in finite sample, namely, it is required
$E(\hat\theta)=\theta$. If it has to be estimated with a biased
estimator, then the bias $b$ of the estimator should be known or can
be obtained by some way.

The efficiency is a measure of the variance of an estimator. Under
the regularity conditions, namely, if the range of $\vec{x}$ is
independent of $\theta$ and the first and second derivatives of the
sample's joint $pdf$ - Likelihood function
$L(\vec{x}|\theta)=\prod_{i=1}^n{f(x_i,\theta)}$ - with respect to
$\theta$ exist, there exists a lower bound on the variance of the
estimates derived from an estimator, which is called the minimum
variance bound MVB, given by Cramer-Rao inequality:
\begin{equation}
MVB=\frac{(1+\frac{\partial b}{\partial \theta})}{I(\theta)},
\label{E1}
\end{equation}
where $I(\theta)$ is the Fisher information:
\begin{equation}
I(\theta)=E[(\frac{\partial lnL}{\partial \theta})^{2}]=\int
(\frac{\partial lnL}{\partial \theta})^{2} \cdot L d\vec{x} \\
         =E[-\frac{\partial^{2} lnL}{\partial \theta^{2}}]=\int
(-\frac{\partial^{2} lnL}{\partial \theta^{2}}) \cdot L d\vec{x}.
\label{E2}
\end{equation}
 The efficiency of an estimator $\hat{\theta}$ is defined as
$e(\hat{\theta})=MVB/V(\hat{\theta})$. Apparently, we hope the
efficiency of the used estimator is close or equal to 1.

The mean-square error (MSE) of an estimator is a convenient quantity
which combine the uncertainties in an estimator due to bias and
variance:
\begin{equation}
MSE=E[(\hat{\theta}-\theta)^{2}]=V(\hat{\theta})+b^{2}. \label{E3}
\end{equation}
\subsection{Estimators for mean and variance}
Suppose we are interested in the expectation $\mu$ of an observable
$x$ (random variable) and its variance, $\sigma^{2}$. We have a set
of $n$ independent measurements $x_i$, which have same unknown
expectation $\mu$ and common unknown variance $\sigma^{2}$. This
corresponds to, for instance, a set of $n$ measurements for an
observable $x$ in an experiment. Then their consistent and unbiased
estimate are the sample mean $\bar{x}$ and sample variance $S^{2}$,
respectively:
\begin{equation}
\hat{\mu}=\bar{x}=\frac{1}{n}\sum_{i=1}^{n} x_{i}, \label{E4}
\end{equation}
\begin{equation}
\hat{\sigma^{2}}=S^{2}=\frac{1}{n-1}\sum_{i=1}^{n}
(x_{i}-\bar{x})^{2}. \label{E5}
\end{equation}
The variance of $\hat{\mu}$ is $\sigma^{2}/n$, while the variance of
$\hat{\sigma^{2}}$ is
\begin{equation}
V(\hat{\sigma^{2}})=\frac{1}{n}(m_{4}-\frac{n-3}{n-1}\sigma^{4}),
\label{E6}
\end{equation}
where $m_{4}$ is the 4th central moment of $x$.

For the known $\mu$, the consistent, unbiased estimator of variance
is
\begin{equation}
\hat{\sigma^{2}}=\frac{1}{n}\sum_{i=1}^{n} (x_{i}-\mu)^{2},
\label{E7}
\end{equation}
which gives a somewhat better estimator of $\sigma^{2}$ compared
with Eq.~\ref{E5} for unknown $\mu$ case. For the binomial, Poisson
and Gaussian variables $x_{i}$, which are often used in data
analysis, the sample mean is an efficient estimator for $\mu$; For
the normal variables, Eq.~\ref{E7} is an efficient estimator of
$\sigma^{2}$ in the case of known $\mu$, and sample variance $S^{2}$
is an asymptotic efficient estimator for $\sigma^{2}$.

For the Gaussian distributed $x_{i}$, Eq.~\ref{E6} becomes
$V(\hat{\sigma^{2}})=2\sigma^{4}/(n-1)$ for any $n\geq2$, and for
large $n$ the standard deviation of $\hat{\sigma}$ (the "error of
the error") is $\sigma/\sqrt{2n}$.

If the $x_{i}$ have different, known variance $\sigma_{i}^{2}$,
which corresponds to the situation that different experiments
measure the same quantity with different uncertainties. Assume
$x_{i}$ can be considered as a measurement of the Gaussian
distributed variable $N(\mu,\sigma_{i}^{2})$, then the unbiased
estimator of the physics quantity $\mu$ is a weighted average
\begin{equation}
\hat{\mu}=\frac{1}{\omega}\sum_{i=1}^{n} \omega_{i} x_{i},
\label{E8}
\end{equation}
where $\omega_{i}=1/\sigma_{i}^{2}$, $\omega=\sum_{i}\omega_{i}$,
and the standard deviation of $\hat{\mu}$ is $1/\sqrt{\omega}$.
\subsection{The method of maximum likelihood(ML)}
From the statistical point of view, the method of maximum likelihood
(ML) is the most important general method of estimation, as the ML
estimator of parameter has many good properties.
\subsubsection{The ML estimators for parameter and its error}
Suppose $x_{i},i=1,...,n$ are the $n$ independent measurements of a
random variable $x$ with the  $pdf$  $f(x,\vec{\theta})$, where
$\vec{\theta}=(\theta_1,...,\theta_k)^T$ are $k$ parameters to be
determined, then the ML estimators $\hat{\vec{\theta}}(x_1,...,x_n)$
are the values of $\vec{\theta}$ that maximize the likelihood
function
\begin{equation}
L(\vec{x}|\hat{\vec{\theta}})=\prod_{i=1}^{n}f(x_{i};\vec{\theta}).
\label{E9}
\end{equation}
Since both $lnL$ and $L$ are maximized for the same parameters
values $\vec{\theta}$ and it is usually easier to work with $lnL$,
therefore, the ML estimators can be found by solving the likelihood
equations
\begin{equation}
\frac{\partial lnL}{\partial \theta_{i}}=0, ~~~~~~i=1,...,k.
\label{E10}
\end{equation}

The ML estimator is invariant under change of parameter, namely,
under an one-to-one change of parameters from $\vec{\theta}$ to
$\vec{\eta}$, the ML estimators $\hat{\vec{\theta}}$ transform to
$\vec{\eta}(\vec{\theta})$  :
$\hat{\vec{\eta}}(\vec{\theta})=\vec{\eta}(\hat{\vec{\theta}})$.
Moreover, the ML estimators are asymptotic unbiased. When the
likelihood function satisfies the regularity conditions, the ML
estimators are consistent estimators. If there exist the efficient
estimators for parameters or their functions, then the efficient
estimators must be the ML estimators, and the likelihood equations
give the unique solutions; while if the efficient estimators do not
exist, the ML estimators give possibly minimum variance for
$\vec{\theta}$. For large size $n$ and the likelihood function
satisfies the regularity conditions, $\hat{\vec{\theta}}$
asymptotically distributed as a normal variable with the mean being
the true values $\vec{\theta}$ and the variances reach the MVB.

The ML estimators give only the values of the parameters. To know
the errors of the parameters, one has to know the variances of
parameters. The expression of the covariance between parameters
$\hat{\theta_{i}}$ and $\hat{\theta_{j}}$ for any size of sample $n$
is
\begin{equation}
V_{ij}(\hat{\vec{\theta}})=\int
(\hat{\theta_{i}}-\theta_{i})(\hat{\theta_{j}}-\theta_{j})
L(\vec{x}|\vec{\theta})d\vec{x},~~~~~~~~ i,j=1,...,k. \label{E11}
\end{equation}
The calculation of this integral is sometimes troublesome, however,
in general, it can be calculated with numerical method in any case.

For the case that $\hat{\theta}$ is an efficient estimator of single
parameter, following equation is applicable for any size of sample
$n$:
\begin{equation}
V(\hat{\theta})=\frac{(1+\frac{\partial b}{\partial
\theta})^{2}}{(-\frac{\partial^{2} lnL}{\partial
\theta^{2}})_{\theta=\hat{\theta}}}; \label{E12}
\end{equation}
In particular, if $\hat{\theta}$ is an unbiased efficient estimator
\begin{equation}
V(\hat{\theta})=\frac{1}{(-\frac{\partial^{2} lnL}{\partial
\theta^{2}})_{\theta=\hat{\theta}}}. \label{E13}
\end{equation}

For multi-parameters and large $n$, if there exists a set of $k$
jointly sufficient statistics $t_1,...,t_k$ for the $k$ parameters
$\theta_1,...,\theta_k$, the inverse of the covariance matrix
$V_{ij}=cov(\hat{\theta_{i}},\hat{\theta_{j}})$ for a set of ML
estimators can be calculated by
\begin{equation}
V_{ij}^{-1}(\hat{\vec{\theta}})=(-\frac{\partial^{2}lnL}{\partial
\theta_{i}\partial
\theta_{j}})_{\vec{\theta}=\hat{\vec{\theta}}}~,~~~~~ i,j=1,...,k.
\label{E14}
\end{equation}
Besides, for large $n$ and the likelihood function satisfying the
regularity conditions, the ML estimators $\hat{\vec{\theta}}$
asymptotically distributed as a multi-dimensional normal variable,
then one has
\begin{equation}
V_{ij}^{-1}(\hat{\vec{\theta}})=
E(-\frac{\partial^{2}lnL(\vec{x}|\vec{\theta})}
{\partial\theta_{i}\partial \theta_{j}})_{\vec{\theta}=
\hat{\vec{\theta}}} \\
=\int (-\frac{\partial^{2}lnL(\vec{x}|\vec{\theta})}{\partial
\theta_{i}\partial
\theta_{j}})_{\vec{\theta}=\hat{\vec{\theta}}}\cdot Ld\vec{x},~~~~
i,j=1,...,k; \label{E15}
\end{equation}
or, one can use the $pdf$ of the random variable $x$ to calculate
the covariance matrix:
\begin{equation}
V_{ij}^{-1}(\hat{\vec{\theta}})=n\int \frac{1}{f}(\frac{\partial
f}{\partial \theta_{i}})(\frac{\partial f}{\partial \theta_{j}})dx,
~~~~~i,j=1,...,k. \label{E16}
\end{equation}
Wherein, the last equation uses only the $pdf$ of the random
variable $x$ and does not need the measured data sample, which is
particularly useful in the design stage of an experiment.

If the observable $x$ is a normal random variable, or the size of
sample $n$ is sufficiently large, then the likelihood function is an
asymptotically normal distribution and $lnL$ is a parabolic
function, a numerically equivalent way of determining
$s$-standard-deviation errors is from the contour given by the
$\vec{\theta}^{\prime}$ such that
\begin{equation}
lnL(\vec{\theta^{\prime}})=lnL_{max}-\frac{s^{2}}{2}, \label{E17}
\end{equation}
where $L_{max}$ is the  value of $lnL$ at the solution point. The
extreme limits of this contour on the $\theta_{i}$ axis give an
$s$-standard-deviation likelihood interval for $\theta_{i}$. In the
case $lnL$ is not a parabolic function, the approximate
1-standard-deviation likelihood interval can also be estimated by
this equation, and it will give an asymmetric positive and negative
errors for each parameter, namely,
$\sigma^{+}(\theta_{i})\neq\sigma^{-}(\theta_{i}), i=1,...,k.$
\subsubsection{The ML method for binned data}
In the case that the size $n$ of data sample
$\vec{x}=(x_{1},...,x_{n})^T$ is sufficiently large, the
measurements often are expressed as a histogram  binned data.  For
constant $n$, the likelihood function (joint $pdf$) of
$n_{i}(i=1,...,m)$ measurement values appearing in i-th bin is
expressed as a multinomial distribution
\begin{equation}
L(n_{1},...,n_{m}|\vec{\theta})=n!\prod_{i=1}^{m}\frac{1}{n_{i}!}p_{i}^{n_{i}}.
\label{E18}
\end{equation}
The probability of one measurement value appearing in i-th bin is
calculated with $pdf$ $f(x|\vec{\theta})$
\begin{equation}
p_{i}=p_{i}(\vec{\theta})=\int_{\triangle x_{i}}f(x|\vec{\theta})dx.
\label{E19}
\end{equation}
Then the likelihood equation becomes
\begin{equation}
(\frac{\partial lnL}{\partial
\theta_{i}})_{\vec{\theta}=\hat{\vec{\theta}}}=\frac{\partial}{\partial
\theta_{i}}[\sum_{i=1}^{m}n_{i}lnp_{i}(\vec{\theta})]_{\vec{\theta}
=\hat{\vec{\theta}}}=0, ~~~~~i=1,...,m, \label{E20}
\end{equation}
Solving this set of equations gives the ML estimators
$\hat{\vec{\theta}}$.
\subsubsection{The extended ML method}
If the size $n$ of data sample is not a constant but a Poisson
random variable with the expectation $\nu$, then the likelihood
function is the product of usual likelihood function and the Poisson
probability of observing $n$ events
\begin{equation}
L(\nu,\vec{\theta})=\frac{\nu^{n}}{n!}e^{-\nu}\prod_{i=1}^{n}f(x_{i},\vec{\theta}),
\label{E21}
\end{equation}
which is called the extended likelihood function~\cite{Cowan}. Then
the solutions of the likelihood equations
\begin{eqnarray}
\frac{\partial lnL(\nu,\vec{\theta})}{\partial \theta_{j}}=0,~~~~~
j=1,...,k, \\
\frac{\partial lnL(\nu,\vec{\theta})}{\partial \nu}=0 \label{E22}
\end{eqnarray}
give the ML estimators $\hat{\vec{\theta}}$.

In the case that $\nu$ is irrelevant to $\vec{\theta}$,
$\frac{\partial lnL(\nu,\vec{\theta})}{\partial \nu}=0$ gives
$\hat{\nu}=n$, the solutions of Eq. A.22 give the same
$\hat{\vec{\theta}}$ as those from Eqs.~\ref{E9},~\ref{E10}. If
$\nu$ is a function of $\vec{\theta}$, the likelihood function
becomes (dropping terms irrelevant to $\vec{\theta}$)
\begin{equation}
lnL(\vec{\theta})=-\nu(\vec{\theta})+\sum_{i=1}^{n}ln[\nu(\vec{\theta})\cdot
f(x_{i},\vec{\theta})]. \label{E24}
\end{equation}
The variances of the ML estimators $\hat{\vec{\theta}}$ derived from
the extended likelihood function are usually smaller than those from
usual likelihood function because the former uses  the information
from both $n$ and $\vec{x}$.

For the binned data, the extended likelihood function is
\begin{equation}
L(n_1,...,n_m|\vec{\theta})=\prod_{i=1}^{m}\frac{1}{n_{i}!}
\nu_{i}^{n_i}e^{-\nu_i}, \label{E25}
\end{equation}
where the expectation of $n_i$, $\nu_i$, is
\begin{equation}
\nu_i=\nu\int_{\triangle x_i}f(x|\vec{\theta})dx,~~~~~
\nu=\sum_{i=1}^{m}\nu_i. \label{E26}
\end{equation}
In the case that $\nu$ is irrelevant to $\vec{\theta}$, the
likelihood equations become
\begin{equation}
\frac{\partial
lnL}{\partial\theta_{j}}|_{\vec{\theta}=\hat{\vec{\theta}}}=
\frac{\partial}{\partial\theta_{j}}
[\sum_{i=1}^{m}n_{i}ln\nu_{i}]_{\vec{\theta}=\hat{\vec{\theta}}}=0,
~~~~~j=1,...,m, \label{E27}
\end{equation}
which has the same form of Eq.~\ref{E20} with $p_{i}(\vec{\theta})$
replaced by $\nu_{i}(\vec{\theta})$, and $\hat{\nu}=n$. If $\nu$ is
a function of $\vec{\theta}$, then  the likelihood equations are
\begin{equation}
\frac{\partial
lnL}{\partial\theta_{j}}|_{\vec{\theta}=\hat{\vec{\theta}}}=
\frac{\partial}{\partial\theta_{j}}
[\sum_{i=1}^{m}n_{i}ln\nu_{i}-\nu]_{\vec{\theta}=\hat{\vec{\theta}}}=0,
~~~~~j=1,...,m. \label{E28}
\end{equation}
The variances of the ML estimators $\hat{\vec{\theta}}$ derived from
these equations are usually smaller than those from usual likelihood
function because the random property of $n$ has been taken into
account here.
\subsubsection{Combining measurements with ML method}
Suppose the observations in two independent experiments are
$\vec{x}=(x_{1},...,x_{l})^T$ and $\vec{y}=(y_{1},...,y_{m})^T$, and
their $pdf$ $f_{x}(x,\vec{\theta})$ and $f_{y}(y,\vec{\theta})$,
depend on same parameters $\vec{\theta}$, which are the quantities
to be measured in the experiments. The joint likelihood function of
these two experiments are
\begin{equation}
L(\vec{x},\vec{y};\vec{\theta})=L(\vec{x};\vec{\theta})\cdot
L(\vec{y};\vec{\theta})=\prod_{i=1}^{l}
f_{x}(x_{i},\vec{\theta})\prod_{j=1}^{m} f_{y}(y_{i},\vec{\theta}).
\label{E29}
\end{equation}
Solving the likelihood equations of this likelihood function with
respect to parameters $\vec{\theta}$ and obtaining the ML estimator
$\hat{\vec{\theta}}$ gives the combined measurement of these two
experiments for parameters $\vec{\theta}$.

In the case $f_{x}(x,\theta)$ and $f_{y}(y,\theta)$ are Gaussians
and the parameter $\theta$ is the mean of Gaussians, the combined
estimator of the parameter and its variance have simple forms:
\begin{eqnarray}
\hat{\theta}=(\frac{\theta_{x}}{\sigma_{x}^{2}}+\frac{\theta_{y}}
{\sigma_{y}^{2}})/
(\frac{1}{\sigma_{x}^{2}}+\frac{1}{\sigma_{y}^{2}}), \\
V(\hat{\theta})=1/(\frac{1}{\sigma_{x}^{2}}+\frac{1}{\sigma_{y}^{2}}),
\label{E30}
\end{eqnarray}
where $\theta_{x}$ and $\sigma_{x}$ are the measured value of
parameter $\theta$ and its error from experiment $x$, respectively.
Above expressions can be directly extended to the situation of
multi-experiments.

If the likelihood function is unknown, and only the results of
parameter $\theta$ and its errors,
$\theta_{i},\sigma_{i}^{+},\sigma_{i}^{-}$, are reported in each
experiment, the combined results for parameter $\theta$  and its
errors of multi-experiments can be deduced with the method suggested
by R.Barlow~\cite{Barlow}. The essence of the method is using the
measured values $\theta_{i},\sigma_{i}^{+},\sigma_{i}^{-}$ to
construct an approximate parametric likelihood function for each
experiment. The $variable ~width ~Gaussians$ are concluded as the
best approximation for our purpose. The likelihood function can be
approximated as
\begin{equation}
lnL(\theta_{i}|\theta)=-\frac{(\theta-\theta_{i})^{2}}{2V_{i}(\theta)},
\label{E32}
\end{equation}
where the true value of parameter is $\theta$, and the measured
value is $\theta_{i}$ in $i$-th experiment. For the $linear~ \sigma
~ parametrization$, we have
\begin{eqnarray}
V_{i}(\theta)=[\sigma_{i}(\theta)]^{2},~~~~~
\sigma_{i}(\theta)=\sigma_{i}+\sigma_{i}^{\prime}(\theta-\theta_{i}),\\
\sigma_{i}=\frac{2\sigma_{i}^{+}\sigma_{i}^{-}}{\sigma_{i}^{+}+\sigma_{i}^{-}},
~~~~~\sigma_{i}^{\prime}=\frac{\sigma_{i}^{+}-\sigma_{i}^{-}}
{\sigma_{i}^{+}+\sigma_{i}^{-}}, \label{E33}
\end{eqnarray}
where $\sigma_{i}^{+},\sigma_{i}^{-}$ are the measured positive and
negative errors of $\theta_{i}$ in $i$-th experiment. For the
$linear ~ V ~ parametrization$,
\begin{eqnarray}
V_{i}(\theta)=V_{i}+V_{i}^{\prime}(\theta-\theta_{i}),\\
V_{i}=\sigma_{i}^{+}\sigma_{i}^{-},~~~~~
V_{i}^{\prime}=\sigma_{i}^{+}-\sigma_{i}^{-}. \label{E35}
\end{eqnarray}
 Thus, the joint likelihood function of
multi-experiments  for parameter $\theta$ is approximately
\begin{equation}
lnL(\theta)=-\frac{1}{2}\sum_{i}\frac{(\theta-\theta_{i})^{2}}{V_{i}(\theta)}.
\label{E37}
\end{equation}
The best estimate of $\theta$, $\hat{\theta}$, is determined by the
maximum of above likelihood function. For the linear $\sigma$ form,
the solution is
\begin{eqnarray}
\hat{\theta}=\Sigma_{i}\omega_{i}\theta_{i}/\Sigma_{i}\omega_{i},\\
\omega_{i}=\frac{\sigma_{i}}{[\sigma_{i}+\sigma_{i}^{\prime}
(\hat{\theta}-\theta_{i})]^{3}}. \label{E38}
\end{eqnarray}
For the linear $V$ form, the solution is
\begin{eqnarray}
\hat{\theta}=\Sigma_{i}\omega_{i}[\theta_{i}-\frac{V_{i}^{\prime}}
{2V_{i}}(\hat{\theta}-\theta_{i})^{2}]/\Sigma_{i}\omega_{i},\\
\omega_{i}=\frac{V_{i}}{[V_{i}+V_{i}^{\prime}
(\hat{\theta}-\theta_{i})]^{2}}. \label{E40}
\end{eqnarray}
Two sets of equations shown above are non-linear for $\hat{\theta}$
and the solution must be found by iteration. The $\triangle lnL$=0.5
points of the likelihood function in Eq.~\ref{E37} are used to
determine the positive and negative errors for $\hat{\theta}$, which
also need to be determined numerically.  The program of combining
results from multi-experiments using parametrization likelihood
function has been coded, and obtainable under
http://www.slac.stanford.edu/~barlow/statistics.html.
\subsection{The method of least squares(LS)}
\subsubsection{The LS estimator for parameter and its error}
Suppose $n$ observations $\vec{y}=(y_{1},...,y_{n})^{T}$ are
measured at $n$ points $\vec{x}=(x_{1},...,x_{n})^{T}$, the
covariance matrix of observations $\vec{y}$ is expressed as
$V_{ij}=cov(y_{i},y_{j})$, and the true values of $\vec{y}$,
$\vec{\eta}$, are described by model
$\eta_{i}=f(x_{i},\vec{\theta})$ ,$i=1,...,n$ , where
$\vec{\theta}=(\theta_{1},...,\theta_{k})^{T}$ are the parameters to
be determined. The least squares (LS) estimators of $\vec{\theta}$,
$\hat{\vec{\theta}}$, can be found by minimizing the LS function
$Q^{2}(\vec{\theta})$ with respect to $\vec{\theta}$:
\begin{equation}
Q^{2}(\vec{\theta})=(\vec{y}-\vec{\eta}(\vec{\theta}))^{T}
V^{-1}(\vec{y}-\vec{\eta}(\vec{\theta}))
=\Sigma_{i=1}^{n}\Sigma_{j=1}^{n}
(y_{i}-\eta_{i})V_{ij}^{-1}(y_{j}-\eta_{j}). \label{E42}
\end{equation}
In the case of $y_{i},i=1,...,n$ being $n$ independent measurements,
the LS function has simple form
\begin{equation}
Q^{2}(\vec{\theta})=\Sigma_{i=1}^{n}
\frac{(y_{i}-\eta_{i})^{2}}{\sigma_{i}^{2}}, \label{E43}
\end{equation}
where $\sigma_{i}$ is the error of $y_{i}$. An usual case is $y_{i}$
is a Poisson variable, then $\sigma_{i}^{2}$ can be approximated by
$y_{i}$ or its predicted value $\eta_{i}$. If $y_{i},i=1,...,n$ are
$n$ independent Gaussians, $y_{i}\sim N(\eta_{i},\sigma_{i}^{2})$,
the likelihood function of $\vec{y}$ is $L(\vec{\theta})\propto
exp[-\frac{1}{2}\sum_{i=1}^{n}(\frac{y_{i}-\eta_{i}}{\sigma_{i}})^{2}]$.
In this case maximizing $L(\vec{\theta})$ with respect to parameters
$\vec{\theta}$ is equivalent to minimizing the LS function
$Q^{2}(\vec{\theta})=\Sigma_{i=1}^{n}
\frac{(y_{i}-\eta_{i})^{2}}{\sigma_{i}^{2}}$, namely, the estimators
of ML and LS methods for $\vec{\theta}$ are identical.

For the linear LS model, $i.e.$ $f(x_{i},\vec{\theta})$ is the
linear function of $\vec{\theta}$:
\begin{equation}
f(x_{i},\vec{\theta})=\Sigma_{j=1}^{k} a_{ij}\theta_{j},
~~~~~i=1,...,n,~~ k<n, \label{E44}
\end{equation}
where $a_{ij}$ equals $x_{i}^{j-1}$ or is the $(j-1)$-th Legendre
polynomial of $x_{i}$, minimizing the LS function
$Q^{2}(\vec{\theta})$ simplified to solve a set of $k$ linear
equations. Define $a_{ij}$ being the elements of a $n\times k$
matrix $A$, minimizing the LS function $Q^{2}(\vec{\theta})$ gives
the LS estimator of parameters $\vec{\theta}$:
\begin{equation}
\hat{\vec{\theta}}=(A^{T}V^{-1}A)^{-1}A^{T}V^{-1}\vec{y},
\label{E45}
\end{equation}
the covariance matrix of $\hat{\vec{\theta}}$ is
\begin{equation}
V(\hat{\vec{\theta}})=(A^{T}V^{-1}A)^{-1}, \label{E46}
\end{equation}
or equivalently
\begin{equation}
(V^{-1}(\hat{\vec{\theta}}))_{ij}=\frac{1}{2}
\frac{\partial^{2}Q^{2}}{\partial\theta_{i}
\partial\theta_{j}}|_{\vec{\theta}=\hat{\vec{\theta}}}
=\Sigma_{l,m=1}^{n}a_{li}a_{mj}(V^{-1})_{lm},~~~~~i,j=1,...,k.
\label{E47}
\end{equation}
If $y_{i},i=1,...,n$ are independent each other, the non-diagonal
elements equal to zeros, then the above equation is simplified to
\begin{equation}
(V^{-1}(\hat{\vec{\theta}}))_{ij}
=\Sigma_{m=1}^{n}a_{mi}a_{mj}/\sigma_{m}^{2},~~~~~ i,j=1,...,k.
\label{E48}
\end{equation}

The linear LS estimators provide the exact solutions for parameters
$\vec{\theta}$, and they are unique and unbiased, and have minimum
variances.

Expanding $Q^{2}(\vec{\theta})$ about $\hat{\vec{\theta}}$, one
finds that the contour in parameter space defined by
\begin{equation}
Q^{2}(\vec{\theta})=Q^{2}(\hat{\vec{\theta}})+s^2=Q_{min}^{2}+s^2
\label{E49}
\end{equation}
has tangent planes located at plus or minus $s$- standard deviation
 from the LS estimates $\hat{\vec{\theta}}$.

For the linear LS model, if the observations $\vec{y}$ are
multi-normal variables, the minimum of the LS function
$Q^{2}(\vec{\theta})$
\begin{equation}
Q_{min}^{2}(\vec{\theta})=\Sigma_{i=1}^{n}\Sigma_{j=1}^{n}
(y_{i}-\hat{\eta_{i}})V_{ij}^{-1}(y_{j}-\hat{\eta_{j}}) \label{E50}
\end{equation}
is a $\chi^{2}$ variable with degree of freedom of $n-k$. This means
the $Q_{min}^{2}$ obtained by LS method is a quantitative measure of
the consistency between the measured values $\vec{y}$ and their
fitted value $\hat{\vec{\eta}}$, $i.e.$, the $Q_{min}^{2}$
represents the goodness of fit (see section A.3.1., goodness of fit
tests).

For the non-linear LS model,  $i.e.$ $f(x_{i},\vec{\theta})$ is the
non-linear function of $\vec{\theta}$, usually the minimizing of the
LS function $Q^{2}(\vec{\theta})$ is implemented via iteration
procedure to obtain an approximate solution of $\hat{\vec{\theta}}$.
The non-linear LS estimator is a biased estimator, its variance does
not reach MVB, and the exact distribution of $Q_{min}^{2}$ is
unknown. However, if $n$ is sufficiently large, the LS estimator is
asymptotically unbiased, and its $Q_{min}^{2}$  is approximately a
$\chi^{2}$ variable.
\subsubsection{The LS method for binned data}
For sufficiently large size $n$ of data sample
$\vec{x}=(x_{1},...,x_{n})^T$ and the measurements expressed as a
histogram  binned data, assuming the observed number of measurements
in $i$-th bin is $n_{i}, i=1,...,m$, and its corresponding
expectation from assumed model is
\begin{equation}
f_{i}(\vec{\theta})=np_{i},~~~~~ p_{i}(\vec{\theta})=\int_{\triangle
x_{i}}g(x|\vec{\theta})dx, \label{E51}
\end{equation}
where $g(x|\vec{\theta})$ is the $pdf$ of observable $x$ and
$\vec{\theta}$ are the parameters to be determined. The
normalization $\Sigma_{i=1}^{m}p_{i}=1$ requires
\begin{equation}
\Sigma_{i=1}^{m}n_{i}=\Sigma_{i=1}^{m}f_{i}(\vec{\theta})=n.
\label{E52}
\end{equation}
It can be proved for a given $n$, the LS function
$Q^{2}(\vec{\theta})$ is of the form
\begin{equation}
Q^{2}(\vec{\theta})=\Sigma_{i=1}^{m}\frac{(n_{i}-np_{i})^{2}}{np_{i}}=
\Sigma_{i=1}^{m}\frac{(n_{i}-f_{i})^{2}}{f_{i}}. \label{E53}
\end{equation}
The $f_{i}$ in the denominator can be approximated by $n_{i}$.
Minimizing this LS function leads to the LS estimates for parameters
$\vec{\theta}$, which usually needs to be implemented by numerical
iteration procedure. The $n_{i}$ is a Poisson variable with the
expectation $np_{i}$, if $n$ is sufficiently large, $n_{i}$ can be
approximated by a Gaussian, then $(n_{i}-f_{i})/\sqrt{f_{i}}$ or
$(n_{i}-f_{i})/\sqrt{n_{i}}$ is approximately a standard normal
variable, therefore, the $Q_{min}^{2}(\vec{\theta})$ distributed
approximately as a $\chi^{2}(m-1)$ variable, where the degree of
freedom of $m-1$ is due to the existence of a constraint equation
~\ref{E52}.
\subsubsection{General LS estimation with constraints}
Often in the estimation problem there exist a set of constraint
equations between the true values of observations $\eta_{i},
i=1,...,n$. The typical example is in the kinematic analysis of a
particle reaction or decay, the momentum and energy conservation
laws constitute a set of restrictions relating the various momenta
and angles for the particle combination defining the kinematic
hypothesis. Some of the quantities have been measured to a certain
accuracy (say, the momenta and angles of curved tracks), and some
are completely unknown (the variables for an unseen particle). The
purpose of the LS estimation is to investigate the kinematic
hypothesis: for a successful minimization the constraint equations
will supply estimates for the unmeasured variables as well as
"improved measurements" for the measured quantities.

Assume $\vec{y}=(y_{1},...,y_{n})^{T}$ are the measured values with
covariant matrix $V(\vec{y})$, let the true values of $\vec{y}$ be
$\vec{\eta}$. In addition, we have a set of J unmeasurable variables
$\vec{\xi}=(\xi_{1},...,\xi_{J})^{T}$. The $n$ measurable and $J$
unmeasurable variables are related and have to satisfy a set of $K$
constraint equations
\begin{displaymath}
f_k(\vec{\eta},\vec{\xi})=0,~~~~~    k=1,...,K.
\end{displaymath}
According to the LS principle, we should adopt as the best estimates
of the unknown $\vec{\eta}$ and $\vec{\xi}$ those values for which
\begin{eqnarray}
Q^{2}(\vec{\eta})=(\vec{y}-\vec{\eta})^{T}V^{-1}(\vec{y})(\vec{y}-\vec{\eta})
=minimum,\\
f(\vec{\eta},\vec{\xi})=\vec{0}. \label{E54}
\end{eqnarray}
Usually the method of the Lagrangian multipliers are used to solve
above equations. We introduce $K$ additional unknowns
$\vec{\lambda}=(\lambda_{1},...,\lambda_{K})^T$ and rephrase the
problem by requiring
\begin{equation}
Q^{2}(\vec{\eta},\vec{\xi},\vec{\lambda})=
(\vec{y}-\vec{\eta})^{T}V^{-1}(\vec{y})(\vec{y}-\vec{\eta})
+2\vec{\lambda}^{T}\vec{f}(\vec{\eta},\vec{\xi})=minimum.
\label{E55}
\end{equation}
We have now a total of $n+J+K$ unknowns. When the derivatives of
$Q^{2}$ with respect to all unknowns are put equal to zero we get
following set of equations
\begin{eqnarray}
V^{-1}(\vec{y})(\vec{\eta}-\vec{y})+F_{\eta}^{T}\vec{\lambda}=\vec{0},\\
F_{\xi}^{T}\vec{\lambda}=\vec{0},\\
\vec{f}(\vec{\eta},\vec{\xi})=\vec{0}, \label{E56}
\end{eqnarray}
where the matrices $F_{\eta}$ (of dimension $K\times N$) and
$F_{\xi}$ (of dimension $K\times J$) are defined by
\begin{equation}
(F_{\eta})_{ki}\equiv\frac{\partial f_{k}}{\partial \eta_{i}},~~~~~
(F_{\xi})_{kj}\equiv\frac{\partial f_{k}}{\partial\xi_{j}}.
\label{E59}
\end{equation}
The solution of this set of equations for the $n+J+K$ unknowns and
their errors must in general case be found by iterations, producing
successively better approximations.

In the linear LS estimation problem for the $n$ measurable and $J$
unmeasurable variables which are related and have to satisfy a set
of $K$ constraint equations, if the measured values
$\vec{y}=(y_{1},...,y_{n})^{T}$ is a multi-normal variable, the
$Q^2_{min}$ is a $\chi^2$ variable with the degree of freedom
$(K-J)$. For the non-linear LS estimation problem of non-linear
constraint equations, and/or $\vec{y}$ is not a multi-normal
variable, the $Q^2_{min}$ may be approximated by $\chi^2(K-J)$.

The momentum-energy conservation laws constitute a set of 4
constraint equations. If all the particle's parameters in a reaction
or a decay process have been measured (no unmeasurable variables)
and the momentum-energy conservation laws are applied to obtain
better values of particles parameters (4C kinematic fit), the
$Q^2_{min}$ of the LS estimator is then an approximate $\chi^2(4)$
variable. If there exist $J$ unmeasurable variables and $r$
intermediate resonances which promptly decayed to observed final
state particles, and the invariant masses of daughter particles of
these resonances are constrained to their mother particles' masses,
then the $Q^2_{min}$ is approximately a $\chi^2(4+r-J)$ variable.
%
%
\section{Interval estimation, confidence interval and upper limit}
The task of the interval estimation is to locate a region which
contains the true value of the parameter $\theta$ to be studied with
a probability $\gamma$. This region is called the confidence
interval with coverage probability $\gamma$. When the goal of an
experiment is to determine a parameter $\theta$, the result is
usually expressed by quoting, in addition to the point estimate,
some sort of confidence interval which reflects the statistical
precision of the measurement. In the simplest case this can be given
by the parameter's estimated value $\hat{\theta}$ plus/minus an
estimate of the standard deviation of $\hat{\theta}$,
$\sigma_{\hat{\theta}}$. If the parameter $\theta$ has boundary
(without losing generality, we assume it is lower boundary with the
value zero throughout this Appendix), and the estimate of $\theta$
in an experiment is close to this boundary, then the determination
of the interval estimation is difficult and needs to be treated in
special way.
\subsection{Frequentist confidence interval}
\subsubsection{Neyman method for confidence interval}
\begin{figure}[hbt]
\centerline{\psfig{file=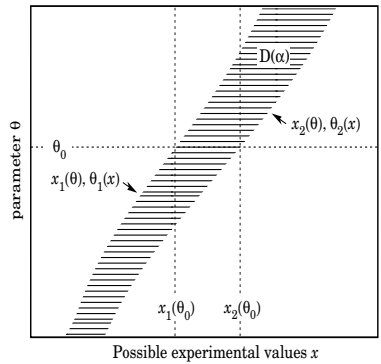,height=5 cm,width=5 cm}}
\caption{\label{F1} Construction of the confidence belt}
\end{figure}

Confidence interval refers to frequentist interval obtained with a
procedure due to Neyman~\cite{Neyman}. Consider a $pdf$
$f(x;\theta)$ where $x$ represents the measurement of the experiment
and $\theta$ the unknown parameter for which we want to construct a
confidence interval. The variable $x$ could (and often does)
represent an estimator of $\theta$. Using $f(x;\theta)$ we can find
for a pre-specified probability $\gamma=1-\alpha$ and for every
value of $\theta$ a set of values $x_{1}(\theta,\alpha)$ and
$x_{2}(\theta,\alpha)$ such that
\begin{equation}
P(x_{1}<x<x_{2};\theta)=1-\alpha\equiv
\gamma=\int_{x_{1}}^{x_{2}}f(x;\theta)dx. \label{E61}
\end{equation}
  This is illustrated in Fig.~\ref{F1}: a horizontal line
segment $[x_{1}(\theta,\alpha),x_{2}(\theta,\alpha)]$ is drawn for
representative values of $\theta$. The union of such intervals for
all values of $\theta$, designated in the figure as $D(\alpha)$, is
known as the confidence belt. Typically the curves
$x_{1}(\theta,\alpha)$ and $x_{2}(\theta,\alpha)$ are monotonic
functions of $\theta$, which we assume for this discussion.

Upon performing an experiment to measure $x$ and obtaining a value
$x_{0}$, one draws a vertical line through $x_{0}$. The confidence
interval for $\theta$ is the set of all values of $\theta$ for which
the corresponding line segment
$[x_{1}(\theta,\alpha),x_{2}(\theta,\alpha)]$ is intercepted by this
vertical line. Such confidence intervals are said to have a
confidence level (CL) equal to $\gamma=1-\alpha$.

Now suppose that the true value of $\theta$ is $\theta_{0}$,
indicated in the figure.We see from the figure that $\theta_{0}$
lies between $\theta_{1}(x)$ and $\theta_{2}(x)$ if and only if $x$
lies between $x_{1}(\theta_{0})$ and $x_{2}(\theta_{0})$. The two
events thus have the same probability, and since this is true for
any value $\theta_{0}$, we can drop the subscript 0 and obtain
\begin{equation}
\gamma=1-\alpha=P(x_{1}(\theta)<x<x_{2}(\theta))=
P(\theta_{2}(x)<\theta<\theta_{1}(x)). \label{E62}
\end{equation}
In this probability statement $\theta_{1}(x)$ and $\theta_{2}(x)$,
$i.e.$, the endpoints of the interval, are the random variables and
$\theta$ is an unknown constant. If the experiment were to be
repeated a large number of times, the interval
$[\theta_{1},\theta_{2}]$ would vary, covering the fixed value
$\theta$ in a fraction $\gamma=1-\alpha$ of the experiments.

The condition of coverage probability does not determine $x_{1}$ and
$x_{2}$ uniquely and additional criteria are needed. The most common
criterion is to choose central intervals such that the probabilities
below $x_{1}$ and above $x_{2}$ are each $\alpha/2$. In other cases
one may want to report only an upper or lower limit, then the
probability excluded below $x_{1}$ or above $x_{2}$ can be set to
zero.

When the observed random variable $x$ is continuous, the coverage
probability obtained with the Neyman construction is
$\gamma=1-\alpha$, regardless of the true value of the parameter. If
$x$ is discrete, however, it is not possible to find segments
$[x_{1}(\theta,\alpha),x_{2}(\theta,\alpha)]$ that satisfy
Eq.~\ref{E62} exactly for all values of $\theta$. By convention one
constructs the confidence belt requiring the probability
$P(x_{1}<x<x_{2})$ to be greater than or equal to $\gamma=1-\alpha$.
This gives confidence intervals that include the true parameter with
a probability greater than or equal to $\gamma=1-\alpha$.
\subsubsection{Gaussian distributed measurements}
An important example of constructing a confidence interval is when
the data consist of a single random variable $x$ that follow a
Gaussian distribution; this is often the case when $x$ represents an
estimator for a parameter and one has a sufficiently large data
sample. If there is more than one parameter being estimated, the
multivariate Gaussian is used. For the univariate case with known
$\sigma$,
\begin{equation}
\gamma=1-\alpha=\frac{1}{\sqrt{2\pi\sigma}}\int_{\mu-\delta}^{\mu+\delta}
e^{-(x-\mu)^2/2\sigma^2}dx=erf(\frac{\delta}{\sqrt{2}\sigma})
\label{E63}
\end{equation}
is the probability that the measured value $x$ will fall within
$\pm\delta$ of the true value $\mu$. From the symmetry of the
Gaussian with respect to $x$ and $\mu$, this is also the probability
for the interval $x\pm\delta$ to include $\mu$. The choice
$\delta=\sigma$ gives an interval called the standard error which
has $\gamma=1-\alpha=68.27\%$ if $\sigma$ is known. Values of
$\alpha$ for other frequently used choices of $\delta$ are given in
Table~\ref{T1}. The relation of $\alpha$ and  $\delta$ can be also
represented by the cumulated distribution function for the $\chi^2$
distribution for $\chi^2=(\delta/\sigma)^2$ and $n=1$
 degree of freedom:
\begin{equation}
\gamma= 1-\alpha=F(\chi^2;n=1). \label{E64}
\end{equation}
\begin{table}[htbp]
\caption{Area of the tails $\alpha$ outside $\pm\delta$ from the
mean of a Gaussian distribution.} \center
\begin{tabular}{|c|c|c|c|}\hline
$\alpha$             & $\delta$ & $\alpha$ &$\delta$      \\\hline
0.3173               & $1\sigma$ &0.2      & $1.28\sigma$ \\
0.0455               & $2\sigma$ &0.1      & $1.64\sigma$ \\
0.0027               & $3\sigma$ &0.05     & $1.96\sigma$ \\
$ 6.3\times 10^{-5}$ & $4\sigma$ &0.01     & $2.58\sigma$ \\
$ 5.7\times 10^{-7}$ & $5\sigma$ &0.001    & $3.29\sigma$ \\
$ 2.0\times 10^{-9}$ & $6\sigma$ &0.0001   & $3.89\sigma$ \\\hline
\end{tabular}
\label{T1}
\end{table}
For multivariate measurements of, say, $n$ parameter estimates
$\hat{\vec{\theta}}=(\hat{\theta_1},...,\hat{\theta_n})^T$, one
requires the full covariance matrix
$V_{ij}=cov(\hat{\theta_i},\hat{\theta_j})$, which can be estimated
by ML or LS method.

If the parameters $\vec{\theta}$ are estimated with the ML method,
for sufficient large $n$ and the likelihood function satisfies the
regularity conditions, the likelihood function distributed
asymptotically as a multi-Gaussian, then we have
\begin{equation}
lnL(\vec{\theta})=lnL_{max}-Q(\vec{\theta})/2, \label{E65}
\end{equation}
where $Q(\vec{\theta})=(\vec{\theta}-\hat{\vec{\theta}})^T
V^{-1}(\vec{\theta})(\vec{\theta}-\hat{\vec{\theta}})$ is
asymptotically a $\chi^2(k)$ variable, and $k$ is the dimension of
$\vec{\theta}$. The intersection contour of super-plane
$lnL=lnL_{max}-Q_{\gamma}/2$ and super-surface $lnL(\vec{\theta})$
forms the boundary of the confidence region of $\vec{\theta}$ with
coverage probability of $\gamma= 1-\alpha$, which is calculated by
the cumulated $\chi^2$ function
\begin{equation}
\gamma= 1-\alpha=P(Q\leq
Q_\gamma)=\int_{0}^{Q_\gamma}\chi^2(Q;\nu=k)dQ=F_\alpha(Q_\gamma;\nu=k).
\label{E66}
\end{equation}

In the case that the parameters $\vec{\theta}$ are estimated with LS
method, for linear LS estimator and multi-Gaussian measurements, we
have
\begin{equation}
Q^2(\vec{\theta})=Q^2_{min}+Q^2_{LS}, \label{E67}
\end{equation}
where $Q^2_{LS}(\vec{\theta})=(\vec{\theta}-\hat{\vec{\theta}})^T
V^{-1}(\vec{\theta})(\vec{\theta}-\hat{\vec{\theta}})$ is a
$\chi^2(k)$ variable and $k$ is the dimension of $\vec{\theta}$ for
non-constraint LS estimation, and the dimension of $\vec{\theta}$
minus the number of independent linear constraint equations for
constraint LS case. The intersection contour of super-plane
$Q^2(\vec{\theta})=Q^2_{min}+Q_{\gamma}$ and super-surface
$Q^2(\vec{\theta})$ forms the boundary of the confidence region of
$\vec{\theta}$ with coverage probability of $\gamma\equiv 1-\alpha$,
which is also calculated by Eq.~\ref{E66}. Values of $Q_\gamma$ for
$k=1,2,3$ are given in Table~\ref{T1p} for several values of the
coverage probability $\gamma=1-\alpha$.
\begin{table}[htbp]
\caption{$Q_\gamma$ for $k=1,2,3$ corresponding to a coverage
probability $\gamma=1-\alpha$ in the large data sample limit.}
\center
\begin{tabular}{|c|ccc|}\hline
$\gamma$(\%) & k=1  & k=2   & k=3    \\\hline
68.27        & 1.00 & 2.30  & 3.53 \\
90.          & 2.71 & 4.61  & 6.25 \\
95.          & 3.84 & 5.99  & 7.82 \\
95.45        & 4.00 & 6.18  & 8.03 \\
99.          & 6.63 & 9.21  &11.34 \\
99.73        & 9.00 &11.83  &14.16 \\\hline
\end{tabular}
\label{T1p}
\end{table}

If the mentioned conditions are not fully satisfied, the confidence
region determined by Eqs.~\ref{E65} and ~\ref{E67} are not exact but
an approximate one.

The ML method has an advantage that is easier to calculate the
confidence region for combining several independent measurements of
same parameters. Assume $N$ independent measurements give likelihood
functions $lnL_i(\vec{\theta}),i=1,...,N$, then the combined
likelihood function is simply
\begin{equation}
lnL(\vec{\theta})=\sum_{i}^{N}lnL_i(\vec{\theta}). \label{E68}
\end{equation}
Then using this likelihood functions in Eq.~\ref{E65} can give the
confidence region with coverage probability $\gamma$ for combined
estimate $\hat{\vec{\theta}}$.
\subsubsection{Poisson distributed measurements}
If $n$ represents the number of events produced in a reaction with
cross section $\sigma$, say in a fixed integrated luminosity $L$,
then it follows a Poisson distribution with mean $s=\sigma L$ in the
case there is no background. Therefore, to determine the cross
section of a reaction or the branching ratio of a decay process in
terms of the number of observed events, the interval estimation of
Poisson distributed data must be met. The probability of observing
$n$ events of the Poisson distribution with the mean $s$ is
\begin{equation}
P(n,s)=\frac{s^{n}e^{-s}}{n!}. \label{E69}
\end{equation}
The upper and lower (one sided) limits on the mean $s$  can be found
from the Neyman procedure to be
\begin{equation}
s_{lo}=\frac{1}{2}F^{-1}_{\chi^{2}}(\alpha_{lo};2n), \label{E70}
\end{equation}
\begin{equation}
s_{up}=\frac{1}{2}F^{-1}_{\chi^{2}}(1-\alpha_{up});2(n+1)),
\label{E71}
\end{equation}
where the upper and lower limits are at confidence levels of
$1-\alpha_{lo}$ and $1-\alpha_{up}$, respectively, and
$F^{-1}_{\chi^{2}}$ is the quantile of the $\chi^{2}$ distribution
(inverse of the cumulative distribution). The quantiles
$F^{-1}_{\chi^{2}}$ can be obtained from standard tables or from the
CERNLIB routine CHISIN. For central confidence intervals at CL
$1-\alpha$, set $\alpha_{lo}=\alpha_{up}=\alpha/2$. Values for
confidence levels of $90\%$ and $95\%$ are shown in Table~\ref{T2}.
\begin{table}[htbp]
\caption{Lower and upper (one-sided) limits for the mean $s$ of a
Poisson variable given $n$ observed events in the absence of
background, for CL of $90\%$ and $95\%$}. \center
\begin{tabular}{|c|cc|cc|}\hline
     &$1-\alpha=90\%$&     &$1-\alpha=95\%$ &     \\\hline
 $n$ &$s_{lo}$&$s_{up}$&$s_{lo}$&$s_{up}$ \\\hline
0    &          & 2.30     &          &  3.00 \\
1    &   0.105  & 3.89     &  0.051   &  4.74 \\
2    &   0.532  & 5.32     &  0.355   &  6.30 \\
3    &   1.10   & 6.68     &  0.818   &  7.75 \\
4    &   1.74   & 7.99     &  1.37    &  9.15 \\
5    &   2.43   & 9.27     &  1.97    & 10.51 \\
6    &   3.15   &10.53     &  2.61    & 11.84 \\
7    &   3.89   &11.77     &  3.29    & 13.15 \\
8    &   4.66   &12.99     &  3.98    & 14.43 \\
9    &   5.43   &14.21     &  4.70    & 15.71 \\
10   &   6.22   &15.41     &  5.43    & 16.96 \\\hline
\end{tabular}
\label{T2}
\end{table}

If the number of observed events $n$ contains both signal and
background events, which are Poisson variables with mean $s$ and
$b$, respectively, then we have
\begin{equation}
P(n,s)=\frac{(s+b)^{n}e^{-(s+b)}}{n!}. \label{E72}
\end{equation}
For a specific value of $s$, the upper and lower limit of the
central confidence region, $[n_{l},n_{u}]$, and the lower limit of
the upper confidence belt, $n_{l_{0}}$, at given confidence level
$\gamma=1-\alpha$ can be determined by
\begin{equation}
\sum_{n=0}^{n_{l}}P(n,s)\leq\frac{\alpha}{2},~~~~~
\sum_{n=n_{u}+1}^{\infty}P(n,s)\leq\frac{\alpha}{2}, \label{E73}
\end{equation}
\begin{equation}
\sum_{n=0}^{n_{lo}} P(n,s)\leq\alpha, \label{E74}
\end{equation}
respectively.  For all $s$ values, such calculations give the
confidence belts for central region and upper confidence belt. The
inequality sign is to ensure the actual coverage greater or equal to
the given coverage in the discrete variable case.
\subsubsection{Confidence interval near the physics boundary}
A number of issues arise in the construction and interpretation of
confidence intervals when the parameter can only take on values in a
restricted range. An important sample is where the mean of a
Gaussian variable is constrained on physical grounds to be
non-negative. This arises, for example, when the square of the
neutrino mass is estimated from $\hat{m}^2=\hat{E}^2-\hat{p}^2$,
where $\hat{E}$ and $\hat{p}$ are independent, Gaussian distributed
estimates of the energy and momentum. Although the true $m^2$ is
constrained to be positive, random errors in $\hat{E}$ and $\hat{p}$
can easily lead to negative values for the estimate  $\hat{m}^2$.

If one uses the prescription given above for Gaussian distributed
measurements, which says to construct the interval by taking the
estimate plus/minus one standard deviation, then this can give
intervals that are partially or entirely in the unphysical region.
In fact, by following strictly the Neyman construction for the
central confidence interval, one finds that the interval is
truncated below zero; nevertheless an extremely small or even a
zero-length interval can result.

An additional important example is where the experiment consists of
counting a certain number of events, $n$, which is assumed to be
Poisson distributed. Suppose the expectation value $E(n)=\mu$ is
equal to $s+b$, where $s$ and $b$ are the means for signal and
background processes, and assume further that $b$ is a known
constant. Then $\hat{s}=n-b$ is an unbiased estimator for $s$.
Depending on true magnitudes of $s$ and $b$, the estimate $\hat{s}$
can easily fall in the negative region. Similar to the Gaussian case
with the positive mean, the central confidence interval or even the
upper limit for $s$ may be of zero length.

An additional difficulty arises when a parameter estimate is not
significantly far away from the boundary, in which case it is
natural to report a one-sided confidence interval (often an upper
limit). It is straightforward to force the Neyman prescription to
produce only an upper limit by setting $x_2=\infty$ in Eq.~\ref{E61}
. Then $x_1$ is uniquely determined and the upper limit can be
obtained. If, however, the data come out such that the parameter
estimate is not so close to the boundary, one might wish to report a
central ($i.e.$, two-sided) confidence interval. As pointed out by
Feldman and Cousins~\cite{FC}, however, if the decision to report an
upper limit or two-sided interval is made by looking at the data
("flip-flopping"), then the resulting intervals will not in general
cover the parameter with the probability $1-\alpha$.

With the confidence intervals suggested by Feldman and
Cousins~\cite{FC}, the prescription determines whether the interval
is one- or two-sided in a way which preserves the coverage
probability. Intervals with this property are said to be unified.
Furthermore, this prescription is such that null intervals do not
occur. For a given choice of $1-\alpha$, if the parameter estimate
is sufficiently close to the boundary, then the method gives an
one-sided limit. In the case of a Poisson variable in the presence
of background, for example, this would occur if the number of
observed events is compatible with the expected background. For
parameter estimates increasingly far away from the boundary, $i.e.$,
for increasing signal significance, the interval makes a smooth
transition from one- to two-sided, and far away from the boundary
one obtains a central interval. The intervals according to this
method for the mean of Poisson variable in the absence of background
are given in Table~\ref{T3}.
\begin{table}[htbp]
\caption{Unified confidence interval $[s_1,s_2]$ for a mean $s$ of a
Poisson variable given $n$ observed events in the absence of
background, for CL of $90\%$ and $95\%$}. \center
\begin{tabular}{|c|cc|cc|}\hline
     &$1-\alpha=90\%$&     &$1-\alpha=95\%$ &     \\\hline
 $n$ &$s_{1 }$&$s_{2}$ &$s_{1}$ &$s_{2}$  \\\hline
0    &   0.00   & 2.44     &  0.00    &  3.09 \\
1    &   0.11   & 4.36     &  0.05    &  5.14 \\
2    &   0.53   & 5.91     &  0.36    &  6.72 \\
3    &   1.10   & 7.42     &  0.82    &  8.25 \\
4    &   1.47   & 8.60     &  1.37    &  9.76 \\
5    &   1.84   & 9.99     &  1.84    & 11.26 \\
6    &   2.21   &11.47     &  2.21    & 12.75 \\
7    &   3.56   &12.53     &  2.58    & 13.81 \\
8    &   3.96   &13.90     &  2.94    & 15.29 \\
9    &   4.36   &15.30     &  4.36    & 16.77 \\
10   &   5.50   &16.50     &  4.75    & 17.82 \\\hline
\end{tabular}
\label{T3}
\end{table}

The intervals constructed according to the unified procedure in
Ref.~\cite{FC} for a Poisson variable $n$ consisting of signal and
background have the property that for $n=0$ observed events, the
upper limit decreases for increasing expected background. This is
counter-intuitive, since it is known that if $n=0$ for the
experiment in question, then no background was observed, and
therefore one may argue that the expected background should not be
relevant. Roe and Woodroofe~\cite{RW} proposed a solution to this
problem by using such a fact that, given an observation $n$, the
background $b$ can not be larger than $n$ in any case. Therefore,
the usual Poisson $pdf$ should be replaced by a conditional $pdf$,
and then this conditional $pdf$ is used to construct the  confidence
intervals following Feldman and Cousins' procedure.
\subsubsection{Confidence interval incorporating
systematic uncertainties}
A modification of the Neyman method incorporating systematic
uncertainty of the signal detection efficiency has been proposed by
Highland and Cousins~\cite{HC}, in which a "semi-Bayesian" approach
is adopted, where an average over the probability of the detection
efficiency is performed. This method is of limited accuracy in the
limit of high relative systematic uncertainties. On the other hand,
an entirely frequentist approach has been proposed for the
uncertainty in the background rate prediction~\cite{RL}. This
approach is based on a two-dimensional confidence belt construction
and likelihood ratio hypothesis testing and treats the uncertainty
in the background as a statistical uncertainty rather than as a
systematic one. Recently, Conrad $etal$ extend the method of
confidence belt construction proposed in~\cite{Conrad} to include
systematic uncertainties  in both the signal and background
efficiencies as well as systematic uncertainty of background
expectation prediction. It takes into account the systematic
uncertainties by assuming a $pdf$ which parameterizes our knowledge
on the uncertainties and integrating over this $pdf$. This method,
combining classical and Bayesian elements, is referred to as
semi-Bayesian approach. A FORTRAN program, POLE, has been coded to
calculate the confidence intervals for a maximum of observed events
of 100 and a maximum signal expectation of 50~\cite{POLE}.
\subsection{Bayesian credible interval}
In Bayesian approach one has to assume a prior $pdf$ of an unknown
parameter and then perform an experiment to update the prior
distribution. The prior $pdf$ reflects the experimenter's subjective
degree of belief about unknown parameter before the measurement was
carried out. The updated prior, called posterior $pdf$, is used to
draw inference on unknown parameter. This updating is done with the
use of Bayes theorem~\cite{Bayes}. Assuming that $n$  represents the
number of observed events, $s$  is the expectation of the number of
signal events which is unknown and to be inferred, $p(n|s)$  is the
conditional $pdf$ of observing $n$ events with given signal $s$,
$\pi(s)$  is the prior $pdf$, the Bayes theorem gives the posterior
$pdf$:
\begin{equation}
h(s|n)=\frac{p(n|s)\pi(s)}{\int_{0}^{\infty}p(n|s)\pi(s)ds}.
\label{E75}
\end{equation}
Here the lower limit of the integral is zero, which is the possible
minimum of the signal expectation. Using this posterior $pdf$, one
can calculate a Bayesian credible interval for the signal
expectation $s$ at given credible level $CL=1-\alpha$ :
\begin{equation}
1-\alpha=\int_{s_L}^{s_U}h(s|n)ds . \label{E76}
\end{equation}
However, such intervals are not uniquely determined. Often, the
highest posterior density (HPD) credible interval $I$  is chosen,
which is determined in following way:
\begin{equation}
1-\alpha=\int_{I}h(s|n)ds,~~~  h(s_1|n)\geq h(s_2|n)~~~for ~any~
s_1\in I~ and~  s_2\notin I. \label{E77}
\end{equation}
The upper limit of the signal expectation $s$ at given credible
level $CL=1-\alpha$, $s_{UP}$, is naturally given by:
\begin{equation}
1-\alpha=\int_{0}^{s_{UP}} h(s|n)ds. \label{E78}
\end{equation}
The nice feature of the Bayesian approach is that the zero value of
an upper limit $s_{UP}$  always corresponds to the zero value of
credible level $CL=1-\alpha$, which is not necessarily true for the
classical approach. The most important issue is to determine a prior
$pdf$ of the parameter. This is an issue which brings most of
controversies into Bayesian methods. An important question is that
if one should use an $informative$ prior, $i.e.$, a prior which
incorporates results of previous experiments, or a $non-informative$
 prior, $i.e.$, a prior which claims total ignorance.  The major
objection against informative prior is based on such argument: if we
assume a prior which incorporates results of previous experiments,
then our measurement will not be independent, hence, we will not be
able to combine our results with previous results by taking a
weighted average.

Thus, we only discuss the Bayesian inference that assumes a
non-informative prior $pdf$ for the non-negative parameter of a
Poisson distribution. For the case that in the "signal region" where
the signal events resides, the number of signal events is a Poisson
variable with unknown expectation $s$ , and the number of background
events is a Poisson variable with expectation $b$, the conditional
$pdf$ of observing  total events $n$, $p(s|n)$, can be represented
by
\begin{equation}
P(s|n)=\frac{(s+b)^{n}e^{-(s+b)}}{n!}. \label{E72p}
\end{equation}
 To deduce the posterior $pdf$, one has to assume a
prior $pdf$. Bayes stated that, the non-informative prior for any
parameter must be flat~\cite{Bayes}. This statement does not based
on any strict mathematical argument, but merely his intuition. The
obvious weakness of Bayes prior $pdf$ is that if one can assume a
flat distribution of an unknown parameter, then one can also assume
a flat distribution for any function of this parameter, and these
two prior functions are apparently not identical.
Jeffreys~\cite{Jeffreys},~\cite{Jeffreys1},Jaynes~\cite{Jaynes}, and
Box $etal$~\cite{Box} derived the non-informative priors from first
principle to resolve this problem, which are proportional to
$1/\theta$ and $1/\sqrt{\theta}$, respectively, where $\theta$ is
the unknown parameter. Comments on these non-informative priors can
be found in Refs.~\cite{Narsky},~\cite{Narsky1}. For the $pdf$ shown
by Eq.~\ref{E72p}, the corresponding priors are proportional to
$1/(s+b)$ and $1/\sqrt{s+b}$. In general, we can use a prior $pdf$
of
\begin{equation}
\pi(s)\propto \frac{1}{(s+b)^m},~~~~~s\geq 0,~~ 0\leq m\leq 1,
\label{E79}
\end{equation}
where $m=0$ corresponds to Bayes prior, $m=0.5$ to $1/\sqrt{s+b}$
prior, and $m=1$ to $1/(s+b)$ prior. One can choose $m$ value as
he/she thinks appropriate, however, it should be always kept in mind
that different $m$ value will give different answer for the credible
interval and upper limit. Expected coverage and length of credible
intervals constructed with these three priors and with the Neyman
construction and unified approach can be found in
Ref.~\cite{Narsky1}. It has been shown that the $1/\sqrt{s+b}$ prior
is the most versatile choice among the Bayesian methods, it provides
a reasonable mean coverage for the credible interval and upper limit
for Poisson observable.

Substituting $p(n|s)$  of Eq.~\ref{E72p} and $\pi(s)$ of
Eq.~\ref{E79} into Eq.~\ref{E75}, the posterior $pdf$ is then given
by
\begin{equation}
h(s|n)=\frac{(s+b)^{n-m}e^{-(s+b)}}{\Gamma(n-m+1,b)}, \label{E80}
\end{equation}
where
\begin{equation}
\Gamma(x,b)=\int_{0}^{\infty}s^{x-1}e^{-x}ds,~~~~~x>0,b>0
\label{E81}
\end{equation}
is an incomplete gamma function.

    In the case that the systematic uncertainties of the signal efficiency
and background expectation can be neglected, the signal expectation
$s$ is an unknown constant and the background expectation $b$ is a
known value. The Bayesian HPD credible interval at given credible
level $1-\alpha$ can be obtained by substituting the posterior $pdf$
of Eq.~\ref{E80} into Eq.~\ref{E77}, while the Bayesian upper limit
can be calculated by substituting the posterior $pdf$ of
Eq.~\ref{E80} into Eq.~\ref{E78}:
\begin{equation}
\alpha=\frac{\Gamma(n-m+1,s_{UP}+b)}{\Gamma(n-m+1,b)}. \label{E82}
\end{equation}
If the flat prior£¨$m=0$£©is used, Eq.~\ref{E82} turns into
\begin{equation}
\alpha=e^{-s_{UP}}\cdot\frac{\sum_{k=0}^{n}\frac{(s_{UP}+b)^k}{k!}
}{\sum_{k=0}^{n}\frac{b^k}{k!}}. \label{E83}
\end{equation}
The upper limit $s_{UP}$  at given credible level $1-\alpha$ can be
acquired by solving Eq.~\ref{E82} or Eq.~\ref{E83} numerically from
measured values of $n$ and $b$. Eq.~\ref{E83} has been recommended
by PDG~\cite{PDG}, therefore, widely used in particle physics
experiments. However, from statistics point of view, the
$1/\sqrt{s+b}$ prior seems to be a more appropriate non-informative
prior as mentioned above, therefore, using Eq.~\ref{E82} with
$m=0.5$ to determine $s_{UP}$ seems a reasonable choice.

 Now we turn to the question of inclusion of systematic uncertainties. First we
consider only the uncertainty of background expectation is present,
and the distribution of the background expectation is represented by
a pdf $f_{b^{\prime}}(b,\sigma_{b})$ with the mean $b$ and standard
deviation $\sigma_{b}$ . The conditional pdf expressed by
Eq.~\ref{E72p} now is modified to
\begin{equation}
q(n|s)_{b}=\int_{0}^{\infty}p(n|s)_{b^{\prime}}\cdot
f_{b^{\prime}}(b,\sigma_{b})db^{\prime}, \label{E84}
\end{equation}
where $p(n|s)_{b^{\prime}}$ has the same expression in
Eq.~\ref{E72p} with $b$ replaced by $b^{\prime}$.

    Next we take into account the uncertainties of the signal
efficiency and background expectation simultaneously, and consider
they are independent each other. The distribution of the signal
relative efficiency $\varepsilon$ (with respect to the nominal
signal detection efficiency) is described by a $pdf$
$f_{\varepsilon}(1,\sigma_{\varepsilon})$   with the mean 1 and
standard deviation $\sigma_{\varepsilon}$. The conditional $pdf$
    described by Eq.~\ref{E72p} is then further modified to
\begin{equation}
q(n|s)_{b}=\int_{0}^{\infty}\int_{0}^{\infty}p(n|s\varepsilon)_{b^{\prime}}
f_{b^{\prime}}(b,\sigma_{b})f_{\varepsilon}(1,\sigma_{\varepsilon})
db^{\prime}d\varepsilon, \label{E85}
\end{equation}
where  $p(n|s\varepsilon)_{b^{\prime}}$ represents that in
Eq.~\ref{E72p} $b$ is replaced by $b^{\prime}$ , and $s$ by
$s\varepsilon$ . One notices that the lower limits of integrals in
Eqs.~\ref{E84},~\ref{E85} are all zeros, which are the possible
minimum value of any efficiencies and number of background events.

Using $q(s|n)_{b}$  in Eqs.~\ref{E84},~\ref{E85} to construct
posterior $pdf$
\begin{equation}
h(s|n)=\frac{q(n|s)_{b}\pi(s)}{\int_{0}^{\infty}q(n|s)_{b}\pi(s)ds},
\label{E86}
\end{equation}
one can calculate the Bayesian HPD credible interval or upper limit
$s_{UP}$ on $s$ at any given credible level with inclusion of
systematic uncertainties in terms of Eq.~\ref{E77} or ~\ref{E78}.

 An method to calculate the Bayesian HPD credible interval or upper
limit at given credible level with or without inclusion of
systematic uncertainties in pure Bayesian approach has been
described in ref.~\cite{ZHU},~\cite{BPOCI}.  It has been implemented
as a FORTRAN program, BPOCI (Bayesian POissonian Credible
Interval)~\cite{BPOCI}.
%
%
\section{Tests of hypotheses}
In addition to estimating parameters, one often wants to assess the
validity of certain statements concerning the data's underlying
distribution. Hypothesis tests provide a rule for accepting or
rejecting hypotheses depending on the outcome of a measurement. We
restrict ourselves here to discuss the Goodness-of-fit tests - one
of the non-parametric tests, which deals with questions of the
functional form for the distribution of the data and gives the
probability to obtain a level of incompatibility with a certain
hypothesis that is greater than or equal to the level observed with
the actual data. Two methods will be stated: Pearson's $\chi^2$ test
and Kolmogorov-Smirnov test, which is applicable for the large and
small size of the measured data sample, respectively. Finally, we
have a section to discuss an important concept in particle physics
experiment-the statistical significance of signal.
\subsection{Goodness-of-fit test}
\subsubsection{Pearson's $\chi^2$ test}
We assume that $n$ observations on the variable $x$ belong to $N$
mutually exclusive classes, such as successive intervals in a
histogram, non-overlapping regions in two-dimentional plot, etc. The
number of events $n_1,n_2,...,n_N$ in the different classes will
then be multinomially distributed, with probabilities $p_i$ for the
individual classes as determined by the underlying distribution
$f(x)$ for continuous variable $x$:
$$p_i=\int_{\triangle x_i} f(x)dx,~~~~~~i=1,2,...,N,$$
or $q_j=P(x=x_j),j=1,2,\cdots$ for discrete variable $x$:
$$p_i=\sum_{j,x_j\in\triangle x_i}q_j,~~~~~~i=1,2,...,N,$$
where $\triangle x_i$ represents the $i$-th interval. The hypothesis
we wish to test specifies the class probabilities according to a
certain prescription,
\begin{equation}
H_0:p_i=p_{0i},~~~~~~i=1,2,...,N, \label{E87}
\end{equation}
where
\begin{equation}
\sum_{i=1}^N p_{0i}=1,\label{E88}
\end{equation}
is the overall normalization and
$$p_{0i}=\int_{\triangle x_i} f_0(x)dx,~~~ or~~~
p_{0i}=\sum_{j,x_j\in\triangle x_i}q_{0j}.$$ Therefore, what we wish
to test is if the distribution of the observation $f(x)$ or $q_j$ is
consistent with the assigned distribution $f_0(x)$ or $q_{0j}$, or
equivalently, if the hypothesis $H_0$ is accepted by the observed
data, given that the total number in all classes is $n$? To test
whether the set of predicted numbers $np_{0i}$ is compatible with
the set of observed numbers $n_i$ we take as our test statistic the
quantity
\begin{equation}
X^2=\sum_{i=1}^N \frac{(n_i-np_{0i})^2}{np_{0i}}=
\frac{1}{n}\sum_{i=1}^N\frac{n_i^2}{p_{0i}}-n.\label{E89}
\end{equation}
When $H_0$ is true this statistic is approximately $\chi^2(N-1)$
distributed.  This is called the Pearson theorem.

If $H_0$ is true and the experiment is repeated many times under the
same conditions with $n$ observations, the actual values obtained
for $X^2$, $X^2_{obs}$, will therefore be distributed nearly like
$\chi^2(N-1)$; in particular, the mean value for $X^2_{obs}$ will be
$\simeq N-1$ and the variance $\simeq 2(N-1)$. If, on the other
hand, $H_0$ is not true, the expectation for each $n_i$ is not
$np_{0i}$, and the sum of terms $(n_i-np_{0i})^2/np_{0i}$ will tend
to become on the average larger than if $H_0$ were true. Hence it
seems reasonable to reject $H_0$ if  $X^2_{obs}$ becomes too large.
The criteria to reject $H_0$ at $significance ~level~ \alpha$ is
\begin{equation}
X^2_{obs}>\chi^2_{\alpha}(N-1), \label{E90}
\end{equation}
where $\chi^2_{\alpha}(N-1)$ is determined by the
$\chi^2(N-1)~~pdf~~f(y;N-1)$ such that
$$\alpha=\int^\infty_{\chi^2_{\alpha}(N-1)} f(y;N-1)dy.$$

Often, the model which to describe the distribution of the measured
data includes $L$ unknown parameters. For a Least-Square estimation
we know that the comparison between data and fitted model is made
using the $\chi^2$ distribution with a number of degrees of freedom
equal to the number of independent observations minus the number of
independent parameters estimated. This procedure is exact only in
the limit of infinitely many observations and with a linear
parameter dependence; otherwise it is an approximation. Thus, if
there are $L$ parameters in $H_0$ which are estimated by the $LS$
method and $N$ classes subject to an overall normalization
condition, Pearson's $\chi^2$ test for goodness-of-fit consists in
comparing the fitted (minimum) value $X^2_{min}$ to the $\chi^2$
distribution with $(N-1-L)$ degrees of freedom.
\subsubsection{Kolmogorov-Smirnov test}
The Kolmogorov-Smirnov (KS) test avoids the binning of individual
observations and may be more sensitive to the data, and is superior
to the $\chi^2$ test in particular for small samples and has many
nice properties when applied to problems in which no parameters are
estimated.

Given $n$ independent observations on the variable $x$ we form an
$ordered ~sample$ by arranging the observations in ascending order
of magnitude, $x_1,x_2,\cdots,x_n$. The cumulative distribution for
this sample of size $n$ is now defined by
\begin{equation}
S_n(x)=\left\{\begin{array}{cc} 0,         & x<x_i,\\
                               \frac{i}{n},& x_i\leq x\leq x_{i+1},\\
                               1,          & x\geq x_n. \end{array}
                               \right.
\label{E91}
\end{equation}
Thus $S_n(x)$ is an increasing step function with a step of height
$1/n$ at each of the observational points $x_1,x_2,\cdots,x_n$.

The KS test involves a comparison between the observed cumulative
distribution function $S_n(x)$ for the data sample and the
cumulative distribution function $F_0(x)$ which is determined by
some theoretical model. We state the null hypothesis as
\begin{equation}
H_0:~~~S_n(x)= F_0(x).  \label{E92}
\end{equation}
For $H_0$ true one expects that the difference between $S_n(x)$ and
$F_0(x)$ at any point should be reasonably small. The KS test looks
at the difference $S_n(x)-F_0(x)$ at all observed points and takes
the maximum of the absolute value of this quantity, $D_n$, as a test
statistic
\begin{equation}
D_n=max|S_n(x)-F_0(x)|.  \label{E93}
\end{equation}
It can be shown that provided no parameter in $F_0(x)$ has been
determined from the data, and assuming $H_0$ true, the variable
$D_n$ has a distribution which is independent of $F_0(x)$, $i.e.$
$D_n$ is $distribution~ free$. This holds irrespective of the sample
size.

For continuous variable $x$ and finite $n$, the $D_n$ has the
distribution of~\cite{Rohatgi}
\begin{equation}
P(D_n<z+\frac{1}{2n})=\left\{\begin{array}{cc} 0,         & z\leq 0,\\
 \int_{\frac{1}{2n}-z}^{\frac{1}{2n}+z}\int_{\frac{3}{2n}-z}^{\frac{3}{2n}+z}
 \cdots\int_{\frac{2n-1}{2n}-z}^{\frac{2n-1}{2n}+z}
 f(y_1,\cdots,y_n)dy_1\cdots dy_n,     & 0<z<1-\frac{1}{2n}\\
                               1,          & z\geq 1-\frac{1}{2n}, \end{array}
                               \right.
\label{E94}
\end{equation}
where
\begin{equation}
f(y_1,\cdots,y_n)=\left\{\begin{array}{cc} n!,  & when~ 0<y_1<\cdots<y_n<1,\\
  0,          & others. \end{array}
                               \right.
\label{E95}
\end{equation}
For large $n$ the $D_n$ has the cumulative distribution of
\begin{equation}
\lim_{n\rightarrow\infty}P(D_n\leq\frac{z}{\sqrt n})=
1-2\sum_{r=1}^{\infty} (-1)^{r-1}e^{-2r^2 z^2},~~~~~(z>0).
\label{E96}
\end{equation}
This relation is approximately valid at $n\simeq 80$.

If $H_0$ is true, the $D_n$ tends to be small, while if $H_0$ is not
true, the $D_n$ tends to be larger than if $H_0$ were true. Hence it
seems reasonable to reject $H_0$ if $D_n$ becomes too large. The
criteria to reject $H_0$ at significance level $\alpha$ is
\begin{equation}
P(D_n>D_{n,\alpha}). \label{E97}
\end{equation}
A table in the Appendix of the book~\cite{Frodesen} or ~\cite{Zhuys}
gives the critical values $D_{n,\alpha}$ at 5 different significance
level $\alpha$ for $n\leq 100$, and the approximate expression for
$n>100$.
\subsection{Statistical significance of signal}
The statistical significance of a signal in an experiment of
particle physics is to quantify the degree of confidence that the
observation in the experiment either confirm or disprove a null
hypothesis $H_0$, in favor of an alternative hypothesis $H_1$.
Usually the $H_0$ stands for known or background processes, while
the alternative hypothesis $H_1$ stands for a new or a signal
process plus background processes with respective production cross
section.   This concept is very useful for usual measurements that
one can have an intuitive estimation, to what extent one can believe
the observed phenomena are due to backgrounds or a signal.  It
becomes crucial for measurements which claim a new discovery or a
new signal.  As a convention in particle physics experiment, the
"$5\sigma$" standard, namely the statistical significance $S\geq 5$
is required to define the sensitivity for discovery; while in the
cases  $S\geq 3$ ($S\geq 2$), one may claim that the observed signal
has strong (weak) evidence.

However, as pointed out in Ref.~\cite{Sinervo}, the concept of the
statistical significance has not been employed consistently in the
most important discoveries made over the last quarter century. Also,
the definitions of the statistical significance in different
measurements differ from each other. Listed below are various
definitions for the statistical significance in counting experiment
(see, for example, refs. ~\cite{Bity1}~\cite{Bity2}~\cite{Narsky}):
\begin{equation}
S_1=(n-b)/\sqrt{b}, \label{E98}\end{equation}
\begin{equation}
S_2=(n-b)/\sqrt{n},\label{E99}
\end{equation}
\begin{equation}
S_{12}=\sqrt{n}/\sqrt{b},\label{E100}
\end{equation}
\begin{equation}
S_{B1}=S_1-k(\alpha)\sqrt{n/b},\label{E101}
\end{equation}
\begin{equation}
S_{B12}=2S_{12}-k(\alpha),\label{E102}
\end{equation}
\begin{equation}
\int_{-\infty}^{S_N}N(0,1)dx=\sum_{i=0}^{n-1}e^{-b}\frac{b^i}{i!},
\label{E103}
\end{equation}
where $n$ is the total number of the observed events, which is the
Poisson variable with the expectation $s+b$, $s$ is the expected
number of signal events to be searched, while $b$ is the known
expected number of Poisson distributed background events. All
numbers are counted in the "signal region" where the searched signal
events are supposed to appear. In equations ~\ref{E101} and
~\ref{E102} the $k(\alpha)$ is a factor related to the $\alpha$ that
the corresponding statistical significance assumes $1-\alpha$
acceptance for positive decision about signal observation, and
$k(0.5)=0,k(0.25)=0.66,k(0.1)=1.28,k(0.05)=1.64$, etc~\cite{Bity2}.
In equation ~\ref{E103}, $N(0,1)$ is a notation for the standard
normal function. On the other hand, the measurements in particle
physics often examine statistical variables that are continuous in
nature. Actually, to identify a sample of events enriched in the
signal process, it is often important to take into account the
entire distribution of a given variable for a set of events , rather
than just to count the events within a given signal region of
values. In this situation, I. Narsky ~\cite{Narsky} gives a
definition of the statistical significance via likelihood function
\begin{equation}
S_{L}=\sqrt{-2\ln L(b)/L(s+b)} \label{E104}
\end{equation}
under the assumption that $-2\ln L(b)/L(s+b)$ distributes as
$\chi^2$ function with degree of freedom of 1.

   Upon above situation, it is clear that we desire to have a
self-consistent definition for statistical significance, which can
avoid the ambiguity that the same $S$ value in different
measurements may imply virtually different statistical significance,
and can be suitable for both counting experiment and continuous test
statistics.
\subsubsection{ Definition of the statistical significance }
  In the PDG~\cite{PDG}, the $p-$value is defined to quantify the level
of agreement between the experimental data and a hypothesis. Assume
an experiment makes a measurement for test statistic $t$ being equal
to $t_{obs}$, and $t$ has a probability density function $g(t|H_0)$
if a null hypothesis $H_0$ is true. We further assume that large $t$
values correspond to poor agreement between the null hypothesis
$H_0$ and data, then the $p-$value of an experiment would be
\begin{equation}
p(t_{obs})=P(t>t_{obs}|H_0)=\int_{t_{obs}}^{\infty}g(t|H_0)dt.
\label{E105}\end{equation} A very small $p-$value tends to reject
the null hypothesis $H_0$.

   Since the $p-$value of an experiment provides a measure of the
consistency between the $H_0$ hypothesis and the measurement,
Zhu~\cite{YS} define the statistical significance $S$ in terms of
the $p-$value in the following form
\begin{equation}    
         \int_{-S}^{S}N(0,1)dx=1-p(t_{obs})
\label{E106}\end{equation}   
under the assumption that the null hypothesis $H_0$ represents that
the observed events can be described merely by background processes.
A small $p-$value (larger $t_{obs}$) corresponds to poor agreement
between $H_0$ and data, in this case one would get a large signal
significance $S$ by this expression. The left side of
equation~\ref{E106} represents the integral probability of the
normal distribution in the region within $S$ standard deviation
($S\sigma$).  In such a definition, some correlated  $S$ and
$p-$values are listed in Table~\ref{pS}.
\begin{table}[htbp]
\caption{\it Statistical Significance $S$ and correlated $p-$value.}
\begin{center}
\begin{tabular}{|c|c|}\hline
$S$  &  $p-$value \\\hline
1 & 0.3173  \\
2 & 0.0455  \\
3 & 0.0027  \\
4 & $ 6.3\times 10^{-5} $ \\
5 & $ 5.7\times 10^{-7} $ \\
6 & $ 2.0\times 10^{-9} $ \\\hline
\end{tabular}
\end{center}
\label{pS}
\end{table}

\subsubsection{Statistical significance in counting experiment}
A group of particle physics experiment involves the search for new
phenomena or signal by observing a unique class of events that can
not be described by background processes. One can address this
problem to that of a "counting experiment", where one identifies a
class of events using well-defined criteria, counts up the number of
observed events, and estimates the average rate of events
contributed by various backgrounds in the signal region, where the
signal events (if exist) will be clustered. Assume in an experiment,
the number of signal events in the signal region is a Poisson
variable with the expectation $s$, while the number of events from
backgrounds is a Poisson variable with a known expectation $b$, then
the observed number of events distributes as the Poisson variable
with the expectation $s+b$. If the experiment observed $n_{obs}$
events in the signal region, then the $p-$value is
\begin{eqnarray}
p(n_{obs}) &=&P(n>n_{obs}|H_0)=\sum_{n=n_{obs}}^{\infty}\frac{b^n}{n!}e^{-b} \\
\nonumber
 &=&  1-\sum_{n=0}^{n_{obs}-1}\frac{b^n}{n!}e^{-b} .
\end{eqnarray}
Substituting this relation to equation~\ref{E106}, one immediately
has
\begin{equation}   
  \int_{-S}^{S}N(0,1)dx=\sum_{n=0}^{n_{obs}-1}\frac{b^n}{n!}e^{-b}
\label{E107}\end{equation}   
Then, the signal significance $S$ can be easily determined.
Comparing this equation with equation ~\ref{E103} given by
Ref.~\cite{Narsky}, we found the lower limit of the integral is
different.
\subsubsection{Statistical significance in continuous test
statistics}
The general problem in this situation can be addressed as follows.
Suppose we identify a class of events using well-defined criteria,
which are characterized by a set of $n$ observations
$x_1,\cdots,x_n$ for a random variable $x$. In addition, one wish to
test a hypothesis which predicts the probability density function of
$x$, say $f(x|\vec{\theta})$, where
$\vec{\theta}=(\theta_1,\theta_2,...,\theta_k)$ is a set of
parameters which need to be estimated from the data. Then the
problem is to define a statistic that gives a measure of the
consistency of the distribution of data with the distribution given
by the hypothesis.

To be concrete, we consider the random variable $x$ is, say, an
invariant mass, and the $n$ observations $x_1,\cdots,x_n$ give an
experimental distribution of $x$. Assuming parameters $\vec{\theta}=
(\theta_1,\cdots,\theta_k)\equiv(\vec{\theta_s};\vec{\theta_b)}$,
where $\vec{\theta_s}$ and $\vec{\theta_b}$ represent the parameters
belong to signal (say, a resonance) and backgrounds contribution,
respectively.  We assume  the null hypothesis $H_0$ stands for that
the experimental distribution of $x$ can be described merely by the
background processes, namely, the null hypothesis $H_0$ specifies
fixed values for a subset of parameters $\vec{\theta_s}$. Therefore,
the parameters $\vec{\theta}$ are restricted to lie in a subspace
$\omega$ of its total space $\Omega$. On the basis of a data sample
of size $n$ from  $f(x|\vec{\theta})$ we want to test the hypothesis
$H_0:\vec{\theta}$ belongs to $\omega$. Given the observations
$x_1,\cdots,x_n$, the likelihood function is $L=\prod_{i=1}^{n}
f(x_i|\vec{\theta})$. The maximum of this function over the total
space $\Omega$ is denoted by $L(\hat{\Omega})$; while within the
subspace $\omega$ the maximum of the likelihood function is denoted
by  $L(\hat{\omega})$, then we define the likelihood-ratio
$\lambda\equiv L(\hat{\omega})/L(\hat{\Omega})$. It can be shown
that for $H_0$ true, the statistic
\begin{equation}   
 t \equiv -2\ln \lambda \equiv 2(\ln L_{max}(s+b)-\ln L_{max}(b))
\label{E108}
\end{equation}    
is distributed as $\chi^2(r)$ where $r$ is the difference in the
number of degrees of freedom between the $H_1$ and $H_0$
hypotheses~\cite{Eadie}. In equation~\ref{E108} we use $\ln
L_{max}(s+b)$ and $\ln L_{max}(b)$ denoting $L(\hat{\Omega})$ and
$L(\hat{\omega})$, respectively.  If $\lambda$ turns out to be in
the neighborhood of 1, the null hypothesis $H_0$ is such that it
renders $L(\hat{\omega})$ close to the maximum $L(\hat{\Omega})$,
and hence $H_0$ will have a large probability of being true. On
the other hand, a small value of $\lambda$ will indicates that
$H_0$ is unlikely. Therefore, the critical region of $\lambda$ is
in the neighborhood of 0, corresponding to large value of
statistic $t$. If the measured value of $t$ in an experiment is
$t_{obs}$, from equation~\ref{E105} we have $p-$value
\begin{equation}   
 p(t_{obs})=\int_{t_{obs}}^{\infty}\chi^2(t;r)dt.
\label{E109}
\end{equation}   
Therefore, in terms of equation~\ref{E106}, one can calculate the
signal
 significance according to following expression:
\begin{equation}    
 \int_{-S}^{S}N(0,1)dx=1-p(t_{obs})=\int_{0}^{t_{obs}}\chi^2(t;r)dt.
\label{E110}
\end{equation}    
For the case of $r=1$, we have
\begin{eqnarray}
 \int_{-S}^{S}N(0,1)dx &=& \int_{0}^{t_{obs}}\chi^2(t;1)dt \\ \nonumber
  &=& 2\int_{0}^{\sqrt{t_{obs}}}N(0,1)dx.
\end{eqnarray}
and immediately obtain
\begin{eqnarray}
 S &=&\sqrt{t_{obs}} \\ \nonumber
   &=&  [2(\ln L_{max}(s+b)- \ln L_{max}(b))]^{1/2},
\end{eqnarray}
which is identical to the equation~\ref{E104} given by
Ref.~\cite{Narsky}.
%
%



\begin{thebibliography}{9}
\bibitem{mixing}
See the Note on $D{^0}\hbox{--}\overline D{^0}$ Mixing by D. Asner
in the Review;
Dalitz plot analysis of the wrong sign rate
$D^0\to K^+\pi^-\pi^0$\cite{cleob} is a candidate processes
\end{thebibliography}
\end{document}